%% file: main.tex
\documentclass[aps,prd,onecolumn,superscriptaddress,nofootinbib,10pt]{revtex4-1}
\usepackage{amsmath,amssymb,bm,latexsym,acronym,float,soul,nicefrac,verbatim,cancel}
\usepackage{array,multirow,dcolumn,longtable,footnote,appendix,graphicx,color,subfigure}
\usepackage[colorlinks,linkcolor=blue,citecolor=blue,urlcolor=blue]{hyperref}
\usepackage{tikz} 
\usepackage[normalem]{ulem}
\usepackage{mathrsfs} 
\allowdisplaybreaks
\newcommand{\be}{\begin{equation}}
\newcommand{\ee}{\end{equation}}
\newcommand{\bea}{\begin{eqnarray}}
\newcommand{\eea}{\end{eqnarray}}
\newcommand{\nn}{\nonumber}
\newcommand{\cA}{{\cal A}}

\newcommand{\cO}{{\cal O}}

\newcommand{\cM}{{\cal M}}

\newcommand{\cQ}{{\cal Q}}
\newcommand{\cS}{{\cal S}}

\newcommand{\cT}{{\cal T}}

\newcommand{\td}{\tilde}

\newcommand{\bk}{{\bf k}}
\newcommand{\hTTij}{{h^{\rm TT}_{ij}}}
\newcommand{\thTTij}{{\tilde h^{\rm TT}_{ij}}}
\newcommand{\tG}{{\tilde G}}
\newcommand{\mSun}{{{\rm~M}_\odot}}
\newcommand{\D}{{\mathrm{d}}}
\graphicspath{{figs/}{tex/}{tex/figs/}}

\def\jcap{Journal of Cosmology and Astroparticle Physics}

\def\nat{Nature}

\def\prd{Physical Review D}

\def\sovast{Sov. Astron.}
\begin{document}

\newcommand{\intitle}{Fundamental Physics and Cosmology with TianQin}
\title{\bf \intitle}

\input{authors}

\begin{abstract}
The exploration of the surrounding world and the universe is an important theme in the legacy of humankind. The detection of gravitational waves is adding a new dimension to this grand effort. What are the fundamental physical laws governing the dynamics of the universe? What is the fundamental composition of the universe? How has the universe evolved in the past and how will it evolve in the future? These are the basic questions that press for answers. The space-based gravitational wave detector TianQin will tune in to gravitational waves in the millihertz frequency range ($10^{-4} \sim 1$ Hz, to be specific), opening a new gravitational wave spectrum window to explore many of the previously hidden sectors of the universe. TianQin will discover many astrophysical systems, populating the universe at different redshifts: some will be of new types that have never been detected before, some will have very high signal-to-noise ratios, and some will have very high parameter estimation precision. The plethora of information collected will bring us to new fronts on which to search for the breaking points of general relativity, the possible violation of established physical laws, the signature of possible new gravitational physics and new fundamental fields, and to improve our knowledge on the expansion history of the universe. In this white paper, we highlight the advances that TianQin can bring to fundamental physics and cosmology.
\end{abstract}


\input{style-coverpage}
\maketitle


\tableofcontents

\newpage

\input{sec0-acronyms}

\input{sec1-intro}
\input{sec2-phys}

\input{sec3-new-phys}
\input{sec4-cosmo}

\input{sec5-sum}

\newpage

\section*{Acknowledgments}

The work has been supported in part by the Guangdong Major Project of Basic and Applied Basic Research (Grant No. 2019B030302001), the Natural Science Foundation of China (Grants No. 12261131504), and the National Key Research and Development Program of China (Grant No. 2023YFC2206700).
L.B. is supported by the National Key Research and Development Program of China (Grant No. 2021YFC2203004), the National Natural Science Foundation of China (Grant No. 12322505, 12347101), Chongqing Talents: Exceptional Young Talents Project (Grant No. cstc2024ycjh-bgzxm0020) and Chongqing Natural Science Foundation (Grant No. CSTB2024NSCQ-JQX0022).
S.-J.H is supported by the Postdoctoral Fellowship Program of CPSF (Grant No. GZC20242112).
S.P is supported in part by the National Key Research and Development Program of China Grant No. 2020YFC2201502, and by National Natural Science Foundation of China No. 12475066 and No. 12047503.
M.S is supported by JSPS KAKENHI grant Nos. 20H05853, and 24K00624 (MS), and also by the World Premier International Research Center Initiative (WPI), MEXT, Japan.
J.-D.Z is supported by Guangdong Basic and Applied Basic Research Foundation (Grant No. 2023A1515030116).

\newpage

\bibliographystyle{apsrev4-1}
\bibliography{ref-sec1,ref-sec2,ref-sec3,ref-sec4}

\end{document}

%% file: authors.tex
%
%
\newcommand{\TRC}{MOE Key Laboratory of TianQin Mission, TianQin Research Center for Gravitational Physics $\&$ School of Physics and Astronomy, Frontiers Science Center for TianQin, Gravitational Wave Research Center of CNSA, Sun Yat-sen University (Zhuhai Campus), Zhuhai 519082, China}
\newcommand{\SPA}{School of Physics and Astronomy, Sun Yat-sen University (Zhuhai Campus), Zhuhai 519082, China}
\newcommand{\SYSUSZ}{School of Science, Shenzhen Campus of Sun Yat-sen University, Shenzhen 518107, China}

\newcommand{\APCTP}{Asia Pacific Center for Theoretical Physics (APCTP), Pohang 37673, Korea}
\newcommand{\AZW}{Department of Physics and Astronomy, Baylor University, Waco, Texas 76798-7316, USA}

\newcommand{\BNU}{School of Physics and Astronomy, Beijing Normal University, Beijing 100875, China}

\newcommand{\CHEPTHU}{Center for High Energy Physics, Tsinghua University, Beijing 100084, China}
\newcommand{\CHEP}{Center for High Energy Physics, Peking University, Beijing 100871, China}
\newcommand{\CQU}{Department of Physics and Chongqing Key Laboratory for Strongly Coupled Physics, Chongqing University, Chongqing 401331, China}

\newcommand{\GLU}{SUPA, School of Physics and Astronomy, University of Glasgow, Glasgow G12 8QQ, United Kingdom}

\newcommand{\HIAS}{School of Fundamental Physics and Mathematical Sciences, Hangzhou Institute for Advanced Study (HIAS), University of Chinese Academy of Sciences (UCAS), Hangzhou 310024, China}
\newcommand{\HUST}{National Gravitation Laboratory, MOE Key Laboratory of Fundamental Physical Quantities Measurement, and School of Physics, Huazhong University of Science and Technology, Wuhan 430074, China}
\newcommand{\HZDR}{Helmholtz-Zentrum Dresden-Rossendorf, Bautzner Landstra{\ss}e 400, 01328 Dresden, Germany}

\newcommand{\IHEP}{Institute of High Energy Physics, Chinese Academy of Sciences, Beijing 100049, China}
\newcommand{\IPMU}{Kavli Institute for the Physics and Mathematics of the Universe (WPI), UTIAS, The University of Tokyo, Kashiwa, Chiba 277-8583, Japan}
\newcommand{\ITP}{CAS Key Laboratory of Theoretical Physics, Institute of Theoretical Physics, Chinese Academy of Sciences, Beijing 100190, China}

\newcommand{\KIAA}{Kavli Institute for Astronomy and Astrophysics, Peking University, Beijing 100871, China}
\newcommand{\KNU}{Department of Physics, Kunsan National University, Kunsan 54150, Korea}

\newcommand{\LZU}{School of Physical Science and Technology, Lanzhou University, Lanzhou 730000, China}

\newcommand{\NBU}{Institute of Fundamental Physics and Quantum Technology, Ningbo University, Ningbo, 315211, China}
\newcommand{\NEU}{Liaoning Key Laboratory of Cosmology and Astrophysics, College of Sciences, Northeastern University, Shenyang 110819, China}
\newcommand{\NJUA}{School of Astronomy and Space Science, Nanjing University, Nanjing,210023, China}
\newcommand{\NJUB}{Key Laboratory of Modern Astronomy and Astrophysics (Nanjing University), Ministry of Education,Nanjing, 210023,  China}

\newcommand{\QMU}{School of Physical Science and Technology, Kunming University, Kunming 650214, China}

\newcommand{\SAI}{Sternberg Astronomical Institute, M.V. Lomonosov Moscow State University, 119234 Moscow, Russia}
\newcommand{\SHAO}{Shanghai Astronomical Obserbvatory, CAS, Shanghai, 200030, China}

\newcommand{\THUP}{Department of Physics, Tsinghua University, Beijing 100084, China}
\newcommand{\THUA}{Department of Astronomy, Tsinghua University, Beijing 100084, China}

\newcommand{\USTC}{Department of Astronomy, University of Science and Technology of China, Hefei, 230026, China}

\newcommand{\WHU}{School of Physics and Technology, Wuhan University, Wuhan 430072, China}

%
%
%
%

\affiliation{\TRC}
\affiliation{\HUST}

\author{Jun Luo}
\affiliation{\TRC}
\author{Haipeng An}
\affiliation{\THUP}
\affiliation{\CHEPTHU}
\author{Ligong Bian}
\affiliation{\CQU}
\author{Rong-Gen Cai}
\affiliation{\NBU}
\affiliation{\ITP}
\affiliation{\HIAS}
\author{Zhoujian Cao}
\affiliation{\BNU}
\author{Wenbiao Han}
\affiliation{\SHAO}
\affiliation{\HIAS}
\author{Jianhua He}
\affiliation{\NJUA}
\affiliation{\NJUB}
\author{Martin A. Hendry}
\affiliation{\GLU}
\author{Bin Hu}
\affiliation{\BNU}
\author{Yi-Ming Hu}
\affiliation{\TRC}
\author{Fa Peng Huang}
\affiliation{\TRC}
\author{Shun-Jia Huang}
\affiliation{\SYSUSZ}
\author{Sang Pyo Kim}
\affiliation{\KNU}
\affiliation{\APCTP}
\author{En-Kun Li}
\affiliation{\TRC}
\author{Yu-Xiao Liu}
\affiliation{\LZU}
\author{Vadim Milyukov}
\affiliation{\SAI}
\author{Shi Pi}
\affiliation{\ITP}
\affiliation{\CHEP}
\affiliation{\IPMU}
\author{Konstantin Postnov}
\affiliation{\SAI}
\author{Misao Sasaki}
\affiliation{\IPMU}
\author{Cheng-Gang Shao}
\affiliation{\HUST}
\author{Lijing Shao}
\affiliation{\KIAA}
\author{Changfu Shi}
\affiliation{\TRC}
\author{Shuo Sun}
\affiliation{\QMU}
\author{Anzhong Wang}
\affiliation{\AZW}
\author{Pan-Pan Wang}
\affiliation{\HUST}
\author{Sai Wang}
\affiliation{\IHEP}
\author{Shao-Jiang Wang}
\affiliation{\ITP}
\affiliation{\APCTP}
\author{Zhong-Zhi Xianyu}
\affiliation{\THUP}
\author{Huan Yang}
\affiliation{\THUA}
\author{Tao Yang}
\affiliation{\WHU}
\author{Jian-dong Zhang}
\affiliation{\TRC}
\author{Xin Zhang}
\affiliation{\NEU}
\author{Wen Zhao}
\affiliation{\USTC}
\author{Liang-Gui Zhu}
\affiliation{\KIAA}
\author{Jianwei Mei}
\email{
\underline{General coordinator}\\
JM:  meijw@sysu.edu.cn\\
\underline{Section coordinators}\\
FPH: huangfp8@sysu.edu.cn\\
LGZ: lianggui.zhu@pku.edu.cn\\
\underline{Subsection coordinators}\\
HPA: anhp@mail.tsinghua.edu.cn\\
LGB: lgbycl@cqu.edu.cn\\
SJH: huangshj69@mail.sysu.edu.cn\\
SPK: sangkim@kunsan.ac.kr\\
CS:  shichf6@mail.sysu.edu.cn\\
SS:  gwyddruid@gmail.com\\
KP:  k.a.postnov@yandex.ru\\
SP:  shi.pi@itp.ac.cn\\
AZW: anzhong_wang@baylor.edu\\
SW:  wangsai@ihep.ac.cn\\
JDZ: zhangjd9@mail.sysu.edu.cn\\
XZ:  zhangxin@mail.neu.edu.cn}
\affiliation{\TRC}
%
\def\ccr#1{{\color{red} #1}}
\def\ccb#1{{\color{blue} #1}}

\newcommand{\LS}[2]{{\color{magenta}{\st{#1} #2}}}
\newcommand{\zhu}[1]{\textcolor{cyan}{\sf{#1}}}
\def\KP#1{{\color{red}KP: ( #1)}}

%% file: style-coverpage.tex
\clearpage
\thispagestyle{empty}
\begin{center}
\begin{tikzpicture}[remember picture, overlay]
 \node(O) at (current page.center) {};
 \node[yshift=-0.02\paperheight] at (O) {%
 \includegraphics[width=\paperwidth]{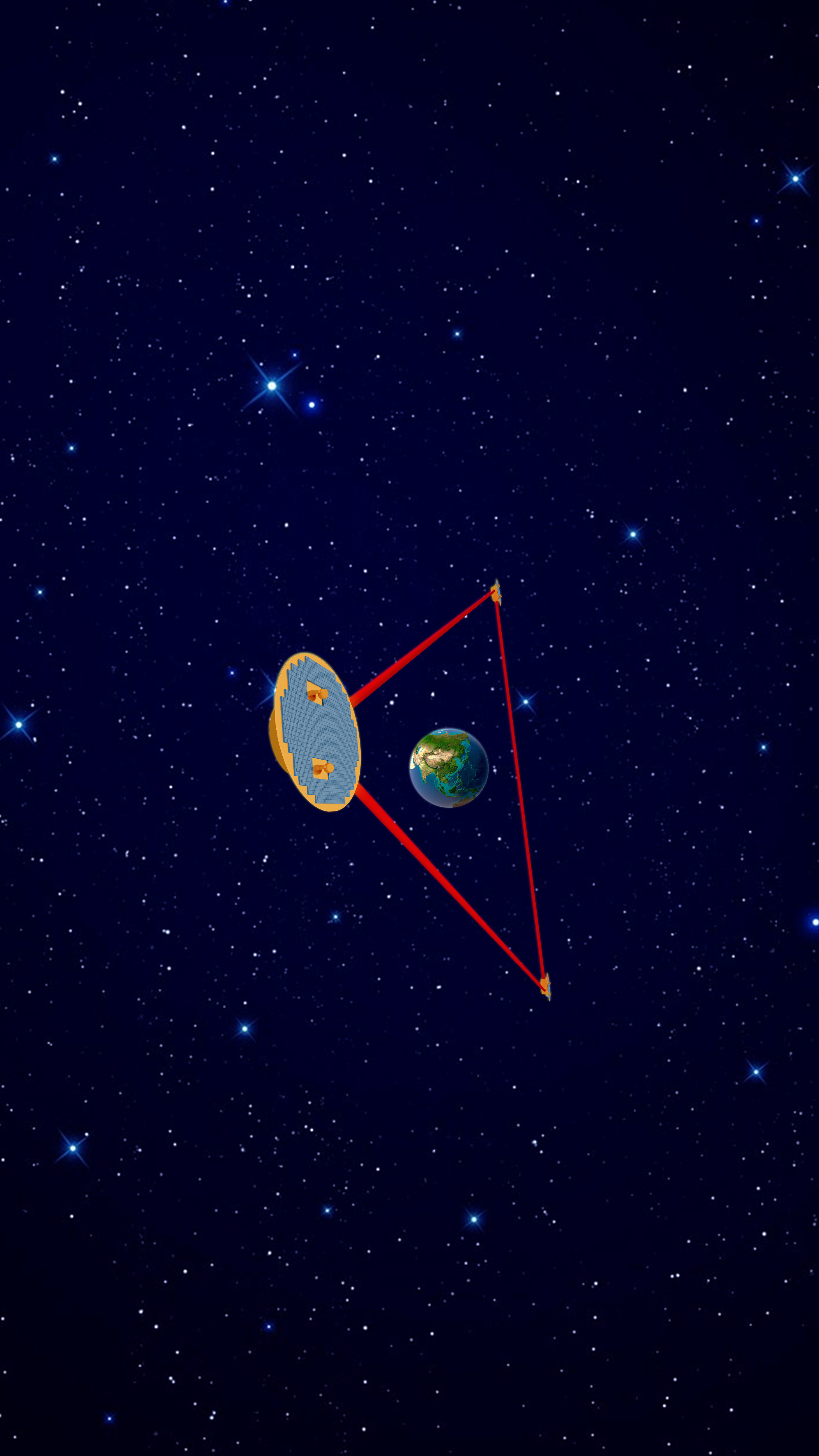}};
 \node[yshift=0.3\paperheight,
 font=\huge\color{white}] at (O) %
 {\parbox[t]{0.9\textwidth}{\baselineskip=2em\centering\bfseries\intitle}};
\end{tikzpicture}
\end{center}
\clearpage
\thispagestyle{empty}

\centerline{This Page Intentionally Left Blank}

\clearpage
\pagestyle{plain}
\setcounter{page}{0}

%% file: sec0-acronyms.tex
\acrodef{EM}{Electromagnetic}
\acrodef{GW}{Gravitational wave}
\acrodef{CW}{Continuous wave}
\acrodef{DWD}{Double white dwarf}
\acrodef{EMRI}{Extreme Mass Ratio Inspiral}
\acrodef{GCB}{Galactic ultra-compact binary}
\acrodef{IMR}{Inspiral-Merger-Ringdown}
\acrodef{IMRI}{Intermediate Mass Ratio Inspiral}
\acrodef{MBHB}{Massive black hole binary}
\acrodef{SBHB}{Stellar mass black hole binary}
\acrodef{SGWB}{Stochastic GW background}
\acrodef{VB}{Verification binary}
\acrodef{CE}{Cosmic Explorer}
\acrodef{ET}{Einstein Telescope}
\acrodef{LIGO}{Laser Interferometer Gravitational-wave Observatory}
\acrodef{LISA}{Laser Interferometer Space Antenna}
\acrodef{LVK}{LIGO $\&$ Virgo $\&$ KAGRA}
\acrodef{PTA}{Pulsar Timing Array}
\acrodef{BHPT}{Black hole perturbation theory}
\acrodef{BSM}{Beyond Standard Model (of Particle Physics)}
\acrodef{dCS}{Dynamical Chern-Simons theory}
\acrodef{EdGB}{Einstein-dilaton Gauss-Bonnet theory}
\acrodef{GR}{General relativity}
\acrodef{MGT}{Modified gravity theory}
\acrodef{NCG}{Noncommutative gravity}
\acrodef{SMPP}{Standard Model of Particle Physics}
\acrodef{BMS}{Bondi-Metzner-Sachs}
\acrodef{CDDR}{Cosmic distance duality relation}
\acrodef{DI}{Diffeomorphism invariance}
\acrodef{EEP}{Einstein Equivalence Principle}
\acrodef{ISL}{Inverse Square Law}
\acrodef{LLI}{Local Lorentz invariance}
\acrodef{LPI}{Local position invariance}
\acrodef{SEP}{Strong Equivalence Principle}
\acrodef{WEP}{Weak equivalence principle}
\acrodef{BAO}{Baryon acoustic oscillation}
\acrodef{BBN}{Big Bang Nucleosynthesis}
\acrodef{CBC}{Compact binary coalescence}
\acrodef{CMB}{Cosmic Microwave Background}
\acrodef{ECO}{Exotic compact object}
\acrodef{EWPT}{Electroweak phase transition}
\acrodef{FLRW}{Friedmann-Lema\^itre-Robertson-Walker}
\acrodef{FoPT}{First order phase transition}
\acrodef{FRB}{Fast radio burst}
\acrodef{GRB}{Gamma-ray burst}
\acrodef{IMBH}{Intermediate mass black hole}
\acrodef{MBH}{Massive black hole}
\acrodef{PBH}{Primordial black hole}
\acrodef{SBH}{Stellar mass black hole}
\acrodef{SN Ia}{Type Ia supernova}
\acrodef{CDF}{Cumulative distribution function}
\acrodef{PDF}{Probability density function}
\acrodef{EoS}{Equation of State}
\acrodef{FIM}{Fisher information matrix}
\acrodef{MCMC}{Markov Chain Monte Carlo}
\acrodef{NFW}{Navarro-Frenk-White}
\acrodef{NR}{Numerical relativity}
\acrodef{PN}{Post-Newtonian}
\acrodef{ppE}{Parameterized post-Einstein}
\acrodef{PPN}{Parameterized post-Newtonian}
\acrodef{QNM}{Quasi-normal mode}
\acrodef{SNR}{Signal-to-noise ratio}
\acrodef{TDI}{Time delay interferometry}

\section*{List of acronyms}

\begin{itemize}
\item Messengers
    \begin{itemize}
    \item \ac{EM}
    \item \ac{GW}
    \end{itemize}
\item GWs and sources
    \begin{itemize}
    \item \ac{CBC}
    \item \ac{CW}
    \item \ac{DWD}
    \item \ac{EMRI}
    \item \ac{GCB}
    \item \ac{IMR}
    \item \ac{IMRI}
    \item \ac{MBHB}
    \item \ac{SBHB}
    \item \ac{SGWB}
    \item \ac{VB}
    \end{itemize}
\item GW Detectors \& Methods
    \begin{itemize}
    \item \ac{CE}
    \item \ac{ET}
    \item \ac{LIGO}
    \item \ac{LISA}
    \item \ac{LVK}
    \item \ac{PTA}
    \end{itemize}
\item Theories
    \begin{itemize}
    \item \ac{BHPT}
    \item \ac{BSM}
    \item \ac{dCS}
    \item \ac{EdGB}
    \item \ac{GR}
    \item \ac{MGT}
    \item \ac{NCG}
    \item \ac{SMPP}
    \end{itemize}
\item Physical laws
    \begin{itemize}
    \item \ac{BMS}
    \item \ac{CDDR}
    \item \ac{DI}
    \item \ac{EEP}
    \item \ac{ISL}
    \item \ac{LLI}
    \item \ac{LPI}
    \item \ac{SEP}
    \item \ac{WEP}
    \end{itemize}
\item Physical entities and process
    \begin{itemize}
    \item \ac{BAO}
    \item \ac{BBN}
    \item \ac{CMB}
    \item \ac{ECO}
    \item \ac{EWPT}
    \item \ac{FLRW}
    \item \ac{FoPT}
    \item \ac{FRB}
    \item \ac{GRB}
    \item \ac{IMBH}
    \item \ac{MBH}
    \item \ac{PBH}
    \item \ac{SBH}
    \item \ac{SN Ia}
    \end{itemize}
\item Methods and Technical names
    \begin{itemize}
    \item \ac{CDF}
    \item \ac{PDF}
    \item \ac{EoS}
    \item \ac{FIM}
    \item \ac{MCMC}
    \item \ac{NFW}
    \item \ac{NR}
    \item \ac{PN}
    \item \ac{ppE}
    \item \ac{PPN}
    \item \ac{QNM}
    \item \ac{SNR}
    \end{itemize}
\end{itemize}

\clearpage

%% file: sec1-intro.tex
\section{Introduction}\label{sec:intro}

The most incomprehensible thing about the Universe is that it is comprehensible \cite{2018Natur.557...30R}.
It is remarkable to note how much is already known about the Universe. However, there are still important open questions related to basically all major epochs of the Universe (see Table \ref{tab:sec1.universe}). One may summarize these into the following three basic questions:
\begin{itemize}
    \item What are the fundamental physical laws governing the dynamics of the Universe?
    \item What are the fundamental compositions of the Universe?
    \item How has the Universe (including everything in it) evolved in the past and how will it evolve in the future?
\end{itemize}

\ac{GR} has been at the basis of our understanding of the dynamics of the Universe for more than a century. But there are good reasons to believe that \ac{GR} has to be extended:
in order to understand the physics of Big Bang, the Planck era, and the singularity of black holes, it is very likely that one must firstly figure out how to quantize gravity and how to unify gravity with other fundamental interactions of nature;
the necessity of introducing dark matter and dark energy could be indications that our understanding of the nature of gravity is incomplete. But going beyond \ac{GR} has proven to be extremely difficult.

On the theoretical side, the search for a consistent quantum gravity theory has led to the development of theoretical frameworks such string theory and loop quantum gravity \cite{Carlip:2015asa,Addazi:2021xuf}, together with the proposal of schools of theoretical insights, such as gauge/gravity duality \cite{Maldacena:1997re}. Unfortunately, none of these can rival \ac{GR} in terms of making quantitative predictions on gravitational phenomena that can be experimentally tested immediately.
As a result, many \acp{MGT} have been constructed to explore possible features foreseen in the so far elusive theory of quantum gravity \cite{Capozziello:2011et,Clifton:2011jh,CANTATA:2021ktz,Shankaranarayanan:2022wbx}.
But \acp{MGT} are not expected to be fully self-consistent or pathology-free. They can be best viewed as phenomenological models parameterizing the possible ways to partially deviate from \ac{GR}.

\begin{figure}[h]
\begin{center}
\includegraphics[width=0.8\textwidth]{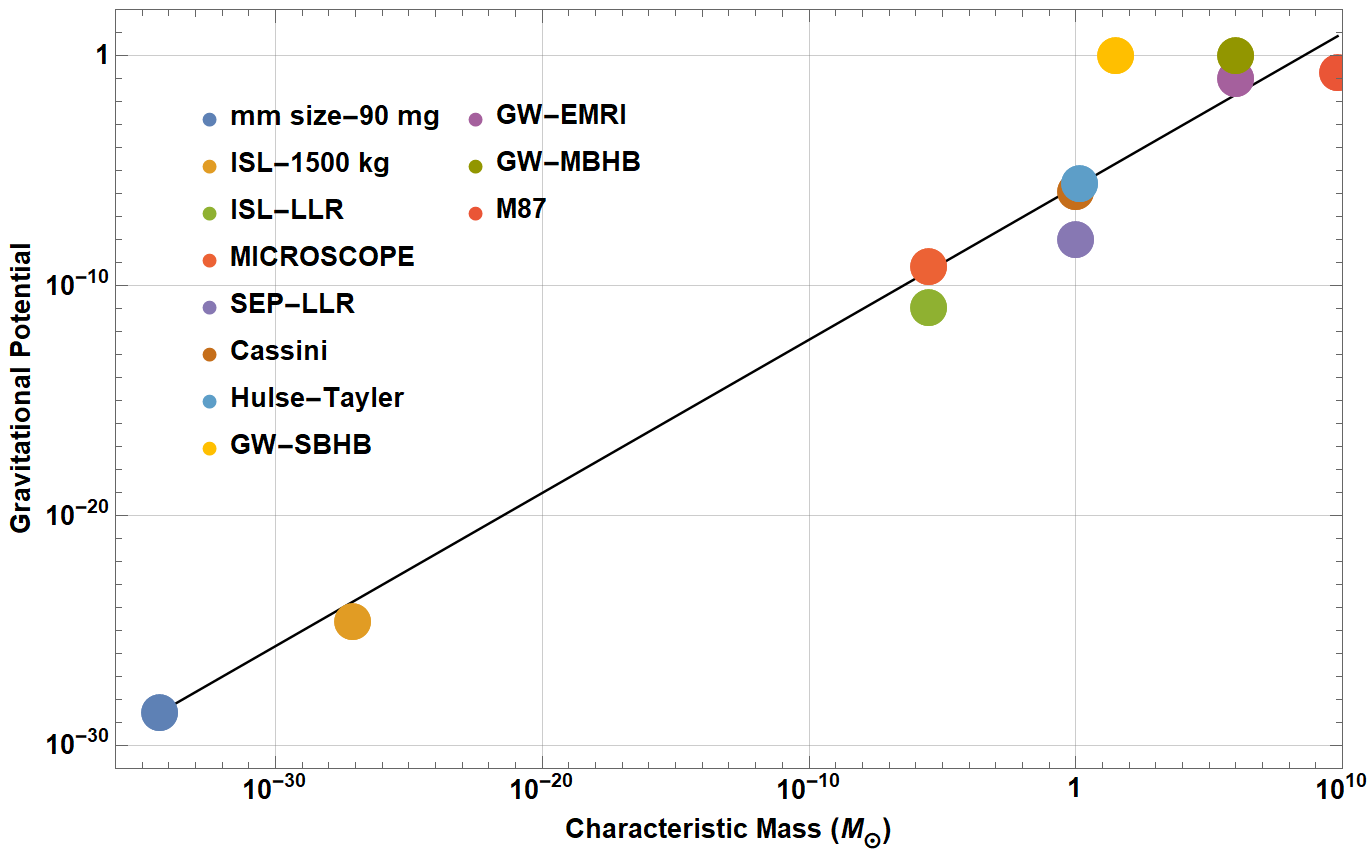}
\caption{The characteristic masses and gravitational potentials involved in typical gravitational experiments. The left-most and right-most points represent the extreme single object masses that have been used to experimentally test \ac{GR}. More discussions can be found in the main text. (References: mm size-90 mg \cite{Westphal:2020okx}, ISL-1500 kg \cite{Moody:1993ir}, ISL-LLR \cite{Adelberger:2003zx}, MICROSCOPE \cite{Touboul:2017grn}, SEP-LLR \cite{Williams:2004qba}, Cassini \cite{Bertotti:2003rm}, Hulse-Taylor \cite{Taylor:1982zz}, M87 \cite{EventHorizonTelescope:2019dse}. Some typical numbers are used for \ac{GW}-SBHB, \ac{GW}-\ac{MBHB} and \ac{GW}-EMRI, but they all can be changed.)}
\label{fig:gr-exps}
\end{center}
\end{figure}

On the experimental side, there has been a great amount of effort. To sort through the complicated landscape of the field, one can ask two basic questions.
\begin{itemize}
\item The first question is what to test. From the perspective of the nature of gravity, past experiments have mainly considered the following \cite{Will:2014kxa}:
\begin{itemize}
    \item Is gravity a purely geometrical property of the spacetime? Here one mainly looks at \ac{EEP}, which implies that gravity should be described by a metric theory and invoking gravity is equivalent to ``replacing the partial derivatives with covariant derivatives" in all the special relativistic theories.
    \item Are the gravitational degrees of freedom the same as the metric field in \ac{GR}? Here one mainly searches for the possible existence of extra interaction-mediating fields that could result in a ``fifth force" and a breaking of \ac{ISL}. One would look at the \ac{SEP}, which indicates that the metric is the only gravitational field in the Universe. One would also look at the polarization and propagation properties of \acp{GW} to see if they are consistent with the prescription of \ac{GR}.
    \item Are the dynamics of gravity the same as the metric field in \ac{GR}? Here one mainly looks at various dynamical processes involving gravity. A large portion of experiments have looked at the weak field and slow motion processes in a static or stationary spacetime background, as is the case for all the gravitational experiments taking place in the solar system. However, \ac{GW} detection has now made it possible to explore highly dynamical and strong field processes. black holes are another consequential prediction of \ac{GR} in the strong field regime, and testing the nature of dark compact objects in the Universe serves as another important test of the nature of gravity and the validity of \ac{GR}.
\end{itemize}

\item The second question is how to test. Given the huge variety of possible gravitational experiments, it is a challenging task trying to organize all the experiments in an intuitively visual way; see \cite{Baker:2014zba} for an example. But all experiments share two basic parameters: the characteristic mass (the mass of the most massive object whose gravitational field is directly tested) and the typical force range in the experiment.
The order of magnitude of the characteristic masses and the gravitational potential (the characteristic mass divided by the force range in natural units) of some typical gravitational experiments are shown in FIG. \ref{fig:gr-exps}.
For experiments based on normal matter, all are within a couple of orders of magnitude from saturating the gravitational potential limit that is achievable for a given characteristic mass. So all of them are distributed around the scaling relation (solid line in FIG. \ref{fig:gr-exps}), $\Phi=\Phi_\odot (M/\mSun)^{2/3}\,$, where $\Phi$ is the gravitational potential, $M$ is the characteristic mass, and $\Phi_\odot$ is the gravitational potential of the Sun near its surface.
The above scaling relation assumes that the density of matter is fixed at the density of the Sun. It fails for $M>1.15\times10^8\mSun$ because so much normal matter at the solar density would collapse into a black hole.
For black holes, the density varies with mass and the gravitational potential can reach $\Phi\approx1\,$ at their horizons.
For experiments at the higher mass end in FIG. \ref{fig:gr-exps}, one not only has normal stars but also black holes, and so the points become more scattered.
As an overall conclusion, one can see that larger gravitational potentials rely on having larger characteristic masses, and it is only with \acp{GW} that it is possible to systematically probe gravity in the genuinely strong field regime. \ac{EM} observations of supermassive black holes such as M87 can also help \cite{Ayzenberg:2023hfw}, but such methods are limited by the number of black holes that can be observed.
\end{itemize}

\ac{SMPP} is currently the best theory to understand the fundamental composition of the Unverse, but it can only account for about $5\%$ of the energy density of the Universe, with the rest being about $27\%$ dark matter and $68\%$ dark energy.
All the $5\%$ of energy density described by \ac{SMPP} is ordinary matter, with almost no antimatter present.
The origin of the matter-antimatter asymmetry remains one of the long-standing mysteries in particle cosmology.
Additionally, the precise shape of the Higgs potential, which imparts mass to fundamental particles, is still unknown.
The origin of matter, namely the origin of the matter-antimatter asymmetry in the Universe, is deeply connected to the shape of the Higgs potential.
One widely studied mechanism for explaining the origin of the matter-antimatter asymmetry is electroweak baryogenesis, which is rooted in the properties of Higgs-related physics, particularly the cosmic phase transition induced by the Higgs potential.
In many new physics models beyond \ac{SMPP}, the Higgs potential can lead to a strong first-order electroweak phase transition, thereby providing the necessary out-of-equilibrium conditions for electroweak baryogenesis \cite{Zhang:1992fs,Grojean:2004xa,Huang:2015izx,Huang:2016odd,Cai:2017tmh}.
Furthermore, the electroweak phase transition process is expected to generate phase-transition \ac{GW} signals.
Dark matter constitutes about $27\%$ of the Universe, yet its true nature eludes us.
Traditional collider experiments and direct detection methods have not yielded the expected signals.
The current status of experimental detection and theoretical research on dark matter suggests the need to explore new mechanisms for dark matter production and novel methods for its detection.
Cosmic phase transitions and the associated \ac{GW} signals offer new perspectives on the production and detection of dark matter \cite{Baker:2019ndr,Chway:2019kft,Jiang:2023nkj, Azatov:2021ifm,Baldes:2021vyz, Krylov:2013qe,Huang:2017kzu,Hong:2020est,Jiang:2023qbm,Jiang:2024zrb}, respectively.
Additionally, \ac{GW} signals from astrophysical sources, arising from their interactions with dark matter, may carry important information about the properties of dark matter.
Typical examples include new mechanisms for the production of superheavy dark matter via cosmic phase transitions and their associated \ac{GW} signals, \ac{PBH} dark matter and the \ac{GW} signals it induces, as well as \ac{GW} signals from ultralight dark matter.
In addition to the issues mentioned above, many other significant problems in particle physics and cosmology may require the introduction of new particles and interactions.
These are often associated with various symmetry-breaking processes in the early universe and the formation of topological defects such as cosmic strings.
\ac{GW} signals have the potential to offer novel approaches to explore these central issues in particle cosmology and probe new physics that may lie beyond the \ac{SMPP}.

Cosmologists have developed a standard picture, the $\Lambda$CDM model, describing the expansion history of the Universe from \ac{BBN} to the present time\cite{Carroll:2000fy,Peebles:2002gy,Bull:2015stt},
by combining data from the global universe, such as \ac{CMB} \cite{WMAP:2012nax, Planck:2018vyg}, \ac{BBN} \cite{Addison:2017fdm, Schoneberg:2019wmt}, and \ac{BAO} \cite{SDSS:2005xqv, Bassett:2009mm, eBOSS:2020yzd},
as well as from the local universe, such as the cosmic distance ladder measurement \cite{Riess:2020fzl, Riess:2021jrx, Freedman:2019jwv, Pesce:2020xfe, Kourkchi:2020iyz} and strong gravitational lensing \cite{Denzel:2020zuq, Wong:2019kwg, DES:2019fny}.
The $\Lambda$CDM model tells us that the Universe is about 13.8 billion years old, the space is almost flat, the ratio of total matter to dark energy of about three to seven, and the current expansion rate is about 70 km/s/Mpc and accelerating \cite{Carroll:2000fy, Peebles:2002gy}.
However, with the accumulation of observational data and improvements in the precision of the cosmological parameters, the status of $\Lambda$CDM as the standard cosmological model has been seriously challenged, with two problems being particularly significant:
\begin{itemize}
\item The Hubble tension, i.e., the inconsistency between early universe measurements of the Hubble-Lema\^itre constant, $H_0$, inferred from \ac{CMB} observations, and late universe measurements of $H_0$, measured with \ac{SN Ia} observations, exceeds a significance level of $4\sigma$ \cite{Freedman:2017yms,Riess:2019qba,DiValentino:2021izs, Schoneberg:2021qvd, Cai:2021weh, Verde:2023lmm};
\item The deviation of the dark energy equation of state relative to the standard model, the cosmological constant $\Lambda$, is of increasing significance \cite{Zhao:2017cud, Zhang:2019jsu, DESI:2024mwx}.
\end{itemize}
Because \ac{GW} detections allow us to directly measure the luminosity distances of \ac{GW} sources without depending on the cosmic distance ladder, they can serve as the so-called ``standard sirens" to independently constrain the cosmic expansion history \cite{Schutz:1986gp, Holz:2005df}.
Using \ac{GW} detection one can also measure the expansion rate of the Universe throughout its entire history, which could be crucial to help resolve the Hubble tension and to explore the nature of dark energy.

Expected to be launched around 2035, the space-based \ac{GW} detector TianQin aims to detect \acp{GW} in the frequency range $10^{-4}$ Hz $\sim$ 1 Hz \cite{Luo:2015ght}.
TianQin will comprise an equilateral triangle constellation, consisting of three drag-free satellites, which orbit the Earth with orbital radii of about $10^5$ km \cite{Ye:2019txh,Hu:2018yqb,Tan:2020xbm}. The detector plane of TianQin will be oriented toward RX J0806.3+1527 (also known as HM Cancri or HM Cnc, hereafter J0806 \cite{Strohmayer:2005uc}) and so will be nearly perpendicular to the ecliptic.
The Sun will pass through the fixed orbital plane of TianQin twice a year, upsetting \ac{GW} detection with complicated thermal load on the satellites and direct Sun light getting into the telescopes.
As a result, TianQin will adopt a consecutive ``three-month on + three-month off" detection scheme, which means that TianQin will firstly observe continuously for three months, and then be put into a safe mode for the next three months before starting observation again \cite{Luo:2015ght}. In this scheme, the total duration of data acquisition will be of 2.5 years for a mission lifetime of 5 years.
The development of TianQin is targeting a sensitivity characterized by the all sky averaged sensitivity \cite{Luo:2015ght,Hu:2018yqb,Lu:2019log},
\be S_n(f) = \frac{10}{3L^2}\left[S_x+\frac{4S_a}{(2\pi f)^4}\left(1+\frac{10^{-4}{\rm Hz}}{f}\right) \right] \times\Big[1+0.6\Big(\frac{f}{f_\ast}\Big)^2\Big]\,,
\label{eq:S_n^SA} \ee
where $L\approx1.7\times10^8$ m is the arm-length of TianQin,
$S_x^{1/2}=1\times10^{-12}$ m/Hz$^{1/2}$ is the displacement measurement noise for each one-way laser link,
$S_a^{1/2}=1\times10^{-15}$ m/s$^2$/Hz$^{1/2}$ is the residual acceleration noise for each test mass along the sensitive direction,
and $f_\ast=c/(2\pi L)\approx0.28$ Hz is the transfer frequency.

TianQin can detect a large variety of important astrophysical and cosmological sources and provide key information to astrophysics, fundamental physics and cosmology \cite{Hu:2017yoc}.
Firstly, TianQin can detect \ac{GW} sources at different times (redshifts) across the cosmic history, for example:
\ac{GCB} systems in the Galaxy \cite{Huang:2020rjf},
\ac{SBHB} systems to redshifts of the order $z\sim\cO(0.1)$ \cite{Liu:2020eko},
\ac{EMRI} systems to redshifts of the order $z\sim\cO(3)$ when the star formation rate was at its peak \cite{Fan:2020zhy},
\ac{MBHB} systems to redshifts of the order $z\sim\cO(20)$ when the first stars and galaxies had just appeared in the Universe \cite{Wang:2019ryf},
and possibly also first order electroweak phase transitions when the Universe was only about $\cO(10^{-10}$ s) old \cite{Liang:2021bde}.
Secondly, TianQin will be able to detect some \ac{GW} signals with extremely high \acp{SNR}. For example, \acp{SNR} for some \ac{MBHB} signals can reach the order of $\cO(10^3)$ \cite{Wang:2019ryf}.
Thirdly, TianQin will be able to measure the source parameters of some \ac{GW} signals to extremely high precision. For example, some of the parameters of \acp{SBHB}, \acp{EMRI} and \acp{MBHB} will be measured to better than $\cO(10^{-6})$ \cite{Liu:2020eko,Fan:2020zhy,Wang:2019ryf}.
These capabilities will enable TianQin to reveal many details of a \ac{GW} event, to test the nature of gravity in the strong field regime to unprecedented precision levels, and to precisely measure cosmological parameters at many different epochs of the cosmic history.

Several other new \ac{GW} detectors are being planned for the mid-2030's \cite{Gong:2021gvw}, for example, the space-based \ac{GW} detectors LISA \cite{LISA:2017pwj} and Taiji \cite{Hu:2017mde}, and the third generation ground-based \ac{GW} detectors \ac{CE} \cite{Evans:2021gyd} and \ac{ET} \cite{Maggiore:2019uih,Branchesi:2023mws}.
Both LISA and Taiji can complement TianQin in the detection of heavier sources while CE and \ac{ET} can join with TianQin to carry out multiband detection of \acp{SBHB}.
There is significant science reward if these detectors can form detector networks.
A systematic study focusing on the TianQin-LISA network can be found in \cite{Torres-Orjuela:2023hfd}.

The purpose of this white paper is to provide a quantitative assessment of the advancement that TianQin can bring to fundamental physics and cosmology.
In preparing for this white paper, we have benefited from many existing review papers. Some of them are listed below for the convenience of the reader:
\begin{itemize}
\item Science overview for \ac{GW} detectors: \cite{eLISA:2013xep,LISA:2017pwj,Hu:2017mde,Evans:2021gyd,Maggiore:2019uih,Branchesi:2023mws,Bailes:2021tot,Gong:2021gvw}
\item Experimental tests of \ac{GR}: \cite{Will:2014kxa,Baker:2014zba,Turyshev:2008dr,Turyshev:2008ur,Will:2010uh,Murata:2014nra,Koyama:2015vza,Sakstein:2017pqi}
\item Fundamental physics with \acp{GW}: \cite{Gair:2012nm,Yunes:2013dva,Berti:2015itd,Yagi:2016jml,Barack:2018yly,Cardoso:2019rvt,Barausse:2020rsu,LISA:2022kgy}
\item Cosmology with \acp{GW}: \cite{LISACosmologyWorkingGroup:2022jok}
\end{itemize}

This paper is organized as follows.
Section \ref{sec:tq_phys} is devoted to discussing how TianQin can help probe the fundamental laws governing the dynamics of the Universe.
This includes verifying key predictions of \ac{GR} in the strong field regime and searching for possible signatures of beyond \ac{GR} effects. Interfering environmental effects and waveform systematics will also be discussed.
Section \ref{sec:tq_new} is devoted to discussing how TianQin can help probe the fundamental composition of the Universe.
This includes detecting a first order phase transition in the early universe, revealing the properties of dark matter particles, studying the matter-anti-matter asymmetry, probing the hidden sector and other new physics beyond the \ac{SMPP}, searching for \acp{PBH}, and so on.
Section \ref{sec:tq_cosmo} is devoted to discussing how TianQin can help probe the expansion history of the Universe.
This includes measuring cosmological parameters across a wide range of redshifts and verifying cosmological laws.
The main results of the paper will be summarized in section \ref{sec:sum}.

\begin{table}[htbp!]
\begin{center}
\caption{Some major epochs, events and open questions in the history of the Universe (see, e.g., \cite{Debono:2016vkp,Baumann:2022mni}).}
\label{tab:sec1.universe}
\begin{tabular}{|m{2cm}<{\centering}|m{2cm}<{\centering}|m{10cm}<{\centering}|m{3cm}<{\centering}|}
\hline
Epochs & Cosmic Age & Major events & Open questions \\
\hline
Big Bang & 0 s
& The Universe is believed to have started from an initial state with extremely high energy density.
Since the size of the observable universe was extremely small at the beginning and then started to increase afterwards, the origin of the Universe is called a Big Bang.
& What is the initial state of the Universe and how did the Universe get into it? \\
\hline
Planck era & $\cO(10^{-43}$ s)
& Due to the high energy density right after the Big Bang, quantum effect was believed to be important even for the gravitational interaction. It stayed so until the characteristic energy of the Universe dropped below the Planck energy, $1.22\times10^{19}$ GeV, corresponding to a temperature of the order $\cO(10^{32}$ K).
& Is gravity quantized? If so, how to quantize gravity? \\
\hline
Separation of fundamental interactions & $\cO(10^{-35}$ s)
& The temperature at the Planck era was so high that all the fundamental interactions are believed to be unified. Gravity, the strong interaction and the electroweak interaction became separated one by one as the temperature dropped to the order $\cO(10^{27}$ K).
& Is it possible to unify all the fundamental interactions? \\
\hline
Inflation & $\cO(10^{-32}$ s)
& After the Planck era the size of the Universe was increased by roughly $\cO(10^{26})$ times through a rapid inflation. The temperature of the Universe dropped during the inflation but was restored to the order $\cO(10^{26}$ K) due to reheating at the end of inflation.
& Did inflation really happen? If yes, what is the driving force? \\
\hline
Electroweak symmetry breaking & $\cO(10^{-11}$ s)
& The electroweak symmetry broke as the temperature dropped to the order $\cO(10^{15}$ K). The universe went through a phase transition, and most particles in the Universe became massive.
& How did the electroweak phase transition happen? \\
\hline
Formation of nuclei and atoms & $\cO(10^{0}$ min) $\sim$ $\cO(10^{5}$ yrs)
& Nuclei became stable and \ac{BBN} took place when the temperature dropped to the order $\cO(10^{9}$ K). Atoms became stable when the temperature dropped to the order $\cO(10^{3}$ K). The photons decoupled to form \ac{CMB} when the Universe was about 380,000 years old.
& When and how did matter became dominant against antimatter? \\
\hline
Dark age, the first stars and galaxies & $\cO(10^{8}$ yrs)
& After the formation of CMB, the Universe was in darkness as the only source of \ac{EM} radiation was the glow of individual atoms. The dark age continued for hundreds of millions of years until the first stars and galaxies started to form. black holes also started to form.
& When and how did the first stars, galaxies and black holes form and evolve? \\
\hline
Cosmic expansion and acceleration & Up to now
& The universe continues expansion after the inflation, governed by the laws of gravity and driven by the energy content of the Universe. It has been found that about 27\% the total energy density of the Universe is dark matter, while about 68\% is dark energy.
& What is the nature of dark matter and dark energy? \\
\hline
\end{tabular}
\end{center}
\end{table}

\clearpage

%% file: sec2-phys.tex
\section{Nature of gravity with TianQin}\label{sec:tq_phys}

{\it Section coordinator: Jianwei Mei}

Among all the fundamental interactions in nature, gravity is the least understood and it is only known at the classical level so far.
\ac{GR} is the current best theory of gravity but it describes gravity in a purely geometrical way.
The theoretical structure of \ac{GR} is completely different from that of \ac{SMPP}, making it difficult for gravity to unify with other fundamental interactions even formally.
Experimentally finding out the breaking points of \ac{GR} may hold the key to answering fundamental questions concerning the quantization of gravity and the unification of gravity with other fundamental {interactions}.

For over a century, \ac{GR} has been passing all sorts of experimental tests with flying colors. But for most experiments, the characteristic masses involved are at the order $\cO(1\mSun)$ or less \cite{Will:2014kxa}.
With the breakthrough of \ac{GW} detection, there start to be tests of \ac{GR} with characteristic masses at the order $\cO(30\mSun)$ \cite{LIGOScientific:2016lio,LIGOScientific:2019fpa,LIGOScientific:2020tif,LIGOScientific:2021sio}, corresponding to the point ``GW-SBHB" in FIG. \ref{fig:gr-exps}.
The most remarkable fact about \acp{GW} is that it can test \ac{GR} in the genuinely strong field regime, i.e., with the {dimensionless} gravitational potential approaching $\cO(1)$ {where nonperturbative gravitational effects might take place}.
Such tests are not possible with non-\ac{GW} experiments and may harbor the chance to making breakthroughs in finding beyond \ac{GR} effects.

TianQin will bring the test of \ac{GR} to a whole new level: some signals will be observed with very high \acp{SNR} so that much details of gravity can be revealed \cite{Shi:2024ttu}, and some source parameters will be measured to very high precision so that various aspects of \ac{GR} can be tested to the corresponding level of precision \cite{Shi:2019hqa,Zi:2021pdp}.
With TianQin, characteristic masses ranging from $\cO(10\mSun)$ to $\cO(10^6\mSun)$ and higher will be employed to test \ac{GR}, significantly populating the parameter space connecting ``GW-SBHB" and ``GW-\ac{MBHB}" in FIG. \ref{fig:gr-exps}.

The effort to test {GR} and to search for possible new physics can be grossly divided into two categories:
\begin{itemize}
\item {Detection}: For characteristic predictions of \ac{GR} that lack experimental verification, one actively searches for the corresponding effect in experiments and tries to establish their existence. The first detection of \ac{GW} itself is a good example in this category, as it robustly establishes the existence of \ac{GW} and firmly verifies a key prediction of \ac{GR} \cite{LIGOScientific:2016aoc}.
\item {Measurement}: For predictions of \ac{GR} that already have been verified by experiments, one tries to improve the precision of measurement to search for possible signatures of beyond \ac{GR} effects. Due to the lack of a complete theory of quantum gravity that is immediately relevant for experiments, one has to use phenomenological parameters to book keep possible deviations from \ac{GR}. {In this sense we} treat all \acp{MGT} as phenomenological parameterization schemes and will group them into this category.
\end{itemize}

In this section, we will use such a classification scheme to discuss how TianQin can help verify key predictions of \ac{GR} in the strong field regime and search for possible signatures of beyond \ac{GR} effects.
We will also discuss the environmental effects that may interfere with such effort.

\subsection{Key predictions of GR in the strong field regime}

The most straightforward way to test \ac{GR} is to see if its predictions can be found in experiments. Many key predictions of \ac{GR} have been verified through classic tests \cite{Will:2014kxa}, such as the perihelion precession of Mercury, the bending of light by {the} Sun, the gravitational redshift of light, and the Shapiro time delay \cite{Shapiro:1964uw}. The detection of \acp{GW} has been another remarkable success for \ac{GR} \cite{LIGOScientific:2016lio}. With the breakthrough in \ac{GW} detection, there come new opportunities to test some key predictions of \ac{GR} in the genuinely strong field regime. For examples:
\begin{itemize}
\item Is the dynamics of gravity still as prescribed by \ac{GR} even at the nonlinear level?
\item Do black holes, which are the most remarkable strong field predictions of \ac{GR}, exist in nature?
\end{itemize}
In this subsection, we highlight how TianQin can be used to help answer these questions.

\subsubsection{Higher modes and nonlinear modes}

{\it Subsection coordinator: Changfu Shi}

Nonlinearity is a characteristic feature of Einstein's equations. On the theoretical side, one can already note a few interesting peculiarities. For examples:
\begin{itemize}
\item Although the Kerr black hole is a solution to Einstein's equations in the strong gravity limit, its metric can be written in the Kerr-Schild form \cite{Kerr:1965wfc}, $g_{\mu\nu}=\eta_{\mu\nu}+fk_\mu k_\nu\,$, where $f$ is a function, $k_\mu$ is a null vector, and the piece $h_{\mu\nu}=fk_\mu k_\nu\,$ satisfies the linearized version of Einstein's equations over the full Kerr background.
\item The nonlinearity of Einstein's equations can be reduced to a finite order by using appropriate variables for the metric components, written in the extended Kerr-Schild form \cite{Harte:2014ooa}.
\item The waveform from the merger of two black holes, which requires Einstein's equations in full to calculate, is unexpectedly simple \cite{Pretorius:2005gq}.
\end{itemize}
All these facts suggest that even the nonlinearity of Einstein's equations itself is something that needs further characterization.
So experimental test of the strong field predictions of Einstein's equations will hopefully help us understand better not only the true dynamics of gravity but also the true nonlinearity of Einstein's equations itself.

\acp{GW} provide an irreplaceable way to study the nonlinearity of gravity under dynamical situations. The \ac{GW} signal from a compact binary {coalescence} event typically includes three stages: inspiral, merger and ringdown. For the {early-inspiral} and ringdown stages, one can rely on perturbative approaches to produce predictions from Einstein's equations. Schematically, the metric can be expanded as
\bea g_{\mu\nu}=\eta_{\mu\nu}+\varepsilon h^{(1)}_{\mu\nu}+\varepsilon^2 h^{(2)}_{\mu\nu}+\cO(\varepsilon^3)\,,\label{sec2.2:expansion-metric}\eea
where $\eta_{\mu\nu}$ is the Minkowski metric and $\varepsilon$ is a bookkeeping parameter indicating the magnitudes of each term in the expansion. Correspondingly, Einstein's equations can be expanded order by order as (focusing on the vacuum case for simplicity),
\bea\cO(\varepsilon)&:&E_{\mu\nu}[h^{(1)}_{\bullet\bullet}]\equiv\frac12\bar\nabla^\rho\Big[\bar\nabla_\nu h^{(1)}_{\mu\rho} +\bar\nabla_\mu h^{(1)}_{\nu\rho} -\bar\nabla_\rho h^{(1)}_{\mu\nu}\Big] -\frac12\bar\nabla_\mu\bar\nabla_\nu h^{(1)}=0\,,\nn\\
\cO(\varepsilon^2)&:&E_{\mu\nu}[h^{(2)}_{\bullet\bullet}]=S^{(2)}[h^{(1)}_{\bullet\bullet},h^{(1)}_{\bullet\bullet}]\,,\label{sec2.2:expansion-EE}\nn\\
&\vdots&\eea
where $h^{(1)}$ is the trace of $h^{(1)}_{\mu\nu}$ and $S^{(2)}[h^{(1)}_{\bullet\bullet},h^{(1)}_{\bullet\bullet}]$ contains the quadratic terms of $h^{(1)}_{\mu\nu}$ acting as the source for the second order perturbations. The metric perturbations at each order can be further expanded. For example, the ringdown signal after the merger of a \ac{MBHB} can be expanded in terms of a series of \acp{QNM} \cite{Kokkotas:1999bd,Berti:2009kk,Konoplya:2011qq},
\bea h=\frac{M_z}{D_L}\sum_{\ell mn}e^{-i\widetilde{\omega}_{\ell mn}(t-t_0)}~_{-2}S_{\ell mn}A_{\ell mn}\,,\label{sec2.2:expansion-h-ringdown} \eea
where we have omitted the indices on $h$, $M_z$ is the red-shifted mass, $D_L$ is the luminosity distance, $\widetilde{\omega}_{lmn} = \omega_{lmn} + i / \tau_{lmn}$ are the complex {\ac{QNM}} frequencies, with $\omega_{lmn}$ and $\tau_{lmn}$ being the oscillation frequency and damping time, respectively, $_{-2}S_{\ell mn}$ are the $-2$ spin-weighted spheroidal harmonics, and $A_{\ell mn}$ are the amplitudes that depend on the parameters of the progenitor binaries.

In practice, one must truncate (\ref{sec2.2:expansion-metric}), (\ref{sec2.2:expansion-EE}) and (\ref{sec2.2:expansion-h-ringdown}) to finite orders to be computationally efficient/feasible. \acp{GW} from each order of the perturbation can be expanded as in (\ref{sec2.2:expansion-h-ringdown}).
So far only \acp{GW} from the linear order have been confirmed in ground-based detectors.
The $(2,2,0)$-mode, which has $\ell=m=2$ and $n=0$, is usually the dominant mode, and all other modes are called higher modes.
All modes with $n=0$ are called the fundamental modes.
The modes with $n\neq0$ are called overtones and all the modes for $h^{(2)}_{\mu\nu}$ and higher are called nonlinear modes.
The detection of a broader spectrum of modes translates into the acquisition of richer information, which in turn can break the degeneracy among certain parameters of \ac{GW} source, thereby enhancing our ability to accurately measure these parameters.
The detection of the higher modes and nonlinear modes also hold the key to finding the possible discrepancy between the nonlinearity of Einstein's equations and the true dynamics of gravity. {It also serves to examine the consistency of GR predictions, and contrast with predictions from alternative gravity theories.}
Detecting the higher modes and nonlinear modes is thus an important aspect of testing \ac{GR}.

Limited by the sensitivity of existing ground-based detectors, most of the detected \ac{GW} events have \acp{SNR} of 30 or less \cite{LIGOScientific:2021sio}, while the \acp{SNR} of the ringdown phase are even weaker.
It has been claimed that there is $(2,2,1)$ {mode} in \ac{GW}150914 with 3.6$\sigma$ confidence \cite{Isi:2019aib}, but the result is still being debated \cite{Carullo:2019flw, Wang:2023mst, Wang:2024yhb}.
The detection of the (3,3,0) mode with a Bayes factor 56 in \ac{GW}190521 has also been claimed \cite{Capano:2022zqm}.
No other higher modes have been {reported} in the existing \ac{GW} data.

\begin{figure*}[htbp!]
	\centering
	\graphicspath{{figs/}}
	\subfigure{
		\includegraphics[scale=0.5]{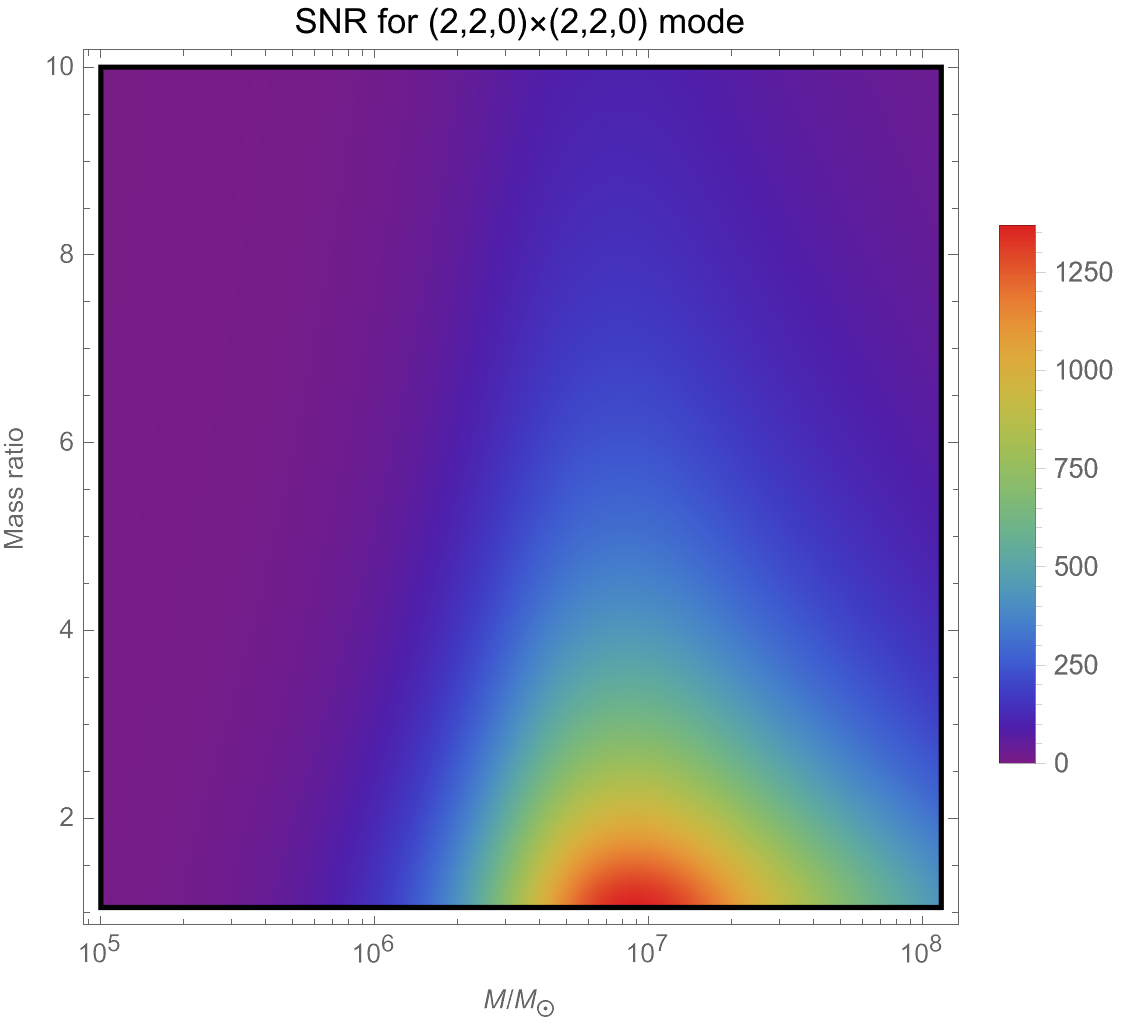}}
	\subfigure{
		\includegraphics[scale=0.5]{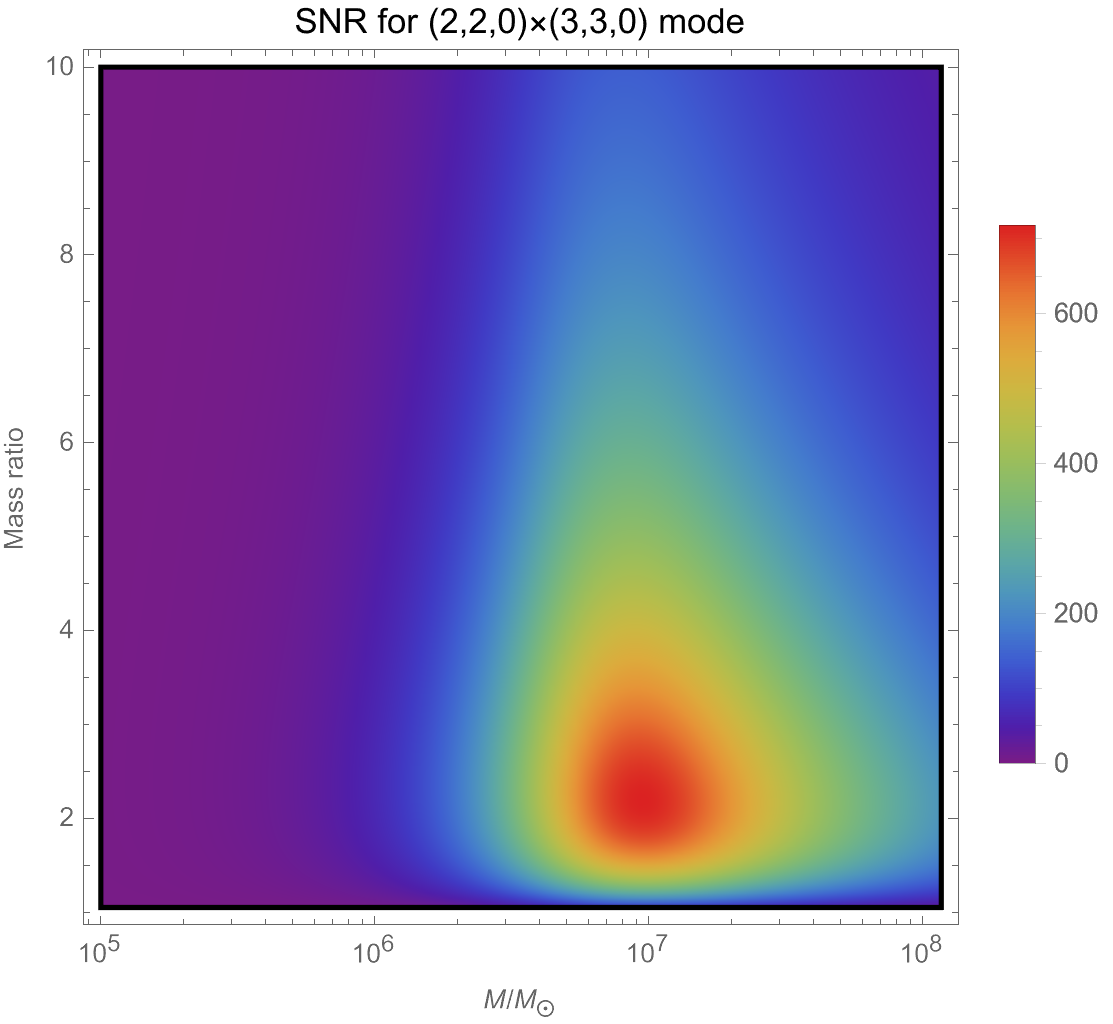}}
	\caption{The dependence of \ac{SNR} for two second-order modes varying with the final mass and mass-ratio. Other parameters used for this plot are: $D_L=1\rm~Gpc,\iota=\pi/3$.}\label{fig:SNR_SM}
\end{figure*}

The space-based \ac{GW} detectors are expected to achieve high \acp{SNR} for \ac{MBHB} signals and can detect a series of higher-order \acp{QNM}.
Berti {\it et al.} have investigated the capability of LISA on detecting the (3,3,0) mode and (4,4,0) mode by assuming that the energy radiated during the ringdown phase is about 3\% of the total mass of the system \cite{Berti:2005ys}.
Baibhav {\it et al.} extended their work to include more fundamental modes, such as (5,5,0), (6,6,0) and even (7,7,0) {modes}, by setting the \ac{SNR} threshold to 8.
They analyzed how the detection horizon of each mode depends on the mass of the remnant black hole \cite{Baibhav:2018rfk}.
Shi {\it et al.} \cite{Shi:2024ttu} analyzed the prospect of using a space-based detector to detect the (2,2,0), (2,1,0), (3,3,0), and (4,4,0) modes, by using an amplitude {fit formula} from numerical waveforms \cite{Kamaretsos:2011um}.

For nonlinear modes, Ioka {\it et al.} have analyzed the metric perturbations to second-order and have found that the second-order \acp{QNM} have frequencies twice those of the first-order and the amplitudes up to about 10\% those of the first-order \cite{Ioka:2007ak}.
They noted that LISA can detect the nonlinear modes by comparing the amplitude of the second-order mode with the amplitude spectral density of the detector.
Approximate analytical formulae for first-order \ac{QNM} amplitudes can be fitted from \ac{NR} waveforms, and this has also been done for the nonlinear (2,2,0)$\times$(2,2,0) and (2,2,0)$\times$(3,3,0) modes \cite{London:2014cma}.
Based on \ac{NR} results for head-on mergers and quasi-circular mergers with non-spinning progenitors, it has been suggested that the second-order modes can be viewed as being sourced by the first-order ones, with the relation \cite{Mitman:2022qdl,Cheung:2022rbm}:
\bea
	\label{eq:srm}
	\omega_{(l_1m_1n_1)(l_2m_2n_2)} &=& \omega_{l_1m_1n_1} + \omega_{l_2m_2n_2},  \nn \\
	\tau_{(l_1m_1n_1)(l_2m_2n_2)}^{-1} &=& \tau_{l_1m_1n_1}^{-1} + \tau_{l_2m_2n_2}^{-1},   \\
	A_{(l_1m_1n_1)(l_2m_2n_2)} &\propto & A_{l_1m_1n_1} A_{l_2m_2n_2}. \nn
\eea
The ratio factor of the amplitudes,
\bea
	\mu_{(l_1m_1n_1)(l_2m_2n_2)} \equiv \frac{A_{(l_1m_1n_1)(l_2m_2n_2)}}{A_{l_1m_1n_1} A_{l_2m_2n_2}},
\eea
is determined solely by the properties of the final black hole. It has been suggested that \cite{Cheung:2022rbm}:
\begin{eqnarray}
	\mu_{(2,2,0)(2,2,0)} = 0.1637,  \nn \\
	\mu_{(2,2,0)(3,3,0)} = 0.4735.
\end{eqnarray}

By using the amplitude formulae fit by London {\it et al.} \cite{London:2014cma}, the prospect of using TianQin to detect 11 different higher modes and nonlinear modes has been studied in \cite{Shi:2024ttu}.
From the definition of the \ac{SNR} and the ringdown waveform \eqref{sec2.2:expansion-h-ringdown}, one can note that the \acp{SNR} of higher modes and nonlinear modes strongly depend on the remnant black hole mass $M_z$, the luminosity distance $D_L$, the mass ratio $q$, and the inclination angle $\iota$.
The dependence on the luminosity distance and the inclination angle is simple: the \ac{SNR} is inversely proportional to the luminosity distance and the effect of inclination angle is fully contained in the spin-weighted harmonics.
The result for the two second-order modes is illustrated in FIG. \ref{fig:SNR_SM}.
For both modes, one can see that the highest \ac{SNR} can be achieved when the total mass is around $10^7\mSun$.
For the (2,2,0)$\times$(2,2,0) mode, the highest \ac{SNR} can be achieved when $q=0$, while for the (2,2,0)$\times$(3,3,0) mode, the highest \ac{SNR} can be achieved around $q=2$.
It is remarkable to note that the \acp{SNR} can reach a few dozens even for {a source at} redshift $z=3$.

The detection numbers of TianQin for each of the 11 higher modes and nonlinear modes are presented in Table \ref{table_QNM_number}.
Three astrophysical models of \acp{MBH} have been considered: pop III, Q3\_d and Q3\_nod.
The result has been obtained by averaging over one thousand sets of data generated from each of the astrophysical models.
One can see that, apart from the (4,3,0) mode, all other modes are expected to be detected in at least one \ac{MBHB} event.

\begin{table}[t]
\caption{The detection number for different QNMs with different astrophysical models.}
\centering
\begin{tabular}{|c|c|c|c|c|c|c|c|c|c|c|c|}
\hline
QNMs      &(2,2,0) &(2,2,1)&(2,1,0)&(3,3,0)&(3,3,1)&(3,2,0)&(4,4,0)&(4,3,0)&(5,5,0)&(2,2,0)$\times$(2,2,0)&(2,2,0)$\times$(3,3,0) \\
\hline
PIII      & 11.6   & 4.8   & 7.2   & 7.7   & 2.5   & 2.9   & 5.0   & 1.0   & 2.8   & 2.2                  & 1.2 \\
\hline
Q3d       & 13.7   & 8.3   & 8.4   & 9.8   & 3.4   & 3.4   & 6.7   & 0.8   & 2.5   & 4.5                  & 1.4 \\
\hline
Q3\_nod   & 168.1  & 59.0  & 78.5  & 110.2 & 16.6  & 16.9  & 46.8  & 4.5   & 16.9  & 15.0                 & 5.6 \\
\hline
\end{tabular}
\label{table_QNM_number}
\end{table}

\subsubsection{Memory effect}

{\it Subsection coordinator: Shuo Sun}

The radiation of \acp{GW} will result in a permanent change to the background spacetime.
Such change is related to the entire history of \ac{GW} radiation, and the phenomenon is referred to as the \ac{GW} memory effect.
The memory effect is one of the direct predictions of \ac{GR} in the nonlinear and strong-field regime, so the detection of the memory effect is a direct test of \ac{GR}.
What's more, the memory effect can be used to search for signatures of \acp{MGT} \cite{Du:2016hww, Seraj:2021qja,Tahura:2021hbk,Hou:2021oxe,Hou:2021bxz,Hou:2023pfz},
to break the degeneracy among \ac{GW} source parameters, such as the inclination angle and luminosity distance \cite{Gasparotto:2023fcg,Sun:2024nut,Xu:2024ybt},
to help distinguish between neutron star-black hole systems and binary black hole systems \cite{Tiwari:2021gfl},
and even aid in detecting the presence of matter around black holes \cite{Lopez:2023aja}.
So far, many studies and waveform models do not consider the memory effect, it is important to know how much systematic error such practice will bring.

In the 1970s, Zel'dovich and Polnarev first discovered the memory effect while studying \acp{GW} emitted by collapsed stars passing each other in galactic nuclei \cite{Zeldovich:1974gvh}. Since the calculations were performed within the framework of linear gravitational theory, this effect is also known as linear memory effect. Because the study focused on unbound systems, it was initially believed that the memory effect was not widely applicable. For an N-body system, the linear memory effect primarily arises from changes in the masses and velocities of the system's components. It can be described by an universal formula \cite{Thorne:1992sdb}
\be
    \Delta h^{\mathrm{TT}}_{jk}=\Delta \sum^{N}_{A=1}\frac{4M_{A}}{R\sqrt{1-v_{A}^{2}}}\left[\frac{v^{j}_{A}v^{k}_{A}}{1-v_{A}\cdot N}\right]^{\mathrm{TT}},
\ee
where the $\Delta$ means the difference before and after radiating \acp{GW}, and TT represents the transverse and traceless (TT) gauge.

In the 1990s, it was discovered that the energy flux radiating outward with the \acp{GW} also produces a memory effect \cite{Christodoulou:1991cr,Blanchet:1992br}. Because this memory effect was discovered in a nonlinear context, it is also known as nonlinear memory effect. Since it is caused by the energy radiated outward from the system, all \ac{GW} sources will produce the nonlinear memory effect. Under the harmonic TT gauge, the nonlinear memory effect can be written as
\be
    \delta h_{j k}^{\mathrm{TT}}=\frac{4}{R} \int_{-\infty}^{T_{R}} d t^{\prime}\left[\int \frac{d E^{\mathrm{gw}}}{d t^{\prime} d \Omega^{\prime}} \frac{n_{j}^{\prime} n_{k}^{\prime}}{\left(1-\boldsymbol{n}^{\prime} \cdot \boldsymbol{N}\right)} d \Omega^{\prime}\right]^{\mathrm{TT}},
\ee
where $T_{R}$ is the retard time and $n_{j}$ is the unit radiation vector.

Due to its relation to the change of the background spacetime, the memory effect is closely related to the \ac{BMS} group \cite{Bondi:1962px,Sachs:1962wk,deBoer:2003vf,Barnich:2009se,Barnich:2010ojg,Kapec:2014opa,Kapec:2016jld,He:2017fsb,Hou:2024exz}, which describes the symmetry of the asymptotically flat spacetime.
In 2014, Strominger and Zhiboedov discovered that the memory effect is related to the \ac{BMS} supertranslation while studying scattering problems in asymptotically flat spacetimes \cite{Strominger:2014pwa}. The effect of a passing \ac{GW} can be viewed as a supertranslation transformation of the spacetime.
The difference between the two spacetimes before and after the supertranslation is the memory effect, also known as the displacement memory effect, for it can permanently change the relative distance between two objects.
Corresponding to the \ac{BMS} superrotation and superboost, there are two more types of memory effects: the spin memory effect \cite{Pasterski:2015tva} and the center-of-mass memory effect \cite{Nichols:2018qac}.
The spin memory effect causes a time delay between two oppositely rotating particles, while the center-of-mass memory effect causes a time delay between two particles moving in opposite parallel directions.
Therefore, the memory effect is related to both the asymptotic symmetries and the Weinberg formula for soft graviton production \cite{Weinberg:1965nx}, forming a triangular relationship, known as the infrared triangle \cite{Strominger:2017zoo}.
Thus, the direct detection of the memory effect can provide an observational mean to study the soft theorem and asymptotic symmetries \cite{Goncharov:2023woe}.

The idea of detecting the memory effect with a ground-based \ac{GW} detector was first proposed by Thorne and Braginsky in the 1980s \cite{Braginsky:1987kwo}.
Lasky {\it et al.} analyzed the data from \ac{GW}50914 and indicated that LIGO is insufficient to directly detect the memory effect generated by sources like \ac{GW}150914.
Instead, approximately $\cO(90)$ \ac{GW}150914 liked events are needed to confidently confirm the memory effect \cite{Lasky:2016knh}.
H\"ubner {\it et al.} and Cheung {\it et al.} analyzed the data from GWTC-1, -2, and -3, and found that LIGO would require approximately $\cO(2000)$ events to confidently confirm the memory effect \cite{Hubner:2019sly,Hubner:2021amk,Cheung:2024zow}.
Zhao {\it et al.} also analyzed data from GWTC-2, and they suggested there is indication of the memory effect in \ac{GW}190814 \cite{Zhao:2021hmx}.
Using pulsar timing array to detect the memory effect has also been considered \cite{Seto:2009nv,vanHaasteren:2009fy,Pshirkov:2009ak,Cordes:2012zz,Madison:2014vca,NANOGrav:2015xuc}.
Recent analysis of 12.5 years of NANOGrav data found that the Bayes factor for the existence of the memory effect is approximately 2.8, which is insufficient to confirm the presence of the memory effect in the data \cite{NANOGrav:2023vfo}.
Space-based \ac{GW} detectors can detect more massive sources of \acp{GW} at cosmological distances, such as \acp{MBHB}.
These sources can radiate more energy during their merger, the resulting memory effect is also more pronounced.
It has been found that LISA could detect approximately 2 to 10 instances of the memory effect produced by \acp{MBHB} during its mission lifetime \cite{Islo:2019qht,Inchauspe:2024ibs}. In addition, a study has explored the prospects of the space-based detector DECIGO in detecting the memory effect generated by \acp{SBHB}. The results suggest that DECIGO could detect approximately 2,258 loud enough memory signals during its 5 years of observation \cite{Hou:2024rgo}.

\begin{table}[t]
\caption{The expected number of \ac{MBHB} events that can be detected by TianQin and LISA with significant memory effect. ``Total" represents the total number of \ac{MBHB} events that can be detected.}
\setlength{\tabcolsep}{2mm}{
\begin{tabular}{lcccccccc}
\hline \hline
& Total &$\cM>\text{Threshold}$ &$\rho_{\text{dis}}>3$ & $\rho_{\text{dis}}>5$ & $\rho_{\text{dis}}>8$ & $\rho_{\text{spin}}>3$ & $\rho_{\text{spin}}>5$ & $\rho_{\text{spin}}>8$\\
\hline
TianQin pop III  & $56.8$ & 0.9&$0.5$ & 0.3 & 0.1 & $\sim0$ & $\sim0$ & $\sim0$ \\
\hline
TianQin Q3\_d   & 18.1 & 0.9&0.6 & 0.3 & 0.2 & $\sim0$ & $\sim0$ & $\sim0$ \\
\hline
TianQin Q3\_nod & 271.4 & 3.6&2.0 & 1.2 & 0.7 & $0.2$ & 0.1 & $\sim0$ \\
\hline
LISA pop III     & $148.35$ &3.3& $1.6$ & 0.7 & 0.4 & $0.1$ & $\sim0$ & $\sim0$ \\
\hline
LISA Q3\_d      & 37.4 &4.9& 2.6 & 1.4 & 0.8 & 0.2 & 0.1 & $\sim0$ \\
\hline
LISA Q3\_nod    & 295.5 &12.2& 5.8 & 2.6 & 1.4 & 0.4 &0.2  &0.1 \\
\hline
\end{tabular}}
\label{pop}
\end{table}

The prospect of using TianQin to detect the memory effect has been studied in \cite{Sun:2022pvh,Sun:2024nut}.
Based on a few astrophysical population models of \acp{MBHB}, the expected detection numbers of \ac{MBHB} events with significant memory effect have been given in Table. \ref{pop}.
One can see that TianQin can detect approximately 0.5 to 2 \ac{MBHB} merger events for which the displacement memory effect can have \acp{SNR} greater than 3.
The chance for TianQin to detect the spin memory effect from a single \ac{MBHB} event is found to be negligible.
LISA is more capable than TianQin in terms of detecting the memory effect for most of the parameter space.
The joint detection of TianQin and LISA can slightly improve over LISA \cite{Sun:2022pvh}.

Apart from detecting the memory effect, the contribution of the memory effect to the waveform systematics is also an issue.
So far, the memory effect can only be calculated in the time domain, and this is an obstacle to the \ac{MBHB} data analysis because calculating waveforms in the frequency domain is usually much faster.
So it is important to know when one can neglect the contribution of the memory effect.
The mismodeling of waveforms and whether it will introduce systematic errors are measured by the mismatch and the mismatch threshold,
\bea
\cM=1-\frac{\langle \tilde{h}_{1}(f),\tilde{h}_{2}(f)\rangle}{\sqrt{\langle \tilde{h}_{1}(f),\tilde{h}_{1}(f)\rangle\langle \tilde{h}_{2}(f),\tilde{h}_{2}(f)\rangle}}\,,\quad
\cT=\frac{D}{2{\rm~SNR}^{2}}\,,\label{mismatch}
\eea
where $<\cdots|\cdots>$ is the inner product.
The factor $D$ is often approximated by the number of intrinsic parameters whose estimation is affected by the waveform accuracy \cite{Chatziioannou:2017tdw}, and can be tuned by calculating the statistical and systematic errors from the posterior distribution of synthetic signals with increasing \acp{SNR} \cite{Purrer:2019jcp}.

\begin{figure}[t]
\subfigure{\includegraphics[scale=0.2]{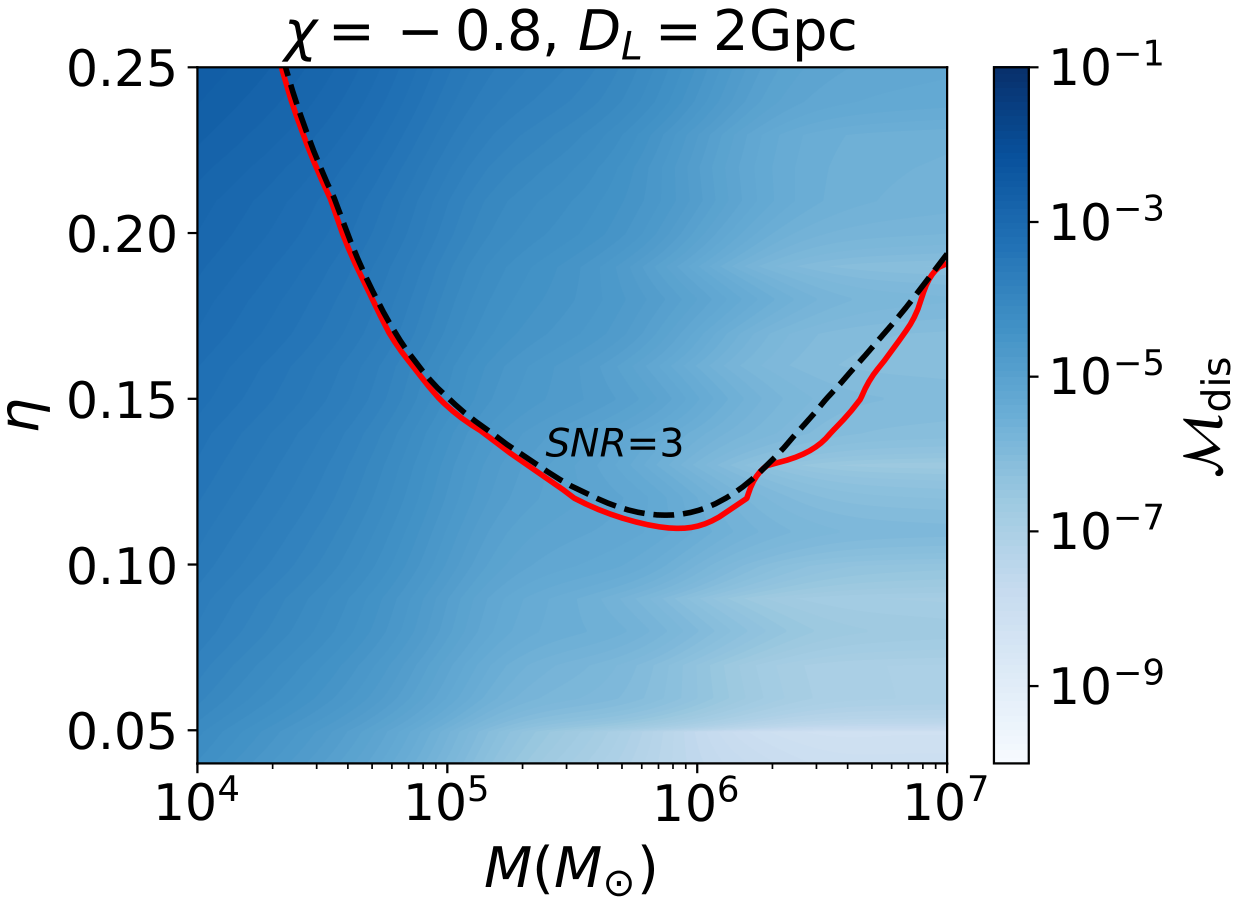}}
~~~~
\subfigure{\includegraphics[scale=0.2]{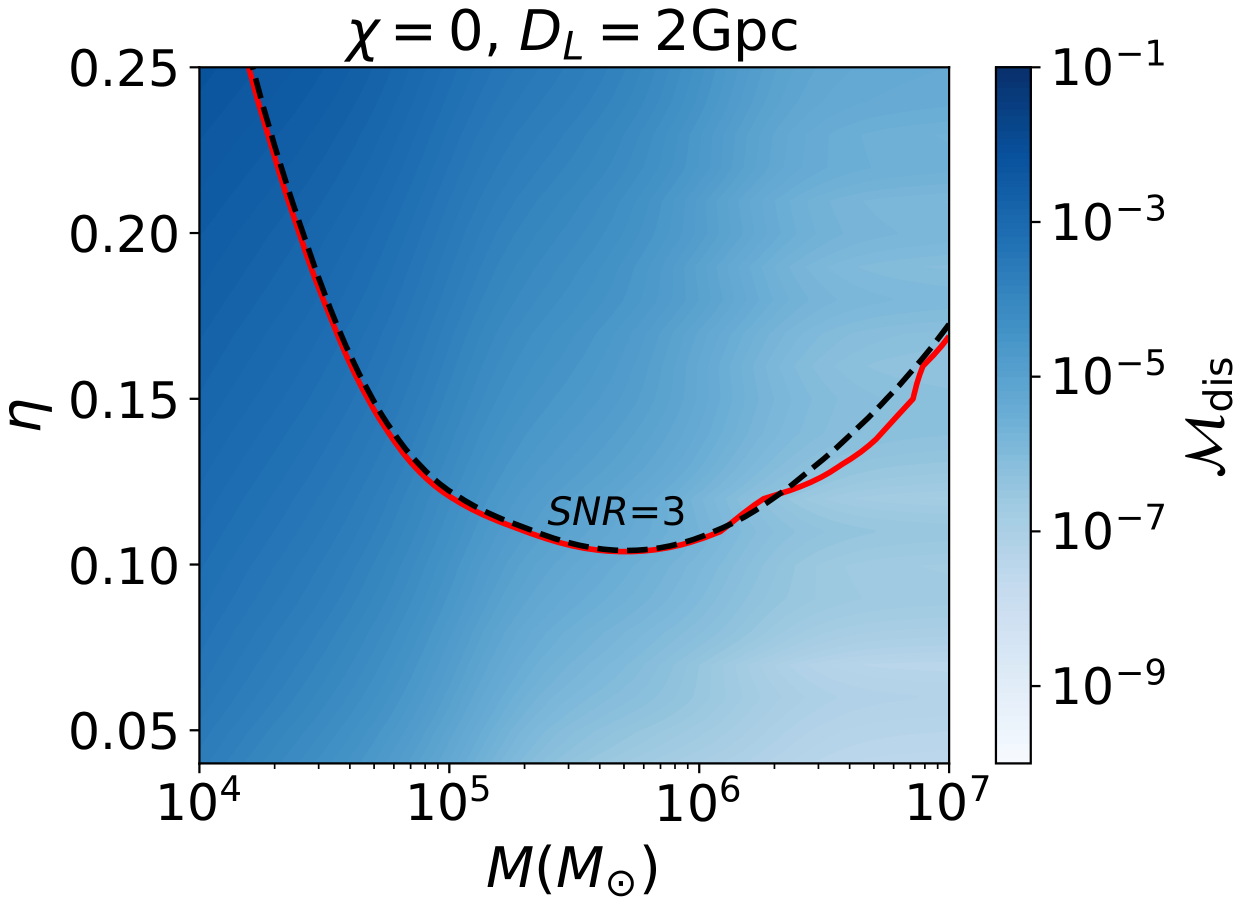}}
\subfigure{\includegraphics[scale=0.2]{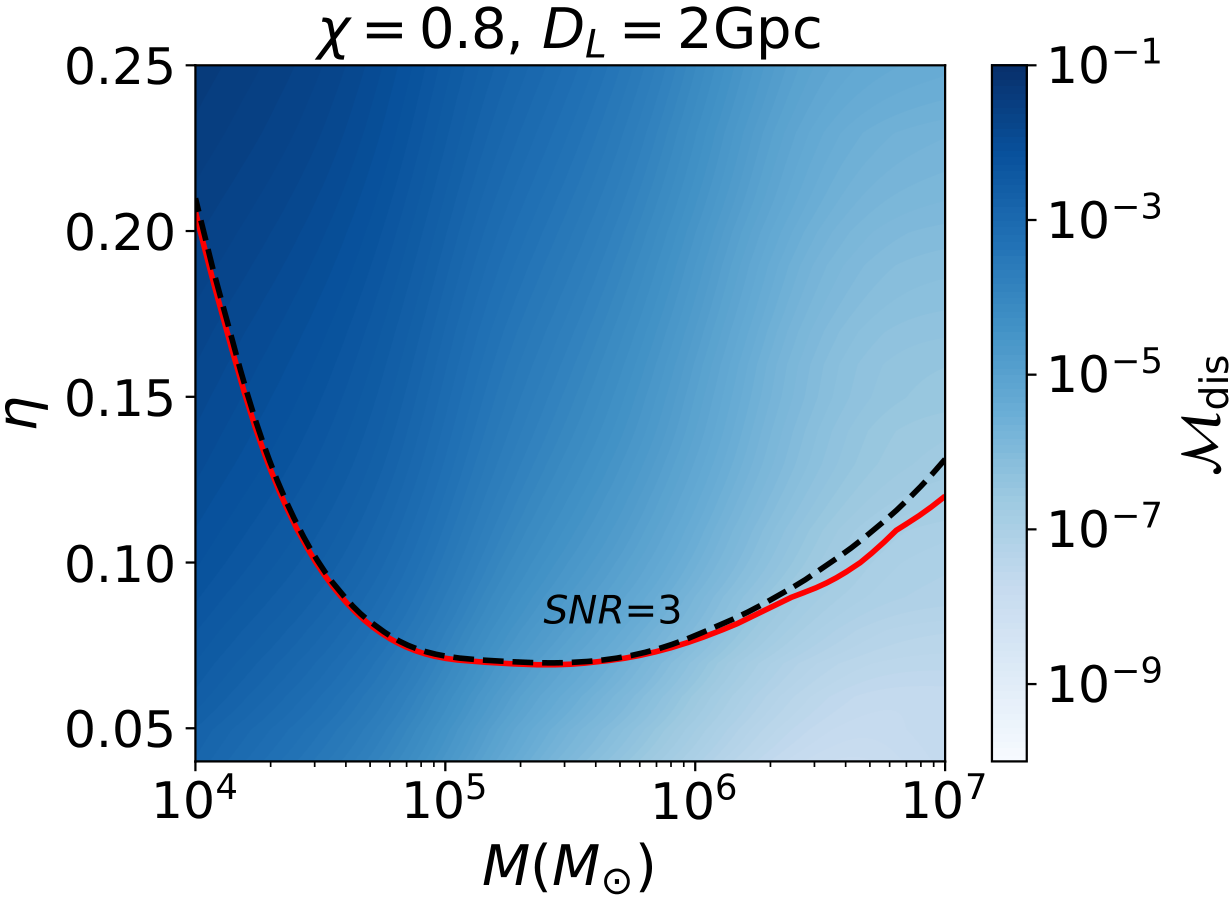}}
~~~~
\subfigure{\includegraphics[scale=0.2]{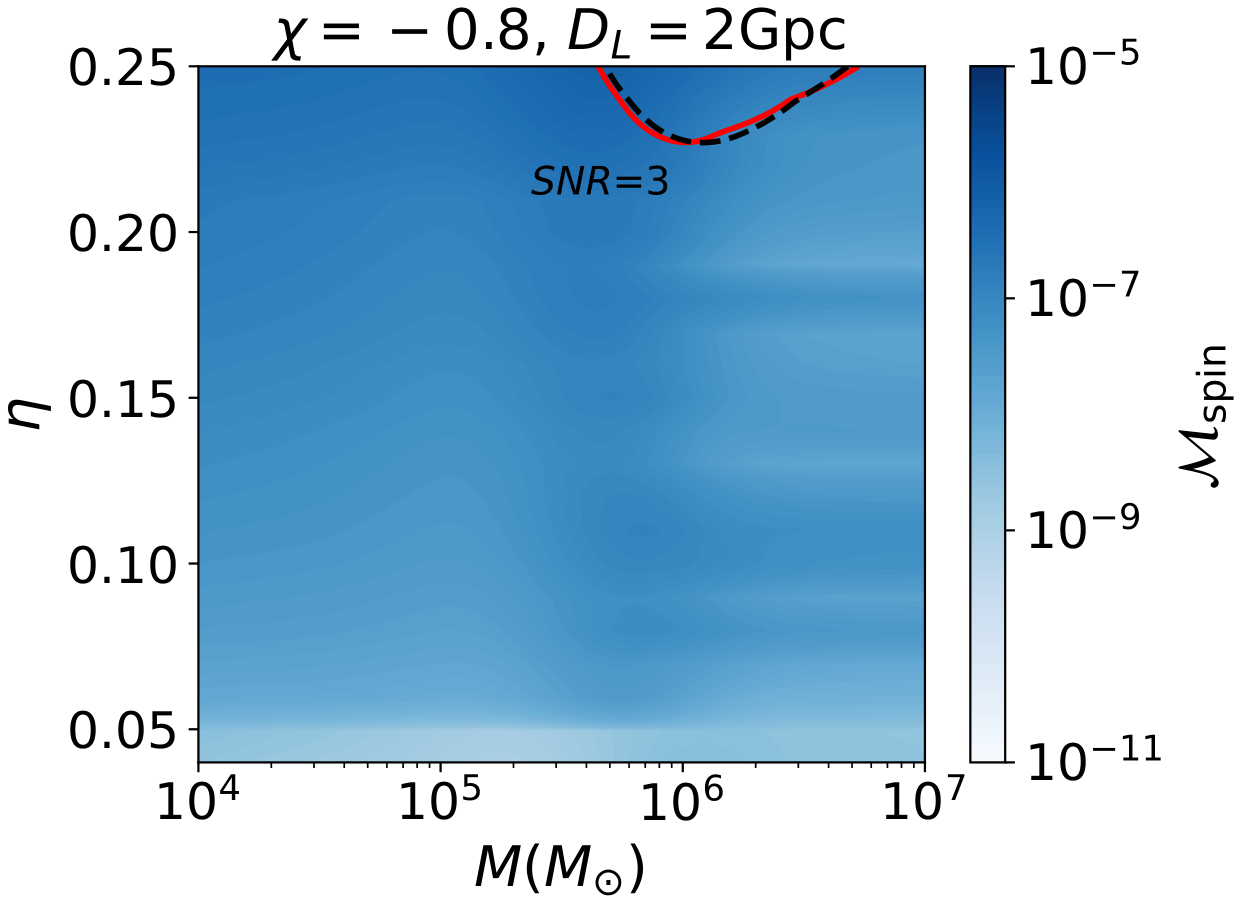}}
\subfigure{\includegraphics[scale=0.2]{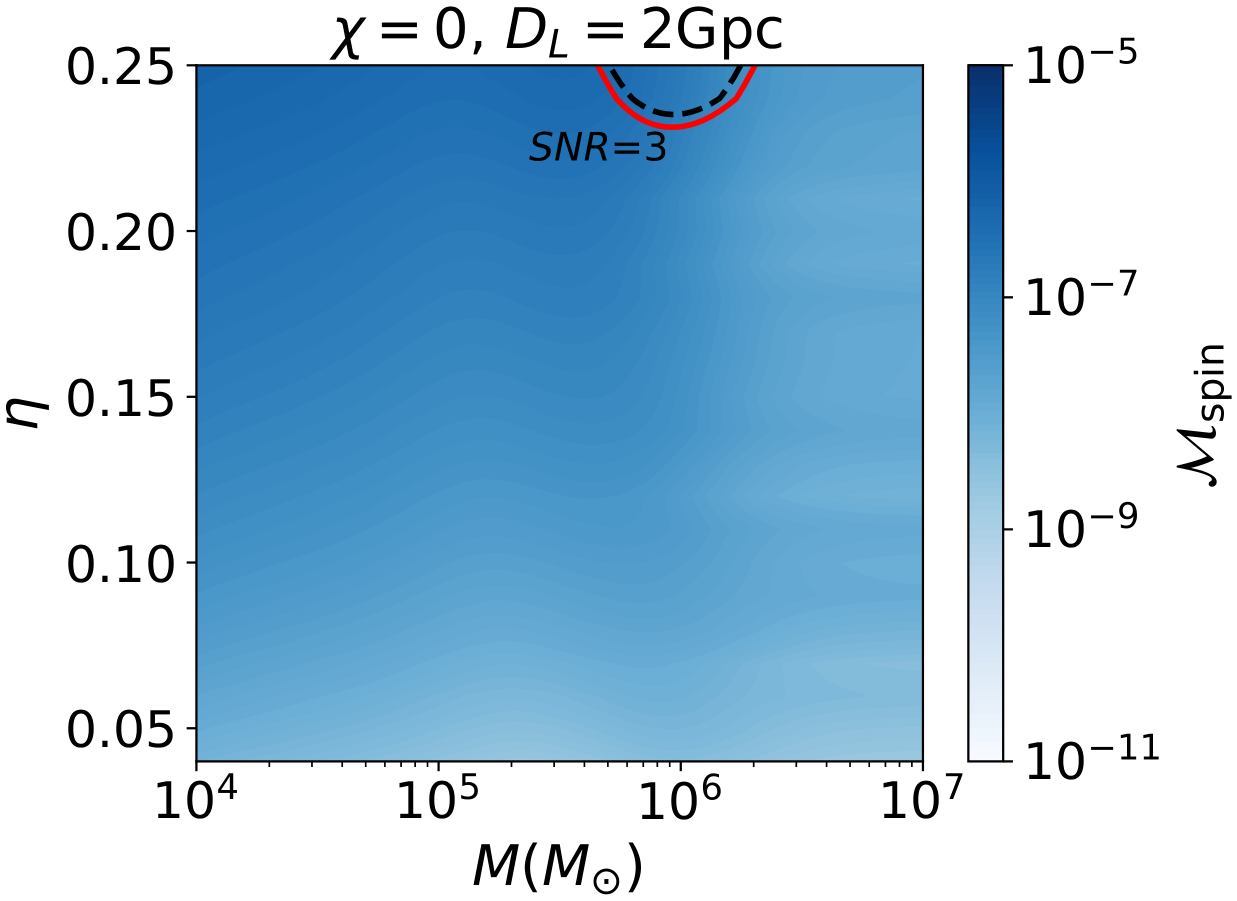}}
~~~~
\subfigure{\includegraphics[scale=0.2]{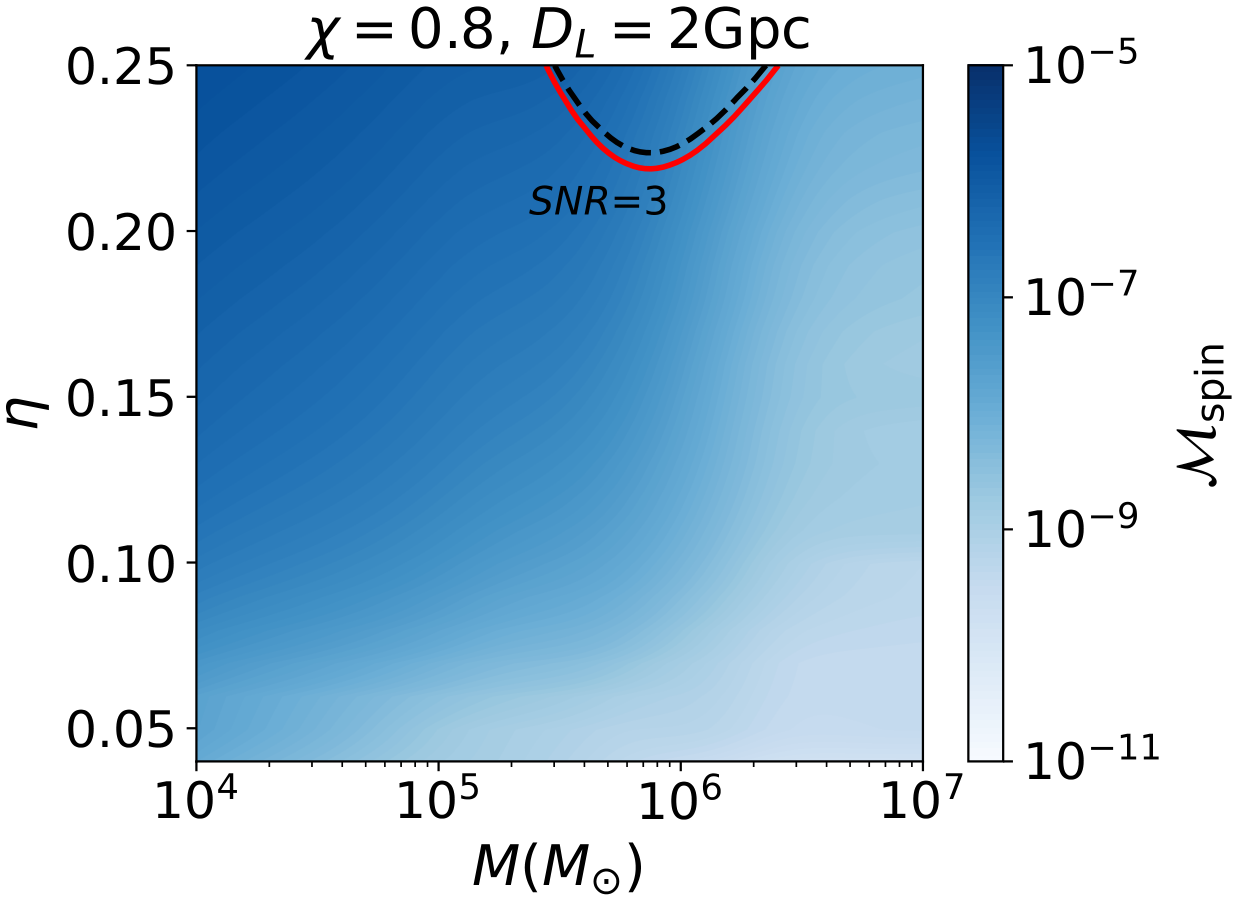}}
\caption{The dependence of the mismatch on \ac{GW} source parameters. The luminosity distance is fixed at $D_{\text{L}}=2$ Gpc. The red contours show where mismatch equals the threshold and the black contours show where \ac{SNR}=3.}
\label{etamassmismatch}
\end{figure}

The dependence of the mismatch on some source parameters has been plotted in FIG. \ref{etamassmismatch}.
The contour of the memory effect with \ac{SNR}=3 has also been plotted.
One can see that the contour of the memory effect with \ac{SNR}=3 always lies above the contour of the mismatch threshold.
This indicates that if the memory effect has an \ac{SNR} of no less than 3, then neglecting the memory effect will introduce systematic errors.
It should be noted that both the mismatch and its threshold in (\ref{mismatch}) are rough estimations.
It has been suggested that the threshold is often too conservative, and when it is violated, biases do not necessarily appear in parameter estimations \cite{Pompili:2023tna,Ossokine:2020kjp}.

Detailed study of parameter estimation with the memory effect shows that the memory effect has very limited impact on the estimation of many parameters, except for the inclination angle and luminosity distance  \cite{Sun:2024nut}.
The memory effect can help to break the degeneracy between the inclination angle and luminosity distance.
By calculating the Bayes factor, it has been found that an \ac{SNR} of approximately 2.36 is sufficient for TianQin to claim the detection of the memory effect.
The calculation also shows that the \ac{SNR} threshold derived from the Bayes factor is close to that derived from the mismatch threshold.

\subsubsection{Kerr hypothesis}\label{subsec:kerr}

{\it Subsection coordinator: Changfu Shi}

The Kerr metric \cite{Kerr:1963ud} is a two-parameter solution of Einstein's equations that describes stationary and rotating black holes.
Astronomical observations have also found ultra-compact objects called black holes, which we {intentionally} call astrophysical black holes so as to distinguish them from the black holes predicted in theory.
Astrophysical black holes are believed to be nearly neutral due to several charge loss and neutralization mechanisms \cite{Gibbons:1975kk,Goldreich:1969sb,Ruderman:1975ju,Blandford:1977ds}, see subsection \ref{sec:3b:ceoc} for more discussion, so one can mainly focus on neutral black holes.
What's remarkable is that, despite their different sizes, masses and surrounding environments, astrophysical black holes are believed to be fully described by the Kerr metric.
This is called the Kerr hypothesis.
In the words of Subrahmanyan Chandrasekhar \cite{Chandrasekhar:1975rye}: {\it ``In my entire scientific life, extending over forty-five years, the most shattering experience has been the realization that an exact solution of Einstein's equations of \ac{GR}, discovered by the New Zealand mathematician, Roy Kerr, provides the absolutely exact representation of untold numbers of massive black holes that populate the universe."}

Although there is no general proof, the theoretical basis for the Kerr hypothesis has been accumulating through various versions of uniqueness and no-hair theorems \cite{Israel:1967wq,Carter:1971zc,Carter:1997im,Robinson:1975bv,Robinson:2004zz,Bekenstein:1996pn,Chrusciel:2012jk}.
For example, a primitive no-hair theorem states that any isolated static black hole in \ac{GR} is necessarily a Schwarzschild black hole, i.e., a Kerr black hole without spin.
This has been generalized by releasing the ``isolated" condition to include black holes in astrophysical environments \cite{Gurlebeck:2015xpa}.
Further generalizations include showing that the tidal Love number of a Kerr black hole is zero \cite{LeTiec:2020spy,LeTiec:2020bos,Chia:2020yla,Charalambous:2021kcz,Charalambous:2021mea}, indicating that a Kerr black hole under the influence of dynamical external tidal force is still a Kerr black hole.

There have been attempts to challenge the Kerr hypothesis \cite{Herdeiro:2015waa,Herdeiro:2022yle,Cardoso:2016ryw}.
But there are very high {standards} for an alternative black hole model to pass \cite{Herdeiro:2022yle}:
\begin{itemize}
    \item Need to be delectably different from the Kerr black hole;
    \item Need to appear in a theory that is no less well motivated or self-consistent than \ac{GR};
    \item Need to be able to universally replace the Kerr black hole as the natural end product of various black hole formation mechanisms, such as {stars' gravitational collapse};
    \item Need to be stable or stable enough so that humans cannot tell the difference {with a reasonable observational time span};
    \item Need to be able to describe black holes with all masses with a single model.
\end{itemize}
So far there appears to be no known model that can satisfy all these conditions. For more discussions of these issues, we refer to \cite{Cardoso:2016ryw}.

Some theories predict the existence of alternative black hole models with particular masses due to the existence of new fundamental fields \cite{Herdeiro:2015waa,Herdeiro:2022yle,Cardoso:2016ryw, Xu:2022frb}. For such cases, a test of the Kerr hypothesis will not only help test \ac{GR}, but also help reveal the possible presence of new fundamental fields at certain mass scales. What's more, in order to properly test the Kerr hypothesis, it is necessary to look at black holes with vastly different masses. In this regard, it is desirable to consider the joint results from TianQin and \ac{GW} detectors targeting {at} other {frequency} bands, such as the third generation ground-based detectors \cite{Evans:2021gyd,Maggiore:2019uih,Branchesi:2023mws}.

High-precision tests of the Kerr hypothesis can be conducted with \acp{GW} through different approaches.
We will discuss two approaches here: detecting the ringdown signal of a black hole \cite{Dreyer:2003bv} and measuring the multi-pole moment of a black hole \cite{Ryan:1995wh}.

\paragraph{Testing the Kerr hypothesis with ringdown signals}~

After the merger of a binary black hole system, the remnant transits from a highly perturbed state to a perfect Kerr black hole, emitting a ringdown signal that damps over time.
The ringdown signal can be modelled via \ac{BHPT} with a series of \acp{QNM}.
If \ac{GR} and the Kerr hypothesis is valid, the oscillation frequencies and damping times of the \acp{QNM} are entirely determined by the mass and spin of the final Kerr black hole.
One can test the Kerr hypothesis by measuring the frequency and damping time of the dominant \ac{QNM}, as well as the frequency or damping time of one of the subdominant modes.
The ringdown waveform can be expanded as in (\ref{sec2.2:expansion-h-ringdown}). To test the Kerr hypothesis, the oscillation frequency $\omega_{lmn}$ and the ringdown times {$\tau_{lmn}$} can be parameterized as:
\bea
	\label{eq:para_qnm}
	&\omega_{lmn}=\omega_{lmn}^{GR}(1+\delta\omega_{lmn}),\nn\\
	&\tau_{lmn}=\tau_{lmn}^{GR}(1+\delta\tau_{lmn})\, ,
\eea
where the deviation parameters $\delta\omega_{lmn}=\delta\tau_{lmn}=0\,$ if \ac{GR} is correct. Due to the lack of analytic result on the relation between $\omega_{lmn}$, {$\tau_{lmn}$} and the black hole parameters, all the deviation parameters $\delta\omega_{lmn}$ and $\delta\tau_{lmn}$ are usually treated as independent parameters.
In the following, we will use $w_{lm}\equiv w_{lm0}$ and $\tau_{lm}\equiv \tau_{lm0}\,$.

For ground-based \ac{GW} detectors, there is a claim that at least one overtone exists in the event \ac{GW}150914 \cite{Isi:2019aib}.
Using Bayesian analysis, the authors found that $\delta f_{221}=-0.05\pm 0.2$ (68\% credible intervals), which establishes agreement with the Kerr hypothesis ($\delta f_{221}=0$) at the 20\% level.
However, due to the limited sensitivity of current ground-based detectors, {as well as the detailed treatments in data analysis (including the sampling rate and detector noise estimation, etc.),} the existence of the overtone is still being debated \cite{Carullo:2019flw, Wang:2023mst, Wang:2024yhb}.

Future space-based \ac{GW} detectors and the third-generation ground-based detectors are expected to achieve high \acp{SNR} for some \ac{GW} signals and can detect a series of higher-order \acp{QNM}.
The first quantitative assessment of the capability of LISA resolving three sub-leading \acp{QNM} has been carried out in  \cite{Berti:2005ys}.
The number \ac{GW} events with which the space-based and third generation ground-based detectors can resolve at least two modes has been estimated in \cite{Berti:2016lat}.
The prospect of eLISA \cite{Amaro-Seoane:2012aqc} and \ac{ET} \cite{Punturo:2010zz} to test the Kerr hypothesis has been studied in \cite{Gossan:2011ha}.
The effect of the ringdown \ac{SNR} and the amplitude ratio on the detectability of the subdominant modes has been investigated in \cite{Bhagwat:2019bwv}. {Tests of charged black holes for current \ac{GW} events and prospects for future ground-based detectors are given in \cite{Gu:2023eaa}.}

\begin{figure}[htbp!]
\centering
\graphicspath{{figs/}}
\subfigure{\includegraphics[scale=0.2]{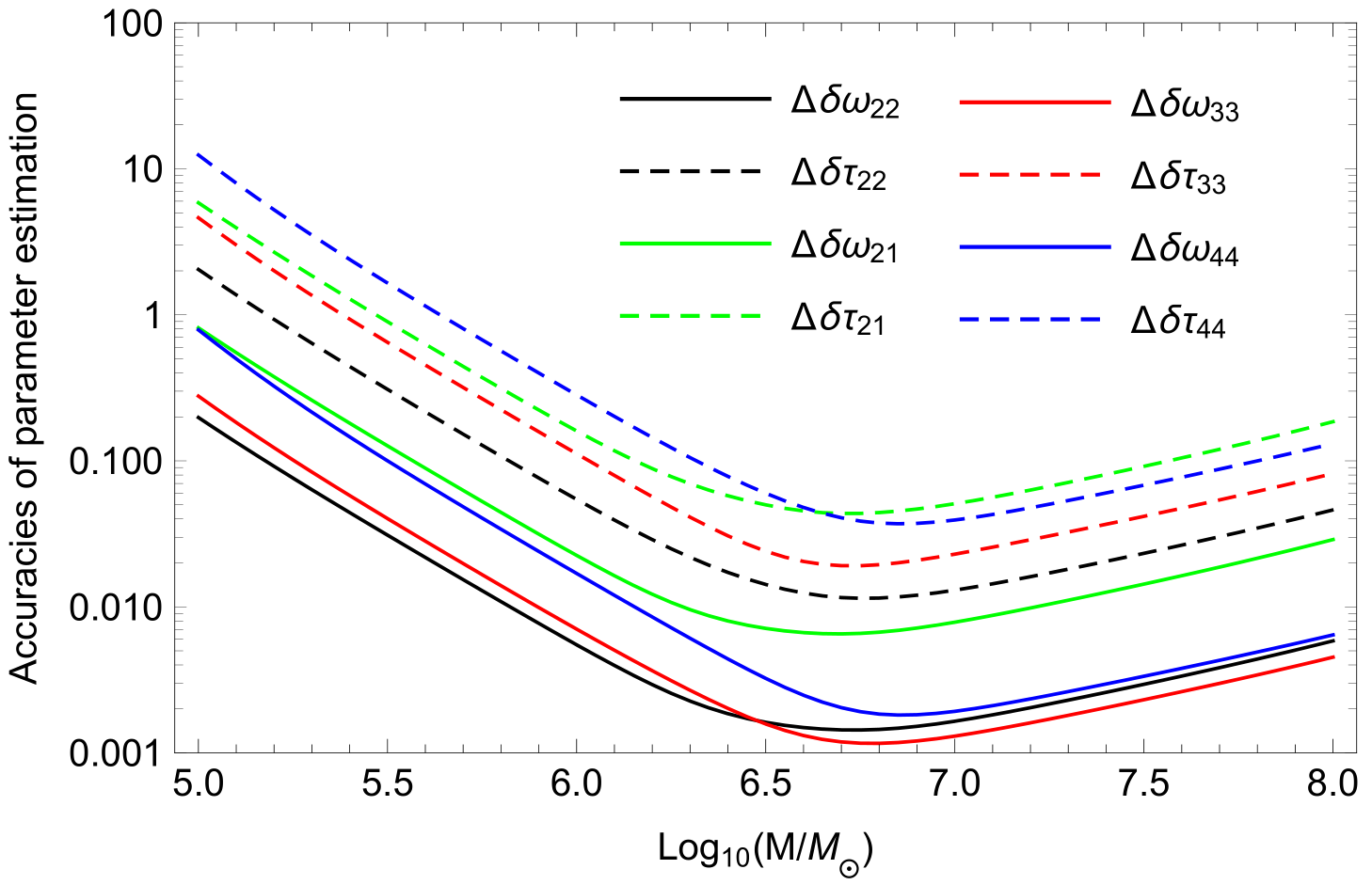}}
~~
\subfigure{\includegraphics[scale=0.2]{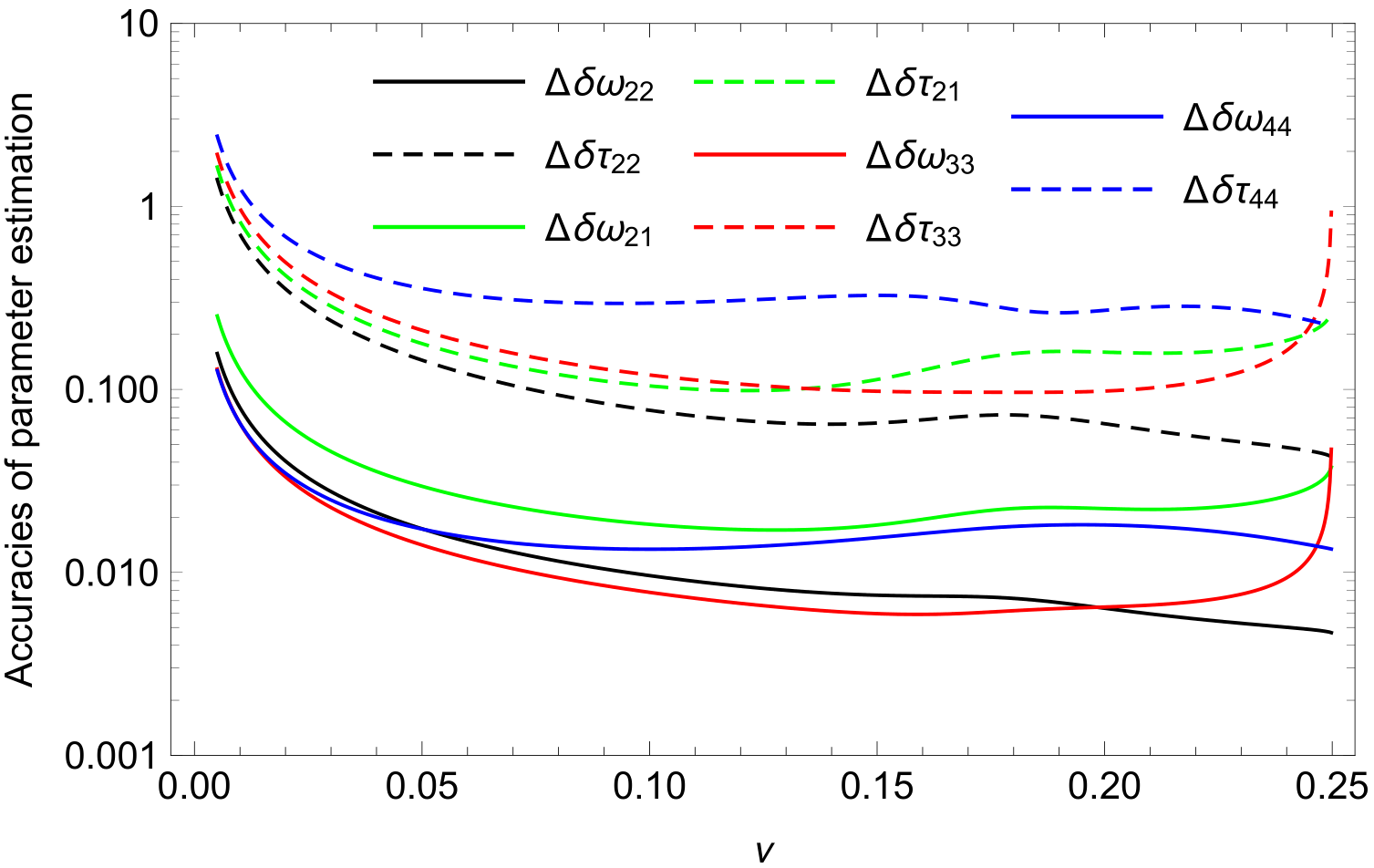}}
\caption{Projected constraints on various GR deviation parameters. For the left plot, the symmetric mass ratio is fixed at $\nu=2/9$. For the right plot, the black hole mass is fixed at $M=10^6\mSun$. Other parameters used in the plot are: $D_L=15$ Gpc, $j=0.76$, and $\chi_{eff}=-0.3$.}
\label{fig:Allmode1}
\end{figure}

The prospect of using TianQin to test the Kerr hypothesis has been studied in \cite{Shi:2019hqa}.
The calculation is based on an \ac{FIM} analysis of the 16-dimensional ringdown signal defined in (\ref{sec2.2:expansion-h-ringdown}) together the sensitivity of TianQin given in \cite{Luo2016}.
The effect of various source parameters on the projected dependence of the constraints on various deviation {parameters} is illustrated in FIG. \ref{fig:Allmode1}.
Among all the damping time related parameters, $\delta \tau_{22}$ is the best constrained, so $\delta\tau_{22}$ and $\delta\omega_{22}$ from the dominant mode can always be chosen as a pair to infer the black hole mass and spin predicted by \ac{GR}.
Among all the other six parameters, $\delta w_{33}$ is the best constrained in the vast majority of cases.
Thus, the combination of $\delta\omega_{22}$, $\delta\tau_{22}$ and $\delta\omega_{33}$ offers the most stringent {test} for vast majority of cases.

The left plot of FIG. \ref{fig:Allmode1} shows that the black hole mass affects the constraints in two ways: through the amplitude and through the detectable frequency range.
For masses below about $2\times10^6\mSun$, the amplitude grows as the mass increases and so the constraints {become} stronger for larger masses.
But when the black hole mass goes above about $4\times10^6\mSun$, the main part of the \ac{GW} signal starts to move away from the sensitivity band of TianQin and the constraints start to get worse as the black hole mass grows.
From the right plot of FIG. \ref{fig:Allmode1}, one can see that the constraints become worse as symmetric mass ratio decreases.
This is because the radiated energy becomes smaller for bigger mass ratio, given a fixed black hole mass.
The amplitude of $(3,3)$ mode trends to zero when the symmetric mass-ratio trends to $1/4$, and so the $(3,3)$ mode will be the worst resolved as the black hole masses become equal.
So one should turn to use $\delta\omega_{21}$ or $\delta\omega_{44}$ to do the test for such cases.

TianQin has the potential to detect many \ac{MBHB} events during its five years of operation time \cite{Wang:2019ryf}.
The constraints on the \ac{GR} deviation parameters can be significantly improved by stacking multiple ringdown signals.
The combined constraints from all the \ac{MBHB} events detectable by TianQin are estimated in Table \ref{tablemr}.
Three astrophyiscal population models, pop III, Q3\_d and Q3\_nod, have been used.
One thousand sets of data for each population model have been produced and the results are the average over these data sets.
The \ac{SNR} threshold is set at 8.
One can see that the selected deviation parameters can always be constrained to the 1.5\% level or better, and some can even reach the $\cO(10^{-4})$ level.

\begin{table}[!htbp]
\caption{Combined constraints on selected GR deviation parameters.}
\label{tablemr}
\begin{tabular}{|c|c|c|c|c|}
\hline
Cases   &Detection number   & $\Delta\delta\omega_{22}$& $\Delta\delta\tau_{22}$& $\Delta\delta\omega_{33}$\\
\hline
pop III &51.7               &$0.0023\pm0.0014$&$0.015\pm0.0092$&$0.0029\pm0.0019$\\
\hline	
Q3\_d   &17.7               &$0.00080\pm0.00041$&$0.0052\pm0.0027$&$0.0014\pm0.00096$\\
\hline
Q3\_nod &274.9              &$0.00044\pm0.00017$&$0.0027\pm0.0011$&$0.00041\pm0.00021$\\
\hline
\end{tabular}
\end{table}

\paragraph{Testing the Kerr hypothesis by measuring the quadrupole moment of black holes}~

For an isolated massive object, its gravitational field can be characterized by summing a series of multipole moments.
In the case of Kerr black hole, its higher multipole moments are fully determined by its mass $M$ and spin $S$:
\bea	
\cM_l+i\cS_l=M(ia)^l\,,\quad l=0,1,2,\cdots\,,
\label{eq:moment}
\eea
where $a\equiv S/M$ is the spin parameter, and $\cM_l$ and $\cS_l$ are the mass and current moments, respectively, with $\cM_0=M$ and $\cS_1=Ma=S$.
Note $\cM_{2m+1}=\cS_{2m}=0\,,\;m=0,1,2,\cdots\,$.

To test the Kerr hypothesis, one can measure one more higher multipole moment with $l\geq2$, in addition to measuring the mass and angular momentum of the black hole.
Although any higher multipole moment can be used, the quadrupole moment is usually the best measured.
So the test of the Kerr hypothesis is mostly done by examining the consistency among the mass, spin, and the quadrupole moment of a black hole.
One can parameterize such test by treating the quadrupole moment as an additional parameter, $Q=-(1+\delta\kappa) a^2 M$, where $\delta\kappa$ is a deviation parameter depending on the internal structure of the object.
For the Kerr black holes in \ac{GR}, $\delta\kappa=0$.
For neutron stars, $\delta\kappa$ can vary between 1--13 \cite{Pappas:2012ns}.
For Boson stars, the range of $\delta\kappa$ is about 10 to 150 \cite{Herdeiro:2014goa,Baumann:2018vus}.
For some other black hole mimickers such as gravastars, the value can be negative.
The presence of $\delta\kappa$ introduces modifications to the  \ac{GW} waveforms.

Due the different method used to calculate the waveforms, the subsequent discussion will be divided into two cases: the low mass ratio systems with $q=m_1/m_2\in (1,15)$ and the large mass ratio systems with $q>10000$.

For the low mass ratio systems, the phase deformation introduced by $\delta\kappa$ for the inspiral stage can be calculated by using the \ac{PN} approximation.
As pointed by Krishnendu \cite{Krishnendu:2017shb}, the leading order correction introduced by $\delta\kappa$ occurs at the 2PN order, with the phase correction:
\bea
	\label{eq:quadru_phase}
	\delta\Psi=\frac{75}{64}\frac{\delta\kappa_1 a_1^2+\delta\kappa_2 a_2^2}{m_1m_2}(\pi M_{tot}f)^{-1/3}.
\eea
Here $a_1$ and $a_2$ are the spin parameters of the two components of the binary system, $m_1$ and $m_2$ are their masses, and $\delta\kappa_1$ and $\delta\kappa_2$ are their deviation parameters.

For ground-based detectors, it has been found that \acp{ECO} are less supported by the data than black holes, $\delta\kappa_1=\delta\kappa_2=\delta\kappa$ is constrained to the level of $\cO(10^2)$, by using the events in GWTC-2 \cite{LIGOScientific:2020ibl,LIGOScientific:2020tif}.
Recent work has also analysed the impact of spin precession and higher modes on the measurement of $\delta\kappa$ with ground-based detectors \cite{Divyajyoti:2023izl}.
The combined Bayesian factor {of} the GWTC events is calculated, $\log^{Kerr}_{\delta\kappa\neq0}=0.9$  in GWTC-3 \cite{KAGRA:2021vkt} and 1.1 in GWTC-2 \cite{LIGOScientific:2020ibl,LIGOScientific:2020tif}.
By using the so-called PSI waveform template constructed with arbitrarily axisymmetric metric \cite{Li:2022fml,Li:2023qcu}, the deviation from the Kerr black hole has been constrained by two LVK events \cite{Li:2023zbm}.
For space-based detectors,  $\delta\kappa$ is expected to be constrained to the level of $\cO(0.1)$ by a sub-population of binary black hole events detectable by LISA and DECIGO \cite{Krishnendu:2019ebd}.

\begin{figure}[!htbp]
	\centering
	\graphicspath{{figs/}}
	\includegraphics[scale=0.25]{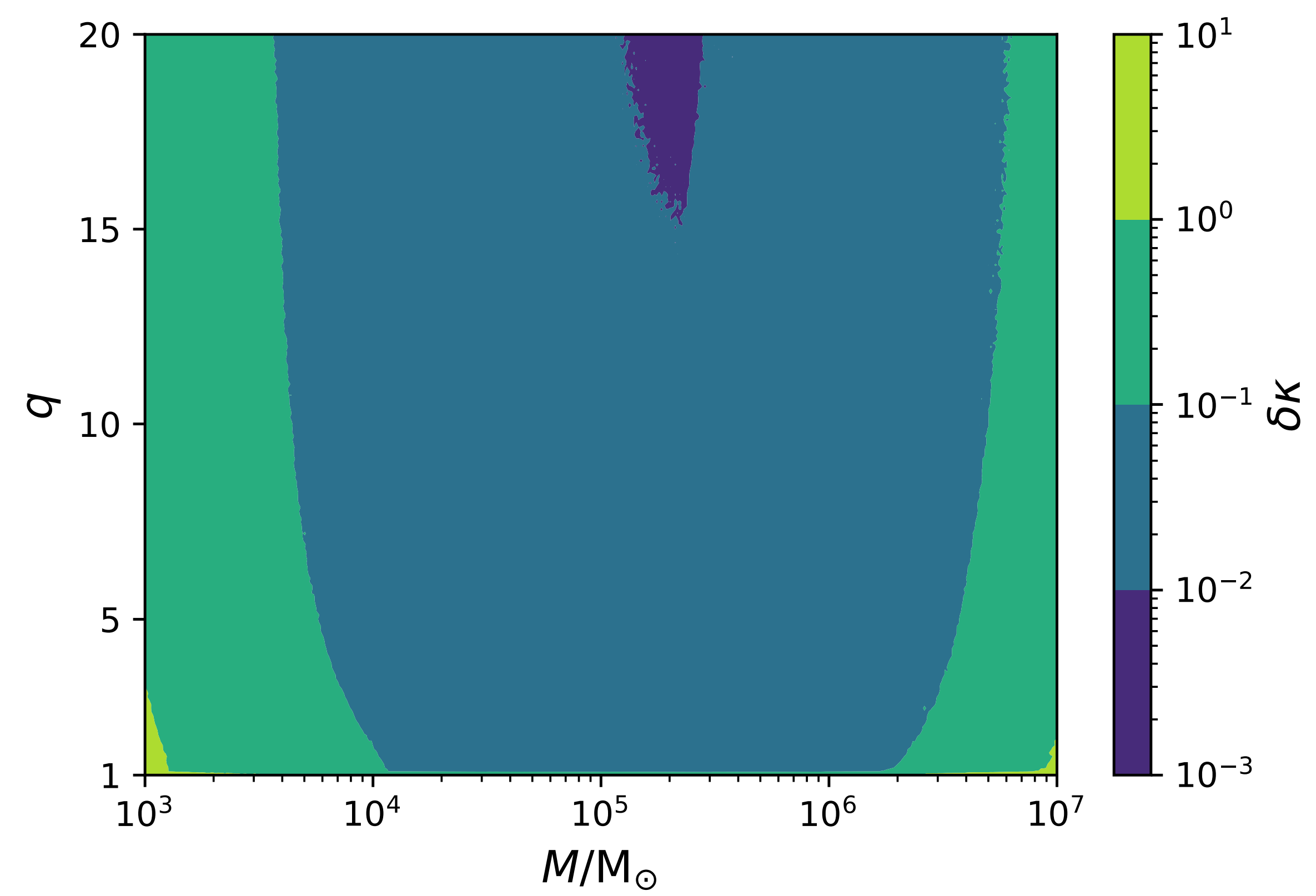}
	\caption{Expected constraints on $\delta\kappa$ with TianQin and \acp{MBHB}.}
 \label{fig:quadru_bbh}
\end{figure}

The capability of TianQin in using \acp{MBHB} to constrain $\delta\kappa$ is shown in FIG. \ref{fig:quadru_bbh}.
All the events are normalized to \ac{SNR}=5000 by changing the luminosity distance.
One can see that the best constraint on $\delta\kappa$  can be achieved for source masses around $10^{5.5}\mSun$, to the level $\delta \kappa \sim \cO(10^{-12})$.
One can also see that higher mass ratio systems can give better constraint on $\delta\kappa$, indicating that \acp{EMRI} should have better constraining power than \acp{MBHB}.

For \acp{EMRI}, a non-zero $\delta\kappa$ will lead to corrections to the orbit of the stellar mass compact object, resulting in corrections to the radiated \ac{GW} waveforms.
Ryan has pioneered the work of using LISA to extract information on the Kerr multipole moments from \ac{EMRI} signals, assuming that the orbit of the stellar mass compact object is circular on the equatorial plane \cite{Ryan:1995wh,Ryan:1997hg}.
When calculating the \ac{EMRI} waveform using the analytic kludge method \cite{Barack:2003fp},  the leading order $\delta\kappa$ correction to the \ac{PN} orbit of the stellar mass object is \cite{Barack:2006pq}:
\bea
\label{eq:quadru_emri}
\delta\frac{d\nu}{dt}&=&-(2\pi M\nu)\delta\kappa\frac{a^2}{M^2}(1-e^2)^{-1}{\frac{33}{16}+\frac{359}{32}e^2-\frac{527}{96}\sin^2 \lambda}\,,\nn\\
\delta\frac{d\gamma}{dt}&=&-\frac{3}{2}\nu\delta\kappa\frac{a^2}{M^2}(2\pi M\nu)^{4/3}(1-e^2)^{-2}(5\cos\lambda-1)\,,\nn\\
\delta\frac{d \alpha}{dt}&=&3\pi \nu \delta\kappa\frac{a^2}{M^2} (2\pi M\nu)^{4/3}(1-e^2)^{-2}\cos\lambda\,,
\eea
where $M$ and $a$ are the mass and rotation parameter of the central black hole, respectively, and the definition of other parameters can be found in \cite{Barack:2003fp}.
By modifying the analytic kludge \ac{EMRI} waveform with a quadrupole moment correction, $Q=-Ma^2 +\Delta Q$, it has been found that LISA can constrain $\Delta\cQ\equiv\Delta Q/M^3=-\delta\kappa a^2/M^2$ to the level $\cO(10^{-4})$ \cite{Barack:2006pq}.
A more detail studied on how LISA can constrain the non-Kerr quadrupole moment by using 12 \acp{EMRI} population models has also been performed \cite{Babak:2017tow}.

\begin{figure}[htbp!]
	\centering
	\graphicspath{{figs/}}
	\includegraphics[scale=0.35]{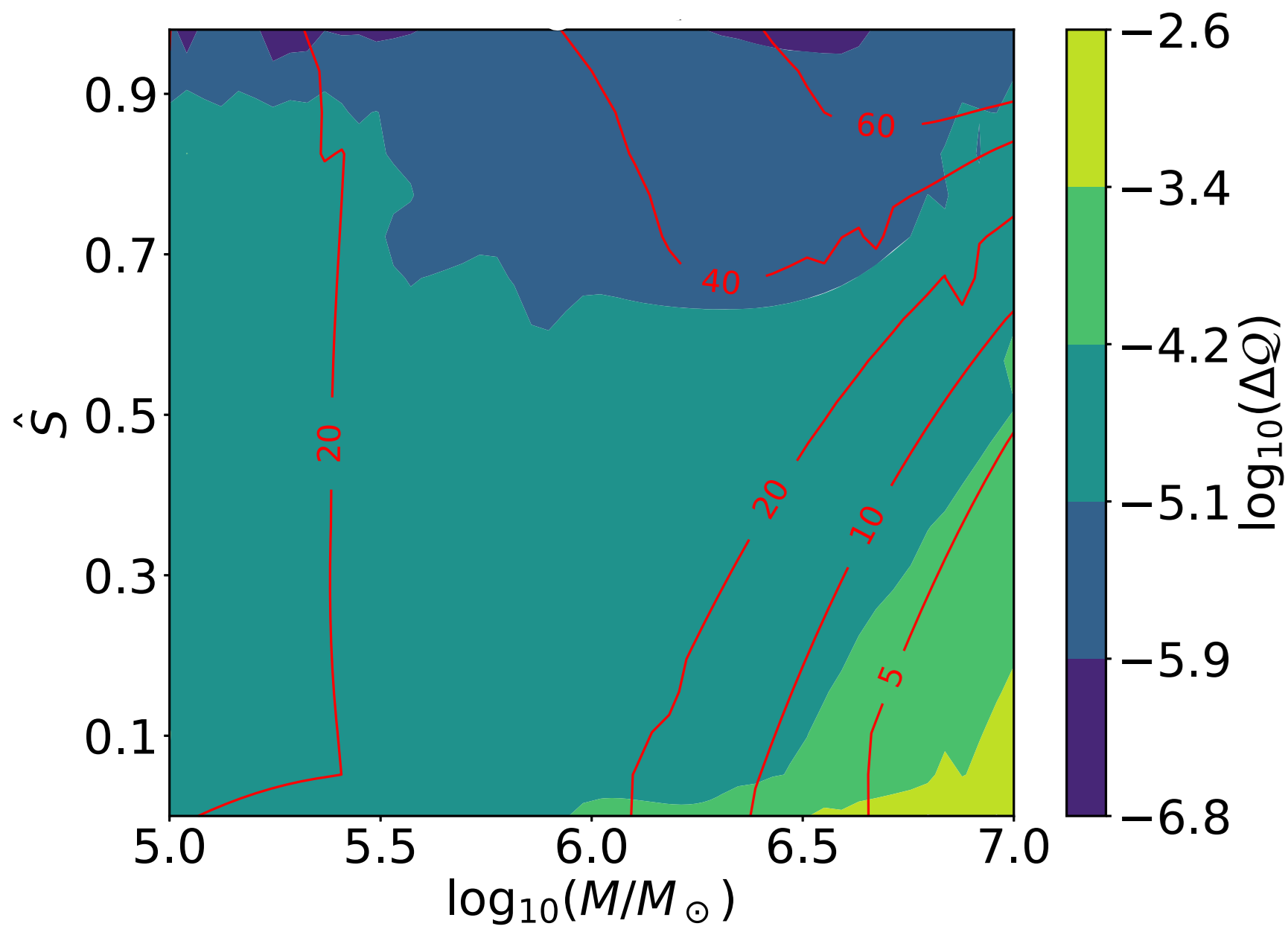}
	\caption{Expected constraints on $\Delta\cQ$ with TianQin and EMRIs (This figure is from the FIG. 1 of \cite{Zi:2021pdp}).}
 \label{fig:quadru_emri}
\end{figure}

The prospect of using TianQin to measure the Kerr quadrupole moment through \acp{EMRI} has been studied in \cite{Zi:2021pdp}.
The main result is illustrated in FIG. \ref{fig:quadru_emri}.
One can see that TianQin can constrain $\Delta\cQ$ to the order $\cO(10^{-6})$ using \acp{EMRI}.
Comparing to the mass, the spin of the central black hole has a more significant impact on the constraints on $\Delta\cQ$, and the larger the spin, the stronger the constraints.

\subsection{Possible signatures of beyond GR effects}

Theoretically, the difference between real gravity and \ac{GR} is encoded in the different ways that one can deviate from \ac{GR}. In this regard, there are three basic questions to ask:
\begin{itemize}
\item How is matter affected by gravity?
\item How is gravity sourced by matter?
\item What is the self-interaction of {the gravitational field}?
\end{itemize}

The anchor point for the first question is \ac{EEP}, which requires \ac{WEP}, \ac{LLI} and \ac{LPI} to be valid for all non-gravitational experiments.
\ac{EEP} mandates the minimal coupling between gravity and matter and it seems that the only theories of gravity that can fully embody EEP are the ``metric theories of gravity", in which the minimal coupling to a symmetric metric is the sole origin of all gravitational interactions {with} matter.
The \ac{DI} is also implied in all these theories.
For more discussions on \ac{EEP}, we refer to  \cite{Will:2014kxa}.
\ac{EEP} does not say anything about how gravity should be sourced by matter or about the self-interaction of the metric, and it also allows the metric to couple to extra gravitational fields that do not directly interact with the matter {in the Jordan frame}. These are exactly the subject of the second and third questions.

For the second question, one usually assumes that the stress energy tensor of matter, $T_{\mu\nu}$, is covariantly conserved and its contribution to Einstein's equations is linear. So a deviation from \ac{GR} is either to give up the covariant conservation of $T_{\mu\nu}$ or to have non-linear contributions of $T_{\mu\nu}$ to Einstein's equations. An example of this can be found in \cite{Pani:2013qfa}.

The anchor point for the third question is two pronged, including \ac{SEP} and the Lovelock theorem. \ac{SEP} requires \ac{WEP}, \ac{LLI} and \ac{LPI} to be valid for all experiments, including those involving significant gravitational self-energy \cite{Baessler:1999iv}. If \ac{SEP} is valid, then the metric is the only gravitational field in the universe \cite{Will:2014kxa}, but the self-interaction of the metric may not be as simple as \ac{GR}. The conditions for the particular type of metric self-interactions used in \ac{GR} are spelled out in the Lovelock theorem \cite{Lovelock:1971yv,Lovelock:1972vz,Sotiriou:2014yhm}, which can be phrased as: {\it In four spacetime dimensions the Einstein tensor and the metric are the only two rank-2 and divergence free tensors constructed solely in terms of the metric and its first two derivatives.}

In view of these, possible deviations from \ac{GR} can come from the following directions, each with gradually more dramatic modifications:
\begin{itemize}
\item Violating \ac{SEP}, such as by coupling the metric to extra gravitational fields.
\item Circumventing the Lovelock theorem \cite{Sotiriou:2014yhm,Berti:2015itd}, such as by involving higher derivatives or higher spacetime dimensions. The latter effectively leads to extra gravitational fields in four spacetime dimensions and can be seen as a violation of \ac{SEP}.
\item Violating \ac{EEP}, such as by violating \ac{WEP} with non-metric theories of gravity or non-minimal coupling between the  metric and matter, violating \ac{LPI} with position dependent coupling and physical constants, or violating \ac{LLI} with a preferred foliation. Since \ac{EEP} implies the \ac{DI}, violating the latter can also be viewed as a violation of \ac{EEP}. Giving the graviton a mass is an example in this direction.
\item Introducing nonlinear contributions of $T_{\mu\nu}$ to Einstein's equations or considering matter contributions that are not covariantly conserved.
\end{itemize}
Some example \acp{MGT} are listed in Table \ref{tab:sec2.B:MGs1} to illustrate the different ways to deviate from \ac{GR}.
The table is by no means complete and more comprehensive reviews can be found in, e.g., \cite{Berti:2015itd,Clifton:2011jh}.

\begin{table}[t]
\begin{center}
\caption{Different ways to deviate from GR.
Blank spaces mean that there is no deviation from \ac{GR} in the corresponding direction.
(``Extra field": having extra gravitational field that may or may not couple to matter directly;
``Higher derivative": involving higher derivatives of the metric;
S: scalar;
$\Gamma$: extra symmetric connection;
V: vector;
T: tensor;
$R_{\rm~GB}^2$: Gauss-Bonnet scalar;
$^\ast RR$: Chern-Simons term;
NC: Noncommutative gravity;
NL: $T_{\mu\nu}$ contributes nonlinearly to the right hand side of Einstein's equations).
Note the reference metric in the dRGT theory is not dynamical. A bimetric theory of gravity can be constructed if the reference metric is made dynamical \cite{Hassan:2011zd}.}
\label{tab:sec2.B:MGs1}
\begin{tabular}{|c|c|c|c|c|}
\hline
\multirow{2}{*}{Theory} & SEP & Lovelock Theorem & \multirow{2}{*}{EEP} & \multirow{2}{*}{$T_{\mu\nu}$} \\
\cline{2-3}
                        & Extra Field & Higher derivative & & \\
\hline
Scalar-Tensor \cite{Bergmann:1968ve,Wagoner:1970vr}
                        & S & & & \\
Metric $f(R)$ \cite{Schmidt:2006jt}
                        & S & & & \\
Horndeski \cite{Horndeski:1974wa}
                        & S & & & \\
Multiscalar \cite{Damour:1992we}
                        & Multiple-S & & & \\
\hline
Palatini $f(R)$ \cite{Olmo:2011uz}
                        & $\Gamma$ & Various powers of $R$ & & \\
EdGB \cite{Kanti:1995vq}
                        & S & $R_{\rm~GB}^2$ & & \\
dCS \cite{Alexander:2009tp}
                        & S & $^\ast RR$ & & \\
Generic Quadratic \cite{Yunes:2011we,Pani:2011gy}
                        & S & $R^2$, $R_{\mu\nu}^2$, $R_{\mu\nu\rho\sigma}^2$, $^\ast RR$ & & \\
\hline
$G(t)$ \cite{Dirac:1937ti}
                        & & & LPI violation & \\
Einstein Aether \cite{Jacobson:2007veq}
                        & V & & LLI violation & \\
Khronometric \cite{Blas:2009qj,Jacobson:2010mx}
                        & S & & LLI violation & \\
Ho$\check{\rm~r}$ava-Lifshitz \cite{Horava:2009uw}
                        & & & LLI violation & \\
$n$-DBI \cite{Herdeiro:2011im}
                        & V & & LLI violation & \\
SME \cite{Kostelecky:2003fs}
                        & T & & LLI violation & \\
\hline
dRGT \cite{deRham:2010ik,deRham:2010kj}
                        & T & & DI violation & \\
Galileon \cite{Nicolis:2008in,Deffayet:2009wt}
                        & S & & DI violation & \\
NC \cite{Calmet:2005qm,Calmet:2006iz,Mukherjee:2006nd,Kobakhidze:2016cqh}
                        & & & DI violation & \\
\hline
PSV \cite{Pani:2013qfa}
         & & & & NL \\
\hline
\end{tabular}
\end{center}
\end{table}

\begin{table}[!htbp]
\begin{center}
\caption{Different predictions on \acp{GW} from some example \acp{MGT}. For propagation, only the results for the $+,\times$ modes at the {\it low energy and flat background} are given, and the references are the same as those for polarization. Predictions with other conditions, such as from a cosmological background could be significantly different, see \cite{Baker:2017hug,Creminelli:2017sry,Sakstein:2017xjx,Ezquiaga:2017ekz} for examples. For results that are too involved, only a reference will be given. A box left blank means that the corresponding result is not known to the authors. For generation, we only present results where the \ac{ppE} parameterization is known \cite{Yunes:2009ke,Tahura:2018zuq}.}
\label{tab:sec2.B:MGs2}
\begin{tabular}{|c|c|c|c|c|c|}
\hline
\multirow{2}{*}{Theory} & \multirow{2}{*}{Polarization} & \multicolumn{3}{c|}{Propagation} & \multirow{2}{*}{Generation} \\
\cline{3-5}
                        & & Speed & Mass & Dispersion & \\
\hline
GR                      & $+,\times$ & 1 & 0 & $w=kv_{+,\times}$ & IMR models \\
\hline
Scalar-Tensor \cite{Bergmann:1968ve,Wagoner:1970vr}
                        & $+,\times,b,l$ \cite{Liang:2017ahj,RizwanaKausar:2016zgi}
                        & 1 & 0 & $w=kv_{+,\times}$ & \ac{ppE} \cite{Zhang:2017sym,Zhang:2017srh} \\
Metric $f(R)$ \cite{Schmidt:2006jt}
                        & $+,\times,b,l$ \cite{Liang:2017ahj}
                        & 1 & 0 & $w=kv_{+,\times}$ & \ac{ppE} \cite{Yunes:2009ke} \\
Horndeski \cite{Horndeski:1974wa}
                        & $+,\times,b,l$ \cite{Hou:2017bqj}
                        & 1 & 0 & $w=kv_{+,\times}$ & \\
Generalized Proca \cite{Heisenberg:2014rta}
			& $+, \times, b, l, x, y$ \cite{Dong:2023xyb} & $(1-\frac{G_{4,X} A^{2}}{G_{4}})^{-1/2}$ & 0 & $w = kv_{+, \times}$ &  \\
\cline{2-5}
General Einstein-Vector \cite{Geng:2015kvs}
			& \multicolumn{4}{c|}{\cite{Lai:2024fza}}&  \\
Most general Scalar-Tensor
			& \multicolumn{4}{c|}{\cite{Dong:2023bgt}}&  \\
Most general second-order Vector-Tensor
			& \multicolumn{4}{c|}{\cite{Dong:2024zal}}&  \\
\cline{2-5}
Multiscalar \cite{Damour:1992we}
			& \multicolumn{4}{c|}{ }&   \\
\hline
Palatini $f(R)$ \cite{Olmo:2011uz}
                        & $+,\times$ \cite{RizwanaKausar:2016zgi}
                        & 1 & 0 & $w=kv_{+,\times}$ & \\
$f(T)$ \cite{Hehl:1976kj,Hayashi:1979qx}& $+, \times$ \cite{Bamba:2013ooa}  & 1 & 0 & $w = kv_{+, \times}$ &  \\
$f(Q)$ \cite{BeltranJimenez:2017tkd,Capozziello:2022zzh} & $+, \times$ \cite{Capozziello:2024vix}  & 1 & 0 & $w = kv_{+, \times}$ &  \\
Palatini-GBD & $+, \times, b$ \cite{Lu:2020eux}  & 1 & 0 & $w = kv_{+, \times}$ &  \\
Palatini-Horndeski
			& $+, \times, b, l$ \cite{Dong:2021jtd} & 1 & 0 & $w = kv_{+, \times}$ &  \\
\cline{2-5}
Horndeski-teleparallel  & \multicolumn{4}{c|}{ \cite{Bahamonde:2021dqn}} & \\
\hline
EdGB \cite{Kanti:1995vq}
                        & $+,\times$ \cite{Wagle:2019mdq}
                        & & & & \ac{ppE} \cite{Yagi:2011xp} \\
dCS \cite{Alexander:2009tp}
                        & $+,\times$ \cite{Wagle:2019mdq,Li:2022grj}
                        & 1 & 0 & & \ac{ppE} \cite{Yagi:2011xp,Zhao:2019xmm} \\
\cline{2-5}
Generic Quadratic \cite{Yunes:2011we,Pani:2011gy}
                        & \multicolumn{4}{c|}{ \cite{Alves:2023rxs}} & \ac{ppE} \cite{Yagi:2011xp} \\
Most general pure metric
			& \multicolumn{4}{c|}{\cite{Dong:2023bgt}}&   \\
\hline
$G(t)$ \cite{Dirac:1937ti}
                        &  &  &  &  &   \\
Einstein Aether \cite{Jacobson:2007veq}
                        & $+,\times,b,l,x,y$ \cite{Jacobson:2004ts,Gong:2018cgj}
                        & $(1-c_{13})^{-1/2}$ & 0 & $w=kv_{+,\times}$ & \ac{ppE} \cite{Hansen:2014ewa} \\
Khronometric \cite{Blas:2009qj,Jacobson:2010mx}
                        & $+,\times,b,l$ \cite{Hansen:2014ewa,Schumacher:2023jxq}
                        & $(1-\beta_{\rm~KG})^{-1/2}$ &  &  & \ac{ppE} \cite{Hansen:2014ewa} \\
Ho$\check{\rm~r}$ava-Lifshitz \cite{Horava:2009uw}
                        & $+,\times,b,l$ \cite{Gong:2018vbo}
                        & $(1+\beta)^{1/2}$ & 0 & $w=kv_{+,\times}$ & \ac{ppE} \cite{Zhao:2019xmm} \\
\cline{2-5}
Spatially covariant gravity\cite{Gao:2014fra}
			& \multicolumn{4}{c|}{\cite{Gao:2014fra,Gao:2019liu}}&  \\
\cline{2-5}
$n$-DBI \cite{Herdeiro:2011im}
                        & \multicolumn{4}{c|}{} & \\
\cline{2-5}
SME \cite{Kostelecky:2003fs}
                        & \multicolumn{4}{c|}{\cite{Hou:2024xbv} (For bumblebee {model}: \cite{Liang:2022hxd})} & \\
\hline
dRGT \cite{deRham:2010ik,deRham:2010kj}
                        & \multicolumn{4}{c|}{} & \\
Galileon \cite{Nicolis:2008in,Deffayet:2009wt}
                        & \multicolumn{4}{c|}{} & \\
\cline{2-5}
NC \cite{Calmet:2005qm,Calmet:2006iz,Mukherjee:2006nd,Kobakhidze:2016cqh}
                        & $+,\times$ \cite{Calmet:2006iz}
                        & & & & \ac{ppE} \cite{Tahura:2018zuq} \\
\cline{2-5}
Scalar-Tensor-Rastall
			& 	 \multicolumn{4}{c|}{\cite{Fan:2024pex}}& \\

\hline
PSV \cite{Pani:2013qfa}
                        & \multicolumn{4}{c|}{} & \\
\hline
\end{tabular}
\end{center}
\end{table}

As is obvious in FIG. \ref{fig:gr-exps}, the advantage of \acp{GW} is the capability to probe gravity in the strong field regime, and the advantage of a space-based detector, such as TianQin, over a ground-based detector is the capability to probe more massive sources. Heavier sources typically means stronger signals, which can allow for more details to be revealed. \acp{GW} predicted from \acp{MGT} will differ from that of \ac{GR} in a number of ways, including:
\begin{itemize}
\item Having different numbers of propagating degrees of freedom;
\item Having different propagation properties;
\item Having different dynamical production characteristics.
\end{itemize}
The predicted \ac{GW} properties from some example \acp{MGT} are listed in Table \ref{tab:sec2.B:MGs2}.

In this subsection, we discuss how TianQin can use all these features to search for possible signatures of beyond \ac{GR} effects.

\subsubsection{GW polarization}

{\it Subsection coordinator: Jian-dong Zhang}

In \ac{GR}, \acp{GW} possess only two tensor polarization modes.
But for a general metric theory of gravity, the metric tensor has 6 propagation d.o.f.,
thus there could exist 6 polarization modes {at most} \cite{Eardley:1973br,Eardley:1973zuo}.
More explicitly, there are two tensor modes, plus ($+$) and cross ($\times$),
two vector modes, x ($x$) and y ($y$), and two scalar modes, the transverse breathing ($b$) and longitudinal ($l$).
Then the general formula for a {spatial-spatial} metric perturbation can be written as:
\be
h_{ij}=\left(
\begin{matrix}
h_++h_b & h_\times & h_x\\
h_\times & -h_++h_b & h_y\\
h_x & h_y & h_l
\end{matrix}
\right)\,.
\ee

These additional polarization modes can be excited by the coupling between the metric and extra gravitational fields.
For example, in the massless Scalar-Tensor theory, \acp{GW} have an additional scalar breathing mode.
In a general Scalar-Tensor theory \cite{Will:2014kxa,Maggiore:1999wm,Capozziello:2006ra} and the $f(R)$ theory \cite{RizwanaKausar:2016zgi,Gong:2017bru,Katsuragawa:2019uto,Moretti:2019yhs}, \acp{GW} have both scalar modes: the breathing mode and the longitude mode.
Generation of additional polarizations of \acp{GW} from a binary system localized on a 3-brane space-time of odd dimensions, associated with the violation of the Huygens principle, and the possibility of registering one of them, the breathing mode, by an observer on the brane have been demonstrated in \cite{2022JCAP...04..014K}.
In many theories with massive scalar modes, there is mode mixing between the breathing and the longitude mode \cite{Liang:2017ahj,Hou:2017bqj,Gong:2018cgj}.
In the Einstein-Aether theory \cite{Gong:2018cgj,Zhang:2019iim}, TeVeS theory \cite{Sagi:2010ei}, bimetric theory \cite{dePaula:2004bc} and so on, all the six polarization modes could exist.
Different polarization modes have different propagation properties.
In the case when there is CPT violation, even the two tensor modes can have birefringence, i.e., they can form the left-handed and right-handed polarization modes that propagate differently \cite{Alexander:2009tp, Kostelecky:2016kfm, Shao:2020shv, Zhao:2019szi, Haegel:2022ymk, Califano:2023aji,ONeal-Ault:2021uwu,Wang:2017igw,Wang:2020pgu,Zhao:2022pun}.
Different types of polarization can also {possess} different masses, leading to different propagation speed.
More examples can be found in Table \ref{tab:sec2.B:MGs2}.

To detect the extra polarization modes, one can either use waveforms including the contribution of extra polarization modes for some specific theories \cite{Will:1994fb,Chatziioannou:2012rf,Sennett:2016klh,Liang:2022hxd}, or one can use a theory agnostic method, such as by using the ``null-stream'' method or using a parameterised waveform model.
The null-stream method relies on multiple detectors to cancel out the tensor modes in the data  \cite{Guersel:1989th,Wen:2005ui,Wen:2007pj,Chatterji:2006nh,Hagihara:2018azu,Hagihara:2019ihn,Hagihara:2019rny,2018Takeda,Hu:2023soi,Liang:2024sfn}.
So if one wants to detect the extra modes with a single detector such as TianQin, an assumption about the waveform is needed.

In general, the detected signal is a linear combination of the responses of all polarizations:
\be\label{eq:signal}
h(t)=\sum_P F_P h_P(t),
\ee
where $P$ stands for different polarization,
$h_P(t)$ is the waveform for each polarization,
and $F_P$ is the antenna pattern function \cite{Poisson:2014book}, which describes the response of the detector {to polarization $P$}.
In the detector frame, the antenna pattern function for each polarization is:
\bea
F_+&=& \frac{\sqrt{3}}{2} \left(\frac{1 + \cos^2\bar{\theta}}{2} \cos2\bar{\phi} \cos2\bar{\psi} - \cos\bar{\theta}\sin2\bar{\phi}  \sin2\bar{\psi} \right),\\
F_\times&=&\frac{\sqrt{3}}{2} \left(\frac{1 + \cos^2\bar{\theta}}{2} \cos2\bar{\phi} \sin2\bar{\psi} + \cos\bar{\theta}\sin2\bar{\phi}  \cos2\bar{\psi} \right),\\
F_x&=& -\frac{\sqrt{3}}{2} \sin\bar{\theta} \left(\cos\bar{\theta}\cos2\bar{\phi} \cos\bar{\psi} - \sin2\bar{\phi} \sin\bar{\psi} \right),\\
F_y&=& -\frac{\sqrt{3}}{2} \sin\bar{\theta} \left(\cos\bar{\theta}\cos2\bar{\phi} \sin\bar{\psi} + \sin2\bar{\phi} \cos\bar{\psi} \right),\\
F_b&=&-F_l= - \frac{\sqrt{3}}{4} \sin^2\bar{\theta}\cos2\bar{\phi}.\label{sec2:Fp}
\eea
where the angles $\bar{\theta}$, $\bar\phi$ and $\bar\psi$ are defined in the detector frame, and more details can be found in \cite{Xie:2022wkx}.
The last equality means that the response of the breathing mode and the longitude mode are degenerate,
\be
F_bh_b(t)+F_lh_l(t)=F_b(h_b(t)-h_l(t))=F_bh_s(t)\,.
\ee

In \ac{GR}, {for a binary system in circular orbit} the leading order contribution comes from the quadrupole radiation,
\bea
h_{+}(t)&=&\cA [(1+\cos^2{\iota})/2] \cos(\Phi),\\
h_{\times}(t)&=&\cA \cos\iota \sin(\Phi)\,,
\eea
where $\cA=\frac{4\cM}{D_L}(\pi\cM f)^{2/3}$ is the amplitude,
$\cM=(m_1m_2)^{3/5}/(m_2+m_2)^{1/5}$ is the chirp mass,
$m_1$ and $m_2$ are the component masses,
$D_L$ is the luminosity distance,
and $\iota$ is the orbital inclination angle.
For a \ac{GW} source located at the direction $(\theta_s, \phi_s)$ in the heliocentric ecliptic coordinate, the \ac{GW} phase $\Phi$ is given by
\be
\Phi=2\pi f t+2\pi f R \sin\theta_s \cos(2\pi f_m t-\phi_s+\phi_m)+\phi_0,
\ee
where the second term is due to the motion of the detector.

The leading order contribution from the extra polarization modes is \cite{Will:2014kxa},
\bea
h_{x}(t)&=&\cA_v \sin(\Phi/2)\,,\quad h_{y}(t)=\cA_v \cos\iota \cos(\Phi/2),\\
h_{b}(t)&=&\cA_b \sin\iota \cos(\Phi/2)\,,\quad h_{l}(t)=\cA_l \sin\iota \cos(\Phi/2).
\eea
Note that the frequency is half that of the tensor modes and the two vector modes have the same amplitude due to rotation symmetry.
The sub-leading contribution from the extra modes is,
\bea
h_{x}(t)&=&\cA'_{v} \sin\iota\sin(\Phi)\,,\quad
h_{y}(t)=\cA'_{v} \sin\iota\cos\iota \cos(\Phi),\\
h_{b}(t)&=&\cA'_{b} \sin^2\iota \cos(\Phi)\,,\quad
h_{l}(t)=\cA'_{l} \sin^2\iota \cos(\Phi)\,.
\eea
Apart from contributing to the waveforms, the extra modes also induce corrections to the {phase evolution of} tensor modes by carrying away extra energy.

For the \ac{GW} events detected by ground-based detectors such as LIGO and Virgo, a Bayesian model selection analysis shows that the data {support} the assumption that the signals are consisted of purely tensor modes, rather than purely vector or scalar modes \cite{LIGOScientific:2017ycc,LIGOScientific:2019fpa}.
The best result is from \ac{GW}170817, which, {using the localization information from optical observation}, gives a Bayes factor that is larger than $10^{20}$ \cite{LIGOScientific:2018dkp}.
The null-stream  method has been used for the O2, O3a and O3b events, and all the data {are} consistent with the pure tensor mode hypothesis \cite{Pang:2020pfz,Wong:2021cmp}.
With the position information of \ac{GW}170817 from \ac{EM} observation, meaningful upper bounds have been obtained on the amplitudes of the vector modes \cite{Hagihara:2019ihn} and the scalar modes \cite{Takeda:2021hgo}.

One can also use \ac{PTA} to search for the extra polarization modes \cite{daSilvaAlves:2011fp,Lee:2008ajo,Niu:2018oox,OBeirne:2019lwp,Boitier:2020xfx,Liang:2024sfn}.
An indication for a purely breathing mode instead of the tensor modes was originally found \cite{Chen:2021wdo} in the 12.5 yr pulsar-timing data in \cite{NANOGrav:2020bcs}.
But the conclusion seems to be strongly related to {a} single pulsar, {PSR} J0030+0451, in the catalogue, and if the pulsar is removed from the data set, the indication for the transverse scalar mode is no longer significant \cite{NANOGrav:2021ini}.
The recent analysis of the 15 yr data set of NANOGrav \cite{Chen:2023uiz} obtains a Bayes factor of 2.5 for the tensor mode relative to the scalar transverse mode.

For space-based detectors, TianQin and LISA are expected to detect about $10^4$ pairs of \acp{GCB} \cite{Lau:2019wzw,Huang:2020rjf}.
The \acp{GW} emitted  by these \acp{GCB} can be regarded as quasi-monochromatic signals \cite{Burdge:2019hgl}.
Because the waveform is simple, the \acp{GCB} are ideal sources for the search of extra polarization modes.

The prospect of using TianQin and \ac{GCB} signals to search for extra polarization modes has been studied in \cite{Xie:2022wkx}.
The dependence of the detection capability on the \ac{GW} frequency, amplitude and observation time is relatively simple.
But because the orientation of the orbital plane of TianQin is nearly fixed in space, the response and the Doppler effect significantly depends on the spatial direction of the source.

\begin{figure}[!htbp]
\includegraphics[width=0.45\linewidth]{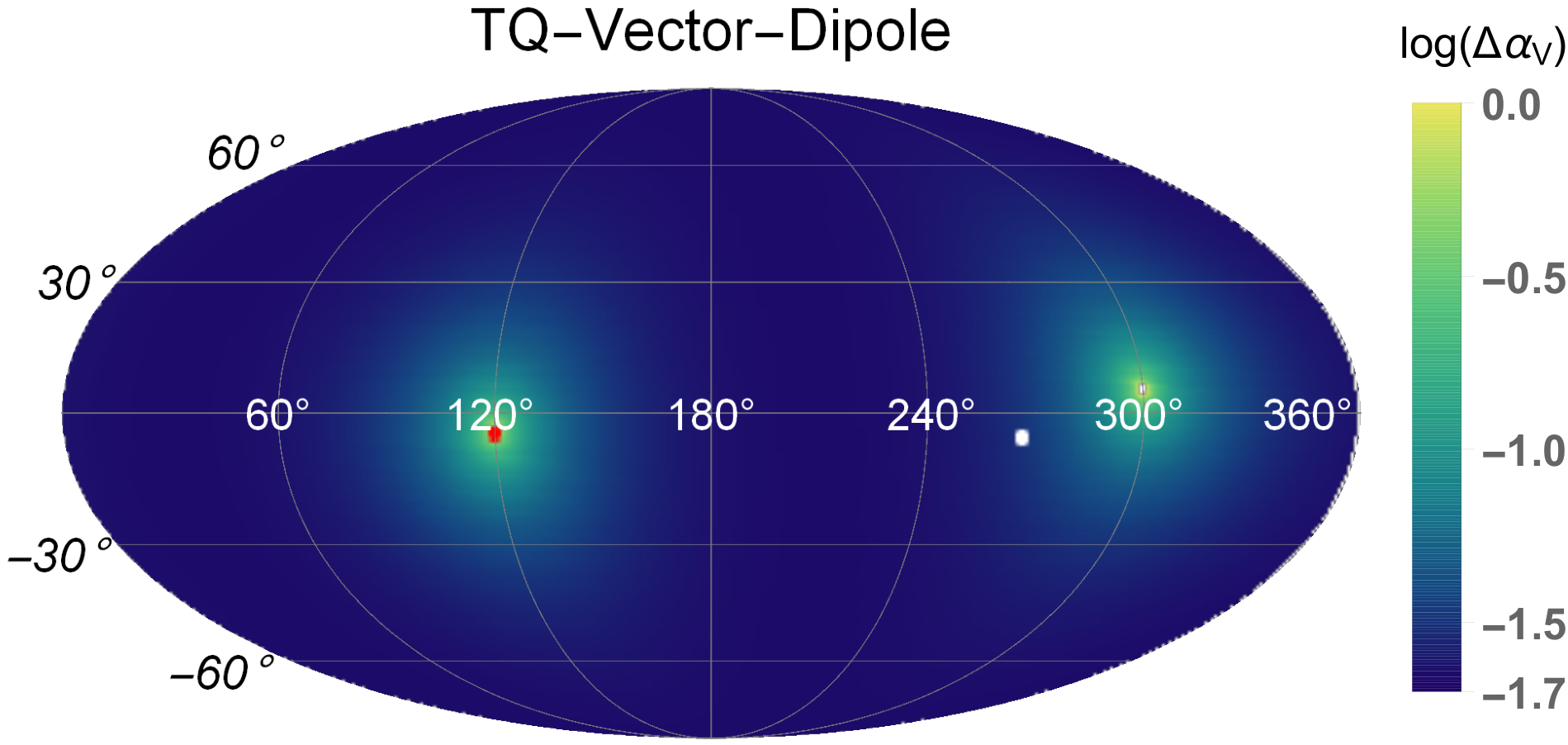}
\includegraphics[width=0.45\linewidth]{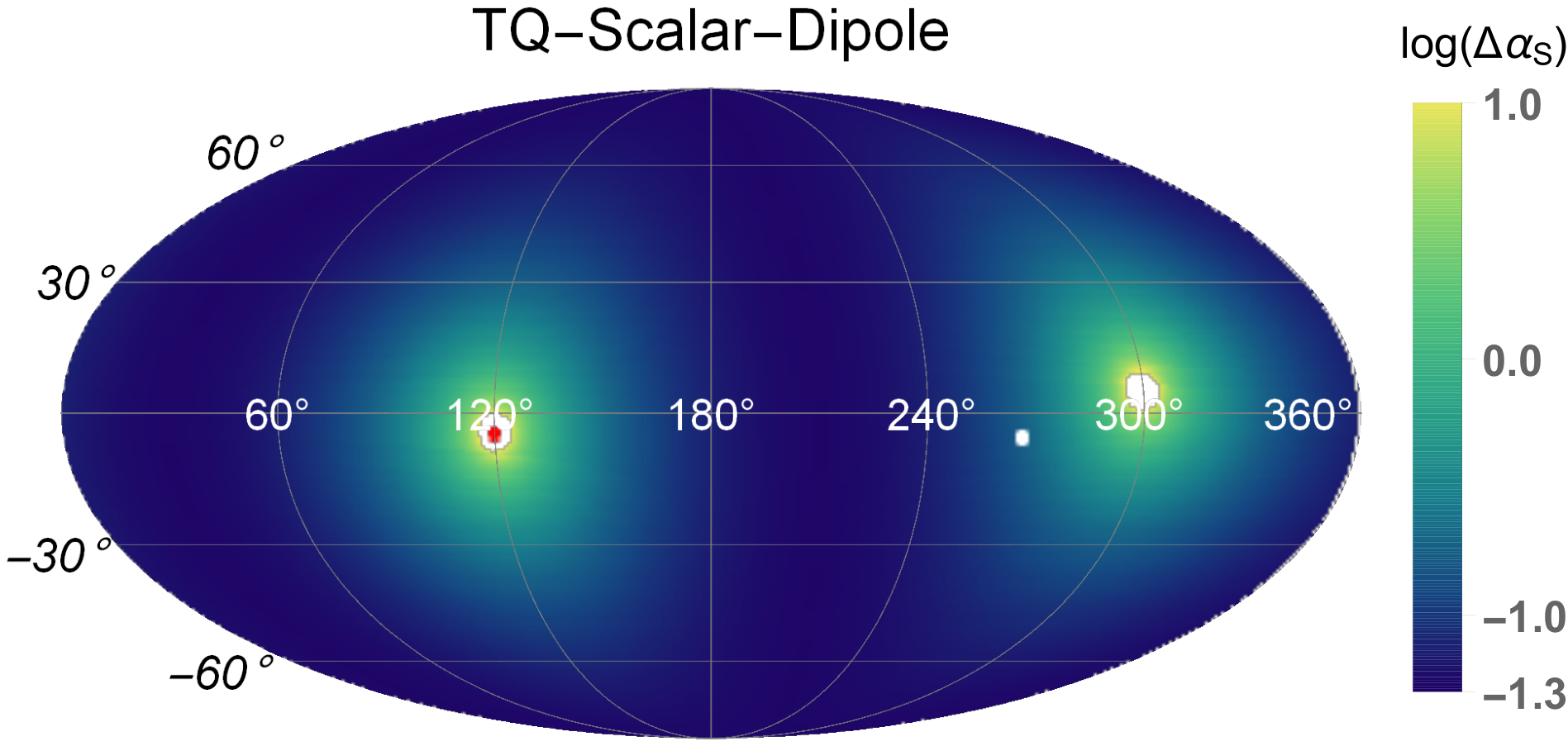}
\caption{The expected precision that TianQin can achieve for $\alpha_v$ and $\alpha_s$. The red (white) dot is J0806 (the Galactic center).	Other parameters: $f=0.02$ Hz, $\cA=10^{-22}$, $\iota=\pi/4$, $\psi_s=\pi/4$. (This plot is the same as FIG. 2 in \cite{Xie:2022wkx}.)}
\label{sec2:dT}
\end{figure}

The capability for TianQin to detect the {leading-order} radiation is illustrated in FIG.~\ref{sec2:dT}.
The result is presented in terms of the amplitude ratios, $\alpha_{v,s}\equiv\cA_{v,s}/\cA$.
Due to the vanishing of the antenna pattern function (\ref{sec2:Fp}) for $\bar{\theta}=0,\pi$, TianQin has no detection power for the vector or scalar modes for sources located in the direction of J0806 and its antipodal point.
For sources located in other directions, the best precision on $\alpha_v$ can reach 2\% level and that for $\alpha_s$ can reach 5\% level.

The capability for TianQin to detect the {subleading-order} radiation is a few times worse than that for the {leading-order} radiation.
Numerical calculation also shows that the angle $\theta_s$ has a strong effect on the result.
If it is treated as an unknown variable, then apart from the divergence along $\bar{\theta}=0,\pi$, there is also divergence along on the ecliptic plane. For the scalar modes, there is also divergence along the detector plane.
If one treats $\theta_s$ as a known parameter, then the divergence along the ecliptic plane can be removed.

Among the thousands of \acp{GCB} that are expected to be detected by TianQin, some have already detected through \ac{EM} observations \cite{Huang:2020rjf}.
These are called the \acp{VB}.
For \acp{VB}, the position and frequency of the \acp{VB} can be determined by \ac{EM} observation,
and the accuracy of the position is much better than the \ac{GW} observations.
As a result, one can hold the angular position parameters ($\theta_s$, $\phi_s$) fixed in the \ac{FIM} analysis.
As mentioned above, this will be helpful in detecting the quadrupole modes.
The prospect of using 14 \acp{VB} of TianQin to search for the extra polarization modes is listed in Table \ref{t:dwd}.
J0806 is not useful for this effort because it is directly facing the detector and so the response is zero for the relevant extra polarization modes.
In contrast, ZTF J1539 is the best among all the \acp{VB}.
But because it is located near the detector plane, the result for the scalar quadrupole mode is not so good.

\begin{table}[!htbp]
\renewcommand{\arraystretch}{1.2}
\caption{The parameters and the parameter estimation accuracy of the extra polarization' amplitude for {\acp{VB}} which can be detected by TianQin within 5 years.}
\label{t:dwd}
\begin{tabular}{lccccccccc}
\hline
\textrm{source} & $\theta$ & $\phi$ & $\iota$ & $f$(mHz) & $\cA (\times 10^{-22})$ & $\Delta\alpha_v$ & $
\Delta\alpha_s$ & $\Delta\alpha'_{v}$ & $\Delta\alpha'_{s}$ \\
 \colrule
 \text{J0806} & 94.7 & 120.4 & 38 & 6.22 & $1.28$ & \textbackslash & \textbackslash & \textbackslash & \textbackslash \\
 \text{ZTF J1539} & 23.8 & 205. & 84 & 4.82 & $3.68$ & 0.026 & 0.037 & 0.007 & 0.651 \\
 \text{V407 Vul} & 43.2 & 295. & 60 & 3.51 & $2.20$ & 0.090 & 0.305 & 0.035 & 0.141 \\
 \text{ES Cet} & 110.3 & 24.6 & 60 & 3.22 & $2.14$ & 0.089 & 0.164 & 0.029 & 0.453 \\
 \text{SDSS J0651} & 94.2 & 101.3 & 87 & 2.61 & $3.24$ & 0.225 & 1.342 & 0.077 & 3.114 \\
 \text{SDSS J1351} & 85.5 & 208.4 & 60 & 2.12 & $1.24$ & 0.360 & 0.658 & 0.114 & 4.142 \\
 \text{AM CVn} & 52.6 & 170.4 & 43 & 1.94 & $5.66$ & 0.090 & 0.285 & 0.043 & 0.172 \\
 \text{SDSS J1908} & 28.5 & 298.2 & 15 & 1.84 & $1.22$ & 0.433 & 4.538 & 0.555 & 6.216 \\
 \text{HP Lib} & 85 & 235.1 & 30 & 1.81 & $3.14$ & 0.173 & 0.784 & 0.115 & 0.635 \\
 \text{SDSS J0935} & 61.9 & 131. & 60 & 1.68 & $5.98$ & 0.172 & 0.712 & 0.065 & 0.317 \\
 \text{SDSS J2322} & 81.5 & 353.4 & 27 & 1.66 & $1.74$ & 0.400 & 2.485 & 0.286 & 1.878 \\
 \text{PTF J0533} & 111.1 & 82.9 & 73 & 1.62 & $1.52$ & 0.686 & 2.065 & 0.230 & 1.132 \\
 \text{CR Boo} & 72.1 & 202.3 & 30 & 1.36 & $2.58$ & 0.362 & 1.363 & 0.236 & 3.229 \\
 \text{V803 Cen} & 120.3 & 216.2 & 14 & 1.25 & $3.20$ & 0.326 & 2.787 & 0.478 & 31.950 \\
\hline
\end{tabular}
\end{table}

The signals for \ac{MBHB} can also be used to constrain the extra polarization modes, and the correction on the phase evolution must be considered in the waveform.
With a preliminary study with Bayes method, the constraint with TianQin on the amplitudes is about a few percent \cite{Ning:2024xxx}.
The possibility of using \ac{SGWB} to constrain the extra polarization modes has been studied in \cite{Hu:2023nfv,Hu:2024toa}.

\subsubsection{GW propagation}

{\it Subsection coordinator: Changfu Shi}

In \ac{GR}, \acp{GW} travel at the speed of light.
But as shown in Table \ref{tab:sec2.B:MGs2}, \acp{GW} in \acp{MGT} can have propagation speeds different from the speed of light even in a flat background.
In cases when there is \ac{EM} counterpart to the \ac{GW} events, it is possible to make a direct comparison between the \ac{GW} and light speeds.
For example, the multi-messenger observations of \ac{GW}170817 \cite{LIGOScientific:2017vwq,LIGOScientific:2017zic} have been used to rule out a significant class of \acp{MGT} \cite{Baker:2017hug,Creminelli:2017sry,Sakstein:2017xjx,Ezquiaga:2017ekz}.

\acp{MGT} also lead to non-trivial dispersion relations for \acp{GW}.
A useful parameterization scheme is
\be
E^2=p^2+\mathbb{A}_\alpha p^\alpha\,,
\ee
where $E$ and $p$ are the graviton energy and momentum, respectively.
$\alpha$ is a power {index} and $\mathbb{A}_\alpha$ is the corresponding magnitude of modification.
In \ac{GR}, $\mathbb{A}_\alpha=0$.
For $\alpha=0$, one usually writes $\mathbb{A}_0=m_g^2$, where $m_g$ corresponds to the graviton mass.
Different \acp{MGT} have different values for $\alpha$ and $\mathbb{A}_\alpha$.
A few examples are listed in Table. \ref{tab:dr_th}.
{Notice that the above equation has assumed the existence of a frame where \acp{GW} are isotropically propagating, but in the presence of \ac{LLI} violation, such a frame may not exist \cite{Kostelecky:2003fs} and then anisotropic propagation would occur \cite{Shao:2020shv}.}

\begin{table}[!htbp]
\renewcommand{\arraystretch}{1.2}
\label{tab:dr_th}
\caption{Some examples of \acp{MGT} with a modified dispersion relation for \acp{GW}. \cite{Yunes:2016jcc}}
\begin{tabular}{|c|c|c|}
\hline
Theory                      &$\alpha$ & $\mathbb{A}_\alpha$\\
\hline
Double Special Relativity   & $3$   &  $\eta_{dsrt}$ \\
\hline	
Extra-Dimensional Theories  &$4$    &$-\alpha_{edt}$\\
\hline
Ho\v{r}ava-Lifshitz         &$4$    &$k^4_{hl}\mu^2_{hl}/16$\\
\hline
Non-Commutative Geometries  &$4$    &$2\alpha_{ncg}/E_p^2$\\
\hline
\end{tabular}
\end{table}

Nontrivial dispersion relation can lead to a dephasing of \acp{GW}.
For example, for massive gravitons there is difference in the propagation speed of \acp{GW} at different frequencies, leading to a phase difference at detection and emission \cite{Will:1997bb}.
Suppose a source emits two massive gravitons of frequencies $f_e$ and $f_e'$ at a time separation $\Delta t_e$, then after propagating through the flat \ac{FLRW} cosmic space, the separation of the arrival time becomes (assuming $\mathbb{A}_\alpha=0$):
\bea
\label{eq:time_dif}
\Delta t_a=(1+z)\Big(\Delta t_e+\frac{D_0}{2\lambda_g^2}\Big)\Big(\frac{1}{f_e^2}-\frac{1}{f_e'^2}\Big)\,,
\eea
where $\lambda_g=E_e/(m_gf_e)=E_{e'}/(m_gf_{e'})$ is the graviton Compton wavelength, $z$ is cosmology red-shift, and $D_0$ is related to the luminosity distance.
The difference between $\Delta t_e$ and $\Delta t_a$ leads to a dephasing of the \acp{GW}:
\bea
\label{eq:mass_phase}
\Delta\Phi=-\frac{\pi^2\cM D_0}{\lambda_g^2(1+z)(\pi \cM f)^{-1}},
\eea
where $\cM$ is the chirp mass of the binary system.
The dephasing corresponding to the general modified dispersion relation (\ref{eq:time_dif}) can be found in \cite{Mirshekari:2011yq}.

Through a Bayesian analysis of the data from the first \ac{GW} event \ac{GW}150914, the \ac{LVK} Collaboration has placed a limit on the graviton mass, $m_g\leq 1.2\times10^{-22}$ eV at the 90\% confidence level \cite{LIGOScientific:2016lio}.
Analysis of the GWTC-1 \cite{LIGOScientific:2018mvr}, GWTC-2 \cite{LIGOScientific:2020ibl} and GWTC-3 \cite{KAGRA:2021vkt} data has {lowered} the bound to $m_g\leq 1.27\times10^{-23}$ eV, which is about 2.5 better than the Solar System bound \cite{Bernus:2020szc}, and about {2 to 3} orders of magnitude better than the bound from from binary-pulsar observations \cite{Finn:2001qi, Miao:2019nhf}.
The current bound on $\mathbb{A}_\alpha$ for $\alpha \in [0,4]$ can be found in \cite{LIGOScientific:2019fpa,LIGOScientific:2020tif,LIGOScientific:2021sio}.
{Analysis of the data of pulsar timing arrays, i.e., NANOGrav \cite{NANOGrav:2023gor} and CPTA \cite{Xu:2023wog}, placed on new limits on the graviton mass, i.e., $m_{g}<8.6\times10^{-24}$ eV and $m_{g}<3.8\times10^{-23}$ eV at the 90\% confidence level \cite{Wang:2023div}.}

Space-based \ac{GW} detectors are expected to do even better in probing the propagation property of \acp{GW}.
It has been shown that \ac{LISA} has the potential to probe the graviton mass to the level $m_g < \cO(10^{-24} \rm~eV)$ by combining about 400 \acp{GCB} with \acp{SNR} greater than 25 \cite{Cooray:2003cv}.
By using the inspiral signal of individual \ac{MBHB} event, it has been shown that \ac{LISA} can probe the graviton mass to the order $m_g < \cO(10^{-25} \rm~eV)$, and the combined bound from 50 events is about ten times better \cite{Berti:2011jz}.
The prospect of using \ac{LISA}, \ac{ET}, and \ac{CE} to probe the generic modified dispersion relation has been studied in \cite{Samajdar:2017mka}.
It has been found that \ac{LISA} can perform better than ground-based detectors for $\alpha\leq1$, and \ac{ET} and \ac{CE} are expected to improve over the current bound on $\mathbb{A}_\alpha$'s by an order of magnitude.

\begin{figure}[htbp!]
\centering
\includegraphics[width=0.45\linewidth]{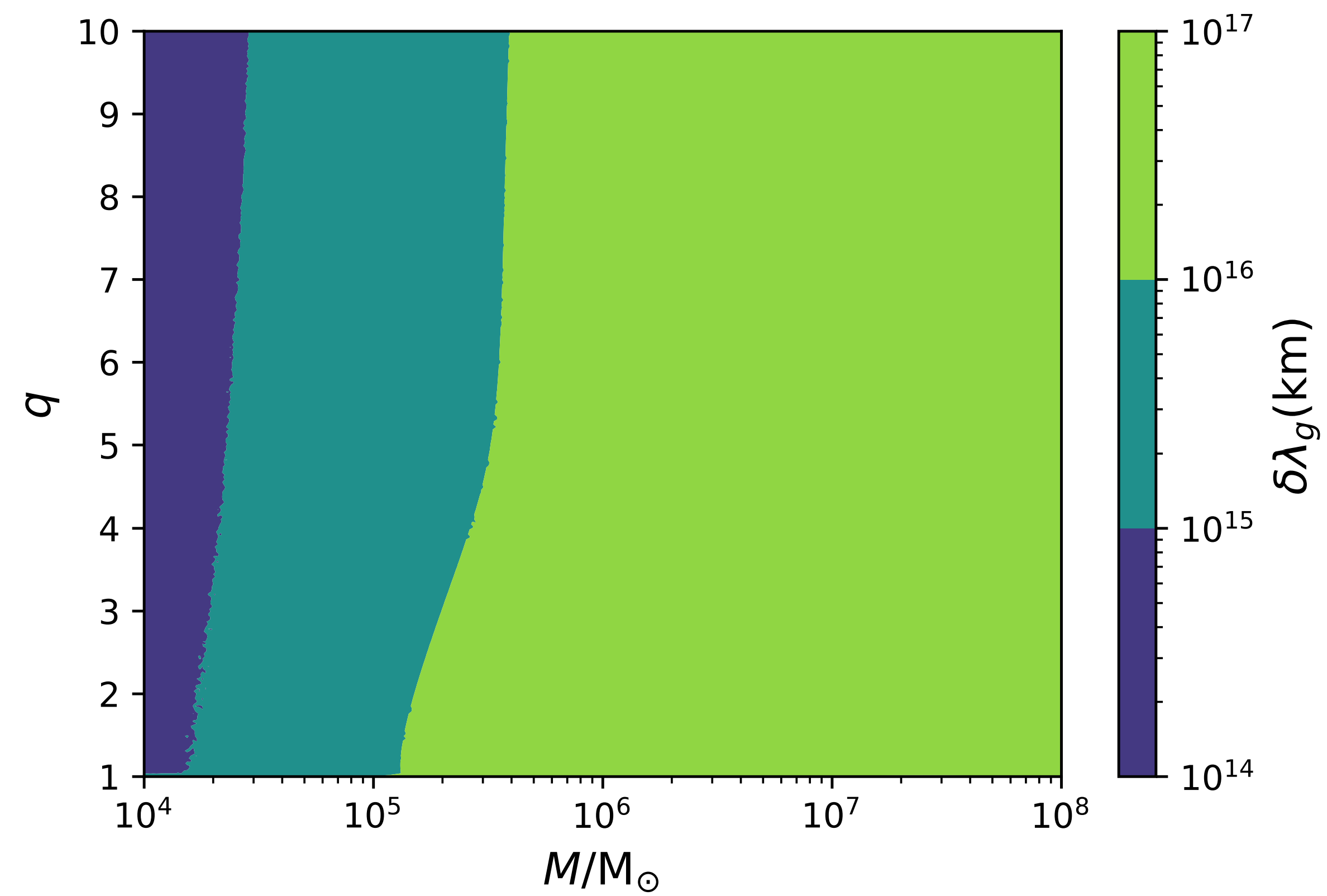}
~~
\includegraphics[width=0.4\linewidth]{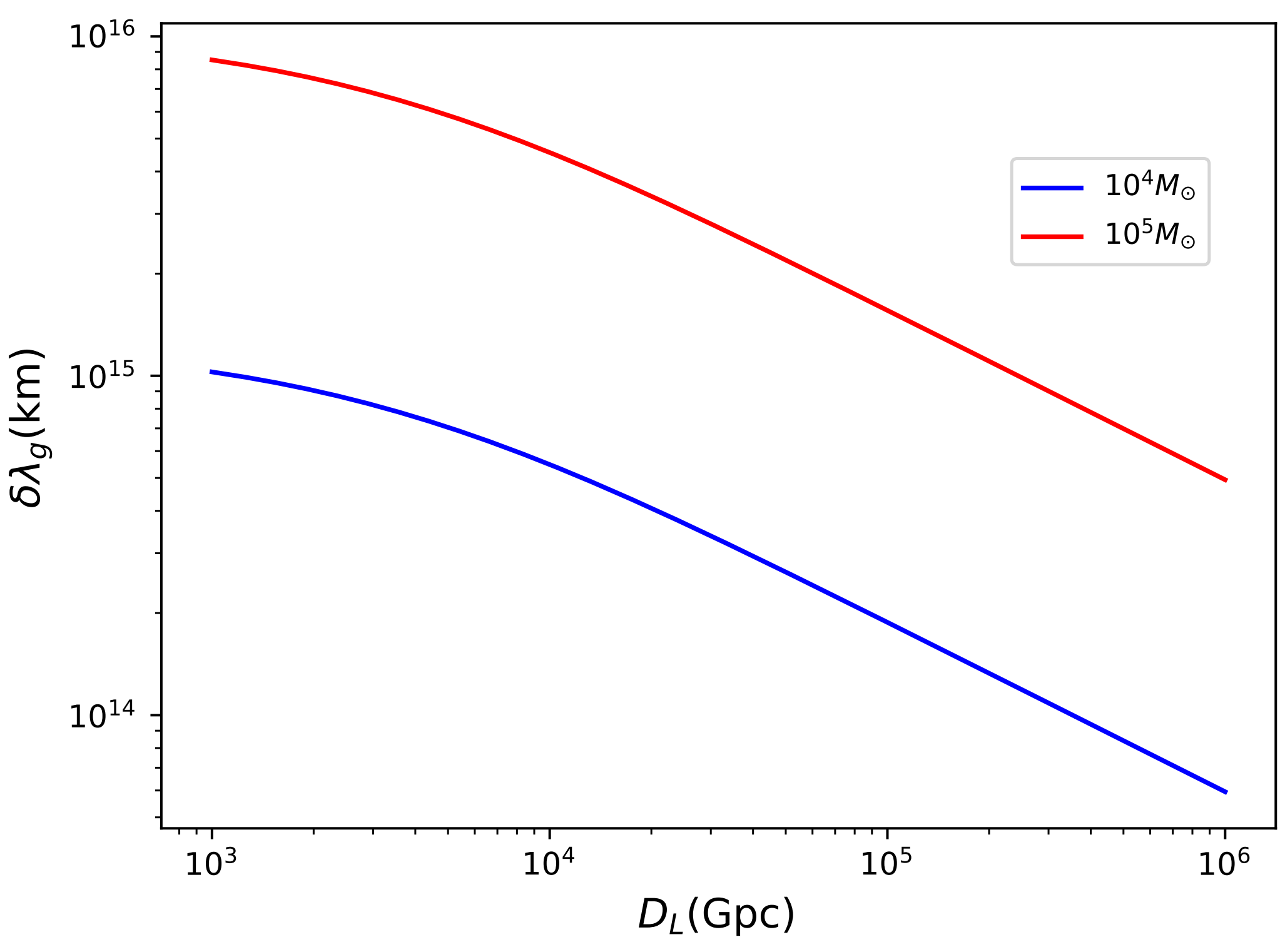}
\caption{The capability of using TianQin to probe the graviton mass. Left: the dependence of $\delta\lambda_g$ on the total redshifted mass $M_z$ and the symmetric mass-ratio $q$, with the luminosity distance $D_L=15\rm~ Gpc$; Right: the dependence of $\delta\lambda_g$ on $D_L$, with $\eta=0.22$, and the indicated masses are measured in the source frame. The dimensionless spin parameters are $s_1=0.4$ and $s_2=0.2$ for both plots.}
\label{fig:mg}
\end{figure}

The prospect of using TianQin to probe the graviton mass is illustrated in FIG. \ref{fig:mg}.
One can see that better probing capability can be obtained with large total mass and less unequal component masses.
With the {chosen} parameters, TianQin can probe the graviton Compton wavelength to better than the order $\cO(10^{17})$ km, corresponding to $m_g<\cO(10^{-27})\rm~eV$, thus improving over the current bound on graviton mass {from \acp{GW}} by four orders {of magnitude}.
One can also see that, if one fixes the total mass in the source frame, then the obtained precision of $\delta\lambda_g$ does not change very significantly with $D_L$.
This is because increasing the luminosity distance {weakens} the signal on the one hand, but it also {increases} the dephasing on the other hand.

The prospect of using TianQin to probe the generic modified dispersion relation is illustrated in Table \ref{tab:dr_cp}, where the result is from an example \ac{MBHB} event with $M_z=10^6\mSun$, $\eta=0.22$, $D_L=15\rm~Gpc$, $s_1=0.4$, and $s_2=0.2$. One can see that TianQin can improve over the current bounds on $\mathbb{A}_0$, $\mathbb{A}_{0.5}$ and $\mathbb{A}_1$ by about eight, five and three orders {of magnitude}, respectively.
For $\alpha>1.5$, TianQin does not have much advantage over the current bounds.

\begin{table}[!htbp]
\label{tab:dr_cp}
\renewcommand{\arraystretch}{1.2}
\caption{The prospect of using TianQin to constrain $\mathbb{A}_\alpha$ with a typical \ac{MBHB} event.}
\centering
\begin{tabular}{|l|l|l|}
\hline
Number & Current & TianQin \\
\hline
$m_g~~[\rm~eV]$ & 1.3$\times10^{-23}$ & 2.1$\times10^{-26}$ \\
$A_0~~[\rm~eV^{2}]$ & 1.9$\times10^{-45}$ & 3.4$\times10^{-53}$ \\
$A_{0.5}[\rm~eV^{1.5}]$ & 5.1$\times10^{-39}$ & 7.3$\times10^{-44}$ \\
$A_1~~[\rm~eV]$ & 1.2$\times10^{-32}$ & 2.7$\times10^{-35}$ \\
$A_{1.5}[\rm~eV^{0.5}]$ & 3.7$\times10^{-26}$ & 8.2$\times10^{-27}$ \\
$A_{2.5}[\rm~eV^{-0.5}]$ & 1.2$\times10^{-14}$ & 6.7$\times10^{-9}$ \\
$A_3~~[\rm~eV^{-1}]$ & 6.6$\times10^{-9}$ & 1.5$\times10^{-1}$ \\
$A_{3.5}[\rm~eV^{-1.5}]$ & 4.5$\times10^{-3}$ & 5.8$\times10^{6}$ \\
$A_4~~[\rm~eV^{-2}]$ & 3.0$\times10^3$ & 4.5$\times10^{13}$  \\
\hline
\end{tabular}
\end{table}

\subsubsection{GW generation}\label{subsec:ppE}

{\it Subsection coordinator: Changfu Shi}

Black holes are ideal laboratories for testing \ac{GR}, as they can provide strong field conditions and are less affected by the environmental contamination that often impacts other astrophysical systems.
The evolution of a black hole binary {coalescence} can be divided into three phases, i.e. inspiral, merger, and ringdown.
The \acp{GW} emitted during the inspiral phase can be accurately modeled using the \ac{PN} approximation, particularly for systems with comparable component masses.
In \ac{GR}, for example, the waveform in the frequency-domain can be written as \cite{Yunes:2009ke}:
\bea
h_{\rm~{GR}}(f)=A(f)e^{i\psi(f)}\,,\quad  \psi(f)=2\pi t_c +\phi_c+\Sigma_{k=0}^\infty\phi_k^{\rm~PN} u^{(k-5)/3}\,,\label{pna}
\eea
where $f$ is frequency, $A(f)$ is the amplitude, $t_c$ and $\phi_c$ are the coalescence time and phase, respectively, $u=(\pi\cM f)^{1/3}$ is a characteristic velocity, $\cM=\eta^{3/5}M$ is the chirp mass,  $M=m_1+m_2$ is the total mass, $\eta=m_1m_2/(m_1+m_2)^2$ is the symmetrical mass ratio, and $\phi^{\rm~PN}_k$ is the phase coefficient at the $(k/2)$ \ac{PN} order. Note that $\phi^{\rm~PN}_k$ is completely determined by the source parameters for the binary black hole system \cite{Blanchet:2002av}.
Pioneering works on using the inspiral signals of black hole binaries to test \ac{GR} can be found in \cite{Will:1994fb}. Pioneering studies of using the signals \LS{}{of} \acp{EMRI} and \acp{IMRI} to test \ac{GR} can be found in \cite{Ryan:1995wh,Ryan:1997hg,Scharre:2001hn}.

Different \acp{MGT} will have different corrections to (\ref{pna}). The \ac{ppE} framework has been developed to enable a theory agnostic probe of possible deviations from \ac{GR} \cite{Berti:2004bd,Arun:2006yw,Arun:2006hn,Yunes:2009ke}.
The basic idea of \ac{ppE} is to focus on the leading \ac{PN}-order corrections to the \ac{PN} waveform,
\bea
h_{\rm~ppE}(f)=h_{\rm~{GR}}(f)(1+\alpha u^a)e^{i\beta u^b}\,,\label{eq:waveform_ppE}
\eea
where $\alpha$ and $\beta$ are the \ac{ppE} parameters, with $\alpha=\beta=0\,$ in \ac{GR}, and $a$ and $b$ are the \ac{PN}-order parameters, with $b=k-5\,$ and $a=b+5$.
Since the initial formulation in \cite{Yunes:2009ke}, which focused on the two polarization modes of \acp{GW} in \ac{GR} and quasi-circular orbits for black hole binaries, the \ac{ppE} framework has been extended to include additional polarization modes \cite{Chatziioannou:2012rf}, time-domain waveform \cite{Huwyler:2014gaa}, eccentricity \cite{Loutrel:2014vja} and environmental effects \cite{Cardoso:2019rou}.
The \ac{ppE} parameters for any given \ac{MGT} can be found by computing the corrections to the orbital evolution of the binary system \cite{Tahura:2018zuq}.
Using this approach, \ac{ppE} parameters have been determined for a spectrum of theoretical models, including Brans-Dicke gravity \cite{Zhang:2017sym}, screened modified gravity \cite{Zhang:2017srh}, parity-violating gravity \cite{Zhao:2019xmm}, Lorentz-violating gravity \cite{PhysRevD.91.082003}, \ac{NCG} \cite{PhysRevD.94.064033}, and quadratic modified gravity \cite{Yagi:2011xp}.
See \cite{Tahura:2018zuq,Chamberlain:2017fjl} for a summary.

Predictions of how advanced LIGO/Virgo and LISA can constrain the \ac{ppE} parameters have been studied in \cite{Cornish:2011ys}.
The prospect of using LISA to constrain the \ac{ppE} phase parameter $\beta$ with \ac{MBHB} signals has been studied in \cite{Huwyler:2014vva}.
The \ac{ppE} formalism has also been extensively applied {in different scenarios} to place constraints on specific \acp{MGT}, such as the Brans-Dicke theory \cite{Zhang:2017sym}, theories involving Lorentz violation in gravity \cite{Hansen:2014ewa}, the {varying-$G$} theory \cite{Yunes:2009bv}, and a range of theories that predict massive gravitons, modified dispersion relations, or the presence of dipole radiation \cite{Keppel:2010qu,Mirshekari:2011yq,Berti:2011jz,Samajdar:2017mka,Arun:2012hf, Shao:2020shv, Zhao:2021bjw, Haegel:2022ymk, Wang:2022yxb}.
Following the direct detection of \acp{GW}, the \ac{GW}190514 and \ac{GW}151226 signals have been used to constrain the \ac{ppE}  phase parameter and the result is then interpreted within the context of specific \acp{MGT} \cite{Yunes:2016jcc}.
The prospect of using future \ac{GW} detectors, including four potential configurations of LISA, aLIGO, A+, Voyager, CE and ET-D, to constrain the \ac{ppE} phase parameter $\beta$ and various \acp{MGT}, including models predicting dipole radiation, extra dimensions, {varying-$G$} theory, Einstein-aether theory, khronometric gravity, and massive graviton theory, has been studied in \cite{Chamberlain:2017fjl}.
The \ac{ppE} formalism has also been used to demonstrate that multiband observation of \acp{SBHB} with LIGO and LISA \cite{Sesana:2016ljz} can significantly enhance the expected constraints on \ac{GW} dipole radiation by an impressive six orders of magnitude \cite{Barausse:2016eii}.
The prospect of {constraining} the \ac{ppE} parameters through multiband observations with CE and an array of space-based detectors, including LISA, TianQin, DECIGO, and B-DECIGO, has been studied in \cite{Carson:2019rda, Liu:2020nwz}.
The multiband enhancement on the constraints on \ac{EdGB} theory and the \ac{IMR} consistency has been studied in \cite{Carson:2020cqb}.

\begin{figure*}[htbp!]
\centering
\subfigure{\includegraphics[width=0.45\linewidth]{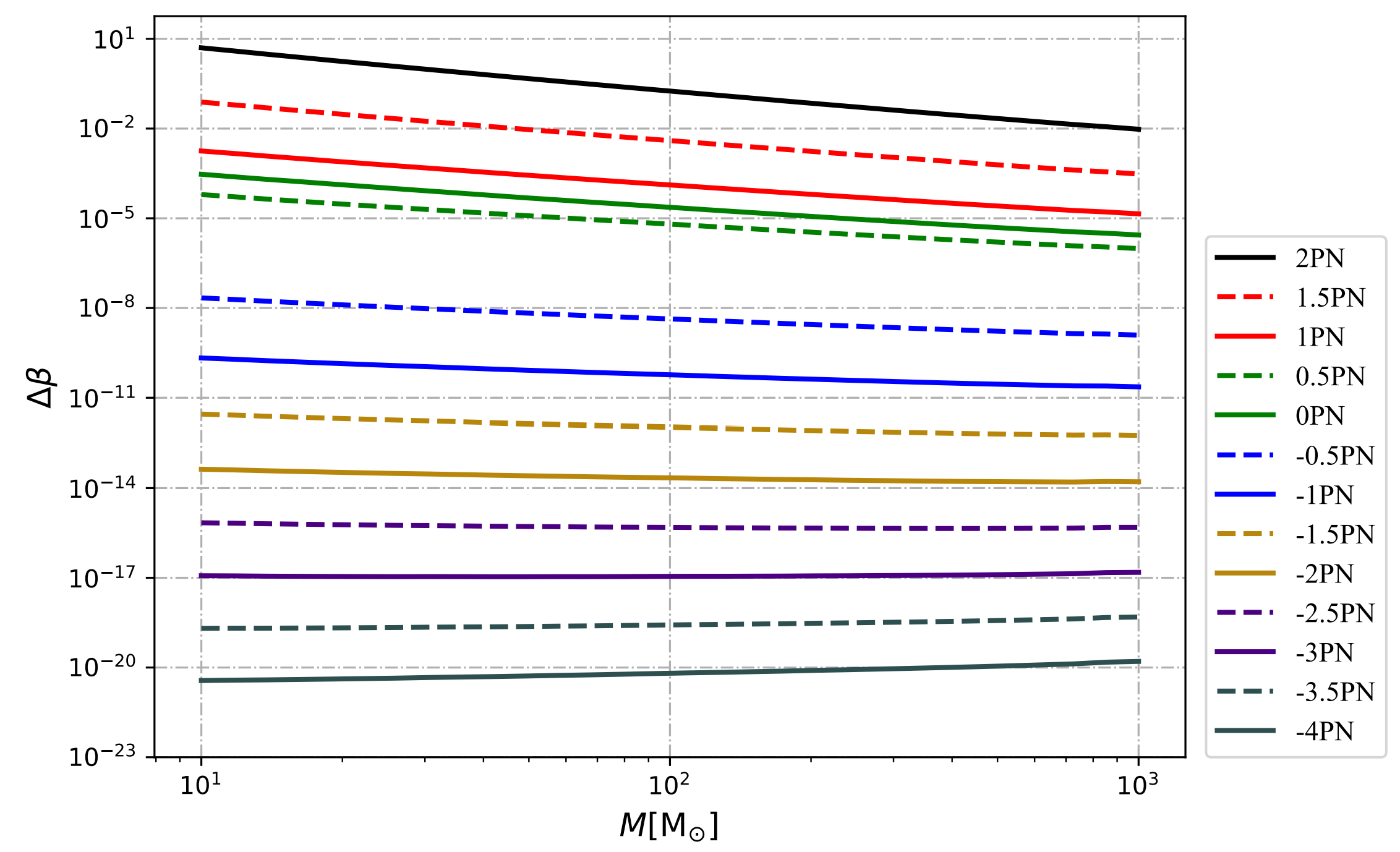}}
\subfigure{\includegraphics[width=0.45\linewidth]{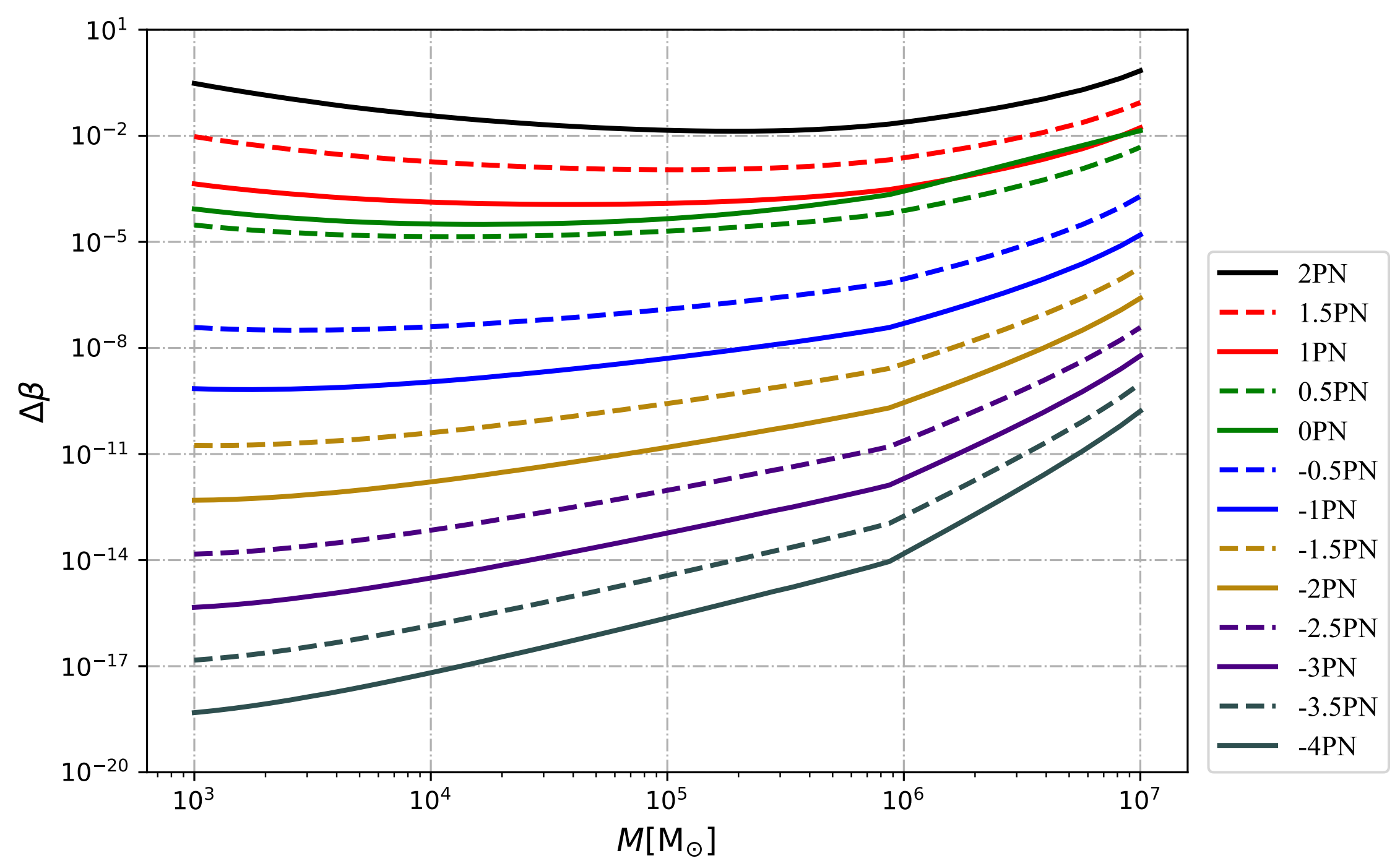}}
\caption{Dependence of $\Delta\beta$ on $M$ at different \ac{PN} orders for TianQin. Other parameters: symmetric mass ratio $\eta=0.22$, and luminosity distance $D_L=$ 500 Mpc (left: low mass region), 15 Gpc (right: high mass region).}
\label{fig:M_ppE}
\end{figure*}

The prospect of using TianQin to test \acp{MGT} with the \ac{ppE} formalism has been studied in \cite{Shi:2022qno}.
The dependence of $\Delta\beta$ on the total mass $M$ at different \ac{PN}  orders is given in FIG. \ref{fig:M_ppE}.
One can see that the total mass has a strong effect on $\Delta\beta$.
For instance, at the 2PN order, the variation of $\Delta\beta$ within the low-mass range can span up to three orders of magnitude, whereas at the {$-4$PN} order, the variation of $\Delta\beta$ within the high-mass range can exceed eight orders of magnitude.
One can also see that $\Delta\beta$ is more tightly constrained by the low-mass sources at the lower \ac{PN} orders, while $\Delta\beta$ at the higher \ac{PN} orders are best constrained with the sources at around $\cO(10^5\mSun)$.
The value of $\Delta\beta$ for a given \ac{PN} order is almost always smaller than those of the higher \ac{PN} orders.
The only exception is $\Delta\beta_{\rm~0PN}$, which is always greater than $\Delta\beta_{\rm~0.5PN}$.
This is due to a strong correlation with the mass parameter \cite{Chamberlain:2017fjl}.

Apart from the theory agnostic method, one can also test specific \acp{MGT}.
Here we consider the following four \acp{MGT} as examples:
\begin{itemize}
\item \ac{EdGB}: The leading-order modification from  \ac{EdGB} starts at the $-1$PN order, corresponding to $b=-7$. The \ac{ppE} phase parameter is \cite{Yagi:2011xp}
\bea
\beta_{\rm EdGB}&=-\frac{5\zeta_{\rm EdGB}}{7168}\frac{(m_1^2\tilde{s}_2-m_2^2\tilde{s}_1)^2}{M^4\eta^{18/5}}\,,
\label{EdGB}
\eea
where $\zeta_{\rm EdGB}\equiv16\pi\bar{\alpha}^2_{\rm EdGB}/M^4\,$, $\bar{\alpha}_{\rm EdGB}$ is the coupling between the scalar field and quadratic curvature term in the theory \cite{Kanti:1995vq}, and $\tilde{s}_n\equiv2(\sqrt{1-\chi_n^2}-1+\chi_n^2)/\chi_n^2\,$, $n=1,2$, is the spin-dependent scalar charge of the $n$th component, with $\chi_n$ being the effective spin. The current best constraint on the theory comes from the observation of \ac{GW}200115, giving $\sqrt{|\bar{\alpha}_{\rm EdGB}|}<1.1$ km \cite{Wang:2023wgv,Lyu:2022gdr,Perkins:2021mhb,Nair:2019iur}.

\item \ac{dCS}: The leading-order modification from \ac{dCS}  starts at the 2PN order, corresponding to $b=-1$. The \ac{ppE} phase parameter is \cite{Yagi:2011xp,Tahura:2018zuq}
\bea
\beta_{\rm~dCS}&=-\frac{1549225\eta^{-14/5}\xi_{\rm~dCS}}{11812864}\Big[\Big(1-\frac{16068\eta}{61969}\Big)\chi_a^2\nn\\
&+\Big(1-\frac{231808\eta}{61969}\Big)\chi_s^2-2\delta_m\chi_a\chi_s\Big]\,,\label{dCS}
\eea
where  $\delta_m\equiv(m_1-m_2)/M\,$, $\chi_s=(\chi_1+\chi_2)/2\,$, $\chi_a=(\chi_1-\chi_2)/2\,$, $\xi_{\rm~dCS}\equiv16\pi\bar{\alpha}^2_{\rm~dCS}/M^4$, and $\bar{\alpha}_{\rm~dCS}$ is the coupling constant of the Chern-Simons correction \cite{Jackiw:2003pm}. The current best constraint on the theory comes from the observation of neutron star systems, giving $\sqrt{\bar{\alpha}_{\rm~dCS}}<8.5$ km \cite{Silva:2020acr}. So far one is unable to place a meaningful constraint on \ac{dCS} theory using \ac{GW} data directly, due to the lack of a viable waveform.

\item \ac{NCG}: A class of theory tried to quantize spacetime by promoting the spacetime coordinates themselves to operators that do not commute \cite{Snyder:1946qz,PMIHES_1985__62__41_0,Chamseddine:1992yx,Landi:1997sh}.. The leading order modification introduced by \ac{NCG} also starts from the 2PN order, and the \ac{ppE} phase parameter is \cite{Kobakhidze:2016cqh}:
\bea
\beta_{NC}=-\frac{75}{256}\eta^{-4/5}(2\eta-1)\td\Lambda^2\,,
\label{NC beta}
\eea
where $\td\Lambda^2=\theta^{0i}\theta_{0i}/(l_p^2t_p^2)$, $t_p$ and $l_p$ are the Planck time and Planck length, respectively, and $\theta^{\alpha\beta}$ is the antisymmetric tensor characterizing the noncommutativity of spacetime
coordinates, $[\hat{x}_\mu,\hat{x}_\nu]=i\theta_{\mu\nu}$. The current best constraint on $\td\Lambda$ comes from \ac{GW}190514, with $\sqrt{\td\Lambda}<3.5$ \cite{Kobakhidze:2016cqh}.

\item $\dot{G}(t)$: A time varying gravitational coupling parameter was suggested in, e.g. \cite{Dirac:1937ti}. The leading order modification from $G(t)$ starts at the  {$-4$PN} order, i.e. $b=-13$, and the \ac{ppE} phase parameter is \cite{Tahura:2019dgr}:
\bea
\beta_{\dot{G}}=-\frac{25}{851968}\dot{G}\eta^{3/5}[(11+3s_1+3s_2)M  -41(s_1m_1+s_2m_2)]\,,
\eea
where $s_1$ and $s_2$ are the sensitivity of the two binary components, and $\dot{G}$ is the time variation of $G(t)$.
The current best constraint comes from lunar laser ranging, giving $|{\dot{G}/G_0}|<\cO(10^{-13})\rm~year^{-1}$ \cite{Hofmann:2010aa}.
For \acp{GW}, \ac{GW}150914 and \ac{GW}151226 have produced the constraints $|{\dot{G}/G_0}|<7.3\times10^6$ $\rm~year^{-1}$ and $|{\dot{G}/G_0}|<2.24\times10^4$ $\rm~year^{-1}$, respectively \cite{Yunes:2016jcc}.
\end{itemize}

\begin{figure*}[htbp!]
\centering
\subfigure{\includegraphics[width=0.45\linewidth]{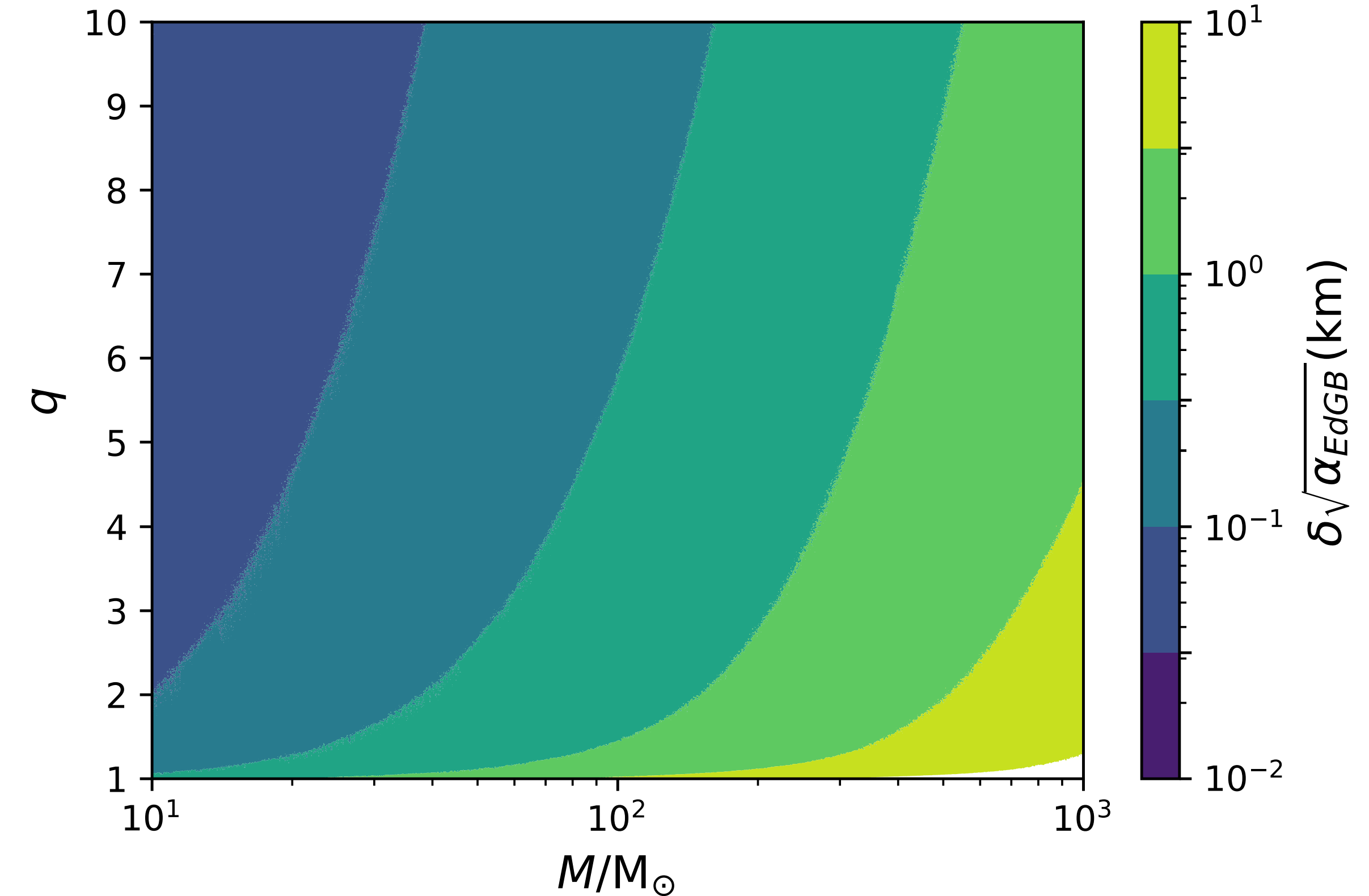}}
\subfigure{\includegraphics[width=0.45\linewidth]{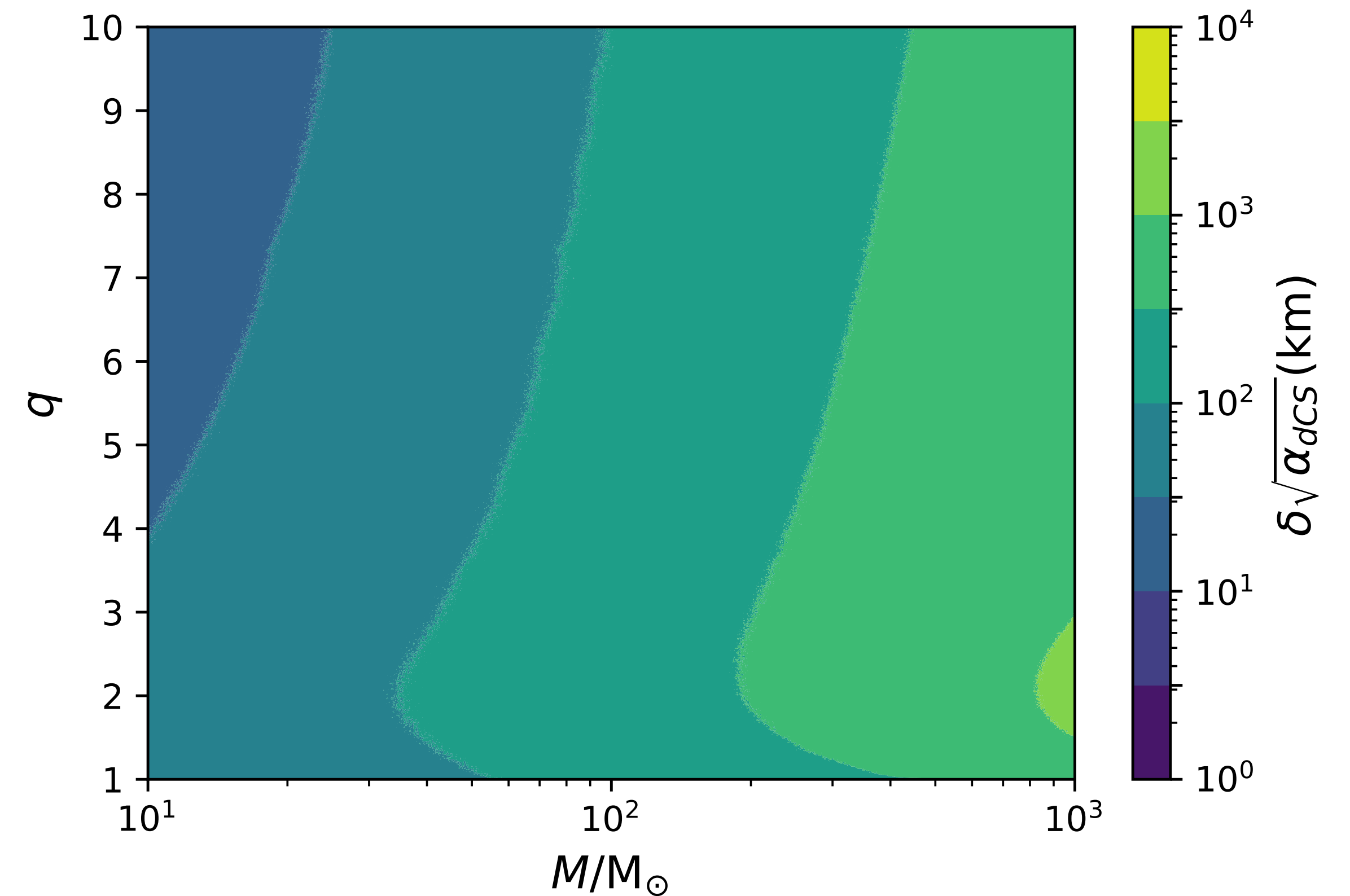}}
\subfigure{\includegraphics[width=0.45\linewidth]{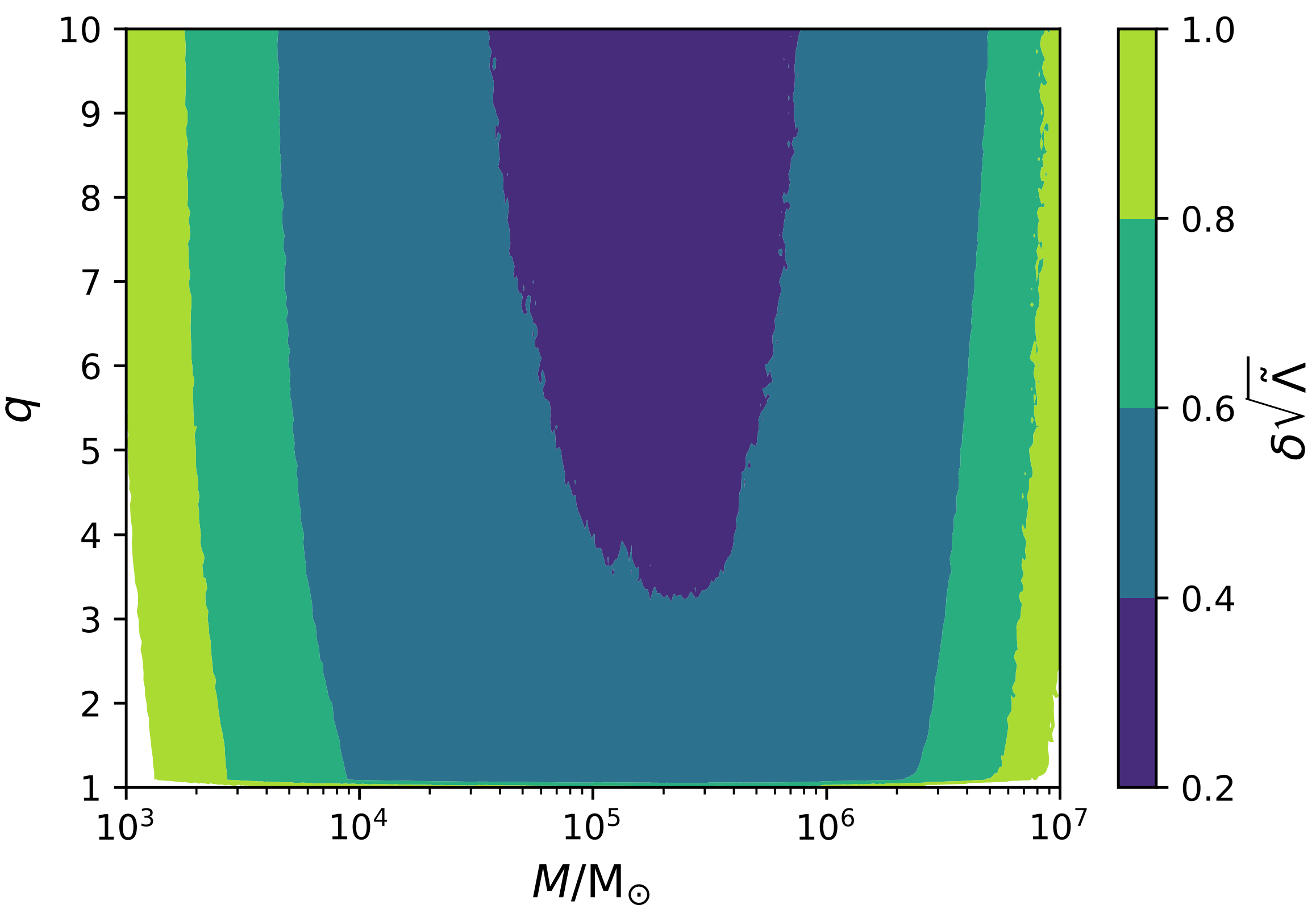}}
\subfigure{\includegraphics[width=0.45\linewidth]{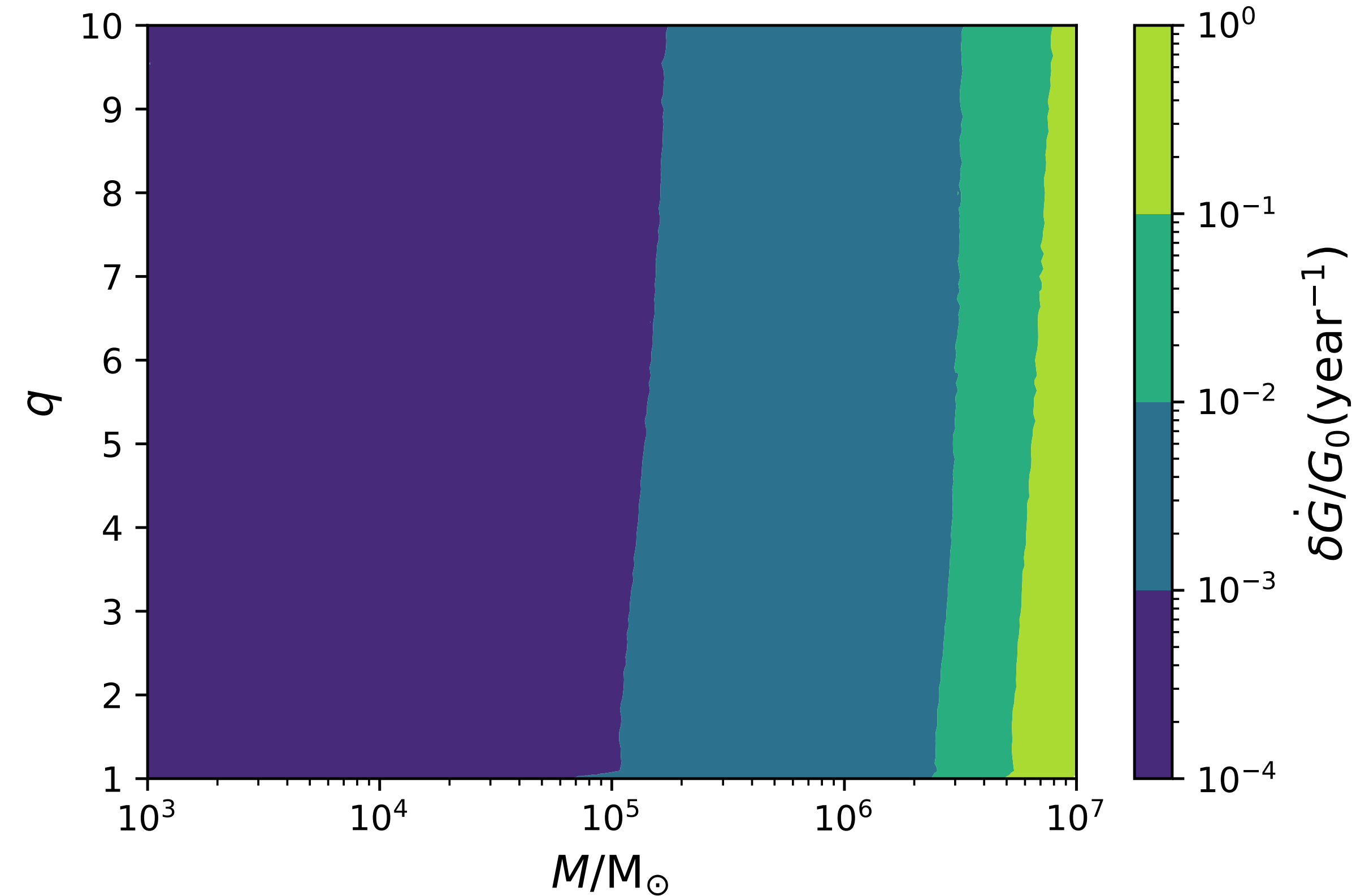}}
\caption{Expected constraint on {theory parameters in} EdGB, dCS, NCG, and $\dot{G}$ with TianQin. The luminosity distance: $D_L=500$ Mpc for low mass sources ($M<10^3\mSun$), and $D_L=15$ Gpc for high mass  sources ($M>10^3\mSun$)}\label{fig:four_theory}
\end{figure*}

The prospect of using TianQin to test these theories is illustrated in FIG. \ref{fig:four_theory}. \ac{EdGB} and \ac{dCS} are better constrained with low mass binaries with disparate masses.
With the detection of \acp{SBHB} with total masses below $M<\cO(10^2\mSun)$, TianQin is expected to constrain EdGB to the order $\sqrt{|\bar{\alpha}_{\rm EdGB}|}<\cO(0.1{\rm~km})$, which is about one order {of magnitude} better than the current best result.
Similarly, TianQin is expected to constrain dCS to the order $\sqrt{\bar{\alpha}_{\rm~dCS}}<\cO(1{\rm~km})$.
However, due to the reliability of existing waveforms being contingent upon the {assumption of} small-coupling limit, this result warrants further scrutiny.
\ac{NCG} and $\dot{G}$ are better constrained with high mass binaries.
In the case of \ac{NCG}, TianQin is expect to improve over the current best bound by an order of magnitude and constrain the theory to the sub-Planckian scale.
For {varying-$G$ theory}, the detection of \acp{MBHB} with total masses ranging from $10^5\mSun$ to $10^6\mSun$ is expected to enable TianQin to constrain the theory to the level of $|{\dot{G}/G_0}|<\cO(10^{-5})\rm~year^{-1}$.

One can also look at the joint detection of \acp{MBHB} with TianQin and LISA, and the multiband observation of \acp{SBHB} with TianQin and ET.
A detection with multiple detectors can help break the correlation between $\beta_{\rm~0PN}$ and mass, enhancing the measurement precision of $\beta_0$ by about three orders of magnitude.
The joint detection of \acp{MBHB} with TianQin and LISA can improve the constraint on \ac{NCG} by about two times.
The multiband observations of \acp{SBHB} with TianQin and \ac{ET} can enhance the constraint on $\sqrt{|\bar{\alpha}_{\rm EdGB}|}$ by about one order of magnitude.
For more details, we refer to \cite{Shi:2022qno}.

\subsection{Environmental effects and waveform systematics}

{\it Subsection coordinator: Jian-dong Zhang}

In probing fundamental physics with \acp{GW}, an important issue is to avoid mistaking false signals for evidence of new physics.
In this subsection, we discuss the correction to \ac{GW} waveforms due to environmental effects.

Around the \ac{GW} radiating binary sources, there may exist accretion disks, dark matter halos, or third gravitational bodies.
The surrounding matter can have three different types of effects on the \ac{GW} source \cite{Barausse:2014pra,Barausse:2014tra}:
\begin{itemize}
\item Change of orbit evolution due to additional conservative forces such as gravitational pull and dissipative forces such as dynamical friction;
\item Change of source mass and spin due to the accretion of surrounding matter;
\item Change of radiation efficiency (lose rate of energy and angular momentum) due to unknown fundamental fields.
\end{itemize}
The existence of matter can also affect the formation and evolution of the binaries before they enter the \ac{GW} dominated region, which determine the initial property of the binaries.

On the path of the \ac{GW} propagation, there can also exist different densities of matter, causing gravitational lensing effect for \acp{GW}.
Depending on the density profile of the lenses, \acp{GW} can be bent, delayed, (de)magnified, and phase-shifted.
Due to the low frequency of \acp{GW}, the wavelength of \acp{GW} maybe comparable to the lens in some cases.
In such cases, the geometric-optic approximation is no longer accurate enough and one may need to consider the wave-optic diffraction effect.
When diffraction happens, the magnification factor may also depend on the frequency of \acp{GW}.

In real \ac{GW} detection, the environment around a \ac{GW} source is {\it a priori} not known.
So the problem is how to distinguish between the environmental effects and possible signatures of new physics.
In the following, we discuss such issues related to TianQin.

\subsubsection{Environmental effects in GW generation}

The most common environmental effects for astrophysical black holes is the gas in the accretion disks.
The gas present in the vicinity of binary black holes affects the orbital evolution and therefore the emission of \acp{GW}.
The main effects include extra gravitational pull on the compact objects \cite{Macedo:2013qea, Barausse:2006vt},
the dynamical friction and planetary migration which are due to the gravitational interactions of the compact objects with their own wake in the matter medium  \cite{Barausse:2007dy,Macedo:2013qea,Barausse:2007ph,Yunes:2011ws,Kocsis:2011dr},
and the accretion which will change the masses and spins of the compact objects \cite{Barausse:2007dy,Macedo:2013qea}.

Dark matter is also a common possibility for the surrounding environment of binary black holes.
Due to its small cross-section, dark matter is collisionless and is hard to form a disk-like configuration.
But black holes can have dynamical interactions with dark matter and can affect its distribution.
In general, there will be a ``spike'' in dark matter halo, i.e., the dark matter density increases significantly near the central black hole \cite{Gondolo:1999ef}.
One common model for the dark matter density profile is the \ac{NFW} model \cite{Navarro:1996gj}, in which the dark matter density varies with the radius as $\rho_{\rm~DM}\sim r^{-7/3}$.
The gravitational pull and the dynamical friction will dominate the interaction between the dark matter halo and the black holes, while the accretion is almost negligible.

Hierarchical triple systems, i.e., systems consisted of a tight inner binary black holes orbiting the third black hole on a wider orbit, are also an interesting possibility in nature \cite{Samsing:2017plz,Samsing:2017xod}.
In these hierarchical triples, the third black hole brings interesting features to the \ac{GW} signal emitted by the inner binary, e.g., the oscillation of eccentricity and inclination of the inner binary orbit due to the Kozai-Lidov mechanism \cite{Kozai:1962zz,Lidov:1962wjn}.
Such oscillations can modify the frequency evolution of the inner binary and this needs to be taken into account in waveform modeling \cite{Chandramouli:2021kts}.

Another possibility for non-trivial environment is the presence of new fundamental fields outside the black holes.
Since this is related to either \acp{MGT} or new physics in the particle physics sector, we will not cover them in this section.
Astrophysical black holes are usually neutral due to various charge neutralization mechanisms \cite{Gibbons:1975kk,Blandford:1977ds}.
However, if there exist external magnetic field generated by the accretion disk, the black holes could be charged.
So a test of the black hole charge is also a test of the black hole environment.

The environment effects can be probed with both the inspiral and the ringdown signals of binary black hole mergers.
For the ringdown signals, the environment can affect the potential of the perturbation equation over the black hole background, and consequently modify the frequencies of \acp{QNM}.
Since the detection of \acp{QNM} has been discussed in detail in the subsection \ref{subsec:kerr}, we will focus on the inspiral signals in the following.

The first problem related to environmental effects is its correction to the \ac{GW} waveforms.
Since the environmental effects are usually small, one usually does not need an extremely precise model of the environment itself.
For example, for dark matter halos, people often use the \ac{NFW} model \cite{Navarro:1996gj} to exemplify the density profile.
For corrections to the waveform, it is usually enough to consider the leading order contribution in the \ac{PN} approximation \cite{Cardoso:2019rou}.
As such, one can firstly employ the \ac{ppE} framework to do theory agnostic study of environment effect, and then to map the result to concrete models black hole environment.
The only exception is \ac{EMRI}, for which the environmental effect can be comparable to the self-force effect \cite{Barausse:2014tra}.
Much work is still needed for precise modelling of \ac{EMRI} waveforms with environmental effect \cite{Kejriwal:2023djc,Jiang:2023xwv,Rahman:2023sof}.
Once this is achieved, \acp{EMRI} will be the best source to study the surrounding environment of black holes \cite{Barausse:2014tra}.

The second problem concerning environmental effects is how to distinguish it from the effect of \acp{MGT}.
Below are some possible ways:
\begin{itemize}
\item Considering existing constraints on the \acp{MGT}.
This method is best used when the detected effect is much larger than the allowed prediction of \acp{MGT}, considering existing constraints from other experiments.
The caveat is if the constraints derived from other experiments on the \acp{MGT} can be extrapolated to the situation of \ac{GW} radiation.
\item Considering the result from multiple \ac{GW} events.
The correction of an \ac{MGT} should be the same for all \ac{GW} events, while the environment of different \ac{GW} sources can be significantly different.
\item Searching for possible \ac{EM} counterparts.
For black hole binaries surrounded with dense matter, there could be \ac{EM} counterparts during merger.
If an \ac{EM} counterpart is observed, then the information can be used to help determine the level of environmental contributions.
\end{itemize}

The capability of TianQin in probing the environmental effect during \ac{GW} generation can be directly obtained from the \ac{ppE} result (see subsection \ref{subsec:ppE}).
For example, the dynamical friction due to a dark matter halo with density profile $\rho_{\rm~DM}=\rho_0(r_0/r)^{3/2}$ affects the inspiral signal at the $-4$ \ac{PN} order, which is the same as the varying-G theory.
From \cite{Yuan:2024duo}, one can see that the precision of relative parameter estimation $\rho_0$ is $\delta\rho_0/\rho_0\sim10^{-2}$.

Because the dark matter halo with $\rho_{\rm~DM}=\rho_0(r_0/r)^{3/2}$ has the same form of \ac{ppE} correction as the varying-G theory, it is interesting to know if these two effects can be distinguished.
For this one can define the following stastistic,
\bea F=\sum_{i=1}^n\frac{(\dot G_i-\bar{\dot {G}})^2}{\sigma_i^2}\,,\eea
where $n$ is the number of detected events and the index $i$ means the $i$-th event,
$\dot G_i$ and $\sigma_i$ are the mean value and variance of $\dot G$ for the $i$-th event.
$\bar{\dot {G}}$ is the mean value of $\dot G_i$ for all the events.
By using a specific astrophysical population model for binary black holes, we find $F$ will be small if the waveform correction is due to the varying-G, while $F$ is very large if the correction is due to dark matter halo \cite{Yuan:2024duo}.
This result shows that it is possible to distinguish these two effects if multiple events are used.

\subsubsection{Environmental effect in GW propagation}

\paragraph{Gravitational lensing effect}\

When \ac{EM} waves pass by a massive object, they can be deflected, delayed and amplified. This is known as the gravitational lensing effect.
Gravitational lensing has a wide range of applications in the study of cosmology, the large scale structure, exoplanets, dark matter and so on \cite{Schneider:2006qyj}.
Similar to \ac{EM} waves, \acp{GW} can also be lensed \cite{Takahashi:2003ix}.
If the lensing effect is not properly included in the analysis of \ac{GW} data, there can be systematic errors in the estimation of source parameters.
What's more, lensed \ac{GW} signals can be used to study the propagation property of \acp{GW}, infer the physical properties of the lensing object, study the nature of dark matter and the expansion of the universe \cite{Fan:2016swi,Liao:2018ofi,Yang:2018bdf,Hannuksela:2020xor,Sereno:2011ty,Liao:2017ioi,Cao:2019kgn,Li:2019rns,Yu:2020agu,Urrutia:2021qak,Chung:2021rcu,Gais:2022xir,Broadhurst:2020cvm}.

If the \ac{GW} wavelength is much shorter than  the gravitational radius of the lens, geometrical optics approximation is applicable to the calculation of the lensing effect.
However, if the \ac{GW} wavelength is comparable or longer than the gravitational radius of the lens, wave optics must be used in the calculation, which requires accurate evaluation of the diffraction integral.
For example, if \acp{GW} in the \ac{LVK} band are lensed by stars, \acp{IMBH} and other objects, they behave much like light diffraction in the wave-optics regime \cite{Ohanian:1974ys,Nakamura:1997sw,Boileau:2020rpg,Leung:2023lmq}.
The wave-optics effect can perturb the plane of \ac{GW} polarization \cite{Ezquiaga:2020dao,Dalang:2021qhu} and cause beat patterns in the time-domain waveform \cite{Yamamoto:2005ea,Hou:2020mpr}.
These effects might allow \ac{LVK} to detect massive stars, \acp{IMBH}, the dense cores of globular clusters, and dark matter halos \cite{Moylan:2007fi,Cao:2014oaa,Takahashi:2016jom,Christian:2018vsi,Dai:2018enj,Jung:2017flg,Liao:2019aqq,Mishra:2021xzz}.

So far the \ac{LVK} collaboration has published 90 \ac{GW} events \cite{LIGOScientific:2016aoc,LIGOScientific:2018mvr,LIGOScientific:2020ibl,LIGOScientific:2021usb,LIGOScientific:2021aug}.
Despite much effort, however, no lensed \ac{GW} signal has been confirmed in these events \cite{Broadhurst:2019ijv,Singer:2019vjs,McIsaac:2019use,Hannuksela:2019kle,Liu:2020par,Dai:2020tpj,LIGOScientific:2021izm,Diego:2021fyd,Baker:2016reh,Fan:2016swi,Goyal:2020bkm,Lai:2018rto,Diego:2019rzc,Oguri:2020ldf,Xu:2021bfn,LIGOScientific:2023bwz}.
Nevertheless, there is a high prospect that many lensed \ac{GW} events will be found using the next-generation \ac{GW} detectors such as \ac{ET} and \ac{CE} \cite{Punturo:2010zza,Reitze:2019iox}.
In the near future, space-based \ac{GW} detectors such as LISA \cite{LISA:2017pwj} and TianQin \cite{TianQin:2015yph} are expected to detect hundreds of \ac{MBHB} merger events \cite{Klein:2015hvg,Wang:2019ryf}.
Some suggest that nearly one percent of the detected events may experience strong gravitational lensing  \cite{Gao:2021sxw}.
It is also possible for wave-optics effects of lensing to be detected \cite{Caliskan:2022hbu,Tambalo:2022wlm}.

The prospect of using TianQin to probe the gravitational lensing effect for \acp{GW} has been studied in \cite{Lin:2023ccz}.
An issue is the calculation of the diffraction integral  \cite{Levin:1992,Takahashi:2004mc,Alfredo:2017,Guo:2020eqw,Tambalo:2022plm}.
Due to the special frequency range, one needs to consider the wave-optics and geometric-optics separately for different parts of the signals.
Three types of lensing models have been used in the study: the point mass lens, the SIS lens and the \ac{NFW} lens.
For each lens model, the amplification factor has been calculated in diffraction limit for lower frequency part, and in geometric optics limit for higher frequency part.
In the geometric optics calculation, the first-order post-geometric optics correction has been included.

\begin{figure}[!htbp]
\centering
\subfigure{\includegraphics[width=0.43\textwidth]{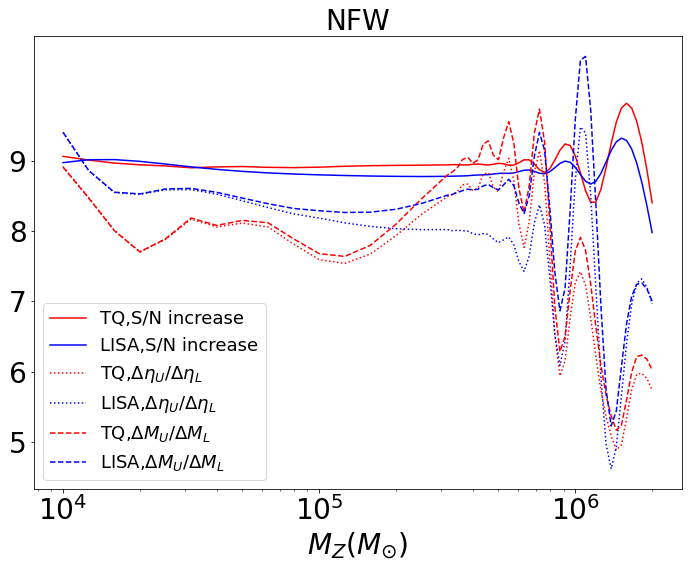}}
\subfigure{\includegraphics[width=0.47\textwidth]{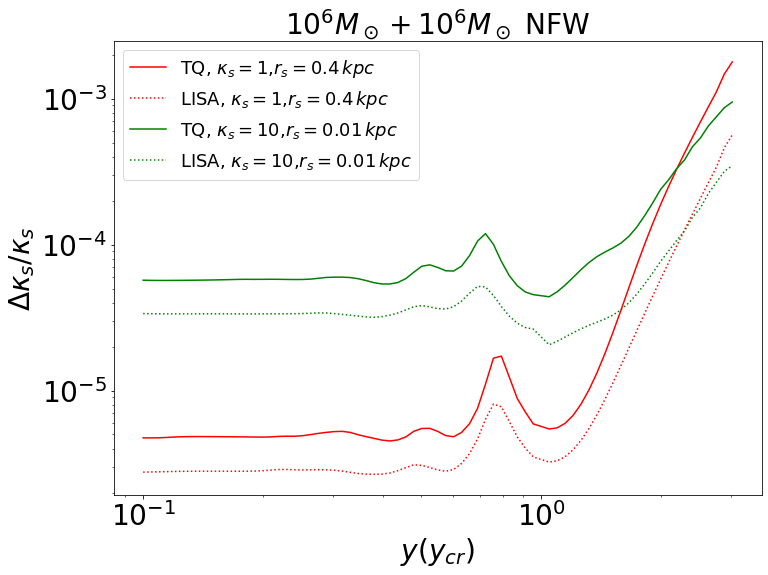}}
\caption{(Left) The increase in the \ac{SNR} and the precision of parameter estimation for \ac{GW} events with an \ac{NFW} lens.
(Right) The expected precision in the estimation of $\kappa_s$.
Both figures are taken from \cite{Lin:2023ccz}.}
\label{fig:NFW}
\end{figure}

Fig. \ref{fig:NFW} (Left) illustrates the effect of lensing on the \ac{SNR} and the precision of parameter estimation for \ac{GW} events, using an \ac{NFW} lens as an example.
One can see that the gravitational lensing increases both the \ac{SNR} and the precision of parameter estimation.
There is a strong correlation between the increase in the \ac{SNR} and the improvement on the precision of parameter estimation.
For low mass sources, for which the geometric optics approximation is valid, the improvement on the \ac{SNR} and the precision of parameter estimation does not change much as the redshifted total mass $M_z$ increases.
For high mass sources, for which the wave optics becomes important, there is notable fluctuation in the improvement on the \ac{SNR} and the precision of parameter estimation as the redshifted total mass $M_z$ increases.
Similar results also hold for the other two lens models.

The lensed gravitational signal can also be used to measure the lens parameters.
Using the \ac{NFW} model as an example, the density profile is given by \cite{Navarro:1994hi}
\bea\rho(r)=\frac{\rho_s}{(r/r_s)(r/r_s+1)^2}\,,\eea
where $\rho_s$ and $r_s$ are two model parameters.
The corresponding lens potential is
\bea\phi(x)=\frac{\kappa_s}2\left\{\begin{matrix}
(\ln\frac{x}2)^2-\Big(\arctan\sqrt{1-x^2}\Big)^2&:&x<1\cr
(\ln\frac{x}2)^2+\Big(\arctan\sqrt{x^2-1}\Big)^2&:&x>1
\end{matrix}\right. , \eea
where $\kappa_s=16\pi\rho_sr_s(d_Ld_{LS}/d_S)$ is the dimensionless surface density.
Fig. \ref{fig:NFW} (Right) illustrates the expected precision in the estimation of $\kappa_s$, varying with the angular separation between the source and the lensing object (see  \cite{Lin:2023ccz} for detail).
Corresponding to the two cases considered, i.e., $r_s=0.4$ kpc for $\kappa_s=1$ and $r_s=0.01$ kpc for $\kappa_s=10$, $M_{200}$ is about $4\times10^9\mSun$ and $5\times10^7\mSun$, respectively.
$M_{200}$ is the quantity to describe the mass of the NFW halo, and it corresponds to the mass enclosed by the radius where the density of the dark matter is 200 times of the density of the universe.
One can see that $\kappa_s$ can be determined to better than $\cO(10^{-5})$ when $\kappa_s=1$ and to the level $\cO(10^{-4})$ when $\kappa_s=10$.

\paragraph{Gravitational Integrated Sachs-Wolfe Effects}\

{\it Subsection coordinator: Anzhong Wang}\

TianQin, together with other space-based detectors, such as   Taiji \cite{Ruan:2018tsw}, LISA \cite{2017arXiv170200786A} and DECIGO \cite{Kawamura:2020pcg},
are able to detect \acp{GW} emitted from  binary systems  as far as the redshift  $z \simeq 100$ \cite{2021NatRP...3..344B}. This
will result in a variety of profound scientific consequences. In particular, \acp{GW} propagating over such long cosmic distances will carry valuable information not only about their sources, but also about
the detail of the cosmological expansion and inhomogeneities of the universe, whereby a completely new window to explore the universe by using \acp{GW} is opened, as so far our understanding of the
universe almost all comes from observations of \ac{EM} waves only (possibly with the important exceptions of  cosmic rays and neutrinos) \cite{Lyth_Liddle_2009}.

One of such effects is the gravitational integrated Sachs-Wolfe (iSW) effect, quite similar to cosmic microwave photons, for which the large-scale structures the photons traverse contribute to the
observed temperature anisotropies.  In the framework of GR, using  the Isaacson geometric
optics approximation \cite{Isaacson:1968hbi}, Luguna {\em et al} \cite{Laguna:2009re} first derived the
\ac{GW} counterpart of this effect for \acp{GW}  propagating on a FRW background with only cosmological scalar perturbations, and later Fier  {\em et al} \cite{Fier:2021fbt} generalized it to the case  with both scalar and  tensor perturbations, and
derived the general formulas for the propagation of \acp{GW} in the inhomogeneous universe. It was showed that the phase, frequency, and amplitude of the \acp{GW} experience
gravitational iSW  effects, in addition to the magnification effects on the amplitude from gravitational lensing. For supermassive black
hole binaries, the iSW effects could account for measurable changes on the frequency, chirp mass, and luminosity
distance of the binary, thus unveiling the presence of inhomogeneities, and potentially dark energy (DE), in the universe. More recently,  
the above studies was further generalized   to scalar-tensor theories \cite{Garoffolo:2019mna}, including Horndeski \cite{Dalang:2020eaj,Ezquiaga:2020dao,Kubota:2022lbn},   and  the Einstein-scalar-Gauss-Bonnet
(EsBG) gravity \cite{Fier:2024}.

For the gravitational iSW effects in GR, all \acp{GW} to be detected by the current and next generations of detectors  can be well approximated  as high frequency \acp{GW},  whereby the geometric optic approximations can be applied \cite{Isaacson:1968hbi}.
For \ac{GW} propagating in the inhomogeneous universe, $g_{\mu\nu} = \gamma_{\mu\nu} + \epsilon h_{\mu\nu}, \;  \gamma_{\mu\nu} = \bar\gamma_{\mu\nu} + \epsilon_c \hat\gamma_{\mu\nu}$, where $\bar\gamma_{\mu\nu}$ and $\hat\gamma_{\mu\nu}$  represent respectively the homogeneous and  inhomogeneous parts of the universe, and  $h_{\mu\nu}$ denotes \acp{GW} first emitted by astrophysical sources (such as binary systems)  and then propagating in the background of $ \gamma_{\mu\nu}$. For the flat FRW universe, we have $\bar\gamma_{\mu\nu} = a^2(\tau)\eta_{\mu\nu}$ and $\hat\gamma_{\mu\nu}dx^{\mu} dx^{\nu} = a^2(\tau)\left(-2\Phi d\tau^2 + \left(2\Psi\delta_{ij} + H_{ij}\right)dx^idx^j\right)$, with $\eta_{\mu\nu} \left[\equiv \text{diag.}\left(-1, 1,  1, 1\right)\right]$ describing the Minkowski spacetime.
It can be found that
\begin{align}
\label{eq4.38}
h_{\mu\nu} = \chi_{\mu\nu} - \frac{1}{2}\chi \gamma_{\mu\nu} =  e_{\mu\nu}\tilde{h},\quad
    \tilde{h} \equiv \mathcal{A}e^{i\varphi} = \frac{(1+z)\mathcal{Q}}{d_L}(1+\xi)e^{i(\varphi_e+\delta\varphi)},
\end{align}
where $\mathcal{Q}$ is a constant, and  ${\cal{R}}  \equiv a r$ denotes the physical distance  between the observer and the source,
while   $r$ the comoving distance, given by
$r^2 \equiv \left(x_e - x_r\right)^2 +\left(y_e - y_r\right)^2 + \left(z_e - z_r\right)^2$, where $x_e^i \equiv (x_e, y_e, z_e)$ and $x_r^i \equiv (x_r, y_r, z_r)$ are the spatial locations of the source and observer, respectively.
The \ac{GW} phase and amplitude are given by
\bea\delta\varphi&=&\varphi-\varphi_e=\int^{\lambda}_{\lambda_e}(\Phi+\Psi)d\lambda' -\frac{1}{2}n^kn^l\int^{\lambda}_{\lambda_e}H_{kl}d\lambda'\,,\nn\\
\xi&=&\left. \left(\Psi-\frac{1}{4}n^kn^lH_{kl}\right)\right|^{\lambda}_{\lambda_e} +\frac{1}{2}I^{(t)}_{iSW} -\frac{1}{2}n^k\int^{\lambda}_{\lambda_e} \partial^lH_{kl}d\lambda'\nn\\
&&-\frac{1}{4}\perp^{ij}\int^{\lambda}_{\lambda_e}\int^{\lambda'}_{\lambda_e}\partial_i\partial_j\Bigg[n^kn^lH_{kl}  -2\left(\Phi+\Psi\right)\Bigg]d\lambda''d\lambda'\,.\eea
Here $I^{(s)}_{iSW}$ represents the gravitational iSW effect due to the cosmological scalar perturbations and was first calculated in \cite{Laguna:2009re}, and the term $I^{(t)}_{iSW}$ is the gravitational integrated  effect due to the cosmological tensor perturbations found in \cite{Fier:2021fbt}. They are given, respectively, by
\begin{align}
I^{(s)}_{iSW} \equiv \int^{\lambda}_{\lambda_e}\partial_{\tau}(\Phi+\Psi)d\lambda', \;\;\;
I^{(t)}_{iSW} \equiv n^kn^l\int^{\lambda}_{\lambda_e}\partial_{\tau}H_{kl}d\lambda'.
\end{align}
To connect the gravitational iSW effects with observations, we first note that observationally we find $\Phi \simeq \Psi$ \cite{Dodelson:2003ft}. Then, in the Fourier space, we have $\Phi_k = \Psi_k = -3H_0^2\delta_k^{(0)} D(t) /(2k^2 a)$, where $D(t)$ is the linear growth function and
$\delta_k^{(0)}$ the Fourier coefficient of the density perturbations at zero redshift with
\begin{align}
\label{eq4.42}
\left<\delta_k^{(0)}\delta_{k'}^{* (0)}\right>  = \frac{4k^4}{25H_0^4}P_k T^2,
\end{align}
where $P_k = 2\pi^2 {\Delta_R^{(0)}}^2/k^3$ is the power spectrum of curvature perturbations and $T(k)$ the transfer function.

Although the detection of \acp{GW} by TianQin or other space-based detectors  will not be affected by matter inhomogeneities, parameter estimation will clearly be affected by them.
TianQin is expected to be sensitive to the chirp mass and the luminosity distance to $\Delta\ln \mathcal{M} \sim 10^{-6}-10^{-2}$ and $\Delta\ln D_L \sim 10^{-4}-10^{-1}$ for low-redshift sources ($z < 0.5$), but also sensitive to high-redshift sources ($z \sim 5 - 10$) up to $\Delta\ln \mathcal{M} \sim 10^{-2}$ and $\Delta\ln D_L \sim 10^{-1}$ \cite{Feng:2019wgq, Wang:2019ryf, Zhu:2021aat}.
Then, it can be shown  that  $\xi$  will be a noise source in the use of standard sirens to measure the equation of state of dark energy through the redshifted luminosity distance.

Alternatively, as first pointed out in  \cite{Laguna:2009re}, we  could also view the above effects as a new link between \ac{EM} (EM)  measurements of density inhomogeneities and \ac{GW} observations. In order to achieve this goal,  we would first have to break the degeneracy between $(\mathcal{M}_z, z, \Upsilon)$ and $(D_L, z, \xi, \Upsilon)$. Given a coincident \ac{EM} and \ac{GW} detection, we could be able to achieve this, by electromagnetically determining the redshift and the component masses via host galaxy identification and correlations between galaxy luminosity and black hole mass. Another possibility is to use large-scale structure observations to measure $\delta_k$ and predict $\Upsilon$. Such measurements would then open up studies of cross-correlations between \acp{GW} and large-scale structure surveys of dark matter  and possibly dark energy. Furthermore, a detection of the cross-correlation between matter distribution and the \ac{GW} iSW effect could potentially be another test of GR since it would show that \acp{GW} propagate in the same metric as \ac{EM} radiation.

\subsubsection{Waveform systematics}

In searching for possible signatures of beyond \ac{GR} effect, an important question is to distinguish possible new physics from other effects \cite{Gupta:2024gun}.
Beside the environmental effect discussed above, there can also be waveform systematics in the modeling.
Such systematics can come from two reasons: the missing of physics and the inaccurate modeling.
The missing of physic can come from the neglect of the eccentricity, the spin and its precession, the recoil after merger, and so on, in the modeling of the dynamics of \ac{GW} sources.
For ground-based \ac{GW} detectors some of these effects are very small, and their impact on the estimation of source parameters can be negligible.
However, due to the high precision that can be achieved with a space-based detector, some of these effects may incorrectly show up as indicators for new physics if not properly included in the \ac{GW} waveform modeling.
The inaccurate modeling corresponds to numerical errors and truncation errors in a perturbative calculation of \ac{GW} waveforms.
To avoid this problem one must model the \ac{GW} waveforms to the required precision, which is for now still a great challenge for the waveform community.

For the detailed review of the currently available waveforms and the requirements on the accuracy of the waveforms, it can be found in \cite{LISAConsortiumWaveformWorkingGroup:2023arg} for LISA, while the situation is similar for TianQin.
For \acp{MBHB} and \acp{SBHB}, there exist several public \ac{NR} catalogs \cite{Boyle:2019kee,Hamilton:2023qkv,Healy:2022wdn,Jani:2016wkt}, and some surrogate models \cite{Blackman:2017pcm}.
But for data analysis, the IMRPhenom \cite{Ajith:2007qp} and EOBNR \cite{Buonanno:2007pf} waveform families are the most widely used waveforms.
However, there still exist {much work} to be done for {fully including} the eccentricity, spin precession, and higher modes {in the relevant portion of the source parameter space}.
For the \acp{EMRI}, the situation is more {complicated}, since the calculation of the second order self-force is very hard.
Beyond the preliminary Kludge waveforms \cite{Chua:2017ujo}, the Black Hole Perturbation Toolkit \cite{BHPToolkit}and the Black Hole Perturbation Club \cite{BHPC} provide a hub for GSF code development, while the Fast EMRI Waveforms package \cite{Katz:2021yft}, {provides} a flexible framework for rapid waveform generation.

\subsection{Summary of the section}

In this section, we discuss the prospect of using TianQin to study the nature of gravity, including detecting the key predictions of \ac{GR} in the strong field regime and search for possible signatures of beyond \ac{GR} effect.

TianQin has the capability to test several key predictions of \ac{GR} in the strong field regime.
This means that for these predictions of \ac{GR}, TianQin can either significantly increases the detection number or improve the precision of parameter measurement to a meaningful level.
For the former, for example, TianQin can increase the number of detectable higher modes and nonlinear modes of \acp{GW} to the level of a dozen, and detect a few \ac{GW} events with recognizable memory effect.
For the later, TianQin can test whether the astrophysical black holes are the Kerr black holes predicted by \ac{GR} to better than $\cO(10^{-2})\,$.

TianQin has the capability to significantly advance the front in searching for possible beyond \ac{GR} effects.
In general, TianQin can either improve on the existing constraint by orders of magnitude or produce a stringent constraint that has not been possible before.
For the former, for example, TianQin can improve the constraints on extra polarization modes, graviton mass, the dispersion coefficients $A_{0.5}$ and $A_1$, some \ac{MGT} couplings such as $\sqrt{\alpha_{\rm EdGB}}$, and so on, by at least an order of magnitude.
For the later, for example, TianQin can constrain possible deviations to the dominant \ac{QNM} frequencies to the 0.1 percent level.

The joint detection with other \ac{GW} detectors will bring significant scientific improvement. For instance, the joint detection with LISA for \acp{MBHB} is expected to improve the constraint on non-commutative theories by a factor of 2 over that of a single detector \cite{Shi:2022qno,Huang:2024ylf};
the multiband detection of \acp{SBHB} with ground-based \ac{GW} detectors such as \ac{ET} and CE is expected to improve the constraint on the \ac{EdGB} theory by about one order of magnitude \cite{Shi:2022qno}.

Some representative quantitative result from this section is listed in Table \ref{tab:sec2.B:MGs3}.

\begin{table}[ht]
\caption{Examples of current and projected constraints.}
\label{tab:sec2.B:MGs3}
\renewcommand{\arraystretch}{1.2}
\begin{center}
\begin{tabular}{|c|l|l|c|c|}
\hline
Category & Check point & Key parameter & Current constraints & TianQin \\
\hline
\multirow{4}{2.4cm}{Key predictions of \ac{GR}} & Higher modes & number of modes & 1 \cite{Isi:2019aib,Capano:2022zqm} & $> 7$ \cite{Shi:2024ttu}\\
\cline{2-5}
& Non-linear modes & number of events & none & $\cO(10)$ \cite{Shi:2024ttu} \\
\cline{2-5}
& Memory effect & number of events & none \cite{Hubner:2019sly,Hubner:2021amk} & $\cO(1)$ \cite{Sun:2022pvh} \\
\cline{2-5}
& Kerr hypothesis &  Quadrupole moment: $\delta\kappa$ & $\cO(10^2)$ \cite{Narikawa:2021pak} & $\cO(10^{-5})$ \cite{Zi:2021pdp,Kong:2024ssa}\\
\hline
\multirow{6}{2.4cm}{Possible signatures of beyond \ac{GR} effects}
& \ac{GW} polarization & {Amplitude ratio:} $\alpha_{v,s}$ & $<\cO(1)$ & $\cO(10^{-2})$  \cite{Xie:2022wkx}\\
\cline{2-5}
& \multirow{3}{2.6cm}{\ac{GW} propagation}
              & Speed: $\delta v_{gw}/c$ & $\cO(10^{-16})$ \cite{Baker:2017hug} & \\
& & Mass: $m_g$ & $<1.3\times10^{-23}\rm~eV/c^2$ \cite{LIGOScientific:2021sio} & $\cO(10^{-26})\rm~eV/c^2$ \\
& & Dispersion: $A_1$  & $<1.2\times10^{-32}\rm~eV$ \cite{LIGOScientific:2021sio} & $\cO(10^{-35})\rm~eV$ \\
\cline{2-5}
& \multirow{2}{2.6cm}{\ac{GW} generation}
              & Inspiral: $\sqrt{\alpha_{\rm EdGB}}$ & $<1.1{\rm~km}$ \cite{Wang:2023wgv} & $<\cO(10^{-1}){\rm~km}$ \cite{Shi:2022qno} \\
& & Ringdown: $\delta\omega_{22}$ & & $<\cO(10^{-3})$ \cite{Shi:2019hqa}\\
\hline
\multirow{2}{2.4cm}{Environmental effect}
& \ac{GW} generation & DM density: $\delta\rho_0/\rho_0$ & & $<\cO(10^{-2})$ \cite{Yuan:2024duo}\\
\cline{2-5}
& \ac{GW} propagation & Lens parameter $\delta\kappa_s/\kappa_s$ & & $<\cO(10^{-5})$ \cite{Lin:2023ccz} \\
\hline
\end{tabular}
\end{center}
\end{table}

\clearpage

%% file: sec3-new-phys.tex
\section{New matter and interaction with TianQin}\label{sec:tq_new}

{\it Section coordinator: Fa Peng Huang}

Apart from probing the nature of gravity as discussed in section \ref{sec:tq_phys}, TianQin also has the potential to probe fundamental physics in the non-gravitational sector, covering a wide range of topics.

\ac{SMPP} is currently the best theory describing the fundamental composition of the Universe.
According to \ac{SMPP}, all matter in the Universe are composed of six types of quarks, six types of leptons, four types of force carriers, and one Higgs boson.
Apart from gravity, there are three fundamental interactions: the strong, weak, and \ac{EM} interaction.
There is indication that the coupling constants of the \ac{EM}, weak, and strong interactions might become unified at an energy scale around the order $\cO(10^{16}$ GeV) \cite{ParticleDataGroup:2024cfk}.
But there are huge problems surrounding \ac{SMPP}.
The most outstanding ones include its lack of explanation for the matter-antimatter asymmetry and its complete omission of dark matter and dark energy: dark energy accounts for about 68\% of the energy density of the universe, dark matter accounts for about 27\% , while the particles known in \ac{SMPP} only account for about 5\%, and they are all matter and not antimatter.
Another notable problem about \ac{SMPP} is its lack of control on its dozens of free parameters and on the detailed shape of the Higgs potential.
Appropriate Higgs potential may explain the matter-antimatter asymmetry, lead to the production of some dark matter, and may also lead to strong first order \ac{EWPT} that generate \ac{GW} signals detectable by TianQin.

\acp{PBH} are hypothetical black holes which formed by gravitational collapse of the density fluctuation directly in the very early universe, long before the recombination.
The mass of such \acp{PBH} depends on the typical size and formation time.
\acp{PBH} from different mass windows are relevant for various aspects of the early universe, such as making up all the dark matter, being responsible for ultra-short timescale microlensing events, contributing a portion to black hole binaries, seeding the early formed galaxies, etc.
Such \acp{PBH} are accompanied by the \ac{SGWB} induced by the same curvature perturbation, of which the frequencies are also determined by the size of the fluctuation.
For instance, \ac{PBH} of $10^{-16}\sim10^{-11} \mSun$ can account for all the dark matter, and the corresponding \ac{SGWB}  peaks at mHz.
Therefore, such concomitant \ac{SGWB} of the \ac{PBH} dark matter is detectable by TianQin.

The slow-roll inflation model remains one of the most compelling frameworks for explaining the epoch of cosmic inflation.
For this model to operate effectively, the inflaton field must traverse a distance comparable to the Planck scale and  transfer its energy into \ac{SMPP} particles at the conclusion of inflation.
This energy transfer necessitates a coupling between the inflaton field and other fields, often referred to as spectator fields.
Consequently, it becomes clear that due to the evolution of the inflaton field, the properties of the spectator fields can undergo significant changes throughout the inflationary period.
These changes can trigger phase transitions \cite{An:2020fff,An:2022cce}, which may give rise to a variety of rich phenomenological consequences.
These include \ac{GW} signals, curvature perturbations, and primordial non-Gaussianities.
Such phase transitions could also play a pivotal role in addressing fundamental questions such as the origin of dark matter and baryogenesis.
The \acp{GW} associated with these phenomena may potentially be detectable by TianQin.

The cosmic string might form in the early Universe after spontaneous symmetry breaking through the Kibble mechanism.
In the post-inflationary scenario, due to the continuous emission of \acp{GW} after which enter the scaling regime, the \ac{GW} signal of the cosmic string can include contributions from radiation and matter dominated Universe.
Therefore, the \ac{GW} spectrum of cosmic string can cover a wide range, i.e., from Nanohertz to kHz.
\ac{GW} detection by TianQin and other detectors can help to reveal possible new physics around even the grand unification scale in the early Universe.

In this section, we discuss the prospect of using TianQin to probe new physics related to all the above topics, mainly focusing on \ac{BSM} particles physics, primordial black holes, phase transition during inflation, and cosmic strings.

\subsection{BSM particle physics}

In this subsection, we discuss the search of \ac{BSM} particle physics with \acp{GW}, focusing on probing the Higgs potential, the origin of matter-antimatter asymmetry and the nature of dark matter by \acp{GW}.
Probing of more \ac{BSM} physics is further discussed with a special type of \acp{ECO}.

\subsubsection{New Higgs potential and FoPT}

The nature of the \ac{EWPT} in the early universe is crucial for understanding the true shape of the Higgs potential. A first-order \ac{EWPT}, characterized by a strong, discontinuous change in the Higgs field, can produce significant \ac{GW} signals which are detectable at space-based interferometers such as TianQin. These signals provide a unique probe into the high-energy physics governing the early universe and the properties of the Higgs potential.

In \ac{SMPP}, the Higgs potential is responsible for spontaneous symmetry breaking, giving rise to the masses of the $W$ and $Z$ bosons and \ac{SMPP} fermions. The shape of the Higgs potential at high temperatures determines the nature of the \ac{EWPT}. In the \ac{SMPP}, the phase transition is a smooth crossover \cite{Kajantie:1996mn,Gurtler:1997hr,Csikor:1998eu}, but many \ac{BSM} theories predict a \ac{FoPT} due to modifications of the Higgs potential.

The Higgs potential in the \ac{SMPP} is given by
\bea V(h) = \frac{1}{2}\mu^2 h^2 + \frac{\lambda}{4} h^4\,,\eea
where $h$ is the Higgs field, $\mu^2$ and $\lambda$ are parameters.

\begin{figure}[htbp!]
\begin{center}
\includegraphics[width=0.5\textwidth]{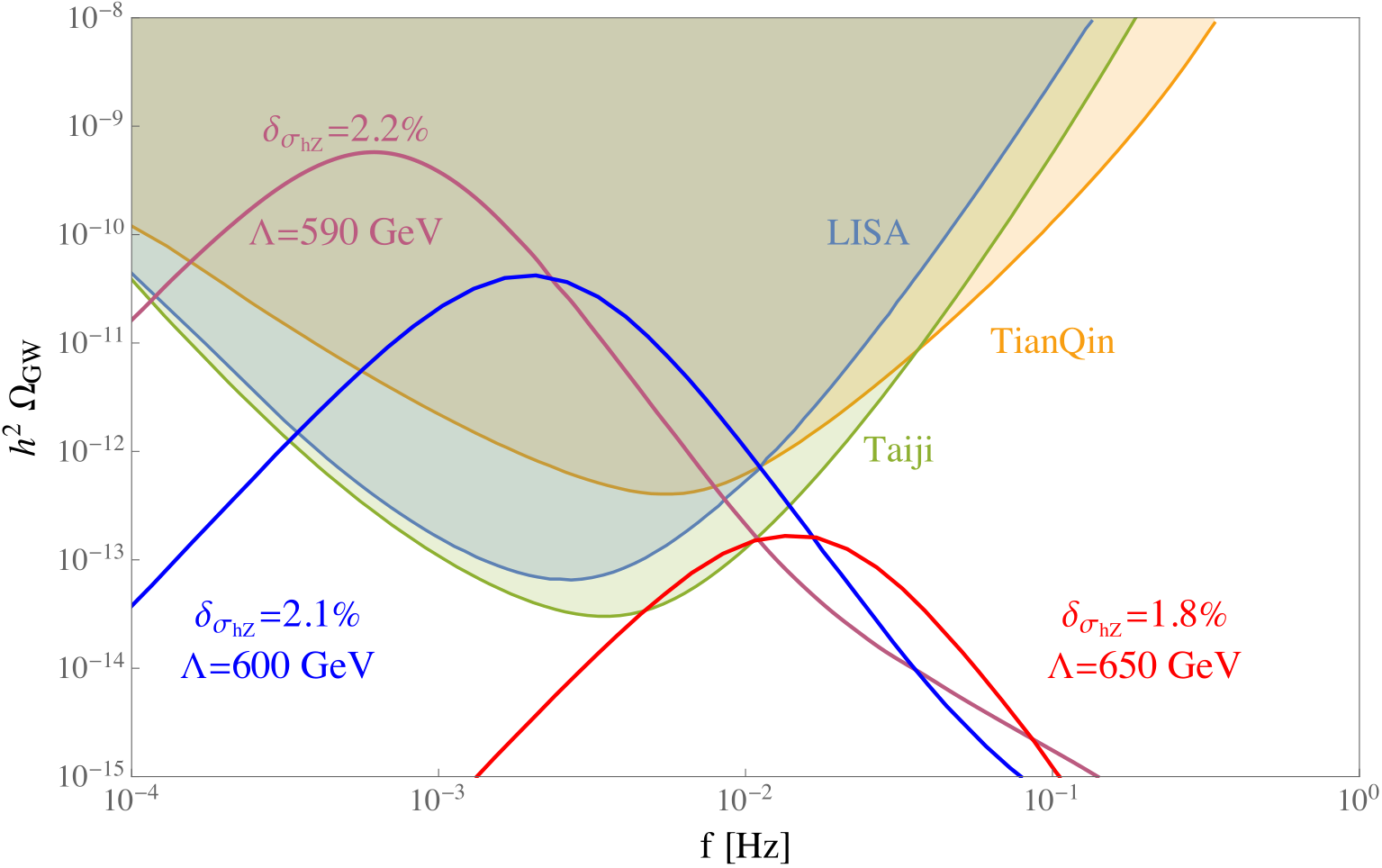}
\caption{General prediction of \ac{GW} signals in the \ac{SMPP} effective field theory under the condition of \ac{EWPT} \cite{Huang:2015izx, Huang:2016odd}.}
\label{fig:gwdim6}
\end{center}
\end{figure}

From the perspective of \ac{SMPP} effective field theory,
a generic Higgs potential can be obtained in the form of the dimension-6 operators \cite{Zhang:1992fs,Grojean:2004xa,Huang:2015izx,Huang:2016odd,Cai:2017tmh},
\bea V(h)=\frac{1}{2} \mu^2 h^2-\frac{\lambda}{4} h^4+\frac{1}{\Lambda^2} h^6\,.\label{vdim6}\eea
This generic Higgs potential usually produces first-order \ac{EWPT} with associated \ac{GW} as shown in FIG.~\ref{fig:gwdim6}.

A first-order \ac{EWPT} involves nucleations \cite{Coleman:1977py,Callan:1977pt,Linde:1980tt,Linde:1981zj}, expansion \cite{Cai:2020djd,Wang:2022txy,Wang:2023kux}, and percolations \cite{Turner:1992tz,Ellis:2018mja} of true vacuum bubbles in the false vacuum. The dynamics of these bubbles produce \acp{GW}, whose characteristic parameters include:
\begin{itemize}
\item {Strength of the Phase Transition ($\alpha$)}: The ratio of the vacuum energy density released during the phase transition to the radiation energy density (note that there are different version of this strength factor when adapting to different fitting templates of numerical simulations), $\alpha = \frac{\Delta \rho_\text{vac}}{\rho_\text{rad}}\,$.
\item {Bubble wall velocity ($v_w$)}: The speed at which the true vacuum bubbles expand, which can be calculated from various approaches (see \cite{Yuwen:2024hme} for the most recent summary).
\item {Duration of the phase transition ($\beta^{-1}$)}: The inverse of the timescale of the phase transition, which is usually measured in the unit of the Hubble time scale at the phase transition,
\bea\frac{\beta}{H_*}=T_*\frac{\mathrm{d}}{\mathrm{d}T}\frac{S_3(T_*)}{T_*}\,.\eea
\end{itemize}

\ac{EWPT} provides three sources of phase transition \acp{GW}: bubble collision, sound wave, and turbulence.
\begin{itemize}
\item Wall collisions \cite{Jinno:2016vai,Jinno:2017fby}
\bea h^2\Omega_\mathrm{env}&=&1.67\times10^{-5}\left(\frac{100}{g_{*}}\right)^\frac13\left(\frac{H_*}{\beta}\right)^2\left(\frac{\kappa_\phi\alpha}{1+\alpha}\right)^2 \frac{0.48v_w^3}{1+5.3v_w^2+5v_w^4}\nn\\
&&\times\left[c_l\left(\frac{f}{f_\mathrm{env}}\right)^{-3}+c_m\left(\frac{f}{f_\mathrm{env}}\right)^{-1}+c_h\left(\frac{f}{f_\mathrm{env}}\right)\right]^{-1}\,,\label{eq:wall1:env}\eea
where $(c_l, c_m, c_h) = (0.064,1-c_l-c_h,0.48)$ and $ \kappa_\phi $ represents the fraction of vacuum energy converted into the scalar-wall gradient energy \cite{Cai:2020djd}. $f_{\mathrm{env}}$ is the peak frequency of bubble wall envelope collisions,
\bea f_\mathrm{env}=1.65\times10^{-5}{\mathrm{Hz}}\left(\frac{g_{*}}{100}\right)^\frac16\left(\frac{T_*}{100\,\mathrm{GeV}}\right)\frac{0.35(\beta/H_*)}{1+0.069v_w+0.69v_w^4}\,,\eea
where $T_\ast$ is the phase transition temperature (see, for example, \cite{Cai:2017tmh} for various definitions of characteristic temperatures).
See also \cite{Zhong:2021hgo} for the Hubble-expansion effect on suppressing the overall amplitude of the \ac{GW} energy density spectrum.
\item Sound waves \cite{Hindmarsh:2013xza,Hindmarsh:2015qta,Hindmarsh:2017gnf}
\bea \Omega_{\mathrm{sw}}h^2 \simeq & 2.65 \times 10^{-6} \Upsilon_{\mathrm{sw}} \left(\frac{H_*}{\beta}\right)\left(\frac{\kappa_v \alpha}{1+\alpha}\right)^2\left(\frac{100}{g_{*}}\right)^{1 / 3}
 v_w \left(f / f_{\mathrm{sw}}\right)^3 \left(\frac{7}{4+3\left(f / f_{\mathrm{sw}}\right)^2}\right)^{7 / 2}\,,\eea
where $\kappa_v$ represents the fraction of vacuum energy that transfers into the acoustic waves \cite{Espinosa:2010hh} (see \cite{Giese:2020znk,Giese:2020rtr,Wang:2020nzm,Wang:2023jto,Wang:2022lyd} for various \ac{EoS} generalizations beyond the simple bag model or even the $\nu$-model, and also \cite{Cai:2018teh,Giombi:2023jqq} for the Hubble-expansion effect on this fluid-motion efficiency factor). The peak frequency of sound waves is red-shifted as
\bea f_{\mathrm{sw}} \simeq 1.9 \times 10^{-5} \mathrm{~Hz} \frac{1}{v_w}\left(\frac{\beta}{H_*}\right)\left(\frac{T_{*}}{100~ \mathrm{GeV}}\right)\left(\frac{g_\star}{100}\right)^{1 / 6}\,.\eea
Here $\Upsilon_{\mathrm{sw}}$ is the suppression factor from the Hubble-expansion effect on the overall amplitude of the \ac{GW} energy density spectrum \cite{Guo:2020grp},
\bea \Upsilon_{\mathrm{sw}} = \left(1-\frac{1}{\sqrt{1+2\tau_{\mathrm{sw}}H_*}}\right),\quad \text{with} \quad \tau_{\mathrm{sw}}H_* \approx (8\pi)^{\frac{1}{3}}\frac{v_w (H_*/\beta)}{ \sqrt{3\kappa_v\alpha/(4+4\alpha)}}\,,\eea
which can be analytically calculated from an analytical model of sound waves called the sound shell model \cite{Hindmarsh:2016lnk,Hindmarsh:2019phv}. Note that the infrared scaling in the original sound shell model was corrected in a hydrodynamical sound shell model \cite{Cai:2023guc}, which was later confirmed in both analytical estimation \cite{RoperPol:2023dzg} and numerical simulation \cite{Sharma:2023mao}.
The contribution from sound waves usually dominates the total \ac{GW} energy density spectrum when most of bubbles collide with each other long after they have approached the terminal wall velocity \cite{Cai:2020djd}.
\item MHD turbulence \cite{Caprini:2009yp,Binetruy:2012ze}
\bea \Omega_{\mathrm{turb}}h^2 \simeq 3.35\times 10^{-4} \left(\frac{H_* v_w}{\beta}\right)\left(\frac{\kappa_{\mathrm{turb}} \alpha}{1+\alpha}\right)^{3 / 2}\left(\frac{100}{g_*}\right)^{1 / 3} \frac{\left (f / f_{\text {turb}}\right)^3}{\left(1+f / f_{\text {turb}}\right)^{11 / 3}\left(1+8 \pi f / H_*\right)}\,,\eea
where $H_*$ is expressed as
\bea H_*=1.65 \times 10^{-5}  ~\mathrm{Hz}\left(\frac{T_{*}}{100~\mathrm{GeV}}\right)\left(\frac{g_*}{100}\right)^{1/6}\,,\eea
and the peak frequency of turbulence processes is
\bea f_{\text {turb}} \simeq 2.7 \times 10^{-5} \mathrm{~Hz}\frac{1}{v_w} \left(\frac{\beta}{H_*}\right)\left(\frac{T_{*}}{100~ \mathrm{GeV}}\right)\left(\frac{g_*}{100}\right)^{1 / 6}\,.\eea
while $\kappa_{\mathrm{turb}}=\tilde{\epsilon} \kappa_v$ represents the efficiency of vacuum energy being converted into turbulent flow, which is usually taken to be negligible with $\tilde{\epsilon}\lesssim5\%\sim10\%$ \cite{Caprini:2015zlo}.
\end{itemize}

TianQin is designed to detect \acp{GW} in the frequency range 0.1 mHz to 1 Hz. The \ac{GW} signals from a first-order \ac{EWPT} typically fall within this range, making these detectors suitable for detecting such signals, enabling them to probe the early-Universe phase transitions.

The shape of the Higgs potential influences the phase transition parameters, and thus the resulting \ac{GW} signal. Different \ac{BSM} scenarios modify the Higgs potential differently, leading to variations in the strength and duration of the phase transition:
\begin{itemize}
\item $\alpha$: A stronger phase transition produces more powerful \ac{GW} signals. The strength is directly related to the height and width of the potential barrier in the Higgs potential.
\item $v_w$: Faster bubble wall velocities can enhance the \ac{GW} signal. The velocity depends on the potential's shape and interactions between the Higgs field and the plasma.
\item $\beta^{-1}$: Shorter phase transitions produce higher frequency \ac{GW} signals. The duration is influenced by the temperature dependence of the Higgs potential and the efficiency of bubble nucleation and growth.
\end{itemize}

Several theoretical models predict modifications to the Higgs potential resulting in a first-order \ac{EWPT}, for examples (see \cite{Cai:2022bcf} for a recent summary):
\begin{itemize}
\item {Supersymmetry}: Predicts additional scalar fields that can lead to a strong first-order \ac{EWPT} \cite{Carena:1996wj,Delepine:1996vn,Carena:2008vj}.
\item {Scalar Extensions}: Adding a singlet, doublet or triplet scalar field modifies the Higgs potential and can induce a first-order \ac{EWPT} \cite{Cao:2017oez,Huang:2017rzf}.
\item {Composite Higgs Models}: In these models, the Higgs is a composite particle, and the dynamics of the strong sector can lead to a \ac{FoPT} \cite{Fujikura:2023fbi}.
\end{itemize}

Different models predict different \ac{GW} signatures that can be tested with TianQin.
For example, a stronger phase transition in supersymmetry models might produce a \ac{GW} signal with a higher amplitude and lower frequency compared to singlet scalar extensions.
Studying \acp{GW} from first-order \ac{EWPT}s provides a unique insight into the early universe's physics and the true shape of the Higgs potential.
TianQin offers the sensitivity required to detect these signals and test \ac{BSM} theories. Correlating the properties of \ac{GW} signals with the Higgs potential's shape can reveal new physics beyond \ac{SMPP}.

Although new physics can be explored by fitting different \ac{EWPT} models to the \ac{GW} data, high degeneracy can be expected among the models.
A more appealing approach is to test an effective field theory description of \ac{EWPT}, which would require a clear separation of scales of new physics \cite{Camargo-Molina:2021zgz,Cai:2022bcf}.

In this section, we discuss the importance of TianQin for advancing our understanding of the Higgs potential and early universe dynamics.
Continued theoretical and experimental efforts are essential for fully exploiting the potential of these observations.

\subsubsection{Matter-antimatter asymmetry}

The matter-antimatter asymmetry of our universe is one of the fundamental puzzles in modern cosmology and particle physics.
According to \ac{SMPP}, matter and antimatter should have been created in equal amounts during the Big Bang.
However, our universe is predominantly composed of matter, with very little antimatter observed \cite{Planck:2015fie}.
This discrepancy suggests new physics beyond \ac{SMPP} that can explain the observed asymmetry.
TianQin can detect \acp{GW} from various cosmic events, including those associated with phase transitions that could produce the necessary conditions for generating the asymmetry, thus offering a novel approach to probing the early universe and unraveling the mechanisms behind the matter-antimatter asymmetry.

Baryogenesis is the theoretical process that describes the generation of a matter-antimatter asymmetry (baryon asymmetry) in the universe. The three Sakharov conditions necessary for baryogenesis are \cite{Sakharov:1967dj}:
1) Baryon number violation;
2) C and CP violation (where C is charge conjugation symmetry, and CP is the combination of C and parity symmetry); and
3) Departure from thermal equilibrium.
Various mechanisms have been proposed to achieve baryogenesis, including electroweak baryogenesis \cite{Kuzmin:1985mm,Cline:2006ts}, leptogenesis \cite{Davidson:2008bu,Buchmuller:2005eh,Huang:2022vkf,Borah:2022cdx,Chun:2023ezg}, and mechanisms involving \acp{FoPT} in the early universe \cite{Azatov:2021irb,Baldes:2021vyz,Baker:2021zsf}.
These mechanisms often offer specific predictions for the \ac{GW} signal that can be tested.

The \ac{GW} signals from phase transitions provide a direct probe of the early universe conditions that could lead to baryogenesis. Different baryogenesis models predict varying \ac{GW} spectra based on the specifics of the phase transition. For instance:
\begin{itemize}
\item {Electroweak Baryogenesis}: This scenario involves a strong first-order \ac{EWPT} which can produce \acp{GW}. The strength and frequency of the \ac{GW} signal depend on the details of the Higgs potential and interactions.
\item {Leptogenesis}: \acp{GW} in this context can arise from phase transitions in models where lepton number violation occurs, later converted to baryon asymmetry through sphaleron processes.
\item {Other \ac{BSM} Models}: Various other models, such as those involving additional scalar fields or symmetry-breaking mechanisms e.g. \cite{1993PhRvD..47.4244D}, can produce distinct \ac{GW} signatures that reflect the dynamics of the phase transition.
\end{itemize}
By analyzing the \ac{GW} signals detected by TianQin, one can constrain the properties of these phase transitions and test the viability of different baryogenesis scenarios. The shape, amplitude, and frequency of the \ac{GW} spectrum provide crucial information about the underlying physics driving the matter-antimatter asymmetry.

\subsubsection{Particle dark matter candidates}

\begin{figure}[htbp!]
\begin{center}
\includegraphics[width=0.45\textwidth]{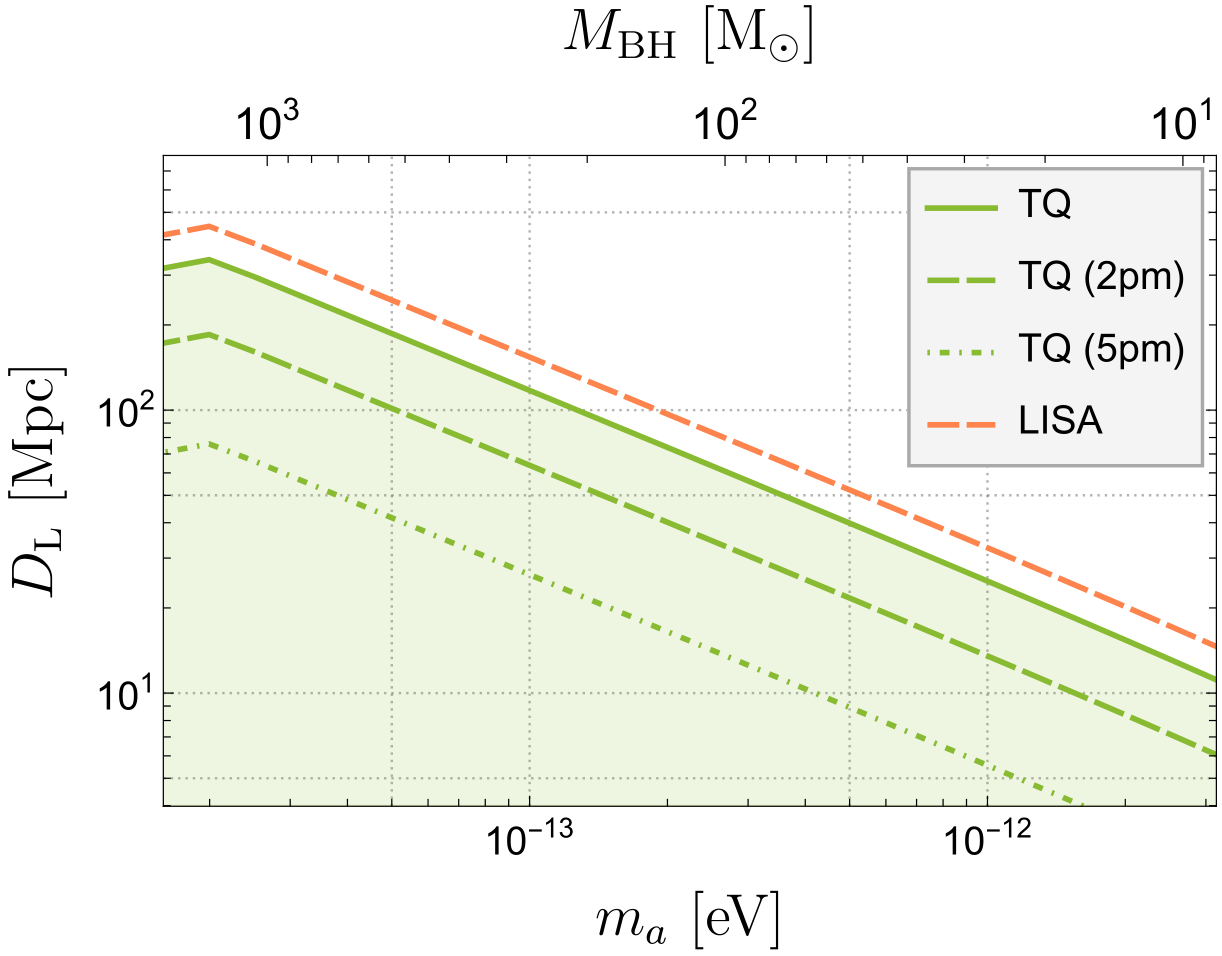}\qquad\qquad
\includegraphics[width=0.45\textwidth]{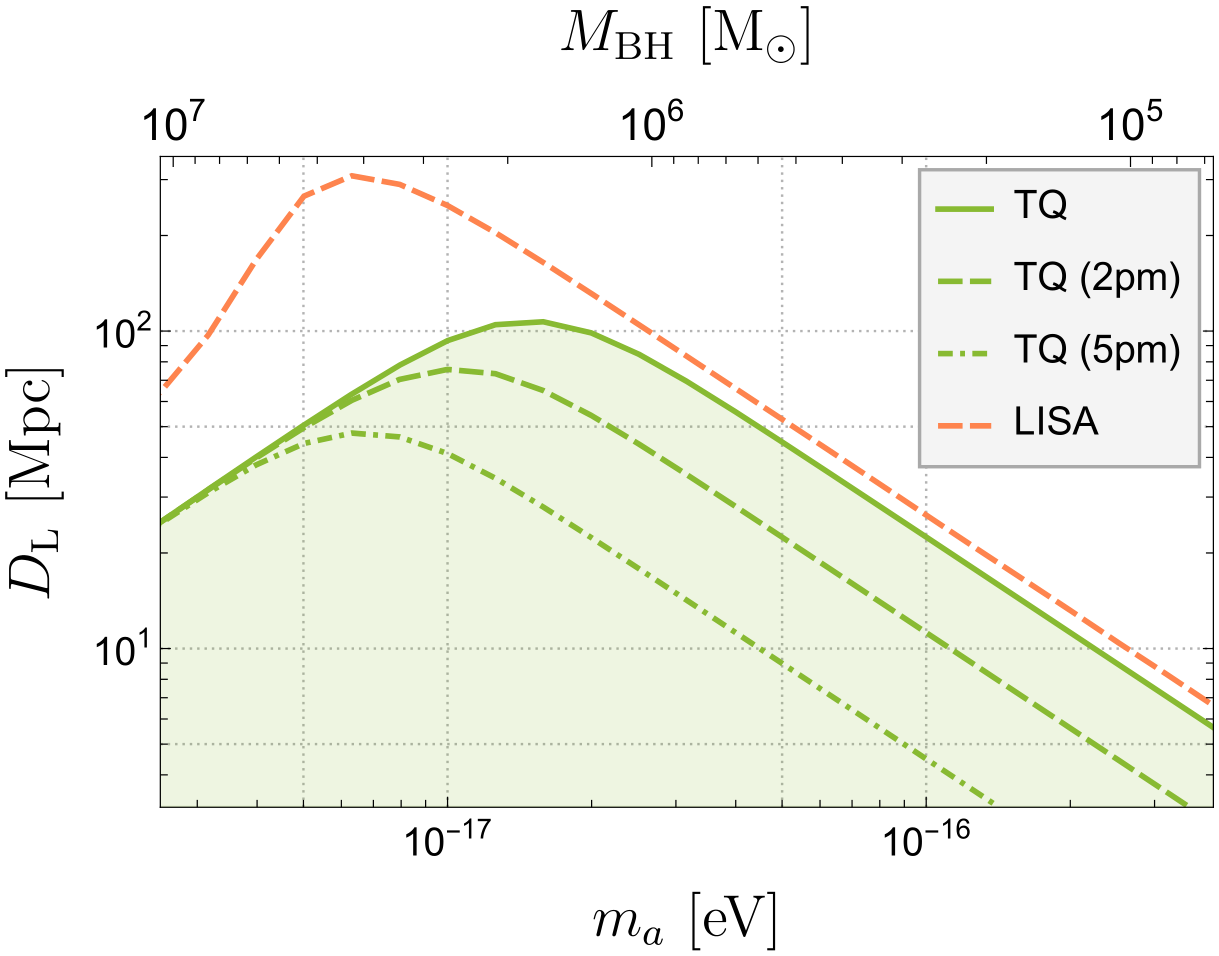}
\caption{The luminosity distance ranges that TianQin can detect:
(Left) the effect of the superradiance cloud on the binary \acp{GW} and
(Right) the \acp{GW} radiated by the superradiance cloud.
The figures are based on the results in \cite{Xie:2022uvp}.}
\label{fig:dm-aedm}
\end{center}
\end{figure}

\acp{GW} are ripples in spacetime produced by violent astrophysical events such as merging black holes or neutron stars.
These waves can carry information about the mass and energy involved in such events, offering potential insights into the dark sector.
Boson clouds formed with dark matter particles such as axions through the superradiance process surrounding the black holes can affect the orbital evolution of binary systems, and hence alter the radiated \acp{GW} \cite{1971SR,1972SR,1973SR,Zouros:1979iw,Detweiler:1980uk,Dolan:2007mj,Arvanitaki:2014wva,2020SRbook,Zhang:2019eid,Xie:2022uvp}.
These novel observational effects can be used to reveal the property of dark matter in the vicinity of black holes \cite{Eda:2013gg,Eda:2014kra,Zhang:2019eid,Xie:2022uvp}.
The macroscopic clouds of scalar and vector boson can also radiate \acp{GW} directly by means of the pair annihilation of the bound-state particles and the energy-level transition in the clouds \cite{Arvanitaki:2010sy,Arvanitaki:2014wva,Arvanitaki:2016qwi}.
The \ac{GW} signal of boson clouds is quasi-monochromatic and the frequency depends on the mass of boson particle.
These \ac{GW} signals can be detected individually with ground-based and space-based \ac{GW} detectors \cite{Brito:2017wnc,Brito:2017zvb,Baryakhtar:2017ngi,Siemonsen:2019ebd,Palomba:2019vxe,LIGOScientific:2021rnv}, and can also contribute to the stochastic background \cite{Brito:2017wnc,Brito:2017zvb,Tsukada:2020lgt,Yang:2023aak}.
Taking the axion cloud as an example, FIG.~\ref{fig:dm-aedm} shows the luminosity distance range in which TianQin can detect the axion effect in modifying binary \ac{GW} radiations and the luminosity distance range in which TianQin can detect the quasi-monochromatic \acp{GW} radiated by the axion cloud with different axion masses.
In the figures here and below, the results for LISA are also shown as comparison.

Some theories propose that dark matter particles could form compact objects or dense clumps. When these clumps interact with regular matter or black holes, they might generate distinctive \ac{GW} signals. For instance, \acp{GW} could be produced by the disruption of a dark matter clump by a neutron star.

\begin{figure}[htbp!]
\begin{center}
\includegraphics[width=0.45\textwidth]{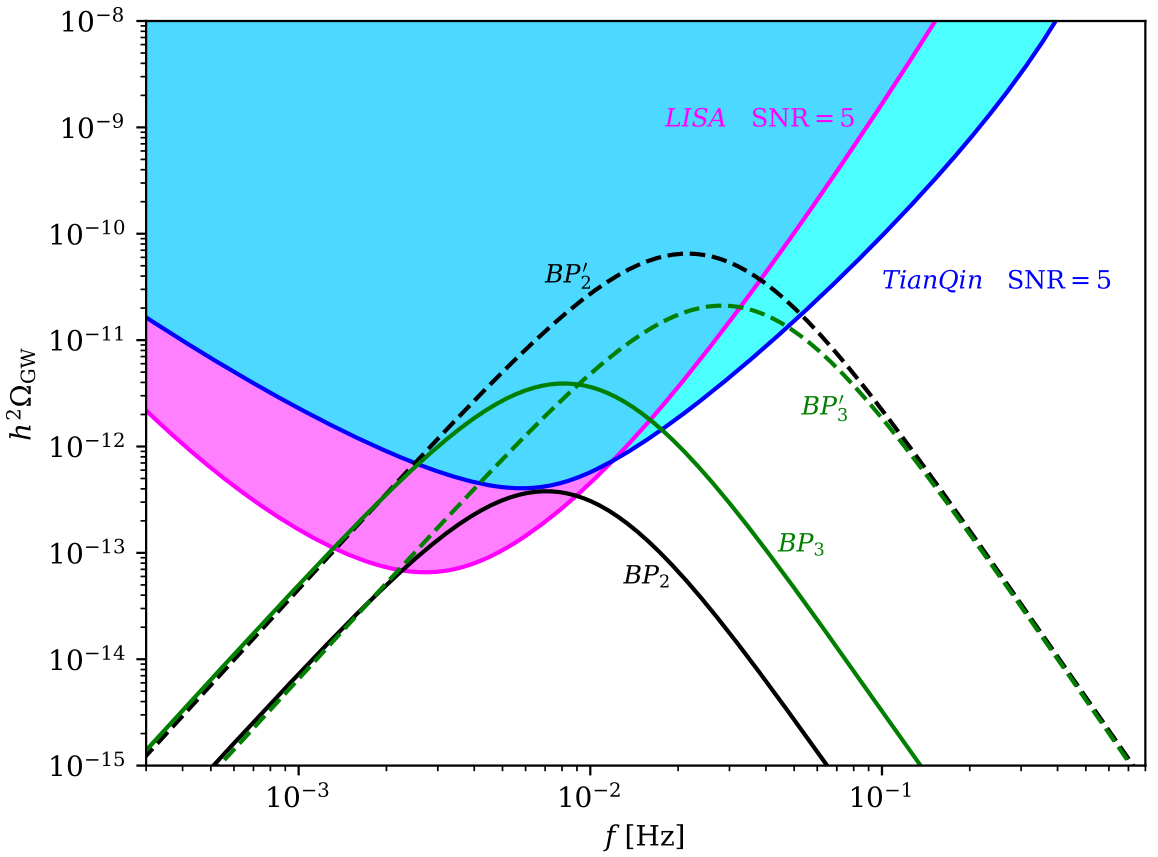}\qquad
\includegraphics[width=0.45\textwidth]{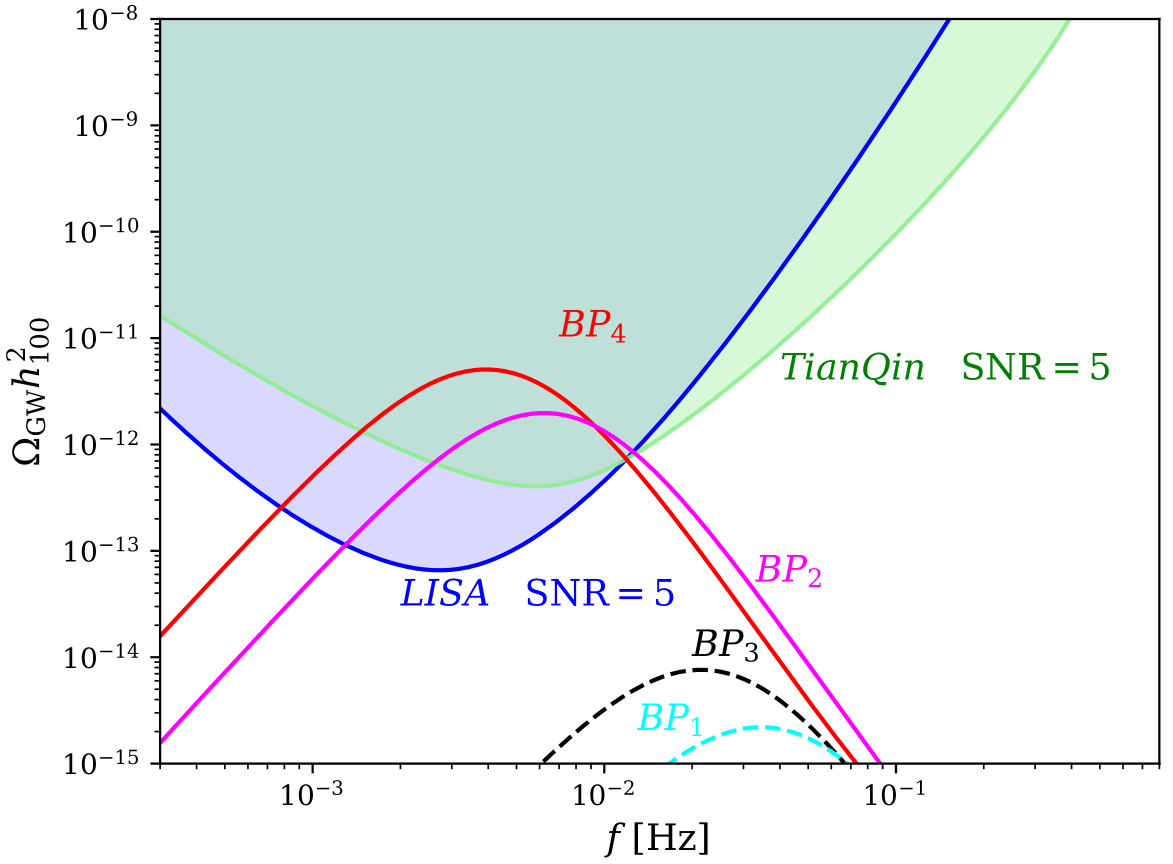}
\caption{GW signals accompanying the production of (Left) filtered dark matter and (Right) soliton dark matter under the condition of first-order \ac{EWPT}. The figures are taken from \cite{Jiang:2023nkj} and \cite{Jiang:2024zrb}, respectively.}
\label{fig:gwfdm}
\end{center}
\end{figure}

Models of dark matter involving new physics at high energy scales often predict phase transitions in the early universe, which could produce a stochastic background of \acp{GW}. The detection of such a background by current or future \ac{GW} detectors would be a groundbreaking confirmation of these models. Different dark matter models give different predictions for phase transition parameters like phase transition strength \cite{Wang:2020jrd}, phase transition duration, and the bubble wall velocity \cite{Moore:1995si,Wang:2020zlf,Jiang:2022btc,Laurent:2022jrs}.
Besides, the phase transitions themselves provide new mechanisms for dark matter production. Typical mechanisms include: filtered dark matter \cite{Baker:2019ndr,Chway:2019kft,Jiang:2023nkj}, production during relativistic bubble expansion \cite{Azatov:2021ifm,Baldes:2021vyz}, soliton dark matter \cite{Krylov:2013qe,Huang:2017kzu,Hong:2020est,Jiang:2023qbm,Jiang:2024zrb}, and so on.
In the left of FIG.~\ref{fig:gwfdm} , we show the \ac{GW} signals accompanying the production of filtered dark matter.
The solid lines represent the original \ac{GW} spectrum and the dashed lines represent the \ac{GW} signals which incorporate the details of bubble dynamics.
In right of FIG.~\ref{fig:gwfdm}, we show \ac{GW} signals accompanying the production of soliton dark matter under the condition of first-order \ac{EWPT}.

Despite the promise to help search for particle dark matter with detectors like TianQin, several challenges remain.
The interpretation of \ac{GW} data in terms of particle dark matter properties is complex and requires careful modeling of both the \ac{GW} sources and the dark matter interactions.
Moreover, distinguishing potential dark matter signals from other astrophysical sources requires highly sensitive and precise measurements.

\subsubsection{Charged exotic compact objects}\label{sec:3b:ceoc}
{\it Subsection coordinator: Sang Pyo Kim}

The Einstein-Maxwell theory has solutions for charged black holes known as Reissner-Nordstr\"{o}m  or rotating Kerr-Newman black holes.
These black holes have electric and/or magnetic charges as parameters (hairs) (for review, see \cite{stephani2009exact,griffiths2009exact}).
The upper bound for black hole charge in the Einstein-Maxwell theory comes from the cosmic censorship: $Q\leq M$ for Reissner-Nordstr\"{o}m  black holes and $Q \leq(M^2 - J^2/M^2)^{1/2}$ for Kerr-Newman black holes. Astrophysical black holes, however, cannot saturate this bound.
The huge disparity between the gravitational and electrostatic interactions in \ac{SMPP} strongly suppresses any significant accretion of charges into black holes formed from gravitational collapse \cite{page2006evidence}.

Black holes under special conditions can acquire charges.
For examples, the rotating black holes placed in magnetic fields, however, can acquire charges \cite{wald1974black}, and, for instance for \acp{MBH}, the accredited charges from selective accretion due to the mass difference between protons and electrons are $Q_{\rm max} =\frac{M}{4 \times 10^6\mSun}\times 3.1 \times 10^{8} {\rm C}$ while the upper bound for induced charges due to a magnetic field on the horizon is $Q_{\rm max} = (\frac{M}{4 \times 10^6\mSun})^2 (\frac{B_H}{10 {\rm G}})\times2.3 \times 10^{15} {\rm C}$ \cite{Zajacek:2018ycb}.
The charged black holes are an essential ingredient to the Blandford-Znajeck process for \acp{GRB} \cite{blandford1977electromagnetic,Lee:1999se} and magnetic Penrose processes for very high energy cosmic rays \cite{wagh1985revival}.
The Einstein-Maxwell theory with Dirac magnetic monopoles has charged Reissner-Nordstr\"{o}m  and Kerr-Newman black holes with electric and magnetic charges \cite{griffiths2009exact}.
The Dirac magnetic monopoles make the Maxwell theory symmetric and explain the quantization of electric charges, and the Yang-Mills theory predicts nonabelian magnetic monopoles \cite{Shnir:2005vvi}.
In \ac{SMPP} 't Hooft-Polyakov monopoles can be produced through phase transition \cite{t1974magnetic,Polyakov:1974ek}.
Dyons in nonabelian gauge theory have both electric and magnetic charges \cite{julia1975poles}.
These hypothetical particles would have been produced during the cosmic phase transition in the early universe and remain as remnants in individual entities or magnetic black holes \cite{lee1992black,Maldacena:2020skw} or dyonic black holes \cite{kasuya1982exact}.
The black holes with electric and magnetic charges may thus viewed as \acp{ECO} signaling the presence of new physics.

Black hole binaries with electric and/or magnetic charges have a distinct feature that \ac{EM} waves can be emitted during the inspiral, not to mention of \acp{GW}.
At the lowest PN order, the orbital motion of binaries with electric charges is governed by the \ac{ISL}of gravitational and electric interactions and emits both \acp{GW} and \ac{EM} waves \cite{Cardoso:2016olt,Liebling:2016orx,Toshmatov:2018tyo,Bai:2019zcd,Allahyari:2019jqz,Liu:2020cds,Christiansen:2020pnv,Wang:2020fra,Bozzola:2020mjx,Kim:2020bhg,Cardoso:2020nst,Cardoso:2020iji, McInnes:2020gxx,Bai:2020ezy,Bozzola:2021elc,McInnes:2021frb,Hou:2021suj,Benavides-Gallego:2021the}, in which the modulation frequency of \ac{EM} signals, twice of that of \acp{GW}, will signal the existence of charged black holes through multimessenger observations \cite{Liu:2020cds}.
Dyonic black hole binaries with both electric and magnetic charges have a generalized angular momentum known as the Laplace-Runge-Lenz vector which fixes the direction of Poincare cone on which the Keplerian orbits reside, and thus exhibit interesting and unique observational characteristics \cite{Liu:2020vsy,Liu:2020bag,liu2022merger}.

Charged black holes become extremal by saturating charges without violating the cosmic censorship. The extremal black holes do not emit Hawking radiation and change the physics of \acp{PBH} \cite{carr2020primordial,carr2021constraints,Carr:2023tpt}.
Near-extremal charged black holes are stable against perturbations \cite{Hod:2012wmy} and have been proposed as a candidate for dark matter \cite{kritos2022mergers}.
Charged black holes emit pairs of charged particles and antiparticles, known as Schwinger mechanism \cite{gibbons1975vacuum,Ruffini:2009hg}.
Even without Hawking radiation of charged pairs, near-extremal charged black holes can still emit charged pairs \cite{Chen:2012zn,Chen:2016caa,Chen:2017mnm} and in the de Sitter space \cite{Chen:2020mqs}.
Near-extremal charged black holes that obey the Breitenlohner-Freedman bound are stable against both Hawking radiation and Schwinger emission and may be dark matter.
In \ac{SMPP} the Breitenlohner-Freedman bound cannot be satisfied unless charged pairs are emitted in high angular momenta, and near-extremal charged black holes are prone to produce charged pairs, the lightest pair being electron-positron. Strong magnetic fields produce
monopole-antimonopole pairs \cite{Affleck:1981ag}, but the emission of monopole pair from magnetic black holes is exponentially suppressed due to heavy monopole mass, and magnetic black holes can be another candidate of dark matter.
The absorption cross section of \ac{EM} waves by extremal Reissner-Nordstr\"{o}m  black holes is $\sigma_{\rm abs} \approx (\frac{M}{10^{15} {\rm g}})^2 (\frac{10^{-13} {\rm cm}}{\lambda})^6 \times 3.1 \times 10^{-23} {\rm cm}^2$ \cite{Oliveira:2011zz}, and thus small mass \acp{ECO} are hard to be optically observed except for very high energy photons but become gravitationally important.

\subsection{Primordial black holes}\label{s:PBH}

{\it Subsection coordinator: Shi Pi, Konstantin Postnov}

\acp{PBH} are hypothetical black holes which form by gravitational collapse of the high peaks of density fluctuations directly in the very early universe.
Such idea was proposed more than 50 years ago \cite{Zeldovich:1967lct,Hawking:1971ei,Carr:1974nx,Meszaros:1974tb,Carr:1975qj,Khlopov:1985jw,1981SvA....25..406P} and has recently attracted much attention after the discovery of \acp{GW} by LIGO.

The conventional mechanism for the formation of \acp{PBH} assumes that the fluctuation of the energy density at the cosmological horizon may accidentally become $\delta \rho/\rho\sim 1$ at the horizon reentry, such that it turns inside its own gravitational radius, decouples from the Hubble flow, and creates a black hole. Therefore, the mass of such a \ac{PBH} is about that comprised inside the Hubble horizon \cite{Carr:2016drx,Green:2020jor},
\begin{equation}
M_\mathrm{PBH}\approx\alpha\frac{4\pi}3\rho H^{-3}\approx4\pi\alpha\frac{M_\mathrm{Pl}^2}{H}\approx 8\pi\alpha M_\mathrm{Pl}^2t
\approx0.95\left(\frac{\alpha}{0.2}\right)\left(\frac{g_*}{106.75}\right)^{-1/2}\left(\frac{T}{100~\mathrm{MeV}}\right)^{-2}\mSun.
\end{equation}
where $\alpha\sim0.2$ is the portion of matter that collapse into \ac{PBH}, $M_\mathrm{Pl}=(8\pi G)^{-1/2}\approx2.435\times 10^{18}$~GeV is the reduced Planck mass, $t$ is the cosmic time at the formation, and $g_*$ is the number of relativistic species at that moment.
One can see that large density perturbations reentering the horizon during QCD phase transition, $T_{QCD}\sim 100-150$~MeV, collapse to solar-mass \acp{PBH}.
Also, large density perturbations reentering at different epochs can give rise to \acp{PBH} with different masses, which leave fruitful observational implications.
For instance, \acp{PBH} can comprise a significant fraction or all of the cold dark matter \cite{Carr:1974nx,Green:2020jor}, can seed supermassive black holes in galactic centers \cite{Blinnikov:2016bxu,Liu:2023pvq} and globular clusters \cite{Dolgov:2017nmh}, can generate cosmic cosmic structure \cite{Carr:2018rid}, and are invoked to solve the present-day questions with early galaxy formation suggested by JWST observations with modest requirements on their abundance \cite{Colazo:2024jmz,Hutsi:2022fzw,Huang:2023chx,Gouttenoire:2023nzr,Huang:2024aog}.
For recent reviews of physics related to \acp{PBH}, see \textit{e.g.} \cite{Sasaki:2018dmp,Carr:2020xqk,Escriva:2022duf,Ozsoy:2023ryl,Carr:2023tpt,Choudhury:2024aji}.

The \acp{PBH} relevance to \ac{GW} in mHz range stems from several facts.
The most important issue is the \ac{PBH} dark matter.
According to the current observations, only \ac{PBH} in the mass range of $\sim 10^{16}-10^{22}$~g (asteroid mass window) can be all the dark matter.
The density fluctuation that collapses into \acp{PBH} can induce a \ac{GW} background at mHz band, which can be probed by TianQin.
Also, the \acp{PBH} can form binaries in the mass range $10^3-10^4 \mSun$ from the tail of an extended \ac{PBH} log-normal mass distribution \cite{Blinnikov:2016bxu}.
The detection of coalescence of such primordial intermediate-mass binaries by TianQin was investigated in \cite{Postnov:2024fra}.
Distinctive features of such primordial I\ac{MBH} mergers are small effective spins, possible high redshifts $z>20$, and lack of association with gas-rich regions or galaxies.

\acp{PBH} can also arise from different subhorizon processes, which leave different \ac{GW} signals.
These include bubble collisions during \acp{FoPT} \cite{Gross:2021qgx,Baker:2021nyl,Kawana:2021tde,Liu:2021svg}, collapses of domain walls \cite{Rubin:2000dq,Deng:2016vzb,Liu:2019lul,Gouttenoire:2023gbn,Li:2024psa,Ferreira:2024eru,Lu:2024ngi}, Q-balls/oscillons \cite{Cotner:2016cvr,Cotner:2017tir,Cotner:2018vug,Cotner:2019ykd}, etc., which we will not talk further here.

\subsubsection{PBH abundance and observational constraints}

The \ac{PBH} mass and its abundance can be estimated with the Press-Schechter-type formalism.
Roughly speaking, the primordial density fluctuation $\delta\equiv\delta\rho/\rho$ has a random distribution among all the Hubble patches, of which $\delta$ might be large in a few rare patches, such that gravitational collapse happens immediately when the over-dense region re-enter the Hubble horizon.
The threshold for the overdensity, estimated by the Jeans instability in \cite{Carr:1974nx}, is approximately the \ac{EoS} parameter $\sim w$.
More accurate calculation shows that $\delta_\mathrm{th}\approx0.41$ \cite{Harada:2013epa}.
Therefore, assuming Gaussian statistics, for a typical overdense region of comoving scale $k_*$, the portion of the energy density that collapses into \acp{PBH} for any given volume is
\begin{equation}\label{3E:beta1}
\beta(M)=\frac{\alpha}{\sqrt{2\pi}\sigma_\delta(k(M))}\int_{\delta_\mathrm{th}}\exp\left(-\frac{\delta^2}{2\sigma_\delta^2(k(M))}\right)\mathrm{d}\delta
=\frac{\alpha}{2\sigma_\delta(M)}\text{erfc}\left(\frac{\delta_\mathrm{th}}{\sqrt2\sigma_\delta(M)}\right)\,,
\end{equation}
where $\alpha\sim0.2$ is the typical portion of the Hubble mass which collapses into \ac{PBH}, and $\sigma_\delta$ is the root-mean-square of the density contrast.
Note that $\sigma_\delta$ depends on the comoving wavenunmber $k$ of the overdensity, which can be transferred to the horizon mass as \acp{PBH} form at horizon reentry.
The integral gives a complementary error function, which approximates a Gaussian suppression in the high-$\sigma$ tail. After the \acp{PBH} form in the very early universe, their energy density decays as $a^{-3}$ while the background radiation decays as $a^{-4}$.
Therefore, a redshift factor must be taken into account when calculating the \ac{PBH} abundance, the fractional \ac{PBH} energy density today normalized by the density of dark matter \cite{Carr:2020gox}
\begin{equation}\label{3E:f-beta}
f_\text{PBH}=3.81\times10^{8}\alpha^{1/2}\left(\frac{g_{*i}}{106.75}\right)^{-1/4}\left(\frac{h}{0.67}\right)^{-2}\beta(M)\left(\frac{M}{\mSun}\right)^{-1/2}\,,
\end{equation}
where $g_*$ is effective number of the relativistic degrees of freedom when the \acp{PBH} form, and $h=H/(100~\mathrm{km/s/Mpc})$. An extrapolation of scale-invariant density perturbation to all scales, $\sigma_\delta\sim10^{-5}$, can only generate negligible amount of \acp{PBH} $\beta\sim\exp(-10^4)$. Abundant \acp{PBH} with $f_\mathrm{PBH}\sim\mathcal{O}(0.1)$ requires $\delta/\sigma_\delta\sim10$, which means that the power spectrum of the curvature perturbation must be enhanced to $\mathcal{P_R}\sim10^{-2}$ on small scales. This is observationally allowed, as the observational constraints on small scales are very weak \cite{Bringmann:2011ut,Chluba:2012we,Green:2018akb,Sato-Polito:2019hws,Gow:2020bzo,Byrnes:2018txb,Inomata:2018epa,Dalianis:2018ymb,Lu:2019sti,Kalaja:2019uju,Ozsoy:2019lyy}. Many inflation models can realize such an enhancement, including the ultra-slow-roll inflation \cite{Yokoyama:1998pt,Garcia-Bellido:2016dkw,Cheng:2016qzb,Garcia-Bellido:2017mdw,Cheng:2018yyr,Dalianis:2018frf,Tada:2019amh,Xu:2019bdp,Mishra:2019pzq,Bhaumik:2019tvl,Liu:2020oqe,Atal:2019erb,Fu:2020lob,Vennin:2020kng,Ragavendra:2020sop,Gao:2021dfi}, multifield inflation \cite{GarciaBellido:1996qt,Kawasaki:1997ju,Frampton:2010sw,Giovannini:2010tk,Clesse:2015wea,Inomata:2017okj,Gong:2017qlj,Inomata:2017vxo,Espinosa:2017sgp,Kawasaki:2019hvt,Palma:2020ejf,Fumagalli:2020adf,Braglia:2020eai,Anguelova:2020nzl,Romano:2020gtn,Gundhi:2020zvb,Gundhi:2020kzm,Wang:2024vfv}, modified gravity \cite{Kannike:2017bxn,Pi:2017gih,Gao:2018pvq,Cheong:2019vzl,Cheong:2020rao,Fu:2019ttf,Dalianis:2019vit,Lin:2020goi,Fu:2019vqc,Aldabergenov:2020bpt,Aldabergenov:2020yok,Yi:2020cut,Gao:2020tsa}, curvaton scenario \cite{Kawasaki:2012wr,Kohri:2012yw,Ando:2017veq,Ando:2018nge,Chen:2019zza}, sound speed resonance and other resonances \cite{Cai:2018tuh,Cai:2019jah,Cai:2019bmk,Chen:2020uhe,Cai:2020ovp,Zhou:2020kkf,Cai:2021yvq,Peng:2021zon,Xie:2024cwp},
phase transition \cite{Hawking:1982ga,Crawford:1982yz,Gross:2021qgx,Baker:2021nyl,Kawana:2021tde,Liu:2021svg,Liu:2022lvz}, oscillon decay \cite{Cotner:2016cvr,Cotner:2017tir,Cotner:2018vug,Cotner:2019ykd,Kusenko:2020pcg}, etc.

Recent developments on numerical relativity and cosmological perturbation theory has renewed our knowledge of \ac{PBH} formation. First of all, numerical relativity shows that the threshold should be put on the ratio of mass excess and the areal radius, the so-called compaction function, $\mathscr{C}=2G\delta M/R$, where $\delta M\equiv M_K(R)-M_H$ is the Kodama mass $M_K$\cite{Kodama:1979vn} inside radius $R$ with the background mass subtracted \cite{Shibata:1999zs}. Using the comoving curvature perturbation $\mathcal{R}$, the compaction function can be written as \cite{Harada:2015yda,Kawasaki:2019mbl,Young:2019yug,DeLuca:2019qsy}
\begin{align}\label{3E:Cnl}
\mathscr{C}=\mathscr{C}_{\ell}-\frac{3}{8} \mathscr{C}_{\ell}^2,\qquad\mathrm{with}~\mathscr{C}_\ell=-\frac43r\frac{\partial\mathcal{R}}{\partial r},
\end{align}
where $r$ is the radial coordinate in the metric of $ds_3^2=a^2e^{2\mathcal{R}}(\mathrm{d} r^2+r^2\mathrm{d}\Omega_2)$, which is related to $R$ by $R=are^{\mathcal{R}(r)}$.
As the high peak is rare, the profile of the curvature perturbation is spherical, and the compaction function $\mathscr{C}(r)$ only depends on the radius.
The threshold $\mathscr{C}_\mathrm{th}$ varies from 0.4 to 0.7, depending on its profile \cite{Musco:2018rwt}.
It is found that, however, the threshold on the compaction function averaged inside the sphere of radius $R_m$ is universal, \textit{i.e.} $\left\langle\mathscr{C}(R)\right\rangle_{R<R_m}>2/5$ \cite{Escriva:2019phb}, which gives the profile-dependent thresholds $\mathscr{C}_{\mathrm{th}}$ and $\mathscr{C}_{\ell,\mathrm{th}}$.
Hypothetical profiles are given by fitting formulas \cite{Musco:2018rwt,Escriva:2019phb,Young:2019osy} or by the theory of peaks \cite{Yoo:2018kvb,Atal:2019cdz,Atal:2019erb,Yoo:2020dkz,Kitajima:2021fpq}.
From the \ac{PDF} of $\mathcal{R}$, one can derive the \ac{PDF} of $\mathscr{C}_\ell$ by the probability conservation, and then integrating it to get the energy density of \acp{PBH} at the collapse \cite{Pi:2024jwt}
\begin{equation}\label{3E:pipeline}
\left.
\begin{matrix}
\left\langle\mathscr{C}(R)\right\rangle_{R<R_m}>2/5\xrightarrow{\mathrm{Eq.~}\eqref{3E:Cnl}}\mathscr{C}_{\mathrm{th}}(\mathrm{profile})
\\
\\
\left(
\begin{matrix}
\delta N~\mathrm{formalism}\\
\mathrm{or~stochastic}
\end{matrix}
\right)
\xrightarrow{}\mathbb{P}(\mathcal{R})\xrightarrow{\mathrm{Prob.}}\mathbb{P}(\mathscr{C}_\ell)
\end{matrix}
\right\}
\xrightarrow[\text{Window function}]{\mathscr{C}_{\mathrm{th}}\mathrm{(profile)}}
\beta=\int^{4/3}_{\mathscr{C}_{\ell,\mathrm{th}}}\mathbb{P}(\mathscr{C}_\ell)\frac{M(\mathscr{C}_\ell)}{M_H}{\mathop{}\!\mathrm{d}}\mathscr{C}_\ell,
\end{equation}
where the mass of the \ac{PBH} obeys a power-law scaling from the critical collapse \cite{Choptuik:1992jv,Evans:1994pj,Koike:1995jm,Niemeyer:1997mt,Hawke:2002rf,Musco:2008hv}
\bea\frac{M(\mathscr{C}_\ell)}{M_H}\sim K\left(\mathscr{C}-\mathscr{C}_\mathrm{th}\right)^\gamma
=K\left[\left(\mathscr{C}_\ell-\frac38\mathscr{C}_\ell^2\right)-\mathscr{C}_\mathrm{th}\right]^\gamma\,,\label{eqn8:M/MH}\eea
with $\gamma\approx0.36$ and $K\sim1$. Then one can use \eqref{3E:f-beta} to calculate $f_\mathrm{PBH}(M)$.
The \ac{PDF} of the linear compaction function $\mathbb{P}(\mathscr{C}_\ell)$ is determined by that of the curvature perturbation $\mathbb{P}(\mathcal{R})$ via probability conservation, \textit{i.e.} $\mathbb{P}(\mathscr{C}_\ell)=\mathbb{P}(\mathcal{R})|\partial\mathcal{R}/\partial\mathscr{C}_\ell|$.
The comoving curvature perturbation originates from the quantum fluctuations of the inflaton field and the metric perturbation, which can be calculated by the classical $\delta N$ formalism or stochastic approach. In the simplest slow-roll case, $\mathcal{R}\approx-(H/\dot\varphi)\delta\varphi$, so
\begin{equation}\label{3E:PDF(R)G}
\mathbb{P}(\mathcal{R})=\frac1{\sqrt{2\pi}\sigma_{\mathcal{R}}(r_w)}\exp\left(-\frac{\mathcal{R}^2}{2\sigma_\mathcal{R}^2(r_w)}\right)
\end{equation}
is a Gaussian \ac{PDF}, where
\begin{equation}\label{3E:def:sigmaR}
\sigma_{\mathcal{R}}^2(r_w)=\int\frac{\mathrm{d}k}{k}\mathcal{P_R}(k)W^2(k;r_w)
\end{equation}
is the variance determined by the power spectrum of $\mathcal{R}$ and the window function $W(k;r_w)$ with a filter scale $r_w$. For thorough discussions of window functions and the smoothing scale, see \textit{e.g.}  \cite{Ando:2018qdb,Young:2019osy,Yoo:2020dkz}. The upper bound $4/3$ of the integral \eqref{3E:pipeline} is the boundary of the type II fluctuation, which is beyond the scope of this paper \cite{Escriva:2022pnz,Escriva:2023uko,Uehara:2024yyp,Inui:2024fgk,Shimada:2024eec}. The \ac{PBH} mass function can also be calculated by the peaks theory, which counts the number density of the peaks of a Gaussian random field, instead of using the \ac{PDF}. Interested readers can check \textit{e.g.}  \cite{Green:2004wb,Yoo:2018kvb,Germani:2018jgr,Atal:2019cdz,Atal:2019erb,Young:2020xmk,Yoo:2020dkz,Taoso:2021uvl,Riccardi:2021rlf,Kitajima:2021fpq,Young:2022phe,Pi:2024ert} for details.

Given a power spectrum of the curvature perturbation $\mathcal{P_R}$, one can calculate the \ac{PBH} mass function and check with observational constraints.
There are many experiments aiming to detect \acp{PBH} of different masses. Small \acp{PBH} around $10^{16}$ g can be probed by the extra-galactic gamma ray or other cosmic rays as these small black holes are approaching their doomsday thus intensely radiate \cite{Carr:2020gox}.
\acp{PBH} larger than $10^{23}$ g can be probed by microlensing experiments like Subaru HSC \cite{Niikura:2017zjd}, EROS \cite{Tisserand:2006zx}, and OGLE \cite{Mroz:2024mse,Mroz:2024wag}.
Recent results combining these experiments show that $f_\mathrm{PBH}\lesssim1\%$ for a wide range from $10^{23}$ g to a few solar mass.
\ac{LVK} direct search for sub-solar-mass black hole binaries puts a constraint $f_\mathrm{PBH}<10\%$ for $M_\mathrm{PBH}=0.4$--$1\mSun$ \cite{LVK:2022ydq}.
Accretion limits from Planck data can constrain \acp{PBH} from a few to ten thousand $\mSun$ \cite{Serpico:2020ehh}, while the \ac{CMB} $\mu$-distortion can exclude \acp{PBH} as seeds of \acp{MBH} \cite{Delabrouille:2019thj,Chluba:2019nxa} unless there is large non-Gaussianity \cite{Nakama:2016kfq,Nakama:2017xvq,Nakama:2019htb,Atal:2020yic,Carr:2018rid,Liu:2022bvr,Biagetti:2022ode,Gouttenoire:2023nzr,Hooper:2023nnl,Hai-LongHuang:2024gtx}.
For a review of all the updated observational constraints, see \textit{e.g.} \cite{Carr:2020xqk,Carr:2023tpt}.

Unfortunately, due to the finite-size effect and wave effect, microlensing can not probe smaller \acp{PBH} of $M\lesssim10^{23}$ g \cite{Sugiyama:2019dgt,Montero-Camacho:2019jte,Smyth:2019whb}, rendering the asteroid-mass range the only open window where \acp{PBH} can be all the dark matter, \textit{i.e.} $f_\mathrm{PBH}=1$. For reviews of this mass window, see \textit{e.g.}  \cite{Green:2020jor,Villanueva-Domingo:2021spv,Green:2024bam,Tinyakov:2024mcy}.
There are some methods to probe asteroid-mass \acp{PBH} in the future, including femtolensing of \acp{GRB} \cite{Katz:2018zrn}, gamma ray telescopes \cite{Ray:2021mxu}, GRB lensing parallax \cite{Jung:2019fcs,Gawade:2023gmt}, X-ray microlensing \cite{Tamta:2024pow}, solar-system capture \cite{Tran:2023jci,Cuadrat-Grzybowski:2024uph,Loeb:2024tcc}, etc.
However, the most effective way of probing \ac{PBH} dark matter is to probe the concomitant induced \acp{GW} at millihertz band \cite{Saito:2008jc,Cai:2018dig,Bartolo:2018evs}.

\subsubsection{Probing induced GWs with TianQin}

The curvature perturbation which drives the density perturbation to collapse into \acp{PBH} can also induce \acp{GW}.
Although there is no directly coupling between scalar and tensor perturbations at linear order, the curvature perturbation can source \acp{GW} at quadratic order, which is usually called secondary \acp{GW} or induced \acp{GW} \cite{Matarrese:1992rp,Matarrese:1993zf,Matarrese:1997ay,Noh:2004bc,Carbone:2004iv,Nakamura:2004rm,Ananda:2006af,Osano:2006ew,Baumann:2007zm} \footnote{Secondary \acp{GW} might also be induced by other couplings like scalar-tensor-tensor or tensor-tensor-tensor. We do not consider such waves as they are usually smaller than the scalar-induced \acp{GW}. See for instance \cite{Chang:2022vlv,Yu:2023lmo,Bari:2023rcw,Picard:2023sbz}. Also, we do not touch the discussion of gauge issue. For recent discussions, see for instance \cite{Hwang:2017oxa,Wang:2019zhj,Gong:2019mui,Tomikawa:2019tvi,DeLuca:2019ufz,Inomata:2019yww,Yuan:2019fwv,Nakamura:2019zbe,Lu:2020diy,Ali:2020sfw,Giovannini:2020qta,Giovannini:2020soq,Chang:2020tji,Chang:2020iji,Chang:2020mky,Domenech:2020xin}. For reviews of scalar-induced \acp{GW}, see  \cite{Domenech:2021ztg,Domenech:2024kmh}.}.
Up to quadratic order in the scalar perturbation, the tensor perturbation $h_{ij}$ in the transverse-traceless gauge has an equation of motion $\square h_{ij}\sim\Lambda^{kl}{}_{ij}\partial_k\mathcal{R}\partial_l\mathcal{R}$, where $\Lambda^{kl}{}_{ij}$ is the transverse-traceless projector, and $\Phi=(2/3)\mathcal{R}$ is the curvature perturbation in longitudinal gauge. By solving this equation of motion, the energy density spectrum of the induced \ac{GW} we observe today is given by \cite{Kohri:2018awv,Pi:2020otn,Domenech:2021ztg}
\bea \Omega_{\text{GW},0}(f)h^2 &=&1.6\times10^{-5}\left(\frac{g_{*s}(\eta_k)}{106.75}\right)^{-1/3}\left(\frac{\Omega_{r,0}h^2}{4.1\times10^{-5}}\right)\nn\\
&&\times3\int^\infty_0\mathrm{d} v\int^{1+v}_{|1-v|}\mathrm{d} u\frac{1}{4u^2v^2}\left[\frac{4v^2-(1+v^2-u^2)^2}{4uv}\right]^2\left(\frac{u^2+v^2-3}{2uv}\right)^4\nn\\
&&\times\left[\left(\ln\left|\frac{3-(u+v)^2}{3-(u-v)^2}\right|-\frac{4uv}{u^2+v^2-3}\right)^2+\pi^2\Theta\left(u+v-\sqrt3\right) \right]
\mathcal{P_R}(uk)\mathcal{P_R}(vk)\,,\label{3E:OmegaGW}\eea
where $\mathcal{P_R}$ is the power spectrum of the comoving curvature perturbation, which also determines $\sigma_\delta$ in \eqref{3E:beta1} and the variance of $\mathcal{R}$ in \eqref{3E:pipeline}. In this manner, the \ac{PBH} abundance and the amplitude of the induced \ac{GW} spectrum are connected.
The integrals in \eqref{3E:OmegaGW} can be calculated analytically for monochromatic \cite{Kohri:2018awv}, lognormal \cite{Pi:2020otn}, and broken-power-law \cite{Li:2024lxx} power spectra. Its spectral shape displays some characteristic features which can be used to distinguish with other \ac{SGWB}s. For instance, in the radiation dominated era, the infrared spectrum scales as $f^2$ for a narrow peak, and as $f^3$ for a broad peak \cite{Cai:2019cdl,Pi:2020otn}. These scalings can be used to probe the thermal history of the early universe, if there are any deviations \cite{Domenech:2019quo,Domenech:2020kqm,Dalianis:2020cla,Hook:2020phx,Brzeminski:2022haa,Franciolini:2023wjm,Domenech:2024wao}. They are distuiguishable from the stochastic \ac{GW} from the incoherent superpositions of the inspiral \acp{MBH}, which goes $\Omega_\mathrm{GW,0}\propto f^{2/3}$ \cite{Ajith:2009bn,Zhu:2011bd,Zhu:2012xw}.

To estimate the peak amplitude, note that such induced \acp{GW} are generated most efficiently at the horizon reentry of the curvature perturbation, which redshift to $\Omega_\mathrm{GW,0}\sim10^{-6}\mathcal{P}_\mathcal{R}^2$ at present. Therefore, a nearly scale-invariant \ac{GW} spectrum of $\Omega_\mathrm{GW,0}\sim10^{-24}$ is ubiquitous on large scales because of $\mathcal{P_R}\sim10^{-9}$ there. This is far beyond our detection, and also much smaller than the primordial \ac{GW} in most of the inflation models, $\Omega_\mathrm{GW,0}^\mathrm{(prim)}\sim r\times10^{-14}$, if the tensor-to-scalar ratio $r>10^{-10}$ \cite{Baumann:2007zm}. However, if on small scales the power spectrum is enhanced to $10^{-2}$, as is required for detectable amount of \acp{PBH}, the accompanying induced \ac{GW} spectrum reaches $\Omega_\mathrm{GW,0}\sim10^{-10}$, which can be probed by many \ac{GW} experiments targeting millihertz and lower frequencies \cite{Saito:2008jc}.

As the \ac{PBH} formation and the induced \ac{GW} generation take place around the horizon reentry of the curvature perturbation, both the \ac{PBH} mass and induced \ac{GW} frequency depend mainly on the Hubble scale of the reentry, which are connected by \cite{Saito:2008jc}
\begin{equation}\label{3E:f-M}
f_\mathrm{IGW}\approx3~\mathrm{Hz}\left(\frac{M_\mathrm{PBH}}{10^{16}~\mathrm{g}}\right)^{-1/2}\left(\frac{g_*}{106.75}\right)^{-1/12}.
\end{equation}
Using this relation, one can make cross checks for \ac{PBH} abundance and \ac{GW} spectrum in most of the mass ranges. For instance, recent pulsar timing array collaborations NANOGrav \cite{NANOGrav:2023gor,NANOGrav:2023hde}, EPTA+InPTA \cite{EPTA:2023fyk,EPTA:2023sfo,EPTA:2023xxk}, PPTA\,\cite{Zic:2023gta,Reardon:2023gzh,Reardon:2023zen}, CPTA\,\cite{Xu:2023wog}, IPTA \cite{InternationalPulsarTimingArray:2023mzf}, and MPTA \cite{Miles:2024seg,Miles:2024rjc} all reported a detection of \ac{SGWB} at $10^{-8}$--$10^{-7}$ Hz. The induced \acp{GW} can fit the data quite well, which should be accompanied by abundant sub-solar-mass \acp{PBH}. Accurate analysis show that sub-solar-mass \ac{PBH} might be overproduced as the amplitude of the reconstructed power spectrum of the curvature perturbation is too large \cite{NANOGrav:2023hvm,Cai:2019elf}. Negative non-Gaussianity \cite{Pi:2022ysn} or stiffer equation of state \cite{Harada:2013epa,Escriva:2022duf} are introduced to suppress the \ac{PBH} formation and reconcile such a problem \cite{Wang:2023ost,Liu:2023pau,Zhu:2023gmx,Domenech:2024rks,Franciolini:2023pbf,Inui:2024fgk}.

As we mentioned before, such a cross check can not be done for the asteroid-mass window from $10^{-16}\mSun$ to $10^{-10}\mSun$ (\textit{i.e.} $10^{16}$ g to $10^{23}$ g), where there is no observational constraints for \acp{PBH}. The asteroid-mass \acp{PBH} can be all the dark matter, which must be accompanied by induced \acp{GW} in a frequency band of $10^{-3}$--1 Hz, given by \eqref{3E:f-M}. This is just the most sensitive frequency band of TianQin \cite{Liang:2021bde} and other space-based \ac{GW} detectors \cite{LISA:2022yao,LISACosmologyWorkingGroup:2022jok,LISA:2022kgy,LISACosmologyWorkingGroup:2023njw,Ren:2023yec}.

For concreteness, we will study the detectability of \ac{PBH} dark matter by TianQin for a monochromatic power spectrum
\begin{equation}\label{3E:monoPS}
\mathcal{P}_\mathcal{R}=\mathcal{A}_\mathcal{R}\delta(\ln k-\ln k_*)
\end{equation}
with Gaussian statistics\footnote{For the discussion of non-Gaussian curvature perturbation and broad power spectrum, see subsection \ref{s3E4}.}. The calculation of induced \ac{GW} spectrum shown in \eqref{3E:OmegaGW} is straightforward and gives a simple analytical expression \cite{Kohri:2018awv}. When changing $\mathcal{A_R}$ and $k_*$, the \ac{GW} spectrum swipes the corresponding parameter space of $\Omega_\mathrm{GW,0}$ and $f_*$. Given the TianQin noise curve of $S_n(f)$ by \eqref{eq:S_n^SA}, one can define
\begin{equation}\label{3E:def:Omega_n}
\Omega_{n}h^2=\frac{4\pi}{3H_{100}^2}f^3S_n(f).
\end{equation}
The detectability of any signal $\Omega_{\mathrm{GW,0}}$ is described by the \ac{SNR}, defined by \cite{Maggiore:2007ulw,Caprini:2018mtu}
\begin{equation}\label{3E:def:SNR}
\mathrm{SNR}=\left[2T\int^\infty_0\left(\frac{\Omega_\mathrm{GW,0}(f)h^2}{\Omega_\mathrm{n}(f)h^2}\right)^2\mathrm{d}f\right]^{1/2},
\end{equation}
where $\Omega_\mathrm{GW,0}(f)$ is the spectrum of the \ac{SGWB} we are searching for, and $T$ is the total duration of the observation. For power-law signals $\Omega_\mathrm{GW,0}(f)\propto f^\beta$, a simple method based on the power-law-integrated sensitivity curve was developed in \cite{Thrane:2013oya}, where the \ac{SNR} can be read directly from comparing the power-law-integrated curve and the power-law \ac{GW} spectrum. It is convenient when we go to the infrared power-law tail of the induced \ac{GW}, $\Omega_\mathrm{GW,0}(f)\propto f^2$ (or $\Omega_\mathrm{GW,0}(f)\propto f^3$ for a broad peak \cite{Pi:2020otn}).
However, this is not convenient when the \ac{GW} spectrum displays a peak in the detectable range, and we should go back to \eqref{3E:def:SNR}.

\begin{figure}[htbp!]
\begin{center}
\includegraphics[width=0.45\textwidth]{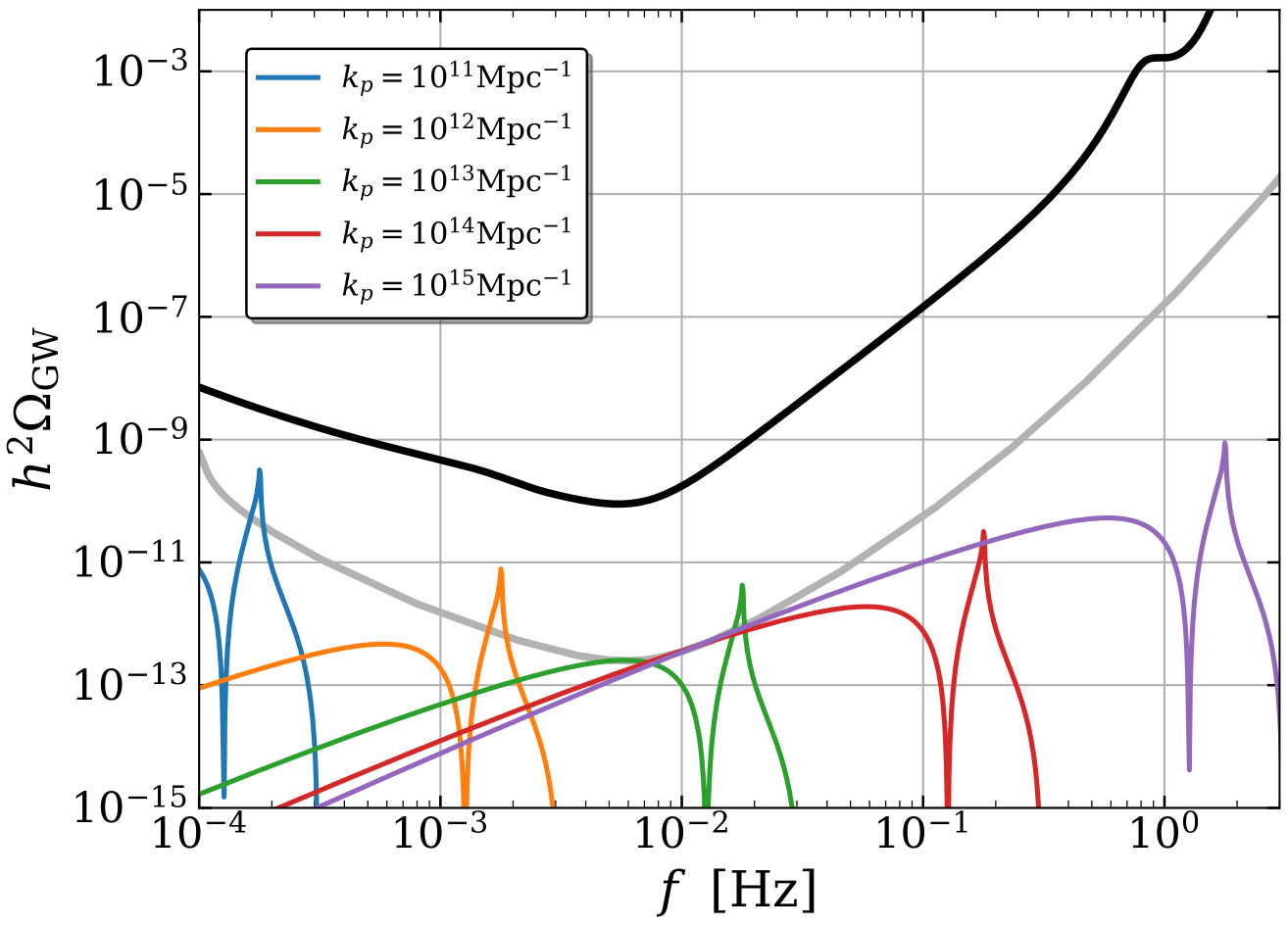}\qquad
\includegraphics[width=0.45\textwidth]{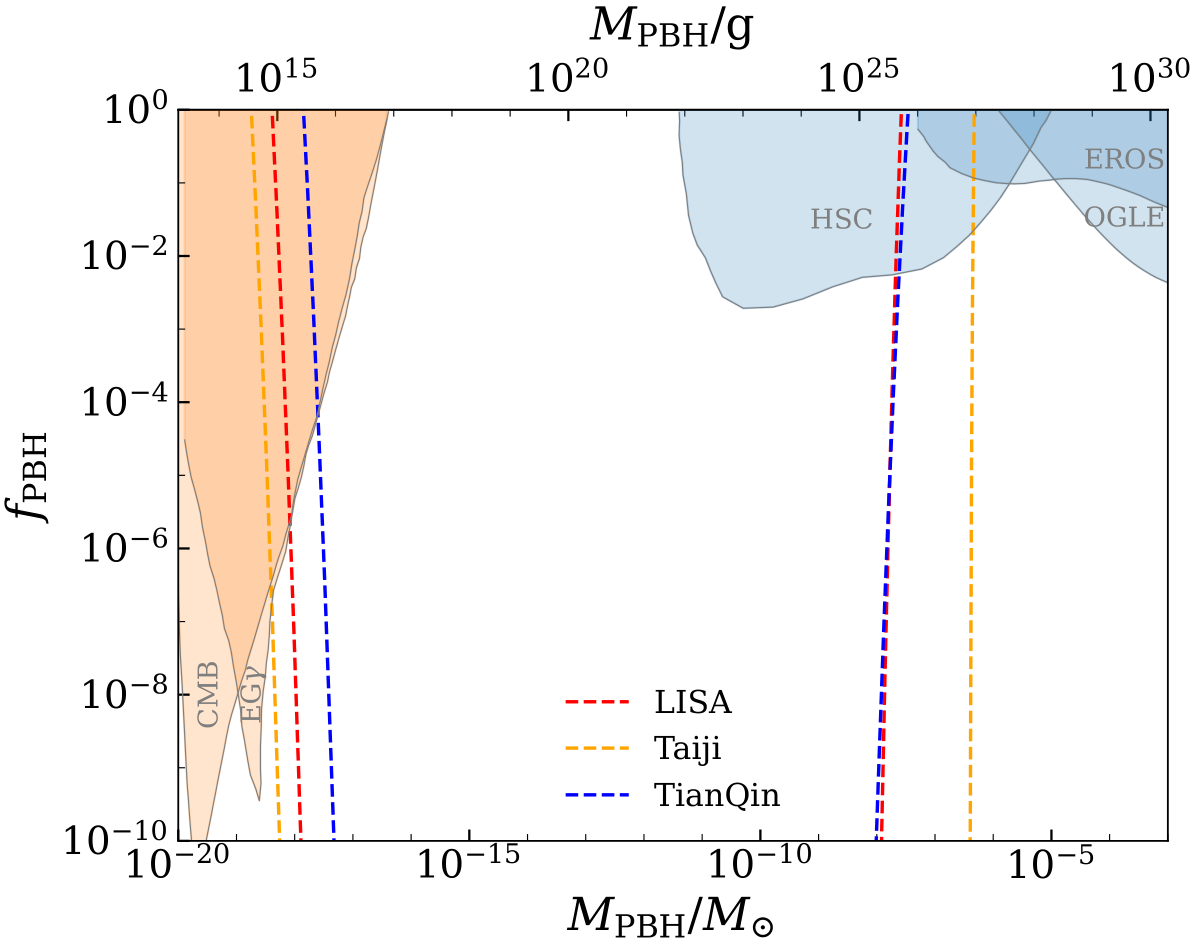}
\caption{(Left) Some possible marginal signals with $\mathrm{\ac{SNR}}=3$. (Right) PBH mass range that can be probed by TianQin, with LISA and Taiji shown as comparison. See main text for more explanation.}
\label{fig:PLI-IGW}
\end{center}
\end{figure}

In FIG.~\ref{fig:PLI-IGW} (Left), we show a few possible marginal signals with $\mathrm{\ac{SNR}}=3$ together with the TianQin noise sensitivity curve transferred to $\Omega_nh^2$ by \eqref{3E:def:Omega_n}, and the power-law-integrated sensitivity curve based on it.
The black curve is the TianQin noise in terms of $\Omega_nh^2$, given by \eqref{3E:def:Omega_n} and \eqref{eq:S_n^SA}.
The \ac{GW} foreground from the white dwarf mergers in our galaxy has been reduced by using the fitting formula given in \cite{Karnesis:2021tsh}, which only leaves a dent around $1.5\times10^{-3}$ Hz.
The gray curve is the power-law-intergrated curve generated by the method proposed in \cite{Thrane:2013oya}.
We also draw some marginal induced \ac{GW} signals with peak wavenumber $k_*=10^{11}$ (blue), $10^{12}$ (orange), $10^{13}$ (green), $10^{14}$ (red), and $10^{15}$ (purple)~$\mathrm{Mpc}^{-1}$.
The power-law-intergrated curve and marginal signals have a total observation duration $T=3~\mathrm{yr}$ and $\mathrm{SNR}=3$.

The TianQin noise curve provides us the detectable region of the amplitude of power spectrum $\mathcal{A_R}$ and the peak frequency $f_*$. By using \eqref{3E:f-M}, $f_*$ can be transferred to the \ac{PBH} mass $M_\mathrm{PBH}$. Also, one can calculate the \ac{PBH} abundance for each $\mathcal{A_R}$. Ignoring the uncertainties from the window function, one can use the Press-Schechter-type formalism shown in \eqref{3E:pipeline} to calculate the \ac{PBH} mass function $\beta(M_\mathrm{PBH})$, and the \ac{PBH} abundance $f_\mathrm{PBH}$ by \eqref{3E:f-beta}. Therefore, one can transfer the detectability of $\mathcal{A_R}(f)$ into the region of $f_\mathrm{PBH}(M_\mathrm{PBH})$, which is shown in FIG.~\ref{fig:PLI-IGW} (Right).
In both figures of FIG. \ref{fig:PLI-IGW}, we see clearly that TianQin can cover the entire asteroid-mass window.
PBH dark matter must leave detectable mHz \ac{SGWB}, thus searching for the signal of \ac{PBH} dark matter is a very important goal for TianQin.

A common misconception of the detectability is that simply transferring the detectable frequency band to \ac{PBH} mass window (related by \eqref{3E:f-M}) leaves a seemingly undetectable gap in the low-mass edge.
For instance, $\Omega_\mathrm{GW,0}\gtrsim10^{-11}$ requires $10^{-4}$--$10^{-1}$ Hz, which corresponds to $10^{19}$--$10^{25}$ g, leaving a undetectable gap $10^{16}$--$10^{19}$ g.
A typical signal in the ``gap'' is shown in FIG. \ref{fig:PLI-IGW} (Left) as the rightmost purple curve.
It is clear that although the peak is undetectable, its infrared tail still lies above the power-law-integrated sensitivity curve.
Therefore, the infrared tail of the induced \ac{GW} helps to close the low-frequency gap, and to probe the entire asteroid-mass window, as is shown explicitly in FIG. \ref{fig:PLI-IGW} (Right).
The detectability of the \ac{PBH} abundance $f_\mathrm{PBH}(M_\mathrm{PBH})$ in the asteroid-mass window, together with some other observational constraints from \ac{PBH} evaporation (yellow shaded) of \ac{CMB} anisotropies \cite{Acharya:2020jbv} and extragalactic $\gamma$-ray\cite{Carr:2009jm}, and microlensing (blue shaded) of Subaru HSC \cite{Niikura:2017zjd}, EROS \cite{EROS-2:2006ryy}, OGLE \cite{Mroz:2024mse}. The region between the blue dashed lines is the parameter space TianQin can probe, with a total duration 3 yr and $\mathrm{\ac{SNR}}=3$.
For comparison, we also draw the detectable region of LISA (red dashed \cite{LISA:2017pwj,Babak:2021mhe}) and Taiji (yellow dashed \cite{Luo:2019zal}) with the same duration and \ac{SNR}.
Although the above conclusion is obtained for a narrow peak with Gaussian statistics, the same conclusion is valid also for broad peak and with local-type non-Gaussianity.
Detailed analysis of the robustness can be found in \cite{Hong:2025tba}.

\subsubsection{Dispersion, Non-Gaussianity and GW Anisotropies}\label{s3E4}

{\it Subsection coordinator: Shi Pi, Sai Wang}

In this subsection we will discuss some recent topics of scalar-induced \acp{GW}. We first show the robustness of the detectability of \ac{PBH} dark matter by TianQin against the dispersion (\textit{i.e.} broadness of the power spectrum) and the non-Gaussianity of the curvature perturbation. We enclose by discussing the \ac{GW} anisotropy as a probe of the non-Gaussianity.

\paragraph{Dispersion and Non-Gaussianity}\

Although the monochromatic power spectrum \eqref{3E:monoPS} is unphysical \cite{Inomata:2024dbr}, it represents the narrow power spectrum which can be generated by, \textit{i.e.}, resonances \cite{Cai:2018tuh,Cai:2019jah,Cai:2019bmk,Chen:2020uhe,Cai:2020ovp,Zhou:2020kkf,Cai:2021yvq,Peng:2021zon,Xie:2024cwp}. Typical power spectrum enhanced by single-field inflation displays a spectral peak from modulated oscilations, of which the width is $\sim1/2\pi$ \cite{Dalianis:2021iig,Pi:2022zxs}, while broader peaks are also possible for mild transitions from slow-roll to ultra-slow-roll \cite{Byrnes:2018txb,Cole:2022xqc,Pi:2022zxs}. Window functions must be taken into account when calculating the \ac{PBH} mass function for such broad power spectra \cite{Ando:2018qdb,Young:2019osy,Yoo:2020dkz,Pi:2024ert}, which leads to extended \ac{PBH} mass functions. All of the observational constraints (microlensing, extragalactic $\gamma$ ray, \textit{etc}.) assume monochromatic mass function, which should be rearranged to generate model-dependent constraints for each extended mass function. The new constraints from extended mass functions are usually more stringent \cite{Carr:2017jsz,Gorton:2024cdm,Green:2017qoa}. On the other hand, the detectable range of \ac{PBH} mass by TianQin only shrinks a little on the small mass edge \cite{Hong:2025tba}, mainly because the infrared scaling becomes steeper (\textit{i.e.} from $k^2$ to $k^3$ \cite{Cai:2019cdl,Pi:2020otn}).
Therefore, our analysis in subsection \ref{s:PBH} on the monochromatic power spectrum is sufficient to show that TianQin can cover all the asteroid-mass window and probe \ac{PBH} dark matter.

We assume Gaussian curvature perturbation in subsection \ref{s:PBH} for simplicity. However, enhancing the power spectrum of the curvature perturbation, especially in the single-field inflation, is usually accompanied with large non-Gaussianities \cite{Byrnes:2012yx,Young:2013oia,Tada:2015noa,Young:2015kda,Young:2015cyn,Franciolini:2018vbk,Ando:2018nge,Atal:2018neu,Passaglia:2018ixg} which significantly increase the \ac{PBH} abundance for the same variance with positive skewness. Traditionally, in \ac{CMB} and large-scale structure, such a local non-Gaussianity is described by the series
\begin{equation}\label{3E:NGseries}
\mathcal{R}=\mathcal{R}_g+\frac35f_\mathrm{NL}\mathcal{R}_g^2+\frac{9}{25}g_\mathrm{NL}\mathcal{R}_g^3+\frac{27}{125}h_\mathrm{NL}\mathcal{R}_g^4+\cdots.
\end{equation}
Following \eqref{3E:pipeline}, such series can be used to calculate the \ac{PBH} abundance, usually only up to quardratic level \cite{Byrnes:2012yx,Young:2013oia,Tada:2015noa,Young:2015kda,Young:2015cyn,Franciolini:2018vbk,Ando:2018nge,Atal:2018neu,Passaglia:2018ixg,Ozsoy:2023ryl}. However, recent studies show that the perturbative series \eqref{3E:NGseries} is not sufficient to fully describe the non-Gaussian impact on the \ac{PBH} formation, as $f_\mathrm{NL}$ goes to $\mathcal{O}(1)$ in ultra-slor-roll and constant-roll models \cite{Namjoo:2012aa,Martin:2012pe,Chen:2013aj,Motohashi:2014ppa,Davies:2021loj,Namjoo:2023rhq,Namjoo:2024ufv}. The \ac{PDF} of $\mathbb{\mathcal{R}}$ usually has an exponential tail $\mathbb{P}(\mathcal{R})\sim\exp[-(6/5)f_\mathrm{NL}\mathcal{R}]$ for $f_\mathrm{NL}>0$ \cite{Pi:2022ysn,Pi:2024jwt}, which greatly enhance the \ac{PBH} abundance and can not be described by perturbative series. Further enhancement caused by a so-called ``heavy tail'' is also possible \cite{Nakama:2016kfq,Hooshangi:2021ubn,Hooshangi:2023kss}, which can be realized in, \textit{e.g.}, the curvaton scenario \cite{Pi:2021dft,Hooper:2023nnl}. As the \ac{PBH} abundance relies sensitively on the \ac{PDF} of $\mathscr{C}_\ell$ around the range of \ac{PBH} formation, $\mathcal{O}(0.5)<\mathscr{C}_\ell<4/3$, such an exponential or heavy tail can enhance the \ac{PBH} abundance significantly.

However, when fixing the \ac{PBH} abundance, say, to be all the dark matter, the change of the required variance of the curvature perturbation with such non-Gaussianities is small, as changes in such far tails only contribute slightly to the variance. This makes it still valid to use the perturbative series \eqref{3E:NGseries} to calculate the induced \ac{GW} spectrum, even if $f_\mathrm{NL}\sim\mathcal{O}(1)$.
As long as the higher order terms stay small, \textit{i.e.} $g_\mathrm{NL}\ll f_\mathrm{NL}/\mathcal{R}$, $h_\mathrm{NL}\ll F_\mathrm{NL}/\mathcal{R}^2$, \textit{etc}., one can always calculate by perturbative series, which was firstly done for quadratic expansion by \cite{Nakama:2016gzw,Garcia-Bellido:2017aan,Cai:2018dig,Unal:2018yaa,Unal:2020mts,Adshead:2021hnm,Ragavendra:2021qdu}, and then extended to cubic (up to $g_\mathrm{NL}$) \cite{Abe:2022xur,Yuan:2023ofl,Li:2023xtl} and quintic (up to $i_\mathrm{NL}$) \cite{Perna:2024ehx} orders. Apparently, when \ac{PBH} abundance is fixed, \textit{i.e.} $f_\mathrm{PBH}=1$, the induced \ac{GW} is reduced when the skewness is positive, as the variance required to generate \ac{PBH} as all the dark matter is smaller.
However, in most of the models, such a reduction is limited, and the large non-Gaussian limits are still above the TianQin sensitivity curves.
Therefore, strong scalar-induced \acp{GW} in millihertz is a robust prediction of \ac{PBH} dark matter, which can be well probed by TianQin.
Besides, non-Gaussianities may leave characteristic features around the peak of the induced \ac{GW} spectrum, making it possible for TianQin to probe primordial non-Gaussianity on small scales. For detailed discussion on non-Gaussianities, see \cite{Pi:2024jwt} and the references therein.

\paragraph{\ac{GW} Anisotropies}\label{3E5}\

The angular power spectrum can be one of the most important observables to identify the existence of induced \acp{GW} and also decode the contained information of the origin and evolution of the Universe.
The definition of angular power spectrum is related with two-point angular correlation of density contrasts of \acp{GW} along two lines-of-sight \cite{Contaldi:2016koz,Bartolo:2019oiq,Bartolo:2019yeu,Schulze:2023ich}.
Hence, it can characterize the anisotropies in the energy density of induced \acp{GW}, thereby encoding the vital information of \ac{GW} sources, e.g., the aforementioned primordial non-Gaussianity.
The relevant research of the angular power spectrum of induced \acp{GW} can be approached similarly to the study of the \ac{CMB} \cite{Bartolo:2019zvb,Li:2023qua,Li:2023xtl,Rey:2024giu,Ruiz:2024weh,Schulze:2023ich,Zhao:2024gan,LISACosmologyWorkingGroup:2022kbp,LISACosmologyWorkingGroup:2022jok,Malhotra:2022ply,Wang:2023ost,Yu:2023jrs}.
It should be noted that the induced \acp{GW} can contain some essential information of the initial inhomogeneities, which are absent for the \ac{CMB} temperature anisotropies and polarization due to the opacity of the early Universe to propagation of photons.
This is one of the most significant advantages of the \ac{GW} probe, compared with other cosmological probes such as the CMB, large-scale structures, etc.
Because of the squeezed primordial non-Gaussianity, the initial inhomogeneities can be produced by couplings between long-wavelength perturbations and short-wavelength perturbations that generated the induced \acp{GW}.
Through a series of detailed analysis in the literature \cite{Bartolo:2019zvb,Li:2023qua,Li:2023xtl,Rey:2024giu,Ruiz:2024weh,Schulze:2023ich,Zhao:2024gan,LISACosmologyWorkingGroup:2022kbp,LISACosmologyWorkingGroup:2022jok,Malhotra:2022ply,Wang:2023ost,Yu:2023jrs}, the angular power spectrum is represented as
$C_{\ell} = 4\pi \int d\ln k\, \mathcal{T}_{\ell}^{2}(q,k,\eta_{0}) P(k)$,
where $P(k)$ stands for the power spectrum of the large-scale primordial curvature perturbations, $\mathcal{T}_{\ell}$ is the transfer function of the density contrasts of induced \acp{GW}, and $\ell$ denotes the angular multipole.
It can be numerically calculated with a modified version of \ac{GW}\_CLASS \cite{Zhao:2024gan}, which is the cosmic linear anisotropy solving system \cite{Schulze:2023ich}.
In particular, since $P(k)$ observed by the \ac{CMB} is nearly scale-invariant at large scales, the angular power spectrum of induced \acp{GW} is roughly inversely-proportional to $\ell (\ell+1)$, exhibiting a similar behavior to that of the \ac{CMB} at low-$\ell$ multipoles.
Depending on specific amplitudes of the squeezed primordial non-Gaussianity, they can give arise in the large anisotropies of induced \acp{GW} \cite{Bartolo:2019zvb,Li:2023qua,Li:2023xtl,Rey:2024giu,Ruiz:2024weh,Wang:2023ost,Yu:2023jrs}, which can be as large as or even larger than the anisotropies of astrophysical \ac{GW} background \cite{Cusin:2018rsq,Cusin:2017fwz, Cusin:2019jhg, Cusin:2019jpv, Jenkins:2018kxc, Jenkins:2018uac, Jenkins:2019nks, Wang:2021djr,Mukherjee:2019oma, Bavera:2021wmw,Bellomo:2021mer,Li:2024qcs}.
The multipole-dependence of induced \acp{GW} is different from that of astrophysical \ac{GW} background, which is roughly inversely-proportional to $(2\ell+1)$ \cite{Cusin:2018rsq,Wang:2021djr}, indicating that one can distinguish astrophysical and cosmological \ac{GW} sources in principle.
In the future, the above theoretical analysis of the anisotropies of induced \acp{GW} might be tested by TianQin \cite{Liang:2023fdf, Li:2024lvt}.
In addition, the study of two-point angular correlations of induced \acp{GW} can be generalized to investigate the n-point angular correlations characterizing the non-Gausianity of induced \acp{GW} \cite{Bartolo:2019zvb, Li:2024zwx}, and the cross-correlations between the induced \acp{GW} and other cosmological probes such as the \ac{CMB} \cite{Dimastrogiovanni:2022eir,Schulze:2023ich,Zhao:2024gan,Cai:2024dya}, which can also be important observables to explore the primordial non-Gaussianity.

\subsection{GWs from phase transitions during inflation}

{\it Subsection coordinator: Haipeng An}

The slow-roll inflation model remains one of the most viable frameworks for explaining cosmic inflation. For this model to function effectively, the inflaton field must undergo an excursion comparable to the Planck scale. Additionally, at the end of inflation, the inflaton field must transfer its energy into the \ac{SMPP} particles.
This necessitates a coupling between the inflaton field and other fields, often referred to as spectator fields. In this sub-section, we denote the inflaton field as $\phi$ and the spectator field as $\sigma$.
Generally, the interaction between $\phi$ and $\sigma$ can be parameterized as
\bea
f(\phi) g(\sigma) \,.
\eea
Thus, it becomes evident that during the evolution of inflaton field $\phi$, the properties of the spectator fields-such as their couplings or masses-may undergo significant changes throughout inflation. These changes can trigger phase transitions \cite{An:2020fff,An:2022cce}, which may lead to a variety of rich phenomenological consequences. These include \ac{GW} signals, curvature perturbations, and primordial non-Gaussianities. Such phase transitions could also play a crucial role in addressing fundamental questions about the nature of our universe.
For instance, the latent energy released during these transitions might contribute to the observed dark matter relic abundance today. Additionally, they could act as a mechanism to generate the matter-antimatter asymmetry in the universe. The \ac{GW} signals produced during inflation could further provide a window into ultrahigh-energy particle physics processes, such as phase transitions related to grand unification theory \cite{Hu:2025xdt}.
These phase transitions can generate two types of \acp{GW}.
The first type, referred to as primary \acp{GW}, arises directly from the phase transition through processes like bubble collisions or the formation of topological defects. The second type, known as secondary \acp{GW}, is induced by curvature perturbations directly produced during the phase transition. In the rest of this subsection, we explore the detailed properties of these gravitational wave (GW) signals and their potential detection using the TianQin observatory.

\subsubsection{Primary GW produced by phase transitions during inflation}

The \ac{GW} produced by classical process, such as \acp{FoPT} can be calculated using the Green's function method. The \ac{GW} wave equation in an expanding universe can be written as
\bea\label{eq:PhInGW1}
\hTTij''(\tau,\bk) + \frac{2a'}{a}\hTTij'(\tau,\bk) - \nabla^2 \hTTij(\tau,\bk) = 16 \pi G_N a^2 \sigma^{\rm TT}_{ij} \,,
\eea
where $\tau$ is the conformal time, $a$ is the scale factor, and $\sigma^{\rm TT}$ is the transverse and traceless part of the energy momentum tensor. The $a'/a$ factor in the second term on the left-hand side is a friction due to the expansion of the universe. The solution to (\ref{eq:PhInGW1}) can be written as
\bea
\thTTij(\tau,\bk) = \int d\tau' \tilde G(\tau,\tau'; k) \tilde\sigma^{\rm TT}_{ij}(\tau',\bk) \,,
\eea
where the retarded Green's function in Fourier space satisfies the equation of motion
\bea
\tG(\tau,\tau';k)'' + \frac{2a'}{a}\tG'(\tau,\tau';k) + k^2 \tG(\tau,\tau';k) = a^{-1} \delta(\tau - \tau') \,.
\eea
During inflation, the expansion of the Universe is accelerating. As a result, the conformal time $\tau$ is finite in the future. This is usually called the future event horizon. By shifting the conformal time, one can set the event horizon at $\tau=0$. Take de Sitter inflation as an instance, we have $a(\tau) = -(H\tau)^{-1}$, and the range of $\tau$ is from $-\infty$ to 0.

\begin{figure}[htbp!]
\begin{center}
\includegraphics[width=0.6\textwidth]{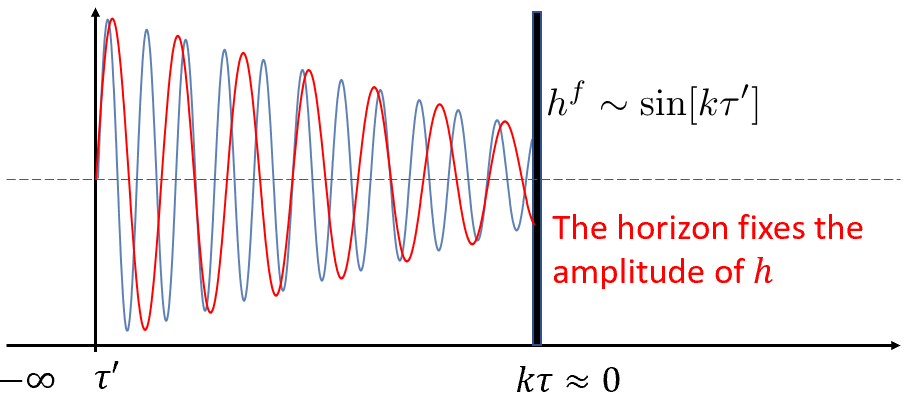}
\caption{Illustration of the evolution of the \ac{GW} mode produced by an instantaneous source during inflation. One can see that modes with different values of the wave number $k$ frozen onto different values $h^f$ at the horizon.}
\label{fig:PhInGW1}
\end{center}
\end{figure}

For a mode, with a certain wave number $\bk$, $\tG(\tau,\tau',k)$ starts to oscillate from zero and then frozen at the future event horizon, as shown in Fig.~\ref{fig:PhInGW1}. The frozen value of $\tG$, denoting as $\tG^f$ will become the amplitude of the oscillation when this mode reenters horizon after inflation, as shown in Fig.~\ref{fig:PhInGW1}. Thus, the amplitude and the energy density of today's \ac{GW} oscillates with the wave number $k$, if the source is instantaneous.

The time scale of a \ac{FoPT} is determined by
\bea
\beta = - \frac{\partial S_B}{d t} \,,
\eea
where $S_B$ is the bounce action of the tunneling process, and $t$ is the physical time.
For a \ac{FoPT} to complete in an expanding universe, the condition $\beta \gg H$ must be satisfied.
Thus, $\beta$ and $H$ separate the wave number domain into three parts. In the region, $k_p > \beta$, where $k_p$ is the physical wave number of the mode at the time of phase transition, the phase transition time scale $\beta^{-1}$ is much larger than the wave length.
As a result, the oscillatory feature of the spectrum is expected to the erased.
In the region $ H < k_p < \beta$, the \ac{GW} source can still be seen as instantaneous.
Therefore, we expect the oscillatory feature of the \ac{GW} spectrum retains in this region.
In the region $k_p < H$, the \ac{GW} mode is already outside the horizon when it is produced.
Therefore, it is already frozen, and does not evolve until reentering the horizon after inflation.

When the \acp{GW} are produced and are inside the horizon $(k_p > H)$, they behave like radiation, and their energy density redshifts as $a^{-4}$.
However, once the mode exits the horizon, the amplitude of the \ac{GW} ceases to decay, and its energy density redshifts only as $a^{-2}$ until it reenters the horizon.
As a result, the \acp{GW} produced during inflation do not experience significant suppression compared to those generated by similar phase transition processes occurring in a radiation dominated universe.

To illustrate this explicitly, consider the case where the universe undergoes instantaneous reheating after inflation.
As shown in \cite{An:2020fff}, the peak value of the \ac{GW} spectrum can be estimated as
\bea
\Omega_{\rm GW}^{\rm peak} \sim \Omega_R \times \left(\frac{H}{\beta}\right)^5 \left(\frac{L}{\rho_{\rm inf}}\right)^2 \,,
\eea
where $\Omega_R \approx 10^{-5}$ represents today's radiation abundance, $L$ is the latent energy density released during the phase transition, and $\rho_{\rm inf}$ is the total energy density of the universe at the time of the phase transition. Compared to \acp{GW} produced during the radiation dominated era, such as those described by (\ref{eq:wall1:env}), the only additional suppression factor here is $(H/\beta)^3$. As discussed in \cite{An:2020fff,An:2022cce}, the typical value of $H/\beta$ for a \ac{FoPT} during inflation is approximately $0.1$.
Therefore, for the instantaneous reheating scenario, the typical peak value of the \ac{GW} signal is around $10^{-10} (L/\rho_{\rm inf})^2$.

\begin{figure}[htbp!]
\begin{center}
\includegraphics[width=0.45\textwidth]{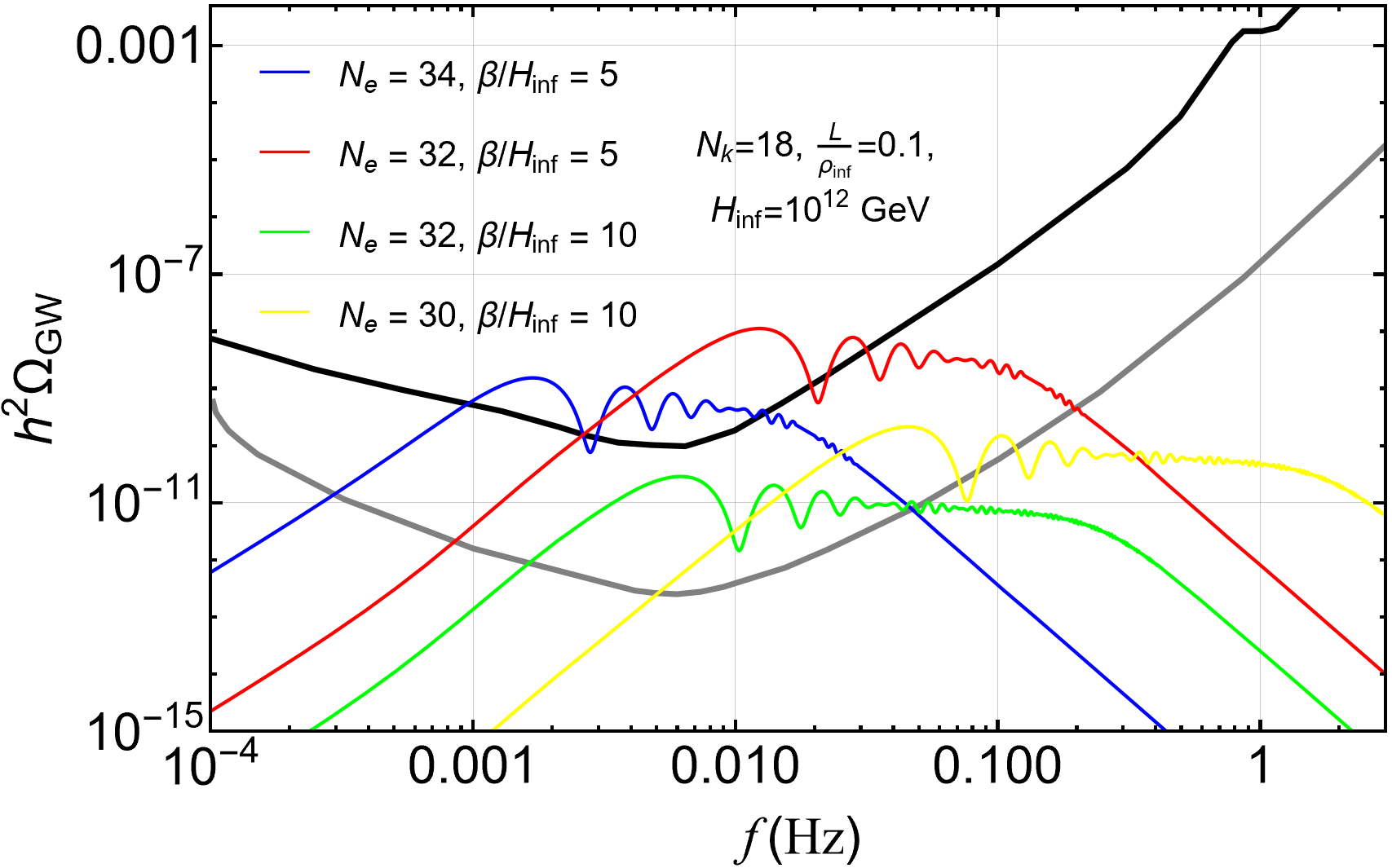}
\caption{Spectra of the primary \acp{GW} produced by \ac{FoPT} during inflation. The black and gray curves are the sensitivity curves of TianQin (see the discussion of Fig.~\ref{fig:PLI-IGW} for more explanation).}
\label{fig:PhInGW3}
\end{center}
\end{figure}

The size of \ac{GW} is also sensitive to the intermediate stage between the end of inflation and the onset of the radiation dominated era.
As detailed in \cite{An:2022cce}, if the \ac{GW} modes reenter the horizon during an intermediate matter dominated era, the present-day value of $\Omega_{\rm GW}$ is relatively suppressed compared to the intermediated reheating case.
Conversely, if the modes reenter during an intermediate kination dominated era, the \ac{GW} signal is significantly enhanced, allowing the peak value of $\Omega_{\rm GW}$ to easily reach $10^{-8}$, as illustrated in Fig.~\ref{fig:PhInGW3}.
Furthermore, not only amplitude but also the slopes of the \ac{GW} spectrum at different frequency regions are influenced by the intermediate stages.
Consequently, once the source of the \ac{GW} is identified through the oscillatory features of the spectrum, the slopes can be utilized to probe the detailed evolution of the universe.

As discussed in \cite{An:2020fff,An:2023idh}, the peak frequency of the primary \ac{GW} produced by the phase transition during inflation is determined by the Hubble parameter at that time.
Then the peak frequency of the \ac{GW} spectrum observed today can be estimated as
\bea\label{eq:PhInGW3}
f^{\rm peak}_{\rm today} \sim H\times e^{ - N_e} \frac{a_{\rm end}}{a_{\rm RH}} \times \left(\frac{T_{\rm CMB}}{T_{\rm RH}}\right)\,,
\eea
where the factor $(T_{\rm CMB}/T_{\rm RH})$ accounts for the redshift after reheating, $(a_{\rm end}/a_{\rm RH})$ represents the redshifts during the intermediate stage, and the factor $e^{-N_e}$ corresponds to the redshift during inflation.
Here $N_e$ denotes the number of e-folds between the phase transition and the end of inflation.
From (\ref{eq:PhInGW3}), it is evident that the peak frequency is strongly dependent on $N_e$.

In Fig.~\ref{fig:PhInGW3}, we present the \ac{GW} spectrum produced by a \ac{FoPT} occurred during inflation.
Assuming a kination domination intermediate stage. We select $N_e$ such that the present-day \ac{GW} frequency lies within the sensitivity range of TianQin.
It is clear with appropriate parameters, the \ac{GW} signal can be sufficiently strong to be detectable by TianQin.

\subsubsection{Primary GW produced by domain walls during inflation}\

The \ac{GW} generated directly during a second-order phase transition is typically too weak to produce any detectable signals.
However, if the phase transition involves symmetry breaking, topological defects are usually formed after the phase transition, which can produce \ac{GW} shortly before the size of these defects exits the horizon.

Immediately after the phase transition, the topological defects are accompanied by small-scale fluctuations.
These fluctuations, however, quickly redshift away, leaving behind a comoving static configuration of the defects.
The results of the detailed simulation of $Z_2$ breaking phase transition can be found in \cite{An:2023idh}.
It is this comovingly static configuration that produces \acp{GW} potentially observable by modern \ac{GW} detectors.

Similar to the \ac{FoPT} scenario, the peak frequency of the \ac{GW} observed today is strongly dependent on the e-fold at which the phase transition occurred, allowing it to be tuned to the frequencies within the detectable range of TianQin.
However, unlike the \ac{FoPT} case, the \ac{GW} spectrum lacks oscillatory features because the source becomes static and persists for a duration much longer than the Hubble time during inflation. As a result, any oscillatory features are expected to be entirely erased.

\subsubsection{Secondary GW induced by phase transition during inflation}\

Since we assume an interaction between the inflaton field $\phi$ and the spectator field $\sigma$, the phase transition in the $\sigma$ sector will inevitably induce a backreaction on the inflaton field. For instance, as discussed in \cite{An:2020fff,An:2022cce,An:2023idh,An:2023jxf}, we consider an interaction of the form
\bea\label{eq:PhIn4}
c \phi^2 \sigma^2 \,,
\eea
where $c$ is the coupling. By expressing $\phi = \phi_0 + \delta\phi$ with $\phi_0$ representing the homogeneous part and $\delta\phi$ the perturbation, the interaction term (\ref{eq:PhIn4}) can be written as
\bea\label{eq:PhIn5}
2 c\phi_0 \delta\phi \sigma^2 = 2 c \phi_0^2 M_{\rm pl}^{-1} \kappa \sigma^2 \delta\phi \,,
\eea
where $\kappa \equiv M_{\rm pl}/\phi_0$. During the phase transition, $c\phi_0^2$ becomes comparable to the mass squared of the spectator field. Thus, it is reasonable to assume that during the phase transition, the combination $c\phi^2_0$ is of the same order as $L^{1/2}$, where $L$ is the latent energy density of the phase transition. From (\ref{eq:PhIn5}), the factor $2 c \kappa \phi_0^2 M_{\rm pl}^{-1}$ can be interpreted as a source term for $\delta\phi$. By calculating this source term for a given phase transition, one can convolve it with the Green's function of $\delta\phi$ to determine the induced curvature perturbation.

The results of a lattice simulation for this source term are presented in \cite{An:2023jxf}. From $\delta\phi$ the curvature perturbation $\zeta$ can be derived. Using the standard methods, the curvature induced \ac{GW}, referred to as the secondary \ac{GW} can then be calculated \cite{Baumann:2007zm,Kohri:2018awv}.

The spectrum of the secondary \ac{GW} can be written as \cite{An:2023jxf}
\bea
\Omega_{\rm GW}^{(2)}(f) = \Omega_R A^2_{\rm ref} {\cal F}_2\left(\frac{k_p}{H}\right) \,,
\eea
where
\bea
A_{\rm ref} = \frac{24 \kappa^2}{\epsilon} \left(\frac{H}{\beta}\right)^3 \left(\frac{L}{\rho_{\rm inf}}\right)^2 \,,
\eea
where $\epsilon$ is the slow-roll parameter of the inflaton field during the phase transition, and ${\cal F}_2$ is a form factor. Its definition can be found in \cite{An:2023jxf}. Compared to the primary \ac{GW}, although the secondary \ac{GW} is suppressed by more powers of $H/\beta$ and $L/\rho_{\rm inf}$, it is enhanced by $\epsilon^{-2}$. Thus, in some parameter space, the size of the secondary \ac{GW} can be larger to the primary \ac{GW}.

\begin{figure}[htbp!]
\begin{center}
\includegraphics[width=0.45\textwidth]{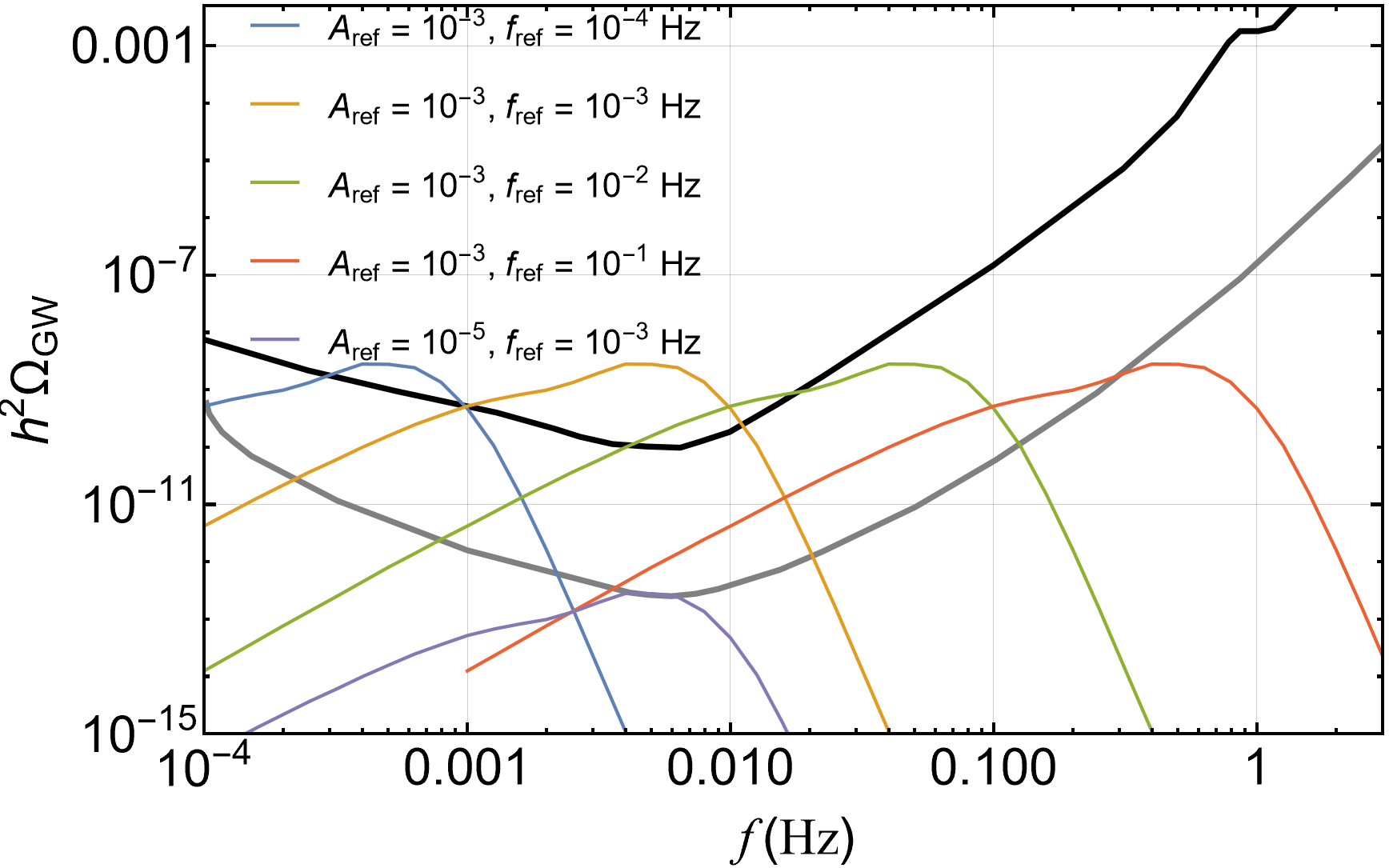}
\caption{Spectra of the secondary \acp{GW} produced by \ac{FoPT} during inflation. The black and gray curves are the sensitivity curves of the TianQin (see the discussion of Fig.~\ref{fig:PLI-IGW} for more explanation).}
\label{fig:PhInGW4}
\end{center}
\end{figure}

Similar to the case of the primary \ac{GW}, the frequency of the secondary \ac{GW} can also be tuned by adjusting $N_e$ to fall within the detectable range of TianQin.
In Fig.~\ref{fig:PhInGW4}, we present the spectrum of the secondary \ac{GW}, assuming instantaneous reheating.
It is evident that with the enhancement of the slow-roll parameter, the relic abundance of secondary \ac{GW} can easily exceed the sensitivity threshold of TianQin.
As discussed in \cite{An:2023jxf}, the infrared slop of the secondary \ac{GW} exhibits a distinctive feature.
The infrared region of ${\cal F}_2$ can be approximated as
\bea
{\cal F}_2^{\rm IR}(x) \approx x^3 \left( \frac{6}{5}\log^2 x + \frac{16}{25}\log x + \frac{28}{125} \right) \,,
\eea
where the logarithmic terms in the bracket arise from the transfer function. These logarithmic factors soften the infrared slope, making the spectrum distinct from that of \ac{GW} induced by a \ac{FoPT} occurring during the radiation-dominated era.

As indicated in \eqref{eq:PhInGW3}, the frequency of the \ac{GW} signal produced by phase transitions during inflation today depends on both the Hubble parameter during inflation and the number of e-folds at which the phase transition occurred. Consequently, in principle, the frequency can span a wide range of values. To detect such a signal, it is essential to utilize a variety of \ac{GW} detectors capable of covering different frequency ranges. The TianQin observatory, with its exceptional sensitivity in the $10^{-4} - 1$ Hz range, will play a crucial role in the search for \ac{GW} signals originating from phase transitions during inflation.

\subsection{SGWB from cosmic string networks}

{\it Subsection coordinator: Ligong Bian}

Cosmic string is one-dimensional topological defect formed after phase transitions when the symmetry gets broken spontaneously \cite{Kibble:1976sj,Hindmarsh:1994re}.
For thin and local strings with no internal structures, the cosmic string dynamics can be described by the Nambu-Goto action. In the scenario, infinite strings would reach scaling regime \cite{Bennett:1987vf,Allen:1990tv,Sakellariadou:1990nd} and go to loops when the intercommutation of intersecting string segments occurs \cite{Vachaspati:1984dz}.
The small loops emit \ac{GW} bursts through the cusps and kinks \cite{Damour:2001bk,Damour:2000wa}, and the \ac{SGWB} is formed through superposition of uncorrelated \ac{GW} bursts from many cosmic strings.
The SGWB is characterized by the string tension ($\mu$) and the loop number density \cite{Vilenkin:2000jqa}, where the dimensionless parameter $G\mu$ ($G$ is the Newtonian constant) parameterizes the gravitational interactions of strings and can be tightly connected with the symmetry breaking scale, and satisfy the relation of $G\mu\sim(\eta/M_{\rm Mpl})^2$ when the cosmic strings are predicted after symmetry breaking at scale of $\eta$.
Therefore, the detection of SGWB emitted from cosmic strings provides an inspiring way to access new physics beyond \ac{SMPP} physics that are inaccessible by high-energy colliders \cite{King:2021gmj,King:2020hyd,Buchmuller:2019gfy,Caldwell:2022qsj}, such as: the seesaw scale and its relation with the leptogenesis  \cite{Dror:2019syi} and some superheavy dark matter scenarios \cite{Bian:2021vmi,Bian:2021dmp}.
Different from local strings, the global strings mostly decay to particles \cite{Saurabh:2020pqe,Baeza-Ballesteros:2023say}.
One of the most motivated global string is the axion string, the axion arises as a pseudo-Nambu-Goldstone boson after the Peccei-Quinn symmetry breaking \cite{DiLuzio:2020wdo}.
Supposing the Peccei-Quinn symmetry is broken in the post-inflationary case, random initial axion field distribution in uncorrelated horizon-size regions can lead to axion strings formation \cite{Kibble:1976sj,Vilenkin:1982ks}.
Axion strings after formation will reach scaling and release their energy by radiating axions and \acp{GW} until QCD phase transition \cite{Yamaguchi:1998gx,Hiramatsu:2010yu}.

The SGWB of cosmic strings comes from three contributions: cusps, kinks, and kink-kink collision, with the SGWB spectrum given by
\begin{equation}\label{eq:sgwb}
\Omega_{\rm GW}(t_{0}, f) = \frac{f}{\rho_{c}}\ \frac{d}{df}\rho_{\rm GW}^{~}(t_{0}, f),
\end{equation}
with $\rho_{c}=\frac{3H_{0}^{2}}{8\pi G}$ being the critical energy density of the universe.
And, $\frac{d}{df}\rho_{\rm GW}^{~}(t_{0}, f)$ is the \ac{GW} energy density per unit frequency at present.
Generally, one have contributions from different modes of loops oscillation ($n$),
\begin{equation}\label{eq:dpf}
\frac{d}{df}\rho_{\rm GW}^{~}(t_0,f)=G\mu^{2}\ \sum_{n}C_{n}(f)\ P_{n}\;.
\end{equation}
Here, $C_n(f)$ is a function of loop distributions, and $P_{n}$ is adopted from numerical simulations to take into account all contributions from cusps, kinks, and kink-kink collisions \cite{Blanco-Pillado:2017oxo}.

In the BOS model, the SGWB from cosmic strings is obtained after the loop production functions for non-self-intersecting loops was obtained from Nambu-Goto simulations of cosmic strings by Blanco-Pillado, Olum, Shlaer \cite{Blanco-Pillado:2013qja,Blanco-Pillado:2011egf}.
With the subscript $r$ to denote radiation dominated era, $rm$ represents the case where loops are formed in radiation dominated era and survive till matter dominated era, and $m$ for the case of matter dominated era, the loop distribution functions are:
\begin{eqnarray}
n_{r}(l, t)&=&\frac{0.18}{t^{3/2}(l + \Gamma G\mu t)^{5/2}}, l/t \leqslant 0.1,\\
n_{rm}(l, t)&=&\frac{0.18t_{\rm eq}^{1/2}}{t^2\ (l + \Gamma G\mu t)^{5/2}},\, l/t < 0.09 t_{\rm eq}/t - \Gamma G\mu,\\
n_{m}(l, t)&=&\frac{0.27-0.45(l/t)^{0.31}}{t^{2}(l + \Gamma G\mu t)^{2}}, l/t < 0.18\;.
\end{eqnarray}
Where, $l/t \leqslant 0.1$ is to consider the cutoff of the maximum size of loops in the radiation dominated era.
And, the constrain on loop size, $l/t < 0.09 t_{\rm eq}/t - \Gamma G\mu$, is to consider that loops formed in the radiation era will survive till the time of radiation-matter equality and continue to emit \acp{GW} in the matter dominated era.
Finally, the loop size should satisfy $l/t < 0.18$ considering loops can also be produced when the cosmic string networks reach the scaling regime in the matter dominated era. With the loop production functions, one has
\begin{equation}
C_{n}(f) = \frac{2n}{f^{2}}\int_{z_{\rm eq}}^{z_{\rm cut}}
\frac{dz}{H_{0}\sqrt{\Omega_{r}}(1+z)^{8}}\ n_{r}(l,t),
\end{equation}
for the SGWB contributions from the radiation era. Here, the redshift in the radiation-matter equality is $z_{\rm eq}$ and the cutoff redshift is $z_{\rm cut}$.
Since both loops that survive to matter dominated era and loops formed in matter dominated era would produce \acp{GW}, the $C_{n}$ will take the form of
\begin{equation}
C_{n}(f) = \frac{2n}{f^{2}}\int_{0}^{z_{\rm eq}}
\frac{dz}{{H_{0}\sqrt{\Omega_{m}}}(1+z)^{15/2}}\ n_{i}(l,t),
\end{equation}
with the subscript $i$ being $rm$ and $m$.

With the distribution of non-self-intersecting scaling loops extracted from simulations rather than the loop production function \cite{Ringeval:2005kr}, the loop distribution functions are obtained analytically by Lorenz, Ringeval, and Sakellariadou \cite{Lorenz:2010sm} (denoted as the LRS model), through which one can also calculate the \ac{GW} spectrum from cosmic strings.
In this model, the \acp{GW} contribution from the radiation dominated era is
\begin{equation}
\Omega_{\rm GW}(f)=\frac{64\pi G^2\mu^2\Omega_r}{3}\sum_n P_n \int dx\ n(x),
\end{equation}
and the \ac{GW} spectrum is given by
\begin{equation}
\Omega_{\rm GW}(f)=\frac{162\pi G^2\mu^2}{\Omega_m^{-2}H_0^{-2}}\frac{1}{f^2}\sum_n\ P_n\ n^2\int_{}^{}dx\ n(x)\,,
\end{equation}
in the matter dominated era. In the model, on the scales $x = l/t \gg \Gamma G\mu \equiv x_d$, the numerical simulation gives
\begin{equation}\label{nx0}
n(x) = \frac{C_0}{x^p},
\end{equation}
with the two constants $C_0$ and $p$ are
\begin{align}
p = 0.60^{+0.21}_{-0.15}\big{|}_r,&\quad p = 0.41^{+0.08}_{-0.07}\big{|}_m\,,&\\
C_0 = 0.21^{-0.12}_{+0.13}\big{|}_r,&\quad C_0 = 0.09^{-0.03}_{+0.03}\big{|}_m\,,&
\end{align}
in the radiation dominated era and matter dominated era, respectively.
The specific distributions are given as \cite{Lorenz:2010sm}
\begin{eqnarray}
n(x > x_d) & \simeq & \frac{C}{(x+x_d)^{3-2\chi}},\\
n(x_c< x < x_d) & \simeq & \frac{C(3\nu-2\chi-1)}{2-2\chi}\frac{1}{x_d}\frac{1}{x^{2(1-\chi)}},\\
n(x < x_c < x_d) & \simeq & \frac{C(3\nu-2\chi-1)}{2-2\chi}\frac{1}{x_c^{2(1-\chi)}}\frac{1}{x_d}\,.
\end{eqnarray}
Here, $C=C_0(1-\nu)^{2-p}$, $\nu=1/2$ and $\nu=2/3$ for radiation and matter dominated era. And, the length scale $x_c \ll x_d$ is the so-called ``gravitational back-reaction scale'', which is given by
$x_c = 20(G\mu)^{1+2\chi}$ after matching the loop distribution on scales $x\gg x_d$
((\ref{nx0})), where $\chi_r=0.2^{+0.07}_{-0.10}$ and
$\chi_m=0.295^{+0.03}_{-0.04}$ for the radiation and matter dominated era \cite{Auclair:2019wcv}.

Another single scale model based on the velocity dependence of long string network motion is the VOS model, its \ac{GW} spectrum is given as \cite{Cui:2018rwi, Gouttenoire:2019rtn,Gouttenoire:2019kij}:
\begin{equation}
\Omega_{\rm GW}^{\rm cs}(f) =\sum_k \Omega_{\rm GW}^{(k)}(f)\;,
\end{equation}
with each $k-$mode's contribution being,
\begin{eqnarray}\label{eq:GWdensity2}
\Omega_{\rm GW}^{(k)}(f) =&&
\frac{1}{\rho_c}
\frac{2k}{f}
\frac{\mathcal{F}_{\alpha}\,\Gamma^{(k)}G\mu^2}
{\alpha_{CS}\left( \alpha_{CS}+\Gamma G\mu\right)}
\int_{t_F}^{t_0}\!d\tilde{t}\;
\frac{C_{\rm eff}(t_i^{(k)})}{t_i^{(k)\,4}}\times\bigg[\frac{a(\tilde{t})}{a(t_0)}\bigg]^5\bigg[\frac{a(t^{(k)}_i)}{a(\tilde{t})}\bigg]^3\Theta(t_i^{(k)} - t_F)~.~~~~
\end{eqnarray}
Where, $\alpha_{CS}$ is the loop size parameter, the parameter $\mathcal{F}_{\alpha}$ characterizes the proportion of the energy released by long strings, which is about $0.1$. The loop production efficiency ($C_{\rm eff}$) is taken as $5.4$ and $0.39$ for the radiation- and matter-dominated Universe \cite{Gouttenoire:2019kij}, and the gravitational emission efficiency of loops is $\Gamma \approx 50$ \cite{Blanco-Pillado:2017oxo}. The Fourier modes of cusps \cite{Olmez:2010bi, Blanco-Pillado:2013qja,Blanco-Pillado:2017oxo} is
$\Gamma^{(k)} = \frac{\Gamma k^{-\frac{4}{3}}}{\sum_{m=1}^{\infty} m^{-\frac{4}{3}}} \;
$ where $\sum_k \Gamma^{(k)}=\Gamma$ and $\sum_{m=1}^{\infty} m^{-\frac{4}{3}} \simeq 3.60$.
The formation time of the $k$-mode loop is estimated through
\begin{equation}\label{eq:ti}
t_i^{(k)}(\tilde{t},f) = \frac{1}{\alpha_{CS}+\Gamma G\mu}\left[
\frac{2 k}{f}\frac{a(\tilde{t})}{a(t_0)} + \Gamma G\mu\;\tilde{t}\;
\right]\,,
\end{equation}
with $\tilde{t}$ being the \ac{GW} emission time.
And the formation temperature of cosmic string network is $t_F$, when cusps dominate the small-scale structure of loops, the high mode and the low mode are related to each other through:
$\Omega_{\rm GW}^{(k)}(f)
= \frac{\Gamma^{(k)}}{\Gamma^{(1)}}\,\Omega_{\rm GW}^{(1)}(f/k)
=k^{-4/3}\,\Omega_{\rm GW}^{(1)}(f/k)$\;.

In the following, we demonstrate the theoretical prediction of the \ac{GW} spectrum for the global string based on recent lattice field simulations \cite{Gorghetto:2021fsn,Jia:2024ejr}. Firstly, the number density of string per Hubble patch, i.e., the scaling parameter and can be extracted from the simulations as
\begin{align}
\xi(t) = \frac{l_s t^2}{a(t)^2 V} \;.
\end{align}
Where, $V$ is the comoving box volume and $l_s$ is the string length, the time t is given by $t = 1/(2H)$ in the radiation dominated Universe. Numerical simulations suggest that the scaling parameter of the axion string enter the linear growth of $\log(m_r/H)$ \cite{Buschmann:2021sdq,Gorghetto:2018myk,Jia:2024ejr}. With the scaling parameter at hand, one can estimate the string energy density through $ \rho_s = \pi f_a^2 \ln(t/(\sqrt{\xi}d_s)) \xi / t^2$ with the $d_s=m_r^{-1}$ being the string width where the $m_r$ is the mass term of the radio mode of the Peccei-Quinn complex field. Generally, axion string would release massless axion and the gravitation wave as \cite{Kawasaki:2014sqa}:
\begin{align}
\frac{ \mathrm{d}\rho_{s}(t)}{ \mathrm{d}t} = - 2 H(t) \rho_{s}(t) - \left[\frac{\mathrm{d} \rho_{s}(t)}{\mathrm{d}t}\right]_{emi} \;, \\
\frac{ \mathrm{d}\rho_{a}(t)}{ \mathrm{d}t} = - 4 H(t) \rho_{a}(t) + \left[\frac{\mathrm{d} \rho_{s}(t)}{\mathrm{d}t}\right]_{emi1} \;, \\ \label{eq:enqx}
\frac{ \mathrm{d}\rho_{gw}(t)}{ \mathrm{d}t} = - 4 H(t) \rho_{gw}(t) + \left[\frac{\mathrm{d} \rho_{s}(t)}{\mathrm{d}t}\right]_{emi2} \;.
\end{align}
For the axion string with the emission rate of \acp{GW} energy density being $\Gamma_{gw} = a^{-4}\frac{\mathrm{d} (a^4 \rho_{gw})}{\mathrm{d}t}$, the instantaneous emission spectrum $F(x,y)$ of the \ac{GW} can be defined as \cite{Gorghetto:2021fsn}:
\begin{align}
F_g(x,y)=\frac{H}{\Gamma_{gw}}\frac{1}{a^3}\frac{\partial}{\partial t} \left(a^3\frac{\partial\rho_{gw}}{\partial k} \right) \;.
\end{align}
Here, $x=k/H$ and $y=m_r/H$. The function $F_g(x,y) \propto x^{-q}$ in the momentum range $[2\pi, m_r/(2H)]$ can be extracted from the numerical simulation results with $q\approx 2$ for axion strings. \cite{Gorghetto:2021fsn}. After taking the theoretical prediction $\Gamma_{gw} \simeq \gamma \Gamma_a G \mu^2/f_a^2 $ with $\Gamma_a = a^{-4}\frac{\mathrm{d} (a^4 \rho_a)}{\mathrm{d}t}$, one can obtain the spectrum at the end time of string evolution after integration of
\begin{align}
\frac{\partial \rho_{gw}}{\partial k}(k,t) = \frac{1}{a^3}\int \mathrm{d} t^\prime \frac{\Gamma_{gw}^\prime}{H^\prime}(a^\prime)^3 F^\prime_g(\frac{k^\prime}{H^\prime},\frac{m_r}{H^\prime}) \;.
\end{align}

\begin{figure}[!htp]
\centering
\includegraphics[width=0.6\textwidth]{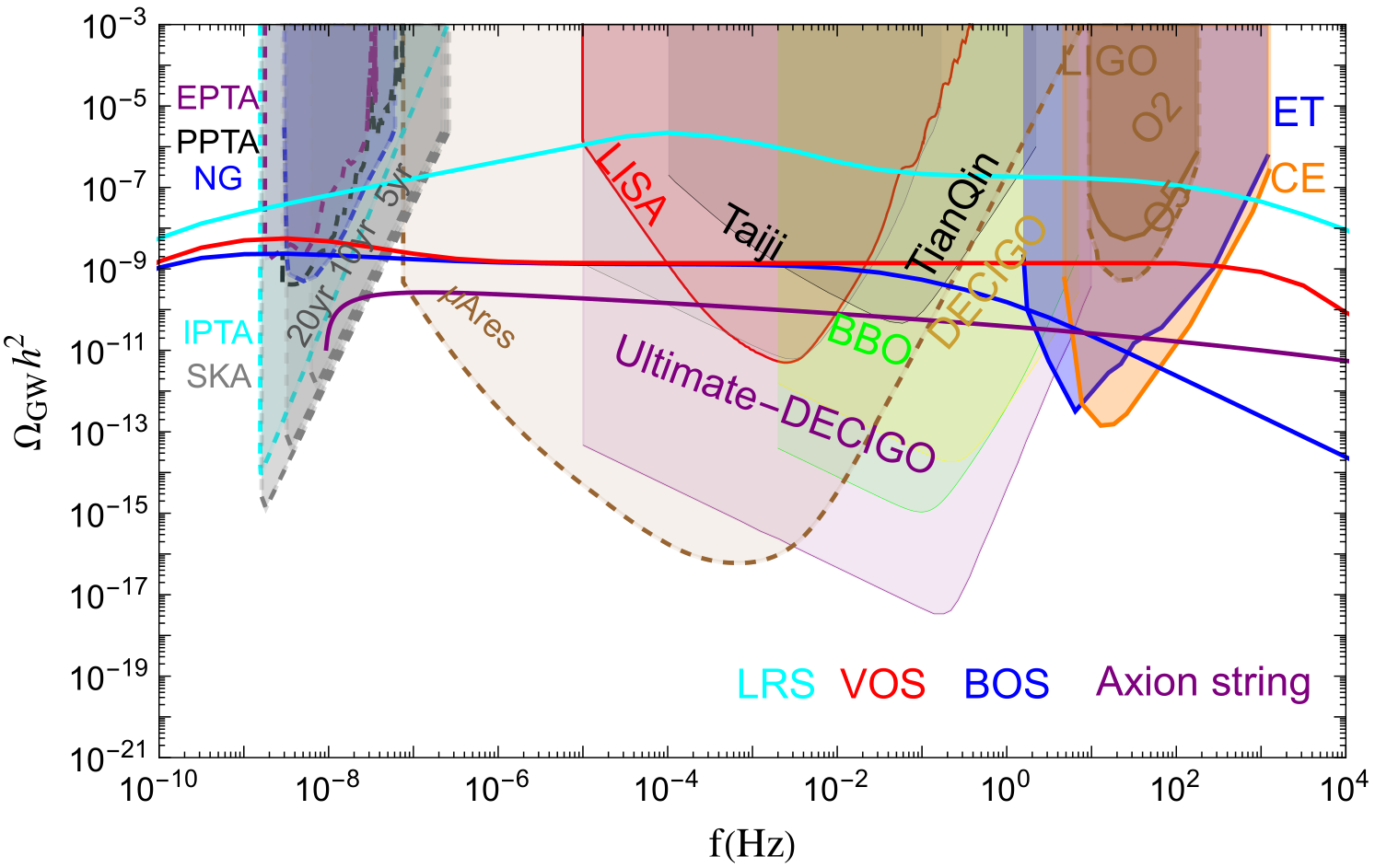}
\caption{Expected spectra of SGWB from cosmic string for the LRS, VOS, and BOS models with $G\mu=
10^{-10}$, which are indicated by cyan, red and blue solid (dashed) lines. The sensitive of TianQin \cite{TianQin:2015yph,Zhou:2023rop} and other \ac{GW} detectors (including LISA \cite{LISA:2017pwj,Baker:2019nia}, Taiji \cite{Hu:2017mde,Ruan:2018tsw}, $\mu Ares$ \cite{Sesana:2019vho}, DECIGO \cite{Seto:2001qf,Kawamura:2011zz,Yagi:2011wg,Isoyama:2018rjb}, BBO \cite{Crowder:2005nr,Corbin:2005ny,Harry:2006fi}, SKA \cite{Janssen:2014dka}, \ac{ET} \cite{Punturo:2010zz,Hild:2010id}, CE \cite{LIGOScientific:2016wof} and LIGO-Virgo-Kigro \cite{LIGOScientific:2014qfs,Thrane:2013oya,LIGOScientific:2019vic}) are presented.}\label{FSGW}
\end{figure}

Different models suggest the \acp{GW} spectra of cosmic strings can span a wide frequency range with a plateau in the high-frequency region.
Therefore, it was expected that the SGWB from cosmic string is one of the most promising target since it can be distinguished from other \ac{GW} sources through a joint detection with ground-based, space-based and \ac{PTA} \ac{GW} detectors \cite{LIGOScientific:2021nrg,LIGOScientific:2017ikf,Auclair:2019wcv,Yonemaru:2020bmr,Sanidas:2012ee}.
Recent study of \cite{Bian:2022tju} shows the complementarity of PTA and LIGO on searching for \acp{GW} from cosmic strings.
In Fig.~\ref{FSGW}, we present the SGWB spectra of the cosmic string-induced SGWB for the local strings through LRS, VOS, and BOS models, and the global string based on lattice simulation of \cite{Jia:2024ejr}. As depicted in the figure,
the LRS model predicts harder spectra, and the BOS and VOS models predict similar spectra in the frequency region of $f\lesssim$mHz.
The magnitude of the \ac{GW} spectrum of the LRS model is much higher at high frequency range because that more small loops in the loop distribution function dominate that range. And, the axion string gives a much lower magnitude \ac{GW} spectrum because that the global strings mostly decay to axions rather than \acp{GW}, see recent numerical studies of Refs.\cite{Jia:2024ejr,Baeza-Ballesteros:2023say} for more detail.

\subsection{Summary of the section}

\ac{GW} astronomy has opened up new avenues for exploring physics beyond \ac{SMPP}, particularly in probing the hidden sectors and other exotic phenomena that do not interact electromagnetically.
These investigations can provide crucial insights into the nature of dark matter, and other novel particles and forces that conventional methods have failed to detect.
In particular, TianQin can detect ultra light dark matter with masses in the ranges $10^{-19} \sim 10^{-15}$ eV and $10^{-13.5}\sim 10^{-11.5}$ eV.
For new physical models and new Higgs potentials that can generate a first-order \ac{EWPT} in the universe, which could be responsible for the origin of matter-antimatter asymmetry and the new mechanism of heavy dark matter production through phase transition, TianQin is expected to explore the parameter space with phase transition strength greater than 0.1.

\acp{PBH} could be produced when the curvature perturbation is strong enough, and the concomitant \ac{SGWB} induced by such enhanced curvature perturbation is also enhanced and could be detectable.
For \ac{SGWB} in the mHz frequency band, the corresponding \acp{PBH} lie in the asteroid-mass window and can account for all the dark matter.
Such induced \acp{GW} in the mHz band is more or less robust against the non-Gaussanity of the curvature perturbation and the dispersion of its power spectrum.
The asteroid-mass window of the \ac{PBH} dark matter can be thoroughly covered by TianQin.
The anisotropies of such induced \ac{SGWB} can be used to probe the primordial non-Gaussianity in the future.

The frequency of the \ac{GW} signal produced by phase transitions during inflation to be observed today depends on both the Hubble parameter during inflation and the number of e-folds at which the phase transition occurred.
As a result, the frequency can, in principle, span a broad range of values. To detect such a signal, it is essential to employ a diverse array of \ac{GW} detectors capable of covering different frequency ranges.
Typically, if the phase transition occurred around 20-30 e-folds after the CMB modes exited the horizon during inflation, the frequency of the corresponding \acp{GW} will fall within the millihertz range and is potentially detectable by TianQin.

TianQin has the capability to probe SGWB radiated from local strings when the \acp{GW} are as predicted by some models with string tension $G\mu\gtrsim \mathcal{O}(10^{-10})$, this corresponds to new physics energy scales above $\sim 10^{14}$ GeV, which is far beyond the current and future colliders' ability.
TianQin can also probe the \acp{GW} emitted from the global string forming after the Peccei-Quinn  symmetry breaking with axion decay constant $f_a\gtrsim \mathcal{O}(10^{15})$ GeV.
A search of the new physics related to the cosmic strings requires joint search of TianQin with detectors covering other frequency band since the GW spectrum from cosmic string has a unique plateau shape range from NanoHertz to kHZ.


\clearpage

%% file: sec4-cosmo.tex
\section{Cosmology with TianQin} \label{sec:tq_cosmo}

{\it Section coordinator: Liang-Gui Zhu}

To reconstruct the cosmic expansion history, i.e., to measure the various parameters
related to the expansion of the universe, it is essential to gather information on the distances and redshifts
of objects at different cosmological distances. Traditional \ac{EM} wave-mediated observations
are very challenging for measuring distances and require the use of a complex cosmic distance ladder system,
however, redshifts are relatively easier to measure,
because each kind of atom possesses a unique spectral signature.
\ac{GW} observations provide complementary capabilities to \ac{EM} observations.
While measuring distances via \ac{GW} observations is straightforward, obtaining redshifts is more challenging.
The frequency and its time derivative of a \ac{GW} signal are entirely determined by the masses of the source binary, while the amplitude of the signal is inversely proportional to the distance \cite{Schutz:1986gp, Cutler:1994ys}.
Once the \ac{GW} signal has propagated through the \ac{FLRW}
geometric background, the observer can directly measure both the redshifted masses and the luminosity
distance to the \ac{GW} source.
However, due to the degeneracy between the mass and redshift \cite{Schutz:1986gp},
extracting the redshift independently remains a challenging task.
Since \ac{GW} detections can directly measure the luminosity distances to the \ac{GW} sources,
when combined with redshift information obtained from \ac{EM} observations,
\acp{GW} can act as \emph{standard sirens}, providing a powerful probe of expansion history of the universe.
 \cite{Schutz:1986gp, Markovic:1993cr, Holz:2005df}.

The use of standard sirens to measure cosmological parameters has been demonstrated in the detection of \ac{GW}170817,
the first binary neutron star merger detected by LIGO and Virgo \cite{LIGOScientific:2017ync, Margutti:2020xbo}.
This detection marked the first realization of the bright standard siren method \cite{LIGOScientific:2017vwq}.
Additionally, detections of other compact object binary mergers without \ac{EM} counterparts
(mainly \acp{SBHB}) realized the first cosmological constraints
that using dark siren method \cite{DES:2019ccw, LIGOScientific:2018gmd, LIGOScientific:2019zcs}.
So far, the first three runs of the \ac{LVK} network have measured the Hubble-Lema\^itre constant $H_0$
with a precision of about 10\% \cite{DES:2020nay, Vasylyev:2020hgb, DESI:2023fij, LIGOScientific:2021aug}.
This precision is still significantly lower than that of traditional \ac{EM} observations,
such as measurements derived from the \ac{CMB} \cite{Planck:2018vyg}
and type Ia supernovae (SNe Ia) \cite{Riess:2020fzl, Freedman:2019jwv}.
However, with anticipated upgrades to ground-based \ac{GW} detectors \cite{KAGRA:2013rdx, Punturo:2010zz, LIGOScientific:2016wof}, the precision of $H_0$ measurements from standard sirens is expected to improve to
better than 1\% \cite{Chen:2020zoq, Zhu:2023jti, Muttoni:2023prw, Song:2022siz, Cai:2016sby}.

Space-based \ac{GW} detectors are primarily designed to detect \ac{GW} sources with frequencies around the millihertz range, thereby
complementing the detection frequency band of ground-based \ac{GW} detectors.
The prospects for standard sirens in space-based \ac{GW} detections were first proposed
for the LISA mission \cite{Holz:2005df}, with early studies primarily focusing on mergers of \acp{MBHB} \cite{Babak:2010ej,
Petiteau:2011we, Tamanini:2016zlh, Caprini:2016qxs, Cai:2017yww}.
As the astrophysical detection capabilities of LISA have been further analyzed
\cite{Amaro-Seoane:2007osp, Klein:2015hvg, Babak:2017tow, Kyutoku:2016ppx, LISA:2017pwj, LISA:2022yao},
the candidate sources for standard sirens have expanded to include \acp{EMRI}
 \cite{MacLeod:2007jd, Laghi:2021pqk} and \ac{SBHB} inspirals \cite{DelPozzo:2017kme,
Muttoni:2021veo} (see the literature \cite{LISACosmologyWorkingGroup:2022jok} for a review).

TianQin, a Chinese space-based \ac{GW} detection plan proposed in 2014 \cite{TianQin:2015yph, Gong:2021gvw},
the probing capability of TianQin in astrophysics is similar to LISA \cite{LISA:2017pwj, Colpi:2024xhw}
and Taiji \cite{Hu:2017mde} but with its own characteristics
 \cite{Wang:2019ryf, Liu:2020eko, Fan:2020zhy, Hu:2017yoc}.
The potential of TianQin for constraining cosmological parameters is the primary focus of this section.
The structure of this section is as follows.
Section \ref{sec:sirens_tq} outlines the basic methodology for using TianQin to constrain the
cosmic expansion history through standard sirens.
Section \ref{sec:LCDM} explores the prospects for TianQin to constrain $\Lambda$CDM
cosmological parameters, including $H_0$, the fractional matter density parameter $\Omega_M$,
and the fractional dark energy density parameter $\Omega_\Lambda$.
Section \ref{sec:cosmo_de} present the potential of TianQin to constrain the \ac{EoS} parameters
of dark energy.
Section \ref{sec:cosmo_TQLISA} present the potential of TianQin combined with other
space-based \ac{GW} detectors on enhancing constraints of the cosmic expansion history.
Section \ref{sec:lensing} addresses the role of gravitational lensing effects of \ac{GW} signals from \acp{MBHB}
in improving constraints on the expansion history of the universe.
Finally, Section \ref{sec:cosmo_summary} summarizes TianQin's prospects for the cosmological inferences.

\subsection{Standard sirens with TianQin} \label{sec:sirens_tq}

\subsubsection{Principles of standard sirens} \label{sec:sirens_principle}
In the framework of GR, the \ac{GW} signal emitted by a compact object binary inspiral system
propagating in a FLRW spacetime background can be expressed as \cite{Colpi:2016fup}
\begin{subequations}
\begin{align}
 h_+ (t) &= \left(\frac{G \mathcal{M}_z}{c^2} \right)^{5/3} \left(\frac{\pi f(t)}{c} \right)^{2/3} \frac{2 (1+ \cos^2 \iota)}{D_L} \cos\! \big[\Psi (t, \mathcal{M}_z, \eta,...) \big], \label{eq:h_plus} \\
 h_{\times} (t) &= \left(\frac{G \mathcal{M}_z}{c^2} \right)^{5/3} \left(\frac{\pi f(t)}{c} \right)^{2/3} \frac{4 \cos \iota}{D_L} \sin\! \big[\Psi (t, \mathcal{M}_z, \eta,...) \big], \label{eq:h_cross}
\end{align}
\end{subequations}
where $h_{+,\times}$ denote the two polarization amplitudes of the \ac{GW} signal,
$c$ is the speed of light in vacuum, $G$ represents Newton's gravitational constant,
$\mathcal{M}_z \equiv (1+z)\mathcal{M}$ represents the redshifted chirp mass,
$\eta$ represents the symmetric mass ratio, $\iota$ is the inclination angle of
the binary orbital angular momentum relative in the line of sight,
$D_L$ is the luminosity distance of the \ac{GW} source to observer, and
$\Psi (t, \mathcal{M}_z, \eta)$ represents the phase of the \ac{GW} signal.
One can see that the phase of the \ac{GW} signal is determined only by parameters related to
the mass of the \ac{GW} source and that the two polarizations of the \ac{GW} signal depend differently
on the cosine of the inclination angle $\cos \iota$.
The masses of the \ac{GW} source can be directly inferred from the measurements of phase of the \ac{GW} signal,
and break the degeneracy between
$\iota$ and $D_L$ by measurements of the two polarizations. So that the estimation of luminosity distance
of the \ac{GW} source can be realized directly by \ac{GW} detections alone, without requiring external calibration.
This property underpins the concept of \ac{GW} signals from compact binary systems as standard sirens \cite{Schutz:1986gp}.

Constraining the cosmic expansion history with standard sirens basically uses the luminosity distance
information provided by the standard sirens to fit the $D_L - z$ relation.
In the FLRW background, the $D_L - z$ relation can be expressed as \cite{Hogg:1999ad}
\begin{equation} \label{eq:DL_z}
\renewcommand\arraystretch{1.5}
D_L = \frac{c(1+z)}{H_0} \left\{\begin{array}{ll}
\frac{1}{\sqrt{\Omega_K}} \sinh\! \big[\sqrt{\Omega_K} \int_{0}^{z} \frac{H_0}{H(z')} \D z' \big], & \textrm{for $\Omega_K > 0$}, \\
 \int_{0}^{z} \frac{H_0}{H(z')} \D z', & \textrm{for $\Omega_K = 0$}, \\
\frac{1}{\sqrt{|\Omega_K|}} \sin\! \big[\sqrt{|\Omega_K|} \int_{0}^{z} \frac{H_0}{H(z')} \D z' \big], & \textrm{for $\Omega_K < 0$},
\end{array} \right.
\end{equation}
where the Hubble-Lema\^itre constant $H_0 \equiv H(z=0)$ describes the current expansion rate of the universe,
$H(z)$ represents the expansion rate of the universe at redshift $z$,
and $\Omega_K$ is the fractional equivalent curvature density with respect to the critical density.
The parameter $H(z)$ can generally be expressed as
\begin{equation} \label{eq:hz_wz}
 H(z) = H_0 \sqrt{ \Omega_M (1 + z)^3 + \Omega_K (1 + z)^2 + \Omega_{\Lambda} f(z)},
\end{equation}
where $\Omega_{M}$ and $\Omega_{\Lambda}$ are the fractional densities for
the total matter and dark energy with respect to the critical density, respectively,
and $f(z)$ is a function that quantifies the effect of dark energy.
The function $f(z)$ can be expressed as \cite{Linder:2002et}
\begin{equation} \label{eq:de_fz}
f(z) = \exp \left( 3\int_{0}^{z} \frac{1 + w(z')}{1 + z'} \D z' \right).
\end{equation}
where $w(z)$ represents the \ac{EoS} of dark energy.
In the standard cosmological model --- $\Lambda$CDM, $w(z) \equiv -1$.
In the widely used Chevallier-Polarski-Linder (CPL) dark energy model \cite{Chevallier:2000qy, Linder:2002et},
\begin{equation} \label{eq:wz_CPL}
w(z) = w_0 + w_a \frac{z}{1+z} ,
\end{equation}
where $w_0$ and $w_a$ represent the corresponding model parameters, respectively.

Fitting the $D_L - z$ relation requires information on both the redshift and luminosity distance of the \ac{GW} source,
see Equation (\ref{eq:DL_z}).
However, extracting the redshift of the \ac{GW} source through \ac{GW} detection alone is challenging
due to the intrinsic degeneracy between the masses of the \ac{GW} source and its redshift,
necessitating the use of additional methods. The literature \cite{Schutz:1986gp} introduces
two primary methods for obtaining redshift information of \ac{GW} sources,
(i) \ac{EM} counterpart searches and (ii) galaxy catalogs.
EM counterparts provide a means of uniquely identifying the host galaxies of \ac{GW} sources,
allowing for the direct measurement of redshifts through spectroscopic observations.
The use of galaxy catalogs to infer redshift of \ac{GW} source relies on the assumption that
all \ac{GW} sources are hosted within galaxies, thus the spatial distribution of galaxies
can serve as a proxy for the probability distribution of the \ac{GW} source's position.
In general, \ac{GW} signals with \ac{EM} counterparts are referred to as bright sirens,
while those without \ac{EM} counterparts are termed dark sirens,
as discussed in Section \ref{sec:sirens_BD} in relation to TianQin's siren observations.
Subsequent studies have proposed additional methods to extract redshift information of \ac{GW} sources, including:
(iii) cross-correlations between \ac{GW} sources and galaxies,
 \cite{Laguna:2009re, Oguri:2016dgk, Diaz:2021pem, Mukherjee:2022afz, LISACosmologyWorkingGroup:2022jok},
(iv) strong gravitational lensing effects \cite{Sereno:2010dr, Liao:2019xug, Wang:2022rvf, Huang:2023prq},
(v) intrinsic redshift distributions of \ac{GW} sources \cite{Ding:2018zrk, Leandro:2021qlc},
(vi) intrinsic distributions of the physical masses of \ac{GW} sources,
also known as spectral sirens for \acp{SBHB}
 \cite{Taylor:2011fs, Taylor:2012db, DelPozzo:2015bna, Farr:2019twy, Mastrogiovanni:2021wsd,
You:2020wju, Mukherjee:2021rtw, Ezquiaga:2022zkx, LIGOScientific:2021aug, Chen:2022fda},
and (vii) tidal deformation phases of neutron stars \cite{Messenger:2011gi, Messenger:2013fya,
Shiralilou:2022urk, Jin:2022qnj, Li:2023gtu}.
The first five methods break the degeneracy between mass and redshift by providing additional redshift
information, whereas the last two methods break the degeneracy by constraining the distribution of
the physical masses of the \ac{GW} sources.
From another perspective, the first three methods rely on additional \ac{EM} observations to obtain
redshift information, whereas the remaining four methods achieve this by introducing additional
astrophysical models, either to provide more detailed mass information or to model the intrinsic
properties of the \ac{GW} sources.

Once the data have been obtained and a cosmological model has been determined, constraints on
the relevant cosmological parameters can be extracted using a Bayesian framework \cite{MacLeod:2007jd,
Babak:2010ej, Petiteau:2011we, DelPozzo:2011vcw, LIGOScientific:2017adf,
Chen:2017rfc, DES:2019ccw, Gray:2019ksv, Zhu:2023jti}.
Let $\mathcal{D}_{\rm GW} \equiv \{ d_{\rm GW}^1, d_{\rm GW}^2, \ldots, d_{\rm GW}^i, \ldots, d_{\rm GW}^N \}$
represent the dataset from \ac{GW} detections, and
let $\mathcal{D}_{\rm EM} \equiv \{ d_{\rm EM}^1, d_{\rm EM}^2, \ldots, d_{\rm EM}^i, \ldots, d_{\rm EM}^N \}$
represent the dataset from corresponding \ac{EM} observations.
The cosmological parameter set of interest is denoted as $\vec{H} \equiv \{H_0, \Omega_M,...\}$,
the posterior probability distribution of $\vec{H}$ can be expressed as
\begin{equation} \label{eq:bayes_theorem}
p(\vec{H} |\mathcal{D}_{\rm GW}, \mathcal{D}_{\rm EM}, T, I) = \frac{\pi(\vec{H}|T, I) \mathcal{L}(\mathcal{D}_{\rm GW}, \mathcal{D}_{\rm EM}|\vec{H}, T, I)}{P(\mathcal{D}_{\rm GW}, \mathcal{D}_{\rm EM}|T, I)} \propto \pi(\vec{H}|T, I) \prod_i^N \mathcal{L}_i (d_{\rm GW}^i, d_{\rm EM}^i|\vec{H}, T, I),
\end{equation}
where $\pi(\vec{H}|T, I)$ represents the prior probability distribution for $\vec{H}$ given the theoretical cosmological model $T$,
$\mathcal{L}(\mathcal{D}_{\rm GW}, \mathcal{D}_{\rm EM}|\vec{H}, T, I)$ represents the likelihood of data,
$P(\mathcal{D}_{\rm GW}, \mathcal{D}_{\rm EM}|T, I) \equiv \!\int\!
\pi(\vec{H}|T, I) \mathcal{L}(\mathcal{D}_{\rm GW}, \mathcal{D}_{\rm EM}|\vec{H}, T, I) \D \vec{H}$
represents the marginal likelihood also called the Bayesian evidence,
and $I$ indicates all the relevant background information.
The individual likelihood $\mathcal{L}_i (d_{\rm GW}^i, d_{\rm EM}^i|\vec{H}, T, I)$ have to
be further decomposed before they can be computed \cite{Chen:2017rfc} and
have to account for the selection effects
on \ac{GW} detections \cite{Mandel:2018mve} and \ac{EM} observations \cite{Chen:2017rfc}.
The exact decomposition forms varies depending on the different redshift extraction methods.
In addition, one can quantitatively compare how well the data
support different cosmological models through a Bayes factor \cite{Jeffreys:1961book, Trotta:2008qt}.
Bayes factor $B_{01}$ is defined as the ratio of the Bayesian evidences obtained for
two theoretical models, $T_0$ and $T_1$, with the same observed data, i.e.,
\begin{equation} \label{eq:bayes_factor}
B_{01} \equiv \frac{P(\mathcal{D}_{\rm GW}, \mathcal{D}_{\rm EM}| T_0, I)}{P(\mathcal{D}_{\rm GW}, \mathcal{D}_{\rm EM}| T_1, I)}.
\end{equation}
A value of $B_{01} > 1$ represents that the observed data is more supportive to the model $T_0$
compared to the model $T_1$ and vice versa. There are strong, moderate, weak or inconclusive
strengths of evidence for data-supported models, which can be referenced to
the ``Jeffreys' scale'' \cite{Jeffreys:1961book, Trotta:2008qt}.

\subsubsection{Candidate sirens of TianQin} \label{sec:sirens_candidateGW}
From the description in the previous section, it follows that \ac{GW} sources
capable of constraining the cosmic expansion history must satisfy two key conditions.
first, they must be individually identifiable and precisely localized to their spatial positions,
and second, these sources must reside at cosmological distances.
Among the five classes of potential \ac{GW} sources detectable by TianQin, three satisfy these criteria:
\acp{SBHB} \cite{Liu:2020eko}, \acp{EMRI} \cite{Fan:2020zhy}, and \acp{MBHB} \cite{Wang:2019ryf}.
These three classes of standard sirens also serve as potential standard sirens for
other milli-hertz space-based \ac{GW} detectors, such as LISA \cite{LISA:2017pwj, Colpi:2024xhw}
and Taiji \cite{Hu:2017mde}. However, the sensitivity and detector configuration of TianQin
exhibit significant differences compared to these other missions \cite{Gong:2021gvw}.

The successful detections of \acp{GW} by the LIGO\&Virgo collaborations have provided direct evidence for
the existence of \acp{SBHB} \cite{LIGOScientific:2016aoc, LIGOScientific:2018mvr,
LIGOScientific:2020ibl, LIGOScientific:2021usb, KAGRA:2021vkt}and have enabled initial investigations
into the properties of the \ac{SBHB} population \cite{LIGOScientific:2018jsj,KAGRA:2021duu}.
By tracing the evolution of compact object binaries backward from the merger phase to the inspiral phase,
it can be inferred that the earlier the time from the binary merger, the lower the frequency of
the \ac{GW} signal radiated from the binary inspiral.
Consequently, these low-frequency \ac{GW} signals from SBHB inspirals are likely to be detectable by
space-based \ac{GW} detectors \cite{Miller:2002vg, Takahashi:2003wm}.
This possibility was first proposed in the literature \cite{Sesana:2016ljz}, where demonstrated that
the \ac{SBHB} inspiral \ac{GW} signal could be detected by space-based \ac{GW} detectors.
As a result, \acp{SBHB} have been recognized as one of the primary candidate \ac{GW} sources
for the space-based \ac{GW} detection missions, such as LISA \cite{Seto:2016wom, Kyutoku:2016ppx,
LISA:2017pwj, Colpi:2024xhw}as well as deci-hertz \ac{GW} detectors \cite{Chen:2017gfm, Isoyama:2018rjb,
Sedda:2019uro}.
The scientific potential of such \ac{GW} sources has been extensively explored in both astrophysical
\cite{Nishizawa:2016jji, Nishizawa:2016eza, Breivik:2016ddj, Randall:2018lnh, Toubiana:2020drf,
Sberna:2022qbn} and cosmological \cite{DelPozzo:2017kme, Muttoni:2021veo} contexts.

The first comprehensive evaluation of TianQin’s capability to detect \ac{SBHB} inspiral
\ac{GW} signals was presented in \cite{Liu:2020eko}.
This study assessed the expected total detection rate and parameter estimation precision,
based on the earliest \ac{SBHB} population models published by the LIGO\&Virgo Collaborations \cite{LIGOScientific:2018jsj}.
A subsequent study \cite{Zhu:2021bpp} updated these predictions by the third version of SBHB population models
published by \ac{LVK} based on the GWTC-3 \cite{KAGRA:2021duu}.
Taking the results from the literature \cite{Zhu:2021bpp} as a reference,
and adopting a detection threshold of ${\rm SNR} = 8$, TianQin is expected to detect about 10 \acp{SBHB},
and TianQin I+II (an alternative configuration of TianQin containing complementary
twin spacecraft constellations, see \cite{Liang:2021bde})
is expected to increase the total detection number by about a factor of three.
The spatial localization errors of the two detector configurations, TianQin and TianQin I+II,
for \acp{SBHB} do not differ much, with sky localization errors of about $0.1-1 ~\!{\rm deg}^2$
and relative luminosity distance errors of about $0.1-0.4$ at $1 \sigma$ confidence level.
Because the \ac{GW} signal radiated by \ac{SBHB} inspiral is a continuous signal for space-based \ac{GW} detectors,
the \ac{SNR} of the inspiral \ac{GW} signal accumulated per unit observation time increases with frequency
 \cite{Cutler:1994ys, Mangiagli:2020rwz, Chen:2023qga}. Given that TianQin has better
sensitivity than LISA in the frequency band that approximately greater than $10^{-2}$ Hz
\cite{TianQin:2015yph, Gong:2021gvw}, it is expected to outperform LISA
in detecting \ac{SBHB} inspirals \cite{Liu:2020eko, Liu:2021yoy, Zhu:2021bpp}.

\acp{EMRI} are extreme relativistic systems consisting of a stellar-mass compact object
orbiting a \ac{MBH}. Given that at least one \ac{MBH}, along with a nuclear star cluster,
is typically present at the centers of galaxies, and that the stellar density within the nuclear star cluster
increases with decreasing distance from the central MBH, it is possible for an central \ac{MBH} to
gravitationally capture a surrounding stellar-mass compact object to form an \ac{EMRI} system
\cite{Barack:2003fp, Amaro-Seoane:2007osp, Babak:2017tow, Amaro-Seoane:2012lgq}.
The standard formation channel for \acp{EMRI} is that an \acp{MBH} at the center of a normal galaxy
gravitationally captures a stellar-mass black holes \cite{Babak:2017tow, Amaro-Seoane:2012lgq}.
The population properties of the resulting \acp{EMRI} are mainly determined by the population properties
of \acp{MBH}, the fraction of \acp{MBH} embedded in dense stellar cusps of nuclear star clusters,
the \ac{EMRI} formation rate per MBH, and the mass and eccentricity distribution of
the inspiralling compact objects \cite{Babak:2017tow, Amaro-Seoane:2012lgq}.
A detailed analysis of TianQin’s capability to detect EMRIs was conducted in the literature \cite{Fan:2020zhy},
utilizing 12 \ac{EMRI} population models presented in the literature \cite{Babak:2017tow}.
Because the \ac{EMRI} rates and properties predicted by the various population models vary greatly,
there are large uncertainties in the detection rates and parameter estimation errors for TianQin's \acp{EMRI}.
TianQin has expected detection rates of a few to a few hundred \acp{EMRI} per year, and TianQin I+II
is expected to increase detection rates by a factor of about two to three; the sky localization errors
of \acp{EMRI} range from about $0.1 ~\!{\rm deg}^2$ to $100 ~\!{\rm deg}^2$, and the relative errors of
the luminosity distance estimations are from about $0.05$ to $0.1$ \cite{Fan:2020zhy}.
The \ac{EMRI} detection rates for TianQin are expected to be approximately one-fifth of that for LISA,
primarily due to the concentration of \ac{EMRI} \ac{GW} signals in the frequency range of $f \lesssim 10^{-2}$
Hz \cite{Babak:2017tow, Fan:2020zhy}.

\acp{MBHB} are the products of galaxy mergers and are expected to be
the loudest \ac{GW} sources in the milli-hertz \ac{GW} band.
Previous study \cite{Klein:2015hvg} have presented three seemingly reliable \ac{MBHB} population models
based on modified cosmological N-body simulations. These simulations employ an extended Press-Schechter
formalism to model the products of dark matter merger trees and utilize semi-analytical
galaxy-formation models to simulate galaxy evolution.
Later works \cite{Wang:2019ryf, Barausse:2020mdt} have obtained similar population results using alternative modified cosmological N-body simulations.
Using the three \ac{MBHB} population models
presented in \cite{Klein:2015hvg}, namely, the light-seeded model \texttt{popIII} and
the heavy-seeded models \texttt{Q3\_d} and \texttt{Q3\_nod}, \cite{Wang:2019ryf} investigates
the early warning capabilities and detection potential of TianQin for \ac{MBHB} mergers.
The detection capability of TianQin for \ac{MBHB} merger \ac{GW} signals varies significantly across the
population models. Expected detection rates range from a few to several tens of \acp{MBHB} per year,
with sky localization errors of approximately $10 ~\!{\rm deg}^2$ and relative errors in luminosity distance estimations of about $3\%$ \cite{Wang:2019ryf}.
The TianQin I+II configuration can enhance the detection rate of \acp{MBHB} by approximately 60\%,
although improvements in spatial localization errors are not significant \cite{Zhu:2021aat}.
Similar to the detections of \acp{EMRI}, the detection capability of TianQin for \acp{MBHB}
is weaker compared to that of LISA \cite{Klein:2015hvg, Wang:2019ryf}.

In summary, the three classes of \ac{GW} sources, namely \acp{SBHB}, \acp{EMRI}, and \acp{MBHB},
provide complementary means to probe the expansion history of the universe.
Because the \acp{SNR} of \acp{SBHB} are typically low, and as a result,
the \acp{SBHB} detectable by TianQin is primarily concentrated in
the low-redshift universe with $z < 0.3$ \cite{Liu:2020eko, Zhu:2021bpp}.
The \acp{SNR} of \acp{EMRI} are generally higher than those of \acp{SBHB}
due to the contributions from the central \acp{MBH}, with the detection horizon for \ac{EMRI} extending to
a redshift of $z \sim 2$ \cite{Fan:2020zhy}.
The \acp{SNR} of \acp{MBHB} are generally the highest of the three, and the detection horizon for \ac{MBHB}
can reach redshifts greater than $z> 10$ \cite{Wang:2019ryf}.
Given the limited event rates of the various class of \ac{GW} sources,
the favored \ac{GW} sources in the low-redshift universe at $z \lesssim 0.3$ are \acp{SBHB},
while \acp{EMRI} dominate the medium-redshift universe of $0.3 \lesssim z \lesssim 2$,
and \acp{MBHB} are the only detectable \ac{GW} sources in the high-redshift universe of $z > 2$.
Consequently, these three classes of \ac{GW} sources form a \emph{probe ladder} system,
enabling space-based \ac{GW} detectors to probe the more complete expansion history of the universe.

\subsubsection{Uncertainties in luminosity distances}

In GR, the amplitude of \ac{GW} is inversely proportional to the luminosity distance, i.e., $h \propto 1/D_L$.
As a result, the precision of luminosity distance that directly estimating
from \ac{GW} detection is generally inversely proportional to the \ac{SNR} of the \ac{GW} signal,
i.e., $\Delta D_L/D_L \propto 1/{\rm SNR}$.
However, when accounting for the various degeneracies between the parameters of \ac{GW}
sources---such as the degeneracy between the inclination angle $\iota$ of the binary orbital angular momentum
relative to the line of sight and the luminosity distance,
as well as the degeneracy between the spatial position and the polarization
angle---the actual error in the luminosity distance becomes more complicated.
Thus, simulation analyses are required to obtain reliable estimates \cite{Finn:1992wt, Cutler:1994ys, Vallisneri:2007ev}.
The error in luminosity distance determined by the detector’s response and noise characteristics is generally referred to as the detection error or uncertainty.

In addition to detection uncertainty, it is important to consider factors that can affect the accuracies
of amplitude measurements of \ac{GW} signals during their emission and propagation,
such as peculiar velocity \cite{Kocsis:2005vv, Gordon:2007zw} and weak gravitational lensing effects
\cite{Markovic:1993cr, Wang:1996as, Takahashi:2003ix, Holz:2005df}.
Each object in the universe exhibits peculiar motion, and the peculiar velocity of a \ac{GW} source
modifies the frequency of the detected \ac{GW} signal due to the Doppler effect.
Since the frequency of a \ac{GW} and its time derivative are key to
estimating the masses of the \ac{GW} source \cite{Cutler:1994ys}, which directly determines the absolute amplitude
of the \ac{GW} signal, the Doppler effect induced by peculiar velocities will impact the accuracy of
the amplitude measurements, and ultimately influence the precision of luminosity distance measurements.
The peculiar velocities of individual \ac{GW} sources are generally difficult to measure directly.
Therefore, the effect of peculiar velocities on luminosity distance estimates can only be accounted for statistically \cite{Kocsis:2005vv}. The additional uncertainty on luminosity distances
due to peculiar velocity effect can be expressed as \cite{Kocsis:2005vv, Gordon:2007zw}
\begin{equation} \label{eq:dDL_pv}
\sigma_{D_L}^{\rm pv}(z) = D_L(z) \times \bigg[ 1 + \frac{c (1+z)^2}{H(z)D_L(z)} \bigg] \frac{\sqrt{\langle v^2 \rangle}}{c},
\end{equation}
where $\sqrt{\langle v^2 \rangle}$ represents the root mean square of the peculiar velocities
of \ac{GW} sources relative to the Hubble flow.
Typically, the peculiar velocities of \ac{GW} sources are approximated by those of galaxies,
with the root mean square taken as
$\sqrt{\langle v^2 \rangle}=500 \ \textrm{km}/\textrm{s}$ \cite{Kocsis:2005vv, He:2019dhl}.

The spatial number density of galaxies in the universe is approximately 0.01 Milky Way-equivalent galaxies
per ${\rm Mpc}^3$ \cite{Kopparapu:2007ib}.
The probability that a \ac{GW} signal emitted from a distant source is affected by
the gravitational potentials of surrounding galaxies during propagation is positively
correlated with the distance of the \ac{GW} source.
Consequently, gravitational lensing effects
will inevitably cause amplification or de-amplification of the \ac{GW} amplitude,
thus impacting the measurement of the luminosity distance to the \ac{GW} source
\cite{Markovic:1993cr, Wang:1996as, Pyne:2003bn, Takahashi:2003ix, Bonvin:2005ps}.
Strong gravitational lensing can be identified and modeled individually
\cite{Sereno:2010dr, Smith:2017mqu}, whereas the weak gravitational
lensing cannot be directly identified. The effects of weak gravitational lensing on luminosity distance
estimations must therefore be accounted for using statistical methods
\cite{Markovic:1993cr, Holz:1997ic, Shapiro:2009sr, Hirata:2010ba}.
The impact of weak gravitational lensing on luminosity distance estimation
can be described by a fitting formula, with the widely adopted expression being \cite{Hirata:2010ba}
\begin{equation} \label{eq:dDL_lens}
 \sigma_{D_L}^{\textrm{lens}}(z) = \frac{1}{2} D_L(z) \times C_l \bigg[ \frac{1 - (1+z)^{-\beta_l}}{\beta_l} \bigg]^{\alpha_l},
\end{equation}
where $C_l = 0.066$, $\alpha_l = 1.8$, and $\beta_l = 0.25$ are fitting parameters.
In comparison to the original formula presented in \cite{Hirata:2010ba},
(\ref{eq:dDL_lens}) introduce an additional factor of $1/2$. This adjustment arises because the original formula
models the effect of weak gravitational lensing on the errors of ${D_L}^{\!2}$ measurements,
whereas \ac{GW} detections directly measure the luminosity distance $D_L$.
Furthermore, the equation has been updated to reflect the growing understanding of
the correlation between \ac{GW} sources \cite{Cusin:2020ezb},
galaxies and the spatial distribution of galaxies \cite{Hilbert:2010am, Wu:2022vrq}.

In addition to the effects of peculiar velocity and weak gravitational lensing,
several other environment factors that host the \ac{GW} source may also influence
the accuracy of luminosity distance measurements.
These include gravitational redshift \cite{Chen:2017xbi, Chen:2020iky},
the presence of gas \cite{Chen:2020lpq},
dark matter \cite{Karydas:2024fcn, Kavanagh:2024lgq},
and the resolution of the \ac{GW} waveform \cite{Jan:2023raq},
as well as the non-stationary \cite{Edy:2021par, Kumar:2022tto} and non-Gaussian \cite{Steltner:2021qjy}
characteristics of \ac{GW} detector noise.
However, these effects are beyond the scope of the present discussion.

\subsubsection{Uncertainties in redshifts} \label{sec:sirens_BD}

In Section \ref{sec:sirens_principle}, seven methods for extracting redshift
information of \ac{GW} sources are summarized. Each method has its own scope of application,
and the uncertainties of redshift derived from different methods vary.
For the characteristics of TianQin and the properties of its candidate standard sirens,
four methods are particularly relevant for extracting redshift information:
EM counterparts, galaxy catalogs, cross-correlations between \ac{GW} sources and galaxies,
and strong gravitational lensing.
This section focuses on the first two methods;
will be addressed in future studies, and the fourth will be discussed in Section \ref{sec:lensing}.

Searching for the \ac{EM} counterpart of a \ac{GW} source is the most desirable method for extracting
the redshift information of \ac{GW} source, as first pointed out by Schutz \cite{Schutz:1986gp}.
EM counterparts of \ac{GW} sources can help observer to uniquely identify their host galaxies,
enabling precise redshift measurements through spectroscopic observations.
In such case, the measurement errors of the redshifts are sufficiently small
to be negligible \cite{Hjorth:2017yza}.
However, the method of identifying \ac{EM} counterparts is not the most universally applicable.
Several factors govern the success of searches for \ac{EM} counterparts
\cite{LIGOScientific:2017ync, FermiGamma-rayBurstMonitorTeam:2018eao}:
(i) whether the \ac{GW} sources emit \ac{EM} radiation (e.g., for black hole binaries);
(ii) whether the \ac{EM} counterparts are sufficiently bright to be detected by
telescopes with limited sensitivities; and
(iii) whether the sky localization information provided by the \ac{GW} detector
is timely and precise enough for telescopes with limited field-of-views to conduct the search within the available time frame.
Among the three classes of candidate sirens for TianQin (also include LISA and Taiji),
the potential bright sirens with observable \ac{EM} counterparts are \acp{MBHB} \cite{Tamanini:2016zlh,
McGee:2018qwb, DeRosa:2019myq, Mangiagli:2020rwz, Bogdanovic:2021aav}.

\begin{figure}[t]
\centering
\includegraphics[width=0.80\textwidth]{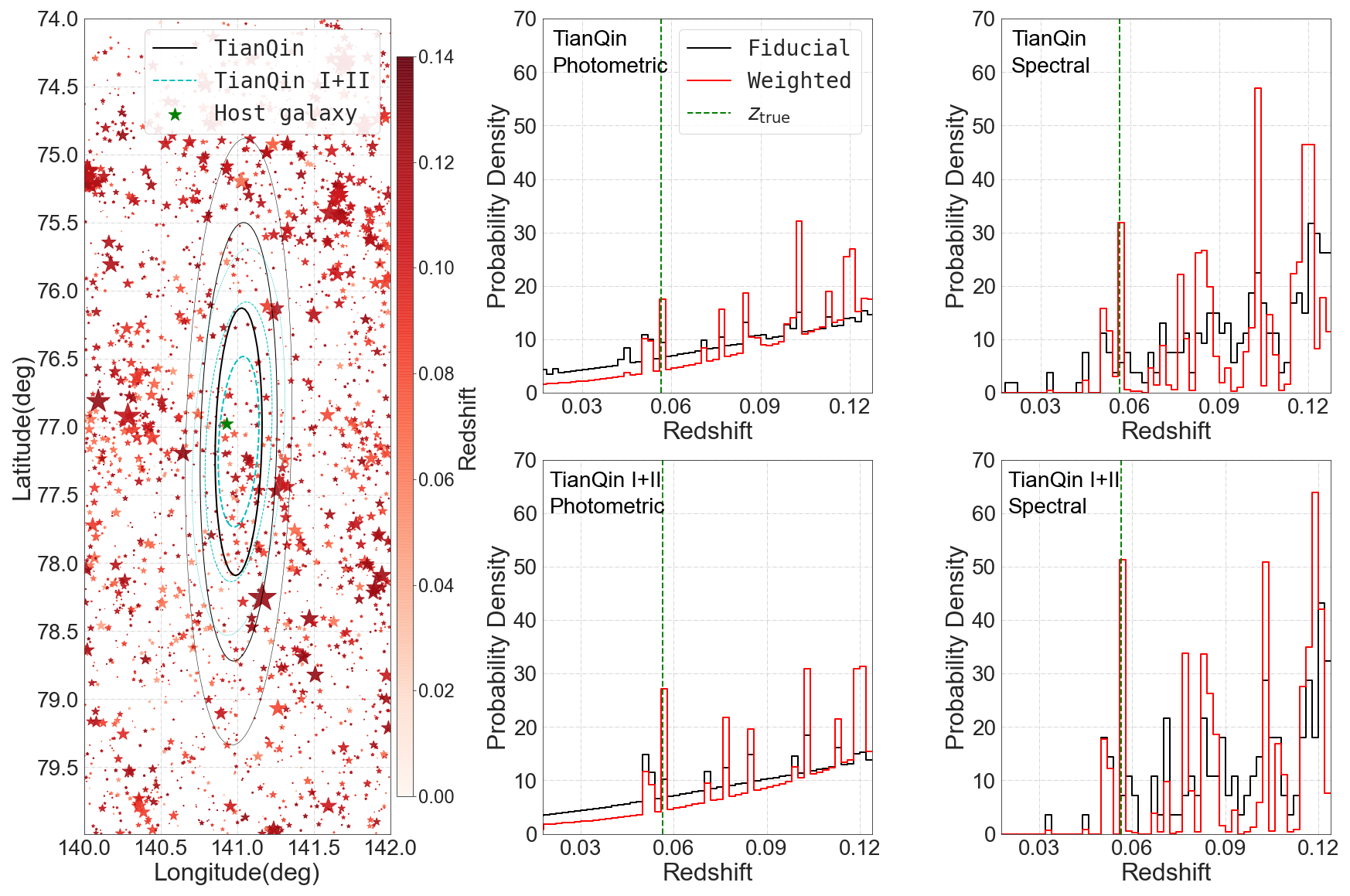}
\caption{Sky localization errors and redshift probability distribution functions
of a \ac{SBHB} \cite{Zhu:2021bpp}.
 The left panel shows sky localization errors from TianQin (solid black contour lines) and TianQin I+II (dashed cyan contour lines), as well as a scatter plot of galaxies within the catalog.
 The green star labels the real host of the \ac{SBHB}, the shades of the other red stars represent their redshifts,
 and the sizes of the stars represent their luminosity-related weights.
 The center panels show the statistical photo-$z$ of TianQin (top) and TianQin I+II (bottom), respectively, using two method of extracting redshift, the fiducial method (black histograms) and the weighted method (red histograms). The vertical green dashed line represents the true redshift of the \ac{SBHB}.
 The right panels show the same as the center panels but with spectroscopic galaxy redshifts.}
\label{fig:z_PDF_SBBH}
\end{figure}

Using galaxy catalogs to extract the redshifts of \ac{GW} sources is the most widely applicable method
for all three classes of candidate standard sirens: \acp{SBHB} \cite{DES:2019ccw, DES:2020nay, Vasylyev:2020hgb,
LIGOScientific:2019zcs, LIGOScientific:2021aug, DelPozzo:2017kme, Zhu:2021bpp},
\acp{EMRI} \cite{MacLeod:2007jd, Laghi:2021pqk, Zhu:2024qpp}, and
\acp{MBHB} \cite{Petiteau:2011we, Wang:2020dkc, Zhu:2021aat}.
Thanks to various galaxy survey projects and plans, such as
the Sloan Digital Sky Survey (SDSS) \cite{SDSS:2000hjo, SDSS:2023tbz},
Dark Energy Survey \cite{DES:2005dhi, LineaScienceServer:2021mgv},
Large Sky Area Multi-Object Fiber Spectroscopic Telescope \cite{2012RAA....12.1197C, 2012RAA....12..723Z},
Dark Energy Spectroscopic Instrument \cite{DESI:2016fyo, DESI:2018ymu},
Rubin Observatory Legacy Survey of Space and Time \cite{LSST:2008ijt},
Euclid \cite{Euclid:2021icp}, and
the Chinese Space Station Telescope \cite{Gong:2019yxt},
our catalog of galaxy information---including sky positions,
magnitudes, spectra (or photometric magnitudes in various bands), and redshifts---continues to improve.
This progress will significantly improve the precision of redshift measurements for TianQin's dark sirens.
The redshift information extracted from galaxy catalogs typically takes the form of
a probability density distribution of redshifts, this method is based on a default assumption
that each galaxy located within the spatial localization error volume of a \ac{GW} source is a potential
host for the \ac{GW} source.
The nature of the clustered distribution of galaxies ensures that the redshift probability
density distribution provided by candidate host galaxies is informative \cite{Schutz:1986gp}.
The errors in the redshifts obtained from galaxy catalogs are primarily determined by:
(i) the spatial localization error volume of the \ac{GW} sources,
(ii) the completeness and redshift precision of the galaxy catalogs, and
(iii) the a priori range of the cosmological parameters
 \cite{MacLeod:2007jd, Petiteau:2011we, DelPozzo:2011vcw, Chen:2017rfc, Zhu:2021aat}.
It is intuitive that reducing the error volume of \ac{GW} sources or increasing the completeness and redshift precision of galaxy catalogs can improve the equivalent precisions of the redshifts.
For example, FIG. \ref{fig:z_PDF_SBBH} shows the probability distribution functions of redshift
extracted from galaxy catalogs for a \ac{SBHB} standard siren detected by TianQin.
The figure demonstrates that improvements in either spatial localization precision or
the redshift precision of galaxies clearly enhance the relative probability of the
\ac{GW} source's true redshift to the redshift probability distribution.
The third factor influencing redshift errors arises from the fact that, when selecting
candidate host galaxies based on the spatial localization error volume of the \ac{GW} source,
the error in luminosity distance must be transformed into a redshift error.
This transformation is dependent on the values of the cosmological parameters.

It should also be noted that the three methods for extracting the redshifts of
\ac{GW} sources, which have not yet been discussed in this section,
are in principle applicable for TianQin's dark sirens.
However, the errors of the extracted redshifts are so large that their practical utility
is almost negligible. The reasons for this are as follows.
The intrinsic redshift distributions as well as the intrinsic distributions of physical masses
of \ac{GW} sources are extremely dependent on astrophysical models of the formation and evolution of
\ac{GW} sources \cite{Ding:2018zrk, Ezquiaga:2022zkx, Mastrogiovanni:2021wsd, Mukherjee:2021rtw}.
There are large uncertainties in our current understanding about the formation and
evolution mechanisms of \acp{EMRI} and \acp{MBHB},
and cannot reasonably predicting their redshifts and masses distributions,
so the two methods are not applicable to \acp{EMRI} and \acp{MBHB}.
For \acp{SBHB}, although one have a preliminary understanding on the population properties of \acp{SBHB}
through \ac{LVK}'s \ac{GW} detections, the detectable \acp{SBHB} of TianQin are all in
the local universe with $z < 0.3$ \cite{Liu:2020eko, Zhu:2021bpp}.
For such low redshift \acp{SBHB}, the errors in the redshifts of the \acp{SBHB} extracted with
the additional information on the redshifts and physical masses are much larger than
the redshift values of the \acp{SBHB} themselves.
Regarding the tidal deformation phases of neutron stars, these additional \ac{GW} phases occur
when binary neutron stars and neutron star-black hole binaries are near merger. Since the frequencies
associated with these phases are relatively high, they fall outside the sensitive band of TianQin.

\subsubsection{Improving the precisions of redshifts} \label{sec:sirens_imporve_z}

Observations of \ac{EM} counterparts provide the most ideal method for improving the precision of
the redshift measurements for \ac{GW} sources.
In this case, extremely accurate redshift information can be obtained.
Near real-time sky localizations of the pre-merger \acp{MBHB} could facilitate
the search for \ac{EM} counterparts to these \ac{GW} sources \cite{Mangiagli:2020rwz, Chen:2023qga}.
However, our current understanding of the \ac{EM} radiation mechanisms and characteristics of \ac{MBHB}
\ac{GW} sources remains highly uncertain \cite{Bogdanovic:2021aav}, and thus, the success rate of
EM counterpart searches is currently unknown.

The method of using galaxy catalogs represent the most widely applicable method for extracting redshift information of \ac{GW} sources.
There are three main ways to improve the precision of the extracted redshifts:
first, decreasing the spatial localization error volume of the \ac{GW} sources;
second, increasing the completeness of galaxy catalogs;
and third, properly assigning weights to each candidate host galaxy.
The first two ways are primarily determined by actual observations from \ac{GW} detectors and
EM telescopes, respectively, and do not require extensive assumptions.
The third way, however, is fundamentally a data processing approach, which necessitates
advance researches and simulation tests.
To enhance clarity, the authors present below the weighting methods for candidate host galaxies
developed for the three classes of candidate standard sirens of TianQin:
\begin{itemize}
\item For \acp{SBHB}, using the multi-band photometric information of galaxies to directly
estimate the total stellar masses of the galaxies, and then assigning the weights of the galaxies
based on the total stellar masses \cite{Zhu:2021bpp};
\item For \acp{MBHB} (also applies to \acp{EMRI}), assigning weights to candidate host galaxies based on their bulge luminosity information and the $M_{\rm MBH}-L_{\rm bulge}$ relation \cite{Zhu:2021aat};
\item For \acp{EMRI} (also applies to \acp{MBHB}), selecting candidate hosts by statistically inferring the formation channels of \acp{EMRI} \cite{Zhu:2024qpp}.
\end{itemize}
The astrophysical foundations and the practical effectiveness of these weighting methods
will be discussed in the following sections.

A proper assignment of weights to candidate host galaxies of \ac{GW} sources must
be based on the astrophysical correlation between the \ac{GW} sources and their host galaxies.
For \acp{SBHB}, there are two potential correlations exist, the \ac{SBHB} event rate of a galaxy
is proportional to the total stellar mass \cite{Phinney:1991ei, Leibler:2010uq, Fong:2013eqa, Rodriguez:2016kxx}
and the star formation rate of the galaxy \cite{Leibler:2010uq, Singer:2016eax}.
The first correlation holds on the basis that galaxies with larger total stellar mass contain more stars
and are therefore likely to generate more stellar-mass black holes.
The second correlation is supported by the fact that the \acp{SBHB} have a similar law of evolution with redshift
as the star formation rate of galaxies \cite{Rodriguez:2018rmd, Yang:2020lhq, Santoliquido:2020bry,
Fishbach:2021mhp, vanSon:2021zpk, KAGRA:2021duu, Vijaykumar:2023bgs}.
Regardless of which correlation is adopted, previous studies have employed single-band luminosities
to approximate the weights $\{ w_i \}$ of candidate host galaxies \cite{DES:2019ccw,
LIGOScientific:2018gmd, LIGOScientific:2019zcs, DES:2020nay, Vasylyev:2020hgb,
LIGOScientific:2021aug, DESI:2023fij}.
For example, the weights be expressed as $w_i \propto L_{K,i}$ based on the total stellar mass \cite{Bell:2003cj, Lin:2004ak}
and $w_i \propto L_{B,i}$ based on the star formation rate \cite{Kennicutt:1998zb, Gehrels:2015uga},
where $L_{K,i}$ and $L_{B,i}$ represent the luminosities of the $i$-th galaxy in the $K$ and $B$ bands, respectively.

\begin{figure}[htbp!]
\centering
\includegraphics[width=0.60\textwidth]{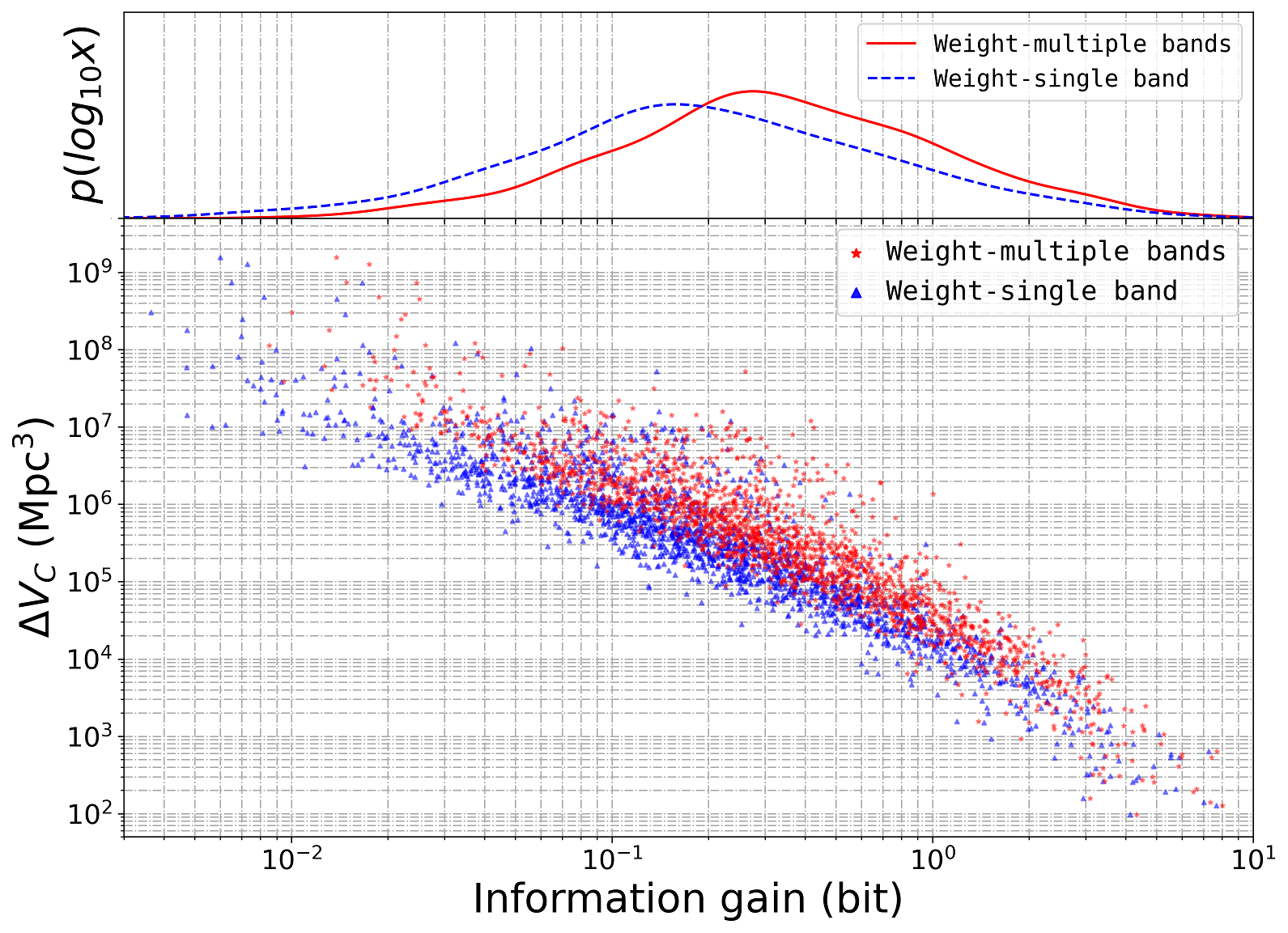}
\caption{Distributions of information gains for the probability distributions of \acp{SBHB}' redshifts.
The blue and red dots or lines represent the results obtained by the single-band and multiple band luminosities weighting methods, respectively.
The top panel shows the distribution of information gains obtained using these two methods, and the bottom panel shows corresponding scatters of information gain $-$ spatial localization error comoving volume $\Delta V_C$.
This figure contains 2000 mock \acp{SBHB} detectable by TianQin \cite{Zhu:2021bpp}.}
\label{fig:InfoGain_SBBH}
\end{figure}

Considering the large scatter in the linear relationship between the luminosities of galaxies
in single $K$ or $B$ bands and their total stellar masses or star formation rates,
such scatters can reduce the effectiveness of the weighting method and may introduce biases
into the weights. A more reasonable approach to weighting is to use photometric luminosities
in multiple bands, as this would improve the accuracy of the estimates for
the total stellar masses or star formation rates of galaxies \cite{Zhu:2021bpp}.
By fitting the spectral energy distribution (SED) of galaxy models using photometric luminosities
from all observed bands, one can estimate the maximum likelihood values for properties of galaxies,
such as total stellar masses and star formation rates \cite{Calzetti:1994vw, Chabrier:2003ki,
Bruzual:2003tq, Arnouts:1999bb, Ilbert:2006dp}.
This multi-band approach allows us to improve the precision of the redshifts extracted from galaxy catalogs.
The information gain can be used to quantitatively describe the effect of the additional multi-band
photometric information on improving redshift precisions of \ac{GW} sources \cite{LIGOScientific:2018gmd, Zhu:2021bpp}.
The information gain is defined as \cite{Sivia:2006book}
\begin{equation} \label{information_gain}
\mathcal{H} = \!\int\! p(z|d_{\rm GW}, d_{\rm gal},\vec{H},I) \log_{2}\!\! \left[ \frac{ p(z|d_{\rm GW}, d_{\rm gal},\vec{H},I)}{ p_0(z| \vec{H}, I)} \right] \D z ,
\end{equation}
where $p(z|d_{\rm GW}, d_{\rm gal},\vec{H},I)$ represents the posterior probability of redshift
from all candidate host galaxies of \ac{GW} sources, $d_{\rm survey}$ represents the galaxy
catalogs, and $p_0(z| \vec{H}, I)$ is the prior probability distribution on redshift.
A larger information gain indicates a stronger constraint on cosmology.
FIG. \ref{fig:InfoGain_SBBH} illustrates a comparison of the information gain distributions
for using single-band and multi-band photometric luminosities to weight candidate host galaxies.
As expected, one can find that multi-band photometric luminosities significantly improves the redshift precision
of the \acp{SBHB} extracted from galaxy catalogs.

\begin{figure*}[htbp!]
\centering
\includegraphics[width=0.439\textwidth]{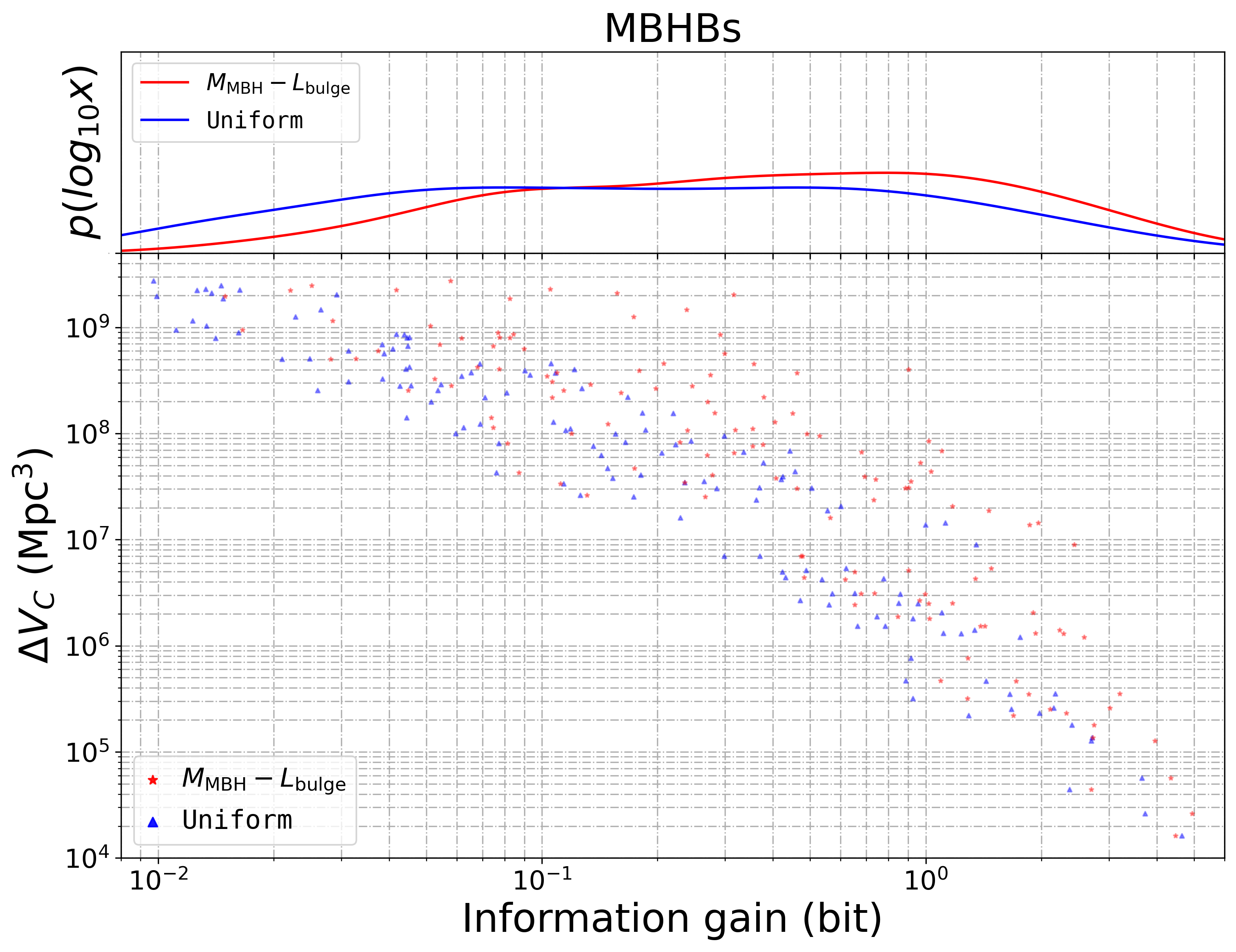}~~~
\includegraphics[width=0.450\textwidth]{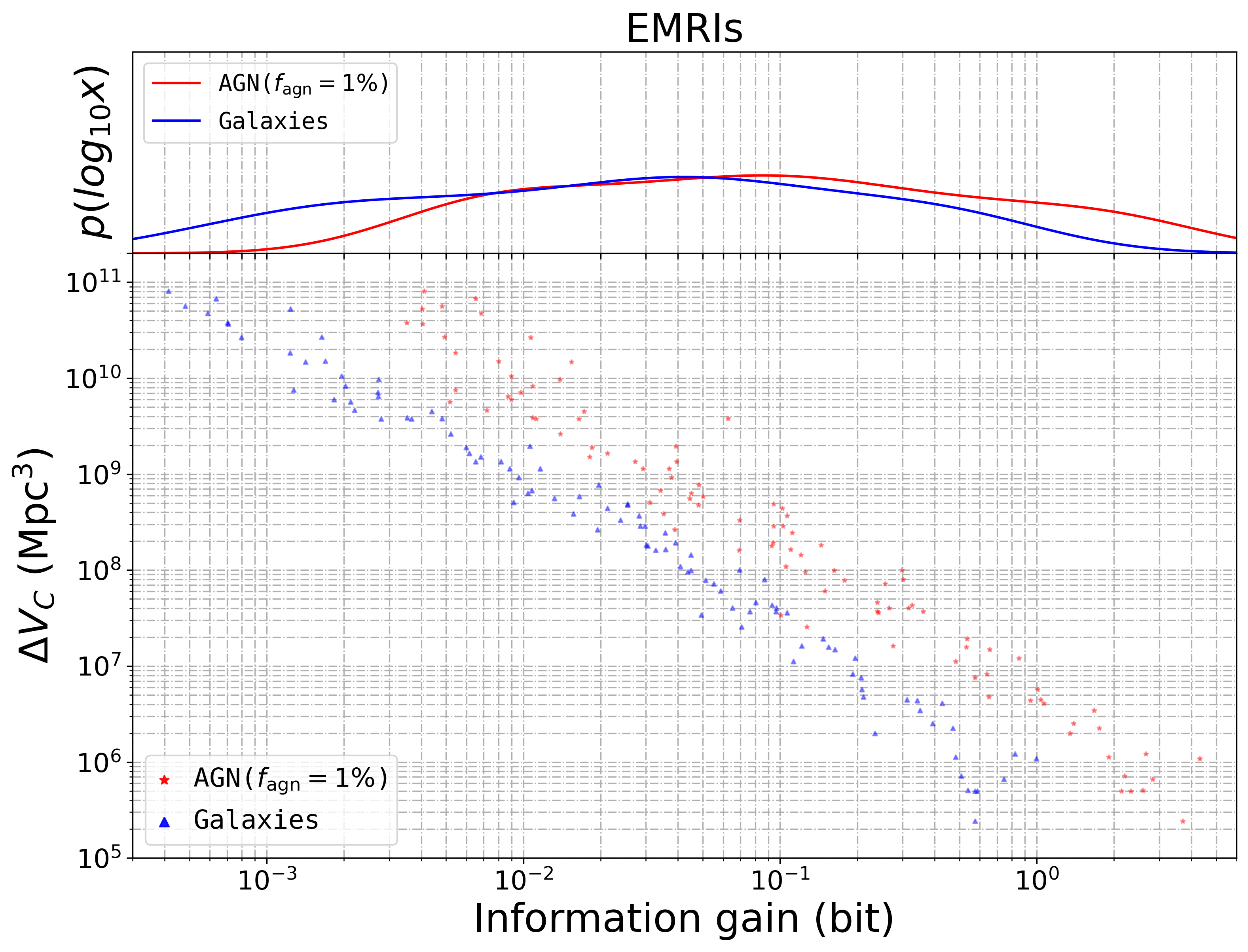}
\caption{Distributions of information gains of redshifts for TianQin's \acp{MBHB} (left panel)
and \acp{EMRI} (right panel).
In left panel, the blue and red dots or lines represent the results obtained by
uniform and $M_{\rm MBH}-L_{\rm bulge}$ weighting methods, respectively, and
the \ac{MBHB} samples are generated based on the pop III population model \cite{Zhu:2021aat}.
In right panel, the blue and red dots or lines represent the results obtained form galaxies and AGNs,
respectively, and the \ac{EMRI} samples are generated based on the M1 population model \cite{Zhu:2024qpp}.
Other sets same as FIG. \ref{fig:InfoGain_SBBH}.}
\label{fig:InfoGain_EMRI&MBHB}
\end{figure*}

For \acp{MBHB} and \acp{EMRI}, the authors identify two methods for weighting their candidate host galaxies.
The first method utilizes the $M_{\rm MBH}-L_{\rm bulge}$ relation \cite{Zhu:2021aat},
while the second method statistically tests the correlation between \ac{GW} sources and AGNs \cite{Zhu:2024qpp, Zhu:2023imz}.
The $M_{\rm MBH}-L_{\rm bulge}$ relation is a well-known linear relationship in logarithmic space between
the bulge luminosities and masses of the central \acp{MBH} of galaxies. The astrophysical basis
for this relation lies in the co-evolution of galaxies and their central \acp{MBH}
\cite{Graham:2007uq, Bentz:2008rt, Jiang:2011bt, Kormendy:2013dxa}.
Since \acp{MBH} are the primary black holes responsible for forming \acp{EMRI} and \acp{MBHB},
\ac{GW} detections can provide precise
estimates of the masses of these sources \cite{Klein:2015hvg, Wang:2019ryf, Babak:2017tow, Fan:2020zhy}.
This allows us to infer the bulge luminosities of the host galaxies using
the $M_{\rm MBH}-L_{\rm bulge}$ relation and the estimated \ac{MBH} masses.
In this context, when extracting the probability distribution of redshift for a \ac{GW} source
like an \ac{EMRI} or \ac{MBHB}, one can assign weights
to each candidate host galaxy based on the degree of consistency between
its bulge luminosity and the value inferred from the $M_{\rm MBH}-L_{\rm bulge}$ relation.
As an example, the authors consider \ac{MBHB} detections by TianQin. The left panel of FIG. \ref{fig:InfoGain_EMRI&MBHB}
illustrates the distributions of information gains for the redshift probability distributions
of candidate host galaxies, where the galaxies are weighted either
using the $M_{\rm MBH}-L_{\rm bulge}$ relation or by uniform weights.
One can see that applying the $M_{\rm MBH}-L_{\rm bulge}$ relation significantly increases the information gain
for the redshift probability distributions of \acp{MBHB}, especially for sources with precise spatial localizations.

The method of statistically testing the correlation between \ac{GW} sources and AGNs is based on
the premise that both \acp{EMRI} and \acp{MBHB} may be correlated with AGNs during their formation.
For \acp{MBHB}, these systems must overcome the so-called ``final-parsec problem''
\cite{Begelman:1980vb, Milosavljevic:2001vi} to enter the phases
dominated by \ac{GW} radiations. The rich gas environments in AGNs can help the \acp{MBHB}
lose orbital angular momentum through dynamical friction, making AGNs potential major hosts
for \acp{MBHB} systems \cite{Begelman:1980vb, Armitage:2002uu, Escala:2004jh, Mayer:2007vk,
Macfadyen:2006jx, Cuadra:2008xn, Goicovic:2015kda, LISA:2022yao}.
For \acp{EMRI}, in addition to the standard channel in normal galaxies \cite{Babak:2017tow, Amaro-Seoane:2012lgq}
discussed in Section \ref{sec:sirens_candidateGW},
some studies suggest that the formation rate of \acp{EMRI} in AGN environments could be several orders of
magnitude higher than in normal galaxies \cite{Levin:2003ej, Levin:2006uc, Yunes:2011ws, Pan:2021ksp,
Pan:2021oob, Pan:2021lyw, Derdzinski:2022ltb}.
Even though AGNs comprise only about $1\%$ of all galaxies (with a maximum of around $10\%$
based on luminosity functions \cite{Dahlen:2005ex, Hopkins:2006fq}), they may still be the
dominant source of \acp{EMRI} detectable by future space-based \ac{GW} detectors \cite{Pan:2021ksp, Pan:2021oob}.
However, there remain significant uncertainties in our current understandings of the formation
mechanisms of both \acp{MBHB} and \acp{EMRI} \cite{LISA:2022yao, Derdzinski:2023qbi},
making it difficult to determine the exact correlation between these two classes of \ac{GW} sources and AGNs.
In cases when the spatial localization precisions of \ac{GW} sources provided by \ac{GW} detections
are much lower than the required precisions to uniquely identify host galaxies of the \ac{GW} sources,
a method to statistically test the correlation between \ac{GW} sources and AGNs is necessary,
as proposed in the literatures \cite{Zhu:2023imz, Zhu:2024qpp}.
Once the correlation between \acp{EMRI} or \acp{MBHB} and AGNs is established statistically,
as outlined in \cite{Zhu:2023imz, Zhu:2024qpp}, one can fully (or partially) extract the redshift
information of these \ac{GW} sources through AGN catalogs \cite{Zhu:2024qpp}.
For example, considering \acp{EMRI} originating from AGNs,
the right panel of FIG. \ref{fig:InfoGain_EMRI&MBHB}
shows the distribution of information gains for the redshift probability distributions of \acp{EMRI}
extracted with AGN catalogs. Compared to redshifts derived from all candidate galaxies,
the information gain from AGN catalogs is significantly larger,
benefiting from the fact that the fraction of AGNs among all galaxies is only about 1\%.

\subsection{TianQin forecast for constraining $\Lambda$CDM} \label{sec:LCDM}

This section will present the expected constraining potentials of TianQin on
various parameters of $\Lambda$CDM cosmological model,
including the Hubble-Lema\^itre constant $H_0$, and
the fractional density parameters $\Omega_M$ and $\Omega_{\Lambda}$.
The standard sirens considered in this section encompass all classes of candidate \ac{GW} sources
introduced in Section \ref{sec:sirens_candidateGW}.
Additionally, since the population properties of \ac{GW} sources can significantly influence the extent to
which \ac{GW} detections are affected by selection effects \cite{LIGOScientific:2017adf,
Chen:2017rfc, Mandel:2018mve, DES:2019ccw},
the capabilities of \ac{GW} sources from different populations to constrain $\Lambda$CDM parameters
will be analyzed separately.

\subsubsection{Hubble tension and standard sirens} \label{sec:Hubble_tension}

The $\Lambda$CDM model, although it fits observations of
the \ac{CMB} \cite{Planck:2015fie, Planck:2018vyg} very well and
describes the evolution of the universe from the era of \ac{BBN}
to the formation of large-scale structures and the accelerated expansion of
the late universe \cite{Carroll:2000fy, Peebles:2002gy, Bull:2015stt},
has faced some challenges in recent years \cite{Bull:2015stt, Perivolaropoulos:2021jda, Verde:2023lmm}.
The most significant of these challenges the discrepancy between the Hubble-Lema\^itre constant $H_0$ measurements
derived from the early universe and those from the late universe.
The Planck satellite's \ac{CMB} anisotropy observations \cite{Planck:2013oqw},
when combined with the flat $\Lambda$CDM model, measure $H_0$ to be
$H_0 = 67.4 \pm 0.5 ~\!{\rm km/s/Mpc}$ \cite{Planck:2018vyg},
a result that is supported by several other observations \cite{Addison:2017fdm, DES:2017txv, ACT:2020gnv}.
Since the \ac{CMB} radiation originates from the early universe,
the $H_0$ values measured through \ac{CMB} observations are referred to as early universe measurements.
On the other hand, the SH0ES project's \ac{SN Ia} observations \cite{Riess:2016jrr},
when combined with the cosmic distance ladder, measure $H_0$ to
be $H_0 = 73.2 \pm 1.3 ~\!{\rm km/s/Mpc}$ \cite{Riess:2020fzl},
and this measurement is supported by other independent observations
\cite{Freedman:2010xv, Soltis:2020gpl, Blakeslee:2021rqi}.
Since SNe Ia primarily occur in the lower redshift universe, the $H_0$ values
derived from \ac{SN Ia} observations are referred to as the late (or local) universe measurements.
The discrepancy between these two classes of measurements has grown to a significance level
greater than $4\sigma$ and is known as the Hubble tension \cite{Freedman:2017yms, Riess:2019qba}.

The Hubble tension represents a fundamental debate regarding the reliability of
the $\Lambda$CDM cosmological model and the cosmic distance ladder system
\cite{DiValentino:2021izs, Schoneberg:2021qvd, Cai:2021weh}.
With new observations from both \ac{CMB} and SNe Ia continuing to reinforce their
respective previous measurements \cite{ACT:2020frw, Riess:2024ohe},
an independent third measurement of $H_0$ would significantly contribute to resolving the Hubble tension.
As first proposed in Schutz \cite{Schutz:1986gp} and detailed
in Section \ref{sec:sirens_principle} of this paper, the measurement of luminosity distances
for \ac{GW} sources is self-calibrating andindependent of
the cosmic distance ladder system. Therefore, the $H_0$ values measured by standard sirens
show considerable promise for clarifying the Hubble tension.
Current and future ground-based and space-based \ac{GW} detectors are well
positioned to contribute to resolving the Hubble tension by providing independent measurements
of $H_0$ through standard sirens \cite{Chen:2017rfc,Feeney:2018mkj,Chen:2020zoq,Califano:2022syd,
Gupta:2022fwd,Zhu:2023jti,Song:2022siz,Muttoni:2023prw,DelPozzo:2017kme,Tamanini:2016zlh,
Wang:2020dkc,Muttoni:2021veo,Laghi:2021pqk,Yang:2021qge,Cai:2021ooo,Yang:2021xox,Yang:2022iwn}.
TianQin, specifically, is expected to detect tens to thousands of standard sirens
 \cite{Liu:2020eko, Wang:2019ryf, Fan:2020zhy}, which could provide valuable insights
 into resolving the Hubble tension \cite{Zhu:2021bpp, Zhu:2021aat, Zhu:2024qpp},
as discussed in the following section.

\subsubsection{Constraints on $H_0$} \label{sec:sec:LCDM_H0}

\begin{figure*}[t]
\centering
\includegraphics[width=0.450\textwidth]{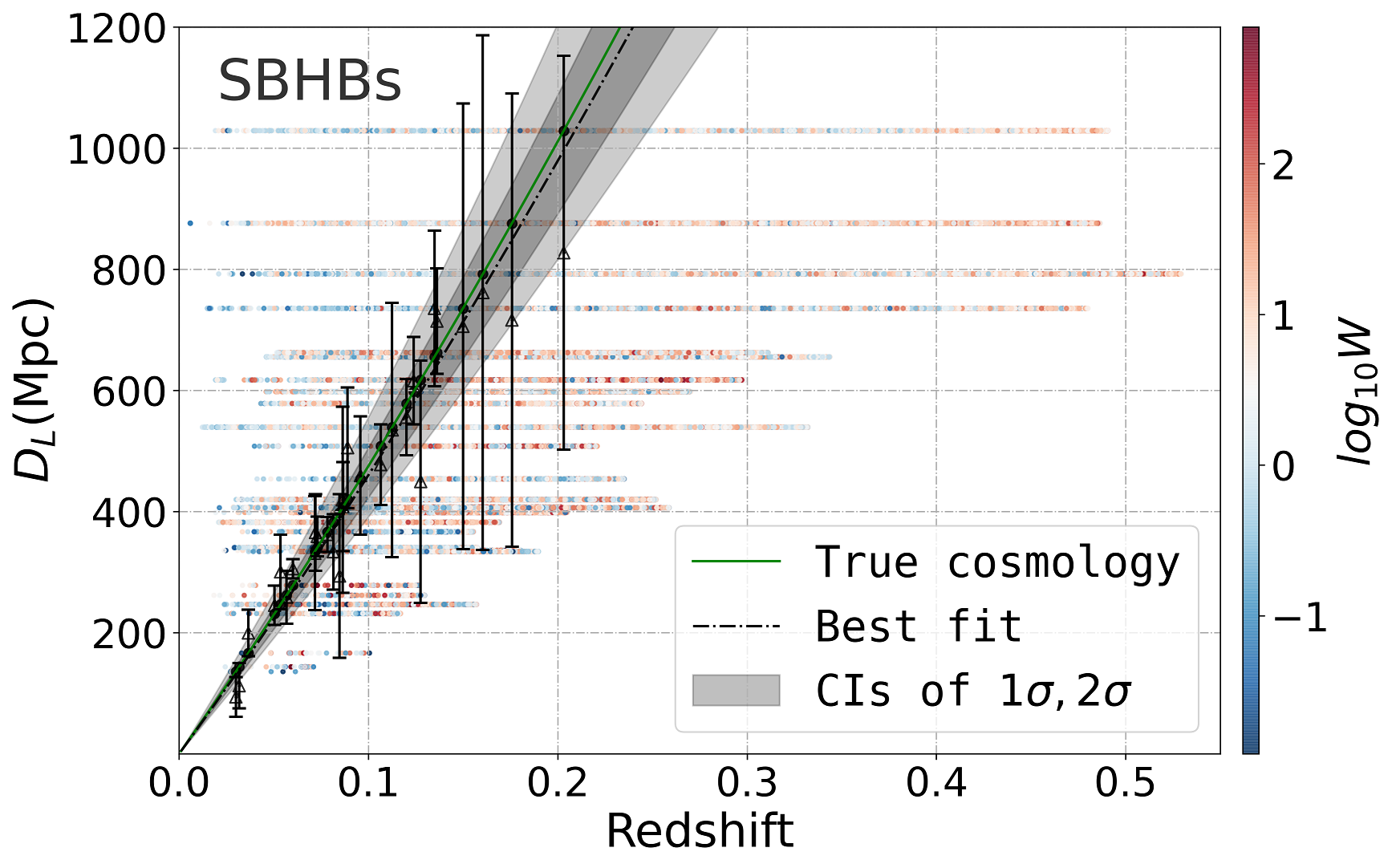}~~~
\includegraphics[width=0.40\textwidth]{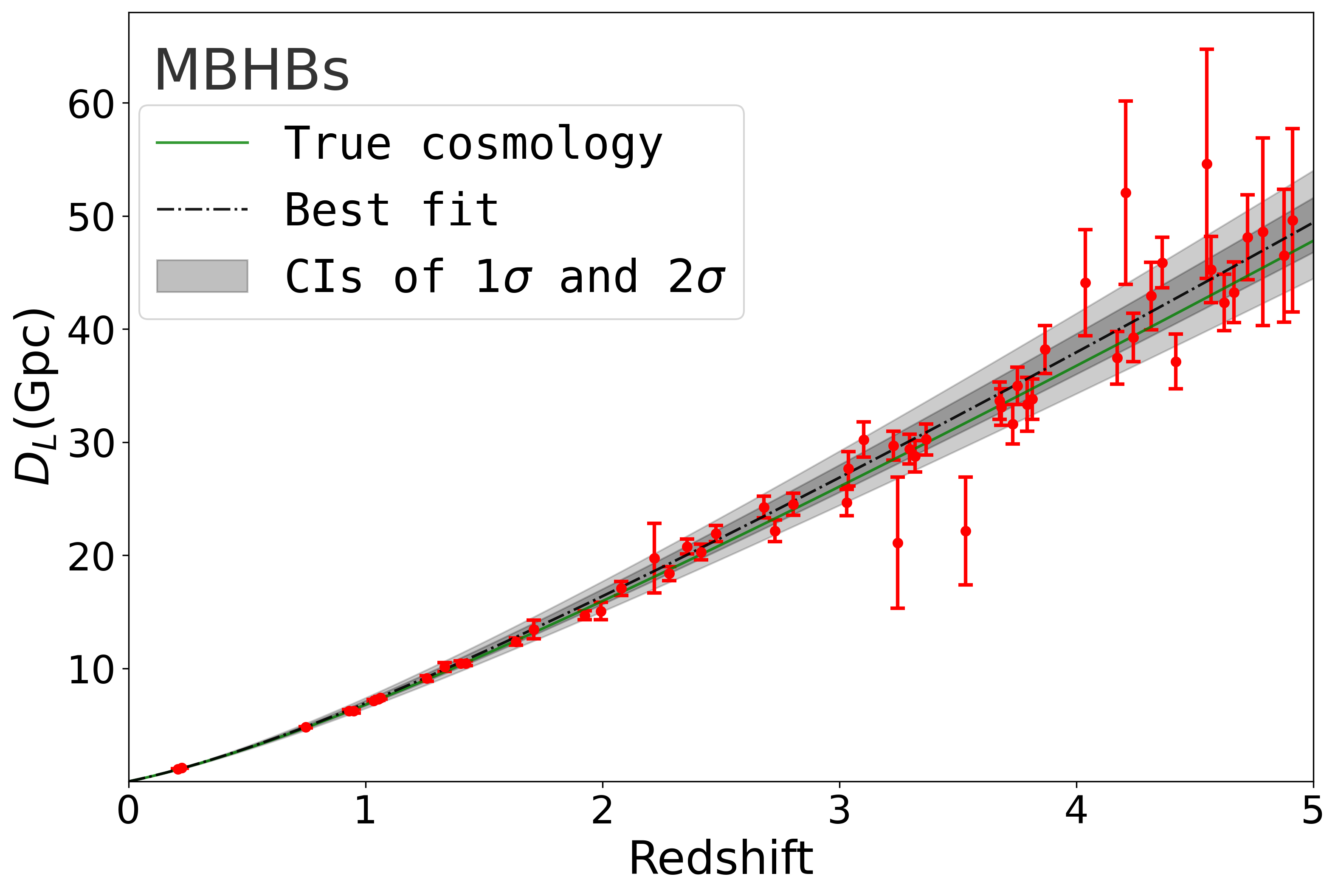} \\
\includegraphics[width=0.90\textwidth]{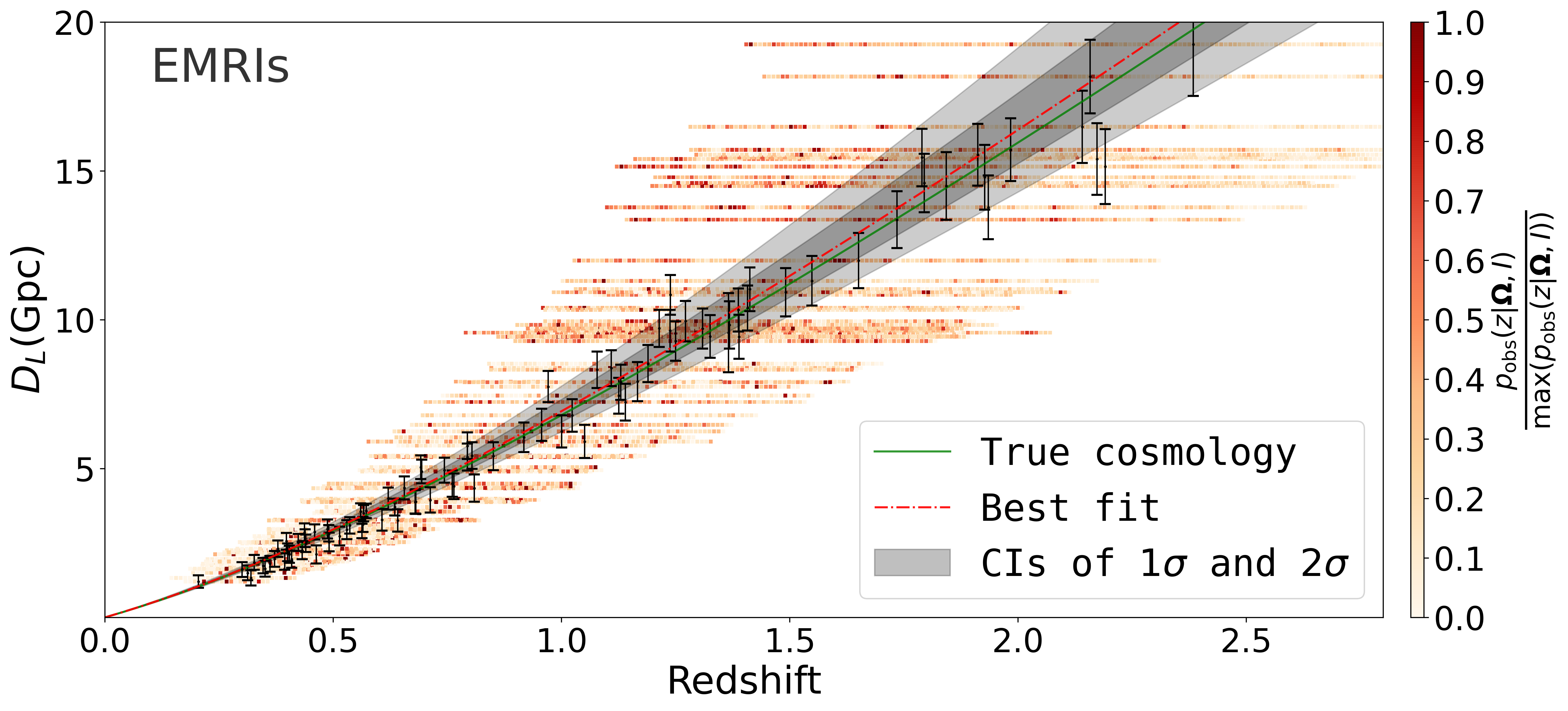}~~~~~~
\caption{Examples of using TianQin standard sirens to fit the $D_L - z$ relation
for different candidate populations. Dark sirens, assumed to be \acp{SBHB} \cite{Zhu:2021bpp}
and \acp{EMRI} \cite{Zhu:2024qpp},
SBHBs are shown in the top-left panel and \acp{EMRI} are shown in the bottom panel.
Bright sirens, assumed to be \acp{MBHB} \cite{Zhu:2021aat}, are shown in the top-right panel.
The error bars represent the total estimation errors of $1\sigma$ confidence interval (CI) on luminosity distances.
Note that for the bright sirens, which can uniquely identify their host galaxies by their \ac{EM} counterparts,
precise redshifts can be obtained from spectroscopic measurements, so the redshift errors are not marked
in the figure. In contrast, the redshift information for the dark sirens are presented as
probability distributions provided by galaxy catalogs, which are indicated by colored horizontal lines
in the figure, with the shade of the color represents the size of probability density.}
\label{fig:fitting_DLz}
\end{figure*}

The process of constraining the cosmological parameters, including $H_0$, using standard sirens
is actually a process of fitting the $D_L - z$ relation by
utilizing the luminosity distance information provided by
\ac{GW} detections and the redshift information obtained from \ac{EM} counterparts, galaxy catalogs, or other means.
Fig. \ref{fig:fitting_DLz} illustrates the processes of fitting the $D_L - z$ relation for TianQin
with dark sirens, such as \acp{SBHB} and \acp{EMRI}, as well as bright sirens, such as \acp{MBHB}.
Bright sirens typically have precise redshift measurements with negligible errors,
whereas the redshift information of dark sirens is usually represented as probability distributions
with large equivalent errors. As a result, bright sirens generally provide stronger constraints on
the $D_L - z$ relation compared to dark sirens,
unless the spatial localizations of dark sirens are sufficiently precise to
uniquely identify their host galaxies.
However, significant uncertainty remains regarding the fraction of actual bright sirens among all standard sirens in future observations, consequently, cosmological prospects from brighter sirens
are often considered an optimistic case. In contrast, redshifts of \ac{GW} sources
extracted from galaxy catalogs can be applied to nearly all detectable \ac{GW} sources, making the cosmological prospects from dark sirens a more conservative case.

\begin{figure*}[t]
\centering
\includegraphics[width=0.450\textwidth]{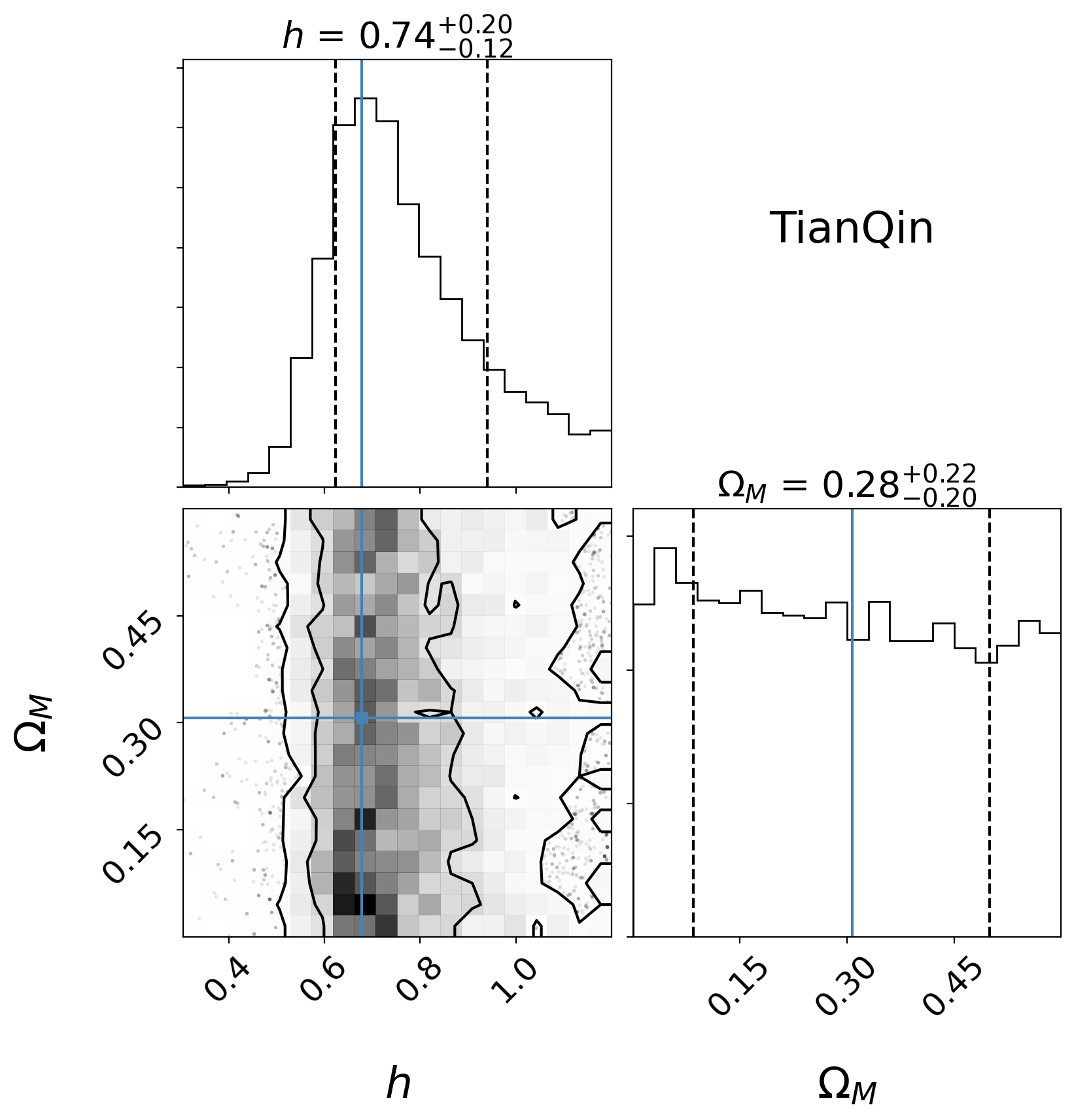} ~~~~
\includegraphics[width=0.450\textwidth]{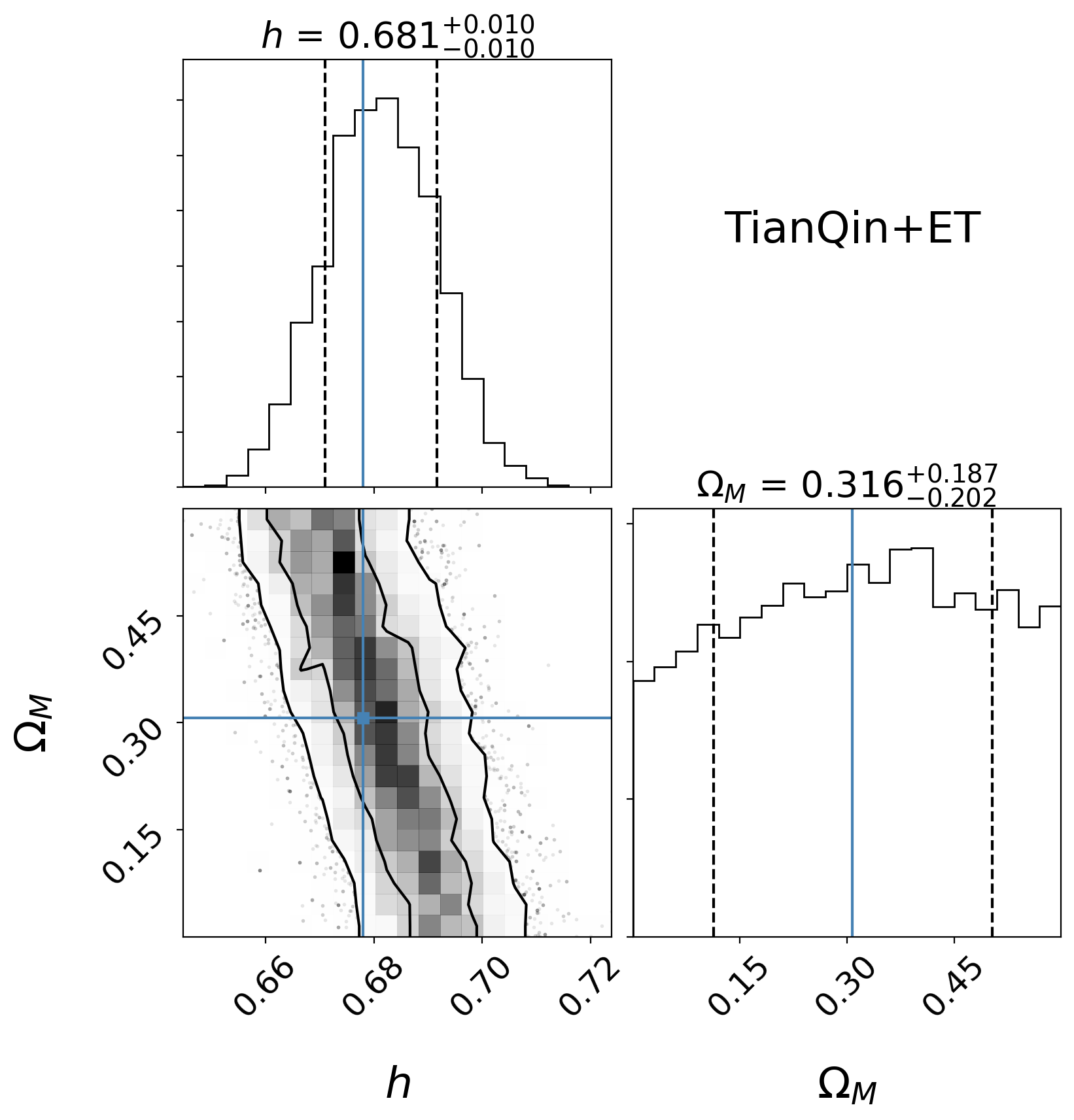}
\caption{Typical posterior probability distributions of the parameters $h$
($h \equiv \frac{H_0}{100 ~\!{\rm km/s/Mpc}}$) and $\Omega_M$ from \acp{SBHB} detected by TianQin (left) and TianQin+ET (right) \cite{Zhu:2021bpp}.
In each subfigure, the lower left panel shows the joint posterior probability of $h$ and $\Omega_M$,
with the contours represent CIs of $1 \sigma$ and $2 \sigma$, respectively;
the upper and right panels show the marginalized posterior probability distributions of the same parameters,
with the dashed lines indicate a $1 \sigma$ CI.
In each panel, the solid cyan lines mark the true values of the parameters.}
\label{fig:cosmo_TQ_ET_H0}
\end{figure*}

By detecting \ac{SBHB} and utilizing the GWENS galaxy catalog
\footnote{GWENS Catalogue: \url{https://astro.ru.nl/catalogs/sdss_gwgalcat/}} derived
from prerelease data of SDSS \cite{SDSS:2000hjo, LIGOScientific:2019zcs},
as well as weighting the candidate host galaxies of \acp{SBHB} with information of
their photometric luminosities in multiple bands, as reported in the literature \cite{Zhu:2021bpp},
TianQin is expected to constrain $H_0$ to a precision of approximately $30\%$.
A twin-constellation configuration of TianQin (i.e., TianQin I+II) could improve this precision to about $15\%$.
At first glance, this level of precisions might appear insufficient for $H_0$ measurements,
especially when compared to the current precision of $H_0$ obtained from dark sirens detected by
\ac{LVK} \cite{LIGOScientific:2021aug}.
There are two main factors for the relatively poor precisions of $H_0$ reported in the literature \cite{Zhu:2021bpp}:
first, the redshifts provided by the GWENS galaxy catalogs contain large photometric redshift errors, and second,
the luminosity distance errors for the detected \acp{SBHB} can be as large as $30\%$.
To address the first issue, spectroscopic redshifts of candidate host galaxies could be introduced.
With galaxy spectroscopic surveys \cite{2012RAA....12.1197C, DESI:2016fyo, Gong:2019yxt},
spectroscopic redshifts for low-redshift galaxies are guaranteed.
For the same \acp{SBHB}, the use of spectroscopic redshifts could improve the precisions of $H_0$ measurements
to about $20\%$ for TianQin and approximately $8\%$ for TianQin I+II \cite{Zhu:2021bpp}.
For the second issue, it can be improved by a joint analysis of multi-band \ac{GW} data.
The \ac{GW} signals of \acp{SBHB} in both the inspiral and merger phases can be detected by space-based
and ground-based \ac{GW} detectors \cite{Sesana:2016ljz}, respectively. With the development of
third-generation ground-based \ac{GW} detectors \cite{Punturo:2010zz, LIGOScientific:2016wof},
multi-band \ac{GW} detections are expected to be realized in the near future.
The third-generation ground-based \ac{GW} detectors will not only contribute high \acp{SNR} to \ac{SBHB} detections
and reduce luminosity distance errors, but will also enhance the detection capacity and horizon of
space-based \ac{GW} detectors for \acp{SBHB} \cite{Wong:2018uwb, Liu:2020nwz, Ewing:2020brd, Zhu:2021bpp}.
As reported in the literature \cite{Zhu:2021bpp}, the precision of $H_0$ can be improved to
about $1\%$ through the joint analysis of multi-band data from \acp{SBHB} provided by TianQin
and \ac{ET}.
Typical results of constraining $H_0$ from single-band \ac{SBHB} data from TianQin alone
and multi-band \ac{SBHB} data from both TianQin and \ac{ET},
are illustrated in FIG. \ref{fig:cosmo_TQ_ET_H0},
where the free cosmological parameters include $H_0$ and $\Omega_M$.

\begin{figure*}[htbp!]
\centering
\includegraphics[width=0.60\textwidth]{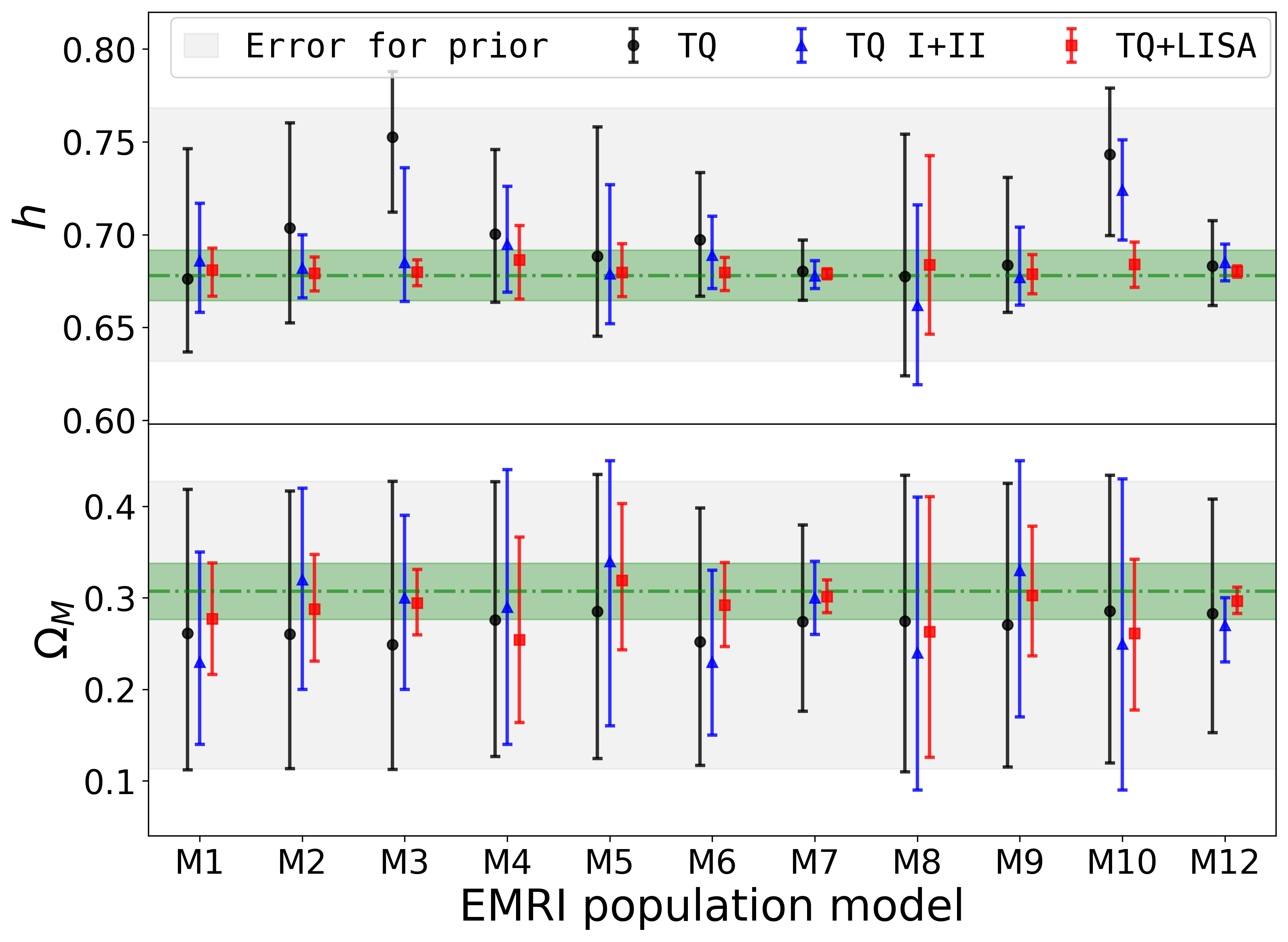} ~~
\includegraphics[width=0.360\textwidth]{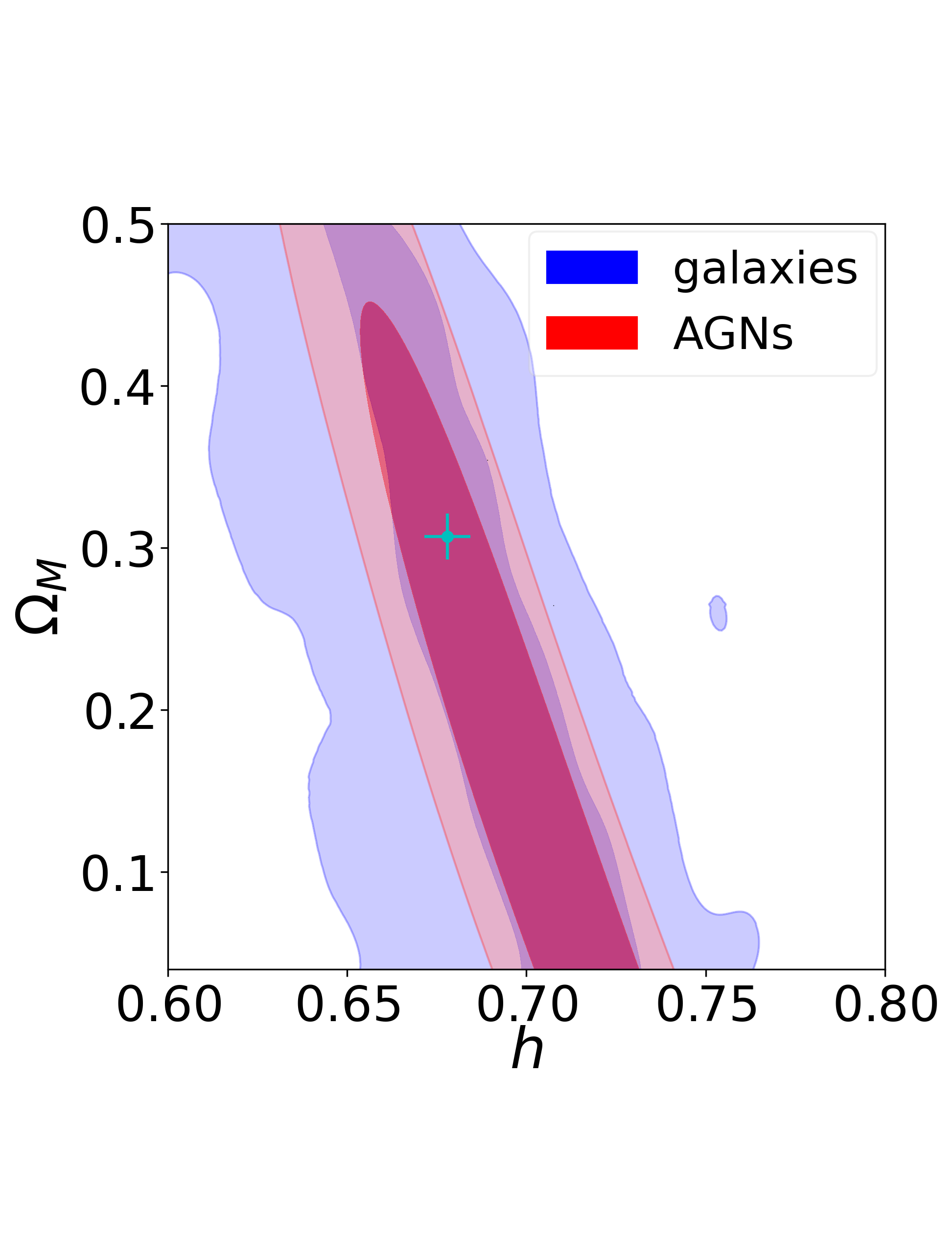}
\caption{Left panel: Constraining errors on $h$ and $\Omega_M$ by TianQin (black), TianQin I+II (blue), and TianQin+LISA (red) under various \ac{EMRI} population models \cite{Zhu:2024qpp}.
The error bars correspond to an $1\sigma$ CI, and the horizontal green dot-dashed line represents
the true value of the corresponding parameter.
The gray-shaded area represents the equivalent error of $1\sigma$ for the prior,
and the green-shaded area represents a referential error scale of $2\%$ ($10\%$) for $h$ ($\Omega_M$) measurements.
Right panel: Typical constraints on $h$ and $\Omega_M$ obtained by TianQin using AGN catalogs (red contours) and galaxy catalogs (blue contours) under the M1 \ac{EMRI} population model \cite{Zhu:2024qpp}, and assuming that both catalogs have spectral redshifts.
Contours of each color represent the $1\sigma$ and $2\sigma$ confidence intervals,
with the cyan cross indicating the true value of the parameters.}
\label{fig:cosmo_TQ_H0M_EMRI}
\end{figure*}

For \ac{EMRI} detections by TianQin, a similar analysis of the expected constraints
on $H_0$ has been done and reported in the literature \cite{Zhu:2024qpp}.
Unlike the case of \acp{SBHB}, there is a great uncertainty in the current understanding of
the population properties of \acp{EMRI}. The literature \cite{Zhu:2024qpp} provides a detailed analysis of TianQin's ability to constrain $H_0$ using 11 population models of \acp{EMRI}, which are
formed through the standard channel presented in the literature \cite{Babak:2017tow},
as shown in FIG. \ref{fig:cosmo_TQ_H0M_EMRI}.
Since the predicted EMRI rates vary significantly between population models,
resulting in significant differences in
the expected precisions of $H_0$ for different population models of \acp{EMRI}.
Under the M1 population model, the expected precision of $H_0$ constrained by TianQin is approximately $8.1\%$,
with TianQin I+II configuration,
the expected precision can only improve to about $4.4\%$.
However, for population models with more optimistic \ac{EMRI} rates,
TianQin can provide constraints on $H_0$ with a precision better than $3\%$,
and TianQin I+II could push the precision to nearly $1\%$ \cite{Zhu:2024qpp}.
In addition, according to the analysis in the literature \cite{Zhu:2024qpp},
if the correlation between the spatial distributions of \acp{EMRI} and AGNs can be established
through statistical methods, TianQin could more accurately extract the redshift information of \acp{EMRI}
using AGN catalogs, as described in Section \ref{sec:sirens_imporve_z}.
For instance, using the same detected \acp{EMRI} under the M1 population model,
TianQin could improve the precision of $H_0$ to approximately $3.2\%$ by utilizing AGN catalogs to
extract redshifts. This improvement arises not only from the lower spatial number density of AGNs
compared to galaxies but also because AGN surveys require spectroscopic observations,
which provide precise spectroscopic redshift information.
Even under the condition where both galaxy and AGN catalogs contain spectroscopic redshift observations,
utilizing AGN catalogs can still significantly enhance the constraint on $H_0$,
as illustrated in the right panel of FIG. \ref{fig:cosmo_TQ_H0M_EMRI}.

\begin{figure*}[t]
\centering
\includegraphics[width=0.70\textwidth]{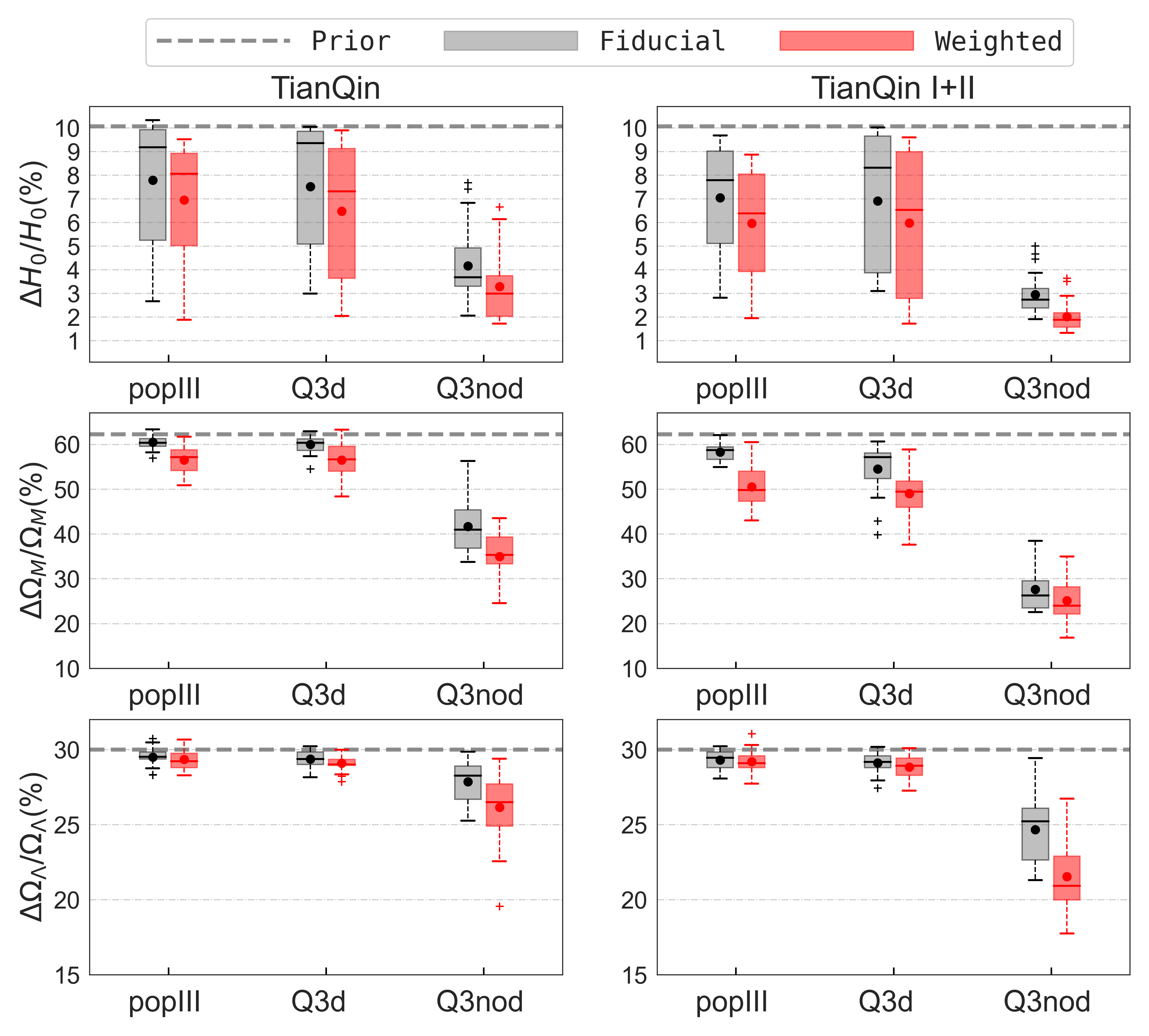}
\caption{Distributions of the errors of $(H_0, \Omega_M, \Omega_{\Lambda})$ constrained by dark \acp{MBHB} under pop III,
Q3\_d and Q3\_nod population models \cite{Zhu:2021aat}.
The left and right columns show the results from TianQin and TianQin I+II, respectively.
The horizontal gray dashed line represents a equivalence error of $1 \sigma$ CI from the prior.
For each result, the box represents the range of $25\%-75\%$ of the data distribution,
the upper limit of the whisker length is 1.5 times the box length, and the crosses are outliers.
In the center of each box, the dot represents the mean value, the short horizontal line represents the median value.
For each box color, gray and red represent the distribution of errors obtained using the fiducial and the weighted methods, respectively.}
\label{fig:cosmo_TQ_H0ML_MBHB}
\end{figure*}

For \ac{MBHB} detections, TianQin is expected to detect a few to hundreds of \acp{MBHB}.
The study \cite{Zhu:2021aat} has analyzed and reported the expected constraint errors on $H_0$ from TianQin's \ac{MBHB} detections,
using the three population models proposed in the literature \cite{Klein:2015hvg}.
Similar to \acp{EMRI}, the expected detection number of \acp{MBHB} varies significantly with population models.
TianQin is projected to achieve sky localization errors better than $10 ~\!{\rm deg}^2$
for approximately half of these detections \cite{Wang:2019ryf}, considering the large uncertainties
in our knowledge of the \acp{MBHB}' \ac{EM} radiation mechanism \cite{Bogdanovic:2021aav},
\acp{MBHB} could be either bright sirens or dark sirens.
In the optimistic scenario where \acp{MBHB} act as bright sirens, the literature \cite{Zhu:2021aat} analyzed
the constraints on various cosmological parameters
under relaxed selection criteria for bright sirens: ${\rm SNR} > 8$, $\Delta \Omega < 10~\!{\rm deg}^2$, and $z<3$.
The expected results are as follows \cite{Zhu:2021aat}: under the three population models pop III, Q3\_d and Q3\_nod,
the constraining errors of $H_0$ by TianQin are about $4.3\%$, $6.2\%$ and $1.9\%$, respectively,
and TianQin I+II can reduce these errors to about 70\%.
In the conservative scenario, where \acp{MBHB} act as dark sirens,
these sources still provide valuable contributions to $H_0$ measurements.
Under the assumption that each candidate
host galaxy has an equal probability of hosting an \ac{MBHB} (denoted as the fiducial method),
the expected errors of $H_0$ measured by TianQin (TianQin I+II) under the pop III, Q3\_d, and
Q3\_nod models are about $7.8\%~(7.0\%)$, $7.5\%~(6.9\%)$,
and $4.2\% ~(2.9\%)$ \cite{Zhu:2021aat}, respectively.
Notably, the constraints on $H_0$ can be significantly improved by the method
of assigning weights to each candidate host galaxy using the $M_{\rm MBH}-L_{\rm bulge}$ relation
(denoted as the weighted method), as described in Section \ref{sec:sirens_imporve_z}.
Fig. \ref{fig:cosmo_TQ_H0ML_MBHB} illustrates the distributions of the constraining errors
on $H_0$ from TianQin's dark \acp{MBHB} using both the fiducial and weighted methods,
obtained from multiple independent simulations.
Applying the weighted method to the same dark \acp{MBHB} under the pop III, Q3\_d and Q3\_nod populations models,
the constraining errors of $H_0$ are expected to be reduced to $6.9\% ~(6.0\%)$, $6.5\% ~(6.0\%)$,
and $3.3\% ~(2.0\%)$ for TianQin (TianQin I+II) \cite{Zhu:2021aat}.
These results demonstrate that the weighted method enhances the precision of $H_0$ constraints
compared to the fiducial approach.

To sum up, there are great uncertainties in the precisions of $H_0$ measurements that
TianQin are expected to provide, and the usefulness for clarifying the Hubble tension is limited by many factors.
For \acp{SBHB}, while the population properties are relatively certain, the realization of precise constraint
on $H_0$ relies on joint multi-band detections with third-generation ground-based \ac{GW} detectors,
and the realization of multi-band \ac{GW} detections is currently subject to a large degree of uncertainty.
For \acp{EMRI} and \acp{MBHB}, the uncertainties on the population properties make the capabilities of TianQin
to utilize them to constrain $H_0$ also subject to significant uncertainties.
However, overall, TianQin allows to contribute helpful $H_0$ measurements,
but there is no guarantee that TianQin will be able to provide a precision
of $H_0$ sufficient to clarify the Hubble tension.
In the end, for clarity, the expected precisions of $H_0$ constrained by TianQin
using three classes of \ac{GW} sources, \acp{SBHB}, \acp{EMRI}, and \acp{MBHB},
are listed in Table \ref{tab:LCDM_errs}.

\subsubsection{Constraints on the fractional densities}

Since the $D_L - z$ relation, i.e. Equation (\ref{eq:DL_z}), can degenerate into
the so-called Hubble-Lema\^itre law at redshifts close to zero \cite{Hubble:1929ig},
in which case the $D_L - z$ relation is determined by only one parameter, $H_0$, the estimations of
the fractional density parameters require relying on \ac{GW} sources at relatively higher redshifts.
In all classes of candidate standard sirens for TianQin, the contributions to constrain
the fractional density parameters $\Omega_M$ and $\Omega_{\Lambda}$ are predominantly
from \acp{EMRI} and \acp{MBHB} because detectable \acp{SBHB} are in the local universe with $z<0.3$.

The literature \cite{Zhu:2024qpp} adopts a flat-$\Lambda$CDM model to analyze the expected
constraining capabilities on $H_0$ and $\Omega_M$ by TianQin's \acp{EMRI}.
The expected errors of $\Omega_M$ constrained by TianQin and TianQin I+II for various
population models is illustrated in the bottom panel of FIG. \ref{fig:cosmo_TQ_H0M_EMRI}.
One can find that TianQin provides little effective constraints on $\Omega_M$ under most
of the population models. Only under the three models with optimistic \ac{EMRI} rates, M6, M7,
and M12, does TianQin show an effective constraint on $\Omega_M$, but the best precision
is only about 33\%.
TianQin I+II configuration is able to achieve an effective constraint on $\Omega_M$
under all but three population models, M4, M8, and M10, and improves the best precision
to about 13\%.

The literature \cite{Zhu:2021aat} adopts a non-flat $\Lambda$CDM model to analyze the joint constraining
capabilities of TianQin on three parameters, $H_0$, $\Omega_M$, and $\Omega_{\Lambda}$, using \acp{MBHB}.
In the optimistic case of \acp{MBHB} as bright sirens, TianQin is able to achieve effective constraints on
$\Omega_M$ and $\Omega_{\Lambda}$ under all population models, the precision of $\Omega_M$ constrained
by TianQin is about 7\% to 27\% and the precision of $\Omega_{\Lambda}$ is about 16\% to 28\%.
TianQin I+II can improve the precision of $\Omega_M$ up to about 5\% to 16\% and
the precision of $\Omega_{\Lambda}$ up to about 11\% to 25\%.
In the conservative case of \acp{MBHB} as dark sirens,
the distribution of the constraining errors on $\Omega_M$ and $\Omega_{\Lambda}$ by TianQin and
TianQin I+II are demonstrated in FIG. \ref{fig:cosmo_TQ_H0ML_MBHB}.
One can find that for both $\Omega_M$ and $\Omega_{\Lambda}$, TianQin can only achieve effective
constraint under Q3\_nod population model with optimistic \ac{MBHB} merger rate. Although the method of
using the $M_{\rm MBH}-L_{\rm bulge}$ relation to weight the candidate host galaxies of \acp{MBHB}
as discussed in Section \ref{sec:sirens_imporve_z} can improve the constraining capability,
the effect is also obvious only under the Q3\_nod model.
TianQin I+II also achieves effective constraints on $\Omega_M$ and $\Omega_{\Lambda}$ only under
the Q3\_nod model. Under the Q3\_nod model, using the weighted method with the
$M_{\rm MBH}-L_{\rm bulge}$ relation, the precisions of $\Omega_M$ and $\Omega_{\Lambda}$
constrained by TianQin are about 35\% and 26\%, respectively, and TianQin I+II is able to improve
these two precisions to about 25\% and 21\%, respectively.

Estimating $\Omega_M$ and $\Omega_{\Lambda}$ can help us measure $H_0$ more accurately,
in addition to helping us understand the fractions of the various constituents of the universe.
This is because when using data from relatively high redshifts to estimate $H_0$, the linear relation
of Hubble-Lema\^itre law breaks down and the full $D_L - z$ relation expressed in Equation (\ref{eq:DL_z})
is needed to accurately estimate $H_0$. Whereas there are certain degree of degeneracies
between $H_0$ and $\Omega_M$ and between $H_0$ and $\Omega_{\Lambda}$, thus constraints on
$\Omega_M$ and $\Omega_{\Lambda}$ can break the degeneracies and improve the accuracy of $H_0$.
Finally, for clarity, the expected precisions of constraining $\Omega_M$ and $\Omega_{\Lambda}$
by TianQin using dark \acp{EMRI} and dark \acp{MBHB} are listed in Table \ref{tab:LCDM_errs}.

\subsection{TianQin forecast for probing dark energy} \label{sec:cosmo_de}

This section will reports the capability of TianQin for constrain the dark energy
\ac{EoS} using standard sirens with a model-dependent method.
The dark energy \ac{EoS} models that one adopted
are present in Section \ref{sec:sirens_principle}.
As in the previous section, the authors also present the potential of
each class of standard sirens separately.

\subsubsection{Dark energy versus the cosmological constant} \label{sec:cosmo_de_wEvolve}

The discoveries of the acceleration of the cosmic expansion \cite{SupernovaCosmologyProject:1998vns,
SupernovaSearchTeam:1998fmf} triggered the proposals for the concept of dark energy and some modified gravity
Theories \cite{Frieman:2008sn, Li:2011sd, Weinberg:2013agg}. Among researchers in astronomy, dark energy are widely accepted.
The standard model as well as the simplest model of dark energy is the cosmological constant
$\Lambda$ \cite{Carroll:2000fy}, which Albert Einstein claimed to be the greatest ``mistake'' he himself made.
The parameter $w$ of the \ac{EoS} of dark energy described by the cosmological constant
is also a constant with the value of $-1$, i.e., $w \equiv -1$.
Although the cosmological constant description of dark energy fits well with many current
cosmological observations \cite{Carroll:2000fy, Planck:2015fie, Planck:2018vyg, DES:2021wwk},
when one attempts to understand the physical nature of dark energy from a quantum field perspective,
one will fall into the so-called ``cosmological constant problem'' \cite{Carroll:2000fy, Frieman:2008sn,
Li:2011sd, Weinberg:2013agg}, i.e., the vacuum energy density predicted by quantum field theory
is about 120 orders of magnitude higher than the equivalent density of dark energy derived from
cosmological observations.

In order to explore the nature of dark energy by means of astronomical observations,
researchers have proposed a number of phenomenological parameterized forms to describe
the deviation of the dark energy \ac{EoS} with respect to the cosmological constant
and its evolution with redshift. The most widely used parametrized form is the CPL
model \cite{Chevallier:2000qy, Linder:2002et}, the form of the CPL \ac{EoS} model
is presented in Equation (\ref{eq:wz_CPL}).
By fitting the CPL model, earlier BAO and SNe Ia datasets showed
some signs of dark energy deviating from the cosmological constant, but not with significant
enough confidence \cite{eBOSS:2020yzd, Brout:2022vxf, DES:2024tys}.
Recently, the latest DESI's BAO data combined with the \ac{CMB} and SNe Ia
datasets obtain evidence of dark energy deviating from the cosmological constant with
greater than $3 \sigma$ confidence \cite{DESI:2024mwx}.
In addition, the reconstructions of the evolution of the dark energy \ac{EoS}
with redshift using the tomographic Alcock-Paczynski method also show signs of dark energy
deviating from the cosmological constant \cite{Zhao:2012aw, Zhao:2017cud, Zhang:2019jsu}.

In conclusion, the nature of dark energy is still a great mystery to be further explored,
either from the theoretical point of view or from the observation point of view.
\ac{GW} standard sirens, as an independent self-calibrating cosmic probe,
is expected to contribute usefulness in probing the nature of dark energy.

\subsubsection{Constraints on dark energy EoS} \label{sec:cosmo_de_w0wa}

\begin{figure*}[t]
\centering
\includegraphics[width=0.60\textwidth]{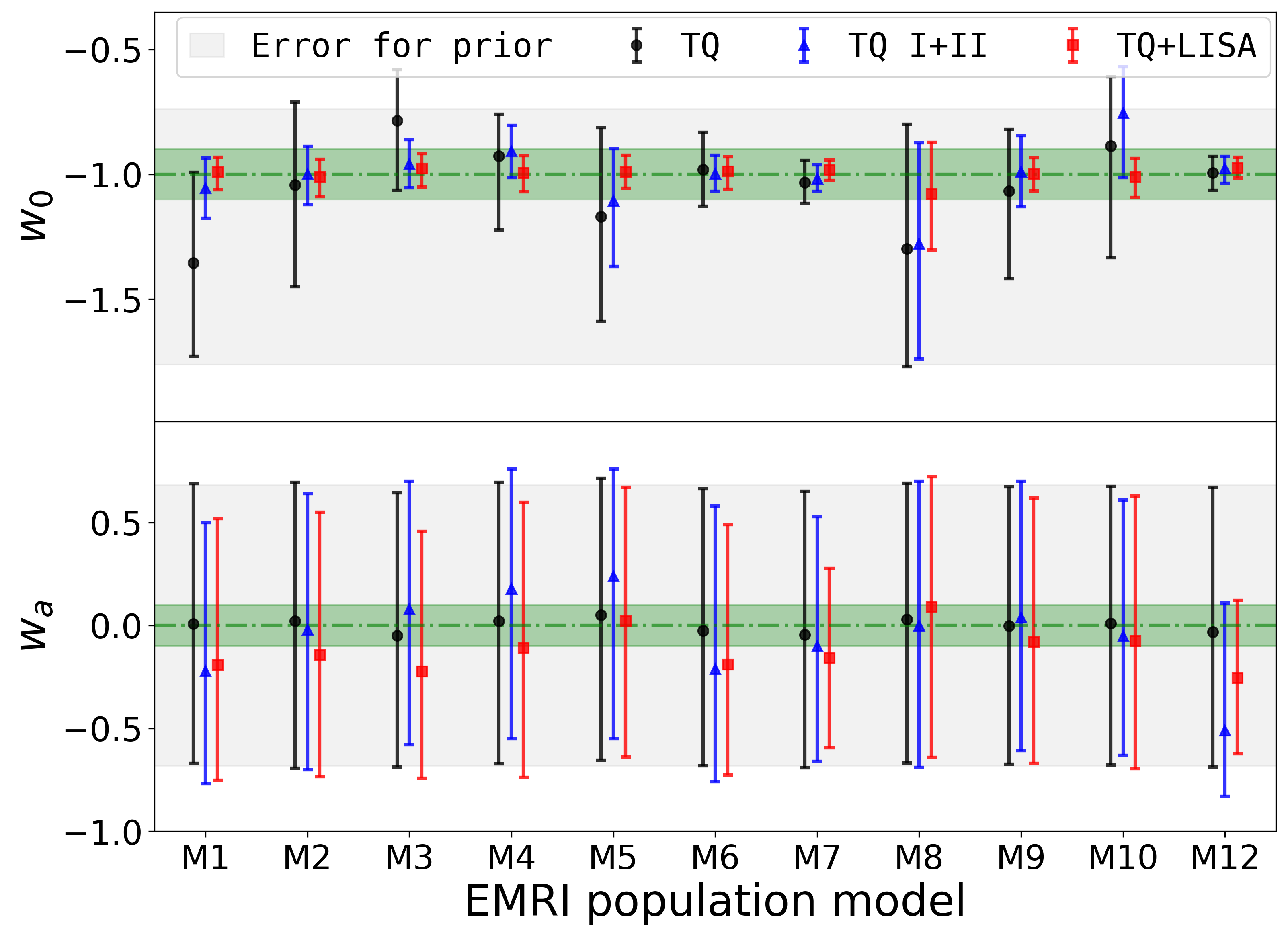}~~
\includegraphics[width=0.360\textwidth]{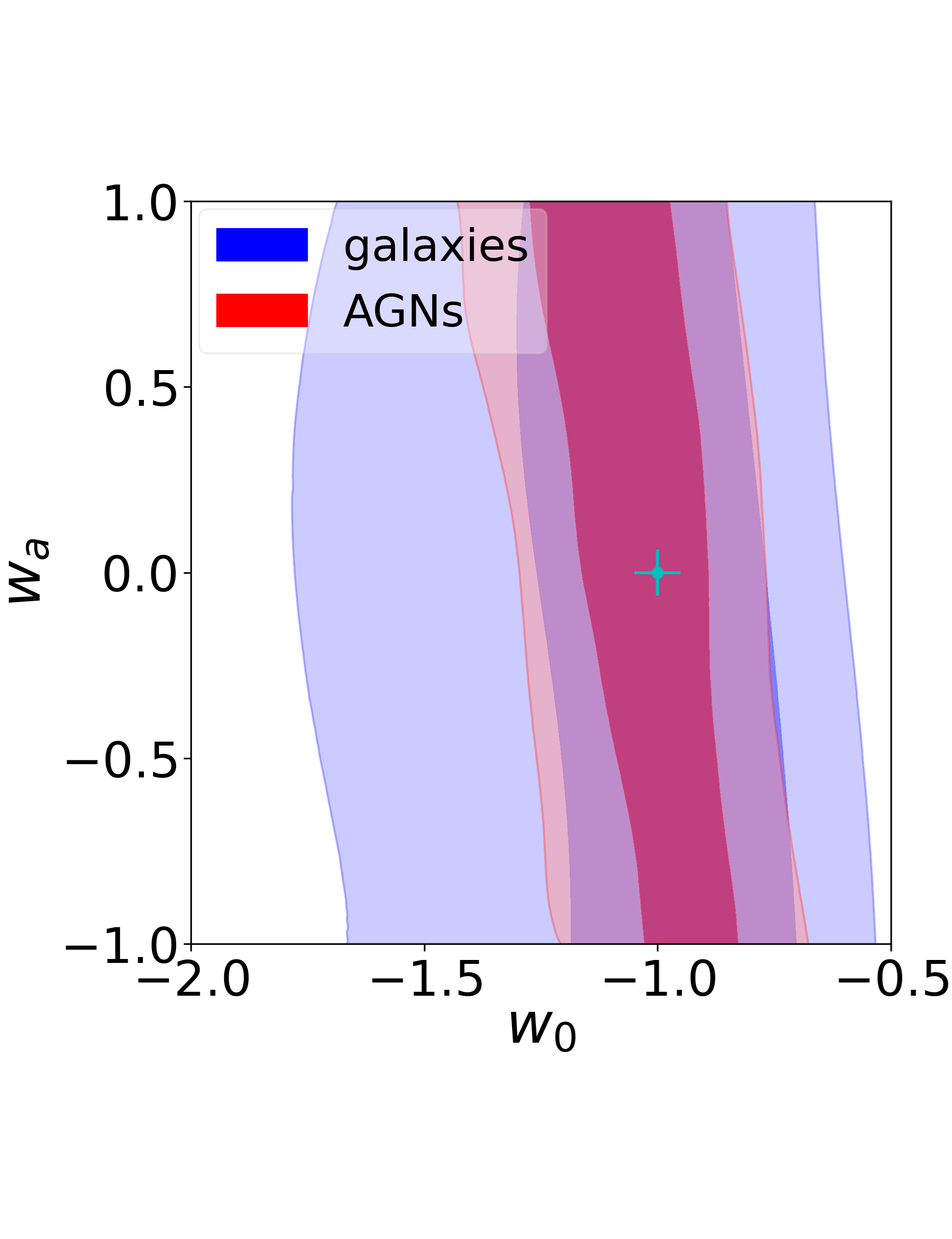}
\caption{Same as FIG. \ref{fig:cosmo_TQ_H0M_EMRI}, but for the constraints from \acp{EMRI} on
dark energy EoS parameters $w_0$ and $w_a$ \cite{Zhu:2024qpp}.
The green-shaded areas represent error scales of $\Delta w_0 = 0.1$ and $\Delta w_a = 0.1$.}
\label{fig:cosmo_TQ_w0wa_EMRI}
\end{figure*}

In this section, the adopted parameterization form of dark energy EoS
is the CPL model \cite{Chevallier:2000qy, Linder:2002et}, while fixing the $\Lambda$CDM parameters,
such as $H_0$, $\Omega_M$ and $\Omega_{\Lambda}$, to their true values.
As reported in the literature \cite{Zhu:2024qpp}, FIG. \ref{fig:cosmo_TQ_w0wa_EMRI} illustrates
the errors of $1\sigma$ CI for $w_0$ and $w_a$ constrained by TianQin and
TianQin I+II under various \ac{EMRI} population models.
TianQin using \ac{EMRI} detections can only achieve effective constraints on the $w_0$ parameter
and cannot effectively constrain $w_a$ under all population models.
The precisions of $w_0$ constraining by TianQin under the various population models vary greatly,
with the precision of $w_0$ being able to be better than $10\%$ under the models with optimistic \ac{EMRI} rates, while under the models with pessimistic \ac{EMRI} rates, the precisions of $w_0$ are only about $40\%$.
As in the case of restricting other cosmological parameters, TianQin I+II can also significantly improve
the constraints on the dark energy EoS. TianQin I+II can improve the precision of $w_0$ to
about $5\%-20\%$ under various population models. And TianQin I+II could realize effective constraints
on the parameter $w_a$ under optimistic \ac{EMRI} population models, such as the M7 and M12 models.
In addition, in particular, if the correlation between the spatial distributions of \acp{EMRI} and AGNs
can be established, AGN catalogs can be used to more precisely extract the redshift information of \acp{EMRI},
thereby improving the precision of constraints on the dark energy EoS.
For example, under the fiducial \ac{EMRI} model, the M1 model, TianQin's constraint on the dark energy EoS parameter $w_0$ using galaxy catalogs is approximately $25\%$, however, once the \ac{EMRI}-AGN correlation is established,
TianQin can improve the precision of $w_0$ to nearly $10\%$ by using AGN catalogs,
as shown in the right panel of FIG. \ref{fig:cosmo_TQ_w0wa_EMRI}.

\begin{figure*}[t]
\centering
\includegraphics[width=0.70\textwidth]{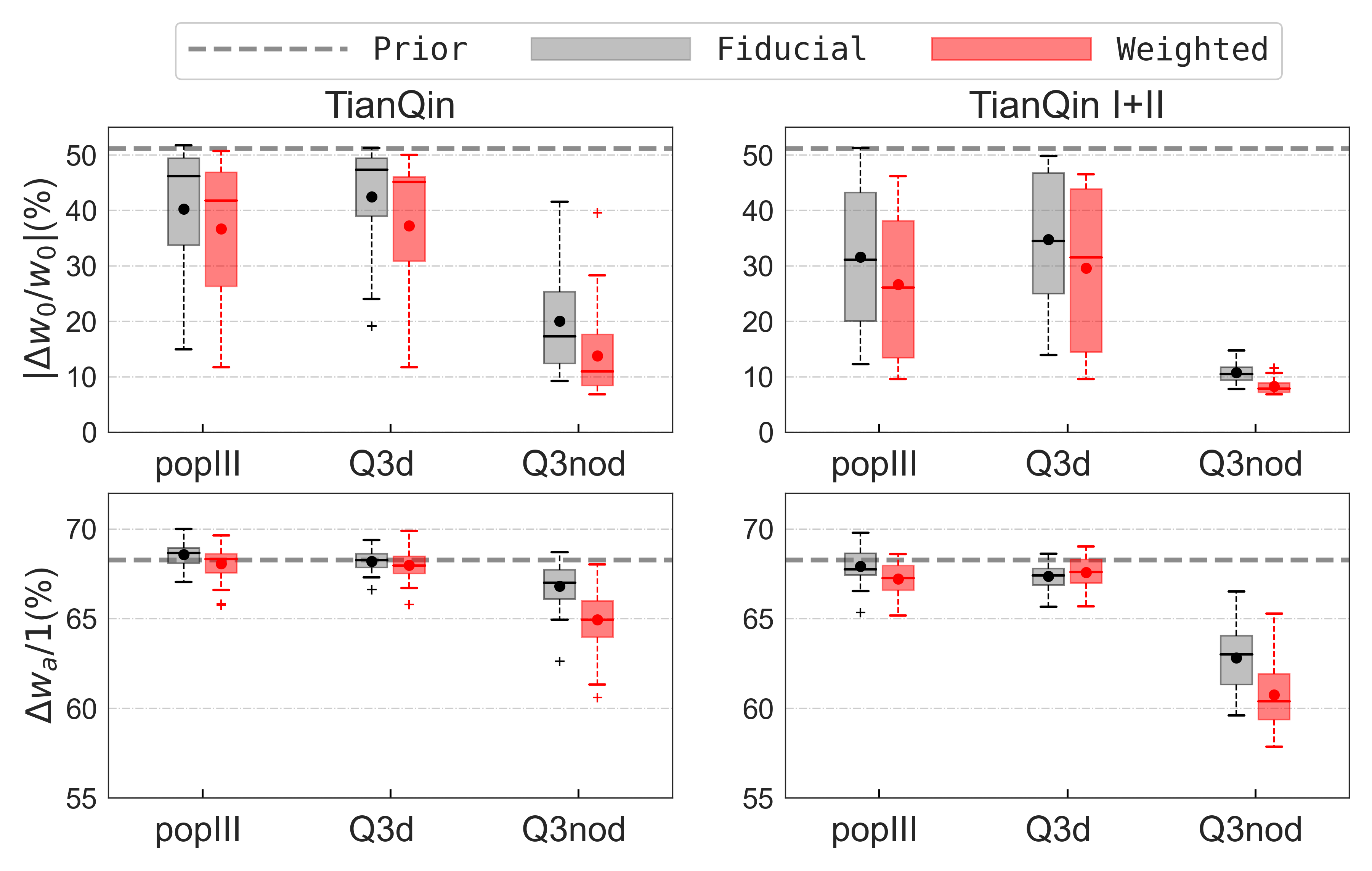}
\caption{Same as FIG. \ref{fig:cosmo_TQ_H0ML_MBHB}, but for the constraints from \acp{MBHB} on
dark energy EoS parameters $w_0$ and $w_a$ \cite{Zhu:2021aat}.}
\label{fig:cosmo_TQ_w0wa_MBHB}
\end{figure*}

For the prospects of constraining the CPL model parameters $w_0$ and $w_a$ with \ac{MBHB} detections,
one follows the scenario setup of Section \ref{sec:LCDM} (also see the literature \cite{Zhu:2021aat})
and present the prospects under the bright and dark siren assumptions separately.
In the optimistic assumption that \acp{MBHB} behave as bright sirens, relying on the precise
luminosity distance information provided by the \ac{MBHB} detections and the precise redshift
information provided by the \ac{EM} counterpart observations, TianQin can achieve effective constraints
on $w_0$ and $w_a$ simultaneously under all three population models, pop III, Q3\_d, and Q3\_nod.
The expected constraining errors on $w_0$ by TianQin are about $12\%$, $14\%$, and $9\%$,
and the expected $\Delta w_a$ are about $0.66$, $0.67$, and $0.49$, respectively;
and TianQin I+II can improve the precisions of $w_0$ to about $11\%$, $12\%$, and $7\%$,
and reduce $\Delta w_a$ to about $0.62$, $0.65$, and $0.39$, respectively.

In the conservative assumption that \acp{MBHB} behave as dark sirens, the constraints
on $w_0$ and $w_a$ are much weaker. The errors of $1\sigma$ CI
for $w_0$ and $w_a$ constrained by TianQin and TianQin I+II
are shown in FIG. \ref{fig:cosmo_TQ_w0wa_MBHB}. One can find that TianQin and
TianQin I+II can only effectively constrain $w_0$ under the pop III and Q3\_d population models,
and effectively constrain $w_a$ only under the Q3\_nod model. This is because the \ac{MBHB} merger rates
predicted by the pop III and Q3\_d models are significantly lower than that of the Q3\_nod model.
Using the weighted method described in Section \ref{sec:sirens_imporve_z}
(also see the literature \cite{Zhu:2021aat}), which uses the $M_{\rm MBH}-L_{\rm bulge}$ relation to
weight the candidate host galaxies of \acp{MBHB}, the expected constraints of $w_0$ are
about $36\%$, $37\%$, and $14\%$ for TianQin, and $27\%$, $30\%$, and $8\%$ for TianQin I+II,
respectively, under the pop III, Q3\_d, and Q3\_nod population models.
Under the Q3\_nod model, TianQin and TianQin I+II can constrain $w_a$ to errors
of $\Delta w_a = 0.65$ and $\Delta w_a = 0.60$, respectively.
Finally, for clarity, the expected precisions for constraining $w_0$ and $w_a$
by TianQin using \acp{EMRI} and \acp{MBHB} without \ac{EM} counterparts are listed in Table \ref{tab:w0wa_errs}.

\subsubsection{Selections on dark energy models} \label{sec:cosmo_de_model}

\begin{figure*}[ht]
\centering
\includegraphics[width=0.460\textwidth]{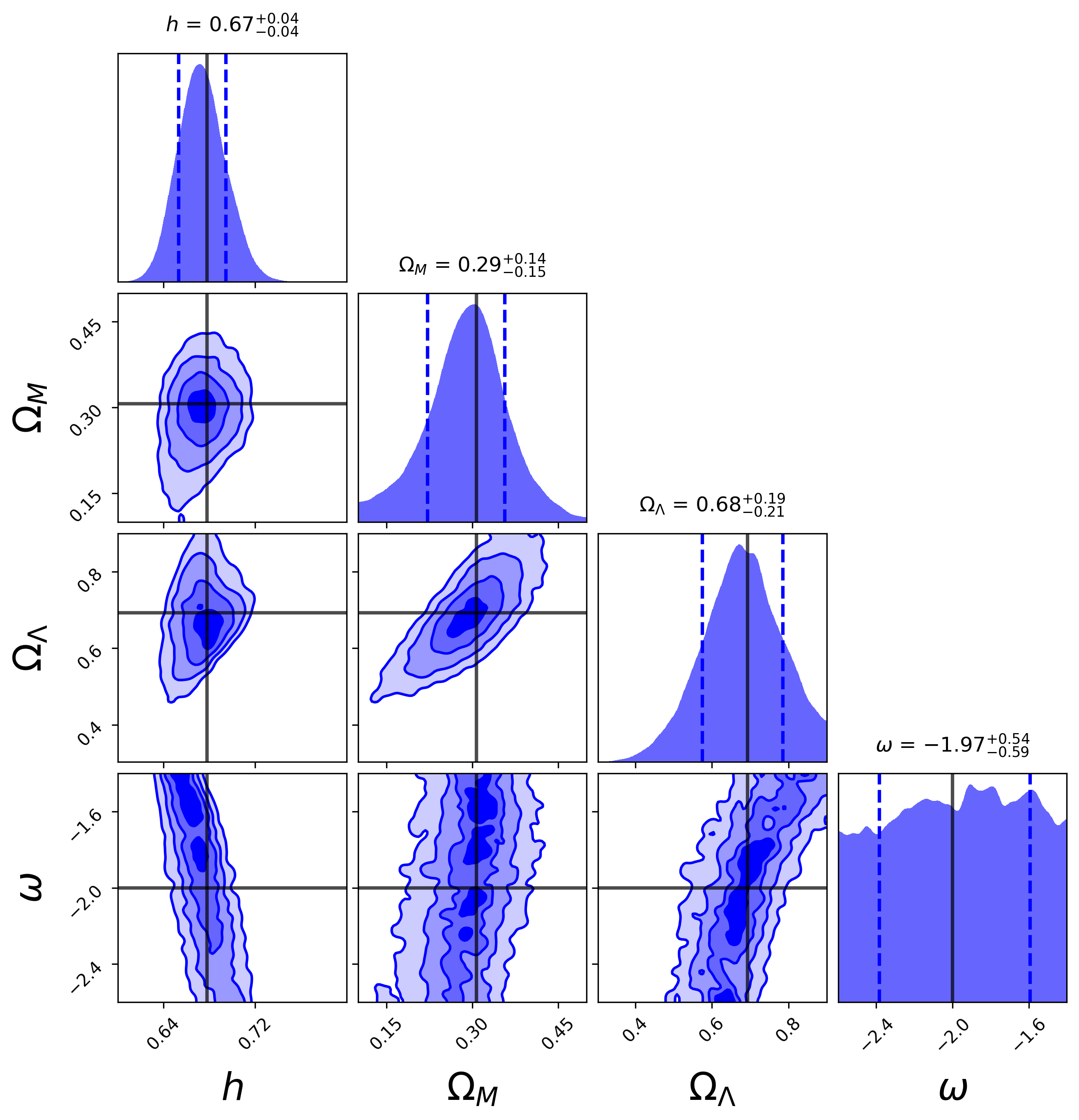} ~~~~
\includegraphics[width=0.460\textwidth]{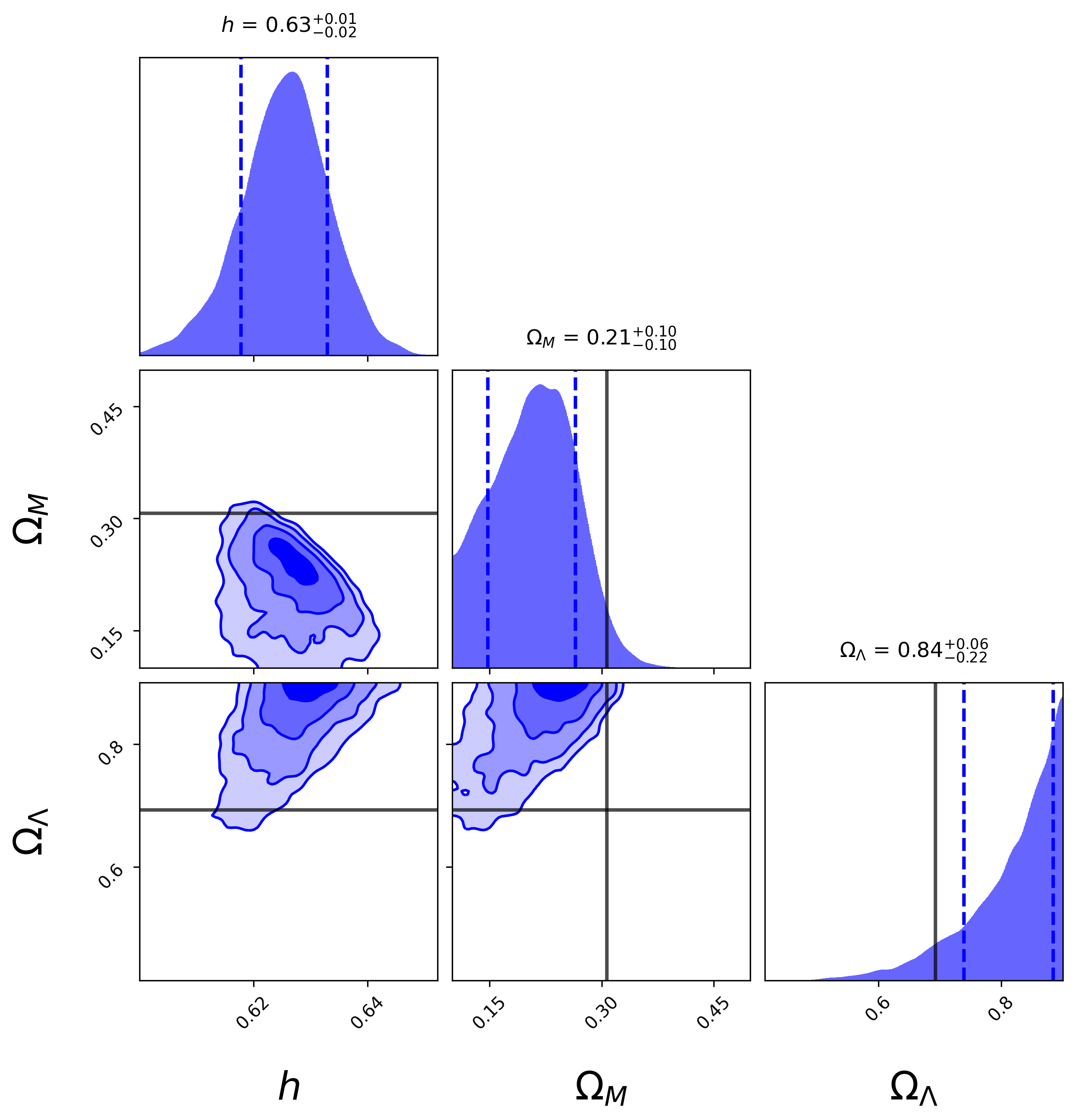}
\caption{Typical constraints on $(h, \Omega_M, \Omega_{\Lambda}, w)$ (left)
and $(h, \Omega_M, \Omega_{\Lambda})$ (right) \cite{Zhu:2022gwc}.
The results in the left and right subfigures constrained from the same mock dataset,
which is the dark \acp{MBHB} detected by TianQin I+II.
Particularly, the true value of the dark energy EoS
injected when simulating the data is $w=-2$.
In each subfigure, the solid black lines mark the true values of the parameters,
and the top panel of each column shows the marginalized result for
the corresponding parameter with an error of $1\sigma$ CI.
The logarithmic Bayes factor
$\ln B_{01} \equiv \ln \!\left[ P(\mathcal{D}_{\rm GW}, \mathcal{D}_{\rm EM}| \Lambda{\rm CDM}, I)
/ P(\mathcal{D}_{\rm GW}, \mathcal{D}_{\rm EM}| w{\rm CDM}, I) \right]$ for the constraints of
right and left subfigures is $\ln B_{01} \approx -2.8$.}
\label{fig:cosmo_TQ2_H0MLw}
\end{figure*}

This section will present the necessity and prospects of TianQin for the analysis of cosmological model selection.
The necessity of conducting a model selection analysis is reflected in two aspects,
the first is the large uncertainty in the precision of the dark energy EoS constrained by TianQin,
and the second is that fitting the data with an improper model
would lead to systematic biases in the parameter estimations.
For the first aspect, the previous section demonstrated that TianQin can achieves
seemingly precise constraint on the dark energy EoS,
but the constraining error varies greatly under different population models.
Furthermore, the results presented in the previous section were all obtained when the $\Lambda$CDM
parameters were fixed to their true values, and if the uncertainties of the $\Lambda$CDM parameters
is taken into account, i.e., $H_0$, $\Omega_M$ and $\Omega_{\Lambda}$ are also taken as free parameters
to be constrained together, then the constraining precision of the dark energy EoS parameters
will become even worse. The left subfigure of FIG. \ref{fig:cosmo_TQ2_H0MLw} illustrates a typical result
of TianQin I+II using \ac{MBHB} dark sirens to simultaneously constrain the four parameters, $\Lambda$CDM parameters
$H_0$, $\Omega_M$, and $\Omega_{\Lambda}$, and dark energy EoS $w$.
One can see that one has almost no constraining capability on $w$ in this case.
For the second aspect, the results of the parameter estimations obtained by fitting the data
with the $\Lambda$CDM model (corresponding to a dark energy EoS of $w \equiv -1$) is
illustrated in the right subfigure of FIG. \ref{fig:cosmo_TQ2_H0MLw}, when the true value of
the dark energy EoS is $w = -2$. One can find that there are significant systematic
biases in the estimations of each $\Lambda$CDM parameter.

\begin{figure*}[htbp!]
\centering
\includegraphics[width=0.60\textwidth]{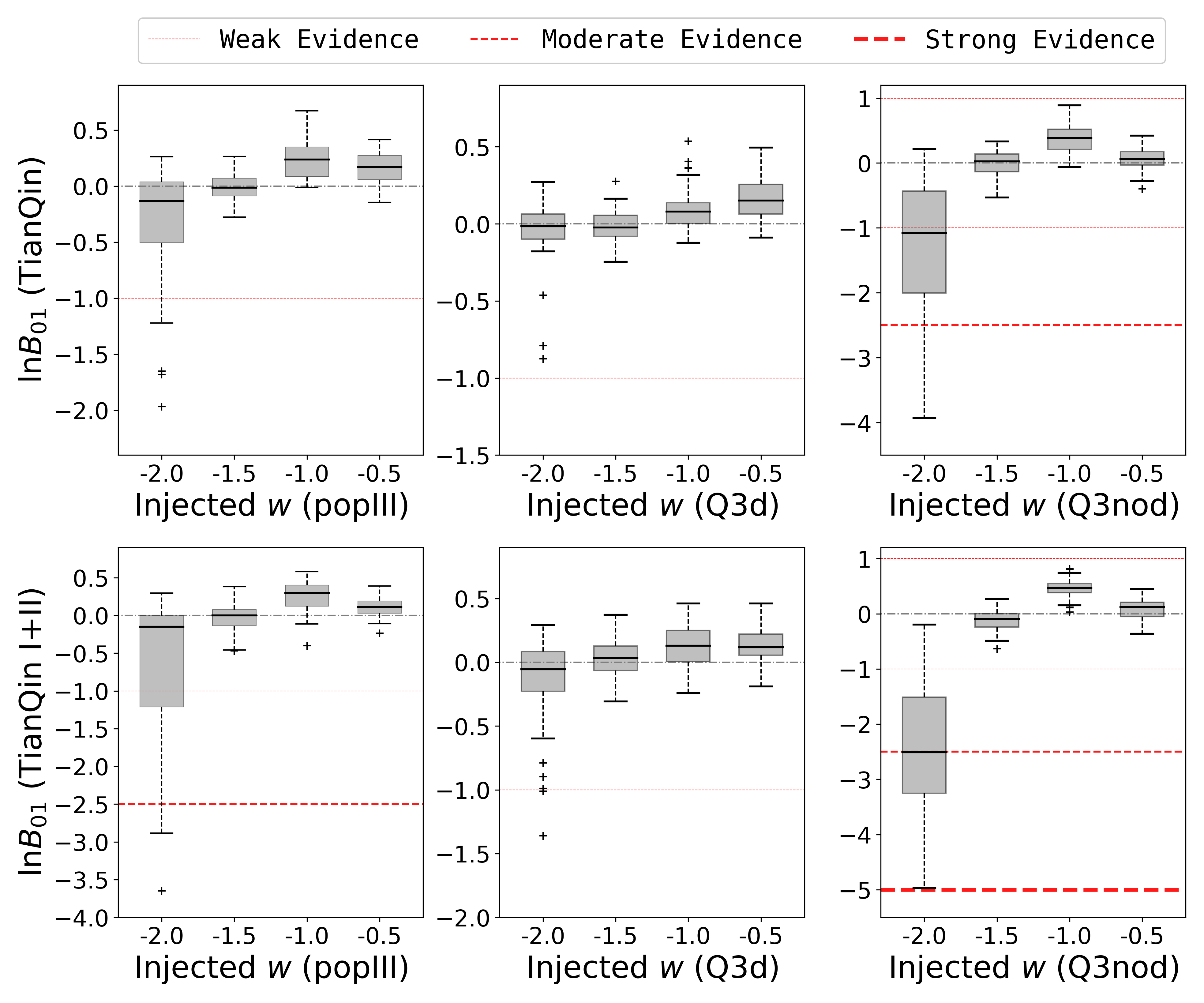}
\caption{Distributions of Bayes factors for $\Lambda$CDM and $w$CDM models
conditional on different values of dark energy EoS $w$ \cite{Zhu:2022gwc}. The data are derived from
\ac{MBHB} detections and assume that \acp{MBHB} are bright sirens. The top and bottom rows
represent the results for TianQin and TianQin I+II, respectively,
and the left, middle, and right columns represent the results in the pop III, Q3\_d,
and Q3\_nod population model conditions, respectively.
The thin, medium, and think dashed red lines represent the weak evidence, moderate evidence,
and strong evidence for the strength of model support, respectively.}
\label{fig:cosmo_TQ_BF_LCDM_MBHB}
\end{figure*}

\begin{figure*}[htbp!]
\centering
\includegraphics[width=0.60\textwidth]{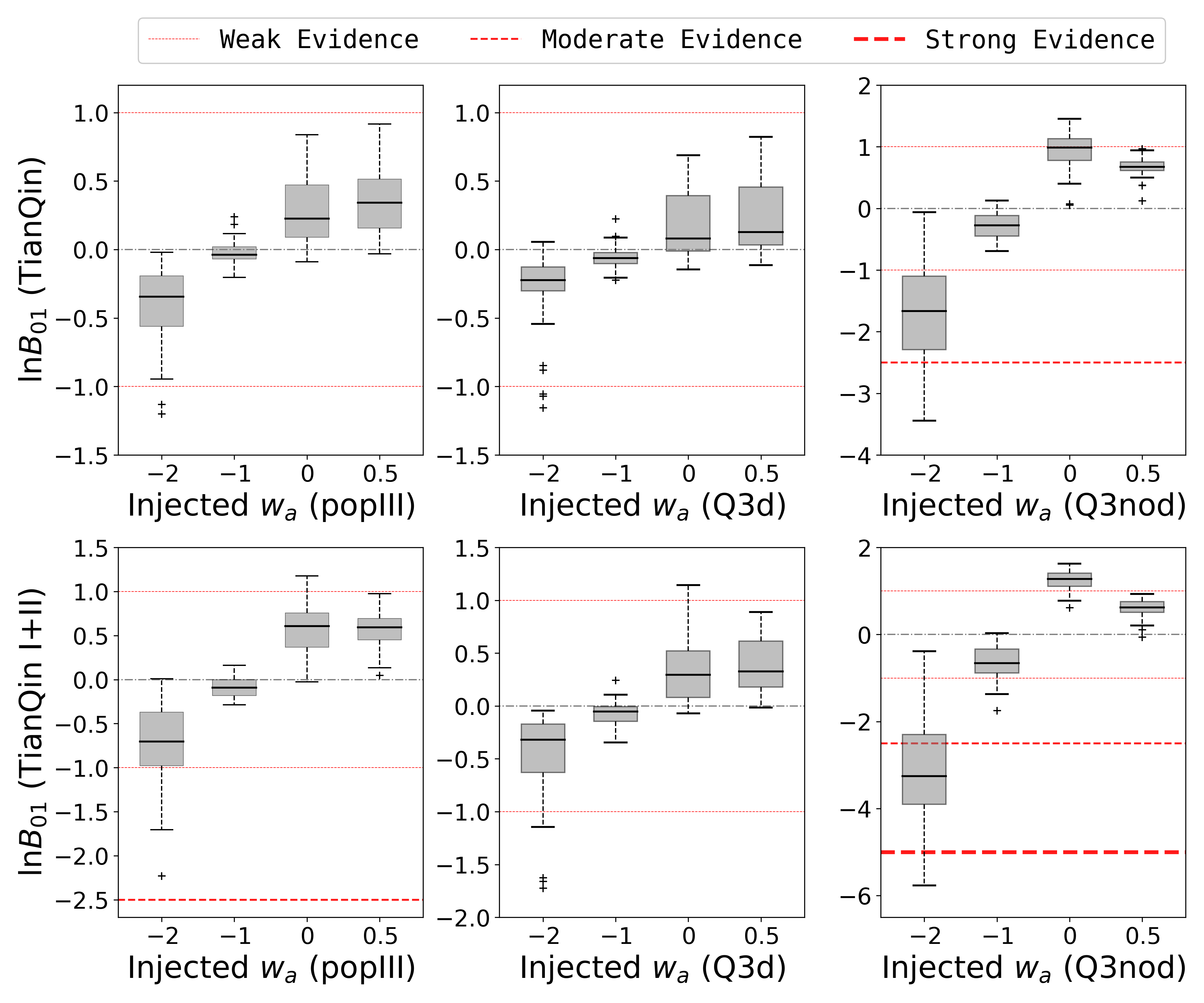}
\caption{Same as FIG. \ref{fig:cosmo_TQ_BF_LCDM_MBHB}, but the logarithmic Bayes factors are
for the $w$CDM and $w_0 w_a$CDM models \cite{Zhu:2022gwc}.}
\label{fig:cosmo_TQ_BF_w0waCDM_MBHB}
\end{figure*}

An optimistic scenario is considered as a preliminary analysis, i.e., \acp{MBHB} as bright sirens.
A Bayes factor described in Section \ref{sec:sirens_principle} is used to quantitatively
compare the degree of support of detection data for different cosmological models.
Following the literature \cite{Zhu:2022gwc}, one carry out the analysis in two steps,
the first focusing on exploring the conditions under which
a variable dark energy EoS parameter $w$ needs to be introduced,
one that is free to vary rather than
being constantly equal to negative one to account for the data, and
and the second focuses on the conditions under which a parameter $w_a$
needs to be introduced, which carves out the evolution of the dark energy EoS
with respect to redshift to account for the data.

In the first step, the two cosmological models one consider are $\Lambda$CDM and $w$CDM.
The free parameters of the $\Lambda$CDM model include three in total, $H_0$, $\Omega_M$,
and $\Omega_{\Lambda}$, and the free parameters of the $w$CDM model include four in total,
$H_0$, $\Omega_M$, $\Omega_{\Lambda}$, and $w$.
When simulating the data one set the fiducial model to be $w$CDM, the simulations are divided into
four groups, each group of simulations to keep $h=0.678$, $\Omega_M=0.307$, and
$\Omega_{\Lambda}=0.693$ unchanged, and only change the value of $w$, the four injected values
of $w$ are set as $w = \{ -2, -1.5, -1, -0.5 \}$.
Setting the logarithmic Bayes factor for this step to
$\ln B_{01} \equiv \ln \!\left[ P(\mathcal{D}_{\rm GW}, \mathcal{D}_{\rm EM}| \Lambda{\rm CDM}, I)
/ P(\mathcal{D}_{\rm GW}, \mathcal{D}_{\rm EM}| w{\rm CDM}, I) \right]$,
and the distributions of Bayes factors for the four groups of simulations are
illustrated in FIG. \ref{fig:cosmo_TQ_BF_LCDM_MBHB}.
One adopts a modified ``Jeffreys' scale'' as a judgment of the support strength of the data
for the models \cite{Jeffreys:1961book, Trotta:2008qt},
the modified Jeffreys' scale has a total of three levels of strength of evidence, namely
``weak evidence'', ``moderate evidence'', and ``strong evidence'',
the critical values of the Bayes factor for the three levels of strength of evidence are
$1 \leq |\ln B_{01}|$, $2.5 \leq |\ln B_{01}|$, and $5 \leq |\ln B_{01}|$, respectively,
and when $|\ln B_{01}| < 1$ represents that the data do not have a clear preference
for the two models \cite{Trotta:2008qt}.
From FIG. \ref{fig:cosmo_TQ_BF_LCDM_MBHB} one can see that TianQin can only obtain a weak evidence
for supporting the $w$CDM model when $w = -2$ and under the Q3\_nod model,
and TianQin I+II can increase the strength of support to a moderate evidence.
When there is a moderate evidence of support for a particular model in the data,
if one tries to fit the data with an unsupported model, there is already a significant
systematic bias, as illustrated in FIG. \ref{fig:cosmo_TQ2_H0MLw}.

In the second step, the two cosmological models one consider are $w$CDM and $w_0 w_a$CDM.
The free parameters of the $w_0 w_a$CDM model include five in total, $H_0$, $\Omega_M$,
$\Omega_{\Lambda}$, and two CPL dark energy EoS parameters $w_0$ and $w_a$.
Here one set the fiducial model as $w_0 w_a$CDM to simulate data, the simulations continues to
be divided into four groups, each group of simulations to keep $h=0.678$, $\Omega_M=0.307$,
$\Omega_{\Lambda}=0.693$ and $w_0 = -1$ unchanged, and only change the value of $w_a$,
the four injected values of $w_a$ are set as $w_a = \{ -2, -1, \!~0, \!~0.5 \}$.
Setting the logarithmic Bayes factor to
$\ln B_{01} \equiv \ln \!\left[ P(\mathcal{D}_{\rm GW}, \mathcal{D}_{\rm EM}| w{\rm CDM}, I)
/ P(\mathcal{D}_{\rm GW}, \mathcal{D}_{\rm EM}| w_0 w_a{\rm CDM}, I) \right]$,
and the distributions of $\ln B_{01}$ for this four groups of simulations are
illustrated in FIG. \ref{fig:cosmo_TQ_BF_w0waCDM_MBHB}.
One can find that, similarly to the first step, TianQin can obtain a weak evidence for
the support on the $w_0 w_a$CDM model only under the conditions of $w_a = -2$ and in the Q3\_nod model,
and TianQin I+II can increase the strength of the support to a moderate evidence.

In general, the potential of TianQin to discriminate between various cosmological models is mainly
limited by the number of \ac{MBHB} detections. TianQin can be expected to go to a better model selection
capability by adding \acp{EMRI}' data. In addition, the inclusion of \acp{SBHB}' data is also expected to
improve the model selection capability, although \acp{SBHB} can only provide effective constraint on $H_0$,
the improvement in $H_0$ precision can break the degeneracy between individual parameters
and thus improve the constraints on the whole cosmological parameters.
The authors leave the analysis of TianQin's model selection capability with the inclusion
of \ac{EMRI} and \ac{SBHB} data for future researches.

\subsection{Improvements from multiple detectors} \label{sec:cosmo_TQLISA}

This section will present the usefulness of multiple space-based \ac{GW} detectors,
such as LISA \cite{LISA:2017pwj, Colpi:2024xhw}, TianQin \cite{TianQin:2015yph, TianQin:2020hid},
and Taiji \cite{Hu:2017mde, TaijiScientific:2021qgx}, to form a network to improve
the constraints on the $\Lambda$CDM model and CPL dark energy model parameters.
This section uses TianQin as the basis to discuss the improvements of cosmological inferences
by a network composed of TianQin and LISA. Since both TianQin and LISA detectors are
planned to be launched around 2035 \cite{LISA:2017pwj, TianQin:2020hid},
it is very likely that the TianQin+LISA network will materialize.
Moreover, considering that LISA and Taiji have similar spatial configurations and sensitivities,
the performance of TianQin+Taiji network \cite{Gong:2021gvw} can be referred to the TianQin+LISA network.
And the cosmological forecasts of a network of LISA plus Taiji are beyond the scope of this paper
and can be found in the literatures \cite{Wang:2020dkc, Wang:2021srv, Jin:2023sfc}.
In addition, this section will present the prospects for cosmological constraints
from TianQin \ac{GW} detections in combination with other cosmological probes.

\subsubsection{LISA forecasts for $\Lambda$CDM and dark energy EoS}

LISA and TianQin have very close detection bands and sensitivities, the prospects presented in
the previous two sections, i.e., Sections \ref{sec:LCDM} and \ref{sec:cosmo_de}, for TianQin to
constrain the parameters of the $\Lambda$CDM cosmological model and CPL dark energy model
were similarly analyzed by LISA at earlier times.
The candidate standard sirens of LISA are the same three classes of \ac{GW} sources, \acp{SBHB}, \acp{EMRI},
and \acp{MBHB}, as for TianQin, except that LISA and TianQin differ in their capabilities to
detect these three classes of \ac{GW} sources and in their prospects for inferring cosmology
using these three classes of \ac{GW} sources
(see the literatures \cite{LISA:2022yao} and \cite{LISACosmologyWorkingGroup:2022jok}
for astrophysical and cosmological reviews of LISA).

For \acp{SBHB}, the literatures \cite{DelPozzo:2017kme} and \cite{Muttoni:2021veo}
report the prospects of measuring $H_0$ when LISA alone and LISA united with \ac{ET} to
form a multi-band network for detections, respectively.
The forecasting precision of $H_0$ from LISA reported in the literature \cite{DelPozzo:2017kme}
can reach the level of a few percent, which looks to be quite a bit better than
the precision of $H_0$ from TianQin reported in the literature \cite{Zhu:2021bpp},
because the literature \cite{DelPozzo:2017kme} used an earlier and
more optimistic \ac{SBHB} population models \cite{Kyutoku:2016ppx}, whereas this analysis of TianQin
was based on the most recent population model from \ac{LVK}'s GWTC-3 \cite{KAGRA:2021duu}.
Based on a similar \ac{SBHB} population model of \ac{LVK},
the literature \cite{Muttoni:2021veo} reports that through a multi-band network composed of LISA and ET,
LISA is able to constrain $H_0$ to a precision of about 2\% as well as
$\Omega_M$ to a precision of about 30\% through 4 years of detection data,
which are similar to the results of TianQin.

For \acp{EMRI} and \acp{MBHB}, the expected precisions of various cosmological parameters constrained by LISA
are reported in the literatures \cite{Laghi:2021pqk} and \cite{Tamanini:2016zlh}, respectively.
According to the reports of the literature \cite{Laghi:2021pqk}, LISA can be expected to achieve
about 2.5\% precision for constraining $H_0$,
about 20\% precision for constraining $\Omega_M$ and
about 10\% precision for constraining dark energy EoS parameter $w_0$
using 4 years of \ac{EMRI} detections under the fiducial population model, M1 model.
As reported in the literature \cite{Tamanini:2016zlh}, LISA through \ac{MBHB} detections holds promise
to achieve constraints close to 1\% precision level for $H_0$,
better than 10\% precision level for $\Omega_M$,
and a precision of about 20\% for $w_0$.
It is likely that the detection capabilities of LISA for
these two classes of \ac{GW} sources (compare \cite{Babak:2017tow} and \cite{Fan:2020zhy} for \acp{EMRI},
and \cite{Klein:2015hvg} and \cite{Wang:2019ryf} for \acp{MBHB}), as well as the capabilities of
utilizing them to constrain $H_0$, $\Omega_M$ and $w_0$, are indeed somewhat better than TianQin.
This is mainly due to the fact that the \ac{GW} signals of \acp{EMRI} and \acp{MBHB} are more concentrated in
the frequency bands where LISA is more sensitive than TianQin.

For clarity, the precisions of $H_0$, $\Omega_M$ and $\Omega_\Lambda$
in Table \ref{tab:LCDM_errs} and the precisions of $w_0$ and $w_a$ are listed in Table \ref{tab:w0wa_errs},
for LISA constraints on the $\Lambda$CDM and CPL dark energy models using \acp{SBHB}, \acp{EMRI} and \acp{MBHB}
without \ac{EM} counterparts.
To summarize, LISA has very impressive capabilities in inferring
the cosmic expansion history using \acp{SBHB}, \acp{EMRI}, and \acp{MBHB}.
However, TianQin will also play crucial roles in \ac{SBHB}, \ac{EMRI}, and \ac{MBHB} detections,
as reported in the literatures \cite{Liu:2020eko, Wang:2019ryf, Fan:2020zhy, Zhu:2021bpp, Zhu:2021aat,
Torres-Orjuela:2023hfd, Zhu:2024qpp} and previous tow sections (i.e., Sections \ref{sec:LCDM} and \ref{sec:cosmo_de}),
and the relationship between TianQin and LISA should not be purely competitive,
but more importantly cooperative, i.e., forming a multi-detector network.
The role of multi-detector networks in improving cosmological inference
will be presented in next two subsections.

\subsubsection{Constraints on $\Lambda$CDM from TianQin+LISA}

According to the reports in the literatures \cite{Zhu:2021bpp, Zhu:2021aat, Zhu:2024qpp},
there are three main reasons why a multi-detector network composed of TianQin and LISA
can significantly improve the capability of constraining the cosmic expansion history.
First, the TianQin+LISA network can significantly improve the spatial localization precisions
of various classes of \ac{GW} sources compared to a single detector detecting alone, e.g.,
the TianQin+LISA can reduce the spatial localization errors of \acp{SBHB}
by several times \cite{Liu:2020eko, Liu:2021yoy, Zhu:2021bpp},
and the spatial localization errors of \acp{EMRI} and \acp{MBHB} by several times or
even one to two orders of magnitude \cite{Zhu:2021aat, Zhu:2024qpp}.
Second, the more precise sky localizations provided by the TianQin+LISA network can improve the chances of finding the \ac{EM} counterparts of \ac{GW} sources.
Third, the TianQin+LISA network can increase the detection number of various classes of \ac{GW} sources, e.g.,
the TianQin+LISA can increase the detection number of \acp{SBHB} by about two times \cite{Liu:2020eko,
Liu:2021yoy, Zhu:2021bpp, Torres-Orjuela:2023hfd}, and
increase the detection numbers of \acp{EMRI} and \acp{MBHB} by several tens
of percent \cite{Zhu:2021aat, Zhu:2024qpp, Torres-Orjuela:2023hfd}.

\begin{figure*}[t]
\centering
\includegraphics[width=0.450\textwidth]{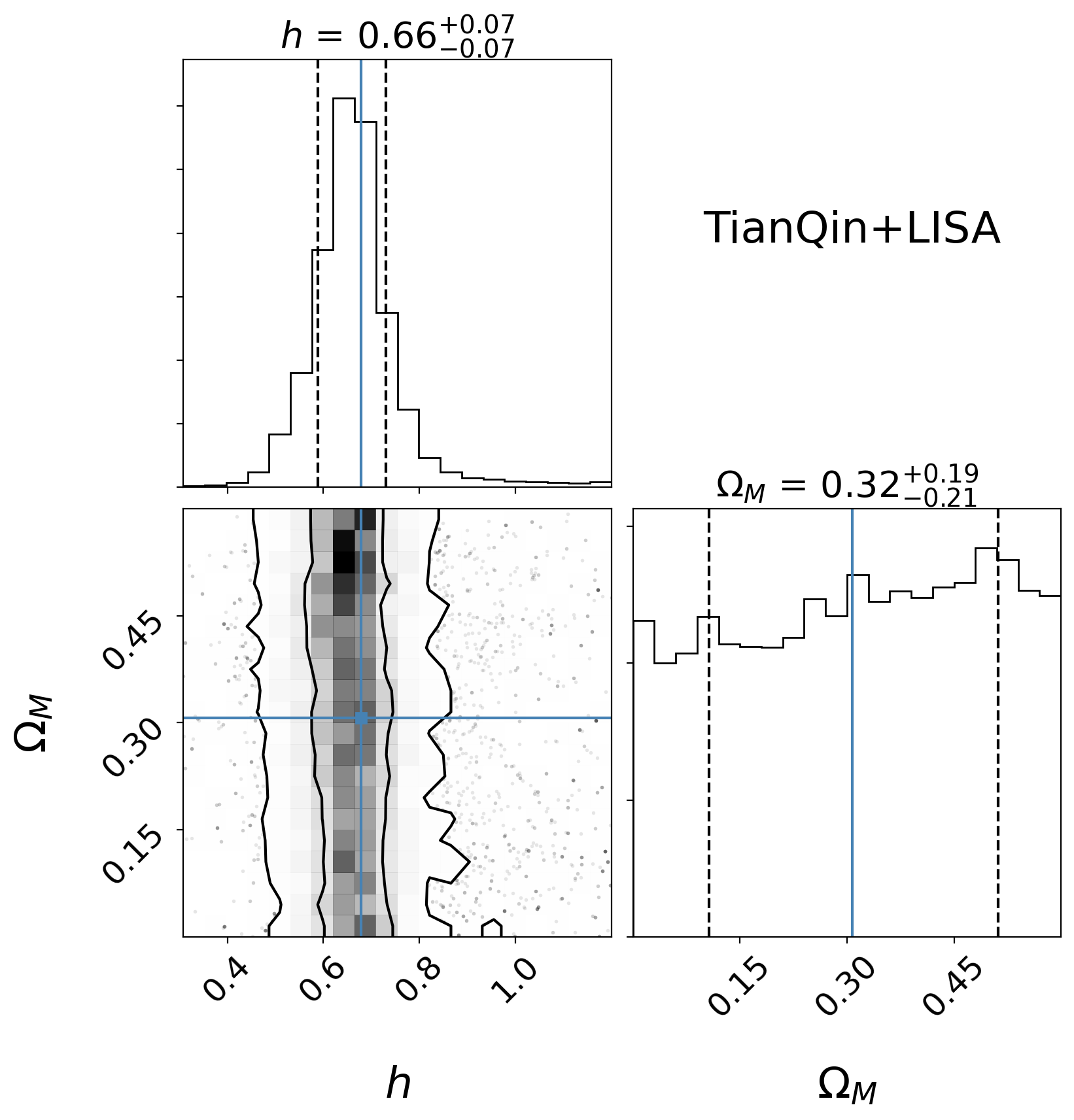} ~~~~
\includegraphics[width=0.450\textwidth]{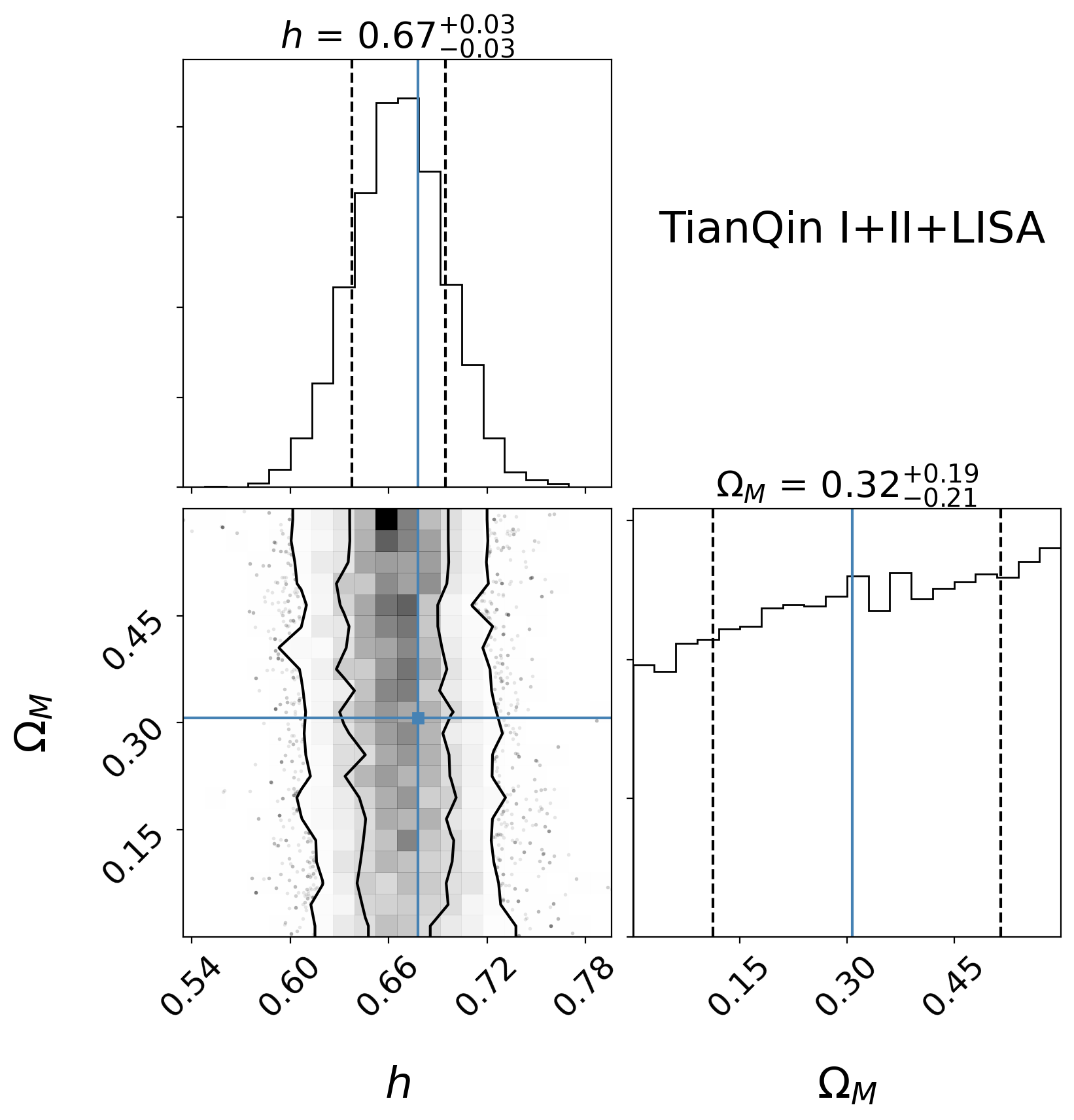}
\caption{Same as FIG. \ref{fig:cosmo_TQ_ET_H0}, but for the TianQin+LISA (left) and TianQin I+II+LISA (right) networks \cite{Zhu:2021bpp}.}
\label{fig:cosmo_TQ_LISA_H0}
\end{figure*}

As reported in the literature \cite{Zhu:2021bpp}, based on the \ac{SBHB} population model from \ac{LVK}'s GWTC-3,
TianQin+LISA network is expected to detect about 11 \ac{SBHB} inspiral \ac{GW} sources with ${\rm SNR} > 12$,
and TianQin I+II+LISA can increase the detection number of \acp{SBHB} with ${\rm SNR} > 12$ to about 18.
On the basis of such improvements, the expected results of constraining $H_0$ by the TianQin+LISA and
TianQin I+II+LISA networks using \ac{SBHB} detections are shown in FIG. \ref{fig:cosmo_TQ_LISA_H0}.
One can find that the TianQin+LISA network is able to constrain $H_0$ to a precision of about $10\%$,
while the TianQin I+II+LISA network can constrain $H_0$ to a precision of about $5\%$ \cite{Zhu:2021bpp}.
In comparison to the precisions from individual TianQin reported in Section \ref{sec:sec:LCDM_H0},
the multi-detector network composed of TianQin (I+II) and LISA reduces the error of $H_0$
by a factor of about three.

By the reports in the literature \cite{Zhu:2024qpp}, the constraining errors of $H_0$ and $\Omega_M$ by
the TianQin+LISA network for various \ac{EMRI} population models are shown in the red error bars
of FIG. \ref{fig:cosmo_TQ_H0M_EMRI}. One can notices that under most population models except M8,
the TianQin+LISA network achieves better than or close to precisions of $2\%$ for the constraints of $H_0$,
especially under the two models, M7 and M12, where the constraints of $H_0$ even achieve precisions of
better than 1\%. These precisions of precisions add confidence to clarifying the Hubble tension.
Besides, the TianQin+LISA network achieves effective constraints on $\Omega_M$ under most population models,
and precisions of better than 10\% are achieved under the M7 and M12 models.

\begin{figure*}[htbp!]
\centering
\includegraphics[width=0.70\textwidth]{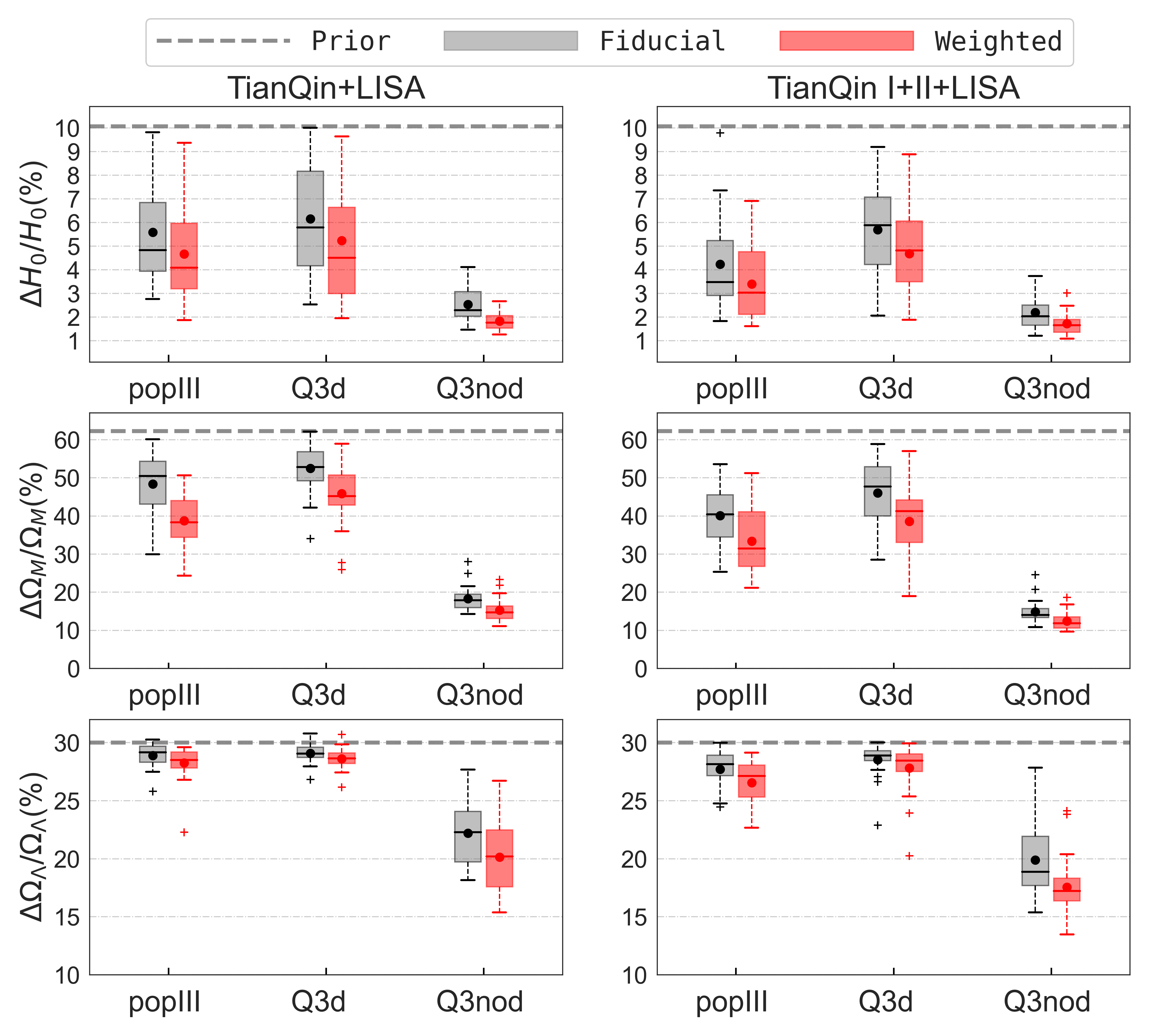}
\caption{Same as FIG. \ref{fig:cosmo_TQ_H0ML_MBHB}, but for the TianQin+LISA (left column) and
TianQin I+II+LISA (right column) networks \cite{Zhu:2021aat}.}
\label{fig:cosmo_TQLISA_H0ML_MBHB}
\end{figure*}

For \acp{MBHB}, considering both the optimistic and conservative scenarios (i.e., with and
without \ac{EM} counterparts) again, as reported in the literature \cite{Zhu:2021aat}.
In the optimistic scenario, the TianQin+LISA network is able to constrain $H_0$, $\Omega_M$ and
$\Omega_\Lambda$ parameters to precisions of about $1.3\% - 3.8\%$, $5\%-15\%$, and $11\% - 26\%$, respectively.
The worst and best precisions are obtained under the Q3\_d and Q3\_nod population models, respectively,
and the precisions obtained under the pop III model lie in the middle.
The TianQin I+II+LISA network can improve the constraints of $H_0$, $\Omega_M$ and $\Omega_\Lambda$
to about $1.2\% - 3.6\%$, $4\%-13\%$, and $10\% - 25\%$ precisions, respectively.
In the conservative scenario, the distributions of precisions on $H_0$, $\Omega_M$ and $\Omega_\Lambda$
constrained by the TianQin+LISA and TianQin I+II+LISA networks under the pop III, Q3\_d, and Q3\_nod models
are illustrated in FIG. \ref{fig:cosmo_TQLISA_H0ML_MBHB}.
One can see that for $H_0$ and $\Omega_M$, the TianQin+LISA or TianQin I+II+LISA network achieves
effective constraints under all three models, but for $\Omega_\Lambda$, TianQin+LISA or TianQin I+II+LISA can
only achieve effective constraint under the Q3\_nod model.
By using the weighted method that the authors describe in Section \ref{sec:sirens_imporve_z}, i.e.,
weighting the candidate host galaxies of \acp{MBHB} with the $M_{\rm MBH}-L_{\rm bulge}$ relation,
the TianQin+LISA (TianQin I+II+LISA) network is able to achieve expected precisions
of about $4.7\% ~\!(3.5\%)$, $5.2\% ~\!(4.7\%)$, and $1.8\% ~\!(1.7\%)$ for $H_0$ constraints,
and about $39\% ~\!(33\%)$, $46\% ~\!(39\%)$, and $15\% ~\!(12\%)$ for $\Omega_M$ constraints
under the pop III, Q3\_d, and Q3\_nod models, respectively.
In particular, the precisions of constraining $\Omega_\Lambda$ for TianQin+LISA and
TianQin I+II+LISA are about 20\% and 18\% under the Q3\_nod model, respectively.
Note that the expected precisions of constraining $H_0$, $\Omega_M$ and $\Omega_{\Lambda}$
by the TianQin+LISA network using \acp{SBHB}, \acp{EMRI} and \acp{MBHB} without \ac{EM} counterparts
are listed in Table \ref{tab:LCDM_errs}, for clearer comparisons with the precisions of
individual detectors on their own.

In summary, regardless of the class of standard sirens, the TianQin+LISA network significantly
improves the capability of constraining the $\Lambda$CDM parameters relative to
the individual detections of TianQin and LISA. For \acp{SBHB}, where the population properties are
relatively certain, the TianQin+LISA network is able to constrain $H_0$ to a precision of about 3\%
without relying on ground-based \ac{GW} detectors, while for the two classes of \ac{GW} sources, \acp{EMRI} and \acp{MBHB},
where the population properties are very uncertain, the TianQin+LISA network significantly increases
the probability of obtaining precise $H_0$ constraint. Therefore the TianQin+LISA network can serve
as a candidate solution for the utilizing space-based \ac{GW} detections to independently provide
$H_0$ measurements that contribute to the clarification of the Hubble tension.

\begin{table*}[ht]
 \caption{Forecasts of constraining $\Lambda$CDM parameters $(H_0, \Omega_M, \Omega_\Lambda)$ with \acp{SBHB}, \acp{EMRI}, \acp{MBHB} for TianQin, LISA and the TianQin+LISA network.
 The upper and lower bounds on the errors are the results obtained in the most optimistic and pessimistic population models, respectively.
 Particularly, here \acp{EMRI} include only the M1, M5, and M6 population models,
 in order to facilitate the comparison between TianQin and LISA.
 The sign ``$\cdots$'' means no effective constraint.}
 \vspace{3pt}
 \renewcommand\arraystretch{1.5}
 \centering
\begin{tabular}{c|c|m{0.25\textwidth}|m{0.25\textwidth}|m{0.25\textwidth}}
 \hline
 \hline
 \multirow{1}*{GW} & \multirow{1}*{Cosmological} & \multicolumn{3}{c}{Expected relative error} \\
 \cline{3-5}
 population & parameter & \multicolumn{1}{c|}{TianQin (+ET)} & \multicolumn{1}{c|}{LISA (+ET)} & \multicolumn{1}{c}{TianQin joint LISA} \\
 \hline
 ~ & \multirow{2}*{$H_0$} & \multicolumn{1}{c|}{$\sim 20\%$ \cite{Zhu:2021bpp}} & \multicolumn{1}{c|}{$\sim 5\%$ \cite{DelPozzo:2017kme}} & \multicolumn{1}{c}{$\sim 3\%$ \cite{Zhu:2021bpp}} \\
 \multirow{2}*{SBHBs} & ~ & \multicolumn{1}{c|}{($\sim 1.5\%$) \cite{Zhu:2021bpp}} & \multicolumn{1}{c|}{($\sim 2\%$) \cite{Muttoni:2021veo}} & \multicolumn{1}{c}{~} \\
 \cline{2-5}
 ~ & $\Omega_M$ & \multicolumn{1}{c|}{$\cdots ~(\sim 40\%)$ \cite{Zhu:2021bpp}} & \multicolumn{1}{c|}{$\cdots ~(\sim 30\%)$ \cite{Muttoni:2021veo}} & \multicolumn{1}{c}{$\cdots$} \\
 \cline{2-5}
 ~ & $\Omega_\Lambda$ & \multicolumn{1}{c|}{$\cdots$} & \multicolumn{1}{c|}{$\cdots$} & \multicolumn{1}{c}{$\cdots$} \\
 \cline{2-5}
 \hline
 \multirow{2}*{EMRIs} & $H_0$ & \multicolumn{1}{c|}{$\sim 5\%-8\%$ \cite{Zhu:2024qpp}} & \multicolumn{1}{c|}{$\sim 1\%-4\%$ \cite{Laghi:2021pqk}} & \multicolumn{1}{c}{$\sim 1\%-2\%$ \cite{Zhu:2024qpp}} \\
 \cline{2-5}
 \multirow{2}*{(M1,M5,M6)} & $\Omega_M$ & \multicolumn{1}{c|}{$\sim 45\% ~- >50\%$ \cite{Zhu:2024qpp}} & \multicolumn{1}{c|}{$\sim 30\%-60\%$ \cite{Laghi:2021pqk}} & \multicolumn{1}{c}{$\sim 15\%-26\%$ \cite{Zhu:2024qpp}} \\
 \cline{2-5}
 ~ & $\Omega_\Lambda$ & \multicolumn{1}{c|}{$\cdots$} & \multicolumn{1}{c|}{$\cdots$} & \multicolumn{1}{c}{$\cdots$} \\
 \hline
 \acp{MBHB} & $H_0$ & \multicolumn{1}{c|}{$\sim 3\%-7\%$ \cite{Zhu:2021aat}} & \multicolumn{1}{c|}{$\sim 2\%-7\%$ \cite{Zhu:2021aat}} & \multicolumn{1}{c}{$\sim 2\%-5\%$ \cite{Zhu:2021aat}} \\
 \cline{2-5}
 (popIII, Q3\_d, & $\Omega_M$ & \multicolumn{1}{c|}{$\sim 35\%-60\%$ \cite{Zhu:2021aat}} & \multicolumn{1}{c|}{$\sim 18\%-50\%$ \cite{Zhu:2021aat}} & \multicolumn{1}{c}{$\sim 12\%-40\%$ \cite{Zhu:2021aat}} \\
 \cline{2-5}
 Q3\_nod, w/o EM) & $\Omega_\Lambda$ & \multicolumn{1}{c|}{$> 30\%$ \cite{Zhu:2021aat}} & \multicolumn{1}{c|}{$\sim 22\%~- >30\%$ \cite{Zhu:2021aat}} & \multicolumn{1}{c}{$\sim 20\%~- >30\%$ \cite{Zhu:2021aat}} \\
 \hline
 \hline
\end{tabular}
 \label{tab:LCDM_errs}
\end{table*}

\subsubsection{Constraints on dark energy from TianQin+LISA}

The improvement effect of the multi-detector network composed of TianQin and LISA on
constraining the dark energy EoS is similar to that shown in the previous subsection
for constraining the $\Lambda$CDM.
The red error bars in FIG. \ref{fig:cosmo_TQ_w0wa_EMRI} show
the constraining errors on the CPL dark energy model parameters $w_0$ and $w_a$
by the TianQin+LISA network for various \ac{EMRI} population models
(also presented in the literature \cite{Zhu:2024qpp}).
One can see that the TianQin+LISA network is able to achieve precisions of better than 10\%
for the constraints on $w_0$ under most of the \ac{EMRI} population models except M8,
among which the precision of $w_0$ reaches better than 5\% under the M7 and
M12 models that have relatively optimistic \ac{EMRI} rates \cite{Babak:2017tow}.
Even under the pessimistic M8 model, the precision of $w_0$ is achieved to about 22\%.
Such level of precision is expected to provide a useful reference for determining
whether the dark energy EoS actually deviates from the cosmological constant.
In addition, for the parameter $w_a$, the TianQin+LISA network also achieves an effective constraint
with a precision of about 40\% under the two optimistic population models, M7 and M12.

\begin{figure*}[t]
\centering
\includegraphics[width=0.70\textwidth]{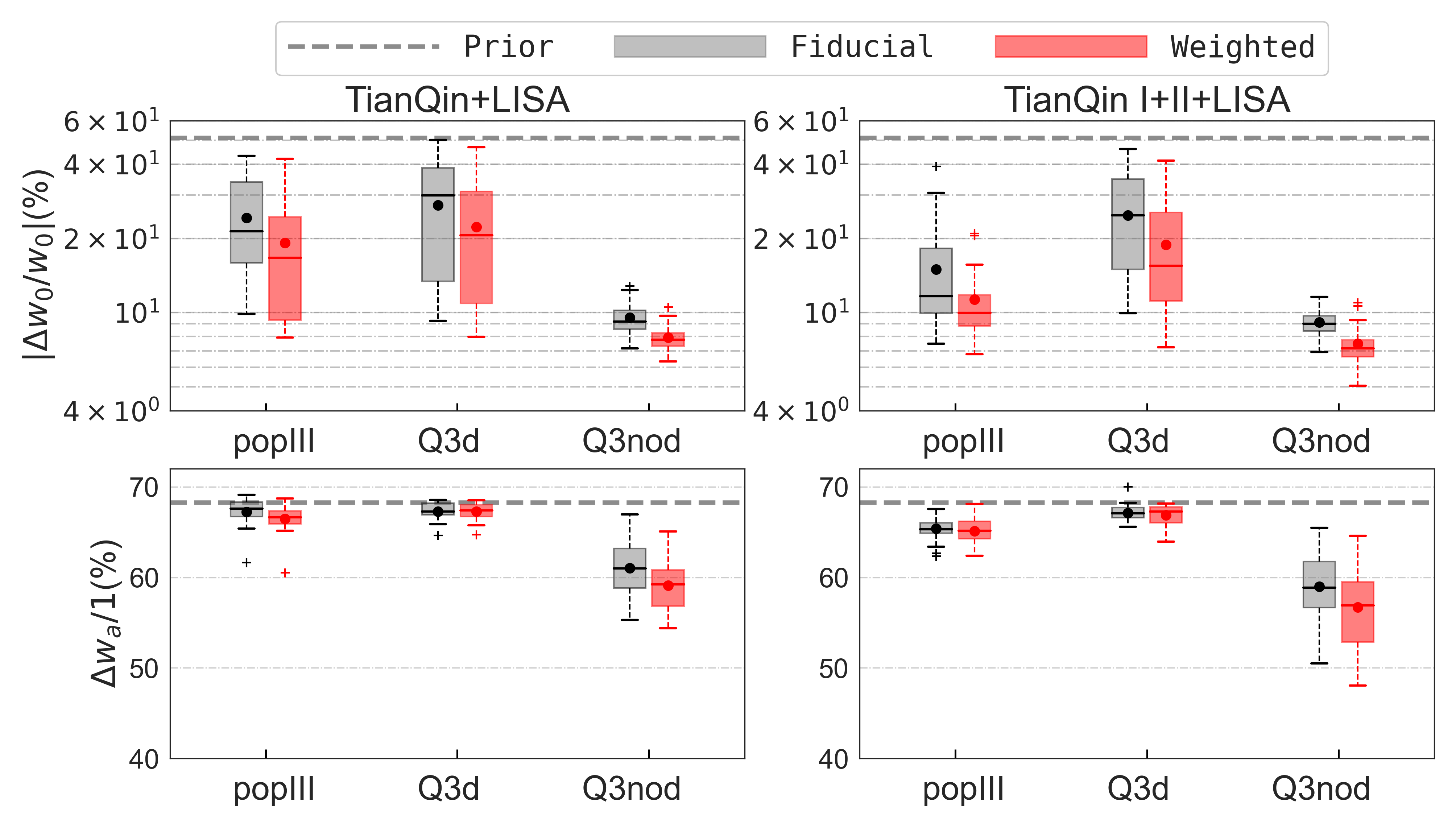}
\caption{Same as FIG. \ref{fig:cosmo_TQ_H0ML_MBHB}, but for the constraints on the CPL
dark energy EoS model parameters $(w_0, w_a)$ by the TianQin+LISA (left column) and
TianQin I+II+LISA (right column) networks \cite{Zhu:2021aat}.}
\label{fig:cosmo_TQLISA_w0wa_MBHB}
\end{figure*}

For \ac{MBHB} constraints on the dark energy equation of state, one follow the setup of
the previous subsection (also presented in the literature \cite{Zhu:2021aat})
and present the prospects in the optimistic and conservative scenarios separately.
In the optimistic scenario, the TianQin+LISA network is able to achieve about 10\%, 12\%,
and 7\% precision for the constraint on $w_0$ under the pop III, Q3\_d, and Q3\_nod population models,
respectively, and the TianQin I+II+LISA network can improve these precisions to about 9\%, 11\%,
and 6\%, respectively.
For the constraint on the parameter wa, the precisions that TianQin+LISA and TianQin I+II+LISA can obtain
are about 60\% under both the pop III and Q3\_d models, and about 35\% under the Q3\_nod model.
In the conservative scenario, both the TianQin+LISA and TianQin I+II+LISA networks can only
effectively constrain the parameter $w_0$ under both the pop III and Q3\_d models,
and only provide effective constraints on the two parameters $w_0$ and $w_a$ simultaneously
under the Q3\_nod model, as illustrated in FIG. \ref{fig:cosmo_TQLISA_w0wa_MBHB}.
Using the weighted method that weights the candidate host galaxies of \acp{MBHB} with
the $M_{\rm MBH}-L_{\rm bulge}$ relation,
Using a weighted approach, the expected precisions of the TianQin+LISA network
for $w_0$ constraints under the pop III, Q3\_d, and Q3\_nod models are about
$19\%$, $22\%$, and $8\%$, respectively, and the TianQin I+II+LISA network can improve
these precisions to about $11\%$, $19\%$, and $7\%$, respectively.
In particular, under the Q3\_nod model, the TianQin+LISA or TianQin I+II+LISA network can
provide constraining precision of about 60\% for $w_a$.

\begin{table*}[t]
 \caption{Forecasts of constraining CPL dark energy EoS parameters $(w_0, w_a)$ with \acp{EMRI} and \acp{MBHB} for TianQin, LISA and the TianQin+LISA network.
 The upper and lower bounds on the errors are the results obtained in the most optimistic and pessimistic population models, respectively.
 Particularly, here \acp{EMRI} include only the M1, M5, and M6 population models,
 in order to facilitate the comparison between TianQin and LISA.
 The sign ``$\cdots$'' means no effective constraint.}
 \vspace{3pt}
 \renewcommand\arraystretch{1.5}
 \centering
\begin{tabular}{c|c|m{0.25\textwidth}|m{0.25\textwidth}|m{0.25\textwidth}}
 \hline
 \hline
 \multirow{1}*{GW} & \multirow{1}*{Dark energy} & \multicolumn{3}{c}{Expected relative error} \\
 \cline{3-5}
 population & EoS parameter & \multicolumn{1}{c|}{TianQin} & \multicolumn{1}{c|}{LISA} & \multicolumn{1}{c}{TianQin joint LISA} \\
 \hline
 \acp{EMRI} & $w_0$ & \multicolumn{1}{c|}{$\sim 15\%-39\%$ \cite{Zhu:2024qpp}} & \multicolumn{1}{c|}{$\sim 7\%-12\%$ \cite{Laghi:2021pqk}} & \multicolumn{1}{c}{$\sim 6\%-7\%$ \cite{Zhu:2024qpp}} \\
 \cline{2-5}
 (M1,M5,M6) & $w_a$ & \multicolumn{1}{c|}{$\cdots$} & \multicolumn{1}{c|}{$\cdots$} & \multicolumn{1}{c}{$\Delta w_a \gtrsim 0.4$ \cite{Zhu:2024qpp}} \\
 \cline{2-5}
 \hline
 \acp{MBHB} (popIII, & $w_0$ & \multicolumn{1}{c|}{$\sim 14\%-37\%$ \cite{Zhu:2021aat}} & \multicolumn{1}{c|}{$\sim 8\%-30\%$ \cite{Zhu:2021aat}} & \multicolumn{1}{c}{$\sim 8\%-22\%$ \cite{Zhu:2021aat}} \\
 \cline{2-5}
 Q3\_d, Q3\_nod, w/o EM) & $w_a$ & \multicolumn{1}{c|}{$\Delta w_a > 0.6$ \cite{Zhu:2021aat}} & \multicolumn{1}{c|}{$\Delta w_a > 0.6$ \cite{Zhu:2021aat}} & \multicolumn{1}{c}{$\Delta w_a \gtrsim 0.6$ \cite{Zhu:2021aat}} \\
 \hline
 \hline
\end{tabular}
 \label{tab:w0wa_errs}
\end{table*}

Finally, for clearer comparisons with the precisions of individual detectors when they are detecting alone,
the expected precisions of constraining $w_0$ and $w_a$ by the TianQin+LISA network using
\acp{EMRI} and \acp{MBHB} without \ac{EM} counterparts are listed in Table \ref{tab:w0wa_errs}.
One can see that the multi-detector network composed of TianQin and LISA always provides a significant improvement in
the constraints on CPL model parameters compared to TianQin or LISA detecting alone.
Thus a multi-detector network can contribute more helpful constraint on dark energy EoS
for probing the nature of dark energy.

\subsubsection{Combination with other cosmological probes}
{\it Subsection coordinator: Xin Zhang}

As mentioned earlier, since \acp{GW} can directly measure absolute distances on cosmological scales, standard siren observations can provide excellent measurements of the Hubble-Lema\^itre constant. However, their effectiveness in measuring other cosmological parameters is somewhat limited. In particular, when using TianQin's \ac{GW} standard siren observations to measure the equation of state of dark energy, the resulting errors are significant. Yet, this does not imply that \ac{GW} standard sirens cannot play an important role in measuring the equation of state of dark energy. On the contrary, \ac{GW} standard siren observations will undoubtedly have a significant impact on dark energy research in the future. This is because \ac{GW} standard siren observations need to be combined with other cosmological probes to play a crucial role in cosmological research.

In terms of measuring dark energy, the best current combination of observational probes is \ac{CMB} + BAO + SN, where BAO represents \ac{BAO} data obtained through galaxy redshift surveys, and SN represents Type Ia supernova observation data. Although \ac{CMB} is the most accurate cosmological probe, as it is an early-universe observation, it cannot effectively measure the equation of state of dark energy alone (because dark energy begins to dominate the evolution of the universe only in the late times). In such cases, cosmological parameter degeneracy is very severe. To break this degeneracy, it is necessary to combine late-universe observations. Therefore, \ac{CMB} must be combined with BAO and SN to effectively constrain the equation of state of dark energy. Currently, using \ac{CMB} + BAO + SN (hereinafter abbreviated as CBS for convenience), the parameter $w$ (assuming the equation of state of dark energy is a constant) can be measured with a precision slightly less than 3\%; for the case of two parameters, $w_0$ and $w_a$, the current precision for $w_0$ is about 8\%, and the absolute error for $w_a$ is approximately 0.3. In the future, \ac{GW} standard siren observations can help significantly improve the measurement accuracy of the equation of state of dark energy.

\begin{figure*}[htb]
\begin{center}
\includegraphics[width=\linewidth]{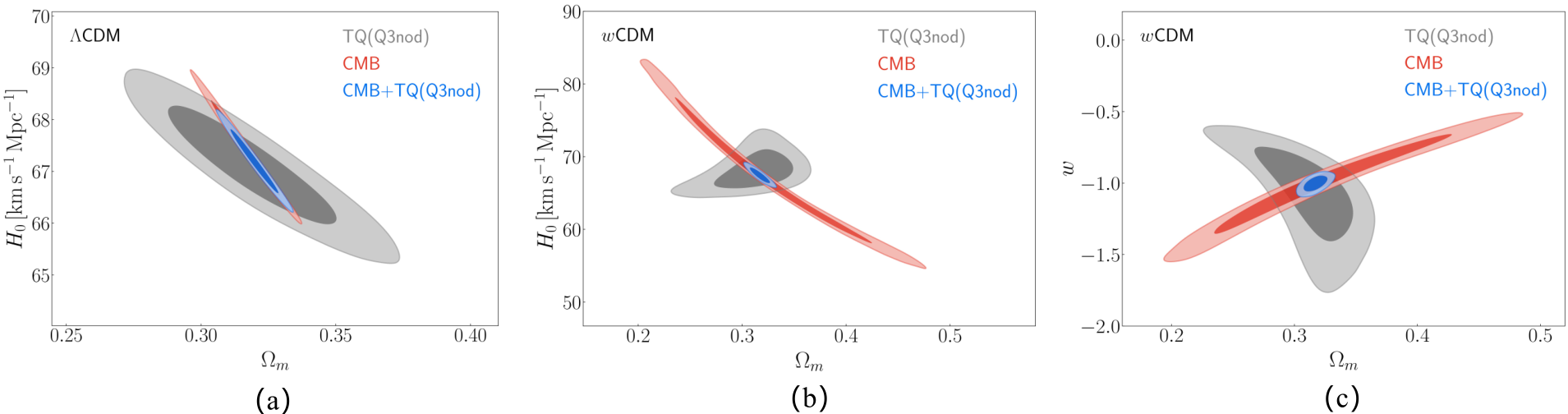}
\end{center}
\caption{Two-dimensional marginalized contours (68.3\% and 95.4\% confidence level) in the $\Omega_{m}$--$H_{0}$ plane for the $\Lambda$CDM model, in the $\Omega_{m}$--$H_{0}$ plane and $\Omega_{m}$--$w$ plane for the $w$CDM model by using the TianQin, CMB, and CMB+TianQin. Here, the TianQin mock data are simulated based on the Q3nod model. This figure is taken from \cite{Wang:2019tto}.} \label{TQCMB}
\end{figure*}

\begin{figure*}[htb]
\begin{center}
\includegraphics[width=\linewidth]{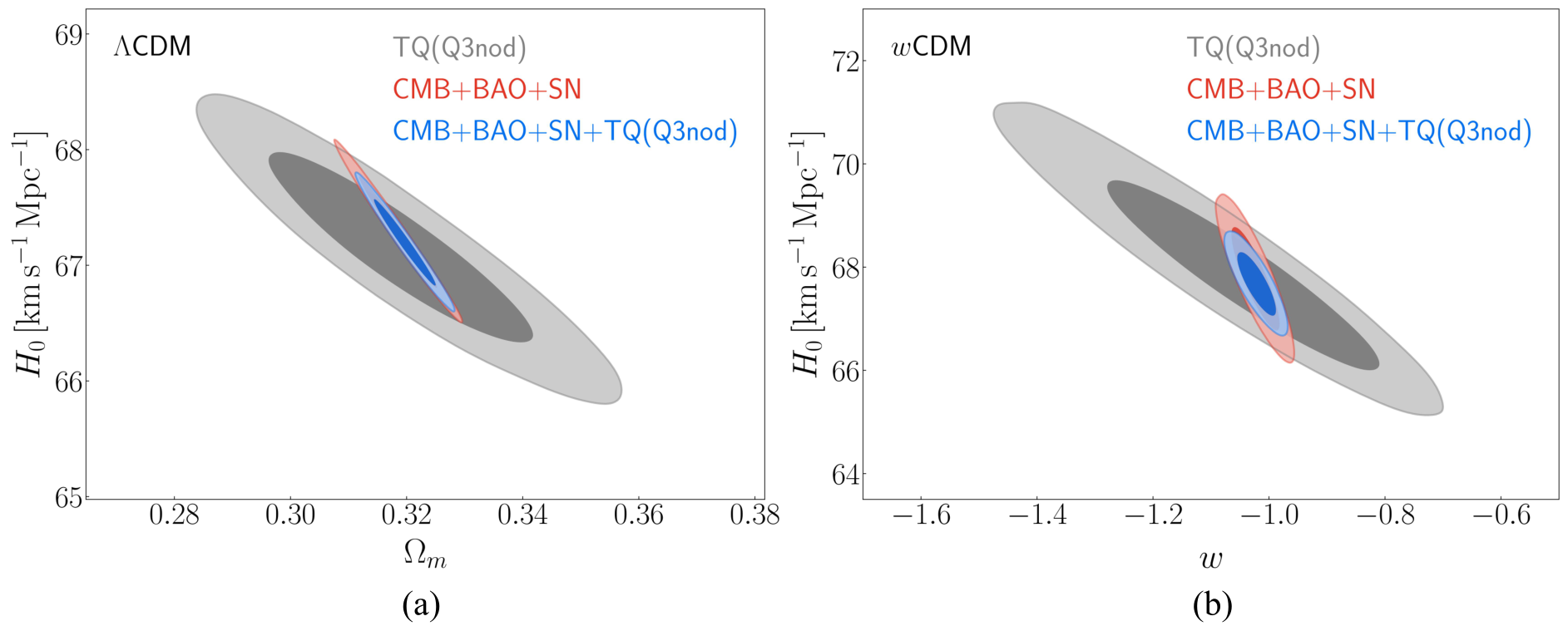}
\end{center}
\caption{Two-dimensional marginalized contours (68.3\% and 95.4\% confidence level) in the $\Omega_{m}$--$H_{0}$ plane for the $\Lambda$CDM model, in the $w$--$H_{0}$ plane for the $w$CDM model by using the TianQin, CMB+BAO+SN, and CMB+BAO+SN+TianQin data combinations. Here, the TianQin mock data are simulated based on the Q3nod model. This figure is taken from \cite{Wang:2019tto}.} \label{TQCBS}
\end{figure*}

The most crucial role of \ac{GW} standard sirens in cosmological research is to help further break cosmological parameter degeneracies. Because standard sirens can play a unique role in measuring the Hubble-Lema\^itre constant, they create unique degeneracy directions in the parameter space of cosmological models. Therefore, when combined with other cosmological probes, their degeneracy directions are often nearly orthogonal in many cases, leading to very effective joint constraints. Research on standard sirens using third-generation ground-based \ac{GW} detectors has clearly demonstrated the role of standard sirens in cosmological research (see  \cite{Zhang:2018byx,Wang:2018lun,Zhang:2019ple,Zhang:2019loq,Li:2019ajo,Jin:2020hmc,Jin:2023zhi,Han:2023exn,Chang:2019xcb}). Compared to higher-frequency standard sirens detected by the ground-based detectors, space-based millihertz band standard sirens have the advantage of much higher signal-to-noise ratios in most cases, so only a small number of standard siren data are needed to effectively contribute to cosmological constraints \cite{Wang:2019tto,Zhao:2019gyk}. Reference \cite{Wang:2019tto} studied the role of TianQin's bright siren observations in cosmological research, which well illustrates this point.

Considering \acp{MBHB} as bright sirens (whose \ac{EM} counterparts can be observed by large optical and radio telescopes currently under construction), TianQin's five-year observations could potentially yield a dozen to two dozen bright siren data (this also depends on the population model, with models such as pop III, Q3d, and Q3nod leading to some differences). Figure~\ref{TQCMB} shows the constraints on the $\Lambda$CDM and $w$CDM models by TianQin's bright siren observations (simulated data) and their joint constraints with CMB, using the Q3nod population model as an example. It is clear that the degeneracy direction of \acp{GW} in parameter space is distinctly different from that of CMB, especially in the case of the $w$CDM model, where they are nearly completely orthogonal, leading to a good breaking of parameter degeneracy. Neither \ac{CMB} nor \acp{GW} alone can effectively constrain the equation of state of dark energy, but their combination can provide excellent constraints, with a measurement precision for $w$ of about 3.6\%. Figure~\ref{TQCBS} shows the combination of TianQin's bright siren observations with CBS, and one find that in this case, standard sirens can further break parameter degeneracy, leading to better constraint results.

The aforementioned example demonstrates that the bright siren observations of TianQin will play a crucial role in the comprehensive measurement of the dark-energy equation of state in the future. They can effectively break the inherent cosmological parameter degeneracies in \ac{CMB} data. In fact, one hopes to accurately measure some key cosmological parameters, including the dark-energy equation of state, solely through late-universe observations. However, this is quite challenging, and without the inclusion of \ac{CMB} data, the cosmological constraints generally yield poor results.

Yet, with the advancement of observational technology, one has new hopes. In the next decade or so, the author is likely to develop new cosmological tools that will make it possible to precisely measure dark energy using only late-universe observations.
For instance, the construction of a new generation of ground-based \ac{GW} detectors and space-based \ac{GW} detectors like TianQin will provide us with abundant \ac{GW} standard siren data. The completion of large radio telescope arrays such as the SKA will enable neutral hydrogen 21-cm intensity mapping (IM) surveys and detect and accurately localize numerous \acp{FRB}, which have the potential to become important late-universe probes \cite{Zhang:2021yof,Wu:2021vfz,Wu:2022jkf,Jin:2021pcv,Zhao:2020ole,Qiu:2021cww,Zhao:2022bpd,Zhang:2023gye}.
\ac{GW} standard sirens have unique advantages in measuring the Hubble-Lema\^itre constant, but their effectiveness in measuring the dark-energy equation of state is limited. In contrast, 21-cm IM and localized \acp{FRB} (with redshift measurements) have advantages in measuring the dark-energy equation of state, but neither can accurately measure the Hubble-Lema\^itre constant alone. The characteristics of these probes indicate that they are highly complementary.
Therefore, when \ac{GW} standard sirens are combined with 21-cm IM, \acp{FRB}, and other probes, they can break parameter degeneracies and yield better joint constraint results \cite{Jin:2021pcv,Zhao:2020ole,Zhang:2023gye}. Reference \cite{Wu:2022dgy} simulated the observational data from four potential late-universe probes in the future: \acp{GW}, 21-cm IM, \acp{FRB}, and strong gravitational lensing.
It found that their combination can accurately measure the Hubble-Lema\^itre constant and the dark-energy equation of state, with constraints superior to those from CBS. Therefore, TianQin's standard siren observations will make significant contributions to the comprehensive development of precise late-universe probes in the future.

\subsection{Potential of gravitational lensing effect} \label{sec:lensing}

{\it Subsection coordinator: Shun-Jia Huang}

If \ac{GW} from a coalescing binary passes near massive objects, gravitational lensing will affect \ac{GW} in the same way as it does for light \cite{Wang:1996as,Nakamura:1997sw,Takahashi:2003ix}, which will influence the strain of \ac{GW}, and, in strong lensing cases, produce multiple images with arrival time delay.
The detection of strongly lensed \ac{GW} signals provide a unique probe for various cosmological studies.
This subsection will introduces the potential contribution to cosmology by using strongly lensed \ac{GW} signals detected with TianQin, including measuring the Hubble-Lema\^itre constant and testing the \ac{CDDR}.

\subsubsection{Measuring the Hubble-Lema\^itre Constant}

\begin{figure*}
\includegraphics[width=0.5\textwidth]{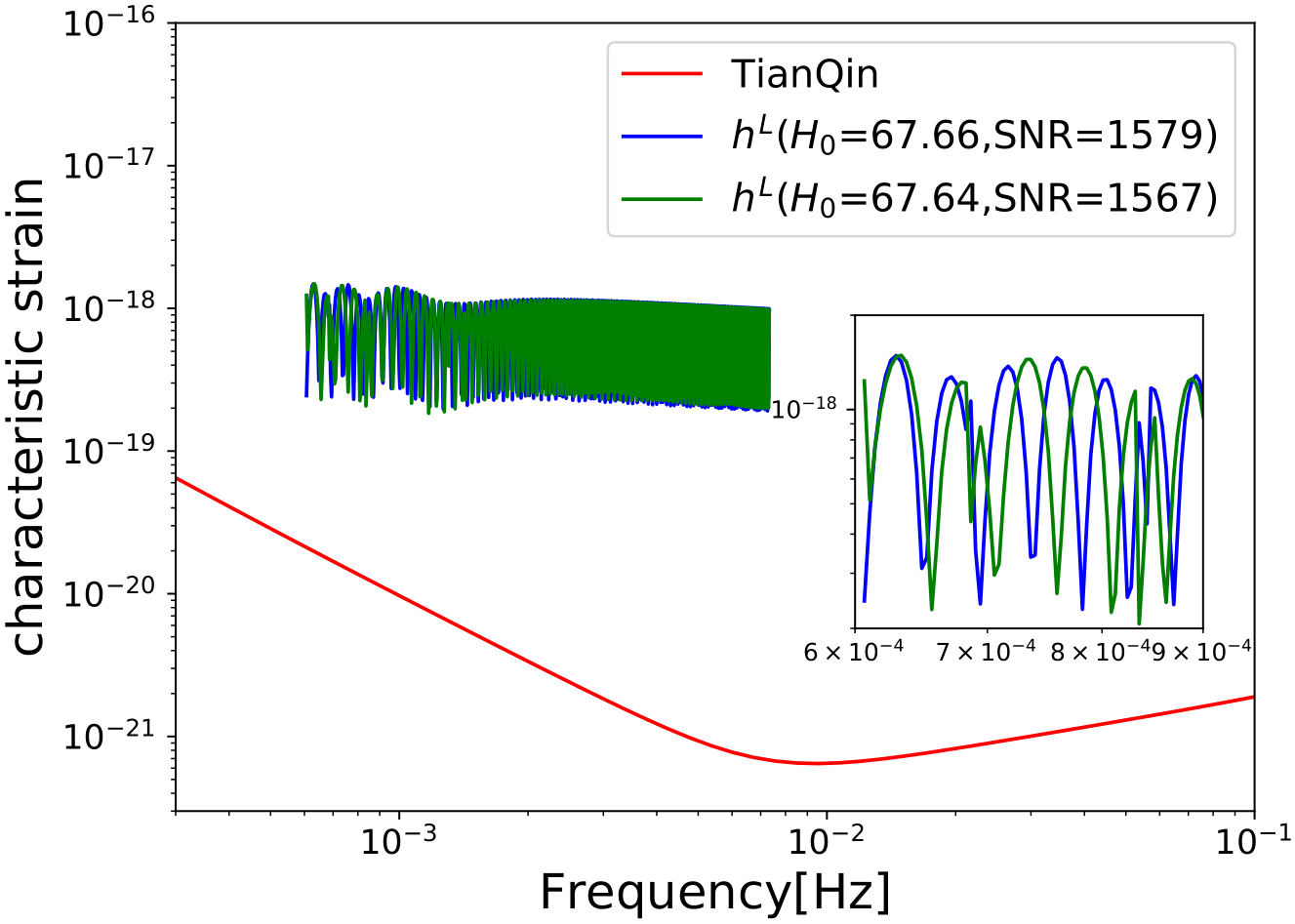}
\caption{Two strongly lensed \ac{GW} waveforms, with all other parameters fixed but $H_0$ slightly different. This figure is taken from  \cite{Huang:2023prq}.}
\label{fig:SLGW_waveform}
\end{figure*}

The phenomenon of strong gravitational lensing results in multiple images of a distant source along different paths, leading to time delays in their arrival at the observer.
This effect is particularly significant for time-varying luminous sources such as quasars and explosive transients, for which these time delays can be measured.
As proposed by Refsdal in 1964, these time delays provide a method for measuring the Hubble-Lema\^itre constant, a technique known as time delay cosmography \cite{Refsdal:1964blz,Treu:2010uj}.
By precisely determining the Einstein radius and time delays, the angular diameter distance ratio of the lensing system can be constrained.
To determine the Hubble-Lema\^itre constant, one also needs accurate knowledge of the lens system's Fermat potential, which depends on the mass model of the lens.
Observations of lensed images help constrain the lens's mass distribution, but additional data, such as extended arcs from the host galaxy or velocity dispersion from spectral observations \cite{Kochanek:2000ue,Treu:2002cb}, may be needed for a complete mass model.
The H0LiCOW project recently used gravitational lensing time delays to measure the Hubble-Lema\^itre constant, obtaining a value of $H_0 = 73.3^{+1.7}_{-1.8}\mathrm{\, km\, s^{-1}\, Mpc^{-1}}$ \cite{Wong:2019kwg}, demonstrating the potential for improving measurement precision with increased observational data.
Furthermore, these time delays can also be used to constrain other cosmological parameters, such as the equation of state of dark energy and the \ac{PPN} parameter of \ac{MGT} \cite{Yang:2020eoh,Yang:2018bdf}.

Gravitational lensing time delays can be used to measure the Hubble-Lema\^itre constant not only through \ac{EM} observations but also via \ac{GW} observations \cite{Sereno:2011ty}.
Although no confirmed lensed \ac{GW} signals have been identified in the \ac{LVK} observations yet \cite{Hannuksela:2019kle,LIGOScientific:2021izm,Diego:2021fyd}, it has been suggested that third-generation ground-based detectors like \ac{ET} could constrain the Hubble-Lema\^itre constant's relative error to 0.68\% using strongly lensed \ac{GW} signals \cite{Liao:2017ioi}, and 50 strongly lensed \ac{GW} events could offer constraints comparable to 580 \ac{SN Ia}, highlighting the method's potential for distinguishing cosmological models \cite{Wei:2017emo}.
With a model-independent approach using the FLRW metric's distance summation rule, it has been shown that 10 strongly lensed \ac{GW} events could provide constraints similar to 300 lensed quasar events, mainly due to the higher precision of \ac{GW} time delay measurements \cite{Li:2019rns}.
If one can combine \ac{GW} and \ac{EM} wave arrival time differences from the same lensed source to constrain cosmological parameters, then even a single event could provide meaningful Hubble-Lema\^itre constant constraints \cite{Cremonese:2019tgb}.
Multi-messenger observations with strongly lensed \acp{GW} could pinpoint the host galaxy of merging black holes, allowing for Hubble-Lema\^itre constant measurements with high precision \cite{Hannuksela:2020xor}.

\begin{figure*}
\includegraphics[width=1\textwidth]{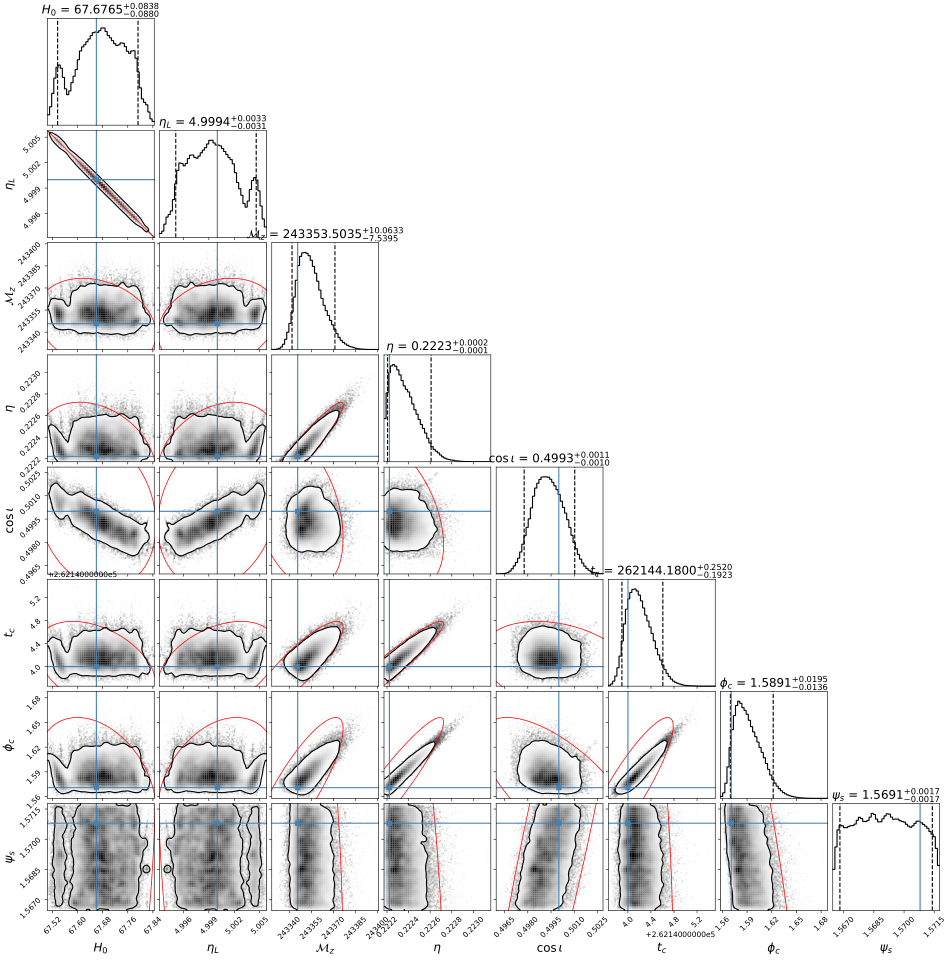}
\caption{Comparison between the posterior distribution sampled by \ac{MCMC} (black) and the ellipses based on the \ac{FIM} (red) with 90\% credible regions.}
\label{fig:SLGW_corner_H0}
\end{figure*}

We have proposed a new method to measure the Hubble-Lema\^itre constant $H_0$ using strongly lensed \ac{GW} signals \cite{Huang:2023prq}.
The waveform is reparameterized to explicitly include $H_0$ in the parameter set.
As shown in Fig. \ref{fig:SLGW_waveform}, the strongly lensed \ac{GW} waveforms are highly sensitivity to $H_0$ and exhibit significant deviation with even a slight difference in the Hubble-Lema\^itre constant.
This was made possible by utilizing the waveform \emph{per se}, instead of just the time delay or the magnification \cite{Sereno:2011ty,Liao:2017ioi,Wei:2017emo,Li:2019rns,Cremonese:2019tgb,Hannuksela:2020xor}.
For the detected \ac{GW} sources, TianQin's sky localization precision can reach the level of 1 deg$^2$ to 0.1 deg$^2$ \cite{Wang:2019ryf,Liu:2020eko,Fan:2020zhy,Huang:2020rjf}, which makes it possible to combine subsequent \ac{EM} observations to implement multi-messenger astronomy.
For an order-of-magnitude estimation of the strongly lensed \ac{GW} detection rate, the detection probability of strongly lensed \ac{GW} from \ac{MBHB} mergers is adopted to be approximately 1\% for space-based \ac{GW} detectors \cite{Gao:2021sxw}, while under an optimistic model, the detection rate of \ac{MBHB} event is expected to be at least on the order of $O(10^2)$ per year \cite{Wang:2019ryf}.
Consequently, the total detection number for strongly lensed \ac{GW} events can be as high as $\gtrsim O(5)$ over the course of a five-year mission lifetime.
It is also assumed that the \ac{EM} counterpart of \ac{GW} source is observed, for if \acp{MBHB} evolve in a gas-rich environment, the accretion of gas could lead to the production of \ac{EM} radiation \cite{DAscoli:2018dbt}, which can then be utilized to ascertain the source redshift \cite{Tamanini:2016zlh}.
Fig. \ref{fig:SLGW_corner_H0} shows that the strongly lensed \ac{GW} waveform method is capable of localizing the Hubble-Lema\^itre constant $H_0$ with a precision of 1\%.

\subsubsection{Testing the Cosmic Distance Duality Relation}

The development of modern cosmology heavily relies on the distance-redshift relation.
The redshift of celestial bodies can be accurately obtained through spectroscopy, making the measurement of distances particularly important.
In general, the key quantities are the luminosity distance and the angular diameter distance.
Theoretically, if the following three conditions are met:
\begin{itemize}
\item Spacetime is described by the metric theory of gravity
\item Photons propagate along null geodesics
\item The number of photons is conserved
\end{itemize}
then the \ac{CDDR} holds \cite{Etherington:1933}:
\bea D_A(z)(1+z)^2/D_L(z)\equiv1\,,\eea
where \( D_A \) and \( D_L \) are the angular diameter distance and luminosity distance, respectively.

Testing the validity of the \ac{CDDR} can not only deepen our understanding of the universe but also reveal possible new physical and astrophysical mechanisms \cite{Bassett:2003vu}.
Testing the \ac{CDDR} requires measurements of both luminosity distance and angular diameter distance at the same redshift.
Common combinations include data from \ac{SN Ia}, which serve as standard candles to provide luminosity distance measurements, and data from galaxy clusters, which provide measurements of angular diameter distance \cite{Holanda:2011hh,Li:2011exa,Hu:2018yah}.
Cosmic opacity, which includes effects such as the dimming of supernovae light by dust \cite{Lima:2011ye} and the transformation of photons into light axions or gravitons, can affect luminosity distance measurements and lead to violations of the \ac{CDDR} \cite{Avgoustidis:2010ju,Liao:2015ccl}.
Angular diameter distances from galaxy clusters are based on observations of X-ray surface brightness and the Sunyaev-Zel'dovich effect \cite{Uzan:2004my}, both of which are affected by cosmic opacity \cite{Li:2013cva}.
Other methods include using \acp{GRB} for luminosity distance measurements \cite{Holanda:2016msr} and \ac{BAO} for angular diameter distance measurements \cite{Wu:2015ixa}.
However, these methods are either affected by cosmic opacity or model-dependent.

\acp{GW} offer direct measurements of luminosity distance that is independent of cosmic opacity or cosmological models \cite{Fu:2019oll}.
Additionally, strong gravitational lensing effects in elliptical galaxies have been utilized for cosmological studies \cite{Chen:2018jcf,Tu:2019vcj,Wong:2019kwg}, which can also provide information on angular diameter distances independent of cosmic opacity \cite{Liao:2015uzb,Yang:2017bkv}.
Projects like H0LiCOW \cite{H0LiCOW:2016xpx} utilize time-delayed gravitational lensing systems to derive angular diameter distances, showing great potential in cosmological research \cite{Rana:2017sfr,Wong:2019kwg,Yildirim:2019vlv}.
Following the first detection of \ac{GW} in 2015, researchers have employed \ac{GW} data to examine the \ac{CDDR}.
In 2019, a study simulated \ac{GW} signals with existing gravitational lensing data to test \ac{CDDR}, setting constraints on the deviation parameter $\eta_0$ at 2.6\% and 4.7\% levels \cite{Liao:2019xug} for the deviation models, $\eta_1(z)=1+\eta_0z$ and $\eta_2(z)=1+\eta_0z/(1+z)$, respectively.

The above tests need to assume that the Universe is isotropy, but anisotropy could invalidate these tests \cite{Li:2017dey}.
In 2020, a method was introduced using strongly lensed \ac{GW} signals to provide both luminosity and angular diameter distances from the same source \cite{Lin:2019mrl}.
Using simulations based on \ac{ET} and methods for error propagation analysis, the deviation parameter $\eta_0$ has been constrained to 1.3\% and 3\% for $\eta_1(z)$ and $\eta_2(z)$, respectively \cite{Lin:2020vqj}.
The potential of using machine learning to constrain $\eta_0$ has been studied in \cite{Arjona:2020axn}.

\begin{figure*}
\includegraphics[width=0.5\textwidth]{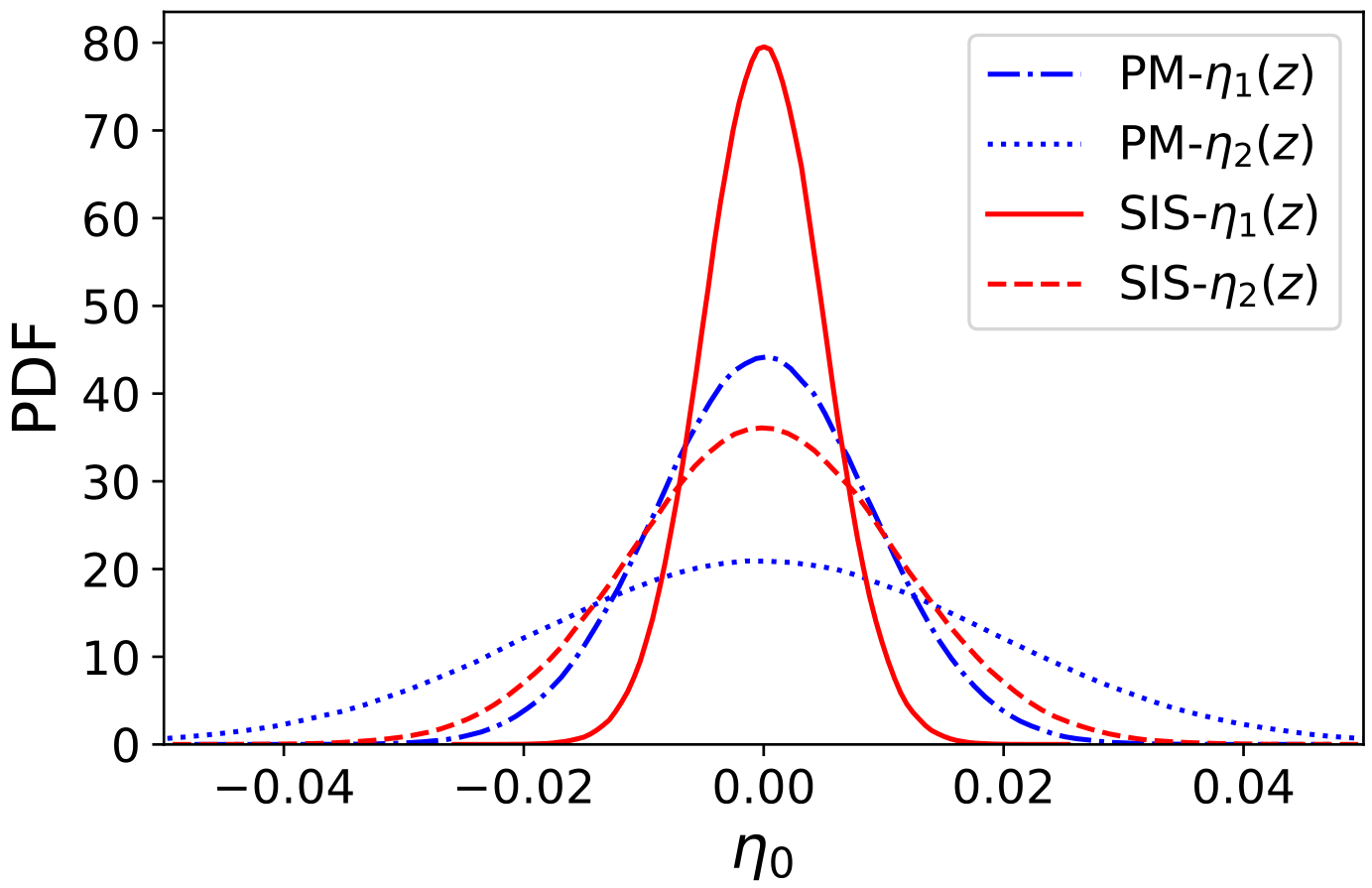}
\caption{The probability density functions of $\eta_0$ constrained with strongly lensed \ac{GW} signals. (PM: point mass; SIS: Singular Isothermal Sphere.) This figure is taken from \cite{Huang:2024zvk}.}
\label{fig:PDF_eta0}
\end{figure*}

Recently, the \ac{CDDR} is tested by incorporating $\eta_0$ into the waveform parameter sets \cite{Huang:2024zvk}.
Assuming the successful detection of the \ac{EM} counterpart of the strongly lensed \ac{GW} signals from \ac{MBHB}, the potential of testing \ac{CDDR} based on TianQin is studied.
The expected measurement precision of $\eta_0$ can reach a considerable 1\% level, as shown in FIG. \ref{fig:PDF_eta0}.
Specifically, for the point mass model for the lens, the measurement precision of $\eta_0$ in the parametric forms $\eta_1(z) = 1 + \eta_0 z$ and $\eta_2(z)=1+\eta_0z/(1+z)$ is 0.9\% and 1.9\%, respectively.
One can see that $\eta_0$ in $\eta_1(z)$ is better constrained than that in $\eta_2(z)$.
This is because $\eta_1(z)$ is more sensitive to changes in $\eta_0$ compared to $\eta_2(z)$.
In contrast, for the Singular Isothermal Sphere model for the lens, the corresponding precision for $\eta_0$ in $\eta_1$ and $\eta_2$ is significantly higher, at 0.5\% and 0.11\%, respectively.
This is because the gravitational lensing effect for the Singular Isothermal Sphere model is stronger than that for the point mass model.

\subsection{Summary of the section} \label{sec:cosmo_summary}

This section reports the prospects of TianQin for constraining the standard cosmological model
$\Lambda$CDM and the dark energy EoS model CPL, using the three classes of candidate standard sirens,
SBHBs, \acp{EMRI}, and \acp{MBHB}. For the $\Lambda$CDM parameters:
\begin{itemize}
\item Using \acp{SBHB}, TianQin can achieve a precision of about $20\%$ for the Hubble-Lema\^itre constant $H_0$ measurement
 when it detects alone, and TianQin is expected to be able to constrain $H_0$ to a precision of
 close to $1\%$ if it can perform a multi-band \ac{GW} analysis in conjunction with the \ac{ET} detections;
\item Using \acp{EMRI}, TianQin can achieve an expected precision of about $3\%-8\%$ for $H_0$ constraint,
 and an expected precision of about $50\%$ for $\Omega_M$ constraint;
\item Using \acp{MBHB}, (i) in the optimistic scenario where their \ac{EM} counterparts can be observed,
 TianQin is expected to achieve a precision of about $2\%-4\%$ for $H_0$ constraint,
 and precisions of about $10\%-30\%$ and about $20\%-30\%$ for $\Omega_M$ and $\Omega_\Lambda$ constraints,
 respectively; and (ii) in the conservative scenario where their \ac{EM} counterparts cannot be observed,
 TianQin can only provide an expected precision of about $3\%-7\%$ for $H_0$ constraint,
 and it is difficult to effectively constrain $\Omega_M$ and $\Omega_\Lambda$;
 besides, (iii) if \ac{MBHB} \ac{GW} signals meet strong gravitational lensing effects, TianQin can ideally
 provide a competitive precision of about $1\%$ for $H_0$ constraint even through a single \ac{MBHB} detection.
\end{itemize}
For the CPL dark energy EoS model parameters, only \acp{EMRI} and \acp{MBHB} can provide effective constraints,
\begin{itemize}
\item Using \acp{EMRI}, TianQin can achieve an expected constraining precision of about $15\%-40\%$ for $w_0$,
 and is unable to effectively constrain $w_a$;
\item Using \acp{MBHB}, (i) in the optimistic scenario, TianQin can be expected to achieve a precision
 of about $10\%-15\%$ for constraining $w_0$, and an expected precision $\Delta w_0 \sim 0.4-0.6$ for $w_a$,
 and (ii) in the conservative scenario, TianQin is expected to achieve a precision of about $15\%-40\%$
 for $w_0$ constraint, but an expected precision of $\Delta w_a > 0.6$ for $w_a$ constraint.
\end{itemize}
Moreover, one note here that the candidate detector configuration of TianQin, i.e., twin constellation ---
TianQin I+II, is expected to significantly improve the constraints for various cosmological parameters.
As constraining capabilities completely independent of the cosmic distance ladder, \ac{CMB} and BAO observations,
the expected constraining capability of the three classes of standard sirens on $H_0$ is promising to
provide an important role in clarifying the Hubble tension,
and on $w_0$ is expected to make an effective contribution to probing the nature of dark energy.

Compared to single detectors, there can be great improvements from the detector networks like TianQin+LISA.
The TianQin+LISA network is expected to improve the constraints for various cosmological parameters,
e.g., $H_0$ and $w_0$, with improvements ranging from a few tens of percent to a factor of a few,
compared to TianQin or LISA detecting alone. These improvements can greatly increase the contribution of
space-based \ac{GW} detections to addressing the Hubble tension and probing dark energy.

Beyond the prospects of constraining cosmological parameters, TianQin can also be used to discriminate dark energy EoS models and test the \ac{CDDR}.
For the selection of dark energy EoS models, utilizing \acp{MBHB} with their \ac{EM} counterparts,
TianQin can provide moderate evidence for the true dark energy model in the best-case scenario.
For testing the \ac{CDDR}, TianQin can measure the deviation parameter $\eta_0$ to a precision of approximately $1\%$.

Finally, we note that space-based \ac{GW} detectors such as TianQin can detect some \ac{GW} events to very high redshifts (such as $z>15$ for \acp{MBHB}), enabling them to probe the global structure of the cosmic space. Such probability has just recently started to be investigated \cite{Shi:2024ula}.


\clearpage

%% file: sec5-sum.tex
\section{Summary}
\label{sec:sum}

In this white paper, we summarize the current understanding of TianQin's capability in answering important questions in fundamental physics and cosmology.

In terms of the nature of gravity, TianQin is expected to make significant progress in multiple directions.
For key predictions of \ac{GR} in the strong field regime,
TianQin can directly detect high-order and nonlinear modes of \ac{MBHB} events, providing critical data for studying nonlinear \ac{GW} phenomena and enabling high-precision test of the Kerr hypothesis;
TianQin is also expected to detect displacement memory from a few \acp{MBHB}, while the detection of spin memory will be challenging.
With its capability to make precision measurements, TianQin can greatly advance the search for possible signatures of beyond \ac{GR} effect, including:
detecting extra polarization modes with amplitudes as low as 1\% of the tensor mode with \ac{GCB} and \ac{MBHB} signals;
improving the measurement accuracy of the quadrupole moment of black holes by approximately 7 orders of magnitude using \ac{EMRI} signals, and improving the constraint on the graviton mass by about 4 orders of magnitude with \ac{MBHB} signals, both compared to the current best \ac{GW} results;
and refining tests of the \ac{EdGB} theory with \ac{SBHB} signals.
Joint detection with other \ac{GW} detectors, such as LISA, \ac{ET} and CE, are expected to yield substantial scientific advancements, such as improved constraints on non-commutative theories and the \ac{EdGB} theory.
Possible environmental effects such as dark matter halos, accretion disks, third-body influences and gravitational lensing can potentially be confused with beyond \ac{GR} effects.
In such cases, statistical analysis of multiple signals is needed to distinguish the different contributions.
Waveform systematics due to model incompleteness and higher-order term inaccuracies could affect parameter estimation and tests of fundamental physics, necessitating comprehensive evaluation to meet the precision requirements.

\ac{GW} detection can also play a significant role in the search for physics beyond the Standard Model, exploring phenomena like ultralight bosons, \acp{PBH}, inflation, and cosmic strings.
There are tantalizing prospects for groundbreaking discoveries that could transform our understanding of the universe.
TianQin is also expected to make significant contributions in this area, some outstanding examples include:
detecting ultra light dark matter with masses in the ranges $10^{-19} \sim 10^{-15}$ eV and $10^{-13.5}\sim 10^{-11.5}$ eV;
exploring the parameter space with phase transition strength greater than 0.1, for new physical models and new Higgs potentials that can generate a first-order \ac{EWPT} related to matter-antimatter asymmetry and heavy dark matter production;
detecting induced \ac{SGWB} accompanying the production of asteroid-mass \acp{PBH}, which could account for all the dark matter;
potentially detecting \acp{GW} produced by phase transitions during inflation and cosmic strings.
But challenges remain, including enhancing detector sensitivity, accurately modeling \ac{GW} sources, and distinguishing signals from noise.

TianQin can detect different types of \ac{GW} sources, \acp{SBHB}, \acp{EMRI}, and \acp{MBHB}, located at vastly different redshifts, making to possible to measure cosmological parameters at different epochs of the universe.
For $\Lambda$CDM parameters, TianQin can achieve about 20\% precision on $H_0$ with \acp{SBHB}, improving to nearly 1\% with multi-band analysis including ET.
\acp{EMRI} can constrain $H_0$ with about 3\%-8\% precision and $\Omega_M$ with 50\% precision.
\acp{MBHB} can provide 2\%-4\% precision on $H_0$ if their \ac{EM} counterparts are observed, but only 3\%-7\% if not; strong gravitational lensing effects could enhance $H_0$ precision to 1\% with single \ac{MBHB} detections.
For dark energy \ac{EoS} parameters, \acp{EMRI} offer 15\%-40\% precision for $w_0$ but not $w_a$.
\acp{MBHB} can provide 10\%-15\% precision on $w_0$ in an optimistic scenario and $\Delta w_a \sim 0.4-0.6$, with 15\%-40\% precision and $\Delta w_a > 0.6$ in a conservative scenario.
Combined with LISA, TianQin could improve precision on $H_0$ and $w_0$ by a factor of a few compared to stand alone detections, thus aiding in resolving the Hubble tension and probing dark energy.
Additionally, TianQin can discriminate dark energy EoS models and test the \ac{CDDR}, with potential to measure the deviation parameter $\eta_0$ to a 1\% level.

In summary, TianQin is poised to make significant contributions to \ac{GW} detection in the future.
Joining other \ac{GW} detectors, TianQin is expect to play a crucial role in advancing our knowledge in fundamental physics, cosmology, and \ac{GW} research as a whole.

%% file: main.bbl
\begin{thebibliography}{1288}%
\makeatletter
\providecommand \@ifxundefined [1]{%
 \@ifx{#1\undefined}
}%
\providecommand \@ifnum [1]{%
 \ifnum #1\expandafter \@firstoftwo
 \else \expandafter \@secondoftwo
 \fi
}%
\providecommand \@ifx [1]{%
 \ifx #1\expandafter \@firstoftwo
 \else \expandafter \@secondoftwo
 \fi
}%
\providecommand \natexlab [1]{#1}%
\providecommand \enquote  [1]{``#1''}%
\providecommand \bibnamefont  [1]{#1}%
\providecommand \bibfnamefont [1]{#1}%
\providecommand \citenamefont [1]{#1}%
\providecommand \href@noop [0]{\@secondoftwo}%
\providecommand \href [0]{\begingroup \@sanitize@url \@href}%
\providecommand \@href[1]{\@@startlink{#1}\@@href}%
\providecommand \@@href[1]{\endgroup#1\@@endlink}%
\providecommand \@sanitize@url [0]{\catcode `\\12\catcode `\$12\catcode
  `\&12\catcode `\#12\catcode `\^12\catcode `\_12\catcode `\%12\relax}%
\providecommand \@@startlink[1]{}%
\providecommand \@@endlink[0]{}%
\providecommand \url  [0]{\begingroup\@sanitize@url \@url }%
\providecommand \@url [1]{\endgroup\@href {#1}{\urlprefix }}%
\providecommand \urlprefix  [0]{URL }%
\providecommand \Eprint [0]{\href }%
\providecommand \doibase [0]{http://dx.doi.org/}%
\providecommand \selectlanguage [0]{\@gobble}%
\providecommand \bibinfo  [0]{\@secondoftwo}%
\providecommand \bibfield  [0]{\@secondoftwo}%
\providecommand \translation [1]{[#1]}%
\providecommand \BibitemOpen [0]{}%
\providecommand \bibitemStop [0]{}%
\providecommand \bibitemNoStop [0]{.\EOS\space}%
\providecommand \EOS [0]{\spacefactor3000\relax}%
\providecommand \BibitemShut  [1]{\csname bibitem#1\endcsname}%
\let\auto@bib@innerbib\@empty
\bibitem [{\citenamefont {{Robinson}}(2018)}]{2018Natur.557...30R}%
  \BibitemOpen
  \bibfield  {author} {\bibinfo {author} {\bibfnamefont {A.}~\bibnamefont
  {{Robinson}}},\ }\href {\doibase 10.1038/d41586-018-05004-4} {\bibfield
  {journal} {\bibinfo  {journal} {\nat}\ }\textbf {\bibinfo {volume} {557}},\
  \bibinfo {pages} {30} (\bibinfo {year} {2018})}\BibitemShut {NoStop}%
\bibitem [{\citenamefont {Carlip}\ \emph {et~al.}(2015)\citenamefont {Carlip},
  \citenamefont {Chiou}, \citenamefont {Ni},\ and\ \citenamefont
  {Woodard}}]{Carlip:2015asa}%
  \BibitemOpen
  \bibfield  {author} {\bibinfo {author} {\bibfnamefont {S.}~\bibnamefont
  {Carlip}}, \bibinfo {author} {\bibfnamefont {D.-W.}\ \bibnamefont {Chiou}},
  \bibinfo {author} {\bibfnamefont {W.-T.}\ \bibnamefont {Ni}}, \ and\ \bibinfo
  {author} {\bibfnamefont {R.}~\bibnamefont {Woodard}},\ }\href {\doibase
  10.1142/S0218271815300281} {\bibfield  {journal} {\bibinfo  {journal} {Int.
  J. Mod. Phys. D}\ }\textbf {\bibinfo {volume} {24}},\ \bibinfo {pages}
  {1530028} (\bibinfo {year} {2015})},\ \Eprint
  {http://arxiv.org/abs/1507.08194} {arXiv:1507.08194 [gr-qc]} \BibitemShut
  {NoStop}%
\bibitem [{\citenamefont {Addazi}\ \emph {et~al.}(2022)\citenamefont {Addazi}
  \emph {et~al.}}]{Addazi:2021xuf}%
  \BibitemOpen
  \bibfield  {author} {\bibinfo {author} {\bibfnamefont {A.}~\bibnamefont
  {Addazi}} \emph {et~al.},\ }\href {\doibase 10.1016/j.ppnp.2022.103948}
  {\bibfield  {journal} {\bibinfo  {journal} {Prog. Part. Nucl. Phys.}\
  }\textbf {\bibinfo {volume} {125}},\ \bibinfo {pages} {103948} (\bibinfo
  {year} {2022})},\ \Eprint {http://arxiv.org/abs/2111.05659} {arXiv:2111.05659
  [hep-ph]} \BibitemShut {NoStop}%
\bibitem [{\citenamefont {Maldacena}(1998)}]{Maldacena:1997re}%
  \BibitemOpen
  \bibfield  {author} {\bibinfo {author} {\bibfnamefont {J.~M.}\ \bibnamefont
  {Maldacena}},\ }\href {\doibase 10.4310/ATMP.1998.v2.n2.a1} {\bibfield
  {journal} {\bibinfo  {journal} {Adv. Theor. Math. Phys.}\ }\textbf {\bibinfo
  {volume} {2}},\ \bibinfo {pages} {231} (\bibinfo {year} {1998})},\ \Eprint
  {http://arxiv.org/abs/hep-th/9711200} {arXiv:hep-th/9711200} \BibitemShut
  {NoStop}%
\bibitem [{\citenamefont {Capozziello}\ and\ \citenamefont
  {De~Laurentis}(2011)}]{Capozziello:2011et}%
  \BibitemOpen
  \bibfield  {author} {\bibinfo {author} {\bibfnamefont {S.}~\bibnamefont
  {Capozziello}}\ and\ \bibinfo {author} {\bibfnamefont {M.}~\bibnamefont
  {De~Laurentis}},\ }\href {\doibase 10.1016/j.physrep.2011.09.003} {\bibfield
  {journal} {\bibinfo  {journal} {Phys. Rept.}\ }\textbf {\bibinfo {volume}
  {509}},\ \bibinfo {pages} {167} (\bibinfo {year} {2011})},\ \Eprint
  {http://arxiv.org/abs/1108.6266} {arXiv:1108.6266 [gr-qc]} \BibitemShut
  {NoStop}%
\bibitem [{\citenamefont {Clifton}\ \emph {et~al.}(2012)\citenamefont
  {Clifton}, \citenamefont {Ferreira}, \citenamefont {Padilla},\ and\
  \citenamefont {Skordis}}]{Clifton:2011jh}%
  \BibitemOpen
  \bibfield  {author} {\bibinfo {author} {\bibfnamefont {T.}~\bibnamefont
  {Clifton}}, \bibinfo {author} {\bibfnamefont {P.~G.}\ \bibnamefont
  {Ferreira}}, \bibinfo {author} {\bibfnamefont {A.}~\bibnamefont {Padilla}}, \
  and\ \bibinfo {author} {\bibfnamefont {C.}~\bibnamefont {Skordis}},\ }\href
  {\doibase 10.1016/j.physrep.2012.01.001} {\bibfield  {journal} {\bibinfo
  {journal} {Phys. Rept.}\ }\textbf {\bibinfo {volume} {513}},\ \bibinfo
  {pages} {1} (\bibinfo {year} {2012})},\ \Eprint
  {http://arxiv.org/abs/1106.2476} {arXiv:1106.2476 [astro-ph.CO]} \BibitemShut
  {NoStop}%
\bibitem [{\citenamefont {Akrami}\ \emph {et~al.}(2021)\citenamefont {Akrami}
  \emph {et~al.}}]{CANTATA:2021ktz}%
  \BibitemOpen
  \bibfield  {author} {\bibinfo {author} {\bibfnamefont {Y.}~\bibnamefont
  {Akrami}} \emph {et~al.} (\bibinfo {collaboration} {CANTATA}),\ }\href
  {\doibase 10.1007/978-3-030-83715-0} {\emph {\bibinfo {title} {{Modified
  Gravity and Cosmology}: {An Update by the CANTATA Network}}}},\ edited by\
  \bibinfo {editor} {\bibfnamefont {E.~N.}\ \bibnamefont {Saridakis}}, \bibinfo
  {editor} {\bibfnamefont {R.}~\bibnamefont {Lazkoz}}, \bibinfo {editor}
  {\bibfnamefont {V.}~\bibnamefont {Salzano}}, \bibinfo {editor} {\bibfnamefont
  {P.}~\bibnamefont {Vargas~Moniz}}, \bibinfo {editor} {\bibfnamefont
  {S.}~\bibnamefont {Capozziello}}, \bibinfo {editor} {\bibfnamefont
  {J.}~\bibnamefont {Beltr\'an~Jim\'enez}}, \bibinfo {editor} {\bibfnamefont
  {M.}~\bibnamefont {De~Laurentis}}, \ and\ \bibinfo {editor} {\bibfnamefont
  {G.~J.}\ \bibnamefont {Olmo}}\ (\bibinfo  {publisher} {Springer},\ \bibinfo
  {year} {2021})\ \Eprint {http://arxiv.org/abs/2105.12582} {arXiv:2105.12582
  [gr-qc]} \BibitemShut {NoStop}%
\bibitem [{\citenamefont {Shankaranarayanan}\ and\ \citenamefont
  {Johnson}(2022)}]{Shankaranarayanan:2022wbx}%
  \BibitemOpen
  \bibfield  {author} {\bibinfo {author} {\bibfnamefont {S.}~\bibnamefont
  {Shankaranarayanan}}\ and\ \bibinfo {author} {\bibfnamefont {J.~P.}\
  \bibnamefont {Johnson}},\ }\href {\doibase 10.1007/s10714-022-02927-2}
  {\bibfield  {journal} {\bibinfo  {journal} {Gen. Rel. Grav.}\ }\textbf
  {\bibinfo {volume} {54}},\ \bibinfo {pages} {44} (\bibinfo {year} {2022})},\
  \Eprint {http://arxiv.org/abs/2204.06533} {arXiv:2204.06533 [gr-qc]}
  \BibitemShut {NoStop}%
\bibitem [{\citenamefont {Will}(2014)}]{Will:2014kxa}%
  \BibitemOpen
  \bibfield  {author} {\bibinfo {author} {\bibfnamefont {C.~M.}\ \bibnamefont
  {Will}},\ }\href {\doibase 10.12942/lrr-2014-4} {\bibfield  {journal}
  {\bibinfo  {journal} {Living Rev. Rel.}\ }\textbf {\bibinfo {volume} {17}},\
  \bibinfo {pages} {4} (\bibinfo {year} {2014})},\ \Eprint
  {http://arxiv.org/abs/1403.7377} {arXiv:1403.7377 [gr-qc]} \BibitemShut
  {NoStop}%
\bibitem [{\citenamefont {Baker}\ \emph {et~al.}(2015)\citenamefont {Baker},
  \citenamefont {Psaltis},\ and\ \citenamefont {Skordis}}]{Baker:2014zba}%
  \BibitemOpen
  \bibfield  {author} {\bibinfo {author} {\bibfnamefont {T.}~\bibnamefont
  {Baker}}, \bibinfo {author} {\bibfnamefont {D.}~\bibnamefont {Psaltis}}, \
  and\ \bibinfo {author} {\bibfnamefont {C.}~\bibnamefont {Skordis}},\ }\href
  {\doibase 10.1088/0004-637X/802/1/63} {\bibfield  {journal} {\bibinfo
  {journal} {Astrophys. J.}\ }\textbf {\bibinfo {volume} {802}},\ \bibinfo
  {pages} {63} (\bibinfo {year} {2015})},\ \Eprint
  {http://arxiv.org/abs/1412.3455} {arXiv:1412.3455 [astro-ph.CO]} \BibitemShut
  {NoStop}%
\bibitem [{\citenamefont {Ayzenberg}\ \emph {et~al.}(2023)\citenamefont
  {Ayzenberg} \emph {et~al.}}]{Ayzenberg:2023hfw}%
  \BibitemOpen
  \bibfield  {author} {\bibinfo {author} {\bibfnamefont {D.}~\bibnamefont
  {Ayzenberg}} \emph {et~al.},\ }\href@noop {} {\  (\bibinfo {year} {2023})},\
  \Eprint {http://arxiv.org/abs/2312.02130} {arXiv:2312.02130 [astro-ph.HE]}
  \BibitemShut {NoStop}%
\bibitem [{\citenamefont {Westphal}\ \emph {et~al.}(2021)\citenamefont
  {Westphal}, \citenamefont {Hepach}, \citenamefont {Pfaff},\ and\
  \citenamefont {Aspelmeyer}}]{Westphal:2020okx}%
  \BibitemOpen
  \bibfield  {author} {\bibinfo {author} {\bibfnamefont {T.}~\bibnamefont
  {Westphal}}, \bibinfo {author} {\bibfnamefont {H.}~\bibnamefont {Hepach}},
  \bibinfo {author} {\bibfnamefont {J.}~\bibnamefont {Pfaff}}, \ and\ \bibinfo
  {author} {\bibfnamefont {M.}~\bibnamefont {Aspelmeyer}},\ }\href {\doibase
  10.1038/s41586-021-03250-7} {\bibfield  {journal} {\bibinfo  {journal}
  {Nature}\ }\textbf {\bibinfo {volume} {591}},\ \bibinfo {pages} {225}
  (\bibinfo {year} {2021})},\ \Eprint {http://arxiv.org/abs/2009.09546}
  {arXiv:2009.09546 [gr-qc]} \BibitemShut {NoStop}%
\bibitem [{\citenamefont {Moody}\ and\ \citenamefont
  {Paik}(1993)}]{Moody:1993ir}%
  \BibitemOpen
  \bibfield  {author} {\bibinfo {author} {\bibfnamefont {M.~V.}\ \bibnamefont
  {Moody}}\ and\ \bibinfo {author} {\bibfnamefont {H.~J.}\ \bibnamefont
  {Paik}},\ }\href {\doibase 10.1103/PhysRevLett.70.1195} {\bibfield  {journal}
  {\bibinfo  {journal} {Phys. Rev. Lett.}\ }\textbf {\bibinfo {volume} {70}},\
  \bibinfo {pages} {1195} (\bibinfo {year} {1993})}\BibitemShut {NoStop}%
\bibitem [{\citenamefont {Adelberger}\ \emph {et~al.}(2003)\citenamefont
  {Adelberger}, \citenamefont {Heckel},\ and\ \citenamefont
  {Nelson}}]{Adelberger:2003zx}%
  \BibitemOpen
  \bibfield  {author} {\bibinfo {author} {\bibfnamefont {E.~G.}\ \bibnamefont
  {Adelberger}}, \bibinfo {author} {\bibfnamefont {B.~R.}\ \bibnamefont
  {Heckel}}, \ and\ \bibinfo {author} {\bibfnamefont {A.~E.}\ \bibnamefont
  {Nelson}},\ }\href {\doibase 10.1146/annurev.nucl.53.041002.110503}
  {\bibfield  {journal} {\bibinfo  {journal} {Ann. Rev. Nucl. Part. Sci.}\
  }\textbf {\bibinfo {volume} {53}},\ \bibinfo {pages} {77} (\bibinfo {year}
  {2003})},\ \Eprint {http://arxiv.org/abs/hep-ph/0307284}
  {arXiv:hep-ph/0307284} \BibitemShut {NoStop}%
\bibitem [{\citenamefont {Touboul}\ \emph {et~al.}(2017)\citenamefont {Touboul}
  \emph {et~al.}}]{Touboul:2017grn}%
  \BibitemOpen
  \bibfield  {author} {\bibinfo {author} {\bibfnamefont {P.}~\bibnamefont
  {Touboul}} \emph {et~al.},\ }\href {\doibase 10.1103/PhysRevLett.119.231101}
  {\bibfield  {journal} {\bibinfo  {journal} {Phys. Rev. Lett.}\ }\textbf
  {\bibinfo {volume} {119}},\ \bibinfo {pages} {231101} (\bibinfo {year}
  {2017})},\ \Eprint {http://arxiv.org/abs/1712.01176} {arXiv:1712.01176
  [astro-ph.IM]} \BibitemShut {NoStop}%
\bibitem [{\citenamefont {Williams}\ \emph {et~al.}(2004)\citenamefont
  {Williams}, \citenamefont {Turyshev},\ and\ \citenamefont
  {Boggs}}]{Williams:2004qba}%
  \BibitemOpen
  \bibfield  {author} {\bibinfo {author} {\bibfnamefont {J.~G.}\ \bibnamefont
  {Williams}}, \bibinfo {author} {\bibfnamefont {S.~G.}\ \bibnamefont
  {Turyshev}}, \ and\ \bibinfo {author} {\bibfnamefont {D.~H.}\ \bibnamefont
  {Boggs}},\ }\href {\doibase 10.1103/PhysRevLett.93.261101} {\bibfield
  {journal} {\bibinfo  {journal} {Phys. Rev. Lett.}\ }\textbf {\bibinfo
  {volume} {93}},\ \bibinfo {pages} {261101} (\bibinfo {year} {2004})},\
  \Eprint {http://arxiv.org/abs/gr-qc/0411113} {arXiv:gr-qc/0411113}
  \BibitemShut {NoStop}%
\bibitem [{\citenamefont {Bertotti}\ \emph {et~al.}(2003)\citenamefont
  {Bertotti}, \citenamefont {Iess},\ and\ \citenamefont
  {Tortora}}]{Bertotti:2003rm}%
  \BibitemOpen
  \bibfield  {author} {\bibinfo {author} {\bibfnamefont {B.}~\bibnamefont
  {Bertotti}}, \bibinfo {author} {\bibfnamefont {L.}~\bibnamefont {Iess}}, \
  and\ \bibinfo {author} {\bibfnamefont {P.}~\bibnamefont {Tortora}},\ }\href
  {\doibase 10.1038/nature01997} {\bibfield  {journal} {\bibinfo  {journal}
  {Nature}\ }\textbf {\bibinfo {volume} {425}},\ \bibinfo {pages} {374}
  (\bibinfo {year} {2003})}\BibitemShut {NoStop}%
\bibitem [{\citenamefont {Taylor}\ and\ \citenamefont
  {Weisberg}(1982)}]{Taylor:1982zz}%
  \BibitemOpen
  \bibfield  {author} {\bibinfo {author} {\bibfnamefont {J.~H.}\ \bibnamefont
  {Taylor}}\ and\ \bibinfo {author} {\bibfnamefont {J.~M.}\ \bibnamefont
  {Weisberg}},\ }\href {\doibase 10.1086/159690} {\bibfield  {journal}
  {\bibinfo  {journal} {Astrophys. J.}\ }\textbf {\bibinfo {volume} {253}},\
  \bibinfo {pages} {908} (\bibinfo {year} {1982})}\BibitemShut {NoStop}%
\bibitem [{\citenamefont {Akiyama}\ \emph {et~al.}(2019)\citenamefont {Akiyama}
  \emph {et~al.}}]{EventHorizonTelescope:2019dse}%
  \BibitemOpen
  \bibfield  {author} {\bibinfo {author} {\bibfnamefont {K.}~\bibnamefont
  {Akiyama}} \emph {et~al.} (\bibinfo {collaboration} {Event Horizon
  Telescope}),\ }\href {\doibase 10.3847/2041-8213/ab0ec7} {\bibfield
  {journal} {\bibinfo  {journal} {Astrophys. J. Lett.}\ }\textbf {\bibinfo
  {volume} {875}},\ \bibinfo {pages} {L1} (\bibinfo {year} {2019})},\ \Eprint
  {http://arxiv.org/abs/1906.11238} {arXiv:1906.11238 [astro-ph.GA]}
  \BibitemShut {NoStop}%
\bibitem [{\citenamefont {Zhang}(1993)}]{Zhang:1992fs}%
  \BibitemOpen
  \bibfield  {author} {\bibinfo {author} {\bibfnamefont {X.-m.}\ \bibnamefont
  {Zhang}},\ }\href {\doibase 10.1103/PhysRevD.47.3065} {\bibfield  {journal}
  {\bibinfo  {journal} {Phys. Rev. D}\ }\textbf {\bibinfo {volume} {47}},\
  \bibinfo {pages} {3065} (\bibinfo {year} {1993})},\ \Eprint
  {http://arxiv.org/abs/hep-ph/9301277} {arXiv:hep-ph/9301277} \BibitemShut
  {NoStop}%
\bibitem [{\citenamefont {Grojean}\ \emph {et~al.}(2005)\citenamefont
  {Grojean}, \citenamefont {Servant},\ and\ \citenamefont
  {Wells}}]{Grojean:2004xa}%
  \BibitemOpen
  \bibfield  {author} {\bibinfo {author} {\bibfnamefont {C.}~\bibnamefont
  {Grojean}}, \bibinfo {author} {\bibfnamefont {G.}~\bibnamefont {Servant}}, \
  and\ \bibinfo {author} {\bibfnamefont {J.~D.}\ \bibnamefont {Wells}},\ }\href
  {\doibase 10.1103/PhysRevD.71.036001} {\bibfield  {journal} {\bibinfo
  {journal} {Phys. Rev. D}\ }\textbf {\bibinfo {volume} {71}},\ \bibinfo
  {pages} {036001} (\bibinfo {year} {2005})},\ \Eprint
  {http://arxiv.org/abs/hep-ph/0407019} {arXiv:hep-ph/0407019} \BibitemShut
  {NoStop}%
\bibitem [{\citenamefont {Huang}\ \emph
  {et~al.}(2016{\natexlab{a}})\citenamefont {Huang}, \citenamefont {Gu},
  \citenamefont {Yin}, \citenamefont {Yu},\ and\ \citenamefont
  {Zhang}}]{Huang:2015izx}%
  \BibitemOpen
  \bibfield  {author} {\bibinfo {author} {\bibfnamefont {F.~P.}\ \bibnamefont
  {Huang}}, \bibinfo {author} {\bibfnamefont {P.-H.}\ \bibnamefont {Gu}},
  \bibinfo {author} {\bibfnamefont {P.-F.}\ \bibnamefont {Yin}}, \bibinfo
  {author} {\bibfnamefont {Z.-H.}\ \bibnamefont {Yu}}, \ and\ \bibinfo {author}
  {\bibfnamefont {X.}~\bibnamefont {Zhang}},\ }\href {\doibase
  10.1103/PhysRevD.93.103515} {\bibfield  {journal} {\bibinfo  {journal} {Phys.
  Rev. D}\ }\textbf {\bibinfo {volume} {93}},\ \bibinfo {pages} {103515}
  (\bibinfo {year} {2016}{\natexlab{a}})},\ \Eprint
  {http://arxiv.org/abs/1511.03969} {arXiv:1511.03969 [hep-ph]} \BibitemShut
  {NoStop}%
\bibitem [{\citenamefont {Huang}\ \emph
  {et~al.}(2016{\natexlab{b}})\citenamefont {Huang}, \citenamefont {Wan},
  \citenamefont {Wang}, \citenamefont {Cai},\ and\ \citenamefont
  {Zhang}}]{Huang:2016odd}%
  \BibitemOpen
  \bibfield  {author} {\bibinfo {author} {\bibfnamefont {F.~P.}\ \bibnamefont
  {Huang}}, \bibinfo {author} {\bibfnamefont {Y.}~\bibnamefont {Wan}}, \bibinfo
  {author} {\bibfnamefont {D.-G.}\ \bibnamefont {Wang}}, \bibinfo {author}
  {\bibfnamefont {Y.-F.}\ \bibnamefont {Cai}}, \ and\ \bibinfo {author}
  {\bibfnamefont {X.}~\bibnamefont {Zhang}},\ }\href {\doibase
  10.1103/PhysRevD.94.041702} {\bibfield  {journal} {\bibinfo  {journal} {Phys.
  Rev. D}\ }\textbf {\bibinfo {volume} {94}},\ \bibinfo {pages} {041702}
  (\bibinfo {year} {2016}{\natexlab{b}})},\ \Eprint
  {http://arxiv.org/abs/1601.01640} {arXiv:1601.01640 [hep-ph]} \BibitemShut
  {NoStop}%
\bibitem [{\citenamefont {Cai}\ \emph {et~al.}(2017{\natexlab{a}})\citenamefont
  {Cai}, \citenamefont {Sasaki},\ and\ \citenamefont {Wang}}]{Cai:2017tmh}%
  \BibitemOpen
  \bibfield  {author} {\bibinfo {author} {\bibfnamefont {R.-G.}\ \bibnamefont
  {Cai}}, \bibinfo {author} {\bibfnamefont {M.}~\bibnamefont {Sasaki}}, \ and\
  \bibinfo {author} {\bibfnamefont {S.-J.}\ \bibnamefont {Wang}},\ }\href
  {\doibase 10.1088/1475-7516/2017/08/004} {\bibfield  {journal} {\bibinfo
  {journal} {JCAP}\ }\textbf {\bibinfo {volume} {08}},\ \bibinfo {pages} {004}
  (\bibinfo {year} {2017}{\natexlab{a}})},\ \Eprint
  {http://arxiv.org/abs/1707.03001} {arXiv:1707.03001 [astro-ph.CO]}
  \BibitemShut {NoStop}%
\bibitem [{\citenamefont {Baker}\ \emph {et~al.}(2020)\citenamefont {Baker},
  \citenamefont {Kopp},\ and\ \citenamefont {Long}}]{Baker:2019ndr}%
  \BibitemOpen
  \bibfield  {author} {\bibinfo {author} {\bibfnamefont {M.~J.}\ \bibnamefont
  {Baker}}, \bibinfo {author} {\bibfnamefont {J.}~\bibnamefont {Kopp}}, \ and\
  \bibinfo {author} {\bibfnamefont {A.~J.}\ \bibnamefont {Long}},\ }\href
  {\doibase 10.1103/PhysRevLett.125.151102} {\bibfield  {journal} {\bibinfo
  {journal} {Phys. Rev. Lett.}\ }\textbf {\bibinfo {volume} {125}},\ \bibinfo
  {pages} {151102} (\bibinfo {year} {2020})},\ \Eprint
  {http://arxiv.org/abs/1912.02830} {arXiv:1912.02830 [hep-ph]} \BibitemShut
  {NoStop}%
\bibitem [{\citenamefont {Chway}\ \emph {et~al.}(2020)\citenamefont {Chway},
  \citenamefont {Jung},\ and\ \citenamefont {Shin}}]{Chway:2019kft}%
  \BibitemOpen
  \bibfield  {author} {\bibinfo {author} {\bibfnamefont {D.}~\bibnamefont
  {Chway}}, \bibinfo {author} {\bibfnamefont {T.~H.}\ \bibnamefont {Jung}}, \
  and\ \bibinfo {author} {\bibfnamefont {C.~S.}\ \bibnamefont {Shin}},\ }\href
  {\doibase 10.1103/PhysRevD.101.095019} {\bibfield  {journal} {\bibinfo
  {journal} {Phys. Rev. D}\ }\textbf {\bibinfo {volume} {101}},\ \bibinfo
  {pages} {095019} (\bibinfo {year} {2020})},\ \Eprint
  {http://arxiv.org/abs/1912.04238} {arXiv:1912.04238 [hep-ph]} \BibitemShut
  {NoStop}%
\bibitem [{\citenamefont {Jiang}\ \emph
  {et~al.}(2023{\natexlab{a}})\citenamefont {Jiang}, \citenamefont {Huang},\
  and\ \citenamefont {Li}}]{Jiang:2023nkj}%
  \BibitemOpen
  \bibfield  {author} {\bibinfo {author} {\bibfnamefont {S.}~\bibnamefont
  {Jiang}}, \bibinfo {author} {\bibfnamefont {F.~P.}\ \bibnamefont {Huang}}, \
  and\ \bibinfo {author} {\bibfnamefont {C.~S.}\ \bibnamefont {Li}},\ }\href
  {\doibase 10.1103/PhysRevD.108.063508} {\bibfield  {journal} {\bibinfo
  {journal} {Phys. Rev. D}\ }\textbf {\bibinfo {volume} {108}},\ \bibinfo
  {pages} {063508} (\bibinfo {year} {2023}{\natexlab{a}})},\ \Eprint
  {http://arxiv.org/abs/2305.02218} {arXiv:2305.02218 [hep-ph]} \BibitemShut
  {NoStop}%
\bibitem [{\citenamefont {Azatov}\ \emph
  {et~al.}(2021{\natexlab{a}})\citenamefont {Azatov}, \citenamefont
  {Vanvlasselaer},\ and\ \citenamefont {Yin}}]{Azatov:2021ifm}%
  \BibitemOpen
  \bibfield  {author} {\bibinfo {author} {\bibfnamefont {A.}~\bibnamefont
  {Azatov}}, \bibinfo {author} {\bibfnamefont {M.}~\bibnamefont
  {Vanvlasselaer}}, \ and\ \bibinfo {author} {\bibfnamefont {W.}~\bibnamefont
  {Yin}},\ }\href {\doibase 10.1007/JHEP03(2021)288} {\bibfield  {journal}
  {\bibinfo  {journal} {JHEP}\ }\textbf {\bibinfo {volume} {03}},\ \bibinfo
  {pages} {288} (\bibinfo {year} {2021}{\natexlab{a}})},\ \Eprint
  {http://arxiv.org/abs/2101.05721} {arXiv:2101.05721 [hep-ph]} \BibitemShut
  {NoStop}%
\bibitem [{\citenamefont {Baldes}\ \emph {et~al.}(2021)\citenamefont {Baldes},
  \citenamefont {Blasi}, \citenamefont {Mariotti}, \citenamefont {Sevrin},\
  and\ \citenamefont {Turbang}}]{Baldes:2021vyz}%
  \BibitemOpen
  \bibfield  {author} {\bibinfo {author} {\bibfnamefont {I.}~\bibnamefont
  {Baldes}}, \bibinfo {author} {\bibfnamefont {S.}~\bibnamefont {Blasi}},
  \bibinfo {author} {\bibfnamefont {A.}~\bibnamefont {Mariotti}}, \bibinfo
  {author} {\bibfnamefont {A.}~\bibnamefont {Sevrin}}, \ and\ \bibinfo {author}
  {\bibfnamefont {K.}~\bibnamefont {Turbang}},\ }\href {\doibase
  10.1103/PhysRevD.104.115029} {\bibfield  {journal} {\bibinfo  {journal}
  {Phys. Rev. D}\ }\textbf {\bibinfo {volume} {104}},\ \bibinfo {pages}
  {115029} (\bibinfo {year} {2021})},\ \Eprint
  {http://arxiv.org/abs/2106.15602} {arXiv:2106.15602 [hep-ph]} \BibitemShut
  {NoStop}%
\bibitem [{\citenamefont {Krylov}\ \emph {et~al.}(2013)\citenamefont {Krylov},
  \citenamefont {Levin},\ and\ \citenamefont {Rubakov}}]{Krylov:2013qe}%
  \BibitemOpen
  \bibfield  {author} {\bibinfo {author} {\bibfnamefont {E.}~\bibnamefont
  {Krylov}}, \bibinfo {author} {\bibfnamefont {A.}~\bibnamefont {Levin}}, \
  and\ \bibinfo {author} {\bibfnamefont {V.}~\bibnamefont {Rubakov}},\ }\href
  {\doibase 10.1103/PhysRevD.87.083528} {\bibfield  {journal} {\bibinfo
  {journal} {Phys. Rev. D}\ }\textbf {\bibinfo {volume} {87}},\ \bibinfo
  {pages} {083528} (\bibinfo {year} {2013})},\ \Eprint
  {http://arxiv.org/abs/1301.0354} {arXiv:1301.0354 [hep-ph]} \BibitemShut
  {NoStop}%
\bibitem [{\citenamefont {Huang}\ and\ \citenamefont
  {Li}(2017)}]{Huang:2017kzu}%
  \BibitemOpen
  \bibfield  {author} {\bibinfo {author} {\bibfnamefont {F.~P.}\ \bibnamefont
  {Huang}}\ and\ \bibinfo {author} {\bibfnamefont {C.~S.}\ \bibnamefont {Li}},\
  }\href {\doibase 10.1103/PhysRevD.96.095028} {\bibfield  {journal} {\bibinfo
  {journal} {Phys. Rev. D}\ }\textbf {\bibinfo {volume} {96}},\ \bibinfo
  {pages} {095028} (\bibinfo {year} {2017})},\ \Eprint
  {http://arxiv.org/abs/1709.09691} {arXiv:1709.09691 [hep-ph]} \BibitemShut
  {NoStop}%
\bibitem [{\citenamefont {Hong}\ \emph {et~al.}(2020)\citenamefont {Hong},
  \citenamefont {Jung},\ and\ \citenamefont {Xie}}]{Hong:2020est}%
  \BibitemOpen
  \bibfield  {author} {\bibinfo {author} {\bibfnamefont {J.-P.}\ \bibnamefont
  {Hong}}, \bibinfo {author} {\bibfnamefont {S.}~\bibnamefont {Jung}}, \ and\
  \bibinfo {author} {\bibfnamefont {K.-P.}\ \bibnamefont {Xie}},\ }\href
  {\doibase 10.1103/PhysRevD.102.075028} {\bibfield  {journal} {\bibinfo
  {journal} {Phys. Rev. D}\ }\textbf {\bibinfo {volume} {102}},\ \bibinfo
  {pages} {075028} (\bibinfo {year} {2020})},\ \Eprint
  {http://arxiv.org/abs/2008.04430} {arXiv:2008.04430 [hep-ph]} \BibitemShut
  {NoStop}%
\bibitem [{\citenamefont {Jiang}\ \emph
  {et~al.}(2024{\natexlab{a}})\citenamefont {Jiang}, \citenamefont {Yang},
  \citenamefont {Ma},\ and\ \citenamefont {Huang}}]{Jiang:2023qbm}%
  \BibitemOpen
  \bibfield  {author} {\bibinfo {author} {\bibfnamefont {S.}~\bibnamefont
  {Jiang}}, \bibinfo {author} {\bibfnamefont {A.}~\bibnamefont {Yang}},
  \bibinfo {author} {\bibfnamefont {J.}~\bibnamefont {Ma}}, \ and\ \bibinfo
  {author} {\bibfnamefont {F.~P.}\ \bibnamefont {Huang}},\ }\href {\doibase
  10.1088/1361-6382/ad24c6} {\bibfield  {journal} {\bibinfo  {journal} {Class.
  Quant. Grav.}\ }\textbf {\bibinfo {volume} {41}},\ \bibinfo {pages} {065009}
  (\bibinfo {year} {2024}{\natexlab{a}})},\ \Eprint
  {http://arxiv.org/abs/2306.17827} {arXiv:2306.17827 [hep-ph]} \BibitemShut
  {NoStop}%
\bibitem [{\citenamefont {Jiang}\ \emph
  {et~al.}(2024{\natexlab{b}})\citenamefont {Jiang}, \citenamefont {Huang},\
  and\ \citenamefont {Ko}}]{Jiang:2024zrb}%
  \BibitemOpen
  \bibfield  {author} {\bibinfo {author} {\bibfnamefont {S.}~\bibnamefont
  {Jiang}}, \bibinfo {author} {\bibfnamefont {F.~P.}\ \bibnamefont {Huang}}, \
  and\ \bibinfo {author} {\bibfnamefont {P.}~\bibnamefont {Ko}},\ }\href
  {\doibase 10.1007/JHEP07(2024)053} {\bibfield  {journal} {\bibinfo  {journal}
  {JHEP}\ }\textbf {\bibinfo {volume} {07}},\ \bibinfo {pages} {053} (\bibinfo
  {year} {2024}{\natexlab{b}})},\ \Eprint {http://arxiv.org/abs/2404.16509}
  {arXiv:2404.16509 [hep-ph]} \BibitemShut {NoStop}%
\bibitem [{\citenamefont {Carroll}(2001)}]{Carroll:2000fy}%
  \BibitemOpen
  \bibfield  {author} {\bibinfo {author} {\bibfnamefont {S.~M.}\ \bibnamefont
  {Carroll}},\ }\href {\doibase 10.12942/lrr-2001-1} {\bibfield  {journal}
  {\bibinfo  {journal} {Living Rev. Rel.}\ }\textbf {\bibinfo {volume} {4}},\
  \bibinfo {pages} {1} (\bibinfo {year} {2001})},\ \Eprint
  {http://arxiv.org/abs/astro-ph/0004075} {arXiv:astro-ph/0004075} \BibitemShut
  {NoStop}%
\bibitem [{\citenamefont {Peebles}\ and\ \citenamefont
  {Ratra}(2003)}]{Peebles:2002gy}%
  \BibitemOpen
  \bibfield  {author} {\bibinfo {author} {\bibfnamefont {P.~J.~E.}\
  \bibnamefont {Peebles}}\ and\ \bibinfo {author} {\bibfnamefont
  {B.}~\bibnamefont {Ratra}},\ }\href {\doibase 10.1103/RevModPhys.75.559}
  {\bibfield  {journal} {\bibinfo  {journal} {Rev. Mod. Phys.}\ }\textbf
  {\bibinfo {volume} {75}},\ \bibinfo {pages} {559} (\bibinfo {year} {2003})},\
  \Eprint {http://arxiv.org/abs/astro-ph/0207347} {arXiv:astro-ph/0207347}
  \BibitemShut {NoStop}%
\bibitem [{\citenamefont {Bull}\ \emph {et~al.}(2016)\citenamefont {Bull} \emph
  {et~al.}}]{Bull:2015stt}%
  \BibitemOpen
  \bibfield  {author} {\bibinfo {author} {\bibfnamefont {P.}~\bibnamefont
  {Bull}} \emph {et~al.},\ }\href {\doibase 10.1016/j.dark.2016.02.001}
  {\bibfield  {journal} {\bibinfo  {journal} {Phys. Dark Univ.}\ }\textbf
  {\bibinfo {volume} {12}},\ \bibinfo {pages} {56} (\bibinfo {year} {2016})},\
  \Eprint {http://arxiv.org/abs/1512.05356} {arXiv:1512.05356 [astro-ph.CO]}
  \BibitemShut {NoStop}%
\bibitem [{\citenamefont {Hinshaw}\ \emph {et~al.}(2013)\citenamefont {Hinshaw}
  \emph {et~al.}}]{WMAP:2012nax}%
  \BibitemOpen
  \bibfield  {author} {\bibinfo {author} {\bibfnamefont {G.}~\bibnamefont
  {Hinshaw}} \emph {et~al.} (\bibinfo {collaboration} {WMAP}),\ }\href
  {\doibase 10.1088/0067-0049/208/2/19} {\bibfield  {journal} {\bibinfo
  {journal} {Astrophys. J. Suppl.}\ }\textbf {\bibinfo {volume} {208}},\
  \bibinfo {pages} {19} (\bibinfo {year} {2013})},\ \Eprint
  {http://arxiv.org/abs/1212.5226} {arXiv:1212.5226 [astro-ph.CO]} \BibitemShut
  {NoStop}%
\bibitem [{\citenamefont {Aghanim}\ \emph {et~al.}(2020)\citenamefont {Aghanim}
  \emph {et~al.}}]{Planck:2018vyg}%
  \BibitemOpen
  \bibfield  {author} {\bibinfo {author} {\bibfnamefont {N.}~\bibnamefont
  {Aghanim}} \emph {et~al.} (\bibinfo {collaboration} {Planck}),\ }\href
  {\doibase 10.1051/0004-6361/201833910} {\bibfield  {journal} {\bibinfo
  {journal} {Astron. Astrophys.}\ }\textbf {\bibinfo {volume} {641}},\ \bibinfo
  {pages} {A6} (\bibinfo {year} {2020})},\ \bibinfo {note} {[Erratum:
  Astron.Astrophys. 652, C4 (2021)]},\ \Eprint
  {http://arxiv.org/abs/1807.06209} {arXiv:1807.06209 [astro-ph.CO]}
  \BibitemShut {NoStop}%
\bibitem [{\citenamefont {Addison}\ \emph {et~al.}(2018)\citenamefont
  {Addison}, \citenamefont {Watts}, \citenamefont {Bennett}, \citenamefont
  {Halpern}, \citenamefont {Hinshaw},\ and\ \citenamefont
  {Weiland}}]{Addison:2017fdm}%
  \BibitemOpen
  \bibfield  {author} {\bibinfo {author} {\bibfnamefont {G.~E.}\ \bibnamefont
  {Addison}}, \bibinfo {author} {\bibfnamefont {D.~J.}\ \bibnamefont {Watts}},
  \bibinfo {author} {\bibfnamefont {C.~L.}\ \bibnamefont {Bennett}}, \bibinfo
  {author} {\bibfnamefont {M.}~\bibnamefont {Halpern}}, \bibinfo {author}
  {\bibfnamefont {G.}~\bibnamefont {Hinshaw}}, \ and\ \bibinfo {author}
  {\bibfnamefont {J.~L.}\ \bibnamefont {Weiland}},\ }\href {\doibase
  10.3847/1538-4357/aaa1ed} {\bibfield  {journal} {\bibinfo  {journal}
  {Astrophys. J.}\ }\textbf {\bibinfo {volume} {853}},\ \bibinfo {pages} {119}
  (\bibinfo {year} {2018})},\ \Eprint {http://arxiv.org/abs/1707.06547}
  {arXiv:1707.06547 [astro-ph.CO]} \BibitemShut {NoStop}%
\bibitem [{\citenamefont {Sch\"oneberg}\ \emph {et~al.}(2019)\citenamefont
  {Sch\"oneberg}, \citenamefont {Lesgourgues},\ and\ \citenamefont
  {Hooper}}]{Schoneberg:2019wmt}%
  \BibitemOpen
  \bibfield  {author} {\bibinfo {author} {\bibfnamefont {N.}~\bibnamefont
  {Sch\"oneberg}}, \bibinfo {author} {\bibfnamefont {J.}~\bibnamefont
  {Lesgourgues}}, \ and\ \bibinfo {author} {\bibfnamefont {D.~C.}\ \bibnamefont
  {Hooper}},\ }\href {\doibase 10.1088/1475-7516/2019/10/029} {\bibfield
  {journal} {\bibinfo  {journal} {JCAP}\ }\textbf {\bibinfo {volume} {10}},\
  \bibinfo {pages} {029} (\bibinfo {year} {2019})},\ \Eprint
  {http://arxiv.org/abs/1907.11594} {arXiv:1907.11594 [astro-ph.CO]}
  \BibitemShut {NoStop}%
\bibitem [{\citenamefont {Eisenstein}\ \emph {et~al.}(2005)\citenamefont
  {Eisenstein} \emph {et~al.}}]{SDSS:2005xqv}%
  \BibitemOpen
  \bibfield  {author} {\bibinfo {author} {\bibfnamefont {D.~J.}\ \bibnamefont
  {Eisenstein}} \emph {et~al.} (\bibinfo {collaboration} {SDSS}),\ }\href
  {\doibase 10.1086/466512} {\bibfield  {journal} {\bibinfo  {journal}
  {Astrophys. J.}\ }\textbf {\bibinfo {volume} {633}},\ \bibinfo {pages} {560}
  (\bibinfo {year} {2005})},\ \Eprint {http://arxiv.org/abs/astro-ph/0501171}
  {arXiv:astro-ph/0501171} \BibitemShut {NoStop}%
\bibitem [{\citenamefont {Bassett}\ and\ \citenamefont
  {Hlozek}(2009)}]{Bassett:2009mm}%
  \BibitemOpen
  \bibfield  {author} {\bibinfo {author} {\bibfnamefont {B.~A.}\ \bibnamefont
  {Bassett}}\ and\ \bibinfo {author} {\bibfnamefont {R.}~\bibnamefont
  {Hlozek}},\ }\href@noop {} {\  (\bibinfo {year} {2009})},\ \Eprint
  {http://arxiv.org/abs/0910.5224} {arXiv:0910.5224 [astro-ph.CO]} \BibitemShut
  {NoStop}%
\bibitem [{\citenamefont {Alam}\ \emph {et~al.}(2021)\citenamefont {Alam} \emph
  {et~al.}}]{eBOSS:2020yzd}%
  \BibitemOpen
  \bibfield  {author} {\bibinfo {author} {\bibfnamefont {S.}~\bibnamefont
  {Alam}} \emph {et~al.} (\bibinfo {collaboration} {eBOSS}),\ }\href {\doibase
  10.1103/PhysRevD.103.083533} {\bibfield  {journal} {\bibinfo  {journal}
  {Phys. Rev. D}\ }\textbf {\bibinfo {volume} {103}},\ \bibinfo {pages}
  {083533} (\bibinfo {year} {2021})},\ \Eprint
  {http://arxiv.org/abs/2007.08991} {arXiv:2007.08991 [astro-ph.CO]}
  \BibitemShut {NoStop}%
\bibitem [{\citenamefont {Riess}\ \emph {et~al.}(2021)\citenamefont {Riess},
  \citenamefont {Casertano}, \citenamefont {Yuan}, \citenamefont {Bowers},
  \citenamefont {Macri}, \citenamefont {Zinn},\ and\ \citenamefont
  {Scolnic}}]{Riess:2020fzl}%
  \BibitemOpen
  \bibfield  {author} {\bibinfo {author} {\bibfnamefont {A.~G.}\ \bibnamefont
  {Riess}}, \bibinfo {author} {\bibfnamefont {S.}~\bibnamefont {Casertano}},
  \bibinfo {author} {\bibfnamefont {W.}~\bibnamefont {Yuan}}, \bibinfo {author}
  {\bibfnamefont {J.~B.}\ \bibnamefont {Bowers}}, \bibinfo {author}
  {\bibfnamefont {L.}~\bibnamefont {Macri}}, \bibinfo {author} {\bibfnamefont
  {J.~C.}\ \bibnamefont {Zinn}}, \ and\ \bibinfo {author} {\bibfnamefont
  {D.}~\bibnamefont {Scolnic}},\ }\href {\doibase 10.3847/2041-8213/abdbaf}
  {\bibfield  {journal} {\bibinfo  {journal} {Astrophys. J. Lett.}\ }\textbf
  {\bibinfo {volume} {908}},\ \bibinfo {pages} {L6} (\bibinfo {year} {2021})},\
  \Eprint {http://arxiv.org/abs/2012.08534} {arXiv:2012.08534 [astro-ph.CO]}
  \BibitemShut {NoStop}%
\bibitem [{\citenamefont {Riess}\ \emph {et~al.}(2022)\citenamefont {Riess}
  \emph {et~al.}}]{Riess:2021jrx}%
  \BibitemOpen
  \bibfield  {author} {\bibinfo {author} {\bibfnamefont {A.~G.}\ \bibnamefont
  {Riess}} \emph {et~al.},\ }\href {\doibase 10.3847/2041-8213/ac5c5b}
  {\bibfield  {journal} {\bibinfo  {journal} {Astrophys. J. Lett.}\ }\textbf
  {\bibinfo {volume} {934}},\ \bibinfo {pages} {L7} (\bibinfo {year} {2022})},\
  \Eprint {http://arxiv.org/abs/2112.04510} {arXiv:2112.04510 [astro-ph.CO]}
  \BibitemShut {NoStop}%
\bibitem [{\citenamefont {Freedman}\ \emph {et~al.}(2019)\citenamefont
  {Freedman} \emph {et~al.}}]{Freedman:2019jwv}%
  \BibitemOpen
  \bibfield  {author} {\bibinfo {author} {\bibfnamefont {W.~L.}\ \bibnamefont
  {Freedman}} \emph {et~al.},\ }\href {\doibase 10.3847/1538-4357/ab2f73}
  {\bibfield  {journal} {\bibinfo  {journal} {Astrophys. J.}\ }\textbf
  {\bibinfo {volume} {882}},\ \bibinfo {pages} {34} (\bibinfo {year} {2019})},\
  \Eprint {http://arxiv.org/abs/1907.05922} {arXiv:1907.05922 [astro-ph.CO]}
  \BibitemShut {NoStop}%
\bibitem [{\citenamefont {Pesce}\ \emph {et~al.}(2020)\citenamefont {Pesce}
  \emph {et~al.}}]{Pesce:2020xfe}%
  \BibitemOpen
  \bibfield  {author} {\bibinfo {author} {\bibfnamefont {D.~W.}\ \bibnamefont
  {Pesce}} \emph {et~al.},\ }\href {\doibase 10.3847/2041-8213/ab75f0}
  {\bibfield  {journal} {\bibinfo  {journal} {Astrophys. J. Lett.}\ }\textbf
  {\bibinfo {volume} {891}},\ \bibinfo {pages} {L1} (\bibinfo {year} {2020})},\
  \Eprint {http://arxiv.org/abs/2001.09213} {arXiv:2001.09213 [astro-ph.CO]}
  \BibitemShut {NoStop}%
\bibitem [{\citenamefont {Kourkchi}\ \emph {et~al.}(2020)\citenamefont
  {Kourkchi}, \citenamefont {Tully}, \citenamefont {Anand}, \citenamefont
  {Courtois}, \citenamefont {Dupuy}, \citenamefont {Neill}, \citenamefont
  {Rizzi},\ and\ \citenamefont {Seibert}}]{Kourkchi:2020iyz}%
  \BibitemOpen
  \bibfield  {author} {\bibinfo {author} {\bibfnamefont {E.}~\bibnamefont
  {Kourkchi}}, \bibinfo {author} {\bibfnamefont {R.~B.}\ \bibnamefont {Tully}},
  \bibinfo {author} {\bibfnamefont {G.~S.}\ \bibnamefont {Anand}}, \bibinfo
  {author} {\bibfnamefont {H.~M.}\ \bibnamefont {Courtois}}, \bibinfo {author}
  {\bibfnamefont {A.}~\bibnamefont {Dupuy}}, \bibinfo {author} {\bibfnamefont
  {J.~D.}\ \bibnamefont {Neill}}, \bibinfo {author} {\bibfnamefont
  {L.}~\bibnamefont {Rizzi}}, \ and\ \bibinfo {author} {\bibfnamefont
  {M.}~\bibnamefont {Seibert}},\ }\href {\doibase 10.3847/1538-4357/ab901c}
  {\bibfield  {journal} {\bibinfo  {journal} {Astrophys. J.}\ }\textbf
  {\bibinfo {volume} {896}},\ \bibinfo {pages} {3} (\bibinfo {year} {2020})},\
  \Eprint {http://arxiv.org/abs/2004.14499} {arXiv:2004.14499 [astro-ph.GA]}
  \BibitemShut {NoStop}%
\bibitem [{\citenamefont {Denzel}\ \emph {et~al.}(2021)\citenamefont {Denzel},
  \citenamefont {Coles}, \citenamefont {Saha},\ and\ \citenamefont
  {Williams}}]{Denzel:2020zuq}%
  \BibitemOpen
  \bibfield  {author} {\bibinfo {author} {\bibfnamefont {P.}~\bibnamefont
  {Denzel}}, \bibinfo {author} {\bibfnamefont {J.~P.}\ \bibnamefont {Coles}},
  \bibinfo {author} {\bibfnamefont {P.}~\bibnamefont {Saha}}, \ and\ \bibinfo
  {author} {\bibfnamefont {L.~L.~R.}\ \bibnamefont {Williams}},\ }\href
  {\doibase 10.1093/mnras/staa3603} {\bibfield  {journal} {\bibinfo  {journal}
  {Mon. Not. Roy. Astron. Soc.}\ }\textbf {\bibinfo {volume} {501}},\ \bibinfo
  {pages} {784} (\bibinfo {year} {2021})},\ \Eprint
  {http://arxiv.org/abs/2007.14398} {arXiv:2007.14398 [astro-ph.CO]}
  \BibitemShut {NoStop}%
\bibitem [{\citenamefont {Wong}\ \emph {et~al.}(2020)\citenamefont {Wong} \emph
  {et~al.}}]{Wong:2019kwg}%
  \BibitemOpen
  \bibfield  {author} {\bibinfo {author} {\bibfnamefont {K.~C.}\ \bibnamefont
  {Wong}} \emph {et~al.},\ }\href {\doibase 10.1093/mnras/stz3094} {\bibfield
  {journal} {\bibinfo  {journal} {Mon. Not. Roy. Astron. Soc.}\ }\textbf
  {\bibinfo {volume} {498}},\ \bibinfo {pages} {1420} (\bibinfo {year}
  {2020})},\ \Eprint {http://arxiv.org/abs/1907.04869} {arXiv:1907.04869
  [astro-ph.CO]} \BibitemShut {NoStop}%
\bibitem [{\citenamefont {Shajib}\ \emph {et~al.}(2020)\citenamefont {Shajib}
  \emph {et~al.}}]{DES:2019fny}%
  \BibitemOpen
  \bibfield  {author} {\bibinfo {author} {\bibfnamefont {A.~J.}\ \bibnamefont
  {Shajib}} \emph {et~al.} (\bibinfo {collaboration} {DES}),\ }\href {\doibase
  10.1093/mnras/staa828} {\bibfield  {journal} {\bibinfo  {journal} {Mon. Not.
  Roy. Astron. Soc.}\ }\textbf {\bibinfo {volume} {494}},\ \bibinfo {pages}
  {6072} (\bibinfo {year} {2020})},\ \Eprint {http://arxiv.org/abs/1910.06306}
  {arXiv:1910.06306 [astro-ph.CO]} \BibitemShut {NoStop}%
\bibitem [{\citenamefont {Freedman}(2017)}]{Freedman:2017yms}%
  \BibitemOpen
  \bibfield  {author} {\bibinfo {author} {\bibfnamefont {W.~L.}\ \bibnamefont
  {Freedman}},\ }\href {\doibase 10.1038/s41550-017-0121} {\bibfield  {journal}
  {\bibinfo  {journal} {Nature Astron.}\ }\textbf {\bibinfo {volume} {1}},\
  \bibinfo {pages} {0121} (\bibinfo {year} {2017})},\ \Eprint
  {http://arxiv.org/abs/1706.02739} {arXiv:1706.02739 [astro-ph.CO]}
  \BibitemShut {NoStop}%
\bibitem [{\citenamefont {Riess}(2019)}]{Riess:2019qba}%
  \BibitemOpen
  \bibfield  {author} {\bibinfo {author} {\bibfnamefont {A.~G.}\ \bibnamefont
  {Riess}},\ }\href {\doibase 10.1038/s42254-019-0137-0} {\bibfield  {journal}
  {\bibinfo  {journal} {Nature Rev. Phys.}\ }\textbf {\bibinfo {volume} {2}},\
  \bibinfo {pages} {10} (\bibinfo {year} {2019})},\ \Eprint
  {http://arxiv.org/abs/2001.03624} {arXiv:2001.03624 [astro-ph.CO]}
  \BibitemShut {NoStop}%
\bibitem [{\citenamefont {Di~Valentino}\ \emph {et~al.}(2021)\citenamefont
  {Di~Valentino}, \citenamefont {Mena}, \citenamefont {Pan}, \citenamefont
  {Visinelli}, \citenamefont {Yang}, \citenamefont {Melchiorri}, \citenamefont
  {Mota}, \citenamefont {Riess},\ and\ \citenamefont
  {Silk}}]{DiValentino:2021izs}%
  \BibitemOpen
  \bibfield  {author} {\bibinfo {author} {\bibfnamefont {E.}~\bibnamefont
  {Di~Valentino}}, \bibinfo {author} {\bibfnamefont {O.}~\bibnamefont {Mena}},
  \bibinfo {author} {\bibfnamefont {S.}~\bibnamefont {Pan}}, \bibinfo {author}
  {\bibfnamefont {L.}~\bibnamefont {Visinelli}}, \bibinfo {author}
  {\bibfnamefont {W.}~\bibnamefont {Yang}}, \bibinfo {author} {\bibfnamefont
  {A.}~\bibnamefont {Melchiorri}}, \bibinfo {author} {\bibfnamefont {D.~F.}\
  \bibnamefont {Mota}}, \bibinfo {author} {\bibfnamefont {A.~G.}\ \bibnamefont
  {Riess}}, \ and\ \bibinfo {author} {\bibfnamefont {J.}~\bibnamefont {Silk}},\
  }\href {\doibase 10.1088/1361-6382/ac086d} {\bibfield  {journal} {\bibinfo
  {journal} {Class. Quant. Grav.}\ }\textbf {\bibinfo {volume} {38}},\ \bibinfo
  {pages} {153001} (\bibinfo {year} {2021})},\ \Eprint
  {http://arxiv.org/abs/2103.01183} {arXiv:2103.01183 [astro-ph.CO]}
  \BibitemShut {NoStop}%
\bibitem [{\citenamefont {Sch\"oneberg}\ \emph {et~al.}(2022)\citenamefont
  {Sch\"oneberg}, \citenamefont {Franco~Abell\'an}, \citenamefont
  {P\'erez~S\'anchez}, \citenamefont {Witte}, \citenamefont {Poulin},\ and\
  \citenamefont {Lesgourgues}}]{Schoneberg:2021qvd}%
  \BibitemOpen
  \bibfield  {author} {\bibinfo {author} {\bibfnamefont {N.}~\bibnamefont
  {Sch\"oneberg}}, \bibinfo {author} {\bibfnamefont {G.}~\bibnamefont
  {Franco~Abell\'an}}, \bibinfo {author} {\bibfnamefont {A.}~\bibnamefont
  {P\'erez~S\'anchez}}, \bibinfo {author} {\bibfnamefont {S.~J.}\ \bibnamefont
  {Witte}}, \bibinfo {author} {\bibfnamefont {V.}~\bibnamefont {Poulin}}, \
  and\ \bibinfo {author} {\bibfnamefont {J.}~\bibnamefont {Lesgourgues}},\
  }\href {\doibase 10.1016/j.physrep.2022.07.001} {\bibfield  {journal}
  {\bibinfo  {journal} {Phys. Rept.}\ }\textbf {\bibinfo {volume} {984}},\
  \bibinfo {pages} {1} (\bibinfo {year} {2022})},\ \Eprint
  {http://arxiv.org/abs/2107.10291} {arXiv:2107.10291 [astro-ph.CO]}
  \BibitemShut {NoStop}%
\bibitem [{\citenamefont {Cai}\ \emph {et~al.}(2022{\natexlab{a}})\citenamefont
  {Cai}, \citenamefont {Guo}, \citenamefont {Wang}, \citenamefont {Yu},\ and\
  \citenamefont {Zhou}}]{Cai:2021weh}%
  \BibitemOpen
  \bibfield  {author} {\bibinfo {author} {\bibfnamefont {R.-G.}\ \bibnamefont
  {Cai}}, \bibinfo {author} {\bibfnamefont {Z.-K.}\ \bibnamefont {Guo}},
  \bibinfo {author} {\bibfnamefont {S.-J.}\ \bibnamefont {Wang}}, \bibinfo
  {author} {\bibfnamefont {W.-W.}\ \bibnamefont {Yu}}, \ and\ \bibinfo {author}
  {\bibfnamefont {Y.}~\bibnamefont {Zhou}},\ }\href {\doibase
  10.1103/PhysRevD.105.L021301} {\bibfield  {journal} {\bibinfo  {journal}
  {Phys. Rev. D}\ }\textbf {\bibinfo {volume} {105}},\ \bibinfo {pages}
  {L021301} (\bibinfo {year} {2022}{\natexlab{a}})},\ \Eprint
  {http://arxiv.org/abs/2107.13286} {arXiv:2107.13286 [astro-ph.CO]}
  \BibitemShut {NoStop}%
\bibitem [{\citenamefont {Verde}\ \emph {et~al.}(2023)\citenamefont {Verde},
  \citenamefont {Sch\"oneberg},\ and\ \citenamefont
  {Gil-Mar\'\i{}n}}]{Verde:2023lmm}%
  \BibitemOpen
  \bibfield  {author} {\bibinfo {author} {\bibfnamefont {L.}~\bibnamefont
  {Verde}}, \bibinfo {author} {\bibfnamefont {N.}~\bibnamefont {Sch\"oneberg}},
  \ and\ \bibinfo {author} {\bibfnamefont {H.}~\bibnamefont {Gil-Mar\'\i{}n}},\
  }\href@noop {} {\  (\bibinfo {year} {2023})},\ \Eprint
  {http://arxiv.org/abs/2311.13305} {arXiv:2311.13305 [astro-ph.CO]}
  \BibitemShut {NoStop}%
\bibitem [{\citenamefont {Zhao}\ \emph {et~al.}(2017)\citenamefont {Zhao} \emph
  {et~al.}}]{Zhao:2017cud}%
  \BibitemOpen
  \bibfield  {author} {\bibinfo {author} {\bibfnamefont {G.-B.}\ \bibnamefont
  {Zhao}} \emph {et~al.},\ }\href {\doibase 10.1038/s41550-017-0216-z}
  {\bibfield  {journal} {\bibinfo  {journal} {Nature Astron.}\ }\textbf
  {\bibinfo {volume} {1}},\ \bibinfo {pages} {627} (\bibinfo {year} {2017})},\
  \Eprint {http://arxiv.org/abs/1701.08165} {arXiv:1701.08165 [astro-ph.CO]}
  \BibitemShut {NoStop}%
\bibitem [{\citenamefont {Zhang}\ \emph
  {et~al.}(2019{\natexlab{a}})\citenamefont {Zhang}, \citenamefont {Gu},
  \citenamefont {Wang}, \citenamefont {Li}, \citenamefont {Sabiu},
  \citenamefont {Park}, \citenamefont {Miao}, \citenamefont {Luo},
  \citenamefont {Fang},\ and\ \citenamefont {Li}}]{Zhang:2019jsu}%
  \BibitemOpen
  \bibfield  {author} {\bibinfo {author} {\bibfnamefont {Z.}~\bibnamefont
  {Zhang}}, \bibinfo {author} {\bibfnamefont {G.}~\bibnamefont {Gu}}, \bibinfo
  {author} {\bibfnamefont {X.}~\bibnamefont {Wang}}, \bibinfo {author}
  {\bibfnamefont {Y.-H.}\ \bibnamefont {Li}}, \bibinfo {author} {\bibfnamefont
  {C.~G.}\ \bibnamefont {Sabiu}}, \bibinfo {author} {\bibfnamefont
  {H.}~\bibnamefont {Park}}, \bibinfo {author} {\bibfnamefont {H.}~\bibnamefont
  {Miao}}, \bibinfo {author} {\bibfnamefont {X.}~\bibnamefont {Luo}}, \bibinfo
  {author} {\bibfnamefont {F.}~\bibnamefont {Fang}}, \ and\ \bibinfo {author}
  {\bibfnamefont {X.-D.}\ \bibnamefont {Li}},\ }\href {\doibase
  10.3847/1538-4357/ab1ea4} {\bibfield  {journal} {\bibinfo  {journal}
  {Astrophys. J.}\ }\textbf {\bibinfo {volume} {878}},\ \bibinfo {pages} {137}
  (\bibinfo {year} {2019}{\natexlab{a}})},\ \Eprint
  {http://arxiv.org/abs/1902.09794} {arXiv:1902.09794 [astro-ph.CO]}
  \BibitemShut {NoStop}%
\bibitem [{\citenamefont {Adame}\ \emph {et~al.}(2024)\citenamefont {Adame}
  \emph {et~al.}}]{DESI:2024mwx}%
  \BibitemOpen
  \bibfield  {author} {\bibinfo {author} {\bibfnamefont {A.~G.}\ \bibnamefont
  {Adame}} \emph {et~al.} (\bibinfo {collaboration} {DESI}),\ }\href@noop {} {\
   (\bibinfo {year} {2024})},\ \Eprint {http://arxiv.org/abs/2404.03002}
  {arXiv:2404.03002 [astro-ph.CO]} \BibitemShut {NoStop}%
\bibitem [{\citenamefont {Schutz}(1986)}]{Schutz:1986gp}%
  \BibitemOpen
  \bibfield  {author} {\bibinfo {author} {\bibfnamefont {B.~F.}\ \bibnamefont
  {Schutz}},\ }\href {\doibase 10.1038/323310a0} {\bibfield  {journal}
  {\bibinfo  {journal} {Nature}\ }\textbf {\bibinfo {volume} {323}},\ \bibinfo
  {pages} {310} (\bibinfo {year} {1986})}\BibitemShut {NoStop}%
\bibitem [{\citenamefont {Holz}\ and\ \citenamefont
  {Hughes}(2005)}]{Holz:2005df}%
  \BibitemOpen
  \bibfield  {author} {\bibinfo {author} {\bibfnamefont {D.~E.}\ \bibnamefont
  {Holz}}\ and\ \bibinfo {author} {\bibfnamefont {S.~A.}\ \bibnamefont
  {Hughes}},\ }\href {\doibase 10.1086/431341} {\bibfield  {journal} {\bibinfo
  {journal} {Astrophys. J.}\ }\textbf {\bibinfo {volume} {629}},\ \bibinfo
  {pages} {15} (\bibinfo {year} {2005})},\ \Eprint
  {http://arxiv.org/abs/astro-ph/0504616} {arXiv:astro-ph/0504616} \BibitemShut
  {NoStop}%
\bibitem [{\citenamefont {Luo}\ \emph {et~al.}(2016{\natexlab{a}})\citenamefont
  {Luo} \emph {et~al.}}]{Luo:2015ght}%
  \BibitemOpen
  \bibfield  {author} {\bibinfo {author} {\bibfnamefont {J.}~\bibnamefont
  {Luo}} \emph {et~al.} (\bibinfo {collaboration} {TianQin}),\ }\href {\doibase
  10.1088/0264-9381/33/3/035010} {\bibfield  {journal} {\bibinfo  {journal}
  {Class. Quant. Grav.}\ }\textbf {\bibinfo {volume} {33}},\ \bibinfo {pages}
  {035010} (\bibinfo {year} {2016}{\natexlab{a}})},\ \Eprint
  {http://arxiv.org/abs/1512.02076} {arXiv:1512.02076 [astro-ph.IM]}
  \BibitemShut {NoStop}%
\bibitem [{\citenamefont {Ye}\ \emph {et~al.}(2019)\citenamefont {Ye},
  \citenamefont {Zhang}, \citenamefont {Zhou}, \citenamefont {Wang},
  \citenamefont {Yuan}, \citenamefont {Gu}, \citenamefont {Ding}, \citenamefont
  {Zhang}, \citenamefont {Mei},\ and\ \citenamefont {Luo}}]{Ye:2019txh}%
  \BibitemOpen
  \bibfield  {author} {\bibinfo {author} {\bibfnamefont {B.-B.}\ \bibnamefont
  {Ye}}, \bibinfo {author} {\bibfnamefont {X.}~\bibnamefont {Zhang}}, \bibinfo
  {author} {\bibfnamefont {M.-Y.}\ \bibnamefont {Zhou}}, \bibinfo {author}
  {\bibfnamefont {Y.}~\bibnamefont {Wang}}, \bibinfo {author} {\bibfnamefont
  {H.-M.}\ \bibnamefont {Yuan}}, \bibinfo {author} {\bibfnamefont
  {D.}~\bibnamefont {Gu}}, \bibinfo {author} {\bibfnamefont {Y.}~\bibnamefont
  {Ding}}, \bibinfo {author} {\bibfnamefont {J.}~\bibnamefont {Zhang}},
  \bibinfo {author} {\bibfnamefont {J.}~\bibnamefont {Mei}}, \ and\ \bibinfo
  {author} {\bibfnamefont {J.}~\bibnamefont {Luo}},\ }\href {\doibase
  10.1142/S0218271819501219} {\bibfield  {journal} {\bibinfo  {journal} {Int.
  J. Mod. Phys. D}\ }\textbf {\bibinfo {volume} {28}},\ \bibinfo {pages}
  {1950121} (\bibinfo {year} {2019})}\BibitemShut {NoStop}%
\bibitem [{\citenamefont {Hu}\ \emph {et~al.}(2018)\citenamefont {Hu},
  \citenamefont {Li}, \citenamefont {Wang}, \citenamefont {Feng}, \citenamefont
  {Zhou}, \citenamefont {Hu}, \citenamefont {Hu}, \citenamefont {Mei},\ and\
  \citenamefont {Shao}}]{Hu:2018yqb}%
  \BibitemOpen
  \bibfield  {author} {\bibinfo {author} {\bibfnamefont {X.-C.}\ \bibnamefont
  {Hu}}, \bibinfo {author} {\bibfnamefont {X.-H.}\ \bibnamefont {Li}}, \bibinfo
  {author} {\bibfnamefont {Y.}~\bibnamefont {Wang}}, \bibinfo {author}
  {\bibfnamefont {W.-F.}\ \bibnamefont {Feng}}, \bibinfo {author}
  {\bibfnamefont {M.-Y.}\ \bibnamefont {Zhou}}, \bibinfo {author}
  {\bibfnamefont {Y.-M.}\ \bibnamefont {Hu}}, \bibinfo {author} {\bibfnamefont
  {S.-C.}\ \bibnamefont {Hu}}, \bibinfo {author} {\bibfnamefont {J.-W.}\
  \bibnamefont {Mei}}, \ and\ \bibinfo {author} {\bibfnamefont {C.-G.}\
  \bibnamefont {Shao}},\ }\href {\doibase 10.1088/1361-6382/aab52f} {\bibfield
  {journal} {\bibinfo  {journal} {Class. Quant. Grav.}\ }\textbf {\bibinfo
  {volume} {35}},\ \bibinfo {pages} {095008} (\bibinfo {year} {2018})},\
  \Eprint {http://arxiv.org/abs/1803.03368} {arXiv:1803.03368 [gr-qc]}
  \BibitemShut {NoStop}%
\bibitem [{\citenamefont {Tan}\ \emph {et~al.}(2020)\citenamefont {Tan},
  \citenamefont {Ye},\ and\ \citenamefont {Zhang}}]{Tan:2020xbm}%
  \BibitemOpen
  \bibfield  {author} {\bibinfo {author} {\bibfnamefont {Z.}~\bibnamefont
  {Tan}}, \bibinfo {author} {\bibfnamefont {B.}~\bibnamefont {Ye}}, \ and\
  \bibinfo {author} {\bibfnamefont {X.}~\bibnamefont {Zhang}},\ }\href
  {\doibase 10.1142/S021827182050056X} {\bibfield  {journal} {\bibinfo
  {journal} {Int. J. Mod. Phys. D}\ }\textbf {\bibinfo {volume} {29}},\
  \bibinfo {pages} {08} (\bibinfo {year} {2020})},\ \Eprint
  {http://arxiv.org/abs/2012.03261} {arXiv:2012.03261 [gr-qc]} \BibitemShut
  {NoStop}%
\bibitem [{\citenamefont {Strohmayer}(2005)}]{Strohmayer:2005uc}%
  \BibitemOpen
  \bibfield  {author} {\bibinfo {author} {\bibfnamefont {T.~E.}\ \bibnamefont
  {Strohmayer}},\ }\href {\doibase 10.1086/430439} {\bibfield  {journal}
  {\bibinfo  {journal} {Astrophys. J.}\ }\textbf {\bibinfo {volume} {627}},\
  \bibinfo {pages} {920} (\bibinfo {year} {2005})},\ \Eprint
  {http://arxiv.org/abs/astro-ph/0504150} {arXiv:astro-ph/0504150 [astro-ph]}
  \BibitemShut {NoStop}%
\bibitem [{\citenamefont {Lu}\ \emph {et~al.}(2019{\natexlab{a}})\citenamefont
  {Lu}, \citenamefont {Tan},\ and\ \citenamefont {Shao}}]{Lu:2019log}%
  \BibitemOpen
  \bibfield  {author} {\bibinfo {author} {\bibfnamefont {X.-Y.}\ \bibnamefont
  {Lu}}, \bibinfo {author} {\bibfnamefont {Y.-J.}\ \bibnamefont {Tan}}, \ and\
  \bibinfo {author} {\bibfnamefont {C.-G.}\ \bibnamefont {Shao}},\ }\href
  {\doibase 10.1103/PhysRevD.100.044042} {\bibfield  {journal} {\bibinfo
  {journal} {Phys. Rev. D}\ }\textbf {\bibinfo {volume} {100}},\ \bibinfo
  {pages} {044042} (\bibinfo {year} {2019}{\natexlab{a}})}\BibitemShut
  {NoStop}%
\bibitem [{\citenamefont {Hu}\ \emph {et~al.}(2017)\citenamefont {Hu},
  \citenamefont {Mei},\ and\ \citenamefont {Luo}}]{Hu:2017yoc}%
  \BibitemOpen
  \bibfield  {author} {\bibinfo {author} {\bibfnamefont {Y.-M.}\ \bibnamefont
  {Hu}}, \bibinfo {author} {\bibfnamefont {J.}~\bibnamefont {Mei}}, \ and\
  \bibinfo {author} {\bibfnamefont {J.}~\bibnamefont {Luo}},\ }\href {\doibase
  10.1093/nsr/nwx115} {\bibfield  {journal} {\bibinfo  {journal} {Natl. Sci.
  Rev.}\ }\textbf {\bibinfo {volume} {4}},\ \bibinfo {pages} {683} (\bibinfo
  {year} {2017})}\BibitemShut {NoStop}%
\bibitem [{\citenamefont {Huang}\ \emph {et~al.}(2020)\citenamefont {Huang},
  \citenamefont {Hu}, \citenamefont {Korol}, \citenamefont {Li}, \citenamefont
  {Liang}, \citenamefont {Lu}, \citenamefont {Wang}, \citenamefont {Yu},\ and\
  \citenamefont {Mei}}]{Huang:2020rjf}%
  \BibitemOpen
  \bibfield  {author} {\bibinfo {author} {\bibfnamefont {S.-J.}\ \bibnamefont
  {Huang}}, \bibinfo {author} {\bibfnamefont {Y.-M.}\ \bibnamefont {Hu}},
  \bibinfo {author} {\bibfnamefont {V.}~\bibnamefont {Korol}}, \bibinfo
  {author} {\bibfnamefont {P.-C.}\ \bibnamefont {Li}}, \bibinfo {author}
  {\bibfnamefont {Z.-C.}\ \bibnamefont {Liang}}, \bibinfo {author}
  {\bibfnamefont {Y.}~\bibnamefont {Lu}}, \bibinfo {author} {\bibfnamefont
  {H.-T.}\ \bibnamefont {Wang}}, \bibinfo {author} {\bibfnamefont
  {S.}~\bibnamefont {Yu}}, \ and\ \bibinfo {author} {\bibfnamefont
  {J.}~\bibnamefont {Mei}},\ }\href@noop {} {\  (\bibinfo {year} {2020})},\
  \Eprint {http://arxiv.org/abs/2005.07889} {arXiv:2005.07889 [astro-ph.HE]}
  \BibitemShut {NoStop}%
\bibitem [{\citenamefont {Liu}\ \emph {et~al.}(2020{\natexlab{a}})\citenamefont
  {Liu}, \citenamefont {Hu}, \citenamefont {Zhang},\ and\ \citenamefont
  {Mei}}]{Liu:2020eko}%
  \BibitemOpen
  \bibfield  {author} {\bibinfo {author} {\bibfnamefont {S.}~\bibnamefont
  {Liu}}, \bibinfo {author} {\bibfnamefont {Y.-M.}\ \bibnamefont {Hu}},
  \bibinfo {author} {\bibfnamefont {J.-d.}\ \bibnamefont {Zhang}}, \ and\
  \bibinfo {author} {\bibfnamefont {J.}~\bibnamefont {Mei}},\ }\href {\doibase
  10.1103/PhysRevD.101.103027} {\bibfield  {journal} {\bibinfo  {journal}
  {Phys. Rev. D}\ }\textbf {\bibinfo {volume} {101}},\ \bibinfo {pages}
  {103027} (\bibinfo {year} {2020}{\natexlab{a}})},\ \Eprint
  {http://arxiv.org/abs/2004.14242} {arXiv:2004.14242 [astro-ph.HE]}
  \BibitemShut {NoStop}%
\bibitem [{\citenamefont {Fan}\ \emph {et~al.}(2020)\citenamefont {Fan},
  \citenamefont {Hu}, \citenamefont {Barausse}, \citenamefont {Sesana},
  \citenamefont {Zhang}, \citenamefont {Zhang}, \citenamefont {Zi},\ and\
  \citenamefont {Mei}}]{Fan:2020zhy}%
  \BibitemOpen
  \bibfield  {author} {\bibinfo {author} {\bibfnamefont {H.-M.}\ \bibnamefont
  {Fan}}, \bibinfo {author} {\bibfnamefont {Y.-M.}\ \bibnamefont {Hu}},
  \bibinfo {author} {\bibfnamefont {E.}~\bibnamefont {Barausse}}, \bibinfo
  {author} {\bibfnamefont {A.}~\bibnamefont {Sesana}}, \bibinfo {author}
  {\bibfnamefont {J.-d.}\ \bibnamefont {Zhang}}, \bibinfo {author}
  {\bibfnamefont {X.}~\bibnamefont {Zhang}}, \bibinfo {author} {\bibfnamefont
  {T.-G.}\ \bibnamefont {Zi}}, \ and\ \bibinfo {author} {\bibfnamefont
  {J.}~\bibnamefont {Mei}},\ }\href {\doibase 10.1103/PhysRevD.102.063016}
  {\bibfield  {journal} {\bibinfo  {journal} {Phys. Rev. D}\ }\textbf {\bibinfo
  {volume} {102}},\ \bibinfo {pages} {063016} (\bibinfo {year} {2020})},\
  \Eprint {http://arxiv.org/abs/2005.08212} {arXiv:2005.08212 [astro-ph.HE]}
  \BibitemShut {NoStop}%
\bibitem [{\citenamefont {Wang}\ \emph
  {et~al.}(2019{\natexlab{a}})\citenamefont {Wang} \emph
  {et~al.}}]{Wang:2019ryf}%
  \BibitemOpen
  \bibfield  {author} {\bibinfo {author} {\bibfnamefont {H.-T.}\ \bibnamefont
  {Wang}} \emph {et~al.},\ }\href {\doibase 10.1103/PhysRevD.100.043003}
  {\bibfield  {journal} {\bibinfo  {journal} {Phys. Rev. D}\ }\textbf {\bibinfo
  {volume} {100}},\ \bibinfo {pages} {043003} (\bibinfo {year}
  {2019}{\natexlab{a}})},\ \Eprint {http://arxiv.org/abs/1902.04423}
  {arXiv:1902.04423 [astro-ph.HE]} \BibitemShut {NoStop}%
\bibitem [{\citenamefont {Liang}\ \emph
  {et~al.}(2022{\natexlab{a}})\citenamefont {Liang}, \citenamefont {Hu},
  \citenamefont {Jiang}, \citenamefont {Cheng}, \citenamefont {Zhang},\ and\
  \citenamefont {Mei}}]{Liang:2021bde}%
  \BibitemOpen
  \bibfield  {author} {\bibinfo {author} {\bibfnamefont {Z.-C.}\ \bibnamefont
  {Liang}}, \bibinfo {author} {\bibfnamefont {Y.-M.}\ \bibnamefont {Hu}},
  \bibinfo {author} {\bibfnamefont {Y.}~\bibnamefont {Jiang}}, \bibinfo
  {author} {\bibfnamefont {J.}~\bibnamefont {Cheng}}, \bibinfo {author}
  {\bibfnamefont {J.-d.}\ \bibnamefont {Zhang}}, \ and\ \bibinfo {author}
  {\bibfnamefont {J.}~\bibnamefont {Mei}},\ }\href {\doibase
  10.1103/PhysRevD.105.022001} {\bibfield  {journal} {\bibinfo  {journal}
  {Phys. Rev. D}\ }\textbf {\bibinfo {volume} {105}},\ \bibinfo {pages}
  {022001} (\bibinfo {year} {2022}{\natexlab{a}})},\ \Eprint
  {http://arxiv.org/abs/2107.08643} {arXiv:2107.08643 [astro-ph.CO]}
  \BibitemShut {NoStop}%
\bibitem [{\citenamefont {Gong}\ \emph {et~al.}(2021)\citenamefont {Gong},
  \citenamefont {Luo},\ and\ \citenamefont {Wang}}]{Gong:2021gvw}%
  \BibitemOpen
  \bibfield  {author} {\bibinfo {author} {\bibfnamefont {Y.}~\bibnamefont
  {Gong}}, \bibinfo {author} {\bibfnamefont {J.}~\bibnamefont {Luo}}, \ and\
  \bibinfo {author} {\bibfnamefont {B.}~\bibnamefont {Wang}},\ }\href {\doibase
  10.1038/s41550-021-01480-3} {\bibfield  {journal} {\bibinfo  {journal}
  {Nature Astron.}\ }\textbf {\bibinfo {volume} {5}},\ \bibinfo {pages} {881}
  (\bibinfo {year} {2021})},\ \Eprint {http://arxiv.org/abs/2109.07442}
  {arXiv:2109.07442 [astro-ph.IM]} \BibitemShut {NoStop}%
\bibitem [{\citenamefont {Amaro-Seoane}\ \emph {et~al.}(2017)\citenamefont
  {Amaro-Seoane} \emph {et~al.}}]{LISA:2017pwj}%
  \BibitemOpen
  \bibfield  {author} {\bibinfo {author} {\bibfnamefont {P.}~\bibnamefont
  {Amaro-Seoane}} \emph {et~al.} (\bibinfo {collaboration} {LISA}),\
  }\href@noop {} {\  (\bibinfo {year} {2017})},\ \Eprint
  {http://arxiv.org/abs/1702.00786} {arXiv:1702.00786 [astro-ph.IM]}
  \BibitemShut {NoStop}%
\bibitem [{\citenamefont {Hu}\ and\ \citenamefont {Wu}(2017)}]{Hu:2017mde}%
  \BibitemOpen
  \bibfield  {author} {\bibinfo {author} {\bibfnamefont {W.-R.}\ \bibnamefont
  {Hu}}\ and\ \bibinfo {author} {\bibfnamefont {Y.-L.}\ \bibnamefont {Wu}},\
  }\href {\doibase 10.1093/nsr/nwx116} {\bibfield  {journal} {\bibinfo
  {journal} {Natl. Sci. Rev.}\ }\textbf {\bibinfo {volume} {4}},\ \bibinfo
  {pages} {685} (\bibinfo {year} {2017})}\BibitemShut {NoStop}%
\bibitem [{\citenamefont {Evans}\ \emph {et~al.}(2021)\citenamefont {Evans}
  \emph {et~al.}}]{Evans:2021gyd}%
  \BibitemOpen
  \bibfield  {author} {\bibinfo {author} {\bibfnamefont {M.}~\bibnamefont
  {Evans}} \emph {et~al.},\ }\href@noop {} {\  (\bibinfo {year} {2021})},\
  \Eprint {http://arxiv.org/abs/2109.09882} {arXiv:2109.09882 [astro-ph.IM]}
  \BibitemShut {NoStop}%
\bibitem [{\citenamefont {Maggiore}\ \emph {et~al.}(2020)\citenamefont
  {Maggiore} \emph {et~al.}}]{Maggiore:2019uih}%
  \BibitemOpen
  \bibfield  {author} {\bibinfo {author} {\bibfnamefont {M.}~\bibnamefont
  {Maggiore}} \emph {et~al.},\ }\href {\doibase 10.1088/1475-7516/2020/03/050}
  {\bibfield  {journal} {\bibinfo  {journal} {JCAP}\ }\textbf {\bibinfo
  {volume} {03}},\ \bibinfo {pages} {050} (\bibinfo {year} {2020})},\ \Eprint
  {http://arxiv.org/abs/1912.02622} {arXiv:1912.02622 [astro-ph.CO]}
  \BibitemShut {NoStop}%
\bibitem [{\citenamefont {Branchesi}\ \emph {et~al.}(2023)\citenamefont
  {Branchesi} \emph {et~al.}}]{Branchesi:2023mws}%
  \BibitemOpen
  \bibfield  {author} {\bibinfo {author} {\bibfnamefont {M.}~\bibnamefont
  {Branchesi}} \emph {et~al.},\ }\href {\doibase 10.1088/1475-7516/2023/07/068}
  {\bibfield  {journal} {\bibinfo  {journal} {JCAP}\ }\textbf {\bibinfo
  {volume} {07}},\ \bibinfo {pages} {068} (\bibinfo {year} {2023})},\ \Eprint
  {http://arxiv.org/abs/2303.15923} {arXiv:2303.15923 [gr-qc]} \BibitemShut
  {NoStop}%
\bibitem [{\citenamefont {Torres-Orjuela}\ \emph {et~al.}(2024)\citenamefont
  {Torres-Orjuela}, \citenamefont {Huang}, \citenamefont {Liang}, \citenamefont
  {Liu}, \citenamefont {Wang}, \citenamefont {Ye}, \citenamefont {Hu},\ and\
  \citenamefont {Mei}}]{Torres-Orjuela:2023hfd}%
  \BibitemOpen
  \bibfield  {author} {\bibinfo {author} {\bibfnamefont {A.}~\bibnamefont
  {Torres-Orjuela}}, \bibinfo {author} {\bibfnamefont {S.-J.}\ \bibnamefont
  {Huang}}, \bibinfo {author} {\bibfnamefont {Z.-C.}\ \bibnamefont {Liang}},
  \bibinfo {author} {\bibfnamefont {S.}~\bibnamefont {Liu}}, \bibinfo {author}
  {\bibfnamefont {H.-T.}\ \bibnamefont {Wang}}, \bibinfo {author}
  {\bibfnamefont {C.-Q.}\ \bibnamefont {Ye}}, \bibinfo {author} {\bibfnamefont
  {Y.-M.}\ \bibnamefont {Hu}}, \ and\ \bibinfo {author} {\bibfnamefont
  {J.}~\bibnamefont {Mei}},\ }\href {\doibase 10.1007/s11433-023-2308-x}
  {\bibfield  {journal} {\bibinfo  {journal} {Sci. China Phys. Mech. Astron.}\
  }\textbf {\bibinfo {volume} {67}},\ \bibinfo {pages} {259511} (\bibinfo
  {year} {2024})},\ \Eprint {http://arxiv.org/abs/2307.16628} {arXiv:2307.16628
  [gr-qc]} \BibitemShut {NoStop}%
\bibitem [{\citenamefont {Seoane}\ \emph {et~al.}(2013)\citenamefont {Seoane}
  \emph {et~al.}}]{eLISA:2013xep}%
  \BibitemOpen
  \bibfield  {author} {\bibinfo {author} {\bibfnamefont {P.~A.}\ \bibnamefont
  {Seoane}} \emph {et~al.} (\bibinfo {collaboration} {eLISA}),\ }\href@noop {}
  {\  (\bibinfo {year} {2013})},\ \Eprint {http://arxiv.org/abs/1305.5720}
  {arXiv:1305.5720 [astro-ph.CO]} \BibitemShut {NoStop}%
\bibitem [{\citenamefont {Bailes}\ \emph {et~al.}(2021)\citenamefont {Bailes}
  \emph {et~al.}}]{Bailes:2021tot}%
  \BibitemOpen
  \bibfield  {author} {\bibinfo {author} {\bibfnamefont {M.}~\bibnamefont
  {Bailes}} \emph {et~al.},\ }\href {\doibase 10.1038/s42254-021-00303-8}
  {\bibfield  {journal} {\bibinfo  {journal} {Nature Rev. Phys.}\ }\textbf
  {\bibinfo {volume} {3}},\ \bibinfo {pages} {344} (\bibinfo {year}
  {2021})}\BibitemShut {NoStop}%
\bibitem [{\citenamefont {Turyshev}(2008)}]{Turyshev:2008dr}%
  \BibitemOpen
  \bibfield  {author} {\bibinfo {author} {\bibfnamefont {S.~G.}\ \bibnamefont
  {Turyshev}},\ }\href {\doibase 10.1146/annurev.nucl.58.020807.111839}
  {\bibfield  {journal} {\bibinfo  {journal} {Ann. Rev. Nucl. Part. Sci.}\
  }\textbf {\bibinfo {volume} {58}},\ \bibinfo {pages} {207} (\bibinfo {year}
  {2008})},\ \Eprint {http://arxiv.org/abs/0806.1731} {arXiv:0806.1731 [gr-qc]}
  \BibitemShut {NoStop}%
\bibitem [{\citenamefont {Turyshev}(2009)}]{Turyshev:2008ur}%
  \BibitemOpen
  \bibfield  {author} {\bibinfo {author} {\bibfnamefont {S.~G.}\ \bibnamefont
  {Turyshev}},\ }\href {\doibase 10.3367/UFNe.0179.200901a.0003} {\bibfield
  {journal} {\bibinfo  {journal} {Usp. Fiz. Nauk}\ }\textbf {\bibinfo {volume}
  {179}},\ \bibinfo {pages} {3034} (\bibinfo {year} {2009})},\ \Eprint
  {http://arxiv.org/abs/0809.3730} {arXiv:0809.3730 [gr-qc]} \BibitemShut
  {NoStop}%
\bibitem [{\citenamefont {Will}(2010)}]{Will:2010uh}%
  \BibitemOpen
  \bibfield  {author} {\bibinfo {author} {\bibfnamefont {C.~M.}\ \bibnamefont
  {Will}},\ }\href {\doibase 10.1119/1.3481700} {\bibfield  {journal} {\bibinfo
   {journal} {Am. J. Phys.}\ }\textbf {\bibinfo {volume} {78}},\ \bibinfo
  {pages} {1240} (\bibinfo {year} {2010})},\ \Eprint
  {http://arxiv.org/abs/1008.0296} {arXiv:1008.0296 [gr-qc]} \BibitemShut
  {NoStop}%
\bibitem [{\citenamefont {Murata}\ and\ \citenamefont
  {Tanaka}(2015)}]{Murata:2014nra}%
  \BibitemOpen
  \bibfield  {author} {\bibinfo {author} {\bibfnamefont {J.}~\bibnamefont
  {Murata}}\ and\ \bibinfo {author} {\bibfnamefont {S.}~\bibnamefont
  {Tanaka}},\ }\href {\doibase 10.1088/0264-9381/32/3/033001} {\bibfield
  {journal} {\bibinfo  {journal} {Class. Quant. Grav.}\ }\textbf {\bibinfo
  {volume} {32}},\ \bibinfo {pages} {033001} (\bibinfo {year} {2015})},\
  \Eprint {http://arxiv.org/abs/1408.3588} {arXiv:1408.3588 [hep-ex]}
  \BibitemShut {NoStop}%
\bibitem [{\citenamefont {Koyama}(2016)}]{Koyama:2015vza}%
  \BibitemOpen
  \bibfield  {author} {\bibinfo {author} {\bibfnamefont {K.}~\bibnamefont
  {Koyama}},\ }\href {\doibase 10.1088/0034-4885/79/4/046902} {\bibfield
  {journal} {\bibinfo  {journal} {Rept. Prog. Phys.}\ }\textbf {\bibinfo
  {volume} {79}},\ \bibinfo {pages} {046902} (\bibinfo {year} {2016})},\
  \Eprint {http://arxiv.org/abs/1504.04623} {arXiv:1504.04623 [astro-ph.CO]}
  \BibitemShut {NoStop}%
\bibitem [{\citenamefont {Sakstein}(2018)}]{Sakstein:2017pqi}%
  \BibitemOpen
  \bibfield  {author} {\bibinfo {author} {\bibfnamefont {J.}~\bibnamefont
  {Sakstein}},\ }\href {\doibase 10.1103/PhysRevD.97.064028} {\bibfield
  {journal} {\bibinfo  {journal} {Phys. Rev. D}\ }\textbf {\bibinfo {volume}
  {97}},\ \bibinfo {pages} {064028} (\bibinfo {year} {2018})},\ \Eprint
  {http://arxiv.org/abs/1710.03156} {arXiv:1710.03156 [astro-ph.CO]}
  \BibitemShut {NoStop}%
\bibitem [{\citenamefont {Gair}\ \emph {et~al.}(2013)\citenamefont {Gair},
  \citenamefont {Vallisneri}, \citenamefont {Larson},\ and\ \citenamefont
  {Baker}}]{Gair:2012nm}%
  \BibitemOpen
  \bibfield  {author} {\bibinfo {author} {\bibfnamefont {J.~R.}\ \bibnamefont
  {Gair}}, \bibinfo {author} {\bibfnamefont {M.}~\bibnamefont {Vallisneri}},
  \bibinfo {author} {\bibfnamefont {S.~L.}\ \bibnamefont {Larson}}, \ and\
  \bibinfo {author} {\bibfnamefont {J.~G.}\ \bibnamefont {Baker}},\ }\href
  {\doibase 10.12942/lrr-2013-7} {\bibfield  {journal} {\bibinfo  {journal}
  {Living Rev. Rel.}\ }\textbf {\bibinfo {volume} {16}},\ \bibinfo {pages} {7}
  (\bibinfo {year} {2013})},\ \Eprint {http://arxiv.org/abs/1212.5575}
  {arXiv:1212.5575 [gr-qc]} \BibitemShut {NoStop}%
\bibitem [{\citenamefont {Yunes}\ and\ \citenamefont
  {Siemens}(2013)}]{Yunes:2013dva}%
  \BibitemOpen
  \bibfield  {author} {\bibinfo {author} {\bibfnamefont {N.}~\bibnamefont
  {Yunes}}\ and\ \bibinfo {author} {\bibfnamefont {X.}~\bibnamefont
  {Siemens}},\ }\href {\doibase 10.12942/lrr-2013-9} {\bibfield  {journal}
  {\bibinfo  {journal} {Living Rev. Rel.}\ }\textbf {\bibinfo {volume} {16}},\
  \bibinfo {pages} {9} (\bibinfo {year} {2013})},\ \Eprint
  {http://arxiv.org/abs/1304.3473} {arXiv:1304.3473 [gr-qc]} \BibitemShut
  {NoStop}%
\bibitem [{\citenamefont {Berti}\ \emph {et~al.}(2015)\citenamefont {Berti}
  \emph {et~al.}}]{Berti:2015itd}%
  \BibitemOpen
  \bibfield  {author} {\bibinfo {author} {\bibfnamefont {E.}~\bibnamefont
  {Berti}} \emph {et~al.},\ }\href {\doibase 10.1088/0264-9381/32/24/243001}
  {\bibfield  {journal} {\bibinfo  {journal} {Class. Quant. Grav.}\ }\textbf
  {\bibinfo {volume} {32}},\ \bibinfo {pages} {243001} (\bibinfo {year}
  {2015})},\ \Eprint {http://arxiv.org/abs/1501.07274} {arXiv:1501.07274
  [gr-qc]} \BibitemShut {NoStop}%
\bibitem [{\citenamefont {Yagi}\ and\ \citenamefont
  {Stein}(2016)}]{Yagi:2016jml}%
  \BibitemOpen
  \bibfield  {author} {\bibinfo {author} {\bibfnamefont {K.}~\bibnamefont
  {Yagi}}\ and\ \bibinfo {author} {\bibfnamefont {L.~C.}\ \bibnamefont
  {Stein}},\ }\href {\doibase 10.1088/0264-9381/33/5/054001} {\bibfield
  {journal} {\bibinfo  {journal} {Class. Quant. Grav.}\ }\textbf {\bibinfo
  {volume} {33}},\ \bibinfo {pages} {054001} (\bibinfo {year} {2016})},\
  \Eprint {http://arxiv.org/abs/1602.02413} {arXiv:1602.02413 [gr-qc]}
  \BibitemShut {NoStop}%
\bibitem [{\citenamefont {Barack}\ \emph {et~al.}(2019)\citenamefont {Barack}
  \emph {et~al.}}]{Barack:2018yly}%
  \BibitemOpen
  \bibfield  {author} {\bibinfo {author} {\bibfnamefont {L.}~\bibnamefont
  {Barack}} \emph {et~al.},\ }\href {\doibase 10.1088/1361-6382/ab0587}
  {\bibfield  {journal} {\bibinfo  {journal} {Class. Quant. Grav.}\ }\textbf
  {\bibinfo {volume} {36}},\ \bibinfo {pages} {143001} (\bibinfo {year}
  {2019})},\ \Eprint {http://arxiv.org/abs/1806.05195} {arXiv:1806.05195
  [gr-qc]} \BibitemShut {NoStop}%
\bibitem [{\citenamefont {Cardoso}\ and\ \citenamefont
  {Pani}(2019)}]{Cardoso:2019rvt}%
  \BibitemOpen
  \bibfield  {author} {\bibinfo {author} {\bibfnamefont {V.}~\bibnamefont
  {Cardoso}}\ and\ \bibinfo {author} {\bibfnamefont {P.}~\bibnamefont {Pani}},\
  }\href {\doibase 10.1007/s41114-019-0020-4} {\bibfield  {journal} {\bibinfo
  {journal} {Living Rev. Rel.}\ }\textbf {\bibinfo {volume} {22}},\ \bibinfo
  {pages} {4} (\bibinfo {year} {2019})},\ \Eprint
  {http://arxiv.org/abs/1904.05363} {arXiv:1904.05363 [gr-qc]} \BibitemShut
  {NoStop}%
\bibitem [{\citenamefont {Barausse}\ \emph
  {et~al.}(2020{\natexlab{a}})\citenamefont {Barausse} \emph
  {et~al.}}]{Barausse:2020rsu}%
  \BibitemOpen
  \bibfield  {author} {\bibinfo {author} {\bibfnamefont {E.}~\bibnamefont
  {Barausse}} \emph {et~al.},\ }\href {\doibase 10.1007/s10714-020-02691-1}
  {\bibfield  {journal} {\bibinfo  {journal} {Gen. Rel. Grav.}\ }\textbf
  {\bibinfo {volume} {52}},\ \bibinfo {pages} {81} (\bibinfo {year}
  {2020}{\natexlab{a}})},\ \Eprint {http://arxiv.org/abs/2001.09793}
  {arXiv:2001.09793 [gr-qc]} \BibitemShut {NoStop}%
\bibitem [{\citenamefont {Arun}\ \emph {et~al.}(2022)\citenamefont {Arun} \emph
  {et~al.}}]{LISA:2022kgy}%
  \BibitemOpen
  \bibfield  {author} {\bibinfo {author} {\bibfnamefont {K.~G.}\ \bibnamefont
  {Arun}} \emph {et~al.} (\bibinfo {collaboration} {LISA}),\ }\href {\doibase
  10.1007/s41114-022-00036-9} {\bibfield  {journal} {\bibinfo  {journal}
  {Living Rev. Rel.}\ }\textbf {\bibinfo {volume} {25}},\ \bibinfo {pages} {4}
  (\bibinfo {year} {2022})},\ \Eprint {http://arxiv.org/abs/2205.01597}
  {arXiv:2205.01597 [gr-qc]} \BibitemShut {NoStop}%
\bibitem [{\citenamefont {Auclair}\ \emph {et~al.}(2023)\citenamefont {Auclair}
  \emph {et~al.}}]{LISACosmologyWorkingGroup:2022jok}%
  \BibitemOpen
  \bibfield  {author} {\bibinfo {author} {\bibfnamefont {P.}~\bibnamefont
  {Auclair}} \emph {et~al.} (\bibinfo {collaboration} {LISA Cosmology Working
  Group}),\ }\href {\doibase 10.1007/s41114-023-00045-2} {\bibfield  {journal}
  {\bibinfo  {journal} {Living Rev. Rel.}\ }\textbf {\bibinfo {volume} {26}},\
  \bibinfo {pages} {5} (\bibinfo {year} {2023})},\ \Eprint
  {http://arxiv.org/abs/2204.05434} {arXiv:2204.05434 [astro-ph.CO]}
  \BibitemShut {NoStop}%
\bibitem [{\citenamefont {Debono}\ and\ \citenamefont
  {Smoot}(2016)}]{Debono:2016vkp}%
  \BibitemOpen
  \bibfield  {author} {\bibinfo {author} {\bibfnamefont {I.}~\bibnamefont
  {Debono}}\ and\ \bibinfo {author} {\bibfnamefont {G.~F.}\ \bibnamefont
  {Smoot}},\ }\href {\doibase 10.3390/universe2040023} {\bibfield  {journal}
  {\bibinfo  {journal} {Universe}\ }\textbf {\bibinfo {volume} {2}},\ \bibinfo
  {pages} {23} (\bibinfo {year} {2016})},\ \Eprint
  {http://arxiv.org/abs/1609.09781} {arXiv:1609.09781 [gr-qc]} \BibitemShut
  {NoStop}%
\bibitem [{\citenamefont {Baumann}(2022)}]{Baumann:2022mni}%
  \BibitemOpen
  \bibfield  {author} {\bibinfo {author} {\bibfnamefont {D.}~\bibnamefont
  {Baumann}},\ }\href {\doibase 10.1017/9781108937092} {\emph {\bibinfo {title}
  {{Cosmology}}}}\ (\bibinfo  {publisher} {Cambridge University Press},\
  \bibinfo {year} {2022})\BibitemShut {NoStop}%
\bibitem [{\citenamefont {Abbott}\ \emph
  {et~al.}(2016{\natexlab{a}})\citenamefont {Abbott} \emph
  {et~al.}}]{LIGOScientific:2016lio}%
  \BibitemOpen
  \bibfield  {author} {\bibinfo {author} {\bibfnamefont {B.~P.}\ \bibnamefont
  {Abbott}} \emph {et~al.} (\bibinfo {collaboration} {LIGO Scientific,
  Virgo}),\ }\href {\doibase 10.1103/PhysRevLett.116.221101} {\bibfield
  {journal} {\bibinfo  {journal} {Phys. Rev. Lett.}\ }\textbf {\bibinfo
  {volume} {116}},\ \bibinfo {pages} {221101} (\bibinfo {year}
  {2016}{\natexlab{a}})},\ \bibinfo {note} {[Erratum: Phys.Rev.Lett. 121,
  129902 (2018)]},\ \Eprint {http://arxiv.org/abs/1602.03841} {arXiv:1602.03841
  [gr-qc]} \BibitemShut {NoStop}%
\bibitem [{\citenamefont {Abbott}\ \emph
  {et~al.}(2019{\natexlab{a}})\citenamefont {Abbott} \emph
  {et~al.}}]{LIGOScientific:2019fpa}%
  \BibitemOpen
  \bibfield  {author} {\bibinfo {author} {\bibfnamefont {B.~P.}\ \bibnamefont
  {Abbott}} \emph {et~al.} (\bibinfo {collaboration} {LIGO Scientific,
  Virgo}),\ }\href {\doibase 10.1103/PhysRevD.100.104036} {\bibfield  {journal}
  {\bibinfo  {journal} {Phys. Rev. D}\ }\textbf {\bibinfo {volume} {100}},\
  \bibinfo {pages} {104036} (\bibinfo {year} {2019}{\natexlab{a}})},\ \Eprint
  {http://arxiv.org/abs/1903.04467} {arXiv:1903.04467 [gr-qc]} \BibitemShut
  {NoStop}%
\bibitem [{\citenamefont {Abbott}\ \emph
  {et~al.}(2021{\natexlab{a}})\citenamefont {Abbott} \emph
  {et~al.}}]{LIGOScientific:2020tif}%
  \BibitemOpen
  \bibfield  {author} {\bibinfo {author} {\bibfnamefont {R.}~\bibnamefont
  {Abbott}} \emph {et~al.} (\bibinfo {collaboration} {LIGO Scientific,
  Virgo}),\ }\href {\doibase 10.1103/PhysRevD.103.122002} {\bibfield  {journal}
  {\bibinfo  {journal} {Phys. Rev. D}\ }\textbf {\bibinfo {volume} {103}},\
  \bibinfo {pages} {122002} (\bibinfo {year} {2021}{\natexlab{a}})},\ \Eprint
  {http://arxiv.org/abs/2010.14529} {arXiv:2010.14529 [gr-qc]} \BibitemShut
  {NoStop}%
\bibitem [{\citenamefont {Abbott}\ \emph
  {et~al.}(2021{\natexlab{b}})\citenamefont {Abbott} \emph
  {et~al.}}]{LIGOScientific:2021sio}%
  \BibitemOpen
  \bibfield  {author} {\bibinfo {author} {\bibfnamefont {R.}~\bibnamefont
  {Abbott}} \emph {et~al.} (\bibinfo {collaboration} {LIGO Scientific, VIRGO,
  KAGRA}),\ }\href@noop {} {\  (\bibinfo {year} {2021}{\natexlab{b}})},\
  \Eprint {http://arxiv.org/abs/2112.06861} {arXiv:2112.06861 [gr-qc]}
  \BibitemShut {NoStop}%
\bibitem [{\citenamefont {Shi}\ \emph {et~al.}(2024)\citenamefont {Shi},
  \citenamefont {Zhang},\ and\ \citenamefont {Mei}}]{Shi:2024ttu}%
  \BibitemOpen
  \bibfield  {author} {\bibinfo {author} {\bibfnamefont {C.}~\bibnamefont
  {Shi}}, \bibinfo {author} {\bibfnamefont {Q.}~\bibnamefont {Zhang}}, \ and\
  \bibinfo {author} {\bibfnamefont {J.}~\bibnamefont {Mei}},\ }\href {\doibase
  10.1103/PhysRevD.110.124007} {\bibfield  {journal} {\bibinfo  {journal}
  {Phys. Rev. D}\ }\textbf {\bibinfo {volume} {110}},\ \bibinfo {pages}
  {124007} (\bibinfo {year} {2024})},\ \Eprint
  {http://arxiv.org/abs/2407.13110} {arXiv:2407.13110 [gr-qc]} \BibitemShut
  {NoStop}%
\bibitem [{\citenamefont {Shi}\ \emph {et~al.}(2019)\citenamefont {Shi},
  \citenamefont {Bao}, \citenamefont {Wang}, \citenamefont {Zhang},
  \citenamefont {Hu}, \citenamefont {Sesana}, \citenamefont {Barausse},
  \citenamefont {Mei},\ and\ \citenamefont {Luo}}]{Shi:2019hqa}%
  \BibitemOpen
  \bibfield  {author} {\bibinfo {author} {\bibfnamefont {C.}~\bibnamefont
  {Shi}}, \bibinfo {author} {\bibfnamefont {J.}~\bibnamefont {Bao}}, \bibinfo
  {author} {\bibfnamefont {H.}~\bibnamefont {Wang}}, \bibinfo {author}
  {\bibfnamefont {J.-d.}\ \bibnamefont {Zhang}}, \bibinfo {author}
  {\bibfnamefont {Y.}~\bibnamefont {Hu}}, \bibinfo {author} {\bibfnamefont
  {A.}~\bibnamefont {Sesana}}, \bibinfo {author} {\bibfnamefont
  {E.}~\bibnamefont {Barausse}}, \bibinfo {author} {\bibfnamefont
  {J.}~\bibnamefont {Mei}}, \ and\ \bibinfo {author} {\bibfnamefont
  {J.}~\bibnamefont {Luo}},\ }\href {\doibase 10.1103/PhysRevD.100.044036}
  {\bibfield  {journal} {\bibinfo  {journal} {Phys. Rev. D}\ }\textbf {\bibinfo
  {volume} {100}},\ \bibinfo {pages} {044036} (\bibinfo {year} {2019})},\
  \Eprint {http://arxiv.org/abs/1902.08922} {arXiv:1902.08922 [gr-qc]}
  \BibitemShut {NoStop}%
\bibitem [{\citenamefont {Zi}\ \emph {et~al.}(2021)\citenamefont {Zi},
  \citenamefont {Zhang}, \citenamefont {Fan}, \citenamefont {Zhang},
  \citenamefont {Hu}, \citenamefont {Shi},\ and\ \citenamefont
  {Mei}}]{Zi:2021pdp}%
  \BibitemOpen
  \bibfield  {author} {\bibinfo {author} {\bibfnamefont {T.-G.}\ \bibnamefont
  {Zi}}, \bibinfo {author} {\bibfnamefont {J.-D.}\ \bibnamefont {Zhang}},
  \bibinfo {author} {\bibfnamefont {H.-M.}\ \bibnamefont {Fan}}, \bibinfo
  {author} {\bibfnamefont {X.-T.}\ \bibnamefont {Zhang}}, \bibinfo {author}
  {\bibfnamefont {Y.-M.}\ \bibnamefont {Hu}}, \bibinfo {author} {\bibfnamefont
  {C.}~\bibnamefont {Shi}}, \ and\ \bibinfo {author} {\bibfnamefont
  {J.}~\bibnamefont {Mei}},\ }\href {\doibase 10.1103/PhysRevD.104.064008}
  {\bibfield  {journal} {\bibinfo  {journal} {Phys. Rev. D}\ }\textbf {\bibinfo
  {volume} {104}},\ \bibinfo {pages} {064008} (\bibinfo {year} {2021})},\
  \Eprint {http://arxiv.org/abs/2104.06047} {arXiv:2104.06047 [gr-qc]}
  \BibitemShut {NoStop}%
\bibitem [{\citenamefont {Abbott}\ \emph
  {et~al.}(2016{\natexlab{b}})\citenamefont {Abbott} \emph
  {et~al.}}]{LIGOScientific:2016aoc}%
  \BibitemOpen
  \bibfield  {author} {\bibinfo {author} {\bibfnamefont {B.~P.}\ \bibnamefont
  {Abbott}} \emph {et~al.} (\bibinfo {collaboration} {LIGO Scientific,
  Virgo}),\ }\href {\doibase 10.1103/PhysRevLett.116.061102} {\bibfield
  {journal} {\bibinfo  {journal} {Phys. Rev. Lett.}\ }\textbf {\bibinfo
  {volume} {116}},\ \bibinfo {pages} {061102} (\bibinfo {year}
  {2016}{\natexlab{b}})},\ \Eprint {http://arxiv.org/abs/1602.03837}
  {arXiv:1602.03837 [gr-qc]} \BibitemShut {NoStop}%
\bibitem [{\citenamefont {Shapiro}(1964)}]{Shapiro:1964uw}%
  \BibitemOpen
  \bibfield  {author} {\bibinfo {author} {\bibfnamefont {I.~I.}\ \bibnamefont
  {Shapiro}},\ }\href {\doibase 10.1103/PhysRevLett.13.789} {\bibfield
  {journal} {\bibinfo  {journal} {Phys. Rev. Lett.}\ }\textbf {\bibinfo
  {volume} {13}},\ \bibinfo {pages} {789} (\bibinfo {year} {1964})}\BibitemShut
  {NoStop}%
\bibitem [{\citenamefont {Kerr}\ and\ \citenamefont
  {Schild}(1965)}]{Kerr:1965wfc}%
  \BibitemOpen
  \bibfield  {author} {\bibinfo {author} {\bibfnamefont {R.~P.}\ \bibnamefont
  {Kerr}}\ and\ \bibinfo {author} {\bibfnamefont {A.}~\bibnamefont {Schild}},\
  }\href@noop {} {\bibfield  {journal} {\bibinfo  {journal} {Proc. Symp. Appl.
  Math.}\ }\textbf {\bibinfo {volume} {17}},\ \bibinfo {pages} {199} (\bibinfo
  {year} {1965})}\BibitemShut {NoStop}%
\bibitem [{\citenamefont {Harte}(2014)}]{Harte:2014ooa}%
  \BibitemOpen
  \bibfield  {author} {\bibinfo {author} {\bibfnamefont {A.~I.}\ \bibnamefont
  {Harte}},\ }\href {\doibase 10.1103/PhysRevLett.113.261103} {\bibfield
  {journal} {\bibinfo  {journal} {Phys. Rev. Lett.}\ }\textbf {\bibinfo
  {volume} {113}},\ \bibinfo {pages} {261103} (\bibinfo {year} {2014})},\
  \Eprint {http://arxiv.org/abs/1409.4674} {arXiv:1409.4674 [gr-qc]}
  \BibitemShut {NoStop}%
\bibitem [{\citenamefont {Pretorius}(2005)}]{Pretorius:2005gq}%
  \BibitemOpen
  \bibfield  {author} {\bibinfo {author} {\bibfnamefont {F.}~\bibnamefont
  {Pretorius}},\ }\href {\doibase 10.1103/PhysRevLett.95.121101} {\bibfield
  {journal} {\bibinfo  {journal} {Phys. Rev. Lett.}\ }\textbf {\bibinfo
  {volume} {95}},\ \bibinfo {pages} {121101} (\bibinfo {year} {2005})},\
  \Eprint {http://arxiv.org/abs/gr-qc/0507014} {arXiv:gr-qc/0507014}
  \BibitemShut {NoStop}%
\bibitem [{\citenamefont {Kokkotas}\ and\ \citenamefont
  {Schmidt}(1999)}]{Kokkotas:1999bd}%
  \BibitemOpen
  \bibfield  {author} {\bibinfo {author} {\bibfnamefont {K.~D.}\ \bibnamefont
  {Kokkotas}}\ and\ \bibinfo {author} {\bibfnamefont {B.~G.}\ \bibnamefont
  {Schmidt}},\ }\href {\doibase 10.12942/lrr-1999-2} {\bibfield  {journal}
  {\bibinfo  {journal} {Living Rev. Rel.}\ }\textbf {\bibinfo {volume} {2}},\
  \bibinfo {pages} {2} (\bibinfo {year} {1999})},\ \Eprint
  {http://arxiv.org/abs/gr-qc/9909058} {arXiv:gr-qc/9909058 [gr-qc]}
  \BibitemShut {NoStop}%
\bibitem [{\citenamefont {Berti}\ \emph {et~al.}(2009)\citenamefont {Berti},
  \citenamefont {Cardoso},\ and\ \citenamefont {Starinets}}]{Berti:2009kk}%
  \BibitemOpen
  \bibfield  {author} {\bibinfo {author} {\bibfnamefont {E.}~\bibnamefont
  {Berti}}, \bibinfo {author} {\bibfnamefont {V.}~\bibnamefont {Cardoso}}, \
  and\ \bibinfo {author} {\bibfnamefont {A.~O.}\ \bibnamefont {Starinets}},\
  }\href {\doibase 10.1088/0264-9381/26/16/163001} {\bibfield  {journal}
  {\bibinfo  {journal} {Class. Quant. Grav.}\ }\textbf {\bibinfo {volume}
  {26}},\ \bibinfo {pages} {163001} (\bibinfo {year} {2009})},\ \Eprint
  {http://arxiv.org/abs/0905.2975} {arXiv:0905.2975 [gr-qc]} \BibitemShut
  {NoStop}%
\bibitem [{\citenamefont {Konoplya}\ and\ \citenamefont
  {Zhidenko}(2011)}]{Konoplya:2011qq}%
  \BibitemOpen
  \bibfield  {author} {\bibinfo {author} {\bibfnamefont {R.~A.}\ \bibnamefont
  {Konoplya}}\ and\ \bibinfo {author} {\bibfnamefont {A.}~\bibnamefont
  {Zhidenko}},\ }\href {\doibase 10.1103/RevModPhys.83.793} {\bibfield
  {journal} {\bibinfo  {journal} {Rev. Mod. Phys.}\ }\textbf {\bibinfo {volume}
  {83}},\ \bibinfo {pages} {793} (\bibinfo {year} {2011})},\ \Eprint
  {http://arxiv.org/abs/1102.4014} {arXiv:1102.4014 [gr-qc]} \BibitemShut
  {NoStop}%
\bibitem [{\citenamefont {Isi}\ \emph {et~al.}(2019)\citenamefont {Isi},
  \citenamefont {Giesler}, \citenamefont {Farr}, \citenamefont {Scheel},\ and\
  \citenamefont {Teukolsky}}]{Isi:2019aib}%
  \BibitemOpen
  \bibfield  {author} {\bibinfo {author} {\bibfnamefont {M.}~\bibnamefont
  {Isi}}, \bibinfo {author} {\bibfnamefont {M.}~\bibnamefont {Giesler}},
  \bibinfo {author} {\bibfnamefont {W.~M.}\ \bibnamefont {Farr}}, \bibinfo
  {author} {\bibfnamefont {M.~A.}\ \bibnamefont {Scheel}}, \ and\ \bibinfo
  {author} {\bibfnamefont {S.~A.}\ \bibnamefont {Teukolsky}},\ }\href {\doibase
  10.1103/PhysRevLett.123.111102} {\bibfield  {journal} {\bibinfo  {journal}
  {Phys. Rev. Lett.}\ }\textbf {\bibinfo {volume} {123}},\ \bibinfo {pages}
  {111102} (\bibinfo {year} {2019})},\ \Eprint
  {http://arxiv.org/abs/1905.00869} {arXiv:1905.00869 [gr-qc]} \BibitemShut
  {NoStop}%
\bibitem [{\citenamefont {Carullo}\ \emph {et~al.}(2019)\citenamefont
  {Carullo}, \citenamefont {Del~Pozzo},\ and\ \citenamefont
  {Veitch}}]{Carullo:2019flw}%
  \BibitemOpen
  \bibfield  {author} {\bibinfo {author} {\bibfnamefont {G.}~\bibnamefont
  {Carullo}}, \bibinfo {author} {\bibfnamefont {W.}~\bibnamefont {Del~Pozzo}},
  \ and\ \bibinfo {author} {\bibfnamefont {J.}~\bibnamefont {Veitch}},\ }\href
  {\doibase 10.1103/PhysRevD.99.123029} {\bibfield  {journal} {\bibinfo
  {journal} {Phys. Rev. D}\ }\textbf {\bibinfo {volume} {99}},\ \bibinfo
  {pages} {123029} (\bibinfo {year} {2019})},\ \bibinfo {note} {[Erratum:
  Phys.Rev.D 100, 089903 (2019)]},\ \Eprint {http://arxiv.org/abs/1902.07527}
  {arXiv:1902.07527 [gr-qc]} \BibitemShut {NoStop}%
\bibitem [{\citenamefont {Wang}\ and\ \citenamefont
  {Shao}(2023)}]{Wang:2023mst}%
  \BibitemOpen
  \bibfield  {author} {\bibinfo {author} {\bibfnamefont {H.-T.}\ \bibnamefont
  {Wang}}\ and\ \bibinfo {author} {\bibfnamefont {L.}~\bibnamefont {Shao}},\
  }\href {\doibase 10.1103/PhysRevD.108.123018} {\bibfield  {journal} {\bibinfo
   {journal} {Phys. Rev. D}\ }\textbf {\bibinfo {volume} {108}},\ \bibinfo
  {pages} {123018} (\bibinfo {year} {2023})},\ \Eprint
  {http://arxiv.org/abs/2311.13300} {arXiv:2311.13300 [gr-qc]} \BibitemShut
  {NoStop}%
\bibitem [{\citenamefont {Wang}\ \emph
  {et~al.}(2024{\natexlab{a}})\citenamefont {Wang}, \citenamefont {Wang},
  \citenamefont {Dong}, \citenamefont {Yim},\ and\ \citenamefont
  {Shao}}]{Wang:2024yhb}%
  \BibitemOpen
  \bibfield  {author} {\bibinfo {author} {\bibfnamefont {H.-T.}\ \bibnamefont
  {Wang}}, \bibinfo {author} {\bibfnamefont {Z.}~\bibnamefont {Wang}}, \bibinfo
  {author} {\bibfnamefont {Y.}~\bibnamefont {Dong}}, \bibinfo {author}
  {\bibfnamefont {G.}~\bibnamefont {Yim}}, \ and\ \bibinfo {author}
  {\bibfnamefont {L.}~\bibnamefont {Shao}},\ }\href@noop {} {\  (\bibinfo
  {year} {2024}{\natexlab{a}})},\ \Eprint {http://arxiv.org/abs/2411.13333}
  {arXiv:2411.13333 [gr-qc]} \BibitemShut {NoStop}%
\bibitem [{\citenamefont {Capano}\ \emph {et~al.}(2024)\citenamefont {Capano},
  \citenamefont {Abedi}, \citenamefont {Kastha}, \citenamefont {Nitz},
  \citenamefont {Westerweck}, \citenamefont {Wang}, \citenamefont {Cabero},
  \citenamefont {Nielsen},\ and\ \citenamefont {Krishnan}}]{Capano:2022zqm}%
  \BibitemOpen
  \bibfield  {author} {\bibinfo {author} {\bibfnamefont {C.~D.}\ \bibnamefont
  {Capano}}, \bibinfo {author} {\bibfnamefont {J.}~\bibnamefont {Abedi}},
  \bibinfo {author} {\bibfnamefont {S.}~\bibnamefont {Kastha}}, \bibinfo
  {author} {\bibfnamefont {A.~H.}\ \bibnamefont {Nitz}}, \bibinfo {author}
  {\bibfnamefont {J.}~\bibnamefont {Westerweck}}, \bibinfo {author}
  {\bibfnamefont {Y.-F.}\ \bibnamefont {Wang}}, \bibinfo {author}
  {\bibfnamefont {M.}~\bibnamefont {Cabero}}, \bibinfo {author} {\bibfnamefont
  {A.~B.}\ \bibnamefont {Nielsen}}, \ and\ \bibinfo {author} {\bibfnamefont
  {B.}~\bibnamefont {Krishnan}},\ }\href {\doibase 10.1088/1361-6382/ad84ae}
  {\bibfield  {journal} {\bibinfo  {journal} {Class. Quant. Grav.}\ }\textbf
  {\bibinfo {volume} {41}},\ \bibinfo {pages} {245009} (\bibinfo {year}
  {2024})},\ \Eprint {http://arxiv.org/abs/2209.00640} {arXiv:2209.00640
  [gr-qc]} \BibitemShut {NoStop}%
\bibitem [{\citenamefont {Berti}\ \emph {et~al.}(2006)\citenamefont {Berti},
  \citenamefont {Cardoso},\ and\ \citenamefont {Will}}]{Berti:2005ys}%
  \BibitemOpen
  \bibfield  {author} {\bibinfo {author} {\bibfnamefont {E.}~\bibnamefont
  {Berti}}, \bibinfo {author} {\bibfnamefont {V.}~\bibnamefont {Cardoso}}, \
  and\ \bibinfo {author} {\bibfnamefont {C.~M.}\ \bibnamefont {Will}},\ }\href
  {\doibase 10.1103/PhysRevD.73.064030} {\bibfield  {journal} {\bibinfo
  {journal} {Phys. Rev. D}\ }\textbf {\bibinfo {volume} {73}},\ \bibinfo
  {pages} {064030} (\bibinfo {year} {2006})},\ \Eprint
  {http://arxiv.org/abs/gr-qc/0512160} {arXiv:gr-qc/0512160} \BibitemShut
  {NoStop}%
\bibitem [{\citenamefont {Baibhav}\ and\ \citenamefont
  {Berti}(2019)}]{Baibhav:2018rfk}%
  \BibitemOpen
  \bibfield  {author} {\bibinfo {author} {\bibfnamefont {V.}~\bibnamefont
  {Baibhav}}\ and\ \bibinfo {author} {\bibfnamefont {E.}~\bibnamefont
  {Berti}},\ }\href {\doibase 10.1103/PhysRevD.99.024005} {\bibfield  {journal}
  {\bibinfo  {journal} {Phys. Rev. D}\ }\textbf {\bibinfo {volume} {99}},\
  \bibinfo {pages} {024005} (\bibinfo {year} {2019})},\ \Eprint
  {http://arxiv.org/abs/1809.03500} {arXiv:1809.03500 [gr-qc]} \BibitemShut
  {NoStop}%
\bibitem [{\citenamefont {Kamaretsos}\ \emph {et~al.}(2012)\citenamefont
  {Kamaretsos}, \citenamefont {Hannam}, \citenamefont {Husa},\ and\
  \citenamefont {Sathyaprakash}}]{Kamaretsos:2011um}%
  \BibitemOpen
  \bibfield  {author} {\bibinfo {author} {\bibfnamefont {I.}~\bibnamefont
  {Kamaretsos}}, \bibinfo {author} {\bibfnamefont {M.}~\bibnamefont {Hannam}},
  \bibinfo {author} {\bibfnamefont {S.}~\bibnamefont {Husa}}, \ and\ \bibinfo
  {author} {\bibfnamefont {B.~S.}\ \bibnamefont {Sathyaprakash}},\ }\href
  {\doibase 10.1103/PhysRevD.85.024018} {\bibfield  {journal} {\bibinfo
  {journal} {Phys. Rev.}\ }\textbf {\bibinfo {volume} {D85}},\ \bibinfo {pages}
  {024018} (\bibinfo {year} {2012})},\ \Eprint {http://arxiv.org/abs/1107.0854}
  {arXiv:1107.0854 [gr-qc]} \BibitemShut {NoStop}%
\bibitem [{\citenamefont {Ioka}\ and\ \citenamefont
  {Nakano}(2007)}]{Ioka:2007ak}%
  \BibitemOpen
  \bibfield  {author} {\bibinfo {author} {\bibfnamefont {K.}~\bibnamefont
  {Ioka}}\ and\ \bibinfo {author} {\bibfnamefont {H.}~\bibnamefont {Nakano}},\
  }\href {\doibase 10.1103/PhysRevD.76.061503} {\bibfield  {journal} {\bibinfo
  {journal} {Phys. Rev. D}\ }\textbf {\bibinfo {volume} {76}},\ \bibinfo
  {pages} {061503} (\bibinfo {year} {2007})},\ \Eprint
  {http://arxiv.org/abs/0704.3467} {arXiv:0704.3467 [astro-ph]} \BibitemShut
  {NoStop}%
\bibitem [{\citenamefont {London}\ \emph {et~al.}(2014)\citenamefont {London},
  \citenamefont {Shoemaker},\ and\ \citenamefont {Healy}}]{London:2014cma}%
  \BibitemOpen
  \bibfield  {author} {\bibinfo {author} {\bibfnamefont {L.}~\bibnamefont
  {London}}, \bibinfo {author} {\bibfnamefont {D.}~\bibnamefont {Shoemaker}}, \
  and\ \bibinfo {author} {\bibfnamefont {J.}~\bibnamefont {Healy}},\ }\href
  {\doibase 10.1103/PhysRevD.90.124032} {\bibfield  {journal} {\bibinfo
  {journal} {Phys. Rev. D}\ }\textbf {\bibinfo {volume} {90}},\ \bibinfo
  {pages} {124032} (\bibinfo {year} {2014})},\ \bibinfo {note} {[Erratum:
  Phys.Rev.D 94, 069902 (2016)]},\ \Eprint {http://arxiv.org/abs/1404.3197}
  {arXiv:1404.3197 [gr-qc]} \BibitemShut {NoStop}%
\bibitem [{\citenamefont {Mitman}\ \emph {et~al.}(2023)\citenamefont {Mitman}
  \emph {et~al.}}]{Mitman:2022qdl}%
  \BibitemOpen
  \bibfield  {author} {\bibinfo {author} {\bibfnamefont {K.}~\bibnamefont
  {Mitman}} \emph {et~al.},\ }\href {\doibase 10.1103/PhysRevLett.130.081402}
  {\bibfield  {journal} {\bibinfo  {journal} {Phys. Rev. Lett.}\ }\textbf
  {\bibinfo {volume} {130}},\ \bibinfo {pages} {081402} (\bibinfo {year}
  {2023})},\ \Eprint {http://arxiv.org/abs/2208.07380} {arXiv:2208.07380
  [gr-qc]} \BibitemShut {NoStop}%
\bibitem [{\citenamefont {Cheung}\ \emph {et~al.}(2023)\citenamefont {Cheung}
  \emph {et~al.}}]{Cheung:2022rbm}%
  \BibitemOpen
  \bibfield  {author} {\bibinfo {author} {\bibfnamefont {M.~H.-Y.}\
  \bibnamefont {Cheung}} \emph {et~al.},\ }\href {\doibase
  10.1103/PhysRevLett.130.081401} {\bibfield  {journal} {\bibinfo  {journal}
  {Phys. Rev. Lett.}\ }\textbf {\bibinfo {volume} {130}},\ \bibinfo {pages}
  {081401} (\bibinfo {year} {2023})},\ \Eprint
  {http://arxiv.org/abs/2208.07374} {arXiv:2208.07374 [gr-qc]} \BibitemShut
  {NoStop}%
\bibitem [{\citenamefont {Du}\ and\ \citenamefont
  {Nishizawa}(2016)}]{Du:2016hww}%
  \BibitemOpen
  \bibfield  {author} {\bibinfo {author} {\bibfnamefont {S.~M.}\ \bibnamefont
  {Du}}\ and\ \bibinfo {author} {\bibfnamefont {A.}~\bibnamefont {Nishizawa}},\
  }\href {\doibase 10.1103/PhysRevD.94.104063} {\bibfield  {journal} {\bibinfo
  {journal} {Phys. Rev. D}\ }\textbf {\bibinfo {volume} {94}},\ \bibinfo
  {pages} {104063} (\bibinfo {year} {2016})},\ \Eprint
  {http://arxiv.org/abs/1609.09825} {arXiv:1609.09825 [gr-qc]} \BibitemShut
  {NoStop}%
\bibitem [{\citenamefont {Seraj}(2021)}]{Seraj:2021qja}%
  \BibitemOpen
  \bibfield  {author} {\bibinfo {author} {\bibfnamefont {A.}~\bibnamefont
  {Seraj}},\ }\href {\doibase 10.1007/JHEP05(2021)283} {\bibfield  {journal}
  {\bibinfo  {journal} {JHEP}\ }\textbf {\bibinfo {volume} {05}},\ \bibinfo
  {pages} {283} (\bibinfo {year} {2021})},\ \Eprint
  {http://arxiv.org/abs/2103.12185} {arXiv:2103.12185 [hep-th]} \BibitemShut
  {NoStop}%
\bibitem [{\citenamefont {Tahura}\ \emph {et~al.}(2021)\citenamefont {Tahura},
  \citenamefont {Nichols},\ and\ \citenamefont {Yagi}}]{Tahura:2021hbk}%
  \BibitemOpen
  \bibfield  {author} {\bibinfo {author} {\bibfnamefont {S.}~\bibnamefont
  {Tahura}}, \bibinfo {author} {\bibfnamefont {D.~A.}\ \bibnamefont {Nichols}},
  \ and\ \bibinfo {author} {\bibfnamefont {K.}~\bibnamefont {Yagi}},\ }\href
  {\doibase 10.1103/PhysRevD.104.104010} {\bibfield  {journal} {\bibinfo
  {journal} {Phys. Rev. D}\ }\textbf {\bibinfo {volume} {104}},\ \bibinfo
  {pages} {104010} (\bibinfo {year} {2021})},\ \Eprint
  {http://arxiv.org/abs/2107.02208} {arXiv:2107.02208 [gr-qc]} \BibitemShut
  {NoStop}%
\bibitem [{\citenamefont {Hou}\ \emph {et~al.}(2022{\natexlab{a}})\citenamefont
  {Hou}, \citenamefont {Zhu},\ and\ \citenamefont {Zhu}}]{Hou:2021oxe}%
  \BibitemOpen
  \bibfield  {author} {\bibinfo {author} {\bibfnamefont {S.}~\bibnamefont
  {Hou}}, \bibinfo {author} {\bibfnamefont {T.}~\bibnamefont {Zhu}}, \ and\
  \bibinfo {author} {\bibfnamefont {Z.-H.}\ \bibnamefont {Zhu}},\ }\href
  {\doibase 10.1103/PhysRevD.105.024025} {\bibfield  {journal} {\bibinfo
  {journal} {Phys. Rev. D}\ }\textbf {\bibinfo {volume} {105}},\ \bibinfo
  {pages} {024025} (\bibinfo {year} {2022}{\natexlab{a}})},\ \Eprint
  {http://arxiv.org/abs/2109.04238} {arXiv:2109.04238 [gr-qc]} \BibitemShut
  {NoStop}%
\bibitem [{\citenamefont {Hou}\ \emph {et~al.}(2022{\natexlab{b}})\citenamefont
  {Hou}, \citenamefont {Zhu},\ and\ \citenamefont {Zhu}}]{Hou:2021bxz}%
  \BibitemOpen
  \bibfield  {author} {\bibinfo {author} {\bibfnamefont {S.}~\bibnamefont
  {Hou}}, \bibinfo {author} {\bibfnamefont {T.}~\bibnamefont {Zhu}}, \ and\
  \bibinfo {author} {\bibfnamefont {Z.-H.}\ \bibnamefont {Zhu}},\ }\href
  {\doibase 10.1088/1475-7516/2022/04/032} {\bibfield  {journal} {\bibinfo
  {journal} {JCAP}\ }\textbf {\bibinfo {volume} {04}},\ \bibinfo {pages} {032}
  (\bibinfo {year} {2022}{\natexlab{b}})},\ \Eprint
  {http://arxiv.org/abs/2112.13049} {arXiv:2112.13049 [gr-qc]} \BibitemShut
  {NoStop}%
\bibitem [{\citenamefont {Hou}\ \emph {et~al.}(2024{\natexlab{a}})\citenamefont
  {Hou}, \citenamefont {Wang},\ and\ \citenamefont {Zhu}}]{Hou:2023pfz}%
  \BibitemOpen
  \bibfield  {author} {\bibinfo {author} {\bibfnamefont {S.}~\bibnamefont
  {Hou}}, \bibinfo {author} {\bibfnamefont {A.}~\bibnamefont {Wang}}, \ and\
  \bibinfo {author} {\bibfnamefont {Z.-H.}\ \bibnamefont {Zhu}},\ }\href
  {\doibase 10.1103/PhysRevD.109.044025} {\bibfield  {journal} {\bibinfo
  {journal} {Phys. Rev. D}\ }\textbf {\bibinfo {volume} {109}},\ \bibinfo
  {pages} {044025} (\bibinfo {year} {2024}{\natexlab{a}})},\ \Eprint
  {http://arxiv.org/abs/2309.01165} {arXiv:2309.01165 [gr-qc]} \BibitemShut
  {NoStop}%
\bibitem [{\citenamefont {Gasparotto}\ \emph {et~al.}(2023)\citenamefont
  {Gasparotto}, \citenamefont {Vicente}, \citenamefont {Blas}, \citenamefont
  {Jenkins},\ and\ \citenamefont {Barausse}}]{Gasparotto:2023fcg}%
  \BibitemOpen
  \bibfield  {author} {\bibinfo {author} {\bibfnamefont {S.}~\bibnamefont
  {Gasparotto}}, \bibinfo {author} {\bibfnamefont {R.}~\bibnamefont {Vicente}},
  \bibinfo {author} {\bibfnamefont {D.}~\bibnamefont {Blas}}, \bibinfo {author}
  {\bibfnamefont {A.~C.}\ \bibnamefont {Jenkins}}, \ and\ \bibinfo {author}
  {\bibfnamefont {E.}~\bibnamefont {Barausse}},\ }\href {\doibase
  10.1103/PhysRevD.107.124033} {\bibfield  {journal} {\bibinfo  {journal}
  {Phys. Rev. D}\ }\textbf {\bibinfo {volume} {107}},\ \bibinfo {pages}
  {124033} (\bibinfo {year} {2023})},\ \Eprint
  {http://arxiv.org/abs/2301.13228} {arXiv:2301.13228 [gr-qc]} \BibitemShut
  {NoStop}%
\bibitem [{\citenamefont {Sun}\ \emph {et~al.}(2024)\citenamefont {Sun},
  \citenamefont {Shi}, \citenamefont {Zhang},\ and\ \citenamefont
  {Mei}}]{Sun:2024nut}%
  \BibitemOpen
  \bibfield  {author} {\bibinfo {author} {\bibfnamefont {S.}~\bibnamefont
  {Sun}}, \bibinfo {author} {\bibfnamefont {C.}~\bibnamefont {Shi}}, \bibinfo
  {author} {\bibfnamefont {J.-d.}\ \bibnamefont {Zhang}}, \ and\ \bibinfo
  {author} {\bibfnamefont {J.}~\bibnamefont {Mei}},\ }\href {\doibase
  10.1103/PhysRevD.110.024050} {\bibfield  {journal} {\bibinfo  {journal}
  {Phys. Rev. D}\ }\textbf {\bibinfo {volume} {110}},\ \bibinfo {pages}
  {024050} (\bibinfo {year} {2024})},\ \Eprint
  {http://arxiv.org/abs/2401.11416} {arXiv:2401.11416 [gr-qc]} \BibitemShut
  {NoStop}%
\bibitem [{\citenamefont {Xu}\ \emph {et~al.}(2024)\citenamefont {Xu},
  \citenamefont {Rossell\'o-Sastre}, \citenamefont {Tiwari}, \citenamefont
  {Ebersold}, \citenamefont {Hamilton}, \citenamefont {Garc\'\i{}a-Quir\'os},
  \citenamefont {Estell\'es},\ and\ \citenamefont {Husa}}]{Xu:2024ybt}%
  \BibitemOpen
  \bibfield  {author} {\bibinfo {author} {\bibfnamefont {Y.}~\bibnamefont
  {Xu}}, \bibinfo {author} {\bibfnamefont {M.}~\bibnamefont
  {Rossell\'o-Sastre}}, \bibinfo {author} {\bibfnamefont {S.}~\bibnamefont
  {Tiwari}}, \bibinfo {author} {\bibfnamefont {M.}~\bibnamefont {Ebersold}},
  \bibinfo {author} {\bibfnamefont {E.~Z.}\ \bibnamefont {Hamilton}}, \bibinfo
  {author} {\bibfnamefont {C.}~\bibnamefont {Garc\'\i{}a-Quir\'os}}, \bibinfo
  {author} {\bibfnamefont {H.}~\bibnamefont {Estell\'es}}, \ and\ \bibinfo
  {author} {\bibfnamefont {S.}~\bibnamefont {Husa}},\ }\href {\doibase
  10.1103/PhysRevD.109.123034} {\bibfield  {journal} {\bibinfo  {journal}
  {Phys. Rev. D}\ }\textbf {\bibinfo {volume} {109}},\ \bibinfo {pages}
  {123034} (\bibinfo {year} {2024})},\ \Eprint
  {http://arxiv.org/abs/2403.00441} {arXiv:2403.00441 [gr-qc]} \BibitemShut
  {NoStop}%
\bibitem [{\citenamefont {Tiwari}\ \emph {et~al.}(2021)\citenamefont {Tiwari},
  \citenamefont {Ebersold},\ and\ \citenamefont {Hamilton}}]{Tiwari:2021gfl}%
  \BibitemOpen
  \bibfield  {author} {\bibinfo {author} {\bibfnamefont {S.}~\bibnamefont
  {Tiwari}}, \bibinfo {author} {\bibfnamefont {M.}~\bibnamefont {Ebersold}}, \
  and\ \bibinfo {author} {\bibfnamefont {E.~Z.}\ \bibnamefont {Hamilton}},\
  }\href {\doibase 10.1103/PhysRevD.104.123024} {\bibfield  {journal} {\bibinfo
   {journal} {Phys. Rev. D}\ }\textbf {\bibinfo {volume} {104}},\ \bibinfo
  {pages} {123024} (\bibinfo {year} {2021})},\ \Eprint
  {http://arxiv.org/abs/2110.11171} {arXiv:2110.11171 [gr-qc]} \BibitemShut
  {NoStop}%
\bibitem [{\citenamefont {Lopez}\ \emph {et~al.}(2024)\citenamefont {Lopez},
  \citenamefont {Tiwari},\ and\ \citenamefont {Ebersold}}]{Lopez:2023aja}%
  \BibitemOpen
  \bibfield  {author} {\bibinfo {author} {\bibfnamefont {D.}~\bibnamefont
  {Lopez}}, \bibinfo {author} {\bibfnamefont {S.}~\bibnamefont {Tiwari}}, \
  and\ \bibinfo {author} {\bibfnamefont {M.}~\bibnamefont {Ebersold}},\ }\href
  {\doibase 10.1103/PhysRevD.109.043039} {\bibfield  {journal} {\bibinfo
  {journal} {Phys. Rev. D}\ }\textbf {\bibinfo {volume} {109}},\ \bibinfo
  {pages} {043039} (\bibinfo {year} {2024})},\ \Eprint
  {http://arxiv.org/abs/2305.04761} {arXiv:2305.04761 [gr-qc]} \BibitemShut
  {NoStop}%
\bibitem [{\citenamefont {Zel'dovich}\ and\ \citenamefont
  {Polnarev}(1974)}]{Zeldovich:1974gvh}%
  \BibitemOpen
  \bibfield  {author} {\bibinfo {author} {\bibfnamefont {Y.~B.}\ \bibnamefont
  {Zel'dovich}}\ and\ \bibinfo {author} {\bibfnamefont {A.~G.}\ \bibnamefont
  {Polnarev}},\ }\href@noop {} {\bibfield  {journal} {\bibinfo  {journal} {Sov.
  Astron.}\ }\textbf {\bibinfo {volume} {18}},\ \bibinfo {pages} {17} (\bibinfo
  {year} {1974})}\BibitemShut {NoStop}%
\bibitem [{\citenamefont {Thorne}(1992)}]{Thorne:1992sdb}%
  \BibitemOpen
  \bibfield  {author} {\bibinfo {author} {\bibfnamefont {K.~S.}\ \bibnamefont
  {Thorne}},\ }\href {\doibase 10.1103/PhysRevD.45.520} {\bibfield  {journal}
  {\bibinfo  {journal} {Phys. Rev. D}\ }\textbf {\bibinfo {volume} {45}},\
  \bibinfo {pages} {520} (\bibinfo {year} {1992})}\BibitemShut {NoStop}%
\bibitem [{\citenamefont {Christodoulou}(1991)}]{Christodoulou:1991cr}%
  \BibitemOpen
  \bibfield  {author} {\bibinfo {author} {\bibfnamefont {D.}~\bibnamefont
  {Christodoulou}},\ }\href {\doibase 10.1103/PhysRevLett.67.1486} {\bibfield
  {journal} {\bibinfo  {journal} {Phys. Rev. Lett.}\ }\textbf {\bibinfo
  {volume} {67}},\ \bibinfo {pages} {1486} (\bibinfo {year}
  {1991})}\BibitemShut {NoStop}%
\bibitem [{\citenamefont {Blanchet}\ and\ \citenamefont
  {Damour}(1992)}]{Blanchet:1992br}%
  \BibitemOpen
  \bibfield  {author} {\bibinfo {author} {\bibfnamefont {L.}~\bibnamefont
  {Blanchet}}\ and\ \bibinfo {author} {\bibfnamefont {T.}~\bibnamefont
  {Damour}},\ }\href {\doibase 10.1103/PhysRevD.46.4304} {\bibfield  {journal}
  {\bibinfo  {journal} {Phys. Rev. D}\ }\textbf {\bibinfo {volume} {46}},\
  \bibinfo {pages} {4304} (\bibinfo {year} {1992})}\BibitemShut {NoStop}%
\bibitem [{\citenamefont {Bondi}\ \emph {et~al.}(1962)\citenamefont {Bondi},
  \citenamefont {van~der Burg},\ and\ \citenamefont {Metzner}}]{Bondi:1962px}%
  \BibitemOpen
  \bibfield  {author} {\bibinfo {author} {\bibfnamefont {H.}~\bibnamefont
  {Bondi}}, \bibinfo {author} {\bibfnamefont {M.~G.~J.}\ \bibnamefont {van~der
  Burg}}, \ and\ \bibinfo {author} {\bibfnamefont {A.~W.~K.}\ \bibnamefont
  {Metzner}},\ }\href {\doibase 10.1098/rspa.1962.0161} {\bibfield  {journal}
  {\bibinfo  {journal} {Proc. Roy. Soc. Lond. A}\ }\textbf {\bibinfo {volume}
  {269}},\ \bibinfo {pages} {21} (\bibinfo {year} {1962})}\BibitemShut
  {NoStop}%
\bibitem [{\citenamefont {Sachs}(1962)}]{Sachs:1962wk}%
  \BibitemOpen
  \bibfield  {author} {\bibinfo {author} {\bibfnamefont {R.~K.}\ \bibnamefont
  {Sachs}},\ }\href {\doibase 10.1098/rspa.1962.0206} {\bibfield  {journal}
  {\bibinfo  {journal} {Proc. Roy. Soc. Lond. A}\ }\textbf {\bibinfo {volume}
  {270}},\ \bibinfo {pages} {103} (\bibinfo {year} {1962})}\BibitemShut
  {NoStop}%
\bibitem [{\citenamefont {de~Boer}\ and\ \citenamefont
  {Solodukhin}(2003)}]{deBoer:2003vf}%
  \BibitemOpen
  \bibfield  {author} {\bibinfo {author} {\bibfnamefont {J.}~\bibnamefont
  {de~Boer}}\ and\ \bibinfo {author} {\bibfnamefont {S.~N.}\ \bibnamefont
  {Solodukhin}},\ }\href {\doibase 10.1016/S0550-3213(03)00494-2} {\bibfield
  {journal} {\bibinfo  {journal} {Nucl. Phys. B}\ }\textbf {\bibinfo {volume}
  {665}},\ \bibinfo {pages} {545} (\bibinfo {year} {2003})},\ \Eprint
  {http://arxiv.org/abs/hep-th/0303006} {arXiv:hep-th/0303006} \BibitemShut
  {NoStop}%
\bibitem [{\citenamefont {Barnich}\ and\ \citenamefont
  {Troessaert}(2010{\natexlab{a}})}]{Barnich:2009se}%
  \BibitemOpen
  \bibfield  {author} {\bibinfo {author} {\bibfnamefont {G.}~\bibnamefont
  {Barnich}}\ and\ \bibinfo {author} {\bibfnamefont {C.}~\bibnamefont
  {Troessaert}},\ }\href {\doibase 10.1103/PhysRevLett.105.111103} {\bibfield
  {journal} {\bibinfo  {journal} {Phys. Rev. Lett.}\ }\textbf {\bibinfo
  {volume} {105}},\ \bibinfo {pages} {111103} (\bibinfo {year}
  {2010}{\natexlab{a}})},\ \Eprint {http://arxiv.org/abs/0909.2617}
  {arXiv:0909.2617 [gr-qc]} \BibitemShut {NoStop}%
\bibitem [{\citenamefont {Barnich}\ and\ \citenamefont
  {Troessaert}(2010{\natexlab{b}})}]{Barnich:2010ojg}%
  \BibitemOpen
  \bibfield  {author} {\bibinfo {author} {\bibfnamefont {G.}~\bibnamefont
  {Barnich}}\ and\ \bibinfo {author} {\bibfnamefont {C.}~\bibnamefont
  {Troessaert}},\ }\href {\doibase 10.22323/1.127.0010} {\bibfield  {journal}
  {\bibinfo  {journal} {PoS}\ }\textbf {\bibinfo {volume} {CNCFG2010}},\
  \bibinfo {pages} {010} (\bibinfo {year} {2010}{\natexlab{b}})},\ \Eprint
  {http://arxiv.org/abs/1102.4632} {arXiv:1102.4632 [gr-qc]} \BibitemShut
  {NoStop}%
\bibitem [{\citenamefont {Kapec}\ \emph {et~al.}(2014)\citenamefont {Kapec},
  \citenamefont {Lysov}, \citenamefont {Pasterski},\ and\ \citenamefont
  {Strominger}}]{Kapec:2014opa}%
  \BibitemOpen
  \bibfield  {author} {\bibinfo {author} {\bibfnamefont {D.}~\bibnamefont
  {Kapec}}, \bibinfo {author} {\bibfnamefont {V.}~\bibnamefont {Lysov}},
  \bibinfo {author} {\bibfnamefont {S.}~\bibnamefont {Pasterski}}, \ and\
  \bibinfo {author} {\bibfnamefont {A.}~\bibnamefont {Strominger}},\ }\href
  {\doibase 10.1007/JHEP08(2014)058} {\bibfield  {journal} {\bibinfo  {journal}
  {JHEP}\ }\textbf {\bibinfo {volume} {08}},\ \bibinfo {pages} {058} (\bibinfo
  {year} {2014})},\ \Eprint {http://arxiv.org/abs/1406.3312} {arXiv:1406.3312
  [hep-th]} \BibitemShut {NoStop}%
\bibitem [{\citenamefont {Kapec}\ \emph {et~al.}(2017)\citenamefont {Kapec},
  \citenamefont {Mitra}, \citenamefont {Raclariu},\ and\ \citenamefont
  {Strominger}}]{Kapec:2016jld}%
  \BibitemOpen
  \bibfield  {author} {\bibinfo {author} {\bibfnamefont {D.}~\bibnamefont
  {Kapec}}, \bibinfo {author} {\bibfnamefont {P.}~\bibnamefont {Mitra}},
  \bibinfo {author} {\bibfnamefont {A.-M.}\ \bibnamefont {Raclariu}}, \ and\
  \bibinfo {author} {\bibfnamefont {A.}~\bibnamefont {Strominger}},\ }\href
  {\doibase 10.1103/PhysRevLett.119.121601} {\bibfield  {journal} {\bibinfo
  {journal} {Phys. Rev. Lett.}\ }\textbf {\bibinfo {volume} {119}},\ \bibinfo
  {pages} {121601} (\bibinfo {year} {2017})},\ \Eprint
  {http://arxiv.org/abs/1609.00282} {arXiv:1609.00282 [hep-th]} \BibitemShut
  {NoStop}%
\bibitem [{\citenamefont {He}\ \emph {et~al.}(2017)\citenamefont {He},
  \citenamefont {Kapec}, \citenamefont {Raclariu},\ and\ \citenamefont
  {Strominger}}]{He:2017fsb}%
  \BibitemOpen
  \bibfield  {author} {\bibinfo {author} {\bibfnamefont {T.}~\bibnamefont
  {He}}, \bibinfo {author} {\bibfnamefont {D.}~\bibnamefont {Kapec}}, \bibinfo
  {author} {\bibfnamefont {A.-M.}\ \bibnamefont {Raclariu}}, \ and\ \bibinfo
  {author} {\bibfnamefont {A.}~\bibnamefont {Strominger}},\ }\href {\doibase
  10.1007/JHEP08(2017)050} {\bibfield  {journal} {\bibinfo  {journal} {JHEP}\
  }\textbf {\bibinfo {volume} {08}},\ \bibinfo {pages} {050} (\bibinfo {year}
  {2017})},\ \Eprint {http://arxiv.org/abs/1701.00496} {arXiv:1701.00496
  [hep-th]} \BibitemShut {NoStop}%
\bibitem [{\citenamefont {Hou}(2024)}]{Hou:2024exz}%
  \BibitemOpen
  \bibfield  {author} {\bibinfo {author} {\bibfnamefont {S.}~\bibnamefont
  {Hou}},\ }\href@noop {} {\  (\bibinfo {year} {2024})},\ \Eprint
  {http://arxiv.org/abs/2411.17318} {arXiv:2411.17318 [gr-qc]} \BibitemShut
  {NoStop}%
\bibitem [{\citenamefont {Strominger}\ and\ \citenamefont
  {Zhiboedov}(2016)}]{Strominger:2014pwa}%
  \BibitemOpen
  \bibfield  {author} {\bibinfo {author} {\bibfnamefont {A.}~\bibnamefont
  {Strominger}}\ and\ \bibinfo {author} {\bibfnamefont {A.}~\bibnamefont
  {Zhiboedov}},\ }\href {\doibase 10.1007/JHEP01(2016)086} {\bibfield
  {journal} {\bibinfo  {journal} {JHEP}\ }\textbf {\bibinfo {volume} {01}},\
  \bibinfo {pages} {086} (\bibinfo {year} {2016})},\ \Eprint
  {http://arxiv.org/abs/1411.5745} {arXiv:1411.5745 [hep-th]} \BibitemShut
  {NoStop}%
\bibitem [{\citenamefont {Pasterski}\ \emph {et~al.}(2016)\citenamefont
  {Pasterski}, \citenamefont {Strominger},\ and\ \citenamefont
  {Zhiboedov}}]{Pasterski:2015tva}%
  \BibitemOpen
  \bibfield  {author} {\bibinfo {author} {\bibfnamefont {S.}~\bibnamefont
  {Pasterski}}, \bibinfo {author} {\bibfnamefont {A.}~\bibnamefont
  {Strominger}}, \ and\ \bibinfo {author} {\bibfnamefont {A.}~\bibnamefont
  {Zhiboedov}},\ }\href {\doibase 10.1007/JHEP12(2016)053} {\bibfield
  {journal} {\bibinfo  {journal} {JHEP}\ }\textbf {\bibinfo {volume} {12}},\
  \bibinfo {pages} {053} (\bibinfo {year} {2016})},\ \Eprint
  {http://arxiv.org/abs/1502.06120} {arXiv:1502.06120 [hep-th]} \BibitemShut
  {NoStop}%
\bibitem [{\citenamefont {Nichols}(2018)}]{Nichols:2018qac}%
  \BibitemOpen
  \bibfield  {author} {\bibinfo {author} {\bibfnamefont {D.~A.}\ \bibnamefont
  {Nichols}},\ }\href {\doibase 10.1103/PhysRevD.98.064032} {\bibfield
  {journal} {\bibinfo  {journal} {Phys. Rev. D}\ }\textbf {\bibinfo {volume}
  {98}},\ \bibinfo {pages} {064032} (\bibinfo {year} {2018})},\ \Eprint
  {http://arxiv.org/abs/1807.08767} {arXiv:1807.08767 [gr-qc]} \BibitemShut
  {NoStop}%
\bibitem [{\citenamefont {Weinberg}(1965)}]{Weinberg:1965nx}%
  \BibitemOpen
  \bibfield  {author} {\bibinfo {author} {\bibfnamefont {S.}~\bibnamefont
  {Weinberg}},\ }\href {\doibase 10.1103/PhysRev.140.B516} {\bibfield
  {journal} {\bibinfo  {journal} {Phys. Rev.}\ }\textbf {\bibinfo {volume}
  {140}},\ \bibinfo {pages} {B516} (\bibinfo {year} {1965})}\BibitemShut
  {NoStop}%
\bibitem [{\citenamefont {Strominger}(2017)}]{Strominger:2017zoo}%
  \BibitemOpen
  \bibfield  {author} {\bibinfo {author} {\bibfnamefont {A.}~\bibnamefont
  {Strominger}},\ }\href@noop {} {\emph {\bibinfo {title} {{Lectures on the
  Infrared Structure of Gravity and Gauge Theory}}}}\ (\bibinfo {year} {2017})\
  \Eprint {http://arxiv.org/abs/1703.05448} {arXiv:1703.05448 [hep-th]}
  \BibitemShut {NoStop}%
\bibitem [{\citenamefont {Goncharov}\ \emph {et~al.}(2024)\citenamefont
  {Goncharov}, \citenamefont {Donnay},\ and\ \citenamefont
  {Harms}}]{Goncharov:2023woe}%
  \BibitemOpen
  \bibfield  {author} {\bibinfo {author} {\bibfnamefont {B.}~\bibnamefont
  {Goncharov}}, \bibinfo {author} {\bibfnamefont {L.}~\bibnamefont {Donnay}}, \
  and\ \bibinfo {author} {\bibfnamefont {J.}~\bibnamefont {Harms}},\ }\href
  {\doibase 10.1103/PhysRevLett.132.241401} {\bibfield  {journal} {\bibinfo
  {journal} {Phys. Rev. Lett.}\ }\textbf {\bibinfo {volume} {132}},\ \bibinfo
  {pages} {241401} (\bibinfo {year} {2024})},\ \Eprint
  {http://arxiv.org/abs/2310.10718} {arXiv:2310.10718 [gr-qc]} \BibitemShut
  {NoStop}%
\bibitem [{\citenamefont {Braginsky}\ and\ \citenamefont
  {Thorne}(1987)}]{Braginsky:1987kwo}%
  \BibitemOpen
  \bibfield  {author} {\bibinfo {author} {\bibfnamefont {V.~B.}\ \bibnamefont
  {Braginsky}}\ and\ \bibinfo {author} {\bibfnamefont {K.~S.}\ \bibnamefont
  {Thorne}},\ }\href {\doibase 10.1038/327123a0} {\bibfield  {journal}
  {\bibinfo  {journal} {Nature}\ }\textbf {\bibinfo {volume} {327}},\ \bibinfo
  {pages} {123} (\bibinfo {year} {1987})}\BibitemShut {NoStop}%
\bibitem [{\citenamefont {Lasky}\ \emph {et~al.}(2016)\citenamefont {Lasky},
  \citenamefont {Thrane}, \citenamefont {Levin}, \citenamefont {Blackman},\
  and\ \citenamefont {Chen}}]{Lasky:2016knh}%
  \BibitemOpen
  \bibfield  {author} {\bibinfo {author} {\bibfnamefont {P.~D.}\ \bibnamefont
  {Lasky}}, \bibinfo {author} {\bibfnamefont {E.}~\bibnamefont {Thrane}},
  \bibinfo {author} {\bibfnamefont {Y.}~\bibnamefont {Levin}}, \bibinfo
  {author} {\bibfnamefont {J.}~\bibnamefont {Blackman}}, \ and\ \bibinfo
  {author} {\bibfnamefont {Y.}~\bibnamefont {Chen}},\ }\href {\doibase
  10.1103/PhysRevLett.117.061102} {\bibfield  {journal} {\bibinfo  {journal}
  {Phys. Rev. Lett.}\ }\textbf {\bibinfo {volume} {117}},\ \bibinfo {pages}
  {061102} (\bibinfo {year} {2016})},\ \Eprint
  {http://arxiv.org/abs/1605.01415} {arXiv:1605.01415 [astro-ph.HE]}
  \BibitemShut {NoStop}%
\bibitem [{\citenamefont {H\"ubner}\ \emph {et~al.}(2020)\citenamefont
  {H\"ubner}, \citenamefont {Talbot}, \citenamefont {Lasky},\ and\
  \citenamefont {Thrane}}]{Hubner:2019sly}%
  \BibitemOpen
  \bibfield  {author} {\bibinfo {author} {\bibfnamefont {M.}~\bibnamefont
  {H\"ubner}}, \bibinfo {author} {\bibfnamefont {C.}~\bibnamefont {Talbot}},
  \bibinfo {author} {\bibfnamefont {P.~D.}\ \bibnamefont {Lasky}}, \ and\
  \bibinfo {author} {\bibfnamefont {E.}~\bibnamefont {Thrane}},\ }\href
  {\doibase 10.1103/PhysRevD.101.023011} {\bibfield  {journal} {\bibinfo
  {journal} {Phys. Rev. D}\ }\textbf {\bibinfo {volume} {101}},\ \bibinfo
  {pages} {023011} (\bibinfo {year} {2020})},\ \Eprint
  {http://arxiv.org/abs/1911.12496} {arXiv:1911.12496 [astro-ph.HE]}
  \BibitemShut {NoStop}%
\bibitem [{\citenamefont {H\"ubner}\ \emph {et~al.}(2021)\citenamefont
  {H\"ubner}, \citenamefont {Lasky},\ and\ \citenamefont
  {Thrane}}]{Hubner:2021amk}%
  \BibitemOpen
  \bibfield  {author} {\bibinfo {author} {\bibfnamefont {M.}~\bibnamefont
  {H\"ubner}}, \bibinfo {author} {\bibfnamefont {P.}~\bibnamefont {Lasky}}, \
  and\ \bibinfo {author} {\bibfnamefont {E.}~\bibnamefont {Thrane}},\ }\href
  {\doibase 10.1103/PhysRevD.104.023004} {\bibfield  {journal} {\bibinfo
  {journal} {Phys. Rev. D}\ }\textbf {\bibinfo {volume} {104}},\ \bibinfo
  {pages} {023004} (\bibinfo {year} {2021})},\ \Eprint
  {http://arxiv.org/abs/2105.02879} {arXiv:2105.02879 [gr-qc]} \BibitemShut
  {NoStop}%
\bibitem [{\citenamefont {Cheung}\ \emph {et~al.}(2024)\citenamefont {Cheung},
  \citenamefont {Lasky},\ and\ \citenamefont {Thrane}}]{Cheung:2024zow}%
  \BibitemOpen
  \bibfield  {author} {\bibinfo {author} {\bibfnamefont {S.~Y.}\ \bibnamefont
  {Cheung}}, \bibinfo {author} {\bibfnamefont {P.~D.}\ \bibnamefont {Lasky}}, \
  and\ \bibinfo {author} {\bibfnamefont {E.}~\bibnamefont {Thrane}},\ }\href
  {\doibase 10.1088/1361-6382/ad3ffe} {\bibfield  {journal} {\bibinfo
  {journal} {Class. Quant. Grav.}\ }\textbf {\bibinfo {volume} {41}},\ \bibinfo
  {pages} {115010} (\bibinfo {year} {2024})},\ \Eprint
  {http://arxiv.org/abs/2404.11919} {arXiv:2404.11919 [gr-qc]} \BibitemShut
  {NoStop}%
\bibitem [{\citenamefont {Zhao}\ \emph
  {et~al.}(2021{\natexlab{a}})\citenamefont {Zhao}, \citenamefont {Liu},
  \citenamefont {Cao},\ and\ \citenamefont {He}}]{Zhao:2021hmx}%
  \BibitemOpen
  \bibfield  {author} {\bibinfo {author} {\bibfnamefont {Z.-C.}\ \bibnamefont
  {Zhao}}, \bibinfo {author} {\bibfnamefont {X.}~\bibnamefont {Liu}}, \bibinfo
  {author} {\bibfnamefont {Z.}~\bibnamefont {Cao}}, \ and\ \bibinfo {author}
  {\bibfnamefont {X.}~\bibnamefont {He}},\ }\href {\doibase
  10.1103/PhysRevD.104.064056} {\bibfield  {journal} {\bibinfo  {journal}
  {Phys. Rev. D}\ }\textbf {\bibinfo {volume} {104}},\ \bibinfo {pages}
  {064056} (\bibinfo {year} {2021}{\natexlab{a}})}\BibitemShut {NoStop}%
\bibitem [{\citenamefont {Seto}(2009)}]{Seto:2009nv}%
  \BibitemOpen
  \bibfield  {author} {\bibinfo {author} {\bibfnamefont {N.}~\bibnamefont
  {Seto}},\ }\href {\doibase 10.1111/j.1745-3933.2009.00758.x} {\bibfield
  {journal} {\bibinfo  {journal} {Mon. Not. Roy. Astron. Soc.}\ }\textbf
  {\bibinfo {volume} {400}},\ \bibinfo {pages} {L38} (\bibinfo {year}
  {2009})},\ \Eprint {http://arxiv.org/abs/0909.1379} {arXiv:0909.1379
  [astro-ph.CO]} \BibitemShut {NoStop}%
\bibitem [{\citenamefont {van Haasteren}\ and\ \citenamefont
  {Levin}(2010)}]{vanHaasteren:2009fy}%
  \BibitemOpen
  \bibfield  {author} {\bibinfo {author} {\bibfnamefont {R.}~\bibnamefont {van
  Haasteren}}\ and\ \bibinfo {author} {\bibfnamefont {Y.}~\bibnamefont
  {Levin}},\ }\href {\doibase 10.1111/j.1365-2966.2009.15885.x} {\bibfield
  {journal} {\bibinfo  {journal} {Mon. Not. Roy. Astron. Soc.}\ }\textbf
  {\bibinfo {volume} {401}},\ \bibinfo {pages} {2372} (\bibinfo {year}
  {2010})},\ \Eprint {http://arxiv.org/abs/0909.0954} {arXiv:0909.0954
  [astro-ph.IM]} \BibitemShut {NoStop}%
\bibitem [{\citenamefont {Pshirkov}\ \emph {et~al.}(2010)\citenamefont
  {Pshirkov}, \citenamefont {Baskaran},\ and\ \citenamefont
  {Postnov}}]{Pshirkov:2009ak}%
  \BibitemOpen
  \bibfield  {author} {\bibinfo {author} {\bibfnamefont {M.~S.}\ \bibnamefont
  {Pshirkov}}, \bibinfo {author} {\bibfnamefont {D.}~\bibnamefont {Baskaran}},
  \ and\ \bibinfo {author} {\bibfnamefont {K.~A.}\ \bibnamefont {Postnov}},\
  }\href {\doibase 10.1111/j.1365-2966.2009.15887.x} {\bibfield  {journal}
  {\bibinfo  {journal} {Mon. Not. Roy. Astron. Soc.}\ }\textbf {\bibinfo
  {volume} {402}},\ \bibinfo {pages} {417} (\bibinfo {year} {2010})},\ \Eprint
  {http://arxiv.org/abs/0909.0742} {arXiv:0909.0742 [astro-ph.CO]} \BibitemShut
  {NoStop}%
\bibitem [{\citenamefont {Cordes}\ and\ \citenamefont
  {Jenet}(2012)}]{Cordes:2012zz}%
  \BibitemOpen
  \bibfield  {author} {\bibinfo {author} {\bibfnamefont {J.~M.}\ \bibnamefont
  {Cordes}}\ and\ \bibinfo {author} {\bibfnamefont {F.~A.}\ \bibnamefont
  {Jenet}},\ }\href {\doibase 10.1088/0004-637X/752/1/54} {\bibfield  {journal}
  {\bibinfo  {journal} {Astrophys. J.}\ }\textbf {\bibinfo {volume} {752}},\
  \bibinfo {pages} {54} (\bibinfo {year} {2012})}\BibitemShut {NoStop}%
\bibitem [{\citenamefont {Madison}\ \emph {et~al.}(2014)\citenamefont
  {Madison}, \citenamefont {Cordes},\ and\ \citenamefont
  {Chatterjee}}]{Madison:2014vca}%
  \BibitemOpen
  \bibfield  {author} {\bibinfo {author} {\bibfnamefont {D.~R.}\ \bibnamefont
  {Madison}}, \bibinfo {author} {\bibfnamefont {J.~M.}\ \bibnamefont {Cordes}},
  \ and\ \bibinfo {author} {\bibfnamefont {S.}~\bibnamefont {Chatterjee}},\
  }\href {\doibase 10.1088/0004-637X/788/2/141} {\bibfield  {journal} {\bibinfo
   {journal} {Astrophys. J.}\ }\textbf {\bibinfo {volume} {788}},\ \bibinfo
  {pages} {141} (\bibinfo {year} {2014})},\ \Eprint
  {http://arxiv.org/abs/1404.5682} {arXiv:1404.5682 [astro-ph.HE]} \BibitemShut
  {NoStop}%
\bibitem [{\citenamefont {Arzoumanian}\ \emph {et~al.}(2015)\citenamefont
  {Arzoumanian} \emph {et~al.}}]{NANOGrav:2015xuc}%
  \BibitemOpen
  \bibfield  {author} {\bibinfo {author} {\bibfnamefont {Z.}~\bibnamefont
  {Arzoumanian}} \emph {et~al.} (\bibinfo {collaboration} {NANOGrav}),\ }\href
  {\doibase 10.1088/0004-637X/810/2/150} {\bibfield  {journal} {\bibinfo
  {journal} {Astrophys. J.}\ }\textbf {\bibinfo {volume} {810}},\ \bibinfo
  {pages} {150} (\bibinfo {year} {2015})},\ \Eprint
  {http://arxiv.org/abs/1501.05343} {arXiv:1501.05343 [astro-ph.GA]}
  \BibitemShut {NoStop}%
\bibitem [{\citenamefont {Agazie}\ \emph {et~al.}(2024)\citenamefont {Agazie}
  \emph {et~al.}}]{NANOGrav:2023vfo}%
  \BibitemOpen
  \bibfield  {author} {\bibinfo {author} {\bibfnamefont {G.}~\bibnamefont
  {Agazie}} \emph {et~al.} (\bibinfo {collaboration} {NANOGrav}),\ }\href
  {\doibase 10.3847/1538-4357/ad0726} {\bibfield  {journal} {\bibinfo
  {journal} {Astrophys. J.}\ }\textbf {\bibinfo {volume} {963}},\ \bibinfo
  {pages} {61} (\bibinfo {year} {2024})},\ \Eprint
  {http://arxiv.org/abs/2307.13797} {arXiv:2307.13797 [gr-qc]} \BibitemShut
  {NoStop}%
\bibitem [{\citenamefont {Islo}\ \emph {et~al.}(2019)\citenamefont {Islo},
  \citenamefont {Simon}, \citenamefont {Burke-Spolaor},\ and\ \citenamefont
  {Siemens}}]{Islo:2019qht}%
  \BibitemOpen
  \bibfield  {author} {\bibinfo {author} {\bibfnamefont {K.}~\bibnamefont
  {Islo}}, \bibinfo {author} {\bibfnamefont {J.}~\bibnamefont {Simon}},
  \bibinfo {author} {\bibfnamefont {S.}~\bibnamefont {Burke-Spolaor}}, \ and\
  \bibinfo {author} {\bibfnamefont {X.}~\bibnamefont {Siemens}},\ }\href@noop
  {} {\  (\bibinfo {year} {2019})},\ \Eprint {http://arxiv.org/abs/1906.11936}
  {arXiv:1906.11936 [astro-ph.HE]} \BibitemShut {NoStop}%
\bibitem [{\citenamefont {Inchausp\'e}\ \emph {et~al.}(2024)\citenamefont
  {Inchausp\'e}, \citenamefont {Gasparotto}, \citenamefont {Blas},
  \citenamefont {Heisenberg}, \citenamefont {Zosso},\ and\ \citenamefont
  {Tiwari}}]{Inchauspe:2024ibs}%
  \BibitemOpen
  \bibfield  {author} {\bibinfo {author} {\bibfnamefont {H.}~\bibnamefont
  {Inchausp\'e}}, \bibinfo {author} {\bibfnamefont {S.}~\bibnamefont
  {Gasparotto}}, \bibinfo {author} {\bibfnamefont {D.}~\bibnamefont {Blas}},
  \bibinfo {author} {\bibfnamefont {L.}~\bibnamefont {Heisenberg}}, \bibinfo
  {author} {\bibfnamefont {J.}~\bibnamefont {Zosso}}, \ and\ \bibinfo {author}
  {\bibfnamefont {S.}~\bibnamefont {Tiwari}},\ }\href@noop {} {\  (\bibinfo
  {year} {2024})},\ \Eprint {http://arxiv.org/abs/2406.09228} {arXiv:2406.09228
  [gr-qc]} \BibitemShut {NoStop}%
\bibitem [{\citenamefont {Hou}\ \emph {et~al.}(2024{\natexlab{b}})\citenamefont
  {Hou}, \citenamefont {Zhao}, \citenamefont {Cao},\ and\ \citenamefont
  {Zhu}}]{Hou:2024rgo}%
  \BibitemOpen
  \bibfield  {author} {\bibinfo {author} {\bibfnamefont {S.}~\bibnamefont
  {Hou}}, \bibinfo {author} {\bibfnamefont {Z.-C.}\ \bibnamefont {Zhao}},
  \bibinfo {author} {\bibfnamefont {Z.}~\bibnamefont {Cao}}, \ and\ \bibinfo
  {author} {\bibfnamefont {Z.-H.}\ \bibnamefont {Zhu}},\ }\href@noop {} {\
  (\bibinfo {year} {2024}{\natexlab{b}})},\ \Eprint
  {http://arxiv.org/abs/2411.18053} {arXiv:2411.18053 [gr-qc]} \BibitemShut
  {NoStop}%
\bibitem [{\citenamefont {Sun}\ \emph {et~al.}(2023)\citenamefont {Sun},
  \citenamefont {Shi}, \citenamefont {Zhang},\ and\ \citenamefont
  {Mei}}]{Sun:2022pvh}%
  \BibitemOpen
  \bibfield  {author} {\bibinfo {author} {\bibfnamefont {S.}~\bibnamefont
  {Sun}}, \bibinfo {author} {\bibfnamefont {C.}~\bibnamefont {Shi}}, \bibinfo
  {author} {\bibfnamefont {J.-d.}\ \bibnamefont {Zhang}}, \ and\ \bibinfo
  {author} {\bibfnamefont {J.}~\bibnamefont {Mei}},\ }\href {\doibase
  10.1103/PhysRevD.107.044023} {\bibfield  {journal} {\bibinfo  {journal}
  {Phys. Rev. D}\ }\textbf {\bibinfo {volume} {107}},\ \bibinfo {pages}
  {044023} (\bibinfo {year} {2023})},\ \Eprint
  {http://arxiv.org/abs/2207.13009} {arXiv:2207.13009 [gr-qc]} \BibitemShut
  {NoStop}%
\bibitem [{\citenamefont {Chatziioannou}\ \emph {et~al.}(2017)\citenamefont
  {Chatziioannou}, \citenamefont {Klein}, \citenamefont {Yunes},\ and\
  \citenamefont {Cornish}}]{Chatziioannou:2017tdw}%
  \BibitemOpen
  \bibfield  {author} {\bibinfo {author} {\bibfnamefont {K.}~\bibnamefont
  {Chatziioannou}}, \bibinfo {author} {\bibfnamefont {A.}~\bibnamefont
  {Klein}}, \bibinfo {author} {\bibfnamefont {N.}~\bibnamefont {Yunes}}, \ and\
  \bibinfo {author} {\bibfnamefont {N.}~\bibnamefont {Cornish}},\ }\href
  {\doibase 10.1103/PhysRevD.95.104004} {\bibfield  {journal} {\bibinfo
  {journal} {Phys. Rev. D}\ }\textbf {\bibinfo {volume} {95}},\ \bibinfo
  {pages} {104004} (\bibinfo {year} {2017})},\ \Eprint
  {http://arxiv.org/abs/1703.03967} {arXiv:1703.03967 [gr-qc]} \BibitemShut
  {NoStop}%
\bibitem [{\citenamefont {P\"urrer}\ and\ \citenamefont
  {Haster}(2020)}]{Purrer:2019jcp}%
  \BibitemOpen
  \bibfield  {author} {\bibinfo {author} {\bibfnamefont {M.}~\bibnamefont
  {P\"urrer}}\ and\ \bibinfo {author} {\bibfnamefont {C.-J.}\ \bibnamefont
  {Haster}},\ }\href {\doibase 10.1103/PhysRevResearch.2.023151} {\bibfield
  {journal} {\bibinfo  {journal} {Phys. Rev. Res.}\ }\textbf {\bibinfo {volume}
  {2}},\ \bibinfo {pages} {023151} (\bibinfo {year} {2020})},\ \Eprint
  {http://arxiv.org/abs/1912.10055} {arXiv:1912.10055 [gr-qc]} \BibitemShut
  {NoStop}%
\bibitem [{\citenamefont {Pompili}\ \emph {et~al.}(2023)\citenamefont {Pompili}
  \emph {et~al.}}]{Pompili:2023tna}%
  \BibitemOpen
  \bibfield  {author} {\bibinfo {author} {\bibfnamefont {L.}~\bibnamefont
  {Pompili}} \emph {et~al.},\ }\href {\doibase 10.1103/PhysRevD.108.124035}
  {\bibfield  {journal} {\bibinfo  {journal} {Phys. Rev. D}\ }\textbf {\bibinfo
  {volume} {108}},\ \bibinfo {pages} {124035} (\bibinfo {year} {2023})},\
  \Eprint {http://arxiv.org/abs/2303.18039} {arXiv:2303.18039 [gr-qc]}
  \BibitemShut {NoStop}%
\bibitem [{\citenamefont {Ossokine}\ \emph {et~al.}(2020)\citenamefont
  {Ossokine} \emph {et~al.}}]{Ossokine:2020kjp}%
  \BibitemOpen
  \bibfield  {author} {\bibinfo {author} {\bibfnamefont {S.}~\bibnamefont
  {Ossokine}} \emph {et~al.},\ }\href {\doibase 10.1103/PhysRevD.102.044055}
  {\bibfield  {journal} {\bibinfo  {journal} {Phys. Rev. D}\ }\textbf {\bibinfo
  {volume} {102}},\ \bibinfo {pages} {044055} (\bibinfo {year} {2020})},\
  \Eprint {http://arxiv.org/abs/2004.09442} {arXiv:2004.09442 [gr-qc]}
  \BibitemShut {NoStop}%
\bibitem [{\citenamefont {Kerr}(1963)}]{Kerr:1963ud}%
  \BibitemOpen
  \bibfield  {author} {\bibinfo {author} {\bibfnamefont {R.~P.}\ \bibnamefont
  {Kerr}},\ }\href {\doibase 10.1103/PhysRevLett.11.237} {\bibfield  {journal}
  {\bibinfo  {journal} {Phys. Rev. Lett.}\ }\textbf {\bibinfo {volume} {11}},\
  \bibinfo {pages} {237} (\bibinfo {year} {1963})}\BibitemShut {NoStop}%
\bibitem [{\citenamefont {Gibbons}(1975{\natexlab{a}})}]{Gibbons:1975kk}%
  \BibitemOpen
  \bibfield  {author} {\bibinfo {author} {\bibfnamefont {G.~W.}\ \bibnamefont
  {Gibbons}},\ }\href {\doibase 10.1007/BF01609829} {\bibfield  {journal}
  {\bibinfo  {journal} {Commun. Math. Phys.}\ }\textbf {\bibinfo {volume}
  {44}},\ \bibinfo {pages} {245} (\bibinfo {year}
  {1975}{\natexlab{a}})}\BibitemShut {NoStop}%
\bibitem [{\citenamefont {Goldreich}\ and\ \citenamefont
  {Julian}(1969)}]{Goldreich:1969sb}%
  \BibitemOpen
  \bibfield  {author} {\bibinfo {author} {\bibfnamefont {P.}~\bibnamefont
  {Goldreich}}\ and\ \bibinfo {author} {\bibfnamefont {W.~H.}\ \bibnamefont
  {Julian}},\ }\href {\doibase 10.1086/150119} {\bibfield  {journal} {\bibinfo
  {journal} {Astrophys. J.}\ }\textbf {\bibinfo {volume} {157}},\ \bibinfo
  {pages} {869} (\bibinfo {year} {1969})}\BibitemShut {NoStop}%
\bibitem [{\citenamefont {Ruderman}\ and\ \citenamefont
  {Sutherland}(1975)}]{Ruderman:1975ju}%
  \BibitemOpen
  \bibfield  {author} {\bibinfo {author} {\bibfnamefont {M.~A.}\ \bibnamefont
  {Ruderman}}\ and\ \bibinfo {author} {\bibfnamefont {P.~G.}\ \bibnamefont
  {Sutherland}},\ }\href {\doibase 10.1086/153393} {\bibfield  {journal}
  {\bibinfo  {journal} {Astrophys. J.}\ }\textbf {\bibinfo {volume} {196}},\
  \bibinfo {pages} {51} (\bibinfo {year} {1975})}\BibitemShut {NoStop}%
\bibitem [{\citenamefont {Blandford}\ and\ \citenamefont
  {Znajek}(1977{\natexlab{a}})}]{Blandford:1977ds}%
  \BibitemOpen
  \bibfield  {author} {\bibinfo {author} {\bibfnamefont {R.~D.}\ \bibnamefont
  {Blandford}}\ and\ \bibinfo {author} {\bibfnamefont {R.~L.}\ \bibnamefont
  {Znajek}},\ }\href {\doibase 10.1093/mnras/179.3.433} {\bibfield  {journal}
  {\bibinfo  {journal} {Mon. Not. Roy. Astron. Soc.}\ }\textbf {\bibinfo
  {volume} {179}},\ \bibinfo {pages} {433} (\bibinfo {year}
  {1977}{\natexlab{a}})}\BibitemShut {NoStop}%
\bibitem [{\citenamefont {Chandrasekhar}(1975)}]{Chandrasekhar:1975rye}%
  \BibitemOpen
  \bibfield  {author} {\bibinfo {author} {\bibfnamefont {S.}~\bibnamefont
  {Chandrasekhar}},\ }\href@noop {} {\bibfield  {journal} {\bibinfo  {journal}
  {Ryerson Lecture, University of Chicago}\ } (\bibinfo {year}
  {1975})}\BibitemShut {NoStop}%
\bibitem [{\citenamefont {Israel}(1967)}]{Israel:1967wq}%
  \BibitemOpen
  \bibfield  {author} {\bibinfo {author} {\bibfnamefont {W.}~\bibnamefont
  {Israel}},\ }\href {\doibase 10.1103/PhysRev.164.1776} {\bibfield  {journal}
  {\bibinfo  {journal} {Phys. Rev.}\ }\textbf {\bibinfo {volume} {164}},\
  \bibinfo {pages} {1776} (\bibinfo {year} {1967})}\BibitemShut {NoStop}%
\bibitem [{\citenamefont {Carter}(1971)}]{Carter:1971zc}%
  \BibitemOpen
  \bibfield  {author} {\bibinfo {author} {\bibfnamefont {B.}~\bibnamefont
  {Carter}},\ }\href {\doibase 10.1103/PhysRevLett.26.331} {\bibfield
  {journal} {\bibinfo  {journal} {Phys. Rev. Lett.}\ }\textbf {\bibinfo
  {volume} {26}},\ \bibinfo {pages} {331} (\bibinfo {year} {1971})}\BibitemShut
  {NoStop}%
\bibitem [{\citenamefont {Carter}(1997)}]{Carter:1997im}%
  \BibitemOpen
  \bibfield  {author} {\bibinfo {author} {\bibfnamefont {B.}~\bibnamefont
  {Carter}},\ }in\ \href@noop {} {\emph {\bibinfo {booktitle} {{8th Marcel
  Grossmann Meeting on Recent Developments in Theoretical and Experimental
  General Relativity, Gravitation and Relativistic Field Theories (MG 8)}}}}\
  (\bibinfo {year} {1997})\ pp.\ \bibinfo {pages} {136--155},\ \Eprint
  {http://arxiv.org/abs/gr-qc/9712038} {arXiv:gr-qc/9712038} \BibitemShut
  {NoStop}%
\bibitem [{\citenamefont {Robinson}(1975)}]{Robinson:1975bv}%
  \BibitemOpen
  \bibfield  {author} {\bibinfo {author} {\bibfnamefont {D.~C.}\ \bibnamefont
  {Robinson}},\ }\href {\doibase 10.1103/PhysRevLett.34.905} {\bibfield
  {journal} {\bibinfo  {journal} {Phys. Rev. Lett.}\ }\textbf {\bibinfo
  {volume} {34}},\ \bibinfo {pages} {905} (\bibinfo {year} {1975})}\BibitemShut
  {NoStop}%
\bibitem [{\citenamefont {Robinson}(2004)}]{Robinson:2004zz}%
  \BibitemOpen
  \bibfield  {author} {\bibinfo {author} {\bibfnamefont {D.}~\bibnamefont
  {Robinson}},\ }in\ \href@noop {} {\emph {\bibinfo {booktitle} {{Kerr Fest:
  Black Holes in Astrophysics, General Relativity and Quantum Gravity}}}}\
  (\bibinfo {year} {2004})\BibitemShut {NoStop}%
\bibitem [{\citenamefont {Bekenstein}(1996)}]{Bekenstein:1996pn}%
  \BibitemOpen
  \bibfield  {author} {\bibinfo {author} {\bibfnamefont {J.~D.}\ \bibnamefont
  {Bekenstein}},\ }in\ \href@noop {} {\emph {\bibinfo {booktitle} {{2nd
  International Sakharov Conference on Physics}}}}\ (\bibinfo {year} {1996})\
  pp.\ \bibinfo {pages} {216--219},\ \Eprint
  {http://arxiv.org/abs/gr-qc/9605059} {arXiv:gr-qc/9605059} \BibitemShut
  {NoStop}%
\bibitem [{\citenamefont {Chrusciel}\ \emph {et~al.}(2012)\citenamefont
  {Chrusciel}, \citenamefont {Lopes~Costa},\ and\ \citenamefont
  {Heusler}}]{Chrusciel:2012jk}%
  \BibitemOpen
  \bibfield  {author} {\bibinfo {author} {\bibfnamefont {P.~T.}\ \bibnamefont
  {Chrusciel}}, \bibinfo {author} {\bibfnamefont {J.}~\bibnamefont
  {Lopes~Costa}}, \ and\ \bibinfo {author} {\bibfnamefont {M.}~\bibnamefont
  {Heusler}},\ }\href {\doibase 10.12942/lrr-2012-7} {\bibfield  {journal}
  {\bibinfo  {journal} {Living Rev. Rel.}\ }\textbf {\bibinfo {volume} {15}},\
  \bibinfo {pages} {7} (\bibinfo {year} {2012})},\ \Eprint
  {http://arxiv.org/abs/1205.6112} {arXiv:1205.6112 [gr-qc]} \BibitemShut
  {NoStop}%
\bibitem [{\citenamefont {G\"urlebeck}(2015)}]{Gurlebeck:2015xpa}%
  \BibitemOpen
  \bibfield  {author} {\bibinfo {author} {\bibfnamefont {N.}~\bibnamefont
  {G\"urlebeck}},\ }\href {\doibase 10.1103/PhysRevLett.114.151102} {\bibfield
  {journal} {\bibinfo  {journal} {Phys. Rev. Lett.}\ }\textbf {\bibinfo
  {volume} {114}},\ \bibinfo {pages} {151102} (\bibinfo {year} {2015})},\
  \Eprint {http://arxiv.org/abs/1503.03240} {arXiv:1503.03240 [gr-qc]}
  \BibitemShut {NoStop}%
\bibitem [{\citenamefont {Le~Tiec}\ and\ \citenamefont
  {Casals}(2021)}]{LeTiec:2020spy}%
  \BibitemOpen
  \bibfield  {author} {\bibinfo {author} {\bibfnamefont {A.}~\bibnamefont
  {Le~Tiec}}\ and\ \bibinfo {author} {\bibfnamefont {M.}~\bibnamefont
  {Casals}},\ }\href {\doibase 10.1103/PhysRevLett.126.131102} {\bibfield
  {journal} {\bibinfo  {journal} {Phys. Rev. Lett.}\ }\textbf {\bibinfo
  {volume} {126}},\ \bibinfo {pages} {131102} (\bibinfo {year} {2021})},\
  \Eprint {http://arxiv.org/abs/2007.00214} {arXiv:2007.00214 [gr-qc]}
  \BibitemShut {NoStop}%
\bibitem [{\citenamefont {Le~Tiec}\ \emph {et~al.}(2021)\citenamefont
  {Le~Tiec}, \citenamefont {Casals},\ and\ \citenamefont
  {Franzin}}]{LeTiec:2020bos}%
  \BibitemOpen
  \bibfield  {author} {\bibinfo {author} {\bibfnamefont {A.}~\bibnamefont
  {Le~Tiec}}, \bibinfo {author} {\bibfnamefont {M.}~\bibnamefont {Casals}}, \
  and\ \bibinfo {author} {\bibfnamefont {E.}~\bibnamefont {Franzin}},\ }\href
  {\doibase 10.1103/PhysRevD.103.084021} {\bibfield  {journal} {\bibinfo
  {journal} {Phys. Rev. D}\ }\textbf {\bibinfo {volume} {103}},\ \bibinfo
  {pages} {084021} (\bibinfo {year} {2021})},\ \Eprint
  {http://arxiv.org/abs/2010.15795} {arXiv:2010.15795 [gr-qc]} \BibitemShut
  {NoStop}%
\bibitem [{\citenamefont {Chia}(2021)}]{Chia:2020yla}%
  \BibitemOpen
  \bibfield  {author} {\bibinfo {author} {\bibfnamefont {H.~S.}\ \bibnamefont
  {Chia}},\ }\href {\doibase 10.1103/PhysRevD.104.024013} {\bibfield  {journal}
  {\bibinfo  {journal} {Phys. Rev. D}\ }\textbf {\bibinfo {volume} {104}},\
  \bibinfo {pages} {024013} (\bibinfo {year} {2021})},\ \Eprint
  {http://arxiv.org/abs/2010.07300} {arXiv:2010.07300 [gr-qc]} \BibitemShut
  {NoStop}%
\bibitem [{\citenamefont {Charalambous}\ \emph
  {et~al.}(2021{\natexlab{a}})\citenamefont {Charalambous}, \citenamefont
  {Dubovsky},\ and\ \citenamefont {Ivanov}}]{Charalambous:2021kcz}%
  \BibitemOpen
  \bibfield  {author} {\bibinfo {author} {\bibfnamefont {P.}~\bibnamefont
  {Charalambous}}, \bibinfo {author} {\bibfnamefont {S.}~\bibnamefont
  {Dubovsky}}, \ and\ \bibinfo {author} {\bibfnamefont {M.~M.}\ \bibnamefont
  {Ivanov}},\ }\href {\doibase 10.1103/PhysRevLett.127.101101} {\bibfield
  {journal} {\bibinfo  {journal} {Phys. Rev. Lett.}\ }\textbf {\bibinfo
  {volume} {127}},\ \bibinfo {pages} {101101} (\bibinfo {year}
  {2021}{\natexlab{a}})},\ \Eprint {http://arxiv.org/abs/2103.01234}
  {arXiv:2103.01234 [hep-th]} \BibitemShut {NoStop}%
\bibitem [{\citenamefont {Charalambous}\ \emph
  {et~al.}(2021{\natexlab{b}})\citenamefont {Charalambous}, \citenamefont
  {Dubovsky},\ and\ \citenamefont {Ivanov}}]{Charalambous:2021mea}%
  \BibitemOpen
  \bibfield  {author} {\bibinfo {author} {\bibfnamefont {P.}~\bibnamefont
  {Charalambous}}, \bibinfo {author} {\bibfnamefont {S.}~\bibnamefont
  {Dubovsky}}, \ and\ \bibinfo {author} {\bibfnamefont {M.~M.}\ \bibnamefont
  {Ivanov}},\ }\href {\doibase 10.1007/JHEP05(2021)038} {\bibfield  {journal}
  {\bibinfo  {journal} {JHEP}\ }\textbf {\bibinfo {volume} {05}},\ \bibinfo
  {pages} {038} (\bibinfo {year} {2021}{\natexlab{b}})},\ \Eprint
  {http://arxiv.org/abs/2102.08917} {arXiv:2102.08917 [hep-th]} \BibitemShut
  {NoStop}%
\bibitem [{\citenamefont {Herdeiro}\ and\ \citenamefont
  {Radu}(2015)}]{Herdeiro:2015waa}%
  \BibitemOpen
  \bibfield  {author} {\bibinfo {author} {\bibfnamefont {C.~A.~R.}\
  \bibnamefont {Herdeiro}}\ and\ \bibinfo {author} {\bibfnamefont
  {E.}~\bibnamefont {Radu}},\ }\href {\doibase 10.1142/S0218271815420146}
  {\bibfield  {journal} {\bibinfo  {journal} {Int. J. Mod. Phys. D}\ }\textbf
  {\bibinfo {volume} {24}},\ \bibinfo {pages} {1542014} (\bibinfo {year}
  {2015})},\ \Eprint {http://arxiv.org/abs/1504.08209} {arXiv:1504.08209
  [gr-qc]} \BibitemShut {NoStop}%
\bibitem [{\citenamefont {Herdeiro}(2023)}]{Herdeiro:2022yle}%
  \BibitemOpen
  \bibfield  {author} {\bibinfo {author} {\bibfnamefont {C.~A.~R.}\
  \bibnamefont {Herdeiro}},\ }\href {\doibase 10.1007/978-3-031-31520-6_8}
  {\bibfield  {journal} {\bibinfo  {journal} {Lect. Notes Phys.}\ }\textbf
  {\bibinfo {volume} {1017}},\ \bibinfo {pages} {315} (\bibinfo {year}
  {2023})},\ \Eprint {http://arxiv.org/abs/2204.05640} {arXiv:2204.05640
  [gr-qc]} \BibitemShut {NoStop}%
\bibitem [{\citenamefont {Cardoso}\ and\ \citenamefont
  {Gualtieri}(2016)}]{Cardoso:2016ryw}%
  \BibitemOpen
  \bibfield  {author} {\bibinfo {author} {\bibfnamefont {V.}~\bibnamefont
  {Cardoso}}\ and\ \bibinfo {author} {\bibfnamefont {L.}~\bibnamefont
  {Gualtieri}},\ }\href {\doibase 10.1088/0264-9381/33/17/174001} {\bibfield
  {journal} {\bibinfo  {journal} {Class. Quant. Grav.}\ }\textbf {\bibinfo
  {volume} {33}},\ \bibinfo {pages} {174001} (\bibinfo {year} {2016})},\
  \Eprint {http://arxiv.org/abs/1607.03133} {arXiv:1607.03133 [gr-qc]}
  \BibitemShut {NoStop}%
\bibitem [{\citenamefont {Xu}\ \emph {et~al.}(2023{\natexlab{a}})\citenamefont
  {Xu}, \citenamefont {Liang},\ and\ \citenamefont {Shao}}]{Xu:2022frb}%
  \BibitemOpen
  \bibfield  {author} {\bibinfo {author} {\bibfnamefont {R.}~\bibnamefont
  {Xu}}, \bibinfo {author} {\bibfnamefont {D.}~\bibnamefont {Liang}}, \ and\
  \bibinfo {author} {\bibfnamefont {L.}~\bibnamefont {Shao}},\ }\href {\doibase
  10.1103/PhysRevD.107.024011} {\bibfield  {journal} {\bibinfo  {journal}
  {Phys. Rev. D}\ }\textbf {\bibinfo {volume} {107}},\ \bibinfo {pages}
  {024011} (\bibinfo {year} {2023}{\natexlab{a}})},\ \Eprint
  {http://arxiv.org/abs/2209.02209} {arXiv:2209.02209 [gr-qc]} \BibitemShut
  {NoStop}%
\bibitem [{\citenamefont {Dreyer}\ \emph {et~al.}(2004)\citenamefont {Dreyer},
  \citenamefont {Kelly}, \citenamefont {Krishnan}, \citenamefont {Finn},
  \citenamefont {Garrison},\ and\ \citenamefont
  {Lopez-Aleman}}]{Dreyer:2003bv}%
  \BibitemOpen
  \bibfield  {author} {\bibinfo {author} {\bibfnamefont {O.}~\bibnamefont
  {Dreyer}}, \bibinfo {author} {\bibfnamefont {B.~J.}\ \bibnamefont {Kelly}},
  \bibinfo {author} {\bibfnamefont {B.}~\bibnamefont {Krishnan}}, \bibinfo
  {author} {\bibfnamefont {L.~S.}\ \bibnamefont {Finn}}, \bibinfo {author}
  {\bibfnamefont {D.}~\bibnamefont {Garrison}}, \ and\ \bibinfo {author}
  {\bibfnamefont {R.}~\bibnamefont {Lopez-Aleman}},\ }\href {\doibase
  10.1088/0264-9381/21/4/003} {\bibfield  {journal} {\bibinfo  {journal}
  {Class. Quant. Grav.}\ }\textbf {\bibinfo {volume} {21}},\ \bibinfo {pages}
  {787} (\bibinfo {year} {2004})},\ \Eprint
  {http://arxiv.org/abs/gr-qc/0309007} {arXiv:gr-qc/0309007 [gr-qc]}
  \BibitemShut {NoStop}%
\bibitem [{\citenamefont {Ryan}(1995)}]{Ryan:1995wh}%
  \BibitemOpen
  \bibfield  {author} {\bibinfo {author} {\bibfnamefont {F.~D.}\ \bibnamefont
  {Ryan}},\ }\href {\doibase 10.1103/PhysRevD.52.5707} {\bibfield  {journal}
  {\bibinfo  {journal} {Phys. Rev. D}\ }\textbf {\bibinfo {volume} {52}},\
  \bibinfo {pages} {5707} (\bibinfo {year} {1995})}\BibitemShut {NoStop}%
\bibitem [{\citenamefont {Berti}\ \emph {et~al.}(2016)\citenamefont {Berti},
  \citenamefont {Sesana}, \citenamefont {Barausse}, \citenamefont {Cardoso},\
  and\ \citenamefont {Belczynski}}]{Berti:2016lat}%
  \BibitemOpen
  \bibfield  {author} {\bibinfo {author} {\bibfnamefont {E.}~\bibnamefont
  {Berti}}, \bibinfo {author} {\bibfnamefont {A.}~\bibnamefont {Sesana}},
  \bibinfo {author} {\bibfnamefont {E.}~\bibnamefont {Barausse}}, \bibinfo
  {author} {\bibfnamefont {V.}~\bibnamefont {Cardoso}}, \ and\ \bibinfo
  {author} {\bibfnamefont {K.}~\bibnamefont {Belczynski}},\ }\href {\doibase
  10.1103/PhysRevLett.117.101102} {\bibfield  {journal} {\bibinfo  {journal}
  {Phys. Rev. Lett.}\ }\textbf {\bibinfo {volume} {117}},\ \bibinfo {pages}
  {101102} (\bibinfo {year} {2016})},\ \Eprint
  {http://arxiv.org/abs/1605.09286} {arXiv:1605.09286 [gr-qc]} \BibitemShut
  {NoStop}%
\bibitem [{\citenamefont {Amaro-Seoane}\ \emph {et~al.}(2013)\citenamefont
  {Amaro-Seoane} \emph {et~al.}}]{Amaro-Seoane:2012aqc}%
  \BibitemOpen
  \bibfield  {author} {\bibinfo {author} {\bibfnamefont {P.}~\bibnamefont
  {Amaro-Seoane}} \emph {et~al.},\ }\href@noop {} {\bibfield  {journal}
  {\bibinfo  {journal} {GW Notes}\ }\textbf {\bibinfo {volume} {6}},\ \bibinfo
  {pages} {4} (\bibinfo {year} {2013})},\ \Eprint
  {http://arxiv.org/abs/1201.3621} {arXiv:1201.3621 [astro-ph.CO]} \BibitemShut
  {NoStop}%
\bibitem [{\citenamefont {Punturo}\ \emph
  {et~al.}(2010{\natexlab{a}})\citenamefont {Punturo} \emph
  {et~al.}}]{Punturo:2010zz}%
  \BibitemOpen
  \bibfield  {author} {\bibinfo {author} {\bibfnamefont {M.}~\bibnamefont
  {Punturo}} \emph {et~al.},\ }\href {\doibase 10.1088/0264-9381/27/19/194002}
  {\bibfield  {journal} {\bibinfo  {journal} {Class. Quant. Grav.}\ }\textbf
  {\bibinfo {volume} {27}},\ \bibinfo {pages} {194002} (\bibinfo {year}
  {2010}{\natexlab{a}})}\BibitemShut {NoStop}%
\bibitem [{\citenamefont {Gossan}\ \emph {et~al.}(2012)\citenamefont {Gossan},
  \citenamefont {Veitch},\ and\ \citenamefont {Sathyaprakash}}]{Gossan:2011ha}%
  \BibitemOpen
  \bibfield  {author} {\bibinfo {author} {\bibfnamefont {S.}~\bibnamefont
  {Gossan}}, \bibinfo {author} {\bibfnamefont {J.}~\bibnamefont {Veitch}}, \
  and\ \bibinfo {author} {\bibfnamefont {B.~S.}\ \bibnamefont
  {Sathyaprakash}},\ }\href {\doibase 10.1103/PhysRevD.85.124056} {\bibfield
  {journal} {\bibinfo  {journal} {Phys. Rev.}\ }\textbf {\bibinfo {volume}
  {D85}},\ \bibinfo {pages} {124056} (\bibinfo {year} {2012})},\ \Eprint
  {http://arxiv.org/abs/1111.5819} {arXiv:1111.5819 [gr-qc]} \BibitemShut
  {NoStop}%
\bibitem [{\citenamefont {Bhagwat}\ \emph {et~al.}(2020)\citenamefont
  {Bhagwat}, \citenamefont {Cabero}, \citenamefont {Capano}, \citenamefont
  {Krishnan},\ and\ \citenamefont {Brown}}]{Bhagwat:2019bwv}%
  \BibitemOpen
  \bibfield  {author} {\bibinfo {author} {\bibfnamefont {S.}~\bibnamefont
  {Bhagwat}}, \bibinfo {author} {\bibfnamefont {M.}~\bibnamefont {Cabero}},
  \bibinfo {author} {\bibfnamefont {C.~D.}\ \bibnamefont {Capano}}, \bibinfo
  {author} {\bibfnamefont {B.}~\bibnamefont {Krishnan}}, \ and\ \bibinfo
  {author} {\bibfnamefont {D.~A.}\ \bibnamefont {Brown}},\ }\href {\doibase
  10.1103/PhysRevD.102.024023} {\bibfield  {journal} {\bibinfo  {journal}
  {Phys. Rev. D}\ }\textbf {\bibinfo {volume} {102}},\ \bibinfo {pages}
  {024023} (\bibinfo {year} {2020})},\ \Eprint
  {http://arxiv.org/abs/1910.13203} {arXiv:1910.13203 [gr-qc]} \BibitemShut
  {NoStop}%
\bibitem [{\citenamefont {Gu}\ \emph {et~al.}(2024)\citenamefont {Gu},
  \citenamefont {Wang},\ and\ \citenamefont {Shao}}]{Gu:2023eaa}%
  \BibitemOpen
  \bibfield  {author} {\bibinfo {author} {\bibfnamefont {H.-P.}\ \bibnamefont
  {Gu}}, \bibinfo {author} {\bibfnamefont {H.-T.}\ \bibnamefont {Wang}}, \ and\
  \bibinfo {author} {\bibfnamefont {L.}~\bibnamefont {Shao}},\ }\href {\doibase
  10.1103/PhysRevD.109.024058} {\bibfield  {journal} {\bibinfo  {journal}
  {Phys. Rev. D}\ }\textbf {\bibinfo {volume} {109}},\ \bibinfo {pages}
  {024058} (\bibinfo {year} {2024})},\ \Eprint
  {http://arxiv.org/abs/2310.10447} {arXiv:2310.10447 [gr-qc]} \BibitemShut
  {NoStop}%
\bibitem [{\citenamefont {Luo}\ \emph {et~al.}(2016{\natexlab{b}})\citenamefont
  {Luo}, \citenamefont {Chen}, \citenamefont {Duan}, \citenamefont {Gong},
  \citenamefont {Hu}, \citenamefont {Ji}, \citenamefont {Liu}, \citenamefont
  {Mei}, \citenamefont {Milyukov}, \citenamefont {Sazhin}, \citenamefont
  {Shao}, \citenamefont {Toth}, \citenamefont {Tu}, \citenamefont {Wang},
  \citenamefont {Wang}, \citenamefont {Yeh}, \citenamefont {Zhan},
  \citenamefont {Zhang}, \citenamefont {Zharov},\ and\ \citenamefont
  {Zhou}}]{Luo2016}%
  \BibitemOpen
  \bibfield  {author} {\bibinfo {author} {\bibfnamefont {J.}~\bibnamefont
  {Luo}}, \bibinfo {author} {\bibfnamefont {L.-S.}\ \bibnamefont {Chen}},
  \bibinfo {author} {\bibfnamefont {H.-Z.}\ \bibnamefont {Duan}}, \bibinfo
  {author} {\bibfnamefont {Y.-G.}\ \bibnamefont {Gong}}, \bibinfo {author}
  {\bibfnamefont {S.}~\bibnamefont {Hu}}, \bibinfo {author} {\bibfnamefont
  {J.}~\bibnamefont {Ji}}, \bibinfo {author} {\bibfnamefont {Q.}~\bibnamefont
  {Liu}}, \bibinfo {author} {\bibfnamefont {J.}~\bibnamefont {Mei}}, \bibinfo
  {author} {\bibfnamefont {V.}~\bibnamefont {Milyukov}}, \bibinfo {author}
  {\bibfnamefont {M.}~\bibnamefont {Sazhin}}, \bibinfo {author} {\bibfnamefont
  {C.-G.}\ \bibnamefont {Shao}}, \bibinfo {author} {\bibfnamefont {V.~T.}\
  \bibnamefont {Toth}}, \bibinfo {author} {\bibfnamefont {H.-B.}\ \bibnamefont
  {Tu}}, \bibinfo {author} {\bibfnamefont {Y.}~\bibnamefont {Wang}}, \bibinfo
  {author} {\bibfnamefont {Y.}~\bibnamefont {Wang}}, \bibinfo {author}
  {\bibfnamefont {H.-C.}\ \bibnamefont {Yeh}}, \bibinfo {author} {\bibfnamefont
  {M.-S.}\ \bibnamefont {Zhan}}, \bibinfo {author} {\bibfnamefont
  {Y.}~\bibnamefont {Zhang}}, \bibinfo {author} {\bibfnamefont
  {V.}~\bibnamefont {Zharov}}, \ and\ \bibinfo {author} {\bibfnamefont {Z.-B.}\
  \bibnamefont {Zhou}},\ }\href {\doibase 10.1088/0264-9381/33/3/035010}
  {\bibfield  {journal} {\bibinfo  {journal} {Classical and Quantum Gravity}\
  }\textbf {\bibinfo {volume} {33}},\ \bibinfo {pages} {035010} (\bibinfo
  {year} {2016}{\natexlab{b}})}\BibitemShut {NoStop}%
\bibitem [{\citenamefont {Pappas}\ and\ \citenamefont
  {Apostolatos}(2012)}]{Pappas:2012ns}%
  \BibitemOpen
  \bibfield  {author} {\bibinfo {author} {\bibfnamefont {G.}~\bibnamefont
  {Pappas}}\ and\ \bibinfo {author} {\bibfnamefont {T.~A.}\ \bibnamefont
  {Apostolatos}},\ }\href {\doibase 10.1103/PhysRevLett.108.231104} {\bibfield
  {journal} {\bibinfo  {journal} {Phys. Rev. Lett.}\ }\textbf {\bibinfo
  {volume} {108}},\ \bibinfo {pages} {231104} (\bibinfo {year} {2012})},\
  \Eprint {http://arxiv.org/abs/1201.6067} {arXiv:1201.6067 [gr-qc]}
  \BibitemShut {NoStop}%
\bibitem [{\citenamefont {Herdeiro}\ and\ \citenamefont
  {Radu}(2014)}]{Herdeiro:2014goa}%
  \BibitemOpen
  \bibfield  {author} {\bibinfo {author} {\bibfnamefont {C.~A.~R.}\
  \bibnamefont {Herdeiro}}\ and\ \bibinfo {author} {\bibfnamefont
  {E.}~\bibnamefont {Radu}},\ }\href {\doibase 10.1103/PhysRevLett.112.221101}
  {\bibfield  {journal} {\bibinfo  {journal} {Phys. Rev. Lett.}\ }\textbf
  {\bibinfo {volume} {112}},\ \bibinfo {pages} {221101} (\bibinfo {year}
  {2014})},\ \Eprint {http://arxiv.org/abs/1403.2757} {arXiv:1403.2757 [gr-qc]}
  \BibitemShut {NoStop}%
\bibitem [{\citenamefont {Baumann}\ \emph {et~al.}(2019)\citenamefont
  {Baumann}, \citenamefont {Chia},\ and\ \citenamefont
  {Porto}}]{Baumann:2018vus}%
  \BibitemOpen
  \bibfield  {author} {\bibinfo {author} {\bibfnamefont {D.}~\bibnamefont
  {Baumann}}, \bibinfo {author} {\bibfnamefont {H.~S.}\ \bibnamefont {Chia}}, \
  and\ \bibinfo {author} {\bibfnamefont {R.~A.}\ \bibnamefont {Porto}},\ }\href
  {\doibase 10.1103/PhysRevD.99.044001} {\bibfield  {journal} {\bibinfo
  {journal} {Phys. Rev. D}\ }\textbf {\bibinfo {volume} {99}},\ \bibinfo
  {pages} {044001} (\bibinfo {year} {2019})},\ \Eprint
  {http://arxiv.org/abs/1804.03208} {arXiv:1804.03208 [gr-qc]} \BibitemShut
  {NoStop}%
\bibitem [{\citenamefont {Krishnendu}\ \emph {et~al.}(2017)\citenamefont
  {Krishnendu}, \citenamefont {Arun},\ and\ \citenamefont
  {Mishra}}]{Krishnendu:2017shb}%
  \BibitemOpen
  \bibfield  {author} {\bibinfo {author} {\bibfnamefont {N.~V.}\ \bibnamefont
  {Krishnendu}}, \bibinfo {author} {\bibfnamefont {K.~G.}\ \bibnamefont
  {Arun}}, \ and\ \bibinfo {author} {\bibfnamefont {C.~K.}\ \bibnamefont
  {Mishra}},\ }\href {\doibase 10.1103/PhysRevLett.119.091101} {\bibfield
  {journal} {\bibinfo  {journal} {Phys. Rev. Lett.}\ }\textbf {\bibinfo
  {volume} {119}},\ \bibinfo {pages} {091101} (\bibinfo {year} {2017})},\
  \Eprint {http://arxiv.org/abs/1701.06318} {arXiv:1701.06318 [gr-qc]}
  \BibitemShut {NoStop}%
\bibitem [{\citenamefont {Abbott}\ \emph
  {et~al.}(2021{\natexlab{c}})\citenamefont {Abbott} \emph
  {et~al.}}]{LIGOScientific:2020ibl}%
  \BibitemOpen
  \bibfield  {author} {\bibinfo {author} {\bibfnamefont {R.}~\bibnamefont
  {Abbott}} \emph {et~al.} (\bibinfo {collaboration} {LIGO Scientific,
  Virgo}),\ }\href {\doibase 10.1103/PhysRevX.11.021053} {\bibfield  {journal}
  {\bibinfo  {journal} {Phys. Rev. X}\ }\textbf {\bibinfo {volume} {11}},\
  \bibinfo {pages} {021053} (\bibinfo {year} {2021}{\natexlab{c}})},\ \Eprint
  {http://arxiv.org/abs/2010.14527} {arXiv:2010.14527 [gr-qc]} \BibitemShut
  {NoStop}%
\bibitem [{\citenamefont {Divyajyoti}\ \emph {et~al.}(2024)\citenamefont
  {Divyajyoti}, \citenamefont {Krishnendu}, \citenamefont {Saleem},
  \citenamefont {Colleoni}, \citenamefont {Vijaykumar}, \citenamefont {Arun},\
  and\ \citenamefont {Mishra}}]{Divyajyoti:2023izl}%
  \BibitemOpen
  \bibfield  {author} {\bibinfo {author} {\bibnamefont {Divyajyoti}}, \bibinfo
  {author} {\bibfnamefont {N.~V.}\ \bibnamefont {Krishnendu}}, \bibinfo
  {author} {\bibfnamefont {M.}~\bibnamefont {Saleem}}, \bibinfo {author}
  {\bibfnamefont {M.}~\bibnamefont {Colleoni}}, \bibinfo {author}
  {\bibfnamefont {A.}~\bibnamefont {Vijaykumar}}, \bibinfo {author}
  {\bibfnamefont {K.~G.}\ \bibnamefont {Arun}}, \ and\ \bibinfo {author}
  {\bibfnamefont {C.~K.}\ \bibnamefont {Mishra}},\ }\href {\doibase
  10.1103/PhysRevD.109.023016} {\bibfield  {journal} {\bibinfo  {journal}
  {Phys. Rev. D}\ }\textbf {\bibinfo {volume} {109}},\ \bibinfo {pages}
  {023016} (\bibinfo {year} {2024})},\ \Eprint
  {http://arxiv.org/abs/2311.05506} {arXiv:2311.05506 [gr-qc]} \BibitemShut
  {NoStop}%
\bibitem [{\citenamefont {Abbott}\ \emph
  {et~al.}(2023{\natexlab{a}})\citenamefont {Abbott} \emph
  {et~al.}}]{KAGRA:2021vkt}%
  \BibitemOpen
  \bibfield  {author} {\bibinfo {author} {\bibfnamefont {R.}~\bibnamefont
  {Abbott}} \emph {et~al.} (\bibinfo {collaboration} {KAGRA, VIRGO, LIGO
  Scientific}),\ }\href {\doibase 10.1103/PhysRevX.13.041039} {\bibfield
  {journal} {\bibinfo  {journal} {Phys. Rev. X}\ }\textbf {\bibinfo {volume}
  {13}},\ \bibinfo {pages} {041039} (\bibinfo {year} {2023}{\natexlab{a}})},\
  \Eprint {http://arxiv.org/abs/2111.03606} {arXiv:2111.03606 [gr-qc]}
  \BibitemShut {NoStop}%
\bibitem [{\citenamefont {Li}\ and\ \citenamefont {Han}(2022)}]{Li:2022fml}%
  \BibitemOpen
  \bibfield  {author} {\bibinfo {author} {\bibfnamefont {S.}~\bibnamefont
  {Li}}\ and\ \bibinfo {author} {\bibfnamefont {W.-B.}\ \bibnamefont {Han}},\
  }\href {\doibase 10.1103/PhysRevD.106.104013} {\bibfield  {journal} {\bibinfo
   {journal} {Phys. Rev. D}\ }\textbf {\bibinfo {volume} {106}},\ \bibinfo
  {pages} {104013} (\bibinfo {year} {2022})},\ \Eprint
  {http://arxiv.org/abs/2204.09367} {arXiv:2204.09367 [gr-qc]} \BibitemShut
  {NoStop}%
\bibitem [{\citenamefont {Li}\ and\ \citenamefont {Han}(2023)}]{Li:2023qcu}%
  \BibitemOpen
  \bibfield  {author} {\bibinfo {author} {\bibfnamefont {S.}~\bibnamefont
  {Li}}\ and\ \bibinfo {author} {\bibfnamefont {W.-B.}\ \bibnamefont {Han}},\
  }\href {\doibase 10.1103/PhysRevD.108.083032} {\bibfield  {journal} {\bibinfo
   {journal} {Phys. Rev. D}\ }\textbf {\bibinfo {volume} {108}},\ \bibinfo
  {pages} {083032} (\bibinfo {year} {2023})},\ \Eprint
  {http://arxiv.org/abs/2307.00797} {arXiv:2307.00797 [gr-qc]} \BibitemShut
  {NoStop}%
\bibitem [{\citenamefont {Li}\ \emph {et~al.}(2024{\natexlab{a}})\citenamefont
  {Li}, \citenamefont {Han},\ and\ \citenamefont {Yang}}]{Li:2023zbm}%
  \BibitemOpen
  \bibfield  {author} {\bibinfo {author} {\bibfnamefont {S.}~\bibnamefont
  {Li}}, \bibinfo {author} {\bibfnamefont {W.-B.}\ \bibnamefont {Han}}, \ and\
  \bibinfo {author} {\bibfnamefont {S.-C.}\ \bibnamefont {Yang}},\ }\href
  {\doibase 10.1088/1475-7516/2024/06/013} {\bibfield  {journal} {\bibinfo
  {journal} {JCAP}\ }\textbf {\bibinfo {volume} {06}},\ \bibinfo {pages} {013}
  (\bibinfo {year} {2024}{\natexlab{a}})},\ \Eprint
  {http://arxiv.org/abs/2312.02841} {arXiv:2312.02841 [gr-qc]} \BibitemShut
  {NoStop}%
\bibitem [{\citenamefont {Krishnendu}\ and\ \citenamefont
  {Yelikar}(2020)}]{Krishnendu:2019ebd}%
  \BibitemOpen
  \bibfield  {author} {\bibinfo {author} {\bibfnamefont {N.~V.}\ \bibnamefont
  {Krishnendu}}\ and\ \bibinfo {author} {\bibfnamefont {A.~B.}\ \bibnamefont
  {Yelikar}},\ }\href {\doibase 10.1088/1361-6382/ababb1} {\bibfield  {journal}
  {\bibinfo  {journal} {Class. Quant. Grav.}\ }\textbf {\bibinfo {volume}
  {37}},\ \bibinfo {pages} {205019} (\bibinfo {year} {2020})},\ \Eprint
  {http://arxiv.org/abs/1904.12712} {arXiv:1904.12712 [gr-qc]} \BibitemShut
  {NoStop}%
\bibitem [{\citenamefont {Ryan}(1997)}]{Ryan:1997hg}%
  \BibitemOpen
  \bibfield  {author} {\bibinfo {author} {\bibfnamefont {F.~D.}\ \bibnamefont
  {Ryan}},\ }\href {\doibase 10.1103/PhysRevD.56.1845} {\bibfield  {journal}
  {\bibinfo  {journal} {Phys. Rev. D}\ }\textbf {\bibinfo {volume} {56}},\
  \bibinfo {pages} {1845} (\bibinfo {year} {1997})}\BibitemShut {NoStop}%
\bibitem [{\citenamefont {Barack}\ and\ \citenamefont
  {Cutler}(2004)}]{Barack:2003fp}%
  \BibitemOpen
  \bibfield  {author} {\bibinfo {author} {\bibfnamefont {L.}~\bibnamefont
  {Barack}}\ and\ \bibinfo {author} {\bibfnamefont {C.}~\bibnamefont
  {Cutler}},\ }\href {\doibase 10.1103/PhysRevD.69.082005} {\bibfield
  {journal} {\bibinfo  {journal} {Phys. Rev. D}\ }\textbf {\bibinfo {volume}
  {69}},\ \bibinfo {pages} {082005} (\bibinfo {year} {2004})},\ \Eprint
  {http://arxiv.org/abs/gr-qc/0310125} {arXiv:gr-qc/0310125} \BibitemShut
  {NoStop}%
\bibitem [{\citenamefont {Barack}\ and\ \citenamefont
  {Cutler}(2007)}]{Barack:2006pq}%
  \BibitemOpen
  \bibfield  {author} {\bibinfo {author} {\bibfnamefont {L.}~\bibnamefont
  {Barack}}\ and\ \bibinfo {author} {\bibfnamefont {C.}~\bibnamefont
  {Cutler}},\ }\href {\doibase 10.1103/PhysRevD.75.042003} {\bibfield
  {journal} {\bibinfo  {journal} {Phys. Rev. D}\ }\textbf {\bibinfo {volume}
  {75}},\ \bibinfo {pages} {042003} (\bibinfo {year} {2007})},\ \Eprint
  {http://arxiv.org/abs/gr-qc/0612029} {arXiv:gr-qc/0612029} \BibitemShut
  {NoStop}%
\bibitem [{\citenamefont {Babak}\ \emph {et~al.}(2017)\citenamefont {Babak},
  \citenamefont {Gair}, \citenamefont {Sesana}, \citenamefont {Barausse},
  \citenamefont {Sopuerta}, \citenamefont {Berry}, \citenamefont {Berti},
  \citenamefont {Amaro-Seoane}, \citenamefont {Petiteau},\ and\ \citenamefont
  {Klein}}]{Babak:2017tow}%
  \BibitemOpen
  \bibfield  {author} {\bibinfo {author} {\bibfnamefont {S.}~\bibnamefont
  {Babak}}, \bibinfo {author} {\bibfnamefont {J.}~\bibnamefont {Gair}},
  \bibinfo {author} {\bibfnamefont {A.}~\bibnamefont {Sesana}}, \bibinfo
  {author} {\bibfnamefont {E.}~\bibnamefont {Barausse}}, \bibinfo {author}
  {\bibfnamefont {C.~F.}\ \bibnamefont {Sopuerta}}, \bibinfo {author}
  {\bibfnamefont {C.~P.~L.}\ \bibnamefont {Berry}}, \bibinfo {author}
  {\bibfnamefont {E.}~\bibnamefont {Berti}}, \bibinfo {author} {\bibfnamefont
  {P.}~\bibnamefont {Amaro-Seoane}}, \bibinfo {author} {\bibfnamefont
  {A.}~\bibnamefont {Petiteau}}, \ and\ \bibinfo {author} {\bibfnamefont
  {A.}~\bibnamefont {Klein}},\ }\href {\doibase 10.1103/PhysRevD.95.103012}
  {\bibfield  {journal} {\bibinfo  {journal} {Phys. Rev. D}\ }\textbf {\bibinfo
  {volume} {95}},\ \bibinfo {pages} {103012} (\bibinfo {year} {2017})},\
  \Eprint {http://arxiv.org/abs/1703.09722} {arXiv:1703.09722 [gr-qc]}
  \BibitemShut {NoStop}%
\bibitem [{\citenamefont {Pani}\ \emph {et~al.}(2013)\citenamefont {Pani},
  \citenamefont {Sotiriou},\ and\ \citenamefont {Vernieri}}]{Pani:2013qfa}%
  \BibitemOpen
  \bibfield  {author} {\bibinfo {author} {\bibfnamefont {P.}~\bibnamefont
  {Pani}}, \bibinfo {author} {\bibfnamefont {T.~P.}\ \bibnamefont {Sotiriou}},
  \ and\ \bibinfo {author} {\bibfnamefont {D.}~\bibnamefont {Vernieri}},\
  }\href {\doibase 10.1103/PhysRevD.88.121502} {\bibfield  {journal} {\bibinfo
  {journal} {Phys. Rev. D}\ }\textbf {\bibinfo {volume} {88}},\ \bibinfo
  {pages} {121502} (\bibinfo {year} {2013})},\ \Eprint
  {http://arxiv.org/abs/1306.1835} {arXiv:1306.1835 [gr-qc]} \BibitemShut
  {NoStop}%
\bibitem [{\citenamefont {Baessler}\ \emph {et~al.}(1999)\citenamefont
  {Baessler}, \citenamefont {Heckel}, \citenamefont {Adelberger}, \citenamefont
  {Gundlach}, \citenamefont {Schmidt},\ and\ \citenamefont
  {Swanson}}]{Baessler:1999iv}%
  \BibitemOpen
  \bibfield  {author} {\bibinfo {author} {\bibfnamefont {S.}~\bibnamefont
  {Baessler}}, \bibinfo {author} {\bibfnamefont {B.~R.}\ \bibnamefont
  {Heckel}}, \bibinfo {author} {\bibfnamefont {E.~G.}\ \bibnamefont
  {Adelberger}}, \bibinfo {author} {\bibfnamefont {J.~H.}\ \bibnamefont
  {Gundlach}}, \bibinfo {author} {\bibfnamefont {U.}~\bibnamefont {Schmidt}}, \
  and\ \bibinfo {author} {\bibfnamefont {H.~E.}\ \bibnamefont {Swanson}},\
  }\href {\doibase 10.1103/PhysRevLett.83.003585} {\bibfield  {journal}
  {\bibinfo  {journal} {Phys. Rev. Lett.}\ }\textbf {\bibinfo {volume} {83}},\
  \bibinfo {pages} {3585} (\bibinfo {year} {1999})}\BibitemShut {NoStop}%
\bibitem [{\citenamefont {Lovelock}(1971)}]{Lovelock:1971yv}%
  \BibitemOpen
  \bibfield  {author} {\bibinfo {author} {\bibfnamefont {D.}~\bibnamefont
  {Lovelock}},\ }\href {\doibase 10.1063/1.1665613} {\bibfield  {journal}
  {\bibinfo  {journal} {J. Math. Phys.}\ }\textbf {\bibinfo {volume} {12}},\
  \bibinfo {pages} {498} (\bibinfo {year} {1971})}\BibitemShut {NoStop}%
\bibitem [{\citenamefont {Lovelock}(1972)}]{Lovelock:1972vz}%
  \BibitemOpen
  \bibfield  {author} {\bibinfo {author} {\bibfnamefont {D.}~\bibnamefont
  {Lovelock}},\ }\href {\doibase 10.1063/1.1666069} {\bibfield  {journal}
  {\bibinfo  {journal} {J. Math. Phys.}\ }\textbf {\bibinfo {volume} {13}},\
  \bibinfo {pages} {874} (\bibinfo {year} {1972})}\BibitemShut {NoStop}%
\bibitem [{\citenamefont {Sotiriou}(2015)}]{Sotiriou:2014yhm}%
  \BibitemOpen
  \bibfield  {author} {\bibinfo {author} {\bibfnamefont {T.~P.}\ \bibnamefont
  {Sotiriou}},\ }\href {\doibase 10.1007/978-3-319-10070-8_1} {\bibfield
  {journal} {\bibinfo  {journal} {Lect. Notes Phys.}\ }\textbf {\bibinfo
  {volume} {892}},\ \bibinfo {pages} {3} (\bibinfo {year} {2015})},\ \Eprint
  {http://arxiv.org/abs/1404.2955} {arXiv:1404.2955 [gr-qc]} \BibitemShut
  {NoStop}%
\bibitem [{\citenamefont {Hassan}\ and\ \citenamefont
  {Rosen}(2012)}]{Hassan:2011zd}%
  \BibitemOpen
  \bibfield  {author} {\bibinfo {author} {\bibfnamefont {S.~F.}\ \bibnamefont
  {Hassan}}\ and\ \bibinfo {author} {\bibfnamefont {R.~A.}\ \bibnamefont
  {Rosen}},\ }\href {\doibase 10.1007/JHEP02(2012)126} {\bibfield  {journal}
  {\bibinfo  {journal} {JHEP}\ }\textbf {\bibinfo {volume} {02}},\ \bibinfo
  {pages} {126} (\bibinfo {year} {2012})},\ \Eprint
  {http://arxiv.org/abs/1109.3515} {arXiv:1109.3515 [hep-th]} \BibitemShut
  {NoStop}%
\bibitem [{\citenamefont {Bergmann}(1968)}]{Bergmann:1968ve}%
  \BibitemOpen
  \bibfield  {author} {\bibinfo {author} {\bibfnamefont {P.~G.}\ \bibnamefont
  {Bergmann}},\ }\href {\doibase 10.1007/BF00668828} {\bibfield  {journal}
  {\bibinfo  {journal} {Int. J. Theor. Phys.}\ }\textbf {\bibinfo {volume}
  {1}},\ \bibinfo {pages} {25} (\bibinfo {year} {1968})}\BibitemShut {NoStop}%
\bibitem [{\citenamefont {Wagoner}(1970)}]{Wagoner:1970vr}%
  \BibitemOpen
  \bibfield  {author} {\bibinfo {author} {\bibfnamefont {R.~V.}\ \bibnamefont
  {Wagoner}},\ }\href {\doibase 10.1103/PhysRevD.1.3209} {\bibfield  {journal}
  {\bibinfo  {journal} {Phys. Rev. D}\ }\textbf {\bibinfo {volume} {1}},\
  \bibinfo {pages} {3209} (\bibinfo {year} {1970})}\BibitemShut {NoStop}%
\bibitem [{\citenamefont {Schmidt}(2006)}]{Schmidt:2006jt}%
  \BibitemOpen
  \bibfield  {author} {\bibinfo {author} {\bibfnamefont {H.-J.}\ \bibnamefont
  {Schmidt}},\ }\href {\doibase 10.1142/S0219887807001977} {\bibfield
  {journal} {\bibinfo  {journal} {eConf}\ }\textbf {\bibinfo {volume}
  {C0602061}},\ \bibinfo {pages} {12} (\bibinfo {year} {2006})},\ \Eprint
  {http://arxiv.org/abs/gr-qc/0602017} {arXiv:gr-qc/0602017} \BibitemShut
  {NoStop}%
\bibitem [{\citenamefont {Horndeski}(1974)}]{Horndeski:1974wa}%
  \BibitemOpen
  \bibfield  {author} {\bibinfo {author} {\bibfnamefont {G.~W.}\ \bibnamefont
  {Horndeski}},\ }\href {\doibase 10.1007/BF01807638} {\bibfield  {journal}
  {\bibinfo  {journal} {Int. J. Theor. Phys.}\ }\textbf {\bibinfo {volume}
  {10}},\ \bibinfo {pages} {363} (\bibinfo {year} {1974})}\BibitemShut
  {NoStop}%
\bibitem [{\citenamefont {Damour}\ and\ \citenamefont
  {Esposito-Farese}(1992)}]{Damour:1992we}%
  \BibitemOpen
  \bibfield  {author} {\bibinfo {author} {\bibfnamefont {T.}~\bibnamefont
  {Damour}}\ and\ \bibinfo {author} {\bibfnamefont {G.}~\bibnamefont
  {Esposito-Farese}},\ }\href {\doibase 10.1088/0264-9381/9/9/015} {\bibfield
  {journal} {\bibinfo  {journal} {Class. Quant. Grav.}\ }\textbf {\bibinfo
  {volume} {9}},\ \bibinfo {pages} {2093} (\bibinfo {year} {1992})}\BibitemShut
  {NoStop}%
\bibitem [{\citenamefont {Olmo}(2011)}]{Olmo:2011uz}%
  \BibitemOpen
  \bibfield  {author} {\bibinfo {author} {\bibfnamefont {G.~J.}\ \bibnamefont
  {Olmo}},\ }\href {\doibase 10.1142/S0218271811018925} {\bibfield  {journal}
  {\bibinfo  {journal} {Int. J. Mod. Phys. D}\ }\textbf {\bibinfo {volume}
  {20}},\ \bibinfo {pages} {413} (\bibinfo {year} {2011})},\ \Eprint
  {http://arxiv.org/abs/1101.3864} {arXiv:1101.3864 [gr-qc]} \BibitemShut
  {NoStop}%
\bibitem [{\citenamefont {Kanti}\ \emph {et~al.}(1996)\citenamefont {Kanti},
  \citenamefont {Mavromatos}, \citenamefont {Rizos}, \citenamefont {Tamvakis},\
  and\ \citenamefont {Winstanley}}]{Kanti:1995vq}%
  \BibitemOpen
  \bibfield  {author} {\bibinfo {author} {\bibfnamefont {P.}~\bibnamefont
  {Kanti}}, \bibinfo {author} {\bibfnamefont {N.~E.}\ \bibnamefont
  {Mavromatos}}, \bibinfo {author} {\bibfnamefont {J.}~\bibnamefont {Rizos}},
  \bibinfo {author} {\bibfnamefont {K.}~\bibnamefont {Tamvakis}}, \ and\
  \bibinfo {author} {\bibfnamefont {E.}~\bibnamefont {Winstanley}},\ }\href
  {\doibase 10.1103/PhysRevD.54.5049} {\bibfield  {journal} {\bibinfo
  {journal} {Phys. Rev. D}\ }\textbf {\bibinfo {volume} {54}},\ \bibinfo
  {pages} {5049} (\bibinfo {year} {1996})},\ \Eprint
  {http://arxiv.org/abs/hep-th/9511071} {arXiv:hep-th/9511071} \BibitemShut
  {NoStop}%
\bibitem [{\citenamefont {Alexander}\ and\ \citenamefont
  {Yunes}(2009)}]{Alexander:2009tp}%
  \BibitemOpen
  \bibfield  {author} {\bibinfo {author} {\bibfnamefont {S.}~\bibnamefont
  {Alexander}}\ and\ \bibinfo {author} {\bibfnamefont {N.}~\bibnamefont
  {Yunes}},\ }\href {\doibase 10.1016/j.physrep.2009.07.002} {\bibfield
  {journal} {\bibinfo  {journal} {Phys. Rept.}\ }\textbf {\bibinfo {volume}
  {480}},\ \bibinfo {pages} {1} (\bibinfo {year} {2009})},\ \Eprint
  {http://arxiv.org/abs/0907.2562} {arXiv:0907.2562 [hep-th]} \BibitemShut
  {NoStop}%
\bibitem [{\citenamefont {Yunes}\ and\ \citenamefont
  {Stein}(2011)}]{Yunes:2011we}%
  \BibitemOpen
  \bibfield  {author} {\bibinfo {author} {\bibfnamefont {N.}~\bibnamefont
  {Yunes}}\ and\ \bibinfo {author} {\bibfnamefont {L.~C.}\ \bibnamefont
  {Stein}},\ }\href {\doibase 10.1103/PhysRevD.83.104002} {\bibfield  {journal}
  {\bibinfo  {journal} {Phys. Rev. D}\ }\textbf {\bibinfo {volume} {83}},\
  \bibinfo {pages} {104002} (\bibinfo {year} {2011})},\ \Eprint
  {http://arxiv.org/abs/1101.2921} {arXiv:1101.2921 [gr-qc]} \BibitemShut
  {NoStop}%
\bibitem [{\citenamefont {Pani}\ \emph {et~al.}(2011)\citenamefont {Pani},
  \citenamefont {Macedo}, \citenamefont {Crispino},\ and\ \citenamefont
  {Cardoso}}]{Pani:2011gy}%
  \BibitemOpen
  \bibfield  {author} {\bibinfo {author} {\bibfnamefont {P.}~\bibnamefont
  {Pani}}, \bibinfo {author} {\bibfnamefont {C.~F.~B.}\ \bibnamefont {Macedo}},
  \bibinfo {author} {\bibfnamefont {L.~C.~B.}\ \bibnamefont {Crispino}}, \ and\
  \bibinfo {author} {\bibfnamefont {V.}~\bibnamefont {Cardoso}},\ }\href
  {\doibase 10.1103/PhysRevD.84.087501} {\bibfield  {journal} {\bibinfo
  {journal} {Phys. Rev. D}\ }\textbf {\bibinfo {volume} {84}},\ \bibinfo
  {pages} {087501} (\bibinfo {year} {2011})},\ \Eprint
  {http://arxiv.org/abs/1109.3996} {arXiv:1109.3996 [gr-qc]} \BibitemShut
  {NoStop}%
\bibitem [{\citenamefont {Dirac}(1937)}]{Dirac:1937ti}%
  \BibitemOpen
  \bibfield  {author} {\bibinfo {author} {\bibfnamefont {P.~A.~M.}\
  \bibnamefont {Dirac}},\ }\href {\doibase 10.1038/139323a0} {\bibfield
  {journal} {\bibinfo  {journal} {Nature}\ }\textbf {\bibinfo {volume} {139}},\
  \bibinfo {pages} {323} (\bibinfo {year} {1937})}\BibitemShut {NoStop}%
\bibitem [{\citenamefont {Jacobson}(2007)}]{Jacobson:2007veq}%
  \BibitemOpen
  \bibfield  {author} {\bibinfo {author} {\bibfnamefont {T.}~\bibnamefont
  {Jacobson}},\ }\href {\doibase 10.22323/1.043.0020} {\bibfield  {journal}
  {\bibinfo  {journal} {PoS}\ }\textbf {\bibinfo {volume} {QG-PH}},\ \bibinfo
  {pages} {020} (\bibinfo {year} {2007})},\ \Eprint
  {http://arxiv.org/abs/0801.1547} {arXiv:0801.1547 [gr-qc]} \BibitemShut
  {NoStop}%
\bibitem [{\citenamefont {Blas}\ \emph {et~al.}(2010)\citenamefont {Blas},
  \citenamefont {Pujolas},\ and\ \citenamefont {Sibiryakov}}]{Blas:2009qj}%
  \BibitemOpen
  \bibfield  {author} {\bibinfo {author} {\bibfnamefont {D.}~\bibnamefont
  {Blas}}, \bibinfo {author} {\bibfnamefont {O.}~\bibnamefont {Pujolas}}, \
  and\ \bibinfo {author} {\bibfnamefont {S.}~\bibnamefont {Sibiryakov}},\
  }\href {\doibase 10.1103/PhysRevLett.104.181302} {\bibfield  {journal}
  {\bibinfo  {journal} {Phys. Rev. Lett.}\ }\textbf {\bibinfo {volume} {104}},\
  \bibinfo {pages} {181302} (\bibinfo {year} {2010})},\ \Eprint
  {http://arxiv.org/abs/0909.3525} {arXiv:0909.3525 [hep-th]} \BibitemShut
  {NoStop}%
\bibitem [{\citenamefont {Jacobson}(2010)}]{Jacobson:2010mx}%
  \BibitemOpen
  \bibfield  {author} {\bibinfo {author} {\bibfnamefont {T.}~\bibnamefont
  {Jacobson}},\ }\href {\doibase 10.1103/PhysRevD.81.101502} {\bibfield
  {journal} {\bibinfo  {journal} {Phys. Rev. D}\ }\textbf {\bibinfo {volume}
  {81}},\ \bibinfo {pages} {101502} (\bibinfo {year} {2010})},\ \bibinfo {note}
  {[Erratum: Phys.Rev.D 82, 129901 (2010)]},\ \Eprint
  {http://arxiv.org/abs/1001.4823} {arXiv:1001.4823 [hep-th]} \BibitemShut
  {NoStop}%
\bibitem [{\citenamefont {Horava}(2009)}]{Horava:2009uw}%
  \BibitemOpen
  \bibfield  {author} {\bibinfo {author} {\bibfnamefont {P.}~\bibnamefont
  {Horava}},\ }\href {\doibase 10.1103/PhysRevD.79.084008} {\bibfield
  {journal} {\bibinfo  {journal} {Phys. Rev. D}\ }\textbf {\bibinfo {volume}
  {79}},\ \bibinfo {pages} {084008} (\bibinfo {year} {2009})},\ \Eprint
  {http://arxiv.org/abs/0901.3775} {arXiv:0901.3775 [hep-th]} \BibitemShut
  {NoStop}%
\bibitem [{\citenamefont {Herdeiro}\ \emph {et~al.}(2011)\citenamefont
  {Herdeiro}, \citenamefont {Hirano},\ and\ \citenamefont
  {Sato}}]{Herdeiro:2011im}%
  \BibitemOpen
  \bibfield  {author} {\bibinfo {author} {\bibfnamefont {C.}~\bibnamefont
  {Herdeiro}}, \bibinfo {author} {\bibfnamefont {S.}~\bibnamefont {Hirano}}, \
  and\ \bibinfo {author} {\bibfnamefont {Y.}~\bibnamefont {Sato}},\ }\href
  {\doibase 10.1103/PhysRevD.84.124048} {\bibfield  {journal} {\bibinfo
  {journal} {Phys. Rev. D}\ }\textbf {\bibinfo {volume} {84}},\ \bibinfo
  {pages} {124048} (\bibinfo {year} {2011})},\ \Eprint
  {http://arxiv.org/abs/1110.0832} {arXiv:1110.0832 [gr-qc]} \BibitemShut
  {NoStop}%
\bibitem [{\citenamefont {Kostelecky}(2004)}]{Kostelecky:2003fs}%
  \BibitemOpen
  \bibfield  {author} {\bibinfo {author} {\bibfnamefont {V.~A.}\ \bibnamefont
  {Kostelecky}},\ }\href {\doibase 10.1103/PhysRevD.69.105009} {\bibfield
  {journal} {\bibinfo  {journal} {Phys. Rev. D}\ }\textbf {\bibinfo {volume}
  {69}},\ \bibinfo {pages} {105009} (\bibinfo {year} {2004})},\ \Eprint
  {http://arxiv.org/abs/hep-th/0312310} {arXiv:hep-th/0312310} \BibitemShut
  {NoStop}%
\bibitem [{\citenamefont {de~Rham}\ and\ \citenamefont
  {Gabadadze}(2010)}]{deRham:2010ik}%
  \BibitemOpen
  \bibfield  {author} {\bibinfo {author} {\bibfnamefont {C.}~\bibnamefont
  {de~Rham}}\ and\ \bibinfo {author} {\bibfnamefont {G.}~\bibnamefont
  {Gabadadze}},\ }\href {\doibase 10.1103/PhysRevD.82.044020} {\bibfield
  {journal} {\bibinfo  {journal} {Phys. Rev. D}\ }\textbf {\bibinfo {volume}
  {82}},\ \bibinfo {pages} {044020} (\bibinfo {year} {2010})},\ \Eprint
  {http://arxiv.org/abs/1007.0443} {arXiv:1007.0443 [hep-th]} \BibitemShut
  {NoStop}%
\bibitem [{\citenamefont {de~Rham}\ \emph {et~al.}(2011)\citenamefont
  {de~Rham}, \citenamefont {Gabadadze},\ and\ \citenamefont
  {Tolley}}]{deRham:2010kj}%
  \BibitemOpen
  \bibfield  {author} {\bibinfo {author} {\bibfnamefont {C.}~\bibnamefont
  {de~Rham}}, \bibinfo {author} {\bibfnamefont {G.}~\bibnamefont {Gabadadze}},
  \ and\ \bibinfo {author} {\bibfnamefont {A.~J.}\ \bibnamefont {Tolley}},\
  }\href {\doibase 10.1103/PhysRevLett.106.231101} {\bibfield  {journal}
  {\bibinfo  {journal} {Phys. Rev. Lett.}\ }\textbf {\bibinfo {volume} {106}},\
  \bibinfo {pages} {231101} (\bibinfo {year} {2011})},\ \Eprint
  {http://arxiv.org/abs/1011.1232} {arXiv:1011.1232 [hep-th]} \BibitemShut
  {NoStop}%
\bibitem [{\citenamefont {Nicolis}\ \emph {et~al.}(2009)\citenamefont
  {Nicolis}, \citenamefont {Rattazzi},\ and\ \citenamefont
  {Trincherini}}]{Nicolis:2008in}%
  \BibitemOpen
  \bibfield  {author} {\bibinfo {author} {\bibfnamefont {A.}~\bibnamefont
  {Nicolis}}, \bibinfo {author} {\bibfnamefont {R.}~\bibnamefont {Rattazzi}}, \
  and\ \bibinfo {author} {\bibfnamefont {E.}~\bibnamefont {Trincherini}},\
  }\href {\doibase 10.1103/PhysRevD.79.064036} {\bibfield  {journal} {\bibinfo
  {journal} {Phys. Rev. D}\ }\textbf {\bibinfo {volume} {79}},\ \bibinfo
  {pages} {064036} (\bibinfo {year} {2009})},\ \Eprint
  {http://arxiv.org/abs/0811.2197} {arXiv:0811.2197 [hep-th]} \BibitemShut
  {NoStop}%
\bibitem [{\citenamefont {Deffayet}\ \emph {et~al.}(2009)\citenamefont
  {Deffayet}, \citenamefont {Esposito-Farese},\ and\ \citenamefont
  {Vikman}}]{Deffayet:2009wt}%
  \BibitemOpen
  \bibfield  {author} {\bibinfo {author} {\bibfnamefont {C.}~\bibnamefont
  {Deffayet}}, \bibinfo {author} {\bibfnamefont {G.}~\bibnamefont
  {Esposito-Farese}}, \ and\ \bibinfo {author} {\bibfnamefont {A.}~\bibnamefont
  {Vikman}},\ }\href {\doibase 10.1103/PhysRevD.79.084003} {\bibfield
  {journal} {\bibinfo  {journal} {Phys. Rev. D}\ }\textbf {\bibinfo {volume}
  {79}},\ \bibinfo {pages} {084003} (\bibinfo {year} {2009})},\ \Eprint
  {http://arxiv.org/abs/0901.1314} {arXiv:0901.1314 [hep-th]} \BibitemShut
  {NoStop}%
\bibitem [{\citenamefont {Calmet}\ and\ \citenamefont
  {Kobakhidze}(2005)}]{Calmet:2005qm}%
  \BibitemOpen
  \bibfield  {author} {\bibinfo {author} {\bibfnamefont {X.}~\bibnamefont
  {Calmet}}\ and\ \bibinfo {author} {\bibfnamefont {A.}~\bibnamefont
  {Kobakhidze}},\ }\href {\doibase 10.1103/PhysRevD.72.045010} {\bibfield
  {journal} {\bibinfo  {journal} {Phys. Rev. D}\ }\textbf {\bibinfo {volume}
  {72}},\ \bibinfo {pages} {045010} (\bibinfo {year} {2005})},\ \Eprint
  {http://arxiv.org/abs/hep-th/0506157} {arXiv:hep-th/0506157} \BibitemShut
  {NoStop}%
\bibitem [{\citenamefont {Calmet}\ and\ \citenamefont
  {Kobakhidze}(2006)}]{Calmet:2006iz}%
  \BibitemOpen
  \bibfield  {author} {\bibinfo {author} {\bibfnamefont {X.}~\bibnamefont
  {Calmet}}\ and\ \bibinfo {author} {\bibfnamefont {A.}~\bibnamefont
  {Kobakhidze}},\ }\href {\doibase 10.1103/PhysRevD.74.047702} {\bibfield
  {journal} {\bibinfo  {journal} {Phys. Rev. D}\ }\textbf {\bibinfo {volume}
  {74}},\ \bibinfo {pages} {047702} (\bibinfo {year} {2006})},\ \Eprint
  {http://arxiv.org/abs/hep-th/0605275} {arXiv:hep-th/0605275} \BibitemShut
  {NoStop}%
\bibitem [{\citenamefont {Mukherjee}\ and\ \citenamefont
  {Saha}(2006)}]{Mukherjee:2006nd}%
  \BibitemOpen
  \bibfield  {author} {\bibinfo {author} {\bibfnamefont {P.}~\bibnamefont
  {Mukherjee}}\ and\ \bibinfo {author} {\bibfnamefont {A.}~\bibnamefont
  {Saha}},\ }\href {\doibase 10.1103/PhysRevD.74.027702} {\bibfield  {journal}
  {\bibinfo  {journal} {Phys. Rev. D}\ }\textbf {\bibinfo {volume} {74}},\
  \bibinfo {pages} {027702} (\bibinfo {year} {2006})},\ \Eprint
  {http://arxiv.org/abs/hep-th/0605287} {arXiv:hep-th/0605287} \BibitemShut
  {NoStop}%
\bibitem [{\citenamefont {Kobakhidze}\ \emph
  {et~al.}(2016{\natexlab{a}})\citenamefont {Kobakhidze}, \citenamefont
  {Lagger},\ and\ \citenamefont {Manning}}]{Kobakhidze:2016cqh}%
  \BibitemOpen
  \bibfield  {author} {\bibinfo {author} {\bibfnamefont {A.}~\bibnamefont
  {Kobakhidze}}, \bibinfo {author} {\bibfnamefont {C.}~\bibnamefont {Lagger}},
  \ and\ \bibinfo {author} {\bibfnamefont {A.}~\bibnamefont {Manning}},\ }\href
  {\doibase 10.1103/PhysRevD.94.064033} {\bibfield  {journal} {\bibinfo
  {journal} {Phys. Rev. D}\ }\textbf {\bibinfo {volume} {94}},\ \bibinfo
  {pages} {064033} (\bibinfo {year} {2016}{\natexlab{a}})},\ \Eprint
  {http://arxiv.org/abs/1607.03776} {arXiv:1607.03776 [gr-qc]} \BibitemShut
  {NoStop}%
\bibitem [{\citenamefont {Baker}\ \emph {et~al.}(2017)\citenamefont {Baker},
  \citenamefont {Bellini}, \citenamefont {Ferreira}, \citenamefont {Lagos},
  \citenamefont {Noller},\ and\ \citenamefont {Sawicki}}]{Baker:2017hug}%
  \BibitemOpen
  \bibfield  {author} {\bibinfo {author} {\bibfnamefont {T.}~\bibnamefont
  {Baker}}, \bibinfo {author} {\bibfnamefont {E.}~\bibnamefont {Bellini}},
  \bibinfo {author} {\bibfnamefont {P.~G.}\ \bibnamefont {Ferreira}}, \bibinfo
  {author} {\bibfnamefont {M.}~\bibnamefont {Lagos}}, \bibinfo {author}
  {\bibfnamefont {J.}~\bibnamefont {Noller}}, \ and\ \bibinfo {author}
  {\bibfnamefont {I.}~\bibnamefont {Sawicki}},\ }\href {\doibase
  10.1103/PhysRevLett.119.251301} {\bibfield  {journal} {\bibinfo  {journal}
  {Phys. Rev. Lett.}\ }\textbf {\bibinfo {volume} {119}},\ \bibinfo {pages}
  {251301} (\bibinfo {year} {2017})},\ \Eprint
  {http://arxiv.org/abs/1710.06394} {arXiv:1710.06394 [astro-ph.CO]}
  \BibitemShut {NoStop}%
\bibitem [{\citenamefont {Creminelli}\ and\ \citenamefont
  {Vernizzi}(2017)}]{Creminelli:2017sry}%
  \BibitemOpen
  \bibfield  {author} {\bibinfo {author} {\bibfnamefont {P.}~\bibnamefont
  {Creminelli}}\ and\ \bibinfo {author} {\bibfnamefont {F.}~\bibnamefont
  {Vernizzi}},\ }\href {\doibase 10.1103/PhysRevLett.119.251302} {\bibfield
  {journal} {\bibinfo  {journal} {Phys. Rev. Lett.}\ }\textbf {\bibinfo
  {volume} {119}},\ \bibinfo {pages} {251302} (\bibinfo {year} {2017})},\
  \Eprint {http://arxiv.org/abs/1710.05877} {arXiv:1710.05877 [astro-ph.CO]}
  \BibitemShut {NoStop}%
\bibitem [{\citenamefont {Sakstein}\ and\ \citenamefont
  {Jain}(2017)}]{Sakstein:2017xjx}%
  \BibitemOpen
  \bibfield  {author} {\bibinfo {author} {\bibfnamefont {J.}~\bibnamefont
  {Sakstein}}\ and\ \bibinfo {author} {\bibfnamefont {B.}~\bibnamefont
  {Jain}},\ }\href {\doibase 10.1103/PhysRevLett.119.251303} {\bibfield
  {journal} {\bibinfo  {journal} {Phys. Rev. Lett.}\ }\textbf {\bibinfo
  {volume} {119}},\ \bibinfo {pages} {251303} (\bibinfo {year} {2017})},\
  \Eprint {http://arxiv.org/abs/1710.05893} {arXiv:1710.05893 [astro-ph.CO]}
  \BibitemShut {NoStop}%
\bibitem [{\citenamefont {Ezquiaga}\ and\ \citenamefont
  {Zumalac\'arregui}(2017)}]{Ezquiaga:2017ekz}%
  \BibitemOpen
  \bibfield  {author} {\bibinfo {author} {\bibfnamefont {J.~M.}\ \bibnamefont
  {Ezquiaga}}\ and\ \bibinfo {author} {\bibfnamefont {M.}~\bibnamefont
  {Zumalac\'arregui}},\ }\href {\doibase 10.1103/PhysRevLett.119.251304}
  {\bibfield  {journal} {\bibinfo  {journal} {Phys. Rev. Lett.}\ }\textbf
  {\bibinfo {volume} {119}},\ \bibinfo {pages} {251304} (\bibinfo {year}
  {2017})},\ \Eprint {http://arxiv.org/abs/1710.05901} {arXiv:1710.05901
  [astro-ph.CO]} \BibitemShut {NoStop}%
\bibitem [{\citenamefont {Yunes}\ and\ \citenamefont
  {Pretorius}(2009)}]{Yunes:2009ke}%
  \BibitemOpen
  \bibfield  {author} {\bibinfo {author} {\bibfnamefont {N.}~\bibnamefont
  {Yunes}}\ and\ \bibinfo {author} {\bibfnamefont {F.}~\bibnamefont
  {Pretorius}},\ }\href {\doibase 10.1103/PhysRevD.80.122003} {\bibfield
  {journal} {\bibinfo  {journal} {Phys.\ Rev.\ D}\ }\textbf {\bibinfo {volume}
  {80}},\ \bibinfo {pages} {122003} (\bibinfo {year} {2009})},\ \Eprint
  {http://arxiv.org/abs/0909.3328} {arXiv:0909.3328 [gr-qc]} \BibitemShut
  {NoStop}%
\bibitem [{\citenamefont {Tahura}\ and\ \citenamefont
  {Yagi}(2018)}]{Tahura:2018zuq}%
  \BibitemOpen
  \bibfield  {author} {\bibinfo {author} {\bibfnamefont {S.}~\bibnamefont
  {Tahura}}\ and\ \bibinfo {author} {\bibfnamefont {K.}~\bibnamefont {Yagi}},\
  }\href {\doibase 10.1103/PhysRevD.98.084042} {\bibfield  {journal} {\bibinfo
  {journal} {Phys. Rev. D}\ }\textbf {\bibinfo {volume} {98}},\ \bibinfo
  {pages} {084042} (\bibinfo {year} {2018})},\ \bibinfo {note} {[Erratum:
  Phys.Rev.D 101, 109902 (2020)]},\ \Eprint {http://arxiv.org/abs/1809.00259}
  {arXiv:1809.00259 [gr-qc]} \BibitemShut {NoStop}%
\bibitem [{\citenamefont {Liang}\ \emph {et~al.}(2017)\citenamefont {Liang},
  \citenamefont {Gong}, \citenamefont {Hou},\ and\ \citenamefont
  {Liu}}]{Liang:2017ahj}%
  \BibitemOpen
  \bibfield  {author} {\bibinfo {author} {\bibfnamefont {D.}~\bibnamefont
  {Liang}}, \bibinfo {author} {\bibfnamefont {Y.}~\bibnamefont {Gong}},
  \bibinfo {author} {\bibfnamefont {S.}~\bibnamefont {Hou}}, \ and\ \bibinfo
  {author} {\bibfnamefont {Y.}~\bibnamefont {Liu}},\ }\href {\doibase
  10.1103/PhysRevD.95.104034} {\bibfield  {journal} {\bibinfo  {journal} {Phys.
  Rev. D}\ }\textbf {\bibinfo {volume} {95}},\ \bibinfo {pages} {104034}
  (\bibinfo {year} {2017})},\ \Eprint {http://arxiv.org/abs/1701.05998}
  {arXiv:1701.05998 [gr-qc]} \BibitemShut {NoStop}%
\bibitem [{\citenamefont {Rizwana~Kausar}\ \emph {et~al.}(2016)\citenamefont
  {Rizwana~Kausar}, \citenamefont {Philippoz},\ and\ \citenamefont
  {Jetzer}}]{RizwanaKausar:2016zgi}%
  \BibitemOpen
  \bibfield  {author} {\bibinfo {author} {\bibfnamefont {H.}~\bibnamefont
  {Rizwana~Kausar}}, \bibinfo {author} {\bibfnamefont {L.}~\bibnamefont
  {Philippoz}}, \ and\ \bibinfo {author} {\bibfnamefont {P.}~\bibnamefont
  {Jetzer}},\ }\href {\doibase 10.1103/PhysRevD.93.124071} {\bibfield
  {journal} {\bibinfo  {journal} {Phys. Rev. D}\ }\textbf {\bibinfo {volume}
  {93}},\ \bibinfo {pages} {124071} (\bibinfo {year} {2016})},\ \Eprint
  {http://arxiv.org/abs/1606.07000} {arXiv:1606.07000 [gr-qc]} \BibitemShut
  {NoStop}%
\bibitem [{\citenamefont {Zhang}\ \emph
  {et~al.}(2017{\natexlab{a}})\citenamefont {Zhang}, \citenamefont {Yu},
  \citenamefont {Liu}, \citenamefont {Zhao},\ and\ \citenamefont
  {Wang}}]{Zhang:2017sym}%
  \BibitemOpen
  \bibfield  {author} {\bibinfo {author} {\bibfnamefont {X.}~\bibnamefont
  {Zhang}}, \bibinfo {author} {\bibfnamefont {J.}~\bibnamefont {Yu}}, \bibinfo
  {author} {\bibfnamefont {T.}~\bibnamefont {Liu}}, \bibinfo {author}
  {\bibfnamefont {W.}~\bibnamefont {Zhao}}, \ and\ \bibinfo {author}
  {\bibfnamefont {A.}~\bibnamefont {Wang}},\ }\href {\doibase
  10.1103/PhysRevD.95.124008} {\bibfield  {journal} {\bibinfo  {journal} {Phys.
  Rev. D}\ }\textbf {\bibinfo {volume} {95}},\ \bibinfo {pages} {124008}
  (\bibinfo {year} {2017}{\natexlab{a}})},\ \Eprint
  {http://arxiv.org/abs/1703.09853} {arXiv:1703.09853 [gr-qc]} \BibitemShut
  {NoStop}%
\bibitem [{\citenamefont {Zhang}\ \emph
  {et~al.}(2017{\natexlab{b}})\citenamefont {Zhang}, \citenamefont {Liu},\ and\
  \citenamefont {Zhao}}]{Zhang:2017srh}%
  \BibitemOpen
  \bibfield  {author} {\bibinfo {author} {\bibfnamefont {X.}~\bibnamefont
  {Zhang}}, \bibinfo {author} {\bibfnamefont {T.}~\bibnamefont {Liu}}, \ and\
  \bibinfo {author} {\bibfnamefont {W.}~\bibnamefont {Zhao}},\ }\href {\doibase
  10.1103/PhysRevD.95.104027} {\bibfield  {journal} {\bibinfo  {journal} {Phys.
  Rev. D}\ }\textbf {\bibinfo {volume} {95}},\ \bibinfo {pages} {104027}
  (\bibinfo {year} {2017}{\natexlab{b}})},\ \Eprint
  {http://arxiv.org/abs/1702.08752} {arXiv:1702.08752 [gr-qc]} \BibitemShut
  {NoStop}%
\bibitem [{\citenamefont {Hou}\ \emph {et~al.}(2018)\citenamefont {Hou},
  \citenamefont {Gong},\ and\ \citenamefont {Liu}}]{Hou:2017bqj}%
  \BibitemOpen
  \bibfield  {author} {\bibinfo {author} {\bibfnamefont {S.}~\bibnamefont
  {Hou}}, \bibinfo {author} {\bibfnamefont {Y.}~\bibnamefont {Gong}}, \ and\
  \bibinfo {author} {\bibfnamefont {Y.}~\bibnamefont {Liu}},\ }\href {\doibase
  10.1140/epjc/s10052-018-5869-y} {\bibfield  {journal} {\bibinfo  {journal}
  {Eur. Phys. J. C}\ }\textbf {\bibinfo {volume} {78}},\ \bibinfo {pages} {378}
  (\bibinfo {year} {2018})},\ \Eprint {http://arxiv.org/abs/1704.01899}
  {arXiv:1704.01899 [gr-qc]} \BibitemShut {NoStop}%
\bibitem [{\citenamefont {Heisenberg}(2014)}]{Heisenberg:2014rta}%
  \BibitemOpen
  \bibfield  {author} {\bibinfo {author} {\bibfnamefont {L.}~\bibnamefont
  {Heisenberg}},\ }\href {\doibase 10.1088/1475-7516/2014/05/015} {\bibfield
  {journal} {\bibinfo  {journal} {JCAP}\ }\textbf {\bibinfo {volume} {05}},\
  \bibinfo {pages} {015} (\bibinfo {year} {2014})},\ \Eprint
  {http://arxiv.org/abs/1402.7026} {arXiv:1402.7026 [hep-th]} \BibitemShut
  {NoStop}%
\bibitem [{\citenamefont {Dong}\ \emph
  {et~al.}(2024{\natexlab{a}})\citenamefont {Dong}, \citenamefont {Liu},\ and\
  \citenamefont {Liu}}]{Dong:2023xyb}%
  \BibitemOpen
  \bibfield  {author} {\bibinfo {author} {\bibfnamefont {Y.-Q.}\ \bibnamefont
  {Dong}}, \bibinfo {author} {\bibfnamefont {Y.-Q.}\ \bibnamefont {Liu}}, \
  and\ \bibinfo {author} {\bibfnamefont {Y.-X.}\ \bibnamefont {Liu}},\ }\href
  {\doibase 10.1103/PhysRevD.109.024014} {\bibfield  {journal} {\bibinfo
  {journal} {Phys. Rev. D}\ }\textbf {\bibinfo {volume} {109}},\ \bibinfo
  {pages} {024014} (\bibinfo {year} {2024}{\natexlab{a}})},\ \Eprint
  {http://arxiv.org/abs/2305.12516} {arXiv:2305.12516 [gr-qc]} \BibitemShut
  {NoStop}%
\bibitem [{\citenamefont {Geng}\ and\ \citenamefont {Lu}(2016)}]{Geng:2015kvs}%
  \BibitemOpen
  \bibfield  {author} {\bibinfo {author} {\bibfnamefont {W.-J.}\ \bibnamefont
  {Geng}}\ and\ \bibinfo {author} {\bibfnamefont {H.}~\bibnamefont {Lu}},\
  }\href {\doibase 10.1103/PhysRevD.93.044035} {\bibfield  {journal} {\bibinfo
  {journal} {Phys. Rev. D}\ }\textbf {\bibinfo {volume} {93}},\ \bibinfo
  {pages} {044035} (\bibinfo {year} {2016})},\ \Eprint
  {http://arxiv.org/abs/1511.03681} {arXiv:1511.03681 [hep-th]} \BibitemShut
  {NoStop}%
\bibitem [{\citenamefont {Lai}\ \emph {et~al.}(2024)\citenamefont {Lai},
  \citenamefont {Dong}, \citenamefont {Liu},\ and\ \citenamefont
  {Liu}}]{Lai:2024fza}%
  \BibitemOpen
  \bibfield  {author} {\bibinfo {author} {\bibfnamefont {X.-B.}\ \bibnamefont
  {Lai}}, \bibinfo {author} {\bibfnamefont {Y.-Q.}\ \bibnamefont {Dong}},
  \bibinfo {author} {\bibfnamefont {Y.-Q.}\ \bibnamefont {Liu}}, \ and\
  \bibinfo {author} {\bibfnamefont {Y.-X.}\ \bibnamefont {Liu}},\ }\href
  {\doibase 10.1103/PhysRevD.110.064073} {\bibfield  {journal} {\bibinfo
  {journal} {Phys. Rev. D}\ }\textbf {\bibinfo {volume} {110}},\ \bibinfo
  {pages} {064073} (\bibinfo {year} {2024})},\ \Eprint
  {http://arxiv.org/abs/2405.20577} {arXiv:2405.20577 [gr-qc]} \BibitemShut
  {NoStop}%
\bibitem [{\citenamefont {Dong}\ \emph
  {et~al.}(2024{\natexlab{b}})\citenamefont {Dong}, \citenamefont {Liu},\ and\
  \citenamefont {Liu}}]{Dong:2023bgt}%
  \BibitemOpen
  \bibfield  {author} {\bibinfo {author} {\bibfnamefont {Y.-Q.}\ \bibnamefont
  {Dong}}, \bibinfo {author} {\bibfnamefont {Y.-Q.}\ \bibnamefont {Liu}}, \
  and\ \bibinfo {author} {\bibfnamefont {Y.-X.}\ \bibnamefont {Liu}},\ }\href
  {\doibase 10.1103/PhysRevD.109.044013} {\bibfield  {journal} {\bibinfo
  {journal} {Phys. Rev. D}\ }\textbf {\bibinfo {volume} {109}},\ \bibinfo
  {pages} {044013} (\bibinfo {year} {2024}{\natexlab{b}})},\ \Eprint
  {http://arxiv.org/abs/2310.11336} {arXiv:2310.11336 [gr-qc]} \BibitemShut
  {NoStop}%
\bibitem [{\citenamefont {Dong}\ \emph
  {et~al.}(2024{\natexlab{c}})\citenamefont {Dong}, \citenamefont {Lai},
  \citenamefont {Liu},\ and\ \citenamefont {Liu}}]{Dong:2024zal}%
  \BibitemOpen
  \bibfield  {author} {\bibinfo {author} {\bibfnamefont {Y.-Q.}\ \bibnamefont
  {Dong}}, \bibinfo {author} {\bibfnamefont {X.-B.}\ \bibnamefont {Lai}},
  \bibinfo {author} {\bibfnamefont {Y.-Q.}\ \bibnamefont {Liu}}, \ and\
  \bibinfo {author} {\bibfnamefont {Y.-X.}\ \bibnamefont {Liu}},\ }\href@noop
  {} {\  (\bibinfo {year} {2024}{\natexlab{c}})},\ \Eprint
  {http://arxiv.org/abs/2409.11838} {arXiv:2409.11838 [gr-qc]} \BibitemShut
  {NoStop}%
\bibitem [{\citenamefont {Hehl}\ \emph {et~al.}(1976)\citenamefont {Hehl},
  \citenamefont {Von Der~Heyde}, \citenamefont {Kerlick},\ and\ \citenamefont
  {Nester}}]{Hehl:1976kj}%
  \BibitemOpen
  \bibfield  {author} {\bibinfo {author} {\bibfnamefont {F.~W.}\ \bibnamefont
  {Hehl}}, \bibinfo {author} {\bibfnamefont {P.}~\bibnamefont {Von Der~Heyde}},
  \bibinfo {author} {\bibfnamefont {G.~D.}\ \bibnamefont {Kerlick}}, \ and\
  \bibinfo {author} {\bibfnamefont {J.~M.}\ \bibnamefont {Nester}},\ }\href
  {\doibase 10.1103/RevModPhys.48.393} {\bibfield  {journal} {\bibinfo
  {journal} {Rev. Mod. Phys.}\ }\textbf {\bibinfo {volume} {48}},\ \bibinfo
  {pages} {393} (\bibinfo {year} {1976})}\BibitemShut {NoStop}%
\bibitem [{\citenamefont {Hayashi}\ and\ \citenamefont
  {Shirafuji}(1979)}]{Hayashi:1979qx}%
  \BibitemOpen
  \bibfield  {author} {\bibinfo {author} {\bibfnamefont {K.}~\bibnamefont
  {Hayashi}}\ and\ \bibinfo {author} {\bibfnamefont {T.}~\bibnamefont
  {Shirafuji}},\ }\href {\doibase 10.1103/PhysRevD.19.3524} {\bibfield
  {journal} {\bibinfo  {journal} {Phys. Rev. D}\ }\textbf {\bibinfo {volume}
  {19}},\ \bibinfo {pages} {3524} (\bibinfo {year} {1979})},\ \bibinfo {note}
  {[Addendum: Phys.Rev.D 24, 3312--3314 (1982)]}\BibitemShut {NoStop}%
\bibitem [{\citenamefont {Bamba}\ \emph {et~al.}(2013)\citenamefont {Bamba},
  \citenamefont {Capozziello}, \citenamefont {De~Laurentis}, \citenamefont
  {Nojiri},\ and\ \citenamefont {S\'aez-G\'omez}}]{Bamba:2013ooa}%
  \BibitemOpen
  \bibfield  {author} {\bibinfo {author} {\bibfnamefont {K.}~\bibnamefont
  {Bamba}}, \bibinfo {author} {\bibfnamefont {S.}~\bibnamefont {Capozziello}},
  \bibinfo {author} {\bibfnamefont {M.}~\bibnamefont {De~Laurentis}}, \bibinfo
  {author} {\bibfnamefont {S.}~\bibnamefont {Nojiri}}, \ and\ \bibinfo {author}
  {\bibfnamefont {D.}~\bibnamefont {S\'aez-G\'omez}},\ }\href {\doibase
  10.1016/j.physletb.2013.10.022} {\bibfield  {journal} {\bibinfo  {journal}
  {Phys. Lett. B}\ }\textbf {\bibinfo {volume} {727}},\ \bibinfo {pages} {194}
  (\bibinfo {year} {2013})},\ \Eprint {http://arxiv.org/abs/1309.2698}
  {arXiv:1309.2698 [gr-qc]} \BibitemShut {NoStop}%
\bibitem [{\citenamefont {Beltr\'an~Jim\'enez}\ \emph
  {et~al.}(2018)\citenamefont {Beltr\'an~Jim\'enez}, \citenamefont
  {Heisenberg},\ and\ \citenamefont {Koivisto}}]{BeltranJimenez:2017tkd}%
  \BibitemOpen
  \bibfield  {author} {\bibinfo {author} {\bibfnamefont {J.}~\bibnamefont
  {Beltr\'an~Jim\'enez}}, \bibinfo {author} {\bibfnamefont {L.}~\bibnamefont
  {Heisenberg}}, \ and\ \bibinfo {author} {\bibfnamefont {T.}~\bibnamefont
  {Koivisto}},\ }\href {\doibase 10.1103/PhysRevD.98.044048} {\bibfield
  {journal} {\bibinfo  {journal} {Phys. Rev. D}\ }\textbf {\bibinfo {volume}
  {98}},\ \bibinfo {pages} {044048} (\bibinfo {year} {2018})},\ \Eprint
  {http://arxiv.org/abs/1710.03116} {arXiv:1710.03116 [gr-qc]} \BibitemShut
  {NoStop}%
\bibitem [{\citenamefont {Capozziello}\ \emph {et~al.}(2022)\citenamefont
  {Capozziello}, \citenamefont {De~Falco},\ and\ \citenamefont
  {Ferrara}}]{Capozziello:2022zzh}%
  \BibitemOpen
  \bibfield  {author} {\bibinfo {author} {\bibfnamefont {S.}~\bibnamefont
  {Capozziello}}, \bibinfo {author} {\bibfnamefont {V.}~\bibnamefont
  {De~Falco}}, \ and\ \bibinfo {author} {\bibfnamefont {C.}~\bibnamefont
  {Ferrara}},\ }\href {\doibase 10.1140/epjc/s10052-022-10823-x} {\bibfield
  {journal} {\bibinfo  {journal} {Eur. Phys. J. C}\ }\textbf {\bibinfo {volume}
  {82}},\ \bibinfo {pages} {865} (\bibinfo {year} {2022})},\ \Eprint
  {http://arxiv.org/abs/2208.03011} {arXiv:2208.03011 [gr-qc]} \BibitemShut
  {NoStop}%
\bibitem [{\citenamefont {Capozziello}\ \emph {et~al.}(2024)\citenamefont
  {Capozziello}, \citenamefont {Capriolo},\ and\ \citenamefont
  {Nojiri}}]{Capozziello:2024vix}%
  \BibitemOpen
  \bibfield  {author} {\bibinfo {author} {\bibfnamefont {S.}~\bibnamefont
  {Capozziello}}, \bibinfo {author} {\bibfnamefont {M.}~\bibnamefont
  {Capriolo}}, \ and\ \bibinfo {author} {\bibfnamefont {S.}~\bibnamefont
  {Nojiri}},\ }\href {\doibase 10.1016/j.physletb.2024.138510} {\bibfield
  {journal} {\bibinfo  {journal} {Phys. Lett. B}\ }\textbf {\bibinfo {volume}
  {850}},\ \bibinfo {pages} {138510} (\bibinfo {year} {2024})},\ \Eprint
  {http://arxiv.org/abs/2401.06424} {arXiv:2401.06424 [gr-qc]} \BibitemShut
  {NoStop}%
\bibitem [{\citenamefont {Lu}\ \emph {et~al.}(2020{\natexlab{a}})\citenamefont
  {Lu}, \citenamefont {Li}, \citenamefont {Guo}, \citenamefont {Zhuang},\ and\
  \citenamefont {Zhao}}]{Lu:2020eux}%
  \BibitemOpen
  \bibfield  {author} {\bibinfo {author} {\bibfnamefont {J.}~\bibnamefont
  {Lu}}, \bibinfo {author} {\bibfnamefont {J.}~\bibnamefont {Li}}, \bibinfo
  {author} {\bibfnamefont {H.}~\bibnamefont {Guo}}, \bibinfo {author}
  {\bibfnamefont {Z.}~\bibnamefont {Zhuang}}, \ and\ \bibinfo {author}
  {\bibfnamefont {X.}~\bibnamefont {Zhao}},\ }\href {\doibase
  10.1016/j.physletb.2020.135985} {\bibfield  {journal} {\bibinfo  {journal}
  {Phys. Lett. B}\ }\textbf {\bibinfo {volume} {811}},\ \bibinfo {pages}
  {135985} (\bibinfo {year} {2020}{\natexlab{a}})},\ \Eprint
  {http://arxiv.org/abs/2012.02343} {arXiv:2012.02343 [gr-qc]} \BibitemShut
  {NoStop}%
\bibitem [{\citenamefont {Dong}\ and\ \citenamefont
  {Liu}(2022)}]{Dong:2021jtd}%
  \BibitemOpen
  \bibfield  {author} {\bibinfo {author} {\bibfnamefont {Y.-Q.}\ \bibnamefont
  {Dong}}\ and\ \bibinfo {author} {\bibfnamefont {Y.-X.}\ \bibnamefont {Liu}},\
  }\href {\doibase 10.1103/PhysRevD.105.064035} {\bibfield  {journal} {\bibinfo
   {journal} {Phys. Rev. D}\ }\textbf {\bibinfo {volume} {105}},\ \bibinfo
  {pages} {064035} (\bibinfo {year} {2022})},\ \Eprint
  {http://arxiv.org/abs/2111.07352} {arXiv:2111.07352 [gr-qc]} \BibitemShut
  {NoStop}%
\bibitem [{\citenamefont {Bahamonde}\ \emph {et~al.}(2021)\citenamefont
  {Bahamonde}, \citenamefont {Caruana}, \citenamefont {Dialektopoulos},
  \citenamefont {Gakis}, \citenamefont {Hohmann}, \citenamefont {Levi~Said},
  \citenamefont {Saridakis},\ and\ \citenamefont
  {Sultana}}]{Bahamonde:2021dqn}%
  \BibitemOpen
  \bibfield  {author} {\bibinfo {author} {\bibfnamefont {S.}~\bibnamefont
  {Bahamonde}}, \bibinfo {author} {\bibfnamefont {M.}~\bibnamefont {Caruana}},
  \bibinfo {author} {\bibfnamefont {K.~F.}\ \bibnamefont {Dialektopoulos}},
  \bibinfo {author} {\bibfnamefont {V.}~\bibnamefont {Gakis}}, \bibinfo
  {author} {\bibfnamefont {M.}~\bibnamefont {Hohmann}}, \bibinfo {author}
  {\bibfnamefont {J.}~\bibnamefont {Levi~Said}}, \bibinfo {author}
  {\bibfnamefont {E.~N.}\ \bibnamefont {Saridakis}}, \ and\ \bibinfo {author}
  {\bibfnamefont {J.}~\bibnamefont {Sultana}},\ }\href {\doibase
  10.1103/PhysRevD.104.084082} {\bibfield  {journal} {\bibinfo  {journal}
  {Phys. Rev. D}\ }\textbf {\bibinfo {volume} {104}},\ \bibinfo {pages}
  {084082} (\bibinfo {year} {2021})},\ \Eprint
  {http://arxiv.org/abs/2105.13243} {arXiv:2105.13243 [gr-qc]} \BibitemShut
  {NoStop}%
\bibitem [{\citenamefont {Wagle}\ \emph {et~al.}(2019)\citenamefont {Wagle},
  \citenamefont {Saffer},\ and\ \citenamefont {Yunes}}]{Wagle:2019mdq}%
  \BibitemOpen
  \bibfield  {author} {\bibinfo {author} {\bibfnamefont {P.}~\bibnamefont
  {Wagle}}, \bibinfo {author} {\bibfnamefont {A.}~\bibnamefont {Saffer}}, \
  and\ \bibinfo {author} {\bibfnamefont {N.}~\bibnamefont {Yunes}},\ }\href
  {\doibase 10.1103/PhysRevD.100.124007} {\bibfield  {journal} {\bibinfo
  {journal} {Phys. Rev. D}\ }\textbf {\bibinfo {volume} {100}},\ \bibinfo
  {pages} {124007} (\bibinfo {year} {2019})},\ \Eprint
  {http://arxiv.org/abs/1910.04800} {arXiv:1910.04800 [gr-qc]} \BibitemShut
  {NoStop}%
\bibitem [{\citenamefont {Yagi}\ \emph {et~al.}(2012)\citenamefont {Yagi},
  \citenamefont {Stein}, \citenamefont {Yunes},\ and\ \citenamefont
  {Tanaka}}]{Yagi:2011xp}%
  \BibitemOpen
  \bibfield  {author} {\bibinfo {author} {\bibfnamefont {K.}~\bibnamefont
  {Yagi}}, \bibinfo {author} {\bibfnamefont {L.~C.}\ \bibnamefont {Stein}},
  \bibinfo {author} {\bibfnamefont {N.}~\bibnamefont {Yunes}}, \ and\ \bibinfo
  {author} {\bibfnamefont {T.}~\bibnamefont {Tanaka}},\ }\href {\doibase
  10.1103/PhysRevD.85.064022} {\bibfield  {journal} {\bibinfo  {journal} {Phys.
  Rev. D}\ }\textbf {\bibinfo {volume} {85}},\ \bibinfo {pages} {064022}
  (\bibinfo {year} {2012})},\ \bibinfo {note} {[Erratum: Phys.Rev.D 93, 029902
  (2016)]},\ \Eprint {http://arxiv.org/abs/1110.5950} {arXiv:1110.5950 [gr-qc]}
  \BibitemShut {NoStop}%
\bibitem [{\citenamefont {Li}\ \emph {et~al.}(2023{\natexlab{a}})\citenamefont
  {Li}, \citenamefont {Qiao}, \citenamefont {Liu}, \citenamefont {Zhu},\ and\
  \citenamefont {Zhao}}]{Li:2022grj}%
  \BibitemOpen
  \bibfield  {author} {\bibinfo {author} {\bibfnamefont {Z.}~\bibnamefont
  {Li}}, \bibinfo {author} {\bibfnamefont {J.}~\bibnamefont {Qiao}}, \bibinfo
  {author} {\bibfnamefont {T.}~\bibnamefont {Liu}}, \bibinfo {author}
  {\bibfnamefont {T.}~\bibnamefont {Zhu}}, \ and\ \bibinfo {author}
  {\bibfnamefont {W.}~\bibnamefont {Zhao}},\ }\href {\doibase
  10.1088/1475-7516/2023/04/006} {\bibfield  {journal} {\bibinfo  {journal}
  {JCAP}\ }\textbf {\bibinfo {volume} {04}},\ \bibinfo {pages} {006} (\bibinfo
  {year} {2023}{\natexlab{a}})},\ \Eprint {http://arxiv.org/abs/2211.12188}
  {arXiv:2211.12188 [gr-qc]} \BibitemShut {NoStop}%
\bibitem [{\citenamefont {Zhao}\ \emph
  {et~al.}(2020{\natexlab{a}})\citenamefont {Zhao}, \citenamefont {Zhu},
  \citenamefont {Qiao},\ and\ \citenamefont {Wang}}]{Zhao:2019xmm}%
  \BibitemOpen
  \bibfield  {author} {\bibinfo {author} {\bibfnamefont {W.}~\bibnamefont
  {Zhao}}, \bibinfo {author} {\bibfnamefont {T.}~\bibnamefont {Zhu}}, \bibinfo
  {author} {\bibfnamefont {J.}~\bibnamefont {Qiao}}, \ and\ \bibinfo {author}
  {\bibfnamefont {A.}~\bibnamefont {Wang}},\ }\href {\doibase
  10.1103/PhysRevD.101.024002} {\bibfield  {journal} {\bibinfo  {journal}
  {Phys. Rev. D}\ }\textbf {\bibinfo {volume} {101}},\ \bibinfo {pages}
  {024002} (\bibinfo {year} {2020}{\natexlab{a}})},\ \Eprint
  {http://arxiv.org/abs/1909.10887} {arXiv:1909.10887 [gr-qc]} \BibitemShut
  {NoStop}%
\bibitem [{\citenamefont {Alves}(2024)}]{Alves:2023rxs}%
  \BibitemOpen
  \bibfield  {author} {\bibinfo {author} {\bibfnamefont {M.~E.~S.}\
  \bibnamefont {Alves}},\ }\href {\doibase 10.1103/PhysRevD.109.104054}
  {\bibfield  {journal} {\bibinfo  {journal} {Phys. Rev. D}\ }\textbf {\bibinfo
  {volume} {109}},\ \bibinfo {pages} {104054} (\bibinfo {year} {2024})},\
  \Eprint {http://arxiv.org/abs/2308.09178} {arXiv:2308.09178 [gr-qc]}
  \BibitemShut {NoStop}%
\bibitem [{\citenamefont {Jacobson}\ and\ \citenamefont
  {Mattingly}(2004)}]{Jacobson:2004ts}%
  \BibitemOpen
  \bibfield  {author} {\bibinfo {author} {\bibfnamefont {T.}~\bibnamefont
  {Jacobson}}\ and\ \bibinfo {author} {\bibfnamefont {D.}~\bibnamefont
  {Mattingly}},\ }\href {\doibase 10.1103/PhysRevD.70.024003} {\bibfield
  {journal} {\bibinfo  {journal} {Phys. Rev. D}\ }\textbf {\bibinfo {volume}
  {70}},\ \bibinfo {pages} {024003} (\bibinfo {year} {2004})},\ \Eprint
  {http://arxiv.org/abs/gr-qc/0402005} {arXiv:gr-qc/0402005} \BibitemShut
  {NoStop}%
\bibitem [{\citenamefont {Gong}\ \emph
  {et~al.}(2018{\natexlab{a}})\citenamefont {Gong}, \citenamefont {Hou},
  \citenamefont {Liang},\ and\ \citenamefont {Papantonopoulos}}]{Gong:2018cgj}%
  \BibitemOpen
  \bibfield  {author} {\bibinfo {author} {\bibfnamefont {Y.}~\bibnamefont
  {Gong}}, \bibinfo {author} {\bibfnamefont {S.}~\bibnamefont {Hou}}, \bibinfo
  {author} {\bibfnamefont {D.}~\bibnamefont {Liang}}, \ and\ \bibinfo {author}
  {\bibfnamefont {E.}~\bibnamefont {Papantonopoulos}},\ }\href {\doibase
  10.1103/PhysRevD.97.084040} {\bibfield  {journal} {\bibinfo  {journal} {Phys.
  Rev. D}\ }\textbf {\bibinfo {volume} {97}},\ \bibinfo {pages} {084040}
  (\bibinfo {year} {2018}{\natexlab{a}})},\ \Eprint
  {http://arxiv.org/abs/1801.03382} {arXiv:1801.03382 [gr-qc]} \BibitemShut
  {NoStop}%
\bibitem [{\citenamefont {Hansen}\ \emph
  {et~al.}(2015{\natexlab{a}})\citenamefont {Hansen}, \citenamefont {Yunes},\
  and\ \citenamefont {Yagi}}]{Hansen:2014ewa}%
  \BibitemOpen
  \bibfield  {author} {\bibinfo {author} {\bibfnamefont {D.}~\bibnamefont
  {Hansen}}, \bibinfo {author} {\bibfnamefont {N.}~\bibnamefont {Yunes}}, \
  and\ \bibinfo {author} {\bibfnamefont {K.}~\bibnamefont {Yagi}},\ }\href
  {\doibase 10.1103/PhysRevD.91.082003} {\bibfield  {journal} {\bibinfo
  {journal} {Phys. Rev. D}\ }\textbf {\bibinfo {volume} {91}},\ \bibinfo
  {pages} {082003} (\bibinfo {year} {2015}{\natexlab{a}})},\ \Eprint
  {http://arxiv.org/abs/1412.4132} {arXiv:1412.4132 [gr-qc]} \BibitemShut
  {NoStop}%
\bibitem [{\citenamefont {Schumacher}\ \emph {et~al.}(2023)\citenamefont
  {Schumacher}, \citenamefont {Yunes},\ and\ \citenamefont
  {Yagi}}]{Schumacher:2023jxq}%
  \BibitemOpen
  \bibfield  {author} {\bibinfo {author} {\bibfnamefont {K.}~\bibnamefont
  {Schumacher}}, \bibinfo {author} {\bibfnamefont {N.}~\bibnamefont {Yunes}}, \
  and\ \bibinfo {author} {\bibfnamefont {K.}~\bibnamefont {Yagi}},\ }\href
  {\doibase 10.1103/PhysRevD.108.104038} {\bibfield  {journal} {\bibinfo
  {journal} {Phys. Rev. D}\ }\textbf {\bibinfo {volume} {108}},\ \bibinfo
  {pages} {104038} (\bibinfo {year} {2023})},\ \Eprint
  {http://arxiv.org/abs/2308.05589} {arXiv:2308.05589 [gr-qc]} \BibitemShut
  {NoStop}%
\bibitem [{\citenamefont {Gong}\ \emph
  {et~al.}(2018{\natexlab{b}})\citenamefont {Gong}, \citenamefont {Hou},
  \citenamefont {Papantonopoulos},\ and\ \citenamefont
  {Tzortzis}}]{Gong:2018vbo}%
  \BibitemOpen
  \bibfield  {author} {\bibinfo {author} {\bibfnamefont {Y.}~\bibnamefont
  {Gong}}, \bibinfo {author} {\bibfnamefont {S.}~\bibnamefont {Hou}}, \bibinfo
  {author} {\bibfnamefont {E.}~\bibnamefont {Papantonopoulos}}, \ and\ \bibinfo
  {author} {\bibfnamefont {D.}~\bibnamefont {Tzortzis}},\ }\href {\doibase
  10.1103/PhysRevD.98.104017} {\bibfield  {journal} {\bibinfo  {journal} {Phys.
  Rev. D}\ }\textbf {\bibinfo {volume} {98}},\ \bibinfo {pages} {104017}
  (\bibinfo {year} {2018}{\natexlab{b}})},\ \Eprint
  {http://arxiv.org/abs/1808.00632} {arXiv:1808.00632 [gr-qc]} \BibitemShut
  {NoStop}%
\bibitem [{\citenamefont {Gao}(2014)}]{Gao:2014fra}%
  \BibitemOpen
  \bibfield  {author} {\bibinfo {author} {\bibfnamefont {X.}~\bibnamefont
  {Gao}},\ }\href {\doibase 10.1103/PhysRevD.90.104033} {\bibfield  {journal}
  {\bibinfo  {journal} {Phys. Rev. D}\ }\textbf {\bibinfo {volume} {90}},\
  \bibinfo {pages} {104033} (\bibinfo {year} {2014})},\ \Eprint
  {http://arxiv.org/abs/1409.6708} {arXiv:1409.6708 [gr-qc]} \BibitemShut
  {NoStop}%
\bibitem [{\citenamefont {Gao}\ and\ \citenamefont {Hong}(2020)}]{Gao:2019liu}%
  \BibitemOpen
  \bibfield  {author} {\bibinfo {author} {\bibfnamefont {X.}~\bibnamefont
  {Gao}}\ and\ \bibinfo {author} {\bibfnamefont {X.-Y.}\ \bibnamefont {Hong}},\
  }\href {\doibase 10.1103/PhysRevD.101.064057} {\bibfield  {journal} {\bibinfo
   {journal} {Phys. Rev. D}\ }\textbf {\bibinfo {volume} {101}},\ \bibinfo
  {pages} {064057} (\bibinfo {year} {2020})},\ \Eprint
  {http://arxiv.org/abs/1906.07131} {arXiv:1906.07131 [gr-qc]} \BibitemShut
  {NoStop}%
\bibitem [{\citenamefont {Hou}\ \emph {et~al.}(2024{\natexlab{c}})\citenamefont
  {Hou}, \citenamefont {Fan}, \citenamefont {Zhu},\ and\ \citenamefont
  {Zhu}}]{Hou:2024xbv}%
  \BibitemOpen
  \bibfield  {author} {\bibinfo {author} {\bibfnamefont {S.}~\bibnamefont
  {Hou}}, \bibinfo {author} {\bibfnamefont {X.-L.}\ \bibnamefont {Fan}},
  \bibinfo {author} {\bibfnamefont {T.}~\bibnamefont {Zhu}}, \ and\ \bibinfo
  {author} {\bibfnamefont {Z.-H.}\ \bibnamefont {Zhu}},\ }\href {\doibase
  10.1103/PhysRevD.109.084011} {\bibfield  {journal} {\bibinfo  {journal}
  {Phys. Rev. D}\ }\textbf {\bibinfo {volume} {109}},\ \bibinfo {pages}
  {084011} (\bibinfo {year} {2024}{\natexlab{c}})},\ \Eprint
  {http://arxiv.org/abs/2401.03474} {arXiv:2401.03474 [gr-qc]} \BibitemShut
  {NoStop}%
\bibitem [{\citenamefont {Liang}\ \emph
  {et~al.}(2022{\natexlab{b}})\citenamefont {Liang}, \citenamefont {Xu},
  \citenamefont {Lu},\ and\ \citenamefont {Shao}}]{Liang:2022hxd}%
  \BibitemOpen
  \bibfield  {author} {\bibinfo {author} {\bibfnamefont {D.}~\bibnamefont
  {Liang}}, \bibinfo {author} {\bibfnamefont {R.}~\bibnamefont {Xu}}, \bibinfo
  {author} {\bibfnamefont {X.}~\bibnamefont {Lu}}, \ and\ \bibinfo {author}
  {\bibfnamefont {L.}~\bibnamefont {Shao}},\ }\href {\doibase
  10.1103/PhysRevD.106.124019} {\bibfield  {journal} {\bibinfo  {journal}
  {Phys. Rev. D}\ }\textbf {\bibinfo {volume} {106}},\ \bibinfo {pages}
  {124019} (\bibinfo {year} {2022}{\natexlab{b}})},\ \Eprint
  {http://arxiv.org/abs/2207.14423} {arXiv:2207.14423 [gr-qc]} \BibitemShut
  {NoStop}%
\bibitem [{\citenamefont {Fan}\ \emph {et~al.}(2024)\citenamefont {Fan},
  \citenamefont {Lai}, \citenamefont {Dong},\ and\ \citenamefont
  {Liu}}]{Fan:2024pex}%
  \BibitemOpen
  \bibfield  {author} {\bibinfo {author} {\bibfnamefont {Y.-Z.}\ \bibnamefont
  {Fan}}, \bibinfo {author} {\bibfnamefont {X.-B.}\ \bibnamefont {Lai}},
  \bibinfo {author} {\bibfnamefont {Y.-Q.}\ \bibnamefont {Dong}}, \ and\
  \bibinfo {author} {\bibfnamefont {Y.-X.}\ \bibnamefont {Liu}},\ }\href@noop
  {} {\  (\bibinfo {year} {2024})},\ \Eprint {http://arxiv.org/abs/2409.18503}
  {arXiv:2409.18503 [gr-qc]} \BibitemShut {NoStop}%
\bibitem [{\citenamefont {Eardley}\ \emph
  {et~al.}(1973{\natexlab{a}})\citenamefont {Eardley}, \citenamefont {Lee},
  \citenamefont {Lightman}, \citenamefont {Wagoner},\ and\ \citenamefont
  {Will}}]{Eardley:1973br}%
  \BibitemOpen
  \bibfield  {author} {\bibinfo {author} {\bibfnamefont {D.~M.}\ \bibnamefont
  {Eardley}}, \bibinfo {author} {\bibfnamefont {D.~L.}\ \bibnamefont {Lee}},
  \bibinfo {author} {\bibfnamefont {A.~P.}\ \bibnamefont {Lightman}}, \bibinfo
  {author} {\bibfnamefont {R.~V.}\ \bibnamefont {Wagoner}}, \ and\ \bibinfo
  {author} {\bibfnamefont {C.~M.}\ \bibnamefont {Will}},\ }\href {\doibase
  10.1103/PhysRevLett.30.884} {\bibfield  {journal} {\bibinfo  {journal} {Phys.
  Rev. Lett.}\ }\textbf {\bibinfo {volume} {30}},\ \bibinfo {pages} {884}
  (\bibinfo {year} {1973}{\natexlab{a}})}\BibitemShut {NoStop}%
\bibitem [{\citenamefont {Eardley}\ \emph
  {et~al.}(1973{\natexlab{b}})\citenamefont {Eardley}, \citenamefont {Lee},\
  and\ \citenamefont {Lightman}}]{Eardley:1973zuo}%
  \BibitemOpen
  \bibfield  {author} {\bibinfo {author} {\bibfnamefont {D.~M.}\ \bibnamefont
  {Eardley}}, \bibinfo {author} {\bibfnamefont {D.~L.}\ \bibnamefont {Lee}}, \
  and\ \bibinfo {author} {\bibfnamefont {A.~P.}\ \bibnamefont {Lightman}},\
  }\href {\doibase 10.1103/PhysRevD.8.3308} {\bibfield  {journal} {\bibinfo
  {journal} {Phys. Rev. D}\ }\textbf {\bibinfo {volume} {8}},\ \bibinfo {pages}
  {3308} (\bibinfo {year} {1973}{\natexlab{b}})}\BibitemShut {NoStop}%
\bibitem [{\citenamefont {Maggiore}\ and\ \citenamefont
  {Nicolis}(2000)}]{Maggiore:1999wm}%
  \BibitemOpen
  \bibfield  {author} {\bibinfo {author} {\bibfnamefont {M.}~\bibnamefont
  {Maggiore}}\ and\ \bibinfo {author} {\bibfnamefont {A.}~\bibnamefont
  {Nicolis}},\ }\href {\doibase 10.1103/PhysRevD.62.024004} {\bibfield
  {journal} {\bibinfo  {journal} {Phys. Rev. D}\ }\textbf {\bibinfo {volume}
  {62}},\ \bibinfo {pages} {024004} (\bibinfo {year} {2000})},\ \Eprint
  {http://arxiv.org/abs/gr-qc/9907055} {arXiv:gr-qc/9907055} \BibitemShut
  {NoStop}%
\bibitem [{\citenamefont {Capozziello}\ and\ \citenamefont
  {Corda}(2006)}]{Capozziello:2006ra}%
  \BibitemOpen
  \bibfield  {author} {\bibinfo {author} {\bibfnamefont {S.}~\bibnamefont
  {Capozziello}}\ and\ \bibinfo {author} {\bibfnamefont {C.}~\bibnamefont
  {Corda}},\ }\href {\doibase 10.1142/S0218271806008814} {\bibfield  {journal}
  {\bibinfo  {journal} {Int. J. Mod. Phys. D}\ }\textbf {\bibinfo {volume}
  {15}},\ \bibinfo {pages} {1119} (\bibinfo {year} {2006})}\BibitemShut
  {NoStop}%
\bibitem [{\citenamefont {Gong}\ and\ \citenamefont
  {Hou}(2018)}]{Gong:2017bru}%
  \BibitemOpen
  \bibfield  {author} {\bibinfo {author} {\bibfnamefont {Y.}~\bibnamefont
  {Gong}}\ and\ \bibinfo {author} {\bibfnamefont {S.}~\bibnamefont {Hou}},\
  }\href {\doibase 10.1051/epjconf/201816801003} {\bibfield  {journal}
  {\bibinfo  {journal} {EPJ Web Conf.}\ }\textbf {\bibinfo {volume} {168}},\
  \bibinfo {pages} {01003} (\bibinfo {year} {2018})},\ \Eprint
  {http://arxiv.org/abs/1709.03313} {arXiv:1709.03313 [gr-qc]} \BibitemShut
  {NoStop}%
\bibitem [{\citenamefont {Katsuragawa}\ \emph {et~al.}(2019)\citenamefont
  {Katsuragawa}, \citenamefont {Nakamura}, \citenamefont {Ikeda},\ and\
  \citenamefont {Capozziello}}]{Katsuragawa:2019uto}%
  \BibitemOpen
  \bibfield  {author} {\bibinfo {author} {\bibfnamefont {T.}~\bibnamefont
  {Katsuragawa}}, \bibinfo {author} {\bibfnamefont {T.}~\bibnamefont
  {Nakamura}}, \bibinfo {author} {\bibfnamefont {T.}~\bibnamefont {Ikeda}}, \
  and\ \bibinfo {author} {\bibfnamefont {S.}~\bibnamefont {Capozziello}},\
  }\href {\doibase 10.1103/PhysRevD.99.124050} {\bibfield  {journal} {\bibinfo
  {journal} {Phys. Rev. D}\ }\textbf {\bibinfo {volume} {99}},\ \bibinfo
  {pages} {124050} (\bibinfo {year} {2019})},\ \Eprint
  {http://arxiv.org/abs/1902.02494} {arXiv:1902.02494 [gr-qc]} \BibitemShut
  {NoStop}%
\bibitem [{\citenamefont {Moretti}\ \emph {et~al.}(2019)\citenamefont
  {Moretti}, \citenamefont {Bombacigno},\ and\ \citenamefont
  {Montani}}]{Moretti:2019yhs}%
  \BibitemOpen
  \bibfield  {author} {\bibinfo {author} {\bibfnamefont {F.}~\bibnamefont
  {Moretti}}, \bibinfo {author} {\bibfnamefont {F.}~\bibnamefont {Bombacigno}},
  \ and\ \bibinfo {author} {\bibfnamefont {G.}~\bibnamefont {Montani}},\ }\href
  {\doibase 10.1103/PhysRevD.100.084014} {\bibfield  {journal} {\bibinfo
  {journal} {Phys. Rev. D}\ }\textbf {\bibinfo {volume} {100}},\ \bibinfo
  {pages} {084014} (\bibinfo {year} {2019})},\ \Eprint
  {http://arxiv.org/abs/1906.01899} {arXiv:1906.01899 [gr-qc]} \BibitemShut
  {NoStop}%
\bibitem [{\citenamefont {{Khlopunov}}\ and\ \citenamefont
  {{Gal'tsov}}(2022)}]{2022JCAP...04..014K}%
  \BibitemOpen
  \bibfield  {author} {\bibinfo {author} {\bibfnamefont {M.}~\bibnamefont
  {{Khlopunov}}}\ and\ \bibinfo {author} {\bibfnamefont {D.~V.}\ \bibnamefont
  {{Gal'tsov}}},\ }\href {\doibase 10.1088/1475-7516/2022/04/014} {\bibfield
  {journal} {\bibinfo  {journal} {\jcap}\ }\textbf {\bibinfo {volume} {2022}},\
  \bibinfo {eid} {014} (\bibinfo {year} {2022})},\ \Eprint
  {http://arxiv.org/abs/2201.11804} {arXiv:2201.11804 [gr-qc]} \BibitemShut
  {NoStop}%
\bibitem [{\citenamefont {Zhang}\ \emph
  {et~al.}(2020{\natexlab{a}})\citenamefont {Zhang}, \citenamefont {Zhao},
  \citenamefont {Wang}, \citenamefont {Wang}, \citenamefont {Yagi},
  \citenamefont {Yunes}, \citenamefont {Zhao},\ and\ \citenamefont
  {Zhu}}]{Zhang:2019iim}%
  \BibitemOpen
  \bibfield  {author} {\bibinfo {author} {\bibfnamefont {C.}~\bibnamefont
  {Zhang}}, \bibinfo {author} {\bibfnamefont {X.}~\bibnamefont {Zhao}},
  \bibinfo {author} {\bibfnamefont {A.}~\bibnamefont {Wang}}, \bibinfo {author}
  {\bibfnamefont {B.}~\bibnamefont {Wang}}, \bibinfo {author} {\bibfnamefont
  {K.}~\bibnamefont {Yagi}}, \bibinfo {author} {\bibfnamefont {N.}~\bibnamefont
  {Yunes}}, \bibinfo {author} {\bibfnamefont {W.}~\bibnamefont {Zhao}}, \ and\
  \bibinfo {author} {\bibfnamefont {T.}~\bibnamefont {Zhu}},\ }\href {\doibase
  10.1103/PhysRevD.104.069905} {\bibfield  {journal} {\bibinfo  {journal}
  {Phys. Rev. D}\ }\textbf {\bibinfo {volume} {101}},\ \bibinfo {pages}
  {044002} (\bibinfo {year} {2020}{\natexlab{a}})},\ \bibinfo {note} {[Erratum:
  Phys.Rev.D 104, 069905 (2021)]},\ \Eprint {http://arxiv.org/abs/1911.10278}
  {arXiv:1911.10278 [gr-qc]} \BibitemShut {NoStop}%
\bibitem [{\citenamefont {Sagi}(2010)}]{Sagi:2010ei}%
  \BibitemOpen
  \bibfield  {author} {\bibinfo {author} {\bibfnamefont {E.}~\bibnamefont
  {Sagi}},\ }\href {\doibase 10.1103/PhysRevD.81.064031} {\bibfield  {journal}
  {\bibinfo  {journal} {Phys. Rev. D}\ }\textbf {\bibinfo {volume} {81}},\
  \bibinfo {pages} {064031} (\bibinfo {year} {2010})},\ \Eprint
  {http://arxiv.org/abs/1001.1555} {arXiv:1001.1555 [gr-qc]} \BibitemShut
  {NoStop}%
\bibitem [{\citenamefont {de~Paula}\ \emph {et~al.}(2004)\citenamefont
  {de~Paula}, \citenamefont {Miranda},\ and\ \citenamefont
  {Marinho}}]{dePaula:2004bc}%
  \BibitemOpen
  \bibfield  {author} {\bibinfo {author} {\bibfnamefont {W.~L.~S.}\
  \bibnamefont {de~Paula}}, \bibinfo {author} {\bibfnamefont {O.~D.}\
  \bibnamefont {Miranda}}, \ and\ \bibinfo {author} {\bibfnamefont {R.~M.}\
  \bibnamefont {Marinho}},\ }\href {\doibase 10.1088/0264-9381/21/19/008}
  {\bibfield  {journal} {\bibinfo  {journal} {Class. Quant. Grav.}\ }\textbf
  {\bibinfo {volume} {21}},\ \bibinfo {pages} {4595} (\bibinfo {year}
  {2004})},\ \Eprint {http://arxiv.org/abs/gr-qc/0409041} {arXiv:gr-qc/0409041}
  \BibitemShut {NoStop}%
\bibitem [{\citenamefont {Kosteleck\'y}\ and\ \citenamefont
  {Mewes}(2016)}]{Kostelecky:2016kfm}%
  \BibitemOpen
  \bibfield  {author} {\bibinfo {author} {\bibfnamefont {V.~A.}\ \bibnamefont
  {Kosteleck\'y}}\ and\ \bibinfo {author} {\bibfnamefont {M.}~\bibnamefont
  {Mewes}},\ }\href {\doibase 10.1016/j.physletb.2016.04.040} {\bibfield
  {journal} {\bibinfo  {journal} {Phys. Lett. B}\ }\textbf {\bibinfo {volume}
  {757}},\ \bibinfo {pages} {510} (\bibinfo {year} {2016})},\ \Eprint
  {http://arxiv.org/abs/1602.04782} {arXiv:1602.04782 [gr-qc]} \BibitemShut
  {NoStop}%
\bibitem [{\citenamefont {Shao}(2020)}]{Shao:2020shv}%
  \BibitemOpen
  \bibfield  {author} {\bibinfo {author} {\bibfnamefont {L.}~\bibnamefont
  {Shao}},\ }\href {\doibase 10.1103/PhysRevD.101.104019} {\bibfield  {journal}
  {\bibinfo  {journal} {Phys. Rev. D}\ }\textbf {\bibinfo {volume} {101}},\
  \bibinfo {pages} {104019} (\bibinfo {year} {2020})},\ \Eprint
  {http://arxiv.org/abs/2002.01185} {arXiv:2002.01185 [hep-ph]} \BibitemShut
  {NoStop}%
\bibitem [{\citenamefont {Zhao}\ \emph
  {et~al.}(2020{\natexlab{b}})\citenamefont {Zhao}, \citenamefont {Liu},
  \citenamefont {Wen}, \citenamefont {Zhu}, \citenamefont {Wang}, \citenamefont
  {Hu},\ and\ \citenamefont {Zhou}}]{Zhao:2019szi}%
  \BibitemOpen
  \bibfield  {author} {\bibinfo {author} {\bibfnamefont {W.}~\bibnamefont
  {Zhao}}, \bibinfo {author} {\bibfnamefont {T.}~\bibnamefont {Liu}}, \bibinfo
  {author} {\bibfnamefont {L.}~\bibnamefont {Wen}}, \bibinfo {author}
  {\bibfnamefont {T.}~\bibnamefont {Zhu}}, \bibinfo {author} {\bibfnamefont
  {A.}~\bibnamefont {Wang}}, \bibinfo {author} {\bibfnamefont {Q.}~\bibnamefont
  {Hu}}, \ and\ \bibinfo {author} {\bibfnamefont {C.}~\bibnamefont {Zhou}},\
  }\href {\doibase 10.1140/epjc/s10052-020-8211-4} {\bibfield  {journal}
  {\bibinfo  {journal} {Eur. Phys. J. C}\ }\textbf {\bibinfo {volume} {80}},\
  \bibinfo {pages} {630} (\bibinfo {year} {2020}{\natexlab{b}})},\ \Eprint
  {http://arxiv.org/abs/1909.13007} {arXiv:1909.13007 [gr-qc]} \BibitemShut
  {NoStop}%
\bibitem [{\citenamefont {Haegel}\ \emph {et~al.}(2023)\citenamefont {Haegel},
  \citenamefont {O'Neal-Ault} \emph {et~al.}}]{Haegel:2022ymk}%
  \BibitemOpen
  \bibfield  {author} {\bibinfo {author} {\bibfnamefont {L.}~\bibnamefont
  {Haegel}}, \bibinfo {author} {\bibfnamefont {K.}~\bibnamefont {O'Neal-Ault}},
   \emph {et~al.},\ }\href {\doibase 10.1103/PhysRevD.107.064031} {\bibfield
  {journal} {\bibinfo  {journal} {Phys. Rev. D}\ }\textbf {\bibinfo {volume}
  {107}},\ \bibinfo {pages} {064031} (\bibinfo {year} {2023})},\ \Eprint
  {http://arxiv.org/abs/2210.04481} {arXiv:2210.04481 [gr-qc]} \BibitemShut
  {NoStop}%
\bibitem [{\citenamefont {Califano}\ \emph {et~al.}(2024)\citenamefont
  {Califano}, \citenamefont {D'Agostino},\ and\ \citenamefont
  {Vernieri}}]{Califano:2023aji}%
  \BibitemOpen
  \bibfield  {author} {\bibinfo {author} {\bibfnamefont {M.}~\bibnamefont
  {Califano}}, \bibinfo {author} {\bibfnamefont {R.}~\bibnamefont
  {D'Agostino}}, \ and\ \bibinfo {author} {\bibfnamefont {D.}~\bibnamefont
  {Vernieri}},\ }\href {\doibase 10.1103/PhysRevD.109.104062} {\bibfield
  {journal} {\bibinfo  {journal} {Phys. Rev. D}\ }\textbf {\bibinfo {volume}
  {109}},\ \bibinfo {pages} {104062} (\bibinfo {year} {2024})},\ \Eprint
  {http://arxiv.org/abs/2311.02161} {arXiv:2311.02161 [gr-qc]} \BibitemShut
  {NoStop}%
\bibitem [{\citenamefont {O'Neal-Ault}\ \emph {et~al.}(2021)\citenamefont
  {O'Neal-Ault}, \citenamefont {Bailey}, \citenamefont {Dumerchat},
  \citenamefont {Haegel},\ and\ \citenamefont {Tasson}}]{ONeal-Ault:2021uwu}%
  \BibitemOpen
  \bibfield  {author} {\bibinfo {author} {\bibfnamefont {K.}~\bibnamefont
  {O'Neal-Ault}}, \bibinfo {author} {\bibfnamefont {Q.~G.}\ \bibnamefont
  {Bailey}}, \bibinfo {author} {\bibfnamefont {T.}~\bibnamefont {Dumerchat}},
  \bibinfo {author} {\bibfnamefont {L.}~\bibnamefont {Haegel}}, \ and\ \bibinfo
  {author} {\bibfnamefont {J.}~\bibnamefont {Tasson}},\ }\href {\doibase
  10.3390/universe7100380} {\bibfield  {journal} {\bibinfo  {journal}
  {Universe}\ }\textbf {\bibinfo {volume} {7}},\ \bibinfo {pages} {380}
  (\bibinfo {year} {2021})},\ \Eprint {http://arxiv.org/abs/2108.06298}
  {arXiv:2108.06298 [gr-qc]} \BibitemShut {NoStop}%
\bibitem [{\citenamefont {Wang}(2020)}]{Wang:2017igw}%
  \BibitemOpen
  \bibfield  {author} {\bibinfo {author} {\bibfnamefont {S.}~\bibnamefont
  {Wang}},\ }\href {\doibase 10.1140/epjc/s10052-020-7812-2} {\bibfield
  {journal} {\bibinfo  {journal} {Eur. Phys. J. C}\ }\textbf {\bibinfo {volume}
  {80}},\ \bibinfo {pages} {342} (\bibinfo {year} {2020})},\ \Eprint
  {http://arxiv.org/abs/1712.06072} {arXiv:1712.06072 [gr-qc]} \BibitemShut
  {NoStop}%
\bibitem [{\citenamefont {Wang}\ and\ \citenamefont
  {Zhao}(2020)}]{Wang:2020pgu}%
  \BibitemOpen
  \bibfield  {author} {\bibinfo {author} {\bibfnamefont {S.}~\bibnamefont
  {Wang}}\ and\ \bibinfo {author} {\bibfnamefont {Z.-C.}\ \bibnamefont
  {Zhao}},\ }\href {\doibase 10.1140/epjc/s10052-020-08628-x} {\bibfield
  {journal} {\bibinfo  {journal} {Eur. Phys. J. C}\ }\textbf {\bibinfo {volume}
  {80}},\ \bibinfo {pages} {1032} (\bibinfo {year} {2020})},\ \Eprint
  {http://arxiv.org/abs/2002.00396} {arXiv:2002.00396 [gr-qc]} \BibitemShut
  {NoStop}%
\bibitem [{\citenamefont {Zhao}\ \emph {et~al.}(2022)\citenamefont {Zhao},
  \citenamefont {Cao},\ and\ \citenamefont {Wang}}]{Zhao:2022pun}%
  \BibitemOpen
  \bibfield  {author} {\bibinfo {author} {\bibfnamefont {Z.-C.}\ \bibnamefont
  {Zhao}}, \bibinfo {author} {\bibfnamefont {Z.}~\bibnamefont {Cao}}, \ and\
  \bibinfo {author} {\bibfnamefont {S.}~\bibnamefont {Wang}},\ }\href {\doibase
  10.3847/1538-4357/ac62d3} {\bibfield  {journal} {\bibinfo  {journal}
  {Astrophys. J.}\ }\textbf {\bibinfo {volume} {930}},\ \bibinfo {pages} {139}
  (\bibinfo {year} {2022})},\ \Eprint {http://arxiv.org/abs/2201.02813}
  {arXiv:2201.02813 [gr-qc]} \BibitemShut {NoStop}%
\bibitem [{\citenamefont {Will}(1994)}]{Will:1994fb}%
  \BibitemOpen
  \bibfield  {author} {\bibinfo {author} {\bibfnamefont {C.~M.}\ \bibnamefont
  {Will}},\ }\href {\doibase 10.1103/PhysRevD.50.6058} {\bibfield  {journal}
  {\bibinfo  {journal} {Phys. Rev. D}\ }\textbf {\bibinfo {volume} {50}},\
  \bibinfo {pages} {6058} (\bibinfo {year} {1994})},\ \Eprint
  {http://arxiv.org/abs/gr-qc/9406022} {arXiv:gr-qc/9406022} \BibitemShut
  {NoStop}%
\bibitem [{\citenamefont {Chatziioannou}\ \emph {et~al.}(2012)\citenamefont
  {Chatziioannou}, \citenamefont {Yunes},\ and\ \citenamefont
  {Cornish}}]{Chatziioannou:2012rf}%
  \BibitemOpen
  \bibfield  {author} {\bibinfo {author} {\bibfnamefont {K.}~\bibnamefont
  {Chatziioannou}}, \bibinfo {author} {\bibfnamefont {N.}~\bibnamefont
  {Yunes}}, \ and\ \bibinfo {author} {\bibfnamefont {N.}~\bibnamefont
  {Cornish}},\ }\href {\doibase 10.1103/PhysRevD.86.022004} {\bibfield
  {journal} {\bibinfo  {journal} {Phys. Rev. D}\ }\textbf {\bibinfo {volume}
  {86}},\ \bibinfo {pages} {022004} (\bibinfo {year} {2012})},\ \bibinfo {note}
  {[Erratum: Phys.Rev.D 95, 129901 (2017)]},\ \Eprint
  {http://arxiv.org/abs/1204.2585} {arXiv:1204.2585 [gr-qc]} \BibitemShut
  {NoStop}%
\bibitem [{\citenamefont {Sennett}\ \emph {et~al.}(2016)\citenamefont
  {Sennett}, \citenamefont {Marsat},\ and\ \citenamefont
  {Buonanno}}]{Sennett:2016klh}%
  \BibitemOpen
  \bibfield  {author} {\bibinfo {author} {\bibfnamefont {N.}~\bibnamefont
  {Sennett}}, \bibinfo {author} {\bibfnamefont {S.}~\bibnamefont {Marsat}}, \
  and\ \bibinfo {author} {\bibfnamefont {A.}~\bibnamefont {Buonanno}},\ }\href
  {\doibase 10.1103/PhysRevD.94.084003} {\bibfield  {journal} {\bibinfo
  {journal} {Phys. Rev. D}\ }\textbf {\bibinfo {volume} {94}},\ \bibinfo
  {pages} {084003} (\bibinfo {year} {2016})},\ \Eprint
  {http://arxiv.org/abs/1607.01420} {arXiv:1607.01420 [gr-qc]} \BibitemShut
  {NoStop}%
\bibitem [{\citenamefont {Guersel}\ and\ \citenamefont
  {Tinto}(1989)}]{Guersel:1989th}%
  \BibitemOpen
  \bibfield  {author} {\bibinfo {author} {\bibfnamefont {Y.}~\bibnamefont
  {Guersel}}\ and\ \bibinfo {author} {\bibfnamefont {M.}~\bibnamefont
  {Tinto}},\ }\href {\doibase 10.1103/PhysRevD.40.3884} {\bibfield  {journal}
  {\bibinfo  {journal} {Phys. Rev. D}\ }\textbf {\bibinfo {volume} {40}},\
  \bibinfo {pages} {3884} (\bibinfo {year} {1989})}\BibitemShut {NoStop}%
\bibitem [{\citenamefont {Wen}\ and\ \citenamefont
  {Schutz}(2005)}]{Wen:2005ui}%
  \BibitemOpen
  \bibfield  {author} {\bibinfo {author} {\bibfnamefont {L.}~\bibnamefont
  {Wen}}\ and\ \bibinfo {author} {\bibfnamefont {B.~F.}\ \bibnamefont
  {Schutz}},\ }\href {\doibase 10.1088/0264-9381/22/18/S46} {\bibfield
  {journal} {\bibinfo  {journal} {Class. Quant. Grav.}\ }\textbf {\bibinfo
  {volume} {22}},\ \bibinfo {pages} {S1321} (\bibinfo {year} {2005})},\ \Eprint
  {http://arxiv.org/abs/gr-qc/0508042} {arXiv:gr-qc/0508042} \BibitemShut
  {NoStop}%
\bibitem [{\citenamefont {Wen}(2008)}]{Wen:2007pj}%
  \BibitemOpen
  \bibfield  {author} {\bibinfo {author} {\bibfnamefont {L.}~\bibnamefont
  {Wen}},\ }\href {\doibase 10.1142/S0218271808012723} {\bibfield  {journal}
  {\bibinfo  {journal} {Int. J. Mod. Phys. D}\ }\textbf {\bibinfo {volume}
  {17}},\ \bibinfo {pages} {1095} (\bibinfo {year} {2008})},\ \Eprint
  {http://arxiv.org/abs/gr-qc/0702096} {arXiv:gr-qc/0702096} \BibitemShut
  {NoStop}%
\bibitem [{\citenamefont {Chatterji}\ \emph {et~al.}(2006)\citenamefont
  {Chatterji}, \citenamefont {Lazzarini}, \citenamefont {Stein}, \citenamefont
  {Sutton}, \citenamefont {Searle},\ and\ \citenamefont
  {Tinto}}]{Chatterji:2006nh}%
  \BibitemOpen
  \bibfield  {author} {\bibinfo {author} {\bibfnamefont {S.}~\bibnamefont
  {Chatterji}}, \bibinfo {author} {\bibfnamefont {A.}~\bibnamefont
  {Lazzarini}}, \bibinfo {author} {\bibfnamefont {L.}~\bibnamefont {Stein}},
  \bibinfo {author} {\bibfnamefont {P.~J.}\ \bibnamefont {Sutton}}, \bibinfo
  {author} {\bibfnamefont {A.}~\bibnamefont {Searle}}, \ and\ \bibinfo {author}
  {\bibfnamefont {M.}~\bibnamefont {Tinto}},\ }\href {\doibase
  10.1103/PhysRevD.74.082005} {\bibfield  {journal} {\bibinfo  {journal} {Phys.
  Rev. D}\ }\textbf {\bibinfo {volume} {74}},\ \bibinfo {pages} {082005}
  (\bibinfo {year} {2006})},\ \Eprint {http://arxiv.org/abs/gr-qc/0605002}
  {arXiv:gr-qc/0605002} \BibitemShut {NoStop}%
\bibitem [{\citenamefont {Hagihara}\ \emph {et~al.}(2018)\citenamefont
  {Hagihara}, \citenamefont {Era}, \citenamefont {Iikawa},\ and\ \citenamefont
  {Asada}}]{Hagihara:2018azu}%
  \BibitemOpen
  \bibfield  {author} {\bibinfo {author} {\bibfnamefont {Y.}~\bibnamefont
  {Hagihara}}, \bibinfo {author} {\bibfnamefont {N.}~\bibnamefont {Era}},
  \bibinfo {author} {\bibfnamefont {D.}~\bibnamefont {Iikawa}}, \ and\ \bibinfo
  {author} {\bibfnamefont {H.}~\bibnamefont {Asada}},\ }\href {\doibase
  10.1103/PhysRevD.98.064035} {\bibfield  {journal} {\bibinfo  {journal} {Phys.
  Rev. D}\ }\textbf {\bibinfo {volume} {98}},\ \bibinfo {pages} {064035}
  (\bibinfo {year} {2018})},\ \Eprint {http://arxiv.org/abs/1807.07234}
  {arXiv:1807.07234 [gr-qc]} \BibitemShut {NoStop}%
\bibitem [{\citenamefont {Hagihara}\ \emph {et~al.}(2019)\citenamefont
  {Hagihara}, \citenamefont {Era}, \citenamefont {Iikawa}, \citenamefont
  {Nishizawa},\ and\ \citenamefont {Asada}}]{Hagihara:2019ihn}%
  \BibitemOpen
  \bibfield  {author} {\bibinfo {author} {\bibfnamefont {Y.}~\bibnamefont
  {Hagihara}}, \bibinfo {author} {\bibfnamefont {N.}~\bibnamefont {Era}},
  \bibinfo {author} {\bibfnamefont {D.}~\bibnamefont {Iikawa}}, \bibinfo
  {author} {\bibfnamefont {A.}~\bibnamefont {Nishizawa}}, \ and\ \bibinfo
  {author} {\bibfnamefont {H.}~\bibnamefont {Asada}},\ }\href {\doibase
  10.1103/PhysRevD.100.064010} {\bibfield  {journal} {\bibinfo  {journal}
  {Phys. Rev. D}\ }\textbf {\bibinfo {volume} {100}},\ \bibinfo {pages}
  {064010} (\bibinfo {year} {2019})},\ \Eprint
  {http://arxiv.org/abs/1904.02300} {arXiv:1904.02300 [gr-qc]} \BibitemShut
  {NoStop}%
\bibitem [{\citenamefont {Hagihara}\ \emph {et~al.}(2020)\citenamefont
  {Hagihara}, \citenamefont {Era}, \citenamefont {Iikawa}, \citenamefont
  {Takeda},\ and\ \citenamefont {Asada}}]{Hagihara:2019rny}%
  \BibitemOpen
  \bibfield  {author} {\bibinfo {author} {\bibfnamefont {Y.}~\bibnamefont
  {Hagihara}}, \bibinfo {author} {\bibfnamefont {N.}~\bibnamefont {Era}},
  \bibinfo {author} {\bibfnamefont {D.}~\bibnamefont {Iikawa}}, \bibinfo
  {author} {\bibfnamefont {N.}~\bibnamefont {Takeda}}, \ and\ \bibinfo {author}
  {\bibfnamefont {H.}~\bibnamefont {Asada}},\ }\href {\doibase
  10.1103/PhysRevD.101.041501} {\bibfield  {journal} {\bibinfo  {journal}
  {Phys. Rev. D}\ }\textbf {\bibinfo {volume} {101}},\ \bibinfo {pages}
  {041501} (\bibinfo {year} {2020})},\ \Eprint
  {http://arxiv.org/abs/1912.06340} {arXiv:1912.06340 [gr-qc]} \BibitemShut
  {NoStop}%
\bibitem [{\citenamefont {Takeda}\ \emph {et~al.}(2018)\citenamefont {Takeda},
  \citenamefont {Nishizawa}, \citenamefont {Michimura}, \citenamefont {Nagano},
  \citenamefont {Komori}, \citenamefont {Ando},\ and\ \citenamefont
  {Hayama}}]{2018Takeda}%
  \BibitemOpen
  \bibfield  {author} {\bibinfo {author} {\bibfnamefont {H.}~\bibnamefont
  {Takeda}}, \bibinfo {author} {\bibfnamefont {A.}~\bibnamefont {Nishizawa}},
  \bibinfo {author} {\bibfnamefont {Y.}~\bibnamefont {Michimura}}, \bibinfo
  {author} {\bibfnamefont {K.}~\bibnamefont {Nagano}}, \bibinfo {author}
  {\bibfnamefont {K.}~\bibnamefont {Komori}}, \bibinfo {author} {\bibfnamefont
  {M.}~\bibnamefont {Ando}}, \ and\ \bibinfo {author} {\bibfnamefont
  {K.}~\bibnamefont {Hayama}},\ }\href {\doibase 10.1103/PhysRevD.98.022008}
  {\bibfield  {journal} {\bibinfo  {journal} {Phys. Rev. D}\ }\textbf {\bibinfo
  {volume} {98}},\ \bibinfo {pages} {022008} (\bibinfo {year} {2018})},\
  \Eprint {http://arxiv.org/abs/1806.02182} {arXiv:1806.02182 [gr-qc]}
  \BibitemShut {NoStop}%
\bibitem [{\citenamefont {Hu}\ \emph {et~al.}(2024{\natexlab{a}})\citenamefont
  {Hu}, \citenamefont {Liang},\ and\ \citenamefont {Shao}}]{Hu:2023soi}%
  \BibitemOpen
  \bibfield  {author} {\bibinfo {author} {\bibfnamefont {J.}~\bibnamefont
  {Hu}}, \bibinfo {author} {\bibfnamefont {D.}~\bibnamefont {Liang}}, \ and\
  \bibinfo {author} {\bibfnamefont {L.}~\bibnamefont {Shao}},\ }\href {\doibase
  10.1103/PhysRevD.109.084023} {\bibfield  {journal} {\bibinfo  {journal}
  {Phys. Rev. D}\ }\textbf {\bibinfo {volume} {109}},\ \bibinfo {pages}
  {084023} (\bibinfo {year} {2024}{\natexlab{a}})},\ \Eprint
  {http://arxiv.org/abs/2310.01249} {arXiv:2310.01249 [gr-qc]} \BibitemShut
  {NoStop}%
\bibitem [{\citenamefont {Liang}\ \emph
  {et~al.}(2024{\natexlab{a}})\citenamefont {Liang}, \citenamefont {Chen},
  \citenamefont {Zhang},\ and\ \citenamefont {Shao}}]{Liang:2024sfn}%
  \BibitemOpen
  \bibfield  {author} {\bibinfo {author} {\bibfnamefont {D.}~\bibnamefont
  {Liang}}, \bibinfo {author} {\bibfnamefont {S.}~\bibnamefont {Chen}},
  \bibinfo {author} {\bibfnamefont {C.}~\bibnamefont {Zhang}}, \ and\ \bibinfo
  {author} {\bibfnamefont {L.}~\bibnamefont {Shao}},\ }\href {\doibase
  10.1103/PhysRevD.110.084040} {\bibfield  {journal} {\bibinfo  {journal}
  {Phys. Rev. D}\ }\textbf {\bibinfo {volume} {110}},\ \bibinfo {pages}
  {084040} (\bibinfo {year} {2024}{\natexlab{a}})},\ \Eprint
  {http://arxiv.org/abs/2404.16680} {arXiv:2404.16680 [gr-qc]} \BibitemShut
  {NoStop}%
\bibitem [{\citenamefont {Poisson}\ and\ \citenamefont
  {Will}(2014)}]{Poisson:2014book}%
  \BibitemOpen
  \bibfield  {author} {\bibinfo {author} {\bibfnamefont {E.}~\bibnamefont
  {Poisson}}\ and\ \bibinfo {author} {\bibfnamefont {C.~M.}\ \bibnamefont
  {Will}},\ }\href@noop {} {\emph {\bibinfo {title} {Gravity: Newtonian,
  post-newtonian, relativistic}}}\ (\bibinfo  {publisher} {Cambridge University
  Press},\ \bibinfo {year} {2014})\BibitemShut {NoStop}%
\bibitem [{\citenamefont {Xie}\ \emph {et~al.}(2022)\citenamefont {Xie},
  \citenamefont {Zhang}, \citenamefont {Huang}, \citenamefont {Hu},\ and\
  \citenamefont {Mei}}]{Xie:2022wkx}%
  \BibitemOpen
  \bibfield  {author} {\bibinfo {author} {\bibfnamefont {N.}~\bibnamefont
  {Xie}}, \bibinfo {author} {\bibfnamefont {J.-d.}\ \bibnamefont {Zhang}},
  \bibinfo {author} {\bibfnamefont {S.-J.}\ \bibnamefont {Huang}}, \bibinfo
  {author} {\bibfnamefont {Y.-M.}\ \bibnamefont {Hu}}, \ and\ \bibinfo {author}
  {\bibfnamefont {J.}~\bibnamefont {Mei}},\ }\href {\doibase
  10.1103/PhysRevD.106.124017} {\bibfield  {journal} {\bibinfo  {journal}
  {Phys. Rev. D}\ }\textbf {\bibinfo {volume} {106}},\ \bibinfo {pages}
  {124017} (\bibinfo {year} {2022})},\ \Eprint
  {http://arxiv.org/abs/2208.10831} {arXiv:2208.10831 [gr-qc]} \BibitemShut
  {NoStop}%
\bibitem [{\citenamefont {Abbott}\ \emph
  {et~al.}(2017{\natexlab{a}})\citenamefont {Abbott} \emph
  {et~al.}}]{LIGOScientific:2017ycc}%
  \BibitemOpen
  \bibfield  {author} {\bibinfo {author} {\bibfnamefont {B.}~\bibnamefont
  {Abbott}} \emph {et~al.} (\bibinfo {collaboration} {LIGO Scientific,
  Virgo}),\ }\href {\doibase 10.1103/PhysRevLett.119.141101} {\bibfield
  {journal} {\bibinfo  {journal} {Phys. Rev. Lett.}\ }\textbf {\bibinfo
  {volume} {119}},\ \bibinfo {pages} {141101} (\bibinfo {year}
  {2017}{\natexlab{a}})},\ \Eprint {http://arxiv.org/abs/1709.09660}
  {arXiv:1709.09660 [gr-qc]} \BibitemShut {NoStop}%
\bibitem [{\citenamefont {Abbott}\ \emph
  {et~al.}(2019{\natexlab{b}})\citenamefont {Abbott} \emph
  {et~al.}}]{LIGOScientific:2018dkp}%
  \BibitemOpen
  \bibfield  {author} {\bibinfo {author} {\bibfnamefont {B.~P.}\ \bibnamefont
  {Abbott}} \emph {et~al.} (\bibinfo {collaboration} {LIGO Scientific,
  Virgo}),\ }\href {\doibase 10.1103/PhysRevLett.123.011102} {\bibfield
  {journal} {\bibinfo  {journal} {Phys. Rev. Lett.}\ }\textbf {\bibinfo
  {volume} {123}},\ \bibinfo {pages} {011102} (\bibinfo {year}
  {2019}{\natexlab{b}})},\ \Eprint {http://arxiv.org/abs/1811.00364}
  {arXiv:1811.00364 [gr-qc]} \BibitemShut {NoStop}%
\bibitem [{\citenamefont {Pang}\ \emph {et~al.}(2020)\citenamefont {Pang},
  \citenamefont {Lo}, \citenamefont {Wong}, \citenamefont {Li},\ and\
  \citenamefont {Van Den~Broeck}}]{Pang:2020pfz}%
  \BibitemOpen
  \bibfield  {author} {\bibinfo {author} {\bibfnamefont {P.~T.~H.}\
  \bibnamefont {Pang}}, \bibinfo {author} {\bibfnamefont {R.~K.~L.}\
  \bibnamefont {Lo}}, \bibinfo {author} {\bibfnamefont {I.~C.~F.}\ \bibnamefont
  {Wong}}, \bibinfo {author} {\bibfnamefont {T.~G.~F.}\ \bibnamefont {Li}}, \
  and\ \bibinfo {author} {\bibfnamefont {C.}~\bibnamefont {Van Den~Broeck}},\
  }\href {\doibase 10.1103/PhysRevD.101.104055} {\bibfield  {journal} {\bibinfo
   {journal} {Phys. Rev. D}\ }\textbf {\bibinfo {volume} {101}},\ \bibinfo
  {pages} {104055} (\bibinfo {year} {2020})},\ \Eprint
  {http://arxiv.org/abs/2003.07375} {arXiv:2003.07375 [gr-qc]} \BibitemShut
  {NoStop}%
\bibitem [{\citenamefont {Wong}\ \emph {et~al.}(2021)\citenamefont {Wong},
  \citenamefont {Pang}, \citenamefont {Lo}, \citenamefont {Li},\ and\
  \citenamefont {Van Den~Broeck}}]{Wong:2021cmp}%
  \BibitemOpen
  \bibfield  {author} {\bibinfo {author} {\bibfnamefont {I.~C.~F.}\
  \bibnamefont {Wong}}, \bibinfo {author} {\bibfnamefont {P.~T.~H.}\
  \bibnamefont {Pang}}, \bibinfo {author} {\bibfnamefont {R.~K.~L.}\
  \bibnamefont {Lo}}, \bibinfo {author} {\bibfnamefont {T.~G.~F.}\ \bibnamefont
  {Li}}, \ and\ \bibinfo {author} {\bibfnamefont {C.}~\bibnamefont {Van
  Den~Broeck}},\ }\href@noop {} {\  (\bibinfo {year} {2021})},\ \Eprint
  {http://arxiv.org/abs/2105.09485} {arXiv:2105.09485 [gr-qc]} \BibitemShut
  {NoStop}%
\bibitem [{\citenamefont {Takeda}\ \emph {et~al.}(2022)\citenamefont {Takeda},
  \citenamefont {Morisaki},\ and\ \citenamefont {Nishizawa}}]{Takeda:2021hgo}%
  \BibitemOpen
  \bibfield  {author} {\bibinfo {author} {\bibfnamefont {H.}~\bibnamefont
  {Takeda}}, \bibinfo {author} {\bibfnamefont {S.}~\bibnamefont {Morisaki}}, \
  and\ \bibinfo {author} {\bibfnamefont {A.}~\bibnamefont {Nishizawa}},\ }\href
  {\doibase 10.1103/PhysRevD.105.084019} {\bibfield  {journal} {\bibinfo
  {journal} {Phys. Rev. D}\ }\textbf {\bibinfo {volume} {105}},\ \bibinfo
  {pages} {084019} (\bibinfo {year} {2022})},\ \Eprint
  {http://arxiv.org/abs/2105.00253} {arXiv:2105.00253 [gr-qc]} \BibitemShut
  {NoStop}%
\bibitem [{\citenamefont {da~Silva~Alves}\ and\ \citenamefont
  {Tinto}(2011)}]{daSilvaAlves:2011fp}%
  \BibitemOpen
  \bibfield  {author} {\bibinfo {author} {\bibfnamefont {M.~E.}\ \bibnamefont
  {da~Silva~Alves}}\ and\ \bibinfo {author} {\bibfnamefont {M.}~\bibnamefont
  {Tinto}},\ }\href {\doibase 10.1103/PhysRevD.83.123529} {\bibfield  {journal}
  {\bibinfo  {journal} {Phys. Rev. D}\ }\textbf {\bibinfo {volume} {83}},\
  \bibinfo {pages} {123529} (\bibinfo {year} {2011})},\ \Eprint
  {http://arxiv.org/abs/1102.4824} {arXiv:1102.4824 [gr-qc]} \BibitemShut
  {NoStop}%
\bibitem [{\citenamefont {Lee}\ \emph {et~al.}(2008)\citenamefont {Lee},
  \citenamefont {Jenet},\ and\ \citenamefont {Price}}]{Lee:2008ajo}%
  \BibitemOpen
  \bibfield  {author} {\bibinfo {author} {\bibfnamefont {K.~J.}\ \bibnamefont
  {Lee}}, \bibinfo {author} {\bibfnamefont {F.~A.}\ \bibnamefont {Jenet}}, \
  and\ \bibinfo {author} {\bibfnamefont {R.~H.}\ \bibnamefont {Price}},\ }\href
  {\doibase 10.1086/591080} {\bibfield  {journal} {\bibinfo  {journal} {The
  Astrophysical Journal}\ }\textbf {\bibinfo {volume} {685}},\ \bibinfo {pages}
  {1304} (\bibinfo {year} {2008})}\BibitemShut {NoStop}%
\bibitem [{\citenamefont {Niu}\ and\ \citenamefont {Zhao}(2019)}]{Niu:2018oox}%
  \BibitemOpen
  \bibfield  {author} {\bibinfo {author} {\bibfnamefont {R.}~\bibnamefont
  {Niu}}\ and\ \bibinfo {author} {\bibfnamefont {W.}~\bibnamefont {Zhao}},\
  }\href {\doibase 10.1007/s11433-018-9340-6} {\bibfield  {journal} {\bibinfo
  {journal} {Sci. China Phys. Mech. Astron.}\ }\textbf {\bibinfo {volume}
  {62}},\ \bibinfo {pages} {970411} (\bibinfo {year} {2019})},\ \Eprint
  {http://arxiv.org/abs/1812.00208} {arXiv:1812.00208 [gr-qc]} \BibitemShut
  {NoStop}%
\bibitem [{\citenamefont {O'Beirne}\ \emph {et~al.}(2019)\citenamefont
  {O'Beirne}, \citenamefont {Cornish}, \citenamefont {Vigeland},\ and\
  \citenamefont {Taylor}}]{OBeirne:2019lwp}%
  \BibitemOpen
  \bibfield  {author} {\bibinfo {author} {\bibfnamefont {L.}~\bibnamefont
  {O'Beirne}}, \bibinfo {author} {\bibfnamefont {N.~J.}\ \bibnamefont
  {Cornish}}, \bibinfo {author} {\bibfnamefont {S.~J.}\ \bibnamefont
  {Vigeland}}, \ and\ \bibinfo {author} {\bibfnamefont {S.~R.}\ \bibnamefont
  {Taylor}},\ }\href {\doibase 10.1103/PhysRevD.99.124039} {\bibfield
  {journal} {\bibinfo  {journal} {Phys. Rev. D}\ }\textbf {\bibinfo {volume}
  {99}},\ \bibinfo {pages} {124039} (\bibinfo {year} {2019})},\ \Eprint
  {http://arxiv.org/abs/1904.02744} {arXiv:1904.02744 [gr-qc]} \BibitemShut
  {NoStop}%
\bibitem [{\citenamefont {Bo\^\i{}tier}\ \emph {et~al.}(2020)\citenamefont
  {Bo\^\i{}tier}, \citenamefont {Tiwari}, \citenamefont {Philippoz},\ and\
  \citenamefont {Jetzer}}]{Boitier:2020xfx}%
  \BibitemOpen
  \bibfield  {author} {\bibinfo {author} {\bibfnamefont {A.}~\bibnamefont
  {Bo\^\i{}tier}}, \bibinfo {author} {\bibfnamefont {S.}~\bibnamefont
  {Tiwari}}, \bibinfo {author} {\bibfnamefont {L.}~\bibnamefont {Philippoz}}, \
  and\ \bibinfo {author} {\bibfnamefont {P.}~\bibnamefont {Jetzer}},\ }\href
  {\doibase 10.1103/PhysRevD.102.064051} {\bibfield  {journal} {\bibinfo
  {journal} {Phys. Rev. D}\ }\textbf {\bibinfo {volume} {102}},\ \bibinfo
  {pages} {064051} (\bibinfo {year} {2020})},\ \Eprint
  {http://arxiv.org/abs/2008.13520} {arXiv:2008.13520 [gr-qc]} \BibitemShut
  {NoStop}%
\bibitem [{\citenamefont {Chen}\ \emph
  {et~al.}(2021{\natexlab{a}})\citenamefont {Chen}, \citenamefont {Yuan},\ and\
  \citenamefont {Huang}}]{Chen:2021wdo}%
  \BibitemOpen
  \bibfield  {author} {\bibinfo {author} {\bibfnamefont {Z.-C.}\ \bibnamefont
  {Chen}}, \bibinfo {author} {\bibfnamefont {C.}~\bibnamefont {Yuan}}, \ and\
  \bibinfo {author} {\bibfnamefont {Q.-G.}\ \bibnamefont {Huang}},\ }\href
  {\doibase 10.1007/s11433-021-1797-y} {\bibfield  {journal} {\bibinfo
  {journal} {Sci. China Phys. Mech. Astron.}\ }\textbf {\bibinfo {volume}
  {64}},\ \bibinfo {pages} {120412} (\bibinfo {year} {2021}{\natexlab{a}})},\
  \Eprint {http://arxiv.org/abs/2101.06869} {arXiv:2101.06869 [astro-ph.CO]}
  \BibitemShut {NoStop}%
\bibitem [{\citenamefont {Arzoumanian}\ \emph {et~al.}(2020)\citenamefont
  {Arzoumanian} \emph {et~al.}}]{NANOGrav:2020bcs}%
  \BibitemOpen
  \bibfield  {author} {\bibinfo {author} {\bibfnamefont {Z.}~\bibnamefont
  {Arzoumanian}} \emph {et~al.} (\bibinfo {collaboration} {NANOGrav}),\ }\href
  {\doibase 10.3847/2041-8213/abd401} {\bibfield  {journal} {\bibinfo
  {journal} {Astrophys. J. Lett.}\ }\textbf {\bibinfo {volume} {905}},\
  \bibinfo {pages} {L34} (\bibinfo {year} {2020})},\ \Eprint
  {http://arxiv.org/abs/2009.04496} {arXiv:2009.04496 [astro-ph.HE]}
  \BibitemShut {NoStop}%
\bibitem [{\citenamefont {Arzoumanian}\ \emph {et~al.}(2021)\citenamefont
  {Arzoumanian} \emph {et~al.}}]{NANOGrav:2021ini}%
  \BibitemOpen
  \bibfield  {author} {\bibinfo {author} {\bibfnamefont {Z.}~\bibnamefont
  {Arzoumanian}} \emph {et~al.} (\bibinfo {collaboration} {NANOGrav}),\ }\href
  {\doibase 10.3847/2041-8213/ac401c} {\bibfield  {journal} {\bibinfo
  {journal} {Astrophys. J. Lett.}\ }\textbf {\bibinfo {volume} {923}},\
  \bibinfo {pages} {L22} (\bibinfo {year} {2021})},\ \Eprint
  {http://arxiv.org/abs/2109.14706} {arXiv:2109.14706 [gr-qc]} \BibitemShut
  {NoStop}%
\bibitem [{\citenamefont {Chen}\ \emph
  {et~al.}(2024{\natexlab{a}})\citenamefont {Chen}, \citenamefont {Wu},
  \citenamefont {Bi},\ and\ \citenamefont {Huang}}]{Chen:2023uiz}%
  \BibitemOpen
  \bibfield  {author} {\bibinfo {author} {\bibfnamefont {Z.-C.}\ \bibnamefont
  {Chen}}, \bibinfo {author} {\bibfnamefont {Y.-M.}\ \bibnamefont {Wu}},
  \bibinfo {author} {\bibfnamefont {Y.-C.}\ \bibnamefont {Bi}}, \ and\ \bibinfo
  {author} {\bibfnamefont {Q.-G.}\ \bibnamefont {Huang}},\ }\href {\doibase
  10.1103/PhysRevD.109.084045} {\bibfield  {journal} {\bibinfo  {journal}
  {Phys. Rev. D}\ }\textbf {\bibinfo {volume} {109}},\ \bibinfo {pages}
  {084045} (\bibinfo {year} {2024}{\natexlab{a}})},\ \Eprint
  {http://arxiv.org/abs/2310.11238} {arXiv:2310.11238 [astro-ph.CO]}
  \BibitemShut {NoStop}%
\bibitem [{\citenamefont {Lau}\ \emph {et~al.}(2020)\citenamefont {Lau},
  \citenamefont {Mandel}, \citenamefont {Vigna-G\'omez}, \citenamefont
  {Neijssel}, \citenamefont {Stevenson},\ and\ \citenamefont
  {Sesana}}]{Lau:2019wzw}%
  \BibitemOpen
  \bibfield  {author} {\bibinfo {author} {\bibfnamefont {M.~Y.~M.}\
  \bibnamefont {Lau}}, \bibinfo {author} {\bibfnamefont {I.}~\bibnamefont
  {Mandel}}, \bibinfo {author} {\bibfnamefont {A.}~\bibnamefont
  {Vigna-G\'omez}}, \bibinfo {author} {\bibfnamefont {C.~J.}\ \bibnamefont
  {Neijssel}}, \bibinfo {author} {\bibfnamefont {S.}~\bibnamefont {Stevenson}},
  \ and\ \bibinfo {author} {\bibfnamefont {A.}~\bibnamefont {Sesana}},\ }\href
  {\doibase 10.1093/mnras/staa002} {\bibfield  {journal} {\bibinfo  {journal}
  {Mon. Not. Roy. Astron. Soc.}\ }\textbf {\bibinfo {volume} {492}},\ \bibinfo
  {pages} {3061} (\bibinfo {year} {2020})},\ \Eprint
  {http://arxiv.org/abs/1910.12422} {arXiv:1910.12422 [astro-ph.HE]}
  \BibitemShut {NoStop}%
\bibitem [{\citenamefont {Burdge}\ \emph {et~al.}(2019)\citenamefont {Burdge}
  \emph {et~al.}}]{Burdge:2019hgl}%
  \BibitemOpen
  \bibfield  {author} {\bibinfo {author} {\bibfnamefont {K.~B.}\ \bibnamefont
  {Burdge}} \emph {et~al.},\ }\href {\doibase 10.1038/s41586-019-1403-0}
  {\bibfield  {journal} {\bibinfo  {journal} {Nature}\ }\textbf {\bibinfo
  {volume} {571}},\ \bibinfo {pages} {528} (\bibinfo {year} {2019})},\ \Eprint
  {http://arxiv.org/abs/1907.11291} {arXiv:1907.11291 [astro-ph.SR]}
  \BibitemShut {NoStop}%
\bibitem [{\citenamefont {Ning}\ and\ \citenamefont {Zhang}()}]{Ning:2024xxx}%
  \BibitemOpen
  \bibfield  {author} {\bibinfo {author} {\bibfnamefont {M.}~\bibnamefont
  {Ning}}\ and\ \bibinfo {author} {\bibfnamefont {J.-d.}\ \bibnamefont
  {Zhang}},\ }\href@noop {} {\bibinfo  {journal} {{In preparation}}\
  }\BibitemShut {NoStop}%
\bibitem [{\citenamefont {Hu}\ \emph {et~al.}(2023)\citenamefont {Hu},
  \citenamefont {Wang}, \citenamefont {Tan},\ and\ \citenamefont
  {Shao}}]{Hu:2023nfv}%
  \BibitemOpen
\bibfield  {journal} {  }\bibfield  {author} {\bibinfo {author} {\bibfnamefont
  {Y.}~\bibnamefont {Hu}}, \bibinfo {author} {\bibfnamefont {P.-P.}\
  \bibnamefont {Wang}}, \bibinfo {author} {\bibfnamefont {Y.-J.}\ \bibnamefont
  {Tan}}, \ and\ \bibinfo {author} {\bibfnamefont {C.-G.}\ \bibnamefont
  {Shao}},\ }\href {\doibase 10.1103/PhysRevD.107.024026} {\bibfield  {journal}
  {\bibinfo  {journal} {Phys. Rev. D}\ }\textbf {\bibinfo {volume} {107}},\
  \bibinfo {pages} {024026} (\bibinfo {year} {2023})}\BibitemShut {NoStop}%
\bibitem [{\citenamefont {Hu}\ \emph {et~al.}(2024{\natexlab{b}})\citenamefont
  {Hu}, \citenamefont {Wang}, \citenamefont {Tan},\ and\ \citenamefont
  {Shao}}]{Hu:2024toa}%
  \BibitemOpen
  \bibfield  {author} {\bibinfo {author} {\bibfnamefont {Y.}~\bibnamefont
  {Hu}}, \bibinfo {author} {\bibfnamefont {P.-P.}\ \bibnamefont {Wang}},
  \bibinfo {author} {\bibfnamefont {Y.-J.}\ \bibnamefont {Tan}}, \ and\
  \bibinfo {author} {\bibfnamefont {C.-G.}\ \bibnamefont {Shao}},\ }\href
  {\doibase 10.3847/1538-4357/ad0cef} {\bibfield  {journal} {\bibinfo
  {journal} {Astrophys. J.}\ }\textbf {\bibinfo {volume} {961}},\ \bibinfo
  {pages} {116} (\bibinfo {year} {2024}{\natexlab{b}})}\BibitemShut {NoStop}%
\bibitem [{\citenamefont {Abbott}\ \emph
  {et~al.}(2017{\natexlab{b}})\citenamefont {Abbott} \emph
  {et~al.}}]{LIGOScientific:2017vwq}%
  \BibitemOpen
  \bibfield  {author} {\bibinfo {author} {\bibfnamefont {B.~P.}\ \bibnamefont
  {Abbott}} \emph {et~al.} (\bibinfo {collaboration} {LIGO Scientific,
  Virgo}),\ }\href {\doibase 10.1103/PhysRevLett.119.161101} {\bibfield
  {journal} {\bibinfo  {journal} {Phys. Rev. Lett.}\ }\textbf {\bibinfo
  {volume} {119}},\ \bibinfo {pages} {161101} (\bibinfo {year}
  {2017}{\natexlab{b}})},\ \Eprint {http://arxiv.org/abs/1710.05832}
  {arXiv:1710.05832 [gr-qc]} \BibitemShut {NoStop}%
\bibitem [{\citenamefont {Abbott}\ \emph
  {et~al.}(2017{\natexlab{c}})\citenamefont {Abbott} \emph
  {et~al.}}]{LIGOScientific:2017zic}%
  \BibitemOpen
  \bibfield  {author} {\bibinfo {author} {\bibfnamefont {B.~P.}\ \bibnamefont
  {Abbott}} \emph {et~al.} (\bibinfo {collaboration} {LIGO Scientific, Virgo,
  Fermi-GBM, INTEGRAL}),\ }\href {\doibase 10.3847/2041-8213/aa920c} {\bibfield
   {journal} {\bibinfo  {journal} {Astrophys. J. Lett.}\ }\textbf {\bibinfo
  {volume} {848}},\ \bibinfo {pages} {L13} (\bibinfo {year}
  {2017}{\natexlab{c}})},\ \Eprint {http://arxiv.org/abs/1710.05834}
  {arXiv:1710.05834 [astro-ph.HE]} \BibitemShut {NoStop}%
\bibitem [{\citenamefont {Yunes}\ \emph {et~al.}(2016)\citenamefont {Yunes},
  \citenamefont {Yagi},\ and\ \citenamefont {Pretorius}}]{Yunes:2016jcc}%
  \BibitemOpen
  \bibfield  {author} {\bibinfo {author} {\bibfnamefont {N.}~\bibnamefont
  {Yunes}}, \bibinfo {author} {\bibfnamefont {K.}~\bibnamefont {Yagi}}, \ and\
  \bibinfo {author} {\bibfnamefont {F.}~\bibnamefont {Pretorius}},\ }\href
  {\doibase 10.1103/PhysRevD.94.084002} {\bibfield  {journal} {\bibinfo
  {journal} {Phys. Rev.}\ }\textbf {\bibinfo {volume} {D94}},\ \bibinfo {pages}
  {084002} (\bibinfo {year} {2016})},\ \Eprint
  {http://arxiv.org/abs/1603.08955} {arXiv:1603.08955 [gr-qc]} \BibitemShut
  {NoStop}%
\bibitem [{\citenamefont {Will}(1998)}]{Will:1997bb}%
  \BibitemOpen
  \bibfield  {author} {\bibinfo {author} {\bibfnamefont {C.~M.}\ \bibnamefont
  {Will}},\ }\href {\doibase 10.1103/PhysRevD.57.2061} {\bibfield  {journal}
  {\bibinfo  {journal} {Phys. Rev. D}\ }\textbf {\bibinfo {volume} {57}},\
  \bibinfo {pages} {2061} (\bibinfo {year} {1998})},\ \Eprint
  {http://arxiv.org/abs/gr-qc/9709011} {arXiv:gr-qc/9709011} \BibitemShut
  {NoStop}%
\bibitem [{\citenamefont {Mirshekari}\ \emph {et~al.}(2012)\citenamefont
  {Mirshekari}, \citenamefont {Yunes},\ and\ \citenamefont
  {Will}}]{Mirshekari:2011yq}%
  \BibitemOpen
  \bibfield  {author} {\bibinfo {author} {\bibfnamefont {S.}~\bibnamefont
  {Mirshekari}}, \bibinfo {author} {\bibfnamefont {N.}~\bibnamefont {Yunes}}, \
  and\ \bibinfo {author} {\bibfnamefont {C.~M.}\ \bibnamefont {Will}},\ }\href
  {\doibase 10.1103/PhysRevD.85.024041} {\bibfield  {journal} {\bibinfo
  {journal} {Phys. Rev. D}\ }\textbf {\bibinfo {volume} {85}},\ \bibinfo
  {pages} {024041} (\bibinfo {year} {2012})},\ \Eprint
  {http://arxiv.org/abs/1110.2720} {arXiv:1110.2720 [gr-qc]} \BibitemShut
  {NoStop}%
\bibitem [{\citenamefont {Abbott}\ \emph
  {et~al.}(2019{\natexlab{c}})\citenamefont {Abbott} \emph
  {et~al.}}]{LIGOScientific:2018mvr}%
  \BibitemOpen
  \bibfield  {author} {\bibinfo {author} {\bibfnamefont {B.~P.}\ \bibnamefont
  {Abbott}} \emph {et~al.} (\bibinfo {collaboration} {LIGO Scientific,
  Virgo}),\ }\href {\doibase 10.1103/PhysRevX.9.031040} {\bibfield  {journal}
  {\bibinfo  {journal} {Phys. Rev. X}\ }\textbf {\bibinfo {volume} {9}},\
  \bibinfo {pages} {031040} (\bibinfo {year} {2019}{\natexlab{c}})},\ \Eprint
  {http://arxiv.org/abs/1811.12907} {arXiv:1811.12907 [astro-ph.HE]}
  \BibitemShut {NoStop}%
\bibitem [{\citenamefont {Bernus}\ \emph {et~al.}(2020)\citenamefont {Bernus},
  \citenamefont {Minazzoli}, \citenamefont {Fienga}, \citenamefont {Gastineau},
  \citenamefont {Laskar}, \citenamefont {Deram},\ and\ \citenamefont
  {Di~Ruscio}}]{Bernus:2020szc}%
  \BibitemOpen
  \bibfield  {author} {\bibinfo {author} {\bibfnamefont {L.}~\bibnamefont
  {Bernus}}, \bibinfo {author} {\bibfnamefont {O.}~\bibnamefont {Minazzoli}},
  \bibinfo {author} {\bibfnamefont {A.}~\bibnamefont {Fienga}}, \bibinfo
  {author} {\bibfnamefont {M.}~\bibnamefont {Gastineau}}, \bibinfo {author}
  {\bibfnamefont {J.}~\bibnamefont {Laskar}}, \bibinfo {author} {\bibfnamefont
  {P.}~\bibnamefont {Deram}}, \ and\ \bibinfo {author} {\bibfnamefont
  {A.}~\bibnamefont {Di~Ruscio}},\ }\href {\doibase
  10.1103/PhysRevD.102.021501} {\bibfield  {journal} {\bibinfo  {journal}
  {Phys. Rev. D}\ }\textbf {\bibinfo {volume} {102}},\ \bibinfo {pages}
  {021501} (\bibinfo {year} {2020})},\ \Eprint
  {http://arxiv.org/abs/2006.12304} {arXiv:2006.12304 [gr-qc]} \BibitemShut
  {NoStop}%
\bibitem [{\citenamefont {Finn}\ and\ \citenamefont
  {Sutton}(2002)}]{Finn:2001qi}%
  \BibitemOpen
  \bibfield  {author} {\bibinfo {author} {\bibfnamefont {L.~S.}\ \bibnamefont
  {Finn}}\ and\ \bibinfo {author} {\bibfnamefont {P.~J.}\ \bibnamefont
  {Sutton}},\ }\href {\doibase 10.1103/PhysRevD.65.044022} {\bibfield
  {journal} {\bibinfo  {journal} {Phys. Rev. D}\ }\textbf {\bibinfo {volume}
  {65}},\ \bibinfo {pages} {044022} (\bibinfo {year} {2002})},\ \Eprint
  {http://arxiv.org/abs/gr-qc/0109049} {arXiv:gr-qc/0109049} \BibitemShut
  {NoStop}%
\bibitem [{\citenamefont {Miao}\ \emph {et~al.}(2019)\citenamefont {Miao},
  \citenamefont {Shao},\ and\ \citenamefont {Ma}}]{Miao:2019nhf}%
  \BibitemOpen
  \bibfield  {author} {\bibinfo {author} {\bibfnamefont {X.}~\bibnamefont
  {Miao}}, \bibinfo {author} {\bibfnamefont {L.}~\bibnamefont {Shao}}, \ and\
  \bibinfo {author} {\bibfnamefont {B.-Q.}\ \bibnamefont {Ma}},\ }\href
  {\doibase 10.1103/PhysRevD.99.123015} {\bibfield  {journal} {\bibinfo
  {journal} {Phys. Rev. D}\ }\textbf {\bibinfo {volume} {99}},\ \bibinfo
  {pages} {123015} (\bibinfo {year} {2019})},\ \Eprint
  {http://arxiv.org/abs/1905.12836} {arXiv:1905.12836 [astro-ph.CO]}
  \BibitemShut {NoStop}%
\bibitem [{\citenamefont {Agazie}\ \emph
  {et~al.}(2023{\natexlab{a}})\citenamefont {Agazie} \emph
  {et~al.}}]{NANOGrav:2023gor}%
  \BibitemOpen
  \bibfield  {author} {\bibinfo {author} {\bibfnamefont {G.}~\bibnamefont
  {Agazie}} \emph {et~al.} (\bibinfo {collaboration} {NANOGrav}),\ }\href
  {\doibase 10.3847/2041-8213/acdac6} {\bibfield  {journal} {\bibinfo
  {journal} {Astrophys. J. Lett.}\ }\textbf {\bibinfo {volume} {951}},\
  \bibinfo {pages} {L8} (\bibinfo {year} {2023}{\natexlab{a}})},\ \Eprint
  {http://arxiv.org/abs/2306.16213} {arXiv:2306.16213 [astro-ph.HE]}
  \BibitemShut {NoStop}%
\bibitem [{\citenamefont {Xu}\ \emph {et~al.}(2023{\natexlab{b}})\citenamefont
  {Xu} \emph {et~al.}}]{Xu:2023wog}%
  \BibitemOpen
  \bibfield  {author} {\bibinfo {author} {\bibfnamefont {H.}~\bibnamefont {Xu}}
  \emph {et~al.},\ }\href {\doibase 10.1088/1674-4527/acdfa5} {\bibfield
  {journal} {\bibinfo  {journal} {Res. Astron. Astrophys.}\ }\textbf {\bibinfo
  {volume} {23}},\ \bibinfo {pages} {075024} (\bibinfo {year}
  {2023}{\natexlab{b}})},\ \Eprint {http://arxiv.org/abs/2306.16216}
  {arXiv:2306.16216 [astro-ph.HE]} \BibitemShut {NoStop}%
\bibitem [{\citenamefont {Wang}\ and\ \citenamefont
  {Zhao}(2024)}]{Wang:2023div}%
  \BibitemOpen
  \bibfield  {author} {\bibinfo {author} {\bibfnamefont {S.}~\bibnamefont
  {Wang}}\ and\ \bibinfo {author} {\bibfnamefont {Z.-C.}\ \bibnamefont
  {Zhao}},\ }\href {\doibase 10.1103/PhysRevD.109.L061502} {\bibfield
  {journal} {\bibinfo  {journal} {Phys. Rev. D}\ }\textbf {\bibinfo {volume}
  {109}},\ \bibinfo {pages} {L061502} (\bibinfo {year} {2024})},\ \Eprint
  {http://arxiv.org/abs/2307.04680} {arXiv:2307.04680 [astro-ph.HE]}
  \BibitemShut {NoStop}%
\bibitem [{\citenamefont {Cooray}\ and\ \citenamefont
  {Seto}(2004)}]{Cooray:2003cv}%
  \BibitemOpen
  \bibfield  {author} {\bibinfo {author} {\bibfnamefont {A.}~\bibnamefont
  {Cooray}}\ and\ \bibinfo {author} {\bibfnamefont {N.}~\bibnamefont {Seto}},\
  }\href {\doibase 10.1103/PhysRevD.69.103502} {\bibfield  {journal} {\bibinfo
  {journal} {Phys. Rev. D}\ }\textbf {\bibinfo {volume} {69}},\ \bibinfo
  {pages} {103502} (\bibinfo {year} {2004})},\ \Eprint
  {http://arxiv.org/abs/astro-ph/0311054} {arXiv:astro-ph/0311054} \BibitemShut
  {NoStop}%
\bibitem [{\citenamefont {Berti}\ \emph {et~al.}(2011)\citenamefont {Berti},
  \citenamefont {Gair},\ and\ \citenamefont {Sesana}}]{Berti:2011jz}%
  \BibitemOpen
  \bibfield  {author} {\bibinfo {author} {\bibfnamefont {E.}~\bibnamefont
  {Berti}}, \bibinfo {author} {\bibfnamefont {J.}~\bibnamefont {Gair}}, \ and\
  \bibinfo {author} {\bibfnamefont {A.}~\bibnamefont {Sesana}},\ }\href
  {\doibase 10.1103/PhysRevD.84.101501} {\bibfield  {journal} {\bibinfo
  {journal} {Phys. Rev. D}\ }\textbf {\bibinfo {volume} {84}},\ \bibinfo
  {pages} {101501} (\bibinfo {year} {2011})},\ \Eprint
  {http://arxiv.org/abs/1107.3528} {arXiv:1107.3528 [gr-qc]} \BibitemShut
  {NoStop}%
\bibitem [{\citenamefont {Samajdar}\ and\ \citenamefont
  {Arun}(2017)}]{Samajdar:2017mka}%
  \BibitemOpen
  \bibfield  {author} {\bibinfo {author} {\bibfnamefont {A.}~\bibnamefont
  {Samajdar}}\ and\ \bibinfo {author} {\bibfnamefont {K.~G.}\ \bibnamefont
  {Arun}},\ }\href {\doibase 10.1103/PhysRevD.96.104027} {\bibfield  {journal}
  {\bibinfo  {journal} {Phys. Rev. D}\ }\textbf {\bibinfo {volume} {96}},\
  \bibinfo {pages} {104027} (\bibinfo {year} {2017})},\ \Eprint
  {http://arxiv.org/abs/1708.00671} {arXiv:1708.00671 [gr-qc]} \BibitemShut
  {NoStop}%
\bibitem [{\citenamefont {Blanchet}(2002)}]{Blanchet:2002av}%
  \BibitemOpen
  \bibfield  {author} {\bibinfo {author} {\bibfnamefont {L.}~\bibnamefont
  {Blanchet}},\ }\href {\doibase 10.12942/lrr-2002-3} {\bibfield  {journal}
  {\bibinfo  {journal} {Living Rev. Rel.}\ }\textbf {\bibinfo {volume} {5}},\
  \bibinfo {pages} {3} (\bibinfo {year} {2002})},\ \Eprint
  {http://arxiv.org/abs/gr-qc/0202016} {arXiv:gr-qc/0202016} \BibitemShut
  {NoStop}%
\bibitem [{\citenamefont {Scharre}\ and\ \citenamefont
  {Will}(2002)}]{Scharre:2001hn}%
  \BibitemOpen
  \bibfield  {author} {\bibinfo {author} {\bibfnamefont {P.~D.}\ \bibnamefont
  {Scharre}}\ and\ \bibinfo {author} {\bibfnamefont {C.~M.}\ \bibnamefont
  {Will}},\ }\href {\doibase 10.1103/PhysRevD.65.042002} {\bibfield  {journal}
  {\bibinfo  {journal} {Phys. Rev. D}\ }\textbf {\bibinfo {volume} {65}},\
  \bibinfo {pages} {042002} (\bibinfo {year} {2002})},\ \Eprint
  {http://arxiv.org/abs/gr-qc/0109044} {arXiv:gr-qc/0109044} \BibitemShut
  {NoStop}%
\bibitem [{\citenamefont {Berti}\ \emph {et~al.}(2005)\citenamefont {Berti},
  \citenamefont {Buonanno},\ and\ \citenamefont {Will}}]{Berti:2004bd}%
  \BibitemOpen
  \bibfield  {author} {\bibinfo {author} {\bibfnamefont {E.}~\bibnamefont
  {Berti}}, \bibinfo {author} {\bibfnamefont {A.}~\bibnamefont {Buonanno}}, \
  and\ \bibinfo {author} {\bibfnamefont {C.~M.}\ \bibnamefont {Will}},\ }\href
  {\doibase 10.1103/PhysRevD.71.084025} {\bibfield  {journal} {\bibinfo
  {journal} {Phys. Rev. D}\ }\textbf {\bibinfo {volume} {71}},\ \bibinfo
  {pages} {084025} (\bibinfo {year} {2005})},\ \Eprint
  {http://arxiv.org/abs/gr-qc/0411129} {arXiv:gr-qc/0411129} \BibitemShut
  {NoStop}%
\bibitem [{\citenamefont {Arun}\ \emph
  {et~al.}(2006{\natexlab{a}})\citenamefont {Arun}, \citenamefont {Iyer},
  \citenamefont {Qusailah},\ and\ \citenamefont {Sathyaprakash}}]{Arun:2006yw}%
  \BibitemOpen
  \bibfield  {author} {\bibinfo {author} {\bibfnamefont {K.~G.}\ \bibnamefont
  {Arun}}, \bibinfo {author} {\bibfnamefont {B.~R.}\ \bibnamefont {Iyer}},
  \bibinfo {author} {\bibfnamefont {M.~S.~S.}\ \bibnamefont {Qusailah}}, \ and\
  \bibinfo {author} {\bibfnamefont {B.~S.}\ \bibnamefont {Sathyaprakash}},\
  }\href {\doibase 10.1088/0264-9381/23/9/L01} {\bibfield  {journal} {\bibinfo
  {journal} {Class. Quant. Grav.}\ }\textbf {\bibinfo {volume} {23}},\ \bibinfo
  {pages} {L37} (\bibinfo {year} {2006}{\natexlab{a}})},\ \Eprint
  {http://arxiv.org/abs/gr-qc/0604018} {arXiv:gr-qc/0604018} \BibitemShut
  {NoStop}%
\bibitem [{\citenamefont {Arun}\ \emph
  {et~al.}(2006{\natexlab{b}})\citenamefont {Arun}, \citenamefont {Iyer},
  \citenamefont {Qusailah},\ and\ \citenamefont {Sathyaprakash}}]{Arun:2006hn}%
  \BibitemOpen
  \bibfield  {author} {\bibinfo {author} {\bibfnamefont {K.~G.}\ \bibnamefont
  {Arun}}, \bibinfo {author} {\bibfnamefont {B.~R.}\ \bibnamefont {Iyer}},
  \bibinfo {author} {\bibfnamefont {M.~S.~S.}\ \bibnamefont {Qusailah}}, \ and\
  \bibinfo {author} {\bibfnamefont {B.~S.}\ \bibnamefont {Sathyaprakash}},\
  }\href {\doibase 10.1103/PhysRevD.74.024006} {\bibfield  {journal} {\bibinfo
  {journal} {Phys. Rev. D}\ }\textbf {\bibinfo {volume} {74}},\ \bibinfo
  {pages} {024006} (\bibinfo {year} {2006}{\natexlab{b}})},\ \Eprint
  {http://arxiv.org/abs/gr-qc/0604067} {arXiv:gr-qc/0604067} \BibitemShut
  {NoStop}%
\bibitem [{\citenamefont {Huwyler}\ \emph
  {et~al.}(2015{\natexlab{a}})\citenamefont {Huwyler}, \citenamefont {Porter},\
  and\ \citenamefont {Jetzer}}]{Huwyler:2014gaa}%
  \BibitemOpen
  \bibfield  {author} {\bibinfo {author} {\bibfnamefont {C.}~\bibnamefont
  {Huwyler}}, \bibinfo {author} {\bibfnamefont {E.~K.}\ \bibnamefont {Porter}},
  \ and\ \bibinfo {author} {\bibfnamefont {P.}~\bibnamefont {Jetzer}},\ }\href
  {\doibase 10.1088/1742-6596/610/1/012046} {\bibfield  {journal} {\bibinfo
  {journal} {J. Phys. Conf. Ser.}\ }\textbf {\bibinfo {volume} {610}},\
  \bibinfo {pages} {012046} (\bibinfo {year} {2015}{\natexlab{a}})},\ \Eprint
  {http://arxiv.org/abs/1410.6687} {arXiv:1410.6687 [gr-qc]} \BibitemShut
  {NoStop}%
\bibitem [{\citenamefont {Loutrel}\ \emph {et~al.}(2014)\citenamefont
  {Loutrel}, \citenamefont {Yunes},\ and\ \citenamefont
  {Pretorius}}]{Loutrel:2014vja}%
  \BibitemOpen
  \bibfield  {author} {\bibinfo {author} {\bibfnamefont {N.}~\bibnamefont
  {Loutrel}}, \bibinfo {author} {\bibfnamefont {N.}~\bibnamefont {Yunes}}, \
  and\ \bibinfo {author} {\bibfnamefont {F.}~\bibnamefont {Pretorius}},\ }\href
  {\doibase 10.1103/PhysRevD.90.104010} {\bibfield  {journal} {\bibinfo
  {journal} {Phys. Rev. D}\ }\textbf {\bibinfo {volume} {90}},\ \bibinfo
  {pages} {104010} (\bibinfo {year} {2014})},\ \Eprint
  {http://arxiv.org/abs/1404.0092} {arXiv:1404.0092 [gr-qc]} \BibitemShut
  {NoStop}%
\bibitem [{\citenamefont {Cardoso}\ and\ \citenamefont
  {Maselli}(2020)}]{Cardoso:2019rou}%
  \BibitemOpen
  \bibfield  {author} {\bibinfo {author} {\bibfnamefont {V.}~\bibnamefont
  {Cardoso}}\ and\ \bibinfo {author} {\bibfnamefont {A.}~\bibnamefont
  {Maselli}},\ }\href {\doibase 10.1051/0004-6361/202037654} {\bibfield
  {journal} {\bibinfo  {journal} {Astron. Astrophys.}\ }\textbf {\bibinfo
  {volume} {644}},\ \bibinfo {pages} {A147} (\bibinfo {year} {2020})},\ \Eprint
  {http://arxiv.org/abs/1909.05870} {arXiv:1909.05870 [astro-ph.HE]}
  \BibitemShut {NoStop}%
\bibitem [{\citenamefont {Hansen}\ \emph
  {et~al.}(2015{\natexlab{b}})\citenamefont {Hansen}, \citenamefont {Yunes},\
  and\ \citenamefont {Yagi}}]{PhysRevD.91.082003}%
  \BibitemOpen
  \bibfield  {author} {\bibinfo {author} {\bibfnamefont {D.}~\bibnamefont
  {Hansen}}, \bibinfo {author} {\bibfnamefont {N.}~\bibnamefont {Yunes}}, \
  and\ \bibinfo {author} {\bibfnamefont {K.}~\bibnamefont {Yagi}},\ }\href
  {\doibase 10.1103/PhysRevD.91.082003} {\bibfield  {journal} {\bibinfo
  {journal} {Phys. Rev. D}\ }\textbf {\bibinfo {volume} {91}},\ \bibinfo
  {pages} {082003} (\bibinfo {year} {2015}{\natexlab{b}})}\BibitemShut
  {NoStop}%
\bibitem [{\citenamefont {Kobakhidze}\ \emph
  {et~al.}(2016{\natexlab{b}})\citenamefont {Kobakhidze}, \citenamefont
  {Lagger},\ and\ \citenamefont {Manning}}]{PhysRevD.94.064033}%
  \BibitemOpen
  \bibfield  {author} {\bibinfo {author} {\bibfnamefont {A.}~\bibnamefont
  {Kobakhidze}}, \bibinfo {author} {\bibfnamefont {C.}~\bibnamefont {Lagger}},
  \ and\ \bibinfo {author} {\bibfnamefont {A.}~\bibnamefont {Manning}},\ }\href
  {\doibase 10.1103/PhysRevD.94.064033} {\bibfield  {journal} {\bibinfo
  {journal} {Phys. Rev. D}\ }\textbf {\bibinfo {volume} {94}},\ \bibinfo
  {pages} {064033} (\bibinfo {year} {2016}{\natexlab{b}})}\BibitemShut
  {NoStop}%
\bibitem [{\citenamefont {Chamberlain}\ and\ \citenamefont
  {Yunes}(2017)}]{Chamberlain:2017fjl}%
  \BibitemOpen
  \bibfield  {author} {\bibinfo {author} {\bibfnamefont {K.}~\bibnamefont
  {Chamberlain}}\ and\ \bibinfo {author} {\bibfnamefont {N.}~\bibnamefont
  {Yunes}},\ }\href {\doibase 10.1103/PhysRevD.96.084039} {\bibfield  {journal}
  {\bibinfo  {journal} {Phys. Rev.}\ }\textbf {\bibinfo {volume} {D96}},\
  \bibinfo {pages} {084039} (\bibinfo {year} {2017})},\ \Eprint
  {http://arxiv.org/abs/1704.08268} {arXiv:1704.08268 [gr-qc]} \BibitemShut
  {NoStop}%
\bibitem [{\citenamefont {Cornish}\ \emph {et~al.}(2011)\citenamefont
  {Cornish}, \citenamefont {Sampson}, \citenamefont {Yunes},\ and\
  \citenamefont {Pretorius}}]{Cornish:2011ys}%
  \BibitemOpen
  \bibfield  {author} {\bibinfo {author} {\bibfnamefont {N.}~\bibnamefont
  {Cornish}}, \bibinfo {author} {\bibfnamefont {L.}~\bibnamefont {Sampson}},
  \bibinfo {author} {\bibfnamefont {N.}~\bibnamefont {Yunes}}, \ and\ \bibinfo
  {author} {\bibfnamefont {F.}~\bibnamefont {Pretorius}},\ }\href {\doibase
  10.1103/PhysRevD.84.062003} {\bibfield  {journal} {\bibinfo  {journal} {Phys.
  Rev. D}\ }\textbf {\bibinfo {volume} {84}},\ \bibinfo {pages} {062003}
  (\bibinfo {year} {2011})},\ \Eprint {http://arxiv.org/abs/1105.2088}
  {arXiv:1105.2088 [gr-qc]} \BibitemShut {NoStop}%
\bibitem [{\citenamefont {Huwyler}\ \emph
  {et~al.}(2015{\natexlab{b}})\citenamefont {Huwyler}, \citenamefont {Porter},\
  and\ \citenamefont {Jetzer}}]{Huwyler:2014vva}%
  \BibitemOpen
  \bibfield  {author} {\bibinfo {author} {\bibfnamefont {C.}~\bibnamefont
  {Huwyler}}, \bibinfo {author} {\bibfnamefont {E.~K.}\ \bibnamefont {Porter}},
  \ and\ \bibinfo {author} {\bibfnamefont {P.}~\bibnamefont {Jetzer}},\ }\href
  {\doibase 10.1103/PhysRevD.91.024037} {\bibfield  {journal} {\bibinfo
  {journal} {Phys. Rev. D}\ }\textbf {\bibinfo {volume} {91}},\ \bibinfo
  {pages} {024037} (\bibinfo {year} {2015}{\natexlab{b}})},\ \Eprint
  {http://arxiv.org/abs/1410.8815} {arXiv:1410.8815 [gr-qc]} \BibitemShut
  {NoStop}%
\bibitem [{\citenamefont {Yunes}\ \emph {et~al.}(2010)\citenamefont {Yunes},
  \citenamefont {Pretorius},\ and\ \citenamefont {Spergel}}]{Yunes:2009bv}%
  \BibitemOpen
  \bibfield  {author} {\bibinfo {author} {\bibfnamefont {N.}~\bibnamefont
  {Yunes}}, \bibinfo {author} {\bibfnamefont {F.}~\bibnamefont {Pretorius}}, \
  and\ \bibinfo {author} {\bibfnamefont {D.}~\bibnamefont {Spergel}},\ }\href
  {\doibase 10.1103/PhysRevD.81.064018} {\bibfield  {journal} {\bibinfo
  {journal} {Phys. Rev. D}\ }\textbf {\bibinfo {volume} {81}},\ \bibinfo
  {pages} {064018} (\bibinfo {year} {2010})},\ \Eprint
  {http://arxiv.org/abs/0912.2724} {arXiv:0912.2724 [gr-qc]} \BibitemShut
  {NoStop}%
\bibitem [{\citenamefont {Keppel}\ and\ \citenamefont
  {Ajith}(2010)}]{Keppel:2010qu}%
  \BibitemOpen
  \bibfield  {author} {\bibinfo {author} {\bibfnamefont {D.}~\bibnamefont
  {Keppel}}\ and\ \bibinfo {author} {\bibfnamefont {P.}~\bibnamefont {Ajith}},\
  }\href {\doibase 10.1103/PhysRevD.82.122001} {\bibfield  {journal} {\bibinfo
  {journal} {Phys. Rev. D}\ }\textbf {\bibinfo {volume} {82}},\ \bibinfo
  {pages} {122001} (\bibinfo {year} {2010})},\ \Eprint
  {http://arxiv.org/abs/1004.0284} {arXiv:1004.0284 [gr-qc]} \BibitemShut
  {NoStop}%
\bibitem [{\citenamefont {Arun}(2012)}]{Arun:2012hf}%
  \BibitemOpen
  \bibfield  {author} {\bibinfo {author} {\bibfnamefont {K.~G.}\ \bibnamefont
  {Arun}},\ }\href {\doibase 10.1088/0264-9381/29/7/075011} {\bibfield
  {journal} {\bibinfo  {journal} {Class. Quant. Grav.}\ }\textbf {\bibinfo
  {volume} {29}},\ \bibinfo {pages} {075011} (\bibinfo {year} {2012})},\
  \Eprint {http://arxiv.org/abs/1202.5911} {arXiv:1202.5911 [gr-qc]}
  \BibitemShut {NoStop}%
\bibitem [{\citenamefont {Zhao}\ \emph
  {et~al.}(2021{\natexlab{b}})\citenamefont {Zhao}, \citenamefont {Shao},
  \citenamefont {Gao}, \citenamefont {Liu}, \citenamefont {Cao},\ and\
  \citenamefont {Ma}}]{Zhao:2021bjw}%
  \BibitemOpen
  \bibfield  {author} {\bibinfo {author} {\bibfnamefont {J.}~\bibnamefont
  {Zhao}}, \bibinfo {author} {\bibfnamefont {L.}~\bibnamefont {Shao}}, \bibinfo
  {author} {\bibfnamefont {Y.}~\bibnamefont {Gao}}, \bibinfo {author}
  {\bibfnamefont {C.}~\bibnamefont {Liu}}, \bibinfo {author} {\bibfnamefont
  {Z.}~\bibnamefont {Cao}}, \ and\ \bibinfo {author} {\bibfnamefont {B.-Q.}\
  \bibnamefont {Ma}},\ }\href {\doibase 10.1103/PhysRevD.104.084008} {\bibfield
   {journal} {\bibinfo  {journal} {Phys. Rev. D}\ }\textbf {\bibinfo {volume}
  {104}},\ \bibinfo {pages} {084008} (\bibinfo {year} {2021}{\natexlab{b}})},\
  \Eprint {http://arxiv.org/abs/2106.04883} {arXiv:2106.04883 [gr-qc]}
  \BibitemShut {NoStop}%
\bibitem [{\citenamefont {Wang}\ \emph
  {et~al.}(2022{\natexlab{a}})\citenamefont {Wang}, \citenamefont {Zhao},
  \citenamefont {An}, \citenamefont {Shao},\ and\ \citenamefont
  {Cao}}]{Wang:2022yxb}%
  \BibitemOpen
  \bibfield  {author} {\bibinfo {author} {\bibfnamefont {Z.}~\bibnamefont
  {Wang}}, \bibinfo {author} {\bibfnamefont {J.}~\bibnamefont {Zhao}}, \bibinfo
  {author} {\bibfnamefont {Z.}~\bibnamefont {An}}, \bibinfo {author}
  {\bibfnamefont {L.}~\bibnamefont {Shao}}, \ and\ \bibinfo {author}
  {\bibfnamefont {Z.}~\bibnamefont {Cao}},\ }\href {\doibase
  10.1016/j.physletb.2022.137416} {\bibfield  {journal} {\bibinfo  {journal}
  {Phys. Lett. B}\ }\textbf {\bibinfo {volume} {834}},\ \bibinfo {pages}
  {137416} (\bibinfo {year} {2022}{\natexlab{a}})},\ \Eprint
  {http://arxiv.org/abs/2208.11913} {arXiv:2208.11913 [gr-qc]} \BibitemShut
  {NoStop}%
\bibitem [{\citenamefont {Sesana}(2016)}]{Sesana:2016ljz}%
  \BibitemOpen
  \bibfield  {author} {\bibinfo {author} {\bibfnamefont {A.}~\bibnamefont
  {Sesana}},\ }\href {\doibase 10.1103/PhysRevLett.116.231102} {\bibfield
  {journal} {\bibinfo  {journal} {Phys. Rev. Lett.}\ }\textbf {\bibinfo
  {volume} {116}},\ \bibinfo {pages} {231102} (\bibinfo {year} {2016})},\
  \Eprint {http://arxiv.org/abs/1602.06951} {arXiv:1602.06951 [gr-qc]}
  \BibitemShut {NoStop}%
\bibitem [{\citenamefont {Barausse}\ \emph {et~al.}(2016)\citenamefont
  {Barausse}, \citenamefont {Yunes},\ and\ \citenamefont
  {Chamberlain}}]{Barausse:2016eii}%
  \BibitemOpen
  \bibfield  {author} {\bibinfo {author} {\bibfnamefont {E.}~\bibnamefont
  {Barausse}}, \bibinfo {author} {\bibfnamefont {N.}~\bibnamefont {Yunes}}, \
  and\ \bibinfo {author} {\bibfnamefont {K.}~\bibnamefont {Chamberlain}},\
  }\href {\doibase 10.1103/PhysRevLett.116.241104} {\bibfield  {journal}
  {\bibinfo  {journal} {Phys. Rev. Lett.}\ }\textbf {\bibinfo {volume} {116}},\
  \bibinfo {pages} {241104} (\bibinfo {year} {2016})},\ \Eprint
  {http://arxiv.org/abs/1603.04075} {arXiv:1603.04075 [gr-qc]} \BibitemShut
  {NoStop}%
\bibitem [{\citenamefont {Carson}\ and\ \citenamefont
  {Yagi}(2020{\natexlab{a}})}]{Carson:2019rda}%
  \BibitemOpen
  \bibfield  {author} {\bibinfo {author} {\bibfnamefont {Z.}~\bibnamefont
  {Carson}}\ and\ \bibinfo {author} {\bibfnamefont {K.}~\bibnamefont {Yagi}},\
  }\href {\doibase 10.1088/1361-6382/ab5c9a} {\bibfield  {journal} {\bibinfo
  {journal} {Class. Quant. Grav.}\ }\textbf {\bibinfo {volume} {37}},\ \bibinfo
  {pages} {02LT01} (\bibinfo {year} {2020}{\natexlab{a}})},\ \Eprint
  {http://arxiv.org/abs/1905.13155} {arXiv:1905.13155 [gr-qc]} \BibitemShut
  {NoStop}%
\bibitem [{\citenamefont {Liu}\ \emph {et~al.}(2020{\natexlab{b}})\citenamefont
  {Liu}, \citenamefont {Shao}, \citenamefont {Zhao},\ and\ \citenamefont
  {Gao}}]{Liu:2020nwz}%
  \BibitemOpen
  \bibfield  {author} {\bibinfo {author} {\bibfnamefont {C.}~\bibnamefont
  {Liu}}, \bibinfo {author} {\bibfnamefont {L.}~\bibnamefont {Shao}}, \bibinfo
  {author} {\bibfnamefont {J.}~\bibnamefont {Zhao}}, \ and\ \bibinfo {author}
  {\bibfnamefont {Y.}~\bibnamefont {Gao}},\ }\href {\doibase
  10.1093/mnras/staa1512} {\bibfield  {journal} {\bibinfo  {journal} {Mon. Not.
  Roy. Astron. Soc.}\ }\textbf {\bibinfo {volume} {496}},\ \bibinfo {pages}
  {182} (\bibinfo {year} {2020}{\natexlab{b}})},\ \Eprint
  {http://arxiv.org/abs/2004.12096} {arXiv:2004.12096 [astro-ph.HE]}
  \BibitemShut {NoStop}%
\bibitem [{\citenamefont {Carson}\ and\ \citenamefont
  {Yagi}(2020{\natexlab{b}})}]{Carson:2020cqb}%
  \BibitemOpen
  \bibfield  {author} {\bibinfo {author} {\bibfnamefont {Z.}~\bibnamefont
  {Carson}}\ and\ \bibinfo {author} {\bibfnamefont {K.}~\bibnamefont {Yagi}},\
  }\href {\doibase 10.1088/1361-6382/aba221} {\bibfield  {journal} {\bibinfo
  {journal} {Class. Quant. Grav.}\ }\textbf {\bibinfo {volume} {37}},\ \bibinfo
  {pages} {215007} (\bibinfo {year} {2020}{\natexlab{b}})},\ \Eprint
  {http://arxiv.org/abs/2002.08559} {arXiv:2002.08559 [gr-qc]} \BibitemShut
  {NoStop}%
\bibitem [{\citenamefont {Shi}\ \emph {et~al.}(2023)\citenamefont {Shi},
  \citenamefont {Ji}, \citenamefont {Zhang},\ and\ \citenamefont
  {Mei}}]{Shi:2022qno}%
  \BibitemOpen
  \bibfield  {author} {\bibinfo {author} {\bibfnamefont {C.}~\bibnamefont
  {Shi}}, \bibinfo {author} {\bibfnamefont {M.}~\bibnamefont {Ji}}, \bibinfo
  {author} {\bibfnamefont {J.-d.}\ \bibnamefont {Zhang}}, \ and\ \bibinfo
  {author} {\bibfnamefont {J.}~\bibnamefont {Mei}},\ }\href {\doibase
  10.1103/PhysRevD.108.024030} {\bibfield  {journal} {\bibinfo  {journal}
  {Phys. Rev. D}\ }\textbf {\bibinfo {volume} {108}},\ \bibinfo {pages}
  {024030} (\bibinfo {year} {2023})},\ \Eprint
  {http://arxiv.org/abs/2210.13006} {arXiv:2210.13006 [gr-qc]} \BibitemShut
  {NoStop}%
\bibitem [{\citenamefont {Wang}\ \emph
  {et~al.}(2023{\natexlab{a}})\citenamefont {Wang}, \citenamefont {Shi},
  \citenamefont {Zhang}, \citenamefont {hu},\ and\ \citenamefont
  {Mei}}]{Wang:2023wgv}%
  \BibitemOpen
  \bibfield  {author} {\bibinfo {author} {\bibfnamefont {B.}~\bibnamefont
  {Wang}}, \bibinfo {author} {\bibfnamefont {C.}~\bibnamefont {Shi}}, \bibinfo
  {author} {\bibfnamefont {J.-d.}\ \bibnamefont {Zhang}}, \bibinfo {author}
  {\bibfnamefont {Y.-M.}\ \bibnamefont {hu}}, \ and\ \bibinfo {author}
  {\bibfnamefont {J.}~\bibnamefont {Mei}},\ }\href {\doibase
  10.1103/PhysRevD.108.044061} {\bibfield  {journal} {\bibinfo  {journal}
  {Phys. Rev. D}\ }\textbf {\bibinfo {volume} {108}},\ \bibinfo {pages}
  {044061} (\bibinfo {year} {2023}{\natexlab{a}})},\ \Eprint
  {http://arxiv.org/abs/2302.10112} {arXiv:2302.10112 [gr-qc]} \BibitemShut
  {NoStop}%
\bibitem [{\citenamefont {Lyu}\ \emph {et~al.}(2022)\citenamefont {Lyu},
  \citenamefont {Jiang},\ and\ \citenamefont {Yagi}}]{Lyu:2022gdr}%
  \BibitemOpen
  \bibfield  {author} {\bibinfo {author} {\bibfnamefont {Z.}~\bibnamefont
  {Lyu}}, \bibinfo {author} {\bibfnamefont {N.}~\bibnamefont {Jiang}}, \ and\
  \bibinfo {author} {\bibfnamefont {K.}~\bibnamefont {Yagi}},\ }\href {\doibase
  10.1103/PhysRevD.105.064001} {\bibfield  {journal} {\bibinfo  {journal}
  {Phys. Rev. D}\ }\textbf {\bibinfo {volume} {105}},\ \bibinfo {pages}
  {064001} (\bibinfo {year} {2022})},\ \Eprint
  {http://arxiv.org/abs/2201.02543} {arXiv:2201.02543 [gr-qc]} \BibitemShut
  {NoStop}%
\bibitem [{\citenamefont {Perkins}\ \emph {et~al.}(2021)\citenamefont
  {Perkins}, \citenamefont {Nair}, \citenamefont {Silva},\ and\ \citenamefont
  {Yunes}}]{Perkins:2021mhb}%
  \BibitemOpen
  \bibfield  {author} {\bibinfo {author} {\bibfnamefont {S.~E.}\ \bibnamefont
  {Perkins}}, \bibinfo {author} {\bibfnamefont {R.}~\bibnamefont {Nair}},
  \bibinfo {author} {\bibfnamefont {H.~O.}\ \bibnamefont {Silva}}, \ and\
  \bibinfo {author} {\bibfnamefont {N.}~\bibnamefont {Yunes}},\ }\href
  {\doibase 10.1103/PhysRevD.104.024060} {\bibfield  {journal} {\bibinfo
  {journal} {Phys. Rev. D}\ }\textbf {\bibinfo {volume} {104}},\ \bibinfo
  {pages} {024060} (\bibinfo {year} {2021})},\ \Eprint
  {http://arxiv.org/abs/2104.11189} {arXiv:2104.11189 [gr-qc]} \BibitemShut
  {NoStop}%
\bibitem [{\citenamefont {Nair}\ \emph {et~al.}(2019)\citenamefont {Nair},
  \citenamefont {Perkins}, \citenamefont {Silva},\ and\ \citenamefont
  {Yunes}}]{Nair:2019iur}%
  \BibitemOpen
  \bibfield  {author} {\bibinfo {author} {\bibfnamefont {R.}~\bibnamefont
  {Nair}}, \bibinfo {author} {\bibfnamefont {S.}~\bibnamefont {Perkins}},
  \bibinfo {author} {\bibfnamefont {H.~O.}\ \bibnamefont {Silva}}, \ and\
  \bibinfo {author} {\bibfnamefont {N.}~\bibnamefont {Yunes}},\ }\href
  {\doibase 10.1103/PhysRevLett.123.191101} {\bibfield  {journal} {\bibinfo
  {journal} {Phys. Rev. Lett.}\ }\textbf {\bibinfo {volume} {123}},\ \bibinfo
  {pages} {191101} (\bibinfo {year} {2019})},\ \Eprint
  {http://arxiv.org/abs/1905.00870} {arXiv:1905.00870 [gr-qc]} \BibitemShut
  {NoStop}%
\bibitem [{\citenamefont {Jackiw}\ and\ \citenamefont
  {Pi}(2003)}]{Jackiw:2003pm}%
  \BibitemOpen
  \bibfield  {author} {\bibinfo {author} {\bibfnamefont {R.}~\bibnamefont
  {Jackiw}}\ and\ \bibinfo {author} {\bibfnamefont {S.}~\bibnamefont {Pi}},\
  }\href {\doibase 10.1103/PhysRevD.68.104012} {\bibfield  {journal} {\bibinfo
  {journal} {Phys. Rev. D}\ }\textbf {\bibinfo {volume} {68}},\ \bibinfo
  {pages} {104012} (\bibinfo {year} {2003})},\ \Eprint
  {http://arxiv.org/abs/gr-qc/0308071} {arXiv:gr-qc/0308071} \BibitemShut
  {NoStop}%
\bibitem [{\citenamefont {Silva}\ \emph {et~al.}(2021)\citenamefont {Silva},
  \citenamefont {Holgado}, \citenamefont {C\'ardenas-Avenda\~no},\ and\
  \citenamefont {Yunes}}]{Silva:2020acr}%
  \BibitemOpen
  \bibfield  {author} {\bibinfo {author} {\bibfnamefont {H.~O.}\ \bibnamefont
  {Silva}}, \bibinfo {author} {\bibfnamefont {A.~M.}\ \bibnamefont {Holgado}},
  \bibinfo {author} {\bibfnamefont {A.}~\bibnamefont {C\'ardenas-Avenda\~no}},
  \ and\ \bibinfo {author} {\bibfnamefont {N.}~\bibnamefont {Yunes}},\ }\href
  {\doibase 10.1103/PhysRevLett.126.181101} {\bibfield  {journal} {\bibinfo
  {journal} {Phys. Rev. Lett.}\ }\textbf {\bibinfo {volume} {126}},\ \bibinfo
  {pages} {181101} (\bibinfo {year} {2021})},\ \Eprint
  {http://arxiv.org/abs/2004.01253} {arXiv:2004.01253 [gr-qc]} \BibitemShut
  {NoStop}%
\bibitem [{\citenamefont {Snyder}(1947)}]{Snyder:1946qz}%
  \BibitemOpen
  \bibfield  {author} {\bibinfo {author} {\bibfnamefont {H.~S.}\ \bibnamefont
  {Snyder}},\ }\href {\doibase 10.1103/PhysRev.71.38} {\bibfield  {journal}
  {\bibinfo  {journal} {Phys. Rev.}\ }\textbf {\bibinfo {volume} {71}},\
  \bibinfo {pages} {38} (\bibinfo {year} {1947})}\BibitemShut {NoStop}%
\bibitem [{\citenamefont {Connes}(1985)}]{PMIHES_1985__62__41_0}%
  \BibitemOpen
  \bibfield  {author} {\bibinfo {author} {\bibfnamefont {A.}~\bibnamefont
  {Connes}},\ }\href {http://www.numdam.org/item/PMIHES_1985__62__41_0/}
  {\bibfield  {journal} {\bibinfo  {journal} {Publications Math\'ematiques de
  l'IH\'ES}\ }\textbf {\bibinfo {volume} {62}},\ \bibinfo {pages} {41}
  (\bibinfo {year} {1985})}\BibitemShut {NoStop}%
\bibitem [{\citenamefont {Chamseddine}\ \emph {et~al.}(1993)\citenamefont
  {Chamseddine}, \citenamefont {Felder},\ and\ \citenamefont
  {Frohlich}}]{Chamseddine:1992yx}%
  \BibitemOpen
  \bibfield  {author} {\bibinfo {author} {\bibfnamefont {A.~H.}\ \bibnamefont
  {Chamseddine}}, \bibinfo {author} {\bibfnamefont {G.}~\bibnamefont {Felder}},
  \ and\ \bibinfo {author} {\bibfnamefont {J.}~\bibnamefont {Frohlich}},\
  }\href {\doibase 10.1007/BF02100059} {\bibfield  {journal} {\bibinfo
  {journal} {Commun. Math. Phys.}\ }\textbf {\bibinfo {volume} {155}},\
  \bibinfo {pages} {205} (\bibinfo {year} {1993})},\ \Eprint
  {http://arxiv.org/abs/hep-th/9209044} {arXiv:hep-th/9209044} \BibitemShut
  {NoStop}%
\bibitem [{\citenamefont {Landi}(1997)}]{Landi:1997sh}%
  \BibitemOpen
  \bibfield  {author} {\bibinfo {author} {\bibfnamefont {G.}~\bibnamefont
  {Landi}},\ }\href {\doibase 10.1007/3-540-14949-X} {\emph {\bibinfo {title}
  {{An Introduction to noncommutative spaces and their geometry}}}},\
  Vol.~\bibinfo {volume} {51}\ (\bibinfo {year} {1997})\ \Eprint
  {http://arxiv.org/abs/hep-th/9701078} {arXiv:hep-th/9701078} \BibitemShut
  {NoStop}%
\bibitem [{\citenamefont {Tahura}\ \emph {et~al.}(2019)\citenamefont {Tahura},
  \citenamefont {Yagi},\ and\ \citenamefont {Carson}}]{Tahura:2019dgr}%
  \BibitemOpen
  \bibfield  {author} {\bibinfo {author} {\bibfnamefont {S.}~\bibnamefont
  {Tahura}}, \bibinfo {author} {\bibfnamefont {K.}~\bibnamefont {Yagi}}, \ and\
  \bibinfo {author} {\bibfnamefont {Z.}~\bibnamefont {Carson}},\ }\href
  {\doibase 10.1103/PhysRevD.100.104001} {\bibfield  {journal} {\bibinfo
  {journal} {Phys. Rev. D}\ }\textbf {\bibinfo {volume} {100}},\ \bibinfo
  {pages} {104001} (\bibinfo {year} {2019})},\ \Eprint
  {http://arxiv.org/abs/1907.10059} {arXiv:1907.10059 [gr-qc]} \BibitemShut
  {NoStop}%
\bibitem [{\citenamefont {{Hofmann, F.}}\ \emph {et~al.}(2010)\citenamefont
  {{Hofmann, F.}}, \citenamefont {{Müller, J.}},\ and\ \citenamefont
  {{Biskupek, L.}}}]{Hofmann:2010aa}%
  \BibitemOpen
  \bibfield  {author} {\bibinfo {author} {\bibnamefont {{Hofmann, F.}}},
  \bibinfo {author} {\bibnamefont {{Müller, J.}}}, \ and\ \bibinfo {author}
  {\bibnamefont {{Biskupek, L.}}},\ }\href {\doibase
  10.1051/0004-6361/201015659} {\bibfield  {journal} {\bibinfo  {journal}
  {A\&A}\ }\textbf {\bibinfo {volume} {522}},\ \bibinfo {pages} {L5} (\bibinfo
  {year} {2010})}\BibitemShut {NoStop}%
\bibitem [{\citenamefont {Barausse}\ \emph {et~al.}(2015)\citenamefont
  {Barausse}, \citenamefont {Cardoso},\ and\ \citenamefont
  {Pani}}]{Barausse:2014pra}%
  \BibitemOpen
  \bibfield  {author} {\bibinfo {author} {\bibfnamefont {E.}~\bibnamefont
  {Barausse}}, \bibinfo {author} {\bibfnamefont {V.}~\bibnamefont {Cardoso}}, \
  and\ \bibinfo {author} {\bibfnamefont {P.}~\bibnamefont {Pani}},\ }\href
  {\doibase 10.1088/1742-6596/610/1/012044} {\bibfield  {journal} {\bibinfo
  {journal} {J. Phys. Conf. Ser.}\ }\textbf {\bibinfo {volume} {610}},\
  \bibinfo {pages} {012044} (\bibinfo {year} {2015})},\ \Eprint
  {http://arxiv.org/abs/1404.7140} {arXiv:1404.7140 [astro-ph.CO]} \BibitemShut
  {NoStop}%
\bibitem [{\citenamefont {Barausse}\ \emph {et~al.}(2014)\citenamefont
  {Barausse}, \citenamefont {Cardoso},\ and\ \citenamefont
  {Pani}}]{Barausse:2014tra}%
  \BibitemOpen
  \bibfield  {author} {\bibinfo {author} {\bibfnamefont {E.}~\bibnamefont
  {Barausse}}, \bibinfo {author} {\bibfnamefont {V.}~\bibnamefont {Cardoso}}, \
  and\ \bibinfo {author} {\bibfnamefont {P.}~\bibnamefont {Pani}},\ }\href
  {\doibase 10.1103/PhysRevD.89.104059} {\bibfield  {journal} {\bibinfo
  {journal} {Phys. Rev. D}\ }\textbf {\bibinfo {volume} {89}},\ \bibinfo
  {pages} {104059} (\bibinfo {year} {2014})},\ \Eprint
  {http://arxiv.org/abs/1404.7149} {arXiv:1404.7149 [gr-qc]} \BibitemShut
  {NoStop}%
\bibitem [{\citenamefont {Macedo}\ \emph {et~al.}(2013)\citenamefont {Macedo},
  \citenamefont {Pani}, \citenamefont {Cardoso},\ and\ \citenamefont
  {Crispino}}]{Macedo:2013qea}%
  \BibitemOpen
  \bibfield  {author} {\bibinfo {author} {\bibfnamefont {C.~F.~B.}\
  \bibnamefont {Macedo}}, \bibinfo {author} {\bibfnamefont {P.}~\bibnamefont
  {Pani}}, \bibinfo {author} {\bibfnamefont {V.}~\bibnamefont {Cardoso}}, \
  and\ \bibinfo {author} {\bibfnamefont {L.~C.~B.}\ \bibnamefont {Crispino}},\
  }\href {\doibase 10.1088/0004-637X/774/1/48} {\bibfield  {journal} {\bibinfo
  {journal} {Astrophys. J.}\ }\textbf {\bibinfo {volume} {774}},\ \bibinfo
  {pages} {48} (\bibinfo {year} {2013})},\ \Eprint
  {http://arxiv.org/abs/1302.2646} {arXiv:1302.2646 [gr-qc]} \BibitemShut
  {NoStop}%
\bibitem [{\citenamefont {Barausse}\ \emph {et~al.}(2007)\citenamefont
  {Barausse}, \citenamefont {Rezzolla}, \citenamefont {Petroff},\ and\
  \citenamefont {Ansorg}}]{Barausse:2006vt}%
  \BibitemOpen
  \bibfield  {author} {\bibinfo {author} {\bibfnamefont {E.}~\bibnamefont
  {Barausse}}, \bibinfo {author} {\bibfnamefont {L.}~\bibnamefont {Rezzolla}},
  \bibinfo {author} {\bibfnamefont {D.}~\bibnamefont {Petroff}}, \ and\
  \bibinfo {author} {\bibfnamefont {M.}~\bibnamefont {Ansorg}},\ }\href
  {\doibase 10.1103/PhysRevD.75.064026} {\bibfield  {journal} {\bibinfo
  {journal} {Phys. Rev. D}\ }\textbf {\bibinfo {volume} {75}},\ \bibinfo
  {pages} {064026} (\bibinfo {year} {2007})},\ \Eprint
  {http://arxiv.org/abs/gr-qc/0612123} {arXiv:gr-qc/0612123} \BibitemShut
  {NoStop}%
\bibitem [{\citenamefont {Barausse}\ and\ \citenamefont
  {Rezzolla}(2008)}]{Barausse:2007dy}%
  \BibitemOpen
  \bibfield  {author} {\bibinfo {author} {\bibfnamefont {E.}~\bibnamefont
  {Barausse}}\ and\ \bibinfo {author} {\bibfnamefont {L.}~\bibnamefont
  {Rezzolla}},\ }\href {\doibase 10.1103/PhysRevD.77.104027} {\bibfield
  {journal} {\bibinfo  {journal} {Phys. Rev. D}\ }\textbf {\bibinfo {volume}
  {77}},\ \bibinfo {pages} {104027} (\bibinfo {year} {2008})},\ \Eprint
  {http://arxiv.org/abs/0711.4558} {arXiv:0711.4558 [gr-qc]} \BibitemShut
  {NoStop}%
\bibitem [{\citenamefont {Barausse}(2007)}]{Barausse:2007ph}%
  \BibitemOpen
  \bibfield  {author} {\bibinfo {author} {\bibfnamefont {E.}~\bibnamefont
  {Barausse}},\ }\href {\doibase 10.1111/j.1365-2966.2007.12408.x} {\bibfield
  {journal} {\bibinfo  {journal} {Mon. Not. Roy. Astron. Soc.}\ }\textbf
  {\bibinfo {volume} {382}},\ \bibinfo {pages} {826} (\bibinfo {year}
  {2007})},\ \Eprint {http://arxiv.org/abs/0709.0211} {arXiv:0709.0211
  [astro-ph]} \BibitemShut {NoStop}%
\bibitem [{\citenamefont {Yunes}\ \emph {et~al.}(2011)\citenamefont {Yunes},
  \citenamefont {Kocsis}, \citenamefont {Loeb},\ and\ \citenamefont
  {Haiman}}]{Yunes:2011ws}%
  \BibitemOpen
  \bibfield  {author} {\bibinfo {author} {\bibfnamefont {N.}~\bibnamefont
  {Yunes}}, \bibinfo {author} {\bibfnamefont {B.}~\bibnamefont {Kocsis}},
  \bibinfo {author} {\bibfnamefont {A.}~\bibnamefont {Loeb}}, \ and\ \bibinfo
  {author} {\bibfnamefont {Z.}~\bibnamefont {Haiman}},\ }\href {\doibase
  10.1103/PhysRevLett.107.171103} {\bibfield  {journal} {\bibinfo  {journal}
  {Phys. Rev. Lett.}\ }\textbf {\bibinfo {volume} {107}},\ \bibinfo {pages}
  {171103} (\bibinfo {year} {2011})},\ \Eprint {http://arxiv.org/abs/1103.4609}
  {arXiv:1103.4609 [astro-ph.CO]} \BibitemShut {NoStop}%
\bibitem [{\citenamefont {Kocsis}\ \emph {et~al.}(2011)\citenamefont {Kocsis},
  \citenamefont {Yunes},\ and\ \citenamefont {Loeb}}]{Kocsis:2011dr}%
  \BibitemOpen
  \bibfield  {author} {\bibinfo {author} {\bibfnamefont {B.}~\bibnamefont
  {Kocsis}}, \bibinfo {author} {\bibfnamefont {N.}~\bibnamefont {Yunes}}, \
  and\ \bibinfo {author} {\bibfnamefont {A.}~\bibnamefont {Loeb}},\ }\href
  {\doibase 10.1103/PhysRevD.86.049907} {\bibfield  {journal} {\bibinfo
  {journal} {Phys. Rev. D}\ }\textbf {\bibinfo {volume} {84}},\ \bibinfo
  {pages} {024032} (\bibinfo {year} {2011})},\ \Eprint
  {http://arxiv.org/abs/1104.2322} {arXiv:1104.2322 [astro-ph.GA]} \BibitemShut
  {NoStop}%
\bibitem [{\citenamefont {Gondolo}\ and\ \citenamefont
  {Silk}(1999)}]{Gondolo:1999ef}%
  \BibitemOpen
  \bibfield  {author} {\bibinfo {author} {\bibfnamefont {P.}~\bibnamefont
  {Gondolo}}\ and\ \bibinfo {author} {\bibfnamefont {J.}~\bibnamefont {Silk}},\
  }\href {\doibase 10.1103/PhysRevLett.83.1719} {\bibfield  {journal} {\bibinfo
   {journal} {Phys. Rev. Lett.}\ }\textbf {\bibinfo {volume} {83}},\ \bibinfo
  {pages} {1719} (\bibinfo {year} {1999})},\ \Eprint
  {http://arxiv.org/abs/astro-ph/9906391} {arXiv:astro-ph/9906391} \BibitemShut
  {NoStop}%
\bibitem [{\citenamefont {Navarro}\ \emph {et~al.}(1997)\citenamefont
  {Navarro}, \citenamefont {Frenk},\ and\ \citenamefont
  {White}}]{Navarro:1996gj}%
  \BibitemOpen
  \bibfield  {author} {\bibinfo {author} {\bibfnamefont {J.~F.}\ \bibnamefont
  {Navarro}}, \bibinfo {author} {\bibfnamefont {C.~S.}\ \bibnamefont {Frenk}},
  \ and\ \bibinfo {author} {\bibfnamefont {S.~D.~M.}\ \bibnamefont {White}},\
  }\href {\doibase 10.1086/304888} {\bibfield  {journal} {\bibinfo  {journal}
  {Astrophys. J.}\ }\textbf {\bibinfo {volume} {490}},\ \bibinfo {pages} {493}
  (\bibinfo {year} {1997})},\ \Eprint {http://arxiv.org/abs/astro-ph/9611107}
  {arXiv:astro-ph/9611107} \BibitemShut {NoStop}%
\bibitem [{\citenamefont {Samsing}\ and\ \citenamefont
  {Ilan}(2018)}]{Samsing:2017plz}%
  \BibitemOpen
  \bibfield  {author} {\bibinfo {author} {\bibfnamefont {J.}~\bibnamefont
  {Samsing}}\ and\ \bibinfo {author} {\bibfnamefont {T.}~\bibnamefont {Ilan}},\
  }\href {\doibase 10.1093/mnras/sty197} {\bibfield  {journal} {\bibinfo
  {journal} {Mon. Not. Roy. Astron. Soc.}\ }\textbf {\bibinfo {volume} {476}},\
  \bibinfo {pages} {1548} (\bibinfo {year} {2018})},\ \Eprint
  {http://arxiv.org/abs/1706.04672} {arXiv:1706.04672 [astro-ph.HE]}
  \BibitemShut {NoStop}%
\bibitem [{\citenamefont {Samsing}\ and\ \citenamefont
  {Ilan}(2019)}]{Samsing:2017xod}%
  \BibitemOpen
  \bibfield  {author} {\bibinfo {author} {\bibfnamefont {J.}~\bibnamefont
  {Samsing}}\ and\ \bibinfo {author} {\bibfnamefont {T.}~\bibnamefont {Ilan}},\
  }\href {\doibase 10.1093/mnras/sty2249} {\bibfield  {journal} {\bibinfo
  {journal} {Mon. Not. Roy. Astron. Soc.}\ }\textbf {\bibinfo {volume} {482}},\
  \bibinfo {pages} {30} (\bibinfo {year} {2019})},\ \Eprint
  {http://arxiv.org/abs/1709.01660} {arXiv:1709.01660 [astro-ph.HE]}
  \BibitemShut {NoStop}%
\bibitem [{\citenamefont {Kozai}(1962)}]{Kozai:1962zz}%
  \BibitemOpen
  \bibfield  {author} {\bibinfo {author} {\bibfnamefont {Y.}~\bibnamefont
  {Kozai}},\ }\href {\doibase 10.1086/108790} {\bibfield  {journal} {\bibinfo
  {journal} {Astron. J.}\ }\textbf {\bibinfo {volume} {67}},\ \bibinfo {pages}
  {591} (\bibinfo {year} {1962})}\BibitemShut {NoStop}%
\bibitem [{\citenamefont {Lidov}(1962)}]{Lidov:1962wjn}%
  \BibitemOpen
  \bibfield  {author} {\bibinfo {author} {\bibfnamefont {M.~L.}\ \bibnamefont
  {Lidov}},\ }\href {\doibase 10.1016/0032-0633(62)90129-0} {\bibfield
  {journal} {\bibinfo  {journal} {Planet. Space Sci.}\ }\textbf {\bibinfo
  {volume} {9}},\ \bibinfo {pages} {719} (\bibinfo {year} {1962})}\BibitemShut
  {NoStop}%
\bibitem [{\citenamefont {Chandramouli}\ and\ \citenamefont
  {Yunes}(2022)}]{Chandramouli:2021kts}%
  \BibitemOpen
  \bibfield  {author} {\bibinfo {author} {\bibfnamefont {R.~S.}\ \bibnamefont
  {Chandramouli}}\ and\ \bibinfo {author} {\bibfnamefont {N.}~\bibnamefont
  {Yunes}},\ }\href {\doibase 10.1103/PhysRevD.105.064009} {\bibfield
  {journal} {\bibinfo  {journal} {Phys. Rev. D}\ }\textbf {\bibinfo {volume}
  {105}},\ \bibinfo {pages} {064009} (\bibinfo {year} {2022})},\ \Eprint
  {http://arxiv.org/abs/2107.00741} {arXiv:2107.00741 [gr-qc]} \BibitemShut
  {NoStop}%
\bibitem [{\citenamefont {Kejriwal}\ \emph {et~al.}(2024)\citenamefont
  {Kejriwal}, \citenamefont {Speri},\ and\ \citenamefont
  {Chua}}]{Kejriwal:2023djc}%
  \BibitemOpen
  \bibfield  {author} {\bibinfo {author} {\bibfnamefont {S.}~\bibnamefont
  {Kejriwal}}, \bibinfo {author} {\bibfnamefont {L.}~\bibnamefont {Speri}}, \
  and\ \bibinfo {author} {\bibfnamefont {A.~J.~K.}\ \bibnamefont {Chua}},\
  }\href {\doibase 10.1103/PhysRevD.110.084060} {\bibfield  {journal} {\bibinfo
   {journal} {Phys. Rev. D}\ }\textbf {\bibinfo {volume} {110}},\ \bibinfo
  {pages} {084060} (\bibinfo {year} {2024})},\ \Eprint
  {http://arxiv.org/abs/2312.13028} {arXiv:2312.13028 [gr-qc]} \BibitemShut
  {NoStop}%
\bibitem [{\citenamefont {Jiang}\ and\ \citenamefont
  {Han}(2023)}]{Jiang:2023xwv}%
  \BibitemOpen
  \bibfield  {author} {\bibinfo {author} {\bibfnamefont {Y.}~\bibnamefont
  {Jiang}}\ and\ \bibinfo {author} {\bibfnamefont {W.-B.}\ \bibnamefont
  {Han}},\ }\href@noop {} {\  (\bibinfo {year} {2023})},\ \Eprint
  {http://arxiv.org/abs/2312.04320} {arXiv:2312.04320 [gr-qc]} \BibitemShut
  {NoStop}%
\bibitem [{\citenamefont {Rahman}\ \emph {et~al.}(2024)\citenamefont {Rahman},
  \citenamefont {Kumar},\ and\ \citenamefont {Bhattacharyya}}]{Rahman:2023sof}%
  \BibitemOpen
  \bibfield  {author} {\bibinfo {author} {\bibfnamefont {M.}~\bibnamefont
  {Rahman}}, \bibinfo {author} {\bibfnamefont {S.}~\bibnamefont {Kumar}}, \
  and\ \bibinfo {author} {\bibfnamefont {A.}~\bibnamefont {Bhattacharyya}},\
  }\href {\doibase 10.1088/1475-7516/2024/01/035} {\bibfield  {journal}
  {\bibinfo  {journal} {JCAP}\ }\textbf {\bibinfo {volume} {01}},\ \bibinfo
  {pages} {035} (\bibinfo {year} {2024})},\ \Eprint
  {http://arxiv.org/abs/2306.14971} {arXiv:2306.14971 [gr-qc]} \BibitemShut
  {NoStop}%
\bibitem [{\citenamefont {Yuan}\ \emph {et~al.}(2024)\citenamefont {Yuan},
  \citenamefont {Zhang},\ and\ \citenamefont {Mei}}]{Yuan:2024duo}%
  \BibitemOpen
  \bibfield  {author} {\bibinfo {author} {\bibfnamefont {X.}~\bibnamefont
  {Yuan}}, \bibinfo {author} {\bibfnamefont {J.-d.}\ \bibnamefont {Zhang}}, \
  and\ \bibinfo {author} {\bibfnamefont {J.}~\bibnamefont {Mei}},\ }\href@noop
  {} {\  (\bibinfo {year} {2024})},\ \Eprint {http://arxiv.org/abs/2412.00915}
  {arXiv:2412.00915 [gr-qc]} \BibitemShut {NoStop}%
\bibitem [{\citenamefont {Schneider}\ and\ \citenamefont
  {Kochanek}(2006)}]{Schneider:2006qyj}%
  \BibitemOpen
  \bibinfo {editor} {\bibfnamefont {P.}~\bibnamefont {Schneider}}\ and\
  \bibinfo {editor} {\bibfnamefont {C.~S.}\ \bibnamefont {Kochanek}},\ eds.,\
  \href@noop {} {\emph {\bibinfo {title} {{Proceedings, 33rd Advanced Saas Fee
  Course on Gravitational Lensing: Strong, Weak, and Micro}: {Les Diablerets,
  Switzerland, April 7-12, 2003}}}},\ Vol.~\bibinfo {volume} {33}\ (\bibinfo
  {publisher} {Springer Berlin Heidelberg},\ \bibinfo {address} {Heidelberg},\
  \bibinfo {year} {2006})\BibitemShut {NoStop}%
\bibitem [{\citenamefont {Takahashi}\ and\ \citenamefont
  {Nakamura}(2003{\natexlab{a}})}]{Takahashi:2003ix}%
  \BibitemOpen
  \bibfield  {author} {\bibinfo {author} {\bibfnamefont {R.}~\bibnamefont
  {Takahashi}}\ and\ \bibinfo {author} {\bibfnamefont {T.}~\bibnamefont
  {Nakamura}},\ }\href {\doibase 10.1086/377430} {\bibfield  {journal}
  {\bibinfo  {journal} {Astrophys. J.}\ }\textbf {\bibinfo {volume} {595}},\
  \bibinfo {pages} {1039} (\bibinfo {year} {2003}{\natexlab{a}})},\ \Eprint
  {http://arxiv.org/abs/astro-ph/0305055} {arXiv:astro-ph/0305055} \BibitemShut
  {NoStop}%
\bibitem [{\citenamefont {Fan}\ \emph {et~al.}(2017)\citenamefont {Fan},
  \citenamefont {Liao}, \citenamefont {Biesiada}, \citenamefont
  {Piorkowska-Kurpas},\ and\ \citenamefont {Zhu}}]{Fan:2016swi}%
  \BibitemOpen
  \bibfield  {author} {\bibinfo {author} {\bibfnamefont {X.-L.}\ \bibnamefont
  {Fan}}, \bibinfo {author} {\bibfnamefont {K.}~\bibnamefont {Liao}}, \bibinfo
  {author} {\bibfnamefont {M.}~\bibnamefont {Biesiada}}, \bibinfo {author}
  {\bibfnamefont {A.}~\bibnamefont {Piorkowska-Kurpas}}, \ and\ \bibinfo
  {author} {\bibfnamefont {Z.-H.}\ \bibnamefont {Zhu}},\ }\href {\doibase
  10.1103/PhysRevLett.118.091102} {\bibfield  {journal} {\bibinfo  {journal}
  {Phys. Rev. Lett.}\ }\textbf {\bibinfo {volume} {118}},\ \bibinfo {pages}
  {091102} (\bibinfo {year} {2017})},\ \Eprint
  {http://arxiv.org/abs/1612.04095} {arXiv:1612.04095 [gr-qc]} \BibitemShut
  {NoStop}%
\bibitem [{\citenamefont {Liao}\ \emph {et~al.}(2018)\citenamefont {Liao},
  \citenamefont {Ding}, \citenamefont {Biesiada}, \citenamefont {Fan},\ and\
  \citenamefont {Zhu}}]{Liao:2018ofi}%
  \BibitemOpen
  \bibfield  {author} {\bibinfo {author} {\bibfnamefont {K.}~\bibnamefont
  {Liao}}, \bibinfo {author} {\bibfnamefont {X.}~\bibnamefont {Ding}}, \bibinfo
  {author} {\bibfnamefont {M.}~\bibnamefont {Biesiada}}, \bibinfo {author}
  {\bibfnamefont {X.-L.}\ \bibnamefont {Fan}}, \ and\ \bibinfo {author}
  {\bibfnamefont {Z.-H.}\ \bibnamefont {Zhu}},\ }\href {\doibase
  10.3847/1538-4357/aae30f} {\bibfield  {journal} {\bibinfo  {journal}
  {Astrophys. J.}\ }\textbf {\bibinfo {volume} {867}},\ \bibinfo {pages} {69}
  (\bibinfo {year} {2018})},\ \Eprint {http://arxiv.org/abs/1809.07079}
  {arXiv:1809.07079 [astro-ph.CO]} \BibitemShut {NoStop}%
\bibitem [{\citenamefont {Yang}\ \emph
  {et~al.}(2019{\natexlab{a}})\citenamefont {Yang}, \citenamefont {Hu},
  \citenamefont {Cai},\ and\ \citenamefont {Wang}}]{Yang:2018bdf}%
  \BibitemOpen
  \bibfield  {author} {\bibinfo {author} {\bibfnamefont {T.}~\bibnamefont
  {Yang}}, \bibinfo {author} {\bibfnamefont {B.}~\bibnamefont {Hu}}, \bibinfo
  {author} {\bibfnamefont {R.-G.}\ \bibnamefont {Cai}}, \ and\ \bibinfo
  {author} {\bibfnamefont {B.}~\bibnamefont {Wang}},\ }\href {\doibase
  10.3847/1538-4357/ab271e} {\bibfield  {journal} {\bibinfo  {journal}
  {Astrophys. J.}\ }\textbf {\bibinfo {volume} {880}},\ \bibinfo {pages} {50}
  (\bibinfo {year} {2019}{\natexlab{a}})},\ \Eprint
  {http://arxiv.org/abs/1810.00164} {arXiv:1810.00164 [astro-ph.CO]}
  \BibitemShut {NoStop}%
\bibitem [{\citenamefont {Hannuksela}\ \emph {et~al.}(2020)\citenamefont
  {Hannuksela}, \citenamefont {Collett}, \citenamefont {\c{C}al\i{}\c{s}kan},\
  and\ \citenamefont {Li}}]{Hannuksela:2020xor}%
  \BibitemOpen
  \bibfield  {author} {\bibinfo {author} {\bibfnamefont {O.~A.}\ \bibnamefont
  {Hannuksela}}, \bibinfo {author} {\bibfnamefont {T.~E.}\ \bibnamefont
  {Collett}}, \bibinfo {author} {\bibfnamefont {M.}~\bibnamefont
  {\c{C}al\i{}\c{s}kan}}, \ and\ \bibinfo {author} {\bibfnamefont {T.~G.~F.}\
  \bibnamefont {Li}},\ }\href {\doibase 10.1093/mnras/staa2577} {\bibfield
  {journal} {\bibinfo  {journal} {Mon. Not. Roy. Astron. Soc.}\ }\textbf
  {\bibinfo {volume} {498}},\ \bibinfo {pages} {3395} (\bibinfo {year}
  {2020})},\ \Eprint {http://arxiv.org/abs/2004.13811} {arXiv:2004.13811
  [astro-ph.HE]} \BibitemShut {NoStop}%
\bibitem [{\citenamefont {Sereno}\ \emph {et~al.}(2011)\citenamefont {Sereno},
  \citenamefont {Jetzer}, \citenamefont {Sesana},\ and\ \citenamefont
  {Volonteri}}]{Sereno:2011ty}%
  \BibitemOpen
  \bibfield  {author} {\bibinfo {author} {\bibfnamefont {M.}~\bibnamefont
  {Sereno}}, \bibinfo {author} {\bibfnamefont {P.}~\bibnamefont {Jetzer}},
  \bibinfo {author} {\bibfnamefont {A.}~\bibnamefont {Sesana}}, \ and\ \bibinfo
  {author} {\bibfnamefont {M.}~\bibnamefont {Volonteri}},\ }\href {\doibase
  10.1111/j.1365-2966.2011.18895.x} {\bibfield  {journal} {\bibinfo  {journal}
  {Mon. Not. Roy. Astron. Soc.}\ }\textbf {\bibinfo {volume} {415}},\ \bibinfo
  {pages} {2773} (\bibinfo {year} {2011})},\ \Eprint
  {http://arxiv.org/abs/1104.1977} {arXiv:1104.1977 [astro-ph.CO]} \BibitemShut
  {NoStop}%
\bibitem [{\citenamefont {Liao}\ \emph {et~al.}(2017)\citenamefont {Liao},
  \citenamefont {Fan}, \citenamefont {Ding}, \citenamefont {Biesiada},\ and\
  \citenamefont {Zhu}}]{Liao:2017ioi}%
  \BibitemOpen
  \bibfield  {author} {\bibinfo {author} {\bibfnamefont {K.}~\bibnamefont
  {Liao}}, \bibinfo {author} {\bibfnamefont {X.-L.}\ \bibnamefont {Fan}},
  \bibinfo {author} {\bibfnamefont {X.-H.}\ \bibnamefont {Ding}}, \bibinfo
  {author} {\bibfnamefont {M.}~\bibnamefont {Biesiada}}, \ and\ \bibinfo
  {author} {\bibfnamefont {Z.-H.}\ \bibnamefont {Zhu}},\ }\href {\doibase
  10.1038/s41467-017-01152-9} {\bibfield  {journal} {\bibinfo  {journal}
  {Nature Commun.}\ }\textbf {\bibinfo {volume} {8}},\ \bibinfo {pages} {1148}
  (\bibinfo {year} {2017})},\ \bibinfo {note} {[Erratum: Nature Commun. 8, 2136
  (2017)]},\ \Eprint {http://arxiv.org/abs/1703.04151} {arXiv:1703.04151
  [astro-ph.CO]} \BibitemShut {NoStop}%
\bibitem [{\citenamefont {Cao}\ \emph {et~al.}(2019)\citenamefont {Cao},
  \citenamefont {Qi}, \citenamefont {Cao}, \citenamefont {Biesiada},
  \citenamefont {Li}, \citenamefont {Pan},\ and\ \citenamefont
  {Zhu}}]{Cao:2019kgn}%
  \BibitemOpen
  \bibfield  {author} {\bibinfo {author} {\bibfnamefont {S.}~\bibnamefont
  {Cao}}, \bibinfo {author} {\bibfnamefont {J.}~\bibnamefont {Qi}}, \bibinfo
  {author} {\bibfnamefont {Z.}~\bibnamefont {Cao}}, \bibinfo {author}
  {\bibfnamefont {M.}~\bibnamefont {Biesiada}}, \bibinfo {author}
  {\bibfnamefont {J.}~\bibnamefont {Li}}, \bibinfo {author} {\bibfnamefont
  {Y.}~\bibnamefont {Pan}}, \ and\ \bibinfo {author} {\bibfnamefont {Z.-H.}\
  \bibnamefont {Zhu}},\ }\href {\doibase 10.1038/s41598-019-47616-4} {\bibfield
   {journal} {\bibinfo  {journal} {Sci. Rep.}\ }\textbf {\bibinfo {volume}
  {9}},\ \bibinfo {pages} {11608} (\bibinfo {year} {2019})},\ \Eprint
  {http://arxiv.org/abs/1910.10365} {arXiv:1910.10365 [astro-ph.CO]}
  \BibitemShut {NoStop}%
\bibitem [{\citenamefont {Li}\ \emph {et~al.}(2019{\natexlab{a}})\citenamefont
  {Li}, \citenamefont {Fan},\ and\ \citenamefont {Gou}}]{Li:2019rns}%
  \BibitemOpen
  \bibfield  {author} {\bibinfo {author} {\bibfnamefont {Y.}~\bibnamefont
  {Li}}, \bibinfo {author} {\bibfnamefont {X.}~\bibnamefont {Fan}}, \ and\
  \bibinfo {author} {\bibfnamefont {L.}~\bibnamefont {Gou}},\ }\href {\doibase
  10.3847/1538-4357/ab037e} {\bibfield  {journal} {\bibinfo  {journal}
  {Astrophys. J.}\ }\textbf {\bibinfo {volume} {873}},\ \bibinfo {pages} {37}
  (\bibinfo {year} {2019}{\natexlab{a}})},\ \Eprint
  {http://arxiv.org/abs/1901.10638} {arXiv:1901.10638 [astro-ph.CO]}
  \BibitemShut {NoStop}%
\bibitem [{\citenamefont {Yu}\ \emph {et~al.}(2020)\citenamefont {Yu},
  \citenamefont {Zhang},\ and\ \citenamefont {Wang}}]{Yu:2020agu}%
  \BibitemOpen
  \bibfield  {author} {\bibinfo {author} {\bibfnamefont {H.}~\bibnamefont
  {Yu}}, \bibinfo {author} {\bibfnamefont {P.}~\bibnamefont {Zhang}}, \ and\
  \bibinfo {author} {\bibfnamefont {F.-Y.}\ \bibnamefont {Wang}},\ }\href
  {\doibase 10.1093/mnras/staa1952} {\bibfield  {journal} {\bibinfo  {journal}
  {Mon. Not. Roy. Astron. Soc.}\ }\textbf {\bibinfo {volume} {497}},\ \bibinfo
  {pages} {204} (\bibinfo {year} {2020})},\ \Eprint
  {http://arxiv.org/abs/2007.00828} {arXiv:2007.00828 [astro-ph.CO]}
  \BibitemShut {NoStop}%
\bibitem [{\citenamefont {Urrutia}\ and\ \citenamefont
  {Vaskonen}(2021)}]{Urrutia:2021qak}%
  \BibitemOpen
  \bibfield  {author} {\bibinfo {author} {\bibfnamefont {J.}~\bibnamefont
  {Urrutia}}\ and\ \bibinfo {author} {\bibfnamefont {V.}~\bibnamefont
  {Vaskonen}},\ }\href {\doibase 10.1093/mnras/stab3118} {\bibfield  {journal}
  {\bibinfo  {journal} {Mon. Not. Roy. Astron. Soc.}\ }\textbf {\bibinfo
  {volume} {509}},\ \bibinfo {pages} {1358} (\bibinfo {year} {2021})},\ \Eprint
  {http://arxiv.org/abs/2109.03213} {arXiv:2109.03213 [astro-ph.CO]}
  \BibitemShut {NoStop}%
\bibitem [{\citenamefont {Chung}\ and\ \citenamefont
  {Li}(2021)}]{Chung:2021rcu}%
  \BibitemOpen
  \bibfield  {author} {\bibinfo {author} {\bibfnamefont {A.~K.-W.}\
  \bibnamefont {Chung}}\ and\ \bibinfo {author} {\bibfnamefont {T.~G.~F.}\
  \bibnamefont {Li}},\ }\href {\doibase 10.1103/PhysRevD.104.124060} {\bibfield
   {journal} {\bibinfo  {journal} {Phys. Rev. D}\ }\textbf {\bibinfo {volume}
  {104}},\ \bibinfo {pages} {124060} (\bibinfo {year} {2021})},\ \Eprint
  {http://arxiv.org/abs/2106.09630} {arXiv:2106.09630 [gr-qc]} \BibitemShut
  {NoStop}%
\bibitem [{\citenamefont {Gais}\ \emph {et~al.}(2022)\citenamefont {Gais},
  \citenamefont {Ng}, \citenamefont {Seo}, \citenamefont {Wong},\ and\
  \citenamefont {Li}}]{Gais:2022xir}%
  \BibitemOpen
  \bibfield  {author} {\bibinfo {author} {\bibfnamefont {J.}~\bibnamefont
  {Gais}}, \bibinfo {author} {\bibfnamefont {K.~K.~Y.}\ \bibnamefont {Ng}},
  \bibinfo {author} {\bibfnamefont {E.}~\bibnamefont {Seo}}, \bibinfo {author}
  {\bibfnamefont {K.~W.~K.}\ \bibnamefont {Wong}}, \ and\ \bibinfo {author}
  {\bibfnamefont {T.~G.~F.}\ \bibnamefont {Li}},\ }\href {\doibase
  10.3847/2041-8213/ac7052} {\bibfield  {journal} {\bibinfo  {journal}
  {Astrophys. J. Lett.}\ }\textbf {\bibinfo {volume} {932}},\ \bibinfo {pages}
  {L4} (\bibinfo {year} {2022})},\ \Eprint {http://arxiv.org/abs/2201.01817}
  {arXiv:2201.01817 [gr-qc]} \BibitemShut {NoStop}%
\bibitem [{\citenamefont {Broadhurst}\ \emph {et~al.}(2020)\citenamefont
  {Broadhurst}, \citenamefont {Diego},\ and\ \citenamefont
  {Smoot}}]{Broadhurst:2020cvm}%
  \BibitemOpen
  \bibfield  {author} {\bibinfo {author} {\bibfnamefont {T.}~\bibnamefont
  {Broadhurst}}, \bibinfo {author} {\bibfnamefont {J.~M.}\ \bibnamefont
  {Diego}}, \ and\ \bibinfo {author} {\bibfnamefont {G.~F.}\ \bibnamefont
  {Smoot}},\ }\href@noop {} {\  (\bibinfo {year} {2020})},\ \Eprint
  {http://arxiv.org/abs/2006.13219} {arXiv:2006.13219 [astro-ph.CO]}
  \BibitemShut {NoStop}%
\bibitem [{\citenamefont {Ohanian}(1974)}]{Ohanian:1974ys}%
  \BibitemOpen
  \bibfield  {author} {\bibinfo {author} {\bibfnamefont {H.~C.}\ \bibnamefont
  {Ohanian}},\ }\href {\doibase 10.1007/BF01810927} {\bibfield  {journal}
  {\bibinfo  {journal} {Int. J. Theor. Phys.}\ }\textbf {\bibinfo {volume}
  {9}},\ \bibinfo {pages} {425} (\bibinfo {year} {1974})}\BibitemShut {NoStop}%
\bibitem [{\citenamefont {Nakamura}(1998)}]{Nakamura:1997sw}%
  \BibitemOpen
  \bibfield  {author} {\bibinfo {author} {\bibfnamefont {T.~T.}\ \bibnamefont
  {Nakamura}},\ }\href {\doibase 10.1103/PhysRevLett.80.1138} {\bibfield
  {journal} {\bibinfo  {journal} {Phys. Rev. Lett.}\ }\textbf {\bibinfo
  {volume} {80}},\ \bibinfo {pages} {1138} (\bibinfo {year}
  {1998})}\BibitemShut {NoStop}%
\bibitem [{\citenamefont {Boileau}\ \emph {et~al.}(2021)\citenamefont
  {Boileau}, \citenamefont {Christensen}, \citenamefont {Meyer},\ and\
  \citenamefont {Cornish}}]{Boileau:2020rpg}%
  \BibitemOpen
  \bibfield  {author} {\bibinfo {author} {\bibfnamefont {G.}~\bibnamefont
  {Boileau}}, \bibinfo {author} {\bibfnamefont {N.}~\bibnamefont
  {Christensen}}, \bibinfo {author} {\bibfnamefont {R.}~\bibnamefont {Meyer}},
  \ and\ \bibinfo {author} {\bibfnamefont {N.~J.}\ \bibnamefont {Cornish}},\
  }\href {\doibase 10.1103/PhysRevD.103.103529} {\bibfield  {journal} {\bibinfo
   {journal} {Phys. Rev. D}\ }\textbf {\bibinfo {volume} {103}},\ \bibinfo
  {pages} {103529} (\bibinfo {year} {2021})},\ \Eprint
  {http://arxiv.org/abs/2011.05055} {arXiv:2011.05055 [gr-qc]} \BibitemShut
  {NoStop}%
\bibitem [{\citenamefont {Leung}\ \emph {et~al.}(2023)\citenamefont {Leung},
  \citenamefont {Jow}, \citenamefont {Saha}, \citenamefont {Dai}, \citenamefont
  {Oguri},\ and\ \citenamefont {Koopmans}}]{Leung:2023lmq}%
  \BibitemOpen
  \bibfield  {author} {\bibinfo {author} {\bibfnamefont {C.}~\bibnamefont
  {Leung}}, \bibinfo {author} {\bibfnamefont {D.}~\bibnamefont {Jow}}, \bibinfo
  {author} {\bibfnamefont {P.}~\bibnamefont {Saha}}, \bibinfo {author}
  {\bibfnamefont {L.}~\bibnamefont {Dai}}, \bibinfo {author} {\bibfnamefont
  {M.}~\bibnamefont {Oguri}}, \ and\ \bibinfo {author} {\bibfnamefont
  {L.~V.~E.}\ \bibnamefont {Koopmans}},\ }\href@noop {} {\  (\bibinfo {year}
  {2023})},\ \Eprint {http://arxiv.org/abs/2304.01202} {arXiv:2304.01202
  [astro-ph.HE]} \BibitemShut {NoStop}%
\bibitem [{\citenamefont {Ezquiaga}\ and\ \citenamefont
  {Zumalac\'arregui}(2020)}]{Ezquiaga:2020dao}%
  \BibitemOpen
  \bibfield  {author} {\bibinfo {author} {\bibfnamefont {J.~M.}\ \bibnamefont
  {Ezquiaga}}\ and\ \bibinfo {author} {\bibfnamefont {M.}~\bibnamefont
  {Zumalac\'arregui}},\ }\href {\doibase 10.1103/PhysRevD.102.124048}
  {\bibfield  {journal} {\bibinfo  {journal} {Phys. Rev. D}\ }\textbf {\bibinfo
  {volume} {102}},\ \bibinfo {pages} {124048} (\bibinfo {year} {2020})},\
  \Eprint {http://arxiv.org/abs/2009.12187} {arXiv:2009.12187 [gr-qc]}
  \BibitemShut {NoStop}%
\bibitem [{\citenamefont {Dalang}\ \emph {et~al.}(2022)\citenamefont {Dalang},
  \citenamefont {Cusin},\ and\ \citenamefont {Lagos}}]{Dalang:2021qhu}%
  \BibitemOpen
  \bibfield  {author} {\bibinfo {author} {\bibfnamefont {C.}~\bibnamefont
  {Dalang}}, \bibinfo {author} {\bibfnamefont {G.}~\bibnamefont {Cusin}}, \
  and\ \bibinfo {author} {\bibfnamefont {M.}~\bibnamefont {Lagos}},\ }\href
  {\doibase 10.1103/PhysRevD.105.024005} {\bibfield  {journal} {\bibinfo
  {journal} {Phys. Rev. D}\ }\textbf {\bibinfo {volume} {105}},\ \bibinfo
  {pages} {024005} (\bibinfo {year} {2022})},\ \Eprint
  {http://arxiv.org/abs/2104.10119} {arXiv:2104.10119 [gr-qc]} \BibitemShut
  {NoStop}%
\bibitem [{\citenamefont {Yamamoto}(2005)}]{Yamamoto:2005ea}%
  \BibitemOpen
  \bibfield  {author} {\bibinfo {author} {\bibfnamefont {K.}~\bibnamefont
  {Yamamoto}},\ }\href {\doibase 10.1103/PhysRevD.71.101301} {\bibfield
  {journal} {\bibinfo  {journal} {Phys. Rev. D}\ }\textbf {\bibinfo {volume}
  {71}},\ \bibinfo {pages} {101301} (\bibinfo {year} {2005})},\ \Eprint
  {http://arxiv.org/abs/astro-ph/0505116} {arXiv:astro-ph/0505116} \BibitemShut
  {NoStop}%
\bibitem [{\citenamefont {Hou}\ \emph {et~al.}(2021{\natexlab{a}})\citenamefont
  {Hou}, \citenamefont {Li}, \citenamefont {Yu}, \citenamefont {Biesiada},
  \citenamefont {Fan}, \citenamefont {Kawamura},\ and\ \citenamefont
  {Zhu}}]{Hou:2020mpr}%
  \BibitemOpen
  \bibfield  {author} {\bibinfo {author} {\bibfnamefont {S.}~\bibnamefont
  {Hou}}, \bibinfo {author} {\bibfnamefont {P.}~\bibnamefont {Li}}, \bibinfo
  {author} {\bibfnamefont {H.}~\bibnamefont {Yu}}, \bibinfo {author}
  {\bibfnamefont {M.}~\bibnamefont {Biesiada}}, \bibinfo {author}
  {\bibfnamefont {X.-L.}\ \bibnamefont {Fan}}, \bibinfo {author} {\bibfnamefont
  {S.}~\bibnamefont {Kawamura}}, \ and\ \bibinfo {author} {\bibfnamefont
  {Z.-H.}\ \bibnamefont {Zhu}},\ }\href {\doibase 10.1103/PhysRevD.103.044005}
  {\bibfield  {journal} {\bibinfo  {journal} {Phys. Rev. D}\ }\textbf {\bibinfo
  {volume} {103}},\ \bibinfo {pages} {044005} (\bibinfo {year}
  {2021}{\natexlab{a}})},\ \Eprint {http://arxiv.org/abs/2009.08116}
  {arXiv:2009.08116 [gr-qc]} \BibitemShut {NoStop}%
\bibitem [{\citenamefont {Moylan}\ \emph {et~al.}(2007)\citenamefont {Moylan},
  \citenamefont {McClelland}, \citenamefont {Scott}, \citenamefont {Searle},\
  and\ \citenamefont {Bicknell}}]{Moylan:2007fi}%
  \BibitemOpen
  \bibfield  {author} {\bibinfo {author} {\bibfnamefont {A.~J.}\ \bibnamefont
  {Moylan}}, \bibinfo {author} {\bibfnamefont {D.~E.}\ \bibnamefont
  {McClelland}}, \bibinfo {author} {\bibfnamefont {S.~M.}\ \bibnamefont
  {Scott}}, \bibinfo {author} {\bibfnamefont {A.~C.}\ \bibnamefont {Searle}}, \
  and\ \bibinfo {author} {\bibfnamefont {G.~V.}\ \bibnamefont {Bicknell}},\
  }in\ \href {\doibase 10.1142/9789812834300_0038} {\emph {\bibinfo {booktitle}
  {{11th Marcel Grossmann Meeting on General Relativity}}}}\ (\bibinfo {year}
  {2007})\ pp.\ \bibinfo {pages} {807--823},\ \Eprint
  {http://arxiv.org/abs/0710.3140} {arXiv:0710.3140 [gr-qc]} \BibitemShut
  {NoStop}%
\bibitem [{\citenamefont {Cao}\ \emph {et~al.}(2014)\citenamefont {Cao},
  \citenamefont {Li},\ and\ \citenamefont {Wang}}]{Cao:2014oaa}%
  \BibitemOpen
  \bibfield  {author} {\bibinfo {author} {\bibfnamefont {Z.}~\bibnamefont
  {Cao}}, \bibinfo {author} {\bibfnamefont {L.-F.}\ \bibnamefont {Li}}, \ and\
  \bibinfo {author} {\bibfnamefont {Y.}~\bibnamefont {Wang}},\ }\href {\doibase
  10.1103/PhysRevD.90.062003} {\bibfield  {journal} {\bibinfo  {journal} {Phys.
  Rev. D}\ }\textbf {\bibinfo {volume} {90}},\ \bibinfo {pages} {062003}
  (\bibinfo {year} {2014})}\BibitemShut {NoStop}%
\bibitem [{\citenamefont {Takahashi}(2017)}]{Takahashi:2016jom}%
  \BibitemOpen
  \bibfield  {author} {\bibinfo {author} {\bibfnamefont {R.}~\bibnamefont
  {Takahashi}},\ }\href {\doibase 10.3847/1538-4357/835/1/103} {\bibfield
  {journal} {\bibinfo  {journal} {Astrophys. J.}\ }\textbf {\bibinfo {volume}
  {835}},\ \bibinfo {pages} {103} (\bibinfo {year} {2017})},\ \Eprint
  {http://arxiv.org/abs/1606.00458} {arXiv:1606.00458 [astro-ph.CO]}
  \BibitemShut {NoStop}%
\bibitem [{\citenamefont {Christian}\ \emph {et~al.}(2018)\citenamefont
  {Christian}, \citenamefont {Vitale},\ and\ \citenamefont
  {Loeb}}]{Christian:2018vsi}%
  \BibitemOpen
  \bibfield  {author} {\bibinfo {author} {\bibfnamefont {P.}~\bibnamefont
  {Christian}}, \bibinfo {author} {\bibfnamefont {S.}~\bibnamefont {Vitale}}, \
  and\ \bibinfo {author} {\bibfnamefont {A.}~\bibnamefont {Loeb}},\ }\href
  {\doibase 10.1103/PhysRevD.98.103022} {\bibfield  {journal} {\bibinfo
  {journal} {Phys. Rev. D}\ }\textbf {\bibinfo {volume} {98}},\ \bibinfo
  {pages} {103022} (\bibinfo {year} {2018})},\ \Eprint
  {http://arxiv.org/abs/1802.02586} {arXiv:1802.02586 [astro-ph.HE]}
  \BibitemShut {NoStop}%
\bibitem [{\citenamefont {Dai}\ \emph {et~al.}(2018)\citenamefont {Dai},
  \citenamefont {Li}, \citenamefont {Zackay}, \citenamefont {Mao},\ and\
  \citenamefont {Lu}}]{Dai:2018enj}%
  \BibitemOpen
  \bibfield  {author} {\bibinfo {author} {\bibfnamefont {L.}~\bibnamefont
  {Dai}}, \bibinfo {author} {\bibfnamefont {S.-S.}\ \bibnamefont {Li}},
  \bibinfo {author} {\bibfnamefont {B.}~\bibnamefont {Zackay}}, \bibinfo
  {author} {\bibfnamefont {S.}~\bibnamefont {Mao}}, \ and\ \bibinfo {author}
  {\bibfnamefont {Y.}~\bibnamefont {Lu}},\ }\href {\doibase
  10.1103/PhysRevD.98.104029} {\bibfield  {journal} {\bibinfo  {journal} {Phys.
  Rev. D}\ }\textbf {\bibinfo {volume} {98}},\ \bibinfo {pages} {104029}
  (\bibinfo {year} {2018})},\ \Eprint {http://arxiv.org/abs/1810.00003}
  {arXiv:1810.00003 [gr-qc]} \BibitemShut {NoStop}%
\bibitem [{\citenamefont {Jung}\ and\ \citenamefont
  {Shin}(2019)}]{Jung:2017flg}%
  \BibitemOpen
  \bibfield  {author} {\bibinfo {author} {\bibfnamefont {S.}~\bibnamefont
  {Jung}}\ and\ \bibinfo {author} {\bibfnamefont {C.~S.}\ \bibnamefont
  {Shin}},\ }\href {\doibase 10.1103/PhysRevLett.122.041103} {\bibfield
  {journal} {\bibinfo  {journal} {Phys. Rev. Lett.}\ }\textbf {\bibinfo
  {volume} {122}},\ \bibinfo {pages} {041103} (\bibinfo {year} {2019})},\
  \Eprint {http://arxiv.org/abs/1712.01396} {arXiv:1712.01396 [astro-ph.CO]}
  \BibitemShut {NoStop}%
\bibitem [{\citenamefont {Liao}\ \emph {et~al.}(2019)\citenamefont {Liao},
  \citenamefont {Biesiada},\ and\ \citenamefont {Fan}}]{Liao:2019aqq}%
  \BibitemOpen
  \bibfield  {author} {\bibinfo {author} {\bibfnamefont {K.}~\bibnamefont
  {Liao}}, \bibinfo {author} {\bibfnamefont {M.}~\bibnamefont {Biesiada}}, \
  and\ \bibinfo {author} {\bibfnamefont {X.-L.}\ \bibnamefont {Fan}},\ }\href
  {\doibase 10.3847/1538-4357/ab1087} {\bibfield  {journal} {\bibinfo
  {journal} {Astrophys. J.}\ }\textbf {\bibinfo {volume} {875}},\ \bibinfo
  {pages} {139} (\bibinfo {year} {2019})},\ \Eprint
  {http://arxiv.org/abs/1903.06612} {arXiv:1903.06612 [gr-qc]} \BibitemShut
  {NoStop}%
\bibitem [{\citenamefont {Mishra}\ \emph {et~al.}(2021)\citenamefont {Mishra},
  \citenamefont {Meena}, \citenamefont {More}, \citenamefont {Bose},\ and\
  \citenamefont {Bagla}}]{Mishra:2021xzz}%
  \BibitemOpen
  \bibfield  {author} {\bibinfo {author} {\bibfnamefont {A.}~\bibnamefont
  {Mishra}}, \bibinfo {author} {\bibfnamefont {A.~K.}\ \bibnamefont {Meena}},
  \bibinfo {author} {\bibfnamefont {A.}~\bibnamefont {More}}, \bibinfo {author}
  {\bibfnamefont {S.}~\bibnamefont {Bose}}, \ and\ \bibinfo {author}
  {\bibfnamefont {J.~S.}\ \bibnamefont {Bagla}},\ }\href {\doibase
  10.1093/mnras/stab2875} {\bibfield  {journal} {\bibinfo  {journal} {Mon. Not.
  Roy. Astron. Soc.}\ }\textbf {\bibinfo {volume} {508}},\ \bibinfo {pages}
  {4869} (\bibinfo {year} {2021})},\ \Eprint {http://arxiv.org/abs/2102.03946}
  {arXiv:2102.03946 [astro-ph.CO]} \BibitemShut {NoStop}%
\bibitem [{\citenamefont {Abbott}\ \emph
  {et~al.}(2024{\natexlab{a}})\citenamefont {Abbott} \emph
  {et~al.}}]{LIGOScientific:2021usb}%
  \BibitemOpen
  \bibfield  {author} {\bibinfo {author} {\bibfnamefont {R.}~\bibnamefont
  {Abbott}} \emph {et~al.} (\bibinfo {collaboration} {LIGO Scientific,
  VIRGO}),\ }\href {\doibase 10.1103/PhysRevD.109.022001} {\bibfield  {journal}
  {\bibinfo  {journal} {Phys. Rev. D}\ }\textbf {\bibinfo {volume} {109}},\
  \bibinfo {pages} {022001} (\bibinfo {year} {2024}{\natexlab{a}})},\ \Eprint
  {http://arxiv.org/abs/2108.01045} {arXiv:2108.01045 [gr-qc]} \BibitemShut
  {NoStop}%
\bibitem [{\citenamefont {Abbott}\ \emph
  {et~al.}(2023{\natexlab{b}})\citenamefont {Abbott} \emph
  {et~al.}}]{LIGOScientific:2021aug}%
  \BibitemOpen
  \bibfield  {author} {\bibinfo {author} {\bibfnamefont {R.}~\bibnamefont
  {Abbott}} \emph {et~al.} (\bibinfo {collaboration} {LIGO Scientific, Virgo,
  KAGRA}),\ }\href {\doibase 10.3847/1538-4357/ac74bb} {\bibfield  {journal}
  {\bibinfo  {journal} {Astrophys. J.}\ }\textbf {\bibinfo {volume} {949}},\
  \bibinfo {pages} {76} (\bibinfo {year} {2023}{\natexlab{b}})},\ \Eprint
  {http://arxiv.org/abs/2111.03604} {arXiv:2111.03604 [astro-ph.CO]}
  \BibitemShut {NoStop}%
\bibitem [{\citenamefont {Broadhurst}\ \emph {et~al.}(2019)\citenamefont
  {Broadhurst}, \citenamefont {Diego},\ and\ \citenamefont
  {Smoot}}]{Broadhurst:2019ijv}%
  \BibitemOpen
  \bibfield  {author} {\bibinfo {author} {\bibfnamefont {T.}~\bibnamefont
  {Broadhurst}}, \bibinfo {author} {\bibfnamefont {J.~M.}\ \bibnamefont
  {Diego}}, \ and\ \bibinfo {author} {\bibfnamefont {G.~F.}\ \bibnamefont
  {Smoot}},\ }\href@noop {} {\  (\bibinfo {year} {2019})},\ \Eprint
  {http://arxiv.org/abs/1901.03190} {arXiv:1901.03190 [astro-ph.CO]}
  \BibitemShut {NoStop}%
\bibitem [{\citenamefont {Singer}\ \emph {et~al.}(2019)\citenamefont {Singer},
  \citenamefont {Goldstein},\ and\ \citenamefont {Bloom}}]{Singer:2019vjs}%
  \BibitemOpen
  \bibfield  {author} {\bibinfo {author} {\bibfnamefont {L.~P.}\ \bibnamefont
  {Singer}}, \bibinfo {author} {\bibfnamefont {D.~A.}\ \bibnamefont
  {Goldstein}}, \ and\ \bibinfo {author} {\bibfnamefont {J.~S.}\ \bibnamefont
  {Bloom}},\ }\href@noop {} {\  (\bibinfo {year} {2019})},\ \Eprint
  {http://arxiv.org/abs/1910.03601} {arXiv:1910.03601 [astro-ph.CO]}
  \BibitemShut {NoStop}%
\bibitem [{\citenamefont {McIsaac}\ \emph {et~al.}(2020)\citenamefont
  {McIsaac}, \citenamefont {Keitel}, \citenamefont {Collett}, \citenamefont
  {Harry}, \citenamefont {Mozzon}, \citenamefont {Edy},\ and\ \citenamefont
  {Bacon}}]{McIsaac:2019use}%
  \BibitemOpen
  \bibfield  {author} {\bibinfo {author} {\bibfnamefont {C.}~\bibnamefont
  {McIsaac}}, \bibinfo {author} {\bibfnamefont {D.}~\bibnamefont {Keitel}},
  \bibinfo {author} {\bibfnamefont {T.}~\bibnamefont {Collett}}, \bibinfo
  {author} {\bibfnamefont {I.}~\bibnamefont {Harry}}, \bibinfo {author}
  {\bibfnamefont {S.}~\bibnamefont {Mozzon}}, \bibinfo {author} {\bibfnamefont
  {O.}~\bibnamefont {Edy}}, \ and\ \bibinfo {author} {\bibfnamefont
  {D.}~\bibnamefont {Bacon}},\ }\href {\doibase 10.1103/PhysRevD.102.084031}
  {\bibfield  {journal} {\bibinfo  {journal} {Phys. Rev. D}\ }\textbf {\bibinfo
  {volume} {102}},\ \bibinfo {pages} {084031} (\bibinfo {year} {2020})},\
  \Eprint {http://arxiv.org/abs/1912.05389} {arXiv:1912.05389 [gr-qc]}
  \BibitemShut {NoStop}%
\bibitem [{\citenamefont {Hannuksela}\ \emph {et~al.}(2019)\citenamefont
  {Hannuksela}, \citenamefont {Haris}, \citenamefont {Ng}, \citenamefont
  {Kumar}, \citenamefont {Mehta}, \citenamefont {Keitel}, \citenamefont {Li},\
  and\ \citenamefont {Ajith}}]{Hannuksela:2019kle}%
  \BibitemOpen
  \bibfield  {author} {\bibinfo {author} {\bibfnamefont {O.~A.}\ \bibnamefont
  {Hannuksela}}, \bibinfo {author} {\bibfnamefont {K.}~\bibnamefont {Haris}},
  \bibinfo {author} {\bibfnamefont {K.~K.~Y.}\ \bibnamefont {Ng}}, \bibinfo
  {author} {\bibfnamefont {S.}~\bibnamefont {Kumar}}, \bibinfo {author}
  {\bibfnamefont {A.~K.}\ \bibnamefont {Mehta}}, \bibinfo {author}
  {\bibfnamefont {D.}~\bibnamefont {Keitel}}, \bibinfo {author} {\bibfnamefont
  {T.~G.~F.}\ \bibnamefont {Li}}, \ and\ \bibinfo {author} {\bibfnamefont
  {P.}~\bibnamefont {Ajith}},\ }\href {\doibase 10.3847/2041-8213/ab0c0f}
  {\bibfield  {journal} {\bibinfo  {journal} {Astrophys. J. Lett.}\ }\textbf
  {\bibinfo {volume} {874}},\ \bibinfo {pages} {L2} (\bibinfo {year} {2019})},\
  \Eprint {http://arxiv.org/abs/1901.02674} {arXiv:1901.02674 [gr-qc]}
  \BibitemShut {NoStop}%
\bibitem [{\citenamefont {Liu}\ \emph {et~al.}(2021{\natexlab{a}})\citenamefont
  {Liu}, \citenamefont {Magana~Hernandez},\ and\ \citenamefont
  {Creighton}}]{Liu:2020par}%
  \BibitemOpen
  \bibfield  {author} {\bibinfo {author} {\bibfnamefont {X.}~\bibnamefont
  {Liu}}, \bibinfo {author} {\bibfnamefont {I.}~\bibnamefont
  {Magana~Hernandez}}, \ and\ \bibinfo {author} {\bibfnamefont
  {J.}~\bibnamefont {Creighton}},\ }\href {\doibase 10.3847/1538-4357/abd7eb}
  {\bibfield  {journal} {\bibinfo  {journal} {Astrophys. J.}\ }\textbf
  {\bibinfo {volume} {908}},\ \bibinfo {pages} {97} (\bibinfo {year}
  {2021}{\natexlab{a}})},\ \Eprint {http://arxiv.org/abs/2009.06539}
  {arXiv:2009.06539 [astro-ph.HE]} \BibitemShut {NoStop}%
\bibitem [{\citenamefont {Dai}\ \emph {et~al.}(2020)\citenamefont {Dai},
  \citenamefont {Zackay}, \citenamefont {Venumadhav}, \citenamefont {Roulet},\
  and\ \citenamefont {Zaldarriaga}}]{Dai:2020tpj}%
  \BibitemOpen
  \bibfield  {author} {\bibinfo {author} {\bibfnamefont {L.}~\bibnamefont
  {Dai}}, \bibinfo {author} {\bibfnamefont {B.}~\bibnamefont {Zackay}},
  \bibinfo {author} {\bibfnamefont {T.}~\bibnamefont {Venumadhav}}, \bibinfo
  {author} {\bibfnamefont {J.}~\bibnamefont {Roulet}}, \ and\ \bibinfo {author}
  {\bibfnamefont {M.}~\bibnamefont {Zaldarriaga}},\ }\href@noop {} {\
  (\bibinfo {year} {2020})},\ \Eprint {http://arxiv.org/abs/2007.12709}
  {arXiv:2007.12709 [astro-ph.HE]} \BibitemShut {NoStop}%
\bibitem [{\citenamefont {Abbott}\ \emph
  {et~al.}(2021{\natexlab{d}})\citenamefont {Abbott} \emph
  {et~al.}}]{LIGOScientific:2021izm}%
  \BibitemOpen
  \bibfield  {author} {\bibinfo {author} {\bibfnamefont {R.}~\bibnamefont
  {Abbott}} \emph {et~al.} (\bibinfo {collaboration} {LIGO Scientific,
  VIRGO}),\ }\href {\doibase 10.3847/1538-4357/ac23db} {\bibfield  {journal}
  {\bibinfo  {journal} {Astrophys. J.}\ }\textbf {\bibinfo {volume} {923}},\
  \bibinfo {pages} {14} (\bibinfo {year} {2021}{\natexlab{d}})},\ \Eprint
  {http://arxiv.org/abs/2105.06384} {arXiv:2105.06384 [gr-qc]} \BibitemShut
  {NoStop}%
\bibitem [{\citenamefont {Diego}\ \emph {et~al.}(2021)\citenamefont {Diego},
  \citenamefont {Broadhurst},\ and\ \citenamefont {Smoot}}]{Diego:2021fyd}%
  \BibitemOpen
  \bibfield  {author} {\bibinfo {author} {\bibfnamefont {J.~M.}\ \bibnamefont
  {Diego}}, \bibinfo {author} {\bibfnamefont {T.}~\bibnamefont {Broadhurst}}, \
  and\ \bibinfo {author} {\bibfnamefont {G.}~\bibnamefont {Smoot}},\ }\href
  {\doibase 10.1103/PhysRevD.104.103529} {\bibfield  {journal} {\bibinfo
  {journal} {Phys. Rev. D}\ }\textbf {\bibinfo {volume} {104}},\ \bibinfo
  {pages} {103529} (\bibinfo {year} {2021})},\ \Eprint
  {http://arxiv.org/abs/2106.06545} {arXiv:2106.06545 [gr-qc]} \BibitemShut
  {NoStop}%
\bibitem [{\citenamefont {Baker}\ and\ \citenamefont
  {Trodden}(2017)}]{Baker:2016reh}%
  \BibitemOpen
  \bibfield  {author} {\bibinfo {author} {\bibfnamefont {T.}~\bibnamefont
  {Baker}}\ and\ \bibinfo {author} {\bibfnamefont {M.}~\bibnamefont
  {Trodden}},\ }\href {\doibase 10.1103/PhysRevD.95.063512} {\bibfield
  {journal} {\bibinfo  {journal} {Phys. Rev. D}\ }\textbf {\bibinfo {volume}
  {95}},\ \bibinfo {pages} {063512} (\bibinfo {year} {2017})},\ \Eprint
  {http://arxiv.org/abs/1612.02004} {arXiv:1612.02004 [astro-ph.CO]}
  \BibitemShut {NoStop}%
\bibitem [{\citenamefont {Goyal}\ \emph {et~al.}(2021)\citenamefont {Goyal},
  \citenamefont {Haris}, \citenamefont {Mehta},\ and\ \citenamefont
  {Ajith}}]{Goyal:2020bkm}%
  \BibitemOpen
  \bibfield  {author} {\bibinfo {author} {\bibfnamefont {S.}~\bibnamefont
  {Goyal}}, \bibinfo {author} {\bibfnamefont {K.}~\bibnamefont {Haris}},
  \bibinfo {author} {\bibfnamefont {A.~K.}\ \bibnamefont {Mehta}}, \ and\
  \bibinfo {author} {\bibfnamefont {P.}~\bibnamefont {Ajith}},\ }\href
  {\doibase 10.1103/PhysRevD.103.024038} {\bibfield  {journal} {\bibinfo
  {journal} {Phys. Rev. D}\ }\textbf {\bibinfo {volume} {103}},\ \bibinfo
  {pages} {024038} (\bibinfo {year} {2021})},\ \Eprint
  {http://arxiv.org/abs/2008.07060} {arXiv:2008.07060 [gr-qc]} \BibitemShut
  {NoStop}%
\bibitem [{\citenamefont {Lai}\ \emph {et~al.}(2018)\citenamefont {Lai},
  \citenamefont {Hannuksela}, \citenamefont {Herrera-Mart\'\i{}n},
  \citenamefont {Diego}, \citenamefont {Broadhurst},\ and\ \citenamefont
  {Li}}]{Lai:2018rto}%
  \BibitemOpen
  \bibfield  {author} {\bibinfo {author} {\bibfnamefont {K.-H.}\ \bibnamefont
  {Lai}}, \bibinfo {author} {\bibfnamefont {O.~A.}\ \bibnamefont {Hannuksela}},
  \bibinfo {author} {\bibfnamefont {A.}~\bibnamefont {Herrera-Mart\'\i{}n}},
  \bibinfo {author} {\bibfnamefont {J.~M.}\ \bibnamefont {Diego}}, \bibinfo
  {author} {\bibfnamefont {T.}~\bibnamefont {Broadhurst}}, \ and\ \bibinfo
  {author} {\bibfnamefont {T.~G.~F.}\ \bibnamefont {Li}},\ }\href {\doibase
  10.1103/PhysRevD.98.083005} {\bibfield  {journal} {\bibinfo  {journal} {Phys.
  Rev. D}\ }\textbf {\bibinfo {volume} {98}},\ \bibinfo {pages} {083005}
  (\bibinfo {year} {2018})},\ \Eprint {http://arxiv.org/abs/1801.07840}
  {arXiv:1801.07840 [gr-qc]} \BibitemShut {NoStop}%
\bibitem [{\citenamefont {Diego}(2020)}]{Diego:2019rzc}%
  \BibitemOpen
  \bibfield  {author} {\bibinfo {author} {\bibfnamefont {J.~M.}\ \bibnamefont
  {Diego}},\ }\href {\doibase 10.1103/PhysRevD.101.123512} {\bibfield
  {journal} {\bibinfo  {journal} {Phys. Rev. D}\ }\textbf {\bibinfo {volume}
  {101}},\ \bibinfo {pages} {123512} (\bibinfo {year} {2020})},\ \Eprint
  {http://arxiv.org/abs/1911.05736} {arXiv:1911.05736 [astro-ph.CO]}
  \BibitemShut {NoStop}%
\bibitem [{\citenamefont {Oguri}\ and\ \citenamefont
  {Takahashi}(2020)}]{Oguri:2020ldf}%
  \BibitemOpen
  \bibfield  {author} {\bibinfo {author} {\bibfnamefont {M.}~\bibnamefont
  {Oguri}}\ and\ \bibinfo {author} {\bibfnamefont {R.}~\bibnamefont
  {Takahashi}},\ }\href {\doibase 10.3847/1538-4357/abafab} {\bibfield
  {journal} {\bibinfo  {journal} {Astrophys. J.}\ }\textbf {\bibinfo {volume}
  {901}},\ \bibinfo {pages} {58} (\bibinfo {year} {2020})},\ \Eprint
  {http://arxiv.org/abs/2007.01936} {arXiv:2007.01936 [astro-ph.CO]}
  \BibitemShut {NoStop}%
\bibitem [{\citenamefont {Xu}\ \emph {et~al.}(2022)\citenamefont {Xu},
  \citenamefont {Ezquiaga},\ and\ \citenamefont {Holz}}]{Xu:2021bfn}%
  \BibitemOpen
  \bibfield  {author} {\bibinfo {author} {\bibfnamefont {F.}~\bibnamefont
  {Xu}}, \bibinfo {author} {\bibfnamefont {J.~M.}\ \bibnamefont {Ezquiaga}}, \
  and\ \bibinfo {author} {\bibfnamefont {D.~E.}\ \bibnamefont {Holz}},\ }\href
  {\doibase 10.3847/1538-4357/ac58f8} {\bibfield  {journal} {\bibinfo
  {journal} {Astrophys. J.}\ }\textbf {\bibinfo {volume} {929}},\ \bibinfo
  {pages} {9} (\bibinfo {year} {2022})},\ \Eprint
  {http://arxiv.org/abs/2105.14390} {arXiv:2105.14390 [astro-ph.CO]}
  \BibitemShut {NoStop}%
\bibitem [{\citenamefont {Abbott}\ \emph
  {et~al.}(2023{\natexlab{c}})\citenamefont {Abbott} \emph
  {et~al.}}]{LIGOScientific:2023bwz}%
  \BibitemOpen
  \bibfield  {author} {\bibinfo {author} {\bibfnamefont {R.}~\bibnamefont
  {Abbott}} \emph {et~al.} (\bibinfo {collaboration} {LIGO Scientific, VIRGO,
  KAGRA}),\ }\href@noop {} {\  (\bibinfo {year} {2023}{\natexlab{c}})},\
  \Eprint {http://arxiv.org/abs/2304.08393} {arXiv:2304.08393 [gr-qc]}
  \BibitemShut {NoStop}%
\bibitem [{\citenamefont {Punturo}\ \emph
  {et~al.}(2010{\natexlab{b}})\citenamefont {Punturo} \emph
  {et~al.}}]{Punturo:2010zza}%
  \BibitemOpen
  \bibfield  {author} {\bibinfo {author} {\bibfnamefont {M.}~\bibnamefont
  {Punturo}} \emph {et~al.},\ }\href {\doibase 10.1088/0264-9381/27/8/084007}
  {\bibfield  {journal} {\bibinfo  {journal} {Class. Quant. Grav.}\ }\textbf
  {\bibinfo {volume} {27}},\ \bibinfo {pages} {084007} (\bibinfo {year}
  {2010}{\natexlab{b}})}\BibitemShut {NoStop}%
\bibitem [{\citenamefont {Reitze}\ \emph {et~al.}(2019)\citenamefont {Reitze}
  \emph {et~al.}}]{Reitze:2019iox}%
  \BibitemOpen
  \bibfield  {author} {\bibinfo {author} {\bibfnamefont {D.}~\bibnamefont
  {Reitze}} \emph {et~al.},\ }\href@noop {} {\bibfield  {journal} {\bibinfo
  {journal} {Bull. Am. Astron. Soc.}\ }\textbf {\bibinfo {volume} {51}},\
  \bibinfo {pages} {035} (\bibinfo {year} {2019})},\ \Eprint
  {http://arxiv.org/abs/1907.04833} {arXiv:1907.04833 [astro-ph.IM]}
  \BibitemShut {NoStop}%
\bibitem [{\citenamefont {Luo}\ \emph {et~al.}(2016{\natexlab{c}})\citenamefont
  {Luo} \emph {et~al.}}]{TianQin:2015yph}%
  \BibitemOpen
  \bibfield  {author} {\bibinfo {author} {\bibfnamefont {J.}~\bibnamefont
  {Luo}} \emph {et~al.} (\bibinfo {collaboration} {TianQin}),\ }\href {\doibase
  10.1088/0264-9381/33/3/035010} {\bibfield  {journal} {\bibinfo  {journal}
  {Class. Quant. Grav.}\ }\textbf {\bibinfo {volume} {33}},\ \bibinfo {pages}
  {035010} (\bibinfo {year} {2016}{\natexlab{c}})},\ \Eprint
  {http://arxiv.org/abs/1512.02076} {arXiv:1512.02076 [astro-ph.IM]}
  \BibitemShut {NoStop}%
\bibitem [{\citenamefont {Klein}\ \emph {et~al.}(2016)\citenamefont {Klein}
  \emph {et~al.}}]{Klein:2015hvg}%
  \BibitemOpen
  \bibfield  {author} {\bibinfo {author} {\bibfnamefont {A.}~\bibnamefont
  {Klein}} \emph {et~al.},\ }\href {\doibase 10.1103/PhysRevD.93.024003}
  {\bibfield  {journal} {\bibinfo  {journal} {Phys. Rev.}\ }\textbf {\bibinfo
  {volume} {D93}},\ \bibinfo {pages} {024003} (\bibinfo {year} {2016})},\
  \Eprint {http://arxiv.org/abs/1511.05581} {arXiv:1511.05581 [gr-qc]}
  \BibitemShut {NoStop}%
\bibitem [{\citenamefont {Gao}\ \emph {et~al.}(2022)\citenamefont {Gao},
  \citenamefont {Chen}, \citenamefont {Hu}, \citenamefont {Zhang},\ and\
  \citenamefont {Huang}}]{Gao:2021sxw}%
  \BibitemOpen
  \bibfield  {author} {\bibinfo {author} {\bibfnamefont {Z.}~\bibnamefont
  {Gao}}, \bibinfo {author} {\bibfnamefont {X.}~\bibnamefont {Chen}}, \bibinfo
  {author} {\bibfnamefont {Y.-M.}\ \bibnamefont {Hu}}, \bibinfo {author}
  {\bibfnamefont {J.-D.}\ \bibnamefont {Zhang}}, \ and\ \bibinfo {author}
  {\bibfnamefont {S.-J.}\ \bibnamefont {Huang}},\ }\href {\doibase
  10.1093/mnras/stac365} {\bibfield  {journal} {\bibinfo  {journal} {Mon. Not.
  Roy. Astron. Soc.}\ }\textbf {\bibinfo {volume} {512}},\ \bibinfo {pages} {1}
  (\bibinfo {year} {2022})},\ \Eprint {http://arxiv.org/abs/2102.10295}
  {arXiv:2102.10295 [astro-ph.CO]} \BibitemShut {NoStop}%
\bibitem [{\citenamefont {\c{C}al\i{}\c{s}kan}\ \emph
  {et~al.}(2022)\citenamefont {\c{C}al\i{}\c{s}kan}, \citenamefont {Ji},
  \citenamefont {Cotesta}, \citenamefont {Berti}, \citenamefont
  {Kamionkowski},\ and\ \citenamefont {Marsat}}]{Caliskan:2022hbu}%
  \BibitemOpen
  \bibfield  {author} {\bibinfo {author} {\bibfnamefont {M.}~\bibnamefont
  {\c{C}al\i{}\c{s}kan}}, \bibinfo {author} {\bibfnamefont {L.}~\bibnamefont
  {Ji}}, \bibinfo {author} {\bibfnamefont {R.}~\bibnamefont {Cotesta}},
  \bibinfo {author} {\bibfnamefont {E.}~\bibnamefont {Berti}}, \bibinfo
  {author} {\bibfnamefont {M.}~\bibnamefont {Kamionkowski}}, \ and\ \bibinfo
  {author} {\bibfnamefont {S.}~\bibnamefont {Marsat}},\ }\href@noop {} {\
  (\bibinfo {year} {2022})},\ \Eprint {http://arxiv.org/abs/2206.02803}
  {arXiv:2206.02803 [astro-ph.CO]} \BibitemShut {NoStop}%
\bibitem [{\citenamefont {Tambalo}\ \emph
  {et~al.}(2022{\natexlab{a}})\citenamefont {Tambalo}, \citenamefont
  {Zumalac\'arregui}, \citenamefont {Dai},\ and\ \citenamefont
  {Cheung}}]{Tambalo:2022wlm}%
  \BibitemOpen
  \bibfield  {author} {\bibinfo {author} {\bibfnamefont {G.}~\bibnamefont
  {Tambalo}}, \bibinfo {author} {\bibfnamefont {M.}~\bibnamefont
  {Zumalac\'arregui}}, \bibinfo {author} {\bibfnamefont {L.}~\bibnamefont
  {Dai}}, \ and\ \bibinfo {author} {\bibfnamefont {M.~H.-Y.}\ \bibnamefont
  {Cheung}},\ }\href@noop {} {\  (\bibinfo {year} {2022}{\natexlab{a}})},\
  \Eprint {http://arxiv.org/abs/2212.11960} {arXiv:2212.11960 [astro-ph.CO]}
  \BibitemShut {NoStop}%
\bibitem [{\citenamefont {Lin}\ \emph {et~al.}(2023)\citenamefont {Lin},
  \citenamefont {Zhang}, \citenamefont {Dai}, \citenamefont {Huang},\ and\
  \citenamefont {Mei}}]{Lin:2023ccz}%
  \BibitemOpen
  \bibfield  {author} {\bibinfo {author} {\bibfnamefont {X.-y.}\ \bibnamefont
  {Lin}}, \bibinfo {author} {\bibfnamefont {J.-d.}\ \bibnamefont {Zhang}},
  \bibinfo {author} {\bibfnamefont {L.}~\bibnamefont {Dai}}, \bibinfo {author}
  {\bibfnamefont {S.-J.}\ \bibnamefont {Huang}}, \ and\ \bibinfo {author}
  {\bibfnamefont {J.}~\bibnamefont {Mei}},\ }\href {\doibase
  10.1103/PhysRevD.108.064020} {\bibfield  {journal} {\bibinfo  {journal}
  {Phys. Rev. D}\ }\textbf {\bibinfo {volume} {108}},\ \bibinfo {pages}
  {064020} (\bibinfo {year} {2023})},\ \Eprint
  {http://arxiv.org/abs/2304.04800} {arXiv:2304.04800 [gr-qc]} \BibitemShut
  {NoStop}%
\bibitem [{\citenamefont {Levin}(1992)}]{Levin:1992}%
  \BibitemOpen
  \bibfield  {author} {\bibinfo {author} {\bibfnamefont {D.}~\bibnamefont
  {Levin}},\ }\href {https://doi.org/10.2307/2007287} {\bibfield  {journal}
  {\bibinfo  {journal} {Mathematics of Computation, Vol. 38, No. 158 (Apr.,
  1982), pp. 531-538 (8 pages)}\ } (\bibinfo {year} {1992})}\BibitemShut
  {NoStop}%
\bibitem [{\citenamefont {Takahashi}(2004)}]{Takahashi:2004mc}%
  \BibitemOpen
  \bibfield  {author} {\bibinfo {author} {\bibfnamefont {R.}~\bibnamefont
  {Takahashi}},\ }\href {\doibase 10.1051/0004-6361:20040212} {\bibfield
  {journal} {\bibinfo  {journal} {Astron. Astrophys.}\ }\textbf {\bibinfo
  {volume} {423}},\ \bibinfo {pages} {787} (\bibinfo {year} {2004})},\ \Eprint
  {http://arxiv.org/abs/astro-ph/0402165} {arXiv:astro-ph/0402165} \BibitemShut
  {NoStop}%
\bibitem [{\citenamefont {Alfredo Dea\~{n}o}(2017)}]{Alfredo:2017}%
  \BibitemOpen
  \bibfield  {author} {\bibinfo {author} {\bibfnamefont {A.~I.}\ \bibnamefont
  {Alfredo Dea\~{n}o}, \bibfnamefont {Daan~Huybrechs}},\ }\href
  {https://doi.org/10.1137/1.9781611975123.ch2} {\emph {\bibinfo {title}
  {{Chapter 2: Asymptotic theory of highly oscillatory integrals}}}}\ (\bibinfo
  {year} {2017})\BibitemShut {NoStop}%
\bibitem [{\citenamefont {Guo}\ and\ \citenamefont {Lu}(2020)}]{Guo:2020eqw}%
  \BibitemOpen
  \bibfield  {author} {\bibinfo {author} {\bibfnamefont {X.}~\bibnamefont
  {Guo}}\ and\ \bibinfo {author} {\bibfnamefont {Y.}~\bibnamefont {Lu}},\
  }\href {\doibase 10.1103/PhysRevD.102.124076} {\bibfield  {journal} {\bibinfo
   {journal} {Phys. Rev. D}\ }\textbf {\bibinfo {volume} {102}},\ \bibinfo
  {pages} {124076} (\bibinfo {year} {2020})},\ \Eprint
  {http://arxiv.org/abs/2012.03474} {arXiv:2012.03474 [gr-qc]} \BibitemShut
  {NoStop}%
\bibitem [{\citenamefont {Tambalo}\ \emph
  {et~al.}(2022{\natexlab{b}})\citenamefont {Tambalo}, \citenamefont
  {Zumalac\'arregui}, \citenamefont {Dai},\ and\ \citenamefont
  {Cheung}}]{Tambalo:2022plm}%
  \BibitemOpen
  \bibfield  {author} {\bibinfo {author} {\bibfnamefont {G.}~\bibnamefont
  {Tambalo}}, \bibinfo {author} {\bibfnamefont {M.}~\bibnamefont
  {Zumalac\'arregui}}, \bibinfo {author} {\bibfnamefont {L.}~\bibnamefont
  {Dai}}, \ and\ \bibinfo {author} {\bibfnamefont {M.~H.-Y.}\ \bibnamefont
  {Cheung}},\ }\href@noop {} {\  (\bibinfo {year} {2022}{\natexlab{b}})},\
  \Eprint {http://arxiv.org/abs/2210.05658} {arXiv:2210.05658 [gr-qc]}
  \BibitemShut {NoStop}%
\bibitem [{\citenamefont {Navarro}\ \emph {et~al.}(1995)\citenamefont
  {Navarro}, \citenamefont {Frenk},\ and\ \citenamefont
  {White}}]{Navarro:1994hi}%
  \BibitemOpen
  \bibfield  {author} {\bibinfo {author} {\bibfnamefont {J.~F.}\ \bibnamefont
  {Navarro}}, \bibinfo {author} {\bibfnamefont {C.~S.}\ \bibnamefont {Frenk}},
  \ and\ \bibinfo {author} {\bibfnamefont {S.~D.~M.}\ \bibnamefont {White}},\
  }\href {\doibase 10.1093/mnras/275.3.720} {\bibfield  {journal} {\bibinfo
  {journal} {Mon. Not. Roy. Astron. Soc.}\ }\textbf {\bibinfo {volume} {275}},\
  \bibinfo {pages} {720} (\bibinfo {year} {1995})},\ \Eprint
  {http://arxiv.org/abs/astro-ph/9408069} {arXiv:astro-ph/9408069} \BibitemShut
  {NoStop}%
\bibitem [{\citenamefont {Ruan}\ \emph {et~al.}(2020)\citenamefont {Ruan},
  \citenamefont {Guo}, \citenamefont {Cai},\ and\ \citenamefont
  {Zhang}}]{Ruan:2018tsw}%
  \BibitemOpen
  \bibfield  {author} {\bibinfo {author} {\bibfnamefont {W.-H.}\ \bibnamefont
  {Ruan}}, \bibinfo {author} {\bibfnamefont {Z.-K.}\ \bibnamefont {Guo}},
  \bibinfo {author} {\bibfnamefont {R.-G.}\ \bibnamefont {Cai}}, \ and\
  \bibinfo {author} {\bibfnamefont {Y.-Z.}\ \bibnamefont {Zhang}},\ }\href
  {\doibase 10.1142/S0217751X2050075X} {\bibfield  {journal} {\bibinfo
  {journal} {Int. J. Mod. Phys. A}\ }\textbf {\bibinfo {volume} {35}},\
  \bibinfo {pages} {2050075} (\bibinfo {year} {2020})},\ \Eprint
  {http://arxiv.org/abs/1807.09495} {arXiv:1807.09495 [gr-qc]} \BibitemShut
  {NoStop}%
\bibitem [{\citenamefont {{Amaro-Seoane}}\ \emph {et~al.}(2017)\citenamefont
  {{Amaro-Seoane}}, \citenamefont {{Audley}}, \citenamefont {{Babak}},
  \citenamefont {{Baker}}, \citenamefont {{Barausse}}, \citenamefont
  {{Bender}}, \citenamefont {{Berti}}, \citenamefont {{Binetruy}},
  \citenamefont {{Born}}, \citenamefont {{Bortoluzzi}}, \citenamefont {{Camp}},
  \citenamefont {{Caprini}}, \citenamefont {{Cardoso}}, \citenamefont
  {{Colpi}}, \citenamefont {{Conklin}}, \citenamefont {{Cornish}},
  \citenamefont {{Cutler}}, \citenamefont {{Danzmann}}, \citenamefont
  {{Dolesi}}, \citenamefont {{Ferraioli}}, \citenamefont {{Ferroni}},
  \citenamefont {{Fitzsimons}}, \citenamefont {{Gair}}, \citenamefont {{Gesa
  Bote}}, \citenamefont {{Giardini}}, \citenamefont {{Gibert}}, \citenamefont
  {{Grimani}}, \citenamefont {{Halloin}}, \citenamefont {{Heinzel}},
  \citenamefont {{Hertog}}, \citenamefont {{Hewitson}}, \citenamefont
  {{Holley-Bockelmann}}, \citenamefont {{Hollington}}, \citenamefont
  {{Hueller}}, \citenamefont {{Inchauspe}}, \citenamefont {{Jetzer}},
  \citenamefont {{Karnesis}}, \citenamefont {{Killow}}, \citenamefont
  {{Klein}}, \citenamefont {{Klipstein}}, \citenamefont {{Korsakova}},
  \citenamefont {{Larson}}, \citenamefont {{Livas}}, \citenamefont {{Lloro}},
  \citenamefont {{Man}}, \citenamefont {{Mance}}, \citenamefont {{Martino}},
  \citenamefont {{Mateos}}, \citenamefont {{McKenzie}}, \citenamefont
  {{McWilliams}}, \citenamefont {{Miller}}, \citenamefont {{Mueller}},
  \citenamefont {{Nardini}}, \citenamefont {{Nelemans}}, \citenamefont
  {{Nofrarias}}, \citenamefont {{Petiteau}}, \citenamefont {{Pivato}},
  \citenamefont {{Plagnol}}, \citenamefont {{Porter}}, \citenamefont
  {{Reiche}}, \citenamefont {{Robertson}}, \citenamefont {{Robertson}},
  \citenamefont {{Rossi}}, \citenamefont {{Russano}}, \citenamefont {{Schutz}},
  \citenamefont {{Sesana}}, \citenamefont {{Shoemaker}}, \citenamefont
  {{Slutsky}}, \citenamefont {{Sopuerta}}, \citenamefont {{Sumner}},
  \citenamefont {{Tamanini}}, \citenamefont {{Thorpe}}, \citenamefont
  {{Troebs}}, \citenamefont {{Vallisneri}}, \citenamefont {{Vecchio}},
  \citenamefont {{Vetrugno}}, \citenamefont {{Vitale}}, \citenamefont
  {{Volonteri}}, \citenamefont {{Wanner}}, \citenamefont {{Ward}},
  \citenamefont {{Wass}}, \citenamefont {{Weber}}, \citenamefont {{Ziemer}},\
  and\ \citenamefont {{Zweifel}}}]{2017arXiv170200786A}%
  \BibitemOpen
  \bibfield  {author} {\bibinfo {author} {\bibfnamefont {P.}~\bibnamefont
  {{Amaro-Seoane}}}, \bibinfo {author} {\bibfnamefont {H.}~\bibnamefont
  {{Audley}}}, \bibinfo {author} {\bibfnamefont {S.}~\bibnamefont {{Babak}}},
  \bibinfo {author} {\bibfnamefont {J.}~\bibnamefont {{Baker}}}, \bibinfo
  {author} {\bibfnamefont {E.}~\bibnamefont {{Barausse}}}, \bibinfo {author}
  {\bibfnamefont {P.}~\bibnamefont {{Bender}}}, \bibinfo {author}
  {\bibfnamefont {E.}~\bibnamefont {{Berti}}}, \bibinfo {author} {\bibfnamefont
  {P.}~\bibnamefont {{Binetruy}}}, \bibinfo {author} {\bibfnamefont
  {M.}~\bibnamefont {{Born}}}, \bibinfo {author} {\bibfnamefont
  {D.}~\bibnamefont {{Bortoluzzi}}}, \bibinfo {author} {\bibfnamefont
  {J.}~\bibnamefont {{Camp}}}, \bibinfo {author} {\bibfnamefont
  {C.}~\bibnamefont {{Caprini}}}, \bibinfo {author} {\bibfnamefont
  {V.}~\bibnamefont {{Cardoso}}}, \bibinfo {author} {\bibfnamefont
  {M.}~\bibnamefont {{Colpi}}}, \bibinfo {author} {\bibfnamefont
  {J.}~\bibnamefont {{Conklin}}}, \bibinfo {author} {\bibfnamefont
  {N.}~\bibnamefont {{Cornish}}}, \bibinfo {author} {\bibfnamefont
  {C.}~\bibnamefont {{Cutler}}}, \bibinfo {author} {\bibfnamefont
  {K.}~\bibnamefont {{Danzmann}}}, \bibinfo {author} {\bibfnamefont
  {R.}~\bibnamefont {{Dolesi}}}, \bibinfo {author} {\bibfnamefont
  {L.}~\bibnamefont {{Ferraioli}}}, \bibinfo {author} {\bibfnamefont
  {V.}~\bibnamefont {{Ferroni}}}, \bibinfo {author} {\bibfnamefont
  {E.}~\bibnamefont {{Fitzsimons}}}, \bibinfo {author} {\bibfnamefont
  {J.}~\bibnamefont {{Gair}}}, \bibinfo {author} {\bibfnamefont
  {L.}~\bibnamefont {{Gesa Bote}}}, \bibinfo {author} {\bibfnamefont
  {D.}~\bibnamefont {{Giardini}}}, \bibinfo {author} {\bibfnamefont
  {F.}~\bibnamefont {{Gibert}}}, \bibinfo {author} {\bibfnamefont
  {C.}~\bibnamefont {{Grimani}}}, \bibinfo {author} {\bibfnamefont
  {H.}~\bibnamefont {{Halloin}}}, \bibinfo {author} {\bibfnamefont
  {G.}~\bibnamefont {{Heinzel}}}, \bibinfo {author} {\bibfnamefont
  {T.}~\bibnamefont {{Hertog}}}, \bibinfo {author} {\bibfnamefont
  {M.}~\bibnamefont {{Hewitson}}}, \bibinfo {author} {\bibfnamefont
  {K.}~\bibnamefont {{Holley-Bockelmann}}}, \bibinfo {author} {\bibfnamefont
  {D.}~\bibnamefont {{Hollington}}}, \bibinfo {author} {\bibfnamefont
  {M.}~\bibnamefont {{Hueller}}}, \bibinfo {author} {\bibfnamefont
  {H.}~\bibnamefont {{Inchauspe}}}, \bibinfo {author} {\bibfnamefont
  {P.}~\bibnamefont {{Jetzer}}}, \bibinfo {author} {\bibfnamefont
  {N.}~\bibnamefont {{Karnesis}}}, \bibinfo {author} {\bibfnamefont
  {C.}~\bibnamefont {{Killow}}}, \bibinfo {author} {\bibfnamefont
  {A.}~\bibnamefont {{Klein}}}, \bibinfo {author} {\bibfnamefont
  {B.}~\bibnamefont {{Klipstein}}}, \bibinfo {author} {\bibfnamefont
  {N.}~\bibnamefont {{Korsakova}}}, \bibinfo {author} {\bibfnamefont {S.~L.}\
  \bibnamefont {{Larson}}}, \bibinfo {author} {\bibfnamefont {J.}~\bibnamefont
  {{Livas}}}, \bibinfo {author} {\bibfnamefont {I.}~\bibnamefont {{Lloro}}},
  \bibinfo {author} {\bibfnamefont {N.}~\bibnamefont {{Man}}}, \bibinfo
  {author} {\bibfnamefont {D.}~\bibnamefont {{Mance}}}, \bibinfo {author}
  {\bibfnamefont {J.}~\bibnamefont {{Martino}}}, \bibinfo {author}
  {\bibfnamefont {I.}~\bibnamefont {{Mateos}}}, \bibinfo {author}
  {\bibfnamefont {K.}~\bibnamefont {{McKenzie}}}, \bibinfo {author}
  {\bibfnamefont {S.~T.}\ \bibnamefont {{McWilliams}}}, \bibinfo {author}
  {\bibfnamefont {C.}~\bibnamefont {{Miller}}}, \bibinfo {author}
  {\bibfnamefont {G.}~\bibnamefont {{Mueller}}}, \bibinfo {author}
  {\bibfnamefont {G.}~\bibnamefont {{Nardini}}}, \bibinfo {author}
  {\bibfnamefont {G.}~\bibnamefont {{Nelemans}}}, \bibinfo {author}
  {\bibfnamefont {M.}~\bibnamefont {{Nofrarias}}}, \bibinfo {author}
  {\bibfnamefont {A.}~\bibnamefont {{Petiteau}}}, \bibinfo {author}
  {\bibfnamefont {P.}~\bibnamefont {{Pivato}}}, \bibinfo {author}
  {\bibfnamefont {E.}~\bibnamefont {{Plagnol}}}, \bibinfo {author}
  {\bibfnamefont {E.}~\bibnamefont {{Porter}}}, \bibinfo {author}
  {\bibfnamefont {J.}~\bibnamefont {{Reiche}}}, \bibinfo {author}
  {\bibfnamefont {D.}~\bibnamefont {{Robertson}}}, \bibinfo {author}
  {\bibfnamefont {N.}~\bibnamefont {{Robertson}}}, \bibinfo {author}
  {\bibfnamefont {E.}~\bibnamefont {{Rossi}}}, \bibinfo {author} {\bibfnamefont
  {G.}~\bibnamefont {{Russano}}}, \bibinfo {author} {\bibfnamefont
  {B.}~\bibnamefont {{Schutz}}}, \bibinfo {author} {\bibfnamefont
  {A.}~\bibnamefont {{Sesana}}}, \bibinfo {author} {\bibfnamefont
  {D.}~\bibnamefont {{Shoemaker}}}, \bibinfo {author} {\bibfnamefont
  {J.}~\bibnamefont {{Slutsky}}}, \bibinfo {author} {\bibfnamefont {C.~F.}\
  \bibnamefont {{Sopuerta}}}, \bibinfo {author} {\bibfnamefont
  {T.}~\bibnamefont {{Sumner}}}, \bibinfo {author} {\bibfnamefont
  {N.}~\bibnamefont {{Tamanini}}}, \bibinfo {author} {\bibfnamefont
  {I.}~\bibnamefont {{Thorpe}}}, \bibinfo {author} {\bibfnamefont
  {M.}~\bibnamefont {{Troebs}}}, \bibinfo {author} {\bibfnamefont
  {M.}~\bibnamefont {{Vallisneri}}}, \bibinfo {author} {\bibfnamefont
  {A.}~\bibnamefont {{Vecchio}}}, \bibinfo {author} {\bibfnamefont
  {D.}~\bibnamefont {{Vetrugno}}}, \bibinfo {author} {\bibfnamefont
  {S.}~\bibnamefont {{Vitale}}}, \bibinfo {author} {\bibfnamefont
  {M.}~\bibnamefont {{Volonteri}}}, \bibinfo {author} {\bibfnamefont
  {G.}~\bibnamefont {{Wanner}}}, \bibinfo {author} {\bibfnamefont
  {H.}~\bibnamefont {{Ward}}}, \bibinfo {author} {\bibfnamefont
  {P.}~\bibnamefont {{Wass}}}, \bibinfo {author} {\bibfnamefont
  {W.}~\bibnamefont {{Weber}}}, \bibinfo {author} {\bibfnamefont
  {J.}~\bibnamefont {{Ziemer}}}, \ and\ \bibinfo {author} {\bibfnamefont
  {P.}~\bibnamefont {{Zweifel}}},\ }\href {\doibase 10.48550/arXiv.1702.00786}
  {\bibfield  {journal} {\bibinfo  {journal} {arXiv e-prints}\ ,\ \bibinfo
  {eid} {arXiv:1702.00786}} (\bibinfo {year} {2017})},\ \Eprint
  {http://arxiv.org/abs/1702.00786} {arXiv:1702.00786 [astro-ph.IM]}
  \BibitemShut {NoStop}%
\bibitem [{\citenamefont {Kawamura}\ \emph {et~al.}(2021)\citenamefont
  {Kawamura} \emph {et~al.}}]{Kawamura:2020pcg}%
  \BibitemOpen
  \bibfield  {author} {\bibinfo {author} {\bibfnamefont {S.}~\bibnamefont
  {Kawamura}} \emph {et~al.},\ }\href {\doibase 10.1093/ptep/ptab019}
  {\bibfield  {journal} {\bibinfo  {journal} {PTEP}\ }\textbf {\bibinfo
  {volume} {2021}},\ \bibinfo {pages} {05A105} (\bibinfo {year} {2021})},\
  \Eprint {http://arxiv.org/abs/2006.13545} {arXiv:2006.13545 [gr-qc]}
  \BibitemShut {NoStop}%
\bibitem [{\citenamefont {{Bailes}}\ \emph {et~al.}(2021)\citenamefont
  {{Bailes}}, \citenamefont {{Berger}}, \citenamefont {{Brady}}, \citenamefont
  {{Branchesi}}, \citenamefont {{Danzmann}}, \citenamefont {{Evans}},
  \citenamefont {{Holley-Bockelmann}}, \citenamefont {{Iyer}}, \citenamefont
  {{Kajita}}, \citenamefont {{Katsanevas}}, \citenamefont {{Kramer}},
  \citenamefont {{Lazzarini}}, \citenamefont {{Lehner}}, \citenamefont
  {{Losurdo}}, \citenamefont {{L{\"u}ck}}, \citenamefont {{McClelland}},
  \citenamefont {{McLaughlin}}, \citenamefont {{Punturo}}, \citenamefont
  {{Ransom}}, \citenamefont {{Raychaudhury}}, \citenamefont {{Reitze}},
  \citenamefont {{Ricci}}, \citenamefont {{Rowan}}, \citenamefont {{Saito}},
  \citenamefont {{Sanders}}, \citenamefont {{Sathyaprakash}}, \citenamefont
  {{Schutz}}, \citenamefont {{Sesana}}, \citenamefont {{Shinkai}},
  \citenamefont {{Siemens}}, \citenamefont {{Shoemaker}}, \citenamefont
  {{Thorpe}}, \citenamefont {{van den Brand}},\ and\ \citenamefont
  {{Vitale}}}]{2021NatRP...3..344B}%
  \BibitemOpen
  \bibfield  {author} {\bibinfo {author} {\bibfnamefont {M.}~\bibnamefont
  {{Bailes}}}, \bibinfo {author} {\bibfnamefont {B.~K.}\ \bibnamefont
  {{Berger}}}, \bibinfo {author} {\bibfnamefont {P.~R.}\ \bibnamefont
  {{Brady}}}, \bibinfo {author} {\bibfnamefont {M.}~\bibnamefont
  {{Branchesi}}}, \bibinfo {author} {\bibfnamefont {K.}~\bibnamefont
  {{Danzmann}}}, \bibinfo {author} {\bibfnamefont {M.}~\bibnamefont {{Evans}}},
  \bibinfo {author} {\bibfnamefont {K.}~\bibnamefont {{Holley-Bockelmann}}},
  \bibinfo {author} {\bibfnamefont {B.~R.}\ \bibnamefont {{Iyer}}}, \bibinfo
  {author} {\bibfnamefont {T.}~\bibnamefont {{Kajita}}}, \bibinfo {author}
  {\bibfnamefont {S.}~\bibnamefont {{Katsanevas}}}, \bibinfo {author}
  {\bibfnamefont {M.}~\bibnamefont {{Kramer}}}, \bibinfo {author}
  {\bibfnamefont {A.}~\bibnamefont {{Lazzarini}}}, \bibinfo {author}
  {\bibfnamefont {L.}~\bibnamefont {{Lehner}}}, \bibinfo {author}
  {\bibfnamefont {G.}~\bibnamefont {{Losurdo}}}, \bibinfo {author}
  {\bibfnamefont {H.}~\bibnamefont {{L{\"u}ck}}}, \bibinfo {author}
  {\bibfnamefont {D.~E.}\ \bibnamefont {{McClelland}}}, \bibinfo {author}
  {\bibfnamefont {M.~A.}\ \bibnamefont {{McLaughlin}}}, \bibinfo {author}
  {\bibfnamefont {M.}~\bibnamefont {{Punturo}}}, \bibinfo {author}
  {\bibfnamefont {S.}~\bibnamefont {{Ransom}}}, \bibinfo {author}
  {\bibfnamefont {S.}~\bibnamefont {{Raychaudhury}}}, \bibinfo {author}
  {\bibfnamefont {D.~H.}\ \bibnamefont {{Reitze}}}, \bibinfo {author}
  {\bibfnamefont {F.}~\bibnamefont {{Ricci}}}, \bibinfo {author} {\bibfnamefont
  {S.}~\bibnamefont {{Rowan}}}, \bibinfo {author} {\bibfnamefont
  {Y.}~\bibnamefont {{Saito}}}, \bibinfo {author} {\bibfnamefont {G.~H.}\
  \bibnamefont {{Sanders}}}, \bibinfo {author} {\bibfnamefont {B.~S.}\
  \bibnamefont {{Sathyaprakash}}}, \bibinfo {author} {\bibfnamefont {B.~F.}\
  \bibnamefont {{Schutz}}}, \bibinfo {author} {\bibfnamefont {A.}~\bibnamefont
  {{Sesana}}}, \bibinfo {author} {\bibfnamefont {H.}~\bibnamefont {{Shinkai}}},
  \bibinfo {author} {\bibfnamefont {X.}~\bibnamefont {{Siemens}}}, \bibinfo
  {author} {\bibfnamefont {D.~H.}\ \bibnamefont {{Shoemaker}}}, \bibinfo
  {author} {\bibfnamefont {J.}~\bibnamefont {{Thorpe}}}, \bibinfo {author}
  {\bibfnamefont {J.~F.~J.}\ \bibnamefont {{van den Brand}}}, \ and\ \bibinfo
  {author} {\bibfnamefont {S.}~\bibnamefont {{Vitale}}},\ }\href {\doibase
  10.1038/s42254-021-00303-8} {\bibfield  {journal} {\bibinfo  {journal}
  {Nature Reviews Physics}\ }\textbf {\bibinfo {volume} {3}},\ \bibinfo {pages}
  {344} (\bibinfo {year} {2021})}\BibitemShut {NoStop}%
\bibitem [{\citenamefont {Lyth}\ and\ \citenamefont
  {Liddle}(2009)}]{Lyth_Liddle_2009}%
  \BibitemOpen
  \bibfield  {author} {\bibinfo {author} {\bibfnamefont {D.~H.}\ \bibnamefont
  {Lyth}}\ and\ \bibinfo {author} {\bibfnamefont {A.~R.}\ \bibnamefont
  {Liddle}},\ }\href@noop {} {\emph {\bibinfo {title} {The Primordial Density
  Perturbation: Cosmology, Inflation and the Origin of Structure}}}\ (\bibinfo
  {publisher} {Cambridge University Press},\ \bibinfo {year}
  {2009})\BibitemShut {NoStop}%
\bibitem [{\citenamefont {Isaacson}(1968)}]{Isaacson:1968hbi}%
  \BibitemOpen
  \bibfield  {author} {\bibinfo {author} {\bibfnamefont {R.~A.}\ \bibnamefont
  {Isaacson}},\ }\href {\doibase 10.1103/PhysRev.166.1263} {\bibfield
  {journal} {\bibinfo  {journal} {Phys. Rev.}\ }\textbf {\bibinfo {volume}
  {166}},\ \bibinfo {pages} {1263} (\bibinfo {year} {1968})}\BibitemShut
  {NoStop}%
\bibitem [{\citenamefont {Laguna}\ \emph {et~al.}(2010)\citenamefont {Laguna},
  \citenamefont {Larson}, \citenamefont {Spergel},\ and\ \citenamefont
  {Yunes}}]{Laguna:2009re}%
  \BibitemOpen
  \bibfield  {author} {\bibinfo {author} {\bibfnamefont {P.}~\bibnamefont
  {Laguna}}, \bibinfo {author} {\bibfnamefont {S.~L.}\ \bibnamefont {Larson}},
  \bibinfo {author} {\bibfnamefont {D.}~\bibnamefont {Spergel}}, \ and\
  \bibinfo {author} {\bibfnamefont {N.}~\bibnamefont {Yunes}},\ }\href
  {\doibase 10.1088/2041-8205/715/1/L12} {\bibfield  {journal} {\bibinfo
  {journal} {Astrophys. J. Lett.}\ }\textbf {\bibinfo {volume} {715}},\
  \bibinfo {pages} {L12} (\bibinfo {year} {2010})},\ \Eprint
  {http://arxiv.org/abs/0905.1908} {arXiv:0905.1908 [gr-qc]} \BibitemShut
  {NoStop}%
\bibitem [{\citenamefont {Fier}\ \emph {et~al.}(2021)\citenamefont {Fier},
  \citenamefont {Fang}, \citenamefont {Li}, \citenamefont {Mukohyama},
  \citenamefont {Wang},\ and\ \citenamefont {Zhu}}]{Fier:2021fbt}%
  \BibitemOpen
  \bibfield  {author} {\bibinfo {author} {\bibfnamefont {J.}~\bibnamefont
  {Fier}}, \bibinfo {author} {\bibfnamefont {X.}~\bibnamefont {Fang}}, \bibinfo
  {author} {\bibfnamefont {B.}~\bibnamefont {Li}}, \bibinfo {author}
  {\bibfnamefont {S.}~\bibnamefont {Mukohyama}}, \bibinfo {author}
  {\bibfnamefont {A.}~\bibnamefont {Wang}}, \ and\ \bibinfo {author}
  {\bibfnamefont {T.}~\bibnamefont {Zhu}},\ }\href {\doibase
  10.1103/PhysRevD.103.123021} {\bibfield  {journal} {\bibinfo  {journal}
  {Phys. Rev. D}\ }\textbf {\bibinfo {volume} {103}},\ \bibinfo {pages}
  {123021} (\bibinfo {year} {2021})},\ \Eprint
  {http://arxiv.org/abs/2102.08968} {arXiv:2102.08968 [astro-ph.CO]}
  \BibitemShut {NoStop}%
\bibitem [{\citenamefont {Garoffolo}\ \emph {et~al.}(2020)\citenamefont
  {Garoffolo}, \citenamefont {Tasinato}, \citenamefont {Carbone}, \citenamefont
  {Bertacca},\ and\ \citenamefont {Matarrese}}]{Garoffolo:2019mna}%
  \BibitemOpen
  \bibfield  {author} {\bibinfo {author} {\bibfnamefont {A.}~\bibnamefont
  {Garoffolo}}, \bibinfo {author} {\bibfnamefont {G.}~\bibnamefont {Tasinato}},
  \bibinfo {author} {\bibfnamefont {C.}~\bibnamefont {Carbone}}, \bibinfo
  {author} {\bibfnamefont {D.}~\bibnamefont {Bertacca}}, \ and\ \bibinfo
  {author} {\bibfnamefont {S.}~\bibnamefont {Matarrese}},\ }\href {\doibase
  10.1088/1475-7516/2020/11/040} {\bibfield  {journal} {\bibinfo  {journal}
  {JCAP}\ }\textbf {\bibinfo {volume} {11}},\ \bibinfo {pages} {040} (\bibinfo
  {year} {2020})},\ \Eprint {http://arxiv.org/abs/1912.08093} {arXiv:1912.08093
  [gr-qc]} \BibitemShut {NoStop}%
\bibitem [{\citenamefont {Dalang}\ \emph {et~al.}(2021)\citenamefont {Dalang},
  \citenamefont {Fleury},\ and\ \citenamefont {Lombriser}}]{Dalang:2020eaj}%
  \BibitemOpen
  \bibfield  {author} {\bibinfo {author} {\bibfnamefont {C.}~\bibnamefont
  {Dalang}}, \bibinfo {author} {\bibfnamefont {P.}~\bibnamefont {Fleury}}, \
  and\ \bibinfo {author} {\bibfnamefont {L.}~\bibnamefont {Lombriser}},\ }\href
  {\doibase 10.1103/PhysRevD.103.064075} {\bibfield  {journal} {\bibinfo
  {journal} {Phys. Rev. D}\ }\textbf {\bibinfo {volume} {103}},\ \bibinfo
  {pages} {064075} (\bibinfo {year} {2021})},\ \Eprint
  {http://arxiv.org/abs/2009.11827} {arXiv:2009.11827 [gr-qc]} \BibitemShut
  {NoStop}%
\bibitem [{\citenamefont {Kubota}\ \emph {et~al.}(2023)\citenamefont {Kubota},
  \citenamefont {Arai},\ and\ \citenamefont {Mukohyama}}]{Kubota:2022lbn}%
  \BibitemOpen
  \bibfield  {author} {\bibinfo {author} {\bibfnamefont {K.-i.}\ \bibnamefont
  {Kubota}}, \bibinfo {author} {\bibfnamefont {S.}~\bibnamefont {Arai}}, \ and\
  \bibinfo {author} {\bibfnamefont {S.}~\bibnamefont {Mukohyama}},\ }\href
  {\doibase 10.1103/PhysRevD.107.064002} {\bibfield  {journal} {\bibinfo
  {journal} {Phys. Rev. D}\ }\textbf {\bibinfo {volume} {107}},\ \bibinfo
  {pages} {064002} (\bibinfo {year} {2023})},\ \Eprint
  {http://arxiv.org/abs/2209.00795} {arXiv:2209.00795 [gr-qc]} \BibitemShut
  {NoStop}%
\bibitem [{\citenamefont {Fier}\ \emph {et~al.}(2024)\citenamefont {Fier},
  \citenamefont {Han}, \citenamefont {Li}, \citenamefont {Lin}, \citenamefont
  {Mukohyama},\ and\ \citenamefont {Wang}}]{Fier:2024}%
  \BibitemOpen
  \bibfield  {author} {\bibinfo {author} {\bibfnamefont {J.}~\bibnamefont
  {Fier}}, \bibinfo {author} {\bibfnamefont {H.}~\bibnamefont {Han}}, \bibinfo
  {author} {\bibfnamefont {B.}~\bibnamefont {Li}}, \bibinfo {author}
  {\bibfnamefont {K.}~\bibnamefont {Lin}}, \bibinfo {author} {\bibfnamefont
  {S.}~\bibnamefont {Mukohyama}}, \ and\ \bibinfo {author} {\bibfnamefont
  {A.}~\bibnamefont {Wang}},\ }\href@noop {} {\bibfield  {journal} {\bibinfo
  {journal} {(in preparation)}\ } (\bibinfo {year} {2024})}\BibitemShut
  {NoStop}%
\bibitem [{\citenamefont {Dodelson}(2003)}]{Dodelson:2003ft}%
  \BibitemOpen
  \bibfield  {author} {\bibinfo {author} {\bibfnamefont {S.}~\bibnamefont
  {Dodelson}},\ }\href@noop {} {\emph {\bibinfo {title} {{Modern Cosmology}}}}\
  (\bibinfo  {publisher} {Academic Press},\ \bibinfo {address} {Amsterdam},\
  \bibinfo {year} {2003})\BibitemShut {NoStop}%
\bibitem [{\citenamefont {Feng}\ \emph {et~al.}(2019)\citenamefont {Feng},
  \citenamefont {Wang}, \citenamefont {Hu}, \citenamefont {Hu},\ and\
  \citenamefont {Wang}}]{Feng:2019wgq}%
  \BibitemOpen
  \bibfield  {author} {\bibinfo {author} {\bibfnamefont {W.-F.}\ \bibnamefont
  {Feng}}, \bibinfo {author} {\bibfnamefont {H.-T.}\ \bibnamefont {Wang}},
  \bibinfo {author} {\bibfnamefont {X.-C.}\ \bibnamefont {Hu}}, \bibinfo
  {author} {\bibfnamefont {Y.-M.}\ \bibnamefont {Hu}}, \ and\ \bibinfo {author}
  {\bibfnamefont {Y.}~\bibnamefont {Wang}},\ }\href {\doibase
  10.1103/PhysRevD.99.123002} {\bibfield  {journal} {\bibinfo  {journal} {Phys.
  Rev. D}\ }\textbf {\bibinfo {volume} {99}},\ \bibinfo {pages} {123002}
  (\bibinfo {year} {2019})},\ \Eprint {http://arxiv.org/abs/1901.02159}
  {arXiv:1901.02159 [astro-ph.IM]} \BibitemShut {NoStop}%
\bibitem [{\citenamefont {Zhu}\ \emph {et~al.}(2022{\natexlab{a}})\citenamefont
  {Zhu}, \citenamefont {Hu}, \citenamefont {Wang}, \citenamefont {Zhang},
  \citenamefont {Li}, \citenamefont {Hendry},\ and\ \citenamefont
  {Mei}}]{Zhu:2021aat}%
  \BibitemOpen
  \bibfield  {author} {\bibinfo {author} {\bibfnamefont {L.-G.}\ \bibnamefont
  {Zhu}}, \bibinfo {author} {\bibfnamefont {Y.-M.}\ \bibnamefont {Hu}},
  \bibinfo {author} {\bibfnamefont {H.-T.}\ \bibnamefont {Wang}}, \bibinfo
  {author} {\bibfnamefont {J.-d.}\ \bibnamefont {Zhang}}, \bibinfo {author}
  {\bibfnamefont {X.-D.}\ \bibnamefont {Li}}, \bibinfo {author} {\bibfnamefont
  {M.}~\bibnamefont {Hendry}}, \ and\ \bibinfo {author} {\bibfnamefont
  {J.}~\bibnamefont {Mei}},\ }\href {\doibase 10.1103/PhysRevResearch.4.013247}
  {\bibfield  {journal} {\bibinfo  {journal} {Phys. Rev. Res.}\ }\textbf
  {\bibinfo {volume} {4}},\ \bibinfo {pages} {013247} (\bibinfo {year}
  {2022}{\natexlab{a}})},\ \Eprint {http://arxiv.org/abs/2104.11956}
  {arXiv:2104.11956 [astro-ph.CO]} \BibitemShut {NoStop}%
\bibitem [{\citenamefont {Gupta}\ \emph {et~al.}(2024)\citenamefont {Gupta}
  \emph {et~al.}}]{Gupta:2024gun}%
  \BibitemOpen
  \bibfield  {author} {\bibinfo {author} {\bibfnamefont {A.}~\bibnamefont
  {Gupta}} \emph {et~al.},\ }\href@noop {} {\  (\bibinfo {year} {2024})},\
  \Eprint {http://arxiv.org/abs/2405.02197} {arXiv:2405.02197 [gr-qc]}
  \BibitemShut {NoStop}%
\bibitem [{\citenamefont {Afshordi}\ \emph {et~al.}(2023)\citenamefont
  {Afshordi} \emph {et~al.}}]{LISAConsortiumWaveformWorkingGroup:2023arg}%
  \BibitemOpen
  \bibfield  {author} {\bibinfo {author} {\bibfnamefont {N.}~\bibnamefont
  {Afshordi}} \emph {et~al.} (\bibinfo {collaboration} {LISA Consortium
  Waveform Working Group}),\ }\href@noop {} {\  (\bibinfo {year} {2023})},\
  \Eprint {http://arxiv.org/abs/2311.01300} {arXiv:2311.01300 [gr-qc]}
  \BibitemShut {NoStop}%
\bibitem [{\citenamefont {Boyle}\ \emph {et~al.}(2019)\citenamefont {Boyle}
  \emph {et~al.}}]{Boyle:2019kee}%
  \BibitemOpen
  \bibfield  {author} {\bibinfo {author} {\bibfnamefont {M.}~\bibnamefont
  {Boyle}} \emph {et~al.},\ }\href {\doibase 10.1088/1361-6382/ab34e2}
  {\bibfield  {journal} {\bibinfo  {journal} {Class. Quant. Grav.}\ }\textbf
  {\bibinfo {volume} {36}},\ \bibinfo {pages} {195006} (\bibinfo {year}
  {2019})},\ \Eprint {http://arxiv.org/abs/1904.04831} {arXiv:1904.04831
  [gr-qc]} \BibitemShut {NoStop}%
\bibitem [{\citenamefont {Hamilton}\ \emph {et~al.}(2024)\citenamefont
  {Hamilton} \emph {et~al.}}]{Hamilton:2023qkv}%
  \BibitemOpen
  \bibfield  {author} {\bibinfo {author} {\bibfnamefont {E.}~\bibnamefont
  {Hamilton}} \emph {et~al.},\ }\href {\doibase 10.1103/PhysRevD.109.044032}
  {\bibfield  {journal} {\bibinfo  {journal} {Phys. Rev. D}\ }\textbf {\bibinfo
  {volume} {109}},\ \bibinfo {pages} {044032} (\bibinfo {year} {2024})},\
  \Eprint {http://arxiv.org/abs/2303.05419} {arXiv:2303.05419 [gr-qc]}
  \BibitemShut {NoStop}%
\bibitem [{\citenamefont {Healy}\ and\ \citenamefont
  {Lousto}(2022)}]{Healy:2022wdn}%
  \BibitemOpen
  \bibfield  {author} {\bibinfo {author} {\bibfnamefont {J.}~\bibnamefont
  {Healy}}\ and\ \bibinfo {author} {\bibfnamefont {C.~O.}\ \bibnamefont
  {Lousto}},\ }\href {\doibase 10.1103/PhysRevD.105.124010} {\bibfield
  {journal} {\bibinfo  {journal} {Phys. Rev. D}\ }\textbf {\bibinfo {volume}
  {105}},\ \bibinfo {pages} {124010} (\bibinfo {year} {2022})},\ \Eprint
  {http://arxiv.org/abs/2202.00018} {arXiv:2202.00018 [gr-qc]} \BibitemShut
  {NoStop}%
\bibitem [{\citenamefont {Jani}\ \emph {et~al.}(2016)\citenamefont {Jani},
  \citenamefont {Healy}, \citenamefont {Clark}, \citenamefont {London},
  \citenamefont {Laguna},\ and\ \citenamefont {Shoemaker}}]{Jani:2016wkt}%
  \BibitemOpen
  \bibfield  {author} {\bibinfo {author} {\bibfnamefont {K.}~\bibnamefont
  {Jani}}, \bibinfo {author} {\bibfnamefont {J.}~\bibnamefont {Healy}},
  \bibinfo {author} {\bibfnamefont {J.~A.}\ \bibnamefont {Clark}}, \bibinfo
  {author} {\bibfnamefont {L.}~\bibnamefont {London}}, \bibinfo {author}
  {\bibfnamefont {P.}~\bibnamefont {Laguna}}, \ and\ \bibinfo {author}
  {\bibfnamefont {D.}~\bibnamefont {Shoemaker}},\ }\href {\doibase
  10.1088/0264-9381/33/20/204001} {\bibfield  {journal} {\bibinfo  {journal}
  {Class. Quant. Grav.}\ }\textbf {\bibinfo {volume} {33}},\ \bibinfo {pages}
  {204001} (\bibinfo {year} {2016})},\ \Eprint
  {http://arxiv.org/abs/1605.03204} {arXiv:1605.03204 [gr-qc]} \BibitemShut
  {NoStop}%
\bibitem [{\citenamefont {Blackman}\ \emph {et~al.}(2017)\citenamefont
  {Blackman}, \citenamefont {Field}, \citenamefont {Scheel}, \citenamefont
  {Galley}, \citenamefont {Ott}, \citenamefont {Boyle}, \citenamefont {Kidder},
  \citenamefont {Pfeiffer},\ and\ \citenamefont
  {Szil\'agyi}}]{Blackman:2017pcm}%
  \BibitemOpen
  \bibfield  {author} {\bibinfo {author} {\bibfnamefont {J.}~\bibnamefont
  {Blackman}}, \bibinfo {author} {\bibfnamefont {S.~E.}\ \bibnamefont {Field}},
  \bibinfo {author} {\bibfnamefont {M.~A.}\ \bibnamefont {Scheel}}, \bibinfo
  {author} {\bibfnamefont {C.~R.}\ \bibnamefont {Galley}}, \bibinfo {author}
  {\bibfnamefont {C.~D.}\ \bibnamefont {Ott}}, \bibinfo {author} {\bibfnamefont
  {M.}~\bibnamefont {Boyle}}, \bibinfo {author} {\bibfnamefont {L.~E.}\
  \bibnamefont {Kidder}}, \bibinfo {author} {\bibfnamefont {H.~P.}\
  \bibnamefont {Pfeiffer}}, \ and\ \bibinfo {author} {\bibfnamefont
  {B.}~\bibnamefont {Szil\'agyi}},\ }\href {\doibase
  10.1103/PhysRevD.96.024058} {\bibfield  {journal} {\bibinfo  {journal} {Phys.
  Rev. D}\ }\textbf {\bibinfo {volume} {96}},\ \bibinfo {pages} {024058}
  (\bibinfo {year} {2017})},\ \Eprint {http://arxiv.org/abs/1705.07089}
  {arXiv:1705.07089 [gr-qc]} \BibitemShut {NoStop}%
\bibitem [{\citenamefont {Ajith}\ \emph {et~al.}(2007)\citenamefont {Ajith}
  \emph {et~al.}}]{Ajith:2007qp}%
  \BibitemOpen
  \bibfield  {author} {\bibinfo {author} {\bibfnamefont {P.}~\bibnamefont
  {Ajith}} \emph {et~al.},\ }\href {\doibase 10.1088/0264-9381/24/19/S31}
  {\bibfield  {journal} {\bibinfo  {journal} {Class. Quant. Grav.}\ }\textbf
  {\bibinfo {volume} {24}},\ \bibinfo {pages} {S689} (\bibinfo {year}
  {2007})},\ \Eprint {http://arxiv.org/abs/0704.3764} {arXiv:0704.3764 [gr-qc]}
  \BibitemShut {NoStop}%
\bibitem [{\citenamefont {Buonanno}\ \emph {et~al.}(2007)\citenamefont
  {Buonanno}, \citenamefont {Pan}, \citenamefont {Baker}, \citenamefont
  {Centrella}, \citenamefont {Kelly}, \citenamefont {McWilliams},\ and\
  \citenamefont {van Meter}}]{Buonanno:2007pf}%
  \BibitemOpen
  \bibfield  {author} {\bibinfo {author} {\bibfnamefont {A.}~\bibnamefont
  {Buonanno}}, \bibinfo {author} {\bibfnamefont {Y.}~\bibnamefont {Pan}},
  \bibinfo {author} {\bibfnamefont {J.~G.}\ \bibnamefont {Baker}}, \bibinfo
  {author} {\bibfnamefont {J.}~\bibnamefont {Centrella}}, \bibinfo {author}
  {\bibfnamefont {B.~J.}\ \bibnamefont {Kelly}}, \bibinfo {author}
  {\bibfnamefont {S.~T.}\ \bibnamefont {McWilliams}}, \ and\ \bibinfo {author}
  {\bibfnamefont {J.~R.}\ \bibnamefont {van Meter}},\ }\href {\doibase
  10.1103/PhysRevD.76.104049} {\bibfield  {journal} {\bibinfo  {journal} {Phys.
  Rev. D}\ }\textbf {\bibinfo {volume} {76}},\ \bibinfo {pages} {104049}
  (\bibinfo {year} {2007})},\ \Eprint {http://arxiv.org/abs/0706.3732}
  {arXiv:0706.3732 [gr-qc]} \BibitemShut {NoStop}%
\bibitem [{\citenamefont {Chua}\ \emph {et~al.}(2017)\citenamefont {Chua},
  \citenamefont {Moore},\ and\ \citenamefont {Gair}}]{Chua:2017ujo}%
  \BibitemOpen
  \bibfield  {author} {\bibinfo {author} {\bibfnamefont {A.~J.~K.}\
  \bibnamefont {Chua}}, \bibinfo {author} {\bibfnamefont {C.~J.}\ \bibnamefont
  {Moore}}, \ and\ \bibinfo {author} {\bibfnamefont {J.~R.}\ \bibnamefont
  {Gair}},\ }\href {\doibase 10.1103/PhysRevD.96.044005} {\bibfield  {journal}
  {\bibinfo  {journal} {Phys. Rev. D}\ }\textbf {\bibinfo {volume} {96}},\
  \bibinfo {pages} {044005} (\bibinfo {year} {2017})},\ \Eprint
  {http://arxiv.org/abs/1705.04259} {arXiv:1705.04259 [gr-qc]} \BibitemShut
  {NoStop}%
\bibitem [{BHP({\natexlab{a}})}]{BHPToolkit}%
  \BibitemOpen
  \href@noop {} {\enquote {\bibinfo {title} {{Black Hole Perturbation
  Toolkit}},}\ }\bibinfo {howpublished}
  {(\href{http://bhptoolkit.org/}{bhptoolkit.org})}
  ({\natexlab{a}})\BibitemShut {NoStop}%
\bibitem [{BHP({\natexlab{b}})}]{BHPC}%
  \BibitemOpen
  \href@noop {} {\enquote {\bibinfo {title} {{Black Hole Perturbation Club}},}\
  }\bibinfo {howpublished}
  {(\href{http://sites.google.com/view/bhpc1996/home}{sites.google.com/view/bhpc1996/home})}
  ({\natexlab{b}})\BibitemShut {NoStop}%
\bibitem [{\citenamefont {Katz}\ \emph {et~al.}(2021)\citenamefont {Katz},
  \citenamefont {Chua}, \citenamefont {Speri}, \citenamefont {Warburton},\ and\
  \citenamefont {Hughes}}]{Katz:2021yft}%
  \BibitemOpen
  \bibfield  {author} {\bibinfo {author} {\bibfnamefont {M.~L.}\ \bibnamefont
  {Katz}}, \bibinfo {author} {\bibfnamefont {A.~J.~K.}\ \bibnamefont {Chua}},
  \bibinfo {author} {\bibfnamefont {L.}~\bibnamefont {Speri}}, \bibinfo
  {author} {\bibfnamefont {N.}~\bibnamefont {Warburton}}, \ and\ \bibinfo
  {author} {\bibfnamefont {S.~A.}\ \bibnamefont {Hughes}},\ }\href {\doibase
  10.1103/PhysRevD.104.064047} {\bibfield  {journal} {\bibinfo  {journal}
  {Phys. Rev. D}\ }\textbf {\bibinfo {volume} {104}},\ \bibinfo {pages}
  {064047} (\bibinfo {year} {2021})},\ \Eprint
  {http://arxiv.org/abs/2104.04582} {arXiv:2104.04582 [gr-qc]} \BibitemShut
  {NoStop}%
\bibitem [{\citenamefont {Huang}\ \emph
  {et~al.}(2024{\natexlab{a}})\citenamefont {Huang}, \citenamefont {Shi},
  \citenamefont {Lyu},\ and\ \citenamefont {Mei}}]{Huang:2024ylf}%
  \BibitemOpen
  \bibfield  {author} {\bibinfo {author} {\bibfnamefont {Z.}~\bibnamefont
  {Huang}}, \bibinfo {author} {\bibfnamefont {C.}~\bibnamefont {Shi}}, \bibinfo
  {author} {\bibfnamefont {X.}~\bibnamefont {Lyu}}, \ and\ \bibinfo {author}
  {\bibfnamefont {J.}~\bibnamefont {Mei}},\ }\href {\doibase
  10.1140/epjc/s10052-024-12873-9} {\bibfield  {journal} {\bibinfo  {journal}
  {Eur. Phys. J. C}\ }\textbf {\bibinfo {volume} {84}},\ \bibinfo {pages} {522}
  (\bibinfo {year} {2024}{\natexlab{a}})},\ \Eprint
  {http://arxiv.org/abs/2401.12940} {arXiv:2401.12940 [gr-qc]} \BibitemShut
  {NoStop}%
\bibitem [{\citenamefont {Narikawa}\ \emph {et~al.}(2021)\citenamefont
  {Narikawa}, \citenamefont {Uchikata},\ and\ \citenamefont
  {Tanaka}}]{Narikawa:2021pak}%
  \BibitemOpen
  \bibfield  {author} {\bibinfo {author} {\bibfnamefont {T.}~\bibnamefont
  {Narikawa}}, \bibinfo {author} {\bibfnamefont {N.}~\bibnamefont {Uchikata}},
  \ and\ \bibinfo {author} {\bibfnamefont {T.}~\bibnamefont {Tanaka}},\ }\href
  {\doibase 10.1103/PhysRevD.104.084056} {\bibfield  {journal} {\bibinfo
  {journal} {Phys. Rev. D}\ }\textbf {\bibinfo {volume} {104}},\ \bibinfo
  {pages} {084056} (\bibinfo {year} {2021})},\ \Eprint
  {http://arxiv.org/abs/2106.09193} {arXiv:2106.09193 [gr-qc]} \BibitemShut
  {NoStop}%
\bibitem [{\citenamefont {Kong}\ and\ \citenamefont
  {Zhang}(2024)}]{Kong:2024ssa}%
  \BibitemOpen
  \bibfield  {author} {\bibinfo {author} {\bibfnamefont {Y.-L.}\ \bibnamefont
  {Kong}}\ and\ \bibinfo {author} {\bibfnamefont {J.-d.}\ \bibnamefont
  {Zhang}},\ }\href {\doibase 10.1103/PhysRevD.110.024059} {\bibfield
  {journal} {\bibinfo  {journal} {Phys. Rev. D}\ }\textbf {\bibinfo {volume}
  {110}},\ \bibinfo {pages} {024059} (\bibinfo {year} {2024})},\ \Eprint
  {http://arxiv.org/abs/2401.12066} {arXiv:2401.12066 [gr-qc]} \BibitemShut
  {NoStop}%
\bibitem [{\citenamefont {Navas}\ \emph {et~al.}(2024)\citenamefont {Navas}
  \emph {et~al.}}]{ParticleDataGroup:2024cfk}%
  \BibitemOpen
  \bibfield  {author} {\bibinfo {author} {\bibfnamefont {S.}~\bibnamefont
  {Navas}} \emph {et~al.} (\bibinfo {collaboration} {Particle Data Group}),\
  }\href {\doibase 10.1103/PhysRevD.110.030001} {\bibfield  {journal} {\bibinfo
   {journal} {Phys. Rev. D}\ }\textbf {\bibinfo {volume} {110}},\ \bibinfo
  {pages} {030001} (\bibinfo {year} {2024})}\BibitemShut {NoStop}%
\bibitem [{\citenamefont {An}\ \emph {et~al.}(2022{\natexlab{a}})\citenamefont
  {An}, \citenamefont {Lyu}, \citenamefont {Wang},\ and\ \citenamefont
  {Zhou}}]{An:2020fff}%
  \BibitemOpen
  \bibfield  {author} {\bibinfo {author} {\bibfnamefont {H.}~\bibnamefont
  {An}}, \bibinfo {author} {\bibfnamefont {K.-F.}\ \bibnamefont {Lyu}},
  \bibinfo {author} {\bibfnamefont {L.-T.}\ \bibnamefont {Wang}}, \ and\
  \bibinfo {author} {\bibfnamefont {S.}~\bibnamefont {Zhou}},\ }\href {\doibase
  10.1088/1674-1137/ac76a7} {\bibfield  {journal} {\bibinfo  {journal} {Chin.
  Phys. C}\ }\textbf {\bibinfo {volume} {46}},\ \bibinfo {pages} {101001}
  (\bibinfo {year} {2022}{\natexlab{a}})},\ \Eprint
  {http://arxiv.org/abs/2009.12381} {arXiv:2009.12381 [astro-ph.CO]}
  \BibitemShut {NoStop}%
\bibitem [{\citenamefont {An}\ \emph {et~al.}(2022{\natexlab{b}})\citenamefont
  {An}, \citenamefont {Lyu}, \citenamefont {Wang},\ and\ \citenamefont
  {Zhou}}]{An:2022cce}%
  \BibitemOpen
  \bibfield  {author} {\bibinfo {author} {\bibfnamefont {H.}~\bibnamefont
  {An}}, \bibinfo {author} {\bibfnamefont {K.-F.}\ \bibnamefont {Lyu}},
  \bibinfo {author} {\bibfnamefont {L.-T.}\ \bibnamefont {Wang}}, \ and\
  \bibinfo {author} {\bibfnamefont {S.}~\bibnamefont {Zhou}},\ }\href {\doibase
  10.1007/JHEP06(2022)050} {\bibfield  {journal} {\bibinfo  {journal} {JHEP}\
  }\textbf {\bibinfo {volume} {06}},\ \bibinfo {pages} {050} (\bibinfo {year}
  {2022}{\natexlab{b}})},\ \Eprint {http://arxiv.org/abs/2201.05171}
  {arXiv:2201.05171 [astro-ph.CO]} \BibitemShut {NoStop}%
\bibitem [{\citenamefont {Kajantie}\ \emph {et~al.}(1996)\citenamefont
  {Kajantie}, \citenamefont {Laine}, \citenamefont {Rummukainen},\ and\
  \citenamefont {Shaposhnikov}}]{Kajantie:1996mn}%
  \BibitemOpen
  \bibfield  {author} {\bibinfo {author} {\bibfnamefont {K.}~\bibnamefont
  {Kajantie}}, \bibinfo {author} {\bibfnamefont {M.}~\bibnamefont {Laine}},
  \bibinfo {author} {\bibfnamefont {K.}~\bibnamefont {Rummukainen}}, \ and\
  \bibinfo {author} {\bibfnamefont {M.~E.}\ \bibnamefont {Shaposhnikov}},\
  }\href {\doibase 10.1103/PhysRevLett.77.2887} {\bibfield  {journal} {\bibinfo
   {journal} {Phys. Rev. Lett.}\ }\textbf {\bibinfo {volume} {77}},\ \bibinfo
  {pages} {2887} (\bibinfo {year} {1996})},\ \Eprint
  {http://arxiv.org/abs/hep-ph/9605288} {arXiv:hep-ph/9605288} \BibitemShut
  {NoStop}%
\bibitem [{\citenamefont {Gurtler}\ \emph {et~al.}(1997)\citenamefont
  {Gurtler}, \citenamefont {Ilgenfritz},\ and\ \citenamefont
  {Schiller}}]{Gurtler:1997hr}%
  \BibitemOpen
  \bibfield  {author} {\bibinfo {author} {\bibfnamefont {M.}~\bibnamefont
  {Gurtler}}, \bibinfo {author} {\bibfnamefont {E.-M.}\ \bibnamefont
  {Ilgenfritz}}, \ and\ \bibinfo {author} {\bibfnamefont {A.}~\bibnamefont
  {Schiller}},\ }\href {\doibase 10.1103/PhysRevD.56.3888} {\bibfield
  {journal} {\bibinfo  {journal} {Phys. Rev. D}\ }\textbf {\bibinfo {volume}
  {56}},\ \bibinfo {pages} {3888} (\bibinfo {year} {1997})},\ \Eprint
  {http://arxiv.org/abs/hep-lat/9704013} {arXiv:hep-lat/9704013} \BibitemShut
  {NoStop}%
\bibitem [{\citenamefont {Csikor}\ \emph {et~al.}(1999)\citenamefont {Csikor},
  \citenamefont {Fodor},\ and\ \citenamefont {Heitger}}]{Csikor:1998eu}%
  \BibitemOpen
  \bibfield  {author} {\bibinfo {author} {\bibfnamefont {F.}~\bibnamefont
  {Csikor}}, \bibinfo {author} {\bibfnamefont {Z.}~\bibnamefont {Fodor}}, \
  and\ \bibinfo {author} {\bibfnamefont {J.}~\bibnamefont {Heitger}},\ }\href
  {\doibase 10.1103/PhysRevLett.82.21} {\bibfield  {journal} {\bibinfo
  {journal} {Phys. Rev. Lett.}\ }\textbf {\bibinfo {volume} {82}},\ \bibinfo
  {pages} {21} (\bibinfo {year} {1999})},\ \Eprint
  {http://arxiv.org/abs/hep-ph/9809291} {arXiv:hep-ph/9809291} \BibitemShut
  {NoStop}%
\bibitem [{\citenamefont {Coleman}(1977)}]{Coleman:1977py}%
  \BibitemOpen
  \bibfield  {author} {\bibinfo {author} {\bibfnamefont {S.~R.}\ \bibnamefont
  {Coleman}},\ }\href {\doibase 10.1103/PhysRevD.16.1248} {\bibfield  {journal}
  {\bibinfo  {journal} {Phys. Rev. D}\ }\textbf {\bibinfo {volume} {15}},\
  \bibinfo {pages} {2929} (\bibinfo {year} {1977})},\ \bibinfo {note}
  {[Erratum: Phys.Rev.D 16, 1248 (1977)]}\BibitemShut {NoStop}%
\bibitem [{\citenamefont {Callan}\ and\ \citenamefont
  {Coleman}(1977)}]{Callan:1977pt}%
  \BibitemOpen
  \bibfield  {author} {\bibinfo {author} {\bibfnamefont {C.~G.}\ \bibnamefont
  {Callan}, \bibfnamefont {Jr.}}\ and\ \bibinfo {author} {\bibfnamefont
  {S.~R.}\ \bibnamefont {Coleman}},\ }\href {\doibase 10.1103/PhysRevD.16.1762}
  {\bibfield  {journal} {\bibinfo  {journal} {Phys. Rev. D}\ }\textbf {\bibinfo
  {volume} {16}},\ \bibinfo {pages} {1762} (\bibinfo {year}
  {1977})}\BibitemShut {NoStop}%
\bibitem [{\citenamefont {Linde}(1981)}]{Linde:1980tt}%
  \BibitemOpen
  \bibfield  {author} {\bibinfo {author} {\bibfnamefont {A.~D.}\ \bibnamefont
  {Linde}},\ }\href {\doibase 10.1016/0370-2693(81)90281-1} {\bibfield
  {journal} {\bibinfo  {journal} {Phys. Lett. B}\ }\textbf {\bibinfo {volume}
  {100}},\ \bibinfo {pages} {37} (\bibinfo {year} {1981})}\BibitemShut
  {NoStop}%
\bibitem [{\citenamefont {Linde}(1983)}]{Linde:1981zj}%
  \BibitemOpen
  \bibfield  {author} {\bibinfo {author} {\bibfnamefont {A.~D.}\ \bibnamefont
  {Linde}},\ }\href {\doibase 10.1016/0550-3213(83)90072-X} {\bibfield
  {journal} {\bibinfo  {journal} {Nucl. Phys. B}\ }\textbf {\bibinfo {volume}
  {216}},\ \bibinfo {pages} {421} (\bibinfo {year} {1983})},\ \bibinfo {note}
  {[Erratum: Nucl.Phys.B 223, 544 (1983)]}\BibitemShut {NoStop}%
\bibitem [{\citenamefont {Cai}\ and\ \citenamefont {Wang}(2021)}]{Cai:2020djd}%
  \BibitemOpen
  \bibfield  {author} {\bibinfo {author} {\bibfnamefont {R.-G.}\ \bibnamefont
  {Cai}}\ and\ \bibinfo {author} {\bibfnamefont {S.-J.}\ \bibnamefont {Wang}},\
  }\href {\doibase 10.1088/1475-7516/2021/03/096} {\bibfield  {journal}
  {\bibinfo  {journal} {JCAP}\ }\textbf {\bibinfo {volume} {03}},\ \bibinfo
  {pages} {096} (\bibinfo {year} {2021})},\ \Eprint
  {http://arxiv.org/abs/2011.11451} {arXiv:2011.11451 [astro-ph.CO]}
  \BibitemShut {NoStop}%
\bibitem [{\citenamefont {Wang}\ and\ \citenamefont
  {Yuwen}(2023)}]{Wang:2022txy}%
  \BibitemOpen
  \bibfield  {author} {\bibinfo {author} {\bibfnamefont {S.-J.}\ \bibnamefont
  {Wang}}\ and\ \bibinfo {author} {\bibfnamefont {Z.-Y.}\ \bibnamefont
  {Yuwen}},\ }\href {\doibase 10.1103/PhysRevD.107.023501} {\bibfield
  {journal} {\bibinfo  {journal} {Phys. Rev. D}\ }\textbf {\bibinfo {volume}
  {107}},\ \bibinfo {pages} {023501} (\bibinfo {year} {2023})},\ \Eprint
  {http://arxiv.org/abs/2205.02492} {arXiv:2205.02492 [hep-ph]} \BibitemShut
  {NoStop}%
\bibitem [{\citenamefont {Wang}\ \emph
  {et~al.}(2024{\natexlab{b}})\citenamefont {Wang}, \citenamefont {Yuwen},
  \citenamefont {Hao},\ and\ \citenamefont {Wang}}]{Wang:2023kux}%
  \BibitemOpen
  \bibfield  {author} {\bibinfo {author} {\bibfnamefont {J.-C.}\ \bibnamefont
  {Wang}}, \bibinfo {author} {\bibfnamefont {Z.-Y.}\ \bibnamefont {Yuwen}},
  \bibinfo {author} {\bibfnamefont {Y.-S.}\ \bibnamefont {Hao}}, \ and\
  \bibinfo {author} {\bibfnamefont {S.-J.}\ \bibnamefont {Wang}},\ }\href
  {\doibase 10.1103/PhysRevD.110.016031} {\bibfield  {journal} {\bibinfo
  {journal} {Phys. Rev. D}\ }\textbf {\bibinfo {volume} {110}},\ \bibinfo
  {pages} {016031} (\bibinfo {year} {2024}{\natexlab{b}})},\ \Eprint
  {http://arxiv.org/abs/2310.07691} {arXiv:2310.07691 [hep-ph]} \BibitemShut
  {NoStop}%
\bibitem [{\citenamefont {Turner}\ \emph {et~al.}(1992)\citenamefont {Turner},
  \citenamefont {Weinberg},\ and\ \citenamefont {Widrow}}]{Turner:1992tz}%
  \BibitemOpen
  \bibfield  {author} {\bibinfo {author} {\bibfnamefont {M.~S.}\ \bibnamefont
  {Turner}}, \bibinfo {author} {\bibfnamefont {E.~J.}\ \bibnamefont
  {Weinberg}}, \ and\ \bibinfo {author} {\bibfnamefont {L.~M.}\ \bibnamefont
  {Widrow}},\ }\href {\doibase 10.1103/PhysRevD.46.2384} {\bibfield  {journal}
  {\bibinfo  {journal} {Phys. Rev. D}\ }\textbf {\bibinfo {volume} {46}},\
  \bibinfo {pages} {2384} (\bibinfo {year} {1992})}\BibitemShut {NoStop}%
\bibitem [{\citenamefont {Ellis}\ \emph {et~al.}(2019)\citenamefont {Ellis},
  \citenamefont {Lewicki},\ and\ \citenamefont {No}}]{Ellis:2018mja}%
  \BibitemOpen
  \bibfield  {author} {\bibinfo {author} {\bibfnamefont {J.}~\bibnamefont
  {Ellis}}, \bibinfo {author} {\bibfnamefont {M.}~\bibnamefont {Lewicki}}, \
  and\ \bibinfo {author} {\bibfnamefont {J.~M.}\ \bibnamefont {No}},\ }\href
  {\doibase 10.1088/1475-7516/2019/04/003} {\bibfield  {journal} {\bibinfo
  {journal} {JCAP}\ }\textbf {\bibinfo {volume} {04}},\ \bibinfo {pages} {003}
  (\bibinfo {year} {2019})},\ \Eprint {http://arxiv.org/abs/1809.08242}
  {arXiv:1809.08242 [hep-ph]} \BibitemShut {NoStop}%
\bibitem [{\citenamefont {Yuwen}\ \emph {et~al.}(2024)\citenamefont {Yuwen},
  \citenamefont {Wang},\ and\ \citenamefont {Wang}}]{Yuwen:2024hme}%
  \BibitemOpen
  \bibfield  {author} {\bibinfo {author} {\bibfnamefont {Z.-Y.}\ \bibnamefont
  {Yuwen}}, \bibinfo {author} {\bibfnamefont {J.-C.}\ \bibnamefont {Wang}}, \
  and\ \bibinfo {author} {\bibfnamefont {S.-J.}\ \bibnamefont {Wang}},\
  }\href@noop {} {\  (\bibinfo {year} {2024})},\ \Eprint
  {http://arxiv.org/abs/2409.20045} {arXiv:2409.20045 [hep-ph]} \BibitemShut
  {NoStop}%
\bibitem [{\citenamefont {Jinno}\ and\ \citenamefont
  {Takimoto}(2017)}]{Jinno:2016vai}%
  \BibitemOpen
  \bibfield  {author} {\bibinfo {author} {\bibfnamefont {R.}~\bibnamefont
  {Jinno}}\ and\ \bibinfo {author} {\bibfnamefont {M.}~\bibnamefont
  {Takimoto}},\ }\href {\doibase 10.1103/PhysRevD.95.024009} {\bibfield
  {journal} {\bibinfo  {journal} {Phys. Rev. D}\ }\textbf {\bibinfo {volume}
  {95}},\ \bibinfo {pages} {024009} (\bibinfo {year} {2017})},\ \Eprint
  {http://arxiv.org/abs/1605.01403} {arXiv:1605.01403 [astro-ph.CO]}
  \BibitemShut {NoStop}%
\bibitem [{\citenamefont {Jinno}\ and\ \citenamefont
  {Takimoto}(2019)}]{Jinno:2017fby}%
  \BibitemOpen
  \bibfield  {author} {\bibinfo {author} {\bibfnamefont {R.}~\bibnamefont
  {Jinno}}\ and\ \bibinfo {author} {\bibfnamefont {M.}~\bibnamefont
  {Takimoto}},\ }\href {\doibase 10.1088/1475-7516/2019/01/060} {\bibfield
  {journal} {\bibinfo  {journal} {JCAP}\ }\textbf {\bibinfo {volume} {01}},\
  \bibinfo {pages} {060} (\bibinfo {year} {2019})},\ \Eprint
  {http://arxiv.org/abs/1707.03111} {arXiv:1707.03111 [hep-ph]} \BibitemShut
  {NoStop}%
\bibitem [{\citenamefont {Zhong}\ \emph {et~al.}(2022)\citenamefont {Zhong},
  \citenamefont {Gong},\ and\ \citenamefont {Qiu}}]{Zhong:2021hgo}%
  \BibitemOpen
  \bibfield  {author} {\bibinfo {author} {\bibfnamefont {H.}~\bibnamefont
  {Zhong}}, \bibinfo {author} {\bibfnamefont {B.}~\bibnamefont {Gong}}, \ and\
  \bibinfo {author} {\bibfnamefont {T.}~\bibnamefont {Qiu}},\ }\href {\doibase
  10.1007/JHEP02(2022)077} {\bibfield  {journal} {\bibinfo  {journal} {JHEP}\
  }\textbf {\bibinfo {volume} {02}},\ \bibinfo {pages} {077} (\bibinfo {year}
  {2022})},\ \Eprint {http://arxiv.org/abs/2107.01845} {arXiv:2107.01845
  [gr-qc]} \BibitemShut {NoStop}%
\bibitem [{\citenamefont {Hindmarsh}\ \emph {et~al.}(2014)\citenamefont
  {Hindmarsh}, \citenamefont {Huber}, \citenamefont {Rummukainen},\ and\
  \citenamefont {Weir}}]{Hindmarsh:2013xza}%
  \BibitemOpen
  \bibfield  {author} {\bibinfo {author} {\bibfnamefont {M.}~\bibnamefont
  {Hindmarsh}}, \bibinfo {author} {\bibfnamefont {S.~J.}\ \bibnamefont
  {Huber}}, \bibinfo {author} {\bibfnamefont {K.}~\bibnamefont {Rummukainen}},
  \ and\ \bibinfo {author} {\bibfnamefont {D.~J.}\ \bibnamefont {Weir}},\
  }\href {\doibase 10.1103/PhysRevLett.112.041301} {\bibfield  {journal}
  {\bibinfo  {journal} {Phys. Rev. Lett.}\ }\textbf {\bibinfo {volume} {112}},\
  \bibinfo {pages} {041301} (\bibinfo {year} {2014})},\ \Eprint
  {http://arxiv.org/abs/1304.2433} {arXiv:1304.2433 [hep-ph]} \BibitemShut
  {NoStop}%
\bibitem [{\citenamefont {Hindmarsh}\ \emph {et~al.}(2015)\citenamefont
  {Hindmarsh}, \citenamefont {Huber}, \citenamefont {Rummukainen},\ and\
  \citenamefont {Weir}}]{Hindmarsh:2015qta}%
  \BibitemOpen
  \bibfield  {author} {\bibinfo {author} {\bibfnamefont {M.}~\bibnamefont
  {Hindmarsh}}, \bibinfo {author} {\bibfnamefont {S.~J.}\ \bibnamefont
  {Huber}}, \bibinfo {author} {\bibfnamefont {K.}~\bibnamefont {Rummukainen}},
  \ and\ \bibinfo {author} {\bibfnamefont {D.~J.}\ \bibnamefont {Weir}},\
  }\href {\doibase 10.1103/PhysRevD.92.123009} {\bibfield  {journal} {\bibinfo
  {journal} {Phys. Rev. D}\ }\textbf {\bibinfo {volume} {92}},\ \bibinfo
  {pages} {123009} (\bibinfo {year} {2015})},\ \Eprint
  {http://arxiv.org/abs/1504.03291} {arXiv:1504.03291 [astro-ph.CO]}
  \BibitemShut {NoStop}%
\bibitem [{\citenamefont {Hindmarsh}\ \emph {et~al.}(2017)\citenamefont
  {Hindmarsh}, \citenamefont {Huber}, \citenamefont {Rummukainen},\ and\
  \citenamefont {Weir}}]{Hindmarsh:2017gnf}%
  \BibitemOpen
  \bibfield  {author} {\bibinfo {author} {\bibfnamefont {M.}~\bibnamefont
  {Hindmarsh}}, \bibinfo {author} {\bibfnamefont {S.~J.}\ \bibnamefont
  {Huber}}, \bibinfo {author} {\bibfnamefont {K.}~\bibnamefont {Rummukainen}},
  \ and\ \bibinfo {author} {\bibfnamefont {D.~J.}\ \bibnamefont {Weir}},\
  }\href {\doibase 10.1103/PhysRevD.96.103520} {\bibfield  {journal} {\bibinfo
  {journal} {Phys. Rev. D}\ }\textbf {\bibinfo {volume} {96}},\ \bibinfo
  {pages} {103520} (\bibinfo {year} {2017})},\ \bibinfo {note} {[Erratum:
  Phys.Rev.D 101, 089902 (2020)]},\ \Eprint {http://arxiv.org/abs/1704.05871}
  {arXiv:1704.05871 [astro-ph.CO]} \BibitemShut {NoStop}%
\bibitem [{\citenamefont {Espinosa}\ \emph {et~al.}(2010)\citenamefont
  {Espinosa}, \citenamefont {Konstandin}, \citenamefont {No},\ and\
  \citenamefont {Servant}}]{Espinosa:2010hh}%
  \BibitemOpen
  \bibfield  {author} {\bibinfo {author} {\bibfnamefont {J.~R.}\ \bibnamefont
  {Espinosa}}, \bibinfo {author} {\bibfnamefont {T.}~\bibnamefont
  {Konstandin}}, \bibinfo {author} {\bibfnamefont {J.~M.}\ \bibnamefont {No}},
  \ and\ \bibinfo {author} {\bibfnamefont {G.}~\bibnamefont {Servant}},\ }\href
  {\doibase 10.1088/1475-7516/2010/06/028} {\bibfield  {journal} {\bibinfo
  {journal} {JCAP}\ }\textbf {\bibinfo {volume} {06}},\ \bibinfo {pages} {028}
  (\bibinfo {year} {2010})},\ \Eprint {http://arxiv.org/abs/1004.4187}
  {arXiv:1004.4187 [hep-ph]} \BibitemShut {NoStop}%
\bibitem [{\citenamefont {Giese}\ \emph {et~al.}(2021)\citenamefont {Giese},
  \citenamefont {Konstandin}, \citenamefont {Schmitz},\ and\ \citenamefont
  {van~de Vis}}]{Giese:2020znk}%
  \BibitemOpen
  \bibfield  {author} {\bibinfo {author} {\bibfnamefont {F.}~\bibnamefont
  {Giese}}, \bibinfo {author} {\bibfnamefont {T.}~\bibnamefont {Konstandin}},
  \bibinfo {author} {\bibfnamefont {K.}~\bibnamefont {Schmitz}}, \ and\
  \bibinfo {author} {\bibfnamefont {J.}~\bibnamefont {van~de Vis}},\ }\href
  {\doibase 10.1088/1475-7516/2021/01/072} {\bibfield  {journal} {\bibinfo
  {journal} {JCAP}\ }\textbf {\bibinfo {volume} {01}},\ \bibinfo {pages} {072}
  (\bibinfo {year} {2021})},\ \Eprint {http://arxiv.org/abs/2010.09744}
  {arXiv:2010.09744 [astro-ph.CO]} \BibitemShut {NoStop}%
\bibitem [{\citenamefont {Giese}\ \emph {et~al.}(2020)\citenamefont {Giese},
  \citenamefont {Konstandin},\ and\ \citenamefont {van~de
  Vis}}]{Giese:2020rtr}%
  \BibitemOpen
  \bibfield  {author} {\bibinfo {author} {\bibfnamefont {F.}~\bibnamefont
  {Giese}}, \bibinfo {author} {\bibfnamefont {T.}~\bibnamefont {Konstandin}}, \
  and\ \bibinfo {author} {\bibfnamefont {J.}~\bibnamefont {van~de Vis}},\
  }\href {\doibase 10.1088/1475-7516/2020/07/057} {\bibfield  {journal}
  {\bibinfo  {journal} {JCAP}\ }\textbf {\bibinfo {volume} {07}},\ \bibinfo
  {pages} {057} (\bibinfo {year} {2020})},\ \Eprint
  {http://arxiv.org/abs/2004.06995} {arXiv:2004.06995 [astro-ph.CO]}
  \BibitemShut {NoStop}%
\bibitem [{\citenamefont {Wang}\ \emph {et~al.}(2021)\citenamefont {Wang},
  \citenamefont {Huang},\ and\ \citenamefont {Zhang}}]{Wang:2020nzm}%
  \BibitemOpen
  \bibfield  {author} {\bibinfo {author} {\bibfnamefont {X.}~\bibnamefont
  {Wang}}, \bibinfo {author} {\bibfnamefont {F.~P.}\ \bibnamefont {Huang}}, \
  and\ \bibinfo {author} {\bibfnamefont {X.}~\bibnamefont {Zhang}},\ }\href
  {\doibase 10.1103/PhysRevD.103.103520} {\bibfield  {journal} {\bibinfo
  {journal} {Phys. Rev. D}\ }\textbf {\bibinfo {volume} {103}},\ \bibinfo
  {pages} {103520} (\bibinfo {year} {2021})},\ \Eprint
  {http://arxiv.org/abs/2010.13770} {arXiv:2010.13770 [astro-ph.CO]}
  \BibitemShut {NoStop}%
\bibitem [{\citenamefont {Wang}\ \emph
  {et~al.}(2023{\natexlab{b}})\citenamefont {Wang}, \citenamefont {Tian},\ and\
  \citenamefont {Huang}}]{Wang:2023jto}%
  \BibitemOpen
  \bibfield  {author} {\bibinfo {author} {\bibfnamefont {X.}~\bibnamefont
  {Wang}}, \bibinfo {author} {\bibfnamefont {C.}~\bibnamefont {Tian}}, \ and\
  \bibinfo {author} {\bibfnamefont {F.~P.}\ \bibnamefont {Huang}},\ }\href
  {\doibase 10.1088/1475-7516/2023/07/006} {\bibfield  {journal} {\bibinfo
  {journal} {JCAP}\ }\textbf {\bibinfo {volume} {07}},\ \bibinfo {pages} {006}
  (\bibinfo {year} {2023}{\natexlab{b}})},\ \Eprint
  {http://arxiv.org/abs/2301.12328} {arXiv:2301.12328 [hep-ph]} \BibitemShut
  {NoStop}%
\bibitem [{\citenamefont {Wang}\ and\ \citenamefont
  {Yuwen}(2022)}]{Wang:2022lyd}%
  \BibitemOpen
  \bibfield  {author} {\bibinfo {author} {\bibfnamefont {S.-J.}\ \bibnamefont
  {Wang}}\ and\ \bibinfo {author} {\bibfnamefont {Z.-Y.}\ \bibnamefont
  {Yuwen}},\ }\href {\doibase 10.1088/1475-7516/2022/10/047} {\bibfield
  {journal} {\bibinfo  {journal} {JCAP}\ }\textbf {\bibinfo {volume} {10}},\
  \bibinfo {pages} {047} (\bibinfo {year} {2022})},\ \Eprint
  {http://arxiv.org/abs/2206.01148} {arXiv:2206.01148 [hep-ph]} \BibitemShut
  {NoStop}%
\bibitem [{\citenamefont {Cai}\ and\ \citenamefont {Wang}(2018)}]{Cai:2018teh}%
  \BibitemOpen
  \bibfield  {author} {\bibinfo {author} {\bibfnamefont {R.-G.}\ \bibnamefont
  {Cai}}\ and\ \bibinfo {author} {\bibfnamefont {S.-J.}\ \bibnamefont {Wang}},\
  }\href {\doibase 10.1007/s11433-018-9216-7} {\bibfield  {journal} {\bibinfo
  {journal} {Sci. China Phys. Mech. Astron.}\ }\textbf {\bibinfo {volume}
  {61}},\ \bibinfo {pages} {080411} (\bibinfo {year} {2018})},\ \Eprint
  {http://arxiv.org/abs/1803.03002} {arXiv:1803.03002 [gr-qc]} \BibitemShut
  {NoStop}%
\bibitem [{\citenamefont {Giombi}\ and\ \citenamefont
  {Hindmarsh}(2024)}]{Giombi:2023jqq}%
  \BibitemOpen
  \bibfield  {author} {\bibinfo {author} {\bibfnamefont {L.}~\bibnamefont
  {Giombi}}\ and\ \bibinfo {author} {\bibfnamefont {M.}~\bibnamefont
  {Hindmarsh}},\ }\href {\doibase 10.1088/1475-7516/2024/03/059} {\bibfield
  {journal} {\bibinfo  {journal} {JCAP}\ }\textbf {\bibinfo {volume} {03}},\
  \bibinfo {pages} {059} (\bibinfo {year} {2024})},\ \Eprint
  {http://arxiv.org/abs/2307.12080} {arXiv:2307.12080 [astro-ph.CO]}
  \BibitemShut {NoStop}%
\bibitem [{\citenamefont {Guo}\ \emph {et~al.}(2021)\citenamefont {Guo},
  \citenamefont {Sinha}, \citenamefont {Vagie},\ and\ \citenamefont
  {White}}]{Guo:2020grp}%
  \BibitemOpen
  \bibfield  {author} {\bibinfo {author} {\bibfnamefont {H.-K.}\ \bibnamefont
  {Guo}}, \bibinfo {author} {\bibfnamefont {K.}~\bibnamefont {Sinha}}, \bibinfo
  {author} {\bibfnamefont {D.}~\bibnamefont {Vagie}}, \ and\ \bibinfo {author}
  {\bibfnamefont {G.}~\bibnamefont {White}},\ }\href {\doibase
  10.1088/1475-7516/2021/01/001} {\bibfield  {journal} {\bibinfo  {journal}
  {JCAP}\ }\textbf {\bibinfo {volume} {01}},\ \bibinfo {pages} {001} (\bibinfo
  {year} {2021})},\ \Eprint {http://arxiv.org/abs/2007.08537} {arXiv:2007.08537
  [hep-ph]} \BibitemShut {NoStop}%
\bibitem [{\citenamefont {Hindmarsh}(2018)}]{Hindmarsh:2016lnk}%
  \BibitemOpen
  \bibfield  {author} {\bibinfo {author} {\bibfnamefont {M.}~\bibnamefont
  {Hindmarsh}},\ }\href {\doibase 10.1103/PhysRevLett.120.071301} {\bibfield
  {journal} {\bibinfo  {journal} {Phys. Rev. Lett.}\ }\textbf {\bibinfo
  {volume} {120}},\ \bibinfo {pages} {071301} (\bibinfo {year} {2018})},\
  \Eprint {http://arxiv.org/abs/1608.04735} {arXiv:1608.04735 [astro-ph.CO]}
  \BibitemShut {NoStop}%
\bibitem [{\citenamefont {Hindmarsh}\ and\ \citenamefont
  {Hijazi}(2019)}]{Hindmarsh:2019phv}%
  \BibitemOpen
  \bibfield  {author} {\bibinfo {author} {\bibfnamefont {M.}~\bibnamefont
  {Hindmarsh}}\ and\ \bibinfo {author} {\bibfnamefont {M.}~\bibnamefont
  {Hijazi}},\ }\href {\doibase 10.1088/1475-7516/2019/12/062} {\bibfield
  {journal} {\bibinfo  {journal} {JCAP}\ }\textbf {\bibinfo {volume} {12}},\
  \bibinfo {pages} {062} (\bibinfo {year} {2019})},\ \Eprint
  {http://arxiv.org/abs/1909.10040} {arXiv:1909.10040 [astro-ph.CO]}
  \BibitemShut {NoStop}%
\bibitem [{\citenamefont {Cai}\ \emph {et~al.}(2023)\citenamefont {Cai},
  \citenamefont {Wang},\ and\ \citenamefont {Yuwen}}]{Cai:2023guc}%
  \BibitemOpen
  \bibfield  {author} {\bibinfo {author} {\bibfnamefont {R.-G.}\ \bibnamefont
  {Cai}}, \bibinfo {author} {\bibfnamefont {S.-J.}\ \bibnamefont {Wang}}, \
  and\ \bibinfo {author} {\bibfnamefont {Z.-Y.}\ \bibnamefont {Yuwen}},\ }\href
  {\doibase 10.1103/PhysRevD.108.L021502} {\bibfield  {journal} {\bibinfo
  {journal} {Phys. Rev. D}\ }\textbf {\bibinfo {volume} {108}},\ \bibinfo
  {pages} {L021502} (\bibinfo {year} {2023})},\ \Eprint
  {http://arxiv.org/abs/2305.00074} {arXiv:2305.00074 [gr-qc]} \BibitemShut
  {NoStop}%
\bibitem [{\citenamefont {Roper~Pol}\ \emph {et~al.}(2024)\citenamefont
  {Roper~Pol}, \citenamefont {Procacci},\ and\ \citenamefont
  {Caprini}}]{RoperPol:2023dzg}%
  \BibitemOpen
  \bibfield  {author} {\bibinfo {author} {\bibfnamefont {A.}~\bibnamefont
  {Roper~Pol}}, \bibinfo {author} {\bibfnamefont {S.}~\bibnamefont {Procacci}},
  \ and\ \bibinfo {author} {\bibfnamefont {C.}~\bibnamefont {Caprini}},\ }\href
  {\doibase 10.1103/PhysRevD.109.063531} {\bibfield  {journal} {\bibinfo
  {journal} {Phys. Rev. D}\ }\textbf {\bibinfo {volume} {109}},\ \bibinfo
  {pages} {063531} (\bibinfo {year} {2024})},\ \Eprint
  {http://arxiv.org/abs/2308.12943} {arXiv:2308.12943 [gr-qc]} \BibitemShut
  {NoStop}%
\bibitem [{\citenamefont {Sharma}\ \emph {et~al.}(2023)\citenamefont {Sharma},
  \citenamefont {Dahl}, \citenamefont {Brandenburg},\ and\ \citenamefont
  {Hindmarsh}}]{Sharma:2023mao}%
  \BibitemOpen
  \bibfield  {author} {\bibinfo {author} {\bibfnamefont {R.}~\bibnamefont
  {Sharma}}, \bibinfo {author} {\bibfnamefont {J.}~\bibnamefont {Dahl}},
  \bibinfo {author} {\bibfnamefont {A.}~\bibnamefont {Brandenburg}}, \ and\
  \bibinfo {author} {\bibfnamefont {M.}~\bibnamefont {Hindmarsh}},\ }\href
  {\doibase 10.1088/1475-7516/2023/12/042} {\bibfield  {journal} {\bibinfo
  {journal} {JCAP}\ }\textbf {\bibinfo {volume} {12}},\ \bibinfo {pages} {042}
  (\bibinfo {year} {2023})},\ \Eprint {http://arxiv.org/abs/2308.12916}
  {arXiv:2308.12916 [gr-qc]} \BibitemShut {NoStop}%
\bibitem [{\citenamefont {Caprini}\ \emph {et~al.}(2009)\citenamefont
  {Caprini}, \citenamefont {Durrer},\ and\ \citenamefont
  {Servant}}]{Caprini:2009yp}%
  \BibitemOpen
  \bibfield  {author} {\bibinfo {author} {\bibfnamefont {C.}~\bibnamefont
  {Caprini}}, \bibinfo {author} {\bibfnamefont {R.}~\bibnamefont {Durrer}}, \
  and\ \bibinfo {author} {\bibfnamefont {G.}~\bibnamefont {Servant}},\ }\href
  {\doibase 10.1088/1475-7516/2009/12/024} {\bibfield  {journal} {\bibinfo
  {journal} {JCAP}\ }\textbf {\bibinfo {volume} {12}},\ \bibinfo {pages} {024}
  (\bibinfo {year} {2009})},\ \Eprint {http://arxiv.org/abs/0909.0622}
  {arXiv:0909.0622 [astro-ph.CO]} \BibitemShut {NoStop}%
\bibitem [{\citenamefont {Binetruy}\ \emph {et~al.}(2012)\citenamefont
  {Binetruy}, \citenamefont {Bohe}, \citenamefont {Caprini},\ and\
  \citenamefont {Dufaux}}]{Binetruy:2012ze}%
  \BibitemOpen
  \bibfield  {author} {\bibinfo {author} {\bibfnamefont {P.}~\bibnamefont
  {Binetruy}}, \bibinfo {author} {\bibfnamefont {A.}~\bibnamefont {Bohe}},
  \bibinfo {author} {\bibfnamefont {C.}~\bibnamefont {Caprini}}, \ and\
  \bibinfo {author} {\bibfnamefont {J.-F.}\ \bibnamefont {Dufaux}},\ }\href
  {\doibase 10.1088/1475-7516/2012/06/027} {\bibfield  {journal} {\bibinfo
  {journal} {JCAP}\ }\textbf {\bibinfo {volume} {06}},\ \bibinfo {pages} {027}
  (\bibinfo {year} {2012})},\ \Eprint {http://arxiv.org/abs/1201.0983}
  {arXiv:1201.0983 [gr-qc]} \BibitemShut {NoStop}%
\bibitem [{\citenamefont {Caprini}\ \emph {et~al.}(2016)\citenamefont {Caprini}
  \emph {et~al.}}]{Caprini:2015zlo}%
  \BibitemOpen
  \bibfield  {author} {\bibinfo {author} {\bibfnamefont {C.}~\bibnamefont
  {Caprini}} \emph {et~al.},\ }\href {\doibase 10.1088/1475-7516/2016/04/001}
  {\bibfield  {journal} {\bibinfo  {journal} {JCAP}\ }\textbf {\bibinfo
  {volume} {04}},\ \bibinfo {pages} {001} (\bibinfo {year} {2016})},\ \Eprint
  {http://arxiv.org/abs/1512.06239} {arXiv:1512.06239 [astro-ph.CO]}
  \BibitemShut {NoStop}%
\bibitem [{\citenamefont {Cai}\ \emph {et~al.}(2022{\natexlab{b}})\citenamefont
  {Cai}, \citenamefont {Hashino}, \citenamefont {Wang},\ and\ \citenamefont
  {Yu}}]{Cai:2022bcf}%
  \BibitemOpen
  \bibfield  {author} {\bibinfo {author} {\bibfnamefont {R.-G.}\ \bibnamefont
  {Cai}}, \bibinfo {author} {\bibfnamefont {K.}~\bibnamefont {Hashino}},
  \bibinfo {author} {\bibfnamefont {S.-J.}\ \bibnamefont {Wang}}, \ and\
  \bibinfo {author} {\bibfnamefont {J.-H.}\ \bibnamefont {Yu}},\ }\href
  {\doibase 10.1088/1572-9494/ad9c3d} {\  (\bibinfo {year}
  {2022}{\natexlab{b}}),\ 10.1088/1572-9494/ad9c3d},\ \Eprint
  {http://arxiv.org/abs/2202.08295} {arXiv:2202.08295 [hep-ph]} \BibitemShut
  {NoStop}%
\bibitem [{\citenamefont {Carena}\ \emph {et~al.}(1996)\citenamefont {Carena},
  \citenamefont {Quiros},\ and\ \citenamefont {Wagner}}]{Carena:1996wj}%
  \BibitemOpen
  \bibfield  {author} {\bibinfo {author} {\bibfnamefont {M.}~\bibnamefont
  {Carena}}, \bibinfo {author} {\bibfnamefont {M.}~\bibnamefont {Quiros}}, \
  and\ \bibinfo {author} {\bibfnamefont {C.~E.~M.}\ \bibnamefont {Wagner}},\
  }\href {\doibase 10.1016/0370-2693(96)00475-3} {\bibfield  {journal}
  {\bibinfo  {journal} {Phys. Lett. B}\ }\textbf {\bibinfo {volume} {380}},\
  \bibinfo {pages} {81} (\bibinfo {year} {1996})},\ \Eprint
  {http://arxiv.org/abs/hep-ph/9603420} {arXiv:hep-ph/9603420} \BibitemShut
  {NoStop}%
\bibitem [{\citenamefont {Delepine}\ \emph {et~al.}(1996)\citenamefont
  {Delepine}, \citenamefont {Gerard}, \citenamefont {Gonzalez~Felipe},\ and\
  \citenamefont {Weyers}}]{Delepine:1996vn}%
  \BibitemOpen
  \bibfield  {author} {\bibinfo {author} {\bibfnamefont {D.}~\bibnamefont
  {Delepine}}, \bibinfo {author} {\bibfnamefont {J.~M.}\ \bibnamefont
  {Gerard}}, \bibinfo {author} {\bibfnamefont {R.}~\bibnamefont
  {Gonzalez~Felipe}}, \ and\ \bibinfo {author} {\bibfnamefont {J.}~\bibnamefont
  {Weyers}},\ }\href {\doibase 10.1016/0370-2693(96)00921-5} {\bibfield
  {journal} {\bibinfo  {journal} {Phys. Lett. B}\ }\textbf {\bibinfo {volume}
  {386}},\ \bibinfo {pages} {183} (\bibinfo {year} {1996})},\ \Eprint
  {http://arxiv.org/abs/hep-ph/9604440} {arXiv:hep-ph/9604440} \BibitemShut
  {NoStop}%
\bibitem [{\citenamefont {Carena}\ \emph {et~al.}(2009)\citenamefont {Carena},
  \citenamefont {Nardini}, \citenamefont {Quiros},\ and\ \citenamefont
  {Wagner}}]{Carena:2008vj}%
  \BibitemOpen
  \bibfield  {author} {\bibinfo {author} {\bibfnamefont {M.}~\bibnamefont
  {Carena}}, \bibinfo {author} {\bibfnamefont {G.}~\bibnamefont {Nardini}},
  \bibinfo {author} {\bibfnamefont {M.}~\bibnamefont {Quiros}}, \ and\ \bibinfo
  {author} {\bibfnamefont {C.~E.~M.}\ \bibnamefont {Wagner}},\ }\href {\doibase
  10.1016/j.nuclphysb.2008.12.014} {\bibfield  {journal} {\bibinfo  {journal}
  {Nucl. Phys. B}\ }\textbf {\bibinfo {volume} {812}},\ \bibinfo {pages} {243}
  (\bibinfo {year} {2009})},\ \Eprint {http://arxiv.org/abs/0809.3760}
  {arXiv:0809.3760 [hep-ph]} \BibitemShut {NoStop}%
\bibitem [{\citenamefont {Cao}\ \emph {et~al.}(2018)\citenamefont {Cao},
  \citenamefont {Huang}, \citenamefont {Xie},\ and\ \citenamefont
  {Zhang}}]{Cao:2017oez}%
  \BibitemOpen
  \bibfield  {author} {\bibinfo {author} {\bibfnamefont {Q.-H.}\ \bibnamefont
  {Cao}}, \bibinfo {author} {\bibfnamefont {F.~P.}\ \bibnamefont {Huang}},
  \bibinfo {author} {\bibfnamefont {K.-P.}\ \bibnamefont {Xie}}, \ and\
  \bibinfo {author} {\bibfnamefont {X.}~\bibnamefont {Zhang}},\ }\href
  {\doibase 10.1088/1674-1137/42/2/023103} {\bibfield  {journal} {\bibinfo
  {journal} {Chin. Phys. C}\ }\textbf {\bibinfo {volume} {42}},\ \bibinfo
  {pages} {023103} (\bibinfo {year} {2018})},\ \Eprint
  {http://arxiv.org/abs/1708.04737} {arXiv:1708.04737 [hep-ph]} \BibitemShut
  {NoStop}%
\bibitem [{\citenamefont {Huang}\ and\ \citenamefont
  {Yu}(2018)}]{Huang:2017rzf}%
  \BibitemOpen
  \bibfield  {author} {\bibinfo {author} {\bibfnamefont {F.~P.}\ \bibnamefont
  {Huang}}\ and\ \bibinfo {author} {\bibfnamefont {J.-H.}\ \bibnamefont {Yu}},\
  }\href {\doibase 10.1103/PhysRevD.98.095022} {\bibfield  {journal} {\bibinfo
  {journal} {Phys. Rev. D}\ }\textbf {\bibinfo {volume} {98}},\ \bibinfo
  {pages} {095022} (\bibinfo {year} {2018})},\ \Eprint
  {http://arxiv.org/abs/1704.04201} {arXiv:1704.04201 [hep-ph]} \BibitemShut
  {NoStop}%
\bibitem [{\citenamefont {Fujikura}\ \emph {et~al.}(2023)\citenamefont
  {Fujikura}, \citenamefont {Nakai}, \citenamefont {Sato},\ and\ \citenamefont
  {Wang}}]{Fujikura:2023fbi}%
  \BibitemOpen
  \bibfield  {author} {\bibinfo {author} {\bibfnamefont {K.}~\bibnamefont
  {Fujikura}}, \bibinfo {author} {\bibfnamefont {Y.}~\bibnamefont {Nakai}},
  \bibinfo {author} {\bibfnamefont {R.}~\bibnamefont {Sato}}, \ and\ \bibinfo
  {author} {\bibfnamefont {Y.}~\bibnamefont {Wang}},\ }\href {\doibase
  10.1007/JHEP09(2023)053} {\bibfield  {journal} {\bibinfo  {journal} {JHEP}\
  }\textbf {\bibinfo {volume} {09}},\ \bibinfo {pages} {053} (\bibinfo {year}
  {2023})},\ \Eprint {http://arxiv.org/abs/2306.01305} {arXiv:2306.01305
  [hep-ph]} \BibitemShut {NoStop}%
\bibitem [{\citenamefont {Camargo-Molina}\ \emph {et~al.}(2021)\citenamefont
  {Camargo-Molina}, \citenamefont {Enberg},\ and\ \citenamefont
  {L\"ofgren}}]{Camargo-Molina:2021zgz}%
  \BibitemOpen
  \bibfield  {author} {\bibinfo {author} {\bibfnamefont {J.~E.}\ \bibnamefont
  {Camargo-Molina}}, \bibinfo {author} {\bibfnamefont {R.}~\bibnamefont
  {Enberg}}, \ and\ \bibinfo {author} {\bibfnamefont {J.}~\bibnamefont
  {L\"ofgren}},\ }\href {\doibase 10.1007/JHEP10(2021)127} {\bibfield
  {journal} {\bibinfo  {journal} {JHEP}\ }\textbf {\bibinfo {volume} {10}},\
  \bibinfo {pages} {127} (\bibinfo {year} {2021})},\ \Eprint
  {http://arxiv.org/abs/2103.14022} {arXiv:2103.14022 [hep-ph]} \BibitemShut
  {NoStop}%
\bibitem [{\citenamefont {Ade}\ \emph {et~al.}(2016)\citenamefont {Ade} \emph
  {et~al.}}]{Planck:2015fie}%
  \BibitemOpen
  \bibfield  {author} {\bibinfo {author} {\bibfnamefont {P.~A.~R.}\
  \bibnamefont {Ade}} \emph {et~al.} (\bibinfo {collaboration} {Planck}),\
  }\href {\doibase 10.1051/0004-6361/201525830} {\bibfield  {journal} {\bibinfo
   {journal} {Astron. Astrophys.}\ }\textbf {\bibinfo {volume} {594}},\
  \bibinfo {pages} {A13} (\bibinfo {year} {2016})},\ \Eprint
  {http://arxiv.org/abs/1502.01589} {arXiv:1502.01589 [astro-ph.CO]}
  \BibitemShut {NoStop}%
\bibitem [{\citenamefont {Sakharov}(1967)}]{Sakharov:1967dj}%
  \BibitemOpen
  \bibfield  {author} {\bibinfo {author} {\bibfnamefont {A.~D.}\ \bibnamefont
  {Sakharov}},\ }\href {\doibase 10.1070/PU1991v034n05ABEH002497} {\bibfield
  {journal} {\bibinfo  {journal} {Pisma Zh. Eksp. Teor. Fiz.}\ }\textbf
  {\bibinfo {volume} {5}},\ \bibinfo {pages} {32} (\bibinfo {year}
  {1967})}\BibitemShut {NoStop}%
\bibitem [{\citenamefont {Kuzmin}\ \emph {et~al.}(1985)\citenamefont {Kuzmin},
  \citenamefont {Rubakov},\ and\ \citenamefont {Shaposhnikov}}]{Kuzmin:1985mm}%
  \BibitemOpen
  \bibfield  {author} {\bibinfo {author} {\bibfnamefont {V.~A.}\ \bibnamefont
  {Kuzmin}}, \bibinfo {author} {\bibfnamefont {V.~A.}\ \bibnamefont {Rubakov}},
  \ and\ \bibinfo {author} {\bibfnamefont {M.~E.}\ \bibnamefont
  {Shaposhnikov}},\ }\href {\doibase 10.1016/0370-2693(85)91028-7} {\bibfield
  {journal} {\bibinfo  {journal} {Phys. Lett. B}\ }\textbf {\bibinfo {volume}
  {155}},\ \bibinfo {pages} {36} (\bibinfo {year} {1985})}\BibitemShut
  {NoStop}%
\bibitem [{\citenamefont {Cline}(2006)}]{Cline:2006ts}%
  \BibitemOpen
  \bibfield  {author} {\bibinfo {author} {\bibfnamefont {J.~M.}\ \bibnamefont
  {Cline}},\ }in\ \href@noop {} {\emph {\bibinfo {booktitle} {{Les Houches
  Summer School - Session 86: Particle Physics and Cosmology: The Fabric of
  Spacetime}}}}\ (\bibinfo {year} {2006})\ \Eprint
  {http://arxiv.org/abs/hep-ph/0609145} {arXiv:hep-ph/0609145} \BibitemShut
  {NoStop}%
\bibitem [{\citenamefont {Davidson}\ \emph {et~al.}(2008)\citenamefont
  {Davidson}, \citenamefont {Nardi},\ and\ \citenamefont
  {Nir}}]{Davidson:2008bu}%
  \BibitemOpen
  \bibfield  {author} {\bibinfo {author} {\bibfnamefont {S.}~\bibnamefont
  {Davidson}}, \bibinfo {author} {\bibfnamefont {E.}~\bibnamefont {Nardi}}, \
  and\ \bibinfo {author} {\bibfnamefont {Y.}~\bibnamefont {Nir}},\ }\href
  {\doibase 10.1016/j.physrep.2008.06.002} {\bibfield  {journal} {\bibinfo
  {journal} {Phys. Rept.}\ }\textbf {\bibinfo {volume} {466}},\ \bibinfo
  {pages} {105} (\bibinfo {year} {2008})},\ \Eprint
  {http://arxiv.org/abs/0802.2962} {arXiv:0802.2962 [hep-ph]} \BibitemShut
  {NoStop}%
\bibitem [{\citenamefont {Buchmuller}\ \emph {et~al.}(2005)\citenamefont
  {Buchmuller}, \citenamefont {Peccei},\ and\ \citenamefont
  {Yanagida}}]{Buchmuller:2005eh}%
  \BibitemOpen
  \bibfield  {author} {\bibinfo {author} {\bibfnamefont {W.}~\bibnamefont
  {Buchmuller}}, \bibinfo {author} {\bibfnamefont {R.~D.}\ \bibnamefont
  {Peccei}}, \ and\ \bibinfo {author} {\bibfnamefont {T.}~\bibnamefont
  {Yanagida}},\ }\href {\doibase 10.1146/annurev.nucl.55.090704.151558}
  {\bibfield  {journal} {\bibinfo  {journal} {Ann. Rev. Nucl. Part. Sci.}\
  }\textbf {\bibinfo {volume} {55}},\ \bibinfo {pages} {311} (\bibinfo {year}
  {2005})},\ \Eprint {http://arxiv.org/abs/hep-ph/0502169}
  {arXiv:hep-ph/0502169} \BibitemShut {NoStop}%
\bibitem [{\citenamefont {Huang}\ and\ \citenamefont
  {Xie}(2022)}]{Huang:2022vkf}%
  \BibitemOpen
  \bibfield  {author} {\bibinfo {author} {\bibfnamefont {P.}~\bibnamefont
  {Huang}}\ and\ \bibinfo {author} {\bibfnamefont {K.-P.}\ \bibnamefont
  {Xie}},\ }\href {\doibase 10.1007/JHEP09(2022)052} {\bibfield  {journal}
  {\bibinfo  {journal} {JHEP}\ }\textbf {\bibinfo {volume} {09}},\ \bibinfo
  {pages} {052} (\bibinfo {year} {2022})},\ \Eprint
  {http://arxiv.org/abs/2206.04691} {arXiv:2206.04691 [hep-ph]} \BibitemShut
  {NoStop}%
\bibitem [{\citenamefont {Borah}\ \emph {et~al.}(2022)\citenamefont {Borah},
  \citenamefont {Dasgupta},\ and\ \citenamefont {Saha}}]{Borah:2022cdx}%
  \BibitemOpen
  \bibfield  {author} {\bibinfo {author} {\bibfnamefont {D.}~\bibnamefont
  {Borah}}, \bibinfo {author} {\bibfnamefont {A.}~\bibnamefont {Dasgupta}}, \
  and\ \bibinfo {author} {\bibfnamefont {I.}~\bibnamefont {Saha}},\ }\href
  {\doibase 10.1007/JHEP11(2022)136} {\bibfield  {journal} {\bibinfo  {journal}
  {JHEP}\ }\textbf {\bibinfo {volume} {11}},\ \bibinfo {pages} {136} (\bibinfo
  {year} {2022})},\ \Eprint {http://arxiv.org/abs/2207.14226} {arXiv:2207.14226
  [hep-ph]} \BibitemShut {NoStop}%
\bibitem [{\citenamefont {Chun}\ \emph {et~al.}(2023)\citenamefont {Chun},
  \citenamefont {Dutka}, \citenamefont {Jung}, \citenamefont {Nagels},\ and\
  \citenamefont {Vanvlasselaer}}]{Chun:2023ezg}%
  \BibitemOpen
  \bibfield  {author} {\bibinfo {author} {\bibfnamefont {E.~J.}\ \bibnamefont
  {Chun}}, \bibinfo {author} {\bibfnamefont {T.~P.}\ \bibnamefont {Dutka}},
  \bibinfo {author} {\bibfnamefont {T.~H.}\ \bibnamefont {Jung}}, \bibinfo
  {author} {\bibfnamefont {X.}~\bibnamefont {Nagels}}, \ and\ \bibinfo {author}
  {\bibfnamefont {M.}~\bibnamefont {Vanvlasselaer}},\ }\href {\doibase
  10.1007/JHEP09(2023)164} {\bibfield  {journal} {\bibinfo  {journal} {JHEP}\
  }\textbf {\bibinfo {volume} {09}},\ \bibinfo {pages} {164} (\bibinfo {year}
  {2023})},\ \Eprint {http://arxiv.org/abs/2305.10759} {arXiv:2305.10759
  [hep-ph]} \BibitemShut {NoStop}%
\bibitem [{\citenamefont {Azatov}\ \emph
  {et~al.}(2021{\natexlab{b}})\citenamefont {Azatov}, \citenamefont
  {Vanvlasselaer},\ and\ \citenamefont {Yin}}]{Azatov:2021irb}%
  \BibitemOpen
  \bibfield  {author} {\bibinfo {author} {\bibfnamefont {A.}~\bibnamefont
  {Azatov}}, \bibinfo {author} {\bibfnamefont {M.}~\bibnamefont
  {Vanvlasselaer}}, \ and\ \bibinfo {author} {\bibfnamefont {W.}~\bibnamefont
  {Yin}},\ }\href {\doibase 10.1007/JHEP10(2021)043} {\bibfield  {journal}
  {\bibinfo  {journal} {JHEP}\ }\textbf {\bibinfo {volume} {10}},\ \bibinfo
  {pages} {043} (\bibinfo {year} {2021}{\natexlab{b}})},\ \Eprint
  {http://arxiv.org/abs/2106.14913} {arXiv:2106.14913 [hep-ph]} \BibitemShut
  {NoStop}%
\bibitem [{\citenamefont {Baker}\ \emph {et~al.}(2022)\citenamefont {Baker},
  \citenamefont {Breitbach}, \citenamefont {Kopp}, \citenamefont {Mittnacht},\
  and\ \citenamefont {Soreq}}]{Baker:2021zsf}%
  \BibitemOpen
  \bibfield  {author} {\bibinfo {author} {\bibfnamefont {M.~J.}\ \bibnamefont
  {Baker}}, \bibinfo {author} {\bibfnamefont {M.}~\bibnamefont {Breitbach}},
  \bibinfo {author} {\bibfnamefont {J.}~\bibnamefont {Kopp}}, \bibinfo {author}
  {\bibfnamefont {L.}~\bibnamefont {Mittnacht}}, \ and\ \bibinfo {author}
  {\bibfnamefont {Y.}~\bibnamefont {Soreq}},\ }\href {\doibase
  10.1007/JHEP08(2022)010} {\bibfield  {journal} {\bibinfo  {journal} {JHEP}\
  }\textbf {\bibinfo {volume} {08}},\ \bibinfo {pages} {010} (\bibinfo {year}
  {2022})},\ \Eprint {http://arxiv.org/abs/2112.08987} {arXiv:2112.08987
  [hep-ph]} \BibitemShut {NoStop}%
\bibitem [{\citenamefont {{Dolgov}}\ and\ \citenamefont
  {{Silk}}(1993)}]{1993PhRvD..47.4244D}%
  \BibitemOpen
  \bibfield  {author} {\bibinfo {author} {\bibfnamefont {A.}~\bibnamefont
  {{Dolgov}}}\ and\ \bibinfo {author} {\bibfnamefont {J.}~\bibnamefont
  {{Silk}}},\ }\href {\doibase 10.1103/PhysRevD.47.4244} {\bibfield  {journal}
  {\bibinfo  {journal} {\prd}\ }\textbf {\bibinfo {volume} {47}},\ \bibinfo
  {pages} {4244} (\bibinfo {year} {1993})}\BibitemShut {NoStop}%
\bibitem [{\citenamefont {{Zel'Dovich}}(1971)}]{1971SR}%
  \BibitemOpen
  \bibfield  {author} {\bibinfo {author} {\bibfnamefont {Y.~B.}\ \bibnamefont
  {{Zel'Dovich}}},\ }\href@noop {} {\bibfield  {journal} {\bibinfo  {journal}
  {Soviet Journal of Experimental and Theoretical Physics Letters}\ }\textbf
  {\bibinfo {volume} {14}},\ \bibinfo {pages} {180} (\bibinfo {year}
  {1971})}\BibitemShut {NoStop}%
\bibitem [{\citenamefont {{Zel'Dovich}}(1972)}]{1972SR}%
  \BibitemOpen
  \bibfield  {author} {\bibinfo {author} {\bibfnamefont {Y.~B.}\ \bibnamefont
  {{Zel'Dovich}}},\ }\href@noop {} {\bibfield  {journal} {\bibinfo  {journal}
  {Soviet Journal of Experimental and Theoretical Physics}\ }\textbf {\bibinfo
  {volume} {35}},\ \bibinfo {pages} {1085} (\bibinfo {year}
  {1972})}\BibitemShut {NoStop}%
\bibitem [{\citenamefont {Starobinsky}(1973)}]{1973SR}%
  \BibitemOpen
  \bibfield  {author} {\bibinfo {author} {\bibfnamefont {A.~A.}\ \bibnamefont
  {Starobinsky}},\ }\href@noop {} {\bibfield  {journal} {\bibinfo  {journal}
  {Soviet Journal of Experimental and Theoretical Physics}\ }\textbf {\bibinfo
  {volume} {37}},\ \bibinfo {pages} {28} (\bibinfo {year} {1973})}\BibitemShut
  {NoStop}%
\bibitem [{\citenamefont {Zouros}\ and\ \citenamefont
  {Eardley}(1979)}]{Zouros:1979iw}%
  \BibitemOpen
  \bibfield  {author} {\bibinfo {author} {\bibfnamefont {T.~J.~M.}\
  \bibnamefont {Zouros}}\ and\ \bibinfo {author} {\bibfnamefont {D.~M.}\
  \bibnamefont {Eardley}},\ }\href {\doibase 10.1016/0003-4916(79)90237-9}
  {\bibfield  {journal} {\bibinfo  {journal} {Annals Phys.}\ }\textbf {\bibinfo
  {volume} {118}},\ \bibinfo {pages} {139} (\bibinfo {year}
  {1979})}\BibitemShut {NoStop}%
\bibitem [{\citenamefont {Detweiler}(1980)}]{Detweiler:1980uk}%
  \BibitemOpen
  \bibfield  {author} {\bibinfo {author} {\bibfnamefont {S.~L.}\ \bibnamefont
  {Detweiler}},\ }\href {\doibase 10.1103/PhysRevD.22.2323} {\bibfield
  {journal} {\bibinfo  {journal} {Phys. Rev. D}\ }\textbf {\bibinfo {volume}
  {22}},\ \bibinfo {pages} {2323} (\bibinfo {year} {1980})}\BibitemShut
  {NoStop}%
\bibitem [{\citenamefont {Dolan}(2007)}]{Dolan:2007mj}%
  \BibitemOpen
  \bibfield  {author} {\bibinfo {author} {\bibfnamefont {S.~R.}\ \bibnamefont
  {Dolan}},\ }\href {\doibase 10.1103/PhysRevD.76.084001} {\bibfield  {journal}
  {\bibinfo  {journal} {Phys. Rev. D}\ }\textbf {\bibinfo {volume} {76}},\
  \bibinfo {pages} {084001} (\bibinfo {year} {2007})},\ \Eprint
  {http://arxiv.org/abs/0705.2880} {arXiv:0705.2880 [gr-qc]} \BibitemShut
  {NoStop}%
\bibitem [{\citenamefont {Arvanitaki}\ \emph {et~al.}(2015)\citenamefont
  {Arvanitaki}, \citenamefont {Baryakhtar},\ and\ \citenamefont
  {Huang}}]{Arvanitaki:2014wva}%
  \BibitemOpen
  \bibfield  {author} {\bibinfo {author} {\bibfnamefont {A.}~\bibnamefont
  {Arvanitaki}}, \bibinfo {author} {\bibfnamefont {M.}~\bibnamefont
  {Baryakhtar}}, \ and\ \bibinfo {author} {\bibfnamefont {X.}~\bibnamefont
  {Huang}},\ }\href {\doibase 10.1103/PhysRevD.91.084011} {\bibfield  {journal}
  {\bibinfo  {journal} {Phys. Rev. D}\ }\textbf {\bibinfo {volume} {91}},\
  \bibinfo {pages} {084011} (\bibinfo {year} {2015})},\ \Eprint
  {http://arxiv.org/abs/1411.2263} {arXiv:1411.2263 [hep-ph]} \BibitemShut
  {NoStop}%
\bibitem [{\citenamefont {{Brito}}\ \emph {et~al.}(2020)\citenamefont
  {{Brito}}, \citenamefont {{Cardoso}},\ and\ \citenamefont
  {{Pani}}}]{2020SRbook}%
  \BibitemOpen
  \bibfield  {author} {\bibinfo {author} {\bibfnamefont {R.}~\bibnamefont
  {{Brito}}}, \bibinfo {author} {\bibfnamefont {V.}~\bibnamefont {{Cardoso}}},
  \ and\ \bibinfo {author} {\bibfnamefont {P.}~\bibnamefont {{Pani}}},\ }\href
  {\doibase 10.1007/978-3-030-46622-0} {\emph {\bibinfo {title}
  {{Superradiance. New Frontiers in Black Hole Physics}}}},\ Vol.\ \bibinfo
  {volume} {971}\ (\bibinfo {year} {2020})\BibitemShut {NoStop}%
\bibitem [{\citenamefont {Zhang}\ and\ \citenamefont
  {Yang}(2020)}]{Zhang:2019eid}%
  \BibitemOpen
  \bibfield  {author} {\bibinfo {author} {\bibfnamefont {J.}~\bibnamefont
  {Zhang}}\ and\ \bibinfo {author} {\bibfnamefont {H.}~\bibnamefont {Yang}},\
  }\href {\doibase 10.1103/PhysRevD.101.043020} {\bibfield  {journal} {\bibinfo
   {journal} {Phys. Rev. D}\ }\textbf {\bibinfo {volume} {101}},\ \bibinfo
  {pages} {043020} (\bibinfo {year} {2020})},\ \Eprint
  {http://arxiv.org/abs/1907.13582} {arXiv:1907.13582 [gr-qc]} \BibitemShut
  {NoStop}%
\bibitem [{\citenamefont {Xie}\ and\ \citenamefont
  {Huang}(2024)}]{Xie:2022uvp}%
  \BibitemOpen
  \bibfield  {author} {\bibinfo {author} {\bibfnamefont {N.}~\bibnamefont
  {Xie}}\ and\ \bibinfo {author} {\bibfnamefont {F.~P.}\ \bibnamefont
  {Huang}},\ }\href {\doibase 10.1007/s11433-023-2172-7} {\bibfield  {journal}
  {\bibinfo  {journal} {Sci. China Phys. Mech. Astron.}\ }\textbf {\bibinfo
  {volume} {67}},\ \bibinfo {pages} {210411} (\bibinfo {year} {2024})},\
  \Eprint {http://arxiv.org/abs/2207.11145} {arXiv:2207.11145 [hep-ph]}
  \BibitemShut {NoStop}%
\bibitem [{\citenamefont {Eda}\ \emph {et~al.}(2013)\citenamefont {Eda},
  \citenamefont {Itoh}, \citenamefont {Kuroyanagi},\ and\ \citenamefont
  {Silk}}]{Eda:2013gg}%
  \BibitemOpen
  \bibfield  {author} {\bibinfo {author} {\bibfnamefont {K.}~\bibnamefont
  {Eda}}, \bibinfo {author} {\bibfnamefont {Y.}~\bibnamefont {Itoh}}, \bibinfo
  {author} {\bibfnamefont {S.}~\bibnamefont {Kuroyanagi}}, \ and\ \bibinfo
  {author} {\bibfnamefont {J.}~\bibnamefont {Silk}},\ }\href {\doibase
  10.1103/PhysRevLett.110.221101} {\bibfield  {journal} {\bibinfo  {journal}
  {Phys. Rev. Lett.}\ }\textbf {\bibinfo {volume} {110}},\ \bibinfo {pages}
  {221101} (\bibinfo {year} {2013})},\ \Eprint {http://arxiv.org/abs/1301.5971}
  {arXiv:1301.5971 [gr-qc]} \BibitemShut {NoStop}%
\bibitem [{\citenamefont {Eda}\ \emph {et~al.}(2015)\citenamefont {Eda},
  \citenamefont {Itoh}, \citenamefont {Kuroyanagi},\ and\ \citenamefont
  {Silk}}]{Eda:2014kra}%
  \BibitemOpen
  \bibfield  {author} {\bibinfo {author} {\bibfnamefont {K.}~\bibnamefont
  {Eda}}, \bibinfo {author} {\bibfnamefont {Y.}~\bibnamefont {Itoh}}, \bibinfo
  {author} {\bibfnamefont {S.}~\bibnamefont {Kuroyanagi}}, \ and\ \bibinfo
  {author} {\bibfnamefont {J.}~\bibnamefont {Silk}},\ }\href {\doibase
  10.1103/PhysRevD.91.044045} {\bibfield  {journal} {\bibinfo  {journal} {Phys.
  Rev. D}\ }\textbf {\bibinfo {volume} {91}},\ \bibinfo {pages} {044045}
  (\bibinfo {year} {2015})},\ \Eprint {http://arxiv.org/abs/1408.3534}
  {arXiv:1408.3534 [gr-qc]} \BibitemShut {NoStop}%
\bibitem [{\citenamefont {Arvanitaki}\ and\ \citenamefont
  {Dubovsky}(2011)}]{Arvanitaki:2010sy}%
  \BibitemOpen
  \bibfield  {author} {\bibinfo {author} {\bibfnamefont {A.}~\bibnamefont
  {Arvanitaki}}\ and\ \bibinfo {author} {\bibfnamefont {S.}~\bibnamefont
  {Dubovsky}},\ }\href {\doibase 10.1103/PhysRevD.83.044026} {\bibfield
  {journal} {\bibinfo  {journal} {Phys. Rev. D}\ }\textbf {\bibinfo {volume}
  {83}},\ \bibinfo {pages} {044026} (\bibinfo {year} {2011})},\ \Eprint
  {http://arxiv.org/abs/1004.3558} {arXiv:1004.3558 [hep-th]} \BibitemShut
  {NoStop}%
\bibitem [{\citenamefont {Arvanitaki}\ \emph {et~al.}(2017)\citenamefont
  {Arvanitaki}, \citenamefont {Baryakhtar}, \citenamefont {Dimopoulos},
  \citenamefont {Dubovsky},\ and\ \citenamefont
  {Lasenby}}]{Arvanitaki:2016qwi}%
  \BibitemOpen
  \bibfield  {author} {\bibinfo {author} {\bibfnamefont {A.}~\bibnamefont
  {Arvanitaki}}, \bibinfo {author} {\bibfnamefont {M.}~\bibnamefont
  {Baryakhtar}}, \bibinfo {author} {\bibfnamefont {S.}~\bibnamefont
  {Dimopoulos}}, \bibinfo {author} {\bibfnamefont {S.}~\bibnamefont
  {Dubovsky}}, \ and\ \bibinfo {author} {\bibfnamefont {R.}~\bibnamefont
  {Lasenby}},\ }\href {\doibase 10.1103/PhysRevD.95.043001} {\bibfield
  {journal} {\bibinfo  {journal} {Phys. Rev. D}\ }\textbf {\bibinfo {volume}
  {95}},\ \bibinfo {pages} {043001} (\bibinfo {year} {2017})},\ \Eprint
  {http://arxiv.org/abs/1604.03958} {arXiv:1604.03958 [hep-ph]} \BibitemShut
  {NoStop}%
\bibitem [{\citenamefont {Brito}\ \emph
  {et~al.}(2017{\natexlab{a}})\citenamefont {Brito}, \citenamefont {Ghosh},
  \citenamefont {Barausse}, \citenamefont {Berti}, \citenamefont {Cardoso},
  \citenamefont {Dvorkin}, \citenamefont {Klein},\ and\ \citenamefont
  {Pani}}]{Brito:2017wnc}%
  \BibitemOpen
  \bibfield  {author} {\bibinfo {author} {\bibfnamefont {R.}~\bibnamefont
  {Brito}}, \bibinfo {author} {\bibfnamefont {S.}~\bibnamefont {Ghosh}},
  \bibinfo {author} {\bibfnamefont {E.}~\bibnamefont {Barausse}}, \bibinfo
  {author} {\bibfnamefont {E.}~\bibnamefont {Berti}}, \bibinfo {author}
  {\bibfnamefont {V.}~\bibnamefont {Cardoso}}, \bibinfo {author} {\bibfnamefont
  {I.}~\bibnamefont {Dvorkin}}, \bibinfo {author} {\bibfnamefont
  {A.}~\bibnamefont {Klein}}, \ and\ \bibinfo {author} {\bibfnamefont
  {P.}~\bibnamefont {Pani}},\ }\href {\doibase 10.1103/PhysRevLett.119.131101}
  {\bibfield  {journal} {\bibinfo  {journal} {Phys. Rev. Lett.}\ }\textbf
  {\bibinfo {volume} {119}},\ \bibinfo {pages} {131101} (\bibinfo {year}
  {2017}{\natexlab{a}})},\ \Eprint {http://arxiv.org/abs/1706.05097}
  {arXiv:1706.05097 [gr-qc]} \BibitemShut {NoStop}%
\bibitem [{\citenamefont {Brito}\ \emph
  {et~al.}(2017{\natexlab{b}})\citenamefont {Brito}, \citenamefont {Ghosh},
  \citenamefont {Barausse}, \citenamefont {Berti}, \citenamefont {Cardoso},
  \citenamefont {Dvorkin}, \citenamefont {Klein},\ and\ \citenamefont
  {Pani}}]{Brito:2017zvb}%
  \BibitemOpen
  \bibfield  {author} {\bibinfo {author} {\bibfnamefont {R.}~\bibnamefont
  {Brito}}, \bibinfo {author} {\bibfnamefont {S.}~\bibnamefont {Ghosh}},
  \bibinfo {author} {\bibfnamefont {E.}~\bibnamefont {Barausse}}, \bibinfo
  {author} {\bibfnamefont {E.}~\bibnamefont {Berti}}, \bibinfo {author}
  {\bibfnamefont {V.}~\bibnamefont {Cardoso}}, \bibinfo {author} {\bibfnamefont
  {I.}~\bibnamefont {Dvorkin}}, \bibinfo {author} {\bibfnamefont
  {A.}~\bibnamefont {Klein}}, \ and\ \bibinfo {author} {\bibfnamefont
  {P.}~\bibnamefont {Pani}},\ }\href {\doibase 10.1103/PhysRevD.96.064050}
  {\bibfield  {journal} {\bibinfo  {journal} {Phys. Rev. D}\ }\textbf {\bibinfo
  {volume} {96}},\ \bibinfo {pages} {064050} (\bibinfo {year}
  {2017}{\natexlab{b}})},\ \Eprint {http://arxiv.org/abs/1706.06311}
  {arXiv:1706.06311 [gr-qc]} \BibitemShut {NoStop}%
\bibitem [{\citenamefont {Baryakhtar}\ \emph {et~al.}(2017)\citenamefont
  {Baryakhtar}, \citenamefont {Lasenby},\ and\ \citenamefont
  {Teo}}]{Baryakhtar:2017ngi}%
  \BibitemOpen
  \bibfield  {author} {\bibinfo {author} {\bibfnamefont {M.}~\bibnamefont
  {Baryakhtar}}, \bibinfo {author} {\bibfnamefont {R.}~\bibnamefont {Lasenby}},
  \ and\ \bibinfo {author} {\bibfnamefont {M.}~\bibnamefont {Teo}},\ }\href
  {\doibase 10.1103/PhysRevD.96.035019} {\bibfield  {journal} {\bibinfo
  {journal} {Phys. Rev. D}\ }\textbf {\bibinfo {volume} {96}},\ \bibinfo
  {pages} {035019} (\bibinfo {year} {2017})},\ \Eprint
  {http://arxiv.org/abs/1704.05081} {arXiv:1704.05081 [hep-ph]} \BibitemShut
  {NoStop}%
\bibitem [{\citenamefont {Siemonsen}\ and\ \citenamefont
  {East}(2020)}]{Siemonsen:2019ebd}%
  \BibitemOpen
  \bibfield  {author} {\bibinfo {author} {\bibfnamefont {N.}~\bibnamefont
  {Siemonsen}}\ and\ \bibinfo {author} {\bibfnamefont {W.~E.}\ \bibnamefont
  {East}},\ }\href {\doibase 10.1103/PhysRevD.101.024019} {\bibfield  {journal}
  {\bibinfo  {journal} {Phys. Rev. D}\ }\textbf {\bibinfo {volume} {101}},\
  \bibinfo {pages} {024019} (\bibinfo {year} {2020})},\ \Eprint
  {http://arxiv.org/abs/1910.09476} {arXiv:1910.09476 [gr-qc]} \BibitemShut
  {NoStop}%
\bibitem [{\citenamefont {Palomba}\ \emph {et~al.}(2019)\citenamefont {Palomba}
  \emph {et~al.}}]{Palomba:2019vxe}%
  \BibitemOpen
  \bibfield  {author} {\bibinfo {author} {\bibfnamefont {C.}~\bibnamefont
  {Palomba}} \emph {et~al.},\ }\href {\doibase 10.1103/PhysRevLett.123.171101}
  {\bibfield  {journal} {\bibinfo  {journal} {Phys. Rev. Lett.}\ }\textbf
  {\bibinfo {volume} {123}},\ \bibinfo {pages} {171101} (\bibinfo {year}
  {2019})},\ \Eprint {http://arxiv.org/abs/1909.08854} {arXiv:1909.08854
  [astro-ph.HE]} \BibitemShut {NoStop}%
\bibitem [{\citenamefont {Abbott}\ \emph
  {et~al.}(2022{\natexlab{a}})\citenamefont {Abbott} \emph
  {et~al.}}]{LIGOScientific:2021rnv}%
  \BibitemOpen
  \bibfield  {author} {\bibinfo {author} {\bibfnamefont {R.}~\bibnamefont
  {Abbott}} \emph {et~al.} (\bibinfo {collaboration} {LIGO Scientific, Virgo,
  KAGRA}),\ }\href {\doibase 10.1103/PhysRevD.105.102001} {\bibfield  {journal}
  {\bibinfo  {journal} {Phys. Rev. D}\ }\textbf {\bibinfo {volume} {105}},\
  \bibinfo {pages} {102001} (\bibinfo {year} {2022}{\natexlab{a}})},\ \Eprint
  {http://arxiv.org/abs/2111.15507} {arXiv:2111.15507 [astro-ph.HE]}
  \BibitemShut {NoStop}%
\bibitem [{\citenamefont {Tsukada}\ \emph {et~al.}(2021)\citenamefont
  {Tsukada}, \citenamefont {Brito}, \citenamefont {East},\ and\ \citenamefont
  {Siemonsen}}]{Tsukada:2020lgt}%
  \BibitemOpen
  \bibfield  {author} {\bibinfo {author} {\bibfnamefont {L.}~\bibnamefont
  {Tsukada}}, \bibinfo {author} {\bibfnamefont {R.}~\bibnamefont {Brito}},
  \bibinfo {author} {\bibfnamefont {W.~E.}\ \bibnamefont {East}}, \ and\
  \bibinfo {author} {\bibfnamefont {N.}~\bibnamefont {Siemonsen}},\ }\href
  {\doibase 10.1103/PhysRevD.103.083005} {\bibfield  {journal} {\bibinfo
  {journal} {Phys. Rev. D}\ }\textbf {\bibinfo {volume} {103}},\ \bibinfo
  {pages} {083005} (\bibinfo {year} {2021})},\ \Eprint
  {http://arxiv.org/abs/2011.06995} {arXiv:2011.06995 [astro-ph.HE]}
  \BibitemShut {NoStop}%
\bibitem [{\citenamefont {Yang}\ \emph {et~al.}(2023)\citenamefont {Yang},
  \citenamefont {Xie},\ and\ \citenamefont {Huang}}]{Yang:2023aak}%
  \BibitemOpen
  \bibfield  {author} {\bibinfo {author} {\bibfnamefont {J.}~\bibnamefont
  {Yang}}, \bibinfo {author} {\bibfnamefont {N.}~\bibnamefont {Xie}}, \ and\
  \bibinfo {author} {\bibfnamefont {F.~P.}\ \bibnamefont {Huang}},\ }\href@noop
  {} {\  (\bibinfo {year} {2023})},\ \Eprint {http://arxiv.org/abs/2306.17113}
  {arXiv:2306.17113 [hep-ph]} \BibitemShut {NoStop}%
\bibitem [{\citenamefont {Wang}\ \emph
  {et~al.}(2020{\natexlab{a}})\citenamefont {Wang}, \citenamefont {Huang},\
  and\ \citenamefont {Zhang}}]{Wang:2020jrd}%
  \BibitemOpen
  \bibfield  {author} {\bibinfo {author} {\bibfnamefont {X.}~\bibnamefont
  {Wang}}, \bibinfo {author} {\bibfnamefont {F.~P.}\ \bibnamefont {Huang}}, \
  and\ \bibinfo {author} {\bibfnamefont {X.}~\bibnamefont {Zhang}},\ }\href
  {\doibase 10.1088/1475-7516/2020/05/045} {\bibfield  {journal} {\bibinfo
  {journal} {JCAP}\ }\textbf {\bibinfo {volume} {05}},\ \bibinfo {pages} {045}
  (\bibinfo {year} {2020}{\natexlab{a}})},\ \Eprint
  {http://arxiv.org/abs/2003.08892} {arXiv:2003.08892 [hep-ph]} \BibitemShut
  {NoStop}%
\bibitem [{\citenamefont {Moore}\ and\ \citenamefont
  {Prokopec}(1995)}]{Moore:1995si}%
  \BibitemOpen
  \bibfield  {author} {\bibinfo {author} {\bibfnamefont {G.~D.}\ \bibnamefont
  {Moore}}\ and\ \bibinfo {author} {\bibfnamefont {T.}~\bibnamefont
  {Prokopec}},\ }\href {\doibase 10.1103/PhysRevD.52.7182} {\bibfield
  {journal} {\bibinfo  {journal} {Phys. Rev. D}\ }\textbf {\bibinfo {volume}
  {52}},\ \bibinfo {pages} {7182} (\bibinfo {year} {1995})},\ \Eprint
  {http://arxiv.org/abs/hep-ph/9506475} {arXiv:hep-ph/9506475} \BibitemShut
  {NoStop}%
\bibitem [{\citenamefont {Wang}\ \emph
  {et~al.}(2020{\natexlab{b}})\citenamefont {Wang}, \citenamefont {Huang},\
  and\ \citenamefont {Zhang}}]{Wang:2020zlf}%
  \BibitemOpen
  \bibfield  {author} {\bibinfo {author} {\bibfnamefont {X.}~\bibnamefont
  {Wang}}, \bibinfo {author} {\bibfnamefont {F.~P.}\ \bibnamefont {Huang}}, \
  and\ \bibinfo {author} {\bibfnamefont {X.}~\bibnamefont {Zhang}},\
  }\href@noop {} {\  (\bibinfo {year} {2020}{\natexlab{b}})},\ \Eprint
  {http://arxiv.org/abs/2011.12903} {arXiv:2011.12903 [hep-ph]} \BibitemShut
  {NoStop}%
\bibitem [{\citenamefont {Jiang}\ \emph
  {et~al.}(2023{\natexlab{b}})\citenamefont {Jiang}, \citenamefont {Huang},\
  and\ \citenamefont {Wang}}]{Jiang:2022btc}%
  \BibitemOpen
  \bibfield  {author} {\bibinfo {author} {\bibfnamefont {S.}~\bibnamefont
  {Jiang}}, \bibinfo {author} {\bibfnamefont {F.~P.}\ \bibnamefont {Huang}}, \
  and\ \bibinfo {author} {\bibfnamefont {X.}~\bibnamefont {Wang}},\ }\href
  {\doibase 10.1103/PhysRevD.107.095005} {\bibfield  {journal} {\bibinfo
  {journal} {Phys. Rev. D}\ }\textbf {\bibinfo {volume} {107}},\ \bibinfo
  {pages} {095005} (\bibinfo {year} {2023}{\natexlab{b}})},\ \Eprint
  {http://arxiv.org/abs/2211.13142} {arXiv:2211.13142 [hep-ph]} \BibitemShut
  {NoStop}%
\bibitem [{\citenamefont {Laurent}\ and\ \citenamefont
  {Cline}(2022)}]{Laurent:2022jrs}%
  \BibitemOpen
  \bibfield  {author} {\bibinfo {author} {\bibfnamefont {B.}~\bibnamefont
  {Laurent}}\ and\ \bibinfo {author} {\bibfnamefont {J.~M.}\ \bibnamefont
  {Cline}},\ }\href {\doibase 10.1103/PhysRevD.106.023501} {\bibfield
  {journal} {\bibinfo  {journal} {Phys. Rev. D}\ }\textbf {\bibinfo {volume}
  {106}},\ \bibinfo {pages} {023501} (\bibinfo {year} {2022})},\ \Eprint
  {http://arxiv.org/abs/2204.13120} {arXiv:2204.13120 [hep-ph]} \BibitemShut
  {NoStop}%
\bibitem [{\citenamefont {Stephani}\ \emph {et~al.}(2009)\citenamefont
  {Stephani}, \citenamefont {Kramer}, \citenamefont {MacCallum}, \citenamefont
  {Hoenselaers},\ and\ \citenamefont {Herlt}}]{stephani2009exact}%
  \BibitemOpen
  \bibfield  {author} {\bibinfo {author} {\bibfnamefont {H.}~\bibnamefont
  {Stephani}}, \bibinfo {author} {\bibfnamefont {D.}~\bibnamefont {Kramer}},
  \bibinfo {author} {\bibfnamefont {M.}~\bibnamefont {MacCallum}}, \bibinfo
  {author} {\bibfnamefont {C.}~\bibnamefont {Hoenselaers}}, \ and\ \bibinfo
  {author} {\bibfnamefont {E.}~\bibnamefont {Herlt}},\ }\href@noop {} {\emph
  {\bibinfo {title} {Exact solutions of Einstein's field equations}}}\
  (\bibinfo  {publisher} {Cambridge university press},\ \bibinfo {year}
  {2009})\BibitemShut {NoStop}%
\bibitem [{\citenamefont {Griffiths}\ and\ \citenamefont
  {Podolsk{\`y}}(2009)}]{griffiths2009exact}%
  \BibitemOpen
  \bibfield  {author} {\bibinfo {author} {\bibfnamefont {J.~B.}\ \bibnamefont
  {Griffiths}}\ and\ \bibinfo {author} {\bibfnamefont {J.}~\bibnamefont
  {Podolsk{\`y}}},\ }\href@noop {} {\emph {\bibinfo {title} {Exact space-times
  in Einstein's general relativity}}}\ (\bibinfo  {publisher} {Cambridge
  University Press},\ \bibinfo {year} {2009})\BibitemShut {NoStop}%
\bibitem [{\citenamefont {Page}(2006)}]{page2006evidence}%
  \BibitemOpen
  \bibfield  {author} {\bibinfo {author} {\bibfnamefont {D.~N.}\ \bibnamefont
  {Page}},\ }\href@noop {} {\bibfield  {journal} {\bibinfo  {journal} {The
  Astrophysical Journal}\ }\textbf {\bibinfo {volume} {653}},\ \bibinfo {pages}
  {1400} (\bibinfo {year} {2006})}\BibitemShut {NoStop}%
\bibitem [{\citenamefont {Wald}(1974)}]{wald1974black}%
  \BibitemOpen
  \bibfield  {author} {\bibinfo {author} {\bibfnamefont {R.~M.}\ \bibnamefont
  {Wald}},\ }\href@noop {} {\bibfield  {journal} {\bibinfo  {journal} {Physical
  Review D}\ }\textbf {\bibinfo {volume} {10}},\ \bibinfo {pages} {1680}
  (\bibinfo {year} {1974})}\BibitemShut {NoStop}%
\bibitem [{\citenamefont {Zaja\v{c}ek}\ \emph {et~al.}(2018)\citenamefont
  {Zaja\v{c}ek}, \citenamefont {Tursunov}, \citenamefont {Eckart},\ and\
  \citenamefont {Britzen}}]{Zajacek:2018ycb}%
  \BibitemOpen
  \bibfield  {author} {\bibinfo {author} {\bibfnamefont {M.}~\bibnamefont
  {Zaja\v{c}ek}}, \bibinfo {author} {\bibfnamefont {A.}~\bibnamefont
  {Tursunov}}, \bibinfo {author} {\bibfnamefont {A.}~\bibnamefont {Eckart}}, \
  and\ \bibinfo {author} {\bibfnamefont {S.}~\bibnamefont {Britzen}},\ }\href
  {\doibase 10.1093/mnras/sty2182} {\bibfield  {journal} {\bibinfo  {journal}
  {Mon. Not. Roy. Astron. Soc.}\ }\textbf {\bibinfo {volume} {480}},\ \bibinfo
  {pages} {4408} (\bibinfo {year} {2018})},\ \Eprint
  {http://arxiv.org/abs/1808.07327} {arXiv:1808.07327 [astro-ph.GA]}
  \BibitemShut {NoStop}%
\bibitem [{\citenamefont {Blandford}\ and\ \citenamefont
  {Znajek}(1977{\natexlab{b}})}]{blandford1977electromagnetic}%
  \BibitemOpen
  \bibfield  {author} {\bibinfo {author} {\bibfnamefont {R.~D.}\ \bibnamefont
  {Blandford}}\ and\ \bibinfo {author} {\bibfnamefont {R.~L.}\ \bibnamefont
  {Znajek}},\ }\href@noop {} {\bibfield  {journal} {\bibinfo  {journal}
  {Monthly Notices of the Royal Astronomical Society}\ }\textbf {\bibinfo
  {volume} {179}},\ \bibinfo {pages} {433} (\bibinfo {year}
  {1977}{\natexlab{b}})}\BibitemShut {NoStop}%
\bibitem [{\citenamefont {Lee}\ \emph {et~al.}(2000)\citenamefont {Lee},
  \citenamefont {Wijers},\ and\ \citenamefont {Brown}}]{Lee:1999se}%
  \BibitemOpen
  \bibfield  {author} {\bibinfo {author} {\bibfnamefont {H.~K.}\ \bibnamefont
  {Lee}}, \bibinfo {author} {\bibfnamefont {R.~A. M.~J.}\ \bibnamefont
  {Wijers}}, \ and\ \bibinfo {author} {\bibfnamefont {G.~E.}\ \bibnamefont
  {Brown}},\ }\href {\doibase 10.1016/S0370-1573(99)00084-8} {\bibfield
  {journal} {\bibinfo  {journal} {Phys. Rept.}\ }\textbf {\bibinfo {volume}
  {325}},\ \bibinfo {pages} {83} (\bibinfo {year} {2000})},\ \Eprint
  {http://arxiv.org/abs/astro-ph/9906213} {arXiv:astro-ph/9906213} \BibitemShut
  {NoStop}%
\bibitem [{\citenamefont {Wagh}\ \emph {et~al.}(1985)\citenamefont {Wagh},
  \citenamefont {Dhurandhar},\ and\ \citenamefont {Dadhich}}]{wagh1985revival}%
  \BibitemOpen
  \bibfield  {author} {\bibinfo {author} {\bibfnamefont {S.}~\bibnamefont
  {Wagh}}, \bibinfo {author} {\bibfnamefont {S.}~\bibnamefont {Dhurandhar}}, \
  and\ \bibinfo {author} {\bibfnamefont {N.}~\bibnamefont {Dadhich}},\
  }\href@noop {} {\bibfield  {journal} {\bibinfo  {journal} {Astrophysical
  Journal}\ }\textbf {\bibinfo {volume} {290}},\ \bibinfo {pages} {12}
  (\bibinfo {year} {1985})}\BibitemShut {NoStop}%
\bibitem [{\citenamefont {Shnir}(2005)}]{Shnir:2005vvi}%
  \BibitemOpen
  \bibfield  {author} {\bibinfo {author} {\bibfnamefont {Y.~M.}\ \bibnamefont
  {Shnir}},\ }\href {\doibase 10.1007/3-540-29082-6} {\emph {\bibinfo {title}
  {{Magnetic Monopoles}}}},\ Text and Monographs in Physics\ (\bibinfo
  {publisher} {Springer},\ \bibinfo {address} {Berlin/Heidelberg},\ \bibinfo
  {year} {2005})\BibitemShut {NoStop}%
\bibitem [{\citenamefont {t~Hooft}(1974)}]{t1974magnetic}%
  \BibitemOpen
  \bibfield  {author} {\bibinfo {author} {\bibfnamefont {G.}~\bibnamefont
  {t~Hooft}},\ }\href@noop {} {\bibfield  {journal} {\bibinfo  {journal} {Nucl.
  Phys. B}\ }\textbf {\bibinfo {volume} {79}},\ \bibinfo {pages} {276}
  (\bibinfo {year} {1974})}\BibitemShut {NoStop}%
\bibitem [{\citenamefont {Polyakov}(1974)}]{Polyakov:1974ek}%
  \BibitemOpen
  \bibfield  {author} {\bibinfo {author} {\bibfnamefont {A.~M.}\ \bibnamefont
  {Polyakov}},\ }\href@noop {} {\bibfield  {journal} {\bibinfo  {journal} {JETP
  Lett.}\ }\textbf {\bibinfo {volume} {20}},\ \bibinfo {pages} {194} (\bibinfo
  {year} {1974})}\BibitemShut {NoStop}%
\bibitem [{\citenamefont {Julia}\ and\ \citenamefont
  {Zee}(1975)}]{julia1975poles}%
  \BibitemOpen
  \bibfield  {author} {\bibinfo {author} {\bibfnamefont {B.}~\bibnamefont
  {Julia}}\ and\ \bibinfo {author} {\bibfnamefont {A.}~\bibnamefont {Zee}},\
  }\href@noop {} {\bibfield  {journal} {\bibinfo  {journal} {Physical Review
  D}\ }\textbf {\bibinfo {volume} {11}},\ \bibinfo {pages} {2227} (\bibinfo
  {year} {1975})}\BibitemShut {NoStop}%
\bibitem [{\citenamefont {Lee}\ \emph {et~al.}(1992)\citenamefont {Lee},
  \citenamefont {Nair},\ and\ \citenamefont {Weinberg}}]{lee1992black}%
  \BibitemOpen
  \bibfield  {author} {\bibinfo {author} {\bibfnamefont {K.}~\bibnamefont
  {Lee}}, \bibinfo {author} {\bibfnamefont {V.}~\bibnamefont {Nair}}, \ and\
  \bibinfo {author} {\bibfnamefont {E.~J.}\ \bibnamefont {Weinberg}},\
  }\href@noop {} {\bibfield  {journal} {\bibinfo  {journal} {Physical Review
  D}\ }\textbf {\bibinfo {volume} {45}},\ \bibinfo {pages} {2751} (\bibinfo
  {year} {1992})}\BibitemShut {NoStop}%
\bibitem [{\citenamefont {Maldacena}(2020)}]{Maldacena:2020skw}%
  \BibitemOpen
  \bibfield  {author} {\bibinfo {author} {\bibfnamefont {J.}~\bibnamefont
  {Maldacena}},\ }\href@noop {} {\  (\bibinfo {year} {2020})},\ \Eprint
  {http://arxiv.org/abs/2004.06084} {arXiv:2004.06084 [hep-th]} \BibitemShut
  {NoStop}%
\bibitem [{\citenamefont {Kasuya}(1982)}]{kasuya1982exact}%
  \BibitemOpen
  \bibfield  {author} {\bibinfo {author} {\bibfnamefont {M.}~\bibnamefont
  {Kasuya}},\ }\href@noop {} {\bibfield  {journal} {\bibinfo  {journal}
  {Physical Review D}\ }\textbf {\bibinfo {volume} {25}},\ \bibinfo {pages}
  {995} (\bibinfo {year} {1982})}\BibitemShut {NoStop}%
\bibitem [{\citenamefont {Cardoso}\ \emph {et~al.}(2016)\citenamefont
  {Cardoso}, \citenamefont {Macedo}, \citenamefont {Pani},\ and\ \citenamefont
  {Ferrari}}]{Cardoso:2016olt}%
  \BibitemOpen
  \bibfield  {author} {\bibinfo {author} {\bibfnamefont {V.}~\bibnamefont
  {Cardoso}}, \bibinfo {author} {\bibfnamefont {C.~F.~B.}\ \bibnamefont
  {Macedo}}, \bibinfo {author} {\bibfnamefont {P.}~\bibnamefont {Pani}}, \ and\
  \bibinfo {author} {\bibfnamefont {V.}~\bibnamefont {Ferrari}},\ }\href
  {\doibase 10.1088/1475-7516/2016/05/054, 10.1088/1475-7516/2020/04/E01}
  {\bibfield  {journal} {\bibinfo  {journal} {JCAP}\ }\textbf {\bibinfo
  {volume} {1605}},\ \bibinfo {pages} {054} (\bibinfo {year} {2016})},\
  \bibinfo {note} {[Erratum: JCAP2004,E01(2020)]},\ \Eprint
  {http://arxiv.org/abs/1604.07845} {arXiv:1604.07845 [hep-ph]} \BibitemShut
  {NoStop}%
\bibitem [{\citenamefont {Liebling}\ and\ \citenamefont
  {Palenzuela}(2016)}]{Liebling:2016orx}%
  \BibitemOpen
  \bibfield  {author} {\bibinfo {author} {\bibfnamefont {S.~L.}\ \bibnamefont
  {Liebling}}\ and\ \bibinfo {author} {\bibfnamefont {C.}~\bibnamefont
  {Palenzuela}},\ }\href {\doibase 10.1103/PhysRevD.94.064046} {\bibfield
  {journal} {\bibinfo  {journal} {Phys. Rev.}\ }\textbf {\bibinfo {volume}
  {D94}},\ \bibinfo {pages} {064046} (\bibinfo {year} {2016})},\ \Eprint
  {http://arxiv.org/abs/1607.02140} {arXiv:1607.02140 [gr-qc]} \BibitemShut
  {NoStop}%
\bibitem [{\citenamefont {Toshmatov}\ \emph {et~al.}(2018)\citenamefont
  {Toshmatov}, \citenamefont {Stuchl\'\i{}k}, \citenamefont {Schee},\ and\
  \citenamefont {Ahmedov}}]{Toshmatov:2018tyo}%
  \BibitemOpen
  \bibfield  {author} {\bibinfo {author} {\bibfnamefont {B.}~\bibnamefont
  {Toshmatov}}, \bibinfo {author} {\bibfnamefont {Z.}~\bibnamefont
  {Stuchl\'\i{}k}}, \bibinfo {author} {\bibfnamefont {J.}~\bibnamefont
  {Schee}}, \ and\ \bibinfo {author} {\bibfnamefont {B.}~\bibnamefont
  {Ahmedov}},\ }\href {\doibase 10.1103/PhysRevD.97.084058} {\bibfield
  {journal} {\bibinfo  {journal} {Phys. Rev. D}\ }\textbf {\bibinfo {volume}
  {97}},\ \bibinfo {pages} {084058} (\bibinfo {year} {2018})},\ \Eprint
  {http://arxiv.org/abs/1805.00240} {arXiv:1805.00240 [gr-qc]} \BibitemShut
  {NoStop}%
\bibitem [{\citenamefont {Bai}\ and\ \citenamefont
  {Orlofsky}(2020)}]{Bai:2019zcd}%
  \BibitemOpen
  \bibfield  {author} {\bibinfo {author} {\bibfnamefont {Y.}~\bibnamefont
  {Bai}}\ and\ \bibinfo {author} {\bibfnamefont {N.}~\bibnamefont {Orlofsky}},\
  }\href {\doibase 10.1103/PhysRevD.101.055006} {\bibfield  {journal} {\bibinfo
   {journal} {Phys. Rev. D}\ }\textbf {\bibinfo {volume} {101}},\ \bibinfo
  {pages} {055006} (\bibinfo {year} {2020})},\ \Eprint
  {http://arxiv.org/abs/1906.04858} {arXiv:1906.04858 [hep-ph]} \BibitemShut
  {NoStop}%
\bibitem [{\citenamefont {Allahyari}\ \emph {et~al.}(2020)\citenamefont
  {Allahyari}, \citenamefont {Khodadi}, \citenamefont {Vagnozzi},\ and\
  \citenamefont {Mota}}]{Allahyari:2019jqz}%
  \BibitemOpen
  \bibfield  {author} {\bibinfo {author} {\bibfnamefont {A.}~\bibnamefont
  {Allahyari}}, \bibinfo {author} {\bibfnamefont {M.}~\bibnamefont {Khodadi}},
  \bibinfo {author} {\bibfnamefont {S.}~\bibnamefont {Vagnozzi}}, \ and\
  \bibinfo {author} {\bibfnamefont {D.~F.}\ \bibnamefont {Mota}},\ }\href
  {\doibase 10.1088/1475-7516/2020/02/003} {\bibfield  {journal} {\bibinfo
  {journal} {JCAP}\ }\textbf {\bibinfo {volume} {02}},\ \bibinfo {pages} {003}
  (\bibinfo {year} {2020})},\ \Eprint {http://arxiv.org/abs/1912.08231}
  {arXiv:1912.08231 [gr-qc]} \BibitemShut {NoStop}%
\bibitem [{\citenamefont {Liu}\ \emph {et~al.}(2020{\natexlab{c}})\citenamefont
  {Liu}, \citenamefont {Guo}, \citenamefont {Cai},\ and\ \citenamefont
  {Kim}}]{Liu:2020cds}%
  \BibitemOpen
  \bibfield  {author} {\bibinfo {author} {\bibfnamefont {L.}~\bibnamefont
  {Liu}}, \bibinfo {author} {\bibfnamefont {Z.-K.}\ \bibnamefont {Guo}},
  \bibinfo {author} {\bibfnamefont {R.-G.}\ \bibnamefont {Cai}}, \ and\
  \bibinfo {author} {\bibfnamefont {S.~P.}\ \bibnamefont {Kim}},\ }\href
  {\doibase 10.1103/PhysRevD.102.043508} {\bibfield  {journal} {\bibinfo
  {journal} {Phys. Rev. D}\ }\textbf {\bibinfo {volume} {102}},\ \bibinfo
  {pages} {043508} (\bibinfo {year} {2020}{\natexlab{c}})},\ \Eprint
  {http://arxiv.org/abs/2001.02984} {arXiv:2001.02984 [astro-ph.CO]}
  \BibitemShut {NoStop}%
\bibitem [{\citenamefont {Christiansen}\ \emph {et~al.}(2021)\citenamefont
  {Christiansen}, \citenamefont {Beltr\'an~Jim\'enez},\ and\ \citenamefont
  {Mota}}]{Christiansen:2020pnv}%
  \BibitemOpen
  \bibfield  {author} {\bibinfo {author} {\bibfnamefont {O.}~\bibnamefont
  {Christiansen}}, \bibinfo {author} {\bibfnamefont {J.}~\bibnamefont
  {Beltr\'an~Jim\'enez}}, \ and\ \bibinfo {author} {\bibfnamefont {D.~F.}\
  \bibnamefont {Mota}},\ }\href {\doibase 10.1088/1361-6382/abdaf5} {\bibfield
  {journal} {\bibinfo  {journal} {Class. Quant. Grav.}\ }\textbf {\bibinfo
  {volume} {38}},\ \bibinfo {pages} {075017} (\bibinfo {year} {2021})},\
  \Eprint {http://arxiv.org/abs/2003.11452} {arXiv:2003.11452 [gr-qc]}
  \BibitemShut {NoStop}%
\bibitem [{\citenamefont {Wang}\ \emph
  {et~al.}(2020{\natexlab{c}})\citenamefont {Wang}, \citenamefont {Li},
  \citenamefont {Jiang}, \citenamefont {Hu},\ and\ \citenamefont
  {Fan}}]{Wang:2020fra}%
  \BibitemOpen
  \bibfield  {author} {\bibinfo {author} {\bibfnamefont {H.-T.}\ \bibnamefont
  {Wang}}, \bibinfo {author} {\bibfnamefont {P.-C.}\ \bibnamefont {Li}},
  \bibinfo {author} {\bibfnamefont {J.-L.}\ \bibnamefont {Jiang}}, \bibinfo
  {author} {\bibfnamefont {Y.-M.}\ \bibnamefont {Hu}}, \ and\ \bibinfo {author}
  {\bibfnamefont {Y.-Z.}\ \bibnamefont {Fan}},\ }\href@noop {} {\  (\bibinfo
  {year} {2020}{\natexlab{c}})},\ \Eprint {http://arxiv.org/abs/2004.12421}
  {arXiv:2004.12421 [gr-qc]} \BibitemShut {NoStop}%
\bibitem [{\citenamefont {Bozzola}\ and\ \citenamefont
  {Paschalidis}(2020)}]{Bozzola:2020mjx}%
  \BibitemOpen
  \bibfield  {author} {\bibinfo {author} {\bibfnamefont {G.}~\bibnamefont
  {Bozzola}}\ and\ \bibinfo {author} {\bibfnamefont {V.}~\bibnamefont
  {Paschalidis}},\ }\href@noop {} {\  (\bibinfo {year} {2020})},\ \Eprint
  {http://arxiv.org/abs/2006.15764} {arXiv:2006.15764 [gr-qc]} \BibitemShut
  {NoStop}%
\bibitem [{\citenamefont {Kim}\ and\ \citenamefont
  {Kobakhidze}(2020)}]{Kim:2020bhg}%
  \BibitemOpen
  \bibfield  {author} {\bibinfo {author} {\bibfnamefont {Y.}~\bibnamefont
  {Kim}}\ and\ \bibinfo {author} {\bibfnamefont {A.}~\bibnamefont
  {Kobakhidze}},\ }\href@noop {} {\  (\bibinfo {year} {2020})},\ \Eprint
  {http://arxiv.org/abs/2008.04506} {arXiv:2008.04506 [gr-qc]} \BibitemShut
  {NoStop}%
\bibitem [{\citenamefont {Cardoso}\ \emph
  {et~al.}(2020{\natexlab{a}})\citenamefont {Cardoso}, \citenamefont {Guo},
  \citenamefont {Macedo},\ and\ \citenamefont {Pani}}]{Cardoso:2020nst}%
  \BibitemOpen
  \bibfield  {author} {\bibinfo {author} {\bibfnamefont {V.}~\bibnamefont
  {Cardoso}}, \bibinfo {author} {\bibfnamefont {W.-d.}\ \bibnamefont {Guo}},
  \bibinfo {author} {\bibfnamefont {C.~F.}\ \bibnamefont {Macedo}}, \ and\
  \bibinfo {author} {\bibfnamefont {P.}~\bibnamefont {Pani}},\ }\href@noop {}
  {\  (\bibinfo {year} {2020}{\natexlab{a}})},\ \Eprint
  {http://arxiv.org/abs/2009.07287} {arXiv:2009.07287 [gr-qc]} \BibitemShut
  {NoStop}%
\bibitem [{\citenamefont {Cardoso}\ \emph
  {et~al.}(2020{\natexlab{b}})\citenamefont {Cardoso}, \citenamefont {Macedo},\
  and\ \citenamefont {Vicente}}]{Cardoso:2020iji}%
  \BibitemOpen
  \bibfield  {author} {\bibinfo {author} {\bibfnamefont {V.}~\bibnamefont
  {Cardoso}}, \bibinfo {author} {\bibfnamefont {C.~F.}\ \bibnamefont {Macedo}},
  \ and\ \bibinfo {author} {\bibfnamefont {R.}~\bibnamefont {Vicente}},\
  }\href@noop {} {\  (\bibinfo {year} {2020}{\natexlab{b}})},\ \Eprint
  {http://arxiv.org/abs/2010.15151} {arXiv:2010.15151 [gr-qc]} \BibitemShut
  {NoStop}%
\bibitem [{\citenamefont {McInnes}(2020)}]{McInnes:2020gxx}%
  \BibitemOpen
  \bibfield  {author} {\bibinfo {author} {\bibfnamefont {B.}~\bibnamefont
  {McInnes}},\ }\href@noop {} {\  (\bibinfo {year} {2020})},\ \Eprint
  {http://arxiv.org/abs/2011.07700} {arXiv:2011.07700 [gr-qc]} \BibitemShut
  {NoStop}%
\bibitem [{\citenamefont {Bai}\ and\ \citenamefont
  {Korwar}(2021)}]{Bai:2020ezy}%
  \BibitemOpen
  \bibfield  {author} {\bibinfo {author} {\bibfnamefont {Y.}~\bibnamefont
  {Bai}}\ and\ \bibinfo {author} {\bibfnamefont {M.}~\bibnamefont {Korwar}},\
  }\href {\doibase 10.1007/JHEP04(2021)119} {\bibfield  {journal} {\bibinfo
  {journal} {JHEP}\ }\textbf {\bibinfo {volume} {04}},\ \bibinfo {pages} {119}
  (\bibinfo {year} {2021})},\ \Eprint {http://arxiv.org/abs/2012.15430}
  {arXiv:2012.15430 [hep-ph]} \BibitemShut {NoStop}%
\bibitem [{\citenamefont {Bozzola}\ and\ \citenamefont
  {Paschalidis}(2021)}]{Bozzola:2021elc}%
  \BibitemOpen
  \bibfield  {author} {\bibinfo {author} {\bibfnamefont {G.}~\bibnamefont
  {Bozzola}}\ and\ \bibinfo {author} {\bibfnamefont {V.}~\bibnamefont
  {Paschalidis}},\ }\href@noop {} {\  (\bibinfo {year} {2021})},\ \Eprint
  {http://arxiv.org/abs/2104.06978} {arXiv:2104.06978 [gr-qc]} \BibitemShut
  {NoStop}%
\bibitem [{\citenamefont {McInnes}(2021)}]{McInnes:2021frb}%
  \BibitemOpen
  \bibfield  {author} {\bibinfo {author} {\bibfnamefont {B.}~\bibnamefont
  {McInnes}},\ }\href {\doibase 10.1016/j.nuclphysb.2021.115525} {\bibfield
  {journal} {\bibinfo  {journal} {Nucl. Phys. B}\ }\textbf {\bibinfo {volume}
  {971}},\ \bibinfo {pages} {115525} (\bibinfo {year} {2021})},\ \Eprint
  {http://arxiv.org/abs/2104.07373} {arXiv:2104.07373 [gr-qc]} \BibitemShut
  {NoStop}%
\bibitem [{\citenamefont {Hou}\ \emph {et~al.}(2021{\natexlab{b}})\citenamefont
  {Hou}, \citenamefont {Tian}, \citenamefont {Cao},\ and\ \citenamefont
  {Zhu}}]{Hou:2021suj}%
  \BibitemOpen
  \bibfield  {author} {\bibinfo {author} {\bibfnamefont {S.}~\bibnamefont
  {Hou}}, \bibinfo {author} {\bibfnamefont {S.}~\bibnamefont {Tian}}, \bibinfo
  {author} {\bibfnamefont {S.}~\bibnamefont {Cao}}, \ and\ \bibinfo {author}
  {\bibfnamefont {Z.-H.}\ \bibnamefont {Zhu}},\ }\href@noop {} {\  (\bibinfo
  {year} {2021}{\natexlab{b}})},\ \Eprint {http://arxiv.org/abs/2110.05084}
  {arXiv:2110.05084 [hep-ph]} \BibitemShut {NoStop}%
\bibitem [{\citenamefont {Benavides-Gallego}\ and\ \citenamefont
  {Han}(2021)}]{Benavides-Gallego:2021the}%
  \BibitemOpen
  \bibfield  {author} {\bibinfo {author} {\bibfnamefont {C.~A.}\ \bibnamefont
  {Benavides-Gallego}}\ and\ \bibinfo {author} {\bibfnamefont {W.-B.}\
  \bibnamefont {Han}},\ }\href@noop {} {\  (\bibinfo {year} {2021})},\ \Eprint
  {http://arxiv.org/abs/2111.04323} {arXiv:2111.04323 [gr-qc]} \BibitemShut
  {NoStop}%
\bibitem [{\citenamefont {Liu}\ \emph {et~al.}(2020{\natexlab{d}})\citenamefont
  {Liu}, \citenamefont {Christiansen}, \citenamefont {Guo}, \citenamefont
  {Cai},\ and\ \citenamefont {Kim}}]{Liu:2020vsy}%
  \BibitemOpen
  \bibfield  {author} {\bibinfo {author} {\bibfnamefont {L.}~\bibnamefont
  {Liu}}, \bibinfo {author} {\bibfnamefont {O.}~\bibnamefont {Christiansen}},
  \bibinfo {author} {\bibfnamefont {Z.-K.}\ \bibnamefont {Guo}}, \bibinfo
  {author} {\bibfnamefont {R.-G.}\ \bibnamefont {Cai}}, \ and\ \bibinfo
  {author} {\bibfnamefont {S.~P.}\ \bibnamefont {Kim}},\ }\href {\doibase
  10.1103/PhysRevD.102.103520} {\bibfield  {journal} {\bibinfo  {journal}
  {Phys. Rev. D}\ }\textbf {\bibinfo {volume} {102}},\ \bibinfo {pages}
  {103520} (\bibinfo {year} {2020}{\natexlab{d}})},\ \Eprint
  {http://arxiv.org/abs/2008.02326} {arXiv:2008.02326 [gr-qc]} \BibitemShut
  {NoStop}%
\bibitem [{\citenamefont {Liu}\ \emph {et~al.}(2021{\natexlab{b}})\citenamefont
  {Liu}, \citenamefont {Christiansen}, \citenamefont {Ruan}, \citenamefont
  {Guo}, \citenamefont {Cai},\ and\ \citenamefont {Kim}}]{Liu:2020bag}%
  \BibitemOpen
  \bibfield  {author} {\bibinfo {author} {\bibfnamefont {L.}~\bibnamefont
  {Liu}}, \bibinfo {author} {\bibfnamefont {O.}~\bibnamefont {Christiansen}},
  \bibinfo {author} {\bibfnamefont {W.-H.}\ \bibnamefont {Ruan}}, \bibinfo
  {author} {\bibfnamefont {Z.-K.}\ \bibnamefont {Guo}}, \bibinfo {author}
  {\bibfnamefont {R.-G.}\ \bibnamefont {Cai}}, \ and\ \bibinfo {author}
  {\bibfnamefont {S.~P.}\ \bibnamefont {Kim}},\ }\href {\doibase
  10.1140/epjc/s10052-021-09849-4} {\bibfield  {journal} {\bibinfo  {journal}
  {Eur. Phys. J. C}\ }\textbf {\bibinfo {volume} {81}},\ \bibinfo {pages}
  {1048} (\bibinfo {year} {2021}{\natexlab{b}})},\ \Eprint
  {http://arxiv.org/abs/2011.13586} {arXiv:2011.13586 [gr-qc]} \BibitemShut
  {NoStop}%
\bibitem [{\citenamefont {Liu}\ and\ \citenamefont
  {Kim}(2022)}]{liu2022merger}%
  \BibitemOpen
  \bibfield  {author} {\bibinfo {author} {\bibfnamefont {L.}~\bibnamefont
  {Liu}}\ and\ \bibinfo {author} {\bibfnamefont {S.~P.}\ \bibnamefont {Kim}},\
  }\href@noop {} {\bibfield  {journal} {\bibinfo  {journal} {Journal of
  Cosmology and Astroparticle Physics}\ }\textbf {\bibinfo {volume} {2022}},\
  \bibinfo {pages} {059} (\bibinfo {year} {2022})}\BibitemShut {NoStop}%
\bibitem [{\citenamefont {Carr}\ and\ \citenamefont
  {K{\"u}hnel}(2020)}]{carr2020primordial}%
  \BibitemOpen
  \bibfield  {author} {\bibinfo {author} {\bibfnamefont {B.}~\bibnamefont
  {Carr}}\ and\ \bibinfo {author} {\bibfnamefont {F.}~\bibnamefont
  {K{\"u}hnel}},\ }\href@noop {} {\bibfield  {journal} {\bibinfo  {journal}
  {Annual Review of Nuclear and Particle Science}\ }\textbf {\bibinfo {volume}
  {70}},\ \bibinfo {pages} {355} (\bibinfo {year} {2020})}\BibitemShut
  {NoStop}%
\bibitem [{\citenamefont {Carr}\ \emph {et~al.}(2021)\citenamefont {Carr},
  \citenamefont {Kohri}, \citenamefont {Sendouda},\ and\ \citenamefont
  {Yokoyama}}]{carr2021constraints}%
  \BibitemOpen
  \bibfield  {author} {\bibinfo {author} {\bibfnamefont {B.}~\bibnamefont
  {Carr}}, \bibinfo {author} {\bibfnamefont {K.}~\bibnamefont {Kohri}},
  \bibinfo {author} {\bibfnamefont {Y.}~\bibnamefont {Sendouda}}, \ and\
  \bibinfo {author} {\bibfnamefont {J.}~\bibnamefont {Yokoyama}},\ }\href@noop
  {} {\bibfield  {journal} {\bibinfo  {journal} {Reports on Progress in
  Physics}\ }\textbf {\bibinfo {volume} {84}},\ \bibinfo {pages} {116902}
  (\bibinfo {year} {2021})}\BibitemShut {NoStop}%
\bibitem [{\citenamefont {Carr}\ \emph {et~al.}(2024)\citenamefont {Carr},
  \citenamefont {Clesse}, \citenamefont {Garcia-Bellido}, \citenamefont
  {Hawkins},\ and\ \citenamefont {Kuhnel}}]{Carr:2023tpt}%
  \BibitemOpen
  \bibfield  {author} {\bibinfo {author} {\bibfnamefont {B.}~\bibnamefont
  {Carr}}, \bibinfo {author} {\bibfnamefont {S.}~\bibnamefont {Clesse}},
  \bibinfo {author} {\bibfnamefont {J.}~\bibnamefont {Garcia-Bellido}},
  \bibinfo {author} {\bibfnamefont {M.}~\bibnamefont {Hawkins}}, \ and\
  \bibinfo {author} {\bibfnamefont {F.}~\bibnamefont {Kuhnel}},\ }\href
  {\doibase 10.1016/j.physrep.2023.11.005} {\bibfield  {journal} {\bibinfo
  {journal} {Phys. Rept.}\ }\textbf {\bibinfo {volume} {1054}},\ \bibinfo
  {pages} {1} (\bibinfo {year} {2024})},\ \Eprint
  {http://arxiv.org/abs/2306.03903} {arXiv:2306.03903 [astro-ph.CO]}
  \BibitemShut {NoStop}%
\bibitem [{\citenamefont {Hod}(2012)}]{Hod:2012wmy}%
  \BibitemOpen
  \bibfield  {author} {\bibinfo {author} {\bibfnamefont {S.}~\bibnamefont
  {Hod}},\ }\href {\doibase 10.1016/j.physletb.2012.06.043} {\bibfield
  {journal} {\bibinfo  {journal} {Phys. Lett. B}\ }\textbf {\bibinfo {volume}
  {713}},\ \bibinfo {pages} {505} (\bibinfo {year} {2012})},\ \Eprint
  {http://arxiv.org/abs/1304.6474} {arXiv:1304.6474 [gr-qc]} \BibitemShut
  {NoStop}%
\bibitem [{\citenamefont {Kritos}\ and\ \citenamefont
  {Silk}(2022)}]{kritos2022mergers}%
  \BibitemOpen
  \bibfield  {author} {\bibinfo {author} {\bibfnamefont {K.}~\bibnamefont
  {Kritos}}\ and\ \bibinfo {author} {\bibfnamefont {J.}~\bibnamefont {Silk}},\
  }\href@noop {} {\bibfield  {journal} {\bibinfo  {journal} {Physical Review
  D}\ }\textbf {\bibinfo {volume} {105}},\ \bibinfo {pages} {063011} (\bibinfo
  {year} {2022})}\BibitemShut {NoStop}%
\bibitem [{\citenamefont {Gibbons}(1975{\natexlab{b}})}]{gibbons1975vacuum}%
  \BibitemOpen
  \bibfield  {author} {\bibinfo {author} {\bibfnamefont {G.~W.}\ \bibnamefont
  {Gibbons}},\ }\href@noop {} {\bibfield  {journal} {\bibinfo  {journal}
  {Communications in Mathematical Physics}\ }\textbf {\bibinfo {volume} {44}},\
  \bibinfo {pages} {245} (\bibinfo {year} {1975}{\natexlab{b}})}\BibitemShut
  {NoStop}%
\bibitem [{\citenamefont {Ruffini}\ \emph {et~al.}(2010)\citenamefont
  {Ruffini}, \citenamefont {Vereshchagin},\ and\ \citenamefont
  {Xue}}]{Ruffini:2009hg}%
  \BibitemOpen
  \bibfield  {author} {\bibinfo {author} {\bibfnamefont {R.}~\bibnamefont
  {Ruffini}}, \bibinfo {author} {\bibfnamefont {G.}~\bibnamefont
  {Vereshchagin}}, \ and\ \bibinfo {author} {\bibfnamefont {S.-S.}\
  \bibnamefont {Xue}},\ }\href {\doibase 10.1016/j.physrep.2009.10.004}
  {\bibfield  {journal} {\bibinfo  {journal} {Phys. Rept.}\ }\textbf {\bibinfo
  {volume} {487}},\ \bibinfo {pages} {1} (\bibinfo {year} {2010})},\ \Eprint
  {http://arxiv.org/abs/0910.0974} {arXiv:0910.0974 [astro-ph.HE]} \BibitemShut
  {NoStop}%
\bibitem [{\citenamefont {Chen}\ \emph {et~al.}(2012)\citenamefont {Chen},
  \citenamefont {Kim}, \citenamefont {Lin}, \citenamefont {Sun},\ and\
  \citenamefont {Wu}}]{Chen:2012zn}%
  \BibitemOpen
  \bibfield  {author} {\bibinfo {author} {\bibfnamefont {C.-M.}\ \bibnamefont
  {Chen}}, \bibinfo {author} {\bibfnamefont {S.~P.}\ \bibnamefont {Kim}},
  \bibinfo {author} {\bibfnamefont {I.-C.}\ \bibnamefont {Lin}}, \bibinfo
  {author} {\bibfnamefont {J.-R.}\ \bibnamefont {Sun}}, \ and\ \bibinfo
  {author} {\bibfnamefont {M.-F.}\ \bibnamefont {Wu}},\ }\href {\doibase
  10.1103/PhysRevD.85.124041} {\bibfield  {journal} {\bibinfo  {journal} {Phys.
  Rev. D}\ }\textbf {\bibinfo {volume} {85}},\ \bibinfo {pages} {124041}
  (\bibinfo {year} {2012})},\ \Eprint {http://arxiv.org/abs/1202.3224}
  {arXiv:1202.3224 [hep-th]} \BibitemShut {NoStop}%
\bibitem [{\citenamefont {Chen}\ \emph {et~al.}(2017)\citenamefont {Chen},
  \citenamefont {Kim}, \citenamefont {Sun},\ and\ \citenamefont
  {Tang}}]{Chen:2016caa}%
  \BibitemOpen
  \bibfield  {author} {\bibinfo {author} {\bibfnamefont {C.-M.}\ \bibnamefont
  {Chen}}, \bibinfo {author} {\bibfnamefont {S.~P.}\ \bibnamefont {Kim}},
  \bibinfo {author} {\bibfnamefont {J.-R.}\ \bibnamefont {Sun}}, \ and\
  \bibinfo {author} {\bibfnamefont {F.-Y.}\ \bibnamefont {Tang}},\ }\href
  {\doibase 10.1103/PhysRevD.95.044043} {\bibfield  {journal} {\bibinfo
  {journal} {Phys. Rev.}\ }\textbf {\bibinfo {volume} {D95}},\ \bibinfo {pages}
  {044043} (\bibinfo {year} {2017})},\ \Eprint
  {http://arxiv.org/abs/1607.02610} {arXiv:1607.02610 [hep-th]} \BibitemShut
  {NoStop}%
\bibitem [{\citenamefont {Chen}\ \emph
  {et~al.}(2018{\natexlab{a}})\citenamefont {Chen}, \citenamefont {Kim},
  \citenamefont {Sun},\ and\ \citenamefont {Tang}}]{Chen:2017mnm}%
  \BibitemOpen
  \bibfield  {author} {\bibinfo {author} {\bibfnamefont {C.-M.}\ \bibnamefont
  {Chen}}, \bibinfo {author} {\bibfnamefont {S.~P.}\ \bibnamefont {Kim}},
  \bibinfo {author} {\bibfnamefont {J.-R.}\ \bibnamefont {Sun}}, \ and\
  \bibinfo {author} {\bibfnamefont {F.-Y.}\ \bibnamefont {Tang}},\ }\href
  {\doibase 10.1016/j.physletb.2018.03.078} {\bibfield  {journal} {\bibinfo
  {journal} {Phys. Lett.}\ }\textbf {\bibinfo {volume} {B781}},\ \bibinfo
  {pages} {129} (\bibinfo {year} {2018}{\natexlab{a}})},\ \Eprint
  {http://arxiv.org/abs/1705.10629} {arXiv:1705.10629 [hep-th]} \BibitemShut
  {NoStop}%
\bibitem [{\citenamefont {Chen}\ and\ \citenamefont
  {Kim}(2020)}]{Chen:2020mqs}%
  \BibitemOpen
  \bibfield  {author} {\bibinfo {author} {\bibfnamefont {C.-M.}\ \bibnamefont
  {Chen}}\ and\ \bibinfo {author} {\bibfnamefont {S.~P.}\ \bibnamefont {Kim}},\
  }\href {\doibase 10.1103/PhysRevD.101.085014} {\bibfield  {journal} {\bibinfo
   {journal} {Phys. Rev. D}\ }\textbf {\bibinfo {volume} {101}},\ \bibinfo
  {pages} {085014} (\bibinfo {year} {2020})},\ \Eprint
  {http://arxiv.org/abs/2002.00394} {arXiv:2002.00394 [hep-th]} \BibitemShut
  {NoStop}%
\bibitem [{\citenamefont {Affleck}\ and\ \citenamefont
  {Manton}(1982)}]{Affleck:1981ag}%
  \BibitemOpen
  \bibfield  {author} {\bibinfo {author} {\bibfnamefont {I.~K.}\ \bibnamefont
  {Affleck}}\ and\ \bibinfo {author} {\bibfnamefont {N.~S.}\ \bibnamefont
  {Manton}},\ }\href {\doibase 10.1016/0550-3213(82)90511-9} {\bibfield
  {journal} {\bibinfo  {journal} {Nucl. Phys. B}\ }\textbf {\bibinfo {volume}
  {194}},\ \bibinfo {pages} {38} (\bibinfo {year} {1982})}\BibitemShut
  {NoStop}%
\bibitem [{\citenamefont {Oliveira}\ \emph {et~al.}(2011)\citenamefont
  {Oliveira}, \citenamefont {Crispino},\ and\ \citenamefont
  {Higuchi}}]{Oliveira:2011zz}%
  \BibitemOpen
  \bibfield  {author} {\bibinfo {author} {\bibfnamefont {E.~S.}\ \bibnamefont
  {Oliveira}}, \bibinfo {author} {\bibfnamefont {L.~C.~B.}\ \bibnamefont
  {Crispino}}, \ and\ \bibinfo {author} {\bibfnamefont {A.}~\bibnamefont
  {Higuchi}},\ }\href {\doibase 10.1103/PhysRevD.84.084048} {\bibfield
  {journal} {\bibinfo  {journal} {Phys. Rev. D}\ }\textbf {\bibinfo {volume}
  {84}},\ \bibinfo {pages} {084048} (\bibinfo {year} {2011})}\BibitemShut
  {NoStop}%
\bibitem [{\citenamefont {Zel'dovich}\ and\ \citenamefont
  {Novikov}(1967)}]{Zeldovich:1967lct}%
  \BibitemOpen
  \bibfield  {author} {\bibinfo {author} {\bibfnamefont {Y.~B.}\ \bibnamefont
  {Zel'dovich}}\ and\ \bibinfo {author} {\bibfnamefont {I.~D.}\ \bibnamefont
  {Novikov}},\ }\href@noop {} {\bibfield  {journal} {\bibinfo  {journal} {Sov.
  Astron.}\ }\textbf {\bibinfo {volume} {10}},\ \bibinfo {pages} {602}
  (\bibinfo {year} {1967})}\BibitemShut {NoStop}%
\bibitem [{\citenamefont {Hawking}(1971)}]{Hawking:1971ei}%
  \BibitemOpen
  \bibfield  {author} {\bibinfo {author} {\bibfnamefont {S.}~\bibnamefont
  {Hawking}},\ }\href {\doibase 10.1093/mnras/152.1.75} {\bibfield  {journal}
  {\bibinfo  {journal} {Mon. Not. Roy. Astron. Soc.}\ }\textbf {\bibinfo
  {volume} {152}},\ \bibinfo {pages} {75} (\bibinfo {year} {1971})}\BibitemShut
  {NoStop}%
\bibitem [{\citenamefont {Carr}\ and\ \citenamefont
  {Hawking}(1974)}]{Carr:1974nx}%
  \BibitemOpen
  \bibfield  {author} {\bibinfo {author} {\bibfnamefont {B.~J.}\ \bibnamefont
  {Carr}}\ and\ \bibinfo {author} {\bibfnamefont {S.~W.}\ \bibnamefont
  {Hawking}},\ }\href {\doibase 10.1093/mnras/168.2.399} {\bibfield  {journal}
  {\bibinfo  {journal} {Mon. Not. Roy. Astron. Soc.}\ }\textbf {\bibinfo
  {volume} {168}},\ \bibinfo {pages} {399} (\bibinfo {year}
  {1974})}\BibitemShut {NoStop}%
\bibitem [{\citenamefont {Meszaros}(1974)}]{Meszaros:1974tb}%
  \BibitemOpen
  \bibfield  {author} {\bibinfo {author} {\bibfnamefont {P.}~\bibnamefont
  {Meszaros}},\ }\href@noop {} {\bibfield  {journal} {\bibinfo  {journal}
  {Astron. Astrophys.}\ }\textbf {\bibinfo {volume} {37}},\ \bibinfo {pages}
  {225} (\bibinfo {year} {1974})}\BibitemShut {NoStop}%
\bibitem [{\citenamefont {Carr}(1975)}]{Carr:1975qj}%
  \BibitemOpen
  \bibfield  {author} {\bibinfo {author} {\bibfnamefont {B.~J.}\ \bibnamefont
  {Carr}},\ }\href {\doibase 10.1086/153853} {\bibfield  {journal} {\bibinfo
  {journal} {Astrophys. J.}\ }\textbf {\bibinfo {volume} {201}},\ \bibinfo
  {pages} {1} (\bibinfo {year} {1975})}\BibitemShut {NoStop}%
\bibitem [{\citenamefont {Khlopov}\ \emph {et~al.}(1985)\citenamefont
  {Khlopov}, \citenamefont {Malomed},\ and\ \citenamefont
  {Zeldovich}}]{Khlopov:1985jw}%
  \BibitemOpen
  \bibfield  {author} {\bibinfo {author} {\bibfnamefont {M.}~\bibnamefont
  {Khlopov}}, \bibinfo {author} {\bibfnamefont {B.}~\bibnamefont {Malomed}}, \
  and\ \bibinfo {author} {\bibfnamefont {I.}~\bibnamefont {Zeldovich}},\
  }\href@noop {} {\bibfield  {journal} {\bibinfo  {journal} {Mon. Not. Roy.
  Astron. Soc.}\ }\textbf {\bibinfo {volume} {215}},\ \bibinfo {pages} {575}
  (\bibinfo {year} {1985})}\BibitemShut {NoStop}%
\bibitem [{\citenamefont {{Polnarev}}\ and\ \citenamefont
  {{Khlopov}}(1981)}]{1981SvA....25..406P}%
  \BibitemOpen
  \bibfield  {author} {\bibinfo {author} {\bibfnamefont {A.~G.}\ \bibnamefont
  {{Polnarev}}}\ and\ \bibinfo {author} {\bibfnamefont {M.~Y.}\ \bibnamefont
  {{Khlopov}}},\ }\href@noop {} {\bibfield  {journal} {\bibinfo  {journal}
  {\sovast}\ }\textbf {\bibinfo {volume} {25}},\ \bibinfo {pages} {406}
  (\bibinfo {year} {1981})}\BibitemShut {NoStop}%
\bibitem [{\citenamefont {Carr}\ \emph {et~al.}(2016)\citenamefont {Carr},
  \citenamefont {Kuhnel},\ and\ \citenamefont {Sandstad}}]{Carr:2016drx}%
  \BibitemOpen
  \bibfield  {author} {\bibinfo {author} {\bibfnamefont {B.}~\bibnamefont
  {Carr}}, \bibinfo {author} {\bibfnamefont {F.}~\bibnamefont {Kuhnel}}, \ and\
  \bibinfo {author} {\bibfnamefont {M.}~\bibnamefont {Sandstad}},\ }\href
  {\doibase 10.1103/PhysRevD.94.083504} {\bibfield  {journal} {\bibinfo
  {journal} {Phys. Rev. D}\ }\textbf {\bibinfo {volume} {94}},\ \bibinfo
  {pages} {083504} (\bibinfo {year} {2016})},\ \Eprint
  {http://arxiv.org/abs/1607.06077} {arXiv:1607.06077 [astro-ph.CO]}
  \BibitemShut {NoStop}%
\bibitem [{\citenamefont {Green}\ and\ \citenamefont
  {Kavanagh}(2021)}]{Green:2020jor}%
  \BibitemOpen
  \bibfield  {author} {\bibinfo {author} {\bibfnamefont {A.~M.}\ \bibnamefont
  {Green}}\ and\ \bibinfo {author} {\bibfnamefont {B.~J.}\ \bibnamefont
  {Kavanagh}},\ }\href {\doibase 10.1088/1361-6471/abc534} {\bibfield
  {journal} {\bibinfo  {journal} {J. Phys. G}\ }\textbf {\bibinfo {volume}
  {48}},\ \bibinfo {pages} {043001} (\bibinfo {year} {2021})},\ \Eprint
  {http://arxiv.org/abs/2007.10722} {arXiv:2007.10722 [astro-ph.CO]}
  \BibitemShut {NoStop}%
\bibitem [{\citenamefont {Blinnikov}\ \emph {et~al.}(2016)\citenamefont
  {Blinnikov}, \citenamefont {Dolgov}, \citenamefont {Porayko},\ and\
  \citenamefont {Postnov}}]{Blinnikov:2016bxu}%
  \BibitemOpen
  \bibfield  {author} {\bibinfo {author} {\bibfnamefont {S.}~\bibnamefont
  {Blinnikov}}, \bibinfo {author} {\bibfnamefont {A.}~\bibnamefont {Dolgov}},
  \bibinfo {author} {\bibfnamefont {N.~K.}\ \bibnamefont {Porayko}}, \ and\
  \bibinfo {author} {\bibfnamefont {K.}~\bibnamefont {Postnov}},\ }\href
  {\doibase 10.1088/1475-7516/2016/11/036} {\bibfield  {journal} {\bibinfo
  {journal} {JCAP}\ }\textbf {\bibinfo {volume} {11}},\ \bibinfo {pages} {036}
  (\bibinfo {year} {2016})},\ \Eprint {http://arxiv.org/abs/1611.00541}
  {arXiv:1611.00541 [astro-ph.HE]} \BibitemShut {NoStop}%
\bibitem [{\citenamefont {Liu}\ and\ \citenamefont
  {Bromm}(2023)}]{Liu:2023pvq}%
  \BibitemOpen
  \bibfield  {author} {\bibinfo {author} {\bibfnamefont {B.}~\bibnamefont
  {Liu}}\ and\ \bibinfo {author} {\bibfnamefont {V.}~\bibnamefont {Bromm}},\
  }\href@noop {} {\  (\bibinfo {year} {2023})},\ \Eprint
  {http://arxiv.org/abs/2312.04085} {arXiv:2312.04085 [astro-ph.GA]}
  \BibitemShut {NoStop}%
\bibitem [{\citenamefont {Dolgov}\ and\ \citenamefont
  {Postnov}(2017)}]{Dolgov:2017nmh}%
  \BibitemOpen
  \bibfield  {author} {\bibinfo {author} {\bibfnamefont {A.}~\bibnamefont
  {Dolgov}}\ and\ \bibinfo {author} {\bibfnamefont {K.}~\bibnamefont
  {Postnov}},\ }\href {\doibase 10.1088/1475-7516/2017/04/036} {\bibfield
  {journal} {\bibinfo  {journal} {JCAP}\ }\textbf {\bibinfo {volume} {04}},\
  \bibinfo {pages} {036} (\bibinfo {year} {2017})},\ \Eprint
  {http://arxiv.org/abs/1702.07621} {arXiv:1702.07621 [astro-ph.CO]}
  \BibitemShut {NoStop}%
\bibitem [{\citenamefont {Carr}\ and\ \citenamefont
  {Silk}(2018)}]{Carr:2018rid}%
  \BibitemOpen
  \bibfield  {author} {\bibinfo {author} {\bibfnamefont {B.}~\bibnamefont
  {Carr}}\ and\ \bibinfo {author} {\bibfnamefont {J.}~\bibnamefont {Silk}},\
  }\href {\doibase 10.1093/mnras/sty1204} {\bibfield  {journal} {\bibinfo
  {journal} {Mon. Not. Roy. Astron. Soc.}\ }\textbf {\bibinfo {volume} {478}},\
  \bibinfo {pages} {3756} (\bibinfo {year} {2018})},\ \Eprint
  {http://arxiv.org/abs/1801.00672} {arXiv:1801.00672 [astro-ph.CO]}
  \BibitemShut {NoStop}%
\bibitem [{\citenamefont {Colazo}\ \emph {et~al.}(2024)\citenamefont {Colazo},
  \citenamefont {Stasyszyn},\ and\ \citenamefont {Padilla}}]{Colazo:2024jmz}%
  \BibitemOpen
  \bibfield  {author} {\bibinfo {author} {\bibfnamefont {P.~E.}\ \bibnamefont
  {Colazo}}, \bibinfo {author} {\bibfnamefont {F.}~\bibnamefont {Stasyszyn}}, \
  and\ \bibinfo {author} {\bibfnamefont {N.}~\bibnamefont {Padilla}},\ }\href
  {\doibase 10.1051/0004-6361/202449565} {\bibfield  {journal} {\bibinfo
  {journal} {Astron. Astrophys.}\ }\textbf {\bibinfo {volume} {685}},\ \bibinfo
  {pages} {L8} (\bibinfo {year} {2024})},\ \Eprint
  {http://arxiv.org/abs/2404.13110} {arXiv:2404.13110 [astro-ph.CO]}
  \BibitemShut {NoStop}%
\bibitem [{\citenamefont {H\"utsi}\ \emph {et~al.}(2023)\citenamefont
  {H\"utsi}, \citenamefont {Raidal}, \citenamefont {Urrutia}, \citenamefont
  {Vaskonen},\ and\ \citenamefont {Veerm\"ae}}]{Hutsi:2022fzw}%
  \BibitemOpen
  \bibfield  {author} {\bibinfo {author} {\bibfnamefont {G.}~\bibnamefont
  {H\"utsi}}, \bibinfo {author} {\bibfnamefont {M.}~\bibnamefont {Raidal}},
  \bibinfo {author} {\bibfnamefont {J.}~\bibnamefont {Urrutia}}, \bibinfo
  {author} {\bibfnamefont {V.}~\bibnamefont {Vaskonen}}, \ and\ \bibinfo
  {author} {\bibfnamefont {H.}~\bibnamefont {Veerm\"ae}},\ }\href {\doibase
  10.1103/PhysRevD.107.043502} {\bibfield  {journal} {\bibinfo  {journal}
  {Phys. Rev. D}\ }\textbf {\bibinfo {volume} {107}},\ \bibinfo {pages}
  {043502} (\bibinfo {year} {2023})},\ \Eprint
  {http://arxiv.org/abs/2211.02651} {arXiv:2211.02651 [astro-ph.CO]}
  \BibitemShut {NoStop}%
\bibitem [{\citenamefont {Huang}\ \emph
  {et~al.}(2024{\natexlab{b}})\citenamefont {Huang}, \citenamefont {Cai},
  \citenamefont {Jiang}, \citenamefont {Zhang},\ and\ \citenamefont
  {Piao}}]{Huang:2023chx}%
  \BibitemOpen
  \bibfield  {author} {\bibinfo {author} {\bibfnamefont {H.-L.}\ \bibnamefont
  {Huang}}, \bibinfo {author} {\bibfnamefont {Y.}~\bibnamefont {Cai}}, \bibinfo
  {author} {\bibfnamefont {J.-Q.}\ \bibnamefont {Jiang}}, \bibinfo {author}
  {\bibfnamefont {J.}~\bibnamefont {Zhang}}, \ and\ \bibinfo {author}
  {\bibfnamefont {Y.-S.}\ \bibnamefont {Piao}},\ }\href {\doibase
  10.1088/1674-4527/ad683d} {\bibfield  {journal} {\bibinfo  {journal} {Res.
  Astron. Astrophys.}\ }\textbf {\bibinfo {volume} {24}},\ \bibinfo {pages}
  {091001} (\bibinfo {year} {2024}{\natexlab{b}})},\ \Eprint
  {http://arxiv.org/abs/2306.17577} {arXiv:2306.17577 [gr-qc]} \BibitemShut
  {NoStop}%
\bibitem [{\citenamefont {Gouttenoire}\ \emph {et~al.}(2023)\citenamefont
  {Gouttenoire}, \citenamefont {Trifinopoulos}, \citenamefont {Valogiannis},\
  and\ \citenamefont {Vanvlasselaer}}]{Gouttenoire:2023nzr}%
  \BibitemOpen
  \bibfield  {author} {\bibinfo {author} {\bibfnamefont {Y.}~\bibnamefont
  {Gouttenoire}}, \bibinfo {author} {\bibfnamefont {S.}~\bibnamefont
  {Trifinopoulos}}, \bibinfo {author} {\bibfnamefont {G.}~\bibnamefont
  {Valogiannis}}, \ and\ \bibinfo {author} {\bibfnamefont {M.}~\bibnamefont
  {Vanvlasselaer}},\ }\href@noop {} {\  (\bibinfo {year} {2023})},\ \Eprint
  {http://arxiv.org/abs/2307.01457} {arXiv:2307.01457 [astro-ph.CO]}
  \BibitemShut {NoStop}%
\bibitem [{\citenamefont {Huang}\ \emph
  {et~al.}(2024{\natexlab{c}})\citenamefont {Huang}, \citenamefont {Jiang},\
  and\ \citenamefont {Piao}}]{Huang:2024aog}%
  \BibitemOpen
  \bibfield  {author} {\bibinfo {author} {\bibfnamefont {H.-L.}\ \bibnamefont
  {Huang}}, \bibinfo {author} {\bibfnamefont {J.-Q.}\ \bibnamefont {Jiang}}, \
  and\ \bibinfo {author} {\bibfnamefont {Y.-S.}\ \bibnamefont {Piao}},\ }\href
  {\doibase 10.1103/PhysRevD.110.103540} {\bibfield  {journal} {\bibinfo
  {journal} {Phys. Rev. D}\ }\textbf {\bibinfo {volume} {110}},\ \bibinfo
  {pages} {103540} (\bibinfo {year} {2024}{\natexlab{c}})},\ \Eprint
  {http://arxiv.org/abs/2407.15781} {arXiv:2407.15781 [astro-ph.CO]}
  \BibitemShut {NoStop}%
\bibitem [{\citenamefont {Sasaki}\ \emph {et~al.}(2018)\citenamefont {Sasaki},
  \citenamefont {Suyama}, \citenamefont {Tanaka},\ and\ \citenamefont
  {Yokoyama}}]{Sasaki:2018dmp}%
  \BibitemOpen
  \bibfield  {author} {\bibinfo {author} {\bibfnamefont {M.}~\bibnamefont
  {Sasaki}}, \bibinfo {author} {\bibfnamefont {T.}~\bibnamefont {Suyama}},
  \bibinfo {author} {\bibfnamefont {T.}~\bibnamefont {Tanaka}}, \ and\ \bibinfo
  {author} {\bibfnamefont {S.}~\bibnamefont {Yokoyama}},\ }\href {\doibase
  10.1088/1361-6382/aaa7b4} {\bibfield  {journal} {\bibinfo  {journal} {Class.
  Quant. Grav.}\ }\textbf {\bibinfo {volume} {35}},\ \bibinfo {pages} {063001}
  (\bibinfo {year} {2018})},\ \Eprint {http://arxiv.org/abs/1801.05235}
  {arXiv:1801.05235 [astro-ph.CO]} \BibitemShut {NoStop}%
\bibitem [{\citenamefont {Carr}\ and\ \citenamefont
  {Kuhnel}(2020)}]{Carr:2020xqk}%
  \BibitemOpen
  \bibfield  {author} {\bibinfo {author} {\bibfnamefont {B.}~\bibnamefont
  {Carr}}\ and\ \bibinfo {author} {\bibfnamefont {F.}~\bibnamefont {Kuhnel}},\
  }\href {\doibase 10.1146/annurev-nucl-050520-125911} {\bibfield  {journal}
  {\bibinfo  {journal} {Ann. Rev. Nucl. Part. Sci.}\ }\textbf {\bibinfo
  {volume} {70}},\ \bibinfo {pages} {355} (\bibinfo {year} {2020})},\ \Eprint
  {http://arxiv.org/abs/2006.02838} {arXiv:2006.02838 [astro-ph.CO]}
  \BibitemShut {NoStop}%
\bibitem [{\citenamefont {Escriv\`a}\ \emph
  {et~al.}(2022{\natexlab{a}})\citenamefont {Escriv\`a}, \citenamefont
  {Kuhnel},\ and\ \citenamefont {Tada}}]{Escriva:2022duf}%
  \BibitemOpen
  \bibfield  {author} {\bibinfo {author} {\bibfnamefont {A.}~\bibnamefont
  {Escriv\`a}}, \bibinfo {author} {\bibfnamefont {F.}~\bibnamefont {Kuhnel}}, \
  and\ \bibinfo {author} {\bibfnamefont {Y.}~\bibnamefont {Tada}},\ }\href@noop
  {} {\  (\bibinfo {year} {2022}{\natexlab{a}})},\ \Eprint
  {http://arxiv.org/abs/2211.05767} {arXiv:2211.05767 [astro-ph.CO]}
  \BibitemShut {NoStop}%
\bibitem [{\citenamefont {\"Ozsoy}\ and\ \citenamefont
  {Tasinato}(2023)}]{Ozsoy:2023ryl}%
  \BibitemOpen
  \bibfield  {author} {\bibinfo {author} {\bibfnamefont {O.}~\bibnamefont
  {\"Ozsoy}}\ and\ \bibinfo {author} {\bibfnamefont {G.}~\bibnamefont
  {Tasinato}},\ }\href {\doibase 10.3390/universe9050203} {\bibfield  {journal}
  {\bibinfo  {journal} {Universe}\ }\textbf {\bibinfo {volume} {9}},\ \bibinfo
  {pages} {203} (\bibinfo {year} {2023})},\ \Eprint
  {http://arxiv.org/abs/2301.03600} {arXiv:2301.03600 [astro-ph.CO]}
  \BibitemShut {NoStop}%
\bibitem [{\citenamefont {Choudhury}\ and\ \citenamefont
  {Sami}(2025)}]{Choudhury:2024aji}%
  \BibitemOpen
  \bibfield  {author} {\bibinfo {author} {\bibfnamefont {S.}~\bibnamefont
  {Choudhury}}\ and\ \bibinfo {author} {\bibfnamefont {M.}~\bibnamefont
  {Sami}},\ }\href {\doibase 10.1016/j.physrep.2024.10.007} {\bibfield
  {journal} {\bibinfo  {journal} {Phys. Rept.}\ }\textbf {\bibinfo {volume}
  {1103}},\ \bibinfo {pages} {1} (\bibinfo {year} {2025})},\ \Eprint
  {http://arxiv.org/abs/2407.17006} {arXiv:2407.17006 [gr-qc]} \BibitemShut
  {NoStop}%
\bibitem [{\citenamefont {Postnov}\ and\ \citenamefont
  {Chekh}(2024)}]{Postnov:2024fra}%
  \BibitemOpen
  \bibfield  {author} {\bibinfo {author} {\bibfnamefont {K.}~\bibnamefont
  {Postnov}}\ and\ \bibinfo {author} {\bibfnamefont {I.}~\bibnamefont
  {Chekh}},\ }in\ \href@noop {} {\emph {\bibinfo {booktitle} {{7th
  International Workshop on the TianQin Science Mission}}}}\ (\bibinfo {year}
  {2024})\ \Eprint {http://arxiv.org/abs/2407.16373} {arXiv:2407.16373
  [astro-ph.CO]} \BibitemShut {NoStop}%
\bibitem [{\citenamefont {Gross}\ \emph {et~al.}(2021)\citenamefont {Gross},
  \citenamefont {Landini}, \citenamefont {Strumia},\ and\ \citenamefont
  {Teresi}}]{Gross:2021qgx}%
  \BibitemOpen
  \bibfield  {author} {\bibinfo {author} {\bibfnamefont {C.}~\bibnamefont
  {Gross}}, \bibinfo {author} {\bibfnamefont {G.}~\bibnamefont {Landini}},
  \bibinfo {author} {\bibfnamefont {A.}~\bibnamefont {Strumia}}, \ and\
  \bibinfo {author} {\bibfnamefont {D.}~\bibnamefont {Teresi}},\ }\href
  {\doibase 10.1007/JHEP09(2021)033} {\bibfield  {journal} {\bibinfo  {journal}
  {JHEP}\ }\textbf {\bibinfo {volume} {09}},\ \bibinfo {pages} {033} (\bibinfo
  {year} {2021})},\ \Eprint {http://arxiv.org/abs/2105.02840} {arXiv:2105.02840
  [hep-ph]} \BibitemShut {NoStop}%
\bibitem [{\citenamefont {Baker}\ \emph {et~al.}(2021)\citenamefont {Baker},
  \citenamefont {Breitbach}, \citenamefont {Kopp},\ and\ \citenamefont
  {Mittnacht}}]{Baker:2021nyl}%
  \BibitemOpen
  \bibfield  {author} {\bibinfo {author} {\bibfnamefont {M.~J.}\ \bibnamefont
  {Baker}}, \bibinfo {author} {\bibfnamefont {M.}~\bibnamefont {Breitbach}},
  \bibinfo {author} {\bibfnamefont {J.}~\bibnamefont {Kopp}}, \ and\ \bibinfo
  {author} {\bibfnamefont {L.}~\bibnamefont {Mittnacht}},\ }\href@noop {} {\
  (\bibinfo {year} {2021})},\ \Eprint {http://arxiv.org/abs/2105.07481}
  {arXiv:2105.07481 [astro-ph.CO]} \BibitemShut {NoStop}%
\bibitem [{\citenamefont {Kawana}\ and\ \citenamefont
  {Xie}(2022)}]{Kawana:2021tde}%
  \BibitemOpen
  \bibfield  {author} {\bibinfo {author} {\bibfnamefont {K.}~\bibnamefont
  {Kawana}}\ and\ \bibinfo {author} {\bibfnamefont {K.-P.}\ \bibnamefont
  {Xie}},\ }\href {\doibase 10.1016/j.physletb.2021.136791} {\bibfield
  {journal} {\bibinfo  {journal} {Phys. Lett. B}\ }\textbf {\bibinfo {volume}
  {824}},\ \bibinfo {pages} {136791} (\bibinfo {year} {2022})},\ \Eprint
  {http://arxiv.org/abs/2106.00111} {arXiv:2106.00111 [astro-ph.CO]}
  \BibitemShut {NoStop}%
\bibitem [{\citenamefont {Liu}\ \emph {et~al.}(2021{\natexlab{c}})\citenamefont
  {Liu}, \citenamefont {Bian}, \citenamefont {Cai}, \citenamefont {Guo},\ and\
  \citenamefont {Wang}}]{Liu:2021svg}%
  \BibitemOpen
  \bibfield  {author} {\bibinfo {author} {\bibfnamefont {J.}~\bibnamefont
  {Liu}}, \bibinfo {author} {\bibfnamefont {L.}~\bibnamefont {Bian}}, \bibinfo
  {author} {\bibfnamefont {R.-G.}\ \bibnamefont {Cai}}, \bibinfo {author}
  {\bibfnamefont {Z.-K.}\ \bibnamefont {Guo}}, \ and\ \bibinfo {author}
  {\bibfnamefont {S.-J.}\ \bibnamefont {Wang}},\ }\href@noop {} {\  (\bibinfo
  {year} {2021}{\natexlab{c}})},\ \Eprint {http://arxiv.org/abs/2106.05637}
  {arXiv:2106.05637 [astro-ph.CO]} \BibitemShut {NoStop}%
\bibitem [{\citenamefont {Rubin}\ \emph {et~al.}(2000)\citenamefont {Rubin},
  \citenamefont {Khlopov},\ and\ \citenamefont {Sakharov}}]{Rubin:2000dq}%
  \BibitemOpen
  \bibfield  {author} {\bibinfo {author} {\bibfnamefont {S.~G.}\ \bibnamefont
  {Rubin}}, \bibinfo {author} {\bibfnamefont {M.~Y.}\ \bibnamefont {Khlopov}},
  \ and\ \bibinfo {author} {\bibfnamefont {A.~S.}\ \bibnamefont {Sakharov}},\
  }\href@noop {} {\bibfield  {journal} {\bibinfo  {journal} {Grav. Cosmol.}\
  }\textbf {\bibinfo {volume} {6}},\ \bibinfo {pages} {51} (\bibinfo {year}
  {2000})},\ \Eprint {http://arxiv.org/abs/hep-ph/0005271}
  {arXiv:hep-ph/0005271} \BibitemShut {NoStop}%
\bibitem [{\citenamefont {Deng}\ \emph {et~al.}(2017)\citenamefont {Deng},
  \citenamefont {Garriga},\ and\ \citenamefont {Vilenkin}}]{Deng:2016vzb}%
  \BibitemOpen
  \bibfield  {author} {\bibinfo {author} {\bibfnamefont {H.}~\bibnamefont
  {Deng}}, \bibinfo {author} {\bibfnamefont {J.}~\bibnamefont {Garriga}}, \
  and\ \bibinfo {author} {\bibfnamefont {A.}~\bibnamefont {Vilenkin}},\ }\href
  {\doibase 10.1088/1475-7516/2017/04/050} {\bibfield  {journal} {\bibinfo
  {journal} {JCAP}\ }\textbf {\bibinfo {volume} {04}},\ \bibinfo {pages} {050}
  (\bibinfo {year} {2017})},\ \Eprint {http://arxiv.org/abs/1612.03753}
  {arXiv:1612.03753 [gr-qc]} \BibitemShut {NoStop}%
\bibitem [{\citenamefont {Liu}\ \emph {et~al.}(2020{\natexlab{e}})\citenamefont
  {Liu}, \citenamefont {Guo},\ and\ \citenamefont {Cai}}]{Liu:2019lul}%
  \BibitemOpen
  \bibfield  {author} {\bibinfo {author} {\bibfnamefont {J.}~\bibnamefont
  {Liu}}, \bibinfo {author} {\bibfnamefont {Z.-K.}\ \bibnamefont {Guo}}, \ and\
  \bibinfo {author} {\bibfnamefont {R.-G.}\ \bibnamefont {Cai}},\ }\href
  {\doibase 10.1103/PhysRevD.101.023513} {\bibfield  {journal} {\bibinfo
  {journal} {Phys. Rev. D}\ }\textbf {\bibinfo {volume} {101}},\ \bibinfo
  {pages} {023513} (\bibinfo {year} {2020}{\natexlab{e}})},\ \Eprint
  {http://arxiv.org/abs/1908.02662} {arXiv:1908.02662 [astro-ph.CO]}
  \BibitemShut {NoStop}%
\bibitem [{\citenamefont {Gouttenoire}\ and\ \citenamefont
  {Vitagliano}(2024)}]{Gouttenoire:2023gbn}%
  \BibitemOpen
  \bibfield  {author} {\bibinfo {author} {\bibfnamefont {Y.}~\bibnamefont
  {Gouttenoire}}\ and\ \bibinfo {author} {\bibfnamefont {E.}~\bibnamefont
  {Vitagliano}},\ }\href {\doibase 10.1103/PhysRevD.109.123507} {\bibfield
  {journal} {\bibinfo  {journal} {Phys. Rev. D}\ }\textbf {\bibinfo {volume}
  {109}},\ \bibinfo {pages} {123507} (\bibinfo {year} {2024})},\ \Eprint
  {http://arxiv.org/abs/2311.07670} {arXiv:2311.07670 [hep-ph]} \BibitemShut
  {NoStop}%
\bibitem [{\citenamefont {Li}\ and\ \citenamefont {Zhou}(2024)}]{Li:2024psa}%
  \BibitemOpen
  \bibfield  {author} {\bibinfo {author} {\bibfnamefont {H.-J.}\ \bibnamefont
  {Li}}\ and\ \bibinfo {author} {\bibfnamefont {Y.-F.}\ \bibnamefont {Zhou}},\
  }\href@noop {} {\  (\bibinfo {year} {2024})},\ \Eprint
  {http://arxiv.org/abs/2401.09138} {arXiv:2401.09138 [hep-ph]} \BibitemShut
  {NoStop}%
\bibitem [{\citenamefont {Ferreira}\ \emph {et~al.}(2024)\citenamefont
  {Ferreira}, \citenamefont {Notari}, \citenamefont {Pujol\`as},\ and\
  \citenamefont {Rompineve}}]{Ferreira:2024eru}%
  \BibitemOpen
  \bibfield  {author} {\bibinfo {author} {\bibfnamefont {R.~Z.}\ \bibnamefont
  {Ferreira}}, \bibinfo {author} {\bibfnamefont {A.}~\bibnamefont {Notari}},
  \bibinfo {author} {\bibfnamefont {O.}~\bibnamefont {Pujol\`as}}, \ and\
  \bibinfo {author} {\bibfnamefont {F.}~\bibnamefont {Rompineve}},\ }\href
  {\doibase 10.1088/1475-7516/2024/06/020} {\bibfield  {journal} {\bibinfo
  {journal} {JCAP}\ }\textbf {\bibinfo {volume} {06}},\ \bibinfo {pages} {020}
  (\bibinfo {year} {2024})},\ \Eprint {http://arxiv.org/abs/2401.14331}
  {arXiv:2401.14331 [astro-ph.CO]} \BibitemShut {NoStop}%
\bibitem [{\citenamefont {Lu}\ \emph {et~al.}(2024)\citenamefont {Lu},
  \citenamefont {Chiang},\ and\ \citenamefont {Li}}]{Lu:2024ngi}%
  \BibitemOpen
  \bibfield  {author} {\bibinfo {author} {\bibfnamefont {B.-Q.}\ \bibnamefont
  {Lu}}, \bibinfo {author} {\bibfnamefont {C.-W.}\ \bibnamefont {Chiang}}, \
  and\ \bibinfo {author} {\bibfnamefont {T.}~\bibnamefont {Li}},\ }\href@noop
  {} {\  (\bibinfo {year} {2024})},\ \Eprint {http://arxiv.org/abs/2409.09986}
  {arXiv:2409.09986 [astro-ph.CO]} \BibitemShut {NoStop}%
\bibitem [{\citenamefont {Cotner}\ and\ \citenamefont
  {Kusenko}(2017{\natexlab{a}})}]{Cotner:2016cvr}%
  \BibitemOpen
  \bibfield  {author} {\bibinfo {author} {\bibfnamefont {E.}~\bibnamefont
  {Cotner}}\ and\ \bibinfo {author} {\bibfnamefont {A.}~\bibnamefont
  {Kusenko}},\ }\href {\doibase 10.1103/PhysRevLett.119.031103} {\bibfield
  {journal} {\bibinfo  {journal} {Phys. Rev. Lett.}\ }\textbf {\bibinfo
  {volume} {119}},\ \bibinfo {pages} {031103} (\bibinfo {year}
  {2017}{\natexlab{a}})},\ \Eprint {http://arxiv.org/abs/1612.02529}
  {arXiv:1612.02529 [astro-ph.CO]} \BibitemShut {NoStop}%
\bibitem [{\citenamefont {Cotner}\ and\ \citenamefont
  {Kusenko}(2017{\natexlab{b}})}]{Cotner:2017tir}%
  \BibitemOpen
  \bibfield  {author} {\bibinfo {author} {\bibfnamefont {E.}~\bibnamefont
  {Cotner}}\ and\ \bibinfo {author} {\bibfnamefont {A.}~\bibnamefont
  {Kusenko}},\ }\href {\doibase 10.1103/PhysRevD.96.103002} {\bibfield
  {journal} {\bibinfo  {journal} {Phys. Rev. D}\ }\textbf {\bibinfo {volume}
  {96}},\ \bibinfo {pages} {103002} (\bibinfo {year} {2017}{\natexlab{b}})},\
  \Eprint {http://arxiv.org/abs/1706.09003} {arXiv:1706.09003 [astro-ph.CO]}
  \BibitemShut {NoStop}%
\bibitem [{\citenamefont {Cotner}\ \emph {et~al.}(2018)\citenamefont {Cotner},
  \citenamefont {Kusenko},\ and\ \citenamefont {Takhistov}}]{Cotner:2018vug}%
  \BibitemOpen
  \bibfield  {author} {\bibinfo {author} {\bibfnamefont {E.}~\bibnamefont
  {Cotner}}, \bibinfo {author} {\bibfnamefont {A.}~\bibnamefont {Kusenko}}, \
  and\ \bibinfo {author} {\bibfnamefont {V.}~\bibnamefont {Takhistov}},\ }\href
  {\doibase 10.1103/PhysRevD.98.083513} {\bibfield  {journal} {\bibinfo
  {journal} {Phys. Rev. D}\ }\textbf {\bibinfo {volume} {98}},\ \bibinfo
  {pages} {083513} (\bibinfo {year} {2018})},\ \Eprint
  {http://arxiv.org/abs/1801.03321} {arXiv:1801.03321 [astro-ph.CO]}
  \BibitemShut {NoStop}%
\bibitem [{\citenamefont {Cotner}\ \emph {et~al.}(2019)\citenamefont {Cotner},
  \citenamefont {Kusenko}, \citenamefont {Sasaki},\ and\ \citenamefont
  {Takhistov}}]{Cotner:2019ykd}%
  \BibitemOpen
  \bibfield  {author} {\bibinfo {author} {\bibfnamefont {E.}~\bibnamefont
  {Cotner}}, \bibinfo {author} {\bibfnamefont {A.}~\bibnamefont {Kusenko}},
  \bibinfo {author} {\bibfnamefont {M.}~\bibnamefont {Sasaki}}, \ and\ \bibinfo
  {author} {\bibfnamefont {V.}~\bibnamefont {Takhistov}},\ }\href {\doibase
  10.1088/1475-7516/2019/10/077} {\bibfield  {journal} {\bibinfo  {journal}
  {JCAP}\ }\textbf {\bibinfo {volume} {10}},\ \bibinfo {pages} {077} (\bibinfo
  {year} {2019})},\ \Eprint {http://arxiv.org/abs/1907.10613} {arXiv:1907.10613
  [astro-ph.CO]} \BibitemShut {NoStop}%
\bibitem [{\citenamefont {Harada}\ \emph {et~al.}(2013)\citenamefont {Harada},
  \citenamefont {Yoo},\ and\ \citenamefont {Kohri}}]{Harada:2013epa}%
  \BibitemOpen
  \bibfield  {author} {\bibinfo {author} {\bibfnamefont {T.}~\bibnamefont
  {Harada}}, \bibinfo {author} {\bibfnamefont {C.-M.}\ \bibnamefont {Yoo}}, \
  and\ \bibinfo {author} {\bibfnamefont {K.}~\bibnamefont {Kohri}},\ }\href
  {\doibase 10.1103/PhysRevD.88.084051} {\bibfield  {journal} {\bibinfo
  {journal} {Phys. Rev. D}\ }\textbf {\bibinfo {volume} {88}},\ \bibinfo
  {pages} {084051} (\bibinfo {year} {2013})},\ \bibinfo {note} {[Erratum: Phys.
  Rev. D 89, 029903 (2014)]},\ \Eprint {http://arxiv.org/abs/1309.4201}
  {arXiv:1309.4201 [astro-ph.CO]} \BibitemShut {NoStop}%
\bibitem [{\citenamefont {Carr}\ \emph {et~al.}(2020)\citenamefont {Carr},
  \citenamefont {Kohri}, \citenamefont {Sendouda},\ and\ \citenamefont
  {Yokoyama}}]{Carr:2020gox}%
  \BibitemOpen
  \bibfield  {author} {\bibinfo {author} {\bibfnamefont {B.}~\bibnamefont
  {Carr}}, \bibinfo {author} {\bibfnamefont {K.}~\bibnamefont {Kohri}},
  \bibinfo {author} {\bibfnamefont {Y.}~\bibnamefont {Sendouda}}, \ and\
  \bibinfo {author} {\bibfnamefont {J.}~\bibnamefont {Yokoyama}},\ }\href
  {\doibase 10.1088/1361-6633/ac1e31} {\bibfield  {journal} {\bibinfo
  {journal} {Rept. Prog. Phys.}\ }\textbf {\bibinfo {volume} {84}},\ \bibinfo
  {pages} {116902} (\bibinfo {year} {2020})},\ \Eprint
  {http://arxiv.org/abs/2002.12778} {arXiv:2002.12778 [astro-ph.CO]}
  \BibitemShut {NoStop}%
\bibitem [{\citenamefont {Bringmann}\ \emph {et~al.}(2012)\citenamefont
  {Bringmann}, \citenamefont {Scott},\ and\ \citenamefont
  {Akrami}}]{Bringmann:2011ut}%
  \BibitemOpen
  \bibfield  {author} {\bibinfo {author} {\bibfnamefont {T.}~\bibnamefont
  {Bringmann}}, \bibinfo {author} {\bibfnamefont {P.}~\bibnamefont {Scott}}, \
  and\ \bibinfo {author} {\bibfnamefont {Y.}~\bibnamefont {Akrami}},\ }\href
  {\doibase 10.1103/PhysRevD.85.125027} {\bibfield  {journal} {\bibinfo
  {journal} {Phys. Rev. D}\ }\textbf {\bibinfo {volume} {85}},\ \bibinfo
  {pages} {125027} (\bibinfo {year} {2012})},\ \Eprint
  {http://arxiv.org/abs/1110.2484} {arXiv:1110.2484 [astro-ph.CO]} \BibitemShut
  {NoStop}%
\bibitem [{\citenamefont {Chluba}\ \emph {et~al.}(2012)\citenamefont {Chluba},
  \citenamefont {Erickcek},\ and\ \citenamefont {Ben-Dayan}}]{Chluba:2012we}%
  \BibitemOpen
  \bibfield  {author} {\bibinfo {author} {\bibfnamefont {J.}~\bibnamefont
  {Chluba}}, \bibinfo {author} {\bibfnamefont {A.~L.}\ \bibnamefont
  {Erickcek}}, \ and\ \bibinfo {author} {\bibfnamefont {I.}~\bibnamefont
  {Ben-Dayan}},\ }\href {\doibase 10.1088/0004-637X/758/2/76} {\bibfield
  {journal} {\bibinfo  {journal} {Astrophys. J.}\ }\textbf {\bibinfo {volume}
  {758}},\ \bibinfo {pages} {76} (\bibinfo {year} {2012})},\ \Eprint
  {http://arxiv.org/abs/1203.2681} {arXiv:1203.2681 [astro-ph.CO]} \BibitemShut
  {NoStop}%
\bibitem [{\citenamefont {Green}(2018)}]{Green:2018akb}%
  \BibitemOpen
  \bibfield  {author} {\bibinfo {author} {\bibfnamefont {A.~M.}\ \bibnamefont
  {Green}},\ }\href {\doibase 10.1103/PhysRevD.98.023529} {\bibfield  {journal}
  {\bibinfo  {journal} {Phys. Rev. D}\ }\textbf {\bibinfo {volume} {98}},\
  \bibinfo {pages} {023529} (\bibinfo {year} {2018})},\ \Eprint
  {http://arxiv.org/abs/1805.05178} {arXiv:1805.05178 [astro-ph.CO]}
  \BibitemShut {NoStop}%
\bibitem [{\citenamefont {Sato-Polito}\ \emph {et~al.}(2019)\citenamefont
  {Sato-Polito}, \citenamefont {Kovetz},\ and\ \citenamefont
  {Kamionkowski}}]{Sato-Polito:2019hws}%
  \BibitemOpen
  \bibfield  {author} {\bibinfo {author} {\bibfnamefont {G.}~\bibnamefont
  {Sato-Polito}}, \bibinfo {author} {\bibfnamefont {E.~D.}\ \bibnamefont
  {Kovetz}}, \ and\ \bibinfo {author} {\bibfnamefont {M.}~\bibnamefont
  {Kamionkowski}},\ }\href {\doibase 10.1103/PhysRevD.100.063521} {\bibfield
  {journal} {\bibinfo  {journal} {Phys. Rev. D}\ }\textbf {\bibinfo {volume}
  {100}},\ \bibinfo {pages} {063521} (\bibinfo {year} {2019})},\ \Eprint
  {http://arxiv.org/abs/1904.10971} {arXiv:1904.10971 [astro-ph.CO]}
  \BibitemShut {NoStop}%
\bibitem [{\citenamefont {Gow}\ \emph {et~al.}(2021)\citenamefont {Gow},
  \citenamefont {Byrnes}, \citenamefont {Cole},\ and\ \citenamefont
  {Young}}]{Gow:2020bzo}%
  \BibitemOpen
  \bibfield  {author} {\bibinfo {author} {\bibfnamefont {A.~D.}\ \bibnamefont
  {Gow}}, \bibinfo {author} {\bibfnamefont {C.~T.}\ \bibnamefont {Byrnes}},
  \bibinfo {author} {\bibfnamefont {P.~S.}\ \bibnamefont {Cole}}, \ and\
  \bibinfo {author} {\bibfnamefont {S.}~\bibnamefont {Young}},\ }\href
  {\doibase 10.1088/1475-7516/2021/02/002} {\bibfield  {journal} {\bibinfo
  {journal} {JCAP}\ }\textbf {\bibinfo {volume} {02}},\ \bibinfo {pages} {002}
  (\bibinfo {year} {2021})},\ \Eprint {http://arxiv.org/abs/2008.03289}
  {arXiv:2008.03289 [astro-ph.CO]} \BibitemShut {NoStop}%
\bibitem [{\citenamefont {Byrnes}\ \emph {et~al.}(2019)\citenamefont {Byrnes},
  \citenamefont {Cole},\ and\ \citenamefont {Patil}}]{Byrnes:2018txb}%
  \BibitemOpen
  \bibfield  {author} {\bibinfo {author} {\bibfnamefont {C.~T.}\ \bibnamefont
  {Byrnes}}, \bibinfo {author} {\bibfnamefont {P.~S.}\ \bibnamefont {Cole}}, \
  and\ \bibinfo {author} {\bibfnamefont {S.~P.}\ \bibnamefont {Patil}},\ }\href
  {\doibase 10.1088/1475-7516/2019/06/028} {\bibfield  {journal} {\bibinfo
  {journal} {JCAP}\ }\textbf {\bibinfo {volume} {06}},\ \bibinfo {pages} {028}
  (\bibinfo {year} {2019})},\ \Eprint {http://arxiv.org/abs/1811.11158}
  {arXiv:1811.11158 [astro-ph.CO]} \BibitemShut {NoStop}%
\bibitem [{\citenamefont {Inomata}\ and\ \citenamefont
  {Nakama}(2019)}]{Inomata:2018epa}%
  \BibitemOpen
  \bibfield  {author} {\bibinfo {author} {\bibfnamefont {K.}~\bibnamefont
  {Inomata}}\ and\ \bibinfo {author} {\bibfnamefont {T.}~\bibnamefont
  {Nakama}},\ }\href {\doibase 10.1103/PhysRevD.99.043511} {\bibfield
  {journal} {\bibinfo  {journal} {Phys. Rev. D}\ }\textbf {\bibinfo {volume}
  {99}},\ \bibinfo {pages} {043511} (\bibinfo {year} {2019})},\ \Eprint
  {http://arxiv.org/abs/1812.00674} {arXiv:1812.00674 [astro-ph.CO]}
  \BibitemShut {NoStop}%
\bibitem [{\citenamefont {Dalianis}(2019)}]{Dalianis:2018ymb}%
  \BibitemOpen
  \bibfield  {author} {\bibinfo {author} {\bibfnamefont {I.}~\bibnamefont
  {Dalianis}},\ }\href {\doibase 10.1088/1475-7516/2019/08/032} {\bibfield
  {journal} {\bibinfo  {journal} {JCAP}\ }\textbf {\bibinfo {volume} {08}},\
  \bibinfo {pages} {032} (\bibinfo {year} {2019})},\ \Eprint
  {http://arxiv.org/abs/1812.09807} {arXiv:1812.09807 [astro-ph.CO]}
  \BibitemShut {NoStop}%
\bibitem [{\citenamefont {Lu}\ \emph {et~al.}(2019{\natexlab{b}})\citenamefont
  {Lu}, \citenamefont {Gong}, \citenamefont {Yi},\ and\ \citenamefont
  {Zhang}}]{Lu:2019sti}%
  \BibitemOpen
  \bibfield  {author} {\bibinfo {author} {\bibfnamefont {Y.}~\bibnamefont
  {Lu}}, \bibinfo {author} {\bibfnamefont {Y.}~\bibnamefont {Gong}}, \bibinfo
  {author} {\bibfnamefont {Z.}~\bibnamefont {Yi}}, \ and\ \bibinfo {author}
  {\bibfnamefont {F.}~\bibnamefont {Zhang}},\ }\href {\doibase
  10.1088/1475-7516/2019/12/031} {\bibfield  {journal} {\bibinfo  {journal}
  {JCAP}\ }\textbf {\bibinfo {volume} {12}},\ \bibinfo {pages} {031} (\bibinfo
  {year} {2019}{\natexlab{b}})},\ \Eprint {http://arxiv.org/abs/1907.11896}
  {arXiv:1907.11896 [gr-qc]} \BibitemShut {NoStop}%
\bibitem [{\citenamefont {Kalaja}\ \emph {et~al.}(2019)\citenamefont {Kalaja},
  \citenamefont {Bellomo}, \citenamefont {Bartolo}, \citenamefont {Bertacca},
  \citenamefont {Matarrese}, \citenamefont {Musco}, \citenamefont
  {Raccanelli},\ and\ \citenamefont {Verde}}]{Kalaja:2019uju}%
  \BibitemOpen
  \bibfield  {author} {\bibinfo {author} {\bibfnamefont {A.}~\bibnamefont
  {Kalaja}}, \bibinfo {author} {\bibfnamefont {N.}~\bibnamefont {Bellomo}},
  \bibinfo {author} {\bibfnamefont {N.}~\bibnamefont {Bartolo}}, \bibinfo
  {author} {\bibfnamefont {D.}~\bibnamefont {Bertacca}}, \bibinfo {author}
  {\bibfnamefont {S.}~\bibnamefont {Matarrese}}, \bibinfo {author}
  {\bibfnamefont {I.}~\bibnamefont {Musco}}, \bibinfo {author} {\bibfnamefont
  {A.}~\bibnamefont {Raccanelli}}, \ and\ \bibinfo {author} {\bibfnamefont
  {L.}~\bibnamefont {Verde}},\ }\href {\doibase 10.1088/1475-7516/2019/10/031}
  {\bibfield  {journal} {\bibinfo  {journal} {JCAP}\ }\textbf {\bibinfo
  {volume} {10}},\ \bibinfo {pages} {031} (\bibinfo {year} {2019})},\ \Eprint
  {http://arxiv.org/abs/1908.03596} {arXiv:1908.03596 [astro-ph.CO]}
  \BibitemShut {NoStop}%
\bibitem [{\citenamefont {\"Ozsoy}\ and\ \citenamefont
  {Tasinato}(2020)}]{Ozsoy:2019lyy}%
  \BibitemOpen
  \bibfield  {author} {\bibinfo {author} {\bibfnamefont {O.}~\bibnamefont
  {\"Ozsoy}}\ and\ \bibinfo {author} {\bibfnamefont {G.}~\bibnamefont
  {Tasinato}},\ }\href {\doibase 10.1088/1475-7516/2020/04/048} {\bibfield
  {journal} {\bibinfo  {journal} {JCAP}\ }\textbf {\bibinfo {volume} {04}},\
  \bibinfo {pages} {048} (\bibinfo {year} {2020})},\ \Eprint
  {http://arxiv.org/abs/1912.01061} {arXiv:1912.01061 [astro-ph.CO]}
  \BibitemShut {NoStop}%
\bibitem [{\citenamefont {Yokoyama}(1998)}]{Yokoyama:1998pt}%
  \BibitemOpen
  \bibfield  {author} {\bibinfo {author} {\bibfnamefont {J.}~\bibnamefont
  {Yokoyama}},\ }\href {\doibase 10.1103/PhysRevD.58.083510} {\bibfield
  {journal} {\bibinfo  {journal} {Phys. Rev. D}\ }\textbf {\bibinfo {volume}
  {58}},\ \bibinfo {pages} {083510} (\bibinfo {year} {1998})},\ \Eprint
  {http://arxiv.org/abs/astro-ph/9802357} {arXiv:astro-ph/9802357} \BibitemShut
  {NoStop}%
\bibitem [{\citenamefont {Garcia-Bellido}\ \emph {et~al.}(2016)\citenamefont
  {Garcia-Bellido}, \citenamefont {Peloso},\ and\ \citenamefont
  {Unal}}]{Garcia-Bellido:2016dkw}%
  \BibitemOpen
  \bibfield  {author} {\bibinfo {author} {\bibfnamefont {J.}~\bibnamefont
  {Garcia-Bellido}}, \bibinfo {author} {\bibfnamefont {M.}~\bibnamefont
  {Peloso}}, \ and\ \bibinfo {author} {\bibfnamefont {C.}~\bibnamefont
  {Unal}},\ }\href {\doibase 10.1088/1475-7516/2016/12/031} {\bibfield
  {journal} {\bibinfo  {journal} {JCAP}\ }\textbf {\bibinfo {volume} {12}},\
  \bibinfo {pages} {031} (\bibinfo {year} {2016})},\ \Eprint
  {http://arxiv.org/abs/1610.03763} {arXiv:1610.03763 [astro-ph.CO]}
  \BibitemShut {NoStop}%
\bibitem [{\citenamefont {Cheng}\ \emph {et~al.}(2017)\citenamefont {Cheng},
  \citenamefont {Lee},\ and\ \citenamefont {Ng}}]{Cheng:2016qzb}%
  \BibitemOpen
  \bibfield  {author} {\bibinfo {author} {\bibfnamefont {S.-L.}\ \bibnamefont
  {Cheng}}, \bibinfo {author} {\bibfnamefont {W.}~\bibnamefont {Lee}}, \ and\
  \bibinfo {author} {\bibfnamefont {K.-W.}\ \bibnamefont {Ng}},\ }\href
  {\doibase 10.1007/JHEP02(2017)008} {\bibfield  {journal} {\bibinfo  {journal}
  {JHEP}\ }\textbf {\bibinfo {volume} {02}},\ \bibinfo {pages} {008} (\bibinfo
  {year} {2017})},\ \Eprint {http://arxiv.org/abs/1606.00206} {arXiv:1606.00206
  [astro-ph.CO]} \BibitemShut {NoStop}%
\bibitem [{\citenamefont {Garcia-Bellido}\ and\ \citenamefont
  {Ruiz~Morales}(2017)}]{Garcia-Bellido:2017mdw}%
  \BibitemOpen
  \bibfield  {author} {\bibinfo {author} {\bibfnamefont {J.}~\bibnamefont
  {Garcia-Bellido}}\ and\ \bibinfo {author} {\bibfnamefont {E.}~\bibnamefont
  {Ruiz~Morales}},\ }\href {\doibase 10.1016/j.dark.2017.09.007} {\bibfield
  {journal} {\bibinfo  {journal} {Phys. Dark Univ.}\ }\textbf {\bibinfo
  {volume} {18}},\ \bibinfo {pages} {47} (\bibinfo {year} {2017})},\ \Eprint
  {http://arxiv.org/abs/1702.03901} {arXiv:1702.03901 [astro-ph.CO]}
  \BibitemShut {NoStop}%
\bibitem [{\citenamefont {Cheng}\ \emph {et~al.}(2018)\citenamefont {Cheng},
  \citenamefont {Lee},\ and\ \citenamefont {Ng}}]{Cheng:2018yyr}%
  \BibitemOpen
  \bibfield  {author} {\bibinfo {author} {\bibfnamefont {S.-L.}\ \bibnamefont
  {Cheng}}, \bibinfo {author} {\bibfnamefont {W.}~\bibnamefont {Lee}}, \ and\
  \bibinfo {author} {\bibfnamefont {K.-W.}\ \bibnamefont {Ng}},\ }\href
  {\doibase 10.1088/1475-7516/2018/07/001} {\bibfield  {journal} {\bibinfo
  {journal} {JCAP}\ }\textbf {\bibinfo {volume} {07}},\ \bibinfo {pages} {001}
  (\bibinfo {year} {2018})},\ \Eprint {http://arxiv.org/abs/1801.09050}
  {arXiv:1801.09050 [astro-ph.CO]} \BibitemShut {NoStop}%
\bibitem [{\citenamefont {Dalianis}\ \emph {et~al.}(2019)\citenamefont
  {Dalianis}, \citenamefont {Kehagias},\ and\ \citenamefont
  {Tringas}}]{Dalianis:2018frf}%
  \BibitemOpen
  \bibfield  {author} {\bibinfo {author} {\bibfnamefont {I.}~\bibnamefont
  {Dalianis}}, \bibinfo {author} {\bibfnamefont {A.}~\bibnamefont {Kehagias}},
  \ and\ \bibinfo {author} {\bibfnamefont {G.}~\bibnamefont {Tringas}},\ }\href
  {\doibase 10.1088/1475-7516/2019/01/037} {\bibfield  {journal} {\bibinfo
  {journal} {JCAP}\ }\textbf {\bibinfo {volume} {01}},\ \bibinfo {pages} {037}
  (\bibinfo {year} {2019})},\ \Eprint {http://arxiv.org/abs/1805.09483}
  {arXiv:1805.09483 [astro-ph.CO]} \BibitemShut {NoStop}%
\bibitem [{\citenamefont {Tada}\ and\ \citenamefont
  {Yokoyama}(2019)}]{Tada:2019amh}%
  \BibitemOpen
  \bibfield  {author} {\bibinfo {author} {\bibfnamefont {Y.}~\bibnamefont
  {Tada}}\ and\ \bibinfo {author} {\bibfnamefont {S.}~\bibnamefont
  {Yokoyama}},\ }\href {\doibase 10.1103/PhysRevD.100.023537} {\bibfield
  {journal} {\bibinfo  {journal} {Phys. Rev. D}\ }\textbf {\bibinfo {volume}
  {100}},\ \bibinfo {pages} {023537} (\bibinfo {year} {2019})},\ \Eprint
  {http://arxiv.org/abs/1904.10298} {arXiv:1904.10298 [astro-ph.CO]}
  \BibitemShut {NoStop}%
\bibitem [{\citenamefont {Xu}\ \emph {et~al.}(2020)\citenamefont {Xu},
  \citenamefont {Liu}, \citenamefont {Gao},\ and\ \citenamefont
  {Guo}}]{Xu:2019bdp}%
  \BibitemOpen
  \bibfield  {author} {\bibinfo {author} {\bibfnamefont {W.-T.}\ \bibnamefont
  {Xu}}, \bibinfo {author} {\bibfnamefont {J.}~\bibnamefont {Liu}}, \bibinfo
  {author} {\bibfnamefont {T.-J.}\ \bibnamefont {Gao}}, \ and\ \bibinfo
  {author} {\bibfnamefont {Z.-K.}\ \bibnamefont {Guo}},\ }\href {\doibase
  10.1103/PhysRevD.101.023505} {\bibfield  {journal} {\bibinfo  {journal}
  {Phys. Rev. D}\ }\textbf {\bibinfo {volume} {101}},\ \bibinfo {pages}
  {023505} (\bibinfo {year} {2020})},\ \Eprint
  {http://arxiv.org/abs/1907.05213} {arXiv:1907.05213 [astro-ph.CO]}
  \BibitemShut {NoStop}%
\bibitem [{\citenamefont {Mishra}\ and\ \citenamefont
  {Sahni}(2020)}]{Mishra:2019pzq}%
  \BibitemOpen
  \bibfield  {author} {\bibinfo {author} {\bibfnamefont {S.~S.}\ \bibnamefont
  {Mishra}}\ and\ \bibinfo {author} {\bibfnamefont {V.}~\bibnamefont {Sahni}},\
  }\href {\doibase 10.1088/1475-7516/2020/04/007} {\bibfield  {journal}
  {\bibinfo  {journal} {JCAP}\ }\textbf {\bibinfo {volume} {04}},\ \bibinfo
  {pages} {007} (\bibinfo {year} {2020})},\ \Eprint
  {http://arxiv.org/abs/1911.00057} {arXiv:1911.00057 [gr-qc]} \BibitemShut
  {NoStop}%
\bibitem [{\citenamefont {Bhaumik}\ and\ \citenamefont
  {Jain}(2020)}]{Bhaumik:2019tvl}%
  \BibitemOpen
  \bibfield  {author} {\bibinfo {author} {\bibfnamefont {N.}~\bibnamefont
  {Bhaumik}}\ and\ \bibinfo {author} {\bibfnamefont {R.~K.}\ \bibnamefont
  {Jain}},\ }\href {\doibase 10.1088/1475-7516/2020/01/037} {\bibfield
  {journal} {\bibinfo  {journal} {JCAP}\ }\textbf {\bibinfo {volume} {01}},\
  \bibinfo {pages} {037} (\bibinfo {year} {2020})},\ \Eprint
  {http://arxiv.org/abs/1907.04125} {arXiv:1907.04125 [astro-ph.CO]}
  \BibitemShut {NoStop}%
\bibitem [{\citenamefont {Liu}\ \emph {et~al.}(2020{\natexlab{f}})\citenamefont
  {Liu}, \citenamefont {Guo},\ and\ \citenamefont {Cai}}]{Liu:2020oqe}%
  \BibitemOpen
  \bibfield  {author} {\bibinfo {author} {\bibfnamefont {J.}~\bibnamefont
  {Liu}}, \bibinfo {author} {\bibfnamefont {Z.-K.}\ \bibnamefont {Guo}}, \ and\
  \bibinfo {author} {\bibfnamefont {R.-G.}\ \bibnamefont {Cai}},\ }\href
  {\doibase 10.1103/PhysRevD.101.083535} {\bibfield  {journal} {\bibinfo
  {journal} {Phys. Rev. D}\ }\textbf {\bibinfo {volume} {101}},\ \bibinfo
  {pages} {083535} (\bibinfo {year} {2020}{\natexlab{f}})},\ \Eprint
  {http://arxiv.org/abs/2003.02075} {arXiv:2003.02075 [astro-ph.CO]}
  \BibitemShut {NoStop}%
\bibitem [{\citenamefont {Atal}\ \emph
  {et~al.}(2020{\natexlab{a}})\citenamefont {Atal}, \citenamefont {Cid},
  \citenamefont {Escriv\`a},\ and\ \citenamefont {Garriga}}]{Atal:2019erb}%
  \BibitemOpen
  \bibfield  {author} {\bibinfo {author} {\bibfnamefont {V.}~\bibnamefont
  {Atal}}, \bibinfo {author} {\bibfnamefont {J.}~\bibnamefont {Cid}}, \bibinfo
  {author} {\bibfnamefont {A.}~\bibnamefont {Escriv\`a}}, \ and\ \bibinfo
  {author} {\bibfnamefont {J.}~\bibnamefont {Garriga}},\ }\href {\doibase
  10.1088/1475-7516/2020/05/022} {\bibfield  {journal} {\bibinfo  {journal}
  {JCAP}\ }\textbf {\bibinfo {volume} {05}},\ \bibinfo {pages} {022} (\bibinfo
  {year} {2020}{\natexlab{a}})},\ \Eprint {http://arxiv.org/abs/1908.11357}
  {arXiv:1908.11357 [astro-ph.CO]} \BibitemShut {NoStop}%
\bibitem [{\citenamefont {Fu}\ \emph {et~al.}(2020{\natexlab{a}})\citenamefont
  {Fu}, \citenamefont {Wu},\ and\ \citenamefont {Yu}}]{Fu:2020lob}%
  \BibitemOpen
  \bibfield  {author} {\bibinfo {author} {\bibfnamefont {C.}~\bibnamefont
  {Fu}}, \bibinfo {author} {\bibfnamefont {P.}~\bibnamefont {Wu}}, \ and\
  \bibinfo {author} {\bibfnamefont {H.}~\bibnamefont {Yu}},\ }\href {\doibase
  10.1103/PhysRevD.102.043527} {\bibfield  {journal} {\bibinfo  {journal}
  {Phys. Rev. D}\ }\textbf {\bibinfo {volume} {102}},\ \bibinfo {pages}
  {043527} (\bibinfo {year} {2020}{\natexlab{a}})},\ \Eprint
  {http://arxiv.org/abs/2006.03768} {arXiv:2006.03768 [astro-ph.CO]}
  \BibitemShut {NoStop}%
\bibitem [{\citenamefont {Vennin}(2020)}]{Vennin:2020kng}%
  \BibitemOpen
  \bibfield  {author} {\bibinfo {author} {\bibfnamefont {V.}~\bibnamefont
  {Vennin}},\ }\emph {\bibinfo {title} {{Stochastic inflation and primordial
  black holes}}},\ \href@noop {} {\bibinfo {type} {Other thesis}},\ \bibinfo
  {school} {U. Paris-Saclay} (\bibinfo {year} {2020}),\ \Eprint
  {http://arxiv.org/abs/2009.08715} {arXiv:2009.08715 [astro-ph.CO]}
  \BibitemShut {NoStop}%
\bibitem [{\citenamefont {Ragavendra}\ \emph {et~al.}(2021)\citenamefont
  {Ragavendra}, \citenamefont {Saha}, \citenamefont {Sriramkumar},\ and\
  \citenamefont {Silk}}]{Ragavendra:2020sop}%
  \BibitemOpen
  \bibfield  {author} {\bibinfo {author} {\bibfnamefont {H.~V.}\ \bibnamefont
  {Ragavendra}}, \bibinfo {author} {\bibfnamefont {P.}~\bibnamefont {Saha}},
  \bibinfo {author} {\bibfnamefont {L.}~\bibnamefont {Sriramkumar}}, \ and\
  \bibinfo {author} {\bibfnamefont {J.}~\bibnamefont {Silk}},\ }\href {\doibase
  10.1103/PhysRevD.103.083510} {\bibfield  {journal} {\bibinfo  {journal}
  {Phys. Rev. D}\ }\textbf {\bibinfo {volume} {103}},\ \bibinfo {pages}
  {083510} (\bibinfo {year} {2021})},\ \Eprint
  {http://arxiv.org/abs/2008.12202} {arXiv:2008.12202 [astro-ph.CO]}
  \BibitemShut {NoStop}%
\bibitem [{\citenamefont {Gao}\ and\ \citenamefont {Yang}(2021)}]{Gao:2021dfi}%
  \BibitemOpen
  \bibfield  {author} {\bibinfo {author} {\bibfnamefont {T.-J.}\ \bibnamefont
  {Gao}}\ and\ \bibinfo {author} {\bibfnamefont {X.-Y.}\ \bibnamefont {Yang}},\
  }\href@noop {} {\  (\bibinfo {year} {2021})},\ \Eprint
  {http://arxiv.org/abs/2101.07616} {arXiv:2101.07616 [astro-ph.CO]}
  \BibitemShut {NoStop}%
\bibitem [{\citenamefont {Garcia-Bellido}\ \emph {et~al.}(1996)\citenamefont
  {Garcia-Bellido}, \citenamefont {Linde},\ and\ \citenamefont
  {Wands}}]{GarciaBellido:1996qt}%
  \BibitemOpen
  \bibfield  {author} {\bibinfo {author} {\bibfnamefont {J.}~\bibnamefont
  {Garcia-Bellido}}, \bibinfo {author} {\bibfnamefont {A.~D.}\ \bibnamefont
  {Linde}}, \ and\ \bibinfo {author} {\bibfnamefont {D.}~\bibnamefont
  {Wands}},\ }\href {\doibase 10.1103/PhysRevD.54.6040} {\bibfield  {journal}
  {\bibinfo  {journal} {Phys. Rev. D}\ }\textbf {\bibinfo {volume} {54}},\
  \bibinfo {pages} {6040} (\bibinfo {year} {1996})},\ \Eprint
  {http://arxiv.org/abs/astro-ph/9605094} {arXiv:astro-ph/9605094} \BibitemShut
  {NoStop}%
\bibitem [{\citenamefont {Kawasaki}\ \emph {et~al.}(1998)\citenamefont
  {Kawasaki}, \citenamefont {Sugiyama},\ and\ \citenamefont
  {Yanagida}}]{Kawasaki:1997ju}%
  \BibitemOpen
  \bibfield  {author} {\bibinfo {author} {\bibfnamefont {M.}~\bibnamefont
  {Kawasaki}}, \bibinfo {author} {\bibfnamefont {N.}~\bibnamefont {Sugiyama}},
  \ and\ \bibinfo {author} {\bibfnamefont {T.}~\bibnamefont {Yanagida}},\
  }\href {\doibase 10.1103/PhysRevD.57.6050} {\bibfield  {journal} {\bibinfo
  {journal} {Phys. Rev. D}\ }\textbf {\bibinfo {volume} {57}},\ \bibinfo
  {pages} {6050} (\bibinfo {year} {1998})},\ \Eprint
  {http://arxiv.org/abs/hep-ph/9710259} {arXiv:hep-ph/9710259} \BibitemShut
  {NoStop}%
\bibitem [{\citenamefont {Frampton}\ \emph {et~al.}(2010)\citenamefont
  {Frampton}, \citenamefont {Kawasaki}, \citenamefont {Takahashi},\ and\
  \citenamefont {Yanagida}}]{Frampton:2010sw}%
  \BibitemOpen
  \bibfield  {author} {\bibinfo {author} {\bibfnamefont {P.~H.}\ \bibnamefont
  {Frampton}}, \bibinfo {author} {\bibfnamefont {M.}~\bibnamefont {Kawasaki}},
  \bibinfo {author} {\bibfnamefont {F.}~\bibnamefont {Takahashi}}, \ and\
  \bibinfo {author} {\bibfnamefont {T.~T.}\ \bibnamefont {Yanagida}},\ }\href
  {\doibase 10.1088/1475-7516/2010/04/023} {\bibfield  {journal} {\bibinfo
  {journal} {JCAP}\ }\textbf {\bibinfo {volume} {04}},\ \bibinfo {pages} {023}
  (\bibinfo {year} {2010})},\ \Eprint {http://arxiv.org/abs/1001.2308}
  {arXiv:1001.2308 [hep-ph]} \BibitemShut {NoStop}%
\bibitem [{\citenamefont {Giovannini}(2010)}]{Giovannini:2010tk}%
  \BibitemOpen
  \bibfield  {author} {\bibinfo {author} {\bibfnamefont {M.}~\bibnamefont
  {Giovannini}},\ }\href {\doibase 10.1103/PhysRevD.82.083523} {\bibfield
  {journal} {\bibinfo  {journal} {Phys. Rev. D}\ }\textbf {\bibinfo {volume}
  {82}},\ \bibinfo {pages} {083523} (\bibinfo {year} {2010})},\ \Eprint
  {http://arxiv.org/abs/1008.1164} {arXiv:1008.1164 [astro-ph.CO]} \BibitemShut
  {NoStop}%
\bibitem [{\citenamefont {Clesse}\ and\ \citenamefont
  {Garc\'\i{}a-Bellido}(2015)}]{Clesse:2015wea}%
  \BibitemOpen
  \bibfield  {author} {\bibinfo {author} {\bibfnamefont {S.}~\bibnamefont
  {Clesse}}\ and\ \bibinfo {author} {\bibfnamefont {J.}~\bibnamefont
  {Garc\'\i{}a-Bellido}},\ }\href {\doibase 10.1103/PhysRevD.92.023524}
  {\bibfield  {journal} {\bibinfo  {journal} {Phys. Rev. D}\ }\textbf {\bibinfo
  {volume} {92}},\ \bibinfo {pages} {023524} (\bibinfo {year} {2015})},\
  \Eprint {http://arxiv.org/abs/1501.07565} {arXiv:1501.07565 [astro-ph.CO]}
  \BibitemShut {NoStop}%
\bibitem [{\citenamefont {Inomata}\ \emph {et~al.}(2017)\citenamefont
  {Inomata}, \citenamefont {Kawasaki}, \citenamefont {Mukaida}, \citenamefont
  {Tada},\ and\ \citenamefont {Yanagida}}]{Inomata:2017okj}%
  \BibitemOpen
  \bibfield  {author} {\bibinfo {author} {\bibfnamefont {K.}~\bibnamefont
  {Inomata}}, \bibinfo {author} {\bibfnamefont {M.}~\bibnamefont {Kawasaki}},
  \bibinfo {author} {\bibfnamefont {K.}~\bibnamefont {Mukaida}}, \bibinfo
  {author} {\bibfnamefont {Y.}~\bibnamefont {Tada}}, \ and\ \bibinfo {author}
  {\bibfnamefont {T.~T.}\ \bibnamefont {Yanagida}},\ }\href {\doibase
  10.1103/PhysRevD.96.043504} {\bibfield  {journal} {\bibinfo  {journal} {Phys.
  Rev. D}\ }\textbf {\bibinfo {volume} {96}},\ \bibinfo {pages} {043504}
  (\bibinfo {year} {2017})},\ \Eprint {http://arxiv.org/abs/1701.02544}
  {arXiv:1701.02544 [astro-ph.CO]} \BibitemShut {NoStop}%
\bibitem [{\citenamefont {Di}\ and\ \citenamefont {Gong}(2018)}]{Gong:2017qlj}%
  \BibitemOpen
  \bibfield  {author} {\bibinfo {author} {\bibfnamefont {H.}~\bibnamefont
  {Di}}\ and\ \bibinfo {author} {\bibfnamefont {Y.}~\bibnamefont {Gong}},\
  }\href {\doibase 10.1088/1475-7516/2018/07/007} {\bibfield  {journal}
  {\bibinfo  {journal} {JCAP}\ }\textbf {\bibinfo {volume} {07}},\ \bibinfo
  {pages} {007} (\bibinfo {year} {2018})},\ \Eprint
  {http://arxiv.org/abs/1707.09578} {arXiv:1707.09578 [astro-ph.CO]}
  \BibitemShut {NoStop}%
\bibitem [{\citenamefont {Inomata}\ \emph {et~al.}(2018)\citenamefont
  {Inomata}, \citenamefont {Kawasaki}, \citenamefont {Mukaida},\ and\
  \citenamefont {Yanagida}}]{Inomata:2017vxo}%
  \BibitemOpen
  \bibfield  {author} {\bibinfo {author} {\bibfnamefont {K.}~\bibnamefont
  {Inomata}}, \bibinfo {author} {\bibfnamefont {M.}~\bibnamefont {Kawasaki}},
  \bibinfo {author} {\bibfnamefont {K.}~\bibnamefont {Mukaida}}, \ and\
  \bibinfo {author} {\bibfnamefont {T.~T.}\ \bibnamefont {Yanagida}},\ }\href
  {\doibase 10.1103/PhysRevD.97.043514} {\bibfield  {journal} {\bibinfo
  {journal} {Phys. Rev. D}\ }\textbf {\bibinfo {volume} {97}},\ \bibinfo
  {pages} {043514} (\bibinfo {year} {2018})},\ \Eprint
  {http://arxiv.org/abs/1711.06129} {arXiv:1711.06129 [astro-ph.CO]}
  \BibitemShut {NoStop}%
\bibitem [{\citenamefont {Espinosa}\ \emph {et~al.}(2018)\citenamefont
  {Espinosa}, \citenamefont {Racco},\ and\ \citenamefont
  {Riotto}}]{Espinosa:2017sgp}%
  \BibitemOpen
  \bibfield  {author} {\bibinfo {author} {\bibfnamefont {J.}~\bibnamefont
  {Espinosa}}, \bibinfo {author} {\bibfnamefont {D.}~\bibnamefont {Racco}}, \
  and\ \bibinfo {author} {\bibfnamefont {A.}~\bibnamefont {Riotto}},\ }\href
  {\doibase 10.1103/PhysRevLett.120.121301} {\bibfield  {journal} {\bibinfo
  {journal} {Phys. Rev. Lett.}\ }\textbf {\bibinfo {volume} {120}},\ \bibinfo
  {pages} {121301} (\bibinfo {year} {2018})},\ \Eprint
  {http://arxiv.org/abs/1710.11196} {arXiv:1710.11196 [hep-ph]} \BibitemShut
  {NoStop}%
\bibitem [{\citenamefont {Kawasaki}\ \emph {et~al.}(2020)\citenamefont
  {Kawasaki}, \citenamefont {Nakatsuka},\ and\ \citenamefont
  {Obata}}]{Kawasaki:2019hvt}%
  \BibitemOpen
  \bibfield  {author} {\bibinfo {author} {\bibfnamefont {M.}~\bibnamefont
  {Kawasaki}}, \bibinfo {author} {\bibfnamefont {H.}~\bibnamefont {Nakatsuka}},
  \ and\ \bibinfo {author} {\bibfnamefont {I.}~\bibnamefont {Obata}},\ }\href
  {\doibase 10.1088/1475-7516/2020/05/007} {\bibfield  {journal} {\bibinfo
  {journal} {JCAP}\ }\textbf {\bibinfo {volume} {05}},\ \bibinfo {pages} {007}
  (\bibinfo {year} {2020})},\ \Eprint {http://arxiv.org/abs/1912.09111}
  {arXiv:1912.09111 [astro-ph.CO]} \BibitemShut {NoStop}%
\bibitem [{\citenamefont {Palma}\ \emph {et~al.}(2020)\citenamefont {Palma},
  \citenamefont {Sypsas},\ and\ \citenamefont {Zenteno}}]{Palma:2020ejf}%
  \BibitemOpen
  \bibfield  {author} {\bibinfo {author} {\bibfnamefont {G.~A.}\ \bibnamefont
  {Palma}}, \bibinfo {author} {\bibfnamefont {S.}~\bibnamefont {Sypsas}}, \
  and\ \bibinfo {author} {\bibfnamefont {C.}~\bibnamefont {Zenteno}},\ }\href
  {\doibase 10.1103/PhysRevLett.125.121301} {\bibfield  {journal} {\bibinfo
  {journal} {Phys. Rev. Lett.}\ }\textbf {\bibinfo {volume} {125}},\ \bibinfo
  {pages} {121301} (\bibinfo {year} {2020})},\ \Eprint
  {http://arxiv.org/abs/2004.06106} {arXiv:2004.06106 [astro-ph.CO]}
  \BibitemShut {NoStop}%
\bibitem [{\citenamefont {Fumagalli}\ \emph {et~al.}(2020)\citenamefont
  {Fumagalli}, \citenamefont {Renaux-Petel}, \citenamefont {Ronayne},\ and\
  \citenamefont {Witkowski}}]{Fumagalli:2020adf}%
  \BibitemOpen
  \bibfield  {author} {\bibinfo {author} {\bibfnamefont {J.}~\bibnamefont
  {Fumagalli}}, \bibinfo {author} {\bibfnamefont {S.}~\bibnamefont
  {Renaux-Petel}}, \bibinfo {author} {\bibfnamefont {J.~W.}\ \bibnamefont
  {Ronayne}}, \ and\ \bibinfo {author} {\bibfnamefont {L.~T.}\ \bibnamefont
  {Witkowski}},\ }\href@noop {} {\  (\bibinfo {year} {2020})},\ \Eprint
  {http://arxiv.org/abs/2004.08369} {arXiv:2004.08369 [hep-th]} \BibitemShut
  {NoStop}%
\bibitem [{\citenamefont {Braglia}\ \emph {et~al.}(2020)\citenamefont
  {Braglia}, \citenamefont {Hazra}, \citenamefont {Finelli}, \citenamefont
  {Smoot}, \citenamefont {Sriramkumar},\ and\ \citenamefont
  {Starobinsky}}]{Braglia:2020eai}%
  \BibitemOpen
  \bibfield  {author} {\bibinfo {author} {\bibfnamefont {M.}~\bibnamefont
  {Braglia}}, \bibinfo {author} {\bibfnamefont {D.~K.}\ \bibnamefont {Hazra}},
  \bibinfo {author} {\bibfnamefont {F.}~\bibnamefont {Finelli}}, \bibinfo
  {author} {\bibfnamefont {G.~F.}\ \bibnamefont {Smoot}}, \bibinfo {author}
  {\bibfnamefont {L.}~\bibnamefont {Sriramkumar}}, \ and\ \bibinfo {author}
  {\bibfnamefont {A.~A.}\ \bibnamefont {Starobinsky}},\ }\href {\doibase
  10.1088/1475-7516/2020/08/001} {\bibfield  {journal} {\bibinfo  {journal}
  {JCAP}\ }\textbf {\bibinfo {volume} {08}},\ \bibinfo {pages} {001} (\bibinfo
  {year} {2020})},\ \Eprint {http://arxiv.org/abs/2005.02895} {arXiv:2005.02895
  [astro-ph.CO]} \BibitemShut {NoStop}%
\bibitem [{\citenamefont {Anguelova}(2021)}]{Anguelova:2020nzl}%
  \BibitemOpen
  \bibfield  {author} {\bibinfo {author} {\bibfnamefont {L.}~\bibnamefont
  {Anguelova}},\ }\href {\doibase 10.1088/1475-7516/2021/06/004} {\bibfield
  {journal} {\bibinfo  {journal} {JCAP}\ }\textbf {\bibinfo {volume} {06}},\
  \bibinfo {pages} {004} (\bibinfo {year} {2021})},\ \Eprint
  {http://arxiv.org/abs/2012.03705} {arXiv:2012.03705 [hep-th]} \BibitemShut
  {NoStop}%
\bibitem [{\citenamefont {Romano}(2020)}]{Romano:2020gtn}%
  \BibitemOpen
  \bibfield  {author} {\bibinfo {author} {\bibfnamefont {A.~E.}\ \bibnamefont
  {Romano}},\ }\href@noop {} {\  (\bibinfo {year} {2020})},\ \Eprint
  {http://arxiv.org/abs/2006.07321} {arXiv:2006.07321 [astro-ph.CO]}
  \BibitemShut {NoStop}%
\bibitem [{\citenamefont {Gundhi}\ and\ \citenamefont
  {Steinwachs}(2021)}]{Gundhi:2020zvb}%
  \BibitemOpen
  \bibfield  {author} {\bibinfo {author} {\bibfnamefont {A.}~\bibnamefont
  {Gundhi}}\ and\ \bibinfo {author} {\bibfnamefont {C.~F.}\ \bibnamefont
  {Steinwachs}},\ }\href {\doibase 10.1140/epjc/s10052-021-09225-2} {\bibfield
  {journal} {\bibinfo  {journal} {Eur. Phys. J. C}\ }\textbf {\bibinfo {volume}
  {81}},\ \bibinfo {pages} {460} (\bibinfo {year} {2021})},\ \Eprint
  {http://arxiv.org/abs/2011.09485} {arXiv:2011.09485 [hep-th]} \BibitemShut
  {NoStop}%
\bibitem [{\citenamefont {Gundhi}\ \emph {et~al.}(2021)\citenamefont {Gundhi},
  \citenamefont {Ketov},\ and\ \citenamefont {Steinwachs}}]{Gundhi:2020kzm}%
  \BibitemOpen
  \bibfield  {author} {\bibinfo {author} {\bibfnamefont {A.}~\bibnamefont
  {Gundhi}}, \bibinfo {author} {\bibfnamefont {S.~V.}\ \bibnamefont {Ketov}}, \
  and\ \bibinfo {author} {\bibfnamefont {C.~F.}\ \bibnamefont {Steinwachs}},\
  }\href {\doibase 10.1103/PhysRevD.103.083518} {\bibfield  {journal} {\bibinfo
   {journal} {Phys. Rev. D}\ }\textbf {\bibinfo {volume} {103}},\ \bibinfo
  {pages} {083518} (\bibinfo {year} {2021})},\ \Eprint
  {http://arxiv.org/abs/2011.05999} {arXiv:2011.05999 [hep-th]} \BibitemShut
  {NoStop}%
\bibitem [{\citenamefont {Wang}\ \emph
  {et~al.}(2024{\natexlab{c}})\citenamefont {Wang}, \citenamefont {Zhang},\
  and\ \citenamefont {Sasaki}}]{Wang:2024vfv}%
  \BibitemOpen
  \bibfield  {author} {\bibinfo {author} {\bibfnamefont {X.}~\bibnamefont
  {Wang}}, \bibinfo {author} {\bibfnamefont {Y.-l.}\ \bibnamefont {Zhang}}, \
  and\ \bibinfo {author} {\bibfnamefont {M.}~\bibnamefont {Sasaki}},\ }\href
  {\doibase 10.1088/1475-7516/2024/07/076} {\bibfield  {journal} {\bibinfo
  {journal} {JCAP}\ }\textbf {\bibinfo {volume} {07}},\ \bibinfo {pages} {076}
  (\bibinfo {year} {2024}{\natexlab{c}})},\ \Eprint
  {http://arxiv.org/abs/2404.02492} {arXiv:2404.02492 [astro-ph.CO]}
  \BibitemShut {NoStop}%
\bibitem [{\citenamefont {Kannike}\ \emph {et~al.}(2017)\citenamefont
  {Kannike}, \citenamefont {Marzola}, \citenamefont {Raidal},\ and\
  \citenamefont {Veerm\"ae}}]{Kannike:2017bxn}%
  \BibitemOpen
  \bibfield  {author} {\bibinfo {author} {\bibfnamefont {K.}~\bibnamefont
  {Kannike}}, \bibinfo {author} {\bibfnamefont {L.}~\bibnamefont {Marzola}},
  \bibinfo {author} {\bibfnamefont {M.}~\bibnamefont {Raidal}}, \ and\ \bibinfo
  {author} {\bibfnamefont {H.}~\bibnamefont {Veerm\"ae}},\ }\href {\doibase
  10.1088/1475-7516/2017/09/020} {\bibfield  {journal} {\bibinfo  {journal}
  {JCAP}\ }\textbf {\bibinfo {volume} {09}},\ \bibinfo {pages} {020} (\bibinfo
  {year} {2017})},\ \Eprint {http://arxiv.org/abs/1705.06225} {arXiv:1705.06225
  [astro-ph.CO]} \BibitemShut {NoStop}%
\bibitem [{\citenamefont {Pi}\ \emph {et~al.}(2018)\citenamefont {Pi},
  \citenamefont {Zhang}, \citenamefont {Huang},\ and\ \citenamefont
  {Sasaki}}]{Pi:2017gih}%
  \BibitemOpen
  \bibfield  {author} {\bibinfo {author} {\bibfnamefont {S.}~\bibnamefont
  {Pi}}, \bibinfo {author} {\bibfnamefont {Y.-l.}\ \bibnamefont {Zhang}},
  \bibinfo {author} {\bibfnamefont {Q.-G.}\ \bibnamefont {Huang}}, \ and\
  \bibinfo {author} {\bibfnamefont {M.}~\bibnamefont {Sasaki}},\ }\href
  {\doibase 10.1088/1475-7516/2018/05/042} {\bibfield  {journal} {\bibinfo
  {journal} {JCAP}\ }\textbf {\bibinfo {volume} {05}},\ \bibinfo {pages} {042}
  (\bibinfo {year} {2018})},\ \Eprint {http://arxiv.org/abs/1712.09896}
  {arXiv:1712.09896 [astro-ph.CO]} \BibitemShut {NoStop}%
\bibitem [{\citenamefont {Gao}\ and\ \citenamefont {Guo}(2018)}]{Gao:2018pvq}%
  \BibitemOpen
  \bibfield  {author} {\bibinfo {author} {\bibfnamefont {T.-J.}\ \bibnamefont
  {Gao}}\ and\ \bibinfo {author} {\bibfnamefont {Z.-K.}\ \bibnamefont {Guo}},\
  }\href {\doibase 10.1103/PhysRevD.98.063526} {\bibfield  {journal} {\bibinfo
  {journal} {Phys. Rev. D}\ }\textbf {\bibinfo {volume} {98}},\ \bibinfo
  {pages} {063526} (\bibinfo {year} {2018})},\ \Eprint
  {http://arxiv.org/abs/1806.09320} {arXiv:1806.09320 [hep-ph]} \BibitemShut
  {NoStop}%
\bibitem [{\citenamefont {Cheong}\ \emph {et~al.}(2021)\citenamefont {Cheong},
  \citenamefont {Lee},\ and\ \citenamefont {Park}}]{Cheong:2019vzl}%
  \BibitemOpen
  \bibfield  {author} {\bibinfo {author} {\bibfnamefont {D.~Y.}\ \bibnamefont
  {Cheong}}, \bibinfo {author} {\bibfnamefont {S.~M.}\ \bibnamefont {Lee}}, \
  and\ \bibinfo {author} {\bibfnamefont {S.~C.}\ \bibnamefont {Park}},\ }\href
  {\doibase 10.1088/1475-7516/2021/01/032} {\bibfield  {journal} {\bibinfo
  {journal} {JCAP}\ }\textbf {\bibinfo {volume} {01}},\ \bibinfo {pages} {032}
  (\bibinfo {year} {2021})},\ \Eprint {http://arxiv.org/abs/1912.12032}
  {arXiv:1912.12032 [hep-ph]} \BibitemShut {NoStop}%
\bibitem [{\citenamefont {Cheong}\ \emph {et~al.}(2020)\citenamefont {Cheong},
  \citenamefont {Lee},\ and\ \citenamefont {Park}}]{Cheong:2020rao}%
  \BibitemOpen
  \bibfield  {author} {\bibinfo {author} {\bibfnamefont {D.~Y.}\ \bibnamefont
  {Cheong}}, \bibinfo {author} {\bibfnamefont {H.~M.}\ \bibnamefont {Lee}}, \
  and\ \bibinfo {author} {\bibfnamefont {S.~C.}\ \bibnamefont {Park}},\ }\href
  {\doibase 10.1016/j.physletb.2020.135453} {\bibfield  {journal} {\bibinfo
  {journal} {Phys. Lett. B}\ }\textbf {\bibinfo {volume} {805}},\ \bibinfo
  {pages} {135453} (\bibinfo {year} {2020})},\ \Eprint
  {http://arxiv.org/abs/2002.07981} {arXiv:2002.07981 [hep-ph]} \BibitemShut
  {NoStop}%
\bibitem [{\citenamefont {Fu}\ \emph {et~al.}(2019{\natexlab{a}})\citenamefont
  {Fu}, \citenamefont {Wu},\ and\ \citenamefont {Yu}}]{Fu:2019ttf}%
  \BibitemOpen
  \bibfield  {author} {\bibinfo {author} {\bibfnamefont {C.}~\bibnamefont
  {Fu}}, \bibinfo {author} {\bibfnamefont {P.}~\bibnamefont {Wu}}, \ and\
  \bibinfo {author} {\bibfnamefont {H.}~\bibnamefont {Yu}},\ }\href {\doibase
  10.1103/PhysRevD.100.063532} {\bibfield  {journal} {\bibinfo  {journal}
  {Phys. Rev. D}\ }\textbf {\bibinfo {volume} {100}},\ \bibinfo {pages}
  {063532} (\bibinfo {year} {2019}{\natexlab{a}})},\ \Eprint
  {http://arxiv.org/abs/1907.05042} {arXiv:1907.05042 [astro-ph.CO]}
  \BibitemShut {NoStop}%
\bibitem [{\citenamefont {Dalianis}\ \emph {et~al.}(2020)\citenamefont
  {Dalianis}, \citenamefont {Karydas},\ and\ \citenamefont
  {Papantonopoulos}}]{Dalianis:2019vit}%
  \BibitemOpen
  \bibfield  {author} {\bibinfo {author} {\bibfnamefont {I.}~\bibnamefont
  {Dalianis}}, \bibinfo {author} {\bibfnamefont {S.}~\bibnamefont {Karydas}}, \
  and\ \bibinfo {author} {\bibfnamefont {E.}~\bibnamefont {Papantonopoulos}},\
  }\href {\doibase 10.1088/1475-7516/2020/06/040} {\bibfield  {journal}
  {\bibinfo  {journal} {JCAP}\ }\textbf {\bibinfo {volume} {06}},\ \bibinfo
  {pages} {040} (\bibinfo {year} {2020})},\ \Eprint
  {http://arxiv.org/abs/1910.00622} {arXiv:1910.00622 [astro-ph.CO]}
  \BibitemShut {NoStop}%
\bibitem [{\citenamefont {Lin}\ \emph {et~al.}(2020)\citenamefont {Lin},
  \citenamefont {Gao}, \citenamefont {Gong}, \citenamefont {Lu}, \citenamefont
  {Zhang},\ and\ \citenamefont {Zhang}}]{Lin:2020goi}%
  \BibitemOpen
  \bibfield  {author} {\bibinfo {author} {\bibfnamefont {J.}~\bibnamefont
  {Lin}}, \bibinfo {author} {\bibfnamefont {Q.}~\bibnamefont {Gao}}, \bibinfo
  {author} {\bibfnamefont {Y.}~\bibnamefont {Gong}}, \bibinfo {author}
  {\bibfnamefont {Y.}~\bibnamefont {Lu}}, \bibinfo {author} {\bibfnamefont
  {C.}~\bibnamefont {Zhang}}, \ and\ \bibinfo {author} {\bibfnamefont
  {F.}~\bibnamefont {Zhang}},\ }\href {\doibase 10.1103/PhysRevD.101.103515}
  {\bibfield  {journal} {\bibinfo  {journal} {Phys. Rev. D}\ }\textbf {\bibinfo
  {volume} {101}},\ \bibinfo {pages} {103515} (\bibinfo {year} {2020})},\
  \Eprint {http://arxiv.org/abs/2001.05909} {arXiv:2001.05909 [gr-qc]}
  \BibitemShut {NoStop}%
\bibitem [{\citenamefont {Fu}\ \emph {et~al.}(2020{\natexlab{b}})\citenamefont
  {Fu}, \citenamefont {Wu},\ and\ \citenamefont {Yu}}]{Fu:2019vqc}%
  \BibitemOpen
  \bibfield  {author} {\bibinfo {author} {\bibfnamefont {C.}~\bibnamefont
  {Fu}}, \bibinfo {author} {\bibfnamefont {P.}~\bibnamefont {Wu}}, \ and\
  \bibinfo {author} {\bibfnamefont {H.}~\bibnamefont {Yu}},\ }\href {\doibase
  10.1103/PhysRevD.101.023529} {\bibfield  {journal} {\bibinfo  {journal}
  {Phys. Rev. D}\ }\textbf {\bibinfo {volume} {101}},\ \bibinfo {pages}
  {023529} (\bibinfo {year} {2020}{\natexlab{b}})},\ \Eprint
  {http://arxiv.org/abs/1912.05927} {arXiv:1912.05927 [astro-ph.CO]}
  \BibitemShut {NoStop}%
\bibitem [{\citenamefont {Aldabergenov}\ \emph {et~al.}(2020)\citenamefont
  {Aldabergenov}, \citenamefont {Addazi},\ and\ \citenamefont
  {Ketov}}]{Aldabergenov:2020bpt}%
  \BibitemOpen
  \bibfield  {author} {\bibinfo {author} {\bibfnamefont {Y.}~\bibnamefont
  {Aldabergenov}}, \bibinfo {author} {\bibfnamefont {A.}~\bibnamefont
  {Addazi}}, \ and\ \bibinfo {author} {\bibfnamefont {S.~V.}\ \bibnamefont
  {Ketov}},\ }\href {\doibase 10.1140/epjc/s10052-020-08506-6} {\bibfield
  {journal} {\bibinfo  {journal} {Eur. Phys. J. C}\ }\textbf {\bibinfo {volume}
  {80}},\ \bibinfo {pages} {917} (\bibinfo {year} {2020})},\ \Eprint
  {http://arxiv.org/abs/2006.16641} {arXiv:2006.16641 [hep-th]} \BibitemShut
  {NoStop}%
\bibitem [{\citenamefont {Aldabergenov}\ \emph {et~al.}(2021)\citenamefont
  {Aldabergenov}, \citenamefont {Addazi},\ and\ \citenamefont
  {Ketov}}]{Aldabergenov:2020yok}%
  \BibitemOpen
  \bibfield  {author} {\bibinfo {author} {\bibfnamefont {Y.}~\bibnamefont
  {Aldabergenov}}, \bibinfo {author} {\bibfnamefont {A.}~\bibnamefont
  {Addazi}}, \ and\ \bibinfo {author} {\bibfnamefont {S.~V.}\ \bibnamefont
  {Ketov}},\ }\href {\doibase 10.1016/j.physletb.2021.136069} {\bibfield
  {journal} {\bibinfo  {journal} {Phys. Lett. B}\ }\textbf {\bibinfo {volume}
  {814}},\ \bibinfo {pages} {136069} (\bibinfo {year} {2021})},\ \Eprint
  {http://arxiv.org/abs/2008.10476} {arXiv:2008.10476 [hep-th]} \BibitemShut
  {NoStop}%
\bibitem [{\citenamefont {Yi}\ \emph {et~al.}(2021)\citenamefont {Yi},
  \citenamefont {Gao}, \citenamefont {Gong},\ and\ \citenamefont
  {Zhu}}]{Yi:2020cut}%
  \BibitemOpen
  \bibfield  {author} {\bibinfo {author} {\bibfnamefont {Z.}~\bibnamefont
  {Yi}}, \bibinfo {author} {\bibfnamefont {Q.}~\bibnamefont {Gao}}, \bibinfo
  {author} {\bibfnamefont {Y.}~\bibnamefont {Gong}}, \ and\ \bibinfo {author}
  {\bibfnamefont {Z.-h.}\ \bibnamefont {Zhu}},\ }\href {\doibase
  10.1103/PhysRevD.103.063534} {\bibfield  {journal} {\bibinfo  {journal}
  {Phys. Rev. D}\ }\textbf {\bibinfo {volume} {103}},\ \bibinfo {pages}
  {063534} (\bibinfo {year} {2021})},\ \Eprint
  {http://arxiv.org/abs/2011.10606} {arXiv:2011.10606 [astro-ph.CO]}
  \BibitemShut {NoStop}%
\bibitem [{\citenamefont {Gao}\ \emph {et~al.}(2020)\citenamefont {Gao},
  \citenamefont {Gong},\ and\ \citenamefont {Yi}}]{Gao:2020tsa}%
  \BibitemOpen
  \bibfield  {author} {\bibinfo {author} {\bibfnamefont {Q.}~\bibnamefont
  {Gao}}, \bibinfo {author} {\bibfnamefont {Y.}~\bibnamefont {Gong}}, \ and\
  \bibinfo {author} {\bibfnamefont {Z.}~\bibnamefont {Yi}},\ }\href@noop {} {\
  (\bibinfo {year} {2020})},\ \Eprint {http://arxiv.org/abs/2012.03856}
  {arXiv:2012.03856 [gr-qc]} \BibitemShut {NoStop}%
\bibitem [{\citenamefont {Kawasaki}\ \emph {et~al.}(2013)\citenamefont
  {Kawasaki}, \citenamefont {Kitajima},\ and\ \citenamefont
  {Yanagida}}]{Kawasaki:2012wr}%
  \BibitemOpen
  \bibfield  {author} {\bibinfo {author} {\bibfnamefont {M.}~\bibnamefont
  {Kawasaki}}, \bibinfo {author} {\bibfnamefont {N.}~\bibnamefont {Kitajima}},
  \ and\ \bibinfo {author} {\bibfnamefont {T.~T.}\ \bibnamefont {Yanagida}},\
  }\href {\doibase 10.1103/PhysRevD.87.063519} {\bibfield  {journal} {\bibinfo
  {journal} {Phys. Rev. D}\ }\textbf {\bibinfo {volume} {87}},\ \bibinfo
  {pages} {063519} (\bibinfo {year} {2013})},\ \Eprint
  {http://arxiv.org/abs/1207.2550} {arXiv:1207.2550 [hep-ph]} \BibitemShut
  {NoStop}%
\bibitem [{\citenamefont {Kohri}\ \emph {et~al.}(2013)\citenamefont {Kohri},
  \citenamefont {Lin},\ and\ \citenamefont {Matsuda}}]{Kohri:2012yw}%
  \BibitemOpen
  \bibfield  {author} {\bibinfo {author} {\bibfnamefont {K.}~\bibnamefont
  {Kohri}}, \bibinfo {author} {\bibfnamefont {C.-M.}\ \bibnamefont {Lin}}, \
  and\ \bibinfo {author} {\bibfnamefont {T.}~\bibnamefont {Matsuda}},\ }\href
  {\doibase 10.1103/PhysRevD.87.103527} {\bibfield  {journal} {\bibinfo
  {journal} {Phys. Rev. D}\ }\textbf {\bibinfo {volume} {87}},\ \bibinfo
  {pages} {103527} (\bibinfo {year} {2013})},\ \Eprint
  {http://arxiv.org/abs/1211.2371} {arXiv:1211.2371 [hep-ph]} \BibitemShut
  {NoStop}%
\bibitem [{\citenamefont {Ando}\ \emph
  {et~al.}(2018{\natexlab{a}})\citenamefont {Ando}, \citenamefont {Inomata},
  \citenamefont {Kawasaki}, \citenamefont {Mukaida},\ and\ \citenamefont
  {Yanagida}}]{Ando:2017veq}%
  \BibitemOpen
  \bibfield  {author} {\bibinfo {author} {\bibfnamefont {K.}~\bibnamefont
  {Ando}}, \bibinfo {author} {\bibfnamefont {K.}~\bibnamefont {Inomata}},
  \bibinfo {author} {\bibfnamefont {M.}~\bibnamefont {Kawasaki}}, \bibinfo
  {author} {\bibfnamefont {K.}~\bibnamefont {Mukaida}}, \ and\ \bibinfo
  {author} {\bibfnamefont {T.~T.}\ \bibnamefont {Yanagida}},\ }\href {\doibase
  10.1103/PhysRevD.97.123512} {\bibfield  {journal} {\bibinfo  {journal} {Phys.
  Rev. D}\ }\textbf {\bibinfo {volume} {97}},\ \bibinfo {pages} {123512}
  (\bibinfo {year} {2018}{\natexlab{a}})},\ \Eprint
  {http://arxiv.org/abs/1711.08956} {arXiv:1711.08956 [astro-ph.CO]}
  \BibitemShut {NoStop}%
\bibitem [{\citenamefont {Ando}\ \emph
  {et~al.}(2018{\natexlab{b}})\citenamefont {Ando}, \citenamefont {Kawasaki},\
  and\ \citenamefont {Nakatsuka}}]{Ando:2018nge}%
  \BibitemOpen
  \bibfield  {author} {\bibinfo {author} {\bibfnamefont {K.}~\bibnamefont
  {Ando}}, \bibinfo {author} {\bibfnamefont {M.}~\bibnamefont {Kawasaki}}, \
  and\ \bibinfo {author} {\bibfnamefont {H.}~\bibnamefont {Nakatsuka}},\ }\href
  {\doibase 10.1103/PhysRevD.98.083508} {\bibfield  {journal} {\bibinfo
  {journal} {Phys. Rev. D}\ }\textbf {\bibinfo {volume} {98}},\ \bibinfo
  {pages} {083508} (\bibinfo {year} {2018}{\natexlab{b}})},\ \Eprint
  {http://arxiv.org/abs/1805.07757} {arXiv:1805.07757 [astro-ph.CO]}
  \BibitemShut {NoStop}%
\bibitem [{\citenamefont {Chen}\ and\ \citenamefont
  {Cai}(2019)}]{Chen:2019zza}%
  \BibitemOpen
  \bibfield  {author} {\bibinfo {author} {\bibfnamefont {C.}~\bibnamefont
  {Chen}}\ and\ \bibinfo {author} {\bibfnamefont {Y.-F.}\ \bibnamefont {Cai}},\
  }\href {\doibase 10.1088/1475-7516/2019/10/068} {\bibfield  {journal}
  {\bibinfo  {journal} {JCAP}\ }\textbf {\bibinfo {volume} {10}},\ \bibinfo
  {pages} {068} (\bibinfo {year} {2019})},\ \Eprint
  {http://arxiv.org/abs/1908.03942} {arXiv:1908.03942 [astro-ph.CO]}
  \BibitemShut {NoStop}%
\bibitem [{\citenamefont {Cai}\ \emph {et~al.}(2018)\citenamefont {Cai},
  \citenamefont {Tong}, \citenamefont {Wang},\ and\ \citenamefont
  {Yan}}]{Cai:2018tuh}%
  \BibitemOpen
  \bibfield  {author} {\bibinfo {author} {\bibfnamefont {Y.-F.}\ \bibnamefont
  {Cai}}, \bibinfo {author} {\bibfnamefont {X.}~\bibnamefont {Tong}}, \bibinfo
  {author} {\bibfnamefont {D.-G.}\ \bibnamefont {Wang}}, \ and\ \bibinfo
  {author} {\bibfnamefont {S.-F.}\ \bibnamefont {Yan}},\ }\href {\doibase
  10.1103/PhysRevLett.121.081306} {\bibfield  {journal} {\bibinfo  {journal}
  {Phys. Rev. Lett.}\ }\textbf {\bibinfo {volume} {121}},\ \bibinfo {pages}
  {081306} (\bibinfo {year} {2018})},\ \Eprint
  {http://arxiv.org/abs/1805.03639} {arXiv:1805.03639 [astro-ph.CO]}
  \BibitemShut {NoStop}%
\bibitem [{\citenamefont {Cai}\ \emph {et~al.}(2019{\natexlab{a}})\citenamefont
  {Cai}, \citenamefont {Chen}, \citenamefont {Tong}, \citenamefont {Wang},\
  and\ \citenamefont {Yan}}]{Cai:2019jah}%
  \BibitemOpen
  \bibfield  {author} {\bibinfo {author} {\bibfnamefont {Y.-F.}\ \bibnamefont
  {Cai}}, \bibinfo {author} {\bibfnamefont {C.}~\bibnamefont {Chen}}, \bibinfo
  {author} {\bibfnamefont {X.}~\bibnamefont {Tong}}, \bibinfo {author}
  {\bibfnamefont {D.-G.}\ \bibnamefont {Wang}}, \ and\ \bibinfo {author}
  {\bibfnamefont {S.-F.}\ \bibnamefont {Yan}},\ }\href {\doibase
  10.1103/PhysRevD.100.043518} {\bibfield  {journal} {\bibinfo  {journal}
  {Phys. Rev. D}\ }\textbf {\bibinfo {volume} {100}},\ \bibinfo {pages}
  {043518} (\bibinfo {year} {2019}{\natexlab{a}})},\ \Eprint
  {http://arxiv.org/abs/1902.08187} {arXiv:1902.08187 [astro-ph.CO]}
  \BibitemShut {NoStop}%
\bibitem [{\citenamefont {Cai}\ \emph {et~al.}(2020{\natexlab{a}})\citenamefont
  {Cai}, \citenamefont {Guo}, \citenamefont {Liu}, \citenamefont {Liu},\ and\
  \citenamefont {Yang}}]{Cai:2019bmk}%
  \BibitemOpen
  \bibfield  {author} {\bibinfo {author} {\bibfnamefont {R.-G.}\ \bibnamefont
  {Cai}}, \bibinfo {author} {\bibfnamefont {Z.-K.}\ \bibnamefont {Guo}},
  \bibinfo {author} {\bibfnamefont {J.}~\bibnamefont {Liu}}, \bibinfo {author}
  {\bibfnamefont {L.}~\bibnamefont {Liu}}, \ and\ \bibinfo {author}
  {\bibfnamefont {X.-Y.}\ \bibnamefont {Yang}},\ }\href {\doibase
  10.1088/1475-7516/2020/06/013} {\bibfield  {journal} {\bibinfo  {journal}
  {JCAP}\ }\textbf {\bibinfo {volume} {06}},\ \bibinfo {pages} {013} (\bibinfo
  {year} {2020}{\natexlab{a}})},\ \Eprint {http://arxiv.org/abs/1912.10437}
  {arXiv:1912.10437 [astro-ph.CO]} \BibitemShut {NoStop}%
\bibitem [{\citenamefont {Chen}\ \emph
  {et~al.}(2020{\natexlab{a}})\citenamefont {Chen}, \citenamefont {Ma},\ and\
  \citenamefont {Cai}}]{Chen:2020uhe}%
  \BibitemOpen
  \bibfield  {author} {\bibinfo {author} {\bibfnamefont {C.}~\bibnamefont
  {Chen}}, \bibinfo {author} {\bibfnamefont {X.-H.}\ \bibnamefont {Ma}}, \ and\
  \bibinfo {author} {\bibfnamefont {Y.-F.}\ \bibnamefont {Cai}},\ }\href
  {\doibase 10.1103/PhysRevD.102.063526} {\bibfield  {journal} {\bibinfo
  {journal} {Phys. Rev. D}\ }\textbf {\bibinfo {volume} {102}},\ \bibinfo
  {pages} {063526} (\bibinfo {year} {2020}{\natexlab{a}})},\ \Eprint
  {http://arxiv.org/abs/2003.03821} {arXiv:2003.03821 [astro-ph.CO]}
  \BibitemShut {NoStop}%
\bibitem [{\citenamefont {Cai}\ \emph {et~al.}(2020{\natexlab{b}})\citenamefont
  {Cai}, \citenamefont {Lin}, \citenamefont {Wang},\ and\ \citenamefont
  {Yan}}]{Cai:2020ovp}%
  \BibitemOpen
  \bibfield  {author} {\bibinfo {author} {\bibfnamefont {Y.-F.}\ \bibnamefont
  {Cai}}, \bibinfo {author} {\bibfnamefont {C.}~\bibnamefont {Lin}}, \bibinfo
  {author} {\bibfnamefont {B.}~\bibnamefont {Wang}}, \ and\ \bibinfo {author}
  {\bibfnamefont {S.-F.}\ \bibnamefont {Yan}},\ }\href@noop {} {\  (\bibinfo
  {year} {2020}{\natexlab{b}})},\ \Eprint {http://arxiv.org/abs/2009.09833}
  {arXiv:2009.09833 [gr-qc]} \BibitemShut {NoStop}%
\bibitem [{\citenamefont {Zhou}\ \emph {et~al.}(2020)\citenamefont {Zhou},
  \citenamefont {Jiang}, \citenamefont {Cai}, \citenamefont {Sasaki},\ and\
  \citenamefont {Pi}}]{Zhou:2020kkf}%
  \BibitemOpen
  \bibfield  {author} {\bibinfo {author} {\bibfnamefont {Z.}~\bibnamefont
  {Zhou}}, \bibinfo {author} {\bibfnamefont {J.}~\bibnamefont {Jiang}},
  \bibinfo {author} {\bibfnamefont {Y.-F.}\ \bibnamefont {Cai}}, \bibinfo
  {author} {\bibfnamefont {M.}~\bibnamefont {Sasaki}}, \ and\ \bibinfo {author}
  {\bibfnamefont {S.}~\bibnamefont {Pi}},\ }\href {\doibase
  10.1103/PhysRevD.102.103527} {\bibfield  {journal} {\bibinfo  {journal}
  {Phys. Rev. D}\ }\textbf {\bibinfo {volume} {102}},\ \bibinfo {pages}
  {103527} (\bibinfo {year} {2020})},\ \Eprint
  {http://arxiv.org/abs/2010.03537} {arXiv:2010.03537 [astro-ph.CO]}
  \BibitemShut {NoStop}%
\bibitem [{\citenamefont {Cai}\ \emph {et~al.}(2021)\citenamefont {Cai},
  \citenamefont {Jiang}, \citenamefont {Sasaki}, \citenamefont {Vardanyan},\
  and\ \citenamefont {Zhou}}]{Cai:2021yvq}%
  \BibitemOpen
  \bibfield  {author} {\bibinfo {author} {\bibfnamefont {Y.-F.}\ \bibnamefont
  {Cai}}, \bibinfo {author} {\bibfnamefont {J.}~\bibnamefont {Jiang}}, \bibinfo
  {author} {\bibfnamefont {M.}~\bibnamefont {Sasaki}}, \bibinfo {author}
  {\bibfnamefont {V.}~\bibnamefont {Vardanyan}}, \ and\ \bibinfo {author}
  {\bibfnamefont {Z.}~\bibnamefont {Zhou}},\ }\href {\doibase
  10.1103/PhysRevLett.127.251301} {\bibfield  {journal} {\bibinfo  {journal}
  {Phys. Rev. Lett.}\ }\textbf {\bibinfo {volume} {127}},\ \bibinfo {pages}
  {251301} (\bibinfo {year} {2021})},\ \Eprint
  {http://arxiv.org/abs/2105.12554} {arXiv:2105.12554 [astro-ph.CO]}
  \BibitemShut {NoStop}%
\bibitem [{\citenamefont {Peng}\ \emph {et~al.}(2021)\citenamefont {Peng},
  \citenamefont {Fu}, \citenamefont {Liu}, \citenamefont {Guo},\ and\
  \citenamefont {Cai}}]{Peng:2021zon}%
  \BibitemOpen
  \bibfield  {author} {\bibinfo {author} {\bibfnamefont {Z.-Z.}\ \bibnamefont
  {Peng}}, \bibinfo {author} {\bibfnamefont {C.}~\bibnamefont {Fu}}, \bibinfo
  {author} {\bibfnamefont {J.}~\bibnamefont {Liu}}, \bibinfo {author}
  {\bibfnamefont {Z.-K.}\ \bibnamefont {Guo}}, \ and\ \bibinfo {author}
  {\bibfnamefont {R.-G.}\ \bibnamefont {Cai}},\ }\href {\doibase
  10.1088/1475-7516/2021/10/050} {\bibfield  {journal} {\bibinfo  {journal}
  {JCAP}\ }\textbf {\bibinfo {volume} {10}},\ \bibinfo {pages} {050} (\bibinfo
  {year} {2021})},\ \Eprint {http://arxiv.org/abs/2106.11816} {arXiv:2106.11816
  [astro-ph.CO]} \BibitemShut {NoStop}%
\bibitem [{\citenamefont {Xie}\ \emph {et~al.}(2024)\citenamefont {Xie},
  \citenamefont {Zhang}, \citenamefont {Jiang}, \citenamefont {Li},
  \citenamefont {Wang},\ and\ \citenamefont {Cai}}]{Xie:2024cwp}%
  \BibitemOpen
  \bibfield  {author} {\bibinfo {author} {\bibfnamefont {T.}~\bibnamefont
  {Xie}}, \bibinfo {author} {\bibfnamefont {D.}~\bibnamefont {Zhang}}, \bibinfo
  {author} {\bibfnamefont {J.}~\bibnamefont {Jiang}}, \bibinfo {author}
  {\bibfnamefont {J.-R.}\ \bibnamefont {Li}}, \bibinfo {author} {\bibfnamefont
  {B.}~\bibnamefont {Wang}}, \ and\ \bibinfo {author} {\bibfnamefont {Y.-F.}\
  \bibnamefont {Cai}},\ }\href {\doibase 10.1103/PhysRevD.109.083529}
  {\bibfield  {journal} {\bibinfo  {journal} {Phys. Rev. D}\ }\textbf {\bibinfo
  {volume} {109}},\ \bibinfo {pages} {083529} (\bibinfo {year} {2024})},\
  \Eprint {http://arxiv.org/abs/2402.02415} {arXiv:2402.02415 [astro-ph.CO]}
  \BibitemShut {NoStop}%
\bibitem [{\citenamefont {Hawking}\ \emph {et~al.}(1982)\citenamefont
  {Hawking}, \citenamefont {Moss},\ and\ \citenamefont
  {Stewart}}]{Hawking:1982ga}%
  \BibitemOpen
  \bibfield  {author} {\bibinfo {author} {\bibfnamefont {S.~W.}\ \bibnamefont
  {Hawking}}, \bibinfo {author} {\bibfnamefont {I.~G.}\ \bibnamefont {Moss}}, \
  and\ \bibinfo {author} {\bibfnamefont {J.~M.}\ \bibnamefont {Stewart}},\
  }\href {\doibase 10.1103/PhysRevD.26.2681} {\bibfield  {journal} {\bibinfo
  {journal} {Phys. Rev. D}\ }\textbf {\bibinfo {volume} {26}},\ \bibinfo
  {pages} {2681} (\bibinfo {year} {1982})}\BibitemShut {NoStop}%
\bibitem [{\citenamefont {Crawford}\ and\ \citenamefont
  {Schramm}(1982)}]{Crawford:1982yz}%
  \BibitemOpen
  \bibfield  {author} {\bibinfo {author} {\bibfnamefont {M.}~\bibnamefont
  {Crawford}}\ and\ \bibinfo {author} {\bibfnamefont {D.~N.}\ \bibnamefont
  {Schramm}},\ }\href {\doibase 10.1038/298538a0} {\bibfield  {journal}
  {\bibinfo  {journal} {Nature}\ }\textbf {\bibinfo {volume} {298}},\ \bibinfo
  {pages} {538} (\bibinfo {year} {1982})}\BibitemShut {NoStop}%
\bibitem [{\citenamefont {Liu}\ \emph {et~al.}(2023{\natexlab{a}})\citenamefont
  {Liu}, \citenamefont {Bian}, \citenamefont {Cai}, \citenamefont {Guo},\ and\
  \citenamefont {Wang}}]{Liu:2022lvz}%
  \BibitemOpen
  \bibfield  {author} {\bibinfo {author} {\bibfnamefont {J.}~\bibnamefont
  {Liu}}, \bibinfo {author} {\bibfnamefont {L.}~\bibnamefont {Bian}}, \bibinfo
  {author} {\bibfnamefont {R.-G.}\ \bibnamefont {Cai}}, \bibinfo {author}
  {\bibfnamefont {Z.-K.}\ \bibnamefont {Guo}}, \ and\ \bibinfo {author}
  {\bibfnamefont {S.-J.}\ \bibnamefont {Wang}},\ }\href {\doibase
  10.1103/PhysRevLett.130.051001} {\bibfield  {journal} {\bibinfo  {journal}
  {Phys. Rev. Lett.}\ }\textbf {\bibinfo {volume} {130}},\ \bibinfo {pages}
  {051001} (\bibinfo {year} {2023}{\natexlab{a}})},\ \Eprint
  {http://arxiv.org/abs/2208.14086} {arXiv:2208.14086 [astro-ph.CO]}
  \BibitemShut {NoStop}%
\bibitem [{\citenamefont {Kusenko}\ \emph {et~al.}(2020)\citenamefont
  {Kusenko}, \citenamefont {Sasaki}, \citenamefont {Sugiyama}, \citenamefont
  {Takada}, \citenamefont {Takhistov},\ and\ \citenamefont
  {Vitagliano}}]{Kusenko:2020pcg}%
  \BibitemOpen
  \bibfield  {author} {\bibinfo {author} {\bibfnamefont {A.}~\bibnamefont
  {Kusenko}}, \bibinfo {author} {\bibfnamefont {M.}~\bibnamefont {Sasaki}},
  \bibinfo {author} {\bibfnamefont {S.}~\bibnamefont {Sugiyama}}, \bibinfo
  {author} {\bibfnamefont {M.}~\bibnamefont {Takada}}, \bibinfo {author}
  {\bibfnamefont {V.}~\bibnamefont {Takhistov}}, \ and\ \bibinfo {author}
  {\bibfnamefont {E.}~\bibnamefont {Vitagliano}},\ }\href {\doibase
  10.1103/PhysRevLett.125.181304} {\bibfield  {journal} {\bibinfo  {journal}
  {Phys. Rev. Lett.}\ }\textbf {\bibinfo {volume} {125}},\ \bibinfo {pages}
  {18} (\bibinfo {year} {2020})},\ \Eprint {http://arxiv.org/abs/2001.09160}
  {arXiv:2001.09160 [astro-ph.CO]} \BibitemShut {NoStop}%
\bibitem [{\citenamefont {Kodama}(1980)}]{Kodama:1979vn}%
  \BibitemOpen
  \bibfield  {author} {\bibinfo {author} {\bibfnamefont {H.}~\bibnamefont
  {Kodama}},\ }\href {\doibase 10.1143/PTP.63.1217} {\bibfield  {journal}
  {\bibinfo  {journal} {Prog. Theor. Phys.}\ }\textbf {\bibinfo {volume}
  {63}},\ \bibinfo {pages} {1217} (\bibinfo {year} {1980})}\BibitemShut
  {NoStop}%
\bibitem [{\citenamefont {Shibata}\ and\ \citenamefont
  {Sasaki}(1999)}]{Shibata:1999zs}%
  \BibitemOpen
  \bibfield  {author} {\bibinfo {author} {\bibfnamefont {M.}~\bibnamefont
  {Shibata}}\ and\ \bibinfo {author} {\bibfnamefont {M.}~\bibnamefont
  {Sasaki}},\ }\href {\doibase 10.1103/PhysRevD.60.084002} {\bibfield
  {journal} {\bibinfo  {journal} {Phys. Rev. D}\ }\textbf {\bibinfo {volume}
  {60}},\ \bibinfo {pages} {084002} (\bibinfo {year} {1999})},\ \Eprint
  {http://arxiv.org/abs/gr-qc/9905064} {arXiv:gr-qc/9905064} \BibitemShut
  {NoStop}%
\bibitem [{\citenamefont {Harada}\ \emph {et~al.}(2015)\citenamefont {Harada},
  \citenamefont {Yoo}, \citenamefont {Nakama},\ and\ \citenamefont
  {Koga}}]{Harada:2015yda}%
  \BibitemOpen
  \bibfield  {author} {\bibinfo {author} {\bibfnamefont {T.}~\bibnamefont
  {Harada}}, \bibinfo {author} {\bibfnamefont {C.-M.}\ \bibnamefont {Yoo}},
  \bibinfo {author} {\bibfnamefont {T.}~\bibnamefont {Nakama}}, \ and\ \bibinfo
  {author} {\bibfnamefont {Y.}~\bibnamefont {Koga}},\ }\href {\doibase
  10.1103/PhysRevD.91.084057} {\bibfield  {journal} {\bibinfo  {journal} {Phys.
  Rev. D}\ }\textbf {\bibinfo {volume} {91}},\ \bibinfo {pages} {084057}
  (\bibinfo {year} {2015})},\ \Eprint {http://arxiv.org/abs/1503.03934}
  {arXiv:1503.03934 [gr-qc]} \BibitemShut {NoStop}%
\bibitem [{\citenamefont {Kawasaki}\ and\ \citenamefont
  {Nakatsuka}(2019)}]{Kawasaki:2019mbl}%
  \BibitemOpen
  \bibfield  {author} {\bibinfo {author} {\bibfnamefont {M.}~\bibnamefont
  {Kawasaki}}\ and\ \bibinfo {author} {\bibfnamefont {H.}~\bibnamefont
  {Nakatsuka}},\ }\href {\doibase 10.1103/PhysRevD.99.123501} {\bibfield
  {journal} {\bibinfo  {journal} {Phys. Rev. D}\ }\textbf {\bibinfo {volume}
  {99}},\ \bibinfo {pages} {123501} (\bibinfo {year} {2019})},\ \Eprint
  {http://arxiv.org/abs/1903.02994} {arXiv:1903.02994 [astro-ph.CO]}
  \BibitemShut {NoStop}%
\bibitem [{\citenamefont {Young}\ \emph {et~al.}(2019)\citenamefont {Young},
  \citenamefont {Musco},\ and\ \citenamefont {Byrnes}}]{Young:2019yug}%
  \BibitemOpen
  \bibfield  {author} {\bibinfo {author} {\bibfnamefont {S.}~\bibnamefont
  {Young}}, \bibinfo {author} {\bibfnamefont {I.}~\bibnamefont {Musco}}, \ and\
  \bibinfo {author} {\bibfnamefont {C.~T.}\ \bibnamefont {Byrnes}},\ }\href
  {\doibase 10.1088/1475-7516/2019/11/012} {\bibfield  {journal} {\bibinfo
  {journal} {JCAP}\ }\textbf {\bibinfo {volume} {11}},\ \bibinfo {pages} {012}
  (\bibinfo {year} {2019})},\ \Eprint {http://arxiv.org/abs/1904.00984}
  {arXiv:1904.00984 [astro-ph.CO]} \BibitemShut {NoStop}%
\bibitem [{\citenamefont {De~Luca}\ \emph {et~al.}(2019)\citenamefont
  {De~Luca}, \citenamefont {Franciolini}, \citenamefont {Kehagias},
  \citenamefont {Peloso}, \citenamefont {Riotto},\ and\ \citenamefont
  {\"Unal}}]{DeLuca:2019qsy}%
  \BibitemOpen
  \bibfield  {author} {\bibinfo {author} {\bibfnamefont {V.}~\bibnamefont
  {De~Luca}}, \bibinfo {author} {\bibfnamefont {G.}~\bibnamefont
  {Franciolini}}, \bibinfo {author} {\bibfnamefont {A.}~\bibnamefont
  {Kehagias}}, \bibinfo {author} {\bibfnamefont {M.}~\bibnamefont {Peloso}},
  \bibinfo {author} {\bibfnamefont {A.}~\bibnamefont {Riotto}}, \ and\ \bibinfo
  {author} {\bibfnamefont {C.}~\bibnamefont {\"Unal}},\ }\href {\doibase
  10.1088/1475-7516/2019/07/048} {\bibfield  {journal} {\bibinfo  {journal}
  {JCAP}\ }\textbf {\bibinfo {volume} {07}},\ \bibinfo {pages} {048} (\bibinfo
  {year} {2019})},\ \Eprint {http://arxiv.org/abs/1904.00970} {arXiv:1904.00970
  [astro-ph.CO]} \BibitemShut {NoStop}%
\bibitem [{\citenamefont {Musco}(2019)}]{Musco:2018rwt}%
  \BibitemOpen
  \bibfield  {author} {\bibinfo {author} {\bibfnamefont {I.}~\bibnamefont
  {Musco}},\ }\href {\doibase 10.1103/PhysRevD.100.123524} {\bibfield
  {journal} {\bibinfo  {journal} {Phys. Rev. D}\ }\textbf {\bibinfo {volume}
  {100}},\ \bibinfo {pages} {123524} (\bibinfo {year} {2019})},\ \Eprint
  {http://arxiv.org/abs/1809.02127} {arXiv:1809.02127 [gr-qc]} \BibitemShut
  {NoStop}%
\bibitem [{\citenamefont {Escriv\`a}\ \emph {et~al.}(2020)\citenamefont
  {Escriv\`a}, \citenamefont {Germani},\ and\ \citenamefont
  {Sheth}}]{Escriva:2019phb}%
  \BibitemOpen
  \bibfield  {author} {\bibinfo {author} {\bibfnamefont {A.}~\bibnamefont
  {Escriv\`a}}, \bibinfo {author} {\bibfnamefont {C.}~\bibnamefont {Germani}},
  \ and\ \bibinfo {author} {\bibfnamefont {R.~K.}\ \bibnamefont {Sheth}},\
  }\href {\doibase 10.1103/PhysRevD.101.044022} {\bibfield  {journal} {\bibinfo
   {journal} {Phys. Rev. D}\ }\textbf {\bibinfo {volume} {101}},\ \bibinfo
  {pages} {044022} (\bibinfo {year} {2020})},\ \Eprint
  {http://arxiv.org/abs/1907.13311} {arXiv:1907.13311 [gr-qc]} \BibitemShut
  {NoStop}%
\bibitem [{\citenamefont {Young}(2019)}]{Young:2019osy}%
  \BibitemOpen
  \bibfield  {author} {\bibinfo {author} {\bibfnamefont {S.}~\bibnamefont
  {Young}},\ }\href {\doibase 10.1142/S0218271820300025} {\bibfield  {journal}
  {\bibinfo  {journal} {Int. J. Mod. Phys. D}\ }\textbf {\bibinfo {volume}
  {29}},\ \bibinfo {pages} {2030002} (\bibinfo {year} {2019})},\ \Eprint
  {http://arxiv.org/abs/1905.01230} {arXiv:1905.01230 [astro-ph.CO]}
  \BibitemShut {NoStop}%
\bibitem [{\citenamefont {Yoo}\ \emph {et~al.}(2018)\citenamefont {Yoo},
  \citenamefont {Harada}, \citenamefont {Garriga},\ and\ \citenamefont
  {Kohri}}]{Yoo:2018kvb}%
  \BibitemOpen
  \bibfield  {author} {\bibinfo {author} {\bibfnamefont {C.-M.}\ \bibnamefont
  {Yoo}}, \bibinfo {author} {\bibfnamefont {T.}~\bibnamefont {Harada}},
  \bibinfo {author} {\bibfnamefont {J.}~\bibnamefont {Garriga}}, \ and\
  \bibinfo {author} {\bibfnamefont {K.}~\bibnamefont {Kohri}},\ }\href
  {\doibase 10.1093/ptep/pty120} {\bibfield  {journal} {\bibinfo  {journal}
  {PTEP}\ }\textbf {\bibinfo {volume} {2018}},\ \bibinfo {pages} {123E01}
  (\bibinfo {year} {2018})},\ \Eprint {http://arxiv.org/abs/1805.03946}
  {arXiv:1805.03946 [astro-ph.CO]} \BibitemShut {NoStop}%
\bibitem [{\citenamefont {Atal}\ \emph {et~al.}(2019)\citenamefont {Atal},
  \citenamefont {Garriga},\ and\ \citenamefont
  {Marcos-Caballero}}]{Atal:2019cdz}%
  \BibitemOpen
  \bibfield  {author} {\bibinfo {author} {\bibfnamefont {V.}~\bibnamefont
  {Atal}}, \bibinfo {author} {\bibfnamefont {J.}~\bibnamefont {Garriga}}, \
  and\ \bibinfo {author} {\bibfnamefont {A.}~\bibnamefont {Marcos-Caballero}},\
  }\href {\doibase 10.1088/1475-7516/2019/09/073} {\bibfield  {journal}
  {\bibinfo  {journal} {JCAP}\ }\textbf {\bibinfo {volume} {09}},\ \bibinfo
  {pages} {073} (\bibinfo {year} {2019})},\ \Eprint
  {http://arxiv.org/abs/1905.13202} {arXiv:1905.13202 [astro-ph.CO]}
  \BibitemShut {NoStop}%
\bibitem [{\citenamefont {Yoo}\ \emph {et~al.}(2021)\citenamefont {Yoo},
  \citenamefont {Harada}, \citenamefont {Hirano},\ and\ \citenamefont
  {Kohri}}]{Yoo:2020dkz}%
  \BibitemOpen
  \bibfield  {author} {\bibinfo {author} {\bibfnamefont {C.-M.}\ \bibnamefont
  {Yoo}}, \bibinfo {author} {\bibfnamefont {T.}~\bibnamefont {Harada}},
  \bibinfo {author} {\bibfnamefont {S.}~\bibnamefont {Hirano}}, \ and\ \bibinfo
  {author} {\bibfnamefont {K.}~\bibnamefont {Kohri}},\ }\href {\doibase
  10.1093/ptep/ptaa155} {\bibfield  {journal} {\bibinfo  {journal} {PTEP}\
  }\textbf {\bibinfo {volume} {2021}},\ \bibinfo {pages} {013E02} (\bibinfo
  {year} {2021})},\ \Eprint {http://arxiv.org/abs/2008.02425} {arXiv:2008.02425
  [astro-ph.CO]} \BibitemShut {NoStop}%
\bibitem [{\citenamefont {Kitajima}\ \emph {et~al.}(2021)\citenamefont
  {Kitajima}, \citenamefont {Tada}, \citenamefont {Yokoyama},\ and\
  \citenamefont {Yoo}}]{Kitajima:2021fpq}%
  \BibitemOpen
  \bibfield  {author} {\bibinfo {author} {\bibfnamefont {N.}~\bibnamefont
  {Kitajima}}, \bibinfo {author} {\bibfnamefont {Y.}~\bibnamefont {Tada}},
  \bibinfo {author} {\bibfnamefont {S.}~\bibnamefont {Yokoyama}}, \ and\
  \bibinfo {author} {\bibfnamefont {C.-M.}\ \bibnamefont {Yoo}},\ }\href
  {\doibase 10.1088/1475-7516/2021/10/053} {\bibfield  {journal} {\bibinfo
  {journal} {JCAP}\ }\textbf {\bibinfo {volume} {10}},\ \bibinfo {pages} {053}
  (\bibinfo {year} {2021})},\ \Eprint {http://arxiv.org/abs/2109.00791}
  {arXiv:2109.00791 [astro-ph.CO]} \BibitemShut {NoStop}%
\bibitem [{\citenamefont {Pi}(2024)}]{Pi:2024jwt}%
  \BibitemOpen
  \bibfield  {author} {\bibinfo {author} {\bibfnamefont {S.}~\bibnamefont
  {Pi}},\ }\href@noop {} {\  (\bibinfo {year} {2024})},\ \Eprint
  {http://arxiv.org/abs/2404.06151} {arXiv:2404.06151 [astro-ph.CO]}
  \BibitemShut {NoStop}%
\bibitem [{\citenamefont {Choptuik}(1993)}]{Choptuik:1992jv}%
  \BibitemOpen
  \bibfield  {author} {\bibinfo {author} {\bibfnamefont {M.~W.}\ \bibnamefont
  {Choptuik}},\ }\href {\doibase 10.1103/PhysRevLett.70.9} {\bibfield
  {journal} {\bibinfo  {journal} {Phys. Rev. Lett.}\ }\textbf {\bibinfo
  {volume} {70}},\ \bibinfo {pages} {9} (\bibinfo {year} {1993})}\BibitemShut
  {NoStop}%
\bibitem [{\citenamefont {Evans}\ and\ \citenamefont
  {Coleman}(1994)}]{Evans:1994pj}%
  \BibitemOpen
  \bibfield  {author} {\bibinfo {author} {\bibfnamefont {C.~R.}\ \bibnamefont
  {Evans}}\ and\ \bibinfo {author} {\bibfnamefont {J.~S.}\ \bibnamefont
  {Coleman}},\ }\href {\doibase 10.1103/PhysRevLett.72.1782} {\bibfield
  {journal} {\bibinfo  {journal} {Phys. Rev. Lett.}\ }\textbf {\bibinfo
  {volume} {72}},\ \bibinfo {pages} {1782} (\bibinfo {year} {1994})},\ \Eprint
  {http://arxiv.org/abs/gr-qc/9402041} {arXiv:gr-qc/9402041} \BibitemShut
  {NoStop}%
\bibitem [{\citenamefont {Koike}\ \emph {et~al.}(1995)\citenamefont {Koike},
  \citenamefont {Hara},\ and\ \citenamefont {Adachi}}]{Koike:1995jm}%
  \BibitemOpen
  \bibfield  {author} {\bibinfo {author} {\bibfnamefont {T.}~\bibnamefont
  {Koike}}, \bibinfo {author} {\bibfnamefont {T.}~\bibnamefont {Hara}}, \ and\
  \bibinfo {author} {\bibfnamefont {S.}~\bibnamefont {Adachi}},\ }\href
  {\doibase 10.1103/PhysRevLett.74.5170} {\bibfield  {journal} {\bibinfo
  {journal} {Phys. Rev. Lett.}\ }\textbf {\bibinfo {volume} {74}},\ \bibinfo
  {pages} {5170} (\bibinfo {year} {1995})},\ \Eprint
  {http://arxiv.org/abs/gr-qc/9503007} {arXiv:gr-qc/9503007} \BibitemShut
  {NoStop}%
\bibitem [{\citenamefont {Niemeyer}\ and\ \citenamefont
  {Jedamzik}(1998)}]{Niemeyer:1997mt}%
  \BibitemOpen
  \bibfield  {author} {\bibinfo {author} {\bibfnamefont {J.~C.}\ \bibnamefont
  {Niemeyer}}\ and\ \bibinfo {author} {\bibfnamefont {K.}~\bibnamefont
  {Jedamzik}},\ }\href {\doibase 10.1103/PhysRevLett.80.5481} {\bibfield
  {journal} {\bibinfo  {journal} {Phys. Rev. Lett.}\ }\textbf {\bibinfo
  {volume} {80}},\ \bibinfo {pages} {5481} (\bibinfo {year} {1998})},\ \Eprint
  {http://arxiv.org/abs/astro-ph/9709072} {arXiv:astro-ph/9709072} \BibitemShut
  {NoStop}%
\bibitem [{\citenamefont {Hawke}\ and\ \citenamefont
  {Stewart}(2002)}]{Hawke:2002rf}%
  \BibitemOpen
  \bibfield  {author} {\bibinfo {author} {\bibfnamefont {I.}~\bibnamefont
  {Hawke}}\ and\ \bibinfo {author} {\bibfnamefont {J.~M.}\ \bibnamefont
  {Stewart}},\ }\href {\doibase 10.1088/0264-9381/19/14/310} {\bibfield
  {journal} {\bibinfo  {journal} {Class. Quant. Grav.}\ }\textbf {\bibinfo
  {volume} {19}},\ \bibinfo {pages} {3687} (\bibinfo {year}
  {2002})}\BibitemShut {NoStop}%
\bibitem [{\citenamefont {Musco}\ \emph {et~al.}(2009)\citenamefont {Musco},
  \citenamefont {Miller},\ and\ \citenamefont {Polnarev}}]{Musco:2008hv}%
  \BibitemOpen
  \bibfield  {author} {\bibinfo {author} {\bibfnamefont {I.}~\bibnamefont
  {Musco}}, \bibinfo {author} {\bibfnamefont {J.~C.}\ \bibnamefont {Miller}}, \
  and\ \bibinfo {author} {\bibfnamefont {A.~G.}\ \bibnamefont {Polnarev}},\
  }\href {\doibase 10.1088/0264-9381/26/23/235001} {\bibfield  {journal}
  {\bibinfo  {journal} {Class. Quant. Grav.}\ }\textbf {\bibinfo {volume}
  {26}},\ \bibinfo {pages} {235001} (\bibinfo {year} {2009})},\ \Eprint
  {http://arxiv.org/abs/0811.1452} {arXiv:0811.1452 [gr-qc]} \BibitemShut
  {NoStop}%
\bibitem [{\citenamefont {Ando}\ \emph
  {et~al.}(2018{\natexlab{c}})\citenamefont {Ando}, \citenamefont {Inomata},\
  and\ \citenamefont {Kawasaki}}]{Ando:2018qdb}%
  \BibitemOpen
  \bibfield  {author} {\bibinfo {author} {\bibfnamefont {K.}~\bibnamefont
  {Ando}}, \bibinfo {author} {\bibfnamefont {K.}~\bibnamefont {Inomata}}, \
  and\ \bibinfo {author} {\bibfnamefont {M.}~\bibnamefont {Kawasaki}},\ }\href
  {\doibase 10.1103/PhysRevD.97.103528} {\bibfield  {journal} {\bibinfo
  {journal} {Phys. Rev. D}\ }\textbf {\bibinfo {volume} {97}},\ \bibinfo
  {pages} {103528} (\bibinfo {year} {2018}{\natexlab{c}})},\ \Eprint
  {http://arxiv.org/abs/1802.06393} {arXiv:1802.06393 [astro-ph.CO]}
  \BibitemShut {NoStop}%
\bibitem [{\citenamefont {Escriv\`a}\ \emph
  {et~al.}(2022{\natexlab{b}})\citenamefont {Escriv\`a}, \citenamefont {Tada},
  \citenamefont {Yokoyama},\ and\ \citenamefont {Yoo}}]{Escriva:2022pnz}%
  \BibitemOpen
  \bibfield  {author} {\bibinfo {author} {\bibfnamefont {A.}~\bibnamefont
  {Escriv\`a}}, \bibinfo {author} {\bibfnamefont {Y.}~\bibnamefont {Tada}},
  \bibinfo {author} {\bibfnamefont {S.}~\bibnamefont {Yokoyama}}, \ and\
  \bibinfo {author} {\bibfnamefont {C.-M.}\ \bibnamefont {Yoo}},\ }\href
  {\doibase 10.1088/1475-7516/2022/05/012} {\bibfield  {journal} {\bibinfo
  {journal} {JCAP}\ }\textbf {\bibinfo {volume} {05}},\ \bibinfo {pages} {012}
  (\bibinfo {year} {2022}{\natexlab{b}})},\ \Eprint
  {http://arxiv.org/abs/2202.01028} {arXiv:2202.01028 [astro-ph.CO]}
  \BibitemShut {NoStop}%
\bibitem [{\citenamefont {Escriv\`a}\ \emph {et~al.}(2023)\citenamefont
  {Escriv\`a}, \citenamefont {Atal},\ and\ \citenamefont
  {Garriga}}]{Escriva:2023uko}%
  \BibitemOpen
  \bibfield  {author} {\bibinfo {author} {\bibfnamefont {A.}~\bibnamefont
  {Escriv\`a}}, \bibinfo {author} {\bibfnamefont {V.}~\bibnamefont {Atal}}, \
  and\ \bibinfo {author} {\bibfnamefont {J.}~\bibnamefont {Garriga}},\ }\href
  {\doibase 10.1088/1475-7516/2023/10/035} {\bibfield  {journal} {\bibinfo
  {journal} {JCAP}\ }\textbf {\bibinfo {volume} {10}},\ \bibinfo {pages} {035}
  (\bibinfo {year} {2023})},\ \Eprint {http://arxiv.org/abs/2306.09990}
  {arXiv:2306.09990 [astro-ph.CO]} \BibitemShut {NoStop}%
\bibitem [{\citenamefont {Uehara}\ \emph {et~al.}(2024)\citenamefont {Uehara},
  \citenamefont {Escriv\`a}, \citenamefont {Harada}, \citenamefont {Saito},\
  and\ \citenamefont {Yoo}}]{Uehara:2024yyp}%
  \BibitemOpen
  \bibfield  {author} {\bibinfo {author} {\bibfnamefont {K.}~\bibnamefont
  {Uehara}}, \bibinfo {author} {\bibfnamefont {A.}~\bibnamefont {Escriv\`a}},
  \bibinfo {author} {\bibfnamefont {T.}~\bibnamefont {Harada}}, \bibinfo
  {author} {\bibfnamefont {D.}~\bibnamefont {Saito}}, \ and\ \bibinfo {author}
  {\bibfnamefont {C.-M.}\ \bibnamefont {Yoo}},\ }\href@noop {} {\  (\bibinfo
  {year} {2024})},\ \Eprint {http://arxiv.org/abs/2401.06329} {arXiv:2401.06329
  [gr-qc]} \BibitemShut {NoStop}%
\bibitem [{\citenamefont {Inui}\ \emph {et~al.}(2024)\citenamefont {Inui},
  \citenamefont {Joana}, \citenamefont {Motohashi}, \citenamefont {Pi},
  \citenamefont {Tada},\ and\ \citenamefont {Yokoyama}}]{Inui:2024fgk}%
  \BibitemOpen
  \bibfield  {author} {\bibinfo {author} {\bibfnamefont {R.}~\bibnamefont
  {Inui}}, \bibinfo {author} {\bibfnamefont {C.}~\bibnamefont {Joana}},
  \bibinfo {author} {\bibfnamefont {H.}~\bibnamefont {Motohashi}}, \bibinfo
  {author} {\bibfnamefont {S.}~\bibnamefont {Pi}}, \bibinfo {author}
  {\bibfnamefont {Y.}~\bibnamefont {Tada}}, \ and\ \bibinfo {author}
  {\bibfnamefont {S.}~\bibnamefont {Yokoyama}},\ }\href@noop {} {\  (\bibinfo
  {year} {2024})},\ \Eprint {http://arxiv.org/abs/2411.07647} {arXiv:2411.07647
  [astro-ph.CO]} \BibitemShut {NoStop}%
\bibitem [{\citenamefont {Shimada}\ \emph {et~al.}(2024)\citenamefont
  {Shimada}, \citenamefont {Escriv\'a}, \citenamefont {Saito}, \citenamefont
  {Uehara},\ and\ \citenamefont {Yoo}}]{Shimada:2024eec}%
  \BibitemOpen
  \bibfield  {author} {\bibinfo {author} {\bibfnamefont {M.}~\bibnamefont
  {Shimada}}, \bibinfo {author} {\bibfnamefont {A.}~\bibnamefont {Escriv\'a}},
  \bibinfo {author} {\bibfnamefont {D.}~\bibnamefont {Saito}}, \bibinfo
  {author} {\bibfnamefont {K.}~\bibnamefont {Uehara}}, \ and\ \bibinfo {author}
  {\bibfnamefont {C.-M.}\ \bibnamefont {Yoo}},\ }\href@noop {} {\  (\bibinfo
  {year} {2024})},\ \Eprint {http://arxiv.org/abs/2411.07648} {arXiv:2411.07648
  [gr-qc]} \BibitemShut {NoStop}%
\bibitem [{\citenamefont {Green}\ \emph {et~al.}(2004)\citenamefont {Green},
  \citenamefont {Liddle}, \citenamefont {Malik},\ and\ \citenamefont
  {Sasaki}}]{Green:2004wb}%
  \BibitemOpen
  \bibfield  {author} {\bibinfo {author} {\bibfnamefont {A.~M.}\ \bibnamefont
  {Green}}, \bibinfo {author} {\bibfnamefont {A.~R.}\ \bibnamefont {Liddle}},
  \bibinfo {author} {\bibfnamefont {K.~A.}\ \bibnamefont {Malik}}, \ and\
  \bibinfo {author} {\bibfnamefont {M.}~\bibnamefont {Sasaki}},\ }\href
  {\doibase 10.1103/PhysRevD.70.041502} {\bibfield  {journal} {\bibinfo
  {journal} {Phys. Rev. D}\ }\textbf {\bibinfo {volume} {70}},\ \bibinfo
  {pages} {041502} (\bibinfo {year} {2004})},\ \Eprint
  {http://arxiv.org/abs/astro-ph/0403181} {arXiv:astro-ph/0403181} \BibitemShut
  {NoStop}%
\bibitem [{\citenamefont {Germani}\ and\ \citenamefont
  {Musco}(2019)}]{Germani:2018jgr}%
  \BibitemOpen
  \bibfield  {author} {\bibinfo {author} {\bibfnamefont {C.}~\bibnamefont
  {Germani}}\ and\ \bibinfo {author} {\bibfnamefont {I.}~\bibnamefont
  {Musco}},\ }\href {\doibase 10.1103/PhysRevLett.122.141302} {\bibfield
  {journal} {\bibinfo  {journal} {Phys. Rev. Lett.}\ }\textbf {\bibinfo
  {volume} {122}},\ \bibinfo {pages} {141302} (\bibinfo {year} {2019})},\
  \Eprint {http://arxiv.org/abs/1805.04087} {arXiv:1805.04087 [astro-ph.CO]}
  \BibitemShut {NoStop}%
\bibitem [{\citenamefont {Young}\ and\ \citenamefont
  {Musso}(2020)}]{Young:2020xmk}%
  \BibitemOpen
  \bibfield  {author} {\bibinfo {author} {\bibfnamefont {S.}~\bibnamefont
  {Young}}\ and\ \bibinfo {author} {\bibfnamefont {M.}~\bibnamefont {Musso}},\
  }\href {\doibase 10.1088/1475-7516/2020/11/022} {\bibfield  {journal}
  {\bibinfo  {journal} {JCAP}\ }\textbf {\bibinfo {volume} {11}},\ \bibinfo
  {pages} {022} (\bibinfo {year} {2020})},\ \Eprint
  {http://arxiv.org/abs/2001.06469} {arXiv:2001.06469 [astro-ph.CO]}
  \BibitemShut {NoStop}%
\bibitem [{\citenamefont {Taoso}\ and\ \citenamefont
  {Urbano}(2021)}]{Taoso:2021uvl}%
  \BibitemOpen
  \bibfield  {author} {\bibinfo {author} {\bibfnamefont {M.}~\bibnamefont
  {Taoso}}\ and\ \bibinfo {author} {\bibfnamefont {A.}~\bibnamefont {Urbano}},\
  }\href@noop {} {\  (\bibinfo {year} {2021})},\ \Eprint
  {http://arxiv.org/abs/2102.03610} {arXiv:2102.03610 [astro-ph.CO]}
  \BibitemShut {NoStop}%
\bibitem [{\citenamefont {Riccardi}\ \emph {et~al.}(2021)\citenamefont
  {Riccardi}, \citenamefont {Taoso},\ and\ \citenamefont
  {Urbano}}]{Riccardi:2021rlf}%
  \BibitemOpen
  \bibfield  {author} {\bibinfo {author} {\bibfnamefont {F.}~\bibnamefont
  {Riccardi}}, \bibinfo {author} {\bibfnamefont {M.}~\bibnamefont {Taoso}}, \
  and\ \bibinfo {author} {\bibfnamefont {A.}~\bibnamefont {Urbano}},\
  }\href@noop {} {\  (\bibinfo {year} {2021})},\ \Eprint
  {http://arxiv.org/abs/2102.04084} {arXiv:2102.04084 [astro-ph.CO]}
  \BibitemShut {NoStop}%
\bibitem [{\citenamefont {Young}(2022)}]{Young:2022phe}%
  \BibitemOpen
  \bibfield  {author} {\bibinfo {author} {\bibfnamefont {S.}~\bibnamefont
  {Young}},\ }\href {\doibase 10.1088/1475-7516/2022/05/037} {\bibfield
  {journal} {\bibinfo  {journal} {JCAP}\ }\textbf {\bibinfo {volume} {05}},\
  \bibinfo {pages} {037} (\bibinfo {year} {2022})},\ \Eprint
  {http://arxiv.org/abs/2201.13345} {arXiv:2201.13345 [astro-ph.CO]}
  \BibitemShut {NoStop}%
\bibitem [{\citenamefont {Pi}\ \emph {et~al.}(2024)\citenamefont {Pi},
  \citenamefont {Sasaki}, \citenamefont {Takhistov},\ and\ \citenamefont
  {Wang}}]{Pi:2024ert}%
  \BibitemOpen
  \bibfield  {author} {\bibinfo {author} {\bibfnamefont {S.}~\bibnamefont
  {Pi}}, \bibinfo {author} {\bibfnamefont {M.}~\bibnamefont {Sasaki}}, \bibinfo
  {author} {\bibfnamefont {V.}~\bibnamefont {Takhistov}}, \ and\ \bibinfo
  {author} {\bibfnamefont {J.}~\bibnamefont {Wang}},\ }\href@noop {} {\
  (\bibinfo {year} {2024})},\ \Eprint {http://arxiv.org/abs/2501.00295}
  {arXiv:2501.00295 [astro-ph.CO]} \BibitemShut {NoStop}%
\bibitem [{\citenamefont {Niikura}\ \emph {et~al.}(2019)\citenamefont {Niikura}
  \emph {et~al.}}]{Niikura:2017zjd}%
  \BibitemOpen
  \bibfield  {author} {\bibinfo {author} {\bibfnamefont {H.}~\bibnamefont
  {Niikura}} \emph {et~al.},\ }\href {\doibase 10.1038/s41550-019-0723-1}
  {\bibfield  {journal} {\bibinfo  {journal} {Nature Astron.}\ }\textbf
  {\bibinfo {volume} {3}},\ \bibinfo {pages} {524} (\bibinfo {year} {2019})},\
  \Eprint {http://arxiv.org/abs/1701.02151} {arXiv:1701.02151 [astro-ph.CO]}
  \BibitemShut {NoStop}%
\bibitem [{\citenamefont {Tisserand}\ \emph
  {et~al.}(2007{\natexlab{a}})\citenamefont {Tisserand} \emph
  {et~al.}}]{Tisserand:2006zx}%
  \BibitemOpen
  \bibfield  {author} {\bibinfo {author} {\bibfnamefont {P.}~\bibnamefont
  {Tisserand}} \emph {et~al.} (\bibinfo {collaboration} {EROS-2}),\ }\href
  {\doibase 10.1051/0004-6361:20066017} {\bibfield  {journal} {\bibinfo
  {journal} {Astron. Astrophys.}\ }\textbf {\bibinfo {volume} {469}},\ \bibinfo
  {pages} {387} (\bibinfo {year} {2007}{\natexlab{a}})},\ \Eprint
  {http://arxiv.org/abs/astro-ph/0607207} {arXiv:astro-ph/0607207} \BibitemShut
  {NoStop}%
\bibitem [{\citenamefont {Mroz}\ \emph
  {et~al.}(2024{\natexlab{a}})\citenamefont {Mroz} \emph
  {et~al.}}]{Mroz:2024mse}%
  \BibitemOpen
  \bibfield  {author} {\bibinfo {author} {\bibfnamefont {P.}~\bibnamefont
  {Mroz}} \emph {et~al.},\ }\href@noop {} {\  (\bibinfo {year}
  {2024}{\natexlab{a}})},\ \Eprint {http://arxiv.org/abs/2403.02386}
  {arXiv:2403.02386 [astro-ph.GA]} \BibitemShut {NoStop}%
\bibitem [{\citenamefont {Mroz}\ \emph
  {et~al.}(2024{\natexlab{b}})\citenamefont {Mroz} \emph
  {et~al.}}]{Mroz:2024wag}%
  \BibitemOpen
  \bibfield  {author} {\bibinfo {author} {\bibfnamefont {P.}~\bibnamefont
  {Mroz}} \emph {et~al.},\ }\href@noop {} {\  (\bibinfo {year}
  {2024}{\natexlab{b}})},\ \Eprint {http://arxiv.org/abs/2403.02398}
  {arXiv:2403.02398 [astro-ph.GA]} \BibitemShut {NoStop}%
\bibitem [{\citenamefont {Abbott}\ \emph
  {et~al.}(2023{\natexlab{d}})\citenamefont {Abbott} \emph
  {et~al.}}]{LVK:2022ydq}%
  \BibitemOpen
  \bibfield  {author} {\bibinfo {author} {\bibfnamefont {R.}~\bibnamefont
  {Abbott}} \emph {et~al.} (\bibinfo {collaboration} {LVK}),\ }\href {\doibase
  10.1093/mnras/stad588} {\bibfield  {journal} {\bibinfo  {journal} {Mon. Not.
  Roy. Astron. Soc.}\ }\textbf {\bibinfo {volume} {524}},\ \bibinfo {pages}
  {5984} (\bibinfo {year} {2023}{\natexlab{d}})},\ \bibinfo {note} {[Erratum:
  Mon.Not.Roy.Astron.Soc. 526, 6234 (2023)]},\ \Eprint
  {http://arxiv.org/abs/2212.01477} {arXiv:2212.01477 [astro-ph.HE]}
  \BibitemShut {NoStop}%
\bibitem [{\citenamefont {Serpico}\ \emph {et~al.}(2020)\citenamefont
  {Serpico}, \citenamefont {Poulin}, \citenamefont {Inman},\ and\ \citenamefont
  {Kohri}}]{Serpico:2020ehh}%
  \BibitemOpen
  \bibfield  {author} {\bibinfo {author} {\bibfnamefont {P.~D.}\ \bibnamefont
  {Serpico}}, \bibinfo {author} {\bibfnamefont {V.}~\bibnamefont {Poulin}},
  \bibinfo {author} {\bibfnamefont {D.}~\bibnamefont {Inman}}, \ and\ \bibinfo
  {author} {\bibfnamefont {K.}~\bibnamefont {Kohri}},\ }\href {\doibase
  10.1103/PhysRevResearch.2.023204} {\bibfield  {journal} {\bibinfo  {journal}
  {Phys. Rev. Res.}\ }\textbf {\bibinfo {volume} {2}},\ \bibinfo {pages}
  {023204} (\bibinfo {year} {2020})},\ \Eprint
  {http://arxiv.org/abs/2002.10771} {arXiv:2002.10771 [astro-ph.CO]}
  \BibitemShut {NoStop}%
\bibitem [{\citenamefont {Delabrouille}\ \emph {et~al.}(2021)\citenamefont
  {Delabrouille} \emph {et~al.}}]{Delabrouille:2019thj}%
  \BibitemOpen
  \bibfield  {author} {\bibinfo {author} {\bibfnamefont {J.}~\bibnamefont
  {Delabrouille}} \emph {et~al.},\ }\href {\doibase 10.1007/s10686-021-09721-z}
  {\bibfield  {journal} {\bibinfo  {journal} {Exper. Astron.}\ }\textbf
  {\bibinfo {volume} {51}},\ \bibinfo {pages} {1471} (\bibinfo {year}
  {2021})},\ \Eprint {http://arxiv.org/abs/1909.01591} {arXiv:1909.01591
  [astro-ph.CO]} \BibitemShut {NoStop}%
\bibitem [{\citenamefont {Chluba}\ \emph {et~al.}(2021)\citenamefont {Chluba}
  \emph {et~al.}}]{Chluba:2019nxa}%
  \BibitemOpen
  \bibfield  {author} {\bibinfo {author} {\bibfnamefont {J.}~\bibnamefont
  {Chluba}} \emph {et~al.},\ }\href {\doibase 10.1007/s10686-021-09729-5}
  {\bibfield  {journal} {\bibinfo  {journal} {Exper. Astron.}\ }\textbf
  {\bibinfo {volume} {51}},\ \bibinfo {pages} {1515} (\bibinfo {year}
  {2021})},\ \Eprint {http://arxiv.org/abs/1909.01593} {arXiv:1909.01593
  [astro-ph.CO]} \BibitemShut {NoStop}%
\bibitem [{\citenamefont {Nakama}\ \emph {et~al.}(2016)\citenamefont {Nakama},
  \citenamefont {Suyama},\ and\ \citenamefont {Yokoyama}}]{Nakama:2016kfq}%
  \BibitemOpen
  \bibfield  {author} {\bibinfo {author} {\bibfnamefont {T.}~\bibnamefont
  {Nakama}}, \bibinfo {author} {\bibfnamefont {T.}~\bibnamefont {Suyama}}, \
  and\ \bibinfo {author} {\bibfnamefont {J.}~\bibnamefont {Yokoyama}},\ }\href
  {\doibase 10.1103/PhysRevD.94.103522} {\bibfield  {journal} {\bibinfo
  {journal} {Phys. Rev. D}\ }\textbf {\bibinfo {volume} {94}},\ \bibinfo
  {pages} {103522} (\bibinfo {year} {2016})},\ \Eprint
  {http://arxiv.org/abs/1609.02245} {arXiv:1609.02245 [gr-qc]} \BibitemShut
  {NoStop}%
\bibitem [{\citenamefont {Nakama}\ \emph {et~al.}(2018)\citenamefont {Nakama},
  \citenamefont {Carr},\ and\ \citenamefont {Silk}}]{Nakama:2017xvq}%
  \BibitemOpen
  \bibfield  {author} {\bibinfo {author} {\bibfnamefont {T.}~\bibnamefont
  {Nakama}}, \bibinfo {author} {\bibfnamefont {B.}~\bibnamefont {Carr}}, \ and\
  \bibinfo {author} {\bibfnamefont {J.}~\bibnamefont {Silk}},\ }\href {\doibase
  10.1103/PhysRevD.97.043525} {\bibfield  {journal} {\bibinfo  {journal} {Phys.
  Rev. D}\ }\textbf {\bibinfo {volume} {97}},\ \bibinfo {pages} {043525}
  (\bibinfo {year} {2018})},\ \Eprint {http://arxiv.org/abs/1710.06945}
  {arXiv:1710.06945 [astro-ph.CO]} \BibitemShut {NoStop}%
\bibitem [{\citenamefont {Nakama}\ \emph {et~al.}(2019)\citenamefont {Nakama},
  \citenamefont {Kohri},\ and\ \citenamefont {Silk}}]{Nakama:2019htb}%
  \BibitemOpen
  \bibfield  {author} {\bibinfo {author} {\bibfnamefont {T.}~\bibnamefont
  {Nakama}}, \bibinfo {author} {\bibfnamefont {K.}~\bibnamefont {Kohri}}, \
  and\ \bibinfo {author} {\bibfnamefont {J.}~\bibnamefont {Silk}},\ }\href
  {\doibase 10.1103/PhysRevD.99.123530} {\bibfield  {journal} {\bibinfo
  {journal} {Phys. Rev. D}\ }\textbf {\bibinfo {volume} {99}},\ \bibinfo
  {pages} {123530} (\bibinfo {year} {2019})},\ \Eprint
  {http://arxiv.org/abs/1905.04477} {arXiv:1905.04477 [astro-ph.CO]}
  \BibitemShut {NoStop}%
\bibitem [{\citenamefont {Atal}\ \emph
  {et~al.}(2020{\natexlab{b}})\citenamefont {Atal}, \citenamefont {Sanglas},\
  and\ \citenamefont {Triantafyllou}}]{Atal:2020yic}%
  \BibitemOpen
  \bibfield  {author} {\bibinfo {author} {\bibfnamefont {V.}~\bibnamefont
  {Atal}}, \bibinfo {author} {\bibfnamefont {A.}~\bibnamefont {Sanglas}}, \
  and\ \bibinfo {author} {\bibfnamefont {N.}~\bibnamefont {Triantafyllou}},\
  }\href@noop {} {\  (\bibinfo {year} {2020}{\natexlab{b}})},\ \Eprint
  {http://arxiv.org/abs/2012.14721} {arXiv:2012.14721 [astro-ph.CO]}
  \BibitemShut {NoStop}%
\bibitem [{\citenamefont {Liu}\ and\ \citenamefont
  {Bromm}(2022)}]{Liu:2022bvr}%
  \BibitemOpen
  \bibfield  {author} {\bibinfo {author} {\bibfnamefont {B.}~\bibnamefont
  {Liu}}\ and\ \bibinfo {author} {\bibfnamefont {V.}~\bibnamefont {Bromm}},\
  }\href {\doibase 10.3847/2041-8213/ac927f} {\bibfield  {journal} {\bibinfo
  {journal} {Astrophys. J. Lett.}\ }\textbf {\bibinfo {volume} {937}},\
  \bibinfo {pages} {L30} (\bibinfo {year} {2022})},\ \Eprint
  {http://arxiv.org/abs/2208.13178} {arXiv:2208.13178 [astro-ph.CO]}
  \BibitemShut {NoStop}%
\bibitem [{\citenamefont {Biagetti}\ \emph {et~al.}(2023)\citenamefont
  {Biagetti}, \citenamefont {Franciolini},\ and\ \citenamefont
  {Riotto}}]{Biagetti:2022ode}%
  \BibitemOpen
  \bibfield  {author} {\bibinfo {author} {\bibfnamefont {M.}~\bibnamefont
  {Biagetti}}, \bibinfo {author} {\bibfnamefont {G.}~\bibnamefont
  {Franciolini}}, \ and\ \bibinfo {author} {\bibfnamefont {A.}~\bibnamefont
  {Riotto}},\ }\href {\doibase 10.3847/1538-4357/acb5ea} {\bibfield  {journal}
  {\bibinfo  {journal} {Astrophys. J.}\ }\textbf {\bibinfo {volume} {944}},\
  \bibinfo {pages} {113} (\bibinfo {year} {2023})},\ \Eprint
  {http://arxiv.org/abs/2210.04812} {arXiv:2210.04812 [astro-ph.CO]}
  \BibitemShut {NoStop}%
\bibitem [{\citenamefont {Hooper}\ \emph {et~al.}(2023)\citenamefont {Hooper},
  \citenamefont {Ireland}, \citenamefont {Krnjaic},\ and\ \citenamefont
  {Stebbins}}]{Hooper:2023nnl}%
  \BibitemOpen
  \bibfield  {author} {\bibinfo {author} {\bibfnamefont {D.}~\bibnamefont
  {Hooper}}, \bibinfo {author} {\bibfnamefont {A.}~\bibnamefont {Ireland}},
  \bibinfo {author} {\bibfnamefont {G.}~\bibnamefont {Krnjaic}}, \ and\
  \bibinfo {author} {\bibfnamefont {A.}~\bibnamefont {Stebbins}},\ }\href@noop
  {} {\  (\bibinfo {year} {2023})},\ \Eprint {http://arxiv.org/abs/2308.00756}
  {arXiv:2308.00756 [astro-ph.CO]} \BibitemShut {NoStop}%
\bibitem [{\citenamefont {Huang}\ \emph
  {et~al.}(2024{\natexlab{d}})\citenamefont {Huang}, \citenamefont {Jiang},
  \citenamefont {He}, \citenamefont {Wang},\ and\ \citenamefont
  {Piao}}]{Hai-LongHuang:2024gtx}%
  \BibitemOpen
  \bibfield  {author} {\bibinfo {author} {\bibfnamefont {H.-L.}\ \bibnamefont
  {Huang}}, \bibinfo {author} {\bibfnamefont {J.-Q.}\ \bibnamefont {Jiang}},
  \bibinfo {author} {\bibfnamefont {J.}~\bibnamefont {He}}, \bibinfo {author}
  {\bibfnamefont {Y.-T.}\ \bibnamefont {Wang}}, \ and\ \bibinfo {author}
  {\bibfnamefont {Y.-S.}\ \bibnamefont {Piao}},\ }\href@noop {} {\  (\bibinfo
  {year} {2024}{\natexlab{d}})},\ \Eprint {http://arxiv.org/abs/2410.20663}
  {arXiv:2410.20663 [astro-ph.GA]} \BibitemShut {NoStop}%
\bibitem [{\citenamefont {Sugiyama}\ \emph {et~al.}(2020)\citenamefont
  {Sugiyama}, \citenamefont {Kurita},\ and\ \citenamefont
  {Takada}}]{Sugiyama:2019dgt}%
  \BibitemOpen
  \bibfield  {author} {\bibinfo {author} {\bibfnamefont {S.}~\bibnamefont
  {Sugiyama}}, \bibinfo {author} {\bibfnamefont {T.}~\bibnamefont {Kurita}}, \
  and\ \bibinfo {author} {\bibfnamefont {M.}~\bibnamefont {Takada}},\ }\href
  {\doibase 10.1093/mnras/staa407} {\bibfield  {journal} {\bibinfo  {journal}
  {Mon. Not. Roy. Astron. Soc.}\ }\textbf {\bibinfo {volume} {493}},\ \bibinfo
  {pages} {3632} (\bibinfo {year} {2020})},\ \Eprint
  {http://arxiv.org/abs/1905.06066} {arXiv:1905.06066 [astro-ph.CO]}
  \BibitemShut {NoStop}%
\bibitem [{\citenamefont {Montero-Camacho}\ \emph {et~al.}(2019)\citenamefont
  {Montero-Camacho}, \citenamefont {Fang}, \citenamefont {Vasquez},
  \citenamefont {Silva},\ and\ \citenamefont
  {Hirata}}]{Montero-Camacho:2019jte}%
  \BibitemOpen
  \bibfield  {author} {\bibinfo {author} {\bibfnamefont {P.}~\bibnamefont
  {Montero-Camacho}}, \bibinfo {author} {\bibfnamefont {X.}~\bibnamefont
  {Fang}}, \bibinfo {author} {\bibfnamefont {G.}~\bibnamefont {Vasquez}},
  \bibinfo {author} {\bibfnamefont {M.}~\bibnamefont {Silva}}, \ and\ \bibinfo
  {author} {\bibfnamefont {C.~M.}\ \bibnamefont {Hirata}},\ }\href {\doibase
  10.1088/1475-7516/2019/08/031} {\bibfield  {journal} {\bibinfo  {journal}
  {JCAP}\ }\textbf {\bibinfo {volume} {08}},\ \bibinfo {pages} {031} (\bibinfo
  {year} {2019})},\ \Eprint {http://arxiv.org/abs/1906.05950} {arXiv:1906.05950
  [astro-ph.CO]} \BibitemShut {NoStop}%
\bibitem [{\citenamefont {Smyth}\ \emph {et~al.}(2020)\citenamefont {Smyth},
  \citenamefont {Profumo}, \citenamefont {English}, \citenamefont {Jeltema},
  \citenamefont {McKinnon},\ and\ \citenamefont
  {Guhathakurta}}]{Smyth:2019whb}%
  \BibitemOpen
  \bibfield  {author} {\bibinfo {author} {\bibfnamefont {N.}~\bibnamefont
  {Smyth}}, \bibinfo {author} {\bibfnamefont {S.}~\bibnamefont {Profumo}},
  \bibinfo {author} {\bibfnamefont {S.}~\bibnamefont {English}}, \bibinfo
  {author} {\bibfnamefont {T.}~\bibnamefont {Jeltema}}, \bibinfo {author}
  {\bibfnamefont {K.}~\bibnamefont {McKinnon}}, \ and\ \bibinfo {author}
  {\bibfnamefont {P.}~\bibnamefont {Guhathakurta}},\ }\href {\doibase
  10.1103/PhysRevD.101.063005} {\bibfield  {journal} {\bibinfo  {journal}
  {Phys. Rev. D}\ }\textbf {\bibinfo {volume} {101}},\ \bibinfo {pages}
  {063005} (\bibinfo {year} {2020})},\ \Eprint
  {http://arxiv.org/abs/1910.01285} {arXiv:1910.01285 [astro-ph.CO]}
  \BibitemShut {NoStop}%
\bibitem [{\citenamefont {Villanueva-Domingo}\ \emph
  {et~al.}(2021)\citenamefont {Villanueva-Domingo}, \citenamefont {Mena},\ and\
  \citenamefont {Palomares-Ruiz}}]{Villanueva-Domingo:2021spv}%
  \BibitemOpen
  \bibfield  {author} {\bibinfo {author} {\bibfnamefont {P.}~\bibnamefont
  {Villanueva-Domingo}}, \bibinfo {author} {\bibfnamefont {O.}~\bibnamefont
  {Mena}}, \ and\ \bibinfo {author} {\bibfnamefont {S.}~\bibnamefont
  {Palomares-Ruiz}},\ }\href {\doibase 10.3389/fspas.2021.681084} {\bibfield
  {journal} {\bibinfo  {journal} {Front. Astron. Space Sci.}\ }\textbf
  {\bibinfo {volume} {8}},\ \bibinfo {pages} {87} (\bibinfo {year} {2021})},\
  \Eprint {http://arxiv.org/abs/2103.12087} {arXiv:2103.12087 [astro-ph.CO]}
  \BibitemShut {NoStop}%
\bibitem [{\citenamefont {Green}(2024)}]{Green:2024bam}%
  \BibitemOpen
  \bibfield  {author} {\bibinfo {author} {\bibfnamefont {A.~M.}\ \bibnamefont
  {Green}},\ }\href {\doibase 10.1016/j.nuclphysb.2024.116494} {\bibfield
  {journal} {\bibinfo  {journal} {Nucl. Phys. B}\ }\textbf {\bibinfo {volume}
  {1003}},\ \bibinfo {pages} {116494} (\bibinfo {year} {2024})},\ \Eprint
  {http://arxiv.org/abs/2402.15211} {arXiv:2402.15211 [astro-ph.CO]}
  \BibitemShut {NoStop}%
\bibitem [{\citenamefont {Tinyakov}(2024)}]{Tinyakov:2024mcy}%
  \BibitemOpen
  \bibfield  {author} {\bibinfo {author} {\bibfnamefont {P.}~\bibnamefont
  {Tinyakov}},\ }\href@noop {} {\  (\bibinfo {year} {2024})},\ \Eprint
  {http://arxiv.org/abs/2406.03114} {arXiv:2406.03114 [astro-ph.CO]}
  \BibitemShut {NoStop}%
\bibitem [{\citenamefont {Katz}\ \emph {et~al.}(2018)\citenamefont {Katz},
  \citenamefont {Kopp}, \citenamefont {Sibiryakov},\ and\ \citenamefont
  {Xue}}]{Katz:2018zrn}%
  \BibitemOpen
  \bibfield  {author} {\bibinfo {author} {\bibfnamefont {A.}~\bibnamefont
  {Katz}}, \bibinfo {author} {\bibfnamefont {J.}~\bibnamefont {Kopp}}, \bibinfo
  {author} {\bibfnamefont {S.}~\bibnamefont {Sibiryakov}}, \ and\ \bibinfo
  {author} {\bibfnamefont {W.}~\bibnamefont {Xue}},\ }\href {\doibase
  10.1088/1475-7516/2018/12/005} {\bibfield  {journal} {\bibinfo  {journal}
  {JCAP}\ }\textbf {\bibinfo {volume} {12}},\ \bibinfo {pages} {005} (\bibinfo
  {year} {2018})},\ \Eprint {http://arxiv.org/abs/1807.11495} {arXiv:1807.11495
  [astro-ph.CO]} \BibitemShut {NoStop}%
\bibitem [{\citenamefont {Ray}\ \emph {et~al.}(2021)\citenamefont {Ray},
  \citenamefont {Laha}, \citenamefont {Mu\~noz},\ and\ \citenamefont
  {Caputo}}]{Ray:2021mxu}%
  \BibitemOpen
  \bibfield  {author} {\bibinfo {author} {\bibfnamefont {A.}~\bibnamefont
  {Ray}}, \bibinfo {author} {\bibfnamefont {R.}~\bibnamefont {Laha}}, \bibinfo
  {author} {\bibfnamefont {J.~B.}\ \bibnamefont {Mu\~noz}}, \ and\ \bibinfo
  {author} {\bibfnamefont {R.}~\bibnamefont {Caputo}},\ }\href {\doibase
  10.1103/PhysRevD.104.023516} {\bibfield  {journal} {\bibinfo  {journal}
  {Phys. Rev. D}\ }\textbf {\bibinfo {volume} {104}},\ \bibinfo {pages}
  {023516} (\bibinfo {year} {2021})},\ \Eprint
  {http://arxiv.org/abs/2102.06714} {arXiv:2102.06714 [astro-ph.CO]}
  \BibitemShut {NoStop}%
\bibitem [{\citenamefont {Jung}\ and\ \citenamefont
  {Kim}(2020)}]{Jung:2019fcs}%
  \BibitemOpen
  \bibfield  {author} {\bibinfo {author} {\bibfnamefont {S.}~\bibnamefont
  {Jung}}\ and\ \bibinfo {author} {\bibfnamefont {T.}~\bibnamefont {Kim}},\
  }\href {\doibase 10.1103/PhysRevResearch.2.013113} {\bibfield  {journal}
  {\bibinfo  {journal} {Phys. Rev. Res.}\ }\textbf {\bibinfo {volume} {2}},\
  \bibinfo {pages} {013113} (\bibinfo {year} {2020})},\ \Eprint
  {http://arxiv.org/abs/1908.00078} {arXiv:1908.00078 [astro-ph.CO]}
  \BibitemShut {NoStop}%
\bibitem [{\citenamefont {Gawade}\ \emph {et~al.}(2023)\citenamefont {Gawade},
  \citenamefont {More},\ and\ \citenamefont {Bhalerao}}]{Gawade:2023gmt}%
  \BibitemOpen
  \bibfield  {author} {\bibinfo {author} {\bibfnamefont {P.}~\bibnamefont
  {Gawade}}, \bibinfo {author} {\bibfnamefont {S.}~\bibnamefont {More}}, \ and\
  \bibinfo {author} {\bibfnamefont {V.}~\bibnamefont {Bhalerao}},\ }\href
  {\doibase 10.1093/mnras/stad3336} {\bibfield  {journal} {\bibinfo  {journal}
  {Mon. Not. Roy. Astron. Soc.}\ }\textbf {\bibinfo {volume} {527}},\ \bibinfo
  {pages} {3306} (\bibinfo {year} {2023})},\ \Eprint
  {http://arxiv.org/abs/2308.01775} {arXiv:2308.01775 [astro-ph.CO]}
  \BibitemShut {NoStop}%
\bibitem [{\citenamefont {Tamta}\ \emph {et~al.}(2024)\citenamefont {Tamta},
  \citenamefont {Raj},\ and\ \citenamefont {Sharma}}]{Tamta:2024pow}%
  \BibitemOpen
  \bibfield  {author} {\bibinfo {author} {\bibfnamefont {M.}~\bibnamefont
  {Tamta}}, \bibinfo {author} {\bibfnamefont {N.}~\bibnamefont {Raj}}, \ and\
  \bibinfo {author} {\bibfnamefont {P.}~\bibnamefont {Sharma}},\ }\href@noop {}
  {\  (\bibinfo {year} {2024})},\ \Eprint {http://arxiv.org/abs/2405.20365}
  {arXiv:2405.20365 [astro-ph.HE]} \BibitemShut {NoStop}%
\bibitem [{\citenamefont {Tran}\ \emph {et~al.}(2024)\citenamefont {Tran},
  \citenamefont {Geller}, \citenamefont {Lehmann},\ and\ \citenamefont
  {Kaiser}}]{Tran:2023jci}%
  \BibitemOpen
  \bibfield  {author} {\bibinfo {author} {\bibfnamefont {T.~X.}\ \bibnamefont
  {Tran}}, \bibinfo {author} {\bibfnamefont {S.~R.}\ \bibnamefont {Geller}},
  \bibinfo {author} {\bibfnamefont {B.~V.}\ \bibnamefont {Lehmann}}, \ and\
  \bibinfo {author} {\bibfnamefont {D.~I.}\ \bibnamefont {Kaiser}},\ }\href
  {\doibase 10.1103/PhysRevD.110.063533} {\bibfield  {journal} {\bibinfo
  {journal} {Phys. Rev. D}\ }\textbf {\bibinfo {volume} {110}},\ \bibinfo
  {pages} {063533} (\bibinfo {year} {2024})},\ \Eprint
  {http://arxiv.org/abs/2312.17217} {arXiv:2312.17217 [astro-ph.CO]}
  \BibitemShut {NoStop}%
\bibitem [{\citenamefont {Cuadrat-Grzybowski}\ \emph
  {et~al.}(2024)\citenamefont {Cuadrat-Grzybowski}, \citenamefont {Clesse},
  \citenamefont {Defraigne}, \citenamefont {Van~Camp},\ and\ \citenamefont
  {Bertrand}}]{Cuadrat-Grzybowski:2024uph}%
  \BibitemOpen
  \bibfield  {author} {\bibinfo {author} {\bibfnamefont {M.}~\bibnamefont
  {Cuadrat-Grzybowski}}, \bibinfo {author} {\bibfnamefont {S.}~\bibnamefont
  {Clesse}}, \bibinfo {author} {\bibfnamefont {P.}~\bibnamefont {Defraigne}},
  \bibinfo {author} {\bibfnamefont {M.}~\bibnamefont {Van~Camp}}, \ and\
  \bibinfo {author} {\bibfnamefont {B.}~\bibnamefont {Bertrand}},\ }\href
  {\doibase 10.1103/PhysRevD.110.063029} {\bibfield  {journal} {\bibinfo
  {journal} {Phys. Rev. D}\ }\textbf {\bibinfo {volume} {110}},\ \bibinfo
  {pages} {063029} (\bibinfo {year} {2024})},\ \Eprint
  {http://arxiv.org/abs/2403.14397} {arXiv:2403.14397 [astro-ph.CO]}
  \BibitemShut {NoStop}%
\bibitem [{\citenamefont {Loeb}(2024)}]{Loeb:2024tcc}%
  \BibitemOpen
  \bibfield  {author} {\bibinfo {author} {\bibfnamefont {A.}~\bibnamefont
  {Loeb}},\ }\href {\doibase 10.3847/2515-5172/ad739e} {\bibfield  {journal}
  {\bibinfo  {journal} {Res. Notes AAS}\ }\textbf {\bibinfo {volume} {8}},\
  \bibinfo {pages} {211} (\bibinfo {year} {2024})},\ \Eprint
  {http://arxiv.org/abs/2408.10799} {arXiv:2408.10799 [hep-ph]} \BibitemShut
  {NoStop}%
\bibitem [{\citenamefont {Saito}\ and\ \citenamefont
  {Yokoyama}(2009)}]{Saito:2008jc}%
  \BibitemOpen
  \bibfield  {author} {\bibinfo {author} {\bibfnamefont {R.}~\bibnamefont
  {Saito}}\ and\ \bibinfo {author} {\bibfnamefont {J.}~\bibnamefont
  {Yokoyama}},\ }\href {\doibase 10.1103/PhysRevLett.102.161101} {\bibfield
  {journal} {\bibinfo  {journal} {Phys. Rev. Lett.}\ }\textbf {\bibinfo
  {volume} {102}},\ \bibinfo {pages} {161101} (\bibinfo {year} {2009})},\
  \bibinfo {note} {[Erratum: Phys. Rev. Lett. 107, 069901 (2011)]},\ \Eprint
  {http://arxiv.org/abs/0812.4339} {arXiv:0812.4339 [astro-ph]} \BibitemShut
  {NoStop}%
\bibitem [{\citenamefont {Cai}\ \emph {et~al.}(2019{\natexlab{b}})\citenamefont
  {Cai}, \citenamefont {Pi},\ and\ \citenamefont {Sasaki}}]{Cai:2018dig}%
  \BibitemOpen
  \bibfield  {author} {\bibinfo {author} {\bibfnamefont {R.-g.}\ \bibnamefont
  {Cai}}, \bibinfo {author} {\bibfnamefont {S.}~\bibnamefont {Pi}}, \ and\
  \bibinfo {author} {\bibfnamefont {M.}~\bibnamefont {Sasaki}},\ }\href
  {\doibase 10.1103/PhysRevLett.122.201101} {\bibfield  {journal} {\bibinfo
  {journal} {Phys. Rev. Lett.}\ }\textbf {\bibinfo {volume} {122}},\ \bibinfo
  {pages} {201101} (\bibinfo {year} {2019}{\natexlab{b}})},\ \Eprint
  {http://arxiv.org/abs/1810.11000} {arXiv:1810.11000 [astro-ph.CO]}
  \BibitemShut {NoStop}%
\bibitem [{\citenamefont {Bartolo}\ \emph
  {et~al.}(2019{\natexlab{a}})\citenamefont {Bartolo}, \citenamefont {De~Luca},
  \citenamefont {Franciolini}, \citenamefont {Lewis}, \citenamefont {Peloso},\
  and\ \citenamefont {Riotto}}]{Bartolo:2018evs}%
  \BibitemOpen
  \bibfield  {author} {\bibinfo {author} {\bibfnamefont {N.}~\bibnamefont
  {Bartolo}}, \bibinfo {author} {\bibfnamefont {V.}~\bibnamefont {De~Luca}},
  \bibinfo {author} {\bibfnamefont {G.}~\bibnamefont {Franciolini}}, \bibinfo
  {author} {\bibfnamefont {A.}~\bibnamefont {Lewis}}, \bibinfo {author}
  {\bibfnamefont {M.}~\bibnamefont {Peloso}}, \ and\ \bibinfo {author}
  {\bibfnamefont {A.}~\bibnamefont {Riotto}},\ }\href {\doibase
  10.1103/PhysRevLett.122.211301} {\bibfield  {journal} {\bibinfo  {journal}
  {Phys. Rev. Lett.}\ }\textbf {\bibinfo {volume} {122}},\ \bibinfo {pages}
  {211301} (\bibinfo {year} {2019}{\natexlab{a}})},\ \Eprint
  {http://arxiv.org/abs/1810.12218} {arXiv:1810.12218 [astro-ph.CO]}
  \BibitemShut {NoStop}%
\bibitem [{\citenamefont {Matarrese}\ \emph {et~al.}(1993)\citenamefont
  {Matarrese}, \citenamefont {Pantano},\ and\ \citenamefont
  {Saez}}]{Matarrese:1992rp}%
  \BibitemOpen
  \bibfield  {author} {\bibinfo {author} {\bibfnamefont {S.}~\bibnamefont
  {Matarrese}}, \bibinfo {author} {\bibfnamefont {O.}~\bibnamefont {Pantano}},
  \ and\ \bibinfo {author} {\bibfnamefont {D.}~\bibnamefont {Saez}},\ }\href
  {\doibase 10.1103/PhysRevD.47.1311} {\bibfield  {journal} {\bibinfo
  {journal} {Phys. Rev. D}\ }\textbf {\bibinfo {volume} {47}},\ \bibinfo
  {pages} {1311} (\bibinfo {year} {1993})}\BibitemShut {NoStop}%
\bibitem [{\citenamefont {Matarrese}\ \emph {et~al.}(1994)\citenamefont
  {Matarrese}, \citenamefont {Pantano},\ and\ \citenamefont
  {Saez}}]{Matarrese:1993zf}%
  \BibitemOpen
  \bibfield  {author} {\bibinfo {author} {\bibfnamefont {S.}~\bibnamefont
  {Matarrese}}, \bibinfo {author} {\bibfnamefont {O.}~\bibnamefont {Pantano}},
  \ and\ \bibinfo {author} {\bibfnamefont {D.}~\bibnamefont {Saez}},\ }\href
  {\doibase 10.1103/PhysRevLett.72.320} {\bibfield  {journal} {\bibinfo
  {journal} {Phys. Rev. Lett.}\ }\textbf {\bibinfo {volume} {72}},\ \bibinfo
  {pages} {320} (\bibinfo {year} {1994})},\ \Eprint
  {http://arxiv.org/abs/astro-ph/9310036} {arXiv:astro-ph/9310036} \BibitemShut
  {NoStop}%
\bibitem [{\citenamefont {Matarrese}\ \emph {et~al.}(1998)\citenamefont
  {Matarrese}, \citenamefont {Mollerach},\ and\ \citenamefont
  {Bruni}}]{Matarrese:1997ay}%
  \BibitemOpen
  \bibfield  {author} {\bibinfo {author} {\bibfnamefont {S.}~\bibnamefont
  {Matarrese}}, \bibinfo {author} {\bibfnamefont {S.}~\bibnamefont
  {Mollerach}}, \ and\ \bibinfo {author} {\bibfnamefont {M.}~\bibnamefont
  {Bruni}},\ }\href {\doibase 10.1103/PhysRevD.58.043504} {\bibfield  {journal}
  {\bibinfo  {journal} {Phys. Rev. D}\ }\textbf {\bibinfo {volume} {58}},\
  \bibinfo {pages} {043504} (\bibinfo {year} {1998})},\ \Eprint
  {http://arxiv.org/abs/astro-ph/9707278} {arXiv:astro-ph/9707278} \BibitemShut
  {NoStop}%
\bibitem [{\citenamefont {Noh}\ and\ \citenamefont {Hwang}(2004)}]{Noh:2004bc}%
  \BibitemOpen
  \bibfield  {author} {\bibinfo {author} {\bibfnamefont {H.}~\bibnamefont
  {Noh}}\ and\ \bibinfo {author} {\bibfnamefont {J.-c.}\ \bibnamefont
  {Hwang}},\ }\href {\doibase 10.1103/PhysRevD.69.104011} {\bibfield  {journal}
  {\bibinfo  {journal} {Phys. Rev. D}\ }\textbf {\bibinfo {volume} {69}},\
  \bibinfo {pages} {104011} (\bibinfo {year} {2004})}\BibitemShut {NoStop}%
\bibitem [{\citenamefont {Carbone}\ and\ \citenamefont
  {Matarrese}(2005)}]{Carbone:2004iv}%
  \BibitemOpen
  \bibfield  {author} {\bibinfo {author} {\bibfnamefont {C.}~\bibnamefont
  {Carbone}}\ and\ \bibinfo {author} {\bibfnamefont {S.}~\bibnamefont
  {Matarrese}},\ }\href {\doibase 10.1103/PhysRevD.71.043508} {\bibfield
  {journal} {\bibinfo  {journal} {Phys. Rev. D}\ }\textbf {\bibinfo {volume}
  {71}},\ \bibinfo {pages} {043508} (\bibinfo {year} {2005})},\ \Eprint
  {http://arxiv.org/abs/astro-ph/0407611} {arXiv:astro-ph/0407611} \BibitemShut
  {NoStop}%
\bibitem [{\citenamefont {Nakamura}(2007)}]{Nakamura:2004rm}%
  \BibitemOpen
  \bibfield  {author} {\bibinfo {author} {\bibfnamefont {K.}~\bibnamefont
  {Nakamura}},\ }\href {\doibase 10.1143/PTP.117.17} {\bibfield  {journal}
  {\bibinfo  {journal} {Prog. Theor. Phys.}\ }\textbf {\bibinfo {volume}
  {117}},\ \bibinfo {pages} {17} (\bibinfo {year} {2007})},\ \Eprint
  {http://arxiv.org/abs/gr-qc/0605108} {arXiv:gr-qc/0605108} \BibitemShut
  {NoStop}%
\bibitem [{\citenamefont {Ananda}\ \emph {et~al.}(2007)\citenamefont {Ananda},
  \citenamefont {Clarkson},\ and\ \citenamefont {Wands}}]{Ananda:2006af}%
  \BibitemOpen
  \bibfield  {author} {\bibinfo {author} {\bibfnamefont {K.~N.}\ \bibnamefont
  {Ananda}}, \bibinfo {author} {\bibfnamefont {C.}~\bibnamefont {Clarkson}}, \
  and\ \bibinfo {author} {\bibfnamefont {D.}~\bibnamefont {Wands}},\ }\href
  {\doibase 10.1103/PhysRevD.75.123518} {\bibfield  {journal} {\bibinfo
  {journal} {Phys. Rev. D}\ }\textbf {\bibinfo {volume} {75}},\ \bibinfo
  {pages} {123518} (\bibinfo {year} {2007})},\ \Eprint
  {http://arxiv.org/abs/gr-qc/0612013} {arXiv:gr-qc/0612013} \BibitemShut
  {NoStop}%
\bibitem [{\citenamefont {Osano}\ \emph {et~al.}(2007)\citenamefont {Osano},
  \citenamefont {Pitrou}, \citenamefont {Dunsby}, \citenamefont {Uzan},\ and\
  \citenamefont {Clarkson}}]{Osano:2006ew}%
  \BibitemOpen
  \bibfield  {author} {\bibinfo {author} {\bibfnamefont {B.}~\bibnamefont
  {Osano}}, \bibinfo {author} {\bibfnamefont {C.}~\bibnamefont {Pitrou}},
  \bibinfo {author} {\bibfnamefont {P.}~\bibnamefont {Dunsby}}, \bibinfo
  {author} {\bibfnamefont {J.-P.}\ \bibnamefont {Uzan}}, \ and\ \bibinfo
  {author} {\bibfnamefont {C.}~\bibnamefont {Clarkson}},\ }\href {\doibase
  10.1088/1475-7516/2007/04/003} {\bibfield  {journal} {\bibinfo  {journal}
  {JCAP}\ }\textbf {\bibinfo {volume} {04}},\ \bibinfo {pages} {003} (\bibinfo
  {year} {2007})},\ \Eprint {http://arxiv.org/abs/gr-qc/0612108}
  {arXiv:gr-qc/0612108} \BibitemShut {NoStop}%
\bibitem [{\citenamefont {Baumann}\ \emph {et~al.}(2007)\citenamefont
  {Baumann}, \citenamefont {Steinhardt}, \citenamefont {Takahashi},\ and\
  \citenamefont {Ichiki}}]{Baumann:2007zm}%
  \BibitemOpen
  \bibfield  {author} {\bibinfo {author} {\bibfnamefont {D.}~\bibnamefont
  {Baumann}}, \bibinfo {author} {\bibfnamefont {P.~J.}\ \bibnamefont
  {Steinhardt}}, \bibinfo {author} {\bibfnamefont {K.}~\bibnamefont
  {Takahashi}}, \ and\ \bibinfo {author} {\bibfnamefont {K.}~\bibnamefont
  {Ichiki}},\ }\href {\doibase 10.1103/PhysRevD.76.084019} {\bibfield
  {journal} {\bibinfo  {journal} {Phys. Rev. D}\ }\textbf {\bibinfo {volume}
  {76}},\ \bibinfo {pages} {084019} (\bibinfo {year} {2007})},\ \Eprint
  {http://arxiv.org/abs/hep-th/0703290} {arXiv:hep-th/0703290} \BibitemShut
  {NoStop}%
\bibitem [{\citenamefont {Chang}\ \emph {et~al.}(2023)\citenamefont {Chang},
  \citenamefont {Zhang},\ and\ \citenamefont {Zhou}}]{Chang:2022vlv}%
  \BibitemOpen
  \bibfield  {author} {\bibinfo {author} {\bibfnamefont {Z.}~\bibnamefont
  {Chang}}, \bibinfo {author} {\bibfnamefont {X.}~\bibnamefont {Zhang}}, \ and\
  \bibinfo {author} {\bibfnamefont {J.-Z.}\ \bibnamefont {Zhou}},\ }\href
  {\doibase 10.1103/PhysRevD.107.063510} {\bibfield  {journal} {\bibinfo
  {journal} {Phys. Rev. D}\ }\textbf {\bibinfo {volume} {107}},\ \bibinfo
  {pages} {063510} (\bibinfo {year} {2023})},\ \Eprint
  {http://arxiv.org/abs/2209.07693} {arXiv:2209.07693 [astro-ph.CO]}
  \BibitemShut {NoStop}%
\bibitem [{\citenamefont {Yu}\ and\ \citenamefont {Wang}(2023)}]{Yu:2023lmo}%
  \BibitemOpen
  \bibfield  {author} {\bibinfo {author} {\bibfnamefont {Y.-H.}\ \bibnamefont
  {Yu}}\ and\ \bibinfo {author} {\bibfnamefont {S.}~\bibnamefont {Wang}},\
  }\href@noop {} {\  (\bibinfo {year} {2023})},\ \Eprint
  {http://arxiv.org/abs/2303.03897} {arXiv:2303.03897 [astro-ph.CO]}
  \BibitemShut {NoStop}%
\bibitem [{\citenamefont {Bari}\ \emph {et~al.}(2024)\citenamefont {Bari},
  \citenamefont {Bartolo}, \citenamefont {Dom\`enech},\ and\ \citenamefont
  {Matarrese}}]{Bari:2023rcw}%
  \BibitemOpen
  \bibfield  {author} {\bibinfo {author} {\bibfnamefont {P.}~\bibnamefont
  {Bari}}, \bibinfo {author} {\bibfnamefont {N.}~\bibnamefont {Bartolo}},
  \bibinfo {author} {\bibfnamefont {G.}~\bibnamefont {Dom\`enech}}, \ and\
  \bibinfo {author} {\bibfnamefont {S.}~\bibnamefont {Matarrese}},\ }\href
  {\doibase 10.1103/PhysRevD.109.023509} {\bibfield  {journal} {\bibinfo
  {journal} {Phys. Rev. D}\ }\textbf {\bibinfo {volume} {109}},\ \bibinfo
  {pages} {023509} (\bibinfo {year} {2024})},\ \Eprint
  {http://arxiv.org/abs/2307.05404} {arXiv:2307.05404 [astro-ph.CO]}
  \BibitemShut {NoStop}%
\bibitem [{\citenamefont {Picard}\ and\ \citenamefont
  {Malik}(2023)}]{Picard:2023sbz}%
  \BibitemOpen
  \bibfield  {author} {\bibinfo {author} {\bibfnamefont {R.}~\bibnamefont
  {Picard}}\ and\ \bibinfo {author} {\bibfnamefont {K.~A.}\ \bibnamefont
  {Malik}},\ }\href@noop {} {\  (\bibinfo {year} {2023})},\ \Eprint
  {http://arxiv.org/abs/2311.14513} {arXiv:2311.14513 [astro-ph.CO]}
  \BibitemShut {NoStop}%
\bibitem [{\citenamefont {Hwang}\ \emph {et~al.}(2017)\citenamefont {Hwang},
  \citenamefont {Jeong},\ and\ \citenamefont {Noh}}]{Hwang:2017oxa}%
  \BibitemOpen
  \bibfield  {author} {\bibinfo {author} {\bibfnamefont {J.-C.}\ \bibnamefont
  {Hwang}}, \bibinfo {author} {\bibfnamefont {D.}~\bibnamefont {Jeong}}, \ and\
  \bibinfo {author} {\bibfnamefont {H.}~\bibnamefont {Noh}},\ }\href {\doibase
  10.3847/1538-4357/aa74be} {\bibfield  {journal} {\bibinfo  {journal}
  {Astrophys. J.}\ }\textbf {\bibinfo {volume} {842}},\ \bibinfo {pages} {46}
  (\bibinfo {year} {2017})},\ \Eprint {http://arxiv.org/abs/1704.03500}
  {arXiv:1704.03500 [astro-ph.CO]} \BibitemShut {NoStop}%
\bibitem [{\citenamefont {Wang}\ and\ \citenamefont
  {Zhang}(2019)}]{Wang:2019zhj}%
  \BibitemOpen
  \bibfield  {author} {\bibinfo {author} {\bibfnamefont {B.}~\bibnamefont
  {Wang}}\ and\ \bibinfo {author} {\bibfnamefont {Y.}~\bibnamefont {Zhang}},\
  }\href {\doibase 10.1103/PhysRevD.99.123008} {\bibfield  {journal} {\bibinfo
  {journal} {Phys. Rev. D}\ }\textbf {\bibinfo {volume} {99}},\ \bibinfo
  {pages} {123008} (\bibinfo {year} {2019})},\ \Eprint
  {http://arxiv.org/abs/1905.03272} {arXiv:1905.03272 [gr-qc]} \BibitemShut
  {NoStop}%
\bibitem [{\citenamefont {Gong}(2019)}]{Gong:2019mui}%
  \BibitemOpen
  \bibfield  {author} {\bibinfo {author} {\bibfnamefont {J.-O.}\ \bibnamefont
  {Gong}},\ }\href@noop {} {\  (\bibinfo {year} {2019})},\ \Eprint
  {http://arxiv.org/abs/1909.12708} {arXiv:1909.12708 [gr-qc]} \BibitemShut
  {NoStop}%
\bibitem [{\citenamefont {Tomikawa}\ and\ \citenamefont
  {Kobayashi}(2020)}]{Tomikawa:2019tvi}%
  \BibitemOpen
  \bibfield  {author} {\bibinfo {author} {\bibfnamefont {K.}~\bibnamefont
  {Tomikawa}}\ and\ \bibinfo {author} {\bibfnamefont {T.}~\bibnamefont
  {Kobayashi}},\ }\href {\doibase 10.1103/PhysRevD.101.083529} {\bibfield
  {journal} {\bibinfo  {journal} {Phys. Rev. D}\ }\textbf {\bibinfo {volume}
  {101}},\ \bibinfo {pages} {083529} (\bibinfo {year} {2020})},\ \Eprint
  {http://arxiv.org/abs/1910.01880} {arXiv:1910.01880 [gr-qc]} \BibitemShut
  {NoStop}%
\bibitem [{\citenamefont {De~Luca}\ \emph {et~al.}(2020)\citenamefont
  {De~Luca}, \citenamefont {Franciolini}, \citenamefont {Kehagias},\ and\
  \citenamefont {Riotto}}]{DeLuca:2019ufz}%
  \BibitemOpen
  \bibfield  {author} {\bibinfo {author} {\bibfnamefont {V.}~\bibnamefont
  {De~Luca}}, \bibinfo {author} {\bibfnamefont {G.}~\bibnamefont
  {Franciolini}}, \bibinfo {author} {\bibfnamefont {A.}~\bibnamefont
  {Kehagias}}, \ and\ \bibinfo {author} {\bibfnamefont {A.}~\bibnamefont
  {Riotto}},\ }\href {\doibase 10.1088/1475-7516/2020/03/014} {\bibfield
  {journal} {\bibinfo  {journal} {JCAP}\ }\textbf {\bibinfo {volume} {03}},\
  \bibinfo {pages} {014} (\bibinfo {year} {2020})},\ \Eprint
  {http://arxiv.org/abs/1911.09689} {arXiv:1911.09689 [gr-qc]} \BibitemShut
  {NoStop}%
\bibitem [{\citenamefont {Inomata}\ and\ \citenamefont
  {Terada}(2020)}]{Inomata:2019yww}%
  \BibitemOpen
  \bibfield  {author} {\bibinfo {author} {\bibfnamefont {K.}~\bibnamefont
  {Inomata}}\ and\ \bibinfo {author} {\bibfnamefont {T.}~\bibnamefont
  {Terada}},\ }\href {\doibase 10.1103/PhysRevD.101.023523} {\bibfield
  {journal} {\bibinfo  {journal} {Phys. Rev. D}\ }\textbf {\bibinfo {volume}
  {101}},\ \bibinfo {pages} {023523} (\bibinfo {year} {2020})},\ \Eprint
  {http://arxiv.org/abs/1912.00785} {arXiv:1912.00785 [gr-qc]} \BibitemShut
  {NoStop}%
\bibitem [{\citenamefont {Yuan}\ \emph {et~al.}(2020)\citenamefont {Yuan},
  \citenamefont {Chen},\ and\ \citenamefont {Huang}}]{Yuan:2019fwv}%
  \BibitemOpen
  \bibfield  {author} {\bibinfo {author} {\bibfnamefont {C.}~\bibnamefont
  {Yuan}}, \bibinfo {author} {\bibfnamefont {Z.-C.}\ \bibnamefont {Chen}}, \
  and\ \bibinfo {author} {\bibfnamefont {Q.-G.}\ \bibnamefont {Huang}},\ }\href
  {\doibase 10.1103/PhysRevD.101.063018} {\bibfield  {journal} {\bibinfo
  {journal} {Phys. Rev. D}\ }\textbf {\bibinfo {volume} {101}},\ \bibinfo
  {pages} {063018} (\bibinfo {year} {2020})},\ \Eprint
  {http://arxiv.org/abs/1912.00885} {arXiv:1912.00885 [astro-ph.CO]}
  \BibitemShut {NoStop}%
\bibitem [{\citenamefont {Nakamura}(2020)}]{Nakamura:2019zbe}%
  \BibitemOpen
  \bibfield  {author} {\bibinfo {author} {\bibfnamefont {K.}~\bibnamefont
  {Nakamura}},\ }\href {\doibase 10.9734/bpi/taps/v3} {\  (\bibinfo {year}
  {2020}),\ 10.9734/bpi/taps/v3},\ \Eprint {http://arxiv.org/abs/1912.12805}
  {arXiv:1912.12805 [gr-qc]} \BibitemShut {NoStop}%
\bibitem [{\citenamefont {Lu}\ \emph {et~al.}(2020{\natexlab{b}})\citenamefont
  {Lu}, \citenamefont {Ali}, \citenamefont {Gong}, \citenamefont {Lin},\ and\
  \citenamefont {Zhang}}]{Lu:2020diy}%
  \BibitemOpen
  \bibfield  {author} {\bibinfo {author} {\bibfnamefont {Y.}~\bibnamefont
  {Lu}}, \bibinfo {author} {\bibfnamefont {A.}~\bibnamefont {Ali}}, \bibinfo
  {author} {\bibfnamefont {Y.}~\bibnamefont {Gong}}, \bibinfo {author}
  {\bibfnamefont {J.}~\bibnamefont {Lin}}, \ and\ \bibinfo {author}
  {\bibfnamefont {F.}~\bibnamefont {Zhang}},\ }\href {\doibase
  10.1103/PhysRevD.102.083503(2020)} {\bibfield  {journal} {\bibinfo  {journal}
  {Phys. Rev. D}\ }\textbf {\bibinfo {volume} {102}},\ \bibinfo {pages}
  {083503} (\bibinfo {year} {2020}{\natexlab{b}})},\ \Eprint
  {http://arxiv.org/abs/2006.03450} {arXiv:2006.03450 [gr-qc]} \BibitemShut
  {NoStop}%
\bibitem [{\citenamefont {Ali}\ \emph {et~al.}(2021)\citenamefont {Ali},
  \citenamefont {Gong},\ and\ \citenamefont {Lu}}]{Ali:2020sfw}%
  \BibitemOpen
  \bibfield  {author} {\bibinfo {author} {\bibfnamefont {A.}~\bibnamefont
  {Ali}}, \bibinfo {author} {\bibfnamefont {Y.}~\bibnamefont {Gong}}, \ and\
  \bibinfo {author} {\bibfnamefont {Y.}~\bibnamefont {Lu}},\ }\href {\doibase
  10.1103/PhysRevD.103.043516} {\bibfield  {journal} {\bibinfo  {journal}
  {Phys. Rev. D}\ }\textbf {\bibinfo {volume} {103}},\ \bibinfo {pages}
  {043516} (\bibinfo {year} {2021})},\ \Eprint
  {http://arxiv.org/abs/2009.11081} {arXiv:2009.11081 [gr-qc]} \BibitemShut
  {NoStop}%
\bibitem [{\citenamefont
  {Giovannini}(2020{\natexlab{a}})}]{Giovannini:2020qta}%
  \BibitemOpen
  \bibfield  {author} {\bibinfo {author} {\bibfnamefont {M.}~\bibnamefont
  {Giovannini}},\ }\href {\doibase 10.1142/S0217751X20501651} {\bibfield
  {journal} {\bibinfo  {journal} {Int. J. Mod. Phys.}\ }\textbf {\bibinfo
  {volume} {A35}},\ \bibinfo {pages} {2050165} (\bibinfo {year}
  {2020}{\natexlab{a}})},\ \Eprint {http://arxiv.org/abs/2005.04962}
  {arXiv:2005.04962 [hep-th]} \BibitemShut {NoStop}%
\bibitem [{\citenamefont
  {Giovannini}(2020{\natexlab{b}})}]{Giovannini:2020soq}%
  \BibitemOpen
  \bibfield  {author} {\bibinfo {author} {\bibfnamefont {M.}~\bibnamefont
  {Giovannini}},\ }\href {\doibase 10.1142/S0218271820501126} {\bibfield
  {journal} {\bibinfo  {journal} {Int. J. Mod. Phys.}\ }\textbf {\bibinfo
  {volume} {D29}},\ \bibinfo {pages} {2050112} (\bibinfo {year}
  {2020}{\natexlab{b}})},\ \Eprint {http://arxiv.org/abs/2007.14956}
  {arXiv:2007.14956 [hep-th]} \BibitemShut {NoStop}%
\bibitem [{\citenamefont {Chang}\ \emph
  {et~al.}(2020{\natexlab{a}})\citenamefont {Chang}, \citenamefont {Wang},\
  and\ \citenamefont {Zhu}}]{Chang:2020tji}%
  \BibitemOpen
  \bibfield  {author} {\bibinfo {author} {\bibfnamefont {Z.}~\bibnamefont
  {Chang}}, \bibinfo {author} {\bibfnamefont {S.}~\bibnamefont {Wang}}, \ and\
  \bibinfo {author} {\bibfnamefont {Q.-H.}\ \bibnamefont {Zhu}},\ }\href@noop
  {} {\  (\bibinfo {year} {2020}{\natexlab{a}})},\ \Eprint
  {http://arxiv.org/abs/2009.11025} {arXiv:2009.11025 [astro-ph.CO]}
  \BibitemShut {NoStop}%
\bibitem [{\citenamefont {Chang}\ \emph
  {et~al.}(2020{\natexlab{b}})\citenamefont {Chang}, \citenamefont {Wang},\
  and\ \citenamefont {Zhu}}]{Chang:2020iji}%
  \BibitemOpen
  \bibfield  {author} {\bibinfo {author} {\bibfnamefont {Z.}~\bibnamefont
  {Chang}}, \bibinfo {author} {\bibfnamefont {S.}~\bibnamefont {Wang}}, \ and\
  \bibinfo {author} {\bibfnamefont {Q.-H.}\ \bibnamefont {Zhu}},\ }\href@noop
  {} {\  (\bibinfo {year} {2020}{\natexlab{b}})},\ \Eprint
  {http://arxiv.org/abs/2009.11994} {arXiv:2009.11994 [gr-qc]} \BibitemShut
  {NoStop}%
\bibitem [{\citenamefont {Chang}\ \emph
  {et~al.}(2020{\natexlab{c}})\citenamefont {Chang}, \citenamefont {Wang},\
  and\ \citenamefont {Zhu}}]{Chang:2020mky}%
  \BibitemOpen
  \bibfield  {author} {\bibinfo {author} {\bibfnamefont {Z.}~\bibnamefont
  {Chang}}, \bibinfo {author} {\bibfnamefont {S.}~\bibnamefont {Wang}}, \ and\
  \bibinfo {author} {\bibfnamefont {Q.-H.}\ \bibnamefont {Zhu}},\ }\href@noop
  {} {\  (\bibinfo {year} {2020}{\natexlab{c}})},\ \Eprint
  {http://arxiv.org/abs/2010.01487} {arXiv:2010.01487 [gr-qc]} \BibitemShut
  {NoStop}%
\bibitem [{\citenamefont {Dom\`{e}nech}\ and\ \citenamefont
  {Sasaki}(2021)}]{Domenech:2020xin}%
  \BibitemOpen
  \bibfield  {author} {\bibinfo {author} {\bibfnamefont {G.}~\bibnamefont
  {Dom\`{e}nech}}\ and\ \bibinfo {author} {\bibfnamefont {M.}~\bibnamefont
  {Sasaki}},\ }\href {\doibase 10.1103/PhysRevD.103.063531} {\bibfield
  {journal} {\bibinfo  {journal} {Phys. Rev.}\ }\textbf {\bibinfo {volume}
  {D103}},\ \bibinfo {pages} {063531} (\bibinfo {year} {2021})},\ \Eprint
  {http://arxiv.org/abs/2012.14016} {arXiv:2012.14016 [gr-qc]} \BibitemShut
  {NoStop}%
\bibitem [{\citenamefont {Dom\`enech}(2021)}]{Domenech:2021ztg}%
  \BibitemOpen
  \bibfield  {author} {\bibinfo {author} {\bibfnamefont {G.}~\bibnamefont
  {Dom\`enech}},\ }\href {\doibase 10.3390/universe7110398} {\bibfield
  {journal} {\bibinfo  {journal} {Universe}\ }\textbf {\bibinfo {volume} {7}},\
  \bibinfo {pages} {398} (\bibinfo {year} {2021})},\ \Eprint
  {http://arxiv.org/abs/2109.01398} {arXiv:2109.01398 [gr-qc]} \BibitemShut
  {NoStop}%
\bibitem [{\citenamefont {Dom\`enech}(2024)}]{Domenech:2024kmh}%
  \BibitemOpen
  \bibfield  {author} {\bibinfo {author} {\bibfnamefont {G.}~\bibnamefont
  {Dom\`enech}},\ }\href@noop {} {\  (\bibinfo {year} {2024})},\ \Eprint
  {http://arxiv.org/abs/2402.17388} {arXiv:2402.17388 [gr-qc]} \BibitemShut
  {NoStop}%
\bibitem [{\citenamefont {Kohri}\ and\ \citenamefont
  {Terada}(2018)}]{Kohri:2018awv}%
  \BibitemOpen
  \bibfield  {author} {\bibinfo {author} {\bibfnamefont {K.}~\bibnamefont
  {Kohri}}\ and\ \bibinfo {author} {\bibfnamefont {T.}~\bibnamefont {Terada}},\
  }\href {\doibase 10.1103/PhysRevD.97.123532} {\bibfield  {journal} {\bibinfo
  {journal} {Phys. Rev. D}\ }\textbf {\bibinfo {volume} {97}},\ \bibinfo
  {pages} {123532} (\bibinfo {year} {2018})},\ \Eprint
  {http://arxiv.org/abs/1804.08577} {arXiv:1804.08577 [gr-qc]} \BibitemShut
  {NoStop}%
\bibitem [{\citenamefont {Pi}\ and\ \citenamefont {Sasaki}(2020)}]{Pi:2020otn}%
  \BibitemOpen
  \bibfield  {author} {\bibinfo {author} {\bibfnamefont {S.}~\bibnamefont
  {Pi}}\ and\ \bibinfo {author} {\bibfnamefont {M.}~\bibnamefont {Sasaki}},\
  }\href {\doibase 10.1088/1475-7516/2020/09/037} {\bibfield  {journal}
  {\bibinfo  {journal} {JCAP}\ }\textbf {\bibinfo {volume} {09}},\ \bibinfo
  {pages} {037} (\bibinfo {year} {2020})},\ \Eprint
  {http://arxiv.org/abs/2005.12306} {arXiv:2005.12306 [gr-qc]} \BibitemShut
  {NoStop}%
\bibitem [{\citenamefont {Li}\ \emph {et~al.}(2024{\natexlab{b}})\citenamefont
  {Li}, \citenamefont {Yuan},\ and\ \citenamefont {Huang}}]{Li:2024lxx}%
  \BibitemOpen
  \bibfield  {author} {\bibinfo {author} {\bibfnamefont {C.-Z.}\ \bibnamefont
  {Li}}, \bibinfo {author} {\bibfnamefont {C.}~\bibnamefont {Yuan}}, \ and\
  \bibinfo {author} {\bibfnamefont {Q.-g.}\ \bibnamefont {Huang}},\ }\href@noop
  {} {\  (\bibinfo {year} {2024}{\natexlab{b}})},\ \Eprint
  {http://arxiv.org/abs/2407.12914} {arXiv:2407.12914 [gr-qc]} \BibitemShut
  {NoStop}%
\bibitem [{\citenamefont {Cai}\ \emph {et~al.}(2020{\natexlab{c}})\citenamefont
  {Cai}, \citenamefont {Pi},\ and\ \citenamefont {Sasaki}}]{Cai:2019cdl}%
  \BibitemOpen
  \bibfield  {author} {\bibinfo {author} {\bibfnamefont {R.-G.}\ \bibnamefont
  {Cai}}, \bibinfo {author} {\bibfnamefont {S.}~\bibnamefont {Pi}}, \ and\
  \bibinfo {author} {\bibfnamefont {M.}~\bibnamefont {Sasaki}},\ }\href
  {\doibase 10.1103/PhysRevD.102.083528} {\bibfield  {journal} {\bibinfo
  {journal} {Phys. Rev.}\ }\textbf {\bibinfo {volume} {D102}},\ \bibinfo
  {pages} {083528} (\bibinfo {year} {2020}{\natexlab{c}})},\ \Eprint
  {http://arxiv.org/abs/1909.13728} {arXiv:1909.13728 [astro-ph.CO]}
  \BibitemShut {NoStop}%
\bibitem [{\citenamefont {Dom\`enech}(2020)}]{Domenech:2019quo}%
  \BibitemOpen
  \bibfield  {author} {\bibinfo {author} {\bibfnamefont {G.}~\bibnamefont
  {Dom\`enech}},\ }\href {\doibase 10.1142/S0218271820500285} {\bibfield
  {journal} {\bibinfo  {journal} {Int. J. Mod. Phys. D}\ }\textbf {\bibinfo
  {volume} {29}},\ \bibinfo {pages} {2050028} (\bibinfo {year} {2020})},\
  \Eprint {http://arxiv.org/abs/1912.05583} {arXiv:1912.05583 [gr-qc]}
  \BibitemShut {NoStop}%
\bibitem [{\citenamefont {Dom\`enech}\ \emph {et~al.}(2020)\citenamefont
  {Dom\`enech}, \citenamefont {Pi},\ and\ \citenamefont
  {Sasaki}}]{Domenech:2020kqm}%
  \BibitemOpen
  \bibfield  {author} {\bibinfo {author} {\bibfnamefont {G.}~\bibnamefont
  {Dom\`enech}}, \bibinfo {author} {\bibfnamefont {S.}~\bibnamefont {Pi}}, \
  and\ \bibinfo {author} {\bibfnamefont {M.}~\bibnamefont {Sasaki}},\ }\href
  {\doibase 10.1088/1475-7516/2020/08/017} {\bibfield  {journal} {\bibinfo
  {journal} {JCAP}\ }\textbf {\bibinfo {volume} {08}},\ \bibinfo {pages} {017}
  (\bibinfo {year} {2020})},\ \Eprint {http://arxiv.org/abs/2005.12314}
  {arXiv:2005.12314 [gr-qc]} \BibitemShut {NoStop}%
\bibitem [{\citenamefont {Dalianis}\ and\ \citenamefont
  {Kritos}(2021)}]{Dalianis:2020cla}%
  \BibitemOpen
  \bibfield  {author} {\bibinfo {author} {\bibfnamefont {I.}~\bibnamefont
  {Dalianis}}\ and\ \bibinfo {author} {\bibfnamefont {K.}~\bibnamefont
  {Kritos}},\ }\href {\doibase 10.1103/PhysRevD.103.023505} {\bibfield
  {journal} {\bibinfo  {journal} {Phys. Rev. D}\ }\textbf {\bibinfo {volume}
  {103}},\ \bibinfo {pages} {023505} (\bibinfo {year} {2021})},\ \Eprint
  {http://arxiv.org/abs/2007.07915} {arXiv:2007.07915 [astro-ph.CO]}
  \BibitemShut {NoStop}%
\bibitem [{\citenamefont {Hook}\ \emph {et~al.}(2020)\citenamefont {Hook},
  \citenamefont {Marques-Tavares},\ and\ \citenamefont {Racco}}]{Hook:2020phx}%
  \BibitemOpen
  \bibfield  {author} {\bibinfo {author} {\bibfnamefont {A.}~\bibnamefont
  {Hook}}, \bibinfo {author} {\bibfnamefont {G.}~\bibnamefont
  {Marques-Tavares}}, \ and\ \bibinfo {author} {\bibfnamefont {D.}~\bibnamefont
  {Racco}},\ }\href@noop {} {\  (\bibinfo {year} {2020})},\ \Eprint
  {http://arxiv.org/abs/2010.03568} {arXiv:2010.03568 [hep-ph]} \BibitemShut
  {NoStop}%
\bibitem [{\citenamefont {Brzeminski}\ \emph {et~al.}(2022)\citenamefont
  {Brzeminski}, \citenamefont {Hook},\ and\ \citenamefont
  {Marques-Tavares}}]{Brzeminski:2022haa}%
  \BibitemOpen
  \bibfield  {author} {\bibinfo {author} {\bibfnamefont {D.}~\bibnamefont
  {Brzeminski}}, \bibinfo {author} {\bibfnamefont {A.}~\bibnamefont {Hook}}, \
  and\ \bibinfo {author} {\bibfnamefont {G.}~\bibnamefont {Marques-Tavares}},\
  }\href {\doibase 10.1007/JHEP11(2022)061} {\bibfield  {journal} {\bibinfo
  {journal} {JHEP}\ }\textbf {\bibinfo {volume} {11}},\ \bibinfo {pages} {061}
  (\bibinfo {year} {2022})},\ \Eprint {http://arxiv.org/abs/2203.13842}
  {arXiv:2203.13842 [hep-ph]} \BibitemShut {NoStop}%
\bibitem [{\citenamefont {Franciolini}\ \emph {et~al.}(2024)\citenamefont
  {Franciolini}, \citenamefont {Racco},\ and\ \citenamefont
  {Rompineve}}]{Franciolini:2023wjm}%
  \BibitemOpen
  \bibfield  {author} {\bibinfo {author} {\bibfnamefont {G.}~\bibnamefont
  {Franciolini}}, \bibinfo {author} {\bibfnamefont {D.}~\bibnamefont {Racco}},
  \ and\ \bibinfo {author} {\bibfnamefont {F.}~\bibnamefont {Rompineve}},\
  }\href {\doibase 10.1103/PhysRevLett.132.081001} {\bibfield  {journal}
  {\bibinfo  {journal} {Phys. Rev. Lett.}\ }\textbf {\bibinfo {volume} {132}},\
  \bibinfo {pages} {081001} (\bibinfo {year} {2024})},\ \bibinfo {note}
  {[Erratum: Phys.Rev.Lett. 133, 189901 (2024)]},\ \Eprint
  {http://arxiv.org/abs/2306.17136} {arXiv:2306.17136 [astro-ph.CO]}
  \BibitemShut {NoStop}%
\bibitem [{\citenamefont {Dom\`enech}\ and\ \citenamefont
  {Tr\"ankle}(2024)}]{Domenech:2024wao}%
  \BibitemOpen
  \bibfield  {author} {\bibinfo {author} {\bibfnamefont {G.}~\bibnamefont
  {Dom\`enech}}\ and\ \bibinfo {author} {\bibfnamefont {J.}~\bibnamefont
  {Tr\"ankle}},\ }\href@noop {} {\  (\bibinfo {year} {2024})},\ \Eprint
  {http://arxiv.org/abs/2409.12125} {arXiv:2409.12125 [gr-qc]} \BibitemShut
  {NoStop}%
\bibitem [{\citenamefont {Ajith}\ \emph {et~al.}(2011)\citenamefont {Ajith}
  \emph {et~al.}}]{Ajith:2009bn}%
  \BibitemOpen
  \bibfield  {author} {\bibinfo {author} {\bibfnamefont {P.}~\bibnamefont
  {Ajith}} \emph {et~al.},\ }\href {\doibase 10.1103/PhysRevLett.106.241101}
  {\bibfield  {journal} {\bibinfo  {journal} {Phys. Rev. Lett.}\ }\textbf
  {\bibinfo {volume} {106}},\ \bibinfo {pages} {241101} (\bibinfo {year}
  {2011})},\ \Eprint {http://arxiv.org/abs/0909.2867} {arXiv:0909.2867 [gr-qc]}
  \BibitemShut {NoStop}%
\bibitem [{\citenamefont {Zhu}\ \emph {et~al.}(2011)\citenamefont {Zhu},
  \citenamefont {Howell}, \citenamefont {Regimbau}, \citenamefont {Blair},\
  and\ \citenamefont {Zhu}}]{Zhu:2011bd}%
  \BibitemOpen
  \bibfield  {author} {\bibinfo {author} {\bibfnamefont {X.-J.}\ \bibnamefont
  {Zhu}}, \bibinfo {author} {\bibfnamefont {E.}~\bibnamefont {Howell}},
  \bibinfo {author} {\bibfnamefont {T.}~\bibnamefont {Regimbau}}, \bibinfo
  {author} {\bibfnamefont {D.}~\bibnamefont {Blair}}, \ and\ \bibinfo {author}
  {\bibfnamefont {Z.-H.}\ \bibnamefont {Zhu}},\ }\href {\doibase
  10.1088/0004-637X/739/2/86} {\bibfield  {journal} {\bibinfo  {journal}
  {Astrophys. J.}\ }\textbf {\bibinfo {volume} {739}},\ \bibinfo {pages} {86}
  (\bibinfo {year} {2011})},\ \Eprint {http://arxiv.org/abs/1104.3565}
  {arXiv:1104.3565 [gr-qc]} \BibitemShut {NoStop}%
\bibitem [{\citenamefont {Zhu}\ \emph {et~al.}(2013)\citenamefont {Zhu},
  \citenamefont {Howell}, \citenamefont {Blair},\ and\ \citenamefont
  {Zhu}}]{Zhu:2012xw}%
  \BibitemOpen
  \bibfield  {author} {\bibinfo {author} {\bibfnamefont {X.-J.}\ \bibnamefont
  {Zhu}}, \bibinfo {author} {\bibfnamefont {E.~J.}\ \bibnamefont {Howell}},
  \bibinfo {author} {\bibfnamefont {D.~G.}\ \bibnamefont {Blair}}, \ and\
  \bibinfo {author} {\bibfnamefont {Z.-H.}\ \bibnamefont {Zhu}},\ }\href
  {\doibase 10.1093/mnras/stt207} {\bibfield  {journal} {\bibinfo  {journal}
  {Mon. Not. Roy. Astron. Soc.}\ }\textbf {\bibinfo {volume} {431}},\ \bibinfo
  {pages} {882} (\bibinfo {year} {2013})},\ \Eprint
  {http://arxiv.org/abs/1209.0595} {arXiv:1209.0595 [gr-qc]} \BibitemShut
  {NoStop}%
\bibitem [{\citenamefont {Agazie}\ \emph
  {et~al.}(2023{\natexlab{b}})\citenamefont {Agazie} \emph
  {et~al.}}]{NANOGrav:2023hde}%
  \BibitemOpen
  \bibfield  {author} {\bibinfo {author} {\bibfnamefont {G.}~\bibnamefont
  {Agazie}} \emph {et~al.} (\bibinfo {collaboration} {NANOGrav}),\ }\href
  {\doibase 10.3847/2041-8213/acda9a} {\bibfield  {journal} {\bibinfo
  {journal} {Astrophys. J. Lett.}\ }\textbf {\bibinfo {volume} {951}},\
  \bibinfo {pages} {L9} (\bibinfo {year} {2023}{\natexlab{b}})},\ \Eprint
  {http://arxiv.org/abs/2306.16217} {arXiv:2306.16217 [astro-ph.HE]}
  \BibitemShut {NoStop}%
\bibitem [{\citenamefont {Antoniadis}\ \emph
  {et~al.}(2023{\natexlab{a}})\citenamefont {Antoniadis} \emph
  {et~al.}}]{EPTA:2023fyk}%
  \BibitemOpen
  \bibfield  {author} {\bibinfo {author} {\bibfnamefont {J.}~\bibnamefont
  {Antoniadis}} \emph {et~al.} (\bibinfo {collaboration} {EPTA}),\ }\href
  {\doibase 10.1051/0004-6361/202346844} {\bibfield  {journal} {\bibinfo
  {journal} {Astron. Astrophys.}\ }\textbf {\bibinfo {volume} {678}},\ \bibinfo
  {pages} {A50} (\bibinfo {year} {2023}{\natexlab{a}})},\ \Eprint
  {http://arxiv.org/abs/2306.16214} {arXiv:2306.16214 [astro-ph.HE]}
  \BibitemShut {NoStop}%
\bibitem [{\citenamefont {Antoniadis}\ \emph
  {et~al.}(2023{\natexlab{b}})\citenamefont {Antoniadis} \emph
  {et~al.}}]{EPTA:2023sfo}%
  \BibitemOpen
  \bibfield  {author} {\bibinfo {author} {\bibfnamefont {J.}~\bibnamefont
  {Antoniadis}} \emph {et~al.} (\bibinfo {collaboration} {EPTA}),\ }\href
  {\doibase 10.1051/0004-6361/202346841} {\  (\bibinfo {year}
  {2023}{\natexlab{b}}),\ 10.1051/0004-6361/202346841},\ \Eprint
  {http://arxiv.org/abs/2306.16224} {arXiv:2306.16224 [astro-ph.HE]}
  \BibitemShut {NoStop}%
\bibitem [{\citenamefont {Antoniadis}\ \emph
  {et~al.}(2023{\natexlab{c}})\citenamefont {Antoniadis} \emph
  {et~al.}}]{EPTA:2023xxk}%
  \BibitemOpen
  \bibfield  {author} {\bibinfo {author} {\bibfnamefont {J.}~\bibnamefont
  {Antoniadis}} \emph {et~al.} (\bibinfo {collaboration} {EPTA}),\ }\href@noop
  {} {\  (\bibinfo {year} {2023}{\natexlab{c}})},\ \Eprint
  {http://arxiv.org/abs/2306.16227} {arXiv:2306.16227 [astro-ph.CO]}
  \BibitemShut {NoStop}%
\bibitem [{\citenamefont {Zic}\ \emph {et~al.}(2023)\citenamefont {Zic} \emph
  {et~al.}}]{Zic:2023gta}%
  \BibitemOpen
  \bibfield  {author} {\bibinfo {author} {\bibfnamefont {A.}~\bibnamefont
  {Zic}} \emph {et~al.},\ }\href@noop {} {\  (\bibinfo {year} {2023})},\
  \Eprint {http://arxiv.org/abs/2306.16230} {arXiv:2306.16230 [astro-ph.HE]}
  \BibitemShut {NoStop}%
\bibitem [{\citenamefont {Reardon}\ \emph
  {et~al.}(2023{\natexlab{a}})\citenamefont {Reardon} \emph
  {et~al.}}]{Reardon:2023gzh}%
  \BibitemOpen
  \bibfield  {author} {\bibinfo {author} {\bibfnamefont {D.~J.}\ \bibnamefont
  {Reardon}} \emph {et~al.},\ }\href {\doibase 10.3847/2041-8213/acdd02}
  {\bibfield  {journal} {\bibinfo  {journal} {Astrophys. J. Lett.}\ }\textbf
  {\bibinfo {volume} {951}},\ \bibinfo {pages} {L6} (\bibinfo {year}
  {2023}{\natexlab{a}})},\ \Eprint {http://arxiv.org/abs/2306.16215}
  {arXiv:2306.16215 [astro-ph.HE]} \BibitemShut {NoStop}%
\bibitem [{\citenamefont {Reardon}\ \emph
  {et~al.}(2023{\natexlab{b}})\citenamefont {Reardon} \emph
  {et~al.}}]{Reardon:2023zen}%
  \BibitemOpen
  \bibfield  {author} {\bibinfo {author} {\bibfnamefont {D.~J.}\ \bibnamefont
  {Reardon}} \emph {et~al.},\ }\href {\doibase 10.3847/2041-8213/acdd03}
  {\bibfield  {journal} {\bibinfo  {journal} {Astrophys. J. Lett.}\ }\textbf
  {\bibinfo {volume} {951}},\ \bibinfo {pages} {L7} (\bibinfo {year}
  {2023}{\natexlab{b}})},\ \Eprint {http://arxiv.org/abs/2306.16229}
  {arXiv:2306.16229 [astro-ph.HE]} \BibitemShut {NoStop}%
\bibitem [{\citenamefont {Agazie}\ \emph
  {et~al.}(2023{\natexlab{c}})\citenamefont {Agazie} \emph
  {et~al.}}]{InternationalPulsarTimingArray:2023mzf}%
  \BibitemOpen
  \bibfield  {author} {\bibinfo {author} {\bibfnamefont {G.}~\bibnamefont
  {Agazie}} \emph {et~al.} (\bibinfo {collaboration} {International Pulsar
  Timing Array}),\ }\href@noop {} {\  (\bibinfo {year} {2023}{\natexlab{c}})},\
  \Eprint {http://arxiv.org/abs/2309.00693} {arXiv:2309.00693 [astro-ph.HE]}
  \BibitemShut {NoStop}%
\bibitem [{\citenamefont {Miles}\ \emph
  {et~al.}(2024{\natexlab{a}})\citenamefont {Miles} \emph
  {et~al.}}]{Miles:2024seg}%
  \BibitemOpen
  \bibfield  {author} {\bibinfo {author} {\bibfnamefont {M.~T.}\ \bibnamefont
  {Miles}} \emph {et~al.},\ }\href {\doibase 10.1093/mnras/stae2571} {\
  (\bibinfo {year} {2024}{\natexlab{a}}),\ 10.1093/mnras/stae2571},\ \Eprint
  {http://arxiv.org/abs/2412.01153} {arXiv:2412.01153 [astro-ph.HE]}
  \BibitemShut {NoStop}%
\bibitem [{\citenamefont {Miles}\ \emph
  {et~al.}(2024{\natexlab{b}})\citenamefont {Miles} \emph
  {et~al.}}]{Miles:2024rjc}%
  \BibitemOpen
  \bibfield  {author} {\bibinfo {author} {\bibfnamefont {M.~T.}\ \bibnamefont
  {Miles}} \emph {et~al.},\ }\href {\doibase 10.1093/mnras/stae2572} {\
  (\bibinfo {year} {2024}{\natexlab{b}}),\ 10.1093/mnras/stae2572},\ \Eprint
  {http://arxiv.org/abs/2412.01148} {arXiv:2412.01148 [astro-ph.HE]}
  \BibitemShut {NoStop}%
\bibitem [{\citenamefont {Afzal}\ \emph {et~al.}(2023)\citenamefont {Afzal}
  \emph {et~al.}}]{NANOGrav:2023hvm}%
  \BibitemOpen
  \bibfield  {author} {\bibinfo {author} {\bibfnamefont {A.}~\bibnamefont
  {Afzal}} \emph {et~al.} (\bibinfo {collaboration} {NANOGrav}),\ }\href
  {\doibase 10.3847/2041-8213/acdc91} {\bibfield  {journal} {\bibinfo
  {journal} {Astrophys. J. Lett.}\ }\textbf {\bibinfo {volume} {951}},\
  \bibinfo {pages} {L11} (\bibinfo {year} {2023})},\ \Eprint
  {http://arxiv.org/abs/2306.16219} {arXiv:2306.16219 [astro-ph.HE]}
  \BibitemShut {NoStop}%
\bibitem [{\citenamefont {Cai}\ \emph {et~al.}(2019{\natexlab{c}})\citenamefont
  {Cai}, \citenamefont {Pi}, \citenamefont {Wang},\ and\ \citenamefont
  {Yang}}]{Cai:2019elf}%
  \BibitemOpen
  \bibfield  {author} {\bibinfo {author} {\bibfnamefont {R.-G.}\ \bibnamefont
  {Cai}}, \bibinfo {author} {\bibfnamefont {S.}~\bibnamefont {Pi}}, \bibinfo
  {author} {\bibfnamefont {S.-J.}\ \bibnamefont {Wang}}, \ and\ \bibinfo
  {author} {\bibfnamefont {X.-Y.}\ \bibnamefont {Yang}},\ }\href {\doibase
  10.1088/1475-7516/2019/10/059} {\bibfield  {journal} {\bibinfo  {journal}
  {JCAP}\ }\textbf {\bibinfo {volume} {10}},\ \bibinfo {pages} {059} (\bibinfo
  {year} {2019}{\natexlab{c}})},\ \Eprint {http://arxiv.org/abs/1907.06372}
  {arXiv:1907.06372 [astro-ph.CO]} \BibitemShut {NoStop}%
\bibitem [{\citenamefont {Pi}\ and\ \citenamefont {Sasaki}(2023)}]{Pi:2022ysn}%
  \BibitemOpen
  \bibfield  {author} {\bibinfo {author} {\bibfnamefont {S.}~\bibnamefont
  {Pi}}\ and\ \bibinfo {author} {\bibfnamefont {M.}~\bibnamefont {Sasaki}},\
  }\href {\doibase 10.1103/PhysRevLett.131.011002} {\bibfield  {journal}
  {\bibinfo  {journal} {Phys. Rev. Lett.}\ }\textbf {\bibinfo {volume} {131}},\
  \bibinfo {pages} {011002} (\bibinfo {year} {2023})},\ \Eprint
  {http://arxiv.org/abs/2211.13932} {arXiv:2211.13932 [astro-ph.CO]}
  \BibitemShut {NoStop}%
\bibitem [{\citenamefont {Wang}\ \emph
  {et~al.}(2024{\natexlab{d}})\citenamefont {Wang}, \citenamefont {Zhao},
  \citenamefont {Li},\ and\ \citenamefont {Zhu}}]{Wang:2023ost}%
  \BibitemOpen
  \bibfield  {author} {\bibinfo {author} {\bibfnamefont {S.}~\bibnamefont
  {Wang}}, \bibinfo {author} {\bibfnamefont {Z.-C.}\ \bibnamefont {Zhao}},
  \bibinfo {author} {\bibfnamefont {J.-P.}\ \bibnamefont {Li}}, \ and\ \bibinfo
  {author} {\bibfnamefont {Q.-H.}\ \bibnamefont {Zhu}},\ }\href {\doibase
  10.1103/PhysRevResearch.6.L012060} {\bibfield  {journal} {\bibinfo  {journal}
  {Phys. Rev. Res.}\ }\textbf {\bibinfo {volume} {6}},\ \bibinfo {pages}
  {L012060} (\bibinfo {year} {2024}{\natexlab{d}})},\ \Eprint
  {http://arxiv.org/abs/2307.00572} {arXiv:2307.00572 [astro-ph.CO]}
  \BibitemShut {NoStop}%
\bibitem [{\citenamefont {Liu}\ \emph {et~al.}(2023{\natexlab{b}})\citenamefont
  {Liu}, \citenamefont {Chen},\ and\ \citenamefont {Huang}}]{Liu:2023pau}%
  \BibitemOpen
  \bibfield  {author} {\bibinfo {author} {\bibfnamefont {L.}~\bibnamefont
  {Liu}}, \bibinfo {author} {\bibfnamefont {Z.-C.}\ \bibnamefont {Chen}}, \
  and\ \bibinfo {author} {\bibfnamefont {Q.-G.}\ \bibnamefont {Huang}},\ }\href
  {\doibase 10.1088/1475-7516/2023/11/071} {\bibfield  {journal} {\bibinfo
  {journal} {JCAP}\ }\textbf {\bibinfo {volume} {11}},\ \bibinfo {pages} {071}
  (\bibinfo {year} {2023}{\natexlab{b}})},\ \Eprint
  {http://arxiv.org/abs/2307.14911} {arXiv:2307.14911 [astro-ph.CO]}
  \BibitemShut {NoStop}%
\bibitem [{\citenamefont {Zhu}\ \emph {et~al.}(2024{\natexlab{a}})\citenamefont
  {Zhu}, \citenamefont {Zhao}, \citenamefont {Wang},\ and\ \citenamefont
  {Zhang}}]{Zhu:2023gmx}%
  \BibitemOpen
  \bibfield  {author} {\bibinfo {author} {\bibfnamefont {Q.-H.}\ \bibnamefont
  {Zhu}}, \bibinfo {author} {\bibfnamefont {Z.-C.}\ \bibnamefont {Zhao}},
  \bibinfo {author} {\bibfnamefont {S.}~\bibnamefont {Wang}}, \ and\ \bibinfo
  {author} {\bibfnamefont {X.}~\bibnamefont {Zhang}},\ }\href {\doibase
  10.1088/1674-1137/ad79d5} {\bibfield  {journal} {\bibinfo  {journal} {Chin.
  Phys. C}\ }\textbf {\bibinfo {volume} {48}},\ \bibinfo {pages} {125105}
  (\bibinfo {year} {2024}{\natexlab{a}})},\ \Eprint
  {http://arxiv.org/abs/2307.13574} {arXiv:2307.13574 [astro-ph.CO]}
  \BibitemShut {NoStop}%
\bibitem [{\citenamefont {Dom\`enech}\ \emph {et~al.}(2024)\citenamefont
  {Dom\`enech}, \citenamefont {Pi}, \citenamefont {Wang},\ and\ \citenamefont
  {Wang}}]{Domenech:2024rks}%
  \BibitemOpen
  \bibfield  {author} {\bibinfo {author} {\bibfnamefont {G.}~\bibnamefont
  {Dom\`enech}}, \bibinfo {author} {\bibfnamefont {S.}~\bibnamefont {Pi}},
  \bibinfo {author} {\bibfnamefont {A.}~\bibnamefont {Wang}}, \ and\ \bibinfo
  {author} {\bibfnamefont {J.}~\bibnamefont {Wang}},\ }\href@noop {} {\
  (\bibinfo {year} {2024})},\ \Eprint {http://arxiv.org/abs/2402.18965}
  {arXiv:2402.18965 [astro-ph.CO]} \BibitemShut {NoStop}%
\bibitem [{\citenamefont {Franciolini}\ \emph {et~al.}(2023)\citenamefont
  {Franciolini}, \citenamefont {Iovino}, \citenamefont {Vaskonen},\ and\
  \citenamefont {Veermae}}]{Franciolini:2023pbf}%
  \BibitemOpen
  \bibfield  {author} {\bibinfo {author} {\bibfnamefont {G.}~\bibnamefont
  {Franciolini}}, \bibinfo {author} {\bibfnamefont {A.}~\bibnamefont {Iovino},
  \bibfnamefont {Junior.}}, \bibinfo {author} {\bibfnamefont {V.}~\bibnamefont
  {Vaskonen}}, \ and\ \bibinfo {author} {\bibfnamefont {H.}~\bibnamefont
  {Veermae}},\ }\href {\doibase 10.1103/PhysRevLett.131.201401} {\bibfield
  {journal} {\bibinfo  {journal} {Phys. Rev. Lett.}\ }\textbf {\bibinfo
  {volume} {131}},\ \bibinfo {pages} {201401} (\bibinfo {year} {2023})},\
  \Eprint {http://arxiv.org/abs/2306.17149} {arXiv:2306.17149 [astro-ph.CO]}
  \BibitemShut {NoStop}%
\bibitem [{\citenamefont {Seoane}\ \emph {et~al.}(2023)\citenamefont {Seoane}
  \emph {et~al.}}]{LISA:2022yao}%
  \BibitemOpen
  \bibfield  {author} {\bibinfo {author} {\bibfnamefont {P.~A.}\ \bibnamefont
  {Seoane}} \emph {et~al.} (\bibinfo {collaboration} {LISA}),\ }\href {\doibase
  10.1007/s41114-022-00041-y} {\bibfield  {journal} {\bibinfo  {journal}
  {Living Rev. Rel.}\ }\textbf {\bibinfo {volume} {26}},\ \bibinfo {pages} {2}
  (\bibinfo {year} {2023})},\ \Eprint {http://arxiv.org/abs/2203.06016}
  {arXiv:2203.06016 [gr-qc]} \BibitemShut {NoStop}%
\bibitem [{\citenamefont {Bagui}\ \emph {et~al.}(2023)\citenamefont {Bagui}
  \emph {et~al.}}]{LISACosmologyWorkingGroup:2023njw}%
  \BibitemOpen
  \bibfield  {author} {\bibinfo {author} {\bibfnamefont {E.}~\bibnamefont
  {Bagui}} \emph {et~al.} (\bibinfo {collaboration} {LISA Cosmology Working
  Group}),\ }\href@noop {} {\  (\bibinfo {year} {2023})},\ \Eprint
  {http://arxiv.org/abs/2310.19857} {arXiv:2310.19857 [astro-ph.CO]}
  \BibitemShut {NoStop}%
\bibitem [{\citenamefont {Ren}\ \emph {et~al.}(2023)\citenamefont {Ren},
  \citenamefont {Zhao}, \citenamefont {Cao}, \citenamefont {Guo}, \citenamefont
  {Han}, \citenamefont {Jin},\ and\ \citenamefont {Wu}}]{Ren:2023yec}%
  \BibitemOpen
  \bibfield  {author} {\bibinfo {author} {\bibfnamefont {Z.}~\bibnamefont
  {Ren}}, \bibinfo {author} {\bibfnamefont {T.}~\bibnamefont {Zhao}}, \bibinfo
  {author} {\bibfnamefont {Z.}~\bibnamefont {Cao}}, \bibinfo {author}
  {\bibfnamefont {Z.-K.}\ \bibnamefont {Guo}}, \bibinfo {author} {\bibfnamefont
  {W.-B.}\ \bibnamefont {Han}}, \bibinfo {author} {\bibfnamefont {H.-B.}\
  \bibnamefont {Jin}}, \ and\ \bibinfo {author} {\bibfnamefont {Y.-L.}\
  \bibnamefont {Wu}},\ }\href {\doibase 10.1007/s11467-023-1318-y} {\bibfield
  {journal} {\bibinfo  {journal} {Front. Phys. (Beijing)}\ }\textbf {\bibinfo
  {volume} {18}},\ \bibinfo {pages} {64302} (\bibinfo {year} {2023})},\ \Eprint
  {http://arxiv.org/abs/2301.02967} {arXiv:2301.02967 [gr-qc]} \BibitemShut
  {NoStop}%
\bibitem [{\citenamefont {Maggiore}(2007)}]{Maggiore:2007ulw}%
  \BibitemOpen
  \bibfield  {author} {\bibinfo {author} {\bibfnamefont {M.}~\bibnamefont
  {Maggiore}},\ }\href {\doibase 10.1093/acprof:oso/9780198570745.001.0001}
  {\emph {\bibinfo {title} {{Gravitational Waves. Vol. 1: Theory and
  Experiments}}}}\ (\bibinfo  {publisher} {Oxford University Press},\ \bibinfo
  {year} {2007})\BibitemShut {NoStop}%
\bibitem [{\citenamefont {Caprini}\ and\ \citenamefont
  {Figueroa}(2018)}]{Caprini:2018mtu}%
  \BibitemOpen
  \bibfield  {author} {\bibinfo {author} {\bibfnamefont {C.}~\bibnamefont
  {Caprini}}\ and\ \bibinfo {author} {\bibfnamefont {D.~G.}\ \bibnamefont
  {Figueroa}},\ }\href {\doibase 10.1088/1361-6382/aac608} {\bibfield
  {journal} {\bibinfo  {journal} {Class. Quant. Grav.}\ }\textbf {\bibinfo
  {volume} {35}},\ \bibinfo {pages} {163001} (\bibinfo {year} {2018})},\
  \Eprint {http://arxiv.org/abs/1801.04268} {arXiv:1801.04268 [astro-ph.CO]}
  \BibitemShut {NoStop}%
\bibitem [{\citenamefont {Thrane}\ and\ \citenamefont
  {Romano}(2013)}]{Thrane:2013oya}%
  \BibitemOpen
  \bibfield  {author} {\bibinfo {author} {\bibfnamefont {E.}~\bibnamefont
  {Thrane}}\ and\ \bibinfo {author} {\bibfnamefont {J.~D.}\ \bibnamefont
  {Romano}},\ }\href {\doibase 10.1103/PhysRevD.88.124032} {\bibfield
  {journal} {\bibinfo  {journal} {Phys. Rev. D}\ }\textbf {\bibinfo {volume}
  {88}},\ \bibinfo {pages} {124032} (\bibinfo {year} {2013})},\ \Eprint
  {http://arxiv.org/abs/1310.5300} {arXiv:1310.5300 [astro-ph.IM]} \BibitemShut
  {NoStop}%
\bibitem [{\citenamefont {Karnesis}\ \emph {et~al.}(2021)\citenamefont
  {Karnesis}, \citenamefont {Babak}, \citenamefont {Pieroni}, \citenamefont
  {Cornish},\ and\ \citenamefont {Littenberg}}]{Karnesis:2021tsh}%
  \BibitemOpen
  \bibfield  {author} {\bibinfo {author} {\bibfnamefont {N.}~\bibnamefont
  {Karnesis}}, \bibinfo {author} {\bibfnamefont {S.}~\bibnamefont {Babak}},
  \bibinfo {author} {\bibfnamefont {M.}~\bibnamefont {Pieroni}}, \bibinfo
  {author} {\bibfnamefont {N.}~\bibnamefont {Cornish}}, \ and\ \bibinfo
  {author} {\bibfnamefont {T.}~\bibnamefont {Littenberg}},\ }\href {\doibase
  10.1103/PhysRevD.104.043019} {\bibfield  {journal} {\bibinfo  {journal}
  {Phys. Rev. D}\ }\textbf {\bibinfo {volume} {104}},\ \bibinfo {pages}
  {043019} (\bibinfo {year} {2021})},\ \Eprint
  {http://arxiv.org/abs/2103.14598} {arXiv:2103.14598 [astro-ph.IM]}
  \BibitemShut {NoStop}%
\bibitem [{\citenamefont {Acharya}\ and\ \citenamefont
  {Khatri}(2020)}]{Acharya:2020jbv}%
  \BibitemOpen
  \bibfield  {author} {\bibinfo {author} {\bibfnamefont {S.~K.}\ \bibnamefont
  {Acharya}}\ and\ \bibinfo {author} {\bibfnamefont {R.}~\bibnamefont
  {Khatri}},\ }\href {\doibase 10.1088/1475-7516/2020/06/018} {\bibfield
  {journal} {\bibinfo  {journal} {JCAP}\ }\textbf {\bibinfo {volume} {06}},\
  \bibinfo {pages} {018} (\bibinfo {year} {2020})},\ \Eprint
  {http://arxiv.org/abs/2002.00898} {arXiv:2002.00898 [astro-ph.CO]}
  \BibitemShut {NoStop}%
\bibitem [{\citenamefont {Carr}\ \emph {et~al.}(2010)\citenamefont {Carr},
  \citenamefont {Kohri}, \citenamefont {Sendouda},\ and\ \citenamefont
  {Yokoyama}}]{Carr:2009jm}%
  \BibitemOpen
  \bibfield  {author} {\bibinfo {author} {\bibfnamefont {B.}~\bibnamefont
  {Carr}}, \bibinfo {author} {\bibfnamefont {K.}~\bibnamefont {Kohri}},
  \bibinfo {author} {\bibfnamefont {Y.}~\bibnamefont {Sendouda}}, \ and\
  \bibinfo {author} {\bibfnamefont {J.}~\bibnamefont {Yokoyama}},\ }\href
  {\doibase 10.1103/PhysRevD.81.104019} {\bibfield  {journal} {\bibinfo
  {journal} {Phys. Rev. D}\ }\textbf {\bibinfo {volume} {81}},\ \bibinfo
  {pages} {104019} (\bibinfo {year} {2010})},\ \Eprint
  {http://arxiv.org/abs/0912.5297} {arXiv:0912.5297 [astro-ph.CO]} \BibitemShut
  {NoStop}%
\bibitem [{\citenamefont {Tisserand}\ \emph
  {et~al.}(2007{\natexlab{b}})\citenamefont {Tisserand} \emph
  {et~al.}}]{EROS-2:2006ryy}%
  \BibitemOpen
  \bibfield  {author} {\bibinfo {author} {\bibfnamefont {P.}~\bibnamefont
  {Tisserand}} \emph {et~al.} (\bibinfo {collaboration} {EROS-2}),\ }\href
  {\doibase 10.1051/0004-6361:20066017} {\bibfield  {journal} {\bibinfo
  {journal} {Astron. Astrophys.}\ }\textbf {\bibinfo {volume} {469}},\ \bibinfo
  {pages} {387} (\bibinfo {year} {2007}{\natexlab{b}})},\ \Eprint
  {http://arxiv.org/abs/astro-ph/0607207} {arXiv:astro-ph/0607207} \BibitemShut
  {NoStop}%
\bibitem [{\citenamefont {Babak}\ \emph {et~al.}(2021)\citenamefont {Babak},
  \citenamefont {Petiteau},\ and\ \citenamefont {Hewitson}}]{Babak:2021mhe}%
  \BibitemOpen
  \bibfield  {author} {\bibinfo {author} {\bibfnamefont {S.}~\bibnamefont
  {Babak}}, \bibinfo {author} {\bibfnamefont {A.}~\bibnamefont {Petiteau}}, \
  and\ \bibinfo {author} {\bibfnamefont {M.}~\bibnamefont {Hewitson}},\
  }\href@noop {} {\  (\bibinfo {year} {2021})},\ \Eprint
  {http://arxiv.org/abs/2108.01167} {arXiv:2108.01167 [astro-ph.IM]}
  \BibitemShut {NoStop}%
\bibitem [{\citenamefont {Luo}\ \emph {et~al.}(2020)\citenamefont {Luo},
  \citenamefont {Guo}, \citenamefont {Jin}, \citenamefont {Wu},\ and\
  \citenamefont {Hu}}]{Luo:2019zal}%
  \BibitemOpen
  \bibfield  {author} {\bibinfo {author} {\bibfnamefont {Z.}~\bibnamefont
  {Luo}}, \bibinfo {author} {\bibfnamefont {Z.}~\bibnamefont {Guo}}, \bibinfo
  {author} {\bibfnamefont {G.}~\bibnamefont {Jin}}, \bibinfo {author}
  {\bibfnamefont {Y.}~\bibnamefont {Wu}}, \ and\ \bibinfo {author}
  {\bibfnamefont {W.}~\bibnamefont {Hu}},\ }\href {\doibase
  10.1016/j.rinp.2019.102918} {\bibfield  {journal} {\bibinfo  {journal}
  {Results Phys.}\ }\textbf {\bibinfo {volume} {16}},\ \bibinfo {pages}
  {102918} (\bibinfo {year} {2020})}\BibitemShut {NoStop}%
\bibitem [{\citenamefont {Hong}\ \emph {et~al.}()\citenamefont {Hong},
  \citenamefont {Kuroyanagi}, \citenamefont {Pi}, \citenamefont {Wang},\ and\
  \citenamefont {Zhang}}]{Hong:2025tba}%
  \BibitemOpen
  \bibfield  {author} {\bibinfo {author} {\bibfnamefont {W.}~\bibnamefont
  {Hong}}, \bibinfo {author} {\bibfnamefont {S.}~\bibnamefont {Kuroyanagi}},
  \bibinfo {author} {\bibfnamefont {S.}~\bibnamefont {Pi}}, \bibinfo {author}
  {\bibfnamefont {A.}~\bibnamefont {Wang}}, \ and\ \bibinfo {author}
  {\bibfnamefont {Z.}~\bibnamefont {Zhang}},\ }\href@noop {} {\bibinfo
  {journal} {To appear}\ }\BibitemShut {NoStop}%
\bibitem [{\citenamefont {Inomata}\ and\ \citenamefont
  {Luo}(2024)}]{Inomata:2024dbr}%
  \BibitemOpen
\bibfield  {journal} {  }\bibfield  {author} {\bibinfo {author} {\bibfnamefont
  {K.}~\bibnamefont {Inomata}}\ and\ \bibinfo {author} {\bibfnamefont
  {X.}~\bibnamefont {Luo}},\ }\href@noop {} {\  (\bibinfo {year} {2024})},\
  \Eprint {http://arxiv.org/abs/2410.07086} {arXiv:2410.07086 [astro-ph.CO]}
  \BibitemShut {NoStop}%
\bibitem [{\citenamefont {Dalianis}\ \emph {et~al.}(2021)\citenamefont
  {Dalianis}, \citenamefont {Kodaxis}, \citenamefont {Stamou}, \citenamefont
  {Tetradis},\ and\ \citenamefont {Tsigkas-Kouvelis}}]{Dalianis:2021iig}%
  \BibitemOpen
  \bibfield  {author} {\bibinfo {author} {\bibfnamefont {I.}~\bibnamefont
  {Dalianis}}, \bibinfo {author} {\bibfnamefont {G.~P.}\ \bibnamefont
  {Kodaxis}}, \bibinfo {author} {\bibfnamefont {I.~D.}\ \bibnamefont {Stamou}},
  \bibinfo {author} {\bibfnamefont {N.}~\bibnamefont {Tetradis}}, \ and\
  \bibinfo {author} {\bibfnamefont {A.}~\bibnamefont {Tsigkas-Kouvelis}},\
  }\href {\doibase 10.1103/PhysRevD.104.103510} {\bibfield  {journal} {\bibinfo
   {journal} {Phys. Rev. D}\ }\textbf {\bibinfo {volume} {104}},\ \bibinfo
  {pages} {103510} (\bibinfo {year} {2021})},\ \Eprint
  {http://arxiv.org/abs/2106.02467} {arXiv:2106.02467 [astro-ph.CO]}
  \BibitemShut {NoStop}%
\bibitem [{\citenamefont {Pi}\ and\ \citenamefont {Wang}(2023)}]{Pi:2022zxs}%
  \BibitemOpen
  \bibfield  {author} {\bibinfo {author} {\bibfnamefont {S.}~\bibnamefont
  {Pi}}\ and\ \bibinfo {author} {\bibfnamefont {J.}~\bibnamefont {Wang}},\
  }\href {\doibase 10.1088/1475-7516/2023/06/018} {\bibfield  {journal}
  {\bibinfo  {journal} {JCAP}\ }\textbf {\bibinfo {volume} {06}},\ \bibinfo
  {pages} {018} (\bibinfo {year} {2023})},\ \Eprint
  {http://arxiv.org/abs/2209.14183} {arXiv:2209.14183 [astro-ph.CO]}
  \BibitemShut {NoStop}%
\bibitem [{\citenamefont {Cole}\ \emph {et~al.}(2024)\citenamefont {Cole},
  \citenamefont {Gow}, \citenamefont {Byrnes},\ and\ \citenamefont
  {Patil}}]{Cole:2022xqc}%
  \BibitemOpen
  \bibfield  {author} {\bibinfo {author} {\bibfnamefont {P.~S.}\ \bibnamefont
  {Cole}}, \bibinfo {author} {\bibfnamefont {A.~D.}\ \bibnamefont {Gow}},
  \bibinfo {author} {\bibfnamefont {C.~T.}\ \bibnamefont {Byrnes}}, \ and\
  \bibinfo {author} {\bibfnamefont {S.~P.}\ \bibnamefont {Patil}},\ }\href
  {\doibase 10.1088/1475-7516/2024/05/022} {\bibfield  {journal} {\bibinfo
  {journal} {JCAP}\ }\textbf {\bibinfo {volume} {05}},\ \bibinfo {pages} {022}
  (\bibinfo {year} {2024})},\ \Eprint {http://arxiv.org/abs/2204.07573}
  {arXiv:2204.07573 [astro-ph.CO]} \BibitemShut {NoStop}%
\bibitem [{\citenamefont {Carr}\ \emph {et~al.}(2017)\citenamefont {Carr},
  \citenamefont {Raidal}, \citenamefont {Tenkanen}, \citenamefont {Vaskonen},\
  and\ \citenamefont {Veerm\"ae}}]{Carr:2017jsz}%
  \BibitemOpen
  \bibfield  {author} {\bibinfo {author} {\bibfnamefont {B.}~\bibnamefont
  {Carr}}, \bibinfo {author} {\bibfnamefont {M.}~\bibnamefont {Raidal}},
  \bibinfo {author} {\bibfnamefont {T.}~\bibnamefont {Tenkanen}}, \bibinfo
  {author} {\bibfnamefont {V.}~\bibnamefont {Vaskonen}}, \ and\ \bibinfo
  {author} {\bibfnamefont {H.}~\bibnamefont {Veerm\"ae}},\ }\href {\doibase
  10.1103/PhysRevD.96.023514} {\bibfield  {journal} {\bibinfo  {journal} {Phys.
  Rev. D}\ }\textbf {\bibinfo {volume} {96}},\ \bibinfo {pages} {023514}
  (\bibinfo {year} {2017})},\ \Eprint {http://arxiv.org/abs/1705.05567}
  {arXiv:1705.05567 [astro-ph.CO]} \BibitemShut {NoStop}%
\bibitem [{\citenamefont {Gorton}\ and\ \citenamefont
  {Green}(2024)}]{Gorton:2024cdm}%
  \BibitemOpen
  \bibfield  {author} {\bibinfo {author} {\bibfnamefont {M.}~\bibnamefont
  {Gorton}}\ and\ \bibinfo {author} {\bibfnamefont {A.~M.}\ \bibnamefont
  {Green}},\ }\href {\doibase 10.21468/SciPostPhys.17.2.032} {\bibfield
  {journal} {\bibinfo  {journal} {SciPost Phys.}\ }\textbf {\bibinfo {volume}
  {17}},\ \bibinfo {pages} {032} (\bibinfo {year} {2024})},\ \Eprint
  {http://arxiv.org/abs/2403.03839} {arXiv:2403.03839 [astro-ph.CO]}
  \BibitemShut {NoStop}%
\bibitem [{\citenamefont {Green}(2017)}]{Green:2017qoa}%
  \BibitemOpen
  \bibfield  {author} {\bibinfo {author} {\bibfnamefont {A.~M.}\ \bibnamefont
  {Green}},\ }\href {\doibase 10.1103/PhysRevD.96.043020} {\bibfield  {journal}
  {\bibinfo  {journal} {Phys. Rev. D}\ }\textbf {\bibinfo {volume} {96}},\
  \bibinfo {pages} {043020} (\bibinfo {year} {2017})},\ \Eprint
  {http://arxiv.org/abs/1705.10818} {arXiv:1705.10818 [astro-ph.CO]}
  \BibitemShut {NoStop}%
\bibitem [{\citenamefont {Byrnes}\ \emph {et~al.}(2012)\citenamefont {Byrnes},
  \citenamefont {Copeland},\ and\ \citenamefont {Green}}]{Byrnes:2012yx}%
  \BibitemOpen
  \bibfield  {author} {\bibinfo {author} {\bibfnamefont {C.~T.}\ \bibnamefont
  {Byrnes}}, \bibinfo {author} {\bibfnamefont {E.~J.}\ \bibnamefont
  {Copeland}}, \ and\ \bibinfo {author} {\bibfnamefont {A.~M.}\ \bibnamefont
  {Green}},\ }\href {\doibase 10.1103/PhysRevD.86.043512} {\bibfield  {journal}
  {\bibinfo  {journal} {Phys. Rev. D}\ }\textbf {\bibinfo {volume} {86}},\
  \bibinfo {pages} {043512} (\bibinfo {year} {2012})},\ \Eprint
  {http://arxiv.org/abs/1206.4188} {arXiv:1206.4188 [astro-ph.CO]} \BibitemShut
  {NoStop}%
\bibitem [{\citenamefont {Young}\ and\ \citenamefont
  {Byrnes}(2013)}]{Young:2013oia}%
  \BibitemOpen
  \bibfield  {author} {\bibinfo {author} {\bibfnamefont {S.}~\bibnamefont
  {Young}}\ and\ \bibinfo {author} {\bibfnamefont {C.~T.}\ \bibnamefont
  {Byrnes}},\ }\href {\doibase 10.1088/1475-7516/2013/08/052} {\bibfield
  {journal} {\bibinfo  {journal} {JCAP}\ }\textbf {\bibinfo {volume} {08}},\
  \bibinfo {pages} {052} (\bibinfo {year} {2013})},\ \Eprint
  {http://arxiv.org/abs/1307.4995} {arXiv:1307.4995 [astro-ph.CO]} \BibitemShut
  {NoStop}%
\bibitem [{\citenamefont {Tada}\ and\ \citenamefont
  {Yokoyama}(2015)}]{Tada:2015noa}%
  \BibitemOpen
  \bibfield  {author} {\bibinfo {author} {\bibfnamefont {Y.}~\bibnamefont
  {Tada}}\ and\ \bibinfo {author} {\bibfnamefont {S.}~\bibnamefont
  {Yokoyama}},\ }\href {\doibase 10.1103/PhysRevD.91.123534} {\bibfield
  {journal} {\bibinfo  {journal} {Phys. Rev. D}\ }\textbf {\bibinfo {volume}
  {91}},\ \bibinfo {pages} {123534} (\bibinfo {year} {2015})},\ \Eprint
  {http://arxiv.org/abs/1502.01124} {arXiv:1502.01124 [astro-ph.CO]}
  \BibitemShut {NoStop}%
\bibitem [{\citenamefont {Young}\ and\ \citenamefont
  {Byrnes}(2015)}]{Young:2015kda}%
  \BibitemOpen
  \bibfield  {author} {\bibinfo {author} {\bibfnamefont {S.}~\bibnamefont
  {Young}}\ and\ \bibinfo {author} {\bibfnamefont {C.~T.}\ \bibnamefont
  {Byrnes}},\ }\href {\doibase 10.1088/1475-7516/2015/04/034} {\bibfield
  {journal} {\bibinfo  {journal} {JCAP}\ }\textbf {\bibinfo {volume} {04}},\
  \bibinfo {pages} {034} (\bibinfo {year} {2015})},\ \Eprint
  {http://arxiv.org/abs/1503.01505} {arXiv:1503.01505 [astro-ph.CO]}
  \BibitemShut {NoStop}%
\bibitem [{\citenamefont {Young}\ \emph {et~al.}(2016)\citenamefont {Young},
  \citenamefont {Regan},\ and\ \citenamefont {Byrnes}}]{Young:2015cyn}%
  \BibitemOpen
  \bibfield  {author} {\bibinfo {author} {\bibfnamefont {S.}~\bibnamefont
  {Young}}, \bibinfo {author} {\bibfnamefont {D.}~\bibnamefont {Regan}}, \ and\
  \bibinfo {author} {\bibfnamefont {C.~T.}\ \bibnamefont {Byrnes}},\ }\href
  {\doibase 10.1088/1475-7516/2016/02/029} {\bibfield  {journal} {\bibinfo
  {journal} {JCAP}\ }\textbf {\bibinfo {volume} {02}},\ \bibinfo {pages} {029}
  (\bibinfo {year} {2016})},\ \Eprint {http://arxiv.org/abs/1512.07224}
  {arXiv:1512.07224 [astro-ph.CO]} \BibitemShut {NoStop}%
\bibitem [{\citenamefont {Franciolini}\ \emph {et~al.}(2018)\citenamefont
  {Franciolini}, \citenamefont {Kehagias}, \citenamefont {Matarrese},\ and\
  \citenamefont {Riotto}}]{Franciolini:2018vbk}%
  \BibitemOpen
  \bibfield  {author} {\bibinfo {author} {\bibfnamefont {G.}~\bibnamefont
  {Franciolini}}, \bibinfo {author} {\bibfnamefont {A.}~\bibnamefont
  {Kehagias}}, \bibinfo {author} {\bibfnamefont {S.}~\bibnamefont {Matarrese}},
  \ and\ \bibinfo {author} {\bibfnamefont {A.}~\bibnamefont {Riotto}},\ }\href
  {\doibase 10.1088/1475-7516/2018/03/016} {\bibfield  {journal} {\bibinfo
  {journal} {JCAP}\ }\textbf {\bibinfo {volume} {03}},\ \bibinfo {pages} {016}
  (\bibinfo {year} {2018})},\ \Eprint {http://arxiv.org/abs/1801.09415}
  {arXiv:1801.09415 [astro-ph.CO]} \BibitemShut {NoStop}%
\bibitem [{\citenamefont {Atal}\ and\ \citenamefont
  {Germani}(2019)}]{Atal:2018neu}%
  \BibitemOpen
  \bibfield  {author} {\bibinfo {author} {\bibfnamefont {V.}~\bibnamefont
  {Atal}}\ and\ \bibinfo {author} {\bibfnamefont {C.}~\bibnamefont {Germani}},\
  }\href {\doibase 10.1016/j.dark.2019.100275} {\bibfield  {journal} {\bibinfo
  {journal} {Phys. Dark Univ.}\ }\textbf {\bibinfo {volume} {24}},\ \bibinfo
  {pages} {100275} (\bibinfo {year} {2019})},\ \Eprint
  {http://arxiv.org/abs/1811.07857} {arXiv:1811.07857 [astro-ph.CO]}
  \BibitemShut {NoStop}%
\bibitem [{\citenamefont {Passaglia}\ \emph {et~al.}(2019)\citenamefont
  {Passaglia}, \citenamefont {Hu},\ and\ \citenamefont
  {Motohashi}}]{Passaglia:2018ixg}%
  \BibitemOpen
  \bibfield  {author} {\bibinfo {author} {\bibfnamefont {S.}~\bibnamefont
  {Passaglia}}, \bibinfo {author} {\bibfnamefont {W.}~\bibnamefont {Hu}}, \
  and\ \bibinfo {author} {\bibfnamefont {H.}~\bibnamefont {Motohashi}},\ }\href
  {\doibase 10.1103/PhysRevD.99.043536} {\bibfield  {journal} {\bibinfo
  {journal} {Phys. Rev. D}\ }\textbf {\bibinfo {volume} {99}},\ \bibinfo
  {pages} {043536} (\bibinfo {year} {2019})},\ \Eprint
  {http://arxiv.org/abs/1812.08243} {arXiv:1812.08243 [astro-ph.CO]}
  \BibitemShut {NoStop}%
\bibitem [{\citenamefont {Namjoo}\ \emph {et~al.}(2013)\citenamefont {Namjoo},
  \citenamefont {Firouzjahi},\ and\ \citenamefont {Sasaki}}]{Namjoo:2012aa}%
  \BibitemOpen
  \bibfield  {author} {\bibinfo {author} {\bibfnamefont {M.~H.}\ \bibnamefont
  {Namjoo}}, \bibinfo {author} {\bibfnamefont {H.}~\bibnamefont {Firouzjahi}},
  \ and\ \bibinfo {author} {\bibfnamefont {M.}~\bibnamefont {Sasaki}},\ }\href
  {\doibase 10.1209/0295-5075/101/39001} {\bibfield  {journal} {\bibinfo
  {journal} {EPL}\ }\textbf {\bibinfo {volume} {101}},\ \bibinfo {pages}
  {39001} (\bibinfo {year} {2013})},\ \Eprint {http://arxiv.org/abs/1210.3692}
  {arXiv:1210.3692 [astro-ph.CO]} \BibitemShut {NoStop}%
\bibitem [{\citenamefont {Martin}\ \emph {et~al.}(2013)\citenamefont {Martin},
  \citenamefont {Motohashi},\ and\ \citenamefont {Suyama}}]{Martin:2012pe}%
  \BibitemOpen
  \bibfield  {author} {\bibinfo {author} {\bibfnamefont {J.}~\bibnamefont
  {Martin}}, \bibinfo {author} {\bibfnamefont {H.}~\bibnamefont {Motohashi}}, \
  and\ \bibinfo {author} {\bibfnamefont {T.}~\bibnamefont {Suyama}},\ }\href
  {\doibase 10.1103/PhysRevD.87.023514} {\bibfield  {journal} {\bibinfo
  {journal} {Phys. Rev. D}\ }\textbf {\bibinfo {volume} {87}},\ \bibinfo
  {pages} {023514} (\bibinfo {year} {2013})},\ \Eprint
  {http://arxiv.org/abs/1211.0083} {arXiv:1211.0083 [astro-ph.CO]} \BibitemShut
  {NoStop}%
\bibitem [{\citenamefont {Chen}\ \emph {et~al.}(2013)\citenamefont {Chen},
  \citenamefont {Firouzjahi}, \citenamefont {Namjoo},\ and\ \citenamefont
  {Sasaki}}]{Chen:2013aj}%
  \BibitemOpen
  \bibfield  {author} {\bibinfo {author} {\bibfnamefont {X.}~\bibnamefont
  {Chen}}, \bibinfo {author} {\bibfnamefont {H.}~\bibnamefont {Firouzjahi}},
  \bibinfo {author} {\bibfnamefont {M.~H.}\ \bibnamefont {Namjoo}}, \ and\
  \bibinfo {author} {\bibfnamefont {M.}~\bibnamefont {Sasaki}},\ }\href
  {\doibase 10.1209/0295-5075/102/59001} {\bibfield  {journal} {\bibinfo
  {journal} {EPL}\ }\textbf {\bibinfo {volume} {102}},\ \bibinfo {pages}
  {59001} (\bibinfo {year} {2013})},\ \Eprint {http://arxiv.org/abs/1301.5699}
  {arXiv:1301.5699 [hep-th]} \BibitemShut {NoStop}%
\bibitem [{\citenamefont {Motohashi}\ \emph {et~al.}(2015)\citenamefont
  {Motohashi}, \citenamefont {Starobinsky},\ and\ \citenamefont
  {Yokoyama}}]{Motohashi:2014ppa}%
  \BibitemOpen
  \bibfield  {author} {\bibinfo {author} {\bibfnamefont {H.}~\bibnamefont
  {Motohashi}}, \bibinfo {author} {\bibfnamefont {A.~A.}\ \bibnamefont
  {Starobinsky}}, \ and\ \bibinfo {author} {\bibfnamefont {J.}~\bibnamefont
  {Yokoyama}},\ }\href {\doibase 10.1088/1475-7516/2015/09/018} {\bibfield
  {journal} {\bibinfo  {journal} {JCAP}\ }\textbf {\bibinfo {volume} {09}},\
  \bibinfo {pages} {018} (\bibinfo {year} {2015})},\ \Eprint
  {http://arxiv.org/abs/1411.5021} {arXiv:1411.5021 [astro-ph.CO]} \BibitemShut
  {NoStop}%
\bibitem [{\citenamefont {Davies}\ \emph {et~al.}(2022)\citenamefont {Davies},
  \citenamefont {Carrilho},\ and\ \citenamefont {Mulryne}}]{Davies:2021loj}%
  \BibitemOpen
  \bibfield  {author} {\bibinfo {author} {\bibfnamefont {M.~W.}\ \bibnamefont
  {Davies}}, \bibinfo {author} {\bibfnamefont {P.}~\bibnamefont {Carrilho}}, \
  and\ \bibinfo {author} {\bibfnamefont {D.~J.}\ \bibnamefont {Mulryne}},\
  }\href {\doibase 10.1088/1475-7516/2022/06/019} {\bibfield  {journal}
  {\bibinfo  {journal} {JCAP}\ }\textbf {\bibinfo {volume} {06}},\ \bibinfo
  {pages} {019} (\bibinfo {year} {2022})},\ \Eprint
  {http://arxiv.org/abs/2110.08189} {arXiv:2110.08189 [astro-ph.CO]}
  \BibitemShut {NoStop}%
\bibitem [{\citenamefont {Namjoo}(2024)}]{Namjoo:2023rhq}%
  \BibitemOpen
  \bibfield  {author} {\bibinfo {author} {\bibfnamefont {M.~H.}\ \bibnamefont
  {Namjoo}},\ }\href {\doibase 10.1088/1475-7516/2024/05/041} {\bibfield
  {journal} {\bibinfo  {journal} {JCAP}\ }\textbf {\bibinfo {volume} {05}},\
  \bibinfo {pages} {041} (\bibinfo {year} {2024})},\ \Eprint
  {http://arxiv.org/abs/2311.12777} {arXiv:2311.12777 [astro-ph.CO]}
  \BibitemShut {NoStop}%
\bibitem [{\citenamefont {Namjoo}\ and\ \citenamefont
  {Nikbakht}(2024)}]{Namjoo:2024ufv}%
  \BibitemOpen
  \bibfield  {author} {\bibinfo {author} {\bibfnamefont {M.~H.}\ \bibnamefont
  {Namjoo}}\ and\ \bibinfo {author} {\bibfnamefont {B.}~\bibnamefont
  {Nikbakht}},\ }\href {\doibase 10.1088/1475-7516/2024/08/005} {\bibfield
  {journal} {\bibinfo  {journal} {JCAP}\ }\textbf {\bibinfo {volume} {08}},\
  \bibinfo {pages} {005} (\bibinfo {year} {2024})},\ \Eprint
  {http://arxiv.org/abs/2401.12958} {arXiv:2401.12958 [astro-ph.CO]}
  \BibitemShut {NoStop}%
\bibitem [{\citenamefont {Hooshangi}\ \emph {et~al.}(2022)\citenamefont
  {Hooshangi}, \citenamefont {Namjoo},\ and\ \citenamefont
  {Noorbala}}]{Hooshangi:2021ubn}%
  \BibitemOpen
  \bibfield  {author} {\bibinfo {author} {\bibfnamefont {S.}~\bibnamefont
  {Hooshangi}}, \bibinfo {author} {\bibfnamefont {M.~H.}\ \bibnamefont
  {Namjoo}}, \ and\ \bibinfo {author} {\bibfnamefont {M.}~\bibnamefont
  {Noorbala}},\ }\href {\doibase 10.1016/j.physletb.2022.137400} {\bibfield
  {journal} {\bibinfo  {journal} {Phys. Lett. B}\ }\textbf {\bibinfo {volume}
  {834}},\ \bibinfo {pages} {137400} (\bibinfo {year} {2022})},\ \Eprint
  {http://arxiv.org/abs/2112.04520} {arXiv:2112.04520 [astro-ph.CO]}
  \BibitemShut {NoStop}%
\bibitem [{\citenamefont {Hooshangi}\ \emph {et~al.}(2023)\citenamefont
  {Hooshangi}, \citenamefont {Namjoo},\ and\ \citenamefont
  {Noorbala}}]{Hooshangi:2023kss}%
  \BibitemOpen
  \bibfield  {author} {\bibinfo {author} {\bibfnamefont {S.}~\bibnamefont
  {Hooshangi}}, \bibinfo {author} {\bibfnamefont {M.~H.}\ \bibnamefont
  {Namjoo}}, \ and\ \bibinfo {author} {\bibfnamefont {M.}~\bibnamefont
  {Noorbala}},\ }\href {\doibase 10.1088/1475-7516/2023/09/023} {\bibfield
  {journal} {\bibinfo  {journal} {JCAP}\ }\textbf {\bibinfo {volume} {09}},\
  \bibinfo {pages} {023} (\bibinfo {year} {2023})},\ \Eprint
  {http://arxiv.org/abs/2305.19257} {arXiv:2305.19257 [astro-ph.CO]}
  \BibitemShut {NoStop}%
\bibitem [{\citenamefont {Pi}\ and\ \citenamefont {Sasaki}(2021)}]{Pi:2021dft}%
  \BibitemOpen
  \bibfield  {author} {\bibinfo {author} {\bibfnamefont {S.}~\bibnamefont
  {Pi}}\ and\ \bibinfo {author} {\bibfnamefont {M.}~\bibnamefont {Sasaki}},\
  }\href {\doibase 10.1103/PhysRevD.108.L101301} {\bibfield  {journal}
  {\bibinfo  {journal} {Phys. Rev. D}\ }\textbf {\bibinfo {volume} {108}},\
  \bibinfo {pages} {L101301} (\bibinfo {year} {2021})},\ \Eprint
  {http://arxiv.org/abs/2112.12680} {arXiv:2112.12680 [astro-ph.CO]}
  \BibitemShut {NoStop}%
\bibitem [{\citenamefont {Nakama}\ \emph {et~al.}(2017)\citenamefont {Nakama},
  \citenamefont {Silk},\ and\ \citenamefont {Kamionkowski}}]{Nakama:2016gzw}%
  \BibitemOpen
  \bibfield  {author} {\bibinfo {author} {\bibfnamefont {T.}~\bibnamefont
  {Nakama}}, \bibinfo {author} {\bibfnamefont {J.}~\bibnamefont {Silk}}, \ and\
  \bibinfo {author} {\bibfnamefont {M.}~\bibnamefont {Kamionkowski}},\ }\href
  {\doibase 10.1103/PhysRevD.95.043511} {\bibfield  {journal} {\bibinfo
  {journal} {Phys. Rev. D}\ }\textbf {\bibinfo {volume} {95}},\ \bibinfo
  {pages} {043511} (\bibinfo {year} {2017})},\ \Eprint
  {http://arxiv.org/abs/1612.06264} {arXiv:1612.06264 [astro-ph.CO]}
  \BibitemShut {NoStop}%
\bibitem [{\citenamefont {Garcia-Bellido}\ \emph {et~al.}(2017)\citenamefont
  {Garcia-Bellido}, \citenamefont {Peloso},\ and\ \citenamefont
  {Unal}}]{Garcia-Bellido:2017aan}%
  \BibitemOpen
  \bibfield  {author} {\bibinfo {author} {\bibfnamefont {J.}~\bibnamefont
  {Garcia-Bellido}}, \bibinfo {author} {\bibfnamefont {M.}~\bibnamefont
  {Peloso}}, \ and\ \bibinfo {author} {\bibfnamefont {C.}~\bibnamefont
  {Unal}},\ }\href {\doibase 10.1088/1475-7516/2017/09/013} {\bibfield
  {journal} {\bibinfo  {journal} {JCAP}\ }\textbf {\bibinfo {volume} {09}},\
  \bibinfo {pages} {013} (\bibinfo {year} {2017})},\ \Eprint
  {http://arxiv.org/abs/1707.02441} {arXiv:1707.02441 [astro-ph.CO]}
  \BibitemShut {NoStop}%
\bibitem [{\citenamefont {Unal}(2019)}]{Unal:2018yaa}%
  \BibitemOpen
  \bibfield  {author} {\bibinfo {author} {\bibfnamefont {C.}~\bibnamefont
  {Unal}},\ }\href {\doibase 10.1103/PhysRevD.99.041301} {\bibfield  {journal}
  {\bibinfo  {journal} {Phys. Rev. D}\ }\textbf {\bibinfo {volume} {99}},\
  \bibinfo {pages} {041301} (\bibinfo {year} {2019})},\ \Eprint
  {http://arxiv.org/abs/1811.09151} {arXiv:1811.09151 [astro-ph.CO]}
  \BibitemShut {NoStop}%
\bibitem [{\citenamefont {\"Unal}\ \emph {et~al.}(2021)\citenamefont {\"Unal},
  \citenamefont {Kovetz},\ and\ \citenamefont {Patil}}]{Unal:2020mts}%
  \BibitemOpen
  \bibfield  {author} {\bibinfo {author} {\bibfnamefont {C.}~\bibnamefont
  {\"Unal}}, \bibinfo {author} {\bibfnamefont {E.~D.}\ \bibnamefont {Kovetz}},
  \ and\ \bibinfo {author} {\bibfnamefont {S.~P.}\ \bibnamefont {Patil}},\
  }\href {\doibase 10.1103/PhysRevD.103.063519} {\bibfield  {journal} {\bibinfo
   {journal} {Phys. Rev. D}\ }\textbf {\bibinfo {volume} {103}},\ \bibinfo
  {pages} {063519} (\bibinfo {year} {2021})},\ \Eprint
  {http://arxiv.org/abs/2008.11184} {arXiv:2008.11184 [astro-ph.CO]}
  \BibitemShut {NoStop}%
\bibitem [{\citenamefont {Adshead}\ \emph {et~al.}(2021)\citenamefont
  {Adshead}, \citenamefont {Lozanov},\ and\ \citenamefont
  {Weiner}}]{Adshead:2021hnm}%
  \BibitemOpen
  \bibfield  {author} {\bibinfo {author} {\bibfnamefont {P.}~\bibnamefont
  {Adshead}}, \bibinfo {author} {\bibfnamefont {K.~D.}\ \bibnamefont
  {Lozanov}}, \ and\ \bibinfo {author} {\bibfnamefont {Z.~J.}\ \bibnamefont
  {Weiner}},\ }\href {\doibase 10.1088/1475-7516/2021/10/080} {\bibfield
  {journal} {\bibinfo  {journal} {JCAP}\ }\textbf {\bibinfo {volume} {10}},\
  \bibinfo {pages} {080} (\bibinfo {year} {2021})},\ \Eprint
  {http://arxiv.org/abs/2105.01659} {arXiv:2105.01659 [astro-ph.CO]}
  \BibitemShut {NoStop}%
\bibitem [{\citenamefont {Ragavendra}(2022)}]{Ragavendra:2021qdu}%
  \BibitemOpen
  \bibfield  {author} {\bibinfo {author} {\bibfnamefont {H.~V.}\ \bibnamefont
  {Ragavendra}},\ }\href {\doibase 10.1103/PhysRevD.105.063533} {\bibfield
  {journal} {\bibinfo  {journal} {Phys. Rev. D}\ }\textbf {\bibinfo {volume}
  {105}},\ \bibinfo {pages} {063533} (\bibinfo {year} {2022})},\ \Eprint
  {http://arxiv.org/abs/2108.04193} {arXiv:2108.04193 [astro-ph.CO]}
  \BibitemShut {NoStop}%
\bibitem [{\citenamefont {Abe}\ \emph {et~al.}(2023)\citenamefont {Abe},
  \citenamefont {Inui}, \citenamefont {Tada},\ and\ \citenamefont
  {Yokoyama}}]{Abe:2022xur}%
  \BibitemOpen
  \bibfield  {author} {\bibinfo {author} {\bibfnamefont {K.~T.}\ \bibnamefont
  {Abe}}, \bibinfo {author} {\bibfnamefont {R.}~\bibnamefont {Inui}}, \bibinfo
  {author} {\bibfnamefont {Y.}~\bibnamefont {Tada}}, \ and\ \bibinfo {author}
  {\bibfnamefont {S.}~\bibnamefont {Yokoyama}},\ }\href {\doibase
  10.1088/1475-7516/2023/05/044} {\bibfield  {journal} {\bibinfo  {journal}
  {JCAP}\ }\textbf {\bibinfo {volume} {05}},\ \bibinfo {pages} {044} (\bibinfo
  {year} {2023})},\ \Eprint {http://arxiv.org/abs/2209.13891} {arXiv:2209.13891
  [astro-ph.CO]} \BibitemShut {NoStop}%
\bibitem [{\citenamefont {Yuan}\ \emph {et~al.}(2023)\citenamefont {Yuan},
  \citenamefont {Meng},\ and\ \citenamefont {Huang}}]{Yuan:2023ofl}%
  \BibitemOpen
  \bibfield  {author} {\bibinfo {author} {\bibfnamefont {C.}~\bibnamefont
  {Yuan}}, \bibinfo {author} {\bibfnamefont {D.-S.}\ \bibnamefont {Meng}}, \
  and\ \bibinfo {author} {\bibfnamefont {Q.-G.}\ \bibnamefont {Huang}},\ }\href
  {\doibase 10.1088/1475-7516/2023/12/036} {\bibfield  {journal} {\bibinfo
  {journal} {JCAP}\ }\textbf {\bibinfo {volume} {12}},\ \bibinfo {pages} {036}
  (\bibinfo {year} {2023})},\ \Eprint {http://arxiv.org/abs/2308.07155}
  {arXiv:2308.07155 [astro-ph.CO]} \BibitemShut {NoStop}%
\bibitem [{\citenamefont {Li}\ \emph {et~al.}(2024{\natexlab{c}})\citenamefont
  {Li}, \citenamefont {Wang}, \citenamefont {Zhao},\ and\ \citenamefont
  {Kohri}}]{Li:2023xtl}%
  \BibitemOpen
  \bibfield  {author} {\bibinfo {author} {\bibfnamefont {J.-P.}\ \bibnamefont
  {Li}}, \bibinfo {author} {\bibfnamefont {S.}~\bibnamefont {Wang}}, \bibinfo
  {author} {\bibfnamefont {Z.-C.}\ \bibnamefont {Zhao}}, \ and\ \bibinfo
  {author} {\bibfnamefont {K.}~\bibnamefont {Kohri}},\ }\href {\doibase
  10.1088/1475-7516/2024/06/039} {\bibfield  {journal} {\bibinfo  {journal}
  {JCAP}\ }\textbf {\bibinfo {volume} {06}},\ \bibinfo {pages} {039} (\bibinfo
  {year} {2024}{\natexlab{c}})},\ \Eprint {http://arxiv.org/abs/2309.07792}
  {arXiv:2309.07792 [astro-ph.CO]} \BibitemShut {NoStop}%
\bibitem [{\citenamefont {Perna}\ \emph {et~al.}(2024)\citenamefont {Perna},
  \citenamefont {Testini}, \citenamefont {Ricciardone},\ and\ \citenamefont
  {Matarrese}}]{Perna:2024ehx}%
  \BibitemOpen
  \bibfield  {author} {\bibinfo {author} {\bibfnamefont {G.}~\bibnamefont
  {Perna}}, \bibinfo {author} {\bibfnamefont {C.}~\bibnamefont {Testini}},
  \bibinfo {author} {\bibfnamefont {A.}~\bibnamefont {Ricciardone}}, \ and\
  \bibinfo {author} {\bibfnamefont {S.}~\bibnamefont {Matarrese}},\ }\href
  {\doibase 10.1088/1475-7516/2024/05/086} {\bibfield  {journal} {\bibinfo
  {journal} {JCAP}\ }\textbf {\bibinfo {volume} {05}},\ \bibinfo {pages} {086}
  (\bibinfo {year} {2024})},\ \Eprint {http://arxiv.org/abs/2403.06962}
  {arXiv:2403.06962 [astro-ph.CO]} \BibitemShut {NoStop}%
\bibitem [{\citenamefont {Contaldi}(2017)}]{Contaldi:2016koz}%
  \BibitemOpen
  \bibfield  {author} {\bibinfo {author} {\bibfnamefont {C.~R.}\ \bibnamefont
  {Contaldi}},\ }\href {\doibase 10.1016/j.physletb.2017.05.020} {\bibfield
  {journal} {\bibinfo  {journal} {Phys. Lett. B}\ }\textbf {\bibinfo {volume}
  {771}},\ \bibinfo {pages} {9} (\bibinfo {year} {2017})},\ \Eprint
  {http://arxiv.org/abs/1609.08168} {arXiv:1609.08168 [astro-ph.CO]}
  \BibitemShut {NoStop}%
\bibitem [{\citenamefont {Bartolo}\ \emph
  {et~al.}(2019{\natexlab{b}})\citenamefont {Bartolo}, \citenamefont
  {Bertacca}, \citenamefont {Matarrese}, \citenamefont {Peloso}, \citenamefont
  {Ricciardone}, \citenamefont {Riotto},\ and\ \citenamefont
  {Tasinato}}]{Bartolo:2019oiq}%
  \BibitemOpen
  \bibfield  {author} {\bibinfo {author} {\bibfnamefont {N.}~\bibnamefont
  {Bartolo}}, \bibinfo {author} {\bibfnamefont {D.}~\bibnamefont {Bertacca}},
  \bibinfo {author} {\bibfnamefont {S.}~\bibnamefont {Matarrese}}, \bibinfo
  {author} {\bibfnamefont {M.}~\bibnamefont {Peloso}}, \bibinfo {author}
  {\bibfnamefont {A.}~\bibnamefont {Ricciardone}}, \bibinfo {author}
  {\bibfnamefont {A.}~\bibnamefont {Riotto}}, \ and\ \bibinfo {author}
  {\bibfnamefont {G.}~\bibnamefont {Tasinato}},\ }\href {\doibase
  10.1103/PhysRevD.100.121501} {\bibfield  {journal} {\bibinfo  {journal}
  {Phys. Rev. D}\ }\textbf {\bibinfo {volume} {100}},\ \bibinfo {pages}
  {121501} (\bibinfo {year} {2019}{\natexlab{b}})},\ \Eprint
  {http://arxiv.org/abs/1908.00527} {arXiv:1908.00527 [astro-ph.CO]}
  \BibitemShut {NoStop}%
\bibitem [{\citenamefont {Bartolo}\ \emph
  {et~al.}(2020{\natexlab{a}})\citenamefont {Bartolo}, \citenamefont
  {Bertacca}, \citenamefont {Matarrese}, \citenamefont {Peloso}, \citenamefont
  {Ricciardone}, \citenamefont {Riotto},\ and\ \citenamefont
  {Tasinato}}]{Bartolo:2019yeu}%
  \BibitemOpen
  \bibfield  {author} {\bibinfo {author} {\bibfnamefont {N.}~\bibnamefont
  {Bartolo}}, \bibinfo {author} {\bibfnamefont {D.}~\bibnamefont {Bertacca}},
  \bibinfo {author} {\bibfnamefont {S.}~\bibnamefont {Matarrese}}, \bibinfo
  {author} {\bibfnamefont {M.}~\bibnamefont {Peloso}}, \bibinfo {author}
  {\bibfnamefont {A.}~\bibnamefont {Ricciardone}}, \bibinfo {author}
  {\bibfnamefont {A.}~\bibnamefont {Riotto}}, \ and\ \bibinfo {author}
  {\bibfnamefont {G.}~\bibnamefont {Tasinato}},\ }\href {\doibase
  10.1103/PhysRevD.102.023527} {\bibfield  {journal} {\bibinfo  {journal}
  {Phys. Rev. D}\ }\textbf {\bibinfo {volume} {102}},\ \bibinfo {pages}
  {023527} (\bibinfo {year} {2020}{\natexlab{a}})},\ \Eprint
  {http://arxiv.org/abs/1912.09433} {arXiv:1912.09433 [astro-ph.CO]}
  \BibitemShut {NoStop}%
\bibitem [{\citenamefont {Schulze}\ \emph {et~al.}(2023)\citenamefont
  {Schulze}, \citenamefont {Valbusa~Dall'Armi}, \citenamefont {Lesgourgues},
  \citenamefont {Ricciardone}, \citenamefont {Bartolo}, \citenamefont
  {Bertacca}, \citenamefont {Fidler},\ and\ \citenamefont
  {Matarrese}}]{Schulze:2023ich}%
  \BibitemOpen
  \bibfield  {author} {\bibinfo {author} {\bibfnamefont {F.}~\bibnamefont
  {Schulze}}, \bibinfo {author} {\bibfnamefont {L.}~\bibnamefont
  {Valbusa~Dall'Armi}}, \bibinfo {author} {\bibfnamefont {J.}~\bibnamefont
  {Lesgourgues}}, \bibinfo {author} {\bibfnamefont {A.}~\bibnamefont
  {Ricciardone}}, \bibinfo {author} {\bibfnamefont {N.}~\bibnamefont
  {Bartolo}}, \bibinfo {author} {\bibfnamefont {D.}~\bibnamefont {Bertacca}},
  \bibinfo {author} {\bibfnamefont {C.}~\bibnamefont {Fidler}}, \ and\ \bibinfo
  {author} {\bibfnamefont {S.}~\bibnamefont {Matarrese}},\ }\href {\doibase
  10.1088/1475-7516/2023/10/025} {\bibfield  {journal} {\bibinfo  {journal}
  {JCAP}\ }\textbf {\bibinfo {volume} {10}},\ \bibinfo {pages} {025} (\bibinfo
  {year} {2023})},\ \Eprint {http://arxiv.org/abs/2305.01602} {arXiv:2305.01602
  [gr-qc]} \BibitemShut {NoStop}%
\bibitem [{\citenamefont {Bartolo}\ \emph
  {et~al.}(2020{\natexlab{b}})\citenamefont {Bartolo}, \citenamefont
  {Bertacca}, \citenamefont {De~Luca}, \citenamefont {Franciolini},
  \citenamefont {Matarrese}, \citenamefont {Peloso}, \citenamefont
  {Ricciardone}, \citenamefont {Riotto},\ and\ \citenamefont
  {Tasinato}}]{Bartolo:2019zvb}%
  \BibitemOpen
  \bibfield  {author} {\bibinfo {author} {\bibfnamefont {N.}~\bibnamefont
  {Bartolo}}, \bibinfo {author} {\bibfnamefont {D.}~\bibnamefont {Bertacca}},
  \bibinfo {author} {\bibfnamefont {V.}~\bibnamefont {De~Luca}}, \bibinfo
  {author} {\bibfnamefont {G.}~\bibnamefont {Franciolini}}, \bibinfo {author}
  {\bibfnamefont {S.}~\bibnamefont {Matarrese}}, \bibinfo {author}
  {\bibfnamefont {M.}~\bibnamefont {Peloso}}, \bibinfo {author} {\bibfnamefont
  {A.}~\bibnamefont {Ricciardone}}, \bibinfo {author} {\bibfnamefont
  {A.}~\bibnamefont {Riotto}}, \ and\ \bibinfo {author} {\bibfnamefont
  {G.}~\bibnamefont {Tasinato}},\ }\href {\doibase
  10.1088/1475-7516/2020/02/028} {\bibfield  {journal} {\bibinfo  {journal}
  {JCAP}\ }\textbf {\bibinfo {volume} {02}},\ \bibinfo {pages} {028} (\bibinfo
  {year} {2020}{\natexlab{b}})},\ \Eprint {http://arxiv.org/abs/1909.12619}
  {arXiv:1909.12619 [astro-ph.CO]} \BibitemShut {NoStop}%
\bibitem [{\citenamefont {Li}\ \emph {et~al.}(2023{\natexlab{b}})\citenamefont
  {Li}, \citenamefont {Wang}, \citenamefont {Zhao},\ and\ \citenamefont
  {Kohri}}]{Li:2023qua}%
  \BibitemOpen
  \bibfield  {author} {\bibinfo {author} {\bibfnamefont {J.-P.}\ \bibnamefont
  {Li}}, \bibinfo {author} {\bibfnamefont {S.}~\bibnamefont {Wang}}, \bibinfo
  {author} {\bibfnamefont {Z.-C.}\ \bibnamefont {Zhao}}, \ and\ \bibinfo
  {author} {\bibfnamefont {K.}~\bibnamefont {Kohri}},\ }\href {\doibase
  10.1088/1475-7516/2023/10/056} {\bibfield  {journal} {\bibinfo  {journal}
  {JCAP}\ }\textbf {\bibinfo {volume} {10}},\ \bibinfo {pages} {056} (\bibinfo
  {year} {2023}{\natexlab{b}})},\ \Eprint {http://arxiv.org/abs/2305.19950}
  {arXiv:2305.19950 [astro-ph.CO]} \BibitemShut {NoStop}%
\bibitem [{\citenamefont {Rey}(2024)}]{Rey:2024giu}%
  \BibitemOpen
  \bibfield  {author} {\bibinfo {author} {\bibfnamefont {J.}~\bibnamefont
  {Rey}},\ }\href@noop {} {\  (\bibinfo {year} {2024})},\ \Eprint
  {http://arxiv.org/abs/2411.08873} {arXiv:2411.08873 [astro-ph.CO]}
  \BibitemShut {NoStop}%
\bibitem [{\citenamefont {Ruiz}\ and\ \citenamefont
  {Rey}(2024)}]{Ruiz:2024weh}%
  \BibitemOpen
  \bibfield  {author} {\bibinfo {author} {\bibfnamefont {J.~A.}\ \bibnamefont
  {Ruiz}}\ and\ \bibinfo {author} {\bibfnamefont {J.}~\bibnamefont {Rey}},\
  }\href@noop {} {\  (\bibinfo {year} {2024})},\ \Eprint
  {http://arxiv.org/abs/2410.09014} {arXiv:2410.09014 [astro-ph.CO]}
  \BibitemShut {NoStop}%
\bibitem [{\citenamefont {Zhao}\ \emph {et~al.}(2024)\citenamefont {Zhao},
  \citenamefont {Wang}, \citenamefont {Li},\ and\ \citenamefont
  {Kohri}}]{Zhao:2024gan}%
  \BibitemOpen
  \bibfield  {author} {\bibinfo {author} {\bibfnamefont {Z.-C.}\ \bibnamefont
  {Zhao}}, \bibinfo {author} {\bibfnamefont {S.}~\bibnamefont {Wang}}, \bibinfo
  {author} {\bibfnamefont {J.-P.}\ \bibnamefont {Li}}, \ and\ \bibinfo {author}
  {\bibfnamefont {K.}~\bibnamefont {Kohri}},\ }\href@noop {} {\  (\bibinfo
  {year} {2024})},\ \Eprint {http://arxiv.org/abs/2412.02500} {arXiv:2412.02500
  [astro-ph.CO]} \BibitemShut {NoStop}%
\bibitem [{\citenamefont {Bartolo}\ \emph {et~al.}(2022)\citenamefont {Bartolo}
  \emph {et~al.}}]{LISACosmologyWorkingGroup:2022kbp}%
  \BibitemOpen
  \bibfield  {author} {\bibinfo {author} {\bibfnamefont {N.}~\bibnamefont
  {Bartolo}} \emph {et~al.} (\bibinfo {collaboration} {LISA Cosmology Working
  Group}),\ }\href {\doibase 10.1088/1475-7516/2022/11/009} {\bibfield
  {journal} {\bibinfo  {journal} {JCAP}\ }\textbf {\bibinfo {volume} {11}},\
  \bibinfo {pages} {009} (\bibinfo {year} {2022})},\ \Eprint
  {http://arxiv.org/abs/2201.08782} {arXiv:2201.08782 [astro-ph.CO]}
  \BibitemShut {NoStop}%
\bibitem [{\citenamefont {Malhotra}\ \emph {et~al.}(2023)\citenamefont
  {Malhotra}, \citenamefont {Dimastrogiovanni}, \citenamefont {Dom\`enech},
  \citenamefont {Fasiello},\ and\ \citenamefont {Tasinato}}]{Malhotra:2022ply}%
  \BibitemOpen
  \bibfield  {author} {\bibinfo {author} {\bibfnamefont {A.}~\bibnamefont
  {Malhotra}}, \bibinfo {author} {\bibfnamefont {E.}~\bibnamefont
  {Dimastrogiovanni}}, \bibinfo {author} {\bibfnamefont {G.}~\bibnamefont
  {Dom\`enech}}, \bibinfo {author} {\bibfnamefont {M.}~\bibnamefont
  {Fasiello}}, \ and\ \bibinfo {author} {\bibfnamefont {G.}~\bibnamefont
  {Tasinato}},\ }\href {\doibase 10.1103/PhysRevD.107.103502} {\bibfield
  {journal} {\bibinfo  {journal} {Phys. Rev. D}\ }\textbf {\bibinfo {volume}
  {107}},\ \bibinfo {pages} {103502} (\bibinfo {year} {2023})},\ \Eprint
  {http://arxiv.org/abs/2212.10316} {arXiv:2212.10316 [gr-qc]} \BibitemShut
  {NoStop}%
\bibitem [{\citenamefont {Yu}\ and\ \citenamefont {Wang}(2024)}]{Yu:2023jrs}%
  \BibitemOpen
  \bibfield  {author} {\bibinfo {author} {\bibfnamefont {Y.-H.}\ \bibnamefont
  {Yu}}\ and\ \bibinfo {author} {\bibfnamefont {S.}~\bibnamefont {Wang}},\
  }\href {\doibase 10.1103/PhysRevD.109.083501} {\bibfield  {journal} {\bibinfo
   {journal} {Phys. Rev. D}\ }\textbf {\bibinfo {volume} {109}},\ \bibinfo
  {pages} {083501} (\bibinfo {year} {2024})},\ \Eprint
  {http://arxiv.org/abs/2310.14606} {arXiv:2310.14606 [astro-ph.CO]}
  \BibitemShut {NoStop}%
\bibitem [{\citenamefont {Cusin}\ \emph {et~al.}(2018)\citenamefont {Cusin},
  \citenamefont {Dvorkin}, \citenamefont {Pitrou},\ and\ \citenamefont
  {Uzan}}]{Cusin:2018rsq}%
  \BibitemOpen
  \bibfield  {author} {\bibinfo {author} {\bibfnamefont {G.}~\bibnamefont
  {Cusin}}, \bibinfo {author} {\bibfnamefont {I.}~\bibnamefont {Dvorkin}},
  \bibinfo {author} {\bibfnamefont {C.}~\bibnamefont {Pitrou}}, \ and\ \bibinfo
  {author} {\bibfnamefont {J.-P.}\ \bibnamefont {Uzan}},\ }\href {\doibase
  10.1103/PhysRevLett.120.231101} {\bibfield  {journal} {\bibinfo  {journal}
  {Phys. Rev. Lett.}\ }\textbf {\bibinfo {volume} {120}},\ \bibinfo {pages}
  {231101} (\bibinfo {year} {2018})},\ \Eprint
  {http://arxiv.org/abs/1803.03236} {arXiv:1803.03236 [astro-ph.CO]}
  \BibitemShut {NoStop}%
\bibitem [{\citenamefont {Cusin}\ \emph {et~al.}(2017)\citenamefont {Cusin},
  \citenamefont {Pitrou},\ and\ \citenamefont {Uzan}}]{Cusin:2017fwz}%
  \BibitemOpen
  \bibfield  {author} {\bibinfo {author} {\bibfnamefont {G.}~\bibnamefont
  {Cusin}}, \bibinfo {author} {\bibfnamefont {C.}~\bibnamefont {Pitrou}}, \
  and\ \bibinfo {author} {\bibfnamefont {J.-P.}\ \bibnamefont {Uzan}},\ }\href
  {\doibase 10.1103/PhysRevD.96.103019} {\bibfield  {journal} {\bibinfo
  {journal} {Phys. Rev. D}\ }\textbf {\bibinfo {volume} {96}},\ \bibinfo
  {pages} {103019} (\bibinfo {year} {2017})},\ \Eprint
  {http://arxiv.org/abs/1704.06184} {arXiv:1704.06184 [astro-ph.CO]}
  \BibitemShut {NoStop}%
\bibitem [{\citenamefont {Cusin}\ \emph {et~al.}(2020)\citenamefont {Cusin},
  \citenamefont {Dvorkin}, \citenamefont {Pitrou},\ and\ \citenamefont
  {Uzan}}]{Cusin:2019jhg}%
  \BibitemOpen
  \bibfield  {author} {\bibinfo {author} {\bibfnamefont {G.}~\bibnamefont
  {Cusin}}, \bibinfo {author} {\bibfnamefont {I.}~\bibnamefont {Dvorkin}},
  \bibinfo {author} {\bibfnamefont {C.}~\bibnamefont {Pitrou}}, \ and\ \bibinfo
  {author} {\bibfnamefont {J.-P.}\ \bibnamefont {Uzan}},\ }\href {\doibase
  10.1093/mnrasl/slz182} {\bibfield  {journal} {\bibinfo  {journal} {Mon. Not.
  Roy. Astron. Soc.}\ }\textbf {\bibinfo {volume} {493}},\ \bibinfo {pages}
  {L1} (\bibinfo {year} {2020})},\ \Eprint {http://arxiv.org/abs/1904.07757}
  {arXiv:1904.07757 [astro-ph.CO]} \BibitemShut {NoStop}%
\bibitem [{\citenamefont {Cusin}\ \emph {et~al.}(2019)\citenamefont {Cusin},
  \citenamefont {Dvorkin}, \citenamefont {Pitrou},\ and\ \citenamefont
  {Uzan}}]{Cusin:2019jpv}%
  \BibitemOpen
  \bibfield  {author} {\bibinfo {author} {\bibfnamefont {G.}~\bibnamefont
  {Cusin}}, \bibinfo {author} {\bibfnamefont {I.}~\bibnamefont {Dvorkin}},
  \bibinfo {author} {\bibfnamefont {C.}~\bibnamefont {Pitrou}}, \ and\ \bibinfo
  {author} {\bibfnamefont {J.-P.}\ \bibnamefont {Uzan}},\ }\href {\doibase
  10.1103/PhysRevD.100.063004} {\bibfield  {journal} {\bibinfo  {journal}
  {Phys. Rev. D}\ }\textbf {\bibinfo {volume} {100}},\ \bibinfo {pages}
  {063004} (\bibinfo {year} {2019})},\ \Eprint
  {http://arxiv.org/abs/1904.07797} {arXiv:1904.07797 [astro-ph.CO]}
  \BibitemShut {NoStop}%
\bibitem [{\citenamefont {Jenkins}\ \emph
  {et~al.}(2019{\natexlab{a}})\citenamefont {Jenkins}, \citenamefont
  {O'Shaughnessy}, \citenamefont {Sakellariadou},\ and\ \citenamefont
  {Wysocki}}]{Jenkins:2018kxc}%
  \BibitemOpen
  \bibfield  {author} {\bibinfo {author} {\bibfnamefont {A.~C.}\ \bibnamefont
  {Jenkins}}, \bibinfo {author} {\bibfnamefont {R.}~\bibnamefont
  {O'Shaughnessy}}, \bibinfo {author} {\bibfnamefont {M.}~\bibnamefont
  {Sakellariadou}}, \ and\ \bibinfo {author} {\bibfnamefont {D.}~\bibnamefont
  {Wysocki}},\ }\href {\doibase 10.1103/PhysRevLett.122.111101} {\bibfield
  {journal} {\bibinfo  {journal} {Phys. Rev. Lett.}\ }\textbf {\bibinfo
  {volume} {122}},\ \bibinfo {pages} {111101} (\bibinfo {year}
  {2019}{\natexlab{a}})},\ \Eprint {http://arxiv.org/abs/1810.13435}
  {arXiv:1810.13435 [astro-ph.CO]} \BibitemShut {NoStop}%
\bibitem [{\citenamefont {Jenkins}\ \emph {et~al.}(2018)\citenamefont
  {Jenkins}, \citenamefont {Sakellariadou}, \citenamefont {Regimbau},\ and\
  \citenamefont {Slezak}}]{Jenkins:2018uac}%
  \BibitemOpen
  \bibfield  {author} {\bibinfo {author} {\bibfnamefont {A.~C.}\ \bibnamefont
  {Jenkins}}, \bibinfo {author} {\bibfnamefont {M.}~\bibnamefont
  {Sakellariadou}}, \bibinfo {author} {\bibfnamefont {T.}~\bibnamefont
  {Regimbau}}, \ and\ \bibinfo {author} {\bibfnamefont {E.}~\bibnamefont
  {Slezak}},\ }\href {\doibase 10.1103/PhysRevD.98.063501} {\bibfield
  {journal} {\bibinfo  {journal} {Phys. Rev. D}\ }\textbf {\bibinfo {volume}
  {98}},\ \bibinfo {pages} {063501} (\bibinfo {year} {2018})},\ \Eprint
  {http://arxiv.org/abs/1806.01718} {arXiv:1806.01718 [astro-ph.CO]}
  \BibitemShut {NoStop}%
\bibitem [{\citenamefont {Jenkins}\ \emph
  {et~al.}(2019{\natexlab{b}})\citenamefont {Jenkins}, \citenamefont {Romano},\
  and\ \citenamefont {Sakellariadou}}]{Jenkins:2019nks}%
  \BibitemOpen
  \bibfield  {author} {\bibinfo {author} {\bibfnamefont {A.~C.}\ \bibnamefont
  {Jenkins}}, \bibinfo {author} {\bibfnamefont {J.~D.}\ \bibnamefont {Romano}},
  \ and\ \bibinfo {author} {\bibfnamefont {M.}~\bibnamefont {Sakellariadou}},\
  }\href {\doibase 10.1103/PhysRevD.100.083501} {\bibfield  {journal} {\bibinfo
   {journal} {Phys. Rev. D}\ }\textbf {\bibinfo {volume} {100}},\ \bibinfo
  {pages} {083501} (\bibinfo {year} {2019}{\natexlab{b}})},\ \Eprint
  {http://arxiv.org/abs/1907.06642} {arXiv:1907.06642 [astro-ph.CO]}
  \BibitemShut {NoStop}%
\bibitem [{\citenamefont {Wang}\ \emph
  {et~al.}(2022{\natexlab{b}})\citenamefont {Wang}, \citenamefont {Vardanyan},\
  and\ \citenamefont {Kohri}}]{Wang:2021djr}%
  \BibitemOpen
  \bibfield  {author} {\bibinfo {author} {\bibfnamefont {S.}~\bibnamefont
  {Wang}}, \bibinfo {author} {\bibfnamefont {V.}~\bibnamefont {Vardanyan}}, \
  and\ \bibinfo {author} {\bibfnamefont {K.}~\bibnamefont {Kohri}},\ }\href
  {\doibase 10.1103/PhysRevD.106.123511} {\bibfield  {journal} {\bibinfo
  {journal} {Phys. Rev. D}\ }\textbf {\bibinfo {volume} {106}},\ \bibinfo
  {pages} {123511} (\bibinfo {year} {2022}{\natexlab{b}})},\ \Eprint
  {http://arxiv.org/abs/2107.01935} {arXiv:2107.01935 [gr-qc]} \BibitemShut
  {NoStop}%
\bibitem [{\citenamefont {Mukherjee}\ and\ \citenamefont
  {Silk}(2020)}]{Mukherjee:2019oma}%
  \BibitemOpen
  \bibfield  {author} {\bibinfo {author} {\bibfnamefont {S.}~\bibnamefont
  {Mukherjee}}\ and\ \bibinfo {author} {\bibfnamefont {J.}~\bibnamefont
  {Silk}},\ }\href {\doibase 10.1093/mnras/stz3226} {\bibfield  {journal}
  {\bibinfo  {journal} {Mon. Not. Roy. Astron. Soc.}\ }\textbf {\bibinfo
  {volume} {491}},\ \bibinfo {pages} {4690} (\bibinfo {year} {2020})},\ \Eprint
  {http://arxiv.org/abs/1912.07657} {arXiv:1912.07657 [gr-qc]} \BibitemShut
  {NoStop}%
\bibitem [{\citenamefont {Bavera}\ \emph {et~al.}(2022)\citenamefont {Bavera},
  \citenamefont {Franciolini}, \citenamefont {Cusin}, \citenamefont {Riotto},
  \citenamefont {Zevin},\ and\ \citenamefont {Fragos}}]{Bavera:2021wmw}%
  \BibitemOpen
  \bibfield  {author} {\bibinfo {author} {\bibfnamefont {S.~S.}\ \bibnamefont
  {Bavera}}, \bibinfo {author} {\bibfnamefont {G.}~\bibnamefont {Franciolini}},
  \bibinfo {author} {\bibfnamefont {G.}~\bibnamefont {Cusin}}, \bibinfo
  {author} {\bibfnamefont {A.}~\bibnamefont {Riotto}}, \bibinfo {author}
  {\bibfnamefont {M.}~\bibnamefont {Zevin}}, \ and\ \bibinfo {author}
  {\bibfnamefont {T.}~\bibnamefont {Fragos}},\ }\href {\doibase
  10.1051/0004-6361/202142208} {\bibfield  {journal} {\bibinfo  {journal}
  {Astron. Astrophys.}\ }\textbf {\bibinfo {volume} {660}},\ \bibinfo {pages}
  {A26} (\bibinfo {year} {2022})},\ \Eprint {http://arxiv.org/abs/2109.05836}
  {arXiv:2109.05836 [astro-ph.CO]} \BibitemShut {NoStop}%
\bibitem [{\citenamefont {Bellomo}\ \emph {et~al.}(2022)\citenamefont
  {Bellomo}, \citenamefont {Bertacca}, \citenamefont {Jenkins}, \citenamefont
  {Matarrese}, \citenamefont {Raccanelli}, \citenamefont {Regimbau},
  \citenamefont {Ricciardone},\ and\ \citenamefont
  {Sakellariadou}}]{Bellomo:2021mer}%
  \BibitemOpen
  \bibfield  {author} {\bibinfo {author} {\bibfnamefont {N.}~\bibnamefont
  {Bellomo}}, \bibinfo {author} {\bibfnamefont {D.}~\bibnamefont {Bertacca}},
  \bibinfo {author} {\bibfnamefont {A.~C.}\ \bibnamefont {Jenkins}}, \bibinfo
  {author} {\bibfnamefont {S.}~\bibnamefont {Matarrese}}, \bibinfo {author}
  {\bibfnamefont {A.}~\bibnamefont {Raccanelli}}, \bibinfo {author}
  {\bibfnamefont {T.}~\bibnamefont {Regimbau}}, \bibinfo {author}
  {\bibfnamefont {A.}~\bibnamefont {Ricciardone}}, \ and\ \bibinfo {author}
  {\bibfnamefont {M.}~\bibnamefont {Sakellariadou}},\ }\href {\doibase
  10.1088/1475-7516/2022/06/030} {\bibfield  {journal} {\bibinfo  {journal}
  {JCAP}\ }\textbf {\bibinfo {volume} {06}},\ \bibinfo {pages} {030} (\bibinfo
  {year} {2022})},\ \Eprint {http://arxiv.org/abs/2110.15059} {arXiv:2110.15059
  [gr-qc]} \BibitemShut {NoStop}%
\bibitem [{\citenamefont {Li}\ \emph {et~al.}(2024{\natexlab{d}})\citenamefont
  {Li}, \citenamefont {Jiang}, \citenamefont {Liu}, \citenamefont {Fan},
  \citenamefont {Gao}, \citenamefont {Chen},\ and\ \citenamefont
  {Xu}}]{Li:2024qcs}%
  \BibitemOpen
  \bibfield  {author} {\bibinfo {author} {\bibfnamefont {Z.}~\bibnamefont
  {Li}}, \bibinfo {author} {\bibfnamefont {Z.}~\bibnamefont {Jiang}}, \bibinfo
  {author} {\bibfnamefont {Y.}~\bibnamefont {Liu}}, \bibinfo {author}
  {\bibfnamefont {X.-L.}\ \bibnamefont {Fan}}, \bibinfo {author} {\bibfnamefont
  {L.}~\bibnamefont {Gao}}, \bibinfo {author} {\bibfnamefont {Y.}~\bibnamefont
  {Chen}}, \ and\ \bibinfo {author} {\bibfnamefont {T.}~\bibnamefont {Xu}},\
  }\href@noop {} {\  (\bibinfo {year} {2024}{\natexlab{d}})},\ \Eprint
  {http://arxiv.org/abs/2412.09956} {arXiv:2412.09956 [astro-ph.HE]}
  \BibitemShut {NoStop}%
\bibitem [{\citenamefont {Liang}\ \emph
  {et~al.}(2024{\natexlab{b}})\citenamefont {Liang}, \citenamefont {Li},
  \citenamefont {Li}, \citenamefont {Zhang},\ and\ \citenamefont
  {Hu}}]{Liang:2023fdf}%
  \BibitemOpen
  \bibfield  {author} {\bibinfo {author} {\bibfnamefont {Z.-C.}\ \bibnamefont
  {Liang}}, \bibinfo {author} {\bibfnamefont {Z.-Y.}\ \bibnamefont {Li}},
  \bibinfo {author} {\bibfnamefont {E.-K.}\ \bibnamefont {Li}}, \bibinfo
  {author} {\bibfnamefont {J.-d.}\ \bibnamefont {Zhang}}, \ and\ \bibinfo
  {author} {\bibfnamefont {Y.-M.}\ \bibnamefont {Hu}},\ }\href {\doibase
  10.1103/PhysRevD.110.043031} {\bibfield  {journal} {\bibinfo  {journal}
  {Phys. Rev. D}\ }\textbf {\bibinfo {volume} {110}},\ \bibinfo {pages}
  {043031} (\bibinfo {year} {2024}{\natexlab{b}})},\ \Eprint
  {http://arxiv.org/abs/2307.01541} {arXiv:2307.01541 [gr-qc]} \BibitemShut
  {NoStop}%
\bibitem [{\citenamefont {Li}\ \emph {et~al.}(2024{\natexlab{e}})\citenamefont
  {Li}, \citenamefont {Liang}, \citenamefont {Li}, \citenamefont {Zhang},\ and\
  \citenamefont {Hu}}]{Li:2024lvt}%
  \BibitemOpen
  \bibfield  {author} {\bibinfo {author} {\bibfnamefont {Z.-Y.}\ \bibnamefont
  {Li}}, \bibinfo {author} {\bibfnamefont {Z.-C.}\ \bibnamefont {Liang}},
  \bibinfo {author} {\bibfnamefont {E.-K.}\ \bibnamefont {Li}}, \bibinfo
  {author} {\bibfnamefont {J.-d.}\ \bibnamefont {Zhang}}, \ and\ \bibinfo
  {author} {\bibfnamefont {Y.-M.}\ \bibnamefont {Hu}},\ }\href@noop {} {\
  (\bibinfo {year} {2024}{\natexlab{e}})},\ \Eprint
  {http://arxiv.org/abs/2409.11245} {arXiv:2409.11245 [gr-qc]} \BibitemShut
  {NoStop}%
\bibitem [{\citenamefont {Li}\ \emph {et~al.}(2024{\natexlab{f}})\citenamefont
  {Li}, \citenamefont {Wang}, \citenamefont {Zhao},\ and\ \citenamefont
  {Kohri}}]{Li:2024zwx}%
  \BibitemOpen
  \bibfield  {author} {\bibinfo {author} {\bibfnamefont {J.-P.}\ \bibnamefont
  {Li}}, \bibinfo {author} {\bibfnamefont {S.}~\bibnamefont {Wang}}, \bibinfo
  {author} {\bibfnamefont {Z.-C.}\ \bibnamefont {Zhao}}, \ and\ \bibinfo
  {author} {\bibfnamefont {K.}~\bibnamefont {Kohri}},\ }\href {\doibase
  10.1088/1475-7516/2024/05/109} {\bibfield  {journal} {\bibinfo  {journal}
  {JCAP}\ }\textbf {\bibinfo {volume} {05}},\ \bibinfo {pages} {109} (\bibinfo
  {year} {2024}{\natexlab{f}})},\ \Eprint {http://arxiv.org/abs/2403.00238}
  {arXiv:2403.00238 [astro-ph.CO]} \BibitemShut {NoStop}%
\bibitem [{\citenamefont {Dimastrogiovanni}\ \emph {et~al.}(2023)\citenamefont
  {Dimastrogiovanni}, \citenamefont {Fasiello}, \citenamefont {Malhotra},\ and\
  \citenamefont {Tasinato}}]{Dimastrogiovanni:2022eir}%
  \BibitemOpen
  \bibfield  {author} {\bibinfo {author} {\bibfnamefont {E.}~\bibnamefont
  {Dimastrogiovanni}}, \bibinfo {author} {\bibfnamefont {M.}~\bibnamefont
  {Fasiello}}, \bibinfo {author} {\bibfnamefont {A.}~\bibnamefont {Malhotra}},
  \ and\ \bibinfo {author} {\bibfnamefont {G.}~\bibnamefont {Tasinato}},\
  }\href {\doibase 10.1088/1475-7516/2023/01/018} {\bibfield  {journal}
  {\bibinfo  {journal} {JCAP}\ }\textbf {\bibinfo {volume} {01}},\ \bibinfo
  {pages} {018} (\bibinfo {year} {2023})},\ \Eprint
  {http://arxiv.org/abs/2205.05644} {arXiv:2205.05644 [astro-ph.CO]}
  \BibitemShut {NoStop}%
\bibitem [{\citenamefont {Cai}\ \emph {et~al.}(2024)\citenamefont {Cai},
  \citenamefont {Wang}, \citenamefont {Yuwen},\ and\ \citenamefont
  {Zeng}}]{Cai:2024dya}%
  \BibitemOpen
  \bibfield  {author} {\bibinfo {author} {\bibfnamefont {R.-G.}\ \bibnamefont
  {Cai}}, \bibinfo {author} {\bibfnamefont {S.-J.}\ \bibnamefont {Wang}},
  \bibinfo {author} {\bibfnamefont {Z.-Y.}\ \bibnamefont {Yuwen}}, \ and\
  \bibinfo {author} {\bibfnamefont {X.-X.}\ \bibnamefont {Zeng}},\ }\href@noop
  {} {\  (\bibinfo {year} {2024})},\ \Eprint {http://arxiv.org/abs/2410.17721}
  {arXiv:2410.17721 [astro-ph.CO]} \BibitemShut {NoStop}%
\bibitem [{\citenamefont {Hu}\ and\ \citenamefont {Zhou}(2025)}]{Hu:2025xdt}%
  \BibitemOpen
  \bibfield  {author} {\bibinfo {author} {\bibfnamefont {X.-H.}\ \bibnamefont
  {Hu}}\ and\ \bibinfo {author} {\bibfnamefont {Y.-L.}\ \bibnamefont {Zhou}},\
  }\href@noop {} {\  (\bibinfo {year} {2025})},\ \Eprint
  {http://arxiv.org/abs/2501.01491} {arXiv:2501.01491 [hep-ph]} \BibitemShut
  {NoStop}%
\bibitem [{\citenamefont {An}\ and\ \citenamefont {Yang}(2024)}]{An:2023idh}%
  \BibitemOpen
  \bibfield  {author} {\bibinfo {author} {\bibfnamefont {H.}~\bibnamefont
  {An}}\ and\ \bibinfo {author} {\bibfnamefont {C.}~\bibnamefont {Yang}},\
  }\href {\doibase 10.1103/PhysRevD.109.123508} {\bibfield  {journal} {\bibinfo
   {journal} {Phys. Rev. D}\ }\textbf {\bibinfo {volume} {109}},\ \bibinfo
  {pages} {123508} (\bibinfo {year} {2024})},\ \Eprint
  {http://arxiv.org/abs/2304.02361} {arXiv:2304.02361 [hep-ph]} \BibitemShut
  {NoStop}%
\bibitem [{\citenamefont {An}\ \emph {et~al.}(2024)\citenamefont {An},
  \citenamefont {Su}, \citenamefont {Tai}, \citenamefont {Wang},\ and\
  \citenamefont {Yang}}]{An:2023jxf}%
  \BibitemOpen
  \bibfield  {author} {\bibinfo {author} {\bibfnamefont {H.}~\bibnamefont
  {An}}, \bibinfo {author} {\bibfnamefont {B.}~\bibnamefont {Su}}, \bibinfo
  {author} {\bibfnamefont {H.}~\bibnamefont {Tai}}, \bibinfo {author}
  {\bibfnamefont {L.-T.}\ \bibnamefont {Wang}}, \ and\ \bibinfo {author}
  {\bibfnamefont {C.}~\bibnamefont {Yang}},\ }\href {\doibase
  10.1103/PhysRevD.109.L121304} {\bibfield  {journal} {\bibinfo  {journal}
  {Phys. Rev. D}\ }\textbf {\bibinfo {volume} {109}},\ \bibinfo {pages}
  {L121304} (\bibinfo {year} {2024})},\ \Eprint
  {http://arxiv.org/abs/2308.00070} {arXiv:2308.00070 [astro-ph.CO]}
  \BibitemShut {NoStop}%
\bibitem [{\citenamefont {Kibble}(1976)}]{Kibble:1976sj}%
  \BibitemOpen
  \bibfield  {author} {\bibinfo {author} {\bibfnamefont {T.~W.~B.}\
  \bibnamefont {Kibble}},\ }\href {\doibase 10.1088/0305-4470/9/8/029}
  {\bibfield  {journal} {\bibinfo  {journal} {J. Phys. A}\ }\textbf {\bibinfo
  {volume} {9}},\ \bibinfo {pages} {1387} (\bibinfo {year} {1976})}\BibitemShut
  {NoStop}%
\bibitem [{\citenamefont {Hindmarsh}\ and\ \citenamefont
  {Kibble}(1995)}]{Hindmarsh:1994re}%
  \BibitemOpen
  \bibfield  {author} {\bibinfo {author} {\bibfnamefont {M.~B.}\ \bibnamefont
  {Hindmarsh}}\ and\ \bibinfo {author} {\bibfnamefont {T.~W.~B.}\ \bibnamefont
  {Kibble}},\ }\href {\doibase 10.1088/0034-4885/58/5/001} {\bibfield
  {journal} {\bibinfo  {journal} {Rept. Prog. Phys.}\ }\textbf {\bibinfo
  {volume} {58}},\ \bibinfo {pages} {477} (\bibinfo {year} {1995})},\ \Eprint
  {http://arxiv.org/abs/hep-ph/9411342} {arXiv:hep-ph/9411342} \BibitemShut
  {NoStop}%
\bibitem [{\citenamefont {Bennett}\ and\ \citenamefont
  {Bouchet}(1988)}]{Bennett:1987vf}%
  \BibitemOpen
  \bibfield  {author} {\bibinfo {author} {\bibfnamefont {D.~P.}\ \bibnamefont
  {Bennett}}\ and\ \bibinfo {author} {\bibfnamefont {F.~R.}\ \bibnamefont
  {Bouchet}},\ }\href {\doibase 10.1103/PhysRevLett.60.257} {\bibfield
  {journal} {\bibinfo  {journal} {Phys. Rev. Lett.}\ }\textbf {\bibinfo
  {volume} {60}},\ \bibinfo {pages} {257} (\bibinfo {year} {1988})}\BibitemShut
  {NoStop}%
\bibitem [{\citenamefont {Allen}\ and\ \citenamefont
  {Shellard}(1990)}]{Allen:1990tv}%
  \BibitemOpen
  \bibfield  {author} {\bibinfo {author} {\bibfnamefont {B.}~\bibnamefont
  {Allen}}\ and\ \bibinfo {author} {\bibfnamefont {E.~P.~S.}\ \bibnamefont
  {Shellard}},\ }\href {\doibase 10.1103/PhysRevLett.64.119} {\bibfield
  {journal} {\bibinfo  {journal} {Phys. Rev. Lett.}\ }\textbf {\bibinfo
  {volume} {64}},\ \bibinfo {pages} {119} (\bibinfo {year} {1990})}\BibitemShut
  {NoStop}%
\bibitem [{\citenamefont {Sakellariadou}\ and\ \citenamefont
  {Vilenkin}(1990)}]{Sakellariadou:1990nd}%
  \BibitemOpen
  \bibfield  {author} {\bibinfo {author} {\bibfnamefont {M.}~\bibnamefont
  {Sakellariadou}}\ and\ \bibinfo {author} {\bibfnamefont {A.}~\bibnamefont
  {Vilenkin}},\ }\href {\doibase 10.1103/PhysRevD.42.349} {\bibfield  {journal}
  {\bibinfo  {journal} {Phys. Rev. D}\ }\textbf {\bibinfo {volume} {42}},\
  \bibinfo {pages} {349} (\bibinfo {year} {1990})}\BibitemShut {NoStop}%
\bibitem [{\citenamefont {Vachaspati}\ and\ \citenamefont
  {Vilenkin}(1984)}]{Vachaspati:1984dz}%
  \BibitemOpen
  \bibfield  {author} {\bibinfo {author} {\bibfnamefont {T.}~\bibnamefont
  {Vachaspati}}\ and\ \bibinfo {author} {\bibfnamefont {A.}~\bibnamefont
  {Vilenkin}},\ }\href {\doibase 10.1103/PhysRevD.30.2036} {\bibfield
  {journal} {\bibinfo  {journal} {Phys. Rev. D}\ }\textbf {\bibinfo {volume}
  {30}},\ \bibinfo {pages} {2036} (\bibinfo {year} {1984})}\BibitemShut
  {NoStop}%
\bibitem [{\citenamefont {Damour}\ and\ \citenamefont
  {Vilenkin}(2001)}]{Damour:2001bk}%
  \BibitemOpen
  \bibfield  {author} {\bibinfo {author} {\bibfnamefont {T.}~\bibnamefont
  {Damour}}\ and\ \bibinfo {author} {\bibfnamefont {A.}~\bibnamefont
  {Vilenkin}},\ }\href {\doibase 10.1103/PhysRevD.64.064008} {\bibfield
  {journal} {\bibinfo  {journal} {Phys. Rev. D}\ }\textbf {\bibinfo {volume}
  {64}},\ \bibinfo {pages} {064008} (\bibinfo {year} {2001})},\ \Eprint
  {http://arxiv.org/abs/gr-qc/0104026} {arXiv:gr-qc/0104026} \BibitemShut
  {NoStop}%
\bibitem [{\citenamefont {Damour}\ and\ \citenamefont
  {Vilenkin}(2000)}]{Damour:2000wa}%
  \BibitemOpen
  \bibfield  {author} {\bibinfo {author} {\bibfnamefont {T.}~\bibnamefont
  {Damour}}\ and\ \bibinfo {author} {\bibfnamefont {A.}~\bibnamefont
  {Vilenkin}},\ }\href {\doibase 10.1103/PhysRevLett.85.3761} {\bibfield
  {journal} {\bibinfo  {journal} {Phys. Rev. Lett.}\ }\textbf {\bibinfo
  {volume} {85}},\ \bibinfo {pages} {3761} (\bibinfo {year} {2000})},\ \Eprint
  {http://arxiv.org/abs/gr-qc/0004075} {arXiv:gr-qc/0004075} \BibitemShut
  {NoStop}%
\bibitem [{\citenamefont {Vilenkin}\ and\ \citenamefont
  {Shellard}(2000)}]{Vilenkin:2000jqa}%
  \BibitemOpen
  \bibfield  {author} {\bibinfo {author} {\bibfnamefont {A.}~\bibnamefont
  {Vilenkin}}\ and\ \bibinfo {author} {\bibfnamefont {E.~P.~S.}\ \bibnamefont
  {Shellard}},\ }\href@noop {} {\emph {\bibinfo {title} {{Cosmic Strings and
  Other Topological Defects}}}}\ (\bibinfo  {publisher} {Cambridge University
  Press},\ \bibinfo {year} {2000})\BibitemShut {NoStop}%
\bibitem [{\citenamefont {King}\ \emph {et~al.}(2021)\citenamefont {King},
  \citenamefont {Pascoli}, \citenamefont {Turner},\ and\ \citenamefont
  {Zhou}}]{King:2021gmj}%
  \BibitemOpen
  \bibfield  {author} {\bibinfo {author} {\bibfnamefont {S.~F.}\ \bibnamefont
  {King}}, \bibinfo {author} {\bibfnamefont {S.}~\bibnamefont {Pascoli}},
  \bibinfo {author} {\bibfnamefont {J.}~\bibnamefont {Turner}}, \ and\ \bibinfo
  {author} {\bibfnamefont {Y.-L.}\ \bibnamefont {Zhou}},\ }\href {\doibase
  10.1007/JHEP10(2021)225} {\bibfield  {journal} {\bibinfo  {journal} {JHEP}\
  }\textbf {\bibinfo {volume} {10}},\ \bibinfo {pages} {225} (\bibinfo {year}
  {2021})},\ \Eprint {http://arxiv.org/abs/2106.15634} {arXiv:2106.15634
  [hep-ph]} \BibitemShut {NoStop}%
\bibitem [{\citenamefont {King}\ \emph {et~al.}(2020)\citenamefont {King},
  \citenamefont {Pascoli}, \citenamefont {Turner},\ and\ \citenamefont
  {Zhou}}]{King:2020hyd}%
  \BibitemOpen
  \bibfield  {author} {\bibinfo {author} {\bibfnamefont {S.~F.}\ \bibnamefont
  {King}}, \bibinfo {author} {\bibfnamefont {S.}~\bibnamefont {Pascoli}},
  \bibinfo {author} {\bibfnamefont {J.}~\bibnamefont {Turner}}, \ and\ \bibinfo
  {author} {\bibfnamefont {Y.-L.}\ \bibnamefont {Zhou}},\ }\href@noop {} {\
  (\bibinfo {year} {2020})},\ \Eprint {http://arxiv.org/abs/2005.13549}
  {arXiv:2005.13549 [hep-ph]} \BibitemShut {NoStop}%
\bibitem [{\citenamefont {Buchmuller}\ \emph {et~al.}(2020)\citenamefont
  {Buchmuller}, \citenamefont {Domcke}, \citenamefont {Murayama},\ and\
  \citenamefont {Schmitz}}]{Buchmuller:2019gfy}%
  \BibitemOpen
  \bibfield  {author} {\bibinfo {author} {\bibfnamefont {W.}~\bibnamefont
  {Buchmuller}}, \bibinfo {author} {\bibfnamefont {V.}~\bibnamefont {Domcke}},
  \bibinfo {author} {\bibfnamefont {H.}~\bibnamefont {Murayama}}, \ and\
  \bibinfo {author} {\bibfnamefont {K.}~\bibnamefont {Schmitz}},\ }\href
  {\doibase 10.1016/j.physletb.2020.135764} {\bibfield  {journal} {\bibinfo
  {journal} {Phys. Lett. B}\ }\textbf {\bibinfo {volume} {809}},\ \bibinfo
  {pages} {135764} (\bibinfo {year} {2020})},\ \Eprint
  {http://arxiv.org/abs/1912.03695} {arXiv:1912.03695 [hep-ph]} \BibitemShut
  {NoStop}%
\bibitem [{\citenamefont {Caldwell}\ \emph {et~al.}(2022)\citenamefont
  {Caldwell} \emph {et~al.}}]{Caldwell:2022qsj}%
  \BibitemOpen
  \bibfield  {author} {\bibinfo {author} {\bibfnamefont {R.}~\bibnamefont
  {Caldwell}} \emph {et~al.},\ }\href@noop {} {\  (\bibinfo {year} {2022})},\
  \Eprint {http://arxiv.org/abs/2203.07972} {arXiv:2203.07972 [gr-qc]}
  \BibitemShut {NoStop}%
\bibitem [{\citenamefont {Dror}\ \emph {et~al.}(2020)\citenamefont {Dror},
  \citenamefont {Hiramatsu}, \citenamefont {Kohri}, \citenamefont {Murayama},\
  and\ \citenamefont {White}}]{Dror:2019syi}%
  \BibitemOpen
  \bibfield  {author} {\bibinfo {author} {\bibfnamefont {J.~A.}\ \bibnamefont
  {Dror}}, \bibinfo {author} {\bibfnamefont {T.}~\bibnamefont {Hiramatsu}},
  \bibinfo {author} {\bibfnamefont {K.}~\bibnamefont {Kohri}}, \bibinfo
  {author} {\bibfnamefont {H.}~\bibnamefont {Murayama}}, \ and\ \bibinfo
  {author} {\bibfnamefont {G.}~\bibnamefont {White}},\ }\href {\doibase
  10.1103/PhysRevLett.124.041804} {\bibfield  {journal} {\bibinfo  {journal}
  {Phys. Rev. Lett.}\ }\textbf {\bibinfo {volume} {124}},\ \bibinfo {pages}
  {041804} (\bibinfo {year} {2020})},\ \Eprint
  {http://arxiv.org/abs/1908.03227} {arXiv:1908.03227 [hep-ph]} \BibitemShut
  {NoStop}%
\bibitem [{\citenamefont {Bian}\ \emph {et~al.}(2021)\citenamefont {Bian},
  \citenamefont {Liu},\ and\ \citenamefont {Xie}}]{Bian:2021vmi}%
  \BibitemOpen
  \bibfield  {author} {\bibinfo {author} {\bibfnamefont {L.}~\bibnamefont
  {Bian}}, \bibinfo {author} {\bibfnamefont {X.}~\bibnamefont {Liu}}, \ and\
  \bibinfo {author} {\bibfnamefont {K.-P.}\ \bibnamefont {Xie}},\ }\href
  {\doibase 10.1007/JHEP11(2021)175} {\bibfield  {journal} {\bibinfo  {journal}
  {JHEP}\ }\textbf {\bibinfo {volume} {11}},\ \bibinfo {pages} {175} (\bibinfo
  {year} {2021})},\ \Eprint {http://arxiv.org/abs/2107.13112} {arXiv:2107.13112
  [hep-ph]} \BibitemShut {NoStop}%
\bibitem [{\citenamefont {Bian}\ \emph
  {et~al.}(2022{\natexlab{a}})\citenamefont {Bian}, \citenamefont {Tang},\ and\
  \citenamefont {Zhou}}]{Bian:2021dmp}%
  \BibitemOpen
  \bibfield  {author} {\bibinfo {author} {\bibfnamefont {L.}~\bibnamefont
  {Bian}}, \bibinfo {author} {\bibfnamefont {Y.-L.}\ \bibnamefont {Tang}}, \
  and\ \bibinfo {author} {\bibfnamefont {R.}~\bibnamefont {Zhou}},\ }\href
  {\doibase 10.1103/PhysRevD.106.035028} {\bibfield  {journal} {\bibinfo
  {journal} {Phys. Rev. D}\ }\textbf {\bibinfo {volume} {106}},\ \bibinfo
  {pages} {035028} (\bibinfo {year} {2022}{\natexlab{a}})},\ \Eprint
  {http://arxiv.org/abs/2111.10608} {arXiv:2111.10608 [hep-ph]} \BibitemShut
  {NoStop}%
\bibitem [{\citenamefont {Saurabh}\ \emph {et~al.}(2020)\citenamefont
  {Saurabh}, \citenamefont {Vachaspati},\ and\ \citenamefont
  {Pogosian}}]{Saurabh:2020pqe}%
  \BibitemOpen
  \bibfield  {author} {\bibinfo {author} {\bibfnamefont {A.}~\bibnamefont
  {Saurabh}}, \bibinfo {author} {\bibfnamefont {T.}~\bibnamefont {Vachaspati}},
  \ and\ \bibinfo {author} {\bibfnamefont {L.}~\bibnamefont {Pogosian}},\
  }\href {\doibase 10.1103/PhysRevD.101.083522} {\bibfield  {journal} {\bibinfo
   {journal} {Phys. Rev. D}\ }\textbf {\bibinfo {volume} {101}},\ \bibinfo
  {pages} {083522} (\bibinfo {year} {2020})},\ \Eprint
  {http://arxiv.org/abs/2001.01030} {arXiv:2001.01030 [hep-ph]} \BibitemShut
  {NoStop}%
\bibitem [{\citenamefont {Baeza-Ballesteros}\ \emph {et~al.}(2024)\citenamefont
  {Baeza-Ballesteros}, \citenamefont {Copeland}, \citenamefont {Figueroa},\
  and\ \citenamefont {Lizarraga}}]{Baeza-Ballesteros:2023say}%
  \BibitemOpen
  \bibfield  {author} {\bibinfo {author} {\bibfnamefont {J.}~\bibnamefont
  {Baeza-Ballesteros}}, \bibinfo {author} {\bibfnamefont {E.~J.}\ \bibnamefont
  {Copeland}}, \bibinfo {author} {\bibfnamefont {D.~G.}\ \bibnamefont
  {Figueroa}}, \ and\ \bibinfo {author} {\bibfnamefont {J.}~\bibnamefont
  {Lizarraga}},\ }\href {\doibase 10.1103/PhysRevD.110.043522} {\bibfield
  {journal} {\bibinfo  {journal} {Phys. Rev. D}\ }\textbf {\bibinfo {volume}
  {110}},\ \bibinfo {pages} {043522} (\bibinfo {year} {2024})},\ \Eprint
  {http://arxiv.org/abs/2308.08456} {arXiv:2308.08456 [astro-ph.CO]}
  \BibitemShut {NoStop}%
\bibitem [{\citenamefont {Di~Luzio}\ \emph {et~al.}(2020)\citenamefont
  {Di~Luzio}, \citenamefont {Giannotti}, \citenamefont {Nardi},\ and\
  \citenamefont {Visinelli}}]{DiLuzio:2020wdo}%
  \BibitemOpen
  \bibfield  {author} {\bibinfo {author} {\bibfnamefont {L.}~\bibnamefont
  {Di~Luzio}}, \bibinfo {author} {\bibfnamefont {M.}~\bibnamefont {Giannotti}},
  \bibinfo {author} {\bibfnamefont {E.}~\bibnamefont {Nardi}}, \ and\ \bibinfo
  {author} {\bibfnamefont {L.}~\bibnamefont {Visinelli}},\ }\href {\doibase
  10.1016/j.physrep.2020.06.002} {\bibfield  {journal} {\bibinfo  {journal}
  {Phys. Rept.}\ }\textbf {\bibinfo {volume} {870}},\ \bibinfo {pages} {1}
  (\bibinfo {year} {2020})},\ \Eprint {http://arxiv.org/abs/2003.01100}
  {arXiv:2003.01100 [hep-ph]} \BibitemShut {NoStop}%
\bibitem [{\citenamefont {Vilenkin}\ and\ \citenamefont
  {Everett}(1982)}]{Vilenkin:1982ks}%
  \BibitemOpen
  \bibfield  {author} {\bibinfo {author} {\bibfnamefont {A.}~\bibnamefont
  {Vilenkin}}\ and\ \bibinfo {author} {\bibfnamefont {A.~E.}\ \bibnamefont
  {Everett}},\ }\href {\doibase 10.1103/PhysRevLett.48.1867} {\bibfield
  {journal} {\bibinfo  {journal} {Phys. Rev. Lett.}\ }\textbf {\bibinfo
  {volume} {48}},\ \bibinfo {pages} {1867} (\bibinfo {year}
  {1982})}\BibitemShut {NoStop}%
\bibitem [{\citenamefont {Yamaguchi}\ \emph {et~al.}(1999)\citenamefont
  {Yamaguchi}, \citenamefont {Kawasaki},\ and\ \citenamefont
  {Yokoyama}}]{Yamaguchi:1998gx}%
  \BibitemOpen
  \bibfield  {author} {\bibinfo {author} {\bibfnamefont {M.}~\bibnamefont
  {Yamaguchi}}, \bibinfo {author} {\bibfnamefont {M.}~\bibnamefont {Kawasaki}},
  \ and\ \bibinfo {author} {\bibfnamefont {J.}~\bibnamefont {Yokoyama}},\
  }\href {\doibase 10.1103/PhysRevLett.82.4578} {\bibfield  {journal} {\bibinfo
   {journal} {Phys. Rev. Lett.}\ }\textbf {\bibinfo {volume} {82}},\ \bibinfo
  {pages} {4578} (\bibinfo {year} {1999})},\ \Eprint
  {http://arxiv.org/abs/hep-ph/9811311} {arXiv:hep-ph/9811311} \BibitemShut
  {NoStop}%
\bibitem [{\citenamefont {Hiramatsu}\ \emph {et~al.}(2011)\citenamefont
  {Hiramatsu}, \citenamefont {Kawasaki}, \citenamefont {Sekiguchi},
  \citenamefont {Yamaguchi},\ and\ \citenamefont
  {Yokoyama}}]{Hiramatsu:2010yu}%
  \BibitemOpen
  \bibfield  {author} {\bibinfo {author} {\bibfnamefont {T.}~\bibnamefont
  {Hiramatsu}}, \bibinfo {author} {\bibfnamefont {M.}~\bibnamefont {Kawasaki}},
  \bibinfo {author} {\bibfnamefont {T.}~\bibnamefont {Sekiguchi}}, \bibinfo
  {author} {\bibfnamefont {M.}~\bibnamefont {Yamaguchi}}, \ and\ \bibinfo
  {author} {\bibfnamefont {J.}~\bibnamefont {Yokoyama}},\ }\href {\doibase
  10.1103/PhysRevD.83.123531} {\bibfield  {journal} {\bibinfo  {journal} {Phys.
  Rev. D}\ }\textbf {\bibinfo {volume} {83}},\ \bibinfo {pages} {123531}
  (\bibinfo {year} {2011})},\ \Eprint {http://arxiv.org/abs/1012.5502}
  {arXiv:1012.5502 [hep-ph]} \BibitemShut {NoStop}%
\bibitem [{\citenamefont {Blanco-Pillado}\ and\ \citenamefont
  {Olum}(2017)}]{Blanco-Pillado:2017oxo}%
  \BibitemOpen
  \bibfield  {author} {\bibinfo {author} {\bibfnamefont {J.~J.}\ \bibnamefont
  {Blanco-Pillado}}\ and\ \bibinfo {author} {\bibfnamefont {K.~D.}\
  \bibnamefont {Olum}},\ }\href {\doibase 10.1103/PhysRevD.96.104046}
  {\bibfield  {journal} {\bibinfo  {journal} {Phys. Rev. D}\ }\textbf {\bibinfo
  {volume} {96}},\ \bibinfo {pages} {104046} (\bibinfo {year} {2017})},\
  \Eprint {http://arxiv.org/abs/1709.02693} {arXiv:1709.02693 [astro-ph.CO]}
  \BibitemShut {NoStop}%
\bibitem [{\citenamefont {Blanco-Pillado}\ \emph {et~al.}(2014)\citenamefont
  {Blanco-Pillado}, \citenamefont {Olum},\ and\ \citenamefont
  {Shlaer}}]{Blanco-Pillado:2013qja}%
  \BibitemOpen
  \bibfield  {author} {\bibinfo {author} {\bibfnamefont {J.~J.}\ \bibnamefont
  {Blanco-Pillado}}, \bibinfo {author} {\bibfnamefont {K.~D.}\ \bibnamefont
  {Olum}}, \ and\ \bibinfo {author} {\bibfnamefont {B.}~\bibnamefont
  {Shlaer}},\ }\href {\doibase 10.1103/PhysRevD.89.023512} {\bibfield
  {journal} {\bibinfo  {journal} {Phys. Rev. D}\ }\textbf {\bibinfo {volume}
  {89}},\ \bibinfo {pages} {023512} (\bibinfo {year} {2014})},\ \Eprint
  {http://arxiv.org/abs/1309.6637} {arXiv:1309.6637 [astro-ph.CO]} \BibitemShut
  {NoStop}%
\bibitem [{\citenamefont {Blanco-Pillado}\ \emph {et~al.}(2011)\citenamefont
  {Blanco-Pillado}, \citenamefont {Olum},\ and\ \citenamefont
  {Shlaer}}]{Blanco-Pillado:2011egf}%
  \BibitemOpen
  \bibfield  {author} {\bibinfo {author} {\bibfnamefont {J.~J.}\ \bibnamefont
  {Blanco-Pillado}}, \bibinfo {author} {\bibfnamefont {K.~D.}\ \bibnamefont
  {Olum}}, \ and\ \bibinfo {author} {\bibfnamefont {B.}~\bibnamefont
  {Shlaer}},\ }\href {\doibase 10.1103/PhysRevD.83.083514} {\bibfield
  {journal} {\bibinfo  {journal} {Phys. Rev. D}\ }\textbf {\bibinfo {volume}
  {83}},\ \bibinfo {pages} {083514} (\bibinfo {year} {2011})},\ \Eprint
  {http://arxiv.org/abs/1101.5173} {arXiv:1101.5173 [astro-ph.CO]} \BibitemShut
  {NoStop}%
\bibitem [{\citenamefont {Ringeval}\ \emph {et~al.}(2007)\citenamefont
  {Ringeval}, \citenamefont {Sakellariadou},\ and\ \citenamefont
  {Bouchet}}]{Ringeval:2005kr}%
  \BibitemOpen
  \bibfield  {author} {\bibinfo {author} {\bibfnamefont {C.}~\bibnamefont
  {Ringeval}}, \bibinfo {author} {\bibfnamefont {M.}~\bibnamefont
  {Sakellariadou}}, \ and\ \bibinfo {author} {\bibfnamefont {F.}~\bibnamefont
  {Bouchet}},\ }\href {\doibase 10.1088/1475-7516/2007/02/023} {\bibfield
  {journal} {\bibinfo  {journal} {JCAP}\ }\textbf {\bibinfo {volume} {02}},\
  \bibinfo {pages} {023} (\bibinfo {year} {2007})},\ \Eprint
  {http://arxiv.org/abs/astro-ph/0511646} {arXiv:astro-ph/0511646} \BibitemShut
  {NoStop}%
\bibitem [{\citenamefont {Lorenz}\ \emph {et~al.}(2010)\citenamefont {Lorenz},
  \citenamefont {Ringeval},\ and\ \citenamefont
  {Sakellariadou}}]{Lorenz:2010sm}%
  \BibitemOpen
  \bibfield  {author} {\bibinfo {author} {\bibfnamefont {L.}~\bibnamefont
  {Lorenz}}, \bibinfo {author} {\bibfnamefont {C.}~\bibnamefont {Ringeval}}, \
  and\ \bibinfo {author} {\bibfnamefont {M.}~\bibnamefont {Sakellariadou}},\
  }\href {\doibase 10.1088/1475-7516/2010/10/003} {\bibfield  {journal}
  {\bibinfo  {journal} {JCAP}\ }\textbf {\bibinfo {volume} {10}},\ \bibinfo
  {pages} {003} (\bibinfo {year} {2010})},\ \Eprint
  {http://arxiv.org/abs/1006.0931} {arXiv:1006.0931 [astro-ph.CO]} \BibitemShut
  {NoStop}%
\bibitem [{\citenamefont {Auclair}\ \emph {et~al.}(2020)\citenamefont {Auclair}
  \emph {et~al.}}]{Auclair:2019wcv}%
  \BibitemOpen
  \bibfield  {author} {\bibinfo {author} {\bibfnamefont {P.}~\bibnamefont
  {Auclair}} \emph {et~al.},\ }\href {\doibase 10.1088/1475-7516/2020/04/034}
  {\bibfield  {journal} {\bibinfo  {journal} {JCAP}\ }\textbf {\bibinfo
  {volume} {04}},\ \bibinfo {pages} {034} (\bibinfo {year} {2020})},\ \Eprint
  {http://arxiv.org/abs/1909.00819} {arXiv:1909.00819 [astro-ph.CO]}
  \BibitemShut {NoStop}%
\bibitem [{\citenamefont {Cui}\ \emph {et~al.}(2019)\citenamefont {Cui},
  \citenamefont {Lewicki}, \citenamefont {Morrissey},\ and\ \citenamefont
  {Wells}}]{Cui:2018rwi}%
  \BibitemOpen
  \bibfield  {author} {\bibinfo {author} {\bibfnamefont {Y.}~\bibnamefont
  {Cui}}, \bibinfo {author} {\bibfnamefont {M.}~\bibnamefont {Lewicki}},
  \bibinfo {author} {\bibfnamefont {D.~E.}\ \bibnamefont {Morrissey}}, \ and\
  \bibinfo {author} {\bibfnamefont {J.~D.}\ \bibnamefont {Wells}},\ }\href
  {\doibase 10.1007/JHEP01(2019)081} {\bibfield  {journal} {\bibinfo  {journal}
  {JHEP}\ }\textbf {\bibinfo {volume} {01}},\ \bibinfo {pages} {081} (\bibinfo
  {year} {2019})},\ \Eprint {http://arxiv.org/abs/1808.08968} {arXiv:1808.08968
  [hep-ph]} \BibitemShut {NoStop}%
\bibitem [{\citenamefont {Gouttenoire}\ \emph
  {et~al.}(2020{\natexlab{a}})\citenamefont {Gouttenoire}, \citenamefont
  {Servant},\ and\ \citenamefont {Simakachorn}}]{Gouttenoire:2019rtn}%
  \BibitemOpen
  \bibfield  {author} {\bibinfo {author} {\bibfnamefont {Y.}~\bibnamefont
  {Gouttenoire}}, \bibinfo {author} {\bibfnamefont {G.}~\bibnamefont
  {Servant}}, \ and\ \bibinfo {author} {\bibfnamefont {P.}~\bibnamefont
  {Simakachorn}},\ }\href {\doibase 10.1088/1475-7516/2020/07/016} {\bibfield
  {journal} {\bibinfo  {journal} {JCAP}\ }\textbf {\bibinfo {volume} {07}},\
  \bibinfo {pages} {016} (\bibinfo {year} {2020}{\natexlab{a}})},\ \Eprint
  {http://arxiv.org/abs/1912.03245} {arXiv:1912.03245 [hep-ph]} \BibitemShut
  {NoStop}%
\bibitem [{\citenamefont {Gouttenoire}\ \emph
  {et~al.}(2020{\natexlab{b}})\citenamefont {Gouttenoire}, \citenamefont
  {Servant},\ and\ \citenamefont {Simakachorn}}]{Gouttenoire:2019kij}%
  \BibitemOpen
  \bibfield  {author} {\bibinfo {author} {\bibfnamefont {Y.}~\bibnamefont
  {Gouttenoire}}, \bibinfo {author} {\bibfnamefont {G.}~\bibnamefont
  {Servant}}, \ and\ \bibinfo {author} {\bibfnamefont {P.}~\bibnamefont
  {Simakachorn}},\ }\href {\doibase 10.1088/1475-7516/2020/07/032} {\bibfield
  {journal} {\bibinfo  {journal} {JCAP}\ }\textbf {\bibinfo {volume} {07}},\
  \bibinfo {pages} {032} (\bibinfo {year} {2020}{\natexlab{b}})},\ \Eprint
  {http://arxiv.org/abs/1912.02569} {arXiv:1912.02569 [hep-ph]} \BibitemShut
  {NoStop}%
\bibitem [{\citenamefont {Olmez}\ \emph {et~al.}(2010)\citenamefont {Olmez},
  \citenamefont {Mandic},\ and\ \citenamefont {Siemens}}]{Olmez:2010bi}%
  \BibitemOpen
  \bibfield  {author} {\bibinfo {author} {\bibfnamefont {S.}~\bibnamefont
  {Olmez}}, \bibinfo {author} {\bibfnamefont {V.}~\bibnamefont {Mandic}}, \
  and\ \bibinfo {author} {\bibfnamefont {X.}~\bibnamefont {Siemens}},\ }\href
  {\doibase 10.1103/PhysRevD.81.104028} {\bibfield  {journal} {\bibinfo
  {journal} {Phys. Rev. D}\ }\textbf {\bibinfo {volume} {81}},\ \bibinfo
  {pages} {104028} (\bibinfo {year} {2010})},\ \Eprint
  {http://arxiv.org/abs/1004.0890} {arXiv:1004.0890 [astro-ph.CO]} \BibitemShut
  {NoStop}%
\bibitem [{\citenamefont {Gorghetto}\ \emph {et~al.}(2021)\citenamefont
  {Gorghetto}, \citenamefont {Hardy},\ and\ \citenamefont
  {Nicolaescu}}]{Gorghetto:2021fsn}%
  \BibitemOpen
  \bibfield  {author} {\bibinfo {author} {\bibfnamefont {M.}~\bibnamefont
  {Gorghetto}}, \bibinfo {author} {\bibfnamefont {E.}~\bibnamefont {Hardy}}, \
  and\ \bibinfo {author} {\bibfnamefont {H.}~\bibnamefont {Nicolaescu}},\
  }\href {\doibase 10.1088/1475-7516/2021/06/034} {\bibfield  {journal}
  {\bibinfo  {journal} {JCAP}\ }\textbf {\bibinfo {volume} {06}},\ \bibinfo
  {pages} {034} (\bibinfo {year} {2021})},\ \Eprint
  {http://arxiv.org/abs/2101.11007} {arXiv:2101.11007 [hep-ph]} \BibitemShut
  {NoStop}%
\bibitem [{\citenamefont {Jia}\ and\ \citenamefont {Bian}(2024)}]{Jia:2024ejr}%
  \BibitemOpen
  \bibfield  {author} {\bibinfo {author} {\bibfnamefont {Y.}~\bibnamefont
  {Jia}}\ and\ \bibinfo {author} {\bibfnamefont {L.}~\bibnamefont {Bian}},\
  }\href@noop {} {\  (\bibinfo {year} {2024})},\ \Eprint
  {http://arxiv.org/abs/2412.04218} {arXiv:2412.04218 [hep-ph]} \BibitemShut
  {NoStop}%
\bibitem [{\citenamefont {Buschmann}\ \emph {et~al.}(2022)\citenamefont
  {Buschmann}, \citenamefont {Foster}, \citenamefont {Hook}, \citenamefont
  {Peterson}, \citenamefont {Willcox}, \citenamefont {Zhang},\ and\
  \citenamefont {Safdi}}]{Buschmann:2021sdq}%
  \BibitemOpen
  \bibfield  {author} {\bibinfo {author} {\bibfnamefont {M.}~\bibnamefont
  {Buschmann}}, \bibinfo {author} {\bibfnamefont {J.~W.}\ \bibnamefont
  {Foster}}, \bibinfo {author} {\bibfnamefont {A.}~\bibnamefont {Hook}},
  \bibinfo {author} {\bibfnamefont {A.}~\bibnamefont {Peterson}}, \bibinfo
  {author} {\bibfnamefont {D.~E.}\ \bibnamefont {Willcox}}, \bibinfo {author}
  {\bibfnamefont {W.}~\bibnamefont {Zhang}}, \ and\ \bibinfo {author}
  {\bibfnamefont {B.~R.}\ \bibnamefont {Safdi}},\ }\href {\doibase
  10.1038/s41467-022-28669-y} {\bibfield  {journal} {\bibinfo  {journal}
  {Nature Commun.}\ }\textbf {\bibinfo {volume} {13}},\ \bibinfo {pages} {1049}
  (\bibinfo {year} {2022})},\ \Eprint {http://arxiv.org/abs/2108.05368}
  {arXiv:2108.05368 [hep-ph]} \BibitemShut {NoStop}%
\bibitem [{\citenamefont {Gorghetto}\ \emph {et~al.}(2018)\citenamefont
  {Gorghetto}, \citenamefont {Hardy},\ and\ \citenamefont
  {Villadoro}}]{Gorghetto:2018myk}%
  \BibitemOpen
  \bibfield  {author} {\bibinfo {author} {\bibfnamefont {M.}~\bibnamefont
  {Gorghetto}}, \bibinfo {author} {\bibfnamefont {E.}~\bibnamefont {Hardy}}, \
  and\ \bibinfo {author} {\bibfnamefont {G.}~\bibnamefont {Villadoro}},\ }\href
  {\doibase 10.1007/JHEP07(2018)151} {\bibfield  {journal} {\bibinfo  {journal}
  {JHEP}\ }\textbf {\bibinfo {volume} {07}},\ \bibinfo {pages} {151} (\bibinfo
  {year} {2018})},\ \Eprint {http://arxiv.org/abs/1806.04677} {arXiv:1806.04677
  [hep-ph]} \BibitemShut {NoStop}%
\bibitem [{\citenamefont {Kawasaki}\ \emph {et~al.}(2015)\citenamefont
  {Kawasaki}, \citenamefont {Saikawa},\ and\ \citenamefont
  {Sekiguchi}}]{Kawasaki:2014sqa}%
  \BibitemOpen
  \bibfield  {author} {\bibinfo {author} {\bibfnamefont {M.}~\bibnamefont
  {Kawasaki}}, \bibinfo {author} {\bibfnamefont {K.}~\bibnamefont {Saikawa}}, \
  and\ \bibinfo {author} {\bibfnamefont {T.}~\bibnamefont {Sekiguchi}},\ }\href
  {\doibase 10.1103/PhysRevD.91.065014} {\bibfield  {journal} {\bibinfo
  {journal} {Phys. Rev. D}\ }\textbf {\bibinfo {volume} {91}},\ \bibinfo
  {pages} {065014} (\bibinfo {year} {2015})},\ \Eprint
  {http://arxiv.org/abs/1412.0789} {arXiv:1412.0789 [hep-ph]} \BibitemShut
  {NoStop}%
\bibitem [{\citenamefont {Zhou}\ \emph {et~al.}(2023)\citenamefont {Zhou},
  \citenamefont {Cheng},\ and\ \citenamefont {Ren}}]{Zhou:2023rop}%
  \BibitemOpen
  \bibfield  {author} {\bibinfo {author} {\bibfnamefont {K.}~\bibnamefont
  {Zhou}}, \bibinfo {author} {\bibfnamefont {J.}~\bibnamefont {Cheng}}, \ and\
  \bibinfo {author} {\bibfnamefont {L.}~\bibnamefont {Ren}},\ }\href@noop {} {\
   (\bibinfo {year} {2023})},\ \Eprint {http://arxiv.org/abs/2306.14439}
  {arXiv:2306.14439 [gr-qc]} \BibitemShut {NoStop}%
\bibitem [{\citenamefont {Baker}\ \emph {et~al.}(2019)\citenamefont {Baker}
  \emph {et~al.}}]{Baker:2019nia}%
  \BibitemOpen
  \bibfield  {author} {\bibinfo {author} {\bibfnamefont {J.}~\bibnamefont
  {Baker}} \emph {et~al.},\ }\href@noop {} {\  (\bibinfo {year} {2019})},\
  \Eprint {http://arxiv.org/abs/1907.06482} {arXiv:1907.06482 [astro-ph.IM]}
  \BibitemShut {NoStop}%
\bibitem [{\citenamefont {Sesana}\ \emph {et~al.}(2021)\citenamefont {Sesana}
  \emph {et~al.}}]{Sesana:2019vho}%
  \BibitemOpen
  \bibfield  {author} {\bibinfo {author} {\bibfnamefont {A.}~\bibnamefont
  {Sesana}} \emph {et~al.},\ }\href {\doibase 10.1007/s10686-021-09709-9}
  {\bibfield  {journal} {\bibinfo  {journal} {Exper. Astron.}\ }\textbf
  {\bibinfo {volume} {51}},\ \bibinfo {pages} {1333} (\bibinfo {year}
  {2021})},\ \Eprint {http://arxiv.org/abs/1908.11391} {arXiv:1908.11391
  [astro-ph.IM]} \BibitemShut {NoStop}%
\bibitem [{\citenamefont {Seto}\ \emph {et~al.}(2001)\citenamefont {Seto},
  \citenamefont {Kawamura},\ and\ \citenamefont {Nakamura}}]{Seto:2001qf}%
  \BibitemOpen
  \bibfield  {author} {\bibinfo {author} {\bibfnamefont {N.}~\bibnamefont
  {Seto}}, \bibinfo {author} {\bibfnamefont {S.}~\bibnamefont {Kawamura}}, \
  and\ \bibinfo {author} {\bibfnamefont {T.}~\bibnamefont {Nakamura}},\ }\href
  {\doibase 10.1103/PhysRevLett.87.221103} {\bibfield  {journal} {\bibinfo
  {journal} {Phys. Rev. Lett.}\ }\textbf {\bibinfo {volume} {87}},\ \bibinfo
  {pages} {221103} (\bibinfo {year} {2001})},\ \Eprint
  {http://arxiv.org/abs/astro-ph/0108011} {arXiv:astro-ph/0108011} \BibitemShut
  {NoStop}%
\bibitem [{\citenamefont {Kawamura}\ \emph {et~al.}(2011)\citenamefont
  {Kawamura} \emph {et~al.}}]{Kawamura:2011zz}%
  \BibitemOpen
  \bibfield  {author} {\bibinfo {author} {\bibfnamefont {S.}~\bibnamefont
  {Kawamura}} \emph {et~al.},\ }\href {\doibase 10.1088/0264-9381/28/9/094011}
  {\bibfield  {journal} {\bibinfo  {journal} {Class. Quant. Grav.}\ }\textbf
  {\bibinfo {volume} {28}},\ \bibinfo {pages} {094011} (\bibinfo {year}
  {2011})}\BibitemShut {NoStop}%
\bibitem [{\citenamefont {Yagi}\ and\ \citenamefont
  {Seto}(2011)}]{Yagi:2011wg}%
  \BibitemOpen
  \bibfield  {author} {\bibinfo {author} {\bibfnamefont {K.}~\bibnamefont
  {Yagi}}\ and\ \bibinfo {author} {\bibfnamefont {N.}~\bibnamefont {Seto}},\
  }\href {\doibase 10.1103/PhysRevD.83.044011} {\bibfield  {journal} {\bibinfo
  {journal} {Phys. Rev. D}\ }\textbf {\bibinfo {volume} {83}},\ \bibinfo
  {pages} {044011} (\bibinfo {year} {2011})},\ \bibinfo {note} {[Erratum:
  Phys.Rev.D 95, 109901 (2017)]},\ \Eprint {http://arxiv.org/abs/1101.3940}
  {arXiv:1101.3940 [astro-ph.CO]} \BibitemShut {NoStop}%
\bibitem [{\citenamefont {Isoyama}\ \emph {et~al.}(2018)\citenamefont
  {Isoyama}, \citenamefont {Nakano},\ and\ \citenamefont
  {Nakamura}}]{Isoyama:2018rjb}%
  \BibitemOpen
  \bibfield  {author} {\bibinfo {author} {\bibfnamefont {S.}~\bibnamefont
  {Isoyama}}, \bibinfo {author} {\bibfnamefont {H.}~\bibnamefont {Nakano}}, \
  and\ \bibinfo {author} {\bibfnamefont {T.}~\bibnamefont {Nakamura}},\ }\href
  {\doibase 10.1093/ptep/pty078} {\bibfield  {journal} {\bibinfo  {journal}
  {PTEP}\ }\textbf {\bibinfo {volume} {2018}},\ \bibinfo {pages} {073E01}
  (\bibinfo {year} {2018})},\ \Eprint {http://arxiv.org/abs/1802.06977}
  {arXiv:1802.06977 [gr-qc]} \BibitemShut {NoStop}%
\bibitem [{\citenamefont {Crowder}\ and\ \citenamefont
  {Cornish}(2005)}]{Crowder:2005nr}%
  \BibitemOpen
  \bibfield  {author} {\bibinfo {author} {\bibfnamefont {J.}~\bibnamefont
  {Crowder}}\ and\ \bibinfo {author} {\bibfnamefont {N.~J.}\ \bibnamefont
  {Cornish}},\ }\href {\doibase 10.1103/PhysRevD.72.083005} {\bibfield
  {journal} {\bibinfo  {journal} {Phys. Rev. D}\ }\textbf {\bibinfo {volume}
  {72}},\ \bibinfo {pages} {083005} (\bibinfo {year} {2005})},\ \Eprint
  {http://arxiv.org/abs/gr-qc/0506015} {arXiv:gr-qc/0506015} \BibitemShut
  {NoStop}%
\bibitem [{\citenamefont {Corbin}\ and\ \citenamefont
  {Cornish}(2006)}]{Corbin:2005ny}%
  \BibitemOpen
  \bibfield  {author} {\bibinfo {author} {\bibfnamefont {V.}~\bibnamefont
  {Corbin}}\ and\ \bibinfo {author} {\bibfnamefont {N.~J.}\ \bibnamefont
  {Cornish}},\ }\href {\doibase 10.1088/0264-9381/23/7/014} {\bibfield
  {journal} {\bibinfo  {journal} {Class. Quant. Grav.}\ }\textbf {\bibinfo
  {volume} {23}},\ \bibinfo {pages} {2435} (\bibinfo {year} {2006})},\ \Eprint
  {http://arxiv.org/abs/gr-qc/0512039} {arXiv:gr-qc/0512039} \BibitemShut
  {NoStop}%
\bibitem [{\citenamefont {Harry}\ \emph {et~al.}(2006)\citenamefont {Harry},
  \citenamefont {Fritschel}, \citenamefont {Shaddock}, \citenamefont
  {Folkner},\ and\ \citenamefont {Phinney}}]{Harry:2006fi}%
  \BibitemOpen
  \bibfield  {author} {\bibinfo {author} {\bibfnamefont {G.~M.}\ \bibnamefont
  {Harry}}, \bibinfo {author} {\bibfnamefont {P.}~\bibnamefont {Fritschel}},
  \bibinfo {author} {\bibfnamefont {D.~A.}\ \bibnamefont {Shaddock}}, \bibinfo
  {author} {\bibfnamefont {W.}~\bibnamefont {Folkner}}, \ and\ \bibinfo
  {author} {\bibfnamefont {E.~S.}\ \bibnamefont {Phinney}},\ }\href {\doibase
  10.1088/0264-9381/23/15/008} {\bibfield  {journal} {\bibinfo  {journal}
  {Class. Quant. Grav.}\ }\textbf {\bibinfo {volume} {23}},\ \bibinfo {pages}
  {4887} (\bibinfo {year} {2006})},\ \bibinfo {note} {[Erratum:
  Class.Quant.Grav. 23, 7361 (2006)]}\BibitemShut {NoStop}%
\bibitem [{\citenamefont {Janssen}\ \emph {et~al.}(2015)\citenamefont {Janssen}
  \emph {et~al.}}]{Janssen:2014dka}%
  \BibitemOpen
  \bibfield  {author} {\bibinfo {author} {\bibfnamefont {G.}~\bibnamefont
  {Janssen}} \emph {et~al.},\ }\href {\doibase 10.22323/1.215.0037} {\bibfield
  {journal} {\bibinfo  {journal} {PoS}\ }\textbf {\bibinfo {volume}
  {AASKA14}},\ \bibinfo {pages} {037} (\bibinfo {year} {2015})},\ \Eprint
  {http://arxiv.org/abs/1501.00127} {arXiv:1501.00127 [astro-ph.IM]}
  \BibitemShut {NoStop}%
\bibitem [{\citenamefont {Hild}\ \emph {et~al.}(2011)\citenamefont {Hild} \emph
  {et~al.}}]{Hild:2010id}%
  \BibitemOpen
  \bibfield  {author} {\bibinfo {author} {\bibfnamefont {S.}~\bibnamefont
  {Hild}} \emph {et~al.},\ }\href {\doibase 10.1088/0264-9381/28/9/094013}
  {\bibfield  {journal} {\bibinfo  {journal} {Class. Quant. Grav.}\ }\textbf
  {\bibinfo {volume} {28}},\ \bibinfo {pages} {094013} (\bibinfo {year}
  {2011})},\ \Eprint {http://arxiv.org/abs/1012.0908} {arXiv:1012.0908 [gr-qc]}
  \BibitemShut {NoStop}%
\bibitem [{\citenamefont {Abbott}\ \emph
  {et~al.}(2017{\natexlab{d}})\citenamefont {Abbott} \emph
  {et~al.}}]{LIGOScientific:2016wof}%
  \BibitemOpen
  \bibfield  {author} {\bibinfo {author} {\bibfnamefont {B.~P.}\ \bibnamefont
  {Abbott}} \emph {et~al.} (\bibinfo {collaboration} {LIGO Scientific}),\
  }\href {\doibase 10.1088/1361-6382/aa51f4} {\bibfield  {journal} {\bibinfo
  {journal} {Class. Quant. Grav.}\ }\textbf {\bibinfo {volume} {34}},\ \bibinfo
  {pages} {044001} (\bibinfo {year} {2017}{\natexlab{d}})},\ \Eprint
  {http://arxiv.org/abs/1607.08697} {arXiv:1607.08697 [astro-ph.IM]}
  \BibitemShut {NoStop}%
\bibitem [{\citenamefont {Aasi}\ \emph {et~al.}(2015)\citenamefont {Aasi} \emph
  {et~al.}}]{LIGOScientific:2014qfs}%
  \BibitemOpen
  \bibfield  {author} {\bibinfo {author} {\bibfnamefont {J.}~\bibnamefont
  {Aasi}} \emph {et~al.} (\bibinfo {collaboration} {LIGO Scientific, VIRGO}),\
  }\href {\doibase 10.1088/0264-9381/32/11/115012} {\bibfield  {journal}
  {\bibinfo  {journal} {Class. Quant. Grav.}\ }\textbf {\bibinfo {volume}
  {32}},\ \bibinfo {pages} {115012} (\bibinfo {year} {2015})},\ \Eprint
  {http://arxiv.org/abs/1410.7764} {arXiv:1410.7764 [gr-qc]} \BibitemShut
  {NoStop}%
\bibitem [{\citenamefont {Abbott}\ \emph
  {et~al.}(2019{\natexlab{d}})\citenamefont {Abbott} \emph
  {et~al.}}]{LIGOScientific:2019vic}%
  \BibitemOpen
  \bibfield  {author} {\bibinfo {author} {\bibfnamefont {B.~P.}\ \bibnamefont
  {Abbott}} \emph {et~al.} (\bibinfo {collaboration} {LIGO Scientific,
  Virgo}),\ }\href {\doibase 10.1103/PhysRevD.100.061101} {\bibfield  {journal}
  {\bibinfo  {journal} {Phys. Rev. D}\ }\textbf {\bibinfo {volume} {100}},\
  \bibinfo {pages} {061101} (\bibinfo {year} {2019}{\natexlab{d}})},\ \Eprint
  {http://arxiv.org/abs/1903.02886} {arXiv:1903.02886 [gr-qc]} \BibitemShut
  {NoStop}%
\bibitem [{\citenamefont {Abbott}\ \emph
  {et~al.}(2021{\natexlab{e}})\citenamefont {Abbott} \emph
  {et~al.}}]{LIGOScientific:2021nrg}%
  \BibitemOpen
  \bibfield  {author} {\bibinfo {author} {\bibfnamefont {R.}~\bibnamefont
  {Abbott}} \emph {et~al.} (\bibinfo {collaboration} {LIGO Scientific, Virgo,
  KAGRA}),\ }\href {\doibase 10.1103/PhysRevLett.126.241102} {\bibfield
  {journal} {\bibinfo  {journal} {Phys. Rev. Lett.}\ }\textbf {\bibinfo
  {volume} {126}},\ \bibinfo {pages} {241102} (\bibinfo {year}
  {2021}{\natexlab{e}})},\ \Eprint {http://arxiv.org/abs/2101.12248}
  {arXiv:2101.12248 [gr-qc]} \BibitemShut {NoStop}%
\bibitem [{\citenamefont {Abbott}\ \emph
  {et~al.}(2018{\natexlab{a}})\citenamefont {Abbott} \emph
  {et~al.}}]{LIGOScientific:2017ikf}%
  \BibitemOpen
  \bibfield  {author} {\bibinfo {author} {\bibfnamefont {B.~P.}\ \bibnamefont
  {Abbott}} \emph {et~al.} (\bibinfo {collaboration} {LIGO Scientific,
  Virgo}),\ }\href {\doibase 10.1103/PhysRevD.97.102002} {\bibfield  {journal}
  {\bibinfo  {journal} {Phys. Rev. D}\ }\textbf {\bibinfo {volume} {97}},\
  \bibinfo {pages} {102002} (\bibinfo {year} {2018}{\natexlab{a}})},\ \Eprint
  {http://arxiv.org/abs/1712.01168} {arXiv:1712.01168 [gr-qc]} \BibitemShut
  {NoStop}%
\bibitem [{\citenamefont {Yonemaru}\ \emph {et~al.}(2021)\citenamefont
  {Yonemaru} \emph {et~al.}}]{Yonemaru:2020bmr}%
  \BibitemOpen
  \bibfield  {author} {\bibinfo {author} {\bibfnamefont {N.}~\bibnamefont
  {Yonemaru}} \emph {et~al.},\ }\href {\doibase 10.1093/mnras/staa3721}
  {\bibfield  {journal} {\bibinfo  {journal} {Mon. Not. Roy. Astron. Soc.}\
  }\textbf {\bibinfo {volume} {501}},\ \bibinfo {pages} {701} (\bibinfo {year}
  {2021})},\ \Eprint {http://arxiv.org/abs/2011.13490} {arXiv:2011.13490
  [gr-qc]} \BibitemShut {NoStop}%
\bibitem [{\citenamefont {Sanidas}\ \emph {et~al.}(2012)\citenamefont
  {Sanidas}, \citenamefont {Battye},\ and\ \citenamefont
  {Stappers}}]{Sanidas:2012ee}%
  \BibitemOpen
  \bibfield  {author} {\bibinfo {author} {\bibfnamefont {S.~A.}\ \bibnamefont
  {Sanidas}}, \bibinfo {author} {\bibfnamefont {R.~A.}\ \bibnamefont {Battye}},
  \ and\ \bibinfo {author} {\bibfnamefont {B.~W.}\ \bibnamefont {Stappers}},\
  }\href {\doibase 10.1103/PhysRevD.85.122003} {\bibfield  {journal} {\bibinfo
  {journal} {Phys. Rev. D}\ }\textbf {\bibinfo {volume} {85}},\ \bibinfo
  {pages} {122003} (\bibinfo {year} {2012})},\ \Eprint
  {http://arxiv.org/abs/1201.2419} {arXiv:1201.2419 [astro-ph.CO]} \BibitemShut
  {NoStop}%
\bibitem [{\citenamefont {Bian}\ \emph
  {et~al.}(2022{\natexlab{b}})\citenamefont {Bian}, \citenamefont {Shu},
  \citenamefont {Wang}, \citenamefont {Yuan},\ and\ \citenamefont
  {Zong}}]{Bian:2022tju}%
  \BibitemOpen
  \bibfield  {author} {\bibinfo {author} {\bibfnamefont {L.}~\bibnamefont
  {Bian}}, \bibinfo {author} {\bibfnamefont {J.}~\bibnamefont {Shu}}, \bibinfo
  {author} {\bibfnamefont {B.}~\bibnamefont {Wang}}, \bibinfo {author}
  {\bibfnamefont {Q.}~\bibnamefont {Yuan}}, \ and\ \bibinfo {author}
  {\bibfnamefont {J.}~\bibnamefont {Zong}},\ }\href {\doibase
  10.1103/PhysRevD.106.L101301} {\bibfield  {journal} {\bibinfo  {journal}
  {Phys. Rev. D}\ }\textbf {\bibinfo {volume} {106}},\ \bibinfo {pages}
  {L101301} (\bibinfo {year} {2022}{\natexlab{b}})},\ \Eprint
  {http://arxiv.org/abs/2205.07293} {arXiv:2205.07293 [hep-ph]} \BibitemShut
  {NoStop}%
\bibitem [{\citenamefont {Cutler}\ and\ \citenamefont
  {Flanagan}(1994)}]{Cutler:1994ys}%
  \BibitemOpen
  \bibfield  {author} {\bibinfo {author} {\bibfnamefont {C.}~\bibnamefont
  {Cutler}}\ and\ \bibinfo {author} {\bibfnamefont {E.~E.}\ \bibnamefont
  {Flanagan}},\ }\href {\doibase 10.1103/PhysRevD.49.2658} {\bibfield
  {journal} {\bibinfo  {journal} {Phys. Rev. D}\ }\textbf {\bibinfo {volume}
  {49}},\ \bibinfo {pages} {2658} (\bibinfo {year} {1994})},\ \Eprint
  {http://arxiv.org/abs/gr-qc/9402014} {arXiv:gr-qc/9402014} \BibitemShut
  {NoStop}%
\bibitem [{\citenamefont {Markovic}(1993)}]{Markovic:1993cr}%
  \BibitemOpen
  \bibfield  {author} {\bibinfo {author} {\bibfnamefont {D.}~\bibnamefont
  {Markovic}},\ }\href {\doibase 10.1103/PhysRevD.48.4738} {\bibfield
  {journal} {\bibinfo  {journal} {Phys. Rev. D}\ }\textbf {\bibinfo {volume}
  {48}},\ \bibinfo {pages} {4738} (\bibinfo {year} {1993})}\BibitemShut
  {NoStop}%
\bibitem [{\citenamefont {Abbott}\ \emph
  {et~al.}(2017{\natexlab{e}})\citenamefont {Abbott} \emph
  {et~al.}}]{LIGOScientific:2017ync}%
  \BibitemOpen
  \bibfield  {author} {\bibinfo {author} {\bibfnamefont {B.~P.}\ \bibnamefont
  {Abbott}} \emph {et~al.} (\bibinfo {collaboration} {LIGO Scientific, Virgo,
  Fermi GBM, INTEGRAL, IceCube, AstroSat Cadmium Zinc Telluride Imager Team,
  IPN, Insight-Hxmt, ANTARES, Swift, AGILE Team, 1M2H Team, Dark Energy Camera
  GW-EM, DES, DLT40, GRAWITA, Fermi-LAT, ATCA, ASKAP, Las Cumbres Observatory
  Group, OzGrav, DWF (Deeper Wider Faster Program), AST3, CAASTRO, VINROUGE,
  MASTER, J-GEM, GROWTH, JAGWAR, CaltechNRAO, TTU-NRAO, NuSTAR, Pan-STARRS,
  MAXI Team, TZAC Consortium, KU, Nordic Optical Telescope, ePESSTO, GROND,
  Texas Tech University, SALT Group, TOROS, BOOTES, MWA, CALET, IKI-GW
  Follow-up, H.E.S.S., LOFAR, LWA, HAWC, Pierre Auger, ALMA, Euro VLBI Team, Pi
  of Sky, Chandra Team at McGill University, DFN, ATLAS Telescopes, High Time
  Resolution Universe Survey, RIMAS, RATIR, SKA South Africa/MeerKAT}),\ }\href
  {\doibase 10.3847/2041-8213/aa91c9} {\bibfield  {journal} {\bibinfo
  {journal} {Astrophys. J. Lett.}\ }\textbf {\bibinfo {volume} {848}},\
  \bibinfo {pages} {L12} (\bibinfo {year} {2017}{\natexlab{e}})},\ \Eprint
  {http://arxiv.org/abs/1710.05833} {arXiv:1710.05833 [astro-ph.HE]}
  \BibitemShut {NoStop}%
\bibitem [{\citenamefont {Margutti}\ and\ \citenamefont
  {Chornock}(2021)}]{Margutti:2020xbo}%
  \BibitemOpen
  \bibfield  {author} {\bibinfo {author} {\bibfnamefont {R.}~\bibnamefont
  {Margutti}}\ and\ \bibinfo {author} {\bibfnamefont {R.}~\bibnamefont
  {Chornock}},\ }\href {\doibase 10.1146/annurev-astro-112420-030742}
  {\bibfield  {journal} {\bibinfo  {journal} {Ann. Rev. Astron. Astrophys.}\
  }\textbf {\bibinfo {volume} {59}},\ \bibinfo {pages} {155} (\bibinfo {year}
  {2021})},\ \Eprint {http://arxiv.org/abs/2012.04810} {arXiv:2012.04810
  [astro-ph.HE]} \BibitemShut {NoStop}%
\bibitem [{\citenamefont {Soares-Santos}\ \emph {et~al.}(2019)\citenamefont
  {Soares-Santos} \emph {et~al.}}]{DES:2019ccw}%
  \BibitemOpen
  \bibfield  {author} {\bibinfo {author} {\bibfnamefont {M.}~\bibnamefont
  {Soares-Santos}} \emph {et~al.} (\bibinfo {collaboration} {DES, LIGO
  Scientific, Virgo}),\ }\href {\doibase 10.3847/2041-8213/ab14f1} {\bibfield
  {journal} {\bibinfo  {journal} {Astrophys. J. Lett.}\ }\textbf {\bibinfo
  {volume} {876}},\ \bibinfo {pages} {L7} (\bibinfo {year} {2019})},\ \Eprint
  {http://arxiv.org/abs/1901.01540} {arXiv:1901.01540 [astro-ph.CO]}
  \BibitemShut {NoStop}%
\bibitem [{\citenamefont {Fishbach}\ \emph {et~al.}(2019)\citenamefont
  {Fishbach} \emph {et~al.}}]{LIGOScientific:2018gmd}%
  \BibitemOpen
  \bibfield  {author} {\bibinfo {author} {\bibfnamefont {M.}~\bibnamefont
  {Fishbach}} \emph {et~al.} (\bibinfo {collaboration} {LIGO Scientific,
  Virgo}),\ }\href {\doibase 10.3847/2041-8213/aaf96e} {\bibfield  {journal}
  {\bibinfo  {journal} {Astrophys. J. Lett.}\ }\textbf {\bibinfo {volume}
  {871}},\ \bibinfo {pages} {L13} (\bibinfo {year} {2019})},\ \Eprint
  {http://arxiv.org/abs/1807.05667} {arXiv:1807.05667 [astro-ph.CO]}
  \BibitemShut {NoStop}%
\bibitem [{\citenamefont {Abbott}\ \emph
  {et~al.}(2021{\natexlab{f}})\citenamefont {Abbott} \emph
  {et~al.}}]{LIGOScientific:2019zcs}%
  \BibitemOpen
  \bibfield  {author} {\bibinfo {author} {\bibfnamefont {B.~P.}\ \bibnamefont
  {Abbott}} \emph {et~al.} (\bibinfo {collaboration} {LIGO Scientific, Virgo,
  VIRGO}),\ }\href {\doibase 10.3847/1538-4357/abdcb7} {\bibfield  {journal}
  {\bibinfo  {journal} {Astrophys. J.}\ }\textbf {\bibinfo {volume} {909}},\
  \bibinfo {pages} {218} (\bibinfo {year} {2021}{\natexlab{f}})},\ \Eprint
  {http://arxiv.org/abs/1908.06060} {arXiv:1908.06060 [astro-ph.CO]}
  \BibitemShut {NoStop}%
\bibitem [{\citenamefont {Palmese}\ \emph {et~al.}(2020)\citenamefont {Palmese}
  \emph {et~al.}}]{DES:2020nay}%
  \BibitemOpen
  \bibfield  {author} {\bibinfo {author} {\bibfnamefont {A.}~\bibnamefont
  {Palmese}} \emph {et~al.} (\bibinfo {collaboration} {DES}),\ }\href {\doibase
  10.3847/2041-8213/abaeff} {\bibfield  {journal} {\bibinfo  {journal}
  {Astrophys. J. Lett.}\ }\textbf {\bibinfo {volume} {900}},\ \bibinfo {pages}
  {L33} (\bibinfo {year} {2020})},\ \Eprint {http://arxiv.org/abs/2006.14961}
  {arXiv:2006.14961 [astro-ph.CO]} \BibitemShut {NoStop}%
\bibitem [{\citenamefont {Vasylyev}\ and\ \citenamefont
  {Filippenko}(2020)}]{Vasylyev:2020hgb}%
  \BibitemOpen
  \bibfield  {author} {\bibinfo {author} {\bibfnamefont {S.}~\bibnamefont
  {Vasylyev}}\ and\ \bibinfo {author} {\bibfnamefont {A.}~\bibnamefont
  {Filippenko}},\ }\href {\doibase 10.3847/1538-4357/abb5f9} {\bibfield
  {journal} {\bibinfo  {journal} {Astrophys. J.}\ }\textbf {\bibinfo {volume}
  {902}},\ \bibinfo {pages} {149} (\bibinfo {year} {2020})},\ \Eprint
  {http://arxiv.org/abs/2007.11148} {arXiv:2007.11148 [astro-ph.CO]}
  \BibitemShut {NoStop}%
\bibitem [{\citenamefont {Ballard}\ \emph {et~al.}(2023)\citenamefont {Ballard}
  \emph {et~al.}}]{DESI:2023fij}%
  \BibitemOpen
  \bibfield  {author} {\bibinfo {author} {\bibfnamefont {W.}~\bibnamefont
  {Ballard}} \emph {et~al.} (\bibinfo {collaboration} {DESI}),\ }\href
  {\doibase 10.3847/2515-5172/ad0eda} {\bibfield  {journal} {\bibinfo
  {journal} {Res. Notes AAS}\ }\textbf {\bibinfo {volume} {7}},\ \bibinfo
  {pages} {250} (\bibinfo {year} {2023})},\ \Eprint
  {http://arxiv.org/abs/2311.13062} {arXiv:2311.13062 [astro-ph.CO]}
  \BibitemShut {NoStop}%
\bibitem [{\citenamefont {Abbott}\ \emph
  {et~al.}(2016{\natexlab{c}})\citenamefont {Abbott} \emph
  {et~al.}}]{KAGRA:2013rdx}%
  \BibitemOpen
  \bibfield  {author} {\bibinfo {author} {\bibfnamefont {B.~P.}\ \bibnamefont
  {Abbott}} \emph {et~al.} (\bibinfo {collaboration} {KAGRA, LIGO Scientific,
  Virgo}),\ }\href {\doibase 10.1007/s41114-020-00026-9} {\bibfield  {journal}
  {\bibinfo  {journal} {Living Rev. Rel.}\ }\textbf {\bibinfo {volume} {19}},\
  \bibinfo {pages} {1} (\bibinfo {year} {2016}{\natexlab{c}})},\ \Eprint
  {http://arxiv.org/abs/1304.0670} {arXiv:1304.0670 [gr-qc]} \BibitemShut
  {NoStop}%
\bibitem [{\citenamefont {Chen}\ \emph
  {et~al.}(2021{\natexlab{b}})\citenamefont {Chen}, \citenamefont
  {Cowperthwaite}, \citenamefont {Metzger},\ and\ \citenamefont
  {Berger}}]{Chen:2020zoq}%
  \BibitemOpen
  \bibfield  {author} {\bibinfo {author} {\bibfnamefont {H.-Y.}\ \bibnamefont
  {Chen}}, \bibinfo {author} {\bibfnamefont {P.~S.}\ \bibnamefont
  {Cowperthwaite}}, \bibinfo {author} {\bibfnamefont {B.~D.}\ \bibnamefont
  {Metzger}}, \ and\ \bibinfo {author} {\bibfnamefont {E.}~\bibnamefont
  {Berger}},\ }\href {\doibase 10.3847/2041-8213/abdab0} {\bibfield  {journal}
  {\bibinfo  {journal} {Astrophys. J. Lett.}\ }\textbf {\bibinfo {volume}
  {908}},\ \bibinfo {pages} {L4} (\bibinfo {year} {2021}{\natexlab{b}})},\
  \Eprint {http://arxiv.org/abs/2011.01211} {arXiv:2011.01211 [astro-ph.CO]}
  \BibitemShut {NoStop}%
\bibitem [{\citenamefont {Zhu}\ and\ \citenamefont {Chen}(2023)}]{Zhu:2023jti}%
  \BibitemOpen
  \bibfield  {author} {\bibinfo {author} {\bibfnamefont {L.-G.}\ \bibnamefont
  {Zhu}}\ and\ \bibinfo {author} {\bibfnamefont {X.}~\bibnamefont {Chen}},\
  }\href {\doibase 10.3847/1538-4357/acc24b} {\bibfield  {journal} {\bibinfo
  {journal} {Astrophys. J.}\ }\textbf {\bibinfo {volume} {948}},\ \bibinfo
  {pages} {26} (\bibinfo {year} {2023})},\ \Eprint
  {http://arxiv.org/abs/2302.10621} {arXiv:2302.10621 [astro-ph.CO]}
  \BibitemShut {NoStop}%
\bibitem [{\citenamefont {Muttoni}\ \emph {et~al.}(2023)\citenamefont
  {Muttoni}, \citenamefont {Laghi}, \citenamefont {Tamanini}, \citenamefont
  {Marsat},\ and\ \citenamefont {Izquierdo-Villalba}}]{Muttoni:2023prw}%
  \BibitemOpen
  \bibfield  {author} {\bibinfo {author} {\bibfnamefont {N.}~\bibnamefont
  {Muttoni}}, \bibinfo {author} {\bibfnamefont {D.}~\bibnamefont {Laghi}},
  \bibinfo {author} {\bibfnamefont {N.}~\bibnamefont {Tamanini}}, \bibinfo
  {author} {\bibfnamefont {S.}~\bibnamefont {Marsat}}, \ and\ \bibinfo {author}
  {\bibfnamefont {D.}~\bibnamefont {Izquierdo-Villalba}},\ }\href {\doibase
  10.1103/PhysRevD.108.043543} {\bibfield  {journal} {\bibinfo  {journal}
  {Phys. Rev. D}\ }\textbf {\bibinfo {volume} {108}},\ \bibinfo {pages}
  {043543} (\bibinfo {year} {2023})},\ \Eprint
  {http://arxiv.org/abs/2303.10693} {arXiv:2303.10693 [astro-ph.CO]}
  \BibitemShut {NoStop}%
\bibitem [{\citenamefont {Song}\ \emph {et~al.}(2024)\citenamefont {Song},
  \citenamefont {Wang}, \citenamefont {Li}, \citenamefont {Zhao}, \citenamefont
  {Zhang}, \citenamefont {Zhao},\ and\ \citenamefont {Zhang}}]{Song:2022siz}%
  \BibitemOpen
  \bibfield  {author} {\bibinfo {author} {\bibfnamefont {J.-Y.}\ \bibnamefont
  {Song}}, \bibinfo {author} {\bibfnamefont {L.-F.}\ \bibnamefont {Wang}},
  \bibinfo {author} {\bibfnamefont {Y.}~\bibnamefont {Li}}, \bibinfo {author}
  {\bibfnamefont {Z.-W.}\ \bibnamefont {Zhao}}, \bibinfo {author}
  {\bibfnamefont {J.-F.}\ \bibnamefont {Zhang}}, \bibinfo {author}
  {\bibfnamefont {W.}~\bibnamefont {Zhao}}, \ and\ \bibinfo {author}
  {\bibfnamefont {X.}~\bibnamefont {Zhang}},\ }\href {\doibase
  10.1007/s11433-023-2260-2} {\bibfield  {journal} {\bibinfo  {journal} {Sci.
  China Phys. Mech. Astron.}\ }\textbf {\bibinfo {volume} {67}},\ \bibinfo
  {pages} {230411} (\bibinfo {year} {2024})},\ \Eprint
  {http://arxiv.org/abs/2212.00531} {arXiv:2212.00531 [astro-ph.CO]}
  \BibitemShut {NoStop}%
\bibitem [{\citenamefont {Cai}\ and\ \citenamefont {Yang}(2017)}]{Cai:2016sby}%
  \BibitemOpen
  \bibfield  {author} {\bibinfo {author} {\bibfnamefont {R.-G.}\ \bibnamefont
  {Cai}}\ and\ \bibinfo {author} {\bibfnamefont {T.}~\bibnamefont {Yang}},\
  }\href {\doibase 10.1103/PhysRevD.95.044024} {\bibfield  {journal} {\bibinfo
  {journal} {Phys. Rev. D}\ }\textbf {\bibinfo {volume} {95}},\ \bibinfo
  {pages} {044024} (\bibinfo {year} {2017})},\ \Eprint
  {http://arxiv.org/abs/1608.08008} {arXiv:1608.08008 [astro-ph.CO]}
  \BibitemShut {NoStop}%
\bibitem [{\citenamefont {Babak}\ \emph {et~al.}(2011)\citenamefont {Babak},
  \citenamefont {Gair}, \citenamefont {Petiteau},\ and\ \citenamefont
  {Sesana}}]{Babak:2010ej}%
  \BibitemOpen
  \bibfield  {author} {\bibinfo {author} {\bibfnamefont {S.}~\bibnamefont
  {Babak}}, \bibinfo {author} {\bibfnamefont {J.~R.}\ \bibnamefont {Gair}},
  \bibinfo {author} {\bibfnamefont {A.}~\bibnamefont {Petiteau}}, \ and\
  \bibinfo {author} {\bibfnamefont {A.}~\bibnamefont {Sesana}},\ }\href
  {\doibase 10.1088/0264-9381/28/11/114001} {\bibfield  {journal} {\bibinfo
  {journal} {Class. Quant. Grav.}\ }\textbf {\bibinfo {volume} {28}},\ \bibinfo
  {pages} {114001} (\bibinfo {year} {2011})},\ \Eprint
  {http://arxiv.org/abs/1011.2062} {arXiv:1011.2062 [gr-qc]} \BibitemShut
  {NoStop}%
\bibitem [{\citenamefont {Petiteau}\ \emph {et~al.}(2011)\citenamefont
  {Petiteau}, \citenamefont {Babak},\ and\ \citenamefont
  {Sesana}}]{Petiteau:2011we}%
  \BibitemOpen
  \bibfield  {author} {\bibinfo {author} {\bibfnamefont {A.}~\bibnamefont
  {Petiteau}}, \bibinfo {author} {\bibfnamefont {S.}~\bibnamefont {Babak}}, \
  and\ \bibinfo {author} {\bibfnamefont {A.}~\bibnamefont {Sesana}},\ }\href
  {\doibase 10.1088/0004-637X/732/2/82} {\bibfield  {journal} {\bibinfo
  {journal} {Astrophys. J.}\ }\textbf {\bibinfo {volume} {732}},\ \bibinfo
  {pages} {82} (\bibinfo {year} {2011})},\ \Eprint
  {http://arxiv.org/abs/1102.0769} {arXiv:1102.0769 [astro-ph.CO]} \BibitemShut
  {NoStop}%
\bibitem [{\citenamefont {Tamanini}\ \emph {et~al.}(2016)\citenamefont
  {Tamanini}, \citenamefont {Caprini}, \citenamefont {Barausse}, \citenamefont
  {Sesana}, \citenamefont {Klein},\ and\ \citenamefont
  {Petiteau}}]{Tamanini:2016zlh}%
  \BibitemOpen
  \bibfield  {author} {\bibinfo {author} {\bibfnamefont {N.}~\bibnamefont
  {Tamanini}}, \bibinfo {author} {\bibfnamefont {C.}~\bibnamefont {Caprini}},
  \bibinfo {author} {\bibfnamefont {E.}~\bibnamefont {Barausse}}, \bibinfo
  {author} {\bibfnamefont {A.}~\bibnamefont {Sesana}}, \bibinfo {author}
  {\bibfnamefont {A.}~\bibnamefont {Klein}}, \ and\ \bibinfo {author}
  {\bibfnamefont {A.}~\bibnamefont {Petiteau}},\ }\href {\doibase
  10.1088/1475-7516/2016/04/002} {\bibfield  {journal} {\bibinfo  {journal}
  {JCAP}\ }\textbf {\bibinfo {volume} {04}},\ \bibinfo {pages} {002} (\bibinfo
  {year} {2016})},\ \Eprint {http://arxiv.org/abs/1601.07112} {arXiv:1601.07112
  [astro-ph.CO]} \BibitemShut {NoStop}%
\bibitem [{\citenamefont {Caprini}\ and\ \citenamefont
  {Tamanini}(2016)}]{Caprini:2016qxs}%
  \BibitemOpen
  \bibfield  {author} {\bibinfo {author} {\bibfnamefont {C.}~\bibnamefont
  {Caprini}}\ and\ \bibinfo {author} {\bibfnamefont {N.}~\bibnamefont
  {Tamanini}},\ }\href {\doibase 10.1088/1475-7516/2016/10/006} {\bibfield
  {journal} {\bibinfo  {journal} {JCAP}\ }\textbf {\bibinfo {volume} {10}},\
  \bibinfo {pages} {006} (\bibinfo {year} {2016})},\ \Eprint
  {http://arxiv.org/abs/1607.08755} {arXiv:1607.08755 [astro-ph.CO]}
  \BibitemShut {NoStop}%
\bibitem [{\citenamefont {Cai}\ \emph {et~al.}(2017{\natexlab{b}})\citenamefont
  {Cai}, \citenamefont {Tamanini},\ and\ \citenamefont {Yang}}]{Cai:2017yww}%
  \BibitemOpen
  \bibfield  {author} {\bibinfo {author} {\bibfnamefont {R.-G.}\ \bibnamefont
  {Cai}}, \bibinfo {author} {\bibfnamefont {N.}~\bibnamefont {Tamanini}}, \
  and\ \bibinfo {author} {\bibfnamefont {T.}~\bibnamefont {Yang}},\ }\href
  {\doibase 10.1088/1475-7516/2017/05/031} {\bibfield  {journal} {\bibinfo
  {journal} {JCAP}\ }\textbf {\bibinfo {volume} {05}},\ \bibinfo {pages} {031}
  (\bibinfo {year} {2017}{\natexlab{b}})},\ \Eprint
  {http://arxiv.org/abs/1703.07323} {arXiv:1703.07323 [astro-ph.CO]}
  \BibitemShut {NoStop}%
\bibitem [{\citenamefont {Amaro-Seoane}\ \emph {et~al.}(2007)\citenamefont
  {Amaro-Seoane}, \citenamefont {Gair}, \citenamefont {Freitag}, \citenamefont
  {Coleman~Miller}, \citenamefont {Mandel}, \citenamefont {Cutler},\ and\
  \citenamefont {Babak}}]{Amaro-Seoane:2007osp}%
  \BibitemOpen
  \bibfield  {author} {\bibinfo {author} {\bibfnamefont {P.}~\bibnamefont
  {Amaro-Seoane}}, \bibinfo {author} {\bibfnamefont {J.~R.}\ \bibnamefont
  {Gair}}, \bibinfo {author} {\bibfnamefont {M.}~\bibnamefont {Freitag}},
  \bibinfo {author} {\bibfnamefont {M.}~\bibnamefont {Coleman~Miller}},
  \bibinfo {author} {\bibfnamefont {I.}~\bibnamefont {Mandel}}, \bibinfo
  {author} {\bibfnamefont {C.~J.}\ \bibnamefont {Cutler}}, \ and\ \bibinfo
  {author} {\bibfnamefont {S.}~\bibnamefont {Babak}},\ }\href {\doibase
  10.1088/0264-9381/24/17/R01} {\bibfield  {journal} {\bibinfo  {journal}
  {Class. Quant. Grav.}\ }\textbf {\bibinfo {volume} {24}},\ \bibinfo {pages}
  {R113} (\bibinfo {year} {2007})},\ \Eprint
  {http://arxiv.org/abs/astro-ph/0703495} {arXiv:astro-ph/0703495} \BibitemShut
  {NoStop}%
\bibitem [{\citenamefont {Kyutoku}\ and\ \citenamefont
  {Seto}(2016)}]{Kyutoku:2016ppx}%
  \BibitemOpen
  \bibfield  {author} {\bibinfo {author} {\bibfnamefont {K.}~\bibnamefont
  {Kyutoku}}\ and\ \bibinfo {author} {\bibfnamefont {N.}~\bibnamefont {Seto}},\
  }\href {\doibase 10.1093/mnras/stw1767} {\bibfield  {journal} {\bibinfo
  {journal} {Mon. Not. Roy. Astron. Soc.}\ }\textbf {\bibinfo {volume} {462}},\
  \bibinfo {pages} {2177} (\bibinfo {year} {2016})},\ \Eprint
  {http://arxiv.org/abs/1606.02298} {arXiv:1606.02298 [astro-ph.HE]}
  \BibitemShut {NoStop}%
\bibitem [{\citenamefont {MacLeod}\ and\ \citenamefont
  {Hogan}(2008)}]{MacLeod:2007jd}%
  \BibitemOpen
  \bibfield  {author} {\bibinfo {author} {\bibfnamefont {C.~L.}\ \bibnamefont
  {MacLeod}}\ and\ \bibinfo {author} {\bibfnamefont {C.~J.}\ \bibnamefont
  {Hogan}},\ }\href {\doibase 10.1103/PhysRevD.77.043512} {\bibfield  {journal}
  {\bibinfo  {journal} {Phys. Rev. D}\ }\textbf {\bibinfo {volume} {77}},\
  \bibinfo {pages} {043512} (\bibinfo {year} {2008})},\ \Eprint
  {http://arxiv.org/abs/0712.0618} {arXiv:0712.0618 [astro-ph]} \BibitemShut
  {NoStop}%
\bibitem [{\citenamefont {Laghi}\ \emph {et~al.}(2021)\citenamefont {Laghi},
  \citenamefont {Tamanini}, \citenamefont {Del~Pozzo}, \citenamefont {Sesana},
  \citenamefont {Gair}, \citenamefont {Babak},\ and\ \citenamefont
  {Izquierdo-Villalba}}]{Laghi:2021pqk}%
  \BibitemOpen
  \bibfield  {author} {\bibinfo {author} {\bibfnamefont {D.}~\bibnamefont
  {Laghi}}, \bibinfo {author} {\bibfnamefont {N.}~\bibnamefont {Tamanini}},
  \bibinfo {author} {\bibfnamefont {W.}~\bibnamefont {Del~Pozzo}}, \bibinfo
  {author} {\bibfnamefont {A.}~\bibnamefont {Sesana}}, \bibinfo {author}
  {\bibfnamefont {J.}~\bibnamefont {Gair}}, \bibinfo {author} {\bibfnamefont
  {S.}~\bibnamefont {Babak}}, \ and\ \bibinfo {author} {\bibfnamefont
  {D.}~\bibnamefont {Izquierdo-Villalba}},\ }\href {\doibase
  10.1093/mnras/stab2741} {\bibfield  {journal} {\bibinfo  {journal} {Mon. Not.
  Roy. Astron. Soc.}\ }\textbf {\bibinfo {volume} {508}},\ \bibinfo {pages}
  {4512} (\bibinfo {year} {2021})},\ \Eprint {http://arxiv.org/abs/2102.01708}
  {arXiv:2102.01708 [astro-ph.CO]} \BibitemShut {NoStop}%
\bibitem [{\citenamefont {Del~Pozzo}\ \emph {et~al.}(2018)\citenamefont
  {Del~Pozzo}, \citenamefont {Sesana},\ and\ \citenamefont
  {Klein}}]{DelPozzo:2017kme}%
  \BibitemOpen
  \bibfield  {author} {\bibinfo {author} {\bibfnamefont {W.}~\bibnamefont
  {Del~Pozzo}}, \bibinfo {author} {\bibfnamefont {A.}~\bibnamefont {Sesana}}, \
  and\ \bibinfo {author} {\bibfnamefont {A.}~\bibnamefont {Klein}},\ }\href
  {\doibase 10.1093/mnras/sty057} {\bibfield  {journal} {\bibinfo  {journal}
  {Mon. Not. Roy. Astron. Soc.}\ }\textbf {\bibinfo {volume} {475}},\ \bibinfo
  {pages} {3485} (\bibinfo {year} {2018})},\ \Eprint
  {http://arxiv.org/abs/1703.01300} {arXiv:1703.01300 [astro-ph.CO]}
  \BibitemShut {NoStop}%
\bibitem [{\citenamefont {Muttoni}\ \emph {et~al.}(2022)\citenamefont
  {Muttoni}, \citenamefont {Mangiagli}, \citenamefont {Sesana}, \citenamefont
  {Laghi}, \citenamefont {Del~Pozzo}, \citenamefont {Izquierdo-Villalba},\ and\
  \citenamefont {Rosati}}]{Muttoni:2021veo}%
  \BibitemOpen
  \bibfield  {author} {\bibinfo {author} {\bibfnamefont {N.}~\bibnamefont
  {Muttoni}}, \bibinfo {author} {\bibfnamefont {A.}~\bibnamefont {Mangiagli}},
  \bibinfo {author} {\bibfnamefont {A.}~\bibnamefont {Sesana}}, \bibinfo
  {author} {\bibfnamefont {D.}~\bibnamefont {Laghi}}, \bibinfo {author}
  {\bibfnamefont {W.}~\bibnamefont {Del~Pozzo}}, \bibinfo {author}
  {\bibfnamefont {D.}~\bibnamefont {Izquierdo-Villalba}}, \ and\ \bibinfo
  {author} {\bibfnamefont {M.}~\bibnamefont {Rosati}},\ }\href {\doibase
  10.1103/PhysRevD.105.043509} {\bibfield  {journal} {\bibinfo  {journal}
  {Phys. Rev. D}\ }\textbf {\bibinfo {volume} {105}},\ \bibinfo {pages}
  {043509} (\bibinfo {year} {2022})},\ \Eprint
  {http://arxiv.org/abs/2109.13934} {arXiv:2109.13934 [astro-ph.CO]}
  \BibitemShut {NoStop}%
\bibitem [{\citenamefont {Colpi}\ \emph {et~al.}(2024)\citenamefont {Colpi}
  \emph {et~al.}}]{Colpi:2024xhw}%
  \BibitemOpen
  \bibfield  {author} {\bibinfo {author} {\bibfnamefont {M.}~\bibnamefont
  {Colpi}} \emph {et~al.},\ }\href@noop {} {\  (\bibinfo {year} {2024})},\
  \Eprint {http://arxiv.org/abs/2402.07571} {arXiv:2402.07571 [astro-ph.CO]}
  \BibitemShut {NoStop}%
\bibitem [{\citenamefont {Colpi}\ and\ \citenamefont
  {Sesana}(2017)}]{Colpi:2016fup}%
  \BibitemOpen
  \bibfield  {author} {\bibinfo {author} {\bibfnamefont {M.}~\bibnamefont
  {Colpi}}\ and\ \bibinfo {author} {\bibfnamefont {A.}~\bibnamefont {Sesana}},\
  }\enquote {\bibinfo {title} {{Gravitational Wave Sources in the Era of
  Multi-Band Gravitational Wave Astronomy}},}\ in\ \href {\doibase
  10.1142/9789813141766_0002} {\emph {\bibinfo {booktitle} {{An Overview of
  Gravitational Waves}: {Theory, Sources and Detection}}}},\ \bibinfo {editor}
  {edited by\ \bibinfo {editor} {\bibfnamefont {G.}~\bibnamefont {Auger}}\ and\
  \bibinfo {editor} {\bibfnamefont {E.}~\bibnamefont {Plagnol}}}\ (\bibinfo
  {year} {2017})\ pp.\ \bibinfo {pages} {43--140},\ \Eprint
  {http://arxiv.org/abs/1610.05309} {arXiv:1610.05309 [astro-ph.HE]}
  \BibitemShut {NoStop}%
\bibitem [{\citenamefont {Hogg}(1999)}]{Hogg:1999ad}%
  \BibitemOpen
  \bibfield  {author} {\bibinfo {author} {\bibfnamefont {D.~W.}\ \bibnamefont
  {Hogg}},\ }\href@noop {} {\  (\bibinfo {year} {1999})},\ \Eprint
  {http://arxiv.org/abs/astro-ph/9905116} {arXiv:astro-ph/9905116} \BibitemShut
  {NoStop}%
\bibitem [{\citenamefont {Linder}(2003)}]{Linder:2002et}%
  \BibitemOpen
  \bibfield  {author} {\bibinfo {author} {\bibfnamefont {E.~V.}\ \bibnamefont
  {Linder}},\ }\href {\doibase 10.1103/PhysRevLett.90.091301} {\bibfield
  {journal} {\bibinfo  {journal} {Phys. Rev. Lett.}\ }\textbf {\bibinfo
  {volume} {90}},\ \bibinfo {pages} {091301} (\bibinfo {year} {2003})},\
  \Eprint {http://arxiv.org/abs/astro-ph/0208512} {arXiv:astro-ph/0208512}
  \BibitemShut {NoStop}%
\bibitem [{\citenamefont {Chevallier}\ and\ \citenamefont
  {Polarski}(2001)}]{Chevallier:2000qy}%
  \BibitemOpen
  \bibfield  {author} {\bibinfo {author} {\bibfnamefont {M.}~\bibnamefont
  {Chevallier}}\ and\ \bibinfo {author} {\bibfnamefont {D.}~\bibnamefont
  {Polarski}},\ }\href {\doibase 10.1142/S0218271801000822} {\bibfield
  {journal} {\bibinfo  {journal} {Int. J. Mod. Phys. D}\ }\textbf {\bibinfo
  {volume} {10}},\ \bibinfo {pages} {213} (\bibinfo {year} {2001})},\ \Eprint
  {http://arxiv.org/abs/gr-qc/0009008} {arXiv:gr-qc/0009008} \BibitemShut
  {NoStop}%
\bibitem [{\citenamefont {Oguri}(2016)}]{Oguri:2016dgk}%
  \BibitemOpen
  \bibfield  {author} {\bibinfo {author} {\bibfnamefont {M.}~\bibnamefont
  {Oguri}},\ }\href {\doibase 10.1103/PhysRevD.93.083511} {\bibfield  {journal}
  {\bibinfo  {journal} {Phys. Rev. D}\ }\textbf {\bibinfo {volume} {93}},\
  \bibinfo {pages} {083511} (\bibinfo {year} {2016})},\ \Eprint
  {http://arxiv.org/abs/1603.02356} {arXiv:1603.02356 [astro-ph.CO]}
  \BibitemShut {NoStop}%
\bibitem [{\citenamefont {Diaz}\ and\ \citenamefont
  {Mukherjee}(2022)}]{Diaz:2021pem}%
  \BibitemOpen
  \bibfield  {author} {\bibinfo {author} {\bibfnamefont {C.~C.}\ \bibnamefont
  {Diaz}}\ and\ \bibinfo {author} {\bibfnamefont {S.}~\bibnamefont
  {Mukherjee}},\ }\href {\doibase 10.1093/mnras/stac208} {\bibfield  {journal}
  {\bibinfo  {journal} {Mon. Not. Roy. Astron. Soc.}\ }\textbf {\bibinfo
  {volume} {511}},\ \bibinfo {pages} {2782} (\bibinfo {year} {2022})},\ \Eprint
  {http://arxiv.org/abs/2107.12787} {arXiv:2107.12787 [astro-ph.CO]}
  \BibitemShut {NoStop}%
\bibitem [{\citenamefont {Mukherjee}\ \emph {et~al.}(2022)\citenamefont
  {Mukherjee}, \citenamefont {Krolewski}, \citenamefont {Wandelt},\ and\
  \citenamefont {Silk}}]{Mukherjee:2022afz}%
  \BibitemOpen
  \bibfield  {author} {\bibinfo {author} {\bibfnamefont {S.}~\bibnamefont
  {Mukherjee}}, \bibinfo {author} {\bibfnamefont {A.}~\bibnamefont
  {Krolewski}}, \bibinfo {author} {\bibfnamefont {B.~D.}\ \bibnamefont
  {Wandelt}}, \ and\ \bibinfo {author} {\bibfnamefont {J.}~\bibnamefont
  {Silk}},\ }\href@noop {} {\  (\bibinfo {year} {2022})},\ \Eprint
  {http://arxiv.org/abs/2203.03643} {arXiv:2203.03643 [astro-ph.CO]}
  \BibitemShut {NoStop}%
\bibitem [{\citenamefont {Sereno}\ \emph {et~al.}(2010)\citenamefont {Sereno},
  \citenamefont {Sesana}, \citenamefont {Bleuler}, \citenamefont {Jetzer},
  \citenamefont {Volonteri},\ and\ \citenamefont {Begelman}}]{Sereno:2010dr}%
  \BibitemOpen
  \bibfield  {author} {\bibinfo {author} {\bibfnamefont {M.}~\bibnamefont
  {Sereno}}, \bibinfo {author} {\bibfnamefont {A.}~\bibnamefont {Sesana}},
  \bibinfo {author} {\bibfnamefont {A.}~\bibnamefont {Bleuler}}, \bibinfo
  {author} {\bibfnamefont {P.}~\bibnamefont {Jetzer}}, \bibinfo {author}
  {\bibfnamefont {M.}~\bibnamefont {Volonteri}}, \ and\ \bibinfo {author}
  {\bibfnamefont {M.~C.}\ \bibnamefont {Begelman}},\ }\href {\doibase
  10.1103/PhysRevLett.105.251101} {\bibfield  {journal} {\bibinfo  {journal}
  {Phys. Rev. Lett.}\ }\textbf {\bibinfo {volume} {105}},\ \bibinfo {pages}
  {251101} (\bibinfo {year} {2010})},\ \Eprint {http://arxiv.org/abs/1011.5238}
  {arXiv:1011.5238 [astro-ph.CO]} \BibitemShut {NoStop}%
\bibitem [{\citenamefont {Liao}(2019)}]{Liao:2019xug}%
  \BibitemOpen
  \bibfield  {author} {\bibinfo {author} {\bibfnamefont {K.}~\bibnamefont
  {Liao}},\ }\href {\doibase 10.3847/1538-4357/ab4819} {\bibfield  {journal}
  {\bibinfo  {journal} {Astrophys. J.}\ }\textbf {\bibinfo {volume} {885}},\
  \bibinfo {pages} {70} (\bibinfo {year} {2019})},\ \Eprint
  {http://arxiv.org/abs/1906.09588} {arXiv:1906.09588 [astro-ph.CO]}
  \BibitemShut {NoStop}%
\bibitem [{\citenamefont {Wang}\ \emph
  {et~al.}(2022{\natexlab{c}})\citenamefont {Wang}, \citenamefont {Qi},
  \citenamefont {Wang}, \citenamefont {Zhang}, \citenamefont {Cui},\ and\
  \citenamefont {Zhang}}]{Wang:2022rvf}%
  \BibitemOpen
  \bibfield  {author} {\bibinfo {author} {\bibfnamefont {Y.-J.}\ \bibnamefont
  {Wang}}, \bibinfo {author} {\bibfnamefont {J.-Z.}\ \bibnamefont {Qi}},
  \bibinfo {author} {\bibfnamefont {B.}~\bibnamefont {Wang}}, \bibinfo {author}
  {\bibfnamefont {J.-F.}\ \bibnamefont {Zhang}}, \bibinfo {author}
  {\bibfnamefont {J.-L.}\ \bibnamefont {Cui}}, \ and\ \bibinfo {author}
  {\bibfnamefont {X.}~\bibnamefont {Zhang}},\ }\href {\doibase
  10.1093/mnras/stac2556} {\bibfield  {journal} {\bibinfo  {journal} {Mon. Not.
  Roy. Astron. Soc.}\ }\textbf {\bibinfo {volume} {516}},\ \bibinfo {pages}
  {5187} (\bibinfo {year} {2022}{\natexlab{c}})},\ \Eprint
  {http://arxiv.org/abs/2201.12553} {arXiv:2201.12553 [astro-ph.CO]}
  \BibitemShut {NoStop}%
\bibitem [{\citenamefont {Huang}\ \emph {et~al.}(2023)\citenamefont {Huang},
  \citenamefont {Hu}, \citenamefont {Chen}, \citenamefont {Zhang},
  \citenamefont {Li}, \citenamefont {Gao},\ and\ \citenamefont
  {Lin}}]{Huang:2023prq}%
  \BibitemOpen
  \bibfield  {author} {\bibinfo {author} {\bibfnamefont {S.-J.}\ \bibnamefont
  {Huang}}, \bibinfo {author} {\bibfnamefont {Y.-M.}\ \bibnamefont {Hu}},
  \bibinfo {author} {\bibfnamefont {X.}~\bibnamefont {Chen}}, \bibinfo {author}
  {\bibfnamefont {J.-d.}\ \bibnamefont {Zhang}}, \bibinfo {author}
  {\bibfnamefont {E.-K.}\ \bibnamefont {Li}}, \bibinfo {author} {\bibfnamefont
  {Z.}~\bibnamefont {Gao}}, \ and\ \bibinfo {author} {\bibfnamefont {X.-Y.}\
  \bibnamefont {Lin}},\ }\href {\doibase 10.1088/1475-7516/2023/08/003}
  {\bibfield  {journal} {\bibinfo  {journal} {JCAP}\ }\textbf {\bibinfo
  {volume} {08}},\ \bibinfo {pages} {003} (\bibinfo {year} {2023})},\ \Eprint
  {http://arxiv.org/abs/2304.10435} {arXiv:2304.10435 [astro-ph.CO]}
  \BibitemShut {NoStop}%
\bibitem [{\citenamefont {Ding}\ \emph {et~al.}(2019)\citenamefont {Ding},
  \citenamefont {Biesiada}, \citenamefont {Zheng}, \citenamefont {Liao},
  \citenamefont {Li},\ and\ \citenamefont {Zhu}}]{Ding:2018zrk}%
  \BibitemOpen
  \bibfield  {author} {\bibinfo {author} {\bibfnamefont {X.}~\bibnamefont
  {Ding}}, \bibinfo {author} {\bibfnamefont {M.}~\bibnamefont {Biesiada}},
  \bibinfo {author} {\bibfnamefont {X.}~\bibnamefont {Zheng}}, \bibinfo
  {author} {\bibfnamefont {K.}~\bibnamefont {Liao}}, \bibinfo {author}
  {\bibfnamefont {Z.}~\bibnamefont {Li}}, \ and\ \bibinfo {author}
  {\bibfnamefont {Z.-H.}\ \bibnamefont {Zhu}},\ }\href {\doibase
  10.1088/1475-7516/2019/04/033} {\bibfield  {journal} {\bibinfo  {journal}
  {JCAP}\ }\textbf {\bibinfo {volume} {04}},\ \bibinfo {pages} {033} (\bibinfo
  {year} {2019})},\ \Eprint {http://arxiv.org/abs/1801.05073} {arXiv:1801.05073
  [astro-ph.CO]} \BibitemShut {NoStop}%
\bibitem [{\citenamefont {Leandro}\ \emph {et~al.}(2022)\citenamefont
  {Leandro}, \citenamefont {Marra},\ and\ \citenamefont
  {Sturani}}]{Leandro:2021qlc}%
  \BibitemOpen
  \bibfield  {author} {\bibinfo {author} {\bibfnamefont {H.}~\bibnamefont
  {Leandro}}, \bibinfo {author} {\bibfnamefont {V.}~\bibnamefont {Marra}}, \
  and\ \bibinfo {author} {\bibfnamefont {R.}~\bibnamefont {Sturani}},\ }\href
  {\doibase 10.1103/PhysRevD.105.023523} {\bibfield  {journal} {\bibinfo
  {journal} {Phys. Rev. D}\ }\textbf {\bibinfo {volume} {105}},\ \bibinfo
  {pages} {023523} (\bibinfo {year} {2022})},\ \Eprint
  {http://arxiv.org/abs/2109.07537} {arXiv:2109.07537 [gr-qc]} \BibitemShut
  {NoStop}%
\bibitem [{\citenamefont {Taylor}\ \emph {et~al.}(2012)\citenamefont {Taylor},
  \citenamefont {Gair},\ and\ \citenamefont {Mandel}}]{Taylor:2011fs}%
  \BibitemOpen
  \bibfield  {author} {\bibinfo {author} {\bibfnamefont {S.~R.}\ \bibnamefont
  {Taylor}}, \bibinfo {author} {\bibfnamefont {J.~R.}\ \bibnamefont {Gair}}, \
  and\ \bibinfo {author} {\bibfnamefont {I.}~\bibnamefont {Mandel}},\ }\href
  {\doibase 10.1103/PhysRevD.85.023535} {\bibfield  {journal} {\bibinfo
  {journal} {Phys. Rev. D}\ }\textbf {\bibinfo {volume} {85}},\ \bibinfo
  {pages} {023535} (\bibinfo {year} {2012})},\ \Eprint
  {http://arxiv.org/abs/1108.5161} {arXiv:1108.5161 [gr-qc]} \BibitemShut
  {NoStop}%
\bibitem [{\citenamefont {Taylor}\ and\ \citenamefont
  {Gair}(2012)}]{Taylor:2012db}%
  \BibitemOpen
  \bibfield  {author} {\bibinfo {author} {\bibfnamefont {S.~R.}\ \bibnamefont
  {Taylor}}\ and\ \bibinfo {author} {\bibfnamefont {J.~R.}\ \bibnamefont
  {Gair}},\ }\href {\doibase 10.1103/PhysRevD.86.023502} {\bibfield  {journal}
  {\bibinfo  {journal} {Phys. Rev. D}\ }\textbf {\bibinfo {volume} {86}},\
  \bibinfo {pages} {023502} (\bibinfo {year} {2012})},\ \Eprint
  {http://arxiv.org/abs/1204.6739} {arXiv:1204.6739 [astro-ph.CO]} \BibitemShut
  {NoStop}%
\bibitem [{\citenamefont {Del~Pozzo}\ \emph {et~al.}(2017)\citenamefont
  {Del~Pozzo}, \citenamefont {Li},\ and\ \citenamefont
  {Messenger}}]{DelPozzo:2015bna}%
  \BibitemOpen
  \bibfield  {author} {\bibinfo {author} {\bibfnamefont {W.}~\bibnamefont
  {Del~Pozzo}}, \bibinfo {author} {\bibfnamefont {T.~G.~F.}\ \bibnamefont
  {Li}}, \ and\ \bibinfo {author} {\bibfnamefont {C.}~\bibnamefont
  {Messenger}},\ }\href {\doibase 10.1103/PhysRevD.95.043502} {\bibfield
  {journal} {\bibinfo  {journal} {Phys. Rev. D}\ }\textbf {\bibinfo {volume}
  {95}},\ \bibinfo {pages} {043502} (\bibinfo {year} {2017})},\ \Eprint
  {http://arxiv.org/abs/1506.06590} {arXiv:1506.06590 [gr-qc]} \BibitemShut
  {NoStop}%
\bibitem [{\citenamefont {Farr}\ \emph {et~al.}(2019)\citenamefont {Farr},
  \citenamefont {Fishbach}, \citenamefont {Ye},\ and\ \citenamefont
  {Holz}}]{Farr:2019twy}%
  \BibitemOpen
  \bibfield  {author} {\bibinfo {author} {\bibfnamefont {W.~M.}\ \bibnamefont
  {Farr}}, \bibinfo {author} {\bibfnamefont {M.}~\bibnamefont {Fishbach}},
  \bibinfo {author} {\bibfnamefont {J.}~\bibnamefont {Ye}}, \ and\ \bibinfo
  {author} {\bibfnamefont {D.}~\bibnamefont {Holz}},\ }\href {\doibase
  10.3847/2041-8213/ab4284} {\bibfield  {journal} {\bibinfo  {journal}
  {Astrophys. J. Lett.}\ }\textbf {\bibinfo {volume} {883}},\ \bibinfo {pages}
  {L42} (\bibinfo {year} {2019})},\ \Eprint {http://arxiv.org/abs/1908.09084}
  {arXiv:1908.09084 [astro-ph.CO]} \BibitemShut {NoStop}%
\bibitem [{\citenamefont {Mastrogiovanni}\ \emph {et~al.}(2021)\citenamefont
  {Mastrogiovanni}, \citenamefont {Leyde}, \citenamefont {Karathanasis},
  \citenamefont {Chassande-Mottin}, \citenamefont {Steer}, \citenamefont
  {Gair}, \citenamefont {Ghosh}, \citenamefont {Gray}, \citenamefont
  {Mukherjee},\ and\ \citenamefont {Rinaldi}}]{Mastrogiovanni:2021wsd}%
  \BibitemOpen
  \bibfield  {author} {\bibinfo {author} {\bibfnamefont {S.}~\bibnamefont
  {Mastrogiovanni}}, \bibinfo {author} {\bibfnamefont {K.}~\bibnamefont
  {Leyde}}, \bibinfo {author} {\bibfnamefont {C.}~\bibnamefont {Karathanasis}},
  \bibinfo {author} {\bibfnamefont {E.}~\bibnamefont {Chassande-Mottin}},
  \bibinfo {author} {\bibfnamefont {D.~A.}\ \bibnamefont {Steer}}, \bibinfo
  {author} {\bibfnamefont {J.}~\bibnamefont {Gair}}, \bibinfo {author}
  {\bibfnamefont {A.}~\bibnamefont {Ghosh}}, \bibinfo {author} {\bibfnamefont
  {R.}~\bibnamefont {Gray}}, \bibinfo {author} {\bibfnamefont {S.}~\bibnamefont
  {Mukherjee}}, \ and\ \bibinfo {author} {\bibfnamefont {S.}~\bibnamefont
  {Rinaldi}},\ }\href {\doibase 10.1103/PhysRevD.104.062009} {\bibfield
  {journal} {\bibinfo  {journal} {Phys. Rev. D}\ }\textbf {\bibinfo {volume}
  {104}},\ \bibinfo {pages} {062009} (\bibinfo {year} {2021})},\ \Eprint
  {http://arxiv.org/abs/2103.14663} {arXiv:2103.14663 [gr-qc]} \BibitemShut
  {NoStop}%
\bibitem [{\citenamefont {You}\ \emph {et~al.}(2021)\citenamefont {You},
  \citenamefont {Zhu}, \citenamefont {Ashton}, \citenamefont {Thrane},\ and\
  \citenamefont {Zhu}}]{You:2020wju}%
  \BibitemOpen
  \bibfield  {author} {\bibinfo {author} {\bibfnamefont {Z.-Q.}\ \bibnamefont
  {You}}, \bibinfo {author} {\bibfnamefont {X.-J.}\ \bibnamefont {Zhu}},
  \bibinfo {author} {\bibfnamefont {G.}~\bibnamefont {Ashton}}, \bibinfo
  {author} {\bibfnamefont {E.}~\bibnamefont {Thrane}}, \ and\ \bibinfo {author}
  {\bibfnamefont {Z.-H.}\ \bibnamefont {Zhu}},\ }\href {\doibase
  10.3847/1538-4357/abd4d4} {\bibfield  {journal} {\bibinfo  {journal}
  {Astrophys. J.}\ }\textbf {\bibinfo {volume} {908}},\ \bibinfo {pages} {215}
  (\bibinfo {year} {2021})},\ \Eprint {http://arxiv.org/abs/2004.00036}
  {arXiv:2004.00036 [astro-ph.CO]} \BibitemShut {NoStop}%
\bibitem [{\citenamefont {Mukherjee}(2022)}]{Mukherjee:2021rtw}%
  \BibitemOpen
  \bibfield  {author} {\bibinfo {author} {\bibfnamefont {S.}~\bibnamefont
  {Mukherjee}},\ }\href {\doibase 10.1093/mnras/stac2152} {\bibfield  {journal}
  {\bibinfo  {journal} {Mon. Not. Roy. Astron. Soc.}\ }\textbf {\bibinfo
  {volume} {515}},\ \bibinfo {pages} {5495} (\bibinfo {year} {2022})},\ \Eprint
  {http://arxiv.org/abs/2112.10256} {arXiv:2112.10256 [astro-ph.CO]}
  \BibitemShut {NoStop}%
\bibitem [{\citenamefont {Ezquiaga}\ and\ \citenamefont
  {Holz}(2022)}]{Ezquiaga:2022zkx}%
  \BibitemOpen
  \bibfield  {author} {\bibinfo {author} {\bibfnamefont {J.~M.}\ \bibnamefont
  {Ezquiaga}}\ and\ \bibinfo {author} {\bibfnamefont {D.~E.}\ \bibnamefont
  {Holz}},\ }\href {\doibase 10.1103/PhysRevLett.129.061102} {\bibfield
  {journal} {\bibinfo  {journal} {Phys. Rev. Lett.}\ }\textbf {\bibinfo
  {volume} {129}},\ \bibinfo {pages} {061102} (\bibinfo {year} {2022})},\
  \Eprint {http://arxiv.org/abs/2202.08240} {arXiv:2202.08240 [astro-ph.CO]}
  \BibitemShut {NoStop}%
\bibitem [{\citenamefont {Chen}\ \emph {et~al.}(2023)\citenamefont {Chen},
  \citenamefont {Du}, \citenamefont {Huang},\ and\ \citenamefont
  {You}}]{Chen:2022fda}%
  \BibitemOpen
  \bibfield  {author} {\bibinfo {author} {\bibfnamefont {Z.-C.}\ \bibnamefont
  {Chen}}, \bibinfo {author} {\bibfnamefont {S.-S.}\ \bibnamefont {Du}},
  \bibinfo {author} {\bibfnamefont {Q.-G.}\ \bibnamefont {Huang}}, \ and\
  \bibinfo {author} {\bibfnamefont {Z.-Q.}\ \bibnamefont {You}},\ }\href
  {\doibase 10.1088/1475-7516/2023/03/024} {\bibfield  {journal} {\bibinfo
  {journal} {JCAP}\ }\textbf {\bibinfo {volume} {03}},\ \bibinfo {pages} {024}
  (\bibinfo {year} {2023})},\ \Eprint {http://arxiv.org/abs/2205.11278}
  {arXiv:2205.11278 [astro-ph.CO]} \BibitemShut {NoStop}%
\bibitem [{\citenamefont {Messenger}\ and\ \citenamefont
  {Read}(2012)}]{Messenger:2011gi}%
  \BibitemOpen
  \bibfield  {author} {\bibinfo {author} {\bibfnamefont {C.}~\bibnamefont
  {Messenger}}\ and\ \bibinfo {author} {\bibfnamefont {J.}~\bibnamefont
  {Read}},\ }\href {\doibase 10.1103/PhysRevLett.108.091101} {\bibfield
  {journal} {\bibinfo  {journal} {Phys. Rev. Lett.}\ }\textbf {\bibinfo
  {volume} {108}},\ \bibinfo {pages} {091101} (\bibinfo {year} {2012})},\
  \Eprint {http://arxiv.org/abs/1107.5725} {arXiv:1107.5725 [gr-qc]}
  \BibitemShut {NoStop}%
\bibitem [{\citenamefont {Messenger}\ \emph {et~al.}(2014)\citenamefont
  {Messenger}, \citenamefont {Takami}, \citenamefont {Gossan}, \citenamefont
  {Rezzolla},\ and\ \citenamefont {Sathyaprakash}}]{Messenger:2013fya}%
  \BibitemOpen
  \bibfield  {author} {\bibinfo {author} {\bibfnamefont {C.}~\bibnamefont
  {Messenger}}, \bibinfo {author} {\bibfnamefont {K.}~\bibnamefont {Takami}},
  \bibinfo {author} {\bibfnamefont {S.}~\bibnamefont {Gossan}}, \bibinfo
  {author} {\bibfnamefont {L.}~\bibnamefont {Rezzolla}}, \ and\ \bibinfo
  {author} {\bibfnamefont {B.~S.}\ \bibnamefont {Sathyaprakash}},\ }\href
  {\doibase 10.1103/PhysRevX.4.041004} {\bibfield  {journal} {\bibinfo
  {journal} {Phys. Rev. X}\ }\textbf {\bibinfo {volume} {4}},\ \bibinfo {pages}
  {041004} (\bibinfo {year} {2014})},\ \Eprint {http://arxiv.org/abs/1312.1862}
  {arXiv:1312.1862 [gr-qc]} \BibitemShut {NoStop}%
\bibitem [{\citenamefont {Shiralilou}\ \emph {et~al.}(2023)\citenamefont
  {Shiralilou}, \citenamefont {Raaiijmakers}, \citenamefont {Duboeuf},
  \citenamefont {Nissanke}, \citenamefont {Foucart}, \citenamefont {Hinderer},\
  and\ \citenamefont {Williamson}}]{Shiralilou:2022urk}%
  \BibitemOpen
  \bibfield  {author} {\bibinfo {author} {\bibfnamefont {B.}~\bibnamefont
  {Shiralilou}}, \bibinfo {author} {\bibfnamefont {G.}~\bibnamefont
  {Raaiijmakers}}, \bibinfo {author} {\bibfnamefont {B.}~\bibnamefont
  {Duboeuf}}, \bibinfo {author} {\bibfnamefont {S.}~\bibnamefont {Nissanke}},
  \bibinfo {author} {\bibfnamefont {F.}~\bibnamefont {Foucart}}, \bibinfo
  {author} {\bibfnamefont {T.}~\bibnamefont {Hinderer}}, \ and\ \bibinfo
  {author} {\bibfnamefont {A.~R.}\ \bibnamefont {Williamson}},\ }\href
  {\doibase 10.3847/1538-4357/acf3dc} {\bibfield  {journal} {\bibinfo
  {journal} {Astrophys. J.}\ }\textbf {\bibinfo {volume} {955}},\ \bibinfo
  {pages} {149} (\bibinfo {year} {2023})},\ \Eprint
  {http://arxiv.org/abs/2207.11792} {arXiv:2207.11792 [astro-ph.CO]}
  \BibitemShut {NoStop}%
\bibitem [{\citenamefont {Jin}\ \emph {et~al.}(2023{\natexlab{a}})\citenamefont
  {Jin}, \citenamefont {Li}, \citenamefont {Zhang},\ and\ \citenamefont
  {Zhang}}]{Jin:2022qnj}%
  \BibitemOpen
  \bibfield  {author} {\bibinfo {author} {\bibfnamefont {S.-J.}\ \bibnamefont
  {Jin}}, \bibinfo {author} {\bibfnamefont {T.-N.}\ \bibnamefont {Li}},
  \bibinfo {author} {\bibfnamefont {J.-F.}\ \bibnamefont {Zhang}}, \ and\
  \bibinfo {author} {\bibfnamefont {X.}~\bibnamefont {Zhang}},\ }\href
  {\doibase 10.1088/1475-7516/2023/08/070} {\bibfield  {journal} {\bibinfo
  {journal} {JCAP}\ }\textbf {\bibinfo {volume} {08}},\ \bibinfo {pages} {070}
  (\bibinfo {year} {2023}{\natexlab{a}})},\ \Eprint
  {http://arxiv.org/abs/2202.11882} {arXiv:2202.11882 [gr-qc]} \BibitemShut
  {NoStop}%
\bibitem [{\citenamefont {Li}\ \emph {et~al.}(2024{\natexlab{g}})\citenamefont
  {Li}, \citenamefont {Jin}, \citenamefont {Li}, \citenamefont {Zhang},\ and\
  \citenamefont {Zhang}}]{Li:2023gtu}%
  \BibitemOpen
  \bibfield  {author} {\bibinfo {author} {\bibfnamefont {T.-N.}\ \bibnamefont
  {Li}}, \bibinfo {author} {\bibfnamefont {S.-J.}\ \bibnamefont {Jin}},
  \bibinfo {author} {\bibfnamefont {H.-L.}\ \bibnamefont {Li}}, \bibinfo
  {author} {\bibfnamefont {J.-F.}\ \bibnamefont {Zhang}}, \ and\ \bibinfo
  {author} {\bibfnamefont {X.}~\bibnamefont {Zhang}},\ }\href {\doibase
  10.3847/1538-4357/ad1bc9} {\bibfield  {journal} {\bibinfo  {journal}
  {Astrophys. J.}\ }\textbf {\bibinfo {volume} {963}},\ \bibinfo {pages} {52}
  (\bibinfo {year} {2024}{\natexlab{g}})},\ \Eprint
  {http://arxiv.org/abs/2310.15879} {arXiv:2310.15879 [astro-ph.CO]}
  \BibitemShut {NoStop}%
\bibitem [{\citenamefont {Del~Pozzo}(2012)}]{DelPozzo:2011vcw}%
  \BibitemOpen
  \bibfield  {author} {\bibinfo {author} {\bibfnamefont {W.}~\bibnamefont
  {Del~Pozzo}},\ }\href {\doibase 10.1103/PhysRevD.86.043011} {\bibfield
  {journal} {\bibinfo  {journal} {Phys. Rev. D}\ }\textbf {\bibinfo {volume}
  {86}},\ \bibinfo {pages} {043011} (\bibinfo {year} {2012})},\ \Eprint
  {http://arxiv.org/abs/1108.1317} {arXiv:1108.1317 [astro-ph.CO]} \BibitemShut
  {NoStop}%
\bibitem [{\citenamefont {Abbott}\ \emph
  {et~al.}(2017{\natexlab{f}})\citenamefont {Abbott} \emph
  {et~al.}}]{LIGOScientific:2017adf}%
  \BibitemOpen
  \bibfield  {author} {\bibinfo {author} {\bibfnamefont {B.~P.}\ \bibnamefont
  {Abbott}} \emph {et~al.} (\bibinfo {collaboration} {LIGO Scientific, Virgo,
  1M2H, Dark Energy Camera GW-E, DES, DLT40, Las Cumbres Observatory, VINROUGE,
  MASTER}),\ }\href {\doibase 10.1038/nature24471} {\bibfield  {journal}
  {\bibinfo  {journal} {Nature}\ }\textbf {\bibinfo {volume} {551}},\ \bibinfo
  {pages} {85} (\bibinfo {year} {2017}{\natexlab{f}})},\ \Eprint
  {http://arxiv.org/abs/1710.05835} {arXiv:1710.05835 [astro-ph.CO]}
  \BibitemShut {NoStop}%
\bibitem [{\citenamefont {Chen}\ \emph
  {et~al.}(2018{\natexlab{b}})\citenamefont {Chen}, \citenamefont {Fishbach},\
  and\ \citenamefont {Holz}}]{Chen:2017rfc}%
  \BibitemOpen
  \bibfield  {author} {\bibinfo {author} {\bibfnamefont {H.-Y.}\ \bibnamefont
  {Chen}}, \bibinfo {author} {\bibfnamefont {M.}~\bibnamefont {Fishbach}}, \
  and\ \bibinfo {author} {\bibfnamefont {D.~E.}\ \bibnamefont {Holz}},\ }\href
  {\doibase 10.1038/s41586-018-0606-0} {\bibfield  {journal} {\bibinfo
  {journal} {Nature}\ }\textbf {\bibinfo {volume} {562}},\ \bibinfo {pages}
  {545} (\bibinfo {year} {2018}{\natexlab{b}})},\ \Eprint
  {http://arxiv.org/abs/1712.06531} {arXiv:1712.06531 [astro-ph.CO]}
  \BibitemShut {NoStop}%
\bibitem [{\citenamefont {Gray}\ \emph {et~al.}(2020)\citenamefont {Gray} \emph
  {et~al.}}]{Gray:2019ksv}%
  \BibitemOpen
  \bibfield  {author} {\bibinfo {author} {\bibfnamefont {R.}~\bibnamefont
  {Gray}} \emph {et~al.},\ }\href {\doibase 10.1103/PhysRevD.101.122001}
  {\bibfield  {journal} {\bibinfo  {journal} {Phys. Rev. D}\ }\textbf {\bibinfo
  {volume} {101}},\ \bibinfo {pages} {122001} (\bibinfo {year} {2020})},\
  \Eprint {http://arxiv.org/abs/1908.06050} {arXiv:1908.06050 [gr-qc]}
  \BibitemShut {NoStop}%
\bibitem [{\citenamefont {Mandel}\ \emph {et~al.}(2019)\citenamefont {Mandel},
  \citenamefont {Farr},\ and\ \citenamefont {Gair}}]{Mandel:2018mve}%
  \BibitemOpen
  \bibfield  {author} {\bibinfo {author} {\bibfnamefont {I.}~\bibnamefont
  {Mandel}}, \bibinfo {author} {\bibfnamefont {W.~M.}\ \bibnamefont {Farr}}, \
  and\ \bibinfo {author} {\bibfnamefont {J.~R.}\ \bibnamefont {Gair}},\ }\href
  {\doibase 10.1093/mnras/stz896} {\bibfield  {journal} {\bibinfo  {journal}
  {Mon. Not. Roy. Astron. Soc.}\ }\textbf {\bibinfo {volume} {486}},\ \bibinfo
  {pages} {1086} (\bibinfo {year} {2019})},\ \Eprint
  {http://arxiv.org/abs/1809.02063} {arXiv:1809.02063 [physics.data-an]}
  \BibitemShut {NoStop}%
\bibitem [{\citenamefont {Jeffreys}(1961)}]{Jeffreys:1961book}%
  \BibitemOpen
  \bibfield  {author} {\bibinfo {author} {\bibfnamefont {H.}~\bibnamefont
  {Jeffreys}},\ }\href@noop {} {\emph {\bibinfo {title} {Theory of
  Probability}}}\ (\bibinfo  {publisher} {Oxford University Press},\ \bibinfo
  {address} {Oxford},\ \bibinfo {year} {1961})\BibitemShut {NoStop}%
\bibitem [{\citenamefont {Trotta}(2008)}]{Trotta:2008qt}%
  \BibitemOpen
  \bibfield  {author} {\bibinfo {author} {\bibfnamefont {R.}~\bibnamefont
  {Trotta}},\ }\href {\doibase 10.1080/00107510802066753} {\bibfield  {journal}
  {\bibinfo  {journal} {Contemp. Phys.}\ }\textbf {\bibinfo {volume} {49}},\
  \bibinfo {pages} {71} (\bibinfo {year} {2008})},\ \Eprint
  {http://arxiv.org/abs/0803.4089} {arXiv:0803.4089 [astro-ph]} \BibitemShut
  {NoStop}%
\bibitem [{\citenamefont {Abbott}\ \emph
  {et~al.}(2019{\natexlab{e}})\citenamefont {Abbott} \emph
  {et~al.}}]{LIGOScientific:2018jsj}%
  \BibitemOpen
  \bibfield  {author} {\bibinfo {author} {\bibfnamefont {B.~P.}\ \bibnamefont
  {Abbott}} \emph {et~al.} (\bibinfo {collaboration} {LIGO Scientific,
  Virgo}),\ }\href {\doibase 10.3847/2041-8213/ab3800} {\bibfield  {journal}
  {\bibinfo  {journal} {Astrophys. J. Lett.}\ }\textbf {\bibinfo {volume}
  {882}},\ \bibinfo {pages} {L24} (\bibinfo {year} {2019}{\natexlab{e}})},\
  \Eprint {http://arxiv.org/abs/1811.12940} {arXiv:1811.12940 [astro-ph.HE]}
  \BibitemShut {NoStop}%
\bibitem [{\citenamefont {Abbott}\ \emph
  {et~al.}(2023{\natexlab{e}})\citenamefont {Abbott} \emph
  {et~al.}}]{KAGRA:2021duu}%
  \BibitemOpen
  \bibfield  {author} {\bibinfo {author} {\bibfnamefont {R.}~\bibnamefont
  {Abbott}} \emph {et~al.} (\bibinfo {collaboration} {KAGRA, VIRGO, LIGO
  Scientific}),\ }\href {\doibase 10.1103/PhysRevX.13.011048} {\bibfield
  {journal} {\bibinfo  {journal} {Phys. Rev. X}\ }\textbf {\bibinfo {volume}
  {13}},\ \bibinfo {pages} {011048} (\bibinfo {year} {2023}{\natexlab{e}})},\
  \Eprint {http://arxiv.org/abs/2111.03634} {arXiv:2111.03634 [astro-ph.HE]}
  \BibitemShut {NoStop}%
\bibitem [{\citenamefont {Miller}(2002)}]{Miller:2002vg}%
  \BibitemOpen
  \bibfield  {author} {\bibinfo {author} {\bibfnamefont {M.~C.}\ \bibnamefont
  {Miller}},\ }\href {\doibase 10.1086/344156} {\bibfield  {journal} {\bibinfo
  {journal} {Astrophys. J.}\ }\textbf {\bibinfo {volume} {581}},\ \bibinfo
  {pages} {438} (\bibinfo {year} {2002})},\ \Eprint
  {http://arxiv.org/abs/astro-ph/0206404} {arXiv:astro-ph/0206404} \BibitemShut
  {NoStop}%
\bibitem [{\citenamefont {Takahashi}\ and\ \citenamefont
  {Nakamura}(2003{\natexlab{b}})}]{Takahashi:2003wm}%
  \BibitemOpen
  \bibfield  {author} {\bibinfo {author} {\bibfnamefont {R.}~\bibnamefont
  {Takahashi}}\ and\ \bibinfo {author} {\bibfnamefont {T.}~\bibnamefont
  {Nakamura}},\ }\href {\doibase 10.1086/379112} {\bibfield  {journal}
  {\bibinfo  {journal} {Astrophys. J. Lett.}\ }\textbf {\bibinfo {volume}
  {596}},\ \bibinfo {pages} {L231} (\bibinfo {year} {2003}{\natexlab{b}})},\
  \Eprint {http://arxiv.org/abs/astro-ph/0307390} {arXiv:astro-ph/0307390}
  \BibitemShut {NoStop}%
\bibitem [{\citenamefont {Seto}(2016)}]{Seto:2016wom}%
  \BibitemOpen
  \bibfield  {author} {\bibinfo {author} {\bibfnamefont {N.}~\bibnamefont
  {Seto}},\ }\href {\doibase 10.1093/mnrasl/slw060} {\bibfield  {journal}
  {\bibinfo  {journal} {Mon. Not. Roy. Astron. Soc.}\ }\textbf {\bibinfo
  {volume} {460}},\ \bibinfo {pages} {L1} (\bibinfo {year} {2016})},\ \Eprint
  {http://arxiv.org/abs/1602.04715} {arXiv:1602.04715 [astro-ph.HE]}
  \BibitemShut {NoStop}%
\bibitem [{\citenamefont {Chen}\ and\ \citenamefont
  {Amaro-Seoane}(2017)}]{Chen:2017gfm}%
  \BibitemOpen
  \bibfield  {author} {\bibinfo {author} {\bibfnamefont {X.}~\bibnamefont
  {Chen}}\ and\ \bibinfo {author} {\bibfnamefont {P.}~\bibnamefont
  {Amaro-Seoane}},\ }\href {\doibase 10.3847/2041-8213/aa74ce} {\bibfield
  {journal} {\bibinfo  {journal} {Astrophys. J. Lett.}\ }\textbf {\bibinfo
  {volume} {842}},\ \bibinfo {pages} {L2} (\bibinfo {year} {2017})},\ \Eprint
  {http://arxiv.org/abs/1702.08479} {arXiv:1702.08479 [astro-ph.HE]}
  \BibitemShut {NoStop}%
\bibitem [{\citenamefont {Sedda}\ \emph {et~al.}(2020)\citenamefont {Sedda}
  \emph {et~al.}}]{Sedda:2019uro}%
  \BibitemOpen
  \bibfield  {author} {\bibinfo {author} {\bibfnamefont {M.~A.}\ \bibnamefont
  {Sedda}} \emph {et~al.},\ }\href {\doibase 10.1088/1361-6382/abb5c1}
  {\bibfield  {journal} {\bibinfo  {journal} {Class. Quant. Grav.}\ }\textbf
  {\bibinfo {volume} {37}},\ \bibinfo {pages} {215011} (\bibinfo {year}
  {2020})},\ \Eprint {http://arxiv.org/abs/1908.11375} {arXiv:1908.11375
  [gr-qc]} \BibitemShut {NoStop}%
\bibitem [{\citenamefont {Nishizawa}\ \emph {et~al.}(2016)\citenamefont
  {Nishizawa}, \citenamefont {Berti}, \citenamefont {Klein},\ and\
  \citenamefont {Sesana}}]{Nishizawa:2016jji}%
  \BibitemOpen
  \bibfield  {author} {\bibinfo {author} {\bibfnamefont {A.}~\bibnamefont
  {Nishizawa}}, \bibinfo {author} {\bibfnamefont {E.}~\bibnamefont {Berti}},
  \bibinfo {author} {\bibfnamefont {A.}~\bibnamefont {Klein}}, \ and\ \bibinfo
  {author} {\bibfnamefont {A.}~\bibnamefont {Sesana}},\ }\href {\doibase
  10.1103/PhysRevD.94.064020} {\bibfield  {journal} {\bibinfo  {journal} {Phys.
  Rev. D}\ }\textbf {\bibinfo {volume} {94}},\ \bibinfo {pages} {064020}
  (\bibinfo {year} {2016})},\ \Eprint {http://arxiv.org/abs/1605.01341}
  {arXiv:1605.01341 [gr-qc]} \BibitemShut {NoStop}%
\bibitem [{\citenamefont {Nishizawa}\ \emph {et~al.}(2017)\citenamefont
  {Nishizawa}, \citenamefont {Sesana}, \citenamefont {Berti},\ and\
  \citenamefont {Klein}}]{Nishizawa:2016eza}%
  \BibitemOpen
  \bibfield  {author} {\bibinfo {author} {\bibfnamefont {A.}~\bibnamefont
  {Nishizawa}}, \bibinfo {author} {\bibfnamefont {A.}~\bibnamefont {Sesana}},
  \bibinfo {author} {\bibfnamefont {E.}~\bibnamefont {Berti}}, \ and\ \bibinfo
  {author} {\bibfnamefont {A.}~\bibnamefont {Klein}},\ }\href {\doibase
  10.1093/mnras/stw2993} {\bibfield  {journal} {\bibinfo  {journal} {Mon. Not.
  Roy. Astron. Soc.}\ }\textbf {\bibinfo {volume} {465}},\ \bibinfo {pages}
  {4375} (\bibinfo {year} {2017})},\ \Eprint {http://arxiv.org/abs/1606.09295}
  {arXiv:1606.09295 [astro-ph.HE]} \BibitemShut {NoStop}%
\bibitem [{\citenamefont {Breivik}\ \emph {et~al.}(2016)\citenamefont
  {Breivik}, \citenamefont {Rodriguez}, \citenamefont {Larson}, \citenamefont
  {Kalogera},\ and\ \citenamefont {Rasio}}]{Breivik:2016ddj}%
  \BibitemOpen
  \bibfield  {author} {\bibinfo {author} {\bibfnamefont {K.}~\bibnamefont
  {Breivik}}, \bibinfo {author} {\bibfnamefont {C.~L.}\ \bibnamefont
  {Rodriguez}}, \bibinfo {author} {\bibfnamefont {S.~L.}\ \bibnamefont
  {Larson}}, \bibinfo {author} {\bibfnamefont {V.}~\bibnamefont {Kalogera}}, \
  and\ \bibinfo {author} {\bibfnamefont {F.~A.}\ \bibnamefont {Rasio}},\ }\href
  {\doibase 10.3847/2041-8205/830/1/L18} {\bibfield  {journal} {\bibinfo
  {journal} {Astrophys. J. Lett.}\ }\textbf {\bibinfo {volume} {830}},\
  \bibinfo {pages} {L18} (\bibinfo {year} {2016})},\ \Eprint
  {http://arxiv.org/abs/1606.09558} {arXiv:1606.09558 [astro-ph.GA]}
  \BibitemShut {NoStop}%
\bibitem [{\citenamefont {Randall}\ and\ \citenamefont
  {Xianyu}(2019)}]{Randall:2018lnh}%
  \BibitemOpen
  \bibfield  {author} {\bibinfo {author} {\bibfnamefont {L.}~\bibnamefont
  {Randall}}\ and\ \bibinfo {author} {\bibfnamefont {Z.-Z.}\ \bibnamefont
  {Xianyu}},\ }\href {\doibase 10.3847/1538-4357/ab20c6} {\bibfield  {journal}
  {\bibinfo  {journal} {Astrophys. J.}\ }\textbf {\bibinfo {volume} {878}},\
  \bibinfo {pages} {75} (\bibinfo {year} {2019})},\ \Eprint
  {http://arxiv.org/abs/1805.05335} {arXiv:1805.05335 [gr-qc]} \BibitemShut
  {NoStop}%
\bibitem [{\citenamefont {Toubiana}\ \emph {et~al.}(2021)\citenamefont
  {Toubiana} \emph {et~al.}}]{Toubiana:2020drf}%
  \BibitemOpen
  \bibfield  {author} {\bibinfo {author} {\bibfnamefont {A.}~\bibnamefont
  {Toubiana}} \emph {et~al.},\ }\href {\doibase 10.1103/PhysRevLett.126.101105}
  {\bibfield  {journal} {\bibinfo  {journal} {Phys. Rev. Lett.}\ }\textbf
  {\bibinfo {volume} {126}},\ \bibinfo {pages} {101105} (\bibinfo {year}
  {2021})},\ \Eprint {http://arxiv.org/abs/2010.06056} {arXiv:2010.06056
  [astro-ph.HE]} \BibitemShut {NoStop}%
\bibitem [{\citenamefont {Sberna}\ \emph {et~al.}(2022)\citenamefont {Sberna}
  \emph {et~al.}}]{Sberna:2022qbn}%
  \BibitemOpen
  \bibfield  {author} {\bibinfo {author} {\bibfnamefont {L.}~\bibnamefont
  {Sberna}} \emph {et~al.},\ }\href {\doibase 10.1103/PhysRevD.106.064056}
  {\bibfield  {journal} {\bibinfo  {journal} {Phys. Rev. D}\ }\textbf {\bibinfo
  {volume} {106}},\ \bibinfo {pages} {064056} (\bibinfo {year} {2022})},\
  \Eprint {http://arxiv.org/abs/2205.08550} {arXiv:2205.08550 [gr-qc]}
  \BibitemShut {NoStop}%
\bibitem [{\citenamefont {Zhu}\ \emph {et~al.}(2022{\natexlab{b}})\citenamefont
  {Zhu}, \citenamefont {Xie}, \citenamefont {Hu}, \citenamefont {Liu},
  \citenamefont {Li}, \citenamefont {Napolitano}, \citenamefont {Tang},
  \citenamefont {Zhang},\ and\ \citenamefont {Mei}}]{Zhu:2021bpp}%
  \BibitemOpen
  \bibfield  {author} {\bibinfo {author} {\bibfnamefont {L.-G.}\ \bibnamefont
  {Zhu}}, \bibinfo {author} {\bibfnamefont {L.-H.}\ \bibnamefont {Xie}},
  \bibinfo {author} {\bibfnamefont {Y.-M.}\ \bibnamefont {Hu}}, \bibinfo
  {author} {\bibfnamefont {S.}~\bibnamefont {Liu}}, \bibinfo {author}
  {\bibfnamefont {E.-K.}\ \bibnamefont {Li}}, \bibinfo {author} {\bibfnamefont
  {N.~R.}\ \bibnamefont {Napolitano}}, \bibinfo {author} {\bibfnamefont
  {B.-T.}\ \bibnamefont {Tang}}, \bibinfo {author} {\bibfnamefont {J.-d.}\
  \bibnamefont {Zhang}}, \ and\ \bibinfo {author} {\bibfnamefont
  {J.}~\bibnamefont {Mei}},\ }\href {\doibase 10.1007/s11433-021-1859-9}
  {\bibfield  {journal} {\bibinfo  {journal} {Sci. China Phys. Mech. Astron.}\
  }\textbf {\bibinfo {volume} {65}},\ \bibinfo {pages} {259811} (\bibinfo
  {year} {2022}{\natexlab{b}})},\ \Eprint {http://arxiv.org/abs/2110.05224}
  {arXiv:2110.05224 [astro-ph.CO]} \BibitemShut {NoStop}%
\bibitem [{\citenamefont {Mangiagli}\ \emph {et~al.}(2020)\citenamefont
  {Mangiagli}, \citenamefont {Klein}, \citenamefont {Bonetti}, \citenamefont
  {Katz}, \citenamefont {Sesana}, \citenamefont {Volonteri}, \citenamefont
  {Colpi}, \citenamefont {Marsat},\ and\ \citenamefont
  {Babak}}]{Mangiagli:2020rwz}%
  \BibitemOpen
  \bibfield  {author} {\bibinfo {author} {\bibfnamefont {A.}~\bibnamefont
  {Mangiagli}}, \bibinfo {author} {\bibfnamefont {A.}~\bibnamefont {Klein}},
  \bibinfo {author} {\bibfnamefont {M.}~\bibnamefont {Bonetti}}, \bibinfo
  {author} {\bibfnamefont {M.~L.}\ \bibnamefont {Katz}}, \bibinfo {author}
  {\bibfnamefont {A.}~\bibnamefont {Sesana}}, \bibinfo {author} {\bibfnamefont
  {M.}~\bibnamefont {Volonteri}}, \bibinfo {author} {\bibfnamefont
  {M.}~\bibnamefont {Colpi}}, \bibinfo {author} {\bibfnamefont
  {S.}~\bibnamefont {Marsat}}, \ and\ \bibinfo {author} {\bibfnamefont
  {S.}~\bibnamefont {Babak}},\ }\href {\doibase 10.1103/PhysRevD.102.084056}
  {\bibfield  {journal} {\bibinfo  {journal} {Phys. Rev. D}\ }\textbf {\bibinfo
  {volume} {102}},\ \bibinfo {pages} {084056} (\bibinfo {year} {2020})},\
  \Eprint {http://arxiv.org/abs/2006.12513} {arXiv:2006.12513 [astro-ph.HE]}
  \BibitemShut {NoStop}%
\bibitem [{\citenamefont {Chen}\ \emph
  {et~al.}(2024{\natexlab{b}})\citenamefont {Chen}, \citenamefont {Lyu},
  \citenamefont {Li},\ and\ \citenamefont {Hu}}]{Chen:2023qga}%
  \BibitemOpen
  \bibfield  {author} {\bibinfo {author} {\bibfnamefont {H.-Y.}\ \bibnamefont
  {Chen}}, \bibinfo {author} {\bibfnamefont {X.-Y.}\ \bibnamefont {Lyu}},
  \bibinfo {author} {\bibfnamefont {E.-K.}\ \bibnamefont {Li}}, \ and\ \bibinfo
  {author} {\bibfnamefont {Y.-M.}\ \bibnamefont {Hu}},\ }\href {\doibase
  10.1007/s11433-023-2377-7} {\bibfield  {journal} {\bibinfo  {journal} {Sci.
  China Phys. Mech. Astron.}\ }\textbf {\bibinfo {volume} {67}},\ \bibinfo
  {pages} {279512} (\bibinfo {year} {2024}{\natexlab{b}})},\ \Eprint
  {http://arxiv.org/abs/2309.06910} {arXiv:2309.06910 [gr-qc]} \BibitemShut
  {NoStop}%
\bibitem [{\citenamefont {Liu}\ \emph {et~al.}(2022)\citenamefont {Liu},
  \citenamefont {Zhu}, \citenamefont {Hu}, \citenamefont {Zhang},\ and\
  \citenamefont {Ji}}]{Liu:2021yoy}%
  \BibitemOpen
  \bibfield  {author} {\bibinfo {author} {\bibfnamefont {S.}~\bibnamefont
  {Liu}}, \bibinfo {author} {\bibfnamefont {L.-G.}\ \bibnamefont {Zhu}},
  \bibinfo {author} {\bibfnamefont {Y.-M.}\ \bibnamefont {Hu}}, \bibinfo
  {author} {\bibfnamefont {J.-d.}\ \bibnamefont {Zhang}}, \ and\ \bibinfo
  {author} {\bibfnamefont {M.-J.}\ \bibnamefont {Ji}},\ }\href {\doibase
  10.1103/PhysRevD.105.023019} {\bibfield  {journal} {\bibinfo  {journal}
  {Phys. Rev. D}\ }\textbf {\bibinfo {volume} {105}},\ \bibinfo {pages}
  {023019} (\bibinfo {year} {2022})},\ \Eprint
  {http://arxiv.org/abs/2110.05248} {arXiv:2110.05248 [astro-ph.HE]}
  \BibitemShut {NoStop}%
\bibitem [{\citenamefont {Amaro-Seoane}(2018)}]{Amaro-Seoane:2012lgq}%
  \BibitemOpen
  \bibfield  {author} {\bibinfo {author} {\bibfnamefont {P.}~\bibnamefont
  {Amaro-Seoane}},\ }\href {\doibase 10.1007/s41114-018-0013-8} {\bibfield
  {journal} {\bibinfo  {journal} {Living Rev. Rel.}\ }\textbf {\bibinfo
  {volume} {21}},\ \bibinfo {pages} {4} (\bibinfo {year} {2018})},\ \Eprint
  {http://arxiv.org/abs/1205.5240} {arXiv:1205.5240 [astro-ph.CO]} \BibitemShut
  {NoStop}%
\bibitem [{\citenamefont {Barausse}\ \emph
  {et~al.}(2020{\natexlab{b}})\citenamefont {Barausse}, \citenamefont
  {Dvorkin}, \citenamefont {Tremmel}, \citenamefont {Volonteri},\ and\
  \citenamefont {Bonetti}}]{Barausse:2020mdt}%
  \BibitemOpen
  \bibfield  {author} {\bibinfo {author} {\bibfnamefont {E.}~\bibnamefont
  {Barausse}}, \bibinfo {author} {\bibfnamefont {I.}~\bibnamefont {Dvorkin}},
  \bibinfo {author} {\bibfnamefont {M.}~\bibnamefont {Tremmel}}, \bibinfo
  {author} {\bibfnamefont {M.}~\bibnamefont {Volonteri}}, \ and\ \bibinfo
  {author} {\bibfnamefont {M.}~\bibnamefont {Bonetti}},\ }\href {\doibase
  10.3847/1538-4357/abba7f} {\bibfield  {journal} {\bibinfo  {journal}
  {Astrophys. J.}\ }\textbf {\bibinfo {volume} {904}},\ \bibinfo {pages} {16}
  (\bibinfo {year} {2020}{\natexlab{b}})},\ \Eprint
  {http://arxiv.org/abs/2006.03065} {arXiv:2006.03065 [astro-ph.GA]}
  \BibitemShut {NoStop}%
\bibitem [{\citenamefont {Finn}(1992)}]{Finn:1992wt}%
  \BibitemOpen
  \bibfield  {author} {\bibinfo {author} {\bibfnamefont {L.~S.}\ \bibnamefont
  {Finn}},\ }\href {\doibase 10.1103/PhysRevD.46.5236} {\bibfield  {journal}
  {\bibinfo  {journal} {Phys. Rev. D}\ }\textbf {\bibinfo {volume} {46}},\
  \bibinfo {pages} {5236} (\bibinfo {year} {1992})},\ \Eprint
  {http://arxiv.org/abs/gr-qc/9209010} {arXiv:gr-qc/9209010} \BibitemShut
  {NoStop}%
\bibitem [{\citenamefont {Vallisneri}(2008)}]{Vallisneri:2007ev}%
  \BibitemOpen
  \bibfield  {author} {\bibinfo {author} {\bibfnamefont {M.}~\bibnamefont
  {Vallisneri}},\ }\href {\doibase 10.1103/PhysRevD.77.042001} {\bibfield
  {journal} {\bibinfo  {journal} {Phys. Rev. D}\ }\textbf {\bibinfo {volume}
  {77}},\ \bibinfo {pages} {042001} (\bibinfo {year} {2008})},\ \Eprint
  {http://arxiv.org/abs/gr-qc/0703086} {arXiv:gr-qc/0703086} \BibitemShut
  {NoStop}%
\bibitem [{\citenamefont {Kocsis}\ \emph {et~al.}(2006)\citenamefont {Kocsis},
  \citenamefont {Frei}, \citenamefont {Haiman},\ and\ \citenamefont
  {Menou}}]{Kocsis:2005vv}%
  \BibitemOpen
  \bibfield  {author} {\bibinfo {author} {\bibfnamefont {B.}~\bibnamefont
  {Kocsis}}, \bibinfo {author} {\bibfnamefont {Z.}~\bibnamefont {Frei}},
  \bibinfo {author} {\bibfnamefont {Z.}~\bibnamefont {Haiman}}, \ and\ \bibinfo
  {author} {\bibfnamefont {K.}~\bibnamefont {Menou}},\ }\href {\doibase
  10.1086/498236} {\bibfield  {journal} {\bibinfo  {journal} {Astrophys. J.}\
  }\textbf {\bibinfo {volume} {637}},\ \bibinfo {pages} {27} (\bibinfo {year}
  {2006})},\ \Eprint {http://arxiv.org/abs/astro-ph/0505394}
  {arXiv:astro-ph/0505394} \BibitemShut {NoStop}%
\bibitem [{\citenamefont {Gordon}\ \emph {et~al.}(2007)\citenamefont {Gordon},
  \citenamefont {Land},\ and\ \citenamefont {Slosar}}]{Gordon:2007zw}%
  \BibitemOpen
  \bibfield  {author} {\bibinfo {author} {\bibfnamefont {C.}~\bibnamefont
  {Gordon}}, \bibinfo {author} {\bibfnamefont {K.}~\bibnamefont {Land}}, \ and\
  \bibinfo {author} {\bibfnamefont {A.}~\bibnamefont {Slosar}},\ }\href
  {\doibase 10.1103/PhysRevLett.99.081301} {\bibfield  {journal} {\bibinfo
  {journal} {Phys. Rev. Lett.}\ }\textbf {\bibinfo {volume} {99}},\ \bibinfo
  {pages} {081301} (\bibinfo {year} {2007})},\ \Eprint
  {http://arxiv.org/abs/0705.1718} {arXiv:0705.1718 [astro-ph]} \BibitemShut
  {NoStop}%
\bibitem [{\citenamefont {Wang}\ \emph {et~al.}(1996)\citenamefont {Wang},
  \citenamefont {Stebbins},\ and\ \citenamefont {Turner}}]{Wang:1996as}%
  \BibitemOpen
  \bibfield  {author} {\bibinfo {author} {\bibfnamefont {Y.}~\bibnamefont
  {Wang}}, \bibinfo {author} {\bibfnamefont {A.}~\bibnamefont {Stebbins}}, \
  and\ \bibinfo {author} {\bibfnamefont {E.~L.}\ \bibnamefont {Turner}},\
  }\href {\doibase 10.1103/PhysRevLett.77.2875} {\bibfield  {journal} {\bibinfo
   {journal} {Phys. Rev. Lett.}\ }\textbf {\bibinfo {volume} {77}},\ \bibinfo
  {pages} {2875} (\bibinfo {year} {1996})},\ \Eprint
  {http://arxiv.org/abs/astro-ph/9605140} {arXiv:astro-ph/9605140} \BibitemShut
  {NoStop}%
\bibitem [{\citenamefont {He}(2019)}]{He:2019dhl}%
  \BibitemOpen
  \bibfield  {author} {\bibinfo {author} {\bibfnamefont {J.-h.}\ \bibnamefont
  {He}},\ }\href {\doibase 10.1103/PhysRevD.100.023527} {\bibfield  {journal}
  {\bibinfo  {journal} {Phys. Rev. D}\ }\textbf {\bibinfo {volume} {100}},\
  \bibinfo {pages} {023527} (\bibinfo {year} {2019})},\ \Eprint
  {http://arxiv.org/abs/1903.11254} {arXiv:1903.11254 [astro-ph.CO]}
  \BibitemShut {NoStop}%
\bibitem [{\citenamefont {Kopparapu}\ \emph {et~al.}(2008)\citenamefont
  {Kopparapu}, \citenamefont {Hanna}, \citenamefont {Kalogera}, \citenamefont
  {O'Shaughnessy}, \citenamefont {Gonz\'alez}, \citenamefont {Brady},\ and\
  \citenamefont {Fairhurst}}]{Kopparapu:2007ib}%
  \BibitemOpen
  \bibfield  {author} {\bibinfo {author} {\bibfnamefont {R.~K.}\ \bibnamefont
  {Kopparapu}}, \bibinfo {author} {\bibfnamefont {C.}~\bibnamefont {Hanna}},
  \bibinfo {author} {\bibfnamefont {V.}~\bibnamefont {Kalogera}}, \bibinfo
  {author} {\bibfnamefont {R.}~\bibnamefont {O'Shaughnessy}}, \bibinfo {author}
  {\bibfnamefont {G.}~\bibnamefont {Gonz\'alez}}, \bibinfo {author}
  {\bibfnamefont {P.~R.}\ \bibnamefont {Brady}}, \ and\ \bibinfo {author}
  {\bibfnamefont {S.}~\bibnamefont {Fairhurst}},\ }\href {\doibase
  10.1086/527348} {\bibfield  {journal} {\bibinfo  {journal} {Astrophys. J.}\
  }\textbf {\bibinfo {volume} {675}},\ \bibinfo {pages} {1459} (\bibinfo {year}
  {2008})},\ \Eprint {http://arxiv.org/abs/0706.1283} {arXiv:0706.1283
  [astro-ph]} \BibitemShut {NoStop}%
\bibitem [{\citenamefont {Pyne}\ and\ \citenamefont
  {Birkinshaw}(2004)}]{Pyne:2003bn}%
  \BibitemOpen
  \bibfield  {author} {\bibinfo {author} {\bibfnamefont {T.}~\bibnamefont
  {Pyne}}\ and\ \bibinfo {author} {\bibfnamefont {M.}~\bibnamefont
  {Birkinshaw}},\ }\href {\doibase 10.1111/j.1365-2966.2004.07362.x} {\bibfield
   {journal} {\bibinfo  {journal} {Mon. Not. Roy. Astron. Soc.}\ }\textbf
  {\bibinfo {volume} {348}},\ \bibinfo {pages} {581} (\bibinfo {year}
  {2004})},\ \Eprint {http://arxiv.org/abs/astro-ph/0310841}
  {arXiv:astro-ph/0310841} \BibitemShut {NoStop}%
\bibitem [{\citenamefont {Bonvin}\ \emph {et~al.}(2006)\citenamefont {Bonvin},
  \citenamefont {Durrer},\ and\ \citenamefont {Gasparini}}]{Bonvin:2005ps}%
  \BibitemOpen
  \bibfield  {author} {\bibinfo {author} {\bibfnamefont {C.}~\bibnamefont
  {Bonvin}}, \bibinfo {author} {\bibfnamefont {R.}~\bibnamefont {Durrer}}, \
  and\ \bibinfo {author} {\bibfnamefont {M.~A.}\ \bibnamefont {Gasparini}},\
  }\href {\doibase 10.1103/PhysRevD.85.029901} {\bibfield  {journal} {\bibinfo
  {journal} {Phys. Rev. D}\ }\textbf {\bibinfo {volume} {73}},\ \bibinfo
  {pages} {023523} (\bibinfo {year} {2006})},\ \bibinfo {note} {[Erratum:
  Phys.Rev.D 85, 029901 (2012)]},\ \Eprint
  {http://arxiv.org/abs/astro-ph/0511183} {arXiv:astro-ph/0511183} \BibitemShut
  {NoStop}%
\bibitem [{\citenamefont {Smith}\ \emph {et~al.}(2018)\citenamefont {Smith},
  \citenamefont {Jauzac}, \citenamefont {Veitch}, \citenamefont {Farr},
  \citenamefont {Massey},\ and\ \citenamefont {Richard}}]{Smith:2017mqu}%
  \BibitemOpen
  \bibfield  {author} {\bibinfo {author} {\bibfnamefont {G.~P.}\ \bibnamefont
  {Smith}}, \bibinfo {author} {\bibfnamefont {M.}~\bibnamefont {Jauzac}},
  \bibinfo {author} {\bibfnamefont {J.}~\bibnamefont {Veitch}}, \bibinfo
  {author} {\bibfnamefont {W.~M.}\ \bibnamefont {Farr}}, \bibinfo {author}
  {\bibfnamefont {R.}~\bibnamefont {Massey}}, \ and\ \bibinfo {author}
  {\bibfnamefont {J.}~\bibnamefont {Richard}},\ }\href {\doibase
  10.1093/mnras/sty031} {\bibfield  {journal} {\bibinfo  {journal} {Mon. Not.
  Roy. Astron. Soc.}\ }\textbf {\bibinfo {volume} {475}},\ \bibinfo {pages}
  {3823} (\bibinfo {year} {2018})},\ \Eprint {http://arxiv.org/abs/1707.03412}
  {arXiv:1707.03412 [astro-ph.HE]} \BibitemShut {NoStop}%
\bibitem [{\citenamefont {Holz}\ and\ \citenamefont
  {Wald}(1998)}]{Holz:1997ic}%
  \BibitemOpen
  \bibfield  {author} {\bibinfo {author} {\bibfnamefont {D.~E.}\ \bibnamefont
  {Holz}}\ and\ \bibinfo {author} {\bibfnamefont {R.~M.}\ \bibnamefont
  {Wald}},\ }\href {\doibase 10.1103/PhysRevD.58.063501} {\bibfield  {journal}
  {\bibinfo  {journal} {Phys. Rev. D}\ }\textbf {\bibinfo {volume} {58}},\
  \bibinfo {pages} {063501} (\bibinfo {year} {1998})},\ \Eprint
  {http://arxiv.org/abs/astro-ph/9708036} {arXiv:astro-ph/9708036} \BibitemShut
  {NoStop}%
\bibitem [{\citenamefont {Shapiro}\ \emph {et~al.}(2010)\citenamefont
  {Shapiro}, \citenamefont {Bacon}, \citenamefont {Hendry},\ and\ \citenamefont
  {Hoyle}}]{Shapiro:2009sr}%
  \BibitemOpen
  \bibfield  {author} {\bibinfo {author} {\bibfnamefont {C.}~\bibnamefont
  {Shapiro}}, \bibinfo {author} {\bibfnamefont {D.}~\bibnamefont {Bacon}},
  \bibinfo {author} {\bibfnamefont {M.}~\bibnamefont {Hendry}}, \ and\ \bibinfo
  {author} {\bibfnamefont {B.}~\bibnamefont {Hoyle}},\ }\href {\doibase
  10.1111/j.1365-2966.2010.16317.x} {\bibfield  {journal} {\bibinfo  {journal}
  {Mon. Not. Roy. Astron. Soc.}\ }\textbf {\bibinfo {volume} {404}},\ \bibinfo
  {pages} {858} (\bibinfo {year} {2010})},\ \Eprint
  {http://arxiv.org/abs/0907.3635} {arXiv:0907.3635 [astro-ph.CO]} \BibitemShut
  {NoStop}%
\bibitem [{\citenamefont {Hirata}\ \emph {et~al.}(2010)\citenamefont {Hirata},
  \citenamefont {Holz},\ and\ \citenamefont {Cutler}}]{Hirata:2010ba}%
  \BibitemOpen
  \bibfield  {author} {\bibinfo {author} {\bibfnamefont {C.~M.}\ \bibnamefont
  {Hirata}}, \bibinfo {author} {\bibfnamefont {D.~E.}\ \bibnamefont {Holz}}, \
  and\ \bibinfo {author} {\bibfnamefont {C.}~\bibnamefont {Cutler}},\ }\href
  {\doibase 10.1103/PhysRevD.81.124046} {\bibfield  {journal} {\bibinfo
  {journal} {Phys. Rev. D}\ }\textbf {\bibinfo {volume} {81}},\ \bibinfo
  {pages} {124046} (\bibinfo {year} {2010})},\ \Eprint
  {http://arxiv.org/abs/1004.3988} {arXiv:1004.3988 [astro-ph.CO]} \BibitemShut
  {NoStop}%
\bibitem [{\citenamefont {Cusin}\ and\ \citenamefont
  {Tamanini}(2021)}]{Cusin:2020ezb}%
  \BibitemOpen
  \bibfield  {author} {\bibinfo {author} {\bibfnamefont {G.}~\bibnamefont
  {Cusin}}\ and\ \bibinfo {author} {\bibfnamefont {N.}~\bibnamefont
  {Tamanini}},\ }\href {\doibase 10.1093/mnras/stab1130} {\bibfield  {journal}
  {\bibinfo  {journal} {Mon. Not. Roy. Astron. Soc.}\ }\textbf {\bibinfo
  {volume} {504}},\ \bibinfo {pages} {3610} (\bibinfo {year} {2021})},\ \Eprint
  {http://arxiv.org/abs/2011.15109} {arXiv:2011.15109 [astro-ph.CO]}
  \BibitemShut {NoStop}%
\bibitem [{\citenamefont {Hilbert}\ \emph {et~al.}(2011)\citenamefont
  {Hilbert}, \citenamefont {Gair},\ and\ \citenamefont
  {King}}]{Hilbert:2010am}%
  \BibitemOpen
  \bibfield  {author} {\bibinfo {author} {\bibfnamefont {S.}~\bibnamefont
  {Hilbert}}, \bibinfo {author} {\bibfnamefont {J.~R.}\ \bibnamefont {Gair}}, \
  and\ \bibinfo {author} {\bibfnamefont {L.~J.}\ \bibnamefont {King}},\ }\href
  {\doibase 10.1111/j.1365-2966.2010.17963.x} {\bibfield  {journal} {\bibinfo
  {journal} {Mon. Not. Roy. Astron. Soc.}\ }\textbf {\bibinfo {volume} {412}},\
  \bibinfo {pages} {1023} (\bibinfo {year} {2011})},\ \Eprint
  {http://arxiv.org/abs/1007.2468} {arXiv:1007.2468 [astro-ph.CO]} \BibitemShut
  {NoStop}%
\bibitem [{\citenamefont {Wu}\ \emph {et~al.}(2023{\natexlab{a}})\citenamefont
  {Wu}, \citenamefont {Chan}, \citenamefont {Hendry},\ and\ \citenamefont
  {Hannuksela}}]{Wu:2022vrq}%
  \BibitemOpen
  \bibfield  {author} {\bibinfo {author} {\bibfnamefont {Z.-F.}\ \bibnamefont
  {Wu}}, \bibinfo {author} {\bibfnamefont {L.~W.~L.}\ \bibnamefont {Chan}},
  \bibinfo {author} {\bibfnamefont {M.}~\bibnamefont {Hendry}}, \ and\ \bibinfo
  {author} {\bibfnamefont {O.~A.}\ \bibnamefont {Hannuksela}},\ }\href
  {\doibase 10.1093/mnras/stad1194} {\bibfield  {journal} {\bibinfo  {journal}
  {Mon. Not. Roy. Astron. Soc.}\ }\textbf {\bibinfo {volume} {522}},\ \bibinfo
  {pages} {4059} (\bibinfo {year} {2023}{\natexlab{a}})},\ \Eprint
  {http://arxiv.org/abs/2211.15160} {arXiv:2211.15160 [astro-ph.CO]}
  \BibitemShut {NoStop}%
\bibitem [{\citenamefont {Chen}\ \emph
  {et~al.}(2019{\natexlab{a}})\citenamefont {Chen}, \citenamefont {Li},\ and\
  \citenamefont {Cao}}]{Chen:2017xbi}%
  \BibitemOpen
  \bibfield  {author} {\bibinfo {author} {\bibfnamefont {X.}~\bibnamefont
  {Chen}}, \bibinfo {author} {\bibfnamefont {S.}~\bibnamefont {Li}}, \ and\
  \bibinfo {author} {\bibfnamefont {Z.}~\bibnamefont {Cao}},\ }\href {\doibase
  10.1093/mnrasl/slz046} {\bibfield  {journal} {\bibinfo  {journal} {Mon. Not.
  Roy. Astron. Soc.}\ }\textbf {\bibinfo {volume} {485}},\ \bibinfo {pages}
  {L141} (\bibinfo {year} {2019}{\natexlab{a}})},\ \Eprint
  {http://arxiv.org/abs/1703.10543} {arXiv:1703.10543 [astro-ph.HE]}
  \BibitemShut {NoStop}%
\bibitem [{\citenamefont {Chen}(2021)}]{Chen:2020iky}%
  \BibitemOpen
  \bibfield  {author} {\bibinfo {author} {\bibfnamefont {X.}~\bibnamefont
  {Chen}},\ }\enquote {\bibinfo {title} {{Distortion of Gravitational-Wave
  Signals by Astrophysical Environments}},}\ \ (\bibinfo {year} {2021})\
  \Eprint {http://arxiv.org/abs/2009.07626} {arXiv:2009.07626 [astro-ph.HE]}
  \BibitemShut {NoStop}%
\bibitem [{\citenamefont {Chen}\ \emph
  {et~al.}(2020{\natexlab{b}})\citenamefont {Chen}, \citenamefont {Xuan},\ and\
  \citenamefont {Peng}}]{Chen:2020lpq}%
  \BibitemOpen
  \bibfield  {author} {\bibinfo {author} {\bibfnamefont {X.}~\bibnamefont
  {Chen}}, \bibinfo {author} {\bibfnamefont {Z.-Y.}\ \bibnamefont {Xuan}}, \
  and\ \bibinfo {author} {\bibfnamefont {P.}~\bibnamefont {Peng}},\ }\href
  {\doibase 10.3847/1538-4357/ab919f} {\bibfield  {journal} {\bibinfo
  {journal} {Astrophys. J.}\ }\textbf {\bibinfo {volume} {896}},\ \bibinfo
  {pages} {171} (\bibinfo {year} {2020}{\natexlab{b}})},\ \Eprint
  {http://arxiv.org/abs/2003.08639} {arXiv:2003.08639 [astro-ph.HE]}
  \BibitemShut {NoStop}%
\bibitem [{\citenamefont {Karydas}\ \emph {et~al.}(2024)\citenamefont
  {Karydas}, \citenamefont {Kavanagh},\ and\ \citenamefont
  {Bertone}}]{Karydas:2024fcn}%
  \BibitemOpen
  \bibfield  {author} {\bibinfo {author} {\bibfnamefont {T.~K.}\ \bibnamefont
  {Karydas}}, \bibinfo {author} {\bibfnamefont {B.~J.}\ \bibnamefont
  {Kavanagh}}, \ and\ \bibinfo {author} {\bibfnamefont {G.}~\bibnamefont
  {Bertone}},\ }\href@noop {} {\  (\bibinfo {year} {2024})},\ \Eprint
  {http://arxiv.org/abs/2402.13053} {arXiv:2402.13053 [gr-qc]} \BibitemShut
  {NoStop}%
\bibitem [{\citenamefont {Kavanagh}\ \emph {et~al.}(2024)\citenamefont
  {Kavanagh}, \citenamefont {Karydas}, \citenamefont {Bertone}, \citenamefont
  {Di~Cintio},\ and\ \citenamefont {Pasquato}}]{Kavanagh:2024lgq}%
  \BibitemOpen
  \bibfield  {author} {\bibinfo {author} {\bibfnamefont {B.~J.}\ \bibnamefont
  {Kavanagh}}, \bibinfo {author} {\bibfnamefont {T.~K.}\ \bibnamefont
  {Karydas}}, \bibinfo {author} {\bibfnamefont {G.}~\bibnamefont {Bertone}},
  \bibinfo {author} {\bibfnamefont {P.}~\bibnamefont {Di~Cintio}}, \ and\
  \bibinfo {author} {\bibfnamefont {M.}~\bibnamefont {Pasquato}},\ }\href@noop
  {} {\  (\bibinfo {year} {2024})},\ \Eprint {http://arxiv.org/abs/2402.13762}
  {arXiv:2402.13762 [gr-qc]} \BibitemShut {NoStop}%
\bibitem [{\citenamefont {Jan}\ \emph {et~al.}(2024)\citenamefont {Jan},
  \citenamefont {Ferguson}, \citenamefont {Lange}, \citenamefont {Shoemaker},\
  and\ \citenamefont {Zimmerman}}]{Jan:2023raq}%
  \BibitemOpen
  \bibfield  {author} {\bibinfo {author} {\bibfnamefont {A.}~\bibnamefont
  {Jan}}, \bibinfo {author} {\bibfnamefont {D.}~\bibnamefont {Ferguson}},
  \bibinfo {author} {\bibfnamefont {J.}~\bibnamefont {Lange}}, \bibinfo
  {author} {\bibfnamefont {D.}~\bibnamefont {Shoemaker}}, \ and\ \bibinfo
  {author} {\bibfnamefont {A.}~\bibnamefont {Zimmerman}},\ }\href {\doibase
  10.1103/PhysRevD.110.024023} {\bibfield  {journal} {\bibinfo  {journal}
  {Phys. Rev. D}\ }\textbf {\bibinfo {volume} {110}},\ \bibinfo {pages}
  {024023} (\bibinfo {year} {2024})},\ \Eprint
  {http://arxiv.org/abs/2312.10241} {arXiv:2312.10241 [gr-qc]} \BibitemShut
  {NoStop}%
\bibitem [{\citenamefont {Edy}\ \emph {et~al.}(2021)\citenamefont {Edy},
  \citenamefont {Lundgren},\ and\ \citenamefont {Nuttall}}]{Edy:2021par}%
  \BibitemOpen
  \bibfield  {author} {\bibinfo {author} {\bibfnamefont {O.}~\bibnamefont
  {Edy}}, \bibinfo {author} {\bibfnamefont {A.}~\bibnamefont {Lundgren}}, \
  and\ \bibinfo {author} {\bibfnamefont {L.~K.}\ \bibnamefont {Nuttall}},\
  }\href {\doibase 10.1103/PhysRevD.103.124061} {\bibfield  {journal} {\bibinfo
   {journal} {Phys. Rev. D}\ }\textbf {\bibinfo {volume} {103}},\ \bibinfo
  {pages} {124061} (\bibinfo {year} {2021})},\ \Eprint
  {http://arxiv.org/abs/2101.07743} {arXiv:2101.07743 [astro-ph.IM]}
  \BibitemShut {NoStop}%
\bibitem [{\citenamefont {Kumar}\ \emph {et~al.}(2022)\citenamefont {Kumar},
  \citenamefont {Nitz},\ and\ \citenamefont {Forteza}}]{Kumar:2022tto}%
  \BibitemOpen
  \bibfield  {author} {\bibinfo {author} {\bibfnamefont {S.}~\bibnamefont
  {Kumar}}, \bibinfo {author} {\bibfnamefont {A.~H.}\ \bibnamefont {Nitz}}, \
  and\ \bibinfo {author} {\bibfnamefont {X.~J.}\ \bibnamefont {Forteza}},\
  }\href@noop {} {\  (\bibinfo {year} {2022})},\ \Eprint
  {http://arxiv.org/abs/2202.12762} {arXiv:2202.12762 [astro-ph.IM]}
  \BibitemShut {NoStop}%
\bibitem [{\citenamefont {Steltner}\ \emph {et~al.}(2022)\citenamefont
  {Steltner}, \citenamefont {Papa},\ and\ \citenamefont
  {Eggenstein}}]{Steltner:2021qjy}%
  \BibitemOpen
  \bibfield  {author} {\bibinfo {author} {\bibfnamefont {B.}~\bibnamefont
  {Steltner}}, \bibinfo {author} {\bibfnamefont {M.~A.}\ \bibnamefont {Papa}},
  \ and\ \bibinfo {author} {\bibfnamefont {H.-B.}\ \bibnamefont {Eggenstein}},\
  }\href {\doibase 10.1103/PhysRevD.105.022005} {\bibfield  {journal} {\bibinfo
   {journal} {Phys. Rev. D}\ }\textbf {\bibinfo {volume} {105}},\ \bibinfo
  {pages} {022005} (\bibinfo {year} {2022})},\ \Eprint
  {http://arxiv.org/abs/2105.09933} {arXiv:2105.09933 [gr-qc]} \BibitemShut
  {NoStop}%
\bibitem [{\citenamefont {Hjorth}\ \emph {et~al.}(2017)\citenamefont {Hjorth},
  \citenamefont {Levan}, \citenamefont {Tanvir}, \citenamefont {Lyman},
  \citenamefont {Wojtak}, \citenamefont {Schr\o{}der}, \citenamefont {Mandel},
  \citenamefont {Gall},\ and\ \citenamefont {Bruun}}]{Hjorth:2017yza}%
  \BibitemOpen
  \bibfield  {author} {\bibinfo {author} {\bibfnamefont {J.}~\bibnamefont
  {Hjorth}}, \bibinfo {author} {\bibfnamefont {A.~J.}\ \bibnamefont {Levan}},
  \bibinfo {author} {\bibfnamefont {N.~R.}\ \bibnamefont {Tanvir}}, \bibinfo
  {author} {\bibfnamefont {J.~D.}\ \bibnamefont {Lyman}}, \bibinfo {author}
  {\bibfnamefont {R.}~\bibnamefont {Wojtak}}, \bibinfo {author} {\bibfnamefont
  {S.~L.}\ \bibnamefont {Schr\o{}der}}, \bibinfo {author} {\bibfnamefont
  {I.}~\bibnamefont {Mandel}}, \bibinfo {author} {\bibfnamefont
  {C.}~\bibnamefont {Gall}}, \ and\ \bibinfo {author} {\bibfnamefont {S.~H.}\
  \bibnamefont {Bruun}},\ }\href {\doibase 10.3847/2041-8213/aa9110} {\bibfield
   {journal} {\bibinfo  {journal} {Astrophys. J. Lett.}\ }\textbf {\bibinfo
  {volume} {848}},\ \bibinfo {pages} {L31} (\bibinfo {year} {2017})},\ \Eprint
  {http://arxiv.org/abs/1710.05856} {arXiv:1710.05856 [astro-ph.GA]}
  \BibitemShut {NoStop}%
\bibitem [{\citenamefont {Burns}\ \emph {et~al.}(2019)\citenamefont {Burns}
  \emph {et~al.}}]{FermiGamma-rayBurstMonitorTeam:2018eao}%
  \BibitemOpen
  \bibfield  {author} {\bibinfo {author} {\bibfnamefont {E.}~\bibnamefont
  {Burns}} \emph {et~al.} (\bibinfo {collaboration} {Fermi Gamma-ray Burst
  Monitor Team, LIGO Scientific, Virgo}),\ }\href {\doibase
  10.3847/1538-4357/aaf726} {\bibfield  {journal} {\bibinfo  {journal}
  {Astrophys. J.}\ }\textbf {\bibinfo {volume} {871}},\ \bibinfo {pages} {90}
  (\bibinfo {year} {2019})},\ \Eprint {http://arxiv.org/abs/1810.02764}
  {arXiv:1810.02764 [astro-ph.HE]} \BibitemShut {NoStop}%
\bibitem [{\citenamefont {McGee}\ \emph {et~al.}(2020)\citenamefont {McGee},
  \citenamefont {Sesana},\ and\ \citenamefont {Vecchio}}]{McGee:2018qwb}%
  \BibitemOpen
  \bibfield  {author} {\bibinfo {author} {\bibfnamefont {S.}~\bibnamefont
  {McGee}}, \bibinfo {author} {\bibfnamefont {A.}~\bibnamefont {Sesana}}, \
  and\ \bibinfo {author} {\bibfnamefont {A.}~\bibnamefont {Vecchio}},\ }\href
  {\doibase 10.1038/s41550-019-0969-7} {\bibfield  {journal} {\bibinfo
  {journal} {Nature Astron.}\ }\textbf {\bibinfo {volume} {4}},\ \bibinfo
  {pages} {26} (\bibinfo {year} {2020})},\ \Eprint
  {http://arxiv.org/abs/1811.00050} {arXiv:1811.00050 [astro-ph.HE]}
  \BibitemShut {NoStop}%
\bibitem [{\citenamefont {De~Rosa}\ \emph {et~al.}(2019)\citenamefont {De~Rosa}
  \emph {et~al.}}]{DeRosa:2019myq}%
  \BibitemOpen
  \bibfield  {author} {\bibinfo {author} {\bibfnamefont {A.}~\bibnamefont
  {De~Rosa}} \emph {et~al.},\ }\href {\doibase 10.1016/j.newar.2020.101525}
  {\bibfield  {journal} {\bibinfo  {journal} {New Astron. Rev.}\ }\textbf
  {\bibinfo {volume} {86}},\ \bibinfo {pages} {101525} (\bibinfo {year}
  {2019})},\ \Eprint {http://arxiv.org/abs/2001.06293} {arXiv:2001.06293
  [astro-ph.GA]} \BibitemShut {NoStop}%
\bibitem [{\citenamefont {Bogdanovic}\ \emph {et~al.}(2022)\citenamefont
  {Bogdanovic}, \citenamefont {Miller},\ and\ \citenamefont
  {Blecha}}]{Bogdanovic:2021aav}%
  \BibitemOpen
  \bibfield  {author} {\bibinfo {author} {\bibfnamefont {T.}~\bibnamefont
  {Bogdanovic}}, \bibinfo {author} {\bibfnamefont {M.~C.}\ \bibnamefont
  {Miller}}, \ and\ \bibinfo {author} {\bibfnamefont {L.}~\bibnamefont
  {Blecha}},\ }\href {\doibase 10.1007/s41114-022-00037-8} {\bibfield
  {journal} {\bibinfo  {journal} {Living Rev. Rel.}\ }\textbf {\bibinfo
  {volume} {25}},\ \bibinfo {pages} {3} (\bibinfo {year} {2022})},\ \Eprint
  {http://arxiv.org/abs/2109.03262} {arXiv:2109.03262 [astro-ph.HE]}
  \BibitemShut {NoStop}%
\bibitem [{\citenamefont {Zhu}\ \emph {et~al.}(2024{\natexlab{b}})\citenamefont
  {Zhu}, \citenamefont {Fan}, \citenamefont {Chen}, \citenamefont {Hu},\ and\
  \citenamefont {Zhang}}]{Zhu:2024qpp}%
  \BibitemOpen
  \bibfield  {author} {\bibinfo {author} {\bibfnamefont {L.-G.}\ \bibnamefont
  {Zhu}}, \bibinfo {author} {\bibfnamefont {H.-M.}\ \bibnamefont {Fan}},
  \bibinfo {author} {\bibfnamefont {X.}~\bibnamefont {Chen}}, \bibinfo {author}
  {\bibfnamefont {Y.-M.}\ \bibnamefont {Hu}}, \ and\ \bibinfo {author}
  {\bibfnamefont {J.-d.}\ \bibnamefont {Zhang}},\ }\href {\doibase
  10.3847/1538-4365/ad5446} {\bibfield  {journal} {\bibinfo  {journal}
  {Astrophys. J. Suppl.}\ }\textbf {\bibinfo {volume} {273}},\ \bibinfo {pages}
  {24} (\bibinfo {year} {2024}{\natexlab{b}})},\ \Eprint
  {http://arxiv.org/abs/2403.04950} {arXiv:2403.04950 [astro-ph.CO]}
  \BibitemShut {NoStop}%
\bibitem [{\citenamefont {Wang}\ \emph
  {et~al.}(2022{\natexlab{d}})\citenamefont {Wang}, \citenamefont {Ruan},
  \citenamefont {Yang}, \citenamefont {Guo}, \citenamefont {Cai},\ and\
  \citenamefont {Hu}}]{Wang:2020dkc}%
  \BibitemOpen
  \bibfield  {author} {\bibinfo {author} {\bibfnamefont {R.}~\bibnamefont
  {Wang}}, \bibinfo {author} {\bibfnamefont {W.-H.}\ \bibnamefont {Ruan}},
  \bibinfo {author} {\bibfnamefont {Q.}~\bibnamefont {Yang}}, \bibinfo {author}
  {\bibfnamefont {Z.-K.}\ \bibnamefont {Guo}}, \bibinfo {author} {\bibfnamefont
  {R.-G.}\ \bibnamefont {Cai}}, \ and\ \bibinfo {author} {\bibfnamefont
  {B.}~\bibnamefont {Hu}},\ }\href {\doibase 10.1093/nsr/nwab054} {\bibfield
  {journal} {\bibinfo  {journal} {Natl. Sci. Rev.}\ }\textbf {\bibinfo {volume}
  {9}},\ \bibinfo {pages} {nwab054} (\bibinfo {year} {2022}{\natexlab{d}})},\
  \Eprint {http://arxiv.org/abs/2010.14732} {arXiv:2010.14732 [astro-ph.CO]}
  \BibitemShut {NoStop}%
\bibitem [{\citenamefont {York}\ \emph {et~al.}(2000)\citenamefont {York} \emph
  {et~al.}}]{SDSS:2000hjo}%
  \BibitemOpen
  \bibfield  {author} {\bibinfo {author} {\bibfnamefont {D.~G.}\ \bibnamefont
  {York}} \emph {et~al.} (\bibinfo {collaboration} {SDSS}),\ }\href {\doibase
  10.1086/301513} {\bibfield  {journal} {\bibinfo  {journal} {Astron. J.}\
  }\textbf {\bibinfo {volume} {120}},\ \bibinfo {pages} {1579} (\bibinfo {year}
  {2000})},\ \Eprint {http://arxiv.org/abs/astro-ph/0006396}
  {arXiv:astro-ph/0006396} \BibitemShut {NoStop}%
\bibitem [{\citenamefont {Almeida}\ \emph {et~al.}(2023)\citenamefont {Almeida}
  \emph {et~al.}}]{SDSS:2023tbz}%
  \BibitemOpen
  \bibfield  {author} {\bibinfo {author} {\bibfnamefont {A.}~\bibnamefont
  {Almeida}} \emph {et~al.} (\bibinfo {collaboration} {SDSS}),\ }\href
  {\doibase 10.3847/1538-4365/acda98} {\bibfield  {journal} {\bibinfo
  {journal} {Astrophys. J. Suppl.}\ }\textbf {\bibinfo {volume} {267}},\
  \bibinfo {pages} {44} (\bibinfo {year} {2023})},\ \Eprint
  {http://arxiv.org/abs/2301.07688} {arXiv:2301.07688 [astro-ph.GA]}
  \BibitemShut {NoStop}%
\bibitem [{\citenamefont {Abbott}\ \emph {et~al.}(2005)\citenamefont {Abbott}
  \emph {et~al.}}]{DES:2005dhi}%
  \BibitemOpen
  \bibfield  {author} {\bibinfo {author} {\bibfnamefont {T.}~\bibnamefont
  {Abbott}} \emph {et~al.} (\bibinfo {collaboration} {DES}),\ }\href@noop {} {\
   (\bibinfo {year} {2005})},\ \Eprint {http://arxiv.org/abs/astro-ph/0510346}
  {arXiv:astro-ph/0510346} \BibitemShut {NoStop}%
\bibitem [{\citenamefont {Abbott}\ \emph
  {et~al.}(2021{\natexlab{g}})\citenamefont {Abbott} \emph
  {et~al.}}]{LineaScienceServer:2021mgv}%
  \BibitemOpen
  \bibfield  {author} {\bibinfo {author} {\bibfnamefont {T.~M.~C.}\
  \bibnamefont {Abbott}} \emph {et~al.} (\bibinfo {collaboration} {Linea
  Science Server, DES}),\ }\href {\doibase 10.3847/1538-4365/ac00b3} {\bibfield
   {journal} {\bibinfo  {journal} {Astrophys. J. Supp.}\ }\textbf {\bibinfo
  {volume} {255}},\ \bibinfo {pages} {20} (\bibinfo {year}
  {2021}{\natexlab{g}})},\ \Eprint {http://arxiv.org/abs/2101.05765}
  {arXiv:2101.05765 [astro-ph.IM]} \BibitemShut {NoStop}%
\bibitem [{\citenamefont {{Cui}}\ \emph {et~al.}(2012)\citenamefont {{Cui}},
  \citenamefont {{Zhao}}, \citenamefont {{Chu}}, \citenamefont {{Li}},
  \citenamefont {{Li}}, \citenamefont {{Zhang}}, \citenamefont {{Su}},
  \citenamefont {{Yao}}, \citenamefont {{Wang}}, \citenamefont {{Xing}},
  \citenamefont {{Li}}, \citenamefont {{Zhu}}, \citenamefont {{Wang}},
  \citenamefont {{Gu}}, \citenamefont {{Luo}}, \citenamefont {{Xu}},
  \citenamefont {{Zhang}}, \citenamefont {{Liu}}, \citenamefont {{Zhang}},
  \citenamefont {{Yang}}, \citenamefont {{Cao}}, \citenamefont {{Chen}},
  \citenamefont {{Chen}}, \citenamefont {{Chen}}, \citenamefont {{Chen}},
  \citenamefont {{Chu}}, \citenamefont {{Feng}}, \citenamefont {{Gong}},
  \citenamefont {{Hou}}, \citenamefont {{Hu}}, \citenamefont {{Hu}},
  \citenamefont {{Hu}}, \citenamefont {{Jia}}, \citenamefont {{Jiang}},
  \citenamefont {{Jiang}}, \citenamefont {{Jiang}}, \citenamefont {{Jin}},
  \citenamefont {{Li}}, \citenamefont {{Li}}, \citenamefont {{Li}},
  \citenamefont {{Liu}}, \citenamefont {{Liu}}, \citenamefont {{Lu}},
  \citenamefont {{Mao}}, \citenamefont {{Men}}, \citenamefont {{Qi}},
  \citenamefont {{Qi}}, \citenamefont {{Shi}}, \citenamefont {{Tang}},
  \citenamefont {{Tao}}, \citenamefont {{Wang}}, \citenamefont {{Wang}},
  \citenamefont {{Wang}}, \citenamefont {{Wang}}, \citenamefont {{Wang}},
  \citenamefont {{Wang}}, \citenamefont {{Wang}}, \citenamefont {{Wang}},
  \citenamefont {{Wang}}, \citenamefont {{Wang}}, \citenamefont {{Wang}},
  \citenamefont {{Wang}}, \citenamefont {{Xu}}, \citenamefont {{Xu}},
  \citenamefont {{Yang}}, \citenamefont {{Yu}}, \citenamefont {{Yuan}},
  \citenamefont {{Yuan}}, \citenamefont {{Zhai}}, \citenamefont {{Zhang}},
  \citenamefont {{Zhang}}, \citenamefont {{Zhang}}, \citenamefont {{Zhao}},
  \citenamefont {{Zhou}}, \citenamefont {{Zhou}}, \citenamefont {{Zhu}},\ and\
  \citenamefont {{Zou}}}]{2012RAA....12.1197C}%
  \BibitemOpen
  \bibfield  {author} {\bibinfo {author} {\bibfnamefont {X.-Q.}\ \bibnamefont
  {{Cui}}}, \bibinfo {author} {\bibfnamefont {Y.-H.}\ \bibnamefont {{Zhao}}},
  \bibinfo {author} {\bibfnamefont {Y.-Q.}\ \bibnamefont {{Chu}}}, \bibinfo
  {author} {\bibfnamefont {G.-P.}\ \bibnamefont {{Li}}}, \bibinfo {author}
  {\bibfnamefont {Q.}~\bibnamefont {{Li}}}, \bibinfo {author} {\bibfnamefont
  {L.-P.}\ \bibnamefont {{Zhang}}}, \bibinfo {author} {\bibfnamefont {H.-J.}\
  \bibnamefont {{Su}}}, \bibinfo {author} {\bibfnamefont {Z.-Q.}\ \bibnamefont
  {{Yao}}}, \bibinfo {author} {\bibfnamefont {Y.-N.}\ \bibnamefont {{Wang}}},
  \bibinfo {author} {\bibfnamefont {X.-Z.}\ \bibnamefont {{Xing}}}, \bibinfo
  {author} {\bibfnamefont {X.-N.}\ \bibnamefont {{Li}}}, \bibinfo {author}
  {\bibfnamefont {Y.-T.}\ \bibnamefont {{Zhu}}}, \bibinfo {author}
  {\bibfnamefont {G.}~\bibnamefont {{Wang}}}, \bibinfo {author} {\bibfnamefont
  {B.-Z.}\ \bibnamefont {{Gu}}}, \bibinfo {author} {\bibfnamefont {A.~L.}\
  \bibnamefont {{Luo}}}, \bibinfo {author} {\bibfnamefont {X.-Q.}\ \bibnamefont
  {{Xu}}}, \bibinfo {author} {\bibfnamefont {Z.-C.}\ \bibnamefont {{Zhang}}},
  \bibinfo {author} {\bibfnamefont {G.-R.}\ \bibnamefont {{Liu}}}, \bibinfo
  {author} {\bibfnamefont {H.-T.}\ \bibnamefont {{Zhang}}}, \bibinfo {author}
  {\bibfnamefont {D.-H.}\ \bibnamefont {{Yang}}}, \bibinfo {author}
  {\bibfnamefont {S.-Y.}\ \bibnamefont {{Cao}}}, \bibinfo {author}
  {\bibfnamefont {H.-Y.}\ \bibnamefont {{Chen}}}, \bibinfo {author}
  {\bibfnamefont {J.-J.}\ \bibnamefont {{Chen}}}, \bibinfo {author}
  {\bibfnamefont {K.-X.}\ \bibnamefont {{Chen}}}, \bibinfo {author}
  {\bibfnamefont {Y.}~\bibnamefont {{Chen}}}, \bibinfo {author} {\bibfnamefont
  {J.-R.}\ \bibnamefont {{Chu}}}, \bibinfo {author} {\bibfnamefont
  {L.}~\bibnamefont {{Feng}}}, \bibinfo {author} {\bibfnamefont {X.-F.}\
  \bibnamefont {{Gong}}}, \bibinfo {author} {\bibfnamefont {Y.-H.}\
  \bibnamefont {{Hou}}}, \bibinfo {author} {\bibfnamefont {H.-Z.}\ \bibnamefont
  {{Hu}}}, \bibinfo {author} {\bibfnamefont {N.-S.}\ \bibnamefont {{Hu}}},
  \bibinfo {author} {\bibfnamefont {Z.-W.}\ \bibnamefont {{Hu}}}, \bibinfo
  {author} {\bibfnamefont {L.}~\bibnamefont {{Jia}}}, \bibinfo {author}
  {\bibfnamefont {F.-H.}\ \bibnamefont {{Jiang}}}, \bibinfo {author}
  {\bibfnamefont {X.}~\bibnamefont {{Jiang}}}, \bibinfo {author} {\bibfnamefont
  {Z.-B.}\ \bibnamefont {{Jiang}}}, \bibinfo {author} {\bibfnamefont
  {G.}~\bibnamefont {{Jin}}}, \bibinfo {author} {\bibfnamefont {A.-H.}\
  \bibnamefont {{Li}}}, \bibinfo {author} {\bibfnamefont {Y.}~\bibnamefont
  {{Li}}}, \bibinfo {author} {\bibfnamefont {Y.-P.}\ \bibnamefont {{Li}}},
  \bibinfo {author} {\bibfnamefont {G.-Q.}\ \bibnamefont {{Liu}}}, \bibinfo
  {author} {\bibfnamefont {Z.-G.}\ \bibnamefont {{Liu}}}, \bibinfo {author}
  {\bibfnamefont {W.-Z.}\ \bibnamefont {{Lu}}}, \bibinfo {author}
  {\bibfnamefont {Y.-D.}\ \bibnamefont {{Mao}}}, \bibinfo {author}
  {\bibfnamefont {L.}~\bibnamefont {{Men}}}, \bibinfo {author} {\bibfnamefont
  {Y.-J.}\ \bibnamefont {{Qi}}}, \bibinfo {author} {\bibfnamefont {Z.-X.}\
  \bibnamefont {{Qi}}}, \bibinfo {author} {\bibfnamefont {H.-M.}\ \bibnamefont
  {{Shi}}}, \bibinfo {author} {\bibfnamefont {Z.-H.}\ \bibnamefont {{Tang}}},
  \bibinfo {author} {\bibfnamefont {Q.-S.}\ \bibnamefont {{Tao}}}, \bibinfo
  {author} {\bibfnamefont {D.-Q.}\ \bibnamefont {{Wang}}}, \bibinfo {author}
  {\bibfnamefont {D.}~\bibnamefont {{Wang}}}, \bibinfo {author} {\bibfnamefont
  {G.-M.}\ \bibnamefont {{Wang}}}, \bibinfo {author} {\bibfnamefont
  {H.}~\bibnamefont {{Wang}}}, \bibinfo {author} {\bibfnamefont {J.-N.}\
  \bibnamefont {{Wang}}}, \bibinfo {author} {\bibfnamefont {J.}~\bibnamefont
  {{Wang}}}, \bibinfo {author} {\bibfnamefont {J.-L.}\ \bibnamefont {{Wang}}},
  \bibinfo {author} {\bibfnamefont {J.-P.}\ \bibnamefont {{Wang}}}, \bibinfo
  {author} {\bibfnamefont {L.}~\bibnamefont {{Wang}}}, \bibinfo {author}
  {\bibfnamefont {S.-Q.}\ \bibnamefont {{Wang}}}, \bibinfo {author}
  {\bibfnamefont {Y.}~\bibnamefont {{Wang}}}, \bibinfo {author} {\bibfnamefont
  {Y.-F.}\ \bibnamefont {{Wang}}}, \bibinfo {author} {\bibfnamefont {L.-Z.}\
  \bibnamefont {{Xu}}}, \bibinfo {author} {\bibfnamefont {Y.}~\bibnamefont
  {{Xu}}}, \bibinfo {author} {\bibfnamefont {S.-H.}\ \bibnamefont {{Yang}}},
  \bibinfo {author} {\bibfnamefont {Y.}~\bibnamefont {{Yu}}}, \bibinfo {author}
  {\bibfnamefont {H.}~\bibnamefont {{Yuan}}}, \bibinfo {author} {\bibfnamefont
  {X.-Y.}\ \bibnamefont {{Yuan}}}, \bibinfo {author} {\bibfnamefont
  {C.}~\bibnamefont {{Zhai}}}, \bibinfo {author} {\bibfnamefont
  {J.}~\bibnamefont {{Zhang}}}, \bibinfo {author} {\bibfnamefont {Y.-X.}\
  \bibnamefont {{Zhang}}}, \bibinfo {author} {\bibfnamefont {Y.}~\bibnamefont
  {{Zhang}}}, \bibinfo {author} {\bibfnamefont {M.}~\bibnamefont {{Zhao}}},
  \bibinfo {author} {\bibfnamefont {F.}~\bibnamefont {{Zhou}}}, \bibinfo
  {author} {\bibfnamefont {G.-H.}\ \bibnamefont {{Zhou}}}, \bibinfo {author}
  {\bibfnamefont {J.}~\bibnamefont {{Zhu}}}, \ and\ \bibinfo {author}
  {\bibfnamefont {S.-C.}\ \bibnamefont {{Zou}}},\ }\href {\doibase
  10.1088/1674-4527/12/9/003} {\bibfield  {journal} {\bibinfo  {journal} {Res.
  Astron. Astrophys.}\ }\textbf {\bibinfo {volume} {12}},\ \bibinfo {pages}
  {1197} (\bibinfo {year} {2012})}\BibitemShut {NoStop}%
\bibitem [{\citenamefont {{Zhao}}\ \emph {et~al.}(2012)\citenamefont {{Zhao}},
  \citenamefont {{Zhao}}, \citenamefont {{Chu}}, \citenamefont {{Jing}},\ and\
  \citenamefont {{Deng}}}]{2012RAA....12..723Z}%
  \BibitemOpen
  \bibfield  {author} {\bibinfo {author} {\bibfnamefont {G.}~\bibnamefont
  {{Zhao}}}, \bibinfo {author} {\bibfnamefont {Y.-H.}\ \bibnamefont {{Zhao}}},
  \bibinfo {author} {\bibfnamefont {Y.-Q.}\ \bibnamefont {{Chu}}}, \bibinfo
  {author} {\bibfnamefont {Y.-P.}\ \bibnamefont {{Jing}}}, \ and\ \bibinfo
  {author} {\bibfnamefont {L.-C.}\ \bibnamefont {{Deng}}},\ }\href {\doibase
  10.1088/1674-4527/12/7/002} {\bibfield  {journal} {\bibinfo  {journal} {Res.
  Astron. Astrophys.}\ }\textbf {\bibinfo {volume} {12}},\ \bibinfo {pages}
  {723} (\bibinfo {year} {2012})}\BibitemShut {NoStop}%
\bibitem [{\citenamefont {Aghamousa}\ \emph {et~al.}(2016)\citenamefont
  {Aghamousa} \emph {et~al.}}]{DESI:2016fyo}%
  \BibitemOpen
  \bibfield  {author} {\bibinfo {author} {\bibfnamefont {A.}~\bibnamefont
  {Aghamousa}} \emph {et~al.} (\bibinfo {collaboration} {DESI}),\ }\href@noop
  {} {\  (\bibinfo {year} {2016})},\ \Eprint {http://arxiv.org/abs/1611.00036}
  {arXiv:1611.00036 [astro-ph.IM]} \BibitemShut {NoStop}%
\bibitem [{\citenamefont {Dey}\ \emph {et~al.}(2019)\citenamefont {Dey} \emph
  {et~al.}}]{DESI:2018ymu}%
  \BibitemOpen
  \bibfield  {author} {\bibinfo {author} {\bibfnamefont {A.}~\bibnamefont
  {Dey}} \emph {et~al.} (\bibinfo {collaboration} {DESI}),\ }\href {\doibase
  10.3847/1538-3881/ab089d} {\bibfield  {journal} {\bibinfo  {journal} {Astron.
  J.}\ }\textbf {\bibinfo {volume} {157}},\ \bibinfo {pages} {168} (\bibinfo
  {year} {2019})},\ \Eprint {http://arxiv.org/abs/1804.08657} {arXiv:1804.08657
  [astro-ph.IM]} \BibitemShut {NoStop}%
\bibitem [{\citenamefont {Ivezi\'c}\ \emph {et~al.}(2019)\citenamefont
  {Ivezi\'c} \emph {et~al.}}]{LSST:2008ijt}%
  \BibitemOpen
  \bibfield  {author} {\bibinfo {author} {\bibfnamefont {v.}~\bibnamefont
  {Ivezi\'c}} \emph {et~al.} (\bibinfo {collaboration} {LSST}),\ }\href
  {\doibase 10.3847/1538-4357/ab042c} {\bibfield  {journal} {\bibinfo
  {journal} {Astrophys. J.}\ }\textbf {\bibinfo {volume} {873}},\ \bibinfo
  {pages} {111} (\bibinfo {year} {2019})},\ \Eprint
  {http://arxiv.org/abs/0805.2366} {arXiv:0805.2366 [astro-ph]} \BibitemShut
  {NoStop}%
\bibitem [{\citenamefont {Scaramella}\ \emph {et~al.}(2022)\citenamefont
  {Scaramella} \emph {et~al.}}]{Euclid:2021icp}%
  \BibitemOpen
  \bibfield  {author} {\bibinfo {author} {\bibfnamefont {R.}~\bibnamefont
  {Scaramella}} \emph {et~al.} (\bibinfo {collaboration} {Euclid}),\ }\href
  {\doibase 10.1051/0004-6361/202141938} {\bibfield  {journal} {\bibinfo
  {journal} {Astron. Astrophys.}\ }\textbf {\bibinfo {volume} {662}},\ \bibinfo
  {pages} {A112} (\bibinfo {year} {2022})},\ \Eprint
  {http://arxiv.org/abs/2108.01201} {arXiv:2108.01201 [astro-ph.CO]}
  \BibitemShut {NoStop}%
\bibitem [{\citenamefont {Gong}\ \emph {et~al.}(2019)\citenamefont {Gong},
  \citenamefont {Liu}, \citenamefont {Cao}, \citenamefont {Chen}, \citenamefont
  {Fan}, \citenamefont {Li}, \citenamefont {Li}, \citenamefont {Li},
  \citenamefont {Zhang},\ and\ \citenamefont {Zhan}}]{Gong:2019yxt}%
  \BibitemOpen
  \bibfield  {author} {\bibinfo {author} {\bibfnamefont {Y.}~\bibnamefont
  {Gong}}, \bibinfo {author} {\bibfnamefont {X.}~\bibnamefont {Liu}}, \bibinfo
  {author} {\bibfnamefont {Y.}~\bibnamefont {Cao}}, \bibinfo {author}
  {\bibfnamefont {X.}~\bibnamefont {Chen}}, \bibinfo {author} {\bibfnamefont
  {Z.}~\bibnamefont {Fan}}, \bibinfo {author} {\bibfnamefont {R.}~\bibnamefont
  {Li}}, \bibinfo {author} {\bibfnamefont {X.-D.}\ \bibnamefont {Li}}, \bibinfo
  {author} {\bibfnamefont {Z.}~\bibnamefont {Li}}, \bibinfo {author}
  {\bibfnamefont {X.}~\bibnamefont {Zhang}}, \ and\ \bibinfo {author}
  {\bibfnamefont {H.}~\bibnamefont {Zhan}},\ }\href {\doibase
  10.3847/1538-4357/ab391e} {\bibfield  {journal} {\bibinfo  {journal}
  {Astrophys. J.}\ }\textbf {\bibinfo {volume} {883}},\ \bibinfo {pages} {203}
  (\bibinfo {year} {2019})},\ \Eprint {http://arxiv.org/abs/1901.04634}
  {arXiv:1901.04634 [astro-ph.CO]} \BibitemShut {NoStop}%
\bibitem [{\citenamefont {Phinney}(1991)}]{Phinney:1991ei}%
  \BibitemOpen
  \bibfield  {author} {\bibinfo {author} {\bibfnamefont {E.~S.}\ \bibnamefont
  {Phinney}},\ }\href {\doibase 10.1086/186163} {\bibfield  {journal} {\bibinfo
   {journal} {Astrophys. J. Lett.}\ }\textbf {\bibinfo {volume} {380}},\
  \bibinfo {pages} {L17} (\bibinfo {year} {1991})}\BibitemShut {NoStop}%
\bibitem [{\citenamefont {Leibler}\ and\ \citenamefont
  {Berger}(2010)}]{Leibler:2010uq}%
  \BibitemOpen
  \bibfield  {author} {\bibinfo {author} {\bibfnamefont {C.~N.}\ \bibnamefont
  {Leibler}}\ and\ \bibinfo {author} {\bibfnamefont {E.}~\bibnamefont
  {Berger}},\ }\href {\doibase 10.1088/0004-637X/725/1/1202} {\bibfield
  {journal} {\bibinfo  {journal} {Astrophys. J.}\ }\textbf {\bibinfo {volume}
  {725}},\ \bibinfo {pages} {1202} (\bibinfo {year} {2010})},\ \Eprint
  {http://arxiv.org/abs/1009.1147} {arXiv:1009.1147 [astro-ph.HE]} \BibitemShut
  {NoStop}%
\bibitem [{\citenamefont {Fong}\ \emph {et~al.}(2013)\citenamefont {Fong} \emph
  {et~al.}}]{Fong:2013eqa}%
  \BibitemOpen
  \bibfield  {author} {\bibinfo {author} {\bibfnamefont {W.-f.}\ \bibnamefont
  {Fong}} \emph {et~al.},\ }\href {\doibase 10.1088/0004-637X/769/1/56}
  {\bibfield  {journal} {\bibinfo  {journal} {Astrophys. J.}\ }\textbf
  {\bibinfo {volume} {769}},\ \bibinfo {pages} {56} (\bibinfo {year} {2013})},\
  \Eprint {http://arxiv.org/abs/1302.3221} {arXiv:1302.3221 [astro-ph.HE]}
  \BibitemShut {NoStop}%
\bibitem [{\citenamefont {Rodriguez}\ \emph {et~al.}(2016)\citenamefont
  {Rodriguez}, \citenamefont {Chatterjee},\ and\ \citenamefont
  {Rasio}}]{Rodriguez:2016kxx}%
  \BibitemOpen
  \bibfield  {author} {\bibinfo {author} {\bibfnamefont {C.~L.}\ \bibnamefont
  {Rodriguez}}, \bibinfo {author} {\bibfnamefont {S.}~\bibnamefont
  {Chatterjee}}, \ and\ \bibinfo {author} {\bibfnamefont {F.~A.}\ \bibnamefont
  {Rasio}},\ }\href {\doibase 10.1103/PhysRevD.93.084029} {\bibfield  {journal}
  {\bibinfo  {journal} {Phys. Rev. D}\ }\textbf {\bibinfo {volume} {93}},\
  \bibinfo {pages} {084029} (\bibinfo {year} {2016})},\ \Eprint
  {http://arxiv.org/abs/1602.02444} {arXiv:1602.02444 [astro-ph.HE]}
  \BibitemShut {NoStop}%
\bibitem [{\citenamefont {Singer}\ \emph {et~al.}(2016)\citenamefont {Singer}
  \emph {et~al.}}]{Singer:2016eax}%
  \BibitemOpen
  \bibfield  {author} {\bibinfo {author} {\bibfnamefont {L.~P.}\ \bibnamefont
  {Singer}} \emph {et~al.},\ }\href {\doibase 10.3847/2041-8205/829/1/L15}
  {\bibfield  {journal} {\bibinfo  {journal} {Astrophys. J. Lett.}\ }\textbf
  {\bibinfo {volume} {829}},\ \bibinfo {pages} {L15} (\bibinfo {year}
  {2016})},\ \Eprint {http://arxiv.org/abs/1603.07333} {arXiv:1603.07333
  [astro-ph.HE]} \BibitemShut {NoStop}%
\bibitem [{\citenamefont {Rodriguez}\ and\ \citenamefont
  {Loeb}(2018)}]{Rodriguez:2018rmd}%
  \BibitemOpen
  \bibfield  {author} {\bibinfo {author} {\bibfnamefont {C.~L.}\ \bibnamefont
  {Rodriguez}}\ and\ \bibinfo {author} {\bibfnamefont {A.}~\bibnamefont
  {Loeb}},\ }\href {\doibase 10.3847/2041-8213/aae377} {\bibfield  {journal}
  {\bibinfo  {journal} {Astrophys. J. Lett.}\ }\textbf {\bibinfo {volume}
  {866}},\ \bibinfo {pages} {L5} (\bibinfo {year} {2018})},\ \Eprint
  {http://arxiv.org/abs/1809.01152} {arXiv:1809.01152 [astro-ph.HE]}
  \BibitemShut {NoStop}%
\bibitem [{\citenamefont {Yang}\ \emph
  {et~al.}(2020{\natexlab{a}})\citenamefont {Yang}, \citenamefont {Bartos},
  \citenamefont {Haiman}, \citenamefont {Kocsis}, \citenamefont {M\'arka},\
  and\ \citenamefont {Tagawa}}]{Yang:2020lhq}%
  \BibitemOpen
  \bibfield  {author} {\bibinfo {author} {\bibfnamefont {Y.}~\bibnamefont
  {Yang}}, \bibinfo {author} {\bibfnamefont {I.}~\bibnamefont {Bartos}},
  \bibinfo {author} {\bibfnamefont {Z.}~\bibnamefont {Haiman}}, \bibinfo
  {author} {\bibfnamefont {B.}~\bibnamefont {Kocsis}}, \bibinfo {author}
  {\bibfnamefont {S.}~\bibnamefont {M\'arka}}, \ and\ \bibinfo {author}
  {\bibfnamefont {H.}~\bibnamefont {Tagawa}},\ }\href {\doibase
  10.3847/1538-4357/ab91b4} {\bibfield  {journal} {\bibinfo  {journal}
  {Astrophys. J.}\ }\textbf {\bibinfo {volume} {896}},\ \bibinfo {pages} {138}
  (\bibinfo {year} {2020}{\natexlab{a}})},\ \Eprint
  {http://arxiv.org/abs/2003.08564} {arXiv:2003.08564 [astro-ph.HE]}
  \BibitemShut {NoStop}%
\bibitem [{\citenamefont {Santoliquido}\ \emph {et~al.}(2020)\citenamefont
  {Santoliquido}, \citenamefont {Mapelli}, \citenamefont {Bouffanais},
  \citenamefont {Giacobbo}, \citenamefont {Di~Carlo}, \citenamefont {Rastello},
  \citenamefont {Artale},\ and\ \citenamefont
  {Ballone}}]{Santoliquido:2020bry}%
  \BibitemOpen
  \bibfield  {author} {\bibinfo {author} {\bibfnamefont {F.}~\bibnamefont
  {Santoliquido}}, \bibinfo {author} {\bibfnamefont {M.}~\bibnamefont
  {Mapelli}}, \bibinfo {author} {\bibfnamefont {Y.}~\bibnamefont {Bouffanais}},
  \bibinfo {author} {\bibfnamefont {N.}~\bibnamefont {Giacobbo}}, \bibinfo
  {author} {\bibfnamefont {U.~N.}\ \bibnamefont {Di~Carlo}}, \bibinfo {author}
  {\bibfnamefont {S.}~\bibnamefont {Rastello}}, \bibinfo {author}
  {\bibfnamefont {M.~C.}\ \bibnamefont {Artale}}, \ and\ \bibinfo {author}
  {\bibfnamefont {A.}~\bibnamefont {Ballone}},\ }\href {\doibase
  10.3847/1538-4357/ab9b78} {\bibfield  {journal} {\bibinfo  {journal}
  {Astrophys. J.}\ }\textbf {\bibinfo {volume} {898}},\ \bibinfo {pages} {152}
  (\bibinfo {year} {2020})},\ \Eprint {http://arxiv.org/abs/2004.09533}
  {arXiv:2004.09533 [astro-ph.HE]} \BibitemShut {NoStop}%
\bibitem [{\citenamefont {Fishbach}\ and\ \citenamefont
  {Kalogera}(2021)}]{Fishbach:2021mhp}%
  \BibitemOpen
  \bibfield  {author} {\bibinfo {author} {\bibfnamefont {M.}~\bibnamefont
  {Fishbach}}\ and\ \bibinfo {author} {\bibfnamefont {V.}~\bibnamefont
  {Kalogera}},\ }\href {\doibase 10.3847/2041-8213/ac05c4} {\bibfield
  {journal} {\bibinfo  {journal} {Astrophys. J. Lett.}\ }\textbf {\bibinfo
  {volume} {914}},\ \bibinfo {pages} {L30} (\bibinfo {year} {2021})},\ \Eprint
  {http://arxiv.org/abs/2105.06491} {arXiv:2105.06491 [astro-ph.HE]}
  \BibitemShut {NoStop}%
\bibitem [{\citenamefont {van Son}\ \emph {et~al.}(2022)\citenamefont {van
  Son}, \citenamefont {de~Mink}, \citenamefont {Callister}, \citenamefont
  {Justham}, \citenamefont {Renzo}, \citenamefont {Wagg}, \citenamefont
  {Broekgaarden}, \citenamefont {Kummer}, \citenamefont {Pakmor},\ and\
  \citenamefont {Mandel}}]{vanSon:2021zpk}%
  \BibitemOpen
  \bibfield  {author} {\bibinfo {author} {\bibfnamefont {L.~A.~C.}\
  \bibnamefont {van Son}}, \bibinfo {author} {\bibfnamefont {S.~E.}\
  \bibnamefont {de~Mink}}, \bibinfo {author} {\bibfnamefont {T.}~\bibnamefont
  {Callister}}, \bibinfo {author} {\bibfnamefont {S.}~\bibnamefont {Justham}},
  \bibinfo {author} {\bibfnamefont {M.}~\bibnamefont {Renzo}}, \bibinfo
  {author} {\bibfnamefont {T.}~\bibnamefont {Wagg}}, \bibinfo {author}
  {\bibfnamefont {F.~S.}\ \bibnamefont {Broekgaarden}}, \bibinfo {author}
  {\bibfnamefont {F.}~\bibnamefont {Kummer}}, \bibinfo {author} {\bibfnamefont
  {R.}~\bibnamefont {Pakmor}}, \ and\ \bibinfo {author} {\bibfnamefont
  {I.}~\bibnamefont {Mandel}},\ }\href {\doibase 10.3847/1538-4357/ac64a3}
  {\bibfield  {journal} {\bibinfo  {journal} {Astrophys. J.}\ }\textbf
  {\bibinfo {volume} {931}},\ \bibinfo {pages} {17} (\bibinfo {year} {2022})},\
  \Eprint {http://arxiv.org/abs/2110.01634} {arXiv:2110.01634 [astro-ph.HE]}
  \BibitemShut {NoStop}%
\bibitem [{\citenamefont {Vijaykumar}\ \emph {et~al.}(2023)\citenamefont
  {Vijaykumar}, \citenamefont {Fishbach}, \citenamefont {Adhikari},\ and\
  \citenamefont {Holz}}]{Vijaykumar:2023bgs}%
  \BibitemOpen
  \bibfield  {author} {\bibinfo {author} {\bibfnamefont {A.}~\bibnamefont
  {Vijaykumar}}, \bibinfo {author} {\bibfnamefont {M.}~\bibnamefont
  {Fishbach}}, \bibinfo {author} {\bibfnamefont {S.}~\bibnamefont {Adhikari}},
  \ and\ \bibinfo {author} {\bibfnamefont {D.~E.}\ \bibnamefont {Holz}},\
  }\href@noop {} {\  (\bibinfo {year} {2023})},\ \Eprint
  {http://arxiv.org/abs/2312.03316} {arXiv:2312.03316 [astro-ph.HE]}
  \BibitemShut {NoStop}%
\bibitem [{\citenamefont {Bell}\ \emph {et~al.}(2003)\citenamefont {Bell},
  \citenamefont {McIntosh}, \citenamefont {Katz},\ and\ \citenamefont
  {Weinberg}}]{Bell:2003cj}%
  \BibitemOpen
  \bibfield  {author} {\bibinfo {author} {\bibfnamefont {E.~F.}\ \bibnamefont
  {Bell}}, \bibinfo {author} {\bibfnamefont {D.~H.}\ \bibnamefont {McIntosh}},
  \bibinfo {author} {\bibfnamefont {N.}~\bibnamefont {Katz}}, \ and\ \bibinfo
  {author} {\bibfnamefont {M.~D.}\ \bibnamefont {Weinberg}},\ }\href {\doibase
  10.1086/378847} {\bibfield  {journal} {\bibinfo  {journal} {Astrophys. J.
  Suppl.}\ }\textbf {\bibinfo {volume} {149}},\ \bibinfo {pages} {289}
  (\bibinfo {year} {2003})},\ \Eprint {http://arxiv.org/abs/astro-ph/0302543}
  {arXiv:astro-ph/0302543} \BibitemShut {NoStop}%
\bibitem [{\citenamefont {Lin}\ \emph {et~al.}(2004)\citenamefont {Lin},
  \citenamefont {Mohr},\ and\ \citenamefont {Stanford}}]{Lin:2004ak}%
  \BibitemOpen
  \bibfield  {author} {\bibinfo {author} {\bibfnamefont {Y.-T.}\ \bibnamefont
  {Lin}}, \bibinfo {author} {\bibfnamefont {J.~J.}\ \bibnamefont {Mohr}}, \
  and\ \bibinfo {author} {\bibfnamefont {S.~A.}\ \bibnamefont {Stanford}},\
  }\href {\doibase 10.1086/421714} {\bibfield  {journal} {\bibinfo  {journal}
  {Astrophys. J.}\ }\textbf {\bibinfo {volume} {610}},\ \bibinfo {pages} {745}
  (\bibinfo {year} {2004})},\ \Eprint {http://arxiv.org/abs/astro-ph/0402308}
  {arXiv:astro-ph/0402308} \BibitemShut {NoStop}%
\bibitem [{\citenamefont {Kennicutt}(1998)}]{Kennicutt:1998zb}%
  \BibitemOpen
  \bibfield  {author} {\bibinfo {author} {\bibfnamefont {R.~C.}\ \bibnamefont
  {Kennicutt}, \bibfnamefont {Jr.}},\ }\href {\doibase
  10.1146/annurev.astro.36.1.189} {\bibfield  {journal} {\bibinfo  {journal}
  {Ann. Rev. Astron. Astrophys.}\ }\textbf {\bibinfo {volume} {36}},\ \bibinfo
  {pages} {189} (\bibinfo {year} {1998})},\ \Eprint
  {http://arxiv.org/abs/astro-ph/9807187} {arXiv:astro-ph/9807187} \BibitemShut
  {NoStop}%
\bibitem [{\citenamefont {Gehrels}\ \emph {et~al.}(2016)\citenamefont
  {Gehrels}, \citenamefont {Cannizzo}, \citenamefont {Kanner}, \citenamefont
  {Kasliwal}, \citenamefont {Nissanke},\ and\ \citenamefont
  {Singer}}]{Gehrels:2015uga}%
  \BibitemOpen
  \bibfield  {author} {\bibinfo {author} {\bibfnamefont {N.}~\bibnamefont
  {Gehrels}}, \bibinfo {author} {\bibfnamefont {J.~K.}\ \bibnamefont
  {Cannizzo}}, \bibinfo {author} {\bibfnamefont {J.}~\bibnamefont {Kanner}},
  \bibinfo {author} {\bibfnamefont {M.~M.}\ \bibnamefont {Kasliwal}}, \bibinfo
  {author} {\bibfnamefont {S.}~\bibnamefont {Nissanke}}, \ and\ \bibinfo
  {author} {\bibfnamefont {L.~P.}\ \bibnamefont {Singer}},\ }\href {\doibase
  10.3847/0004-637X/820/2/136} {\bibfield  {journal} {\bibinfo  {journal}
  {Astrophys. J.}\ }\textbf {\bibinfo {volume} {820}},\ \bibinfo {pages} {136}
  (\bibinfo {year} {2016})},\ \Eprint {http://arxiv.org/abs/1508.03608}
  {arXiv:1508.03608 [astro-ph.HE]} \BibitemShut {NoStop}%
\bibitem [{\citenamefont {Calzetti}\ \emph {et~al.}(1994)\citenamefont
  {Calzetti}, \citenamefont {Kinney},\ and\ \citenamefont
  {Storchi-Bergmann}}]{Calzetti:1994vw}%
  \BibitemOpen
  \bibfield  {author} {\bibinfo {author} {\bibfnamefont {D.}~\bibnamefont
  {Calzetti}}, \bibinfo {author} {\bibfnamefont {A.~L.}\ \bibnamefont
  {Kinney}}, \ and\ \bibinfo {author} {\bibfnamefont {T.}~\bibnamefont
  {Storchi-Bergmann}},\ }\href {\doibase 10.1086/174346} {\bibfield  {journal}
  {\bibinfo  {journal} {Astrophys. J.}\ }\textbf {\bibinfo {volume} {429}},\
  \bibinfo {pages} {582} (\bibinfo {year} {1994})}\BibitemShut {NoStop}%
\bibitem [{\citenamefont {Chabrier}(2003)}]{Chabrier:2003ki}%
  \BibitemOpen
  \bibfield  {author} {\bibinfo {author} {\bibfnamefont {G.}~\bibnamefont
  {Chabrier}},\ }\href {\doibase 10.1086/376392} {\bibfield  {journal}
  {\bibinfo  {journal} {Publ. Astron. Soc. Pac.}\ }\textbf {\bibinfo {volume}
  {115}},\ \bibinfo {pages} {763} (\bibinfo {year} {2003})},\ \Eprint
  {http://arxiv.org/abs/astro-ph/0304382} {arXiv:astro-ph/0304382} \BibitemShut
  {NoStop}%
\bibitem [{\citenamefont {Bruzual}\ and\ \citenamefont
  {Charlot}(2003)}]{Bruzual:2003tq}%
  \BibitemOpen
  \bibfield  {author} {\bibinfo {author} {\bibfnamefont {G.}~\bibnamefont
  {Bruzual}}\ and\ \bibinfo {author} {\bibfnamefont {S.}~\bibnamefont
  {Charlot}},\ }\href {\doibase 10.1046/j.1365-8711.2003.06897.x} {\bibfield
  {journal} {\bibinfo  {journal} {Mon. Not. Roy. Astron. Soc.}\ }\textbf
  {\bibinfo {volume} {344}},\ \bibinfo {pages} {1000} (\bibinfo {year}
  {2003})},\ \Eprint {http://arxiv.org/abs/astro-ph/0309134}
  {arXiv:astro-ph/0309134} \BibitemShut {NoStop}%
\bibitem [{\citenamefont {Arnouts}\ \emph {et~al.}(1999)\citenamefont
  {Arnouts}, \citenamefont {Cristiani}, \citenamefont {Moscardini},
  \citenamefont {Matarrese}, \citenamefont {Lucchin}, \citenamefont {Fontana},\
  and\ \citenamefont {Giallongo}}]{Arnouts:1999bb}%
  \BibitemOpen
  \bibfield  {author} {\bibinfo {author} {\bibfnamefont {S.}~\bibnamefont
  {Arnouts}}, \bibinfo {author} {\bibfnamefont {S.}~\bibnamefont {Cristiani}},
  \bibinfo {author} {\bibfnamefont {L.}~\bibnamefont {Moscardini}}, \bibinfo
  {author} {\bibfnamefont {S.}~\bibnamefont {Matarrese}}, \bibinfo {author}
  {\bibfnamefont {F.}~\bibnamefont {Lucchin}}, \bibinfo {author} {\bibfnamefont
  {A.}~\bibnamefont {Fontana}}, \ and\ \bibinfo {author} {\bibfnamefont
  {E.}~\bibnamefont {Giallongo}},\ }\href {\doibase
  10.1046/j.1365-8711.1999.02978.x} {\bibfield  {journal} {\bibinfo  {journal}
  {Mon. Not. Roy. Astron. Soc.}\ }\textbf {\bibinfo {volume} {310}},\ \bibinfo
  {pages} {540} (\bibinfo {year} {1999})},\ \Eprint
  {http://arxiv.org/abs/astro-ph/9902290} {arXiv:astro-ph/9902290} \BibitemShut
  {NoStop}%
\bibitem [{\citenamefont {Ilbert}\ \emph {et~al.}(2006)\citenamefont {Ilbert}
  \emph {et~al.}}]{Ilbert:2006dp}%
  \BibitemOpen
  \bibfield  {author} {\bibinfo {author} {\bibfnamefont {O.}~\bibnamefont
  {Ilbert}} \emph {et~al.},\ }\href {\doibase 10.1051/0004-6361:20065138}
  {\bibfield  {journal} {\bibinfo  {journal} {Astron. Astrophys.}\ }\textbf
  {\bibinfo {volume} {457}},\ \bibinfo {pages} {841} (\bibinfo {year}
  {2006})},\ \Eprint {http://arxiv.org/abs/astro-ph/0603217}
  {arXiv:astro-ph/0603217} \BibitemShut {NoStop}%
\bibitem [{\citenamefont {Sivia}(2006)}]{Sivia:2006book}%
  \BibitemOpen
  \bibfield  {author} {\bibinfo {author} {\bibfnamefont {D.~S.}\ \bibnamefont
  {Sivia}},\ }\enquote {\bibinfo {title} {{Information gain: quantifying the
  worth of an experiment}},}\ in\ \href {\doibase
  10.1093/oso/9780198568315.001.0001} {\emph {\bibinfo {booktitle} {{Data
  Analysis: A Bayesian Tutorial}}}}\ (\bibinfo  {publisher} {Oxford University
  Press},\ \bibinfo {year} {2006})\BibitemShut {NoStop}%
\bibitem [{\citenamefont {Zhu}\ and\ \citenamefont {Chen}(2024)}]{Zhu:2023imz}%
  \BibitemOpen
  \bibfield  {author} {\bibinfo {author} {\bibfnamefont {L.-G.}\ \bibnamefont
  {Zhu}}\ and\ \bibinfo {author} {\bibfnamefont {X.}~\bibnamefont {Chen}},\
  }\href {\doibase 10.3847/1538-4357/ad0cf2} {\bibfield  {journal} {\bibinfo
  {journal} {Astrophys. J.}\ }\textbf {\bibinfo {volume} {960}},\ \bibinfo
  {pages} {43} (\bibinfo {year} {2024})},\ \Eprint
  {http://arxiv.org/abs/2308.12499} {arXiv:2308.12499 [astro-ph.HE]}
  \BibitemShut {NoStop}%
\bibitem [{\citenamefont {Graham}(2007)}]{Graham:2007uq}%
  \BibitemOpen
  \bibfield  {author} {\bibinfo {author} {\bibfnamefont {A.~W.}\ \bibnamefont
  {Graham}},\ }\href {\doibase 10.1111/j.1365-2966.2007.11950.x} {\bibfield
  {journal} {\bibinfo  {journal} {Mon. Not. Roy. Astron. Soc.}\ }\textbf
  {\bibinfo {volume} {379}},\ \bibinfo {pages} {711} (\bibinfo {year}
  {2007})},\ \Eprint {http://arxiv.org/abs/0705.0618} {arXiv:0705.0618
  [astro-ph]} \BibitemShut {NoStop}%
\bibitem [{\citenamefont {Bentz}\ \emph {et~al.}(2009)\citenamefont {Bentz},
  \citenamefont {Peterson}, \citenamefont {Pogge},\ and\ \citenamefont
  {Vestergaard}}]{Bentz:2008rt}%
  \BibitemOpen
  \bibfield  {author} {\bibinfo {author} {\bibfnamefont {M.~C.}\ \bibnamefont
  {Bentz}}, \bibinfo {author} {\bibfnamefont {B.~M.}\ \bibnamefont {Peterson}},
  \bibinfo {author} {\bibfnamefont {R.~W.}\ \bibnamefont {Pogge}}, \ and\
  \bibinfo {author} {\bibfnamefont {M.}~\bibnamefont {Vestergaard}},\ }\href
  {\doibase 10.1088/0004-637X/694/2/L166} {\bibfield  {journal} {\bibinfo
  {journal} {Astrophys. J. Lett.}\ }\textbf {\bibinfo {volume} {694}},\
  \bibinfo {pages} {L166} (\bibinfo {year} {2009})},\ \Eprint
  {http://arxiv.org/abs/0812.2284} {arXiv:0812.2284 [astro-ph]} \BibitemShut
  {NoStop}%
\bibitem [{\citenamefont {Jiang}\ \emph {et~al.}(2011)\citenamefont {Jiang},
  \citenamefont {Greene},\ and\ \citenamefont {Ho}}]{Jiang:2011bt}%
  \BibitemOpen
  \bibfield  {author} {\bibinfo {author} {\bibfnamefont {Y.}~\bibnamefont
  {Jiang}}, \bibinfo {author} {\bibfnamefont {J.}~\bibnamefont {Greene}}, \
  and\ \bibinfo {author} {\bibfnamefont {L.}~\bibnamefont {Ho}},\ }\href
  {\doibase 10.1088/2041-8205/737/2/L45} {\bibfield  {journal} {\bibinfo
  {journal} {Astrophys. J. Lett.}\ }\textbf {\bibinfo {volume} {737}},\
  \bibinfo {pages} {L45} (\bibinfo {year} {2011})},\ \Eprint
  {http://arxiv.org/abs/1107.4103} {arXiv:1107.4103 [astro-ph.CO]} \BibitemShut
  {NoStop}%
\bibitem [{\citenamefont {Kormendy}\ and\ \citenamefont
  {Ho}(2013)}]{Kormendy:2013dxa}%
  \BibitemOpen
  \bibfield  {author} {\bibinfo {author} {\bibfnamefont {J.}~\bibnamefont
  {Kormendy}}\ and\ \bibinfo {author} {\bibfnamefont {L.~C.}\ \bibnamefont
  {Ho}},\ }\href {\doibase 10.1146/annurev-astro-082708-101811} {\bibfield
  {journal} {\bibinfo  {journal} {Ann. Rev. Astron. Astrophys.}\ }\textbf
  {\bibinfo {volume} {51}},\ \bibinfo {pages} {511} (\bibinfo {year} {2013})},\
  \Eprint {http://arxiv.org/abs/1304.7762} {arXiv:1304.7762 [astro-ph.CO]}
  \BibitemShut {NoStop}%
\bibitem [{\citenamefont {Begelman}\ \emph {et~al.}(1980)\citenamefont
  {Begelman}, \citenamefont {Blandford},\ and\ \citenamefont
  {Rees}}]{Begelman:1980vb}%
  \BibitemOpen
  \bibfield  {author} {\bibinfo {author} {\bibfnamefont {M.~C.}\ \bibnamefont
  {Begelman}}, \bibinfo {author} {\bibfnamefont {R.~D.}\ \bibnamefont
  {Blandford}}, \ and\ \bibinfo {author} {\bibfnamefont {M.~J.}\ \bibnamefont
  {Rees}},\ }\href {\doibase 10.1038/287307a0} {\bibfield  {journal} {\bibinfo
  {journal} {Nature}\ }\textbf {\bibinfo {volume} {287}},\ \bibinfo {pages}
  {307} (\bibinfo {year} {1980})}\BibitemShut {NoStop}%
\bibitem [{\citenamefont {Milosavljevic}\ and\ \citenamefont
  {Merritt}(2001)}]{Milosavljevic:2001vi}%
  \BibitemOpen
  \bibfield  {author} {\bibinfo {author} {\bibfnamefont {M.}~\bibnamefont
  {Milosavljevic}}\ and\ \bibinfo {author} {\bibfnamefont {D.}~\bibnamefont
  {Merritt}},\ }\href {\doibase 10.1086/323830} {\bibfield  {journal} {\bibinfo
   {journal} {Astrophys. J.}\ }\textbf {\bibinfo {volume} {563}},\ \bibinfo
  {pages} {34} (\bibinfo {year} {2001})},\ \Eprint
  {http://arxiv.org/abs/astro-ph/0103350} {arXiv:astro-ph/0103350} \BibitemShut
  {NoStop}%
\bibitem [{\citenamefont {Armitage}\ and\ \citenamefont
  {Natarajan}(2002)}]{Armitage:2002uu}%
  \BibitemOpen
  \bibfield  {author} {\bibinfo {author} {\bibfnamefont {P.~J.}\ \bibnamefont
  {Armitage}}\ and\ \bibinfo {author} {\bibfnamefont {P.}~\bibnamefont
  {Natarajan}},\ }\href {\doibase 10.1086/339770} {\bibfield  {journal}
  {\bibinfo  {journal} {Astrophys. J. Lett.}\ }\textbf {\bibinfo {volume}
  {567}},\ \bibinfo {pages} {L9} (\bibinfo {year} {2002})},\ \Eprint
  {http://arxiv.org/abs/astro-ph/0201318} {arXiv:astro-ph/0201318} \BibitemShut
  {NoStop}%
\bibitem [{\citenamefont {Escala}\ \emph {et~al.}(2005)\citenamefont {Escala},
  \citenamefont {Larson}, \citenamefont {Coppi},\ and\ \citenamefont
  {Mardones}}]{Escala:2004jh}%
  \BibitemOpen
  \bibfield  {author} {\bibinfo {author} {\bibfnamefont {A.}~\bibnamefont
  {Escala}}, \bibinfo {author} {\bibfnamefont {R.~B.}\ \bibnamefont {Larson}},
  \bibinfo {author} {\bibfnamefont {P.~S.}\ \bibnamefont {Coppi}}, \ and\
  \bibinfo {author} {\bibfnamefont {D.}~\bibnamefont {Mardones}},\ }\href
  {\doibase 10.1086/431747} {\bibfield  {journal} {\bibinfo  {journal}
  {Astrophys. J.}\ }\textbf {\bibinfo {volume} {630}},\ \bibinfo {pages} {152}
  (\bibinfo {year} {2005})},\ \Eprint {http://arxiv.org/abs/astro-ph/0406304}
  {arXiv:astro-ph/0406304} \BibitemShut {NoStop}%
\bibitem [{\citenamefont {Mayer}\ \emph {et~al.}(2007)\citenamefont {Mayer},
  \citenamefont {Kazantzidis}, \citenamefont {Madau}, \citenamefont {Colpi},
  \citenamefont {Quinn},\ and\ \citenamefont {Wadsley}}]{Mayer:2007vk}%
  \BibitemOpen
  \bibfield  {author} {\bibinfo {author} {\bibfnamefont {L.}~\bibnamefont
  {Mayer}}, \bibinfo {author} {\bibfnamefont {S.}~\bibnamefont {Kazantzidis}},
  \bibinfo {author} {\bibfnamefont {P.}~\bibnamefont {Madau}}, \bibinfo
  {author} {\bibfnamefont {M.}~\bibnamefont {Colpi}}, \bibinfo {author}
  {\bibfnamefont {T.~R.}\ \bibnamefont {Quinn}}, \ and\ \bibinfo {author}
  {\bibfnamefont {J.}~\bibnamefont {Wadsley}},\ }\href {\doibase
  10.1126/science.1141858} {\bibfield  {journal} {\bibinfo  {journal}
  {Science}\ }\textbf {\bibinfo {volume} {316}},\ \bibinfo {pages} {1874}
  (\bibinfo {year} {2007})},\ \Eprint {http://arxiv.org/abs/0706.1562}
  {arXiv:0706.1562 [astro-ph]} \BibitemShut {NoStop}%
\bibitem [{\citenamefont {Macfadyen}\ and\ \citenamefont
  {Milosavljevic}(2008)}]{Macfadyen:2006jx}%
  \BibitemOpen
  \bibfield  {author} {\bibinfo {author} {\bibfnamefont {A.~I.}\ \bibnamefont
  {Macfadyen}}\ and\ \bibinfo {author} {\bibfnamefont {M.}~\bibnamefont
  {Milosavljevic}},\ }\href {\doibase 10.1086/523869} {\bibfield  {journal}
  {\bibinfo  {journal} {Astrophys. J.}\ }\textbf {\bibinfo {volume} {672}},\
  \bibinfo {pages} {83} (\bibinfo {year} {2008})},\ \Eprint
  {http://arxiv.org/abs/astro-ph/0607467} {arXiv:astro-ph/0607467} \BibitemShut
  {NoStop}%
\bibitem [{\citenamefont {Cuadra}\ \emph {et~al.}(2009)\citenamefont {Cuadra},
  \citenamefont {Armitage}, \citenamefont {Alexander},\ and\ \citenamefont
  {Begelman}}]{Cuadra:2008xn}%
  \BibitemOpen
  \bibfield  {author} {\bibinfo {author} {\bibfnamefont {J.}~\bibnamefont
  {Cuadra}}, \bibinfo {author} {\bibfnamefont {P.~J.}\ \bibnamefont
  {Armitage}}, \bibinfo {author} {\bibfnamefont {R.~D.}\ \bibnamefont
  {Alexander}}, \ and\ \bibinfo {author} {\bibfnamefont {M.~C.}\ \bibnamefont
  {Begelman}},\ }\href {\doibase 10.1111/j.1365-2966.2008.14147.x} {\bibfield
  {journal} {\bibinfo  {journal} {Mon. Not. Roy. Astron. Soc.}\ }\textbf
  {\bibinfo {volume} {393}},\ \bibinfo {pages} {1423} (\bibinfo {year}
  {2009})},\ \Eprint {http://arxiv.org/abs/0809.0311} {arXiv:0809.0311
  [astro-ph]} \BibitemShut {NoStop}%
\bibitem [{\citenamefont {Goicovic}\ \emph {et~al.}(2016)\citenamefont
  {Goicovic}, \citenamefont {Cuadra}, \citenamefont {Sesana}, \citenamefont
  {Stasyszyn}, \citenamefont {Amaro-Seoane},\ and\ \citenamefont
  {Tanaka}}]{Goicovic:2015kda}%
  \BibitemOpen
  \bibfield  {author} {\bibinfo {author} {\bibfnamefont {F.~G.}\ \bibnamefont
  {Goicovic}}, \bibinfo {author} {\bibfnamefont {J.}~\bibnamefont {Cuadra}},
  \bibinfo {author} {\bibfnamefont {A.}~\bibnamefont {Sesana}}, \bibinfo
  {author} {\bibfnamefont {F.}~\bibnamefont {Stasyszyn}}, \bibinfo {author}
  {\bibfnamefont {P.}~\bibnamefont {Amaro-Seoane}}, \ and\ \bibinfo {author}
  {\bibfnamefont {T.~L.}\ \bibnamefont {Tanaka}},\ }\href {\doibase
  10.1093/mnras/stv2470} {\bibfield  {journal} {\bibinfo  {journal} {Mon. Not.
  Roy. Astron. Soc.}\ }\textbf {\bibinfo {volume} {455}},\ \bibinfo {pages}
  {1989} (\bibinfo {year} {2016})},\ \Eprint {http://arxiv.org/abs/1507.05596}
  {arXiv:1507.05596 [astro-ph.HE]} \BibitemShut {NoStop}%
\bibitem [{\citenamefont {Levin}(2003)}]{Levin:2003ej}%
  \BibitemOpen
  \bibfield  {author} {\bibinfo {author} {\bibfnamefont {Y.}~\bibnamefont
  {Levin}},\ }\href@noop {} {\  (\bibinfo {year} {2003})},\ \Eprint
  {http://arxiv.org/abs/astro-ph/0307084} {arXiv:astro-ph/0307084} \BibitemShut
  {NoStop}%
\bibitem [{\citenamefont {Levin}(2007)}]{Levin:2006uc}%
  \BibitemOpen
  \bibfield  {author} {\bibinfo {author} {\bibfnamefont {Y.}~\bibnamefont
  {Levin}},\ }\href {\doibase 10.1111/j.1365-2966.2006.11155.x} {\bibfield
  {journal} {\bibinfo  {journal} {Mon. Not. Roy. Astron. Soc.}\ }\textbf
  {\bibinfo {volume} {374}},\ \bibinfo {pages} {515} (\bibinfo {year}
  {2007})},\ \Eprint {http://arxiv.org/abs/astro-ph/0603583}
  {arXiv:astro-ph/0603583} \BibitemShut {NoStop}%
\bibitem [{\citenamefont {Pan}\ and\ \citenamefont {Yang}(2021)}]{Pan:2021ksp}%
  \BibitemOpen
  \bibfield  {author} {\bibinfo {author} {\bibfnamefont {Z.}~\bibnamefont
  {Pan}}\ and\ \bibinfo {author} {\bibfnamefont {H.}~\bibnamefont {Yang}},\
  }\href {\doibase 10.1103/PhysRevD.103.103018} {\bibfield  {journal} {\bibinfo
   {journal} {Phys. Rev. D}\ }\textbf {\bibinfo {volume} {103}},\ \bibinfo
  {pages} {103018} (\bibinfo {year} {2021})},\ \Eprint
  {http://arxiv.org/abs/2101.09146} {arXiv:2101.09146 [astro-ph.HE]}
  \BibitemShut {NoStop}%
\bibitem [{\citenamefont {Pan}\ \emph {et~al.}(2021)\citenamefont {Pan},
  \citenamefont {Lyu},\ and\ \citenamefont {Yang}}]{Pan:2021oob}%
  \BibitemOpen
  \bibfield  {author} {\bibinfo {author} {\bibfnamefont {Z.}~\bibnamefont
  {Pan}}, \bibinfo {author} {\bibfnamefont {Z.}~\bibnamefont {Lyu}}, \ and\
  \bibinfo {author} {\bibfnamefont {H.}~\bibnamefont {Yang}},\ }\href {\doibase
  10.1103/PhysRevD.104.063007} {\bibfield  {journal} {\bibinfo  {journal}
  {Phys. Rev. D}\ }\textbf {\bibinfo {volume} {104}},\ \bibinfo {pages}
  {063007} (\bibinfo {year} {2021})},\ \Eprint
  {http://arxiv.org/abs/2104.01208} {arXiv:2104.01208 [astro-ph.HE]}
  \BibitemShut {NoStop}%
\bibitem [{\citenamefont {Pan}\ \emph {et~al.}(2022)\citenamefont {Pan},
  \citenamefont {Lyu},\ and\ \citenamefont {Yang}}]{Pan:2021lyw}%
  \BibitemOpen
  \bibfield  {author} {\bibinfo {author} {\bibfnamefont {Z.}~\bibnamefont
  {Pan}}, \bibinfo {author} {\bibfnamefont {Z.}~\bibnamefont {Lyu}}, \ and\
  \bibinfo {author} {\bibfnamefont {H.}~\bibnamefont {Yang}},\ }\href {\doibase
  10.1103/PhysRevD.105.083005} {\bibfield  {journal} {\bibinfo  {journal}
  {Phys. Rev. D}\ }\textbf {\bibinfo {volume} {105}},\ \bibinfo {pages}
  {083005} (\bibinfo {year} {2022})},\ \Eprint
  {http://arxiv.org/abs/2112.10237} {arXiv:2112.10237 [astro-ph.HE]}
  \BibitemShut {NoStop}%
\bibitem [{\citenamefont {Derdzinski}\ and\ \citenamefont
  {Mayer}(2022)}]{Derdzinski:2022ltb}%
  \BibitemOpen
  \bibfield  {author} {\bibinfo {author} {\bibfnamefont {A.}~\bibnamefont
  {Derdzinski}}\ and\ \bibinfo {author} {\bibfnamefont {L.}~\bibnamefont
  {Mayer}},\ }\href {\doibase 10.1093/mnras/stad749} {\  (\bibinfo {year}
  {2022}),\ 10.1093/mnras/stad749},\ \Eprint {http://arxiv.org/abs/2205.10382}
  {arXiv:2205.10382 [astro-ph.GA]} \BibitemShut {NoStop}%
\bibitem [{\citenamefont {Dahlen}\ \emph {et~al.}(2005)\citenamefont {Dahlen},
  \citenamefont {Mobasher}, \citenamefont {Somerville}, \citenamefont
  {Moustakas}, \citenamefont {Dickinson}, \citenamefont {Ferguson},\ and\
  \citenamefont {Giavalisco}}]{Dahlen:2005ex}%
  \BibitemOpen
  \bibfield  {author} {\bibinfo {author} {\bibfnamefont {T.}~\bibnamefont
  {Dahlen}}, \bibinfo {author} {\bibfnamefont {B.}~\bibnamefont {Mobasher}},
  \bibinfo {author} {\bibfnamefont {R.~S.}\ \bibnamefont {Somerville}},
  \bibinfo {author} {\bibfnamefont {L.~A.}\ \bibnamefont {Moustakas}}, \bibinfo
  {author} {\bibfnamefont {M.}~\bibnamefont {Dickinson}}, \bibinfo {author}
  {\bibfnamefont {H.~C.}\ \bibnamefont {Ferguson}}, \ and\ \bibinfo {author}
  {\bibfnamefont {M.}~\bibnamefont {Giavalisco}},\ }\href {\doibase
  10.1086/432027} {\bibfield  {journal} {\bibinfo  {journal} {Astrophys. J.}\
  }\textbf {\bibinfo {volume} {631}},\ \bibinfo {pages} {126} (\bibinfo {year}
  {2005})},\ \Eprint {http://arxiv.org/abs/astro-ph/0505297}
  {arXiv:astro-ph/0505297} \BibitemShut {NoStop}%
\bibitem [{\citenamefont {Hopkins}\ \emph {et~al.}(2007)\citenamefont
  {Hopkins}, \citenamefont {Richards},\ and\ \citenamefont
  {Hernquist}}]{Hopkins:2006fq}%
  \BibitemOpen
  \bibfield  {author} {\bibinfo {author} {\bibfnamefont {P.~F.}\ \bibnamefont
  {Hopkins}}, \bibinfo {author} {\bibfnamefont {G.~T.}\ \bibnamefont
  {Richards}}, \ and\ \bibinfo {author} {\bibfnamefont {L.}~\bibnamefont
  {Hernquist}},\ }\href {\doibase 10.1086/509629} {\bibfield  {journal}
  {\bibinfo  {journal} {Astrophys. J.}\ }\textbf {\bibinfo {volume} {654}},\
  \bibinfo {pages} {731} (\bibinfo {year} {2007})},\ \Eprint
  {http://arxiv.org/abs/astro-ph/0605678} {arXiv:astro-ph/0605678} \BibitemShut
  {NoStop}%
\bibitem [{\citenamefont {Derdzinski}\ and\ \citenamefont
  {Zwick}(2023)}]{Derdzinski:2023qbi}%
  \BibitemOpen
  \bibfield  {author} {\bibinfo {author} {\bibfnamefont {A.}~\bibnamefont
  {Derdzinski}}\ and\ \bibinfo {author} {\bibfnamefont {L.}~\bibnamefont
  {Zwick}}\ }(\bibinfo {year} {2023})\ \Eprint
  {http://arxiv.org/abs/2310.16900} {arXiv:2310.16900 [astro-ph.HE]}
  \BibitemShut {NoStop}%
\bibitem [{\citenamefont {Perivolaropoulos}\ and\ \citenamefont
  {Skara}(2022)}]{Perivolaropoulos:2021jda}%
  \BibitemOpen
  \bibfield  {author} {\bibinfo {author} {\bibfnamefont {L.}~\bibnamefont
  {Perivolaropoulos}}\ and\ \bibinfo {author} {\bibfnamefont {F.}~\bibnamefont
  {Skara}},\ }\href {\doibase 10.1016/j.newar.2022.101659} {\bibfield
  {journal} {\bibinfo  {journal} {New Astron. Rev.}\ }\textbf {\bibinfo
  {volume} {95}},\ \bibinfo {pages} {101659} (\bibinfo {year} {2022})},\
  \Eprint {http://arxiv.org/abs/2105.05208} {arXiv:2105.05208 [astro-ph.CO]}
  \BibitemShut {NoStop}%
\bibitem [{\citenamefont {Ade}\ \emph {et~al.}(2014)\citenamefont {Ade} \emph
  {et~al.}}]{Planck:2013oqw}%
  \BibitemOpen
  \bibfield  {author} {\bibinfo {author} {\bibfnamefont {P.~A.~R.}\
  \bibnamefont {Ade}} \emph {et~al.} (\bibinfo {collaboration} {Planck}),\
  }\href {\doibase 10.1051/0004-6361/201321529} {\bibfield  {journal} {\bibinfo
   {journal} {Astron. Astrophys.}\ }\textbf {\bibinfo {volume} {571}},\
  \bibinfo {pages} {A1} (\bibinfo {year} {2014})},\ \Eprint
  {http://arxiv.org/abs/1303.5062} {arXiv:1303.5062 [astro-ph.CO]} \BibitemShut
  {NoStop}%
\bibitem [{\citenamefont {Abbott}\ \emph
  {et~al.}(2018{\natexlab{b}})\citenamefont {Abbott} \emph
  {et~al.}}]{DES:2017txv}%
  \BibitemOpen
  \bibfield  {author} {\bibinfo {author} {\bibfnamefont {T.~M.~C.}\
  \bibnamefont {Abbott}} \emph {et~al.} (\bibinfo {collaboration} {DES}),\
  }\href {\doibase 10.1093/mnras/sty1939} {\bibfield  {journal} {\bibinfo
  {journal} {Mon. Not. Roy. Astron. Soc.}\ }\textbf {\bibinfo {volume} {480}},\
  \bibinfo {pages} {3879} (\bibinfo {year} {2018}{\natexlab{b}})},\ \Eprint
  {http://arxiv.org/abs/1711.00403} {arXiv:1711.00403 [astro-ph.CO]}
  \BibitemShut {NoStop}%
\bibitem [{\citenamefont {Aiola}\ \emph {et~al.}(2020)\citenamefont {Aiola}
  \emph {et~al.}}]{ACT:2020gnv}%
  \BibitemOpen
  \bibfield  {author} {\bibinfo {author} {\bibfnamefont {S.}~\bibnamefont
  {Aiola}} \emph {et~al.} (\bibinfo {collaboration} {ACT}),\ }\href {\doibase
  10.1088/1475-7516/2020/12/047} {\bibfield  {journal} {\bibinfo  {journal}
  {JCAP}\ }\textbf {\bibinfo {volume} {12}},\ \bibinfo {pages} {047} (\bibinfo
  {year} {2020})},\ \Eprint {http://arxiv.org/abs/2007.07288} {arXiv:2007.07288
  [astro-ph.CO]} \BibitemShut {NoStop}%
\bibitem [{\citenamefont {Riess}\ \emph {et~al.}(2016)\citenamefont {Riess}
  \emph {et~al.}}]{Riess:2016jrr}%
  \BibitemOpen
  \bibfield  {author} {\bibinfo {author} {\bibfnamefont {A.~G.}\ \bibnamefont
  {Riess}} \emph {et~al.},\ }\href {\doibase 10.3847/0004-637X/826/1/56}
  {\bibfield  {journal} {\bibinfo  {journal} {Astrophys. J.}\ }\textbf
  {\bibinfo {volume} {826}},\ \bibinfo {pages} {56} (\bibinfo {year} {2016})},\
  \Eprint {http://arxiv.org/abs/1604.01424} {arXiv:1604.01424 [astro-ph.CO]}
  \BibitemShut {NoStop}%
\bibitem [{\citenamefont {Freedman}\ and\ \citenamefont
  {Madore}(2010)}]{Freedman:2010xv}%
  \BibitemOpen
  \bibfield  {author} {\bibinfo {author} {\bibfnamefont {W.~L.}\ \bibnamefont
  {Freedman}}\ and\ \bibinfo {author} {\bibfnamefont {B.~F.}\ \bibnamefont
  {Madore}},\ }\href {\doibase 10.1146/annurev-astro-082708-101829} {\bibfield
  {journal} {\bibinfo  {journal} {Ann. Rev. Astron. Astrophys.}\ }\textbf
  {\bibinfo {volume} {48}},\ \bibinfo {pages} {673} (\bibinfo {year} {2010})},\
  \Eprint {http://arxiv.org/abs/1004.1856} {arXiv:1004.1856 [astro-ph.CO]}
  \BibitemShut {NoStop}%
\bibitem [{\citenamefont {Soltis}\ \emph {et~al.}(2021)\citenamefont {Soltis},
  \citenamefont {Casertano},\ and\ \citenamefont {Riess}}]{Soltis:2020gpl}%
  \BibitemOpen
  \bibfield  {author} {\bibinfo {author} {\bibfnamefont {J.}~\bibnamefont
  {Soltis}}, \bibinfo {author} {\bibfnamefont {S.}~\bibnamefont {Casertano}}, \
  and\ \bibinfo {author} {\bibfnamefont {A.~G.}\ \bibnamefont {Riess}},\ }\href
  {\doibase 10.3847/2041-8213/abdbad} {\bibfield  {journal} {\bibinfo
  {journal} {Astrophys. J. Lett.}\ }\textbf {\bibinfo {volume} {908}},\
  \bibinfo {pages} {L5} (\bibinfo {year} {2021})},\ \Eprint
  {http://arxiv.org/abs/2012.09196} {arXiv:2012.09196 [astro-ph.GA]}
  \BibitemShut {NoStop}%
\bibitem [{\citenamefont {Blakeslee}\ \emph {et~al.}(2021)\citenamefont
  {Blakeslee}, \citenamefont {Jensen}, \citenamefont {Ma}, \citenamefont
  {Milne},\ and\ \citenamefont {Greene}}]{Blakeslee:2021rqi}%
  \BibitemOpen
  \bibfield  {author} {\bibinfo {author} {\bibfnamefont {J.~P.}\ \bibnamefont
  {Blakeslee}}, \bibinfo {author} {\bibfnamefont {J.~B.}\ \bibnamefont
  {Jensen}}, \bibinfo {author} {\bibfnamefont {C.-P.}\ \bibnamefont {Ma}},
  \bibinfo {author} {\bibfnamefont {P.~A.}\ \bibnamefont {Milne}}, \ and\
  \bibinfo {author} {\bibfnamefont {J.~E.}\ \bibnamefont {Greene}},\ }\href
  {\doibase 10.3847/1538-4357/abe86a} {\bibfield  {journal} {\bibinfo
  {journal} {Astrophys. J.}\ }\textbf {\bibinfo {volume} {911}},\ \bibinfo
  {pages} {65} (\bibinfo {year} {2021})},\ \Eprint
  {http://arxiv.org/abs/2101.02221} {arXiv:2101.02221 [astro-ph.CO]}
  \BibitemShut {NoStop}%
\bibitem [{\citenamefont {Choi}\ \emph {et~al.}(2020)\citenamefont {Choi} \emph
  {et~al.}}]{ACT:2020frw}%
  \BibitemOpen
  \bibfield  {author} {\bibinfo {author} {\bibfnamefont {S.~K.}\ \bibnamefont
  {Choi}} \emph {et~al.} (\bibinfo {collaboration} {ACT}),\ }\href {\doibase
  10.1088/1475-7516/2020/12/045} {\bibfield  {journal} {\bibinfo  {journal}
  {JCAP}\ }\textbf {\bibinfo {volume} {12}},\ \bibinfo {pages} {045} (\bibinfo
  {year} {2020})},\ \Eprint {http://arxiv.org/abs/2007.07289} {arXiv:2007.07289
  [astro-ph.CO]} \BibitemShut {NoStop}%
\bibitem [{\citenamefont {Riess}\ \emph {et~al.}(2024)\citenamefont {Riess},
  \citenamefont {Anand}, \citenamefont {Yuan}, \citenamefont {Casertano},
  \citenamefont {Dolphin}, \citenamefont {Macri}, \citenamefont {Breuval},
  \citenamefont {Scolnic}, \citenamefont {Perrin},\ and\ \citenamefont
  {Anderson}}]{Riess:2024ohe}%
  \BibitemOpen
  \bibfield  {author} {\bibinfo {author} {\bibfnamefont {A.~G.}\ \bibnamefont
  {Riess}}, \bibinfo {author} {\bibfnamefont {G.~S.}\ \bibnamefont {Anand}},
  \bibinfo {author} {\bibfnamefont {W.}~\bibnamefont {Yuan}}, \bibinfo {author}
  {\bibfnamefont {S.}~\bibnamefont {Casertano}}, \bibinfo {author}
  {\bibfnamefont {A.}~\bibnamefont {Dolphin}}, \bibinfo {author} {\bibfnamefont
  {L.~M.}\ \bibnamefont {Macri}}, \bibinfo {author} {\bibfnamefont
  {L.}~\bibnamefont {Breuval}}, \bibinfo {author} {\bibfnamefont
  {D.}~\bibnamefont {Scolnic}}, \bibinfo {author} {\bibfnamefont
  {M.}~\bibnamefont {Perrin}}, \ and\ \bibinfo {author} {\bibfnamefont {I.~R.}\
  \bibnamefont {Anderson}},\ }\href {\doibase 10.3847/2041-8213/ad1ddd}
  {\bibfield  {journal} {\bibinfo  {journal} {Astrophys. J. Lett.}\ }\textbf
  {\bibinfo {volume} {962}},\ \bibinfo {pages} {L17} (\bibinfo {year}
  {2024})},\ \Eprint {http://arxiv.org/abs/2401.04773} {arXiv:2401.04773
  [astro-ph.CO]} \BibitemShut {NoStop}%
\bibitem [{\citenamefont {Feeney}\ \emph {et~al.}(2019)\citenamefont {Feeney},
  \citenamefont {Peiris}, \citenamefont {Williamson}, \citenamefont {Nissanke},
  \citenamefont {Mortlock}, \citenamefont {Alsing},\ and\ \citenamefont
  {Scolnic}}]{Feeney:2018mkj}%
  \BibitemOpen
  \bibfield  {author} {\bibinfo {author} {\bibfnamefont {S.~M.}\ \bibnamefont
  {Feeney}}, \bibinfo {author} {\bibfnamefont {H.~V.}\ \bibnamefont {Peiris}},
  \bibinfo {author} {\bibfnamefont {A.~R.}\ \bibnamefont {Williamson}},
  \bibinfo {author} {\bibfnamefont {S.~M.}\ \bibnamefont {Nissanke}}, \bibinfo
  {author} {\bibfnamefont {D.~J.}\ \bibnamefont {Mortlock}}, \bibinfo {author}
  {\bibfnamefont {J.}~\bibnamefont {Alsing}}, \ and\ \bibinfo {author}
  {\bibfnamefont {D.}~\bibnamefont {Scolnic}},\ }\href {\doibase
  10.1103/PhysRevLett.122.061105} {\bibfield  {journal} {\bibinfo  {journal}
  {Phys. Rev. Lett.}\ }\textbf {\bibinfo {volume} {122}},\ \bibinfo {pages}
  {061105} (\bibinfo {year} {2019})},\ \Eprint
  {http://arxiv.org/abs/1802.03404} {arXiv:1802.03404 [astro-ph.CO]}
  \BibitemShut {NoStop}%
\bibitem [{\citenamefont {Califano}\ \emph {et~al.}(2023)\citenamefont
  {Califano}, \citenamefont {de~Martino}, \citenamefont {Vernieri},\ and\
  \citenamefont {Capozziello}}]{Califano:2022syd}%
  \BibitemOpen
  \bibfield  {author} {\bibinfo {author} {\bibfnamefont {M.}~\bibnamefont
  {Califano}}, \bibinfo {author} {\bibfnamefont {I.}~\bibnamefont
  {de~Martino}}, \bibinfo {author} {\bibfnamefont {D.}~\bibnamefont
  {Vernieri}}, \ and\ \bibinfo {author} {\bibfnamefont {S.}~\bibnamefont
  {Capozziello}},\ }\href {\doibase 10.1103/PhysRevD.107.123519} {\bibfield
  {journal} {\bibinfo  {journal} {Phys. Rev. D}\ }\textbf {\bibinfo {volume}
  {107}},\ \bibinfo {pages} {123519} (\bibinfo {year} {2023})},\ \Eprint
  {http://arxiv.org/abs/2208.13999} {arXiv:2208.13999 [astro-ph.CO]}
  \BibitemShut {NoStop}%
\bibitem [{\citenamefont {Gupta}(2023)}]{Gupta:2022fwd}%
  \BibitemOpen
  \bibfield  {author} {\bibinfo {author} {\bibfnamefont {I.}~\bibnamefont
  {Gupta}},\ }\href {\doibase 10.1093/mnras/stad2115} {\bibfield  {journal}
  {\bibinfo  {journal} {Mon. Not. Roy. Astron. Soc.}\ }\textbf {\bibinfo
  {volume} {524}},\ \bibinfo {pages} {3537} (\bibinfo {year} {2023})},\ \Eprint
  {http://arxiv.org/abs/2212.00163} {arXiv:2212.00163 [gr-qc]} \BibitemShut
  {NoStop}%
\bibitem [{\citenamefont {Yang}(2021)}]{Yang:2021qge}%
  \BibitemOpen
  \bibfield  {author} {\bibinfo {author} {\bibfnamefont {T.}~\bibnamefont
  {Yang}},\ }\href {\doibase 10.1088/1475-7516/2021/05/044} {\bibfield
  {journal} {\bibinfo  {journal} {JCAP}\ }\textbf {\bibinfo {volume} {05}},\
  \bibinfo {pages} {044} (\bibinfo {year} {2021})},\ \Eprint
  {http://arxiv.org/abs/2103.01923} {arXiv:2103.01923 [astro-ph.CO]}
  \BibitemShut {NoStop}%
\bibitem [{\citenamefont {Cai}\ and\ \citenamefont {Yang}(2021)}]{Cai:2021ooo}%
  \BibitemOpen
  \bibfield  {author} {\bibinfo {author} {\bibfnamefont {R.-G.}\ \bibnamefont
  {Cai}}\ and\ \bibinfo {author} {\bibfnamefont {T.}~\bibnamefont {Yang}},\
  }\href {\doibase 10.1088/1475-7516/2021/12/017} {\bibfield  {journal}
  {\bibinfo  {journal} {JCAP}\ }\textbf {\bibinfo {volume} {12}},\ \bibinfo
  {pages} {017} (\bibinfo {year} {2021})},\ \Eprint
  {http://arxiv.org/abs/2107.13919} {arXiv:2107.13919 [gr-qc]} \BibitemShut
  {NoStop}%
\bibitem [{\citenamefont {Yang}\ \emph
  {et~al.}(2022{\natexlab{a}})\citenamefont {Yang}, \citenamefont {Lee},
  \citenamefont {Cai}, \citenamefont {Choi},\ and\ \citenamefont
  {Jung}}]{Yang:2021xox}%
  \BibitemOpen
  \bibfield  {author} {\bibinfo {author} {\bibfnamefont {T.}~\bibnamefont
  {Yang}}, \bibinfo {author} {\bibfnamefont {H.~M.}\ \bibnamefont {Lee}},
  \bibinfo {author} {\bibfnamefont {R.-G.}\ \bibnamefont {Cai}}, \bibinfo
  {author} {\bibfnamefont {H.~G.}\ \bibnamefont {Choi}}, \ and\ \bibinfo
  {author} {\bibfnamefont {S.}~\bibnamefont {Jung}},\ }\href {\doibase
  10.1088/1475-7516/2022/01/042} {\bibfield  {journal} {\bibinfo  {journal}
  {JCAP}\ }\textbf {\bibinfo {volume} {01}},\ \bibinfo {pages} {042} (\bibinfo
  {year} {2022}{\natexlab{a}})},\ \Eprint {http://arxiv.org/abs/2110.09967}
  {arXiv:2110.09967 [gr-qc]} \BibitemShut {NoStop}%
\bibitem [{\citenamefont {Yang}\ \emph
  {et~al.}(2022{\natexlab{b}})\citenamefont {Yang}, \citenamefont {Cai},\ and\
  \citenamefont {Lee}}]{Yang:2022iwn}%
  \BibitemOpen
  \bibfield  {author} {\bibinfo {author} {\bibfnamefont {T.}~\bibnamefont
  {Yang}}, \bibinfo {author} {\bibfnamefont {R.-G.}\ \bibnamefont {Cai}}, \
  and\ \bibinfo {author} {\bibfnamefont {H.~M.}\ \bibnamefont {Lee}},\ }\href
  {\doibase 10.1088/1475-7516/2022/10/061} {\bibfield  {journal} {\bibinfo
  {journal} {JCAP}\ }\textbf {\bibinfo {volume} {10}},\ \bibinfo {pages} {061}
  (\bibinfo {year} {2022}{\natexlab{b}})},\ \Eprint
  {http://arxiv.org/abs/2208.10998} {arXiv:2208.10998 [gr-qc]} \BibitemShut
  {NoStop}%
\bibitem [{\citenamefont {Wong}\ \emph {et~al.}(2018)\citenamefont {Wong},
  \citenamefont {Kovetz}, \citenamefont {Cutler},\ and\ \citenamefont
  {Berti}}]{Wong:2018uwb}%
  \BibitemOpen
  \bibfield  {author} {\bibinfo {author} {\bibfnamefont {K.~W.~K.}\
  \bibnamefont {Wong}}, \bibinfo {author} {\bibfnamefont {E.~D.}\ \bibnamefont
  {Kovetz}}, \bibinfo {author} {\bibfnamefont {C.}~\bibnamefont {Cutler}}, \
  and\ \bibinfo {author} {\bibfnamefont {E.}~\bibnamefont {Berti}},\ }\href
  {\doibase 10.1103/PhysRevLett.121.251102} {\bibfield  {journal} {\bibinfo
  {journal} {Phys. Rev. Lett.}\ }\textbf {\bibinfo {volume} {121}},\ \bibinfo
  {pages} {251102} (\bibinfo {year} {2018})},\ \Eprint
  {http://arxiv.org/abs/1808.08247} {arXiv:1808.08247 [astro-ph.HE]}
  \BibitemShut {NoStop}%
\bibitem [{\citenamefont {Ewing}\ \emph {et~al.}(2021)\citenamefont {Ewing},
  \citenamefont {Sachdev}, \citenamefont {Borhanian},\ and\ \citenamefont
  {Sathyaprakash}}]{Ewing:2020brd}%
  \BibitemOpen
  \bibfield  {author} {\bibinfo {author} {\bibfnamefont {B.}~\bibnamefont
  {Ewing}}, \bibinfo {author} {\bibfnamefont {S.}~\bibnamefont {Sachdev}},
  \bibinfo {author} {\bibfnamefont {S.}~\bibnamefont {Borhanian}}, \ and\
  \bibinfo {author} {\bibfnamefont {B.~S.}\ \bibnamefont {Sathyaprakash}},\
  }\href {\doibase 10.1103/PhysRevD.103.023025} {\bibfield  {journal} {\bibinfo
   {journal} {Phys. Rev. D}\ }\textbf {\bibinfo {volume} {103}},\ \bibinfo
  {pages} {023025} (\bibinfo {year} {2021})},\ \Eprint
  {http://arxiv.org/abs/2011.03036} {arXiv:2011.03036 [gr-qc]} \BibitemShut
  {NoStop}%
\bibitem [{\citenamefont {Hubble}(1929)}]{Hubble:1929ig}%
  \BibitemOpen
  \bibfield  {author} {\bibinfo {author} {\bibfnamefont {E.}~\bibnamefont
  {Hubble}},\ }\href {\doibase 10.1073/pnas.15.3.168} {\bibfield  {journal}
  {\bibinfo  {journal} {Proc. Nat. Acad. Sci.}\ }\textbf {\bibinfo {volume}
  {15}},\ \bibinfo {pages} {168} (\bibinfo {year} {1929})}\BibitemShut
  {NoStop}%
\bibitem [{\citenamefont {Perlmutter}\ \emph {et~al.}(1999)\citenamefont
  {Perlmutter} \emph {et~al.}}]{SupernovaCosmologyProject:1998vns}%
  \BibitemOpen
  \bibfield  {author} {\bibinfo {author} {\bibfnamefont {S.}~\bibnamefont
  {Perlmutter}} \emph {et~al.} (\bibinfo {collaboration} {Supernova Cosmology
  Project}),\ }\href {\doibase 10.1086/307221} {\bibfield  {journal} {\bibinfo
  {journal} {Astrophys. J.}\ }\textbf {\bibinfo {volume} {517}},\ \bibinfo
  {pages} {565} (\bibinfo {year} {1999})},\ \Eprint
  {http://arxiv.org/abs/astro-ph/9812133} {arXiv:astro-ph/9812133} \BibitemShut
  {NoStop}%
\bibitem [{\citenamefont {Riess}\ \emph {et~al.}(1998)\citenamefont {Riess}
  \emph {et~al.}}]{SupernovaSearchTeam:1998fmf}%
  \BibitemOpen
  \bibfield  {author} {\bibinfo {author} {\bibfnamefont {A.~G.}\ \bibnamefont
  {Riess}} \emph {et~al.} (\bibinfo {collaboration} {Supernova Search Team}),\
  }\href {\doibase 10.1086/300499} {\bibfield  {journal} {\bibinfo  {journal}
  {Astron. J.}\ }\textbf {\bibinfo {volume} {116}},\ \bibinfo {pages} {1009}
  (\bibinfo {year} {1998})},\ \Eprint {http://arxiv.org/abs/astro-ph/9805201}
  {arXiv:astro-ph/9805201} \BibitemShut {NoStop}%
\bibitem [{\citenamefont {Frieman}\ \emph {et~al.}(2008)\citenamefont
  {Frieman}, \citenamefont {Turner},\ and\ \citenamefont
  {Huterer}}]{Frieman:2008sn}%
  \BibitemOpen
  \bibfield  {author} {\bibinfo {author} {\bibfnamefont {J.}~\bibnamefont
  {Frieman}}, \bibinfo {author} {\bibfnamefont {M.}~\bibnamefont {Turner}}, \
  and\ \bibinfo {author} {\bibfnamefont {D.}~\bibnamefont {Huterer}},\ }\href
  {\doibase 10.1146/annurev.astro.46.060407.145243} {\bibfield  {journal}
  {\bibinfo  {journal} {Ann. Rev. Astron. Astrophys.}\ }\textbf {\bibinfo
  {volume} {46}},\ \bibinfo {pages} {385} (\bibinfo {year} {2008})},\ \Eprint
  {http://arxiv.org/abs/0803.0982} {arXiv:0803.0982 [astro-ph]} \BibitemShut
  {NoStop}%
\bibitem [{\citenamefont {Li}\ \emph {et~al.}(2011{\natexlab{a}})\citenamefont
  {Li}, \citenamefont {Li}, \citenamefont {Wang},\ and\ \citenamefont
  {Wang}}]{Li:2011sd}%
  \BibitemOpen
  \bibfield  {author} {\bibinfo {author} {\bibfnamefont {M.}~\bibnamefont
  {Li}}, \bibinfo {author} {\bibfnamefont {X.-D.}\ \bibnamefont {Li}}, \bibinfo
  {author} {\bibfnamefont {S.}~\bibnamefont {Wang}}, \ and\ \bibinfo {author}
  {\bibfnamefont {Y.}~\bibnamefont {Wang}},\ }\href {\doibase
  10.1088/0253-6102/56/3/24} {\bibfield  {journal} {\bibinfo  {journal}
  {Commun. Theor. Phys.}\ }\textbf {\bibinfo {volume} {56}},\ \bibinfo {pages}
  {525} (\bibinfo {year} {2011}{\natexlab{a}})},\ \Eprint
  {http://arxiv.org/abs/1103.5870} {arXiv:1103.5870 [astro-ph.CO]} \BibitemShut
  {NoStop}%
\bibitem [{\citenamefont {Weinberg}\ \emph {et~al.}(2013)\citenamefont
  {Weinberg}, \citenamefont {Mortonson}, \citenamefont {Eisenstein},
  \citenamefont {Hirata}, \citenamefont {Riess},\ and\ \citenamefont
  {Rozo}}]{Weinberg:2013agg}%
  \BibitemOpen
  \bibfield  {author} {\bibinfo {author} {\bibfnamefont {D.~H.}\ \bibnamefont
  {Weinberg}}, \bibinfo {author} {\bibfnamefont {M.~J.}\ \bibnamefont
  {Mortonson}}, \bibinfo {author} {\bibfnamefont {D.~J.}\ \bibnamefont
  {Eisenstein}}, \bibinfo {author} {\bibfnamefont {C.}~\bibnamefont {Hirata}},
  \bibinfo {author} {\bibfnamefont {A.~G.}\ \bibnamefont {Riess}}, \ and\
  \bibinfo {author} {\bibfnamefont {E.}~\bibnamefont {Rozo}},\ }\href {\doibase
  10.1016/j.physrep.2013.05.001} {\bibfield  {journal} {\bibinfo  {journal}
  {Phys. Rept.}\ }\textbf {\bibinfo {volume} {530}},\ \bibinfo {pages} {87}
  (\bibinfo {year} {2013})},\ \Eprint {http://arxiv.org/abs/1201.2434}
  {arXiv:1201.2434 [astro-ph.CO]} \BibitemShut {NoStop}%
\bibitem [{\citenamefont {Abbott}\ \emph
  {et~al.}(2022{\natexlab{b}})\citenamefont {Abbott} \emph
  {et~al.}}]{DES:2021wwk}%
  \BibitemOpen
  \bibfield  {author} {\bibinfo {author} {\bibfnamefont {T.~M.~C.}\
  \bibnamefont {Abbott}} \emph {et~al.} (\bibinfo {collaboration} {DES}),\
  }\href {\doibase 10.1103/PhysRevD.105.023520} {\bibfield  {journal} {\bibinfo
   {journal} {Phys. Rev. D}\ }\textbf {\bibinfo {volume} {105}},\ \bibinfo
  {pages} {023520} (\bibinfo {year} {2022}{\natexlab{b}})},\ \Eprint
  {http://arxiv.org/abs/2105.13549} {arXiv:2105.13549 [astro-ph.CO]}
  \BibitemShut {NoStop}%
\bibitem [{\citenamefont {Brout}\ \emph {et~al.}(2022)\citenamefont {Brout}
  \emph {et~al.}}]{Brout:2022vxf}%
  \BibitemOpen
  \bibfield  {author} {\bibinfo {author} {\bibfnamefont {D.}~\bibnamefont
  {Brout}} \emph {et~al.},\ }\href {\doibase 10.3847/1538-4357/ac8e04}
  {\bibfield  {journal} {\bibinfo  {journal} {Astrophys. J.}\ }\textbf
  {\bibinfo {volume} {938}},\ \bibinfo {pages} {110} (\bibinfo {year}
  {2022})},\ \Eprint {http://arxiv.org/abs/2202.04077} {arXiv:2202.04077
  [astro-ph.CO]} \BibitemShut {NoStop}%
\bibitem [{\citenamefont {Abbott}\ \emph
  {et~al.}(2024{\natexlab{b}})\citenamefont {Abbott} \emph
  {et~al.}}]{DES:2024tys}%
  \BibitemOpen
  \bibfield  {author} {\bibinfo {author} {\bibfnamefont {T.~M.~C.}\
  \bibnamefont {Abbott}} \emph {et~al.} (\bibinfo {collaboration} {DES}),\
  }\href@noop {} {\  (\bibinfo {year} {2024}{\natexlab{b}})},\ \Eprint
  {http://arxiv.org/abs/2401.02929} {arXiv:2401.02929 [astro-ph.CO]}
  \BibitemShut {NoStop}%
\bibitem [{\citenamefont {Zhao}\ \emph {et~al.}(2012)\citenamefont {Zhao},
  \citenamefont {Crittenden}, \citenamefont {Pogosian},\ and\ \citenamefont
  {Zhang}}]{Zhao:2012aw}%
  \BibitemOpen
  \bibfield  {author} {\bibinfo {author} {\bibfnamefont {G.-B.}\ \bibnamefont
  {Zhao}}, \bibinfo {author} {\bibfnamefont {R.~G.}\ \bibnamefont
  {Crittenden}}, \bibinfo {author} {\bibfnamefont {L.}~\bibnamefont
  {Pogosian}}, \ and\ \bibinfo {author} {\bibfnamefont {X.}~\bibnamefont
  {Zhang}},\ }\href {\doibase 10.1103/PhysRevLett.109.171301} {\bibfield
  {journal} {\bibinfo  {journal} {Phys. Rev. Lett.}\ }\textbf {\bibinfo
  {volume} {109}},\ \bibinfo {pages} {171301} (\bibinfo {year} {2012})},\
  \Eprint {http://arxiv.org/abs/1207.3804} {arXiv:1207.3804 [astro-ph.CO]}
  \BibitemShut {NoStop}%
\bibitem [{\citenamefont {Zhu}(2022)}]{Zhu:2022gwc}%
  \BibitemOpen
  \bibfield  {author} {\bibinfo {author} {\bibfnamefont {L.-G.}\ \bibnamefont
  {Zhu}},\ }\emph {\bibinfo {title} {{Researching Cosmic Expansion with
  TianQin}}},\ \href@noop {} {Ph.D. thesis},\ \bibinfo  {school} {Sun Yat-sen
  University} (\bibinfo {year} {2022})\BibitemShut {NoStop}%
\bibitem [{\citenamefont {Mei}\ \emph {et~al.}(2021)\citenamefont {Mei} \emph
  {et~al.}}]{TianQin:2020hid}%
  \BibitemOpen
  \bibfield  {author} {\bibinfo {author} {\bibfnamefont {J.}~\bibnamefont
  {Mei}} \emph {et~al.} (\bibinfo {collaboration} {TianQin}),\ }\href {\doibase
  10.1093/ptep/ptaa114} {\bibfield  {journal} {\bibinfo  {journal} {PTEP}\
  }\textbf {\bibinfo {volume} {2021}},\ \bibinfo {pages} {05A107} (\bibinfo
  {year} {2021})},\ \Eprint {http://arxiv.org/abs/2008.10332} {arXiv:2008.10332
  [gr-qc]} \BibitemShut {NoStop}%
\bibitem [{\citenamefont {Wu}\ \emph {et~al.}(2021)\citenamefont {Wu} \emph
  {et~al.}}]{TaijiScientific:2021qgx}%
  \BibitemOpen
  \bibfield  {author} {\bibinfo {author} {\bibfnamefont {Y.-L.}\ \bibnamefont
  {Wu}} \emph {et~al.} (\bibinfo {collaboration} {Taiji Scientific}),\ }\href
  {\doibase 10.1038/s42005-021-00529-z} {\bibfield  {journal} {\bibinfo
  {journal} {Commun. Phys.}\ }\textbf {\bibinfo {volume} {4}},\ \bibinfo
  {pages} {34} (\bibinfo {year} {2021})}\BibitemShut {NoStop}%
\bibitem [{\citenamefont {Wang}\ \emph
  {et~al.}(2022{\natexlab{e}})\citenamefont {Wang}, \citenamefont {Jin},
  \citenamefont {Zhang},\ and\ \citenamefont {Zhang}}]{Wang:2021srv}%
  \BibitemOpen
  \bibfield  {author} {\bibinfo {author} {\bibfnamefont {L.-F.}\ \bibnamefont
  {Wang}}, \bibinfo {author} {\bibfnamefont {S.-J.}\ \bibnamefont {Jin}},
  \bibinfo {author} {\bibfnamefont {J.-F.}\ \bibnamefont {Zhang}}, \ and\
  \bibinfo {author} {\bibfnamefont {X.}~\bibnamefont {Zhang}},\ }\href
  {\doibase 10.1007/s11433-021-1736-6} {\bibfield  {journal} {\bibinfo
  {journal} {Sci. China Phys. Mech. Astron.}\ }\textbf {\bibinfo {volume}
  {65}},\ \bibinfo {pages} {210411} (\bibinfo {year} {2022}{\natexlab{e}})},\
  \Eprint {http://arxiv.org/abs/2101.11882} {arXiv:2101.11882 [gr-qc]}
  \BibitemShut {NoStop}%
\bibitem [{\citenamefont {Jin}\ \emph {et~al.}(2024)\citenamefont {Jin},
  \citenamefont {Zhang}, \citenamefont {Song}, \citenamefont {Zhang},\ and\
  \citenamefont {Zhang}}]{Jin:2023sfc}%
  \BibitemOpen
  \bibfield  {author} {\bibinfo {author} {\bibfnamefont {S.-J.}\ \bibnamefont
  {Jin}}, \bibinfo {author} {\bibfnamefont {Y.-Z.}\ \bibnamefont {Zhang}},
  \bibinfo {author} {\bibfnamefont {J.-Y.}\ \bibnamefont {Song}}, \bibinfo
  {author} {\bibfnamefont {J.-F.}\ \bibnamefont {Zhang}}, \ and\ \bibinfo
  {author} {\bibfnamefont {X.}~\bibnamefont {Zhang}},\ }\href {\doibase
  10.1007/s11433-023-2276-1} {\bibfield  {journal} {\bibinfo  {journal} {Sci.
  China Phys. Mech. Astron.}\ }\textbf {\bibinfo {volume} {67}},\ \bibinfo
  {pages} {220412} (\bibinfo {year} {2024})},\ \Eprint
  {http://arxiv.org/abs/2305.19714} {arXiv:2305.19714 [astro-ph.CO]}
  \BibitemShut {NoStop}%
\bibitem [{\citenamefont {Wang}\ \emph
  {et~al.}(2019{\natexlab{b}})\citenamefont {Wang}, \citenamefont {Zhao},
  \citenamefont {Zhang},\ and\ \citenamefont {Zhang}}]{Wang:2019tto}%
  \BibitemOpen
  \bibfield  {author} {\bibinfo {author} {\bibfnamefont {L.-F.}\ \bibnamefont
  {Wang}}, \bibinfo {author} {\bibfnamefont {Z.-W.}\ \bibnamefont {Zhao}},
  \bibinfo {author} {\bibfnamefont {J.-F.}\ \bibnamefont {Zhang}}, \ and\
  \bibinfo {author} {\bibfnamefont {X.}~\bibnamefont {Zhang}},\ }\href@noop {}
  {\  (\bibinfo {year} {2019}{\natexlab{b}})},\ \Eprint
  {http://arxiv.org/abs/1907.01838} {arXiv:1907.01838 [astro-ph.CO]}
  \BibitemShut {NoStop}%
\bibitem [{\citenamefont {Zhang}\ \emph
  {et~al.}(2019{\natexlab{b}})\citenamefont {Zhang}, \citenamefont {Wang},
  \citenamefont {Zhang},\ and\ \citenamefont {Zhang}}]{Zhang:2018byx}%
  \BibitemOpen
  \bibfield  {author} {\bibinfo {author} {\bibfnamefont {X.-N.}\ \bibnamefont
  {Zhang}}, \bibinfo {author} {\bibfnamefont {L.-F.}\ \bibnamefont {Wang}},
  \bibinfo {author} {\bibfnamefont {J.-F.}\ \bibnamefont {Zhang}}, \ and\
  \bibinfo {author} {\bibfnamefont {X.}~\bibnamefont {Zhang}},\ }\href
  {\doibase 10.1103/PhysRevD.99.063510} {\bibfield  {journal} {\bibinfo
  {journal} {Phys. Rev. D}\ }\textbf {\bibinfo {volume} {99}},\ \bibinfo
  {pages} {063510} (\bibinfo {year} {2019}{\natexlab{b}})},\ \Eprint
  {http://arxiv.org/abs/1804.08379} {arXiv:1804.08379 [astro-ph.CO]}
  \BibitemShut {NoStop}%
\bibitem [{\citenamefont {Wang}\ \emph {et~al.}(2018)\citenamefont {Wang},
  \citenamefont {Zhang}, \citenamefont {Zhang},\ and\ \citenamefont
  {Zhang}}]{Wang:2018lun}%
  \BibitemOpen
  \bibfield  {author} {\bibinfo {author} {\bibfnamefont {L.-F.}\ \bibnamefont
  {Wang}}, \bibinfo {author} {\bibfnamefont {X.-N.}\ \bibnamefont {Zhang}},
  \bibinfo {author} {\bibfnamefont {J.-F.}\ \bibnamefont {Zhang}}, \ and\
  \bibinfo {author} {\bibfnamefont {X.}~\bibnamefont {Zhang}},\ }\href
  {\doibase 10.1016/j.physletb.2018.05.027} {\bibfield  {journal} {\bibinfo
  {journal} {Phys. Lett. B}\ }\textbf {\bibinfo {volume} {782}},\ \bibinfo
  {pages} {87} (\bibinfo {year} {2018})},\ \Eprint
  {http://arxiv.org/abs/1802.04720} {arXiv:1802.04720 [astro-ph.CO]}
  \BibitemShut {NoStop}%
\bibitem [{\citenamefont {Zhang}\ \emph
  {et~al.}(2020{\natexlab{b}})\citenamefont {Zhang}, \citenamefont {Dong},
  \citenamefont {Qi},\ and\ \citenamefont {Zhang}}]{Zhang:2019ple}%
  \BibitemOpen
  \bibfield  {author} {\bibinfo {author} {\bibfnamefont {J.-F.}\ \bibnamefont
  {Zhang}}, \bibinfo {author} {\bibfnamefont {H.-Y.}\ \bibnamefont {Dong}},
  \bibinfo {author} {\bibfnamefont {J.-Z.}\ \bibnamefont {Qi}}, \ and\ \bibinfo
  {author} {\bibfnamefont {X.}~\bibnamefont {Zhang}},\ }\href {\doibase
  10.1140/epjc/s10052-020-7767-3} {\bibfield  {journal} {\bibinfo  {journal}
  {Eur. Phys. J. C}\ }\textbf {\bibinfo {volume} {80}},\ \bibinfo {pages} {217}
  (\bibinfo {year} {2020}{\natexlab{b}})},\ \Eprint
  {http://arxiv.org/abs/1906.07504} {arXiv:1906.07504 [astro-ph.CO]}
  \BibitemShut {NoStop}%
\bibitem [{\citenamefont {Zhang}\ \emph
  {et~al.}(2019{\natexlab{c}})\citenamefont {Zhang}, \citenamefont {Zhang},
  \citenamefont {Jin}, \citenamefont {Qi},\ and\ \citenamefont
  {Zhang}}]{Zhang:2019loq}%
  \BibitemOpen
  \bibfield  {author} {\bibinfo {author} {\bibfnamefont {J.-F.}\ \bibnamefont
  {Zhang}}, \bibinfo {author} {\bibfnamefont {M.}~\bibnamefont {Zhang}},
  \bibinfo {author} {\bibfnamefont {S.-J.}\ \bibnamefont {Jin}}, \bibinfo
  {author} {\bibfnamefont {J.-Z.}\ \bibnamefont {Qi}}, \ and\ \bibinfo {author}
  {\bibfnamefont {X.}~\bibnamefont {Zhang}},\ }\href {\doibase
  10.1088/1475-7516/2019/09/068} {\bibfield  {journal} {\bibinfo  {journal}
  {JCAP}\ }\textbf {\bibinfo {volume} {09}},\ \bibinfo {pages} {068} (\bibinfo
  {year} {2019}{\natexlab{c}})},\ \Eprint {http://arxiv.org/abs/1907.03238}
  {arXiv:1907.03238 [astro-ph.CO]} \BibitemShut {NoStop}%
\bibitem [{\citenamefont {Li}\ \emph {et~al.}(2020)\citenamefont {Li},
  \citenamefont {He}, \citenamefont {Zhang},\ and\ \citenamefont
  {Zhang}}]{Li:2019ajo}%
  \BibitemOpen
  \bibfield  {author} {\bibinfo {author} {\bibfnamefont {H.-L.}\ \bibnamefont
  {Li}}, \bibinfo {author} {\bibfnamefont {D.-Z.}\ \bibnamefont {He}}, \bibinfo
  {author} {\bibfnamefont {J.-F.}\ \bibnamefont {Zhang}}, \ and\ \bibinfo
  {author} {\bibfnamefont {X.}~\bibnamefont {Zhang}},\ }\href {\doibase
  10.1088/1475-7516/2020/06/038} {\bibfield  {journal} {\bibinfo  {journal}
  {JCAP}\ }\textbf {\bibinfo {volume} {06}},\ \bibinfo {pages} {038} (\bibinfo
  {year} {2020})},\ \Eprint {http://arxiv.org/abs/1908.03098} {arXiv:1908.03098
  [astro-ph.CO]} \BibitemShut {NoStop}%
\bibitem [{\citenamefont {Jin}\ \emph {et~al.}(2020)\citenamefont {Jin},
  \citenamefont {He}, \citenamefont {Xu}, \citenamefont {Zhang},\ and\
  \citenamefont {Zhang}}]{Jin:2020hmc}%
  \BibitemOpen
  \bibfield  {author} {\bibinfo {author} {\bibfnamefont {S.-J.}\ \bibnamefont
  {Jin}}, \bibinfo {author} {\bibfnamefont {D.-Z.}\ \bibnamefont {He}},
  \bibinfo {author} {\bibfnamefont {Y.}~\bibnamefont {Xu}}, \bibinfo {author}
  {\bibfnamefont {J.-F.}\ \bibnamefont {Zhang}}, \ and\ \bibinfo {author}
  {\bibfnamefont {X.}~\bibnamefont {Zhang}},\ }\href {\doibase
  10.1088/1475-7516/2020/03/051} {\bibfield  {journal} {\bibinfo  {journal}
  {JCAP}\ }\textbf {\bibinfo {volume} {03}},\ \bibinfo {pages} {051} (\bibinfo
  {year} {2020})},\ \Eprint {http://arxiv.org/abs/2001.05393} {arXiv:2001.05393
  [astro-ph.CO]} \BibitemShut {NoStop}%
\bibitem [{\citenamefont {Jin}\ \emph {et~al.}(2023{\natexlab{b}})\citenamefont
  {Jin}, \citenamefont {Xing}, \citenamefont {Shao}, \citenamefont {Zhang},\
  and\ \citenamefont {Zhang}}]{Jin:2023zhi}%
  \BibitemOpen
  \bibfield  {author} {\bibinfo {author} {\bibfnamefont {S.-J.}\ \bibnamefont
  {Jin}}, \bibinfo {author} {\bibfnamefont {S.-S.}\ \bibnamefont {Xing}},
  \bibinfo {author} {\bibfnamefont {Y.}~\bibnamefont {Shao}}, \bibinfo {author}
  {\bibfnamefont {J.-F.}\ \bibnamefont {Zhang}}, \ and\ \bibinfo {author}
  {\bibfnamefont {X.}~\bibnamefont {Zhang}},\ }\href {\doibase
  10.1088/1674-1137/acc8be} {\bibfield  {journal} {\bibinfo  {journal} {Chin.
  Phys. C}\ }\textbf {\bibinfo {volume} {47}},\ \bibinfo {pages} {065104}
  (\bibinfo {year} {2023}{\natexlab{b}})},\ \Eprint
  {http://arxiv.org/abs/2301.06722} {arXiv:2301.06722 [astro-ph.CO]}
  \BibitemShut {NoStop}%
\bibitem [{\citenamefont {Han}\ \emph {et~al.}(2024)\citenamefont {Han},
  \citenamefont {Jin}, \citenamefont {Zhang},\ and\ \citenamefont
  {Zhang}}]{Han:2023exn}%
  \BibitemOpen
  \bibfield  {author} {\bibinfo {author} {\bibfnamefont {T.}~\bibnamefont
  {Han}}, \bibinfo {author} {\bibfnamefont {S.-J.}\ \bibnamefont {Jin}},
  \bibinfo {author} {\bibfnamefont {J.-F.}\ \bibnamefont {Zhang}}, \ and\
  \bibinfo {author} {\bibfnamefont {X.}~\bibnamefont {Zhang}},\ }\href
  {\doibase 10.1140/epjc/s10052-024-12999-w} {\bibfield  {journal} {\bibinfo
  {journal} {Eur. Phys. J. C}\ }\textbf {\bibinfo {volume} {84}},\ \bibinfo
  {pages} {663} (\bibinfo {year} {2024})},\ \Eprint
  {http://arxiv.org/abs/2309.14965} {arXiv:2309.14965 [astro-ph.CO]}
  \BibitemShut {NoStop}%
\bibitem [{\citenamefont {Chang}\ \emph {et~al.}(2019)\citenamefont {Chang},
  \citenamefont {Huang}, \citenamefont {Wang},\ and\ \citenamefont
  {Zhao}}]{Chang:2019xcb}%
  \BibitemOpen
  \bibfield  {author} {\bibinfo {author} {\bibfnamefont {Z.}~\bibnamefont
  {Chang}}, \bibinfo {author} {\bibfnamefont {Q.-G.}\ \bibnamefont {Huang}},
  \bibinfo {author} {\bibfnamefont {S.}~\bibnamefont {Wang}}, \ and\ \bibinfo
  {author} {\bibfnamefont {Z.-C.}\ \bibnamefont {Zhao}},\ }\href {\doibase
  10.1140/epjc/s10052-019-6664-0} {\bibfield  {journal} {\bibinfo  {journal}
  {Eur. Phys. J. C}\ }\textbf {\bibinfo {volume} {79}},\ \bibinfo {pages} {177}
  (\bibinfo {year} {2019})}\BibitemShut {NoStop}%
\bibitem [{\citenamefont {Zhao}\ \emph
  {et~al.}(2020{\natexlab{c}})\citenamefont {Zhao}, \citenamefont {Wang},
  \citenamefont {Zhang},\ and\ \citenamefont {Zhang}}]{Zhao:2019gyk}%
  \BibitemOpen
  \bibfield  {author} {\bibinfo {author} {\bibfnamefont {Z.-W.}\ \bibnamefont
  {Zhao}}, \bibinfo {author} {\bibfnamefont {L.-F.}\ \bibnamefont {Wang}},
  \bibinfo {author} {\bibfnamefont {J.-F.}\ \bibnamefont {Zhang}}, \ and\
  \bibinfo {author} {\bibfnamefont {X.}~\bibnamefont {Zhang}},\ }\href
  {\doibase 10.1016/j.scib.2020.04.032} {\bibfield  {journal} {\bibinfo
  {journal} {Sci. Bull.}\ }\textbf {\bibinfo {volume} {65}},\ \bibinfo {pages}
  {1340} (\bibinfo {year} {2020}{\natexlab{c}})},\ \Eprint
  {http://arxiv.org/abs/1912.11629} {arXiv:1912.11629 [astro-ph.CO]}
  \BibitemShut {NoStop}%
\bibitem [{\citenamefont {Zhang}\ \emph {et~al.}(2021)\citenamefont {Zhang},
  \citenamefont {Wang}, \citenamefont {Wu}, \citenamefont {Qi}, \citenamefont
  {Xu}, \citenamefont {Zhang},\ and\ \citenamefont {Zhang}}]{Zhang:2021yof}%
  \BibitemOpen
  \bibfield  {author} {\bibinfo {author} {\bibfnamefont {M.}~\bibnamefont
  {Zhang}}, \bibinfo {author} {\bibfnamefont {B.}~\bibnamefont {Wang}},
  \bibinfo {author} {\bibfnamefont {P.-J.}\ \bibnamefont {Wu}}, \bibinfo
  {author} {\bibfnamefont {J.-Z.}\ \bibnamefont {Qi}}, \bibinfo {author}
  {\bibfnamefont {Y.}~\bibnamefont {Xu}}, \bibinfo {author} {\bibfnamefont
  {J.-F.}\ \bibnamefont {Zhang}}, \ and\ \bibinfo {author} {\bibfnamefont
  {X.}~\bibnamefont {Zhang}},\ }\href {\doibase 10.3847/1538-4357/ac0ef5}
  {\bibfield  {journal} {\bibinfo  {journal} {Astrophys. J.}\ }\textbf
  {\bibinfo {volume} {918}},\ \bibinfo {pages} {56} (\bibinfo {year} {2021})},\
  \Eprint {http://arxiv.org/abs/2102.03979} {arXiv:2102.03979 [astro-ph.CO]}
  \BibitemShut {NoStop}%
\bibitem [{\citenamefont {Wu}\ and\ \citenamefont {Zhang}(2022)}]{Wu:2021vfz}%
  \BibitemOpen
  \bibfield  {author} {\bibinfo {author} {\bibfnamefont {P.-J.}\ \bibnamefont
  {Wu}}\ and\ \bibinfo {author} {\bibfnamefont {X.}~\bibnamefont {Zhang}},\
  }\href {\doibase 10.1088/1475-7516/2022/01/060} {\bibfield  {journal}
  {\bibinfo  {journal} {JCAP}\ }\textbf {\bibinfo {volume} {01}},\ \bibinfo
  {pages} {060} (\bibinfo {year} {2022})},\ \Eprint
  {http://arxiv.org/abs/2108.03552} {arXiv:2108.03552 [astro-ph.CO]}
  \BibitemShut {NoStop}%
\bibitem [{\citenamefont {Wu}\ \emph {et~al.}(2023{\natexlab{b}})\citenamefont
  {Wu}, \citenamefont {Li}, \citenamefont {Zhang},\ and\ \citenamefont
  {Zhang}}]{Wu:2022jkf}%
  \BibitemOpen
  \bibfield  {author} {\bibinfo {author} {\bibfnamefont {P.-J.}\ \bibnamefont
  {Wu}}, \bibinfo {author} {\bibfnamefont {Y.}~\bibnamefont {Li}}, \bibinfo
  {author} {\bibfnamefont {J.-F.}\ \bibnamefont {Zhang}}, \ and\ \bibinfo
  {author} {\bibfnamefont {X.}~\bibnamefont {Zhang}},\ }\href {\doibase
  10.1007/s11433-022-2104-7} {\bibfield  {journal} {\bibinfo  {journal} {Sci.
  China Phys. Mech. Astron.}\ }\textbf {\bibinfo {volume} {66}},\ \bibinfo
  {pages} {270413} (\bibinfo {year} {2023}{\natexlab{b}})},\ \Eprint
  {http://arxiv.org/abs/2212.07681} {arXiv:2212.07681 [astro-ph.CO]}
  \BibitemShut {NoStop}%
\bibitem [{\citenamefont {Jin}\ \emph {et~al.}(2021)\citenamefont {Jin},
  \citenamefont {Wang}, \citenamefont {Wu}, \citenamefont {Zhang},\ and\
  \citenamefont {Zhang}}]{Jin:2021pcv}%
  \BibitemOpen
  \bibfield  {author} {\bibinfo {author} {\bibfnamefont {S.-J.}\ \bibnamefont
  {Jin}}, \bibinfo {author} {\bibfnamefont {L.-F.}\ \bibnamefont {Wang}},
  \bibinfo {author} {\bibfnamefont {P.-J.}\ \bibnamefont {Wu}}, \bibinfo
  {author} {\bibfnamefont {J.-F.}\ \bibnamefont {Zhang}}, \ and\ \bibinfo
  {author} {\bibfnamefont {X.}~\bibnamefont {Zhang}},\ }\href {\doibase
  10.1103/PhysRevD.104.103507} {\bibfield  {journal} {\bibinfo  {journal}
  {Phys. Rev. D}\ }\textbf {\bibinfo {volume} {104}},\ \bibinfo {pages}
  {103507} (\bibinfo {year} {2021})},\ \Eprint
  {http://arxiv.org/abs/2106.01859} {arXiv:2106.01859 [astro-ph.CO]}
  \BibitemShut {NoStop}%
\bibitem [{\citenamefont {Zhao}\ \emph
  {et~al.}(2020{\natexlab{d}})\citenamefont {Zhao}, \citenamefont {Li},
  \citenamefont {Qi}, \citenamefont {Gao}, \citenamefont {Zhang},\ and\
  \citenamefont {Zhang}}]{Zhao:2020ole}%
  \BibitemOpen
  \bibfield  {author} {\bibinfo {author} {\bibfnamefont {Z.-W.}\ \bibnamefont
  {Zhao}}, \bibinfo {author} {\bibfnamefont {Z.-X.}\ \bibnamefont {Li}},
  \bibinfo {author} {\bibfnamefont {J.-Z.}\ \bibnamefont {Qi}}, \bibinfo
  {author} {\bibfnamefont {H.}~\bibnamefont {Gao}}, \bibinfo {author}
  {\bibfnamefont {J.-F.}\ \bibnamefont {Zhang}}, \ and\ \bibinfo {author}
  {\bibfnamefont {X.}~\bibnamefont {Zhang}},\ }\href {\doibase
  10.3847/1538-4357/abb8ce} {\bibfield  {journal} {\bibinfo  {journal}
  {Astrophys. J.}\ }\textbf {\bibinfo {volume} {903}},\ \bibinfo {pages} {83}
  (\bibinfo {year} {2020}{\natexlab{d}})},\ \Eprint
  {http://arxiv.org/abs/2006.01450} {arXiv:2006.01450 [astro-ph.CO]}
  \BibitemShut {NoStop}%
\bibitem [{\citenamefont {Qiu}\ \emph {et~al.}(2022)\citenamefont {Qiu},
  \citenamefont {Zhao}, \citenamefont {Wang}, \citenamefont {Zhang},\ and\
  \citenamefont {Zhang}}]{Qiu:2021cww}%
  \BibitemOpen
  \bibfield  {author} {\bibinfo {author} {\bibfnamefont {X.-W.}\ \bibnamefont
  {Qiu}}, \bibinfo {author} {\bibfnamefont {Z.-W.}\ \bibnamefont {Zhao}},
  \bibinfo {author} {\bibfnamefont {L.-F.}\ \bibnamefont {Wang}}, \bibinfo
  {author} {\bibfnamefont {J.-F.}\ \bibnamefont {Zhang}}, \ and\ \bibinfo
  {author} {\bibfnamefont {X.}~\bibnamefont {Zhang}},\ }\href {\doibase
  10.1088/1475-7516/2022/02/006} {\bibfield  {journal} {\bibinfo  {journal}
  {JCAP}\ }\textbf {\bibinfo {volume} {02}},\ \bibinfo {pages} {006} (\bibinfo
  {year} {2022})},\ \Eprint {http://arxiv.org/abs/2108.04127} {arXiv:2108.04127
  [astro-ph.CO]} \BibitemShut {NoStop}%
\bibitem [{\citenamefont {Zhao}\ \emph {et~al.}(2023)\citenamefont {Zhao},
  \citenamefont {Wang}, \citenamefont {Zhang}, \citenamefont {Zhang},\ and\
  \citenamefont {Zhang}}]{Zhao:2022bpd}%
  \BibitemOpen
  \bibfield  {author} {\bibinfo {author} {\bibfnamefont {Z.-W.}\ \bibnamefont
  {Zhao}}, \bibinfo {author} {\bibfnamefont {L.-F.}\ \bibnamefont {Wang}},
  \bibinfo {author} {\bibfnamefont {J.-G.}\ \bibnamefont {Zhang}}, \bibinfo
  {author} {\bibfnamefont {J.-F.}\ \bibnamefont {Zhang}}, \ and\ \bibinfo
  {author} {\bibfnamefont {X.}~\bibnamefont {Zhang}},\ }\href {\doibase
  10.1088/1475-7516/2023/04/022} {\bibfield  {journal} {\bibinfo  {journal}
  {JCAP}\ }\textbf {\bibinfo {volume} {04}},\ \bibinfo {pages} {022} (\bibinfo
  {year} {2023})},\ \Eprint {http://arxiv.org/abs/2210.07162} {arXiv:2210.07162
  [astro-ph.CO]} \BibitemShut {NoStop}%
\bibitem [{\citenamefont {Zhang}\ \emph {et~al.}(2023)\citenamefont {Zhang},
  \citenamefont {Zhao}, \citenamefont {Li}, \citenamefont {Zhang},
  \citenamefont {Li},\ and\ \citenamefont {Zhang}}]{Zhang:2023gye}%
  \BibitemOpen
  \bibfield  {author} {\bibinfo {author} {\bibfnamefont {J.-G.}\ \bibnamefont
  {Zhang}}, \bibinfo {author} {\bibfnamefont {Z.-W.}\ \bibnamefont {Zhao}},
  \bibinfo {author} {\bibfnamefont {Y.}~\bibnamefont {Li}}, \bibinfo {author}
  {\bibfnamefont {J.-F.}\ \bibnamefont {Zhang}}, \bibinfo {author}
  {\bibfnamefont {D.}~\bibnamefont {Li}}, \ and\ \bibinfo {author}
  {\bibfnamefont {X.}~\bibnamefont {Zhang}},\ }\href {\doibase
  10.1007/s11433-023-2212-9} {\bibfield  {journal} {\bibinfo  {journal} {Sci.
  China Phys. Mech. Astron.}\ }\textbf {\bibinfo {volume} {66}},\ \bibinfo
  {pages} {120412} (\bibinfo {year} {2023})},\ \Eprint
  {http://arxiv.org/abs/2307.01605} {arXiv:2307.01605 [astro-ph.CO]}
  \BibitemShut {NoStop}%
\bibitem [{\citenamefont {Wu}\ \emph {et~al.}(2023{\natexlab{c}})\citenamefont
  {Wu}, \citenamefont {Shao}, \citenamefont {Jin},\ and\ \citenamefont
  {Zhang}}]{Wu:2022dgy}%
  \BibitemOpen
  \bibfield  {author} {\bibinfo {author} {\bibfnamefont {P.-J.}\ \bibnamefont
  {Wu}}, \bibinfo {author} {\bibfnamefont {Y.}~\bibnamefont {Shao}}, \bibinfo
  {author} {\bibfnamefont {S.-J.}\ \bibnamefont {Jin}}, \ and\ \bibinfo
  {author} {\bibfnamefont {X.}~\bibnamefont {Zhang}},\ }\href {\doibase
  10.1088/1475-7516/2023/06/052} {\bibfield  {journal} {\bibinfo  {journal}
  {JCAP}\ }\textbf {\bibinfo {volume} {06}},\ \bibinfo {pages} {052} (\bibinfo
  {year} {2023}{\natexlab{c}})},\ \Eprint {http://arxiv.org/abs/2202.09726}
  {arXiv:2202.09726 [astro-ph.CO]} \BibitemShut {NoStop}%
\bibitem [{\citenamefont {Refsdal}(1964)}]{Refsdal:1964blz}%
  \BibitemOpen
  \bibfield  {author} {\bibinfo {author} {\bibfnamefont {S.}~\bibnamefont
  {Refsdal}},\ }\href {\doibase 10.1093/mnras/128.4.307} {\bibfield  {journal}
  {\bibinfo  {journal} {Mon. Not. Roy. Astron. Soc.}\ }\textbf {\bibinfo
  {volume} {128}},\ \bibinfo {pages} {307} (\bibinfo {year}
  {1964})}\BibitemShut {NoStop}%
\bibitem [{\citenamefont {Treu}(2010)}]{Treu:2010uj}%
  \BibitemOpen
  \bibfield  {author} {\bibinfo {author} {\bibfnamefont {T.}~\bibnamefont
  {Treu}},\ }\href {\doibase 10.1146/annurev-astro-081309-130924} {\bibfield
  {journal} {\bibinfo  {journal} {Ann. Rev. Astron. Astrophys.}\ }\textbf
  {\bibinfo {volume} {48}},\ \bibinfo {pages} {87} (\bibinfo {year} {2010})},\
  \Eprint {http://arxiv.org/abs/1003.5567} {arXiv:1003.5567 [astro-ph.CO]}
  \BibitemShut {NoStop}%
\bibitem [{\citenamefont {Kochanek}\ \emph {et~al.}(2001)\citenamefont
  {Kochanek}, \citenamefont {Keeton},\ and\ \citenamefont
  {McLeod}}]{Kochanek:2000ue}%
  \BibitemOpen
  \bibfield  {author} {\bibinfo {author} {\bibfnamefont {C.~S.}\ \bibnamefont
  {Kochanek}}, \bibinfo {author} {\bibfnamefont {C.~R.}\ \bibnamefont
  {Keeton}}, \ and\ \bibinfo {author} {\bibfnamefont {B.~A.}\ \bibnamefont
  {McLeod}},\ }\href {\doibase 10.1086/318350} {\bibfield  {journal} {\bibinfo
  {journal} {Astrophys. J.}\ }\textbf {\bibinfo {volume} {547}},\ \bibinfo
  {pages} {50} (\bibinfo {year} {2001})},\ \Eprint
  {http://arxiv.org/abs/astro-ph/0006116} {arXiv:astro-ph/0006116} \BibitemShut
  {NoStop}%
\bibitem [{\citenamefont {Treu}\ and\ \citenamefont
  {Koopmans}(2002)}]{Treu:2002cb}%
  \BibitemOpen
  \bibfield  {author} {\bibinfo {author} {\bibfnamefont {T.}~\bibnamefont
  {Treu}}\ and\ \bibinfo {author} {\bibfnamefont {L.~V.~E.}\ \bibnamefont
  {Koopmans}},\ }\href {\doibase 10.1046/j.1365-8711.2002.06107.x} {\bibfield
  {journal} {\bibinfo  {journal} {Mon. Not. Roy. Astron. Soc.}\ }\textbf
  {\bibinfo {volume} {337}},\ \bibinfo {pages} {L6} (\bibinfo {year} {2002})},\
  \Eprint {http://arxiv.org/abs/astro-ph/0210002} {arXiv:astro-ph/0210002}
  \BibitemShut {NoStop}%
\bibitem [{\citenamefont {Yang}\ \emph
  {et~al.}(2020{\natexlab{b}})\citenamefont {Yang}, \citenamefont {Birrer},\
  and\ \citenamefont {Hu}}]{Yang:2020eoh}%
  \BibitemOpen
  \bibfield  {author} {\bibinfo {author} {\bibfnamefont {T.}~\bibnamefont
  {Yang}}, \bibinfo {author} {\bibfnamefont {S.}~\bibnamefont {Birrer}}, \ and\
  \bibinfo {author} {\bibfnamefont {B.}~\bibnamefont {Hu}},\ }\href {\doibase
  10.1093/mnrasl/slaa107} {\bibfield  {journal} {\bibinfo  {journal} {Mon. Not.
  Roy. Astron. Soc.}\ }\textbf {\bibinfo {volume} {497}},\ \bibinfo {pages}
  {L56} (\bibinfo {year} {2020}{\natexlab{b}})},\ \Eprint
  {http://arxiv.org/abs/2003.03277} {arXiv:2003.03277 [astro-ph.CO]}
  \BibitemShut {NoStop}%
\bibitem [{\citenamefont {Wei}\ and\ \citenamefont {Wu}(2017)}]{Wei:2017emo}%
  \BibitemOpen
  \bibfield  {author} {\bibinfo {author} {\bibfnamefont {J.-J.}\ \bibnamefont
  {Wei}}\ and\ \bibinfo {author} {\bibfnamefont {X.-F.}\ \bibnamefont {Wu}},\
  }\href {\doibase 10.1093/mnras/stx2210} {\bibfield  {journal} {\bibinfo
  {journal} {Mon. Not. Roy. Astron. Soc.}\ }\textbf {\bibinfo {volume} {472}},\
  \bibinfo {pages} {2906} (\bibinfo {year} {2017})},\ \Eprint
  {http://arxiv.org/abs/1707.04152} {arXiv:1707.04152 [astro-ph.CO]}
  \BibitemShut {NoStop}%
\bibitem [{\citenamefont {Cremonese}\ and\ \citenamefont
  {Salzano}(2020)}]{Cremonese:2019tgb}%
  \BibitemOpen
  \bibfield  {author} {\bibinfo {author} {\bibfnamefont {P.}~\bibnamefont
  {Cremonese}}\ and\ \bibinfo {author} {\bibfnamefont {V.}~\bibnamefont
  {Salzano}},\ }\href {\doibase 10.1016/j.dark.2020.100517} {\bibfield
  {journal} {\bibinfo  {journal} {Phys. Dark Univ.}\ }\textbf {\bibinfo
  {volume} {28}},\ \bibinfo {pages} {100517} (\bibinfo {year} {2020})},\
  \Eprint {http://arxiv.org/abs/1911.11786} {arXiv:1911.11786 [astro-ph.CO]}
  \BibitemShut {NoStop}%
\bibitem [{\citenamefont {D'Ascoli}\ \emph {et~al.}(2018)\citenamefont
  {D'Ascoli}, \citenamefont {Noble}, \citenamefont {Bowen}, \citenamefont
  {Campanelli}, \citenamefont {Krolik},\ and\ \citenamefont
  {Mewes}}]{DAscoli:2018dbt}%
  \BibitemOpen
  \bibfield  {author} {\bibinfo {author} {\bibfnamefont {S.}~\bibnamefont
  {D'Ascoli}}, \bibinfo {author} {\bibfnamefont {S.~C.}\ \bibnamefont {Noble}},
  \bibinfo {author} {\bibfnamefont {D.~B.}\ \bibnamefont {Bowen}}, \bibinfo
  {author} {\bibfnamefont {M.}~\bibnamefont {Campanelli}}, \bibinfo {author}
  {\bibfnamefont {J.~H.}\ \bibnamefont {Krolik}}, \ and\ \bibinfo {author}
  {\bibfnamefont {V.}~\bibnamefont {Mewes}},\ }\href {\doibase
  10.3847/1538-4357/aad8b4} {\bibfield  {journal} {\bibinfo  {journal}
  {Astrophys. J.}\ }\textbf {\bibinfo {volume} {865}},\ \bibinfo {pages} {140}
  (\bibinfo {year} {2018})},\ \Eprint {http://arxiv.org/abs/1806.05697}
  {arXiv:1806.05697 [astro-ph.HE]} \BibitemShut {NoStop}%
\bibitem [{\citenamefont {{Etherington}}(1933)}]{Etherington:1933}%
  \BibitemOpen
  \bibfield  {author} {\bibinfo {author} {\bibfnamefont {I.~M.~H.}\
  \bibnamefont {{Etherington}}},\ }\href@noop {} {\bibfield  {journal}
  {\bibinfo  {journal} {Philosophical Magazine}\ }\textbf {\bibinfo {volume}
  {15}},\ \bibinfo {pages} {761} (\bibinfo {year} {1933})}\BibitemShut
  {NoStop}%
\bibitem [{\citenamefont {Bassett}\ and\ \citenamefont
  {Kunz}(2004)}]{Bassett:2003vu}%
  \BibitemOpen
  \bibfield  {author} {\bibinfo {author} {\bibfnamefont {B.~A.}\ \bibnamefont
  {Bassett}}\ and\ \bibinfo {author} {\bibfnamefont {M.}~\bibnamefont {Kunz}},\
  }\href {\doibase 10.1103/PhysRevD.69.101305} {\bibfield  {journal} {\bibinfo
  {journal} {Phys. Rev. D}\ }\textbf {\bibinfo {volume} {69}},\ \bibinfo
  {pages} {101305} (\bibinfo {year} {2004})},\ \Eprint
  {http://arxiv.org/abs/astro-ph/0312443} {arXiv:astro-ph/0312443} \BibitemShut
  {NoStop}%
\bibitem [{\citenamefont {Holanda}\ \emph {et~al.}(2012)\citenamefont
  {Holanda}, \citenamefont {Lima},\ and\ \citenamefont
  {Ribeiro}}]{Holanda:2011hh}%
  \BibitemOpen
  \bibfield  {author} {\bibinfo {author} {\bibfnamefont {R.~F.~L.}\
  \bibnamefont {Holanda}}, \bibinfo {author} {\bibfnamefont {J.~A.~S.}\
  \bibnamefont {Lima}}, \ and\ \bibinfo {author} {\bibfnamefont {M.~B.}\
  \bibnamefont {Ribeiro}},\ }\href {\doibase 10.1051/0004-6361/201118343}
  {\bibfield  {journal} {\bibinfo  {journal} {Astron. Astrophys.}\ }\textbf
  {\bibinfo {volume} {538}},\ \bibinfo {pages} {A131} (\bibinfo {year}
  {2012})},\ \Eprint {http://arxiv.org/abs/1104.3753} {arXiv:1104.3753
  [astro-ph.CO]} \BibitemShut {NoStop}%
\bibitem [{\citenamefont {Li}\ \emph {et~al.}(2011{\natexlab{b}})\citenamefont
  {Li}, \citenamefont {Wu},\ and\ \citenamefont {Yu}}]{Li:2011exa}%
  \BibitemOpen
  \bibfield  {author} {\bibinfo {author} {\bibfnamefont {Z.}~\bibnamefont
  {Li}}, \bibinfo {author} {\bibfnamefont {P.}~\bibnamefont {Wu}}, \ and\
  \bibinfo {author} {\bibfnamefont {H.~W.}\ \bibnamefont {Yu}},\ }\href
  {\doibase 10.1088/2041-8205/729/1/L14} {\bibfield  {journal} {\bibinfo
  {journal} {Astrophys. J. Lett.}\ }\textbf {\bibinfo {volume} {729}},\
  \bibinfo {pages} {L14} (\bibinfo {year} {2011}{\natexlab{b}})},\ \Eprint
  {http://arxiv.org/abs/1101.5255} {arXiv:1101.5255 [astro-ph.CO]} \BibitemShut
  {NoStop}%
\bibitem [{\citenamefont {Hu}\ and\ \citenamefont {Wang}(2018)}]{Hu:2018yah}%
  \BibitemOpen
  \bibfield  {author} {\bibinfo {author} {\bibfnamefont {J.}~\bibnamefont
  {Hu}}\ and\ \bibinfo {author} {\bibfnamefont {F.~Y.}\ \bibnamefont {Wang}},\
  }\href {\doibase 10.1093/mnras/sty955} {\bibfield  {journal} {\bibinfo
  {journal} {Mon. Not. Roy. Astron. Soc.}\ }\textbf {\bibinfo {volume} {477}},\
  \bibinfo {pages} {5064} (\bibinfo {year} {2018})},\ \Eprint
  {http://arxiv.org/abs/1804.06606} {arXiv:1804.06606 [astro-ph.CO]}
  \BibitemShut {NoStop}%
\bibitem [{\citenamefont {Lima}\ \emph {et~al.}(2011)\citenamefont {Lima},
  \citenamefont {Cunha},\ and\ \citenamefont {Zanchin}}]{Lima:2011ye}%
  \BibitemOpen
  \bibfield  {author} {\bibinfo {author} {\bibfnamefont {J.~A.~S.}\
  \bibnamefont {Lima}}, \bibinfo {author} {\bibfnamefont {J.~V.}\ \bibnamefont
  {Cunha}}, \ and\ \bibinfo {author} {\bibfnamefont {V.~T.}\ \bibnamefont
  {Zanchin}},\ }\href {\doibase 10.1088/2041-8205/742/2/L26} {\bibfield
  {journal} {\bibinfo  {journal} {Astrophys. J. Lett.}\ }\textbf {\bibinfo
  {volume} {742}},\ \bibinfo {pages} {L26} (\bibinfo {year} {2011})},\ \Eprint
  {http://arxiv.org/abs/1110.5065} {arXiv:1110.5065 [astro-ph.CO]} \BibitemShut
  {NoStop}%
\bibitem [{\citenamefont {Avgoustidis}\ \emph {et~al.}(2010)\citenamefont
  {Avgoustidis}, \citenamefont {Burrage}, \citenamefont {Redondo},
  \citenamefont {Verde},\ and\ \citenamefont {Jimenez}}]{Avgoustidis:2010ju}%
  \BibitemOpen
  \bibfield  {author} {\bibinfo {author} {\bibfnamefont {A.}~\bibnamefont
  {Avgoustidis}}, \bibinfo {author} {\bibfnamefont {C.}~\bibnamefont
  {Burrage}}, \bibinfo {author} {\bibfnamefont {J.}~\bibnamefont {Redondo}},
  \bibinfo {author} {\bibfnamefont {L.}~\bibnamefont {Verde}}, \ and\ \bibinfo
  {author} {\bibfnamefont {R.}~\bibnamefont {Jimenez}},\ }\href {\doibase
  10.1088/1475-7516/2010/10/024} {\bibfield  {journal} {\bibinfo  {journal}
  {JCAP}\ }\textbf {\bibinfo {volume} {10}},\ \bibinfo {pages} {024} (\bibinfo
  {year} {2010})},\ \Eprint {http://arxiv.org/abs/1004.2053} {arXiv:1004.2053
  [astro-ph.CO]} \BibitemShut {NoStop}%
\bibitem [{\citenamefont {Liao}\ \emph {et~al.}(2015)\citenamefont {Liao},
  \citenamefont {Avgoustidis},\ and\ \citenamefont {Li}}]{Liao:2015ccl}%
  \BibitemOpen
  \bibfield  {author} {\bibinfo {author} {\bibfnamefont {K.}~\bibnamefont
  {Liao}}, \bibinfo {author} {\bibfnamefont {A.}~\bibnamefont {Avgoustidis}}, \
  and\ \bibinfo {author} {\bibfnamefont {Z.}~\bibnamefont {Li}},\ }\href
  {\doibase 10.1103/PhysRevD.92.123539} {\bibfield  {journal} {\bibinfo
  {journal} {Phys. Rev. D}\ }\textbf {\bibinfo {volume} {92}},\ \bibinfo
  {pages} {123539} (\bibinfo {year} {2015})},\ \Eprint
  {http://arxiv.org/abs/1512.01861} {arXiv:1512.01861 [astro-ph.CO]}
  \BibitemShut {NoStop}%
\bibitem [{\citenamefont {Uzan}\ \emph {et~al.}(2004)\citenamefont {Uzan},
  \citenamefont {Aghanim},\ and\ \citenamefont {Mellier}}]{Uzan:2004my}%
  \BibitemOpen
  \bibfield  {author} {\bibinfo {author} {\bibfnamefont {J.-P.}\ \bibnamefont
  {Uzan}}, \bibinfo {author} {\bibfnamefont {N.}~\bibnamefont {Aghanim}}, \
  and\ \bibinfo {author} {\bibfnamefont {Y.}~\bibnamefont {Mellier}},\ }\href
  {\doibase 10.1103/PhysRevD.70.083533} {\bibfield  {journal} {\bibinfo
  {journal} {Phys. Rev. D}\ }\textbf {\bibinfo {volume} {70}},\ \bibinfo
  {pages} {083533} (\bibinfo {year} {2004})},\ \Eprint
  {http://arxiv.org/abs/astro-ph/0405620} {arXiv:astro-ph/0405620} \BibitemShut
  {NoStop}%
\bibitem [{\citenamefont {Li}\ \emph {et~al.}(2013)\citenamefont {Li},
  \citenamefont {Wu}, \citenamefont {Yu},\ and\ \citenamefont
  {Zhu}}]{Li:2013cva}%
  \BibitemOpen
  \bibfield  {author} {\bibinfo {author} {\bibfnamefont {Z.}~\bibnamefont
  {Li}}, \bibinfo {author} {\bibfnamefont {P.}~\bibnamefont {Wu}}, \bibinfo
  {author} {\bibfnamefont {H.}~\bibnamefont {Yu}}, \ and\ \bibinfo {author}
  {\bibfnamefont {Z.-H.}\ \bibnamefont {Zhu}},\ }\href {\doibase
  10.1103/PhysRevD.87.103013} {\bibfield  {journal} {\bibinfo  {journal} {Phys.
  Rev. D}\ }\textbf {\bibinfo {volume} {87}},\ \bibinfo {pages} {103013}
  (\bibinfo {year} {2013})},\ \Eprint {http://arxiv.org/abs/1304.7317}
  {arXiv:1304.7317 [astro-ph.CO]} \BibitemShut {NoStop}%
\bibitem [{\citenamefont {Holanda}\ \emph {et~al.}(2017)\citenamefont
  {Holanda}, \citenamefont {Busti}, \citenamefont {Lima},\ and\ \citenamefont
  {Alcaniz}}]{Holanda:2016msr}%
  \BibitemOpen
  \bibfield  {author} {\bibinfo {author} {\bibfnamefont {R.~F.~L.}\
  \bibnamefont {Holanda}}, \bibinfo {author} {\bibfnamefont {V.~C.}\
  \bibnamefont {Busti}}, \bibinfo {author} {\bibfnamefont {F.~S.}\ \bibnamefont
  {Lima}}, \ and\ \bibinfo {author} {\bibfnamefont {J.~S.}\ \bibnamefont
  {Alcaniz}},\ }\href {\doibase 10.1088/1475-7516/2017/09/039} {\bibfield
  {journal} {\bibinfo  {journal} {JCAP}\ }\textbf {\bibinfo {volume} {09}},\
  \bibinfo {pages} {039} (\bibinfo {year} {2017})},\ \Eprint
  {http://arxiv.org/abs/1611.09426} {arXiv:1611.09426 [astro-ph.CO]}
  \BibitemShut {NoStop}%
\bibitem [{\citenamefont {Wu}\ \emph {et~al.}(2015)\citenamefont {Wu},
  \citenamefont {Li}, \citenamefont {Liu},\ and\ \citenamefont
  {Yu}}]{Wu:2015ixa}%
  \BibitemOpen
  \bibfield  {author} {\bibinfo {author} {\bibfnamefont {P.}~\bibnamefont
  {Wu}}, \bibinfo {author} {\bibfnamefont {Z.}~\bibnamefont {Li}}, \bibinfo
  {author} {\bibfnamefont {X.}~\bibnamefont {Liu}}, \ and\ \bibinfo {author}
  {\bibfnamefont {H.}~\bibnamefont {Yu}},\ }\href {\doibase
  10.1103/PhysRevD.92.023520} {\bibfield  {journal} {\bibinfo  {journal} {Phys.
  Rev. D}\ }\textbf {\bibinfo {volume} {92}},\ \bibinfo {pages} {023520}
  (\bibinfo {year} {2015})}\BibitemShut {NoStop}%
\bibitem [{\citenamefont {Fu}\ \emph {et~al.}(2019{\natexlab{b}})\citenamefont
  {Fu}, \citenamefont {Zhou},\ and\ \citenamefont {Chen}}]{Fu:2019oll}%
  \BibitemOpen
  \bibfield  {author} {\bibinfo {author} {\bibfnamefont {X.}~\bibnamefont
  {Fu}}, \bibinfo {author} {\bibfnamefont {L.}~\bibnamefont {Zhou}}, \ and\
  \bibinfo {author} {\bibfnamefont {J.}~\bibnamefont {Chen}},\ }\href {\doibase
  10.1103/PhysRevD.99.083523} {\bibfield  {journal} {\bibinfo  {journal} {Phys.
  Rev. D}\ }\textbf {\bibinfo {volume} {99}},\ \bibinfo {pages} {083523}
  (\bibinfo {year} {2019}{\natexlab{b}})},\ \Eprint
  {http://arxiv.org/abs/1903.09913} {arXiv:1903.09913 [gr-qc]} \BibitemShut
  {NoStop}%
\bibitem [{\citenamefont {Chen}\ \emph
  {et~al.}(2019{\natexlab{b}})\citenamefont {Chen}, \citenamefont {Li},
  \citenamefont {Shu},\ and\ \citenamefont {Cao}}]{Chen:2018jcf}%
  \BibitemOpen
  \bibfield  {author} {\bibinfo {author} {\bibfnamefont {Y.}~\bibnamefont
  {Chen}}, \bibinfo {author} {\bibfnamefont {R.}~\bibnamefont {Li}}, \bibinfo
  {author} {\bibfnamefont {Y.}~\bibnamefont {Shu}}, \ and\ \bibinfo {author}
  {\bibfnamefont {X.}~\bibnamefont {Cao}},\ }\href {\doibase
  10.1093/mnras/stz1902} {\bibfield  {journal} {\bibinfo  {journal} {Mon. Not.
  Roy. Astron. Soc.}\ }\textbf {\bibinfo {volume} {488}},\ \bibinfo {pages}
  {3745} (\bibinfo {year} {2019}{\natexlab{b}})},\ \Eprint
  {http://arxiv.org/abs/1809.09845} {arXiv:1809.09845 [astro-ph.CO]}
  \BibitemShut {NoStop}%
\bibitem [{\citenamefont {Tu}\ \emph {et~al.}(2019)\citenamefont {Tu},
  \citenamefont {Hu},\ and\ \citenamefont {Wang}}]{Tu:2019vcj}%
  \BibitemOpen
  \bibfield  {author} {\bibinfo {author} {\bibfnamefont {Z.~L.}\ \bibnamefont
  {Tu}}, \bibinfo {author} {\bibfnamefont {J.}~\bibnamefont {Hu}}, \ and\
  \bibinfo {author} {\bibfnamefont {F.~Y.}\ \bibnamefont {Wang}},\ }\href
  {\doibase 10.1093/mnras/stz286} {\bibfield  {journal} {\bibinfo  {journal}
  {Mon. Not. Roy. Astron. Soc.}\ }\textbf {\bibinfo {volume} {484}},\ \bibinfo
  {pages} {4337} (\bibinfo {year} {2019})},\ \Eprint
  {http://arxiv.org/abs/1901.09144} {arXiv:1901.09144 [astro-ph.CO]}
  \BibitemShut {NoStop}%
\bibitem [{\citenamefont {Liao}\ \emph {et~al.}(2016)\citenamefont {Liao},
  \citenamefont {Li}, \citenamefont {Cao}, \citenamefont {Biesiada},
  \citenamefont {Zheng},\ and\ \citenamefont {Zhu}}]{Liao:2015uzb}%
  \BibitemOpen
  \bibfield  {author} {\bibinfo {author} {\bibfnamefont {K.}~\bibnamefont
  {Liao}}, \bibinfo {author} {\bibfnamefont {Z.}~\bibnamefont {Li}}, \bibinfo
  {author} {\bibfnamefont {S.}~\bibnamefont {Cao}}, \bibinfo {author}
  {\bibfnamefont {M.}~\bibnamefont {Biesiada}}, \bibinfo {author}
  {\bibfnamefont {X.}~\bibnamefont {Zheng}}, \ and\ \bibinfo {author}
  {\bibfnamefont {Z.-H.}\ \bibnamefont {Zhu}},\ }\href {\doibase
  10.3847/0004-637X/822/2/74} {\bibfield  {journal} {\bibinfo  {journal}
  {Astrophys. J.}\ }\textbf {\bibinfo {volume} {822}},\ \bibinfo {pages} {74}
  (\bibinfo {year} {2016})},\ \Eprint {http://arxiv.org/abs/1511.01318}
  {arXiv:1511.01318 [astro-ph.CO]} \BibitemShut {NoStop}%
\bibitem [{\citenamefont {Yang}\ \emph
  {et~al.}(2019{\natexlab{b}})\citenamefont {Yang}, \citenamefont {Holanda},\
  and\ \citenamefont {Hu}}]{Yang:2017bkv}%
  \BibitemOpen
  \bibfield  {author} {\bibinfo {author} {\bibfnamefont {T.}~\bibnamefont
  {Yang}}, \bibinfo {author} {\bibfnamefont {R.~F.~L.}\ \bibnamefont
  {Holanda}}, \ and\ \bibinfo {author} {\bibfnamefont {B.}~\bibnamefont {Hu}},\
  }\href {\doibase 10.1016/j.astropartphys.2019.01.005} {\bibfield  {journal}
  {\bibinfo  {journal} {Astropart. Phys.}\ }\textbf {\bibinfo {volume} {108}},\
  \bibinfo {pages} {57} (\bibinfo {year} {2019}{\natexlab{b}})},\ \Eprint
  {http://arxiv.org/abs/1710.10929} {arXiv:1710.10929 [astro-ph.CO]}
  \BibitemShut {NoStop}%
\bibitem [{\citenamefont {Suyu}\ \emph {et~al.}(2017)\citenamefont {Suyu} \emph
  {et~al.}}]{H0LiCOW:2016xpx}%
  \BibitemOpen
  \bibfield  {author} {\bibinfo {author} {\bibfnamefont {S.~H.}\ \bibnamefont
  {Suyu}} \emph {et~al.} (\bibinfo {collaboration} {H0LiCOW}),\ }\href
  {\doibase 10.1093/mnras/stx483} {\bibfield  {journal} {\bibinfo  {journal}
  {Mon. Not. Roy. Astron. Soc.}\ }\textbf {\bibinfo {volume} {468}},\ \bibinfo
  {pages} {2590} (\bibinfo {year} {2017})},\ \Eprint
  {http://arxiv.org/abs/1607.00017} {arXiv:1607.00017 [astro-ph.CO]}
  \BibitemShut {NoStop}%
\bibitem [{\citenamefont {Rana}\ \emph {et~al.}(2017)\citenamefont {Rana},
  \citenamefont {Jain}, \citenamefont {Mahajan}, \citenamefont {Mukherjee},\
  and\ \citenamefont {Holanda}}]{Rana:2017sfr}%
  \BibitemOpen
  \bibfield  {author} {\bibinfo {author} {\bibfnamefont {A.}~\bibnamefont
  {Rana}}, \bibinfo {author} {\bibfnamefont {D.}~\bibnamefont {Jain}}, \bibinfo
  {author} {\bibfnamefont {S.}~\bibnamefont {Mahajan}}, \bibinfo {author}
  {\bibfnamefont {A.}~\bibnamefont {Mukherjee}}, \ and\ \bibinfo {author}
  {\bibfnamefont {R.~F.~L.}\ \bibnamefont {Holanda}},\ }\href {\doibase
  10.1088/1475-7516/2017/07/010} {\bibfield  {journal} {\bibinfo  {journal}
  {JCAP}\ }\textbf {\bibinfo {volume} {07}},\ \bibinfo {pages} {010} (\bibinfo
  {year} {2017})},\ \Eprint {http://arxiv.org/abs/1705.04549} {arXiv:1705.04549
  [astro-ph.CO]} \BibitemShut {NoStop}%
\bibitem [{\citenamefont {Y\i{}ld\i{}r\i{}m}\ \emph {et~al.}(2020)\citenamefont
  {Y\i{}ld\i{}r\i{}m}, \citenamefont {Suyu},\ and\ \citenamefont
  {Halkola}}]{Yildirim:2019vlv}%
  \BibitemOpen
  \bibfield  {author} {\bibinfo {author} {\bibfnamefont {A.}~\bibnamefont
  {Y\i{}ld\i{}r\i{}m}}, \bibinfo {author} {\bibfnamefont {S.~H.}\ \bibnamefont
  {Suyu}}, \ and\ \bibinfo {author} {\bibfnamefont {A.}~\bibnamefont
  {Halkola}},\ }\href {\doibase 10.1093/mnras/staa498} {\bibfield  {journal}
  {\bibinfo  {journal} {Mon. Not. Roy. Astron. Soc.}\ }\textbf {\bibinfo
  {volume} {493}},\ \bibinfo {pages} {4783} (\bibinfo {year} {2020})},\ \Eprint
  {http://arxiv.org/abs/1904.07237} {arXiv:1904.07237 [astro-ph.CO]}
  \BibitemShut {NoStop}%
\bibitem [{\citenamefont {Li}\ \emph {et~al.}(2019{\natexlab{b}})\citenamefont
  {Li}, \citenamefont {Tang},\ and\ \citenamefont {Lin}}]{Li:2017dey}%
  \BibitemOpen
  \bibfield  {author} {\bibinfo {author} {\bibfnamefont {X.}~\bibnamefont
  {Li}}, \bibinfo {author} {\bibfnamefont {L.}~\bibnamefont {Tang}}, \ and\
  \bibinfo {author} {\bibfnamefont {H.-N.}\ \bibnamefont {Lin}},\ }\href
  {\doibase 10.1093/mnras/sty3116} {\bibfield  {journal} {\bibinfo  {journal}
  {Mon. Not. Roy. Astron. Soc.}\ }\textbf {\bibinfo {volume} {482}},\ \bibinfo
  {pages} {5678} (\bibinfo {year} {2019}{\natexlab{b}})},\ \Eprint
  {http://arxiv.org/abs/1707.00390} {arXiv:1707.00390 [gr-qc]} \BibitemShut
  {NoStop}%
\bibitem [{\citenamefont {Lin}\ and\ \citenamefont {Li}(2020)}]{Lin:2019mrl}%
  \BibitemOpen
  \bibfield  {author} {\bibinfo {author} {\bibfnamefont {H.-N.}\ \bibnamefont
  {Lin}}\ and\ \bibinfo {author} {\bibfnamefont {X.}~\bibnamefont {Li}},\
  }\href {\doibase 10.1088/1674-1137/44/7/075101} {\bibfield  {journal}
  {\bibinfo  {journal} {Chin. Phys. C}\ }\textbf {\bibinfo {volume} {44}},\
  \bibinfo {pages} {075101} (\bibinfo {year} {2020})},\ \Eprint
  {http://arxiv.org/abs/1911.00263} {arXiv:1911.00263 [gr-qc]} \BibitemShut
  {NoStop}%
\bibitem [{\citenamefont {Lin}\ \emph {et~al.}(2021)\citenamefont {Lin},
  \citenamefont {Li},\ and\ \citenamefont {Tang}}]{Lin:2020vqj}%
  \BibitemOpen
  \bibfield  {author} {\bibinfo {author} {\bibfnamefont {H.-N.}\ \bibnamefont
  {Lin}}, \bibinfo {author} {\bibfnamefont {X.}~\bibnamefont {Li}}, \ and\
  \bibinfo {author} {\bibfnamefont {L.}~\bibnamefont {Tang}},\ }\href {\doibase
  10.1088/1674-1137/abc53a} {\bibfield  {journal} {\bibinfo  {journal} {Chin.
  Phys. C}\ }\textbf {\bibinfo {volume} {45}},\ \bibinfo {pages} {015109}
  (\bibinfo {year} {2021})},\ \Eprint {http://arxiv.org/abs/2010.03754}
  {arXiv:2010.03754 [gr-qc]} \BibitemShut {NoStop}%
\bibitem [{\citenamefont {Arjona}\ \emph {et~al.}(2021)\citenamefont {Arjona},
  \citenamefont {Lin}, \citenamefont {Nesseris},\ and\ \citenamefont
  {Tang}}]{Arjona:2020axn}%
  \BibitemOpen
  \bibfield  {author} {\bibinfo {author} {\bibfnamefont {R.}~\bibnamefont
  {Arjona}}, \bibinfo {author} {\bibfnamefont {H.-N.}\ \bibnamefont {Lin}},
  \bibinfo {author} {\bibfnamefont {S.}~\bibnamefont {Nesseris}}, \ and\
  \bibinfo {author} {\bibfnamefont {L.}~\bibnamefont {Tang}},\ }\href {\doibase
  10.1103/PhysRevD.103.103513} {\bibfield  {journal} {\bibinfo  {journal}
  {Phys. Rev. D}\ }\textbf {\bibinfo {volume} {103}},\ \bibinfo {pages}
  {103513} (\bibinfo {year} {2021})},\ \Eprint
  {http://arxiv.org/abs/2011.02718} {arXiv:2011.02718 [astro-ph.CO]}
  \BibitemShut {NoStop}%
\bibitem [{\citenamefont {Huang}\ \emph {et~al.}(2025)\citenamefont {Huang},
  \citenamefont {Li}, \citenamefont {Zhang}, \citenamefont {Chen},
  \citenamefont {Gao}, \citenamefont {Lin},\ and\ \citenamefont
  {Hu}}]{Huang:2024zvk}%
  \BibitemOpen
  \bibfield  {author} {\bibinfo {author} {\bibfnamefont {S.-J.}\ \bibnamefont
  {Huang}}, \bibinfo {author} {\bibfnamefont {E.-K.}\ \bibnamefont {Li}},
  \bibinfo {author} {\bibfnamefont {J.-d.}\ \bibnamefont {Zhang}}, \bibinfo
  {author} {\bibfnamefont {X.}~\bibnamefont {Chen}}, \bibinfo {author}
  {\bibfnamefont {Z.}~\bibnamefont {Gao}}, \bibinfo {author} {\bibfnamefont
  {X.-y.}\ \bibnamefont {Lin}}, \ and\ \bibinfo {author} {\bibfnamefont
  {Y.-M.}\ \bibnamefont {Hu}},\ }\href {\doibase 10.1016/j.dark.2025.101810}
  {\bibfield  {journal} {\bibinfo  {journal} {Phys. Dark Univ.}\ }\textbf
  {\bibinfo {volume} {47}},\ \bibinfo {pages} {101810} (\bibinfo {year}
  {2025})},\ \Eprint {http://arxiv.org/abs/2402.17349} {arXiv:2402.17349
  [astro-ph.CO]} \BibitemShut {NoStop}%
\bibitem [{\citenamefont {Shi}\ \emph {et~al.}(2025)\citenamefont {Shi},
  \citenamefont {Che}, \citenamefont {Huang}, \citenamefont {Hu},\ and\
  \citenamefont {Mei}}]{Shi:2024ula}%
  \BibitemOpen
  \bibfield  {author} {\bibinfo {author} {\bibfnamefont {C.}~\bibnamefont
  {Shi}}, \bibinfo {author} {\bibfnamefont {X.}~\bibnamefont {Che}}, \bibinfo
  {author} {\bibfnamefont {Z.}~\bibnamefont {Huang}}, \bibinfo {author}
  {\bibfnamefont {Y.-M.}\ \bibnamefont {Hu}}, \ and\ \bibinfo {author}
  {\bibfnamefont {J.}~\bibnamefont {Mei}},\ }\href {\doibase
  10.1103/PhysRevD.111.023022} {\bibfield  {journal} {\bibinfo  {journal}
  {Phys. Rev. D}\ }\textbf {\bibinfo {volume} {111}},\ \bibinfo {pages}
  {023022} (\bibinfo {year} {2025})},\ \Eprint
  {http://arxiv.org/abs/2411.17177} {arXiv:2411.17177 [gr-qc]} \BibitemShut
  {NoStop}%
\end{thebibliography}%
